\documentclass[a4paper,twoside,twocolumn,headsepline]{scrbook}
\usepackage{wallpaper}
\usepackage{import}    

\usepackage[T1]{fontenc} 
\usepackage[utf8]{inputenc}
\usepackage{chapterbib}
\usepackage{scrpage2}
\usepackage{calc}
\usepackage{graphicx}
\usepackage{subfig}
\usepackage{booktabs}
\usepackage{longtable}
\usepackage[compact]{titlesec} 
\usepackage{multirow}
\usepackage{amsmath} 
\usepackage{appendix}
\usepackage{eurosym}
\usepackage{afterpage} 
\usepackage{placeins}
\usepackage{flafter}
\usepackage{float}
\usepackage{multicol}
\usepackage{caption}
\usepackage[figuresright]{rotating}

\usepackage{balance}

\usepackage{color}
\definecolor{light-gray}{gray}{0.7}

\usepackage{hyperref}         
\usepackage[all]{hypcap}

\hypersetup{
  pdftitle = {SuperB Detector Technical Design Report},
  colorlinks=true,
  breaklinks=true,
  linkcolor=blue,
  anchorcolor=blue,
  citecolor=blue,
  filecolor=blue,
  menucolor=blue,
  pagecolor=blue,
  urlcolor=blue,
  pdfpagelabels
}

\input{babarsym}

\newcommand{\EE}[1] {\ensuremath{\times 10^{#1}}\xspace}

\def\ze#1   {\ensuremath{\zeta_{#1}}\xspace}

\def\amp    {\ensuremath{{\rm \,A}}\xspace}
\def\mA     {\ensuremath{{\rm \,mA}}\xspace}
\def\muA     {\ensuremath{\mu{\rm A}}\xspace}
\def\kA	    {\ensuremath{{\rm \,kA}}\xspace}

\def\pC     {\ensuremath{{\rm \,pC}}\xspace}
\def\Ohm    {\ensuremath{{\,\rm \Omega}}\xspace}
\def\mOhm   {\ensuremath{{\rm \,m\Omega}}\xspace}

\def\V     {\ensuremath{{\rm \,V}}\xspace}
\def\kV     {\ensuremath{{\rm \,kV}}\xspace}

\def\mV     {\ensuremath{{\rm \,mV}}\xspace}	
\def\Tesla  {\ensuremath{{\rm \,T}}\xspace}

\def\Henry  {\ensuremath{{\rm \,H}}\xspace}

\def\MJ     {\ensuremath{{\rm \,MJ}}\xspace}

\def\Jpgm   {\ensuremath{{\rm \,J/g}}\xspace}

\def\Watt      {\ensuremath{{\rm \,W}}\xspace}

\def\kW     {\ensuremath{{\rm \,kW}}\xspace}

\def\Hz     {\ensuremath{{\rm \,Hz}}\xspace}
\def\MHz    {\ensuremath{{\rm \,MHz}}\xspace}
\def\GHz    {\ensuremath{{\rm \,GHz}}\xspace}
\def\kHz    {\ensuremath{{\rm \,kHz}}\xspace}

\def\kHzpmma {\ensuremath{{\rm \,kHz/mm^2}}\xspace}

\def\Hzpcma {\ensuremath{{\rm \,Hz/cm^2}}\xspace}
\def\kHzpcma {\ensuremath{{\rm \,kHz/cm^2}}\xspace}

\def\neqpcma {\ensuremath{{\,{\rm n}_{eq}/{\rm cm^2}}}\xspace}
\def\neqpcmaps  {\ensuremath{{\,{\rm n}_{eq}/{\rm cm^2/s}}}\xspace}

\def\MPa    {\ensuremath{{\rm \,MPa}}\xspace}

\def\kN    {\ensuremath{{\rm \,kN}}\xspace} 
\def\kNpmm {\ensuremath{{\rm \,kN/mm}}\xspace}

\def\pepii    {\pep2 \xspace}

\def\CO2  {$\mathrm{CO}_2$\xspace}

\def\Bruno        {{\texttt Bruno}\xspace}

\newcommand{\tdrchap}[1]
{
  \chapter{#1}\nobalance
  \label{\ChapLabel} 
}

\iftrue
\titleformat{\section}%
  {\raggedsection\normalfont\sectfont\Large\bfseries}%
  {\thesection}%
  {0.4em}%
  {\phantomsection}%
  [\vspace{0.5ex}\centerline{\rule[0.15in]{0.9\columnwidth}{2pt}}\vspace*{-1.5ex}]
\fi
\newcommand{\tdrsec}[1]
{
  \section{#1}
}

\newcommand{\tdrsubsec}[1]
{
  \subsection{#1}
}

\newcommand{\tdrsubsubsec}[1]
{
  \subsubsection{#1}
}

\newcommand{\tdrparagraph}[1]
{
  \paragraph{#1}
}

\newcommand{\ChapLabel}{chap:empty}

\long\def\ignore#1{\relax}

\makeatletter
\def\url@tdrstyle{%
  \@ifundefined{selectfont}{\def\UrlFont{\normalfont}}{\def\UrlFont{\normalfont\smaller}}}
\makeatother
\makeatletter
\renewcommand*\bib@heading{%
  \section{\bibname}
  \@mkboth{\bibname}{\bibname}
}
\makeatother

\renewcommand{\topfraction}{.85} 
\renewcommand{\bottomfraction}{.70} 
\renewcommand{\textfraction}{.15}
\renewcommand{\floatpagefraction}{.66} 

\makeatletter
\DeclareRobustCommand\bfseries{%
  \not@math@alphabet\bfseries\mathbf
  \fontseries\bfdefault\selectfont\boldmath}

\makeatother

\setcapindent{0pt} 

\newcommand{\tdrbibstyle}{abuser-unsrt-alu}

\ifdefined\inclCmds\inclCmds\fi

\areaset[12mm]{165mm}{240mm}
\ifx\footfont\undefined
\else
\renewcommand{\footfont}{\normalfont\rmfamily\scshape\large}
\fi

\begin{document}


%
%
 





\thispagestyle{empty}


\ThisCenterWallPaper{1.015}{coverfront.pdf}
\phantom{test}

\cleardoublepage

\pagestyle{empty}
\onecolumn
\titlespacing*{\chapter}{0pt}{3.5ex plus 1ex minus .2ex}{2.3ex plus .2ex}
\setcounter{page}{-5}


\begin{flushright}
INFN-13-01/PI,
LAL 13-01, 
SLAC-R-1003
\end{flushright}
\vspace*{0.20\textheight}

\begin{center}\Huge\bfseries
  Super$B$ Detector\\
  Technical Design Report
\end{center}

\begin{center}\large\bfseries
  Abstract
\end{center}
In this Technical Design Report (TDR) we describe the \superb detector
to be installed on the \superb \epem high luminosity collider. The
\superb asymmetric collider, foreseen to be constructed on the Tor
Vergata campus near the INFN Frascati National Laboratory, is designed
to operate both at the \FourS energy in the center of mass with a
luminosity of $10^{36}\cms$ and at the $\tau/$charm production
threshold with a luminosity of $10^{35}\cms$. This high luminosity,
producing a data sample about a factor 100 larger than present $B$
Factories would allow investigation of new physics effects in rare
decays, CP Violation and Lepton Flavour Violaion. This document
details the detector design presented in the Conceptual Design Report
(CDR) in 2007. The R\&D and engineering studies perfomed to arrive at
the full detector design are described, and an updated cost estimate
is presented .

\clearpage
\null
\vfill
\begin{center}

\end{center}


\begin{center}
\chapter*{Preface}
\end{center}

Flavour physics not only provides  insight in the physics of the standard model 
but also offers great discovery potential for new physics processes, as  the B-Factories 
experiments,  \babar\ at SLAC and Belle at KEK,  have demonstrated very effectively.
Increasing the luminosity has been identified from the very beginning as the key 
element to extend the physics reach of these machines. Since 2003 a group of physicists
began to explore the physics potential of very high luminosity B-Factory machines.
An upgrade of the the PEP-II accelerator was initially investigated; then the \babar and Belle 
community started in 2004 a series of joint workshops in Hawaii.

The \superb Project was formally born in 2005 when INFN inserted in its three-years planning
 document the intention of building a high luminosity flavour factory in the Frascati area. 
In the course of the years \superb has evolved from an intention into a full-fledged project, 
with a Conceptual Design Report published in 2007, progress reports in 2009, and a 
formal collaboration structure setup in 2010 with hundreds of members from several countries.
All aspects of the  project, physics potential, accelerator design, detector design,
successfully passed several international reviews setup by INFN. 
In 2010  \superb was inserted  in the Italian Research Ministry National 
Research Plan as Flagship Project, and a good fraction of the required funds
were allocated, although not the full amount. 
The decision to build \superb on 
the land of the University of Rome Tor Vergata led, in 2011, to the formation
of the Cabibbo Laboratory consortium between INFN and TorVergata, with the explicit
mission of constructing and managing a new research infrastructure for flavour physics. 
A ministerial cost and schedule review of the accelerator project was held in fall 2012.
A combination of a more realistic cost estimates and the unavailability of funds due of the
global economic climate led to a formal cancelation of the project on Nov 27, 2012.

The community who had been committed to the project for so long, although devastated by
the sudden cancelation, decided to try to preserve and document as much as possible
of the work done in \superb, both to retain a lasting trace of the committment of the group
and, more importantly, to provide a written basis of the technical achievements for the
use of future scientific endavours. 
It is in this spirit that this Detector Technical Design Report, 
whose preparation was quite advanced 
at the time of the cancelation of the project, has been completed and is being published.
We felt that the tone and grammar of the text should remain that of a project that will be 
built rather than that of a project that would have been built. 
Therefore we kept the assertiveness
and optimism of a community that was expecting to start constructing the machine and
experiment within a few months. We sincerely hope that it can be of use to the scientific
community.

\clearpage
\null
\vfill
\begin{center}\large\bfseries
  Acknowledgements
\end{center}
Work supported in part by: the Italian Institute for Nuclear Physics, INFN; 
the U.S. Department of Energy under contract number DE-AC02-76SF00515 and other grants.

The TDR covers have been designed by Bruno Mazoyer, graphics designer at the Laboratoire de l'Accélérateur Linéaire (Orsay, France).

The authors would like to acknowledge the continuing support of Lucia Lilli, INFN-Pisa, as the project scientific secretary.

\cleardoublepage
\pagenumbering{roman}
\makeatletter
\newcommand\authorfootnotemark{%
  \begingroup
  \renewcommand{\@makefnmark}{\@thefnmark}%
  \footnotemark%
  \endgroup
}
\makeatother

\begin{center}
M.~Baszczyk,
P.~Dorosz,
J.~Kolodziej,
W.~Kucewicz,
M.~Sapor\\*
{\it
AGH University of Science and Technology,
Kraków, Poland
}
\end{center}
\begin{center}
A.~Jeremie\\*
{\it
LAPP - Annecy,
Annecy-le-Vieux, France
}
\end{center}
\begin{center}
E.~Grauges Pous\\*
{\it
Universitat de Barcelona,
Barcelona, Spain
}
\end{center}
\begin{center}
\mbox{G.~E.~Bruno$^{ab}$},
\mbox{G.~De Robertis$^{b}$},
\mbox{D.~Diacono$^{b}$},
\mbox{G.~Donvito$^{b}$},
\mbox{P.~Fusco$^{ab}$},
\mbox{F.~Gargano$^{b}$},
\mbox{F.~Giordano$^{ab}$},
\mbox{F.~Loddo$^{b}$},
\mbox{F.~Loparco$^{ab}$},
\mbox{G.~P.~Maggi$^{b}$},
\mbox{V.~Manzari$^{b}$},
\mbox{M.~N.~Mazziotta$^{b}$},
\mbox{E.~Nappi$^{b}$},
\mbox{A.~Palano$^{b}$},
\mbox{B.~Santeramo$^{ab}$},
\mbox{I.~Sgura$^{b}$},
\mbox{L.~Silvestris$^{b}$},
\mbox{V.~Spinoso$^{b}$}\\*
{\it
{}$^a$Dipartimento di Fisica dell’Università e del Politecnico di Bari;
{}$^b$INFN - Sezione di Bari,
Bari, Italy
}
\end{center}
\begin{center}
% ...
\end{center}

\cleardoublepage
\tableofcontents

\twocolumn
\automark[section]{chapter}
\ohead{\pagemark}
\ihead{\headmark}
\ofoot{Super{\large\it B} Detector Technical Design Report}

\pagestyle{scrheadings}

\pagenumbering{arabic}



\renewcommand{\ChapLabel}{chap:Introduction} 
\tdrchap{Introduction}

\tdrsec{The Physics Motivation} The Standard Model successfully
explains the wide variety of experimental data that has been gathered
over several decades with energies ranging from under a \gev up to
several hundred \gev. At the start of the millennium, the flavor
sector was perhaps less explored than the gauge sector, but the PEP-II
and KEK-B asymmetric $B$ Factories, and their associated experiments
\babar\ and Belle, have produced a wealth of important flavor physics
highlights during the past decade~\cite{pdg_2012}. The most notable
experimental objective, the establishment of the
Cabibbo-Kobayashi-Maskawa phase as consistent with experimentally
observed CP-violating asymmetries in $B$ meson decay, was cited in the
award of the 2008 Nobel Prize to Kobayashi and
Maskawa~\cite{nobel_2008}.

The $B$ Factories have provided a set of unique, over-constrained tests of
the Unitarity Triangle. These have, in the main, been found to be
consistent with Standard Model predictions. The $B$ factories have done
far more physics than originally envisioned; \babar\ alone has published
more than 480 papers in refereed journals to date.  
Some
examples of important experimental results are
measurements of all
three angles of the Unitarity Triangle, including $\alpha$ and $\gamma$ in
addition to sin $2\beta$; the establishment of  $D^0\bar{D^0}$ mixing;
the uncovering of intriguing clues for potential New Physics in  
$B \rightarrow K^{(\star)} l^+l^-$, $B \rightarrow K\pi$,  
$B \rightarrow D\tau\nu$, and $B \rightarrow D^{(\star)}\tau\nu$
decays; and the unveiling of an entirely unexpected new spectroscopy.

All present measurements confirm the Standard Model, reducing the
parameter space for New Physics Models, whose signals remain
elusive. With the LHC in full operation, we expect major experimental
discoveries during the next few years at the energy frontier. We
would hope not only to complete the Standard Model by confirming that
the exciting recent observations by ATLAS and CMS around 126~\gevcc
indeed come from the Higgs boson, but also to find signals of New Physics,
which are widely expected to lie around the 1~\tev energy scale. If
found, the New Physics phenomena will need data from very sensitive
heavy flavor experiments if they are to be understood in
detail. Determining the flavor structure of the New Physics involved
requires the information on rare $b$, $c$, and $\tau$ decays, and on $CP$
violation in $b$ and $c$ quark decays that only a very high luminosity
asymmetric $B$ Factory can provide~\cite{Hitlin:2008gf}. On the other
hand, if such signatures of New Physics are not observed at the LHC,
then the excellent sensitivity provided at the luminosity frontier by
a next generation super $B$ factory provides another avenue to
observing New Physics at mass scales up to 10~\tev or more through
observation of rare processes involving $B$ and $D$ mesons and studies
of lepton flavour violation in $\tau$ decays.

\tdrsec{The \superb\ Project Elements}
It is generally agreed that the physics being addressed by a next-generation $B$ 
factory requires a data sample that is some 50--100 times larger than the
existing combined sample from \babar\ and Belle, or at least 
50--75~\invab. Acquiring such an integrated luminosity in a 5 year time
frame requires that the collider run at a luminosity of at least
$10^{36} \cm^{-2}\rm{s}^{-1}$.

For a number of years, an Italian led, INFN hosted, collaboration of
scientists from Canada, Italy, Israel, France, Norway, Spain, Poland,
UK, and USA has worked to design and  propose a high
luminosity $10^{36}$ asymmetric $B$ Factory project, called \superb, to be
built at or near the Frascati laboratory~\cite{Bona:2007qt}.

The accelerator portion of the project employs lessons learned from
modern low-emittance synchrotron light sources and ILC/CLIC R\&D, and
an innovative new idea for the intersection region of the storage
rings~\cite{panta_2006_1,panta_2006_2}, called crab waist, to reach
luminosities over 50 times greater than those obtained by earlier $B$
factories at KEK and SLAC. There is now an attractive, cost-effective
accelerator design, including polarized beams, which is being further
refined and optimized~\cite{Biagini:2010cc}. It is designed to
incorporate many PEP-II components. This facility promises to deliver
fundamental discovery-level science at the luminosity frontier.

There is also an active international collaboration working
effectively on the design of the detector. The detector team draws
heavily on its deep experience with the \babar\ detector, which has
performed in an outstanding manner both in terms of scientific
productivity and operational efficiency, and which serves as the
foundation of the design of the \superb\ detector. 
Substantial R\&D has been undertaking to understand the
modifications and enhancements of the detector components required to 
deal with the much
higher luminosity and the different physics and machine backgrounds.

The \superb\ project has been developed over a substantial time
frame. In 2010, the design and development of all elements of the
project were described in the ``Design Progress Reports'', a set of
several descriptive documents that summarized developments and status
of each major division of the project: Physics~\cite{O'Leary:2010af},
Accelerator~\cite{Biagini:2010cc}, and Detector
~\cite{Grauges:2010fi}. These documents present a snapshot of the
entire project at an intermediate stage between the Conceptual Design
Report~\cite{Bona:2007qt}, which was written in 2007, and the
Technical Design Reports (TDRs).

\tdrsec{The Detector Technical Design Report} 

This TDR describes the design and development of 
the \superb\ detector, which is based on a 
major upgrade of \babar. It begins
with overviews of each of the elements of the project (Accelerator,
Detector, and the Physics). In particular, the Detector section
summarizes the base detector design, the challenges involved in
detector operations at the luminosity frontier, the approach being
taken to optimize remaining general design options, and the R\&D
program that is underway to validate the system and subsystem
designs. A chapter detailing the many challenging issues involved in
the machine detector interface at this high luminosity machine
follows. Each of the detector subsystems and the general detector
systems (especially electronics and Data Acquisition) are then
described in more detail, followed by a discussion of the computing
and software. The detector hall facilities are then described in the
context of the integration and assembly of the full detector. Finally,
the paper concludes with a discussion of project management, detector
costs, and a schedule overview.

\bibliographystyle{\tdrbibstyle}
\balance
\bibliography{dtdr-bib,Physics/tau-bib}


\renewcommand{\ChapLabel}{chap:AccOverview} 
\tdrchap{Collider Overview}

The main goal of the \superb facility is to produce \FourS mesons at a rate of 
$\sim 1\kHz$ by colliding electron on positron beams exploiting
the process $\epem \to \FourS$ whose maximum cross section ($\sim 1\nb$) 
is reached at $\sqrt{s} \sim 10.58\gev$. 
This requirement set the design luminosity of the collider at \designlumi,
a two orders of magnitude increase with respect to the present 
generation $B$ factories.
\superb will be able to operate in a $\sqrt{s}$ range spanning from below 
the $\b \bbar$ threshold up the $\FiveS$, moreover \superb will be also able 
to run at the $\tau$/charm threshold with a luminosity in the range of $10^{35}\lumi$.

The \FourS will be produced with a
sizeable boost in the laboratory frame ($\beta \gamma \sim 0.28$) to
allow for an accurate measurement of the longitudinal separation of
the decay vertices of the two \B mesons originating from the \FourS.
The electron colliding beam will be longitudinally polarized at the
Interaction Point (IP) to allow for a precise determination of the
electroweak parameter $\sin \vartheta_\mathrm{W}$ and to provide
additional kinematical informations useful to the searches of lepton
flavour violating (LFV) decays of the $\tau$ leptons. These additional
informations will be useful both for rejecting the dominant backgrounds
minimising these decays and, in case of observation, for determining the
Lorentz structure of the LFV coupling. 

The followings sections will provide an overview of the novel crab-waist collision scheme
with a brief summary of the test performed at DA$\Phi$NE with this new  scheme.
The accelerator complex and its main parameters are also presented.

\tdrsec{The \superb collision scheme}
A \superb factory can provide a uniquely sensitive probe of New Physics in the flavour sector
of the Standard Model. 

The two proposed second-generation $B$~factory designs, \superb\ and
SuperKEKB, come after the successes of the first generation $B$~factories PEP-II at SLAC and KEKB at KEK
which, in turn, came after three former colliders  which operated 
at the  \FourS resonance (DORIS-II at DESY, VEPP-4M at BINP and CESR at Cornell). \par
The two $B$ factories PEP-II  and KEKB were conceived in the 90's with very ambitious design parameters,
that --- compared to typical accelerator commissioning schedules --- were reached very rapidly. Both colliders
achieved very high peak luminosities with very good reliability, assuring their success. 
Both colliders  are based on a similar design: double rings with asymmetric energies, flat beams and a
multibunch regime.
Key features of their design are  also a low $\beta^{*}$ at the IP with $\beta^{*}_{y}$ 
between 1 and 2\cm. PEP-II adopted a head-on collision scheme and beams were separated off the IP by 
dipole magnets, while in KEKB beams were separated by a  finite-angle crossing scheme with +/-11\mrad.\par
 High luminosity is  obtained essentially by  high beam currents pushed to their maximum values.
Of course, beam current cannot be increased infinitely, due to  
beam-beam interaction, but also  to beam instabilities and to synchrotron radiation emission.
As beam density is increased, beam-beam interaction gets stronger 
and  other effects, such as electron cloud and fast ion,  that cause beam instabilities get stronger. 
Also synchrotron radiation emission in the interaction region sets a limit to stored current.
Bunch-by-bunch feedback systems have played a primary role for damping  instabilities.
Moreover, in $B$ factories  high currents induced also high Interaction Region (IR) backgrounds, that have been 
 well handled, together with lifetime effects.
First generation $B$ factories PEP-II and KEKB,
as for all the  factories, namely the Frascati $\Phi$ Factory  DA$\Phi$NE and the Chinese Tau-Charm factory BTCF,
 are electron-positron colliders based on standard collision schemes.
Some of the main parameters of the two $B$ factories are reported in Table~\ref{table:accelerator_parameters},
with a comparison with one of the former colliders running at the  \FourS resonance CESR-Phase III.\par

A new collision scheme named {\it crab-waist}~\cite{panta_2006_1} opened  the possibility of
enhancing the luminosity by about two orders of magnitudes with no currents increase,
making a breaktrough in the colliders design. 
This launched a new generation of $B$ factories, namely \superb\ and SuperKEKB,  based on the new scheme
proposed by Pantaleo Raimondi, that allows to step forward from factories to super factories.
The {\it crab-waist} (CW) scheme combines several ingredients that together are very powerful
for a luminosity increase: one is the tighter focus on beams at IP,  a second one is the large Piwinski angle and the third and very
important one  is the twist of the beam waists at the IP obtained by placing two proper sextupoles per ring. 
 \superb\ is designed to fully accomplish the CW scheme.\par


First of all \superb\ relies on two low emittance storage rings to provide small 
bunches (see the emittances in Table~\ref{table:accelerator_parameters}).
The vertical dimension of the bunches will be demagnified down to 36\nm
by a strong final focus doublet located half a meter from the IP. 
Higher luminosity will be obtained  as a simple consequence that this scheme provides a smaller spot size at the IP, 
thus by exploiting the recent advances in damping rings and final focus design techniques.\par

The second ingredient of the CW scheme is the large Piwinski angle, with flat beams and small horizontal
crossing angle, obtained by decreasing the horizontal beam size $\sigma_{x}$ and increasing the crossing angle $\theta$.
This, in turn, brings a luminosity increase with a horizontal tune shift decrease, but, most importantly,
at the same time the overlap area of the colliding bunches is decreased. 
So, the vertical beta function $\beta_{y}^{*}$ can be of the same order of this overlap area, 
therefore  much smaller than the longitudinal bunch length $\sigma_{z}$ as in the standard collision scheme,
where the bunch length is, for this reason, needed as small as possible.
This is a good gain, as with longer bunches high order modes (HOM) heating is decreased together with coherent syncrotron radiation emission,
power consumption and beam instabilities are less severe. Moreover,  a lower $\beta_{y}^{*}$ allows also
for lower vertical beam-beam tune shift.\par
The last ingredient of the CW collision scheme is  the crab waist transformation by means of two sextupole lenses
 ($\it crab~sextupoles$) at the opposite sides of the IP
 that provide a waist rotation of one bunch to the axis of the other one. This translates into a luminosity gain due to
a better overlap area, where both beams collide in the minimum $\beta_{y}$ region.
In addition, most importantly, these two crab sextupoles  suppress very efficiently betatron and synchrobetatron resonances. 
They must have  a $\pi$  phase advance horizontally and $\pi/2$ vertically with respect to the IP, and a proper strength.\par
 
This new  CW collision scheme has been successfully tested at DA$\Phi$NE after a shut-down in 2007
for the installation of the new experimental detector SIDDHARTA and for the 
major changes in the interaction region necessary to implement the new collision scheme~\cite{Zobov:2010zza}.
The best peak luminosity of  $4.53 \cdot 10^{32}\lumi$ was obtained in June 2009,
showing an increase of a factor 3 with respect to the peak luminosity reached before the upgrade.
The luminosity gain came partly from the smaller vertical beta function with the high Piwinski angle,
and partly from the beam-beam resonance suppression that happens with the crab sextupoles.
The vertical beam-beam tune-shift has improved  by about a factor 1.5, going to 0.044.
Numerical beam-beam simulations are in good agreement (within 15-20\%) with experimental data. 

\begin{table*}[t]
\caption{ \label{table:accelerator_parameters} 
Main parameters of \superb. PEP-II, KEKB and CESR best achieved luminosity 
parameters are also reported.}
\scriptsize
\begin{center}
\begin{tabular}{l c c c c c c c}
\hline 
Parameter & \multicolumn{2}{c}{\superb} & \multicolumn{2}{c}{PEP-II} 
& \multicolumn{2}{c}{KEKB} & CESR \\
& HER($e^+$) & LER($e^-$) &  HER($e^-$) & LER($e^+$)  & HER($e^-$) & LER($e^+$)&\\
\hline
\hline\\[-0.7em]  
Luminosity ($10^{33}\lumi$) &
\multicolumn{2}{c}{$ 1000 $} & 
\multicolumn{2}{c}{ $ 12 $} & 
\multicolumn{2}{c}{ $17 $} & 
$1.2 $ \\

Circumference (m) & 
\multicolumn{2}{c}{1200} & 
\multicolumn{2}{c}{2200} & 
\multicolumn{2}{c}{3014} & 
768 \\
Energy (GeV) & 
6.7 & 4.18 & 
8.97 & 3.12 & 
8.0 & 3.5 & 
5.29 \\
Lorentz $\beta \gamma$ factor & 
\multicolumn{2}{c}{ 0.238 } & 
\multicolumn{2}{c}{ 0.553 } & 
\multicolumn{2}{c}{ 0.426 } & 
 ---    
\\
Beam current (mA) & 
1900 & 2440 & 
1875 & 2990 & 
1330 & 1650 & 
370    
\\
Particle per bunch ($\times 10^{10}$) & 
5.1 & 6.6 & 
5.0 & 7.9 & 
6.0 & 7.5 & 
  12  
\\
h/v emittance (nm / pm) & 
2.0/5.0 & 2.5/6.2 & 
73/~1000 & 36/~1000 & 
24/~600 & 18/~600 & 
400/~4500 
\\
h/v $\beta^\star$ (mm) & 
26/0.26 & 32/0.21& 
740/11  & 210/10 & 
560/6.5  & 590/5.9 & 
450/21    
\\
h/v beam size @ IP (\mum / nm) & 
7.2/36 & 8.9/36& 
\multicolumn{2}{c}{147/5000}  & 
116/1900 & 103/1900 & 
420/~4000 
\\

\hline 
\end{tabular}
\end{center}
\end{table*}

\begin{sidewaysfigure*}
\centering
\includegraphics[width=22cm]{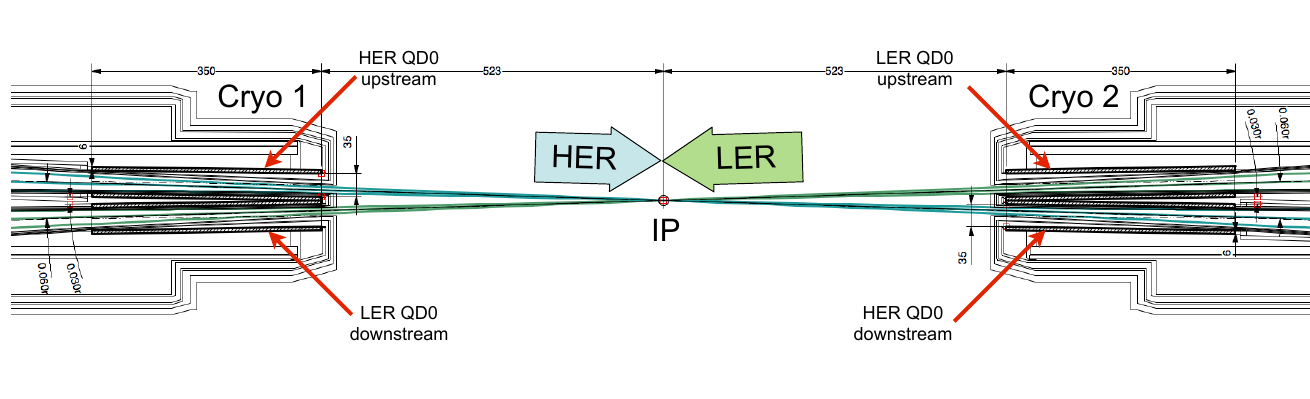}
\caption{The \superb\ interaction region layout.}
\label{fig:IR_layout}
\end{sidewaysfigure*}

\tdrsec{The accelerator complex}

\begin{figure*}[t]
\centering
\includegraphics[height=12cm]{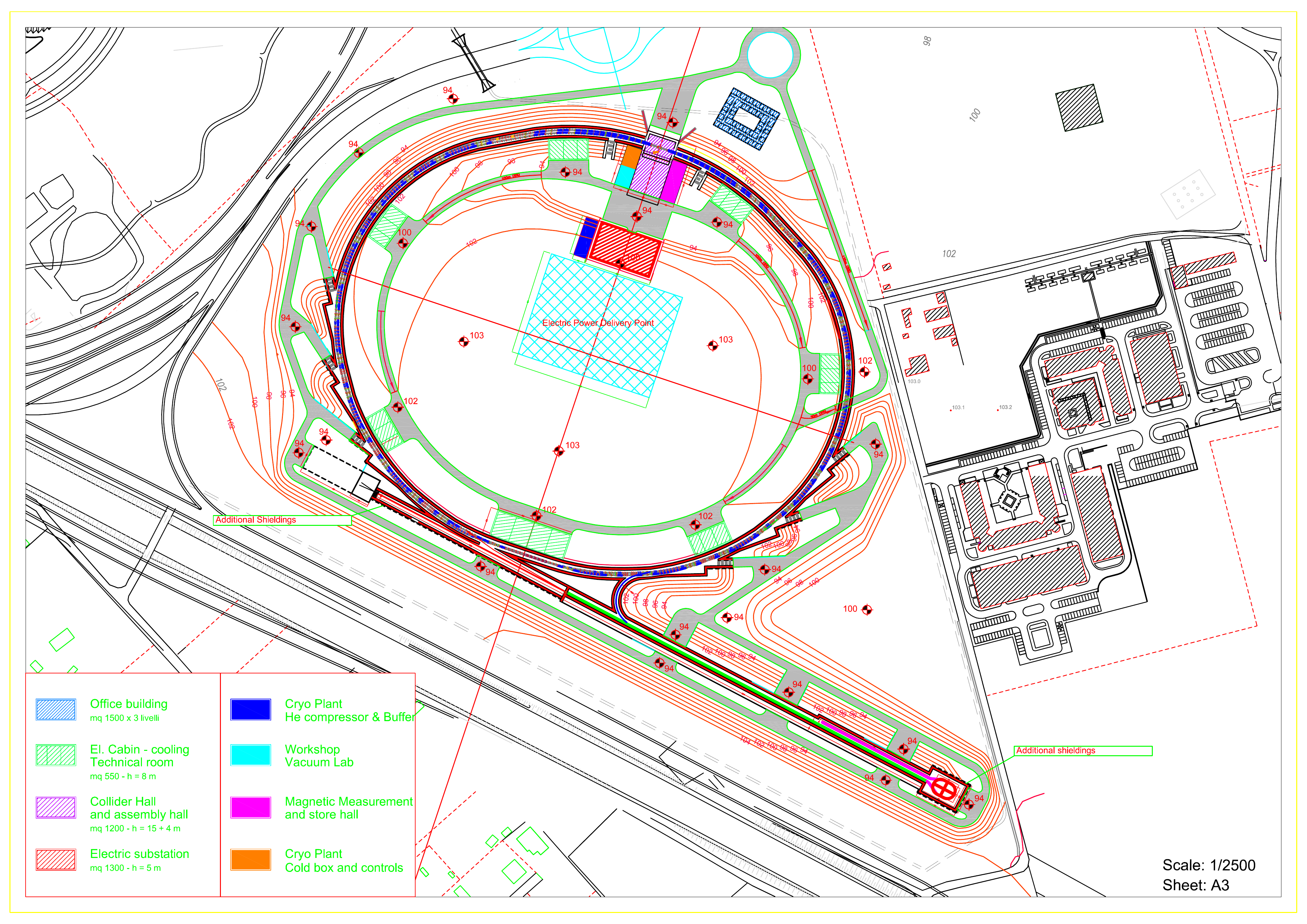}
\caption{The \superb\ site layout. The two main rings, the injection system
and the positrons damping rings are represented together with the main services.}
\label{fig:site_layout}
\end{figure*}

\begin{figure*}[t]
\centering
\includegraphics[width=15cm]{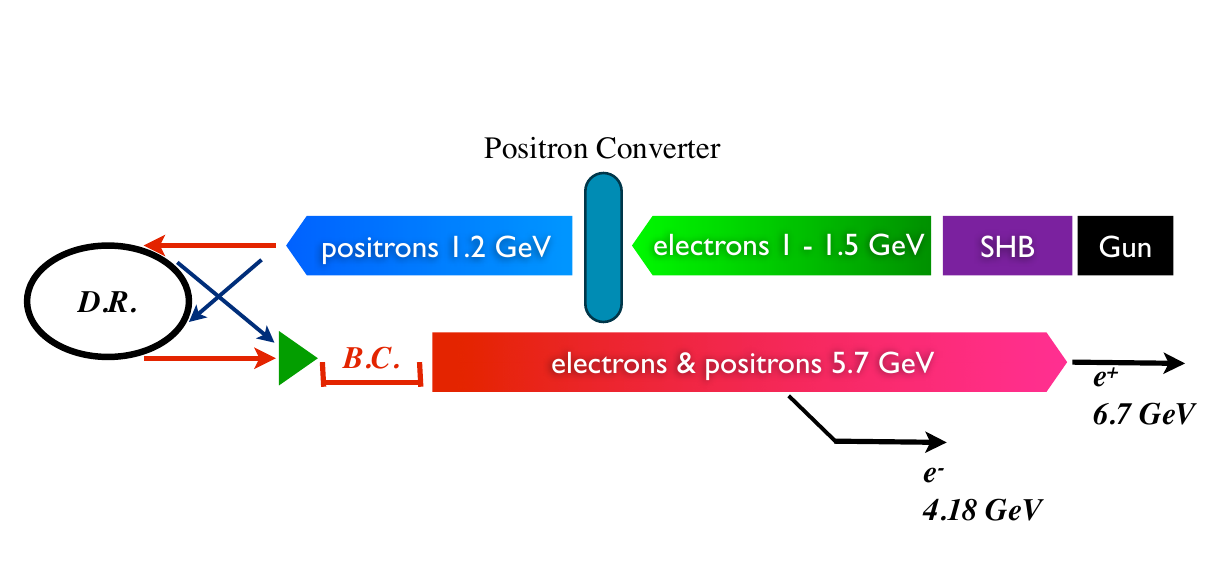}
\caption{The \superb\ injection system conceptual layout.
\label{fig:injection_layout}}
\end{figure*}

The \superb\ accelerators complex consists of an injection system that will inject electrons and positrons into their
respective rings at full energies.  Electrons and positrons have a nominal energy of $4.2\gev$and $6.7\gev$, respectively.
In addition, electrons will be polarized. The two main rings will have a circumference of  about 1200\m.\par
The injection complex (Fig.~\ref{fig:injection_layout})  will consist of a polarized electron gun, a positron production system, electron and
positron linac sections, a damping ring and the transfer lines connecting these systems to the collider main rings.
Electrons will be accelerated in the linac and then into the damping ring;
positrons, produced by accelerated electrons impinging on a positron converter target, will be accelerated and injected in the damping ring as for the elctrons.\par
To keep the ultra high luminosity nearly constant, continuous injection in both rings is necessary. 
The charge required for injection into the main rings is 300\pC/bunch in five bunches. All the three linac sections 
are based on S-band accelerating sections operating at a repetition frequency of 100\Hz. The injection repetition cycle is 30\ms for each beam; this provides a third time slot with  30\ms repetition cycle that can be used to accelerate an ultra-low emittance beam to drive a SASE-FEL facility. 
Electrons will be produced by a polarized gun similar to the SLC gun, where 80\% polarization has been routinely achieved.
At full luminosity and beam currents the lifetime of the beams in the two main rings is expected to be between 3 and 8 minutes. Therefore, to keep beam currents and luminosity nearly constant, the injection process must be continuous with top-up injection. \par
The \superb\ HER and LER lattice designs have several constraints, like an extremely low emittance
and small IP beam sizes needed for the high luminosity, as well as reasonable beam lifetimes and high electron beam polarization.\par
After intense  work on the optimization of the lattice and beam parameters,  asymmetric emittances and asymmetric beam currents 
for the two rings were chosen. 
The \superb\ rings consist of two main lattice systems: arcs and final focus. 
The arcs bend the beams back into collisions and generate the  horizontal
emittance. The final focus consists of an extremely low-$\beta$ insertion and a crab-waist scheme with special optics.
The \superb\ rings can be considered as two damping rings, similar to the ILC and CLIC ones, with the constraint of
including a final focus section for collisions. So, the challenge is not only to achieve low emittance beams, but also to choose all the beam parameters
to reach a very high luminosity with acceptable lifetimes and small beam degradation.\par

The Final Focus is the most crucial system for achieving the \superb\
performance. Reaching the luminosity goal relies on the capability to
de-magnify the vertical beam size down to 35\nm at the IP.  Therefore, the
Final Focus design has to provide small beta functions at the IP:
$\beta^{*}_{x}$ is 2.6\cm for the HER and 3.2\cm for the LER, while
$\beta^{*}_{y}$ is 0.23\mm for the HER and 0.20\mm for the LER.  These
small beta functions at the IP can be reached if the IR magnets are as
close as possible to the IP.

The total length of the Final Focus is about $\pm$~300\m around the IP
and it includes the IR, which is defined as a region of $\pm$~5\m
around the IP.

The IR is designed to have a total crossing angle at the IP of
60\mrad; this is the lowest value that allows for sufficient
seperation of the beams to account for parasitic collisions, and to
avoid synchrotron radiation backgrounds from large transverse orbits
hitting the detector beam pipe.  In the IR the two beams will be
focused by two independent magnetic systems; no quadrupole will be
shared among the LER and the HER. The advantage of this design choice
with respect to most conventional designs is the smaller dispersion
attainable inside of QD0. This is a bonus for both the machine
(smaller equilibrium emittance) and the detector (lower background
rates).  In the present design the HER and LER permanent magnet
quadrupoles will provide the first IP vertical focusing. The remaining
vertical focusing strength will be provided by two superconducting
quadrupoles for the HER and one for the LER. The HER and LER
superconducting quadrupoles are installed in separate warm bore
cryostats, respectively.  Extra focusing for the HER is provided by
two additional quadrupoles, named QD0H and QF1H.  Soft upstream bend
magnets are used to minimize as much as possible the synchrotron
radiation power in the IP area. \par

The design of the superconducting quadrupoles (SCQ) is particularly
challenging. The thermal load by synchrotron radiation on the beam
pipe inside the SCQs will be \~200\W, so a cold bore design is not
suitable.  The horizontal beam clearance and the crossing angle at the
IP fix both the warm bore diameter to 24\mm and the maximum thickness
allowed for the cryostat and the SCQ cold mass to 22\mm.  The SCQs
will be built using a double helical winding scheme and a novel
compensation technique to cancel out the cross talk between adjacent
quadrupoles (i.e., LER QD0 upstream and HER QD0 downstream). A sketch
of the IR layout is shown in in Fig.\ref{fig:IR_layout}.

In summary, the \superb\ accelerator complex has been carefully
studied and a mechanical layout provided, allowing for a rigorous cost
estimate.

\bibliographystyle{\tdrbibstyle}
\balance
\bibliography{dtdr-bib}


\renewcommand{\ChapLabel}{chap:DetOverview} 
\tdrchap{Detector Overview}
\label{chap:detoverview}
  
The \superb\ detector concept is based on the \babar\ detector, with
those modifications required to operate at a luminosity of $10^{36}\cms$
or more, and with a reduced center-of-mass boost. 
Higher luminosity and machine-related backgrounds, as well
as the need to  improve detector hermeticity
and performance, require significant R\&D 
to implement this upgrade. 

The \babar\ detector, which operated on \pep2\ from 1999 to 2008,
consists of a tracking system with a five
layer double-sided silicon strip vertex tracker (SVT) and a 40 layer
drift chamber (DCH) inside a 1.5\Tesla magnetic field, a Cherenkov detector
with fused silica bar radiators (DIRC), an electromagnetic calorimeter
(EMC) consisting of 6580 CsI(Tl) crystals and an instrumented
flux-return (IFR)  for $\KL$ detection and $\mu$
identification.

The \superb\ detector concept reuses a number of components from
\babar: the flux-return steel, the superconducting coil, the barrel of
the EMC and the fused silica bars of the DIRC. The flux-return will be
augmented with additional absorber to increase the number of
interaction lengths for muons to roughly $7\lambda$.  The DIRC camera
will be replaced by a twelve-fold modular camera using multi-anode
photo multipliers (MAPMT) photon detectors in a focusing configuration
using fused silica optics to reduce the impact of beam related
backgrounds and improve performance. The readout electronics will also
be replaced with a much faster system.  For the forward EMC, subject
to higher particle rates and background, a replacement of the CsI(Tl)
crystals with faster devices is required at least where the rates are
highest.  A solution based on cerium-doped LYSO (lutetium yttrium
orthosilicate) crystals, which have a much shorter scintillation time
constant, a lower Moli\`ere radius and much better radiation hardness
than the current CsI(Tl) crystals, would give the ultimate detector
performance.  The high cost of this material has prompted the
collaboration to also explore different solutions, and the present
design is a hybrid of the existing CsI(Tl) in the outermost rings,
plus LYSO in the inner rings where the rates are highest.  Another
attractive possibility is based on pure CsI crystals which, while
featuring a fast response time, suffer from a low light yield,
requiring R\&D on the readout device.

\begin{figure*}[tbp]
 \begin{center}
 \includegraphics[angle=90,width=0.8\linewidth]{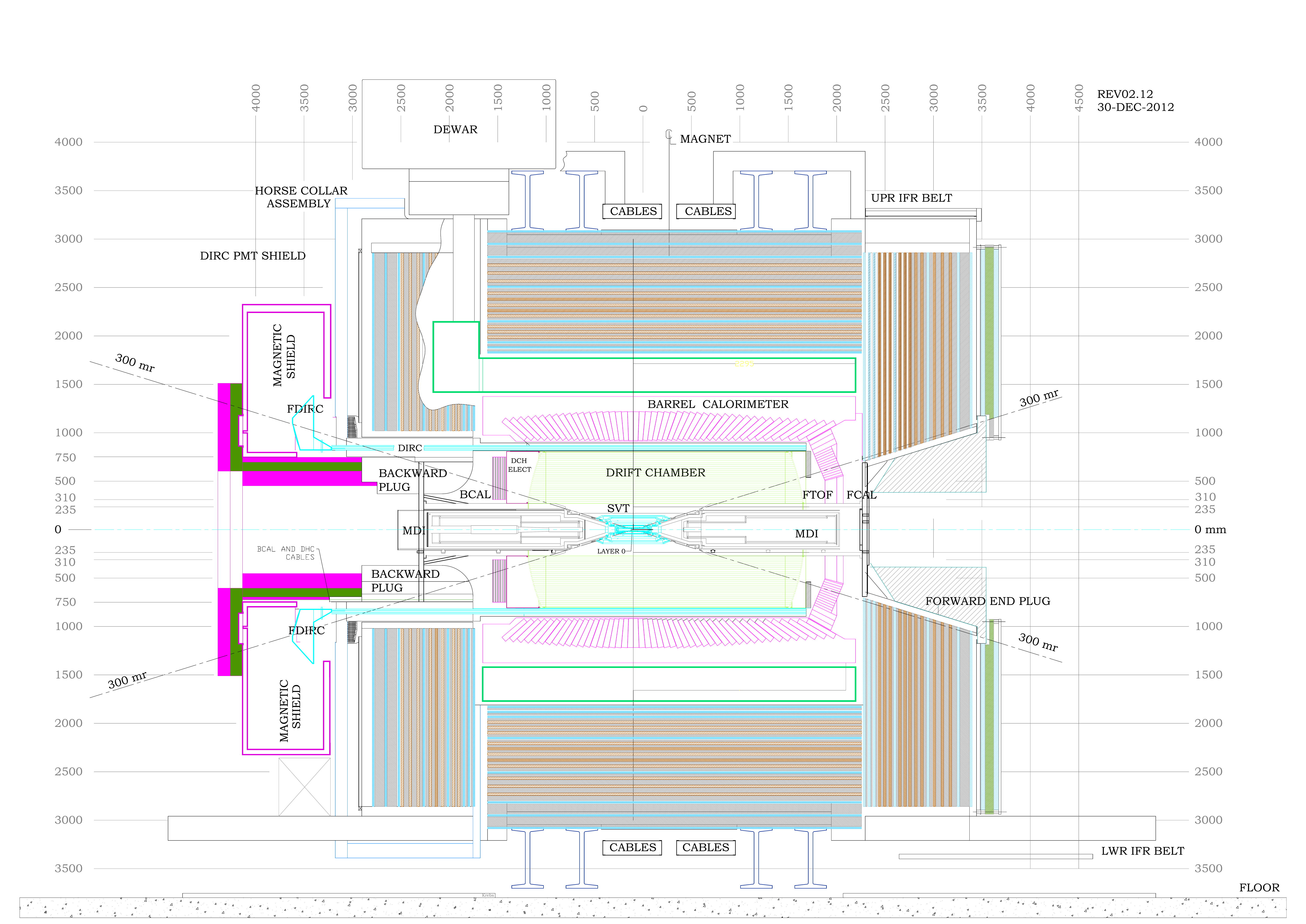}
 \caption{Concept for the \superb\ detector. Space is reserved for the
   backwards calorimeter (BCAL) and the forward time-of-flight system
   (FTOF).}
 \label{fig:det:superb}
 \end{center}
\end{figure*}

The tracking detectors for \superb\ will be new. The current SVT
cannot operate at $\mathcal{L} = 10^{36}\cms$, and the DCH has reached
the end of its design lifetime and must be replaced. To maintain
sufficient proper-time difference ($\deltat$) resolution for
time-dependent \CP violation measurements with the \superb\ boost of
$\beta\gamma=0.24$, the vertex resolution will be improved by reducing
the radius of the beam pipe, placing the innermost layer of the SVT at
a radius of roughly $1.5\cm$.  This innermost layer of the SVT will be
initially constructed of silicon striplets, with an upgrade path to
Monolithic Active Pixel Sensors (MAPS) or other pixelated sensors,
depending on the estimated background levels and the success of the
connected R\&D.  Likewise, the design of the cell size and geometry of
the DCH will be driven by occupancy considerations.

It is desirable to improve the hermeticity of the \superb\ detector,
and thereby its performance for certain physics channels, by including
a backward ``veto-quality'' EMC detector comprising a
lead-scintillator stack and a forward PID based on a fast Cherenkov
light time-of-flight system with multichannel plate PMT readout.
Currently a final decision on the inclusion of these two devices has
not been taken, but enough space has been reserved to make their
installation possible.

The \superb\ detector concept is shown in Figure~\ref{fig:det:superb}.

\tdrsec{Physics Performance}
\label{subsec:Overview_Perf}

The \superb\ detector design, as described in the Conceptual Design
Report~\cite{Bona:2007qt}, left open a number of issues that have a
large impact on the overall detector geometry. These include the
physics impact of a PID device in front of the forward EMC; the
need for an EMC in the backward region; the position of the innermost
layer of the SVT; the SVT internal geometry and the SVT-DCH transition
radius; and the amount and distribution of absorber in the IFR.

These issues have been addressed by evaluating the performance of
different detector configurations in reconstructing charged and
neutral particles as well as the overall sensitivity of each
configuration to a set of benchmark decay channels. To accomplish this
task, a fast simulation code specifically developed for the \superb\
detector has been used (see Section~\ref{sec:computing}), combined with
a complete set of analysis tools inherited, for the most part, from
the \babar\ experiment.  \geantFour-based code has been used to simulate
the primary sources of backgrounds---including both machine-induced
and physics processes---in order to estimate the rates and
occupancies of various sub-detectors as a function of position.  The
main results from these ongoing studies are summarized in this
section.

Time-dependent measurements are an important part of the \superb\
physics program.  In order to achieve a $\deltat$ resolution
comparable to that at \babar, the reduced boost at \superb\ must be
compensated by improving the vertex resolution. This requires a thin
beam pipe plus a SVT Layer0 that is placed as close as possible to the
IP.  The main factor limiting the minimum radius of Layer0 is the hit
rate from $e^+e^-\to e^+e^-e^+e^-$ background events.  Two candidate
detector technologies with appropriate characteristics, especially in
radiation length (\Xrad) and hit resolution, for application in Layer0
are (1) a hybrid pixel detector with 1.08\% \Xrad and 14\mum hit
resolution, and (2) striplets with 0.40\% \Xrad and 8\mum hit
resolution.  Simulation studies of $B^0\to\phi\KS$ decays have shown
that with a boost of $\beta\gamma=0.28$ the hybrid pixels (the
striplets) reach a $\sin 2\beta_{\rm eff}$ per event error equal to
\babar\ at an inner radius of 1.5\cm (2.0\cm). With $\beta\gamma=0.24$
the error increases by 7--8\%. Similar conclusions also apply to
$B^0\to\pi^+\pi^-$ decays.

The \babar\ SVT five-layer design was motivated both by the need for
standalone tracking for low-$p_T$ tracks as well as the need for
redundancy in case several modules failed during operation. The
default \superb\ SVT design, consisting of a \babar-like SVT detector
and an additional Layer0, has been compared with two alternative
configurations with a total of either five or four layers.  These
simulation studies, which used the decay $B\to D^*K$ as the benchmark
channel, focused on the impact of the detector configuration on track
quality as well as on the reconstruction efficiency for low $p_T$
tracks.  The studies have shown that, as expected, the low-$p_T$
tracking efficiency is significantly decreased for configurations with
reduced numbers of SVT layers, while the track quality is basically
unaffected.  Given the importance of low momentum tracking efficiency
for the \superb\ physics program, these results support a six-layer
layout.

Studies have also shown that the best overall SVT+DCH tracking
performance is achieved if the outer radius of the SVT is kept small
(14\cm as in \babar\ or even less) and the inner wall of the DCH is as
close to the SVT as possible. However, as some space between the SVT
and DCH is needed for the cryostats that contain the super-conducting
magnets in the interaction region, the minimum DCH inner radius is
expected to be about 27\cm.

\begin{figure}[tbh] 
\includegraphics[width=0.5\textwidth]{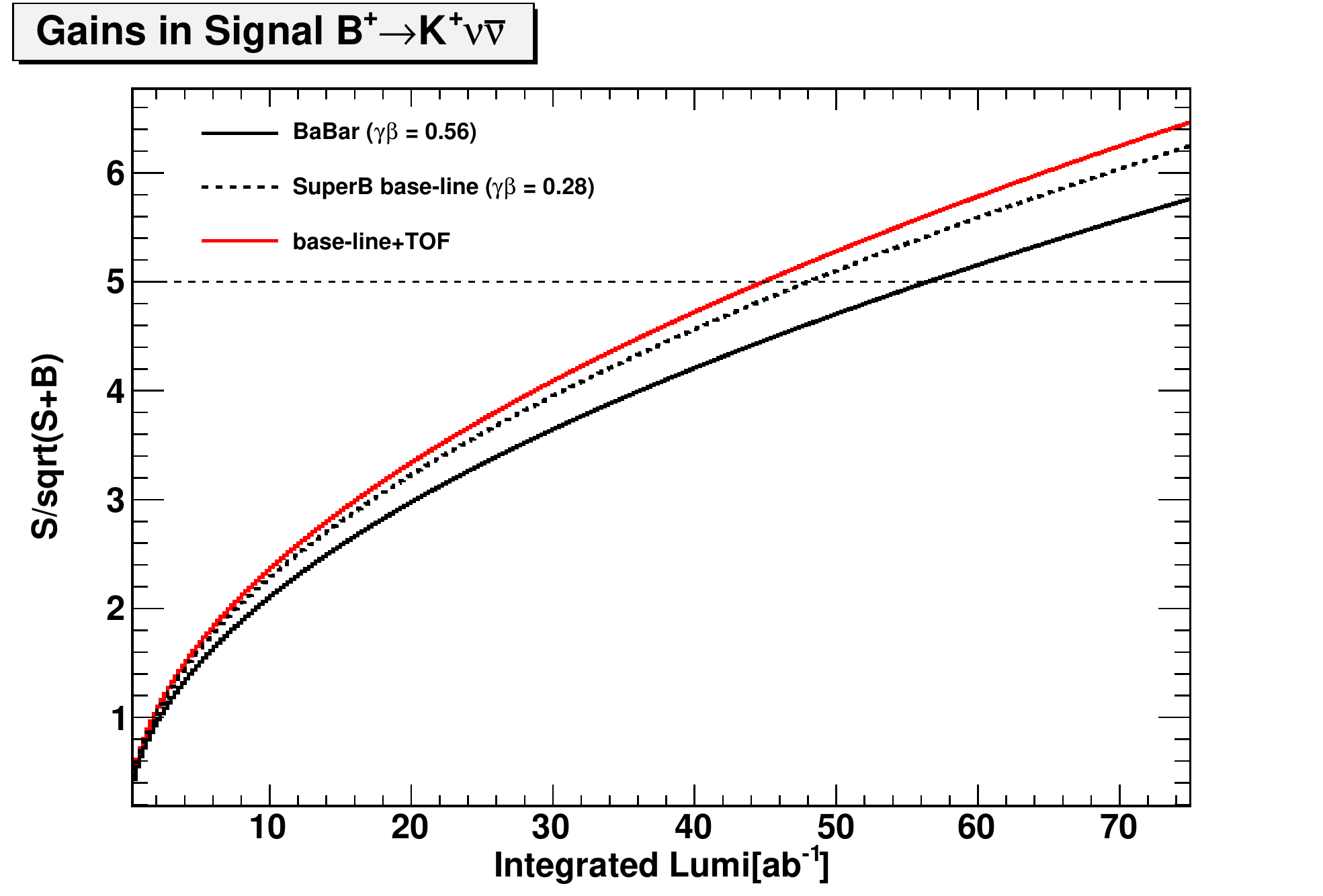} 
\caption{ ${\rm S}/\sqrt{{\rm S}+{\rm B}}$ of $B\to K\nu\bar\nu$ as a
  function of the integrated luminosity in three different detector
  configurations.}
\label{fig:BtoKnunubarBF} 
\end{figure}

The impact of a forward PID device is estimated using benchmark modes
such as $B\to K^{(*)}\nu\bar\nu$, balancing the advantage of having
better PID information in the forward region with the drawbacks
arising from more material in front of the EMC and a slightly shorter
DCH.  Three detector configurations have been compared in these
simulation studies: \babar, the \superb\ baseline (with no forward PID
device), and a configuration that includes a time-of-flight (TOF)
detector between the DCH and the forward EMC.  The results, presented
in terms of ${\rm S}/\sqrt{{\rm S}+{\rm B}}$, for the decay mode $B\to
K\nu\bar\nu$ with the tag side reconstructed in the semileptonic
modes, are shown in Figure~\ref{fig:BtoKnunubarBF}.  In summary, while
the default \superb\ design leads to an improvement of about 7--8\% in
${\rm S}/\sqrt{{\rm S}+{\rm B}}$, primarily due the reduced boost in
\superb, the configuration with the forward TOF provides an
additional 5--6\% improved sensitivity for this channel. Machine
backgrounds have not been included in these simulations.
 
\begin{figure}[tbh]
\includegraphics[clip=true,trim=20 30 55 40,width=0.5\textwidth]{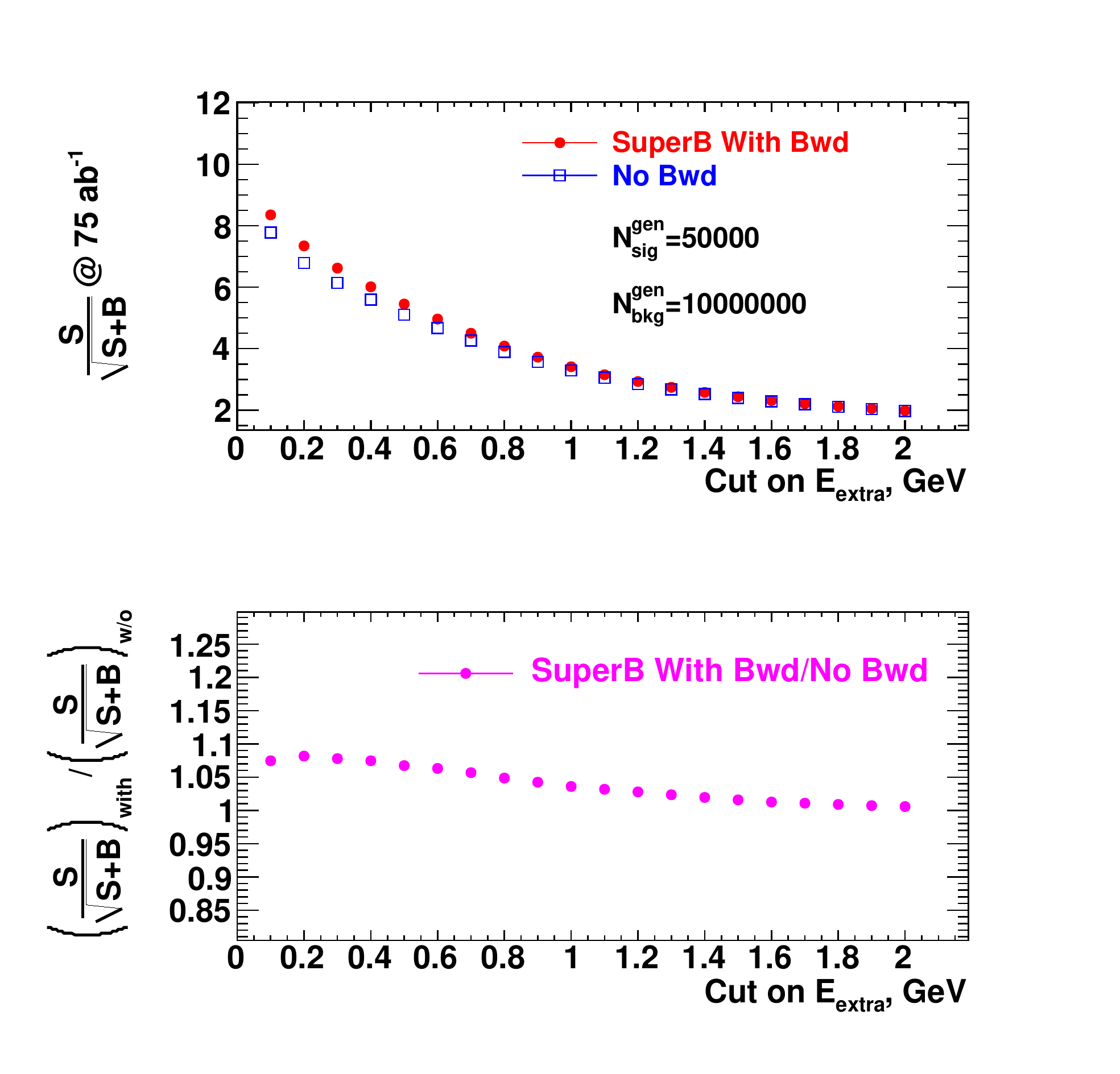}
\caption{Top: ${\rm S}/\sqrt{{\rm S}+{\rm B}}$ as a function of the
  cut on $E_{\rm extra}$ with (circles) and without (squares) the
  backward EMC. Bottom: ratio of ${\rm S}/\sqrt{{\rm S}+{\rm B}}$ for
  detector configurations with or without a backward EMC as a function
  of the $E_{\rm extra}$ cut.}
\label{fig:bwdEMC_btotaunu}
\end{figure}

The backward calorimeter under consideration is designed to be used in
veto mode. Its impact on physics can be estimated by studying the
sensitivity of rare $B$ decays with one or more neutrinos in the final
state, which benefit from having more hermetic detection of neutrals
to reduce the background contamination.  One of the most important
benchmark channels of this kind is $B\to\tau\nu$.  Preliminary
studies, not including the machine backgrounds, indicate that, when
the backward calorimeter is installed, the statistical precision ${\rm
  S}/\sqrt{{\rm S}+{\rm B}}$ is enhanced by about 8\%.  The results
are summarized in Figure~\ref{fig:bwdEMC_btotaunu}. The top plot shows
how ${\rm S}/\sqrt{{\rm S}+{\rm B}}$ changes as a function of the cut
on $E_{\rm extra}$ (the total energy of charged and neutral particles
that cannot be directly associated with the reconstructed daughters of
the signal or tag $B$) with or without the backward EMC. The signal is
peaked at zero.  The bottom plot shows the ratio of ${\rm
  S}/\sqrt{{\rm S}+{\rm B}}$ for detector configurations with and
without a backward EMC, again as a function of the $E_{\rm extra}$
cut.  This analysis did not include machine backgrounds, which could
affect the $E_{\rm extra}$ distributions significantly.  It is
possible that the backward calorimeter could also be used as a PID
time-of-flight device.

The presence of a forward PID or backward EMC affects the maximum
extension of the DCH and therefore the tracking and $dE/dx$
performance in those regions. The impact of a TOF PID detector is
practically negligible because it only takes a few centimeters from
the DCH. On the other hand, the effect of a forward RICH device ($\sim
20\cm$ DCH length reduction) or the backward EMC ($\sim 30\cm$) is
somewhat larger. For example, for tracks with polar angles $<23^\circ$
and $>150^\circ$, there is an increase in $\sigma_p/p$ of 25\% and
35\%, respectively. Even in this case, however, the overall impact on
the physics is generally quite limited because only a small fraction
of tracks cross the extreme forward and backward regions.

The IFR system will be upgraded by replacing the \babar\ RPCs and LSTs
with layers of much faster extruded plastic scintillator coupled to
WLS fibers read out by APDs operated in Geiger mode. The
identification of muons and \KL is optimized with a \geantFour simulation
by tuning the amount of iron absorber and the distribution of the
active detector layers. The current baseline design has a total iron
thickness of 92\cm interspersed with eight layers of scintillator.
Preliminary estimates indicate a muon efficiency larger than 90\% for
$p>1.5$\gevc at a pion misidentification rate of about 2\%.

\tdrsec{Challenges on Detector Design} 
\label{subsec:Overview_challenges}

Machine background is one of the leading challenges of 
the \superb\ project: 
each subsystem must be designed so that its performance is
minimally degraded because of the occupancy produced by background
hits.  Moreover, the detectors must be protected against deterioration
arising from radiation damage.  In effect, what is required is that
each detector perform as well or better than \babar, with similar
operational lifetimes, but for two orders of magnitude higher
luminosity.  An extensive \geantFour-based background simulation effort
has been undertaken to provide the parameters for detector and
electronics design. The simulation takes into account both the overlap
of background generated particles on physics events, with the
consequent performance deterioration, and the overall ionizing and
non-ionizing radiation levels (in particular neutrons) in the various
locations of the detector.

Single beam backgrounds can be classified as follows:
\begin{itemize} 
\item synchrotron radiation (SR) produced near the interaction region
  (IR). Although no bending is foreseen near the IR, the strong
  focusing required to obtain the high luminosity causes significant
  synchrotron radiation levels, which must be directed to absorber
  masks strategically located very close to the interaction point.
  Residual low energy photons, mainly generated in the backscattering
  of SR are absorbed by a coating of the beam pipe with high-Z
  material.

\item beam-gas coulomb scattering and Touschek intra-beam scattering;
  these cause particle losses all around the rings, and require a
  careful collimator design to prevent these particles from reaching
  the interaction region or showering in the nearby machine elements.

\end{itemize} 

Other very relevant backgrounds are related to the interaction of the
primary particles and thus proportional to luminosity:
\begin{itemize} 
\item radiative Bhabha scattering (\ie\ $\epem \to \epem \gamma$)
  where one of the incoming beam particle looses a significant
  fraction of its energy by the emission of a photon.  Both the photon
  and the radiating beam particles emerge from the IP traveling almost
  collinearly with respect to the beam-line. The magnetic elements
  downstream of the IP over-steer these primary particles into the
  vacuum chamber walls producing electromagnetic showers, whose final
  products are the background particles seen by the subsystems. The
  particles of these electromagnetic showers can also excite giant
  nuclear resonances in the material around the beam line expelling
  neutrons from the nucleus.  Careful optimization of the mechanical
  apertures of the vacuum chambers and the optical elements is needed
  to keep a large stay-clear for the off-energy primary particles,
  hence reducing the background rate.  The simulations indicate that a
  mechanically challenging 4.5\cm tungsten shield all around the
  machine elements near the IR is necessary to reduce this background
  source to an acceptable level.

\item electron-positron pair production by the two photon process
  $\epem \to \epem \epem$, whose total cross section evaluated at
  leading order with the Monte Carlo generator DIAG36
  \cite{Berends:1986ig} is $7.3\mbarn$ corresponding, at nominal
  luminosity, to a rate of $7.3\GHz$.  The pairs produced by this
  process are characterized by very soft transverse momenta particles
  which are for the most part confined inside the beam pipe by the
  $1.5\Tesla$ solenoidal field.  Those
  particles having a transverse momentum large enough to reach the
  beam pipe ( $p_T > 2.5 \mevc$) and with a polar angle inside the SVT
  Layer0 acceptance are expected to give a track rate of
  $2\,\mathrm{MHz/cm^2}$ at a radius of 1.5\cm, and are the driving
  element in the design of the segmentation and the read-out
  architecture for SVT Layer0.

\item The elastic Bhabha process has also been considered. The cross
  section for producing a primary particle reconstructed by the
  detector via this process is $\sim 100 \nb$ corresponding to a rate
  of about $100 \kHz$. This is expected to be the driving term for the
  level one trigger rate, which is nominally set at $150 \kHz$, more
  than an order of magnitude larger than in \babar.

\end{itemize}

The \superb detector design poses therefore many challenges and
requires some detector R\&D to be finalized and optimized, although a
baseline solution exists for all subsystems, as can been seen in
Table~\ref{tab:detrd }, where the baseline solution, the main design
challenges and required detector R\&D are detailed for each subsystem.

\begin{table*}
\caption{Baseline solution, main design challenges and required detector R\&D for each 
subsystem }
\label{tab:detrd	}
\begin{center}
\begin{tabular}{|l|l|p{8cm}|}
\hline
{\bf System} & {\bf Baseline} & {\bf Challenges and R\&D} \\
\hline
\multicolumn{ 1}{|l|}{MDI} & \multicolumn{ 1}{|l|}{IR baseline designed} & Magnetic elements and radiation masks.  \\

\multicolumn{ 1}{|l|}{} & \multicolumn{ 1}{|l|}{} & Design of tungsten shields. Cryostat radius. \\

\multicolumn{ 1}{|l|}{} & \multicolumn{ 1}{|l|}{} & Background simulations: global map, detector occupancy. \\
\hline
\multicolumn{ 1}{|l|}{SVT} & \multicolumn{ 1}{|l|}{6-layer silicon} & Technology for Layer 0: striplets or pixels. \\

\multicolumn{ 1}{|l|}{} & \multicolumn{ 1}{|l|}{Striplets Layer 0} & Thin pixels R\&D.  Readout chip for strips. Readout architecture.  Mechanical design. \\
\hline
\multicolumn{ 1}{|l|}{DCH} & \multicolumn{ 1}{|l|}{Stereo-axial He-based} & Dimensions (inner radius, length).  Mechanical structure. Cluster counting option. \\
\hline
\multicolumn{ 1}{|l|}{PID} & \multicolumn{ 1}{|l|}{DIRC w/ FBLOCK} & Focusing Block design. Photon detection. Mechanical structure. \\

\multicolumn{ 1}{|l|}{} & \multicolumn{ 1}{|l|}{} & Forward PID: cost/benefit analysis. Prove TOF technology. \\

\hline
\multicolumn{ 1}{|l|}{EMC} & Barrel: CsI(Tl) & Electronics and trigger.  Mechanical structure. Transport and refurbishing. \\

\multicolumn{ 1}{|l|}{} & Forw: LYSO+CsI(Tl) & Crystal choice: LYSO, LYSO + CsI(Tl), pure CsI; photon detetector. \\

\multicolumn{ 1}{|l|}{} &            & Backward EMC: cost/benefit analysis. \\
\hline
       IFR & Scintillator+ fibers & 8 vs 9 layers. SiPM radiation damage and location. Extra 10~cm iron. Mechanical design and yoke reuse. \\
\hline
       ETD & Synchronous const. latency & Fast link rad hardness.  L1 Trigger (jitter and rate). ROM design. Link to computing for HLT. \\
\hline
\end{tabular}  
\end{center}

\end{table*}

\tdrsec{Detector R\&D}
\label{subsec:Overview_RandD}

\label{sec:rd}

The basic geometry, structure and physics performance of the \superb\
detector is mainly predetermined by the retention of the solenoidal
magnet, the return steel, and the support structure from the \babar\
detector, and a number of its largest and most expensive subsystems.
Even though this fixes both the basic geometry, and much of the
physics performance, it does not really constrain the expected
performance of the \superb\ detector in any important respect. \babar\
was already an optimized $B$-factory detector for physics, and any
improvements in performance that could come from changing the overall
layout or rebuilding the large subsystems would be modest overall. The
primary challenge for \superb\ is to retain physics performance
similar to \babar\ in the higher background environment while
operating at much higher ($\sim \times 50$) data taking rates.

Subdetector envelopes have been defined for the most part, although
some adjustements especially regarding the DCH length and inner radius
might be necessary.  The benefits of a backward EMC and forward PID
have been evaluated and enough space has been left in the general
layout to allow the installation of these devices, provided that
enough resources can be found to build them. The TOF technology for
the forward PID is particularly challenging and requires further R\&D
to arrive at a robust detector design.  The inner radius of the DCH is
determined by the dimensions of the cryostat for the final focus
superconducting quadrupoles and of the tungsten shields.

The \superb\ detector concept rests, for the most part, on
well-validated basic detector technology.  Nonetheless, because of the
high rates and demanding performance requirements, each of the
sub-detectors has many challenges to tackle, with several R\&D
initiatives ongoing to improve the specific performance and optimize
the overall detector design.  These are described in more detail in
each subsystem section.

\tdrsubsec{Machine Detector Interface (MDI) and Backgrounds Simulation}
\label{MDI} 

The machine detector interface is a crucial element of both the
accelerator and detector design. It contains the final focus, composed
by several important magnetic elements: the focusing quadrupoles, the
crab waist sextupoles, and the anti-solenoids compensating for the
detector field. These elements are for the most part implemented with
superconducting magnets contained in a single cryostat. Some permanent
magnets very close to the interaction point are also required.

\begin{figure*}
\includegraphics[width=\textwidth]{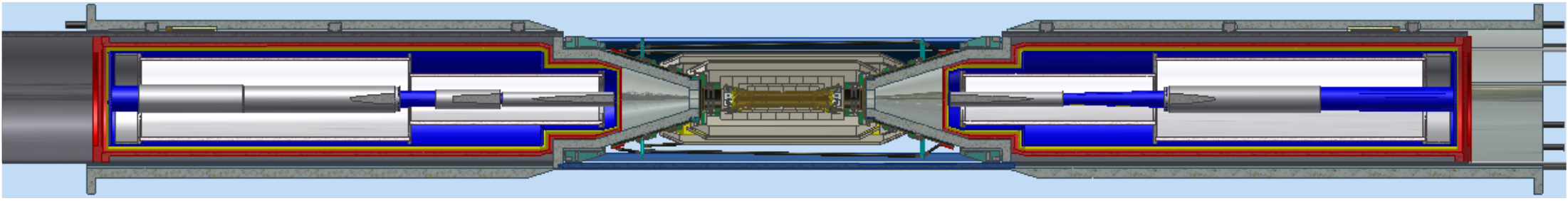}
\caption{Cross section of the interaction region and machine-detector interface area.}
\label{fig:sbir}
\end{figure*}

The Interaction Region (IR) design is also crucial to keep backgrounds
in the detector under control. The most notable elements in this
respect are:
\begin{itemize} 
\item the synchrotron radiation masks, carefully placed inside the
  primary vacuum to absorb radiation emitted during the steering and
  focusing of the beams
\item the final focus quadrupoles (QD0), whose design has required a
  great deal of R\&D.  In fact, radiative Bhabha scattering will
  produce on-axis particles with reduced momentum that will be
  deviated by any residual dipole field component on the nominal
  trajectory. This implies that QD0 cannot be shared by the two beams,
  and complex twin quad design had to be developed with a distance of
  only a few \cm between the two beam axes.
\item the tungsten radiation shields absorb the showers produced by
  the particles that even in a careful QD0 design do get deviated out
  of the beam pipe in the vicinity of the IR. Simulation studies
  indicate that a 4.5\cm thick shield is adequate.  The mechanical
  design of the support system for the shields and of the assembly
  system for the entire IR (including the SVT) is extremely
  challenging.
\end{itemize} 
A cross section of the IR, including the tungsten shields, the
cryostat, and the SVT is shown in Figure~\ref{fig:sbir}.

Backgrounds simulation is essential to provide guidance for both the
machine and detector design. A \geantFour-based background simulation
framework has been developed that can provide input to detector
performance studies, provide radiation maps throughout the detector
and experimental hall, as well as background frames to be superimposed
to physics events for physics studies. Validation of these simulation
results is very difficult, and has been realized through a combination
of extrapolation from \babar\ information, comparison with lower
energy Da$\Phi$ne running, and comparison with analogous simulations
performed by the Belle-II collaboration~\cite{bib:vienna-workshop}.
In addition, some of the backgrounds critically depend on the
placement accuracy of machine elements that might not be at their
nominal position in reality. To account for all these uncertainties we
require the detector to operate with a $\times 5$ safety factor over
the simulated nominal background level.

\tdrsubsec{Silicon Vertex Tracker (SVT)}
\label{SVT}
The SVT baseline design is based on double-sided silicon microstrip
detectors for all layers.  The current Layer0 baseline configuration
for the beginning of data taking is based on the double-sided silicon
striplets technology, which is mature and has been shown to provide
good physics performance thanks to the low material budget associated
(\ie\ no readout electronics in the active area).  However, an upgrade
to pixel sensors, hybrid pixels or thinner CMOS Monolithic Active
Pixel Sensor (MAPS), more robust against background occupancy, is
planned for the full luminosity run.  Specific R\&D programs are
ongoing in order to meet the Layer0 requirements, which include low
pitch and material budget, high readout speed and radiation hardness.
If successful, this will allow the replacement of the Layer0 striplet
modules in a second phase of the experiment.  For this purpose, the
\superb\ interaction region and the SVT mechanics will be designed to
ensure a rapid access to the detector to ease a replacement of the
Layer0.

The expected $2\,\mathrm{MHz/cm^2}$ track rate at 1.5\cm translates
into a much larger hit rate once the hit multiplicity, loopers, and
safety factor are taken into account, setting a requirement of
$100\,\mathrm{MHz/cm^2}$ hit rate with a required module readout
bandwidth of 5\gbitps.  Material budget is the other
essential element of the Layer0 design, with a requirement of about
$0.5\%\,\mathrm{X_0}$.

The hybrid pixels technology represents a mature and viable solution
but reduction in the front-end pitch and in the total material budget
with respect to pixel systems developed for LHC experiments is
required to meet the Layer0 requirements.  The spatial resolution
constraints of 10--15\mum set a limit to the area of the elementary
readout cell and, as a consequence, to the amount of functionalities
that can be included in the front-end electronics.  For a
50$\times$50$\mum^2$  pixel cell a planar 130\nm CMOS technology may guarantee
the required density to implement in-pixel data sparsification and
fast time stamping ($< 1\mus$), meeting the Layer0 requirements.

Denser CMOS technologies (90 or 65\nm technology) can
be used to increase the functional density in the readout electronics
and include such functions as gain calibration, local threshold
adjustment and amplitude measurement and storage.  In this case, costs
for R\&D (and, eventually, production) would increase significantly.

Vertical integration (or 3D) CMOS technologies may represent a lower
cost alternative to sub-100\nm CMOS processes to increase the
functional density in the pixel cell~\cite{Yarema:2008zza, vipix}.

CMOS MAPS are very appealing for applications where the material
budget is critical since in this technology the sensor and readout
electronics share the same substrate that can be thinned down to
several tens of microns.  As the readout speed is another crucial
aspect for application in Layer0 a new Deep NWell MAPS design approach
has been developed in the last few years. This allowed for the first
time the implementation of thin CMOS sensors with similar
functionalities as in hybrid pixels, such as pixel-level
sparsification and fast time stamping~\cite{Gabrielli:2011zz}. A
limiting factor of this technology, though, is the presence of
competitive N-Wells inside the pixel cell, that can subtract charge from
the main collecting electrode leading to an efficiency of the order of
90\%.  A significant degradation of the charge collected (about 50\%)
has also been measured after neutron irradiation 
of $7 \times 10^{12}$ n/cm$^2$, corresponding to about 1.5 years of
operation in the Layer0~\cite{ratti:radecs2009}.

Further MAPS performance improvements are currently under
investigation with two different approaches: the use of INMAPS CMOS
process, featuring a quadruple well and a high resistivity substrate,
and 3D CMOS MAPS, realized with vertical integration technology.

In order to increase the charge collection efficiency, the INMAPS 180\nm CMOS process is being explored: a deep P-well implant, deposited
beneath the competitive N-Wells, can prevent them from stealing charge
from the main collecting electrode.  Moreover, the use of a high
resistivity substrate, also available in this process, can further
improve charge collection and radiation resistance with respect to
standard CMOS devices.

The realization of 3D MAPS, using two CMOS layers interconnected with
vertical integration technology, also offers several advantages with
respect to standard 2D MAPS.  In these devices, one CMOS tier is
hosting the sensor with the analog front-end and the second tier is
dedicated to the in-pixel digital front-end and the peripheral readout
logic. With this splitting of functionalities the collection
efficiency can be improved, significantly reducing the N-Well
competitive area in the sensor layer, allowing the implementation of a
more performant readout architecture, and minimizing the cross-talk
between analog and digital blocks.

To keep material at a minimum it is essential to also develop low mass
cooling systems as well as low mass cables and data transmission
systems.  A vigorous R\&D program on single- and double-phase
microchannel cooling is ongoing, with the final goal of integrating
the microchannels directly on the silicon substrate, thus
significantly reducing the extra material needed.

\tdrsubsec{Drift Chamber (DCH)}
\label{DCH}

In the DCH, many parameters must be optimized for \superb\ running,
such as the gas mixture and the cell layout.  Precision measurements
of fundamental gas parameters are ongoing, as well as studies with
small cell chamber prototypes and simulation of the properties of
different gas mixtures and cell layouts.

An improvement of the performance of the DCH could be obtained by
using the innovative ``Cluster Counting'' method. Signals in drift
chambers are usually split to an analog chain that integrates the
charge and a digital chain that records the arrival time of the first
electron, discriminated with a given threshold. The cluster counting
technique consists instead in digitizing the full waveform to count
and measure the time of all individual peaks. On the assumption that
these peaks can be associated to the primary ionization acts along the
track, the energy loss and to some extent the spatial coordinate
measurements can be substantially improved.  In counting the
individual cluster, one indeed removes the sensitivity of the specific
energy loss measurement to fluctuations in the amplification gain and
in the number of electrons produced in each cluster, fluctuations
which significantly limit the intrinsic resolution of conventional
\dedx measurements.

The ability to count the individual ionization clusters and measure
their drift times strongly depends on the average time separation
between them, which is, in general, relatively large in He-based gas
mixtures thanks to their low primary yield and slow drift velocity.
Other requirements for efficient cluster conting include good
signal-to noise ratio but no or limited gas-gain saturation, high
preamplifier bandwidth, and digitization of the signal with a sampling
speed of the order of $1\,\mathrm{Gs/sec}$. Finally, it is necessary
to extract online the relevant signal features (\ie\ the cluster
times), because the DAQ system of the experiment would hardly be able
to manage the enormous amount of data from the digitized waveforms of
the about 10,000 drift chamber channels.

In order to optimize the gas mixture for the \superb\ environment, and
to assess both the feasibility and the operational improvements for
the cluster counting technique an R\&D program has been ongoing.  The
program includes both beam tests and cosmic ray stands to monitor
performances of {\it ad hoc} built prototypes.  While the \dedx
resolution gain of the cluster counting method is in principle quite
sizeable compared to the traditional total charge collection, the
actual capability of the measured number of clusters might not retain
the same analyzing power, due to a plethora of experimental effects
that should be studied in detail so that the energy loss measurement
derating should be assessed and, if possible, cured.  Although this
technique requires significant R\&D, the cable plant is being designed
to allow the upgrade of preamplifiers and readout system if cluster
counting is proven feasible in the experiment.

\tdrsubsec{Particle Identification (PID)}
\label{PID}

The DIRC (Detector of Internally Reflected Cherenkov light) is an
example of innovative detector technology that has been crucial to the
performance of the \babar\ science program.  The DIRC main performance
parameters are the following: (a) a measured time resolution per
photon of $\sim$1.7\ns, close to the photomultiplier (PMT) transit
time spread of 1.5\ns, (b) a single photon Cherenkov angle resolution
of 9.6\mrad for tracks from di-muon events, (c) a Cherenkov angle
resolution per track of 2.5\mrad in di-muon events, and (d) a $\pi/K$
separation greater than $2.5\sigma$ over the entire track momentum
range, from the pion Cherenkov threshold up to 4.2\gevc.

A new Cherenkov ring imaging detector is being planned for the
\superb\ barrel, called the Focusing DIRC, or FDIRC. It will reuse the
existing \babar\ bar boxes and mechanical support structure. This
structure will be attached to a new photon ``camera'', which will be
optically coupled to the bar box exit window. The new camera design
combines a small modular focusing structure that images the photons
onto a focal plane instrumented with very fast, highly pixelated,
photon detectors (MaPMTs)~\cite{Vavra:2010zz}.  The readout
electronics will be completely redesigned, bringing the time accuracy
of the full readout chain to $200-250\ps$, which helps reduce the
background and allows a correction for chromatic aberrations.  These
elements should combine to attain the same PID performance as the
\babar\ DIRC while being at least 100 times less sensitive to
backgrounds.

Two FDIRC prototypes have been built.  The first prototype, which
used a spherical mirror design not suitable for \superb, was tested
on a beam and with cosmic rays, giving the proof of principle of the
FDIRC concept.  The second is a full-scale prototype of a \superb\ FDIRC
sector. Being able to purchase parts matching the
specifications and to build it is already a great achievement. The
test of this final prototype will validate the detector and
electronics design.  Parallel to this effort, there are many detailed
studies of the MaPMT ongoing in all major institutions involved in the
FDIRC, addressing issues such as the the timing resolution and pixel
cross-talk. The results of this complex R\&D program will allow the
finalization of the design of this novel and performant device.

The possibility of increasing the hermeticity of the detector by including
a PID device in the forward area has been considered. The physics gain
from such a device is about 4--5\%, which is considered significant
enough, provided that the device is sufficiently performant and does
not degrade the performance of the rest of the apparatus. Different
technologies have been examined: a Focusing Aerogel RICH, a
scintillating detector coupled to SiPMT arrays, and a DIRC-like time of
flight detector based on fused silica and MCP-PMT (FTOF). Given the
space constraints and the performance requirements the FTOF is the
chosen technology.

The FTOF~\cite{jjv_toflike1, jjv_toflike2} belongs to the family of
the DIRC detectors.  Charged particles crossing a thin layer of fused
silica emit Cherenkov light along their trajectories, provided that
their momentum is high enough. A fraction of these photons are trapped by
total internal reflection and propagate inside the quartz until an
array of Multi-Channel Plate Photomultipliers (MCP-PMT), where they are
detected.  Unlike the (F)DIRC, no attempt is made to reconstruct
Cherenkov angles: K/$\pi$ separation is provided by \tof.  Given the
short flight distance between the IP and the FTOF (about 2~meters,
hence a 90\ps difference for 3~\gevc particles), the whole detector
chain must make measurements accurate at the 30\ps level or better.
This is one of the main challenges of this design, the others being
the photon yield, the sensitivity to the background (including ageing
effects due to the charge integrated over time by the photon
detectors) and the event reconstruction.  Regarding the latter point,
one should emphasize that the FTOF is a two-dimensional
device: the PID separation uses both the timing and spatial
distributions of the photons detected in the MCP-PMT arrays.  A first
prototype was built and tested with cosmic rays, giving an initial
proof of principle of the method, but R\&D is ongoing to build a full
scale prototype and to study various design options and their performance
in magnetic field, as well as their robustness against background and
radiation damage.

\tdrsubsec{Electromagnetic Calorimeter (EMC)}
\label{EMC}

The \superb\ EMC reuses the barrel part of the \babar\ EMC detector
consisting of 5760 CsI(Tl) crystals readout by PIN-diodes.  The
increased event and background rate requires a replacement of the
readout electronics and a careful assessment of the effect of
radiation damage on the \babar\ crystals.  In addition, there are
several important technical issues related to the EMC barrel
transportation and refurbishing.

On the other hand, the \babar\ forward calorimeter will need to be
partially replaced, due to the higher radiation and higher rates at
\superb compared with \pep2.  At least the innermost rings of the
forward endcap will have to be replaced by new scintillating crystals,
designed to work well in this new environment.  After an intensive
R$\&$D program the baseline option for the inner rings of the forward
EMC is to use the faster and more radiation resistant Ce-doped LYSO
crystals readout by avalanche photo diodes (APD). This is the best
choice in terms of performance and radiation hardness over the
alternatives we have considered because of the faster response time
and shorter Moli\`ere radius that address the higher event and
background rates.  LYSO is a fast scintillator largely used in medical
applications with crystals of small size.  The R\&D was concentrated
on the optimization of performance for large crystals ($2\cm \times
2\cm \times 20\cm$) with good light yield uniformity and optimized Ce
doping, in order to have the best possible light output.  Thanks to
this effort, more than one producer is able to grow LYSO crystals of
good quality that can be used in high energy physics applications. The
high cost of LYSO has prompted the collaboration to examine other
lower cost options, such as BGO and pure CsI. The latter is
particularly attractive for its very fast response (30\ns compared to
40\ns of LYSO), while at the same time posing significant readout
problems with a light yield a factor 20 less than LYSO and a factor 45
less than CsI(Tl). A vigorous R\&D program has started to examine
photo-pentodes and large area APDs as readout options for pure CsI.
Neither solution at this time guarantees the required performance, and
additional studies are needed. R\&D is also needed to verify the
radiation hardness of pure CsI.

A lead-scintillator-sandwich backward endcap calorimeter readout with
SiPM/MPPC improves the hermeticity of the detector.  The main purpose
of this component is to detect energy in the backward endcap region,
as a veto of events with ``extra'' energy.  This is particularly
important for studying channels with neutrinos in the final state.
Because of the fast time response, the backward EMC may also have a
role in particle identification by providing time-of-flight for the
relatively slow backward-going charged particles.

\tdrsubsec{Instrumented Flux Return (IFR)}
\label{IFR}

The Instrumented Flux Return (IFR) is designed primarily to identify
muons, and, in conjunction with the electromagnetic calorimeter, to
identify neutral hadrons, such as \KL.  The iron yoke of the detector
magnet, providing the large amount of material needed to absorb
hadrons, is segmented in depth, with large area extruded scintillator
bars readout with Silicon Photo Multipliers.

The detector consists of a central part, the barrel, and of two
endcaps, to cover the forward and backward regions.  Both the barrel
and the two endcaps, are made of $\sim 92\cm$ of iron interleaved by 9
layers of segmented scintillator planes (with a thickness of $2\cm$).
The use of plastic scintillator as active material ensures fast
response, reliability, robustness and long term stability while the
high granularity guarantees a good space resolution, very important to
cope with the expected high particle flux.  Each active plane provides
both coordinates simultaneously, since it is made of two layers of
orthogonal scintillator bars: the longitudinal bars are 5\cm wide
while the transversal bars are 10\cm wide. The scintillation light is
collected by three Wave Length Shifting fibers (Kuraray Y11,
$1.2\mm$), housed in surface grooves machined on each bar, and guided
to the SiPM placed at one end of each bar.

The general layout outlined above is the result of extensive R\&D
studies, which concerned all the active parts of the detector.  We
compared fibers from different producers (Saint- Gobain and Kuraray)
and with different diameters ($1.0\mm$ and $1.2\mm$), and we studied
the optimal number of fibers to collect light from each scintillator
bar.  On the photodetectors side, we are still comparing the
performance of devices from various producers (FBK-Advansid,
Hamamatsu, SensL), with different pixel size (e.g. $25\times 25$,
$50\times 50$, and $100\times 100\mum^2$).  A final decision has not
been taken yet, as the performances of these recently developed
devices are improving very rapidly, and more R\&D studies are needed.

The performances of the detector have been studied through a full
scale prototype, i.e. a prototype with the same iron-scintillator
structure of the detector. It has been tested at the Fermilab Test
Beam Facility, where muons and pions of energy in the range of
interest for \superb are available.

Given the very high luminosity and, as a consequence, the high level
of background radiation, one of the most critical issues is to
understand the radiation damage of all the sensitive detector
components. For semiconductor devices like SiPMs and front end
electronics, it is of particular importance the effect of neutrons,
which are known to produce an increase in the dark count and a possible
malfunctioning of FPGAs.  A R\&D program is ongoing to characterize
devices from different vendors as well as to identify mitigation
strategies, both in the SiPM design, and in their location and
shielding.

\tdrsubsec{Electronics, Trigger, and DAQ (ETD)}
\label{ETD}

The \superb Electronics, Trigger, Data acquisition and Online system
(ETD) comprises the Level-1 trigger, the event data chain and their
support systems. Event data corresponding to accepted Level-1 triggers
move from the Front-End Electronics (FEE) through the Read-Out Modules
(ROMs), the network event builder and the High Level Trigger (HLT) to
a data logging buffer where they are handed over to the offline for
archival and further processing. ETD also includes the hardware and
software components that control and monitor the detector and data
acquisition systems and perform near-real-time data quality monitoring
and online calibration.

At present, the Electronics, DAQ, and Trigger (ETD), have been
designed for the base luminosity of $1 \times 10^{36}
\cm^{-2}\sec^{-1}$, with adequate headroom.  Further R\&D is needed to
understand the requirements at a luminosity up to 4 times greater, and
to insure that there is a smooth upgrade path when the present design
becomes inadequate.  In particular, radiation levels in various areas
of the detector are driving elements of many of the technical choices.

Data Links are the most subject to radiation and require R\&D to
qualify components and understand the possibility of using new
generation FPGAs. A strong connection with LHC technology development
is in place.

For the L1 trigger, the achievable minimum latency and physics
performance will need to be studied. The studies require to address
many factors including (1) improved time resolution and trigger-level
granularity of the EMC and a faster DCH; (2) potential inclusion of
SVT information at L1; (3) the possibility of a L1 Bhabha veto; (4)
possibilities for handling pile-up and overlapping (spatially and
temporally) events at L1; and (5) opportunities created by modern
FPGAs to improve the trigger algorithms.

The main R\&D topics for the Event Builder and HLT Farm are (1) the
applicability of existing tools and frameworks for constructing the
event builder; (2) the HLT farm framework; and (3), event building
protocols and how they map onto the network hardware.

To provide the most efficient use of resources, it is important to
investigate how much of the software infrastructure, frameworks and
code implementation can be shared with Offline computing. This
requires us to determine the level of reliability-engineering required
in such a shared approach. We also must develop frameworks to take
advantage of multi-core CPUs.

\bibliographystyle{\tdrbibstyle}
\balance
\bibliography{DetOverview/det-esg,Physics/tau-bib}


\renewcommand{\ChapLabel}{chap:Physics} 
\tdrchap{Physics with \superb }

\section{Introduction}
\label{sec:intro}

The \superb project consists of an \epem collider with a general
purpose detector dedicated to the study of precision flavor physics.
This will be built at the Cabibbo Laboratory to commence taking data
later this decade.  \superb is designed to accumulate: 75\invab of
data at the \FourS, more than 100 times the \babar data sample;
1000\invfb at the \psiprpr, 100 times the expected data sample from
BES III; and several \invab at the \FiveS, 15 times the data sample
accumulated by \belle; as well as running at other center of mass
energies.  Using these data one will be able to search for New Physics
(NP) and test the Standard Model (SM) in a multitude of ways.

This document is divided into summaries of the heavy quark flavour
physics programme, i.e. $B_{u,d,s}$ and $D$ decays
(section~\ref{sec:BandD}), $\tau$ physics (section~\ref{sec:tau}),
precision electroweak physics (section~\ref{sec:precisionew}), exotic
spectroscopy (section~\ref{sec:spectroscopy}), and direct searches
(section~\ref{sec:searches}).  Finally section~\ref{sec:summary} gives
a broad overview of the physics programme and how this fits into the
context of furthering our understanding of nature at a fundamental
level.  The ensemble of precision measurements, studies of rare
decays, \CP and other symmetry violating observables all play a role
in the process of trying to elucidate the nature of physics beyond the
Standard Model.  These are augmented by searches for processes
forbidden within the SM, such as charged lepton flavour violation,
Dark Matter, and Dark Forces.  More detailed discussions of the
\superb project and physics programme can be found in
Refs.~\cite{Biagini:2010cc,Grauges:2010fi,Bona:2007qt,Hitlin:2008gf,O'Leary:2010af,Meadows:2011bk}.

The recent discovery of a Higgs-like particle by ATLAS and CMS
(presumed here to be the spin-zero SM Higgs particle) heralds a new
era in particle physics.  There are two ramifications of this
discovery that relate to the \superb physics programme, the first is
that the SM Higgs is related to a number of flavour processes that
will be studied at \superb, and so now that we know this particle
exists, and have started to learn about its properties --- we are
compelled to check how this fits in with our understanding of the
electroweak sector.  The second ramification is that many scenarios of
physics beyond the SM predict other Higgs particles, so while this
discovery is profound, it is not the final missing piece of the puzzle
required to explain the sub-atomic world.  Some of these additional
Higgs particles may be hard to study directly at the energy frontier,
and \superb is in a position to test such theories to constrain the
presence of light ($<10\gevcc$) Higgs particles and charged Higgs
particles that may be present in nature.

The measurements relating to the SM Higgs particle that are of
particular interest to this programme are determinations of
$\sin^2\theta_W$, which at the center of mass energy corresponding to
the \FourS is devoid of hadronisation uncertainties that affect
interpretation of the LEP/SLC measurements at the $Z^0$ pole in the
$\epem\to b\bbar$ transition.  The weak angle can be measured for
$\epem$, $\mu^+\mu^-$, $c\cbar$ and $b\bbar$ final states at \superb.
The interest in $\sin^2\theta_W$ is in inverting precision electroweak
fits that have been used to infer the Higgs mass since the LEP-days to
use the now-known mass to predict the weak angle as a function of
$\sqrt{s}$.  Hence one can test the running of the weak mixing angle.
Looking at the measurements made at the $Z^0$ pole, it is the
$e^+e^-\to b\overline{b}$ measurement that is furthest from
expectation (and also has a limiting hadronisation uncertainty).  A
similar physics potential, albeit with larger uncertainty and not
using the $b\bbar$ mode, exists for a run at the $\psi(3770)$ with a
polarized electron beam.  In terms of constraining charged Higgs
particles, \superb can study rare tree decays such as $B\to
(D^{(*)})\tau \nu$, which are mediated by a $W$ boson.  Beyond the SM
these decays could have additional contributions arising from the
charged Higgs sector.  Similarly \CP violation can be introduced into
$\tau$ decays via the presence of a charged Higgs, and could alter the
expected asymmetry in decay modes such as $\tau \to \KS \pi\nu$.
Loop-mediated Higgs transitions can also be detected using final
states such as $b\to s \gamma$. Hence comparison of the decay rates
and \CP asymmetries measured in data, with those expected in the SM
can be used to constrain $m_{H^+}$ as a function of the parameters of
a given NP model.  Many rare decay constraints on NP have direct
parallels between \B and Charm decays, and so larger samples of
$D_{(s)}$ mesons accumulated at the $\psi(3770)$ and the \FourS
provide alternate ways to test the SM.

For example if one considers a type-II two-Higgs doublet model (2HDM),
the decay $b\to s \gamma$ already excludes $m_{H^+}$ below about 300
\gevcc for all values of the Higgs vacuum expectation value
($\tan\beta$), $B\to \tau \nu$ excludes masses below $\sim 1$
$\tev/c^2$ for large $\tan\beta$, and the combination of $B\to \tau
\nu$, $B\to D\tau \nu$ and $B\to D^*\tau \nu$ disfavor the type II
2HDM model for any value of $\tan\beta$.  Just as with this example,
the combination of measurements of rare decays at \superb will play a
role in elucidating the charge partners of the SM Higgs. The results
of these searches are expected to be available before the high
luminosity running of the SLHC commences early next decade, and could
shed light on the SM Higgs sector and beyond.

\section{$B$ and $D$ decays}
\label{sec:BandD}

\subsection{Rare $B$ decays}
\label{sec:rareB}

Rare $B$ decays provide an excellent opportunity to search for NP at a
flavour factory. Processes highly suppressed in the SM, such as
Flavour Changing Neutral Currents (FCNCs), could receive large NP
contributions, often resulting in measurable deviations in the
branching fractions and other observables.

The \superb detector is designed to be well equipped to study a wide
variety of rare $B$ meson decays. Excellent reconstruction of charged
and neutral particles, with very good particle identification and
virtually fully efficient triggering for $\B\Bbar$ events, allow the
study of complex decay chains that involve several charged and neutral
particles. In addition, decay modes with missing neutrinos can be
studied effectively by employing {\em hadronic} or {\em semileptonic
  tagging} techniques. This approach, which has been used very
successfully at the current \B Factories, entails fully reconstructing
one of the two $B$ meson decays in the event, thus assigning the
remaining particles (including missing particles) to the signal decay
under study. This powerful technique is used to identify decays with
missing neutrinos. We discuss selected rare $B$ decays in the
following paragraphs, most of which can only be studied in an $e^+e^-$
environment.  These are important parts of the \superb physics
programme.

\begin{figure}[!ht]
  \begin{center}
  \includegraphics[width=0.50\textwidth]{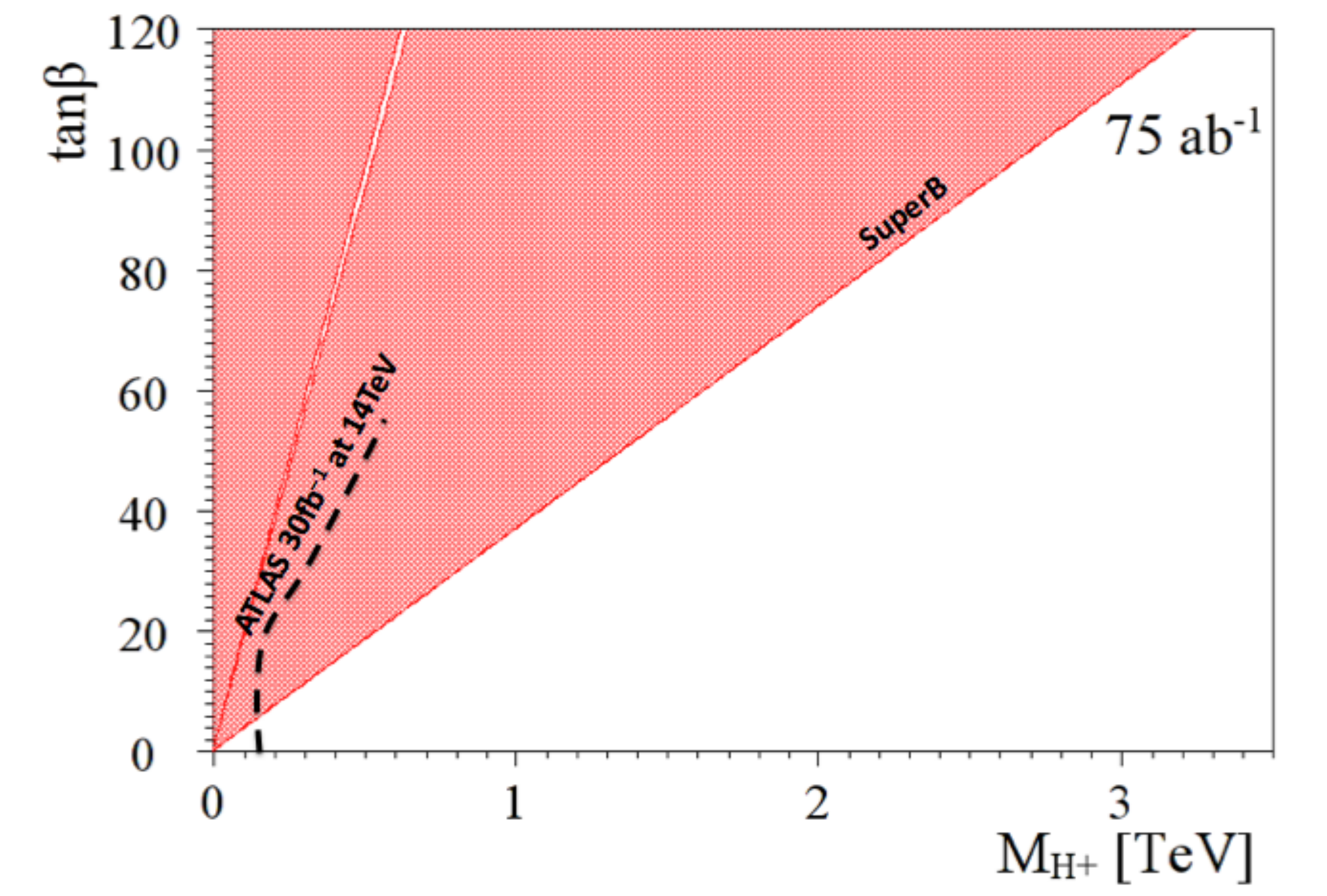}
  \caption{\label{fig:btaunu}
The constraint on the mass of a charged Higgs boson as a function of $\tan\beta$ in a
2HDM-II (95\% C.L.). The constraint anticipated from the LHC with 30\invfb of data
collected at a center of mass
energy of 14 TeV is also shown for comparison.  The \superb constraint is dominated
by $B\to \tau \nu$ up to luminosities $\sim 30\invab$, and the $B\to \mu \nu$
contribution dominates above this.  The ATLAS constraint is taken from arXiv:0901.0512.}
  \end{center}
\end{figure}

The decay mode $B^+\to\tau^+\nu_\tau$ is sensitive to NP scenarios
that include a charged Higgs boson, such as SUSY and two-Higgs doublet
models. Currently, the branching fraction for this decay measured at
\babar and Belle, while current compatible with SM expectations, has
at times been in tension with the SM value obtained from global fits
to CKM quantities~\cite{Bona:2009cj,Charles:2011va}. The lower bound
on the branching fraction provides the most stringent constraint on a
charged Higgs (corresponding to the SM $W^\pm$ and non-SM $H^\pm$
amplitudes destructively interfering).  In fact, the technique of
hadronic or semileptonic tagging, where the non-signal $B$ meson in
the event is fully reconstructed, allows the search on the signal side
of a single charged track and missing energy. With this method,
\superb will measure ${\cal B}(B^+\to\tau^+\nu_\tau)$ with a precision
of approximately 3\%.  The related channel $B^+\to\mu^+\nu_\mu$, will
be measured to about 5\% at \superb, and will increase the overall
sensitivity to the presence of charged Higgs bosons. Belle have
recently updated their $B^+\to\tau^+\nu_\tau$ measurement and obtain a
result compatible with the SM~\cite{BelleICHEP2012TauNu}, however the
corresponding \babar update remains in tension with the
SM~\cite{Lees:2012ju}, prompting the need for further experimental
investigation.  Figure~\ref{fig:btaunu} shows the constraint on the
charged Higgs mass vs $\tan\beta$ plane that would be obtained from a
\superb measurement of $B\to \tau \nu$ and $B\to \mu \nu$ consistent
with the SM.  A similar constraint would be obtained in the MSSM.

Semileptonic $B$ decays with a $\tau$ lepton in the final state are
sensitive to NP effects, in particular those arising from the presence
of a charged Higgs boson. The \babar experiment recently reported an
excess of $B\to D^{(*)}\tau\nu$ decays with respect to SM
expectations, with a combined statistical significance of
3.4$\sigma$~\cite{Lees:2012xj}. Interestingly, the results seem to
exclude the simplest NP model with a charged Higgs, the Type II Two
Higgs Doublet Model. More precise measurements of these channels are
needed to determine if the discrepancy is more than a statistical
fluctuation. The large data sample of \superb will lead to greatly
reduced uncertainties on these branching fractions and the
determination of the $q^2$-dependence of the decay rate, an important
measurement for unraveling the nature of possible NP effects.

\begin{figure}[!ht]
  \begin{center}
  \includegraphics[width=0.50\textwidth]{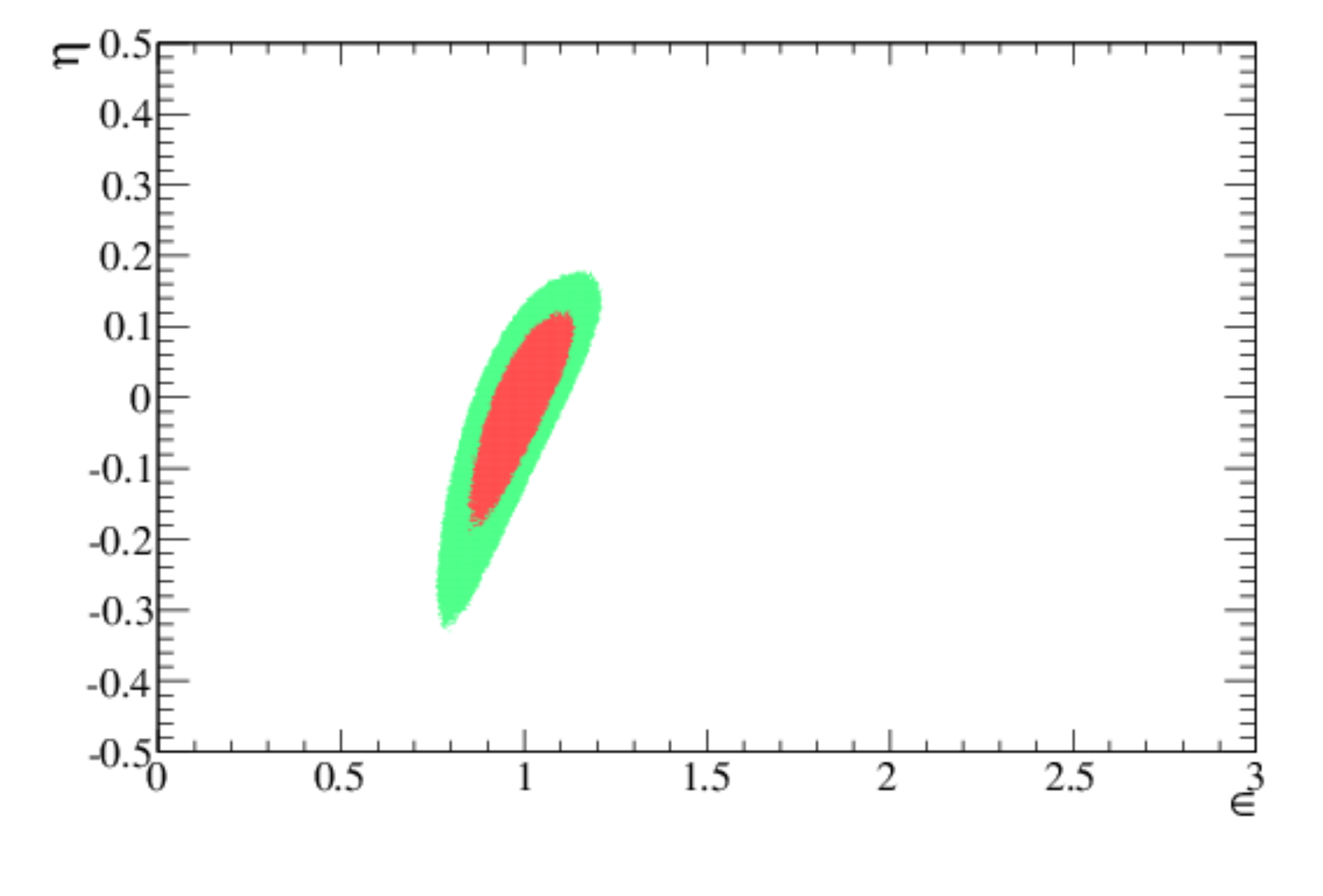}
  \caption{\label{fig:bknunu} Expected constraints on the model-independent parameters
          $\varepsilon$ and $\eta$ describing NP contributions to
          $B\to K^{(*)}\nu\nub$~\cite{Altmannshofer:2009ma} from \superb measurements.  
       }
  \end{center}
\end{figure}

The decays $\B\to K^{(*)}\nu\overline{\nu}$ are interesting probes of
NP, since they allow one to study $Z$ penguin and other electroweak
penguin effects in the absence of other operators that can dominate in
$b\to s\ell^+\ell^-$ decays. Here again hadronic and semileptonic
tagging will be used to isolate the non-signal $B$ meson in the event.
The signal signature then becomes a single kaon (or $K^*$) candidate
accompanied by a momentum imbalance, and nothing else. \superb will
maximize sensitivity to NP by independently measuring the $\B\to
K\nu\overline\nu$ and $\B\to K^{*}\nu\overline\nu$ branching
fractions, along with $F_L$, the $K^*$ longitudinal polarization
fraction. Furthermore, it may be possible to measure the fully
inclusive mode $\B\to X_s\nu\overline\nu$ at \superb, increasing the
sensitivity to NP effects in $Z$ penguin processes. The expected
constraints on the model-independent NP parameters $\varepsilon$ and
$\eta$ introduced in~\cite{Altmannshofer:2009ma} are shown in
Figure~\ref{fig:bknunu}.

One fruitful area to look for NP is in $b\to s\ell^+\ell^-$
transitions, which occur in the SM via electroweak penguin loops and
box diagrams. This transition represents a rather broad field of study
since it comprises several different exclusive decays of charged and
neutral $B$ mesons, along with the fully inclusive channel, usually
denoted $B\to X_s\ell^+\ell^-$. \superb is expected to accumulate
$10-15K$ events in each of the exclusive channels, including final
states with an $e^+e^-$ pair. The ratio $R^{(*)}_\mu = B(B\to
K^{(*)}\mu^+\mu^-)/B(B\to K^{(*)}e^+e^-)$, an important observable
sensitive to neutral Higgs particles in SUSY models, will only be
measured precisely at future Super Flavour Factories.  The event
sample accumulated at \superb will complement the current studies
being performed on $B^0 \to K^{*0}\mu^+\mu^-$ at
LHC~\cite{Aaij:2011aa}.  Additionally, \superb can make precise
measurements of rates, asymmetries and angular distributions of the
inclusive channel $B\to X_s\ell^+\ell^-$, exploiting the possibility
of performing hadronic and semileptonic tagging in the $e^+e^-$
environment. The inclusive channel is of interest because theoretical
calculations of inclusive observables generally have lower
uncertainties compared to the corresponding exclusive channels.

The study of $b \to s\gamma$ transitions has been one of the
cornerstones of the $B$ physics programmes at \babar and Belle. The
combination of precise measurements of the inclusive branching
fraction along with its precise SM calculation, both at the level of
about 7\%, leads to strong constraints on Flavour Changing Neutral
Currents in NP models. \superb can measure this branching fraction to
a precision of about 3\%, increasing the constraints imposed by this
transition.  There is no consensus whether the theoretical uncertainty
could match this precision or not.  Additionally, the high statistics
of the \superb data sample will permit the measurement of the
Cabibbo-suppressed $\B \to X_d\gamma$ channel to good precision.

\subsection{Rare $D$ decays}
\label{sec:rareD}

For a long time the HEP community has worked on $K \to \pi \nu \nub$
transitions and plans new projects to study this mode in greater
detail.  Here SM dynamics are greatly suppressed, and one can deal
with the impact of long distance (LD) dynamics by using correlations
with semileptonic decays.  This situation is different for $D \to X_u
\nu \nub$: short distance dynamics are truly tiny with BR$_{\rm SD}(D
\to X_u \nu \nub) \sim O(10^{-15})$, and impact of LD contributions on
inclusive final states is negligible in the SM.  This rare charm decay
mode is a good place to look for signs of NP which could enhance the
branching fraction by many orders of magnitude above the SM
expectation.  For general models of NP like Little Higgs with T parity
(LHT) or non-minimal charged Higgs models, one can have enhancements
up to $10^4$.  More exotic models can lead to larger enhancements.
This decay can be studied using fully reconstructed hadronically
tagged $D$ events at the $\psi(3770)$.  Other advantages of studying
rare charm decays at \superb include final states with $\pi$, $\eta$,
$\eta ^{\prime}$ and $2\pi$ which can contribute around 10--20\% for
the inclusive width. This is sufficient to probe enhancements from NP.
Direct and indirect \CP violation can be compared using $D^{\pm}
\to \pi^{\pm} \nu \nub$ and $D^0/\Dbar^0 \to \pi^0 \nu \nub$.  At the
$\psi (3770)$ one can use quantum correlations between $D\Dbar$ pairs,
e.g.  at threshold one can analyze $D_L \to (\pi^0, \eta, \eta
^{\prime}) \nu \nub$ due to correlation with $D \to K^+K^-/\pi^+\pi^-$
to probe direct vs indirect \CP violation in the case of large NP
manifestations.

Another rare decay \D that can be used to obtain an insight into
physics beyond the SM is $D\to \gamma\gamma$.  This channel is
dominated by long distance transitions, however it is possible that NP
could enhance the rate well above the SM expectation of a few $\times
10^{-8}$.  \superb is expected to be able to achieve a sensitivity
that will enable a rudimentary measurement of this mode at the SM
rate.  If NP is not manifest, then this rate will provide a stringent
constraint on the LD contribution to the decay $D\to \mu^+\mu^-$,
which itself is a mode very sensitive to NP.  The $D\to \mu^+\mu^-$
amplitude is dominated by the LD component, and so in order to
interpret if any measured rate is compatible with the SM or not, one
has to have a good understanding of $D\to \gamma\gamma$.

It is also possible to probe NP using the set of decays $D\to h
\ell^+\ell^-$, where $h=\pi, \rho$.  Even though the total rate is
dominated by LD contributions, the shape of the $q^2$ distribution of
the di-lepton pair is sensitive to NP, and in particular one may be
able to learn about the underlying dynamics of these final states by
comparing resonant to off-resonant features within the di-lepton
system.

Finally one can search for light dark matter particles by studying $D$
decays to invisible, or invisible $+\gamma$ final states.
Measurements made at the $\psi(3770)$ can complement the corresponding
studies of rare $B_{d,s}$ mesons performed at the \FourS and \FiveS.

\subsection{CKM matrix and unitarity triangle}
\label{sec:ckmsym:elements}

Broadly speaking improved determinations of CKM matrix elements will
lead to a better understanding of the SM, and in particular more
precise tests of unitarity.  It should be noted however that some of
the rare decay searches that are sensitive probes of NP have
significant theoretical uncertainties arising from the lack of
precision CKM parameter determinations.  Hence not only NP could show
up in precision CKM matrix element measurements, but these
measurements also play a significant role in the search for NP
elsewhere.  \superb can improve the determination of four of the CKM
matrix elements via measurements of \Vub, \Vcb, \Vus and \Vcd as
briefly summarized below.

At the end of the \B Factory era there is still tension between
inclusive and exclusive measurements of \Vub and \Vcb.  In
anticipation of larger data samples theorists have started working
with experimentalists to understand how one can make measurements in
the future that are both experimentally and theoretically more robust.
Several promising ideas have been put forward that will require data
samples comparable to that achievable at \superb.  Unitarity tests are
just one of the reasons why it is important to improve our
understanding of these parameters.  Precision measurements of these
quantities will be beneficial to the Lattice community and enable a
number of rare decay searches for NP.

The measurement of \Vus is currently dominated by the interpretation
of semi-leptonic kaon decays.  However the limiting factor in those
measurements comes from theory and unless significant progress is made
on that front there will be essentially no improvement on the
precision of \Vus from more precise kaon measurements.  If one wishes
to make a significant improvement in the measurement of \Vus one has
to resort to the use of $\tau$ leptons.  \superb with about 75 times
the existing worlds sample of $\tau$ lepton pairs will be able to make
a significant improvement on the determination of this parameter.  A
high statistics run at charm threshold would also lead to an improved
constraint on \Vus from $\tau$ decays.

Data accumulated at the \psiprpr will enable one to determine \Vcd via
the study of semi-leptonic \D decays.  This CKM factor is currently
known to a precision of 4.8\%.  It is expected that BES III will be
able to improve the precision on this quantity before \superb starts
to take data.  However with an expected 100 times the data sample of
BES III, \superb will be able to perform a necessary cross check, and
improve our understanding of this CKM element.

It should be noted that the ratio of $|\vtd/\vts|$ can be determined
using radiative $B$ decays via $b\to d$ and $b\to s$ transitions.
Theoretical uncertainties that would otherwise limit the
interpretation of these results cancel in the ratio of rates.

\superb can improve the measurements of the unitarity triangle angles
$\alpha$, $\beta$ and $\gamma$ to expected precisions of $1^\circ$,
$0.2^\circ$, and $1^\circ$, respectively.  These estimates are for the
most sensitive modes and are sufficient to perform a percent level
determination of the apex of the unitarity triangle
($\sigma_{\etabar}/\etabar \sim 1\%$, $\sigma_{\rhobar}/\rhobar \sim
3\%$) using only angles, and will provide a precision input to global
fits.  The primary measurements of $\gamma$ and $\beta$ are expected
to be theoretically clean in the context of the full \superb data
sample.

Current theory results in an uncertainty on $\alpha$ of about
$1^\circ$ from SU(2) symmetry breaking.  There are four sets of modes
that will be used to extract this angle: $B\to \pi\pi$, $B\to
\rho\rho$, $B\to \pi^+\pi^-\pi^0$, and $B\to a_1\pi$, where SU(3)
breaking affects the latter measurement.  Consistency between modes
will allow one to constrain NP entering through loop contributions, as
any large difference in the measured central values would not be
consistent with the SM.  Theoretical uncertainties on $\alpha$
determination will be comparable in magnitude to the experimental
precision attainable at \superb.

Current measurements of $\gamma$ from the \B Factories and LHCb
highlight the (well known) need for diversity in channels used to
measure $\gamma$. Dalitz methods can access the weak phase directly,
however the most popular methods implemented (ADS and GLW) rely on
reconstructing quasi-two-body $D^{(*)}K^{(*)}$ final states whereby
the weak phase is measured as a term multiplied by a Cabibbo
suppression factor $r_B$.  This number is small ($1-2\%$) and poorly
determined, and so the constraint on $\gamma$ is highly sensitive to
the extracted value of this nuisance parameter.  For that reason we
expect that it will be necessary to continue to average together
different sets of constraints on $\gamma$ in order to obtain a robust
measurement for the foreseeable future.  With the full data sample
from \superb we expect to be able to measure $\gamma$ to a precision
of $1^\circ$ in a given channel.  The GGSZ (Dalitz) method for
$\gamma$ was expected to have a significant systematic uncertainty
arising from limited knowledge of the strong phase difference as a
function of position in the $D\to \KS h^+h^-$ Dalitz plot.  The
\superb run at charm threshold is expected to provide 100 times the
data available from BES III, which can be used to effectively remove
this uncertainty and enable the continued use of this method.  The
GGSZ method currently dominates our knowledge of $\gamma$, so it is
important that one retain this powerful approach in the determination
of the angles.  Also the assumption that \CP violation in charm decays
is negligible in the determination of $\gamma$ is something that
should be revisited in light of recent results from LHCb, CDF and
Belle.

\subsection{\CP violation in \B decays}
\label{sec:ckmsym:cpv}

It is important to compare tree- and loop-level processes which
measure the same unitarity triangle angles in the SM. NP, more likely
to affect loop-dominated processes, can then be revealed by a
difference in the extracted angle for different modes.

Typical examples are the loop-dominated $b\to s$ transitions.  These
are a set of modes that are able to measure $\beta_{eff}$ in the SM,
where $\beta_{eff}\sim \beta$, up to hadronic uncertainties from
higher order contributions.  By testing the consistency with $\beta$
from tree dominated charmonium + \Kz modes, one can search for signs
of heavy particles with non-trivial quark flavour transitions in
loops.  The most promising ``{\em golden}'' modes are those that are
both experimentally precise and theoretically clean.  A full list of
the golden modes is given in Ref.~\cite{O'Leary:2010af}, and the
highlights are (for $b\to s$ penguin transitions) $B\to \eta^\prime
\Kz$ ($\sigma_S =0.007$), $B\to \phi \Kz$ ($\sigma_S =0.02$), and (for
$b\to d$ transitions) $B\to J/\psi \piz$ ($\sigma_S =0.02$).  The
latter mode is used to bound theoretical uncertainties in the tree
dominated $B\to J/\psi \Kz$ channel.

A set of \CP violating observables that are central to the $B$ physics
programme at \superb are \CP asymmetries in radiative decays. These
include $a_{CP}(B\to X_s\gamma)$ and $a_{CP}(B\to X_{s+d}\gamma)$
which are predicted to be small in the SM. While the former seems
theoretically limited, the latter has quite small theoretical
uncertainty~\cite{O'Leary:2010af}.  The experimental error on
$a_{CP}(B\to X_{s+d}\gamma)$ at \superb will be about 2\%.

Furthermore, a time-dependent analysis of the \CP asymmetry in the
exclusive mode $B^0\to K_s^0 \pi^0\gamma$ will be very sensitive to
the presence of right-handed currents arising from NP.  For this
decay, we estimate an uncertainty of 0.03 on the \CP violation
parameter $C$.

\subsection{\CP violation in \D decays}

The recent evidence for direct \CP violation in charm decays using a
time-integrated asymmetry $\Delta A_{CP}$ is extremely
interesting\footnote{This is a difference in time-integrated rates of
neutral $D$ mesons decaying into $K^+K^-$ and $\pi^+\pi^-$ final
states.}.  It is clear from the response of the theoretical community
that one needs to perform a detailed analysis of hadronic uncertainty
before heralding this result as a sign of NP, since various
conclusions have been reached either in favor of NP or of being
compatible with the SM depending on how one treats the problem.

Shortly before the $\Delta A_{CP}$ result was announced a systematic
evaluation of time-dependent \CP asymmetries in charm was proposed
using data from \psiprpr, \FourS and at hadron
machines~\cite{Bevan:2011up}.  There are at least thirty five neutral
modes that provide interesting tests of the SM, and a number will only
be accessible at $e^+e^-$ machines, and will complement the work
underway at the LHC.  Unlike \B decays where golden channels can be
readily identified in time-dependent measurements, hadronic effects
cloud ones ability to interpret any small (percent level or less)
result given the data accessible today.  To resolve the issue of
hadronic uncertainties one needs to measure a multitude of final
states, and once again \superb can contribute to the endeavor.  An
important step forward would be to perform a time-dependent
measurement of a single mode.  Currently, in order to control of
systematic uncertainties at the LHC, a difference is taken between two
modes.  One presumes it will soon be possible to extract the TDCP
parameters from a single mode at LHCb without resorting to taking a
difference.  In the case of $\D\to \pi^+\pi^-$ (which is more
interesting in the context of the SM than the $KK$ final state also
used for $\Delta A_{CP}$ as it can be used to measure an angle of a
unitarity triangle) one needs to control penguin contributions, and an
Isospin analysis will result in an error of a few degrees.  It has
been suggested for example that a time-dependent analysis of $\D\to
\pi^+\pi^-$ and $\D\to \piz\piz$ channels can be used to resolve this
problem up to Isospin breaking from electroweak penguins.  This
residual contribution can be controlled using the direct \CP violation
measurement from $\D\to \pi^+\pi^0$ decays.\footnote{For this mode in
particular, the required precision of parts per mille is probably only
possible with data from charm threshold.}  It should also be noted
that if a relatively large non-zero value of $\Delta A$ remains, then
it is possible that TDCP parameters would be clearly non-SM when
measured. Many of these issues can be resolved by \superb, and it
would be possible to explore this new area in the full range of final
states.  It is also important to quantify \CP violation in charm
decays as the methods to measure $\gamma$ using \B decays assume this
is negligible, and one must verify this is the case for precision
measurements.  A run at the $\psi(3770)$ lasting a year is expected to
provide independent measurements of TDCP asymmetries with statistical
precisions comparable to a five year run at the \FourS for \superb or
using the full data sample expected at LHCb according to
Ref.~\cite{Bevan:2011up}.

It is also possible to measure direct \CP violation in charged $D$
decays, however as with \B and kaon manifestations of direct \CP
violation, detailed understanding of hadronic uncertainties and strong
phase differences will be required in order to interpret any results
in the context of the SM.  Four body final states such as $D\to
K^+K^-\pi^+\pi^-$ can be used to measure triple product asymmetries
related to $T$ odd, \CP violating phenomena.

It should be possible for the low energy machine to reach 4.005 GeV,
on doing so one could measure the \CP asymmetry in $e^+e^- \to D D^*
\to D \Dbar \gamma$.  It is possible to use the method proposed by
Bondar {\em et al.} to measure charm mixing parameters using
time-integrated observables at this energy~\cite{Bondar:2010qs}.

\subsection{Other symmetry tests}
\label{sec:ckmsym:other}

\CPT is conserved in relativistic quantum field theories, and
underpins the SM.  \superb will be able to test \CPT using neutral \B
and \D mesons created in collisions at the \FourS and \psiprpr,
respectively.  The most recent test of \CPT from \babar yields a
deviation from the SM expectation of $2.8\sigma$~\cite{Aubert:2007bp}.
While this is not significant it is important to repeat this
measurement to understand if this is an out lier, or if there is some
contribution from NP.

If \CPT is conserved, then given that \CP is violated one can infer
that \T must also be violated. \babar recently discovered \T violation
via a time-dependent asymmetry constructed from T conjugated pairs of
neutral \B meson decays~\cite{Lees:2012kn}.  This established the
existence of T violation using a self-consistent approach, without
invoking \CPT to relate \CP violation to \T violation.  This is a
significant step forward in the field, and should be studied in
greater detail. \superb will be able to significantly improve the
precision on this measurement, which can then be compared with
complementary tests of \CP violation and \CPT to test the SM.

\subsection{Charm mixing}
\label{sec:charmmixing}

Whatever the outcome of studies of direct and indirect \CPV, an open
question remains whether mixing-induced \CP violation can also be
observed.  An obvious sign of this phenomenon would be a mode
dependence of the mixing parameters $x_{D}$ or $y_D$, respectively
related to mass and decay widths for the underlying $D$ mass
eigenstates.  \CP violation in mixing itself would show up in
$D-\Dbar$ differences in mixing rates $|q/p|$ and in the phase of this
quantity.  We anticipate~\cite{O'Leary:2010af} that \superb should be
sensitive to a value for $|q/p|_D$ differing from 1.0 by a few
percent, or to a mixing phase $\arg (q/p)$ differing from zero by
$\sim 1^{\circ}$.  Furthermore, we expect that such results can also
be obtained in a variety of decay modes.  We also anticipate
measurements of $x_D$ and $y_D$ with precisions close to $\sim 1\times
10^{-4}$, an improvement of an order of magnitude from current world
averages.  This precision compares well with expectations from \lhcb
but the comparison will depend on the ultimate \lhcb trigger, whose
present design limits their yield of decays to the ``golden'' \CP
self-conjugate modes (primarily $\KS\pip\pim$ or $\KS\Kp\Km$).  The
\epem data should, however, cover a considerably wider range of modes
that will complement well the results from \lhcb.

The \superb estimates are derived from a robust extrapolation from charm mixing
parameters measured by \babar~\cite{O'Leary:2010af}.  Specifically, we assume that sample sizes at the $\Upsilon(4S)$ will
be larger, thereby improving statistical precision by a factor at least $\sim 12$
\footnote{Simulation studies indicate that the improved time-resolution in \superb could introduce further gains that we did not include in the estimates.}.
We expect most systematic effects to
scale with square root of luminosity, but we do note that the time-dependent Dalitz plot analyses of decays
to the golden modes, which are crucial, will be limited in precision by an irreducible systematic
uncertainty in the nature of the Dalitz plot decay model.  We estimate this limits precision in $x_D$ to
$7\times 10^{-4}$, only a factor 4 improvement over \babar.  Replacing the Dalitz models with
measurements of \DzDzb phases obtained from quantum correlations observed in $\psi(3770)$ data from
BES III, should improve the precision on $x_D (y_D)$ to $4 (2)\times 10^{-4}$.  This model
uncertainty would be virtually eliminated by the use of a $1~\invab$ \superb run at charm threshold
to reach precisions $\sim 2(1)\times 10^{-4}$.

\begin{figure}[tbp]
  \begin{center}
    \includegraphics[width=0.50\textwidth]{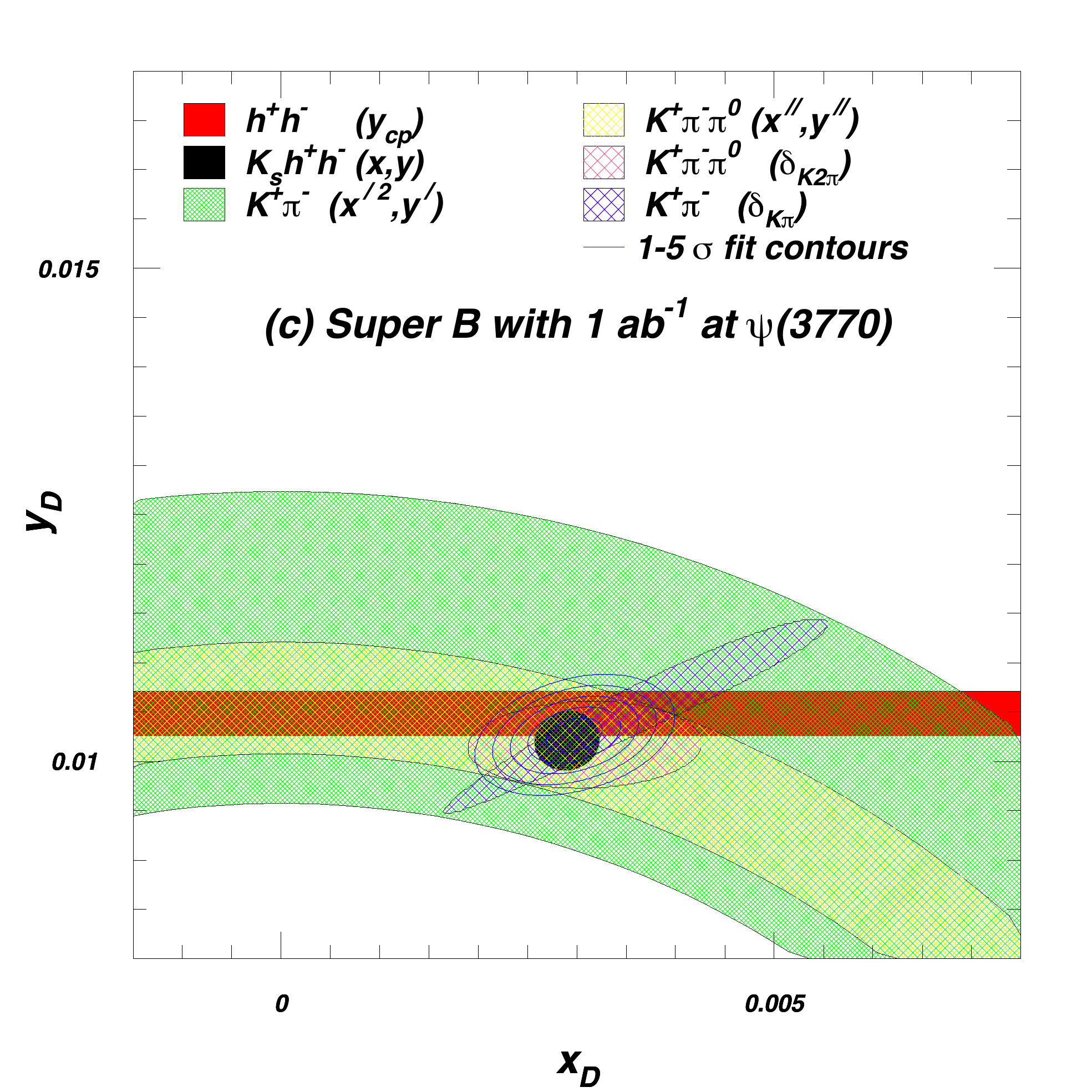}
   \caption{\label{fig:physics:charmmixing}
     The expected constraint from \superb on the charm mixing parameters $x$ and $y$ using the full 75\invab sample
     from the \FourS, and 1\invab from \psiprpr.
           }
  \end{center}
\end{figure}

The anticipated integrated luminosity at charm threshold from \superb
is expected to be 1\invab. Previous estimates of the sensitivity to
charm mixing have assumed half of this data sample in order to
determine the anticipated precision on $x$ and $y$.
Figure~\ref{fig:physics:charmmixing} shows the corresponding
constraint obtained when one uses the full 1\invab sample.

\subsection{\B physics at the \FiveS}
\label{sec:fives}

It is expected that \superb will accumulate several \invab of data at the \FiveS.  
The aim of the physics programme at this center of mass energy is to search for physics 
beyond the SM via rare decays and to perform precision tests of the SM.  An example of 
a decay channel that may be affected by NP is $B_s \to \gamma\gamma$.  Various NP scenarios
suggest correlations between this $B_s$ decay
and $B_d\to X_s \gamma$.  Hence, a branching fraction measurement of this decay is 
an important cross check of searches for NP performed using data collected at the \FourS.
A number of rare decays with neutral final states or missing energy can be 
studied in this environment.

The semi-leptonic asymmetry $a_{SL}$ in $B_s$ decays is of interest as this is sensitive to 
NP and can be correlated with flavor observables measured at the \FourS.  \superb
is expected to be able to measure $a_{SL}$ to a precision of 0.006 with 1\invab of data
at the \FiveS.


It will be possible to study $B_s$ decays inclusively using data from the \FiveS,
and one can also search for the $B_s\to \tau^+\tau^-$ channel using kinematics
of the final state to suppress background.

Finally, compared to $B_d$ and $B_u$ decays, knowledge of the decay modes of $B_s$
is limited. With a data sample of this size, \superb will be able to contribute to 
the general level of understanding of absolute branching fractions of $B_s$ decays,
which will enable theorists to improve their calculations and may reduce uncertainties
on normalization channels for hadronic experiments to use (e.g. a precise measurement of
the absolute value of the branching fraction of $B\to J/\psi \phi$ would be of particular 
interest as a normalization mode for LHC measurements of the rate of $B_s \to \mu^+\mu^-$).

\section{\texorpdfstring{$\tau$ physics at \superb}{tau physics at \superb}}\label{sec:tau}

Measurements on $\tau$ leptons at the intensity frontier provide a
powerful tool to search for NP effects in a way that is competitive
and complementary with respect to LHC and other existing or planned
experiments like MEG and Mu2e. Lepton Flavor violation (LFV) in $\tau$
decay is a measurable unambiguous signal for many NP models, and the
expected sensitivities of the planned Super Flavor Factories is
unrivaled. The large and clean dataset of $\epem \to \tau^+\tau^-$
events will also permit additional NP searches based on precision
measurements of \CP{}V in $\tau$ decays, and of the $\tau$ $g{-}2$ and
EDM form factors, with impressive improvements on the present
experimental accuracy. Furthermore, \superb offers the unique features
of the largest design luminosity and of an 80\% polarized electron
beam, which provides an extra handle to identify sources of NP in LFV
and to improve the experimental precision on the $\tau$ $g{-}2$ and
EDM form factors.

In addition to being a powerful probe for NP, a high
luminosity $\tau$ factory at the \FourS peak represents an ideal
facility for most $\tau$ physics measurements because the data
analysis is facilitated from the fact that events consist in a
correlated $\tau^+\tau^-$ pair in a well defined initial
state with no superposition of background.
And in the \superb project, the information of the initial
state is further refined in a useful way using a polarized electron
beam. In these conditions, a large number of precision measurements
are possible, such as $|V_{us}|$ from $\tau$ decays into strange final
states, $\alpha_s$, the $(g{-}2)_\mu$ hadronic contribution, and
leptonic charged weak current couplings.  

While the statistics of a 1\invab run at the $\psi(3770)$ would 
be smaller than that obtained using a 75\invab run at the \FourS, 
there are several instances where data collected at the lower energy 
would have a distinct advantage.  This may become relevant should
one discover new physics, and wish to perform precise cross checks 
in a cleaner environment than would be possible at the \FourS, 
or perform a high luminosity run at low energy, or at a high
luminosity $\tau-$charm factory.

\subsection{\texorpdfstring{%
    Lepton flavor violation in $\tau$ decay}{%
    Lepton flavor violation in tau decay}\label{sec:tau:lfv}}

The SM with neutrino mixing includes Lepton Flavor Violation (LFV)
in $\tau$ decays, although at rates too small to be detected. On the
other hand, measurable LFV rates can naturally arise for a variety of
NP models, for instance MSSM~\cite{Antusch:2006vw,Arganda:2005ji},
MSSM with flavor symmetries~\cite{Calibbi:2008qt,Calibbi:2009ja}, Non
Universal Higgs Masses (NUHM)
SUSY~\cite{Paradisi:2005tk,Arganda:2008jj,Herrero:2009tm}, MSSM with
R-parity violation (RPV)~\cite{Dreiner:2006gu},
LHT~\cite{Blanke:2007db}, $Z^{'}$
models~\cite{Raidal:2008jk}).

In many cases, the most sensitive mode is \taumg, and some important
complementary modes are $\tau\to e\gamma$, $\tau\to\ell\ell\ell$,
$\tau\to\mu\eta$. The planned Super Flavor Factories, \superb and
\belletwo, have no significant competition in these NP searches, and
\superb has the advantage of a larger design luminosity and electron
beam polarization, which can be used to distinguish between left and
right handed NP currents.

LFV in muon decay is related to $\tau$ LFV in a way that is determined
by the specific details of a NP model, on which there are no
experimental clues so far. Therefore, LFV searches on $\mu\to e
\gamma$ and $\mu\to e$ conversion in material are complementary in
nature with the $\tau$ ones, and in order to understand NP in
the long term one has to understand
all three sets of measurements in detail. While for some specific
NP models, with specific choices of parameters, the experimental sensitivity for LFV is
larger for muon decays than for $\tau$ decays, in general
searches for LFV $\tau$ decays remain important in understanding
the underlying dynamics of NP models.

\begin{figure*}[tb]
\includegraphics[width=\linewidth]{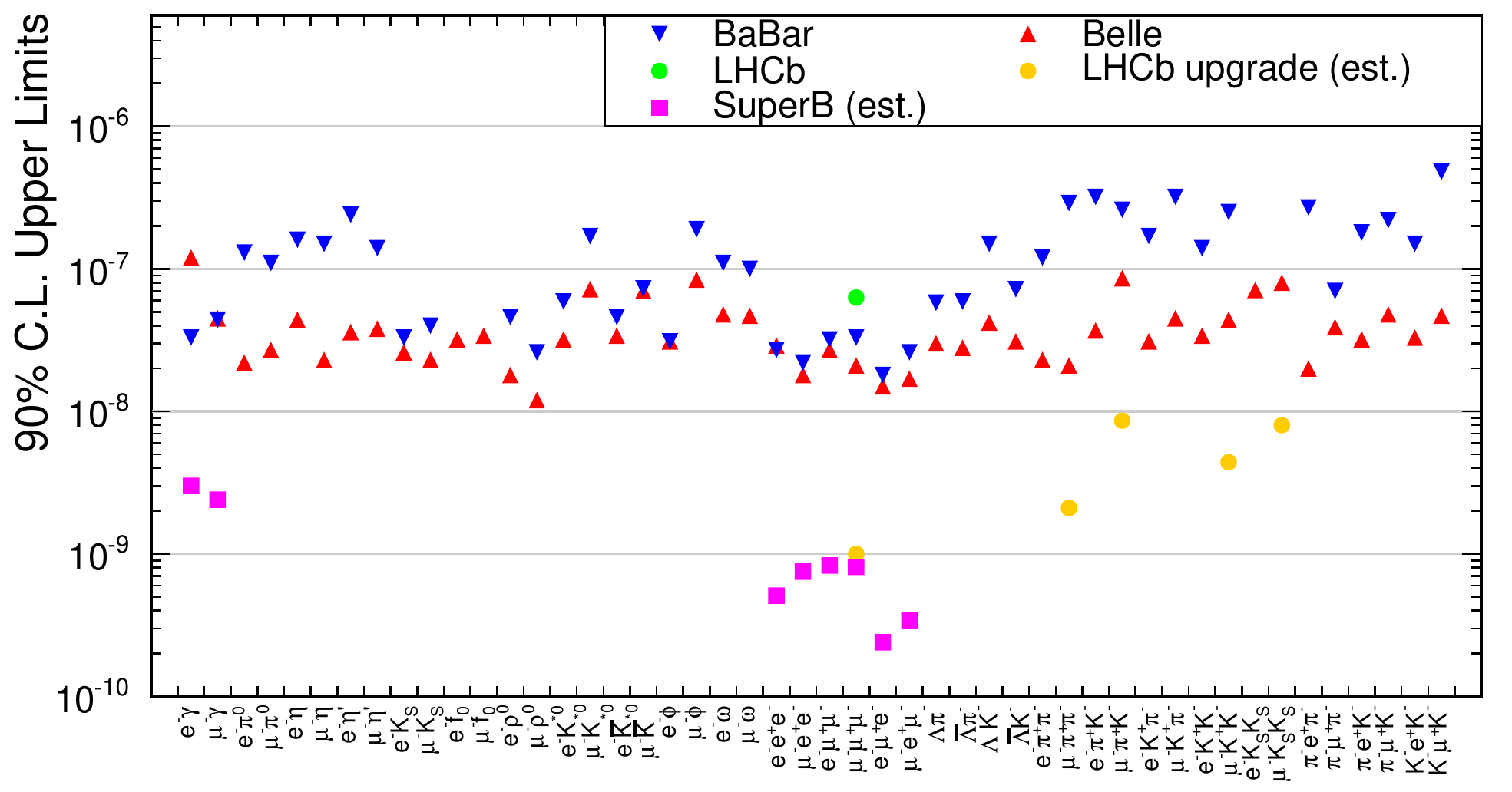}
\caption{\label{fig:tau:lfvsumm}
  Summary of present upper limits on $\tau$ LFV compared with expected
  upper limits from \superb and the LHCb upgrade. The latter estimates are
  the order-of-magnitude estimates listed in the letter of
  intent~\cite{lhcbupgradeloi}. A naive LHCb-upgrade sensitivity estimate for $\tau\to 3\mu$ based on
  the 2012 LHCb 90\% expected upper limit~\cite{LHCb-CONF-2012-015}
  extrapolated from 1\invfb to 50\invfb with the square root of the
  luminosity gives $1.5\EE{-8}$, to be compared with $1.0\EE{-9}$ in the
  plot.  In some cases, the \babar results are hidden by very similar results
  by Belle, which are superposed in the plot.}
\end{figure*}

The \superb design integrated luminosity of 75\invab, in combination
with moderate improvements in the detector efficiency and resolution,
provides a LFV sensitivity 10 times better than the \BFs\ for
background-limited $\tau$ decay modes like \taumg and up to 100 times
better for the cleanest $\tau\to\ell\ell\ell$ searches.
Figure~\ref{fig:tau:lfvsumm} summarizes the expected LFV
sensitivities from \superb and the LHCb upgrade~\cite{lhcbupgradeloi}.

If one considers the search for $\tau \to \mu\gamma$: for data collected at the \FourS,
this has an irreducible background from events with initial state radiation.
Data collected at low energy (at or below about 4.0\gev) can be used to isolate
a clean (background free) sample of events.  An extended run of several \invab at the $\psi(3770)$ 
would be competitive with limits achievable at the \FourS using \superb.
Such a run would be justified in the event of a discovery at the \FourS.

\subsection{\texorpdfstring{%
    \CP violation in $\tau$ decay}{%
    CP violation in tau decay}\label{sec:tau:cpv}}

The SM predicts vanishingly small \CP violation effects in $\tau$
decays, for instance, the $\CP$ asymmetry rate of $\tau^{\pm}\to
K^{\pm}\pi^0\nu$ is estimated to be of order ${\cal
O}(10^{-12})$~\cite{Delepine:2005tw}. For the decay $\tau^{\pm}\to K_S
\pi^\pm \nu$ the SM predicts, with a 2\% relative precision, a \CP\
asymmetry of $3.3\EE{-3}$ due to the known \CP-violating phase of the
$\KzKzb$ mixing amplitude~\cite{Bigi:2005ts}.  Any observed deviation
from this clean, and precisely defined, picture would be a clear sign of
NP.  Specific NP models such as RPV
SUSY~\cite{Delepine:2007qg,Dreiner:2006gu} and non-SUSY
multi-Higgs
models~\cite{Datta:2006kd,Kiers:2008mv,Kimura:2008gh,Kimura:2009pm}
can accommodate \CP violation in $\tau$ decays up to $10^{-1}$ or the
present experimental limits~\cite{Bischofberger:2011pw}, while
satisfying other experimental constraints.  Examples of decay modes
with potentially sizable asymmetries are $\tau \to K \pi \nu_\tau$,
$\tau \to K \eta^{(\prime)} \nu_\tau$, and $\tau \to K \pi \pi
\nu_\tau$~\cite{Delepine:2007qg,Datta:2006kd,Kiers:2008mv,Kimura:2008gh,Kimura:2009pm}.

The Belle and \babar Collaborations have searched for \CP violation in
$\tau$ decay~\cite{Bischofberger:2011pw,BABAR:2011aa} with different
analyses on the same decay mode. \babar measured the \CP asymmetry in the
total branching ratio, which is affected by the $K_S$ \CP violation
amplitude and found a $2.8\sigma$ discrepancy with respect to the SM.
Belle measured an angular dependent \CP asymmetry where any detectable
effect would correspond to NP, with a sensitivity of order
$10^{-3}$. \superb can conceivably improve by an order of magnitude. As \CP
violation is an interference effect that is linear in the NP amplitude, a
$10^{-4}$ precision on a \CP asymmetry corresponds very roughly to measuring
the $\tau\to\mu\gamma$ branching fraction with $10^{-10}$
precision~\cite{Bigi:2008qr}.

\subsection{\texorpdfstring{%
    Measurement of the $\tau$ $g{-}2$ and EDM form factors}{%
    Measurement of the tau g-2 and EDM form factors}\label{sec:pol:edm}}

The anomalous magnetic and the electric dipole moments of the $\tau$
are fundamental properties of the heaviest known lepton and are
predicted with extreme precision within the SM both at zero and
non-zero transferred momentum. They can be measured at
transferred momenta of the order of the $\tau$ mass from the
angular distributions of the $\tau$ decay products in
$\epem\to\tau^+\tau^-$ events. For this purpose having polarized
beams represents a clear
advantage~\cite{GonzalezSprinberg:2007qj,Bernabeu:2006wf}.

Within the SM, the $\tau$ $g{-}2$ has been computed with very good
precision, $\Delta a_\tau = \Delta (g{-}2)_\tau/2 =
5\EE{-8}$~\cite{Eidelman:2007sb}; its size is mainly determined by the
first perturbative term which is of order $\alpha$ and is the
same for all leptons.

If one assumes that the present muon $g{-}2$ discrepancy is caused by
NP corresponding to the MSSM with $\tan\beta \gsim 10$, one
expects a much larger NP effect on the $\tau$ $g{-}2$, at the level of
$10^{-6}$.

There is limited experimental information on the $\tau$ $g{-}2$, the
OPAL collaboration obtained a limit on the size of $F_2(q)$ [where
$F_2(0) = (g{-}2)/2$] averaging transferred momenta between
approximately the $\tau$ and the $Z$ mass at the level of
$7\%$~\cite{Ackerstaff:1998mt}. For \superb with 75\invab of data and
a 100\% polarized electron beam, the real and imaginary part of the
$g{-}2$ form factor $F_2$ can be measured assuming perfect reconstruction
with a resolution in the range $[0.75 - 1.7]\EE{-6}$~\cite{Bernabeu:2007rr}.
When the design 80\% beam polarization and realistic experimental
reconstruction effects are included, the resolution is be estimated
at about $2.4\EE{-6}$, which is comparable to the size of possible
NP MSSM contributions compatible with the muon $g{-}2$ discrepancy.

The $\tau$ EDM is expected to be vanishing small, and the existing
upper limit on the electron EDM imposes severe constraints also on
contributions from NP to about $d_{\tau}\lsim 10^{-22} e\,{\rm cm}$.
Experimentally, from a Belle analysis on $29.5\invfb$ of
data~\cite{Inami:2002ah} (unpolarized beams), we know that
the experimental resolution on the real and imaginary parts of the
$\tau$ EDM is between $0.9\EE{-17}\,e\,\text{cm}$ and
$1.7\EE{-17}\,e\,\text{cm}$, including systematic effects.

Recent studies have provided an estimate of the \superb upper limit
sensitivity for the real part of the $\tau$ EDM
$\left|\text{Re}\{d_\tau^\gamma\}\right| \le 7.2\EE{-20}\,e\,\text{cm}$
with 75\invab~\cite{GonzalezSprinberg:2007qj}. The result assumes a $100\%$
polarized electron beam colliding with unpolarized positrons at the
\FourS resonance. Uncertainty on the polarization is neglected, and a
perfect reconstruction of the decay $\tau \to \pi\nu$ is assumed. A
refined sensitivity estimate has been obtained assuming an electron
beam with a polarization of $80\% \pm 1\%$ and realistic experimental
reconstruction efficiency and uncertainties at $\approx 10\EE{-20}
e\,\text{cm}$.  Such a measurement would constitute a remarkable experimental
improvement over current constraints.

\def\Babar{{\it B}{\it\footnotesize A}{\it B}{\it \footnotesize A}{\it R}}
\def\SM{SM}

\section{\texorpdfstring{\superb\ Neutral Current Electroweak Physics Programme}{\superb\ Neutral Current Electroweak Physics Programme}}
\label{sec:precisionew}

\begin{figure}[tbp]
  \begin{center}
  \includegraphics[width=0.50\textwidth]{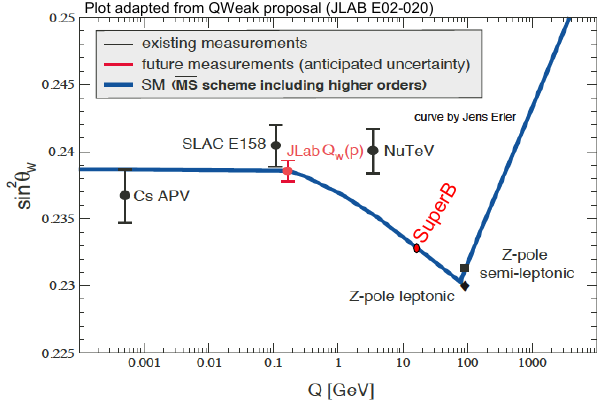}
  \caption{\label{fig:sin2thetaW} Determination of $\sin^2\theta_W$ at present and future experimental facilities.
                                  }
  \end{center}
\end{figure}

With its ultra-high luminosity, polarized beam, and the ability to run at both the $b$-quark and $c$-quark thresholds,
\superb\ is a unique and versatile facility and no other machine with all of these capabilities is planned.
In particular, the polarization of the electron beam enables \superb\ to measure the weak neutral current vector coupling
 constants of the $b$-quark, $c$-quark and muon at significantly higher precision than any previous experiment. The precision of
 the vector coupling to the tau and electron will be measured with a precision comparable to that attained at SLC and LEP.
 Within the framework of the SM these measurements of $g_V^f$ can be used to determine the weak mixing angle, $\theta_W$, 
through the relation: $g_V^f = T_3^f - 2Q_f \sin^2\theta_W$, where $T_3^f$ is the $3^{rd}$ component of weak isospin of fermion $f$,
 $Q_f$ is its electric charge in units of electron charge and higher order corrections are ignored here for simplicity.
Figure~\ref{fig:sin2thetaW} shows the determinations of $\sin^2\theta_W$ at present and future experimental
facilities including \superb.  If one were to run with a polarized electron beam 
at the $\psi(3770)$, then a similar programme
could be exploited using $c$-quark and muon pairs.  An interesting level of precision would be achievable
in order to test the SM at an energy similar to the NuTeV result.

\superb\ determines $g_V^f$ by measuring the left-right asymmetry, $A_{LR}^f$, for each identified final-state fermion-pair in the process
$e^+e^- \rightarrow f \overline{f}$.
\begin{equation}
 A_{LR}^f = \frac{ \sigma_L - \sigma_R}{\sigma_L + \sigma_R} = \frac{s G_F }{\sqrt{2} \pi \alpha Q_f} g_A^e g_V^f \langle Pol \rangle
\end{equation}
where $g_A^e$ is the neutral current axial coupling of the electron, $g_A^e = T_3^e$, $G_F$ is the Fermi coupling constant,
$s$ is the square of the center-of-mass energy,
and 
\begin{equation}
\langle Pol \rangle =0.5 \left[ \left(N_{asym}\right)_{RPB} -
 \left(N_{asym}\right)_{LPB} \right],
\end{equation}
is the average beam polarization, where $N_{asym}$ is the asymmetry between $N_{eR}$ and $N_{eL}$,
and $LPB$ and $RPB$ indicate left or right polarized beams.

These asymmetries arise from $\gamma-Z$ interference and although the
 SM asymmetries are small ($-3\times 10^{-4}$ for the leptons, $-3\times 10^{-3}$ for charm and
 $-1.3\%$ for the $b$-quarks), the unprecedented precision is possible because of
 the high luminosity of \superb\ together with a $70\%$ beam polarization measured to $\pm 0.5\%$.
 To achieve the required precision all detector-related systematic errors are made to cancel by flipping
 the polarization from $R$ to $L$ in a random, but known, pattern through a run.
 $\langle Pol \rangle$ is measured with both a Compton Polarimeter, which has an uncertainty at the interaction point of $<1\%$ and
 by measuring the forward-backward polarization asymmetry of the $\tau$-pairs using the kinematic distributions of the
 decay products of the $\tau$. 
The latter can be used to determine $\langle Pol \rangle$ to $0.5\%$ at the interaction point in a manner entirely independent
of the Compton Polarimeter.

The $3 \sigma$ discrepancy between the LEP measurements of the
right-handed $b$-quark couplings to the $Z$ will be experimentally
probed with higher precision at \superb.  Also, measurements with the
projected precision will enable \superb\ to probe parity violation
induced by the exchange of heavy particles such as $Z$-boson or
hypothetical TeV-scale $Z^\prime$ boson(s). If such bosons only couple
to leptons they will not be produced at the LHC.  Moreover, the
\superb\ machine will have a unique possibility to probe parity
violation in the lepton sector mediated by light and very weakly
coupled particles often referred to as ``dark forces''
(Sec.~\ref{sec:spectroscopy}). Recently, such forces have been
entertained as a possible connecting link between normal and dark
matter \cite{Pospelov:2008jd,ArkaniHamed:2008qn}. The enhancement of
parity violation in the muon sector has been an automatic consequence
of some models \cite{Batell:2011qq} that aim at explaining the
unexpected result for the recent Lamb shift measurement in muonic
hydrogen \cite{Pohl:2010zza}. The left-right asymmetry of the $e^-e^+
\to \mu^-\mu^+$ in such models is expected to be enhanced at
low-to-intermediate energies, and the \superb\ facility may provide a
conclusive test of such models, as well as impose new constraints on
parity-violating dark sector.

\section{Exotic Spectroscopy in \superb}
\label{sec:spectroscopy}

Although the SM is well-established, QCD, the fundamental theory of
strong interactions, provides a quantitative comprehension only of
phenomena at very high energy scales, where perturbation theory is
effective due to asymptotic freedom.  The description of hadron
dynamics below the QCD dimensional transmutation scale is therefore
far from being under full theoretical control.

Systems that include heavy quark-antiquark pairs (quarkonia) are a
unique and, in fact, ideal laboratories for probing both the high
energy regimes of QCD, where an expansion in terms of the coupling
constant is possible, and the low energy regimes, where
non-perturbative effects dominate.  For this reason, quarkonia have
been studied for decades in great detail.  The detailed level of
understanding of the quarkonia mass spectra is such that a particle
mimicking quarkonium properties, but not fitting any quarkonium level,
is most likely to be considered to be of a different nature.

In particular, in the past few years the $B$ Factories and the
Tevatron have provided evidence for a large number of states that do
not admit the conventional mesonic interpretation and that instead
could be made of a larger number of constituents. While this
possibility has been considered since the beginning of the quark
model~\cite{ GellMann:1964nj}, the actual identification of such
states would represent a major revolution in our understanding of
elementary particles. It would also imply the existence of a large
number of additional states that have not yet been observed.

For an overview of the observed states and their interpretation we
refer to~\cite{Drenska:2010kg}, here we will discuss the \superb
potential and its interplay with the other planned experiments.

Regardless of the theoretical ansatz in which the observed states are interpreted, the existence of a new fundamental 
type of state implies the existence of a number of new states that is much 
larger than the set of states observed so far. For instance, as detailed in Ref.~\cite{Drenska:2010kg}, under the assumption of the  tetraquark nature of the observed new state there are at least 18 states predicted for each $J^{PC}$; similarly in the molecular model, the number of thresholds close to which a new state could be is extremely large.

The next generation of experiments ought therefore to make the leap from the discovery phase to the 
systematic investigation of the whole spectroscopy. As an example one should search for a 
charged state decaying into $J/\psi K$ and produced in $B$ decays in conjunction to a $\omega$. 
This search was not performed in \babar and Belle both because of lack of a certain theoretical 
model, but above all because of lack of statistics.

In this process of building the complete picture \superb plays a critical role because of the following characteristics:
\begin{itemize}
\item {\em large statistics}. The largest limitation of the current data sample is the low statistics. As a consequence, signals are still not well established, precision analyses for the determination of the quantum numbers is impossible and several states are likely not to have been observed because of a low branching fraction. In particular all final states with $D$ and $D_s$ mesons will be accessible for the first time at \superb.
\item {\em clean environment}. Albeit the high luminosity, that will slightly degrade the detector performance, the experiment 
is a multipurpose one, like \babar. It will therefore have detection capabilities for all neutral and charged particles. Detection efficiencies are not expected to vary significantly from the \babar ones.
\item {\em center-of-mass scan potential}. The cleanest way to study a particle is to produce it directly, and this is possible, for $J^{PC}=1^{--}$ states if the machine can adjust its energy to match the mass of the state. This is the case for \superb over a large range of energy, in the vicinity of the charm threshold and for energies above the $\Upsilon(4\rm S)$, thus allowing for direct studies on any $J^{PC}=1^{--}$  exotic state.
\end{itemize}

\superb\ is not the only next generation experiment capable of investigating heavy quark spectroscopy.

The LHCb experiment is starting to investigate its potential in the
field.  The larger number of mesons produced allows detailed studies
of the decay modes with final states made of charged particles. All
other modes are best investigated by $e^+ e^-$ machines.

The only other next generation experiments at an $e^+ e^-$ machine are
BES-III and Belle-II.  BES-III is a $\tau$-charm factory planning to
run below the energies of interest, at the
$\psi(3770)$~\cite{Asner:2008nq}, where they expect to collect
$5\invfb$ per year.  Concerning Belle-II, the Super-KEKB accelerator
is not designed to have the scan capabilities of \superb and would not
therefore be able to cover the corresponding physics.

A separate mention is deserved by the PANDA experiment at FAIR~\cite{Lutz:2009ff}, a proton-antiproton collider which could produce the exotic resonances at  threshold (i.e. $e^+e^- \to X,Y$). This innovative production mechanism allows for copious production without the hindrance of fragmentation products. 

The complementarity of the two experiments is guaranteed by the fact that the final states that can be 
studied by the two experiments  are different and that the PANDA experiment can more easily access the narrow states while \superb can study in detail larger states if the production mechanism is favorable. Furthermore, in case the center-of-mass-energy of \superb is changed to  the $Y(4260)$ mass, assuming a factor 10 loss in luminosity with respect to  running at the \FourS, the number of events produced in a few weeks would be equivalent to the PANDA dataset.
Finally, PANDA can only reach center-of-mass energies as high as 5 GeV and therefore has no access to bottomonium spectroscopy.

\section{Direct searches}
\label{sec:searches}

Bottomonium decays can also be very sensitive to new particles that might have escaped direct searches at lower luminosity and higher energy facilities because of small couplings with ordinary matter. Among such searches we consider here the quest for dark matter and dark forces.

Two promising reactions for searching {\bf {low-mass dark matter}}  ($\chi$) are the invisible 
and radiative decays of the $\Upsilon(1S)$ mesons, which can be measured in the mode
$\Upsilon(3S) \to \pi^+ \pi^- invisible$~\cite{McElrath:2005bp}. The current best sensitivity
to this process has been achieved by the \babar
experiment \cite{:2009qd};
however, this result is still an order of magnitude above
the SM prediction. 
The sensitivity is
limited by the amount of background that needs to
be subtracted, primarily due to undetected leptons
from $\Upsilon(1S) \to \ell^+ \ell^-$ decays.
Achieving a $3-5\sigma$ sensitivity to the SM 
will require active background tagging down to
5-10 degrees above the beam-line in both the forward
and backward directions.
Also the searches for bino-like neutralino in radiative $\Upsilon(1S) \to \gamma + invisible$ 
decays require extended detector coverage in the forward and
backward directions to reduce the dominant radiative QED backgrounds.

Concerning the search for {\bf {dark forces}}, recent cosmic ray
measurements of the electron and positron flux
\cite{atic,fermi,pamela} have spectra which are not well described by
galactic cosmic ray models and that lead to the introduction of a
class of theories containing a new ``dark force'' and a light, hidden
sector.  In this model, the observed signals are due to dark matter
particles with mass $\sim 400-800\gevcc$ annihilating into the gauge
boson force carrier with mass $\sim 1\gevcc$, dubbed the $A^\prime$ (a
``dark photon''), which subsequently decays to SM particles.  If the
$A^\prime$ mass is below twice the proton mass, decays to
$p\overline{p}$ are kinematically forbidden allowing only decays to
states like $\epem$, $\mu^+\mu^-$, and $\pi\pi$.

The dark sector couples to the SM through kinetic mixing with the
photon with a mixing strength $\epsilon$.  Low-energy, high luminosity
$\epem$ experiments like the \B Factories are in excellent position to
probe these theories~\cite{batell,essig} by searching the $A^\prime$
either in direct production, or in rare \B mesons and light mesons
decays.

It is possible to search for dark photon production in data from \epem 
collisions (see \cite{Bjorken:2009mm} and references therein) where current limits on the kinetic mixing 
parameter $\epsilon$ are a few $\times 10^{-3}$.  SuperB 
should be able to improve upon these limits by an order of 
magnitude.


Other possible search channels are: 
\begin{itemize}

\item the dark Higgs-strahlung process: the combined production with a
  dark Higgs ($h'$) (e.g. $\epem \to A^\prime h^\prime, h^\prime \to
  A^\prime A^\prime$) where \babar set a limit on the dark sector
  coupling constant and the mixing parameter at the level of $10^{-10}
  - 10^{-8}$ for $0.8 < m_{h'} < 10.0 \gev$ and $0.25 < m_{A'} < 3.0
  \gev$ with the constraint $m_{h'} > 2 m_{A'}$ and where \superb
  could improve these limits by two orders of magnitude;

\item very rare decays of the \B meson, as resonances in the dilepton
  spectrum of $B\to K l^+l^-$, $B^0\to l^+l^- l^+l^-$ or $B\to K
  l^+l^- l^+l^-$, where \superb could probe couplings between a dark
  pseudoscalar and the top quark at the level of $1000-10000 \tev$,
  much larger than the scale accessible at LHC in direct collisions;

\item rare meson decays~\cite{Reece:2009un} where the huge samples of
  lighter mesons such as $\pi^0$, $\eta$, $K$, $\phi$, and $J\psi$ can
  be used to search for the channel $\pi^0/\eta\to\gamma
  A^\prime\to\gamma l^+ l^-$.
 
\end{itemize}

\section{Executive Summary}
\label{sec:summary}

Up to now the SM managed to pass all the experimental challenges
unscathed, providing an overall good description of particle physics up to the energy
scales probed in experiments, which is now approaching the TeV scale.

In spite of its phenomenological success, however, the SM is not
satisfactory for several theoretical reasons, including the
instability of the Fermi scale against radiative corrections, the lack
of an explanation for the origin of flavour and \CP violation and the
non-unified description of the fundamental interactions, with gravity
totally unaccounted for. In addition, the SM does have
phenomenological problems when confronted with astrophysical and
cosmological data: it cannot provide a viable dark matter candidate or
an amount of \CP violation large enough to account for the observed
matter-antimatter asymmetry in the universe.

For these reasons, the SM is regarded as a low-energy theory bound to
fail at some energy scale larger than the Fermi scale, where New
Physics (NP) effects become important. The search for these effects is
the main goal of particle physics in the next decades, both at present
and future experimental facilities.

Flavor physics is the best candidate as a tool for indirect NP
searches for several well-known reasons: FCNCs, neutral
meson-antimeson mixing and \CP violation occur at the loop level in
the SM and therefore are potentially subject to large virtual
corrections from new heavy particles. In addition, flavour violation
in the SM is governed by weak interactions and, in the case of quarks,
suppressed by small mixing angles.  All these features are not
necessarily shared by NP which could then produce very large effects.

In spite of its potential, so far flavour physics has not provided a
clear NP signal, although deviations with low significance are present
here and there. This failure, however, should not be considered
discouraging. Broadly speaking, flavour physics probed and excluded so
far only the region of small NP scales and/or large NP flavour
couplings. On the other hand, we should not forget that flavour
physics successfully anticipated the existence and sometimes the mass
range of all the heavy quarks which are nowadays part of the
SM~\cite{Glashow:1970gm,Kobayashi:1973fv,Bigi:1988jr}. This may well
happen again for heavy particles beyond the SM.

\begin{figure}[tbp]
  \begin{center}
    \includegraphics[width=0.50\textwidth]{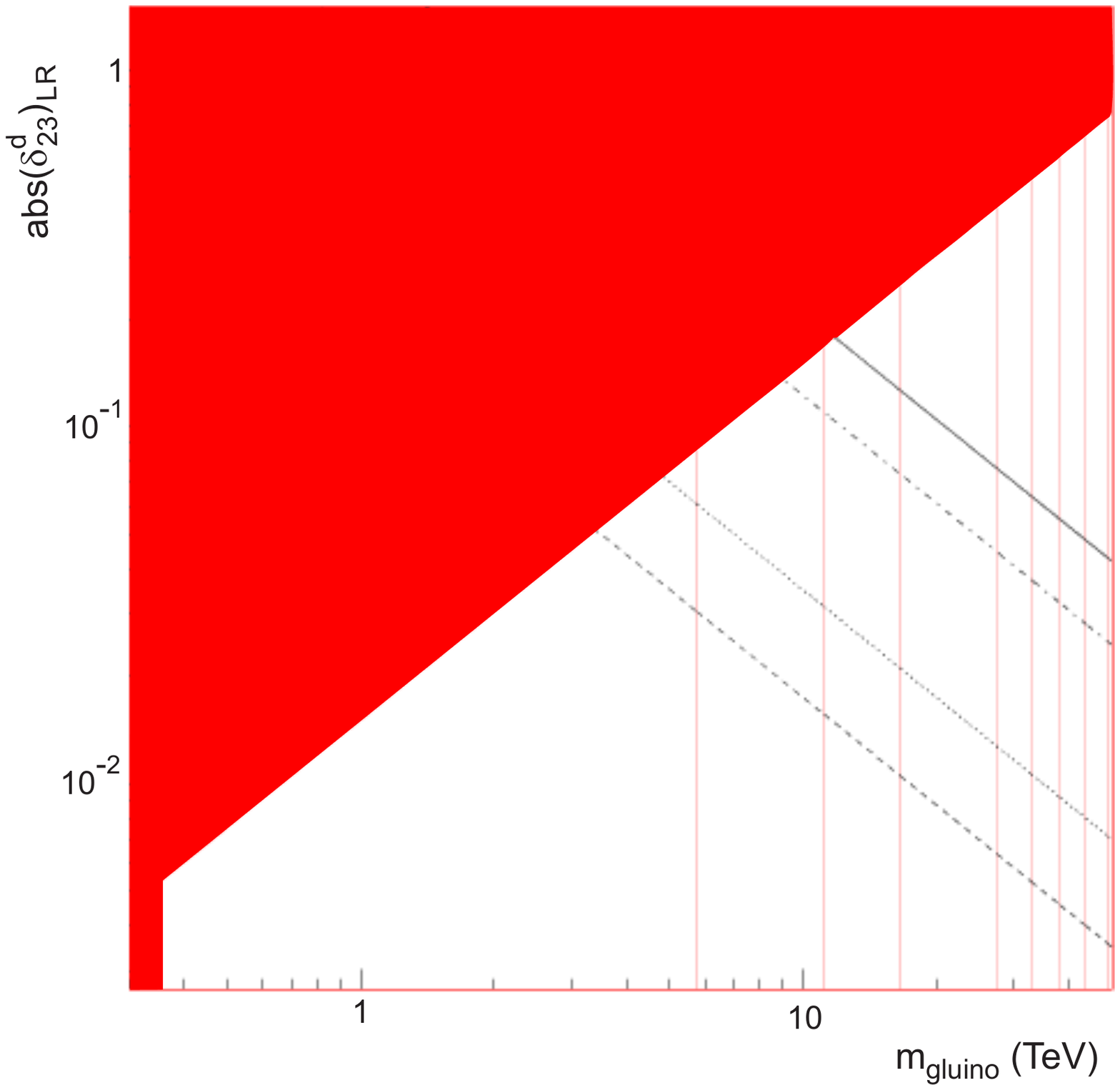}
   \caption{\label{fig:MI2}
           Region of the MSSM parameter space where a non-vanishing squark mass insertion parameter
           $(\delta_{23}^d)_{LR}$ can be extracted with at least $3\sigma$ significance (in red) using
           $a_\mathrm{CP}(B_d\to X_s\gamma)$, \BR($B_d\to X_s\gamma$) and \BR($B_d\to X_s\ell^+\ell^-$) as
           constraints and assuming dominance of gluino-mediated contributions.
           For $m_{\widetilde g}\sim 1$ TeV, \superb is sensitive to $(\delta_{23}^d)_{LR}\sim 10^{-2}$.
           For larger values of $(\delta_{23}^d)_{LR}$, \superb sensitivity extends to $m_{\widetilde g}>10$ TeV.
           }
  \end{center}
\end{figure}

\begin{figure}[tbp]
  \begin{center}
   \includegraphics[width=0.50\textwidth]{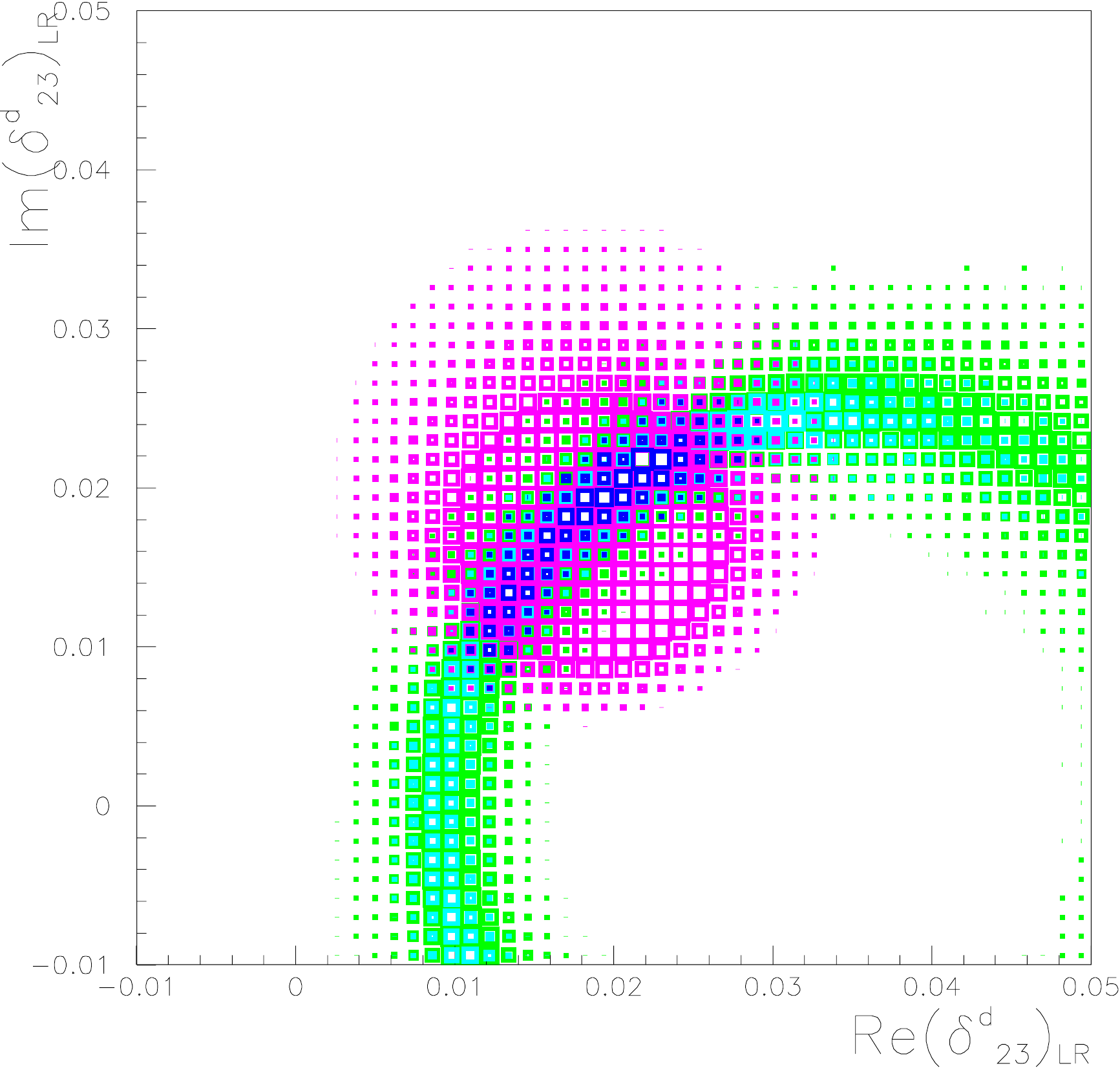}
   \caption{\label{fig:MI}
           Extraction of $(\delta_{23}^d)_{LR}$ from the measurements
           of $a_\mathrm{CP}(B_d\to X_s\gamma)$ (magenta), \BR($B_d\to X_s\gamma$) (green) and
           \BR($B_d\to X_s\ell^+\ell^-$) (cyan) with the errors expected at \superb.
           Central values are generated using $(\delta_{23}^d)_{LR}=0.028\,e^{i\pi/4}$
           and squark and gluino masses at $1$ TeV.
           }
  \end{center}
\end{figure}

From a phenomenological point of view, flavour physics can be used in
two modes. In the ``NP discovery'' mode, deviations from the
SM expectation are looked for in each NP-sensitive measurement
irrespective of their origin. The name of the game in this case is
high precision and a control of the SM contributions which matches it.
Depending on the NP at work, measurements in the $B$ and $D$ sectors
at \superb can be sensitive to contribution of new particles with
masses up to tens or hundreds TeV~\cite{Bona:2007qt,O'Leary:2010af}
(see the MSSM example in Figure~\ref{fig:MI2}). Although in general a
precise determination of the NP masses is hindered by the unknown NP
flavour couplings, even a broad indication of the NP mass range would
be a most valuable information in the unfortunate case direct searches
at the \lhc will be unsuccessful.  The ``NP
characterization'' mode entails measuring significant deviations in
several observables from different flavour sectors and studying the
correlations among them. If some information on the NP mass spectrum
is already available, this kind of analysis could provide crucial
information to disentangle different NP contributions, elucidate the
NP flavour structure and measure the NP flavour
couplings~\cite{Bona:2007qt,O'Leary:2010af} (see the MSSM example in
Figure~\ref{fig:MI}).  It is worth recalling that, on top of the
intrinsic phenomenological interest, flavour is intimately related to
NP symmetry properties in many extensions of the SM. For example, new
flavour structures appearing in the MSSM originate from the soft
SUSY-breaking sector. By measuring many NP-sensitive observables in
different flavour sectors, \superb can give a major contribution to
this ambitious program, which needs combining all available flavour
information, from kaon fixed-target experiments to hadronic collider
results.

The \superb physics programme sketched in this document extends over a
large set of rare decays, FCNCs, \CP-violation in the $B_d$/$B_u$, $D$
and $\tau$ sectors, measured at the \FourS and the \psiprpr
resonances~\cite{Bona:2007qt,O'Leary:2010af}.  The full set of
measurements performed at the $B$ factories can be repeated taking
advantage of the much larger data samples which typically allows one
to increase experimental precision/sensitivity by a factor 5 to 10.
These include the unitarity triangle angles measured with various
techniques, inclusive and exclusive semileptonic decays, $B\to
X_s\gamma$, $B\to\tau\nu$, $B\to D\tau\nu$, several penguin-dominated
$b\to s$ decays, $\tau$ flavour-violating decays, $D$--$\Dbar$ mixing
parameters, and many other processes.  Indeed, with very few
exceptions, these measurements are not severely limited by systematic
or theoretical uncertainties. In addition, several new measurements,
which were not feasible before, become possible: these include various
observables in the decays $B \to K^{(*)} \nu\nub$, $B\to X_s
\ell^+\ell^-$, $B\to X_d\gamma$, \CP violation in $D$ mixing and
decays, $\tau$ form factors, including EDM and $(g-2)_\tau$.  Some LFV
$\tau$ decay searches are expected to remain background free at
\superb, and the corresponding limits are expected to be improved by
two orders of magnitude on these modes (e.g. $\tau \to 3\mu$) relative
to constraints from the \B Factories.  Some of these rare decays can
be used to probe the nature of the Higgs sector, related to the
recently discovered particle at CERN, by constraining charged and
neutral partners of this Higgs which have been predicted by many
postulated NP scenarios.

Within its physics programme, \superb will explore to unprecedented
precision a number of recent developments in the field of $B$ and $D$
physics that could be a hint of physics beyond the SM.  The large
\BR($B\to\tau\nu$) measured at the \BFs\ and the recent evidence for
an excess also in $\B \to D^{(*)}\tau\nu$ decays reported by \babar is
something that needs to be verified and studied to high precision. The
same holds true for the \CP asymmetry of $\tau\to K_s\pi\nu$ which
\babar measured almost 3$\sigma$ away from the SM and for the direct
\CP asymmetry in singly Cabibbo-suppressed $D$ decay measured by
\lhcb. For the latter, an open issue is the actual SM prediction which
is hindered by the theoretical inability to compute the relevant $D$
decay amplitudes. \superb will be able to perform time-dependent \CP
asymmetry studies in neutral $D$ decays at the \psiprpr and \FourS,
enabling one to exploit the different statistical data samples using
complementary experimental techniques, as well as measuring hadronic
modes required to constrain theoretical uncertainties relating to the
interpretation of the measured observables.

\begin{figure}[tbp]
  \begin{center}
  \includegraphics[width=0.50\textwidth]{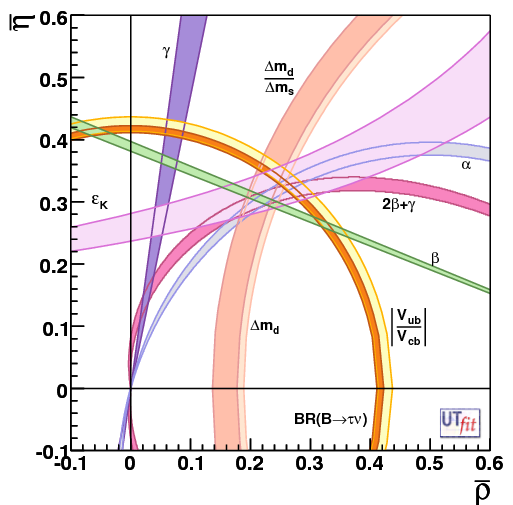}
  \caption{\label{fig:precisionCKM} 
           Constraints on the $(\bar \rho, \bar \eta)$ plane using measurements from \superb.
           Existing central
	   values are extrapolated to sensitivities expected from \superb with 75 ab$^{-1}$
           and one can see that the constraints are not consistent with the CKM scenario.
           This highlights the importance of performing a precision CKM test with \superb.}
  \end{center}
\end{figure}

Given the importance of controlling precisely the SM contributions, it
is crucial to improve the determination of the CKM parameters $\rho$
and $\eta$ in the presence of NP, using fit strategies already applied
successfully to \BF\ data. By combining data taken at the \psiprpr as
well as \FourS one will be able to make competitive measurements of
all the angles of the unitarity triangle ($\alpha$, $\beta$, and
$\gamma$) and the magnitude of several matrix elements (\vub, \vcb,
\vus, and \vcd), as well as the ratio $\Vtd/\Vts$.  In particular, the
crucial information on \vub and \vcb coming from semileptonic decays
can be improved only at flavour factories, where in addition the large
data set can be used to control theoretical uncertainties in inclusive
decays. Using this information, one can determine $\rho$ and $\eta$
with percent precision irrespective of possible NP contributions to
$\Delta F=2$ amplitudes. The reduced uncertainty on the CKM parameters
allows one to boost the NP sensitivity of several observables measured
at \superb and elsewhere.  One should not forget that the CKM fit
itself is a test of NP in $\Delta F=2$ amplitudes which could produce
an unmistakable signal at \superb, like the one shown in
Figure~\ref{fig:precisionCKM}.

$\tau$ physics is another prominent topic of the \superb physics
programme. Flavor-violating $\tau$ decays would be a clear and clean
mark of new physics~\cite{Bona:2007qt,Hitlin:2008gf,O'Leary:2010af}.
\superb extends sensitivity for \BR($\tau\to\mu\gamma$) down to the
$10^{-9}$ region, where many NP models produce detectable signals. In
addition, this region is complementary to the MEG sensitivity for
$\mu\to e\gamma$ ($10^{-13}$) in some SM extensions, allowing one to
probe the mechanism responsible for the LFV. In addition, the
excellent performance of \superb in measuring \BR($\tau\to 3\ell$) not
only gives an additional chance to detect LFV, but could also allow us
to tell apart models where LFV is generated through magnetic dipole
operators (such as the MSSM) from those using different mechanisms
(for example little Higgs models). Finally, the $\tau$ physics
programme does not only include LFV: a complete characterization of
the $\tau$ properties to high precision can be carried out, including
\CP violation and form factor measurements, such as the EDM and
$(g-2)_\tau$. Interestingly enough, assuming that the present
deviation of $(g-2)_\mu$ from its SM prediction is a NP effect, then
one expects the corresponding deviation in $(g-2)_\tau$ to be
measurable at \superb in many SM extensions.

In spite of the superior performance of hadronic colliders in the
$B_s$ sector, \superb can still provide few complementary $B_s$
measurements running at the \FiveS. Examples are
\BR($B_s\to\gamma\gamma$) and the semileptonic asymmetry $a_{SL}$
which can be used, together with $\Delta\Gamma_s$ and $\Delta m_s$, to
determine the NP-sensitive phase $\phi_s=\arg(-M_{12}/\Gamma_{12})$.
In addition, a precise measurement of the branching ratio of reference
$B_s$ decay modes would provide a valuable input to the \lhcb-upgrade
physics programme, allowing for a reduction of systematic
uncertainties on high-precision measurements of absolute branching
fractions (for example a precise measurement of the branching fraction
of $B_s \to J/\psi \phi$ will be useful to control systematic
uncertainties in the measured branching fraction of $B_s\to
\mu^+\mu^-$ from measurements made in the future at the LHC).

Polarization of the electron beam, a unique feature of \superb, makes
it possible to measure the $e^+e^-$ LR asymmetry in various channels.
The bottom and charm NC vector couplings will be determined with
unprecedented precision and $\sin^2\theta_W$ will be measured at the
\FourS using leptons with a precision comparable to that at the $Z$
peak~\cite{O'Leary:2010af}.  These unique measurements provide
important EW precision tests which, after the measurement of the Higgs
boson mass, become another powerful tool for indirect NP searches.
Furthermore, polarization provides new handles for $\tau$ physics
studies, in particular FV and form factor measurements.

Finally, \superb can perform direct NP searches as well: elusive light
particles like low-mass dark matter or light vector bosons, such as
the ``dark'' photon, can be looked for in bottomonium decays and using
data collected from the low energy run.  In addition, a rich
spectroscopy programme can be carried out at \superb. It is aimed at
finding further unconventional states in the quarkonia mass spectra
and possibly shedding light on the strong interaction mechanism
leading to their formation. In this respect, the \superb scan
capabilities offer unique opportunities to study both the charmonium
and bottomonium spectra with high statistics.

While there is a subset of \superb measurements which can be
interpreted with little or no theoretical input (for example: the CKM
angles $\gamma$ and $\alpha$, flavour-violating $\tau$ decays, many
\CP violating $D$ transitions), in order to exploit the full
phenomenological potential of the \superb program, one needs
theoretical uncertainties to match the experimental precision to allow
one to reveal and disentangling genuine NP effects. Extrapolation of
the present systematics indicates that lattice QCD could reach the
required precision on the relevant hadronic parameters in few
years~\cite{Bona:2007qt,O'Leary:2010af}. It seems also possible to
control theoretical uncertainties at the required level in inclusive
semileptonic decays using data-driven methods based on the heavy quark
expansion and the huge \superb data set. Inclusive radiative decays,
most notably the $B\to X_s\gamma$ rate, could be limited by theory
uncertainties, although there is no general consensus on the ultimate
theoretical error which could be obtained in the next years. Only for
some two-body non-leptonic decays, in particular the penguin-dominated
$b\to s$ transitions, theory seems not able, in the absence of a
breakthrough, to match the experimental precision. Yet, even in this
case, data and flavour symmetries can be sometimes used to bound the
theoretical uncertainties and allow for the extraction of the NP
signal.

In summary, we have shown that there are a significant number of
golden modes for which \superb, collecting 75\invab of data at the
\FourS and 1\invab of data at the $\psi(3770)$, has the best expected
sensitivities~\cite{Meadows:2011bk}.  The physics potential of \superb
and \belletwo are similar, but \superb provides additional
opportunities owing to its charm threshold running capability and beam
polarization.  As with \babar and \belle, and ATLAS and CMS, there is
a significant benefit in having two super flavour factories, as it
will be possible to verify any discoveries made by one of the
experiments.  The super flavour factories and the \lhcb upgrade
physics programs are largely complementary and the overlap is confined
to $B^0\to K^{0*} \mu^+\mu^-$ and the charm mixing parameters, which
the \lhcb upgrade will eventually improve upon~\cite{Meadows:2011bk}.
All together, \superb, \belletwo, \lhcb upgrade and experiments
measuring ultra-rare $K$ decays and searching for muon flavor
violation, will be able to provide full and redundant coverage of all
quark and charged lepton flavour sectors with high-precision
measurements of rare decays, FCNCs, flavour and \CP violating
processes.  These will be a key tool to characterize NP at the TeV
scale or to start probing the multi-TeV region.

\bibliographystyle{\tdrbibstyle}
\balance
\bibliography{Physics/BandD-bib,Physics/EWPhysics-bib,Physics/SG-spectroscopy,Physics/tau-bib,Physics/summary}


\renewcommand{\ChapLabel}{chap:MDI} 
\tdrchap{Machine Detector Interface and Backgrounds}
\label{chap:mdi}

\tdrsec{Introduction}

The \superb\ experiment consists of a detector and a collider that
must be properly interfaced to guarantee optimal operations.

The set of constraints and requirements from both sides is managed
by the Machine Detector Interface (MDI).

One of the leading challenges in designing the detector 
is the background model of the environment in which it will
operate since both the large luminosity boost and the 
substantial emittance decrease with respect to the present generation
facilities will put the detector in a dangerous unexplored realm.

In addition to the particles that are the main object of measurement,
the detector will also be exposed to particles produced away
from the IP that are an unpleasant but unavoidable side effect of the
accelerator operations. These particles constitute the 
machine background.
\superb\ will also detect the
particles produced at the IP by QED scattering 
processes that have cross sections of a few mbarn.

The machine background affecting \superb is mainly the result of an
electromagnetic shower initiated by a beam particle hitting the beam
pipe somewhere along the ring. The machine lattice can provide stable
orbits only for a limited region in phase space around the reference
orbit. Either scattering processes occurring at the IP, or Touschek
and beam-gas scattering taking place all around the ring can bring a
beam particle outside the stability region making it eventually 
collides with the beam pipe.

In addition to these backgrounds the detector will be also exposed
to the risks connected to the intense synchrotron radiation produced 
by the beams passing through the magnetic elements upstream the IP 
whose total power (in the kW range) and energy spectrum (spanning 
from the soft to the hard  X-rays) constitute a serious concern 
for the inner detectors.

Background considerations influence several aspects of the design:
readout segmentation, electronics shaping time, data transmission rate, 
triggering and radiation hardness. 

The following sections will describe how these sources are simulated,
how their effects on the detectors are mitigated and what is their impact 
on the subsystems.

\tdrsec{Backgrounds Sources Overview}

With the proposed collider design, the primary sources of background
are scattering of the following type: radiative Bhabha, pairs
production via 2-photon QED interactions, Touschek, and beam-gas.
Photons from synchrotron radiation, if properly handled and kept away
from sensitive areas, give smaller contributions limited to the
innermost detectors (SVT).  The heat load due to synchrotron radiation
photons striking masks and primaries and secondaries particles hitting
the super-conducting magnets must be also carefully evaluated.  We
have developed a detailed Geant4 based software called \Bruno that is
able to
simulate the detector and the beam-line, described from -16 to 16
meters from the IP. \Bruno has been used to estimate the impact of
the different background sources on the operation of the detector.
The relevant magnetic and physical elements used in the configuration
are the quadrupoles of the final focus doublets closest to the IP and
the dipoles of the first bends near the detector. Tungsten shields
placed around the cryostat of the final focus super conducting doublet
protect the detector on both sides of the interaction region 
(Sec.~\ref{sec:mdi_RadBhaBha_bkg_shields}).

The background sources described below have been simulated separately
with \Bruno. The details of the simulated samples are reported in
Table \ref{tab:fullsim_samples}.  Touschek scattering, beam scattering
with residual gas, and photons from synchrotron radiation do not have
an equivalent time because each event is rescaled with a statistical
weight obtained from the dedicated simulation used to produce them, as 
described in Sec.~\ref{sec:mdi_touschek} and \ref{sec:mdi_beamgas}.

\begin{table*}
  \caption{
    Details of simulated samples for the the background sources: source name, 
    description with parameters, number of events (or bunch-crossing), and 
    equivalent time, if applicable.
  }
\vspace{0.2cm} \centering
\begin{tabular}{l p{0.4\textwidth} c c}
\hline \hline 
Name & Brief description & \# of events  & Equivalent time \\
\hline \hline 
RadBhabha & Radiative Bhabha with $\Delta E / E > 30\%$ & 15k & $66\mu \rm s$\\
\hline 
RadBhabhaLowDE & Radiative Bhabha with $5\% < \Delta E / E < 30\%$ & 20k & $88\mu \rm s$ \\
\hline 
Pairs & 2-photon & 100k & $442\mu \rm s$\\
\hline 
Touschek HER & Touschek scattering from HER & 90k & N/A \\
\hline 
Touschek LER & Touschek scattering from LER & 198k & N/A \\
\hline 
Beam gas HER & Beam gas scattering from HER & 285k & N/A \\
\hline 
Beam gas LER & Beam gas scattering from LER & 283k & N/A \\
\hline 
SyncRad HER & Synchrotron radiation from HER & 9.8k & N/A \\
\hline 
SyncRad LER & Synchrotron radiation from HER & 9.6k & N/A\\
\hline \hline 
\end{tabular}
\label{tab:fullsim_samples}
\end{table*}
\tdrsec{ Radiative Bhabha}

\label{sec:mdi_RadBhaBha_bkg}

Radiative Bhabha scattering $e^+e^- \rightarrow e^+e^-\gamma$
is expected to be one the dominant background sources for the
detector since it is, along with Touschek scattering,
the main mechanism by which beam particles are lost along the machine.
Since the rings can provide stable orbits only for particles in
a small range of energies (roughly $\pm 1\%$ from the nominal 
beam energy) the photon radiated by one of the incoming leptons
can carry away enough energy to lead to the lepton loss at some
point downstream of the IP.
The particle loss cross section for this process in \superb is 
roughly $170\,\mathrm{mb}$  \cite{Bona:2007qt,Kotkin:2009ju}, that is
order of $170\,\mathrm{GHz}$ per ring at nominal luminosity.
This huge loss rate requires an accurate design of the final 
focus in order to transport most of these particles
far away from the IP and minimize the detrimental effects 
on the detector.
These reasons are a clear motivation to develop a reliable 
and accurate simulation tool for this kind of process.

\tdrsubsec{Simulation tools}
\label{sec:mdi_RadBhaBha_bkg_simTools}

The effects on the detector of particles scattered by the radiative
Bhabha processes is studied with \Bruno, 
a Geant4~\cite{Agostinelli:2002hh} based simulation program used in conjunction
with the BBBREM generator~\cite{Kleiss:1994wq}. BBBREM is a Monte
Carlo generator which simulates at tree level single
bremsstrahlung in Bhabha scattering in the very forward direction.
BBBREM correctly takes into account the finite lepton masses, thus
providing an accurate description of the angular distribution of the
final products.  BBBREM takes into account in its matrix element
evaluation only the two t-channel Feynman graphs describing the photon
emission by initial or final state radiation.  This approximation is
meaningful because the amplitude of the process is very large for
photons emitted nearly parallel to the radiating beam. The typical
photon emission angle with respect to the radiating lepton is order of
$1/\gamma$ where $\gamma$ is the Lorentz factor of the incoming
radiating lepton.  Hence, one expects negligible contributions to the
total amplitude coming both from the s-channel Feynman graphs and from
the interference terms among graphs describing the emission from the
electron and the positron.

The BBBREM generator takes also into account the finite bunch density
effect that reduces the radiative Bhabha cross-section. The effect is
modeled by setting up an infrared cutoff on the transferred momentum
squared ($Q^2_0$) that depends on the average distance among two
electrons/positrons in the colliding section of the bunch.

The BBBREM generator uses the following input parameters:

\begin{itemize}
\item Total center-of-mass energy of the incoming $e^+e^-$ system
  (including the energy spread of the electrons/positrons)
\item The photon energy cutoff fraction $\kappa_0$. This is a minimum
  value of the energy fraction (in the CM) carried away by the
  radiated photon ($\kappa = \DeltaE/E$). As the total radiative Bhabha
  cross-section is infrared divergent a sensitive minimum value needs
  to be set.
\item The number of interactions to be simulated ($N_{\rm int}$).
\end{itemize}

The BBBREM code is embedded in \Bruno to generate the primaries. The
beams nominal parameters at the IP are used to accurately simulate the
statistical distribution of the primary vertex position and of the
incoming particle momentum spread.

These parameters are also used to calculate the luminosity (${\mathcal
  L}$) per bunch crossing.  It has been checked that the main
backgrounds come from radiative Bhabha events with $\kappa > 0.3$
(High-$\kappa$ radiative Bhabha). Events with $0.005 < \kappa < 0.3$
(Low-$\kappa$ radiative Bhabha) give smaller but non-negligible
contribution.  Both of them have been simulated and considered in the
detector studies.

\tdrsubsec{Losses at the beam-pipe}
\label{sec:mdi_RadBhaBha_bkg_losses}

The impact of this source of
background extends to the whole detector. A large fraction of the
off-energy electrons and positrons hit the downstream beam-line
elements producing electromagnetic showers.  Low energy particles from
these showers hit all of the detector subsystems. An accurate evaluation
requires a careful modeling of the final focus which includes the
geometry (beam-pipe, magnets, shields, cryostat) and the magnetic
field. The interaction of the beams with the material in the final
focus produces mainly low-energy electrons and photons. Neutrons with a
mean kinetic energy around $1~{\rm MeV}$ are also produced via
photo-nuclear reactions.  Neutrons are produced along the beam line
both near the IP and near the bending magnets downstream of the IP.
These neutrons build up a neutron cloud that thermalize
inside the detector hall for times up to tens of $\mu {\rm s}$, 
causing delayed background hits at the EMC and IFR, and also affecting
the front-end-electronics.

\begin{figure}[tbh]
 \begin{center}
 \includegraphics[width=0.5\textwidth]{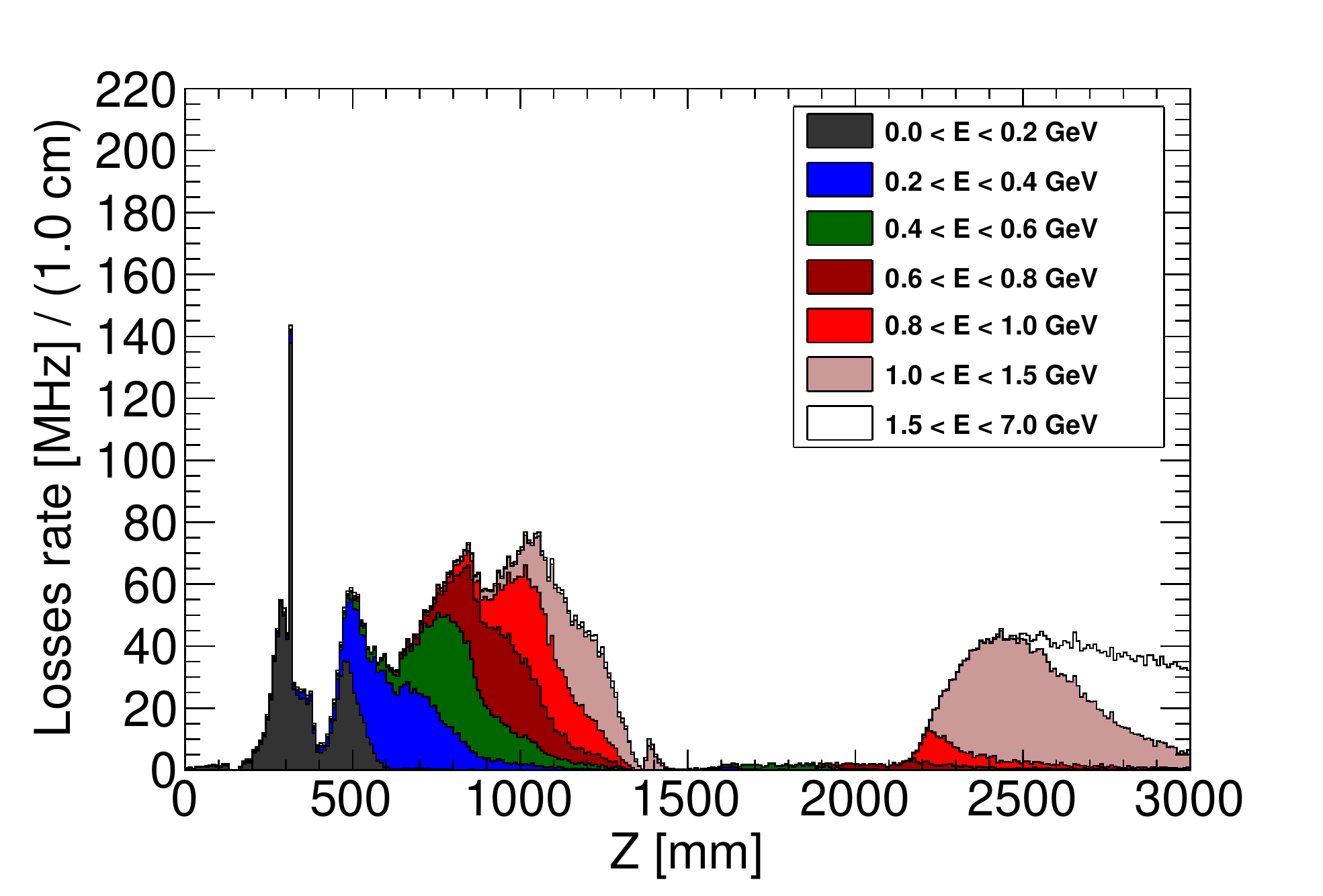}
 \includegraphics[width=0.5\textwidth]{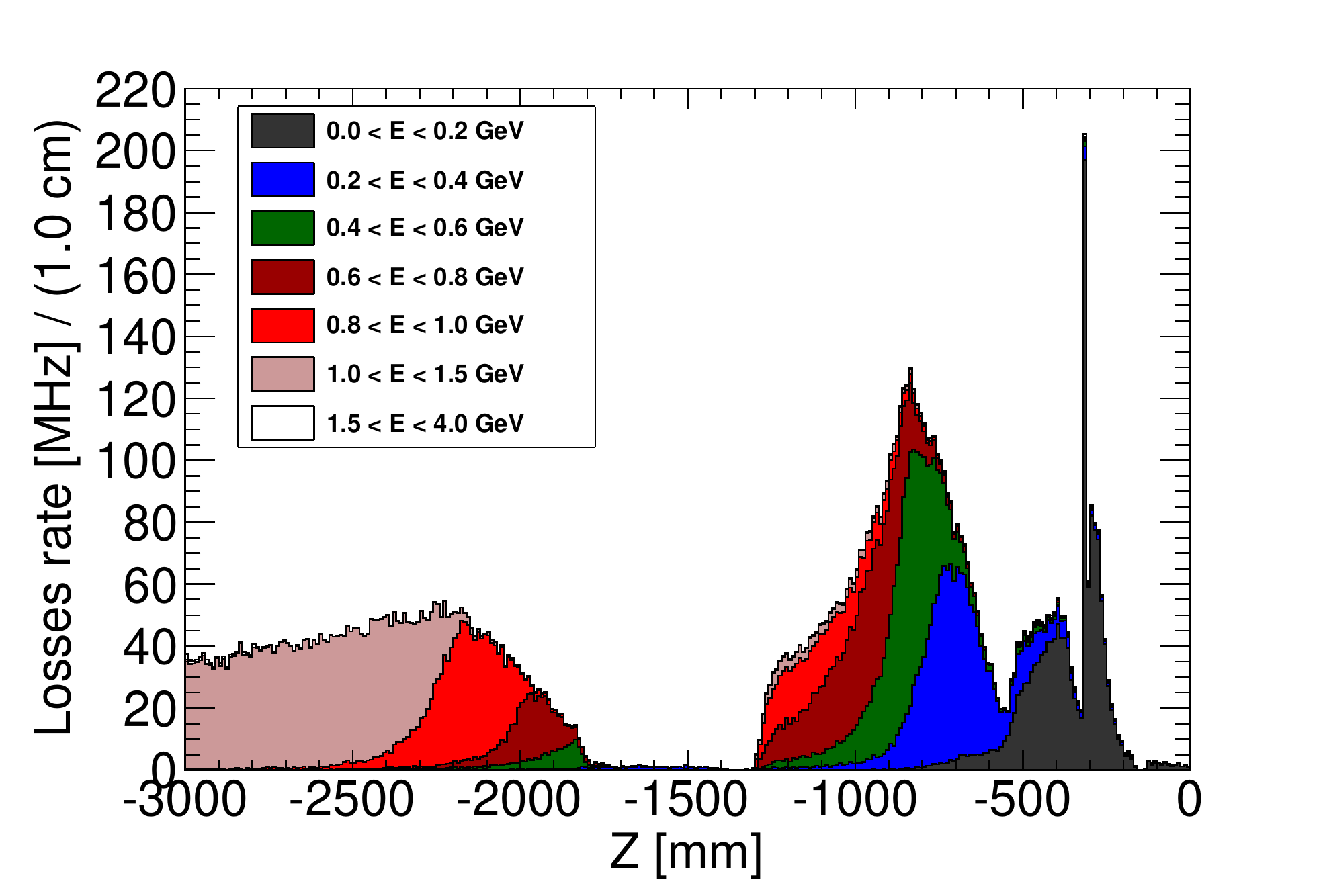}
 \caption{Rad-Bhabha loss rate at the beam-pipe due to positrons 
(top) and due to electrons (bottom) as a function of the Z coordinate.}
 \label{fig:radbhabha_losses}
 \end{center}
\end{figure}

A very useful and efficient approach to understand how the radiative
Bhabha background affects the detector is to estimate the particle loss
rate along the beam pipe. This information is very valuable to decide
where protective shields for the detector (see
Sec.~\ref{sec:mdi_RadBhaBha_bkg_shields}) have to be placed before
proceeding with more time consuming simulations.  The loss rate due to
positrons (high-energy-ring or HER) and due to electrons
(low-energy-ring or LER) as a function of the Z-coordinate is shown in
Figure~\ref{fig:radbhabha_losses} in the region $-3{\rm m} < Z < 3{\rm
  m}$ (inside the detector).

Both plots show similar patterns. Only very low-energy (off-energy)
particles hit the beam pipe near the interaction region. The
contribution to the losses from higher energy particles increases far
away the IP. The loss structure (the different peaks) is due to two
effects: the beam-pipe profile (the beam pipe gets wider for higher
$|Z|$); and the location of the magnets along the beam-line. Even for
the losses occurring inside the \superb\ detector, the lost particles
are quite collimated along the beam-pipe, so the debris generated from
them are the ones expected to affect the outer sub-detectors.  Some
debris produced outside of the detector ($|Z| > 3{\rm m}$) 
produces background by back-scattering with the material around the
final focus. Even though the back-scattering probability is small, the
large loss cross-section makes this component non-negligible.  This
justifies the necessity of a model of the final focus up to 16~m
from the IP.

Without any shield or mask the \superb\ detector would be overwhelmed
mainly by radiate Bhabha backgrounds. Some shields have been included
to reduce this background source, and will be briefly described in the
next section.

\tdrsubsec{Shield System}
\label{sec:mdi_RadBhaBha_bkg_shields}

Several types of shields are needed around the hot regions.  A
tungsten shield (W-shield) surrounds the final focus regions inside
the detector. The aim of this shield is to reduce the amount of
secondary electrons and photon coming from the interaction of the
primary electrons/positrons with the final focus material around IP.
The shield consists of two sections.  The conical section is $3~{\rm
  cm}$ thick and is as close as possible to the IP ($22 < |Z| <
80~{\rm cm}$).  It does not interfere with the SVT angular acceptance
($\pm 300~{\rm mrad}$). The cylindrical section has a $4.5~{\rm cm}$
thickness, which is mainly limited by the geometrical constraints
coming from the DCH inner radius and the final focus cryostats outer
radius.  The cylinder covers the regions $80 < Z < 280~{\rm cm}$ and
$-220 < Z < -80~{\rm cm}$. This thickness is sufficient to reduce
backgrounds on the EMC/DCH up to acceptable levels with a reasonable
safety margin.

A pair of lead-plugs in the forward and backward regions have been
included following a similar design to \babar in order to shield the
detector from secondaries produced far away from the IP.  All these
shields reduce the radiative Bhabha background by a factor of 2--5 in
most of the subsystems. The shields tailored for the FDIRC and IFR 
are described in the corresponding sections. 

\tdrsec{Pairs production}

During high energy beam-beam interaction, electromagnetic secondary
backgrounds are produced from Incoherent Pair Creation (IPC)
processes~\cite{pairsYokoya}, where $e^+e^-$ pairs
arise from interaction of both real or virtual photons of each beam
with individual particles of the other beam. Three processes are
responsible for IPC: the Breit-Wheeler process, in which two real
beamstrahlung photons interact, the Bethe-Heitler one, where a real
and a virtual photon collide and finally the Landau-Lifschitz (LL)
process with two virtual photons. Hence, although almost no
beamstrahlung is produced during beam-beam collision at \superb, pairs
can be created through the processes involving virtual photons, in
particular the Landau-Lifschitz one.  

A fraction of the pairs are produced with transverse momentum large enough to
reach the beam pipe. The large cross section combined with a high
collision frequency can lead to an significant background rate that it
is necessary to predict wll.
Several simulation tools have been used to evaluate the production
rate of the $e^+e^-$ pairs and estimate backgrounds in the detector:
two versions of the BDK four fermion generator~\cite{pairsBDK},
and the GUINEA-PIG++ (GP++)~\cite{pairsDaniel,pairsGP} beam-beam
interaction code, based on the Weizaecker-Williams approximation for
the calculation of the processes involving virtual photons. 

\tdrsubsec{Cross section and spectrum comparison}
BDK is a Monte-Carlo generator for four-fermion processes in $e^+e^-$
interactions.  It is based on complete calculations with
leading-order massive matrix elements for all relevant electro-weak
diagrams involved. Several adaptations of the initial code have been
developed within lepton collider collaborations. For the present
study, two of them have been used and compared, one retrieved from the Delphi
repository and the other, DIAG36, from the \babar\ one. The generations are
performed in the center of mass of the two mother particles, each with
an energy of 5.29~\gev. 

GP++ is a beam-beam interaction program developed for high energy
$e^+e^-$ linear collider, including several secondary production
processes. For the IPC, the three processes described above are
calculated using the equivalent photon approximation for the processes
involving virtual photons. This approximation treats virtual photons
as real ones by convolving an equivalent spectrum for the virtual
photons with the cross section for the real-real case. These photons
are treated as being real as long as their virtuality remains below an
upper limit, above which they are ignored. This limit is set to the
half of the invariant mass squared. Once they are produced, the pairs
are tracked through the electromagnetic field of the beams until the
end of the interaction.

The cross section prediction is about 7.30~mbarn for both BDK
simulations, and about 7.74~mbarn for the Landau-Lifshitz process
in GP++, which represents 97\% of the total IPC, the remainder being
produced by the Bethe-Heitler process.

\begin{figure}
\centering
\includegraphics[width=\linewidth,clip=true,trim = 0 10 40 30]{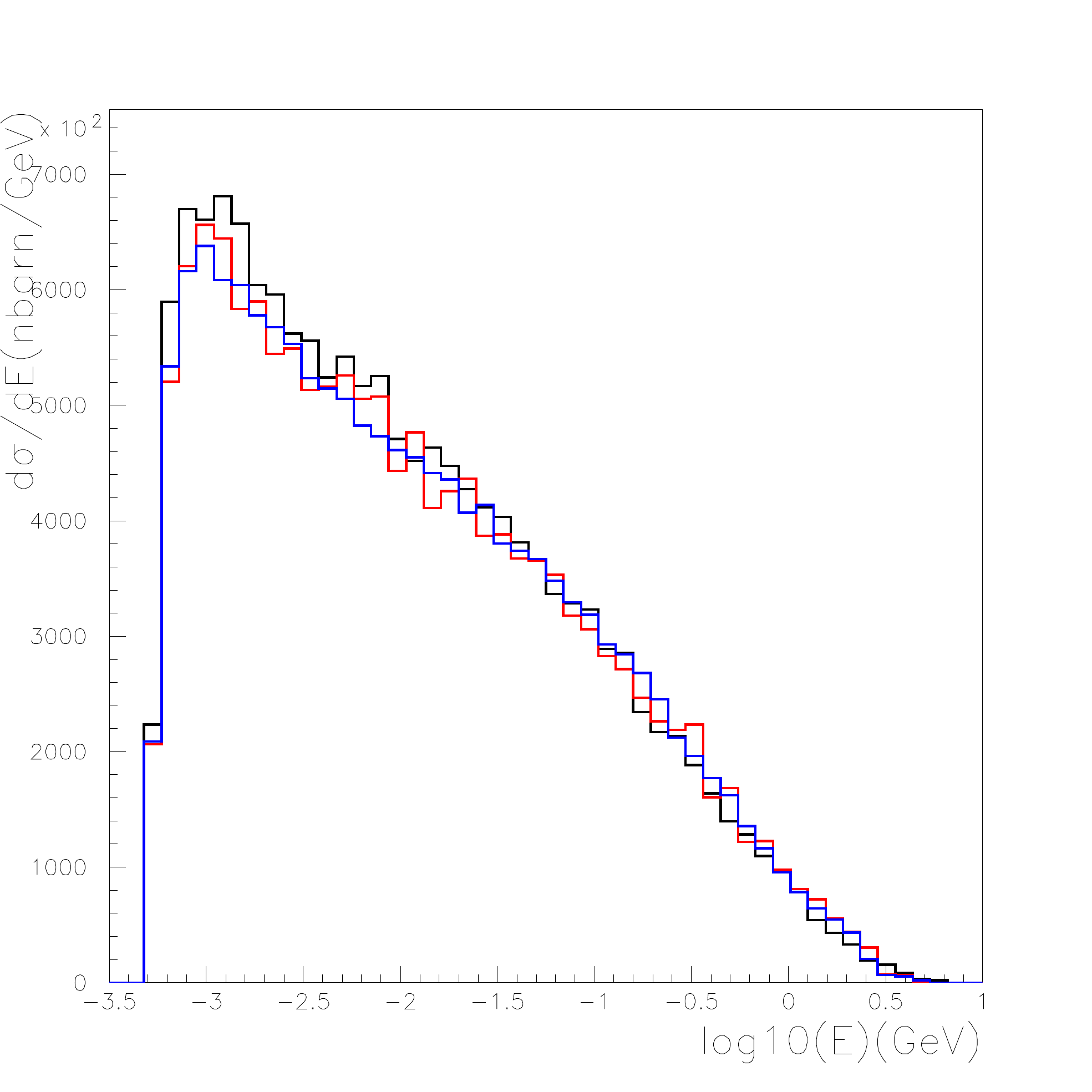}
\caption{Comparison of energy distributions generated with BDK in blue, DIAG36 in red and GP++ in black.}
\label{figpairs1_energy}
\end{figure}

\begin{figure}
\centering
\includegraphics[width=\linewidth,clip=true,trim = 0 10 40 30]{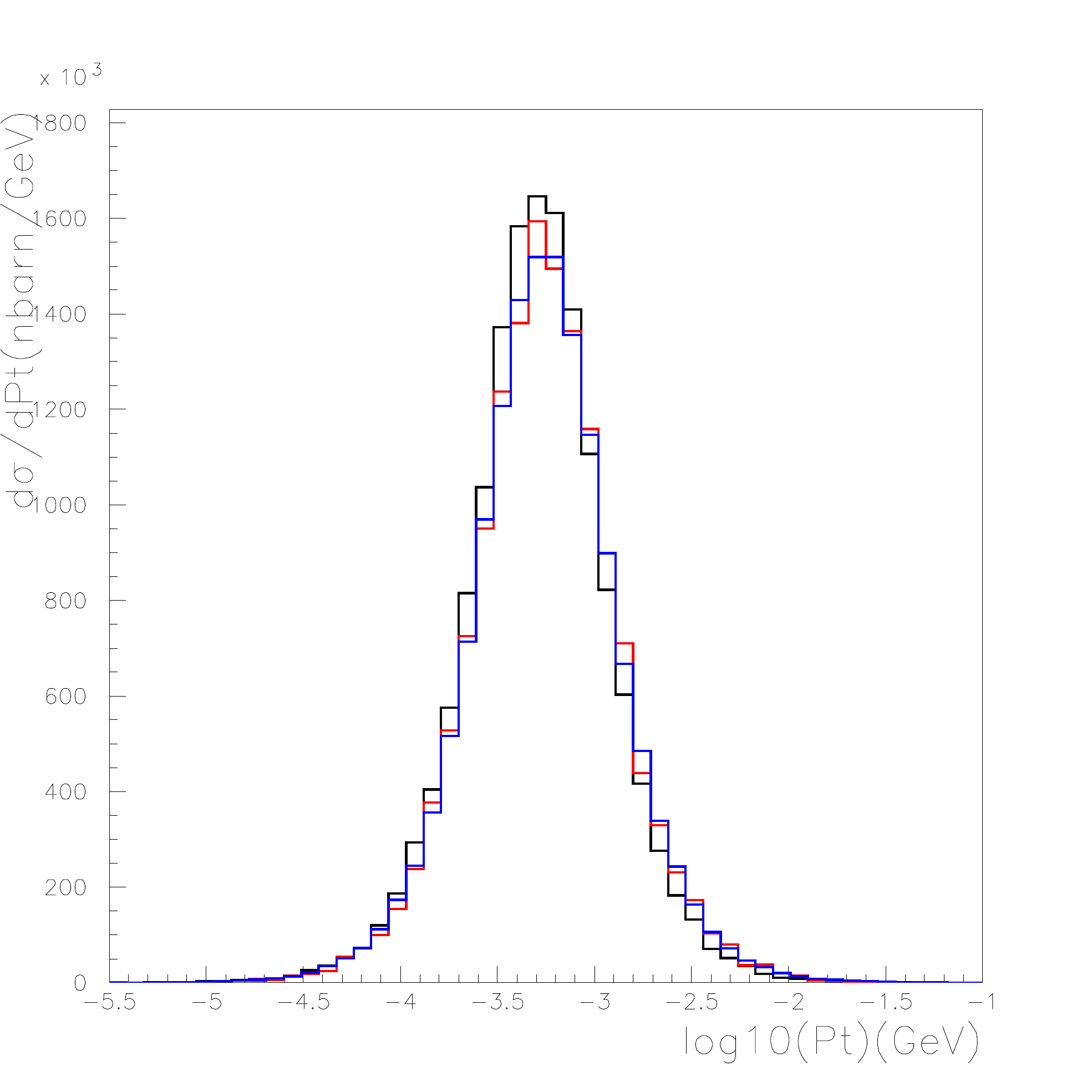}
\caption{Comparison of transverse momentum distributions generated with BDK in blue, DIAG36 in red and GP++ in black.}
\label{figpairs1_pt}
\end{figure}

\begin{figure}
\centering
\includegraphics[width=\linewidth,clip=true,trim = 0 10 40 30]{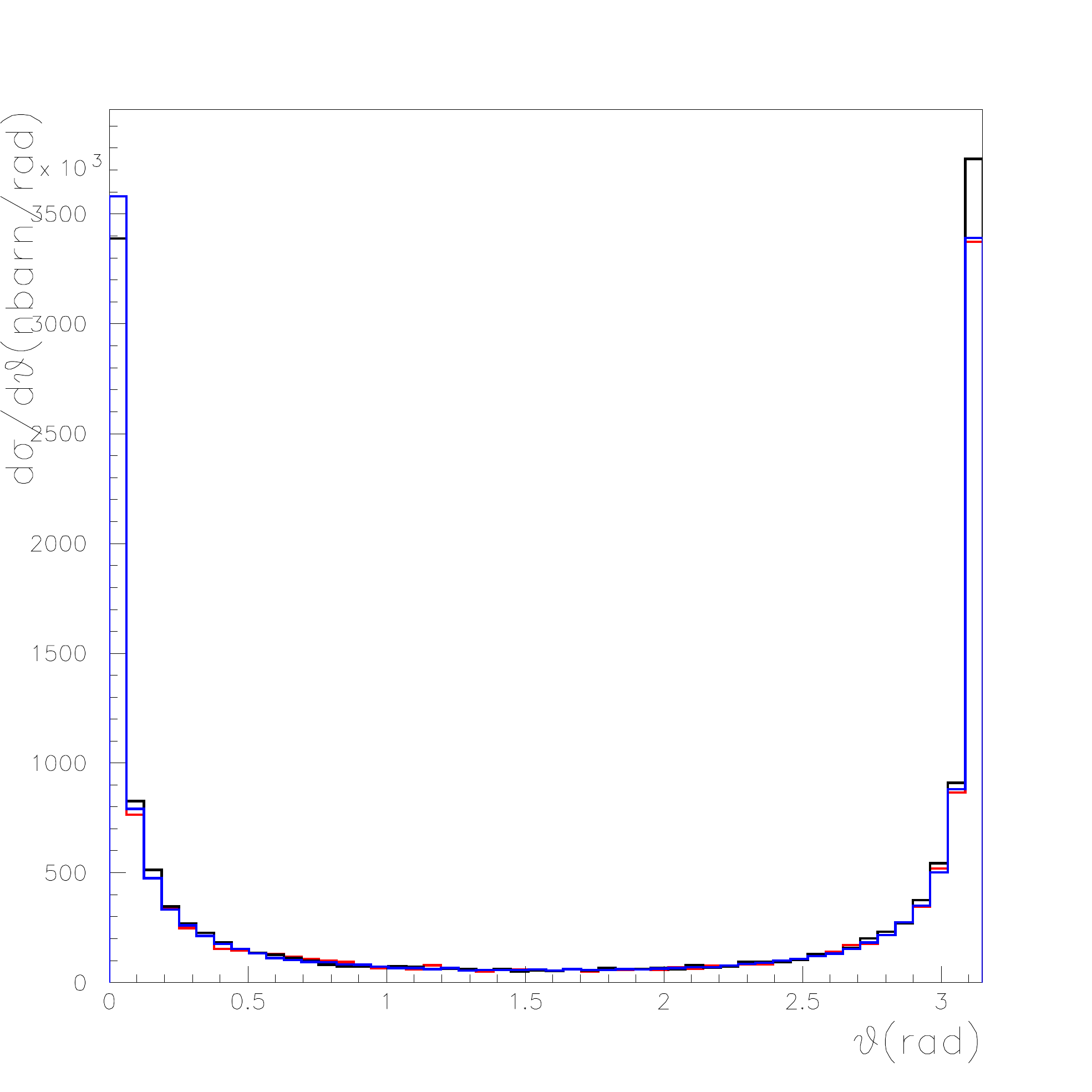}
\caption{Comparison of angular pair distributions generated with BDK in blue, DIAG36 in red and GP++ in black.}
\label{figpairs1_angular}
\end{figure}

Figures~\ref{figpairs1_energy}--\ref{figpairs1_angular} 
show the energy, transverse momentum  and
angular distributions of the lepton pairs produced by BDK, DIAG36 and GP++
at the production time. Good agreement can be observed between the
three generators. The pairs are produced with a mean energy of 120 MeV
and a mean transverse momentum of $\approx 1.5$~\mev.

\tdrsubsec{Impact of beam-beam effect on secondaries}
\begin{figure}[htpb]
\centering
\includegraphics[width=\linewidth]{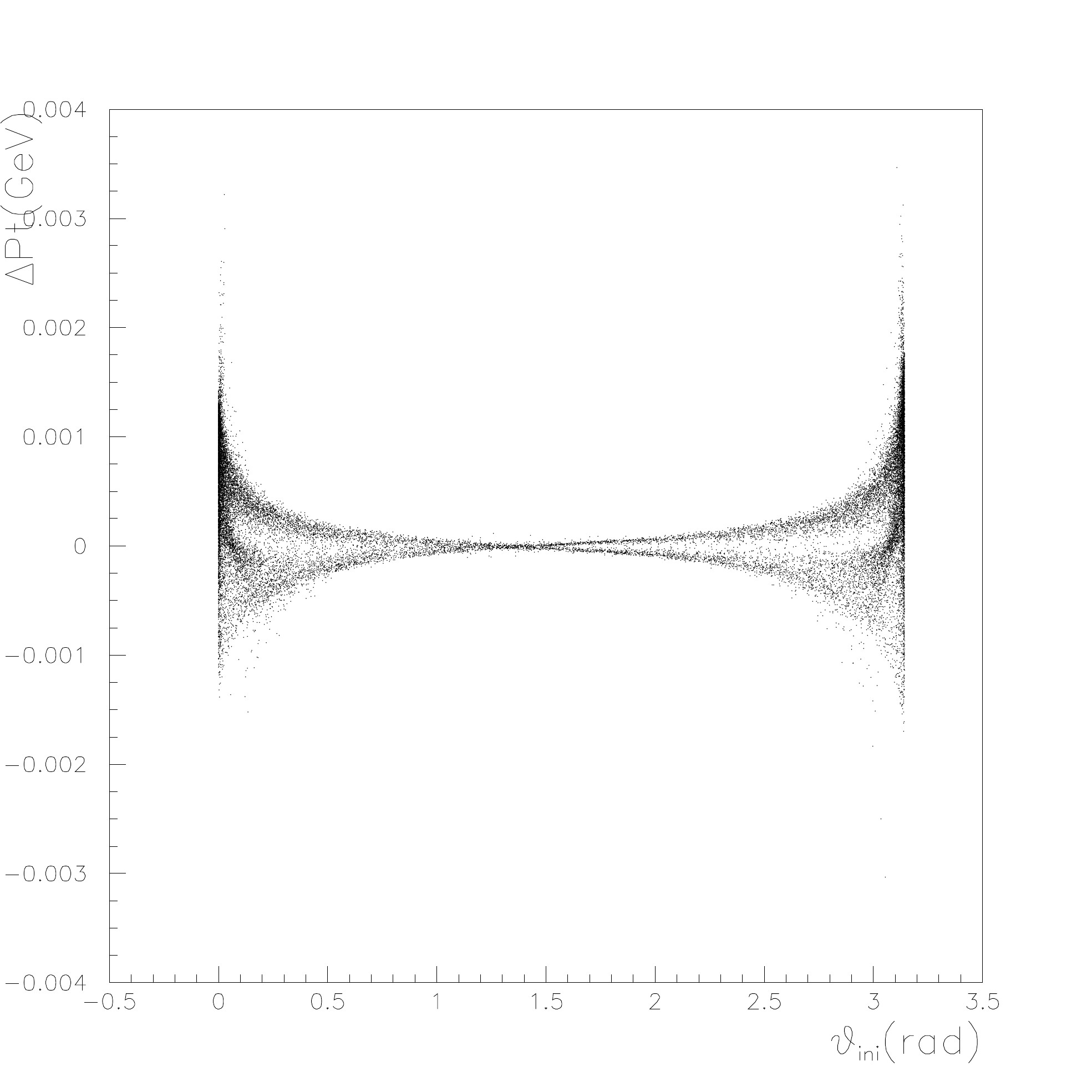}
\caption{Difference between the final and initial transverse momentum as a function of initial angle of the pair particles.}
\label{figpairs2}
\end{figure}
\begin{figure}[htpb]
\centering
\includegraphics[width=\linewidth]{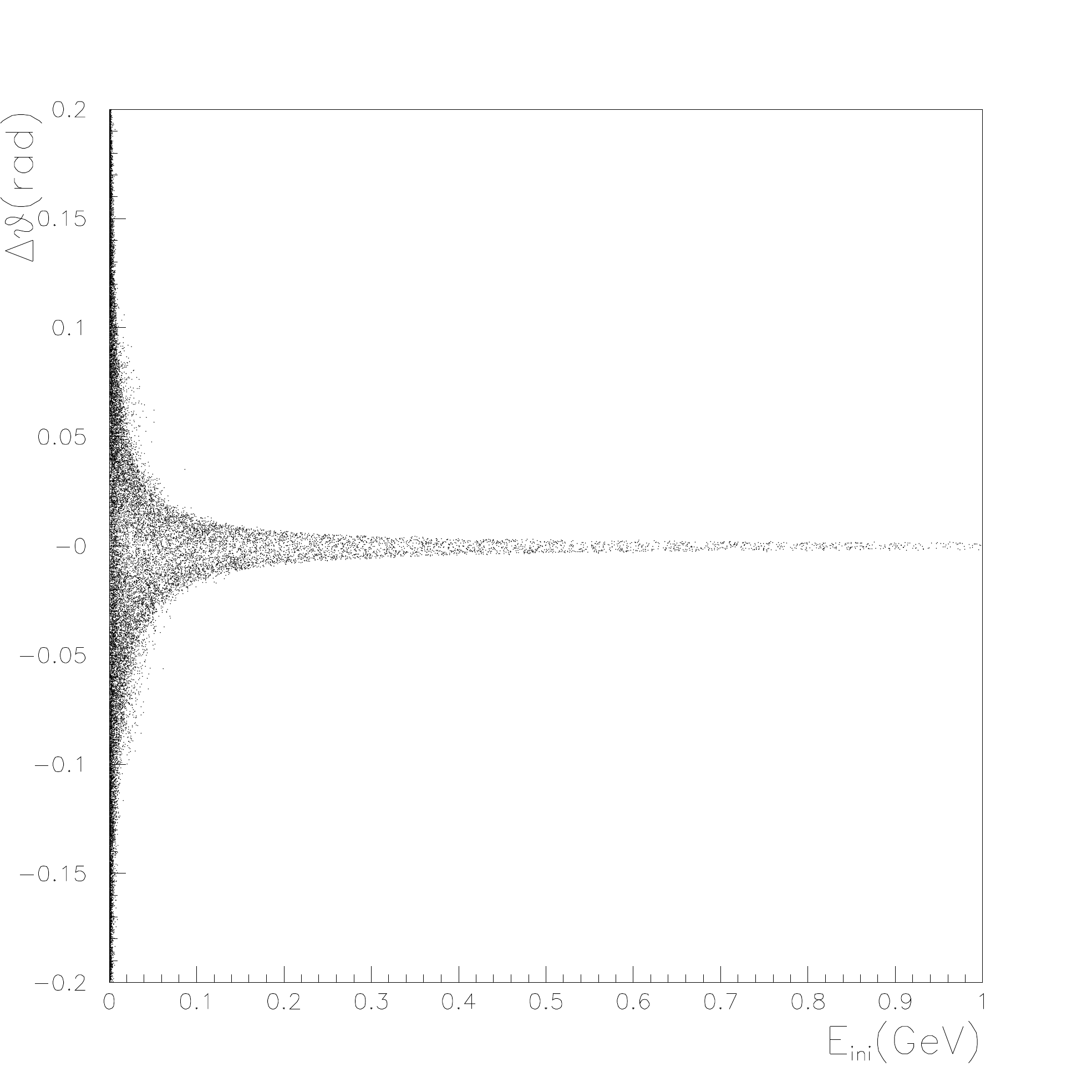}
\caption{Deflection angle as a function of the pair particle energy.}
\label{figpairs3}
\end{figure}

\begin{figure}[htpb]
\centering
\includegraphics[width=\linewidth]{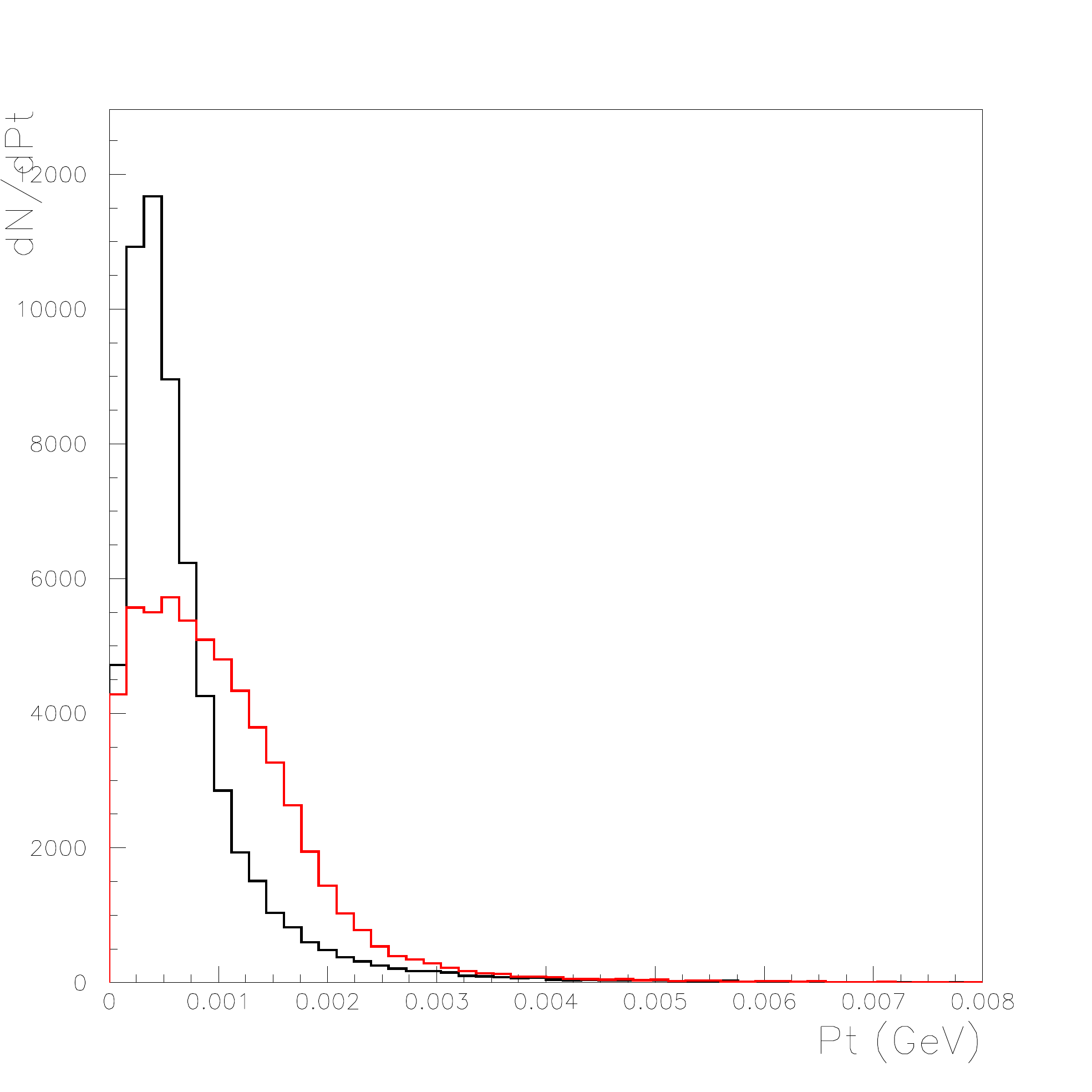}
\caption{Comparison of transverse momentum distributions generated with GP++ at the creation time in black and the end of beam-beam interaction in red.}
\label{figpairs4}
\end{figure}

Since the pairs are mostly produced at low energy and small angle,
they are affected by the electromagnetic field of the oncoming beam,
being either focused or defocused according to their charge. The most
deflected particles are those produced at very low angle
(Figure~\ref{figpairs2}) and very low energy (Figure~\ref{figpairs3}).

The global impact of electromagnetic deflections is shown on
Figure~\ref{figpairs4}. The transverse momentum is on average slightly
increased of a few hundreds keV. 

\tdrsubsec{Beam pipe losses}
A fraction of the pair particles can reach the beam pipe (BP), whose
inner radius is $10\mm$.

In a magnetic field, the charged particles move along an helix of a
radius $r_0 = 3.33p_t/B$, $r_0$ being in meter, $p_t$ in \gevc and $B$ in
Tesla. The rates of particles reaching the beam pipe for a magnetic
field of 1.5\;T are presented in Table~\ref{tabpairs1} as obtained
from DIAG36 and GP++ simulations, without angular cuts and within the
angular acceptance of $100~\mrad < \theta < \pi - 100~\mrad$, corresponding to
a length of 20\;cm. In addition to the Landau-Lifshitz process, the
total Incoherent Pairs Creation cross-sections are also listed. The
losses in the 20\;cm long beam pipe are of the order of 400 MHz. The
electromagnetic deflections, affecting the low energy particles
produced at low angle, lead to an increase in the rate 
of around 10\%.
 
\begin{table*}[htb]
\caption{Pair cross sections and estimation of the losses in the beam pipe for  a sample of $10^6$ events generated with DIAG36 and $3\times 10^5$ with GP++.}
\begin{center}
\begin{tabular}{cccc}
\hline
\hline 
&$\sigma_{tot}$& $\sigma_{BP}$ & $\sigma_{BP}$ in $\pm100\;mrad$\\
&(mbarn)(GHz)& ($\mu$barn)(MHz) & ($\mu$barn)(MHz)\\
\hline 
DIAG36 &$7.30\pm 0.03$&$814\pm 23$&$430\pm 17$\\
GP++ LL $@$ creation& $7.74\pm 0.29$ & $618\pm40$ &$346\pm 29$\\
GP++ LL after deflection&$7.74\pm0.29$&$1027\pm52$&$389\pm31$\\
\hline
GP++ IPC $@$ creation&$7.98\pm0.29$&$627\pm40$&$347\pm29$\\
GP++ IPC after deflection&$7.98\pm0.29$&$1052\pm53$&$390\pm31$\\

\hline
\hline
\end{tabular}
\end{center}
\label{tabpairs1}
\end{table*}

\tdrsec{Touschek background}
\label{sec:mdi_touschek}

The same ingredients that characterize the crab waist scheme directly
affect the Touschek scattering, which is for this reason expected to
be relevant for the \superb\ rings, particularly for the LER.  
The Touschek
effect is Coulomb scattering of charged particles in a stored
bunch that induces energy exchange between transverse and
longitudinal motions. Small transverse momentum fluctuations lead to
larger longitudinal variations, as the effect is amplified by the
Lorentz factor.  After that, these off-momentum particles can exceed
the RF momentum acceptance or they may hit the aperture when displaced
by dispersion. In both cases they get lost.  The features of the
crab waist scheme on which \superb\ design is based are---with respect to
conventional collision schemes---the smaller emittance and the smaller
transverse beam sizes, especially at the IP.  But the achievement of
the high luminosity that this collision scheme can provide is not
enough: its use depends also on how well the backgrounds can be
controlled.  However, some remedies to counteract this effect can be
undertaken, such as implementing proper shielding to decrease the impact
of backgrounds on the detector performance.

Several studies have been performed to control and reduce the
background induced by the Touschek effect together with the beam
related lifetime, by means of a numerical code named STAR~\cite{PhysRevSTAB.15.104201} which has
been developed for $DA\Phi NE$\cite{Vignola:1993dy} and tested with the KLOE data and, more
recently, on the crab waist collision scheme~\cite{Milardi:2009zz} . \par
 At the LER and HER of
the \superb\ accelerator, the Touschek Interaction Region (IR) particle losses and lifetime
have been optimized numerically by means of a trade-off between
critical parameters, such as emittance, bunch current and bunch
length.  The Touschek scattered particles are simulated from their
generation to their loss point in the beam pipe; secondary particles
are then propagated through the IR up to all the sub-detectors.  LER
and HER Touschek simulations have been performed following the lattice
evolutions; loss rates and relative lifetime for each new lattice have
changed consistently with the lattice developments and the parameters
changes.  We present here loss rates and lifetimes for the optics design 
 named V12. This lattice has been studied in detail, especially the interaction region, which is essential for the
particle losses simulations.

Synchrotron radiation evaluation, beam stay clear constraints, and,
mostly important, physical aperture constraints given by the QD0 and
QF1 design have been implemented in the IR design, which, in turn, has
been implemented in the Touschek simulation.  The primary losses at
the beam pipe induced by this effect are used for tracking
secondaries using a Geant4
detailed model of the \superb\ detector and IR.

\begin{table}[tbp]
\caption{Summary of parameters used for the Touschek simulations for
the V12 lattice. Results depend upon horizontal emittance, which is
expected to be enlarged by Intra Beam Scattering (IBS) about 40\%}
\begin{center}			
\hspace{-2em}%
\begin{tabular}{lcc}
  \toprule
  &HER & LER \\
  \midrule
  Beam energy (GeV) & 6.7 & 4.18 \\ 
  nominal $\epsilon_{x}$ (nm) & 1.97 & 1.80 \\ 
  $\epsilon_{x}$ with IBS (nm) & 2.00 & 2.46 \\ 
  nominal $\epsilon_{y}$ (pm) & 5.00 & 6.15 \\ 
  particles per bunch & $5.08\times 10^{10}$ & $6.56 \times 10^{10}$ \\
  number of bunches & 978 & 978 \\ 
  bunch length (mm) & 5 & 5 \\ 
  RF acceptance & 0.03 & 0.04 \\ 
  coupling at full current & 0.25\% & 0.25\% \\ 
  \bottomrule
\end{tabular}%
\hspace*{-2em}\mbox{}%
\end{center}
\label{table_parameters}
\end{table}

\begin{figure*}
  \begin{center}
    \includegraphics[trim=20 20 40 40,clip=true,width=0.8\linewidth]{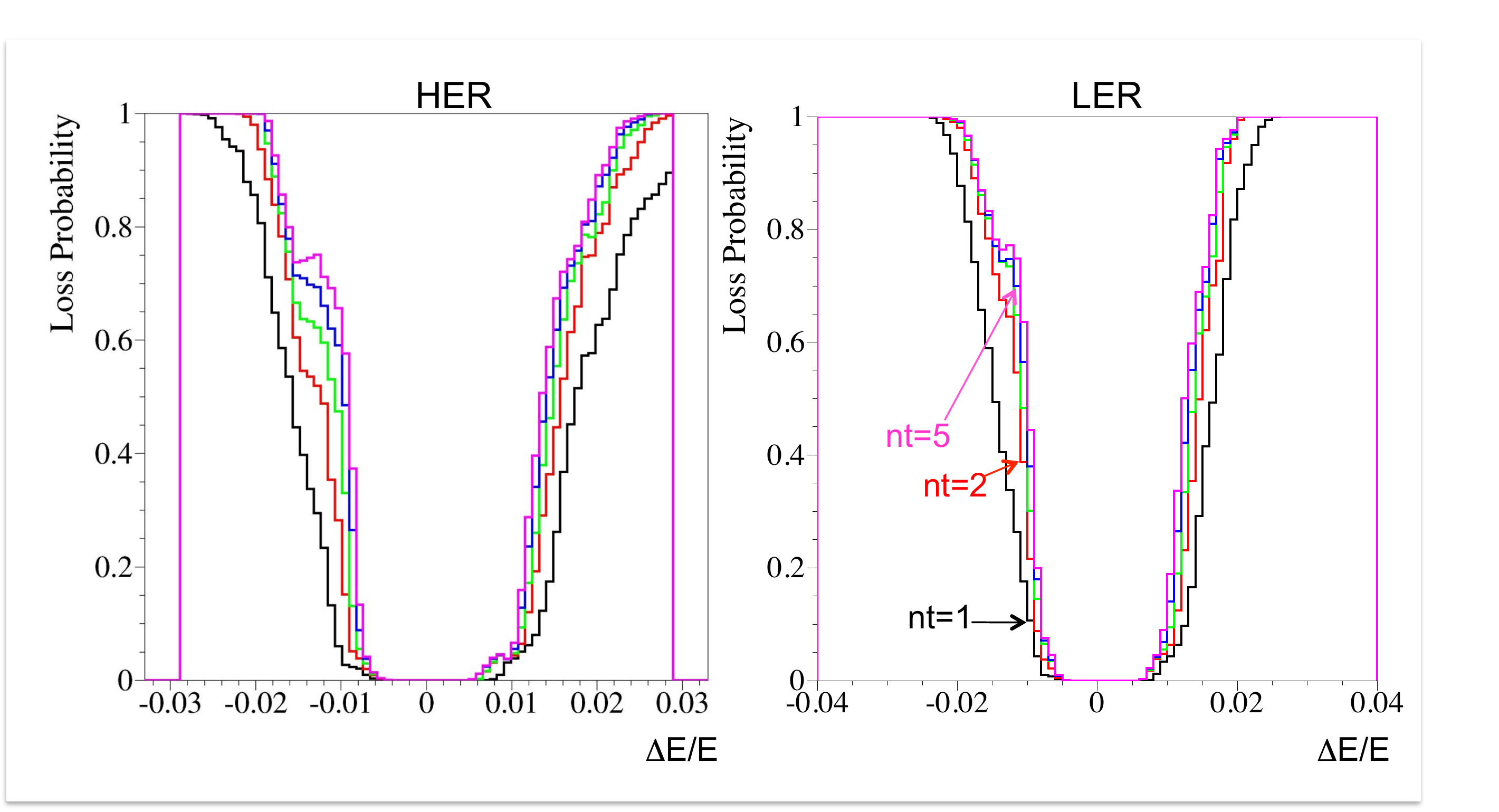}
  \end{center}
  \caption{Left: Energy acceptance of Touschek scattered particles
    for the HER (left) and LER (right) for the first five machine turns
    (nt).}
  \label{energyacceptance} \vfill
\end{figure*}

The Touschek simulation code is based on the Monte Carlo technique.
It has been developed for handling both lifetime and particle losses,
which has been very useful for understanding the critical beam parameters
and optics knobs. An accurate analysis of the critical positions where
Touschek particles are generated---mainly dispersive regions---can be
performed, together with the optimization of collimators, both for
finding the optimal longitudinal position along the ring and the
optimal radial jaw position. IR losses can be studied in detail, like
transverse phase space and energy deviation of these off-energy
particle losses as a function of different beam parameters, of
different optics and for different sets of movable collimators.  The
generation of the scattering events in the simulation code is done
continuously all over the ring. By properly slicing all the
elements in the lattice and evaluating the transverse beam size,
we obtain a good estimate of the Touschek probability density function
along the ring. We have verified that good accuracy is obtained, even in
regions where the optical functions change rapidly.  We track about
$10^{6}$ Touschek scattered macroparticles for a few machine turns; we
have checked that five are enough to have stable results.  The
Touschek lifetime can be evaluated directly from the ratio between the
initial number of particles per bunch N and the particles loss rate,
$\tau \equiv N/ \dot{N}$. A realistic tracking of the off-energy
particles includes the main non-linear terms present in the magnetic
lattice.  Figure~\ref{energyacceptance} shows the energy acceptance of
the Touschek particles for the two rings for the V12 lattice. The
plots represent the loss probability of the Touschek particles as a
function of their energy deviation, and each coloured line stands for
each machine turn.  The energy acceptance found from this technique is
in agreement with the one calculated with the MAD~\cite{Grote:1991zp} program.

\begin{figure}[tbp]
 \begin{center}
   \includegraphics[width=\linewidth]{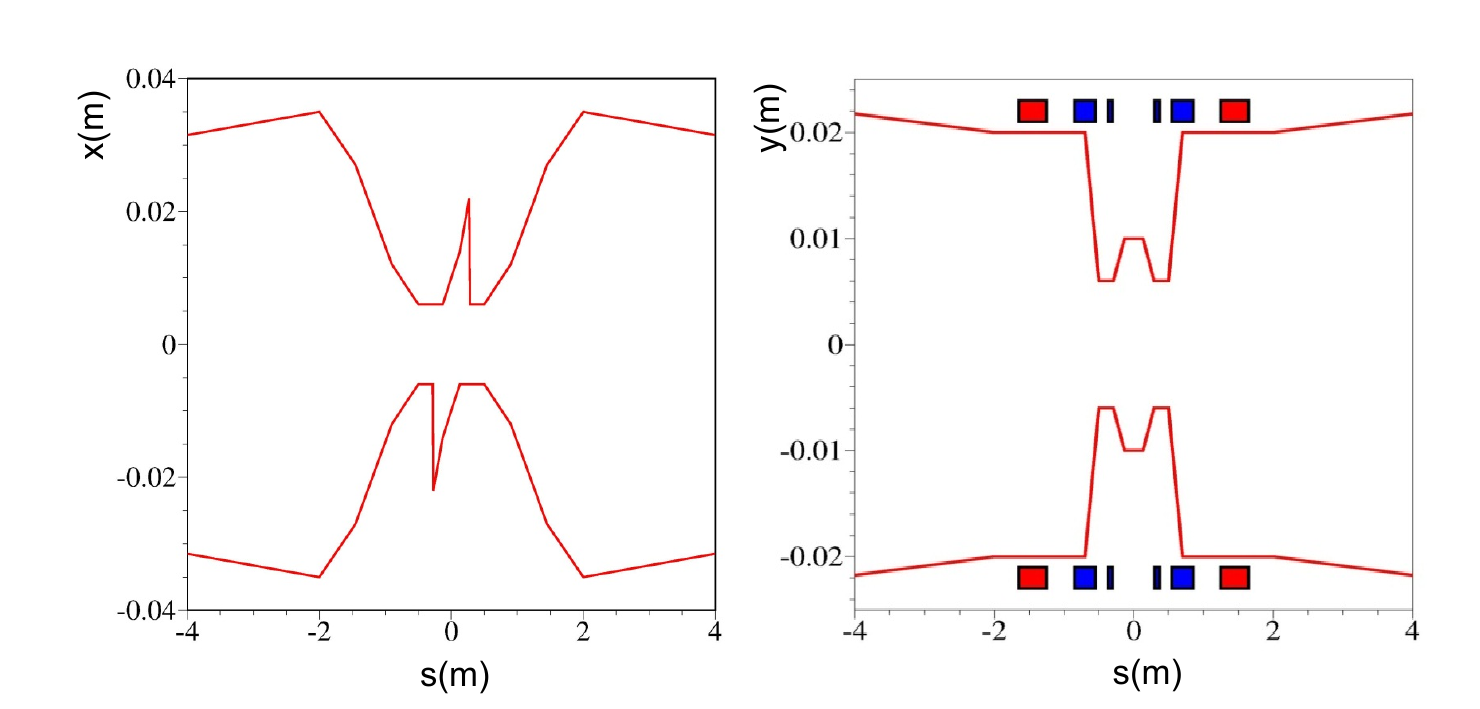}
 \end{center}
\caption{The IR dimension experienced by the HER beam moving from left to right.}
    \label{HER_beam} \vfill
\end{figure}

The physical aperture is modeled in the simulation code by a circular beam pipe with a radius
of 2.5~\cm everywhere but at the IR, where it is considered elliptical. 
In fact, in the IR the horizontal and vertical apertures have been carefully evaluated at the QD0 and at the QF1
to allow for a beam-stay-clear of 30~$\sigma_{x}$ and 100~$\sigma_{y}$ for a full coupled beam. 
In the  simulation code the beam pipe dimensions are given in terms of the aperture that the tracking particles experience
through the elements (in figure~\ref{HER_beam} is reported an example for the HER).

\begin{figure}[tbp]
 \begin{center}
   \includegraphics[width=\linewidth]{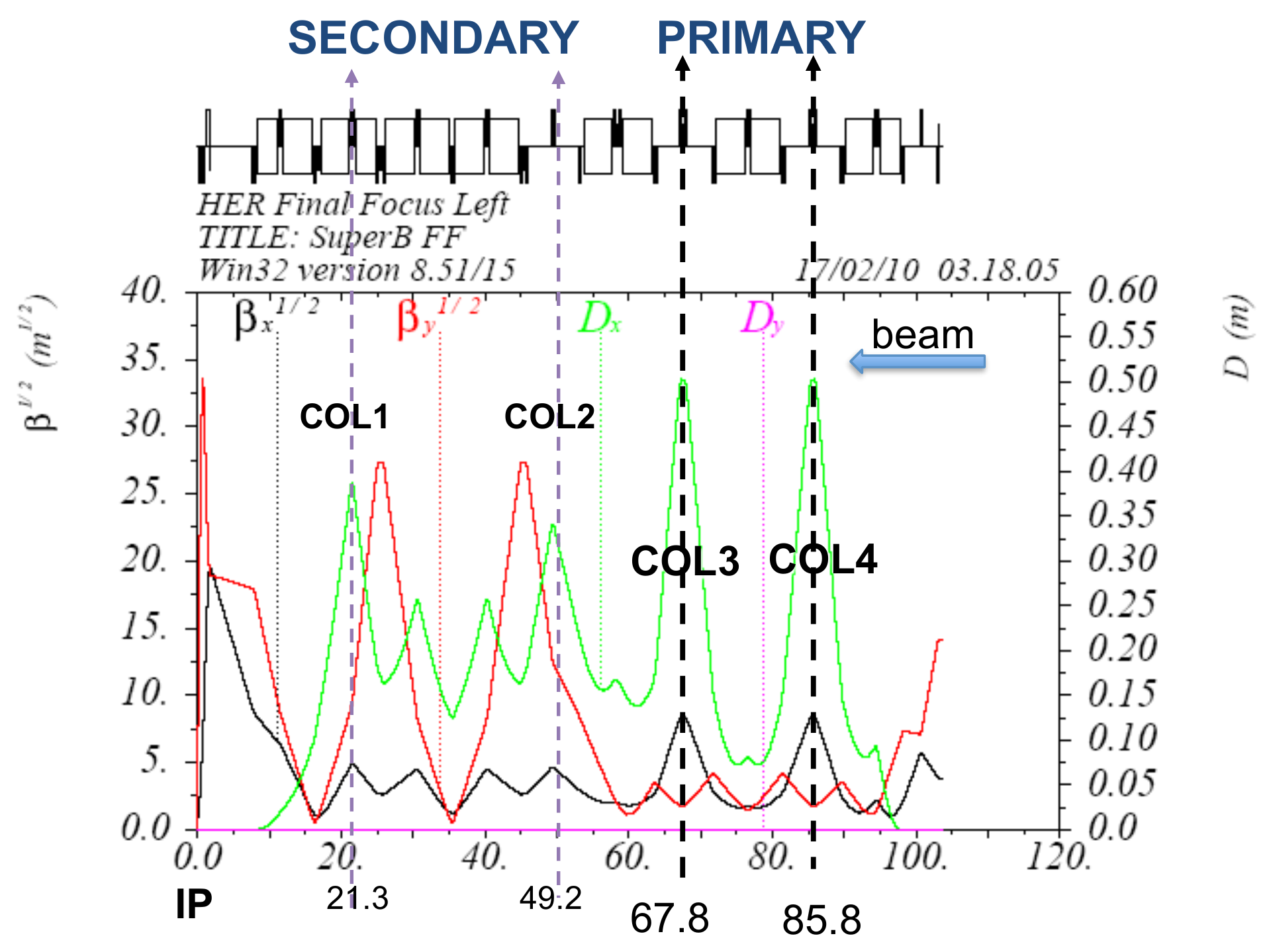}
 \end{center}
\caption{Longitudinal position of horizontal collimators upstream
the Final Focus. The beam direction is from right to left.}
    \label{collimators} \vfill
\end{figure}

 The collimator system has been designed to reduce very efficiently IR
particle losses to minimize Touschek backgrounds in the experiment.
For this source of backgrounds only horizontal collimators are
necessary.  The collimator system is shown in Figure~\ref{collimators}
together with the final focus optical functions. The beam direction is
from right to left, and 0~\m is the IP.  

Collimators are modeled in the simulation as perfectly absorbing, and
infinitely thin. This is a good approximation, as the closest
collimator upstream IP is at $-20\m$.  The primary collimators are in
the Final Focus upstream of the IR at $-85.8\m$ and $-67.8\m$, stopping
most of the energy-deviated particles that otherwise would be lost at
the IR. These two primary collimators are close to two horizontal
focusing sextupoles and their location corresponds to a maximum of the
dispersion and of the $\beta_{x}$ function.  The secondary collimators
are positioned at $-49.2\m$ and $-21.3\m$.  
Simulation indicate that 
this system is very
efficient, reducing IR losses by a factor of
about 350 for the HER and about 230 for the LER.  This happens at the
expenses of Touschek lifetime, which is reduced correspondingly by a
factor about 1.2 for the HER and 1.4 for the LER.  However, the two
horizontal jaws of the collimators have been optimized by simulations
to obtain a good trade-off between IR losses and lifetime. 
Note that the correct position of the jaws will be found
experimentally. Also note that in the simulation the orbit is
perfectly centered on the beam pipe.

\begin{figure*}[tbp]
 \begin{center}
   \includegraphics[width=0.85\linewidth]{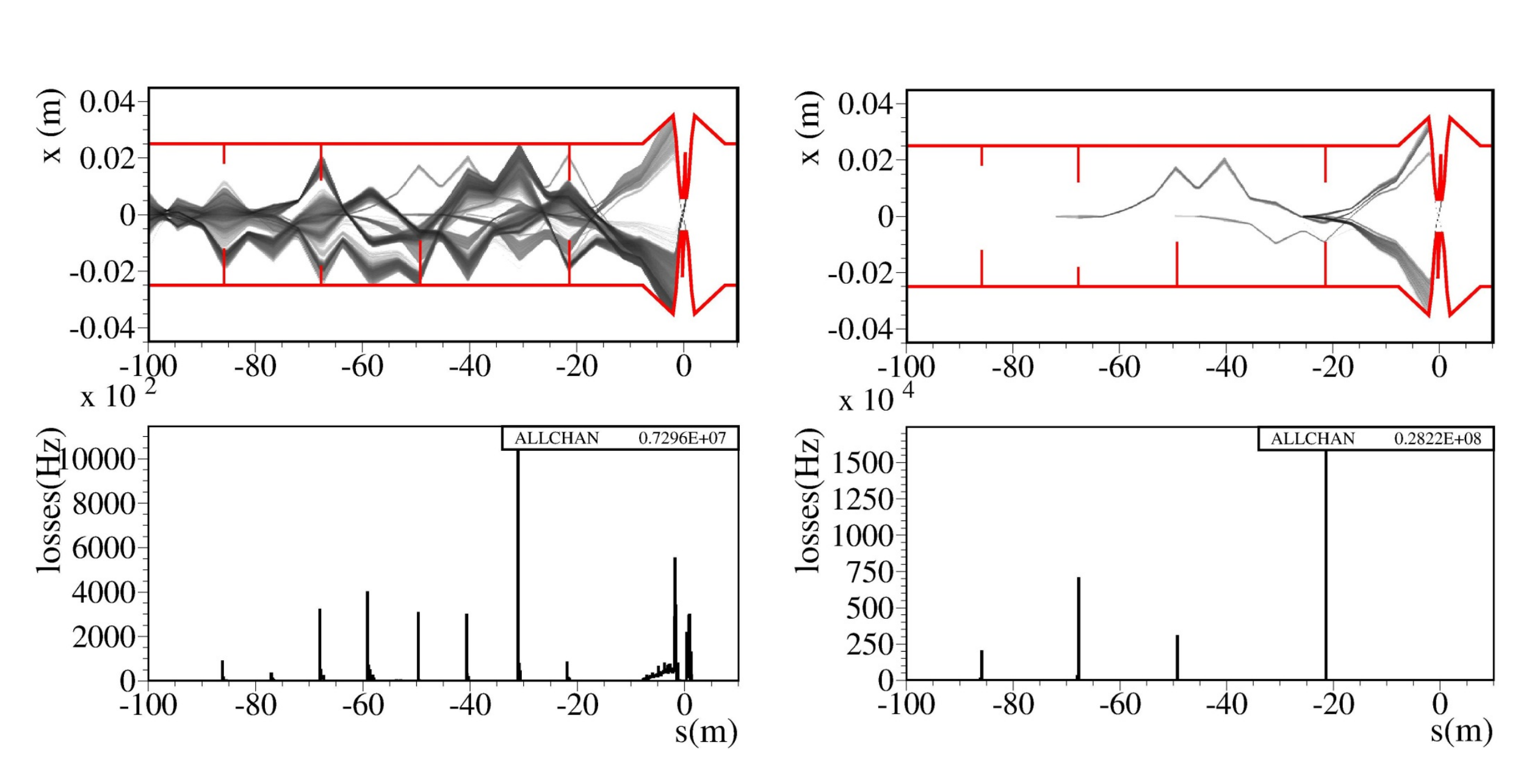}
 \end{center}
 \caption{Upper plots: HER Touschek trajectories of particles
   eventually lost at IR without (left) and with (right) collimators.
   Lower plots: distribution of particle losses in the same region,
   showing the collimators effectiveness. IP is at 0~m.}
 \label{her_traj}
\end{figure*}

The effectiveness of the
collimator system is illustrated in the two lower plots of
Figure~\ref{her_traj}, where HER loss distributions are shown without
and with the insertion of collimators. The trajectories of the particles
that are eventually lost at the IR are shown in the
upper plots of Figure~\ref{her_traj}.

The technical solution we propose for the two primary collimators
design is to place masks at the two horizontal sextupoles, to
intercept the Touschek energy deviated particles that would be lost at
the QF1.  This is obtained simply by placing masks that give at these
locations the same beam stay clear present at QF1.  We then propose to
add two movable jaws close to these sextupoles as further knobs to
tune IR backgrounds.

\begin{figure}[tbp]
  \begin{center}
    \includegraphics[trim=0 40 70 70,clip=true,width=0.9\linewidth]{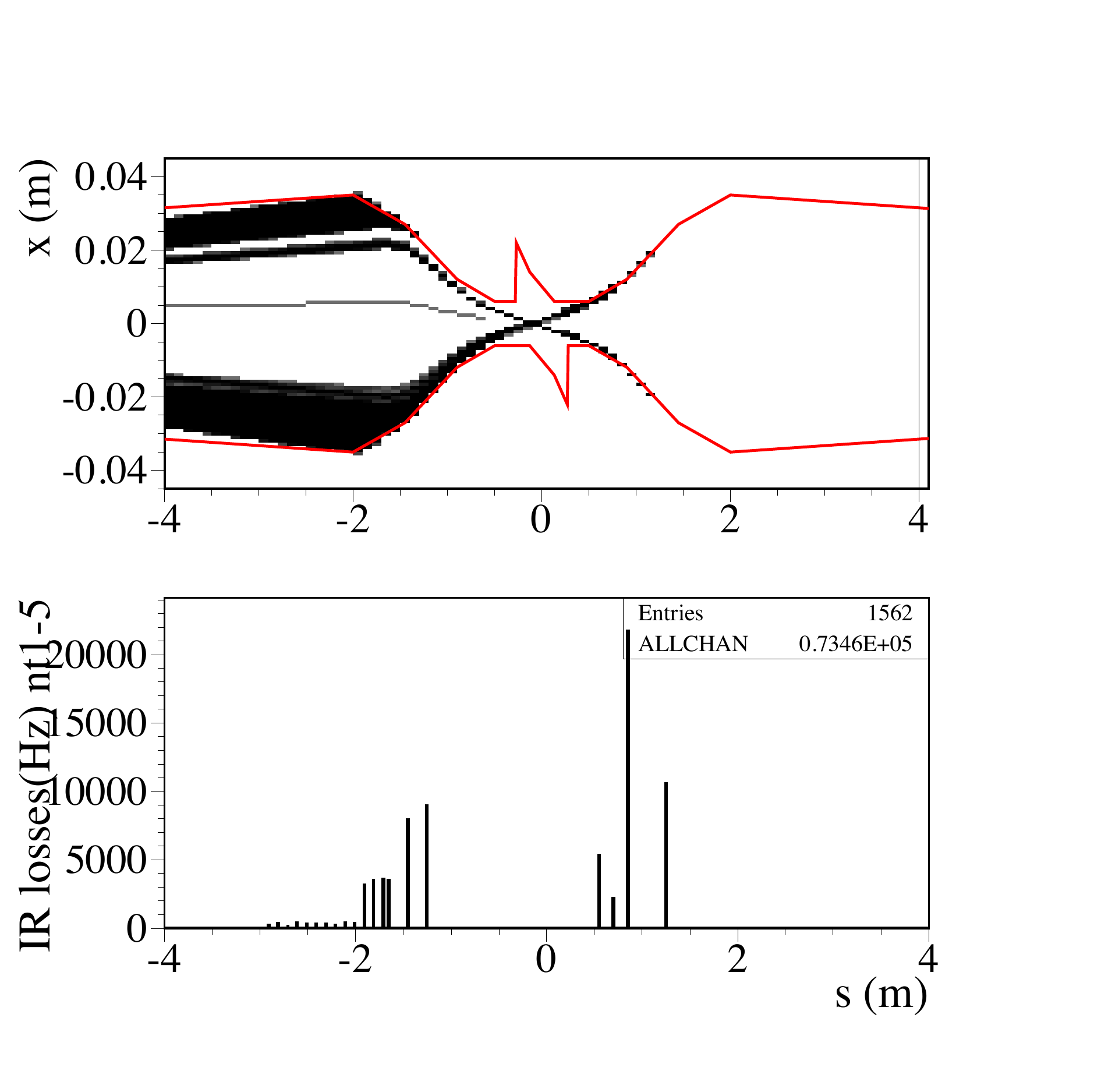}
  \end{center}
  \caption{LER IR Touschek trajectories with collimators at the
    optimized locations. Lower plot: corresponding distribution of
    particle losses at the IR.}
  \label{ler_ir_traj} 
\end{figure}

\begin{figure*}[tbp]
 \begin{center}
   \includegraphics[width=0.85\linewidth]{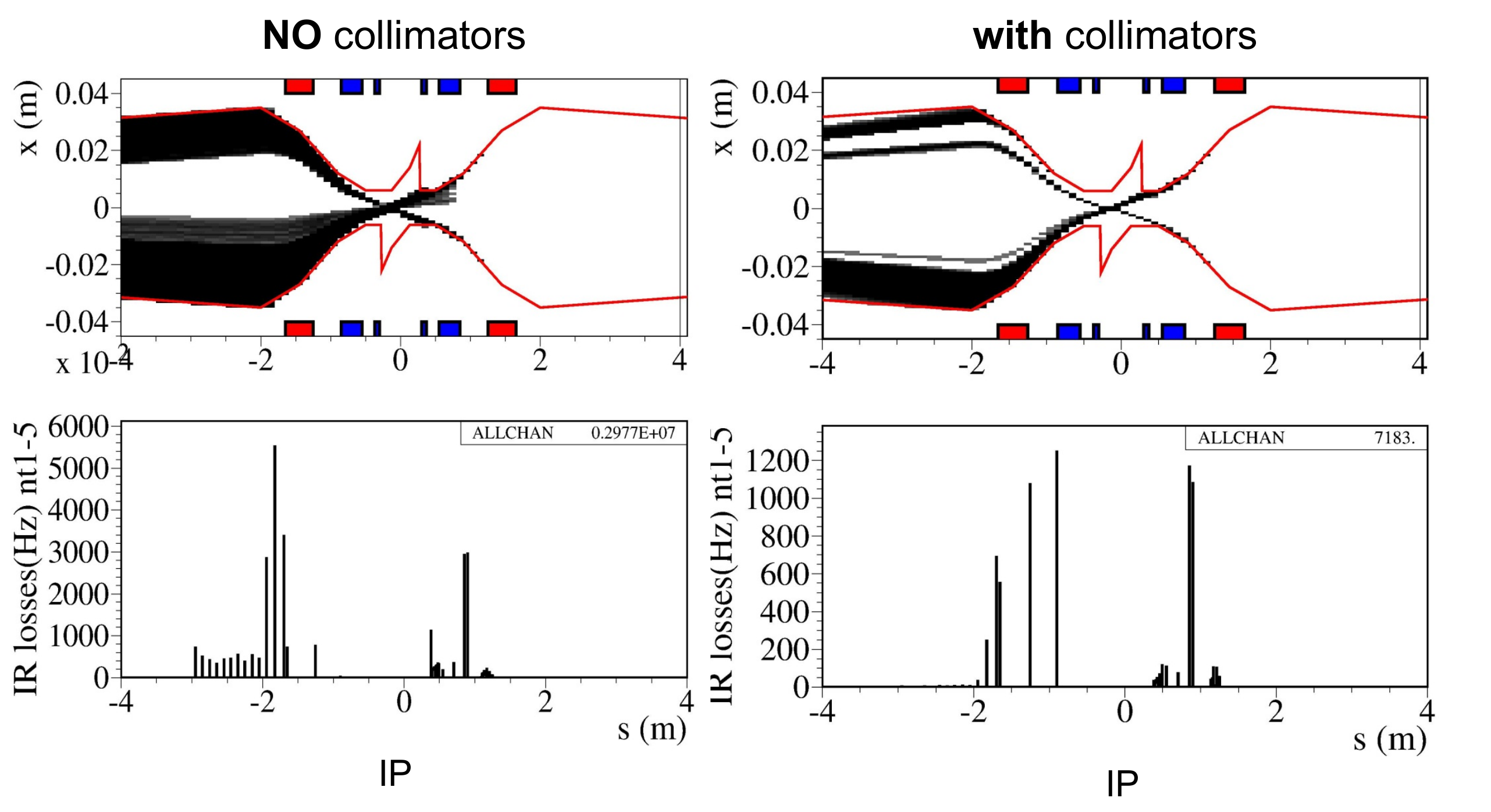}
 \end{center}
\caption{Upper plots: HER Touschek trajectories zoomed at the IR
without (left plots) and with (right plots) collimators.  Lower plots:
corresponding distribution of particle losses. IP is at $0\m$}
    \label{her_ir_traj} \vfill
\end{figure*}

The trajectories of the Touschek particles that are eventually lost at
the IR are shown for the HER (Figure~\ref{her_ir_traj}) and for the
LER (Figure~\ref{ler_ir_traj}), with a zoomed IR view between $-4\m$
and $+4\m$ the IP.  Both beams direction in the plots is from left to
right, and the physical aperture is the one actually seen by the beam.
In this way the actual beam stay clear is taken into account in the
tracking simulation.  For the LER, the figure shows only the results with 
collimators.  The HER figure shows results both and without to illustrate
their effectiveness.  The simulations also include the vertical collimators
optimized for stopping beam gas scattered particles (Sec.~\ref{sec:mdi_beamgas}).

\begin{figure}[tbp]
  \begin{center}
    {\includegraphics[trim=0 20 70 10,clip=true,width=\linewidth]{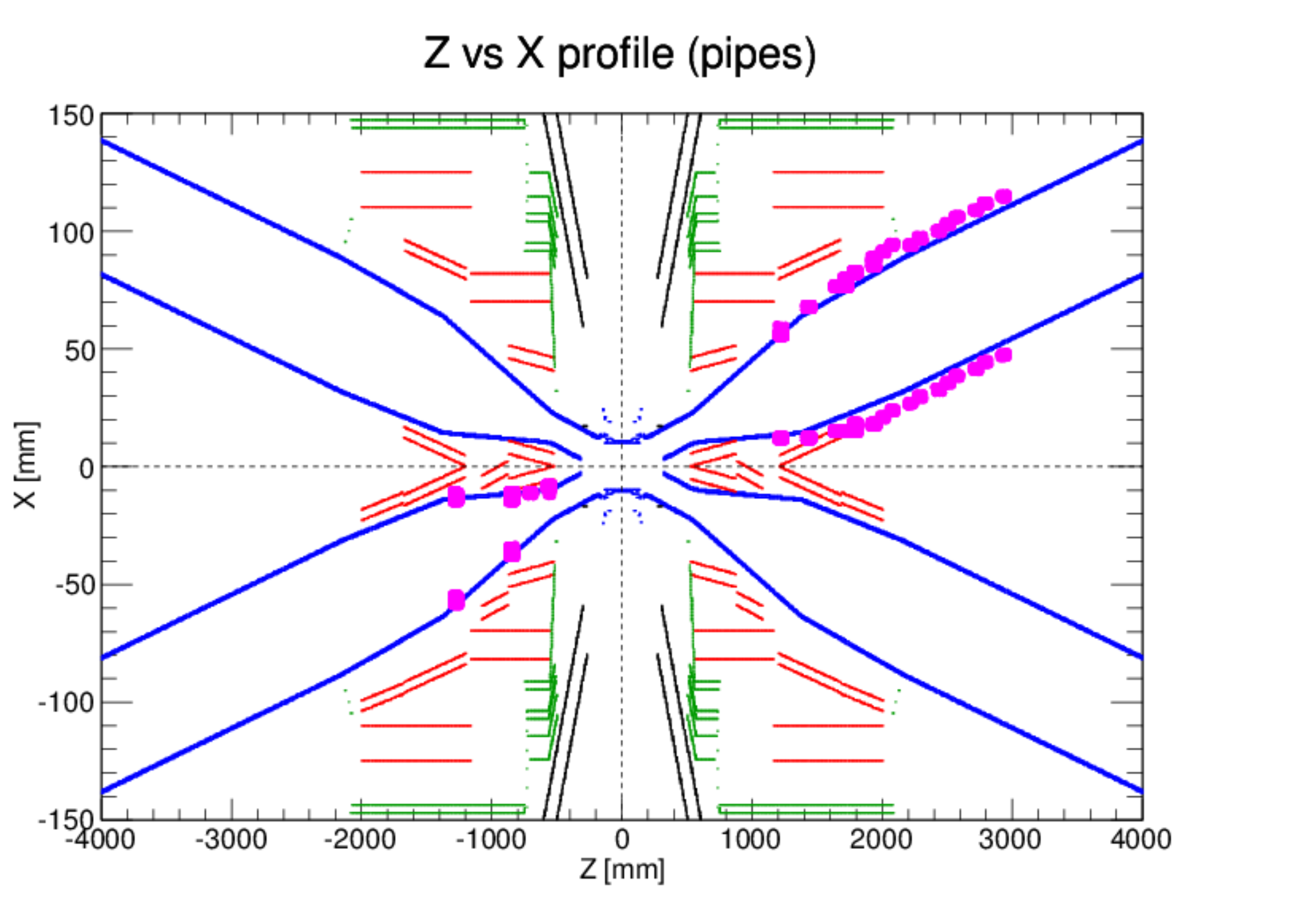}}
  \end{center}
  \caption{x-z profile of the beam pipe and IR with the LER loss
    position of the Touschek primaries.}
  \label{touschek_primaries} \vfill
\end{figure}

The two figures also show the corresponding
losses distribution at the IR.  These particle losses are tracked into the
detector for further investigations. These positions correspond to the
locations where Touschek primaries get lost and secondary particles
are produced. Figure~\ref{touschek_primaries} shows the
locations in the LER where secondaries are generated.
 
The background rates seen by the detectors are evaluated by
tracking the secondary particles produced by the interaction of the
lost primary particles using the full Geant4 simulation.
HER and LER particle losses are summarized for the different
cases in Table~\ref{table_tou_rates}.
Corresponding lifetime predictions are in
Table~\ref{table_tou_lifetimes}.

\begin{table*}[ht]
\caption{Summary of Touschek particle losses predicted by numerical
simulations for the V12 lattice for the HER and LER in the IR within
$2 \m $ for the nominal beam current.}  \vspace{0.2cm} \centering
			\begin{tabular}{l c c }\hline & HER & LER \\
\hline \hline 
 & Loss rate/beam/4m (GHz) & Loss rate/beam/4m (GHz) \\ 
no collimators, $\epsilon_{x}$ with IBS & 2.4 GHz & 17 GHz \\ 
with collimators, $\epsilon_{x}$ with IBS & 6.8 MHz & 72 MHz \\ 
\hline \hline
		\end{tabular}
	\label{table_tou_rates}
\end{table*}

\begin{table*}[hbt]
\caption{Summary of Touschek lifetimes predicted by numerical
simulations for the V12 lattice for the HER and LER.}  \vspace{0.2cm}
\centering
			\begin{tabular}{|c|c|c|}\hline & HER & LER \\
\hline & Touschek lifetime (minutes) & Touschek lifetime(minutes) \\
\hline no collimators, nominal $\epsilon_{x}$ (no IBS)& 26 & 7.4 \\
\hline no collimators, $\epsilon_{x}$ with IBS & 26 & 10.2 \\ \hline
with collimators, $\epsilon_{x}$ with IBS & 22 & 7.0 \\ \hline

		\end{tabular}
	\label{table_tou_lifetimes}
\end{table*}

KLOE experiment data and numerical studies have been compared, 
showing an overall agreement within a factor of
two. The shapes of the distributions are well described. 
The agreement between measured and
calculated lifetime with scrapers inserted is good, while the one
without scrapers show a disagreement of a factor of two. This 
could be the result of beam scraping in the opposite ring crossing section; the simulation
assumes the beam is perfectly aligned and centered. 

\begin{figure}[tbp]
 \begin{center}
   {\includegraphics[trim=20 140 30 140,clip=true,width=\linewidth]{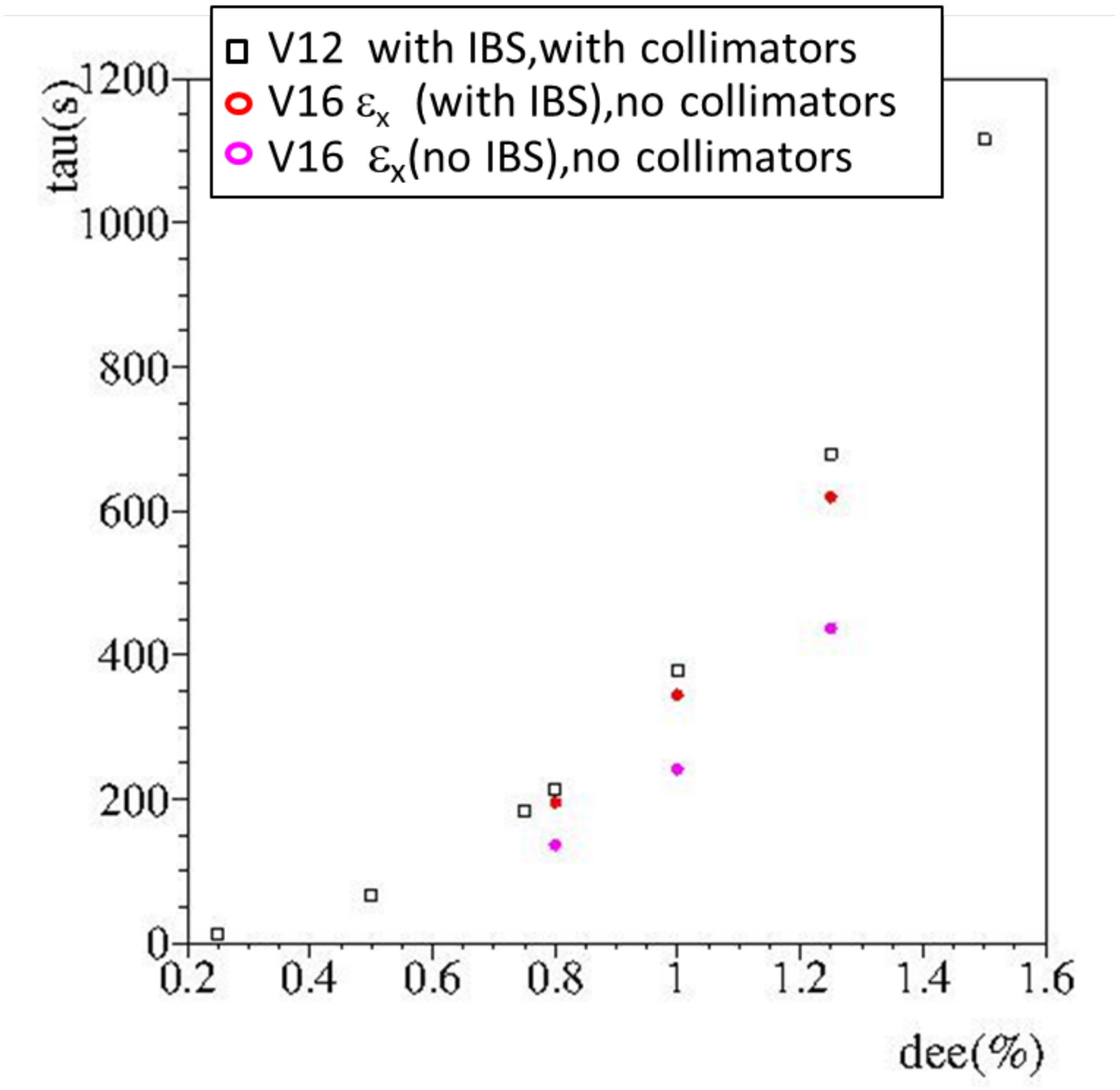}}
 \end{center}
 \caption{LER Touschek lifetime as a function of beam energy
   acceptance for the V12 lattice (black squares) and V16 (circles).}
 \label{tau_vs_dee}
\end{figure}

To conclude, background rates at the IR due to Touschek scattering are
under control with the proposed horizontal collimation system in the
final focus.  The radiative Bhabha remains the limiting effect for the
lifetime. However, LER Touschek lifetime is close to this lower limit,
so care has to be paid to this effect and, in particular, to the
dynamic aperture.  Dedicated simulations have been performed to see
the impact of the energy acceptance on the Touschek lifetime.  These
results have been obtained without tracking, evaluating lifetime while
imposing the energy acceptance as input.  Figure~\ref{tau_vs_dee}
shows the prediction of the LER Touschek lifetime as a function of the
energy acceptance for the V12 lattice with emittance enlarged by IBS
and with collimators installed. A similar result is obtained for the
newer lattice V16.  When the IBS effect on emittance is not taken into
account, lifetime drops. To have a lifetime larger than 5 minutes, the
energy acceptance is required to be larger than 0.9\%. This energy
acceptance is required even when crab sextupoles are switched on.  We
remark that earlier optical models have had the crab
sextupoles switched off.  These calculations show that LER dynamical
aperture and energy acceptance are crucial for the LER Touschek
lifetime, and that, if not large enough, this single beam effect could
be the lower limit for the beam lifetime.

\tdrsec{Beam gas background}
\label{sec:mdi_beamgas}

Beam gas scattering is a single beam effect that affects lifetime
and induces backgrounds in the detectors.  At \superb\ this effect has
been thoroughly studied with the same Monte Carlo approach used for the
Touschek effect.  Bunch particles are scattered with the residual gas
ions that are present in the vacuum chamber and can get lost.  The
better the vacuum, the lower will be this effect. However, as a
perfect vacuum is not feasible, we need to determine the requirement
on the vacuum in order to control lifetime and particle losses at a
sustainable level.  This beam gas interaction can be elastic or
inelastic.  In the first case the bunch particle is transversally
deflected without losing energy; it increases its betatron oscillation
amplitude with the possibility of getting lost, if the gained
scattering angle is large enough to exceed the physical aperture or
the dynamic aperture. We call this process Coulomb beam gas
scattering.  In the second case during the scattering with a rest gas
atom, the bunch particle also loses energy. Two processes can take
place in this case, either Bremsstrahlung, in which the particles
emits a photon, or inelastic scattering, in which the bunch particle
transfers energy to the atom of the rest gas.  Both processes have
been included in the inelastic beam gas simulation.

Dedicated Monte Carlo simulations have been performed to evaluate
elastic and inelastic scattering, and particle losses and relative
lifetime have been found for the V12 lattice. Beam gas scattering is
generally estimated from the integrated cross-section, but we have
developed a Monte Carlo simulation code 
that uses the process differential cross-sections and tracks the
scattered particles.  This technique enables the detailed prediction of
the locations where losses occur by hitting the beam pipe, the
prediction at which tracking turn these particles are lost and the
optimization of a collimator set that intercepts them before they get
lost at the IR.  As for Touschek simulation, the physical
aperture is modeled by a circular beam pipe with a radius of $2.5\cm$
everywhere but at the IR.  In the simulations
presented here a constant pressure of $1\,\mathrm{nTorr}$ has been
assumed, with a gas composition of $Z=8$.
 
The bunch particles scattered for the gas Bremsstrahlung process have
a behaviour similar to the Touschek particles that lose energy in the
scattering.  The particles undergoing inelastic scattering are lost by
exceeding either the RF bucket or the physical and/or dynamic
aperture.  We note that when using the integrated cross-section, only
the first loss source is taken into account.  Although most of the
losses happen in the first tracking turn, ten turns have been
considered for completeness.  The bunch particle that undergoes an
elastic scattering increases its betatron oscillation amplitude, so
the particle may get lost after some machine turns, as confirmed by
the Monte Carlo tracking simulation.  The calculated lifetime for the
inelastic scattering is in good agreement with the standard integrated
cross-section method. 

\begin{table*}
\caption{\label{table_beamgas_rates} Summary of elastic and inelastic beam gas particle losses for
the nominal beam current predicted by numerical simulations for the
V12 lattice for the HER and LER in the IR (from $-2~m$ to $+2~m$ from
IP .}  \vspace{0.2cm} 
\centering
\begin{tabular}{c c c}
\hline 
       & HER & LER \\
\hline \hline
  & Loss rate/beam (MHz) & Loss rate/beam (MHz) \\   
Coulomb no collimators, $\epsilon_{x}$ with IBS & $1.05\cdot 10^{4}$ & $2.5\cdot
10^{4}$ \\ 
\hline 
Coulomb with collimators, $\epsilon_{x}$ with IBS & 3.7 & 20 \\ 
Inelastic no collimators, $\epsilon_{x}$ with IBS & 0.6 & 1.0 \\ 
Inelastic with collimators, $\epsilon_{x}$ with IBS & 0.1 & 0.4 \\ 
\hline \hline
\end{tabular}
\end{table*}

The expected particle loss rates due to elastic beam gas scattering
are much higher than the inelastic one, and, accordingly, the lifetime
is much shorter (Tables~\ref{table_beamgas_rates} and
\ref{table_beamgas_lifetimes}). This result is an outcome of the crab
waist collision scheme, particularly for the small vertical beam size
in QD0 at the IR, which is the hottest location for this
process.

\begin{table*}
\caption{Summary of elastic and inelastic beam gas lifetimes predicted
by numerical simulations for the V12 lattice for the HER and LER.}
\vspace{0.2cm} \centering
			
\begin{tabular}{c c c}
\hline 
   & HER & LER \\
\hline \hline
   & lifetime (seconds) & lifetime (seconds) \\ 
Coulomb no collimators, $\epsilon_{x}$ with IBS & 4590 & 2520 \\ 
Coulomb with collimators, $\epsilon_{x}$ with IBS & 3040 & 1420 \\ 
Inelastic no collimators, $\epsilon_{x}$ with IBS & $\sim 72$hours & $\sim 77$ hours \\ 
Inelastic with collimators, $\epsilon_{x}$ with IBS & $\sim 72$ hours & $\sim 77$ hours \\ 
\hline \hline 
		\end{tabular}
	\label{table_beamgas_lifetimes}
\end{table*}

Beam gas Coulomb background is proportional to vacuum pressure and to
beam current, but also to the betatron function values in the
scattering location, as the scattering angle gained by the bunch
particles interacting with a residual ion is directly proportional to
these values.  Therefore, a large fraction of the losses is found to be at
QD0 of the IR, corresponding to the maximum $\beta_{y}$ position.  To
cope with this loss source, two vertical collimators have been
implemented in the final focus upstream the IR.  
Only vertical collimators are necessary for this
source of backgrounds. 
However, for a realistic simulation, we have also included the
horizontal collimators that intercept Touschek particles.

\begin{figure}
 \begin{center} 
   {\includegraphics[trim=45 10 40 35,clip=true,width=\linewidth]{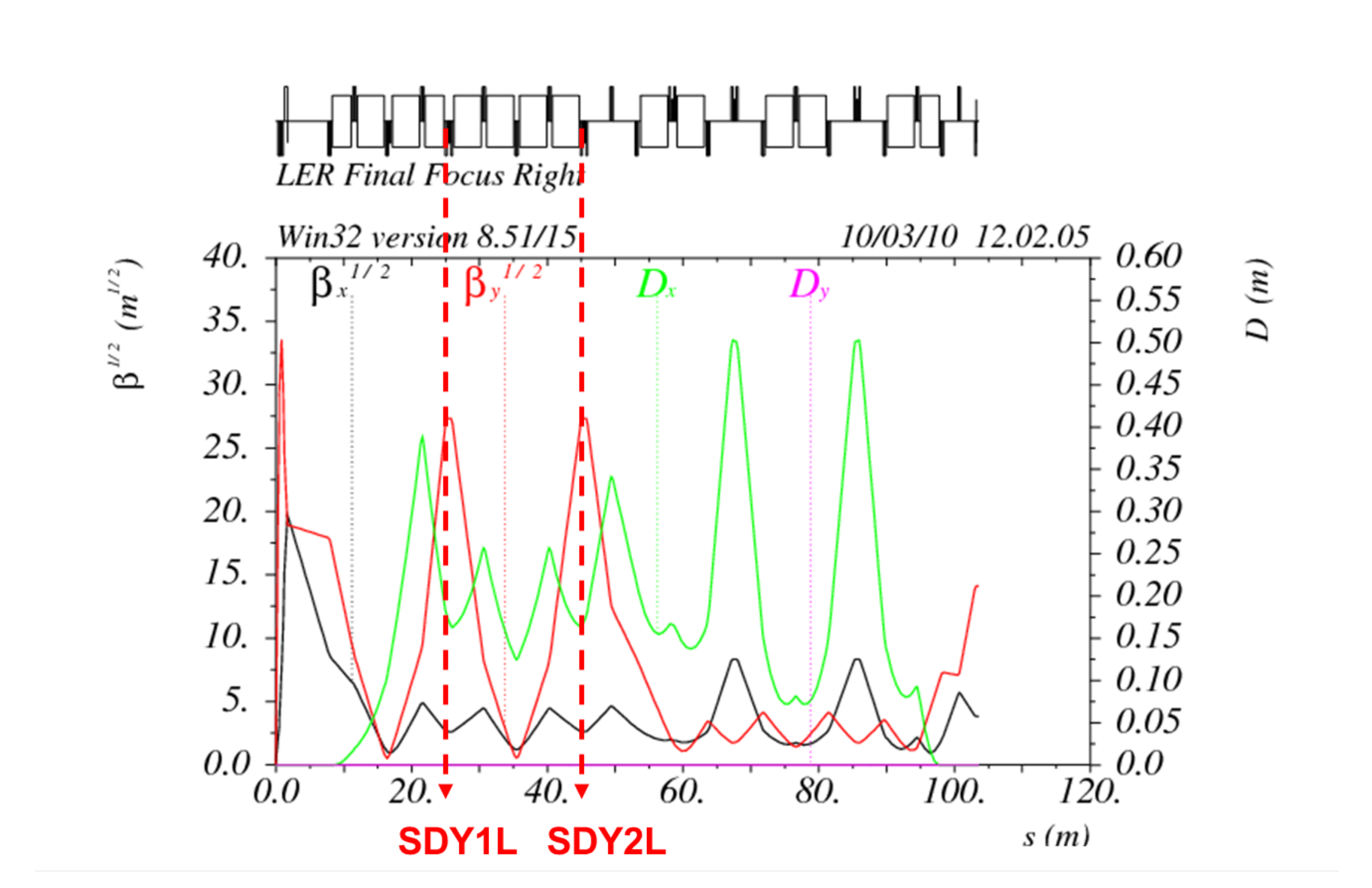}}
 \end{center}
\caption{Longitudinal position of vertical collimators upstream
the Final Focus (red dashed lines). The beam direction is from right to left.}
    \label{verticalcollimators} \vfill
\end{figure}

\begin{figure}
 \begin{center} \includegraphics[clip=true,trim=40 230 40 230,width=\linewidth]{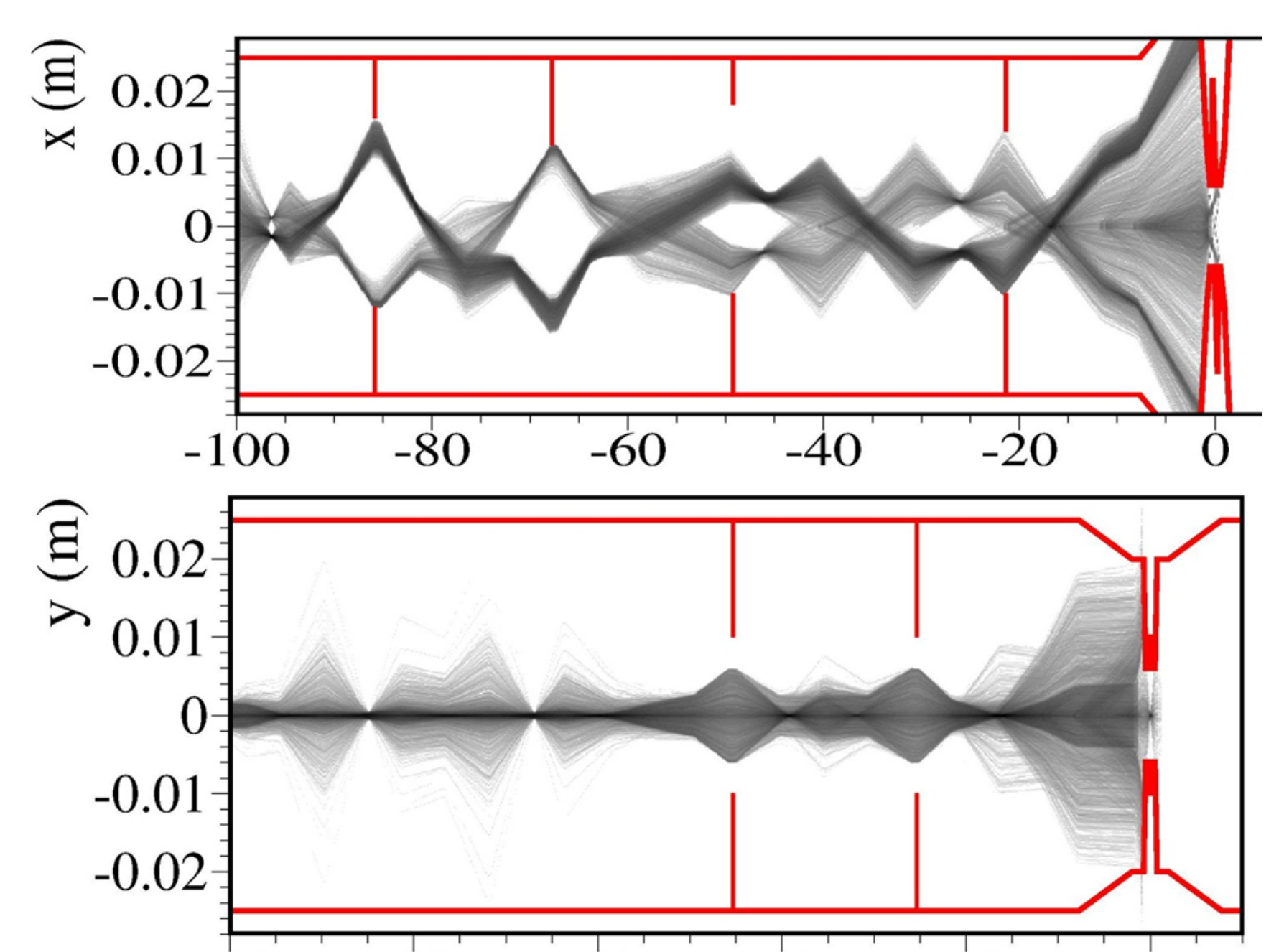}
 \end{center} 
\caption{Horizontal (upper) and vertical (lower) trajectories of
Coulomb gas scattered eventually lost at IR; horizontal (for Touschek)
and vertical collimators are inserted}
    \label{coulomb_traj_x_y} 
\end{figure}

The collimator system is shown in Figure~\ref{verticalcollimators},
together with the final focus optical functions. The beam direction is
from right to left, and the IP is at $0\m$. Collimators are modeled in the
simulation as perfectly absorbing, and infinitely thin.  These two
collimators are close to two vertical focusing sextupoles and their
location corresponds to a maximum of the $\beta_{y}$ function, at
about $-45\m$ and $-25\m$ far from IP.  
Simulation indicates that this system is very 
efficient, reducing IR beam gas losses by a 
large factor (Table~\ref{table_beamgas_rates}).  This happens at
the expenses of Touschek lifetime, which is reduced by
about a factor 1.5 for the HER and 1.8 for the LER.  The technical
solution we propose for the two vertical collimators is to
place masks at the two vertical sextupoles, which intercept the
scattered particles that otherwise would be lost at QD0.  This is
achieved by  masks that give at these locations the same
beam stay clear that is present at QD0.  We also envisage two movable jaws
close to these sextupoles as further knobs to tune IR backgrounds.
Examples of the HER trajectories of the beam gas Coulomb scattered
particles that are eventually lost at the IR are shown in
Figure~\ref{coulomb_traj_x_y}. 

\begin{figure}[tbp]
 \begin{center} \includegraphics[clip=true,trim=20 20 50 45,width=\linewidth]{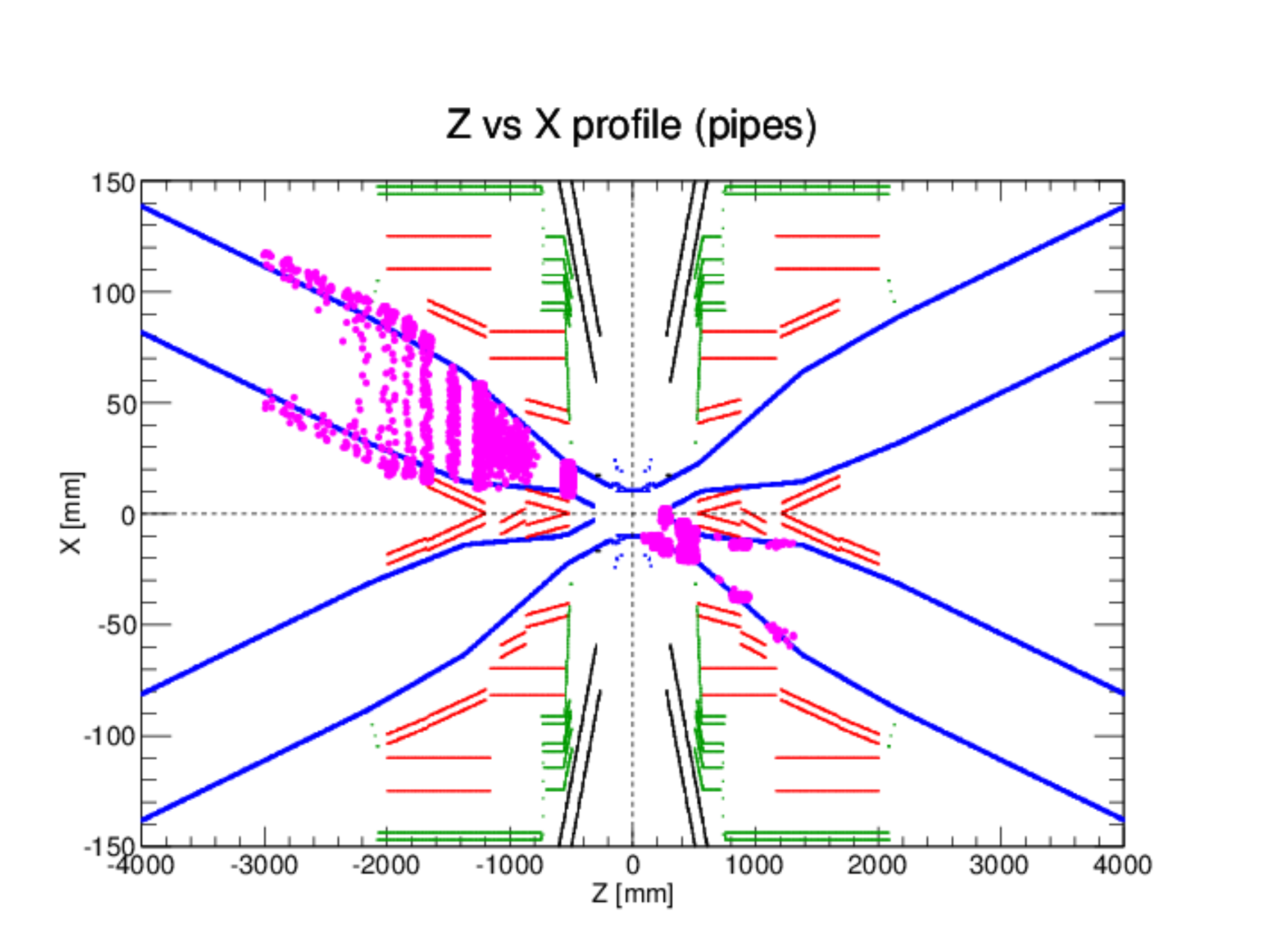}
 \end{center}
\caption{HER Coulomb loss particles along the beam pipe at the IR,
shown in the horizontal-longitudinal dimensions.}
    \label{her_coulomb_primaries} 
\end{figure}

A zoom of the IR (Figure~\ref{her_coulomb_primaries}) shows the
the loss position on the beam pipe of these particles.
The full Geant simulation is then used to track the resulting secondaries. 

In conclusion, elastic and inelastic beam gas backgrounds have been
analyzed in detail, showing that these effects are not the main
source. As well, the related lifetime is not the limiting one.
However, attention has to be paid especially to LER beam gas
Coulomb scattering, as the high loss rates expected at QD0 could
be dangerous.  

\tdrsec{SVT background overview}
\label{sec:mdi_svt_bkg}

The background in the detector environment affect the SVT in two
separate ways. First, a hit rate higher than the design value can
deteriorate the general performances of the tracking system.  Second,
the silicon layers can be progressively damaged when being exposed to
a significant rate of particle for a long period of time.  In both
cases, accurate simulation of the rates is needed to estimate the
performance and the life-span of the SVT.

The main tool used to estimate the rate in the SVT is the simulation
of the various background sources using the full simulation described
in Sec.~\ref{sec:mdi_RadBhaBha_bkg_simTools}.

In the full simulation, the silicon layers are implemented as
homogeneous volumes of silicon, and each particle passing through
those volumes is recorded, together with additional information about
the type of particle, the position, the incident energy, the vector
momentum, and the deposited energy.  The simulation is not aware of
the detector segmentation (strips or striplets).  This choice, even if
intensive from the data output point of view, gives us some
flexibility in testing different configurations without running the
full simulation multiple times.  The raw simulation output needs to be
further processed to obtain the quantities useful for the tracker
design and performance evaluation, such as the rate of ``fired'' strip
per \cma, the radiation dose, and the 1~\mev neutron equivalent
fluence.

\begin{figure}[tbh]
 \begin{center}
   \includegraphics[width=\linewidth]{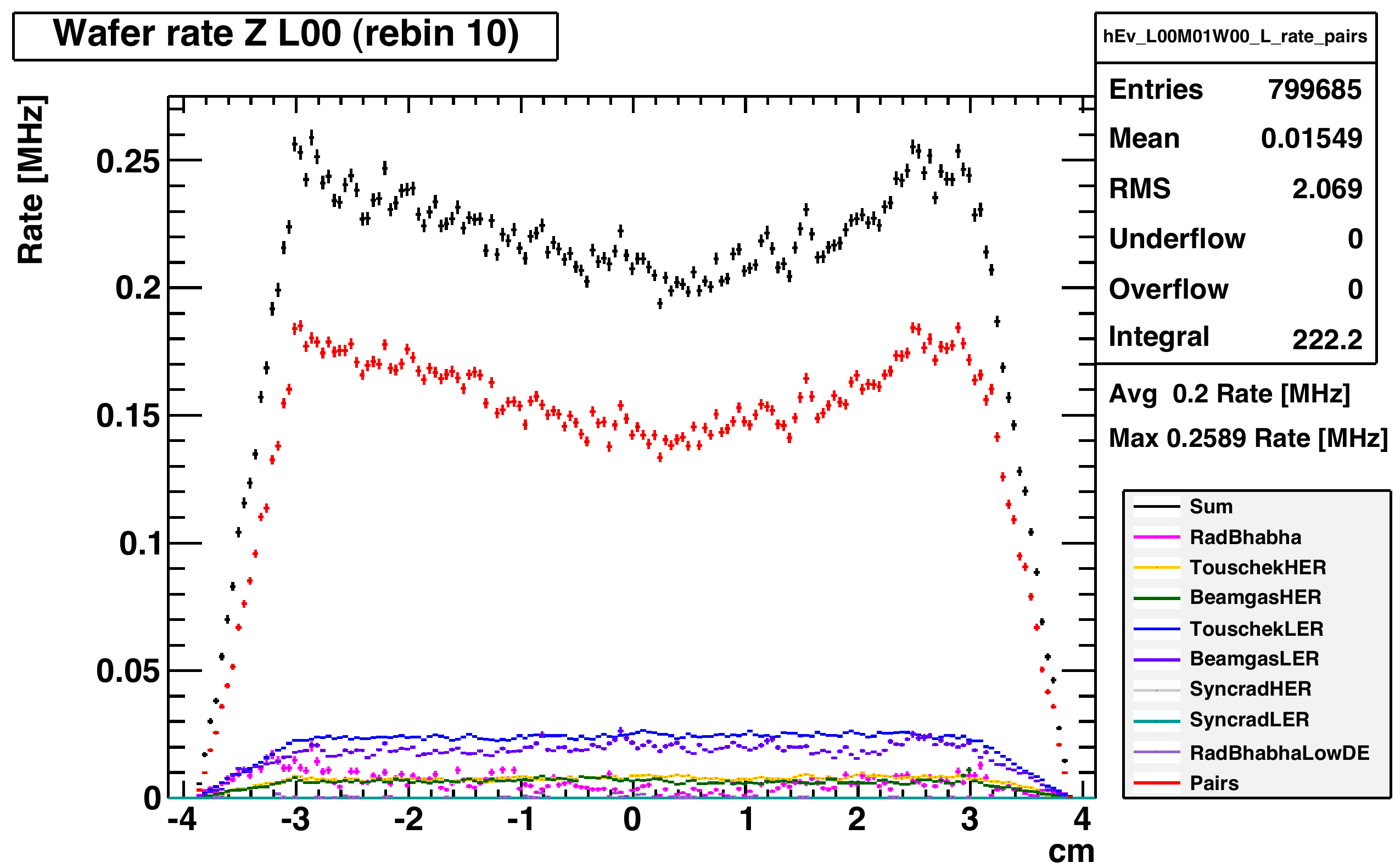}	
 \caption{Hit rate for the striplet Layer0 option as a funcion
of the longitudinal position. The contributions from different
background is represented in color code and explained in the legenda. }
 \label{fig:mdi_svt_localRate}
 \end{center}
\end{figure}

A specific application has been develop to analyze the simulation
output and produce the relevant quantities and plots. Due to the
similarities with other sub-systems, this application has been also
extended using multiple modules to produce the same or similar
quantities for other sub-detectors.  The basic information in the
simulation output corresponds to a single step of the particle in
GEANT4.  A crossing of the volume is defined as a particle entering
and exiting the volume once, or as a particle entering the volume and
destroyed inside, or as a particle created inside the volume and then
exiting, or as a particle created and destroyed inside.  Each crossing
has a well-defined starting and end point, an incident energy, and a
deposited energy.  Assuming a specific pitch of the strips (or
striplets) in each layer, and approximating the trajectory of the
particle with a straight line from the starting point to the end
point, we can compute the number of strips (or striplets) crossed by
the particle in both directions.  Counting the number of strips (or
striplets) crossed in all the simulated events, and rescaling by the
event frequency, gives a measurement of the rate for each silicon
module.  Figure~\ref{fig:mdi_svt_localRate} shows the 
rate in Layer0, with the 
different
background sources reported separately and summed together.

The rate variations observed as a function of the position can be
explained with geometrical or topological considerations.  For all the
layers the largest contribution to the background in the tracker is
given by the pairs production process, where the particles hitting the
silicon are coming directly from the IP.  This contribution ranges
from $\approx 80\%$ for the innermost layer to $\approx 50\%$ for the
outermost one. The production of
secondaries with low transverse momentum is highly favoured for this
process, and the detector magnetic field prevents those particles
from reaching the outer layers.  The rate is quite high, but it is
substantially irreducible, given that it scales
with the luminosity and no shielding can be added in between the IP
and the SVT.

Together with the strip rate, we can easily compute also the crossing
multiplicity and the distribution of deposited energy per strip. All
those data are a fundamental input for getting a reliable simulation
of the readout electronics, as detailed in
Sec.~\ref{sec:svt_rd_strip_readout_chips}.  A similar approach can be
used also for the other relevant variables: summing together the
deposited energy by all the crossings and dividing by the module mass,
we have an estimation of the total ionizing radiation dose; knowing
the path length in the volume and the incident energy for each
particle, we weight each crossing to get the equivalent damage
produced by 1 \mev neutrons~\cite{niel:rose}.  Then, using the
appropriate rescaling, we compute the integrated radiation dose, and
the fluence by equivalent 1 \mev neutrons after 10 \invab of collected
data.  Using previous data or irradiating in a short time a test
module with an equivalent amount of particles, we can estimate the
deterioration of the tracker performance after a tenth of its designed
running period.

Additional details on the impact of the background on the tracker
performance can be found in Chap.~\ref{chap:svt}.

\tdrsec{ DCH background overview }
\label{sec:mdi_dch_bkg}

Background processes are responsible for the largest part of the DCH
rate.  A rate higher than the design level would affect the charged
track reconstruction efficiency and resolution.  Additionally, the
DAQ electronics can be progressively damaged when being exposed to a
significant rate of particle for a long period of time.  In both
cases, accurate simulation of the rates is needed to estimate the real
performances and the life-span of the DCH electronics.

The main tool used to estimate the rate in the DCH is the simulation
of the various background sources, described above, using the full
simulation described in detail in
Sec.~\ref{sec:mdi_RadBhaBha_bkg_simTools}.  In the full simulation,
the gas chamber is implemented as an homogeneous volume with a density
corresponding to a weighted value including the gas and the wires.
The chamber structure is described accurately including carbon fiber
inner and outer wall plus the end-caps with appropriate density. Three
silicon plates are included on the backwards side to simulate the
amount of material corresponding to the front-end electronics.  Each
particle going through the gas volume and the silicon plates is
recorded, together with additional information about the type of
particle, the position, the incident energy, the vector momentum, and
the deposited energy.  The simulation is not aware of the internal
wire structure: this allows us to simulates background events faster,
and rates can be computed in each layer for different wire
configurations in a post-simulation analysis.  The
post-simulation analysis also evaluates the radiation dose, the
equivalent fluence of 1~\mev neutrons, and fluences of different kind
of particles in the plates representing the DAQ electronics.

A specific application has been develop to analyze the simulation
output and produce the relevant quantities and plots. Due to the
similarities with other sub-systems, this application has been also
extended using multiple modules that produce the same or similar
quantities, as already described in Sec.~\ref{sec:mdi_svt_bkg}.  A
given wire structure is provided to the application, with all the
information on the cell size and the eventual stereo angles.  The
application automatically creates a table where the deposited energy
is stored for each cell.  Given the long path, the curvature of
charged tracks in the solenoidal magnetic field cannot be neglected.
For each charged track crossing the gas chamber, the helix parameters
are computed and the deposited energy in each crossed cell is added in
the table.  After going through all of the charged tracks
in the event, the table is used to fill a plot with one entry for each
layer. The number of cells with deposited energy greater than zero in
each layer is added into the corresponding entry. The plot is produced
for all the background sources and then rescaled for the
cross-section/frequency of each process.  At the transverse radius of
the DCH the tracks produced by background processes are with good
approximation isotropic in the azimuthal angle, so no information is
lost when averaging the cells over the same layer.

The impact of the background processes on
the DCH performances is discussed in
Sec.~\ref{subsec:DCH_backgrounds}.
\tdrsec{ FTOF background overview }

The FTOF background has been studied using the \superb\ full
simulation framework \Bruno in which the quartz tiles (including the
Cherenkov physics) and the MCP-PMTs have been simulated in details.
The FTOF layout used for these studies is the one optimized during the
selection process of the forward PID device and which was driven by
the knowledge to date of the \superb\ forward region.  The background
samples generated in each Monte-Carlo production have been analyzed
and the rates computed in kHz/\cma, averaged sector by sector. By
convention, the sectors are labelled clockwise seen from the IP, with
the sector 0 at the top.

The implementation of the Cherenkov physics using \geant4 has been
tested by comparing the computed photon rates with an independent
computation based only on the particle flux incoming to the FTOF and
an estimation from \babar data of the optical photon yield per track.
The estimates are in good agreement, given the geometry
differences between the \babar\ DIRC and the \superb\ FTOF.

\begin{figure*}[!tbp]
\centering
\includegraphics[clip=true,trim = 10 0 40 0,width=0.38\linewidth]{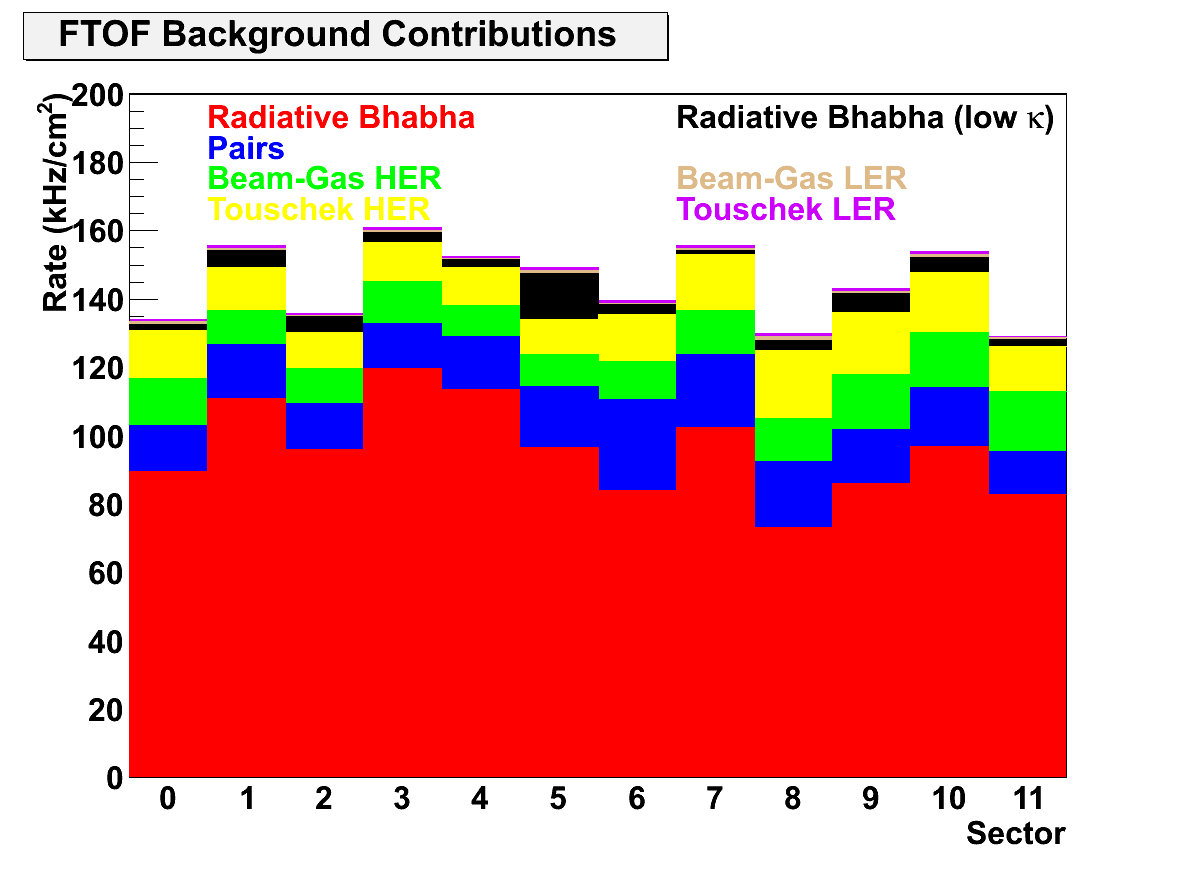}\hfill
\includegraphics[clip=true,trim = 20 0 50 0,width=0.61\linewidth]{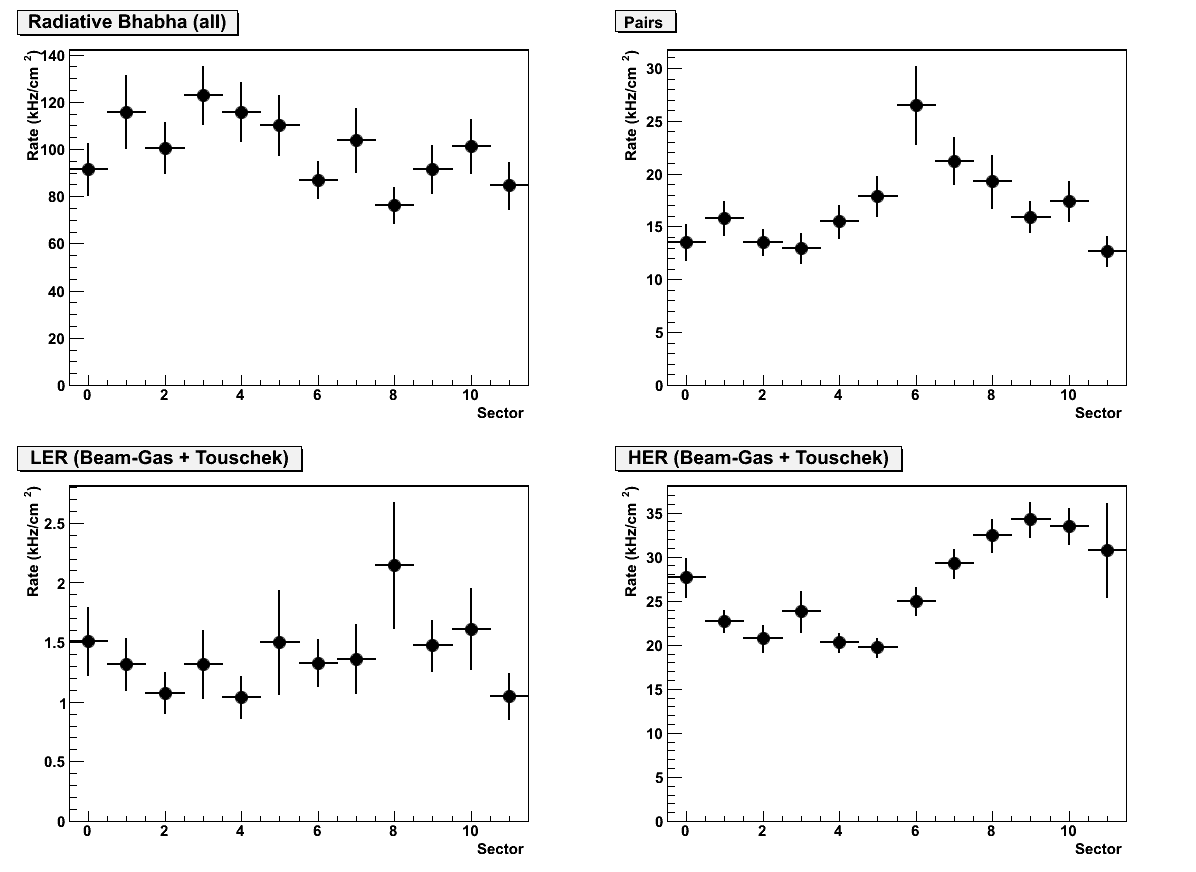}
\caption{Left: contributions of the different background sources to
  the FTOF rate in each sector; the radiative Bhabha sample (red and
  black histograms) is clearly dominant.  Right: The background rates
  per sector associated with the four dominant background
  contributions (radiative Bhabha, Touschek HER, pairs and Touschek
  LER) are displayed separately.}
\label{fig:FTOF-sectorRates}
\end{figure*}

\begin{figure*}[tbp]
\begin{center}
\includegraphics[width=0.8\linewidth]{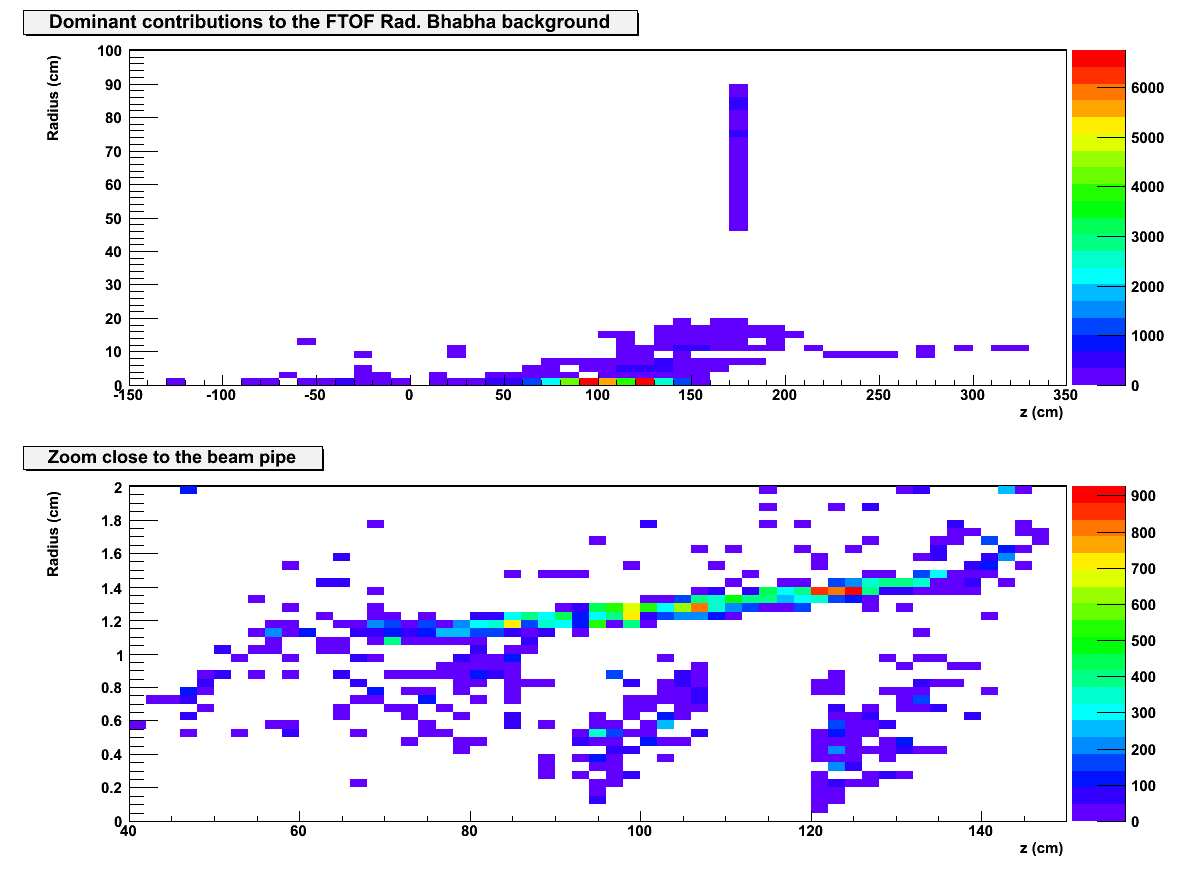}
\caption{Top: distribution in the radius vs.\ $z$ plane of the decay
  vertices of the initial radiative Bhabha particles which contribute
  to the FTOF background. Bottom: zoom of the top map in the low
  radius region. While about 10\% of the background hits originate
  from radiative Bhabha particles which hit directly the FTOF, the
  dominant contribution (more than 50\%) comes from off-energy
  positrons which hit the beam-pipe around 1~m away from the IP on the
  forward side and which generate an electromagnetic shower.}
\label{fig:FTOF-BhabhaDaughter}
\end{center}
\end{figure*}

Figure~\ref{fig:FTOF-sectorRates} shows the contributions of the
different background samples sector by sector. The radiative Bhabha is
clearly dominant, representing 70--80\% of the total rate.  Therefore,
a dedicated study based on Monte-Carlo truth information has been done
to find the origin of this background. The parent genealogy of each
Cherenkov photon hitting the FTOF has been followed up to the first
daughter of the original radiative Bhabha particle. Its creation
vertex has been histogrammed in the plane $R=\sqrt{x^2+y^2}$ vs.\ $z$
(Figure~\ref{fig:FTOF-BhabhaDaughter}). The rectangle at large radius
and $z {\sim} 175$~cm corresponds to the FTOF location: about 10\% of
the background comes from positrons from a radiative Bhabha
interaction which hit directly the detector. But the largest
contribution (${\sim} 50\%$) comes from particles generated at very
low radius about 1 meter away from the IP. They correspond to
positrons that have lost a significant fraction of their energy by
radiating a photon.  They leave the HER nominal trajectory and hit the
beam-pipe, producing electromagnetic showers whose end products
finally reach the FTOF.

\begin{figure}[tbp]
\begin{center}
\includegraphics[width=\linewidth]{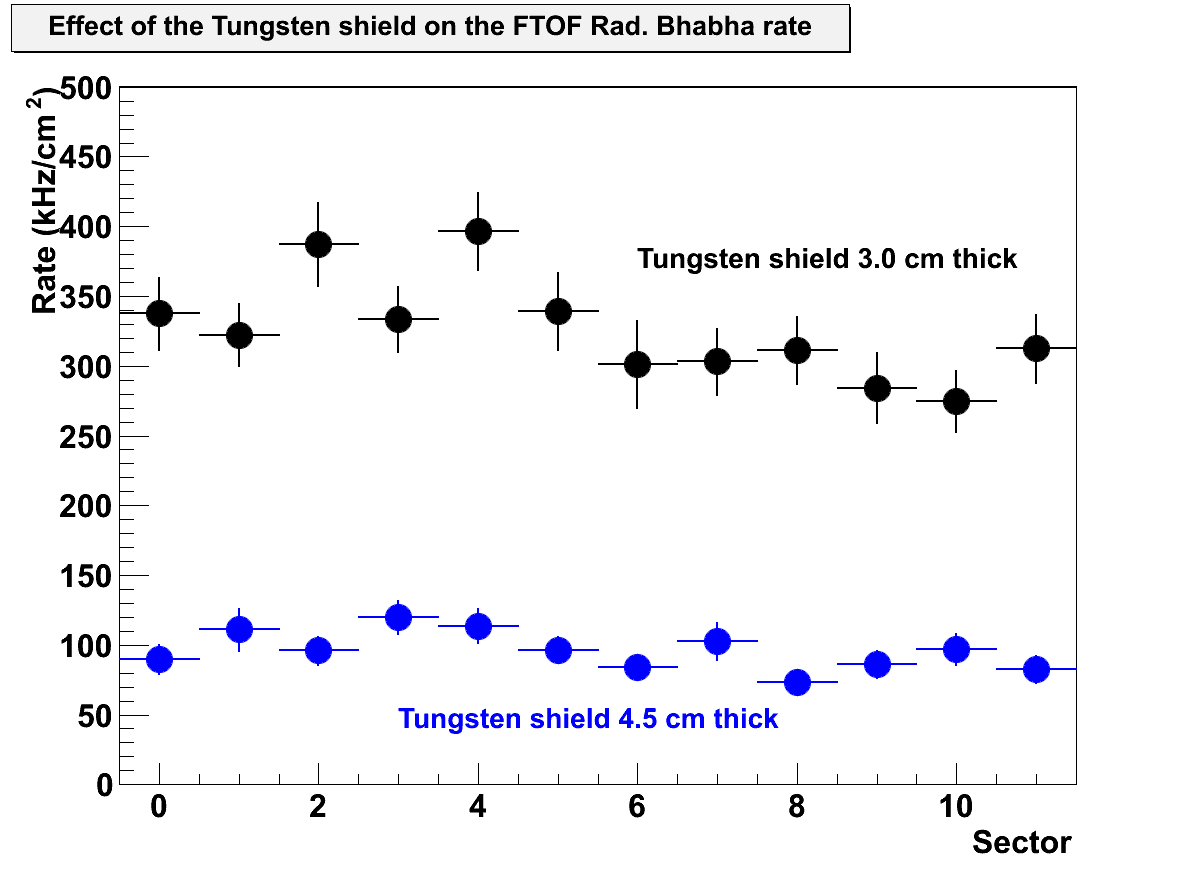}
\caption{Impact of the increase of the width of the Tungsten shield
  around the beam pipe.  Moving from 3~cm (black dots) to 4.5~cm (blue
  dots) of W made the Rad. Bhabha dominant background rate decrease by
  almost a factor 3.}
\label{fig:FTOF-WShield}
\end{center}
\end{figure}

As the dominant source of background is well-localized in space,
studies have been done to mitigate its impact. In particular,
increasing the width of the Tungsten shield from 3 to 4.5~cm has
reduced this contribution by a factor 3
(Figure~\ref{fig:FTOF-WShield}).

The dose absorbed by the electronics and the corresponding neutron
flux have been estimated as 185~Rad and $1.8 \times 10^{11}$n/\cma
respectively, for $10^7$~s at the \superb\ nominal luminosity of
$10^{36}$\cms. The 1~MeV neutron equivalent fluency is $1.0 \times
10^{11}$. These values are quite high but still manageable for front-end
boards that have been properly designed.
\tdrsec{ FDIRC background overview}
\label{sec:mdi_FDIRC_backgrounds}

The FDIRC background is mainly due to low energy electrons and
positrons that produce Cherenkov light either in the fused silica bars
or in the FBLOCK.  Some of these photons reach the MaPMT plane where
they are detected and combine with the hits coming from the high
momentum tracks of the physics events.  These low momentum particles
are the result of complicated interactions between background
primaries and components of the \superb\ detector or machine elements.
The primaries come both from luminosity-driven backgrounds, such as
radiative Bhabha, and from beam-induced backgrounds, such as Touschek
and beam-gas scattering.  

As for the other \superb\ systems, the background hits impact the
FDIRC in two ways. First, they decrease the performance of the
reconstruction algorithm by making the identification of the Cherenkov
cones more difficult. More hits mean smaller signal-to-background
ratios and more combinatorials.  Moreover, real signal photons can be
masked by background hits close in time in the same pixel.  In
addition, the background induces ageing of the FDIRC MaPMTs and
front-end electronics.

Therefore, estimating the FDIRC background rates is crucial, not only
to validate the choices made at the hardware level but also to define
the requirements on the software algorithms.  The past experience
of the \babar\ DIRC makes this goal even more important.  Less
than a year after the beginning of  \babar\ data taking, the DIRC
background was significantly exceeding the predictions. The dominant
component was few \mev neutrons produced in electromagnetic showers 
from radiative Bhabha scattering events.
These interacted with the SOB water, generating Compton
photons which then scattered, producing electrons and positrons
finally giving Cherenkov light. A dedicated shield had to be designed
to cover the beam line around the \babar\ backward end-cap. This
shield provided a direct reduction of the background by a factor 3 and
efficiently protected the DIRC as the \pep2\ luminosity rose.

The simulation of the FDIRC background is based on \Bruno
(Sec.~\ref{sec:mdi_RadBhaBha_bkg_simTools}), the \superb\ Geant4-based
full simulation software. The implementation of the FDIRC hardware
components matches the latest design of the detector and has been
extensively tested in a dedicated standalone simulation, also based on
the \geant\ framework. The fused silica bars and FBLOCKs have the correct
optical properties; the MaPMTs (including the quantum efficiency as
a function of the wavelength), mirrors (at the forward bar ends and in
the FBLOCKs), and glue joints are simulated as well. Hits on the
photon detectors are recorded and stored in ROOT format, along with
some truth information about the particles that produced the
Cherenkov light. The generation, propagation and detection of the
optical photons are completely driven by \geant.  Offline, during the
analysis phase, the locations of the hits are converted into MaPMT and
pixel numbers, taking into account dead areas and the space lost
between neighboring photon detectors.

The FDIRC simulation is validated before each major production and
dedicated quality checks are run over the simulated events to test
their consistency and track unphysical hits due to mis-modeling of
detector components or software issues. This monitoring found, for
example, that the simulation of the MaPMT photo-cathode using
BK7 glass was responsible for 10--20\% excess hits for the radiative
Bhabha background. Changing this material to aluminium
made those hits disappear.

\tdrsubsec{Shielding the FDIRC}

As explained above, experience based on the \babar\ DIRC has shown
that the FDIRC requires a dedicated shielding to reduce the hit rate.
This comes in addition to the design improvements (smaller camera made
of quartz instead of water, faster photon detectors and front-end
electronics) that aim at mitigating the FDIRC background.

Looking at a side view of the \superb\ detector, the FDIRC can be split into
three main regions:
\begin{enumerate}
\item $-160\cm < z <  220\cm$: inside the magnet,
\item $-280\cm < z < -160\cm$: within the flux return steel, and
\item $-400\cm < z < -280\cm$: outside the magnet.
\end{enumerate}

The first region cannot include a dedicated FDIRC shielding for
obvious reasons.  Therefore, its background protection only relies on
the W-shield around the beam pipe. The thickness (45\mm) is as
large as possible given the maximum allowed weight 
and the constraints coming from detector integration.

With the W-shielding, simulations show that the dominant background
comes from the third region where there is space for a dedicated
shield. As the flux of particles entering this area is dominated by
photons with energy below 200\mev, a detailed optimization procedure
has lead to the choice of a steel-lead-steel sandwich shielding
(2.5-10-2.5\cm-width respectively) which reduces the background by
about one order of magnitude. A side effect of this shielding is to
roughly double the number of neutrons in this area. Therefore, it is
complemented by a 10\cm layer of Boron-loaded (5\%) polyethylene.

Without this shielding, the radiative Bhabha events are about one
order of magnitude larger than the other background components. This
is not the case with the new design, as shown in the
following section: most background sources give similar contributions
to the total rate.

\begin{figure}[tbh]
 \begin{center}
 \includegraphics[clip=true,trim=40 70 20 100,width=\linewidth]{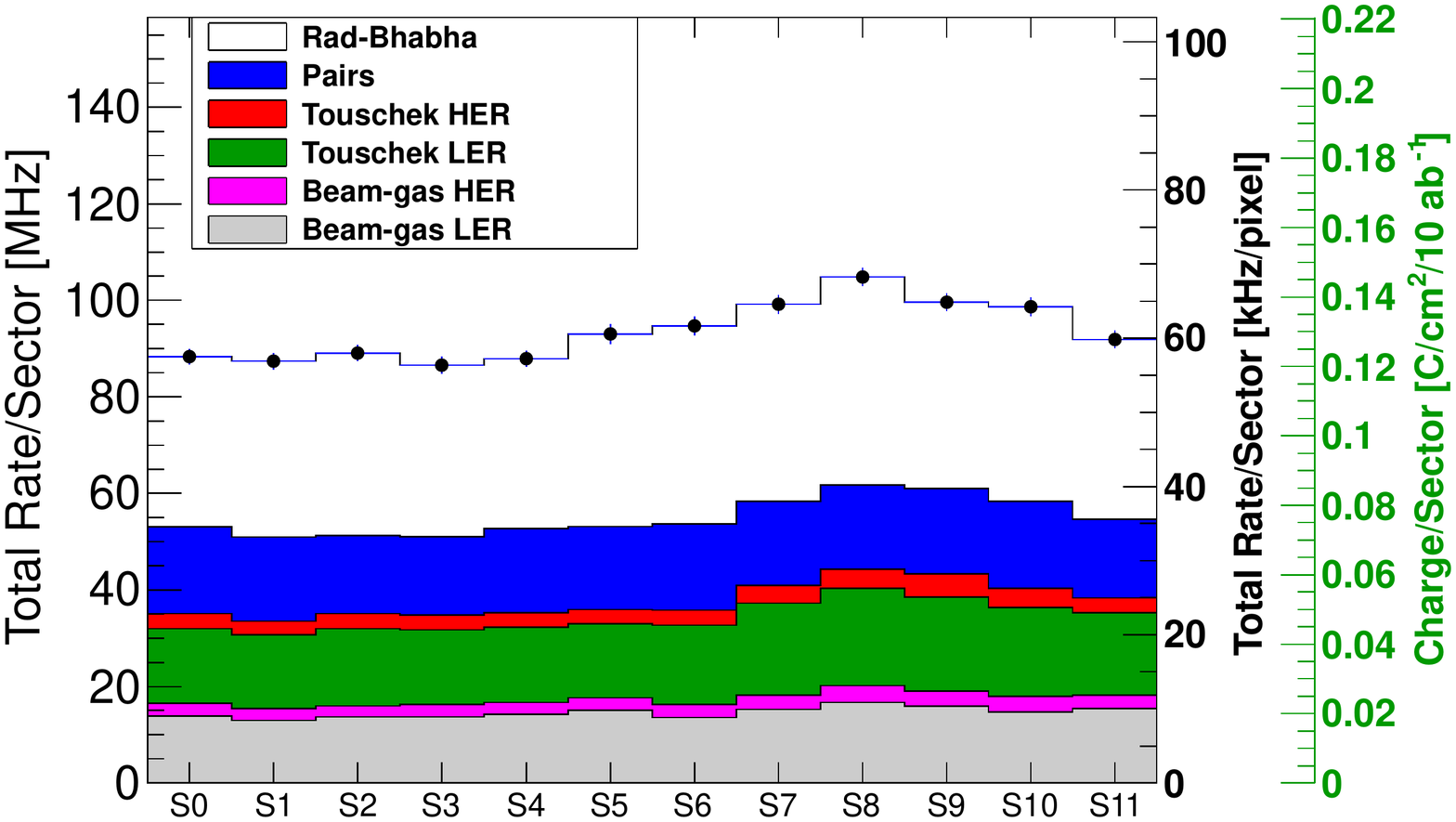}
 \caption{Total rates and integrated charge versus sector number on
   the FDIRC.  These rates are the predicted ones and do not include
   any safety factor.}
 \label{fig:Total_FDIRC_rates}
 \end{center}
\end{figure}

\tdrsubsec{Background rates in the FDIRC}

The backgrounds rates are estimated for each pixel on the focal plane
of the FDIRC photo-camera. Simulation shows that the distribution of
hits per bunch-crossing within a given sector is rather uniform on the
azimuthal angle, but has a small variation with the radius of the
pixel location $(R = \sqrt{x^2 + y^2})$, with a maximum around the
center of the focal plane.  Figure~\ref{fig:Total_FDIRC_rates} shows
the total rate of background hits on the FDIRC photo-camera.  The
horizontal axis labels the FDIRC sectors, counted clockwise as seen
from the interaction point.  The left-vertical axis shows the total
rate on sector and the black right-vertical scale shows the average
hit rate per pixel. As can be seen, radiative Bhabha gives the highest
background contribution but the other background sources give
significant contributions of the same order of magnitude. Also
reported on the green right-scale of the plot is the charge integrated
on a time period of $10^{7}\sec$, which is the time needed to collect
$10\invab$ of data at the nominal luminosity ($10^{36}{\rm Hz}/\cma$).
The estimated average hit rate per pixel including a safety factor of
$5$ is $(301.9 \pm 1.7) \, {\rm kHz}/{\rm pixel}$, which corresponds
to an integrated charge of $(3.2 \pm 0.02) \, {\rm
  C}/\cma/(50\invab)$.

\tdrsubsec{Integrated charges and doses}

Given the high radiation environment of \superb, it is crucial to
estimate the amount of ionizing and non-ionizing energy loss (NIEL) on
the front-end-electronics (FEE) boards of the FDIRC
(Sec.~\ref{sec:mdi_ETD}). In order to accomplish this goal the FDIRC
GEANT-4 model includes the FEE boards, which are implemented as
$200\mum$-thick silicon boards of $34\times7\cma$ area, as specified
by the design. Each FDIRC sector has 21 of such FEE boards.
Simulations show that there is no variation on the radiation levels
within a sector. The numbers reported below correspond then to the
total radiation level for each FDIRC sector.

\begin{figure}[tbh]
 \begin{center}
 \includegraphics[clip=true,trim=40 70 100 120,width=\linewidth]{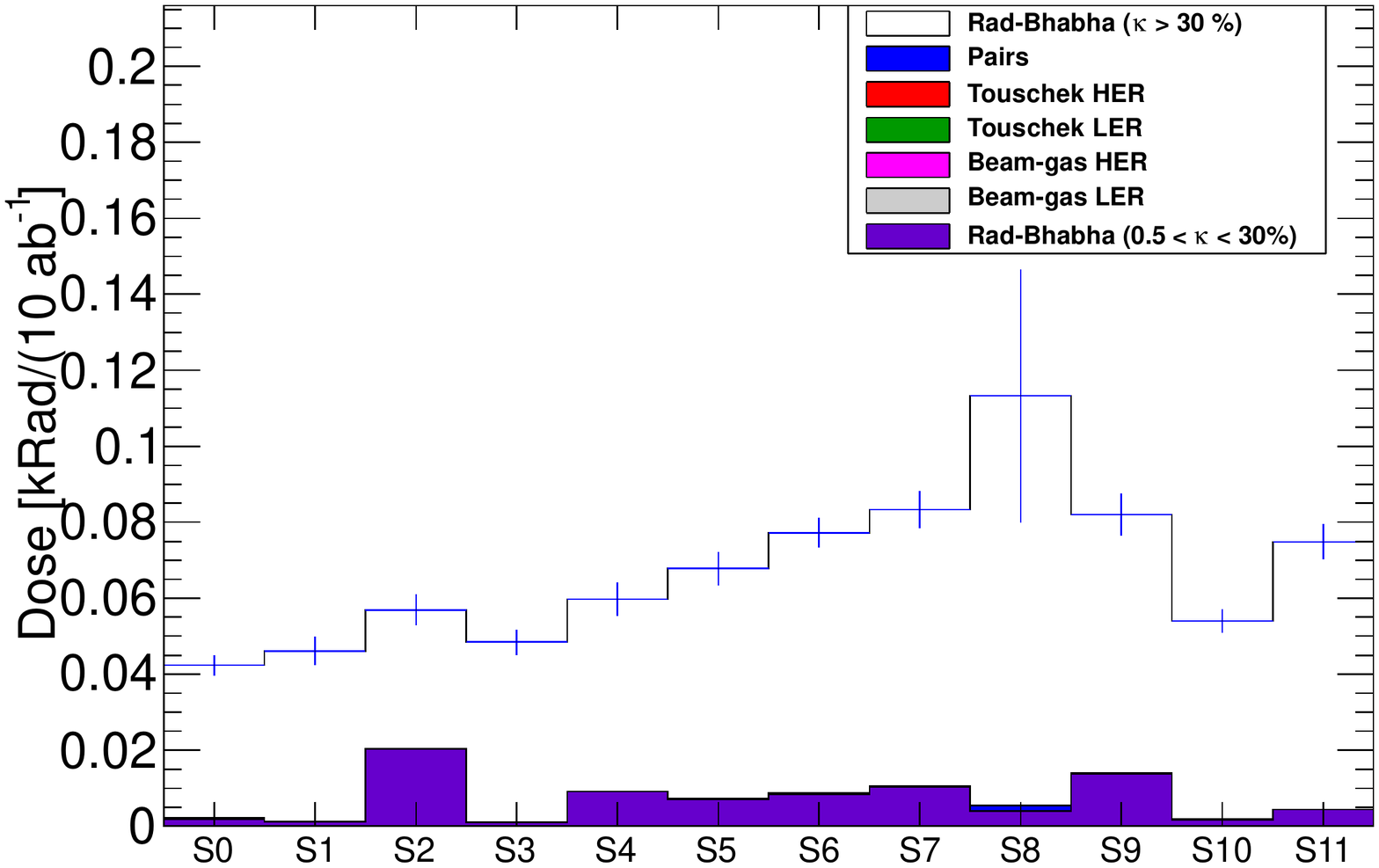}
 \includegraphics[clip=true,trim=40 70 100 120,width=\linewidth]{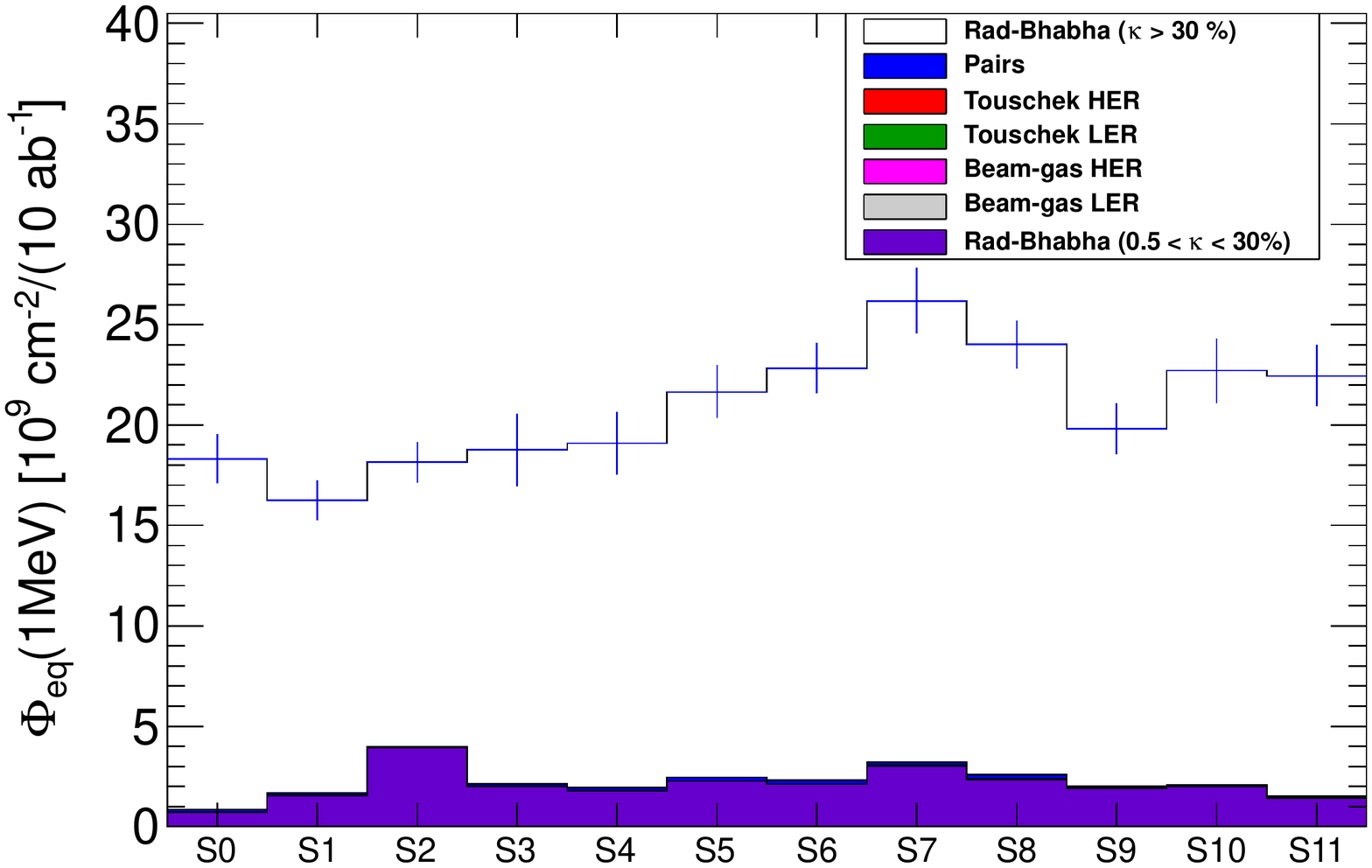}
 \caption{Total dose (top) and total equivalent $1\mev$ neutron
   fluency (bottom) per sector on the FDIRC FEE.  These rates are the
   predicted ones and do not include the safety factor of 5.}
 \label{fig:FDIRC_Dose_and_NIEL}
 \end{center}
\end{figure}

The deposited energy per unit of mass due to ionizing radiation, or
dose, is shown on the top plot of
Figure~\ref{fig:FDIRC_Dose_and_NIEL}. The horizontal axis labels the
FDIRC sector number and the vertical axis is the accumulated dose in
${\rm krad}$ for every $10\invab$ of integrated luminosity. The color
code shows the contribution from the different background sources:
radiative Bhabha (white and purple), Touschek from HER/LER (red and
green) and beam-gas from HER/LER (magenta and gray). As can be seen,
the main radiation source of this kind is radiative Bhabha, the other
components giving a negligible contribution. The average dose per
sector including a safety factor of 5 is $(291.2 \pm 5.8) \, {\rm
  rad}/(10\invab)$, which seems to be a bearable level for boards
properly designed for the full lifetime of the experiment.

The NIEL is measured in $1\mev$ neutron equivalent fluency ($\Phi_{\rm
  eq}(1\mev n^{0})$). This quantity is the amount of $1\mev$ neutrons
which would produce the same amount of damage as the actual
radiation estimated on the detector, to which particles of different
types and energies are contributing.  The bottom plot of
Figure~\ref{fig:FDIRC_Dose_and_NIEL} shows the total $\Phi_{\rm
  eq}(1\mev n^{0})$ for the different FDIRC sectors (horizontal axis).
The vertical axis units are $10^9 \, {\rm cm}^{-2}/(10\invab)$.
Again, the dominant contribution to the NIEL are Radiative Bhabha
interactions. The average NIEL per sector including a safety factor of
5 is $(10.2 \pm 0.2)\times 10^{10} {\rm cm}^{-2}/(10\invab)$, a level
within the acceptable range of properly-designed boards.
\tdrsec{ EMC background overview. }
\label{sec:MDI:EMCbkg}

\begin{figure}[tbh]
 \begin{center}
 \includegraphics[width=0.5\textwidth]{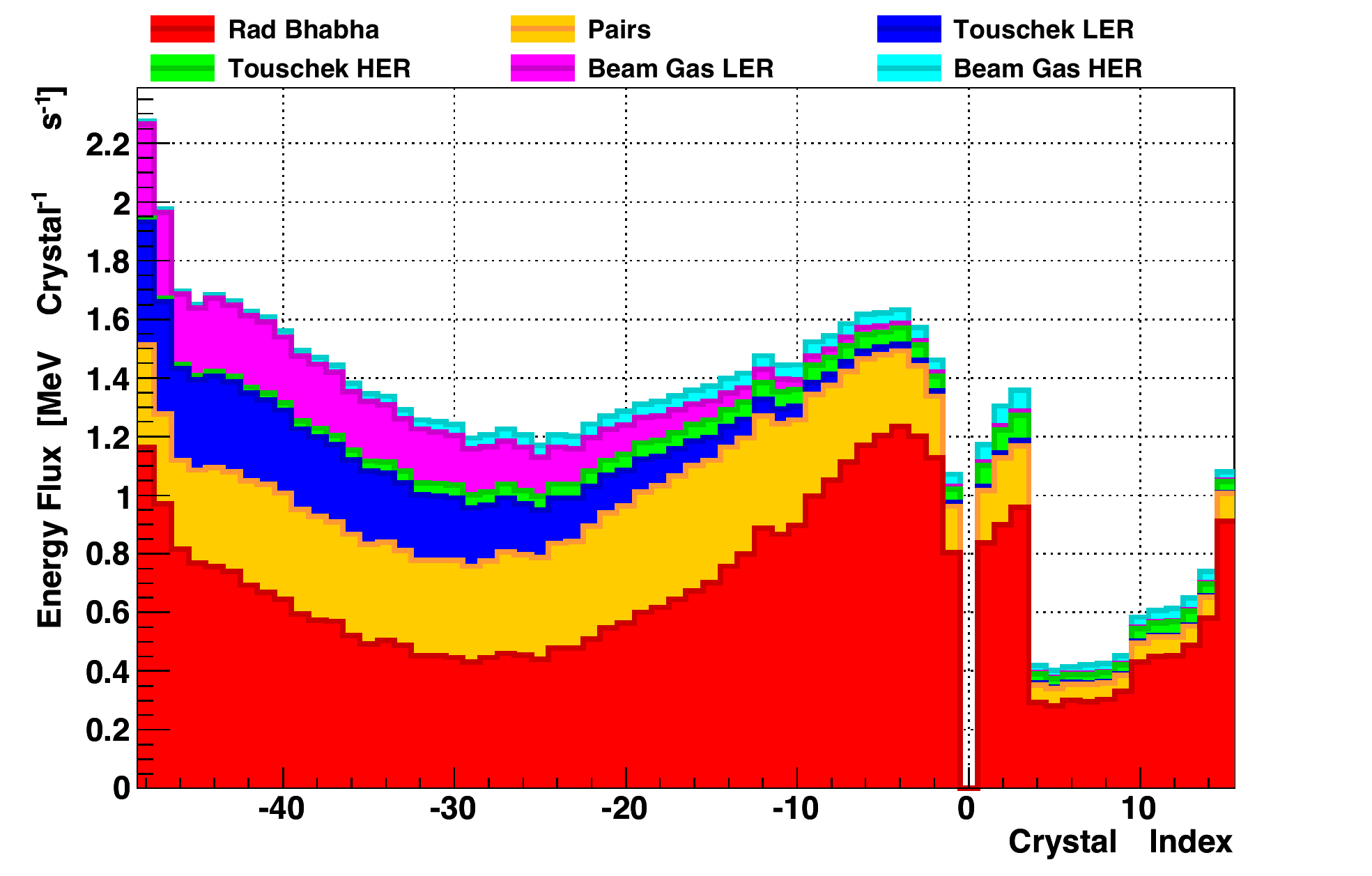}
 \caption{Contributions to the total energy deposit rate per crystal
   vs polar angle ($\theta$) position. Negative indexes are for the
   CsI(Tl) Barrel crystals, positive indexes are for the forward
   Endcap which includes three rings of CsI(Tl) crystals in the region
   close to the Barrel and LYSO crystals in the remaining part; the 0
   index corresponds to the Barrel-Endcap transition.  The energy
   deposit shows no dependence on the azimuthal angle.}
 \label{fig:emc_edep_ring_stack}
 \end{center}
\end{figure}

The main effect of machine background in the EMC is a diffuse energy
deposit all over the detector, mainly the result of several low energy
deposits per crystal which, given the slow crystal decay time
($\tau\simeq 1.2 \mu s$ for the CsI(Tl)) compared to the bunch
crossing rate, may result in a significant integrated contribution.
The angular ($\theta$) dependence and the overall level for the
deposited energy varies with the background source, while the
cumulative effect peaks in the forward and backward direction
(Figure~\ref{fig:emc_edep_ring_stack}).

The Radiative Bhabha background is dominant and determines the shape
of the distribution.  The sharp change in average deposited energy for
the forward EMC (positive $\theta$ indexes) between the third and the
fourth ring, which is shown by all the background sources, is related
to the smaller size of the LYSO crystals whose front face is about
$1/4-1/5$ with respect to the CsI(Tl) ones in the Barrel and in the
outer part of the Endcap.  Another geometry related feature is the
decrease in energy deposit per crystal for the Barrel-Endcap
transition region.  This is explained with the fact that the
background particles are coming from the beam pipe direction but not
directly from the interaction point, so the forward Endcap acts as a
shield protecting the above mentioned section of the calorimeter 
(see Figure~\ref{EMC:fig:xsecview} for a cross section of the relative
barrel and Endcap position).  Touschek and Beam Gas backgrounds show a
similar behaviour.

The sum of all the energy deposit contributions is of the same order
as the single Radiative Bhabha one. The maximum difference is the
central part of the barrel, the maximum of the Pairs contribution,
where a factor of about 2.5 is reached with respect to the single
Radiative Bhabha contribution. The mean energy deposit per crystal
keeps the peaked shape in the forward and backward directions but,
including all the background sources, the maximum of the deposit is in
the backward region of the barrel.  This shape change is mainly
related to the very asymmetric behaviour shown by the Touschek and
Beam Gas backgrounds and to the very different rates between the low
energy and the high energy rings.  It needs to be stressed about the
Touschek and Beam Gas backgrounds that, even if their average energy
deposit contribution seems almost negligible, the incoming particles
energy spectrum is much harder compared to both Pairs and Radiative
Bhabha so that, considering only the single energy deposits above 10
MeV, Touschek and Beam Gas rates are comparable with the Radiative
Bhabha one. This feature is particularly evident in the forward and
backward regions of the barrel and becomes more and more conspicuous
for higher deposited energy.

Concerning the effects related to machine backgrounds on the EMC
performance, the energy deposited by any particle of physical interest
will be superimposed with the diffuse contribution from the
backgrounds, whose fluctuations will degrade both the angular and
energetic resolution.  In addition to this effect, machine backgrounds
can create clustered energy deposits which in the reconstruction
process may be considered as real particle clusters and therefore
spoil the event reconstruction.

In addition to the instantaneous effect of machine background on the
event reconstruction, the integrated radiation dose causes a loss of
light yield for the scintillating crystals which, if significant, may
result in a degradation of the performance. The radiation dose (see
Figure~\ref{fig:EMC:dose}) and the actual impact of the background on
the EMC performance which, apart from the overall background level,
depends also on the crystal signal processing and on the
reconstruction process, are discussed in more details in the dedicated
EMC chapter \ref{chap:Calorimeter}.

\tdrsec{ IFR background overview}

To study the machine related background on the Instrumented Flux
Return, the IFR detector has been simulated using \Bruno
(Sec.~\ref{sec:mdi_RadBhaBha_bkg_simTools}.) Radiative Bhabha
scattering is the dominant background source for the IFR detector.

All of the background processes produce a high-energy primary $e^{\pm}$ or
$\gamma$ that strikes a beam line element within a few meters of the
IP and produces showering secondaries that can be charged or neutral
particles with energies ranging from sub-MeV to several tens of MeV.
These particles hit the IFR and may affect the detector performance
and reduce the photodetector lifetimes.

\begin{figure}[htb]
  \begin{center}
   \includegraphics[width=0.5\textwidth]{EnergyBarrelNeutronsNorm_nt.png}
   \caption{Neutron energy spectrum on IFR barrel, the neutron are
     classified according to their kinetic energy range as high energy
     neutron, fast, epithermal or thermal.}
\label{fig:nspectrum}
\end{center}
\end{figure}

\tdrsubsec{Neutron Background}
\label{neutronsection}

In this context there is a non-negligible production of neutrons via
giant resonance formation~\cite{giantresonance}.  The neutrons
produced with this mechanism have energy of some MeV. These neutrons
are moderated by the interaction with the detector material: they
scatter back and forth on the nuclei both elastically and
inelastically losing energy until they come into thermal equilibrium
with the surrounding atoms. At this point they will diffuse through
matter until they are finally captured by a nucleus. For this reason
the neutron energy spectrum in the \superb\ environment has a very
wide range as shown in Figure \ref{fig:nspectrum}, where the neutrons
are classified according to their kinetic energy.

Since the neutron spectrum is so wide in energy and the interaction
with the matter strongly depends on it, it is customary to
characterize the neutron effects in terms of an equivalent mono
energetic neutron
source~\cite{NeutronEq,Lindstrom:2000gd}.
For this reason all the neutron rates in this section are normalized
to the equivalent of a 1~MeV neutron.

The neutron background not only contributes to the radiation dose of
the IFR detector, but is also particularly dangerous for the SiPM,
the devices used for the detector readout, which are quite
sensitive to radiation.  
A fluence of  $10^{11}n_{eq}/cm^{2}$ ($n_{eq}$ is the equivalent
number of 1 MeV neutrons on silicon) can severely damaged these 
devices, producing a loss of efficiency, increasing of dark current, 
and singles rate~\cite{villaolmo}). 
Figure~\ref{fig:nrateL0} shows the neutron
rate on layer 0 of the Barrel due to the different background
sources, including a safety factor of 5 that takes into account
the fact that the simulation may not perfectly reproduce the reality.
 
The main background source is Radiative Bhabhas. The rate on Barrel
layer 0 corresponds to $\sim 6 \times 10^{9}$ neutrons/\cma for a
year. This rate is acceptable for a 10 year run and is the
result of the complex shielding system described in Sec.~\ref{sec:ifrshield}.

\begin{figure}[htb]
  \begin{center}
   \includegraphics[width=0.5\textwidth]{TotBkg_RateL0Barrel_nt_Pisa2012.png}
   \caption{Neutron rate on IFR Barrel Layer 0 due to different
     background sources, normalized to 1~MeV equivalent and multiplied
     by a safety factor of 5.  The different background sources are
     stacked.}
\label{fig:nrateL0}
\end{center}
\end{figure}

\tdrsubsec{Charged Particles}

The background due to charged particles is particularly interesting
since it can affect directly the detector performances. It can give
fake signals in the detector and affect the track reconstruction and
consequently the muon ID. Figure~\ref{fig:erateL0} shows the rate for
electrons and positrons coming from different background sources. As
for the neutron case, the dominant source of the background are
radiative Bhabha events.  The rate shown in Figure~\ref{fig:erateL0}
is for electrons/positrons with energy deposited greater than 150~keV
the nominal threshold for the detector.

\begin{figure}[htb]
  \begin{center}
   \includegraphics[width=0.5\textwidth]{TotBkg_RateL0Barrel_ele_Pisa2012_150thr.png}
   \caption{Electrons/positron rate for the IFR Barrel layer 0 due to
     different background sources, multiplied by a safety factor of 5.
     The deposited energy of the particle is $>150$~keV. The different
     background sources are stacked.}
\label{fig:erateL0}
\end{center}
\end{figure}
 
A high proton rate is observed in the IFR. These protons are produced
inside the scintillator in an (n,p) reaction in which the neutron is
captured by the scintillator material and a proton is emitted. The
cross section for this process falls as $1/v$, so it is more likely to
happen when the neutron has low energy. The resulting proton has an
energy below threshold, and can therefore be neglected.

\tdrsubsec{Background remediation}
\label{sec:ifrshield}

The rates shown on the previous section are the result of a complex
shielding system implemented primarily to significantly reduce the
number of neutrons crossing the IFR. Adding shielding material also
has the effect of reducing the electron and photon rate.  We added a
shielding system to the external structure of the IFR and a inner
shield to protect the layer 0. The shields are made of Boron
Loaded (5\%) Polyethylene ($(C_{2}H_{4})_{n}H_{2}$). The polyethylene
has a high hydrogen density, which slows neutron particles down so
they can be absorbed by the Boron.  One of the isotopes that comprise
natural Boron, $B^{10}$, has a very high cross section for neutron
capture.  This shielding design, when implemented in the \superb\ Full
Sim, reduced the neutron rate by an order of magnitude.

\tdrsec{ ETD background overview}
\label{sec:mdi_ETD} 

The high luminosity of the machine and the presence of numerous
massive elements close to the interaction region will generate much
higher levels of radiation than in \babar, where radiation effects on
digital read-out electronics had only been observed after the
introduction of FPGAs on the end plate of the drift chamber.

In \superb\ a large flux of charged particles and photons originating
from the interaction point and the beam pipes will cross the detector,
and generate large numbers of secondary neutrons as well.  Therefore a
common general radiation policy has been put in place at the ETD
level.

Long term radiation effects are of two types (ionizing and
non-ionizing), while short term effects are linked to instantaneous
ionization (Single Events). Radiation levels have been simulated, and
Total Ionizing Doses (TID) range from 5 kGy down to a few Gy depending
on the electronics location.  The neutron fluence is required to
estimate the effect of the Non-Ionizing Energy Loss (NIEL) and is
still under study.

Shielding of sensitive parts of the detector like electronics is a key
point of the design, which has been carefully studied. However, all
electronics located on the detector have to be able to handle not only
the damages linked to TID and NIEL, but also to present minimal
sensitivity to Single Event Upsets (SEUs), thanks to the intensive use
of mitigation techniques like triple modular redundancy (TMR) for the
latches and flip-flops and of safety codes (like parity bits) for data
stored in memories. All the components used will be validated for
their proven capacity to handle the integrated dose and NIEL
foreseen. Power supplies are also designed specifically in order to
perform safely in the radiative environment. The architecture of the
system has been designed in such a way to reduce as much as possible
the risk of failure, especially the critical elements
linked to experiment control and readout. In case of failure despite
all these precautions, malfunctioning will be detected, experiment
control system will be immediately warned, and a fast recovery
strategy will be deployed in order to limit as much as possible the
dead-time due to these radiation effects.

\tdrsec{ SVT radiation monitor }
\label{sec:svt_rad_mon} 

A radiation monitor near the interaction point has been designed to
protect the SVT from a high radiation dose and to abort the beam in
the presence of a current spike or a prolonged radiation dose.  A
radiation monitor addresses the following more general issues:

\begin{itemize}
  \item Allow the protection of the equipment during beam
instabilities / accidents;
  \item Give feedback to the machine, thereby helping it
provide optimum conditions; and 
  \item Monitor the instantaneous dose during operation
\end{itemize} 

The primary goal of the SVT radiation-monitoring
detector is to measure the interaction rate and the background level
in an high radiation environment in order to provide inputs to the
background alarm. In addition, it can be used for luminosity
measurements and detailed background characterizations during both
stable beams and the setting up for collisions.  The requirements for
a full use of this detector therefore are the following:
\begin{itemize}
  \item A simple DC (or slow amplification) readout to measure the
beam induced DC current.
   \item Fast electronics with very low noise for detecting minimum
ionizing particles and allowing more sophisticated logic coincidences
(timing measurements).
\end{itemize} 

The chosen baseline technology is Chemical Vapour
Deposition (CVD) diamond sensors\cite{Adam:2004ce}. In recent years CVD
diamond sensors have been successfully employed as beam monitoring in
several experiments: \babar~\cite{Edwards:2005hi}, Belle~\cite{Wallny:2007zz}, 
CDF~\cite{Sfyrla:2007ng}, ATLAS~\cite{Gorisek:2007cv}, 
CMS~\cite{FernandezHernando:2005hj}, LHCb~\cite{Ilgner:2010vu}, 
and ALICE~\cite{Lechman:2011}. A recent idea is to implement
time-of-flight measurements in order to distinguish collision events
from background. Very high time resolution (of the order of 100 ps) is
required to achieve this goal.

\paragraph{CVD diamond properties}

Chemical Vapour Deposition (CVD) diamond has a large band-gap (5.5 eV)
and a large displacement energy (42 eV/atom) that make it inherently
radiation tolerant. It has
been extensively tested by the RD42 Diamond
Collaboration~\cite{RD42:2008}.
The main features of Chemical Vapour Deposition diamonds are:

\begin{itemize}
  \item Fast timing
  \item High radiation tolerance ($> 1$ MGy)
  \item Single-particle detection and current monitoring capabilities
  \item Efficiency for charged-particle detection very close to
100$\%$
  \item 100 pA absorption current for a 500 V applied voltage (low
power consumption)
  \item Leakage current of a few pA, which does not increase with the
accumulated dose
  \item Insensitivity to temperature variations
\end{itemize}

CVD diamond can be grown on substrate wafers. The deposited films are
typically polycrystalline (pCVD), which have the following drawbacks:
\begin{itemize}
  \item Many grain boundaries originating defects in the diamond
itself
  \item Non-uniformity of charge-collection properties
\end{itemize} 
Poly-cystalline CVD diamond quality for radiation
sensors is typically measured by its charge collection distance
(CCD). The CCD is the average separation distance of electron hole
pairs under an electric field before they recombine. Current
technology allows the growth of pCVD diamonds with CCD of several
hundreds of microns. A diamond sensor CCD is dependent on the bias
voltage applied to it and increases with the applied bias voltage. For
pCVD diamonds, the CCD typically saturates at approximately 1~V per
micron.  

The development of special substrates allows the homoepitaxial growth
of CVD diamond with a mono-crystalline structure
(scCVD)\cite{Isberg:2004}.  These diamonds offer two major advantages
over pCVD diamond. The CCD is substantially larger, ensuring that a
nearly fully efficient charge collection is observed, and the maximum
charge collection is reached at an applied electric field of
0.2~V/\mum, an order of magnitude lower than poly-crystalline
material.

\paragraph{Location of the SVT radiation monitoring} 

\begin{figure} \centering
\includegraphics[width=0.99\linewidth]{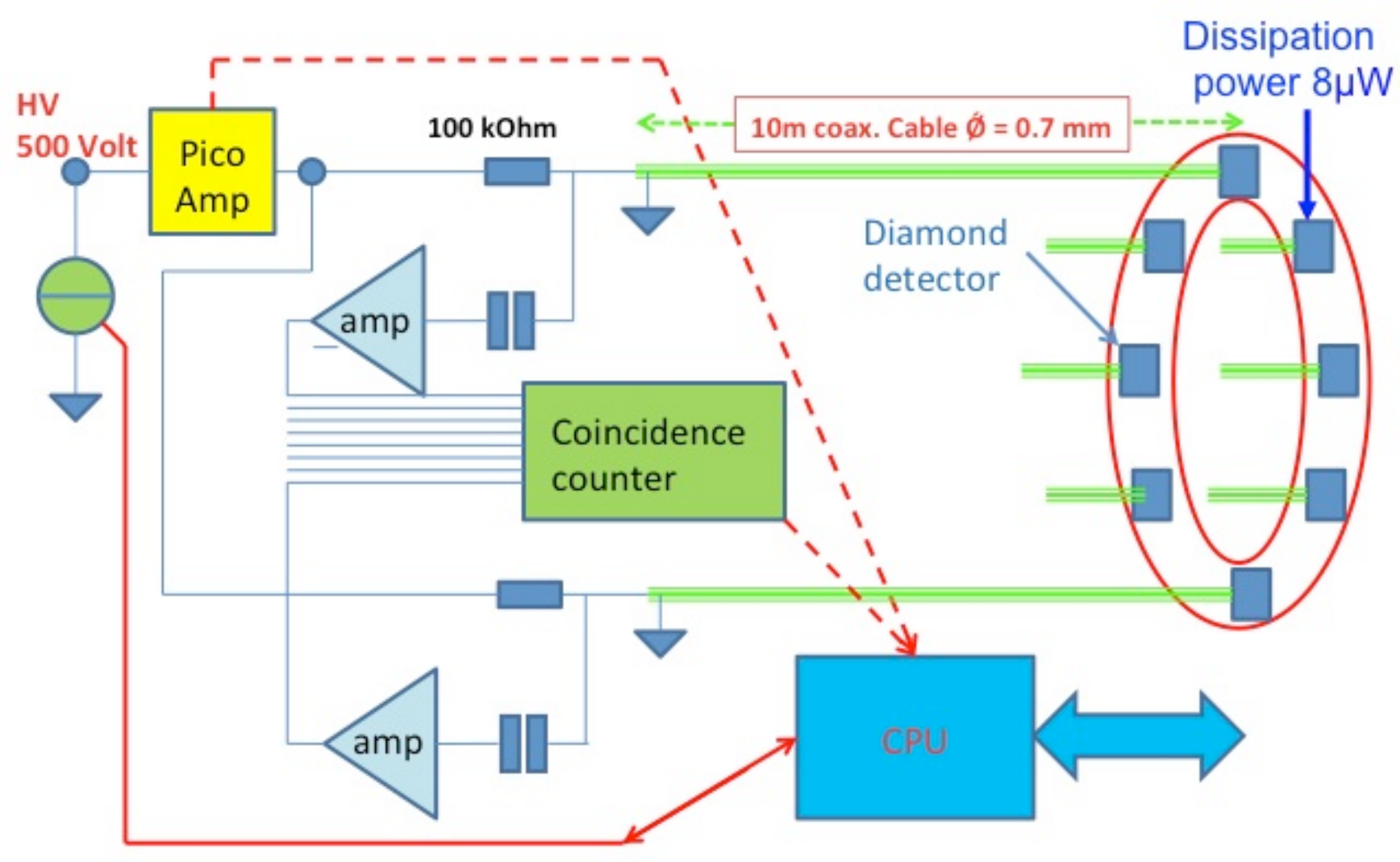}
\caption{Read-out scheme for the SuperB diamond radiation monitor.}
\label{fig:readout}
\end{figure}

The protection system will be made of two rings of CVD diamond sensors
placed on opposite sides of the interaction point. Each ring has eight
sensors of $4\times 4$\mma dimension. The location of the SVT
radiation monitor, between the first silicon layer and the beam pipe,
will be between 8 to 12~cm from the interaction point and at about
12~mm from the beam pipe. The available region is very narrow and the
read-out electronics need to be placed a few meters away from the
detector, making the design of the preamplifier.
Figure~\ref{fig:readout} shows a schematic drawing of the read-out
system.

\begin{figure} \centering
\includegraphics[width=0.99\linewidth]{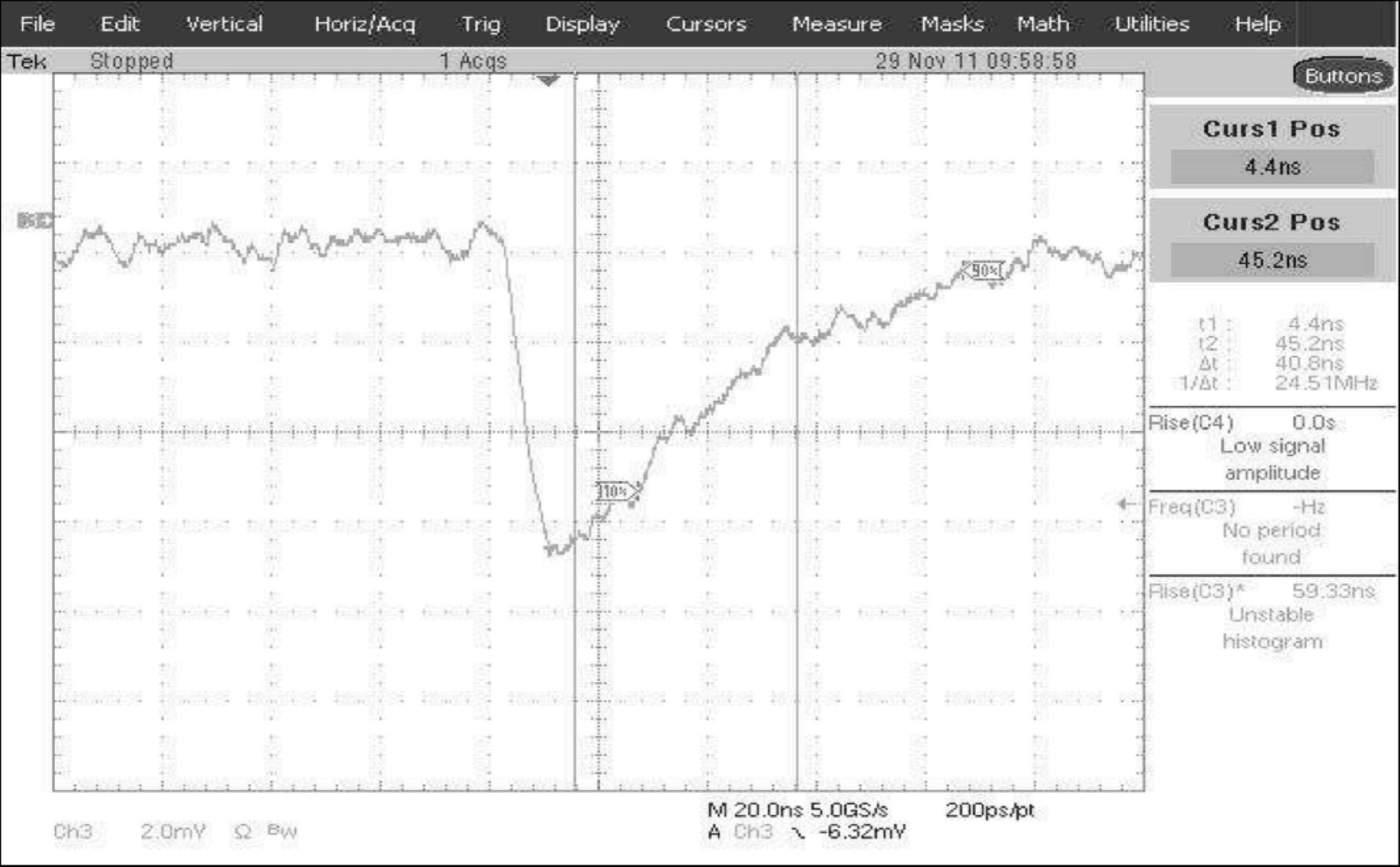}
  \caption{Typical signal generated by a minimum-ionizing particle in
a CVD diamond detector.}
\label{fig:oscilloscope}
\end{figure}


\begin{figure} \centering
\includegraphics[width=0.99\linewidth]{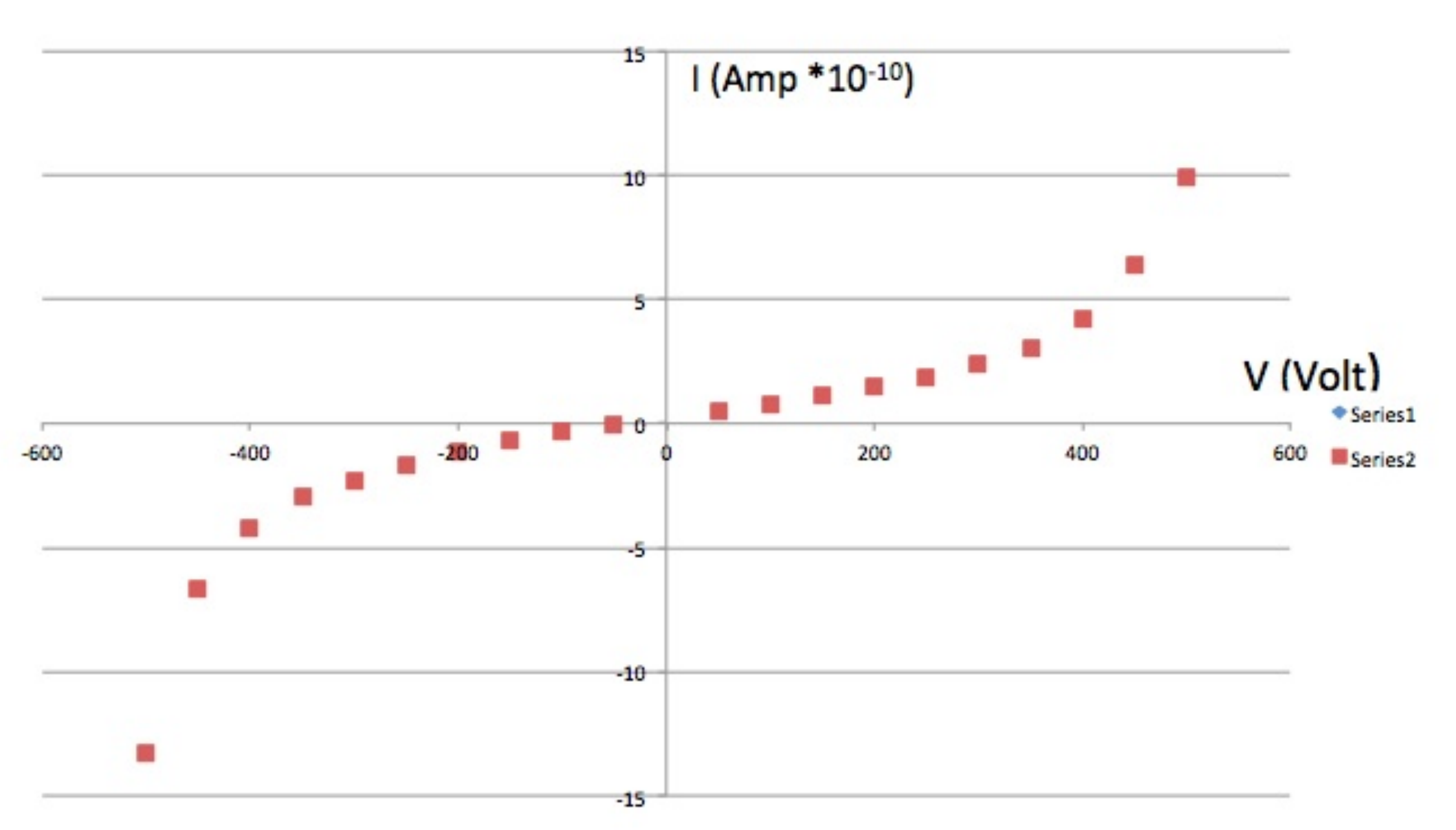}
\caption{ I-V curve of a scCVD diamond detector measured with a
picoamperometer.}
\label{fig:IV}
\end{figure}

\begin{figure} \centering
\includegraphics[width=0.85\linewidth]{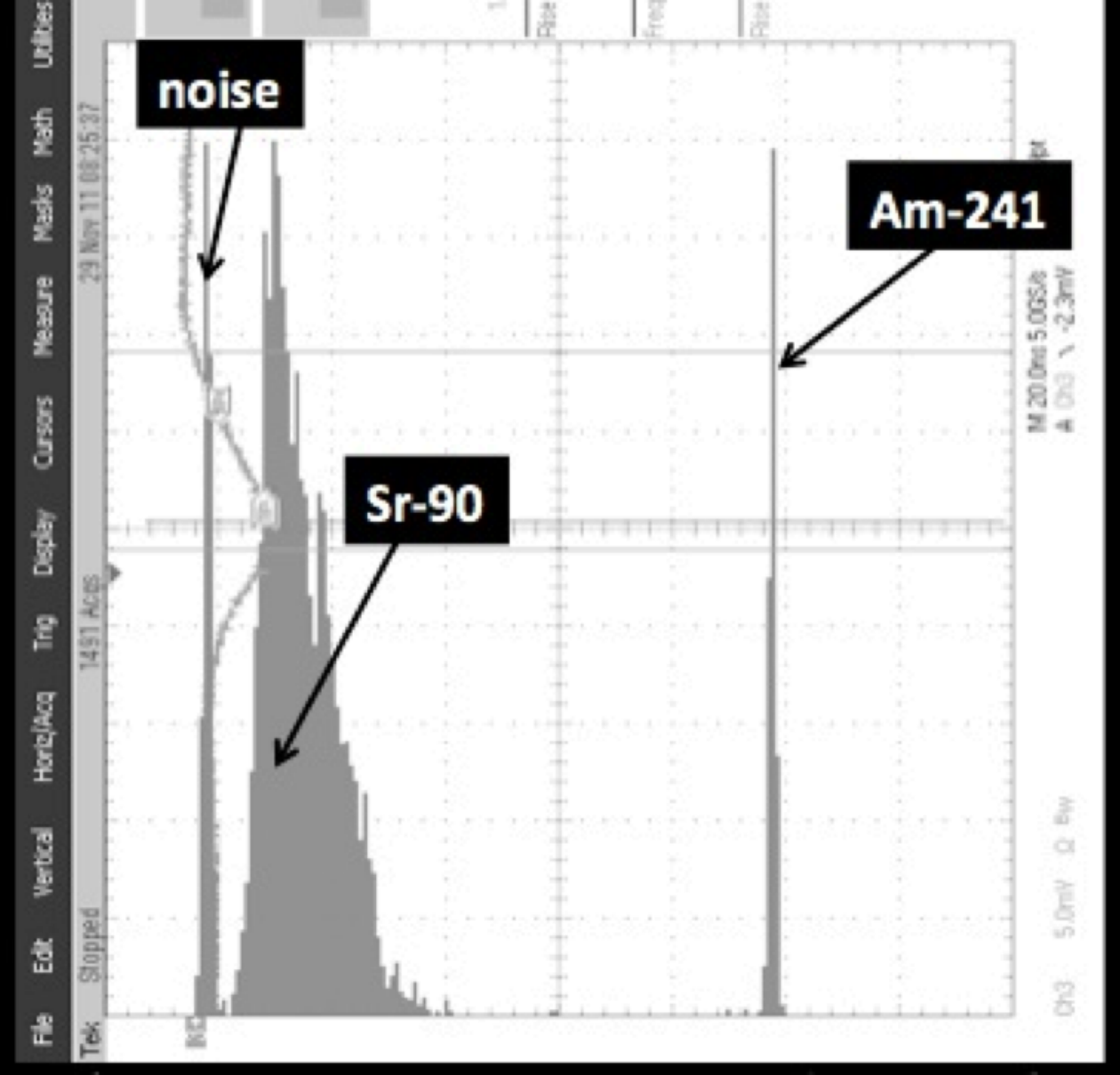}
\caption{5.5 MeV alpha and electrons signals from a $^{241}$Am source
and from a $^{90}$Str radioactive source respectively, measured with a
scCVD diamond sensor.}
\label{fig:signal}
\end{figure}

\paragraph{Tests at the Tor Vergata INFN laboratory}

Mono and polycrystalline CVD diamond sensors, purchased by Element Six
Ltd\footnote{Kings Ride Park Kings Ride, Ascot, Berkshire, SL5 8BP,
U.K.}  have been tested at the Roma Tor Vergata INFN laboratory. The
sensors were read-out with three different amplifiers, placed about 4
m away:
\begin{itemize}
  \item Two-stage amplifier, AC, (BJT Si BFQ67)
  \item AC, (BJT SiGe, BFP650)
  \item AC, (BJT SiGe, BFP740)
\end{itemize} 
The SiGe transistors have a 
short transit time in the base because they take advantage of the
ballistic effect due to the electric field. This produces an increase
in the performance, particularly in the transition frequency,
amplification, and noise reduction. The features
of the BFP740 amplifier used in the test are:
\begin{itemize}
  \item Voltage supply: 5 V
  \item Sensitivity: 6 mV/fC
  \item Noise: 500 e$^{-}$ RMS
  \item Input impedance: 50 $\Omega$
  \item Band width: 30 MHz
  \item Power consumption: 10 mW/ch
  \item Cost: a few {\euro}/ch
  \item Radiation hardness: 50 Mrad, $10^{15}$ n/cm$^{-2}$
\end{itemize}

Figure \ref{fig:oscilloscope} shows an example of a signal from a CVD
diamond detector induced by a minimum-ionizing particle.  Laboratory
tests with $^{241}$Am and $^{90}$Sr radioactive sources of both
mono-crystalline and poly-crystalline CVD diamond sensors gave very
promising results in terms of signal-to-noise separation. These
amplifiers have already been employed in measurements to read-out CVD
diamond sensors for neutron spectroscopy application at the ITER
Tokamak project\cite{Angelone:2008}. The detector's signal from
radioactive sources has been transported up to the fast charge
pre-amplifier by means of a high frequency, single, low attenuation
four meters long cable. The preamplifier is able to read, stretch (up
to 100~ns) and amplify the small and ultra fast ($<$100~ps wide)
signal produced by the radiation in the diamond detector and has a
measured noise of about 500 e$^-$ r.m.s. Figure \ref{fig:IV} shows the
characteristic I-V curve of a scCVD diamond detector measured with a
picoammeter. Figure \ref{fig:signal} shows the signals induced in
the scCVD diamond sensor by 5.5~MeV alpha particles from a
$^{241}$Am radioactive source and electrons from a $^{90}$Sr
radioactive source. These measurements have demonstrated that adequate
performance in terms of SNR can be achieved with the fast high
transition frequency SiGe transistor amplifiers.

\bibliographystyle{\tdrbibstyle}
\balance
\bibliography{MDI/MDI-bib,IFR/IFR-bib}


\renewcommand{\ChapLabel}{chap:SVT} 
\tdrchap{Silicon Vertex Tracker}
\label{chap:svt}

\tdrsec{Overview}
\label{svt:overview}


The Silicon Vertex Tracker (SVT), as in \babar,  together with the drift chamber 
(DCH) and the solenoidal magnet provide both track and vertex reconstruction  
capability for the \superb\ detector. 
Precise vertex information, primarily extracted from precise position 
measurements near the interaction point (IP) by the SVT, is crucial to the measurement of 
time-dependent \CP asymmetries in $B^0$ decays, which remain a key element 
of the \superb\ physics program. In addition, charged particles with 
transverse momenta lower than 100\mevc  will not reach the DCH, so for these particles the SVT must
provide the complete tracking information.

\tdrsubsec{SVT and Layer0}

\begin{figure*}[tbp]
\begin{center}
\includegraphics[clip=true,trim=0cm 0cm 0cm 21cm, width=\textwidth]{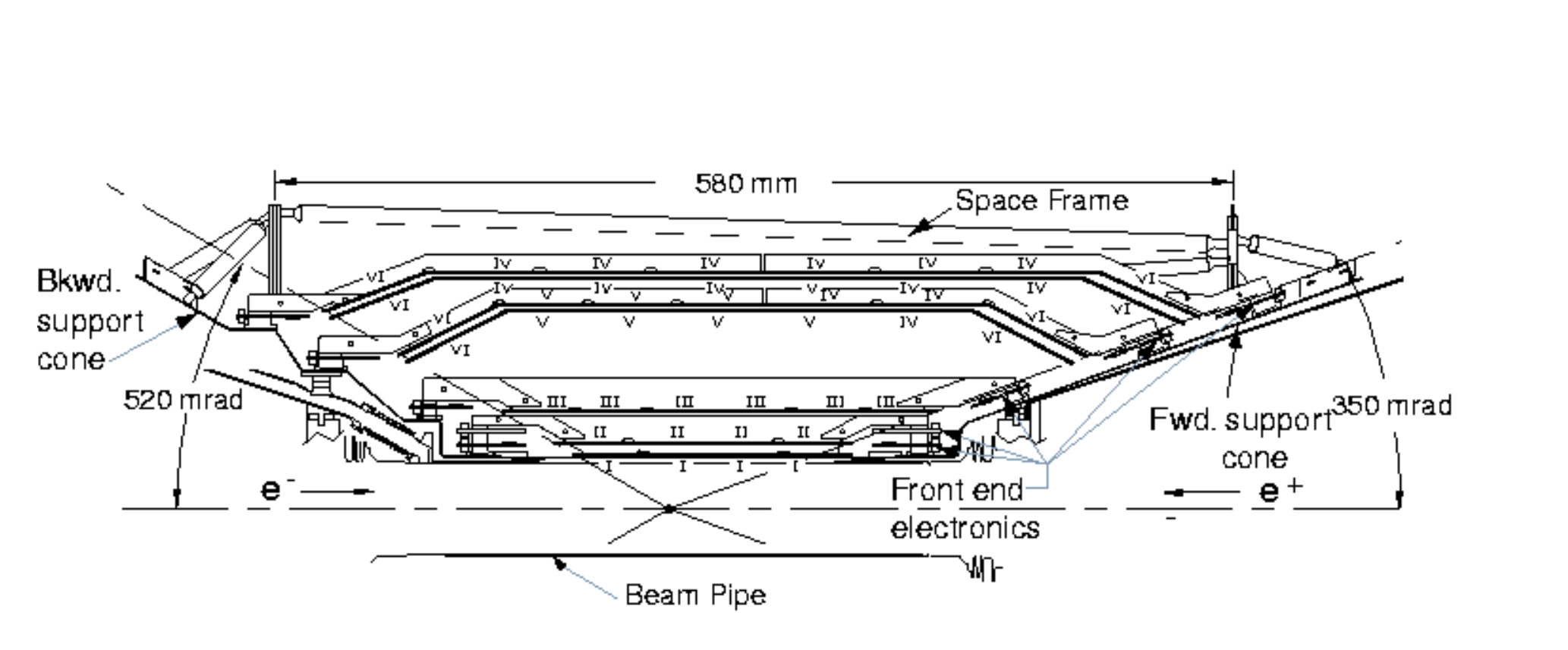}
\end{center}
\caption{Longitudinal section of the \babar\ SVT.}
\label{fig:babar_svt}
\end{figure*}

The above goals where achieved in the \babar\ detector with a five-layer  
silicon strip detector with a 
low mass design, shown schematically in Figure~\ref{fig:babar_svt}.
The \babar\ SVT 
provided excellent performance for the whole life of the
experiment, thanks to a robust design~\cite{Bozzi:2000} that took 
into account the physics 
requirements as well as a sufficient safety margin, to cope with the machine 
background, and redundancy considerations. 
%
\begin{figure*}[tbp]
\begin{center}
\includegraphics[width=\textwidth]{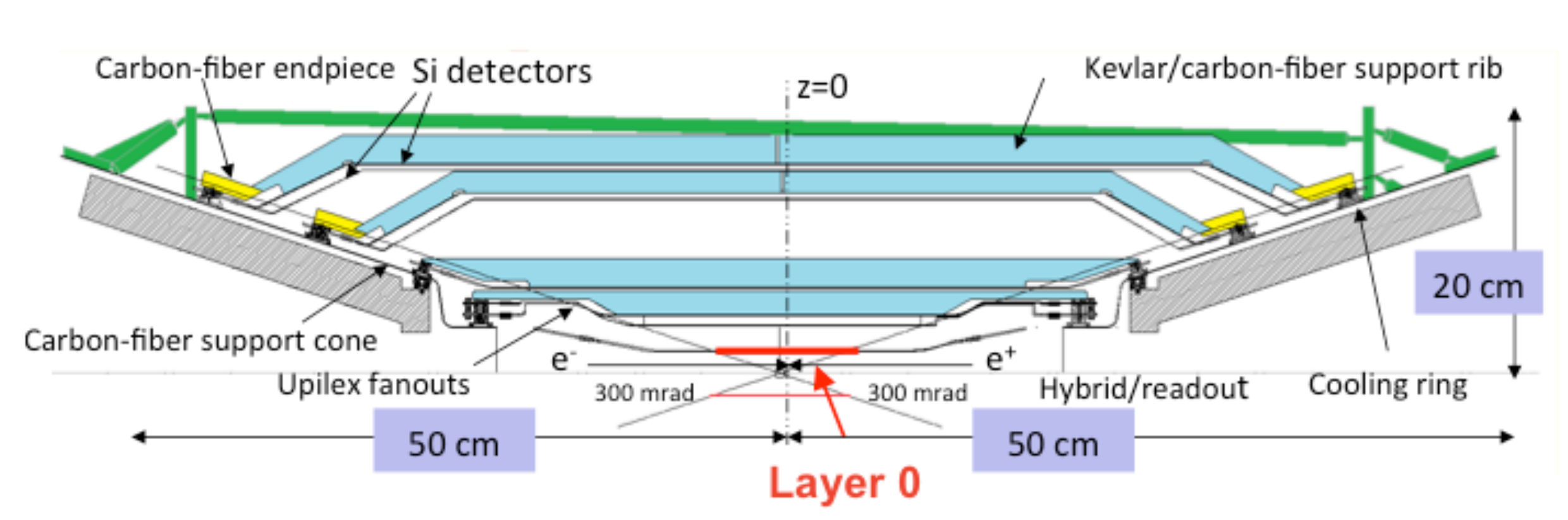}
\end{center}
\caption{Longitudinal section of the \superb\ SVT.}
\label{fig:superb_svt}
\end{figure*}

The \superb\ SVT design, shown schematically in Figure~\ref{fig:superb_svt}, is 
based on the \babar\ vertex detector layout 
with those modifications needed to operate at an instantaneous luminosity of $10^{36}$~cm$^{-2}$~s$^{-1}$ 
or more, and with a reduced center-of-mass boost.
In particular the SVT will be equipped  with an innermost layer closer 
to the IP (Layer0) to improve  
vertex resolution and compensate for the reduced boost at the \superb 
accelerator, thus retaining an adequate   
$\deltat$ resolution for \B decays for time-dependent \CP asymmetries.
Preliminary studies indicate the same is true for time-dependent mixing and 
\CP asymmetry measurements of charm decays.

Physics studies and background conditions, as explained in detail 
in the next sections, set stringent
requirements on the Layer0 design: radius of about 1.5\cm; resolution of 
10-15$\mum$ in both coordinates; 
low material budget (about 1\% \Xrad); and adequate radiation resistance.

Several options are under study for the Layer0
technology, with different levels of maturity, expected performance and safety 
margin against background conditions. These include striplet modules based on 
a high resistivity double-sided 
silicon detector with short strips, 
hybrid pixels and other thin pixel sensors based on CMOS Monolithic Active
Pixel Sensors (MAPS).

The current baseline configuration of the SVT Layer0 is based on the
striplet technology, which has been shown to provide the better physics 
performance, as
detailed in the next sections. However,  options based on pixel sensors, 
which are more robust in high
background conditions, are still being developed with specific R\&D programs 
in order to meet
the Layer0 requirements (see Section~\ref{sec:pixel_upgrade}). 
These include low pitch and material budget, 
high readout speed and
radiation hardness. If successful, this will allow the replacement of the 
Layer0 striplets modules in
a ``second phase" of the experiment. For this purpose, the \superb\ 
interaction region and the SVT mechanics will be designed to ensure
relatively rapid access to the detector for a replacement of Layer0.

The external SVT layers (1-5), with a radius between 3 and 15\cm, 
will be built with the same technology used for the \babar\ SVT (double
sided silicon strip sensors), which is adequate for the machine 
background conditions expected in the \superb\ accelerator scheme
(\ie\ with low beam currents). 
Although the SVT module design for layers 1 through 5 will be very similar to that of \babar, 
albeit with larger solid angle coverage, a completely 
new readout electronics chain needs to be developed to cope with the 
higher background rates expected 
in \superb\ .


A review of the main SVT requirements will be given in the next 
sections, followed by an overview of the general detector layout. 
A detailed discussion of all the specific design aspects will be covered 
in the rest of the chapter.

\tdrsubsec{SVT Requirements}

\tdrsubsubsec{Resolution}

Without the measurement of the \B decay vertex, no useful time-dependent \CP asymmetries 
can be extracted at the \FourS. 
Therefore one of the main goals of the SVT is the determination of the 
\B decay positions, especially along the beam direction ($z$). 
Measurements performed in \babar, where the mean separation between 
\B vertices is \deltaz $\simeq \beta\gamma c \tau_{B} = 250 \mum$, 
demonstrated that good sensitivity to time dependent measurement can 
be achieved with typical vertex resolution of 50-80 $\mum$ in the $z$ 
coordinate for exclusively reconstructed modes, 
and 100-150 $\mum$ for inclusively modes (tag side in CPV measurements).
The reduced \superb\ boost  ($\beta\gamma = 0.24$) with respect to  PEP-II  
($\beta\gamma = 0.55$) requires an improved vertex resolution, by about 
a factor of two, in order to maintain a suitable $\deltat$ resolution for 
time dependent analyses.

The \babar\ resolution was achieved thanks to an intrinsic detector 
resolution of about 10 $\mum$ in the first measured point of 
the SVT, taken at a  radius of about 3 cm, and keeping to the minimum 
the amount of material between the IP and the first measurement. 
Multiple scattering affects the resolution on the impact parameter 
especially for low momentum tracks, and sets a lower limit on the useful 
intrinsic resolution on the various SVT layers. These correspond to a point 
resolution of about 10-15 $\mum$ for 
measurements made close to the IP and 30-40 $\mum$ for the outer 
layers \cite{babar:TDR}.


The required improved track impact parameter and vertex resolution 
can be reached in \superb\, with the same 
intrinsic resolution used in \babar, by reducing the radius of the first 
measured SVT point by a factor of 2 
(Layer0 radius at about 1.5 cm)
and keeping a very low mass design for the beam pipe and the detector itself. 

\tdrsubsubsec{Acceptance}

The coverage of the SVT must be as complete as technically feasible, 
given the constraints of the machine components close to the IP.
The SVT angular acceptance, constrained by the \superb\ interaction 
region design, will be 300\mrad in both the forward and backward direction. 
This corresponds to a solid angle coverage of 95\% in the e$^+e^-$ center-of-mass  (CM)  
frame, thus increasing the acceptance with 
respect to the \babar\ SVT.

There should be as little material as possible within the active 
tracking volume;
minimizing the material between the IP and the first measurement 
is crucial to reducing the multiple 
scattering and preserving the impact parameter resolution. 
The small beam pipe (1 cm radius) requires active cooling with liquid coolant in the detector acceptance
to remove the large power deposition due to 
image beam currents. The total amount of radial material for the 
actual design of this new beryllium pipe is estimated to be less 
than 0.5\% $\X_0$. 
Material located beyond the inner layers does not significantly degrade 
the measurement of track impact parameters, 
but does affect the performance of the overall tracking system and
increases photon conversions in the active region.

\tdrsubsubsec{Efficiency}

Our goal is to achieve
close-to-perfect track reconstruction efficiency within the active
volume of the tracking detectors when information from
both the DCH and the SVT is used.
The pattern recognition capabilities of the combined tracking system 
must be robust enough to tolerate background levels up to 5 times the
nominal rates, as reported in Table~\ref{tab:background_summary}.
Low momentum particles
that do not traverse many drift chamber planes,
such as many of the charged
pions from $D^{*}$ decays, 
must be reconstructed in the SVT alone. 
For this category of tracks, with transverse momentum \pt\ less than 100\mevc, we  
want to achieve reconstruction efficiencies of at least 80--90\%. 
The SVT must also be efficient for particles such as
$\KS$s that decay within the active volume.
  
The \babar\ SVT design with 5 layers was optimized to ensure enough 
redundancy to keep a high tracking efficiency even in the case of failure 
of some modules or inefficient detectors.
The robustness of this choice was demonstrated with the good detector 
performance over the entire life of the \babar\ experiment.
The \superb\ SVT design with 6 layers (inserting the Layer0) 
is inspired by the same philosophy. 
Simulation studies \cite{walsh:frascati09} of a non-perfect/real detector in high background running conditions, indicated that a reduction 
in the number of layers, from 6 to 5 or 4, yields very modest gains in tracking 
performance, while showing a sizeable reduction in the reconstruction efficiency 
for low momentum tracks in $D^{*+}$ decays.

\tdrsubsubsec{Background \& Radiation Tolerance} 

The expected background influences several aspects of the SVT 
design (segmentation, shaping time, data transmission rates), and 
sets the requirements for the 
radiation resistance of the SVT components.
The system has been designed to withstand at least 5 times 
the total expected background rates. 
Whenever possible, detectors and 
front-end electronics are specified to be able to survive the entire 
life of the experiment including a safety factor of 5 on the total 
dose and equivalent neutron 
fluence: 7.5 $n_{eq}$/cm$^2$/yr$\times$5 yrs at nominal peak luminosity of 10$^{36}$~cm$^{-2}$~s$^{-1}$.
 
As described in Section~\ref{svt_background}, 
the effect of background depends steeply on 
radius, as shown in Tab.~\ref{tab:background_summary}.

With the high strip rates expected, especially in the inner layers (0-3), 
the front-end 
electronics should be fast enough to avoid pulse overlap and consequent hit 
inefficiency. This requires shaping time in the range 25-150 ns. Furthermore a good 
time stamp (TS) resolution (30 MHz TS clock) is needed to get a good hit time 
resolution. The hit time resolution determines the width of the time window cut used in the reconstruction 
code  to accept/reject a hit (offline time window $\simeq 10 \sigma_{t-hit}$), and it is then crucial to 
reduce the offline occupancy  to  levels acceptable for efficient reconstruction.
The offline cluster occupancy in the x5 nominal background scenario, defined as the detector strip occupancy after applying the 
offline time window cut, divided by the hit multiplicity in a cluster, is on 
average 2\% in each SVT layer.
This corresponds to offline time windows of 100-150 ns in layers 0-3, 
and about 500 ns in layers 4-5, where the longer shaping time 
is dominating the hit time resolution. See Sections~\ref{sec:performance_and_background} and 
\ref{sec:chips_requirements} for more details.

In Layer0 the expected integrated dose is about 3\,Mrad/yr and the 
equivalent neutron fluence is about 
$5\times 10^{12}$ $n_{eq}$/cm$^2$/yr in the sensor area. 
In the other SVT layers radiation
levels are at least one order of magnitude lower: 
in Layer1 a Total Integrated Dose (TID) $\simeq$ 0.3 Mrad/yr and an equivalent neutron fluence of about 
$8\times 10^{11}$ $n_{eq}$/cm$^2$/yr are expected.

With this expected background in Layer1-5, sensors are proven 
to be sufficiently radiation hard to survive the entire life of 
the experiment (with a safety factor 5 included), maintaining acceptable
performance, as discussed in Section~\ref{par:SVT:SensorDesign} and 
\ref{sec:chips_requirements}.

For Layer0, where the radiation is an order of magnitude higher, a quick
replacement of the entire layer is foreseen, as frequently as necessary, 
depending on the actual background and the radiation hardness of the 
technology chosen.

\tdrsubsubsec{Reliability}

Although the \superb interaction region and the SVT mechanics will be designed 
to ensure a relatively rapid access to the detector for replacement of Layer0,
access to the SVT is not possible without a major shutdown.
The reliability requirements for the SVT are therefore more
stringent than usual for such a device, with implications for 
engineering design at all levels. The detector layout must provide redundant
measurements wherever possible; the electronic readout must be
very robust, and the functionality of all components must not
be compromised by exposure to the expected radiation levels.
The detector monitoring and interlock system must serve as a
safeguard against major failure in the
event of a component malfunction or a human error.

\tdrsubsec{Baseline Detector Concept}

\tdrsubsubsec{Technology}

The SVT baseline design is based on double-sided silicon microstrip
detectors for all layers.  The characteristics of this technology that make it
attractive for the \superb\ detector are: high precision for measuring
the location of charged-particles tracks, tolerance to high background
levels, and reduction in mass made possible through double-sided
readout. Double-sided silicon detectors have already been employed with success 
in \babar\ and in several other large-scale applications, meeting
 the performance standards outlined above.

\tdrsubsubsec{Layout}

%
%

\begin{figure}[!bh]
\begin{center}
\includegraphics[clip=true,trim=3cm 9cm 3.5cm 9cm, angle=90, width=0.5\textwidth]
{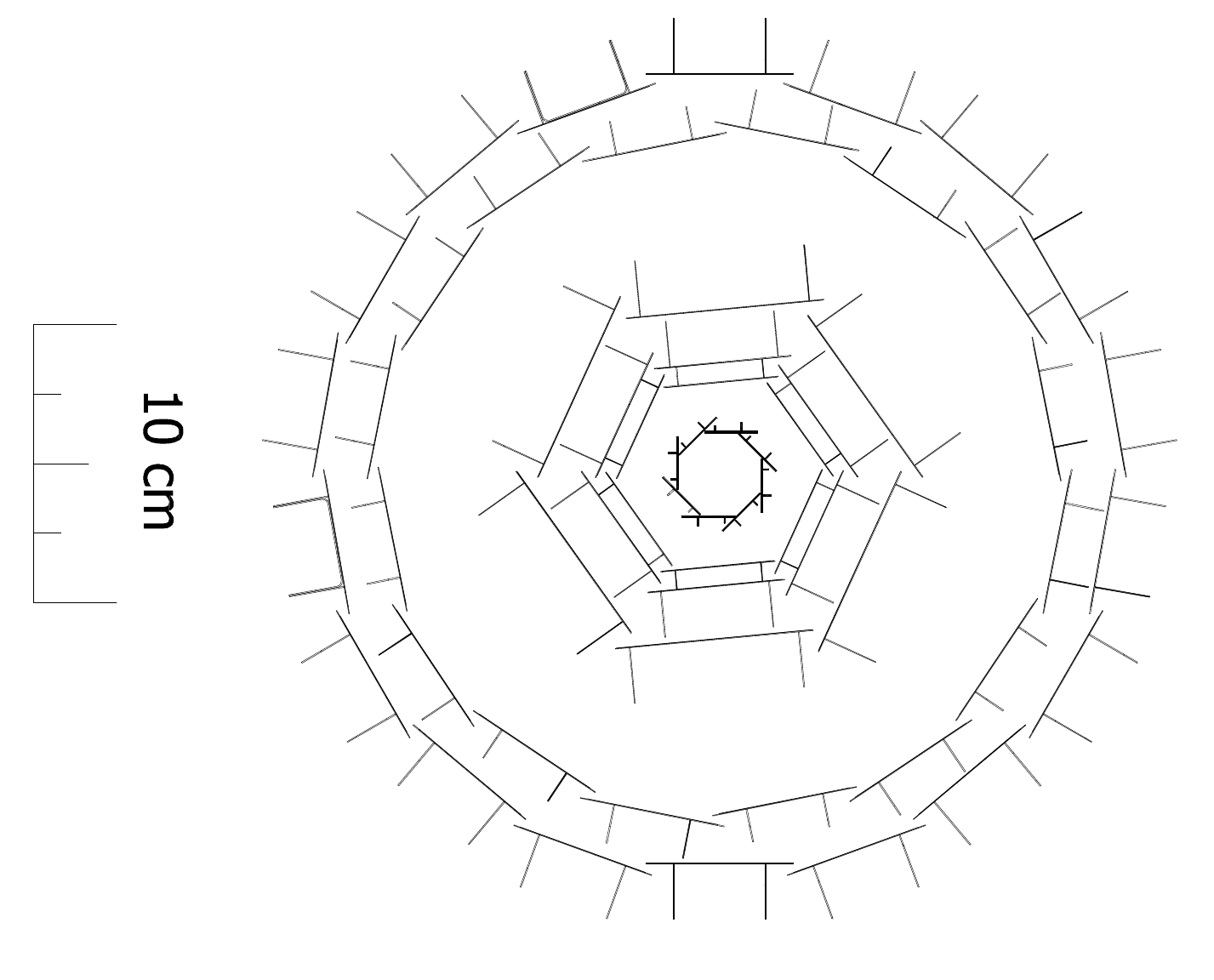}
\end{center}
\caption{Cross section of the SVT in the plane perpendicular to the 
beam axis. The lines perpendicular to the detectors represent structural 
support beams.}
\label{fig:svt_rphi}
\end{figure}

\begin{figure*}[tbp]
\begin{center}
\includegraphics[clip=true,trim=5cm 6cm 5cm 5cm, angle=90, width=\textwidth]{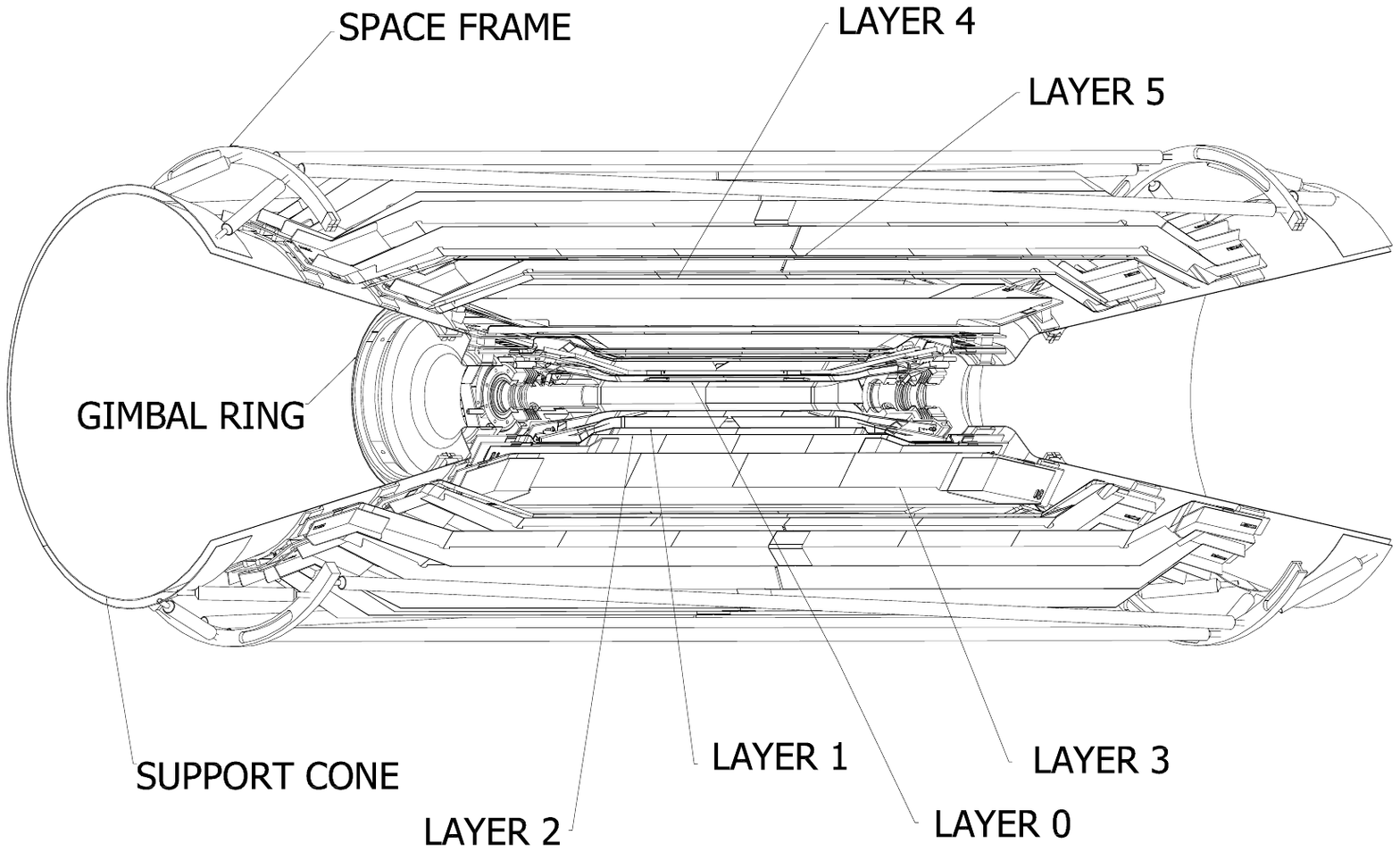}
\end{center}
\caption{Three dimensional cutaway of the SVT.}
\label{fig:svt_3D}
\end{figure*}

The SVT will provide six measurements,
in two orthogonal directions, of the positions of all charged
particles with polar angles in the region  
17\degrees $< \theta < $167 \degrees .
Layer0 has eight detector modules; 
the rest of the detector 
keeps the same number modules/layer as in \babar: layers 1-2-3  have
six detector modules,  arrayed azimuthally around the beam pipe,
while the outer two layers
consist of 16 and 18 detector
modules, respectively.  A side view of the detector is shown in
Figure~\ref{fig:superb_svt}, and 
an end view is shown in Figure~\ref{fig:svt_rphi}.  A three-dimensional cut-away view of the SVT is shown 
in Figure~\ref{fig:svt_3D}.  

\begin{figure*}[tbp]
\begin{center}
\includegraphics[clip=true,trim=0cm 0cm 0cm 0cm, width=\textwidth]
{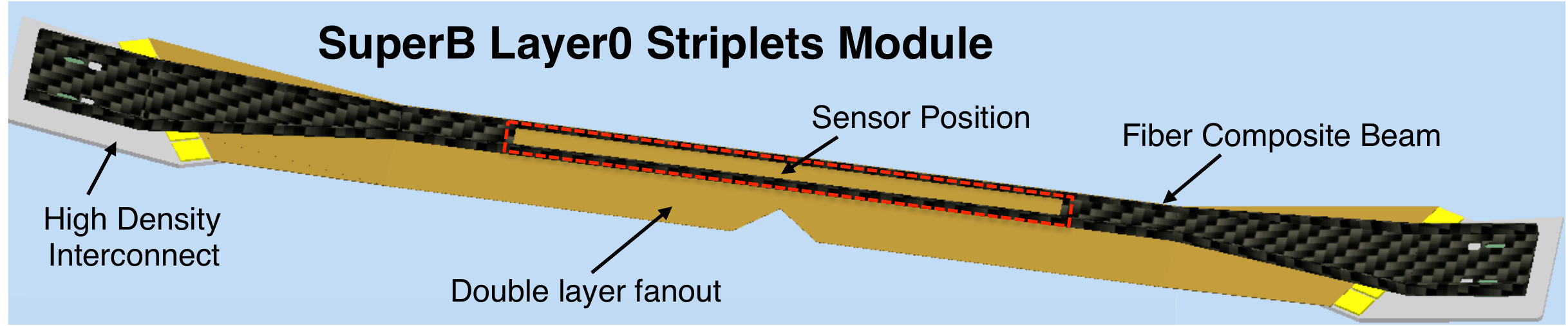}
\end{center}
\caption{Schematic drawing of the Layer0 striplet module.}
\label{fig:layer0_striplets_module}
\end{figure*}
\begin{figure*}[tbp]
\begin{center}
\includegraphics[clip=true,trim=0.5cm 0.1cm 0cm 0cm, width=0.9\textwidth]
{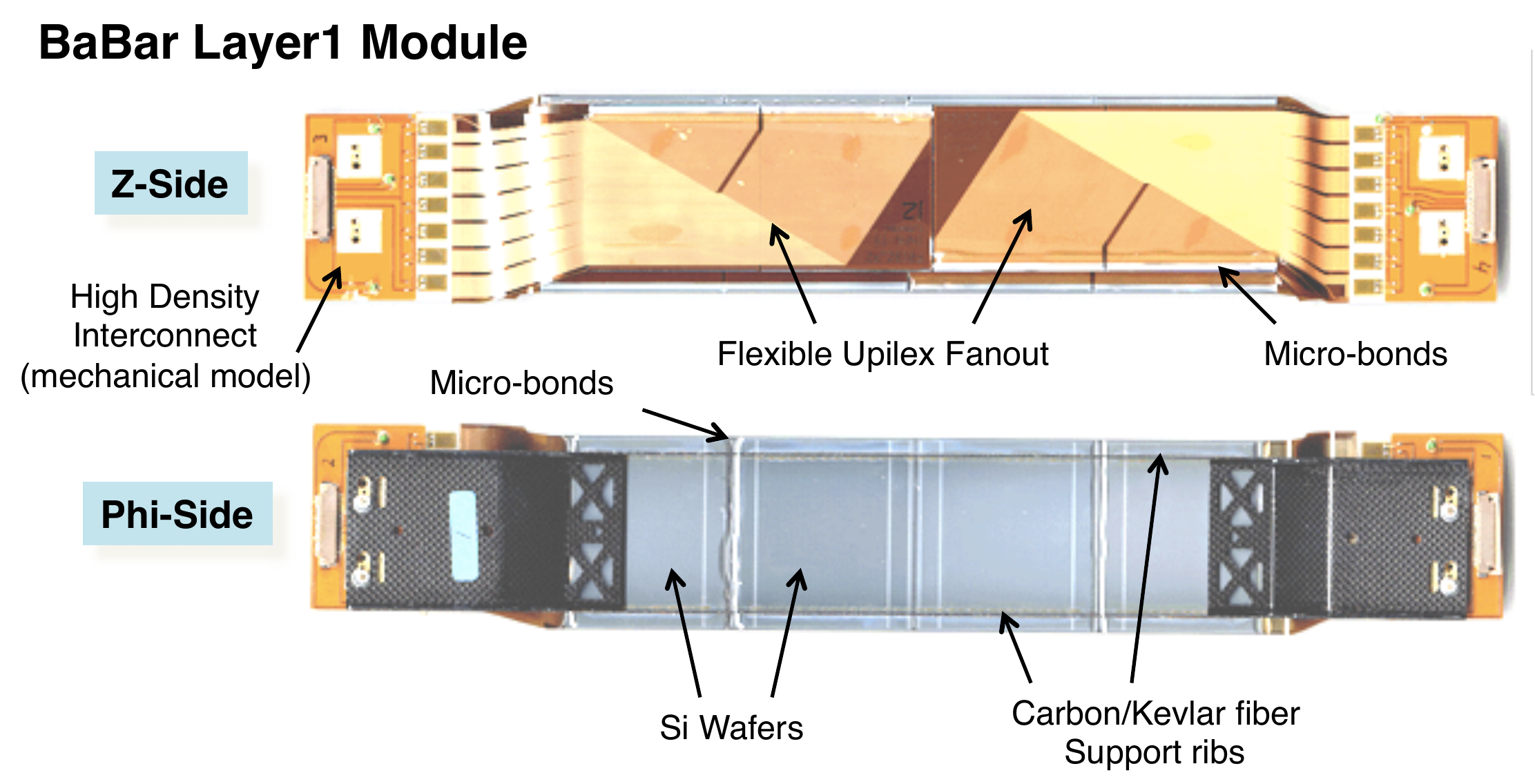}
\end{center}
\caption{Details of the \babar\ SVT Layer1 module.}
\label{fig:babar_layer1_module}
\end{figure*}
\begin{figure}
\begin{center}
\includegraphics[clip=true,trim=0cm 0cm 0cm 0cm, width=0.4\textwidth]
{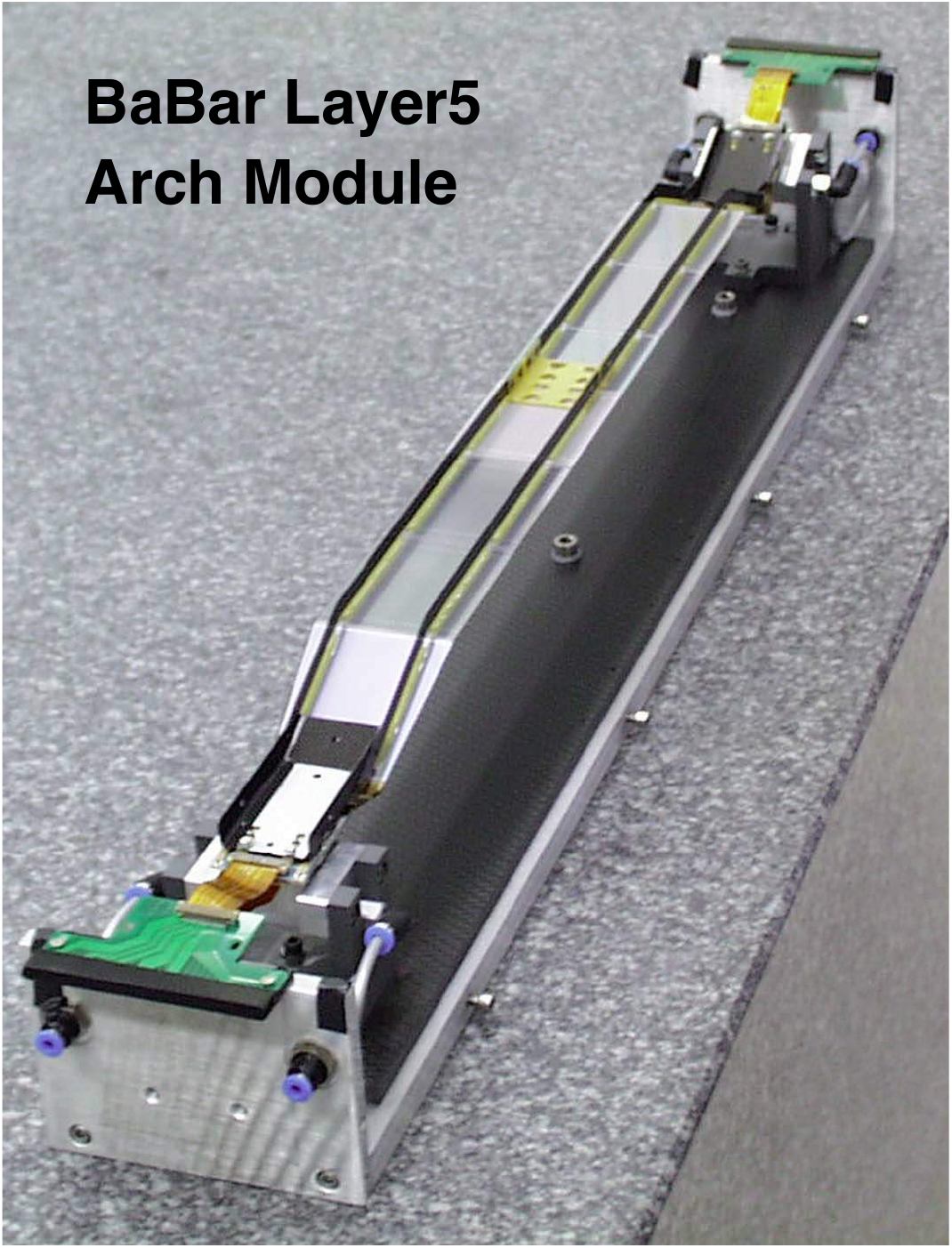}
\end{center}
\caption{Details of the \babar\ SVT Layer5 arch module.}
\label{fig:babar_arch_module}
\end{figure}

The design of the Layer0 striplets module is completely new, and it has
quite a complex shape, as shown in Figure~\ref{fig:layer0_striplets_module}.
This is necessary to  
fit within the 300 mrad acceptance and 
the very limited space available between 
Layer1 and the beam pipe.
The layout of the other five layers is very similar to the \babar\ 
SVT strip modules, shown as a reference in 
Figure~\ref{fig:babar_layer1_module} and Figure~\ref{fig:babar_arch_module}.

The inner detector modules (layers 0 through 3) are traditional barrel-style structures,
while the outer detector modules (layers 4 and 5) employ an arch structure,
in which the detectors are electrically connected across an angle.
The bends in  the arch modules, proven to be functional in \babar,
minimize the area of silicon required to cover the
solid angle and also  
avoid very large track incident angles. 

In order to satisfy the requirement of minimizing material in the  
detector acceptance region,
one of the main features of the SVT design is the mounting  
of the readout electronics
entirely outside the active detector volume. 
For this reason signals from the silicon strips 
are carried to the front-end chips by flexible fanout circuits.

There is a 1\cm\ space  
between the 300\mrad\ stay-clear in the forward and backward directions 
and the first element of the IR ({\it i.e.}, the tungsten shield cones) and 
all of the electronics are mounted here.  
In both directions, space is very tight,  and
the electronic and mechanical designs are closely coupled in the narrow
region available.

The layout specifications for this six-layer design are given in 
Table~\ref{tab:svt_layout} and described in more detail in the text.

For Layer0 short strips, oriented at 45 degrees with respect to the detector 
edges ($u$, $v$ strips), are adopted on both faces of the sensor 
in order to reduce the strip length and the related 
background occupancy to reasonable levels.
For layers 1 to 5 the strips on the two sides of the rectangular detectors 
in the barrel regions are oriented parallel
($\phi$ strips) or perpendicular ($z$ strips) to the beam line.
In the forward and backward regions of the two outer layers,
the angle between the strips on the two sides of the trapezoidal
detectors is approximately 90\degrees , and
the $\phi$ strips are tapered.
Floating strips are used to improve the position resolution
for near-perpendicular angles of incidence;
the capacitive coupling between the floating strip and
the neighboring strips results in 
increased charge sharing and better interpolation.
For larger incident angles the wider readout pitch minimizes the 
degradation in resolution that occurs because of the limited track path  
length associated with each strip.
These issues are discussed in more detail in section~\ref{par:SVT:SensorDesign}.

\begin{table*}[tbp]
\caption{See text for more detail
on the meaning of the different quantities. The intrinsic resolution, quoted for tracks at 90\degrees 
incidence angle, is taken from \babar\ SVT data for Layers 1-5 (pitches are identical). For Layer0, data from a recent
beam test on a striplet module prototype are used.
The $z$-ganging/pairing numbers  
represent the percentage of readout channels connected to the specified strip configuration.}
\begin{tabular}{|l|c|c|c|c|c|c|c|c|} \hline
 Quantity 
 & Layer & Layer & Layer & Layer & Layer & Layer & Layer & Layer \\
 & 0 & 1 & 2 & 3 & 4a & 4b & 5a & 5b  \\
\hline\hline
 Radius (mm) & 15.6 & 33 & 40 & 59 & 120 & 124& 140 & 144     \\
 Wafers/Module & 1 & 2 & 4 &4 & 6 &6  &8 &8 \\  
Modules/Layer  &8 &6 &6 & 6 & 8 & 8 & 9 & 9\\
Silicon Area (cm$^2$)  & 127 & 554  & 787 & 1655 & 2459 & 2548 & 3502 & 3610 \\
Overlap in $\phi$ (\%) & 2.0 & 2.4 & 1.8 & 1.8 & 4.0 & 4.0  & 2.0 & 2.0 \\   
\hline
Readout pitch ($\mu$m): & & & & &\multicolumn{2}{|c|}{} &  
\multicolumn{2}{|c|}{} \\
 $\phi$ ($u$ for Layer 0) &54 &50 &55 &100 &\multicolumn{2}{|c|}{82--100} &  
\multicolumn{2}{|c|}{82--100} \\
 $z$ ($v$ for Layer 0)   &54&100&100&110 &\multicolumn{2}{|c|}{210} &  
\multicolumn{2}{|c|}{210} \\
\hline
Floating Strips: & & & & & \multicolumn{2}{|c|}{} & \multicolumn{2}{|c|}{} \\
 $\phi$ ($u$ for Layer 0) &--- &--- &--- &1 &\multicolumn{2}{|c|}{1} & \multicolumn{2}{|c|}{1} \\
 $z$ ($v$ for Layer 0)   &--- &1 &1 &1 &\multicolumn{2}{|c|}{1} & \multicolumn{2}{|c|}{1} \\
\hline
Intrinsic  & & & & & \multicolumn{2}{|c|}{} & \multicolumn{2}{|c|}{} \\
Resolution ($\mu$m): & & & & & \multicolumn{2}{|c|}{} & \multicolumn{2}{|c|}{} \\
 $\phi$ ($u$ for Layer 0) &10 &10 &10 & 15 & \multicolumn{2}{|c|}{15} &  
\multicolumn{2}{|c|}{15} \\
 $z$ ($v$ for Layer 0)   &10 &14 &14 & 15 & \multicolumn{2}{|c|}{25} &  
\multicolumn{2}{|c|}{25} \\
Readout Section & & & &  &  \multicolumn{2}{|c|}{} & \multicolumn{2}{|c|}{} \\
ROS/Module &4 &4 &4 & 4 &  \multicolumn{2}{|c|}{4} & \multicolumn{2}{|c|}{4} \\
ICs/ROS ($\phi$-$z$) \hfill & 6-6  & 7-7  &  7-7  & 6-10 &   \multicolumn{2}{|c|}{4-5}&
 \multicolumn{2}{|c|}{4-5} \\
 Readout Channels  & 24576 & 21504 & 21504 & 24576 &\multicolumn{2}{|c|}{36864}  
& 
\multicolumn{2}{|c|}{41472} \\
\hline
Strip Length \hfill & & & & & & & & \\
Half Module (mm):\hfill & & & & & & & & \\
$\phi$ ($u$ for Layer 0)  & 20 & 110 & 130 &  190 & 293 & 303 & 369 & 380 \\
 $z$ ($v$ for Layer 0) & 20 & 40 & 48  & 70   & 51--103& 103--154 & 103--154 & 103--154 \\
\hline
Fraction of $z$-side   &  &  &  &  &   &   &   & \\
readout channels with &  &  &  &  &   &   &   & \\
Pairing/Ganging:   &       &       &      &     &      &       &      & \\
None &   & 77\%  & 55\% & 65\%  & 4\% &     &      &      \\
Pairing $\times$2  &    &  23\%  &  45\% &   35\%  &     &     &     &  \\
Ganging $\times$2 &     &     &     &      & 73\% & 74\%  & 25\% & 16\% \\
Gang.$\times$2 + Pair.$\times$2 &       &       &      &     & 23\% & 24\%  & 41\%  & 43\%  \\
Ganging $\times$3 &       &       &      &      &     & 2\%  & 34\%  &41\%  \\
\hline

\end{tabular}
\label{tab:svt_layout}
\end{table*}


The design has a total of 308 silicon detectors of nine different  
types.
The total silicon area in the SVT is about 1.5\m$^2$, and
the  number of readout channels is $\sim$170,000.

\tdrsubsubsec{Readout Electronics }

As emphasized above, all readout electronics are located outside the 
active volume, below 300\mrad\ in the forward and backward regions.
To accomplish this, $\phi$ strips on the forward 
or backward half of a detector module
are electrically connected with wire bonds.
This results in total strip lengths associated with a single readout  
channel
of up to 
$\sim$19\cm\ in the inner layers and up to $\sim$38\cm\ in the outer two
layers.  

The signals from striplets for the Layer0 ($u$ and $v$ strips) and the $z$ 
strips for all the other layers are brought to the readout electronics 
using fanout circuits consisting of 
conductive traces on a thin flexible insulator (for example, copper traces on  
Upilex, as in \babar).
The fanout traces are wire-bonded to the ends of the silicon strips.
  
On the $z$ side of the modules 
the number of readout strips 
exceed the number of available 
electronic channels, constrained by the number of chips that can fit in 
the limited space available. To reduce the number of readout channels 
needed,    
the connection scheme for the $z$ 
fanout circuits includes ``pairing'' and ``ganging'' 
(described in Section~\ref{par:SVT:SensorDesign}) with two or three 
 strips bonded to a single fanout/readout channel.
The length of the $z$ strips is much shorter than $\phi$ strips, 
typically 4-7\cm\ in the
inner layers and either 10 or 15\cm\ in the outer layers, where 
there is either $\times$2 or $\times$3 ganging.

Front-end signal processing is performed by ICs mounted on the 
High-Density Interconnect (HDI), a thick-film hybrid circuit 
fabricated on aluminum nitride (AlN) substrate.
The HDI provides the physical support; it distributes power and signals, 
and thermally 
interfaces the ICs to the cooling system.

New front-end custom-design ICs are currently under development for the 
\superb\ SVT \cite{giorgi:elba2012} since none of the existing chips 
matches all 
the requirements (Sec.~{\ref{sec:chips_requirements}).  
The signals from the readout strips, after amplification and shaping,
are compared to a preset threshold.
The time interval during which they exceed the threshold (time over
threshold, or ToT) is related to
the charge induced on the strip.
The ToT technique has a nonlinear 
input-to-output relationship which is approximately logarithmic.
This is an advantage, since it compresses the dynamic range and
allows one to achieve good position resolution
and large dynamic range with a minimum number of bits.
ToT readout has been successfully employed in the front-end chip of the 
\babar\ SVT ({\it i.e.}, Atom chip~\cite{Bozzi:2000})
providing good analog resolution
for position interpolation, time-walk correction, and background
rejection.

For each channel with a signal
above threshold,
the ToT information together with the hit time stamp will be buffered until 
a trigger is received; 
it will then be transferred, with the strip number, to an output interface,
where data will be serialized and 
transmitted off chip on output LVDS lines. 

The readout IC is expected to be about 6x4 mm$^2$ and to dissipate
about 4.0 mW per channel.
The total power that will be generated by the SVT readout chips 
is $\sim$700 W.

There are four readout sections (ROS) per detector module: the module
is electrically divided in half along $z$, and the $\phi$ and $z$ strips are 
read out separately. The data from one-half of a detector module 
will be transmitted from the hybrid
on a flexible cable to a transition card located approximately
50\cm\ away, where the signals are converted and transmitted 
to optical fibers.

\tdrsubsubsec{Module design and Mechanical Support}

The silicon detectors and the associated readout electronics are
assembled into mechanical units called detector modules.
Each module contains from 1 to 8 silicon detectors, the flex circuits
to bring the signal from strip to the front-end chips, and a low-mass beam 
constructed of carbon and Kevlar fiber-epoxy laminates ({\it i.e.}, ribs) to stiffen 
the module structure.
The ribs are attached through a carbon fiber end-piece to the HDI hybrid 
circuit.
An aluminum nitride substrate for the HDI
provides precise mechanical mounting surfaces and heat sink for
the electronics.

The Layer0 module has certain unique features (see Figure~\ref{fig:layer0_striplets_module}). 
Only one sensor is used; the strips are still orthogonal on both sides 
but are tilted at a 45$^\circ$ angle with respect to the detector edge. 
Layer0 is arranged in an octagonal geometry ({\it i.e.}, not hexagonal. as in layers 1-3) to have 
shorter strip length. 
Due to the large number of strips on each side, the fanout circuits must have 
an extension almost twice the width of the detector and a stack of two fanout 
layers is required (see details on section~\ref{sec:L0fanout}). 
The HDI of Layer0 lies in a plane inclined by 10$^{\circ}$ with respect to the sensor plane, 
positioned outside the active region, fitting the constraints on available space.
The HDI can fit the necessary 6 read-out chips into its available width by tilting them with a 30$^\circ$ angle
with respect to the edges.

The module material budget in the active region is very limited 
(about 0.45\% $X_0$ per layer) and is similar to \babar.
For layers 1 to 5, this is dominated by the silicon 
sensor thickness of 300 $\mu$m; a contribution 
of about 0.1\% $X_0$ is due to the composite ribs, with about 0.05\% $X_0$ 
for the $z$ fanout, which isin the active area.
For the Layer0 striplets, the contribution of the flex circuit is 
considerably higher: the total material for the two multilayer flex circuits, now under development, 
is about 0.15\% $X_0$, while  the carbon fiber support structure contributes about 0.1\%~$X_0$.
Since the sensor thickness of the striplets is only 200 $\mu$m, the total material 
budget for Layer0 is also about 0.45\% $X_0$. 

Layer0 modules are supported on the cold flanges, directly coupled to 
the Be beam-pipe, allowing minimization of the distance to the beam pipe, 
and quick access to the Layer0 without demounting the other SVT layers. 
The other five SVT layers are mounted on support 
cones coupled with the conical tungsten shields with kinematic mounts 
({\it i.e.}, the gimbal rings) that allow relative motion of the 
forward and backward shielding cones without placing stress 
on the silicon detectors. 

The detector modules from Layers 1 and 2 are 
glued together with rigid beams, forming
sextants that are then mounted from
the support cones in the forward and backward directions.  
Each detector module of layers 3, 4 and 5  is mounted  
on the support cones independently of the other modules.
For layer 4 and 5, there are
two different types of modules in each layer,  an inner one, labeled  
{\em a}, 
and an outer one, labeled {\em b}, 
occupying slightly different radial positions.
Thus there are eight different types of detector modules.  


The support cones are laminated carbon fiber structures which are
mounted on the conical tungsten shields, present in the IR region for background reduction. 
Cooling water
flows in brass cooling rings surrounding the outer surface of the support cone. 
Mounting pins in the hybrid structure provide the alignment between the
modules and the brass mounts on the cone, and thermal contact is made to provide
cooling for the front-end electronics located on the hybrid. The  
support cones are divided to allow the vertex
detector to be assembled in two halves and then mounted on the shielding cones 
and the beam pipeby clam-shelling the pieces together.
During the assembly/disassembly procedure the splitting of the support cones 
with the five SVT layers mounted, allows easy 
access to Layer0 without the need to disassemble the entire SVT.

The stiffness of the overall SVT structure is provided by a very low mass  
space frame, constructed of carbon fiber tubes, connecting the forward and 
backward support cones, similar to the one designed for the \babar\ SVT.
The motivation for this space frame stems mainly from the 
possible relative motion of the two support cones during the assembly/disassembly
procedure or in case of an earthquake. 
Cooling water, power, and signal lines are routed along the support cones 
to points outside the active region where manifolds for the cooling 
water, transition cards for wire to optical transition, and power regulation and
distribution are located.

\tdrsubsec{Layer0 Pixel Upgrade}
\label{sec:layer0_upgrade_overview}
\tdrsubsubsec{Motivations}

With the machine operating at full luminosity, Layer0 of the SVT 
may benefit from an upgrade to a pixellated detector that has 
better performance in high background conditions, thanks to 
a lower expected background rate per channel.
A background rate of about 1 MHz/strip ($\times$5 safety factor included) is expected 
with a striplets length of about 2 cm and 50~$\mu$m pitch,  
while only 2.5 kHz/pixel are expected for pixels with a 50x50~$\mu$m$^2$ pixel area.

Possible effects of background hits on performance are: the reduction of the 
hit reconstruction efficiency (due to pileup), 
the increase of the effective hit resolution, 
the reduction of efficiency of the pattern recognition for 
charged tracks along with the increase of fake tracks.
Most of these effects have been included in specific simulation studies 
performed to evaluate the SVT performance in the high background scenario
({\it i.e.}, full luminosity including a $\times$5 safety factor on the nominal background).  
The results, described in more detail in
Sec.~\ref{sec:SVT:performance:striplets_sensitivity_studies_background}
and \ref{sec:SVT:performance:Layer0_pixel_performance},
showed a significant degradation in the striplet
performance with high background occupancy, while  
the pixel solutions explored showed more stable performance 
against background conditions. The pixel occupancy is reduced  by afactor of at 
least 200 w.r.t. the striplets, due to
the smaller electrode dimensions,
even when the possible worse time resolution of the pixels w.r.t. striplets is included.  

An example of these studies is shown in Figure~\ref{fig:s_perevent_error}.
The impact of machine background on the SVT performance 
has been studied evaluating the per-event error 
on the physics parameter $S$, adding background hits to signal events.
$S$ is measured in time-dependent analyses (this corresponds to sin(2$\beta$) for 
$B^0 \to J\psi K^0_S$ decays, or, in the Standard Model, to $B^0 \to \phi K^0_S$ decays) and the 
per-event error on $S$ is defined as the error on the parameter S normalized 
to the number of signal events.
In Figure~\ref{fig:s_perevent_error} the impact of 
background on the physics parameter $S$ is shown for the striplet and pixel solutions, 
for the case of nominal background and with $\times$ nominal background rates.
For the striplets, the reduction in the sensitivity of $S$ w.r.t. \babar\
is small with the nominal background, only  
about 3\%, but it is up to about 15\% with the 5$\times$ nominal background. 
On the contrary, with a pixel option, as the effect of 
background occupancy is negligible, the reduction to the sensitivity on $S$ is 
only 3\%, even in the high background scenario, and is largely related to the 
effect of the background in the rest of the SVT.
\begin{figure}[tbp]
\begin{center}
\includegraphics[clip=true,trim=0cm 1.3cm 1.3cm 0cm,width=0.5\textwidth]{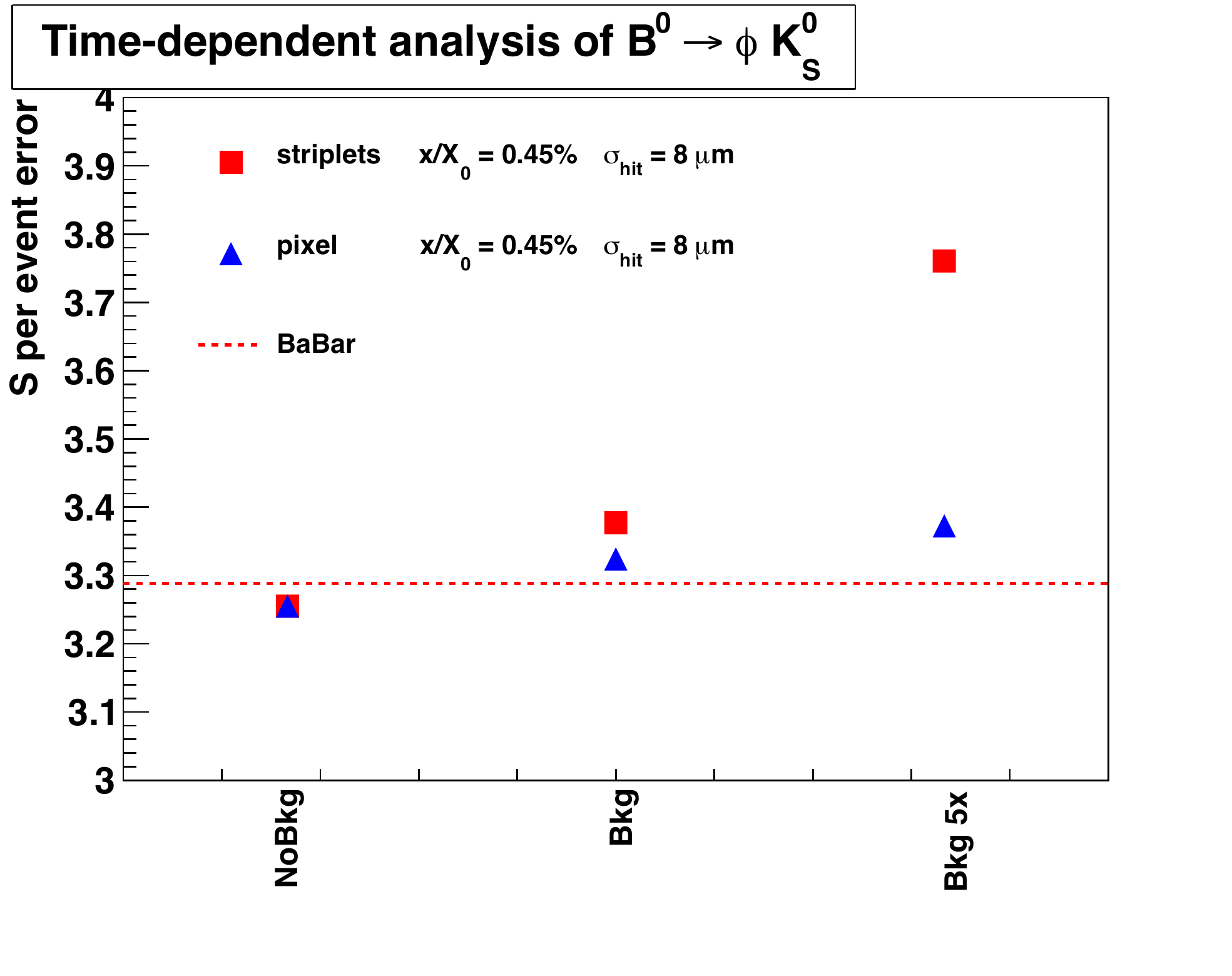}
\end{center}
\caption{Variation of the per-event error $S$ in $\Bz\to\phi\KS$ time-dependent
 analysis in presence of background events, for a Layer0 based on striplets 
or pixels with the same radius ($r$=1.6~cm) and material budget. 
Efficiency and resolution deterioration are both included in the 
simulation study.}
\label{fig:s_perevent_error}
\end{figure}

It is important to stress that in the study reported here the pixel option 
has the same radius and material budget used for the striplets ($r$=1.6 cm, 0.45\% $X_0$), and has the
same performance without background included. 
Of course, the advantage of pixels over striplets in a high background environment 
is reduced if the material budget of the pixel solution is 
significantly higher.
Conversely, if one can achieve a very low material budget with a thin 
pixel option, below the 
striplet material budget, then the upgrade to a pixel solution for Layer0 is 
well-motivated even in nominal background conditions
(see for example Figure~\ref{fig:SVT:FOM_PixelX0Radius}).

While for strip modules most of the material budget is due to the silicon 
of the sensor itself, in pixel modules there are several other important 
contributions in the active area. Including the  
readout electronics, cooling, and the pixel bus 
for the connection of the front-end chips to the periphery of the module,  
one can easily reach a total material budget for a pixel solution above 1\%  $X_0$. 
A discussion on the material budget for various Layer0 pixel options 
is presented in the following sections.

\tdrsubsubsec{Technology Options for Layer0 pixel upgrade}

Two main technologies are under evaluation for the upgrade of Layer0:
hybrid pixel and thinner CMOS Monolithic Active Pixel Sensor (MAPS).
Specific R\&D programs are ongoing on these options 
to meet the Layer0 requirements, such as low 
pitch and material budget, more critical for hybrid pixels, and high readout speed and radiation hardness, 
challenging aspects for CMOS MAPS sensors. 

A short summary of the current status of the R\&D on 
the different pixel options is given below, while a more detailed review
is presented in Sec.\ref{sec:pixel_upgrade}.

\paragraph{Hybrid Pixel} 
 technology represents a mature and viable solution.
Reduction in the front-end pitch and in the total material
budget, with respect to pixel systems
developed for LHC experiments, is required for application in Layer0.

The spatial resolution requirement of \hbox{10-15}~$\mu m$ sets a limit to the area 
of the elementary readout cell 
and, as a consequence, to the amount of functionalities that can be included 
in the front-end electronics. 
For a pixel cell of 50$\times$50~$\mu$m$^2$, a planar 
130~nm CMOS technology may guarantee the required density to implement
in-pixel data sparsification and fast time stamping ($< 1~\mu s$). This is  
required for the maximal hit rate in Layer0 of 100 MHz/cm$^2$ in order 
to keep the module bandwidth to an acceptable level ($<$5 Gbit/s). 

Denser CMOS technologies, such as the 65~nm technology, can be 
used to increase the functional density in the readout electronics and include 
such functions as local threshold adjustment and amplitude measurement 
and storage. 
In this case, costs for R\&D and production would increase 
significantly. 
Vertical integration (or 3D) CMOS technologies may represent a lower cost
alternative to sub-100~nm CMOS processes to increase the functional density 
in the pixel cell~\cite{yarema:twepp2008,vipix}.

A front-end chip for high resistivity pixel sensors with $50 \times50$ 
$\mu$m$^2$ pitch is under development for applications to \superb\ .
A first prototype chip with 4k pixels has been produced 
with the ST Microelectronics 130 nm process 
adopting the same readout architecture, with in-pixel sparsification and 
timestamping, 
developed within the SLIM5 Collaboration~\cite{slim5} for CMOS
Deep NWell MAPS~\cite{rizzo:ieee08, gabrielli:nim2007}.
The chip bump bonded to a high resistivity sensor matrix has been fully 
characterized, with charged particle beams, with good results~\cite{giorgi:nss2011}. 

In this first prototype, only basic functionalities have been implemented.
The readout architecture has been recently optimized 
to efficiently sustain the target Layer0 hit rate of 100 MHz/cm$^2$ on 
matrices larger than 50k pixels. The new architecture requires a more 
complex in-pixel logic and implements a data push and a triggered version of 
the readout~\cite{gabrielli:nim2011}.

The design of a 3D front-end chip for hybrid pixels with 
this new readout architecture, and some improved features,
is now in progress  with the vertical integration 
CMOS technology offered by the 130 nm Chartered/Tezzaron process.

\paragraph{CMOS MAPS}
 are very appealing for applications where the material budget is 
critical:   
in this technology the sensor and readout electronics share the same substrate 
that can be thinned down to several tens of microns.
Since a fast readout is another crucial aspect for Layer0,  
a new Deep N-well MAPS design approach has been developed by the SLIM5 
Collaboration\cite{slim5} to improve readout speed in CMOS MAPS sensors. 
This approach allowed 
for the first time the implementation of thin CMOS sensors with similar 
functionalities as in 
hybrid pixels, such as pixel-level sparsification and fast time 
stamping~\cite{rizzo:ieee08,gabrielli:nim2011}  

Thanks to an intense R\&D program the development of DNW CMOS MAPS 
(with the ST Microelectronics 130 nm process) has reached a good level of 
maturity. A limiting factor in this design is 
the presence of competitive N-wells, 
inside the pixel cell, that can subtract charge from the main collecting 
electrode. 
The last prototype realized, the APSEL4D chip, a 4k pixel matrix with 
$50 \times 50 \mu$m$^2$ pitch has been tested with 
beams~\cite{bettarini:slim5_testbeam} 
reporting a hit 
efficiency of 92\%. This is related to the pixel cell fill 
factor (ratio of the DNW area to the total area 
of N-wells) which is about 90\% in the APSEL design.
Another critical issue for the application of CMOS MAPS 
in  Layer0 is their radiation hardness, 
especially as it relates to bulk damage effects. 
A significant degradation of the charge collected (about 50\%) 
has been measured after irradiation with neutrons up to a fluence of 
about $7$x$10^{12}$ $n$/cm$^2$,   
corresponding to about 1.5 years of Layer0 operation~\cite{ratti:radecs2011}.   

Improvements of MAPS performance are currently under investigation 
with two different 
approaches: the use of INMAPS CMOS process, featuring a quadruple well and 
a high resistivity substrate, 
and 3D CMOS MAPS, realized with vertical integration technology.

In order to increase the charge collection efficiency, the INMAPS 180 nm 
CMOS process is being explored:
a deep P-well implant, deposited beneath the competitive N-wells, 
can prevent them from subtracting charge from the main collecting electrode.
Moreover, the use of a high resistivity substrate 
available 
in this process can further improve charge collection and radiation 
resistance with respect to 
standard CMOS devices.
First prototype INMAPS matrices have been realized with the improved readout 
architecture suitable for the application in the \superb 
Layer0~\cite{giorgi:nss2011}. The devices, currently under test, have shown 
promising results, with measured Signal-to-Noise of about 27 for MIPs~\cite{rizzo:elba2012}. 
Radiation hardness of these devices, at the level required for a safe operation 
in Layer0
is currently under investigation with encouraging results, 
as shown in Sec.~\ref{sec:inmaps_radhard}.

\begin{table*}[htbp]
\caption{Layer0 module material budget for the different 
technologies under evaluation.}
\label{table:material_budget}
\begin{center}
\begin{tabular}{|l|c|c|c|} 
\hline \multicolumn{4}{|c|}{Layer0 Module Material Budget ($X_0$) }   \\
\hline \hline
         &    Striplets      & Hybrid       & CMOS      \\
         &        & Pixel      & MAPS     \\
\hline 
Sensor &    0.21\%  & 0.11-0.21\%  & 0.05\% \\
FE-chip+bump bonding &       & 0.14-0.19\%  &   \\
Multilayer bus or fanout  &    0.15\%   & 0.15-0.30\%  &  0.15-0.30\%    \\
Module Support \& ground plane  &    0.09\%    & 0.15\%   & 0.15\%     \\
 (include cooling for pixels) & & & \\
\hline
Total Material  Budget (X$_0$)  & 0.45\%  & 0.55-0.85\%  & 0.35-0.50\%    \\
\hline
\end{tabular}
\end{center}
\end{table*}

The realization of 3D MAPS, using two CMOS layers interconnected with 
vertical integration technology, 
also offer several advantages with respect to standard 2D MAPS. 
In these devices one CMOS tier is 
hosting the sensor with the analog front-end and the second tier is 
dedicated to the in-pixel 
digital front-end and the peripheral readout logic. With this splitting 
of functionality the collection efficiency can be improved, significantly 
reducing the N-well competitve area in the sensor 
layer. Having more room for the in-pixel logic allows 
the implementation of a more performant 
readout architecture.  
Finally, in 3D MAPS the cross-talk between analog and digital blocks can be 
minimized.

The characterization of the first 3D MAPS prototypes realized with the 130 nm  
Chartered/Tezzaron 3D process is under way; first beam test results on the 
MAPS layer implementing the sensor and the analog front end showed very 
good hit efficiency (above 98\%).


\paragraph{Conclusions}

Very promising results have been achieved with both evolution of the Apsel DNW MAPS: 
DNW 3D MAPS with vertical integration and MAPS with quadruple-well process.
In particular the encouraging results on Apsel4well sensor seems to 
suggest that the INMAPS 180~nm CMOS process with high resistivity epitaxial layer
might overcome the radiation tolerance issues found in DNW MAPS and become the 
candidate technology for the Layer0 pixel upgrade. 
Other developments of CMOS MAPS with quadruple-well 180~nm CMOS process can also be 
considered for Layer0 application~\cite{alice_upgrade,rsmdd12,stanitzki07,crooks07,Ballin08,Mylroie12} 


\tdrsubsubsec{Pixel Module \& Material Budget}

The schematic drawing of the full Layer0 made of 8 pixel modules mounted 
around the beam pipe is shown in Figure~\ref{fig:layer0_beampipe}.
With the pixel option the radius of Layer0 can be slightly reduced ($r$=1.39~cm), with respect to the one used 
for the striplets option ($r$=1.56~cm), thanks to the simpler geometry of the pixel module.  

\begin{figure}[tbp]
\begin{center}
\includegraphics[clip=true,trim=0cm 0cm 0cm 0cm, width=0.5\textwidth]
{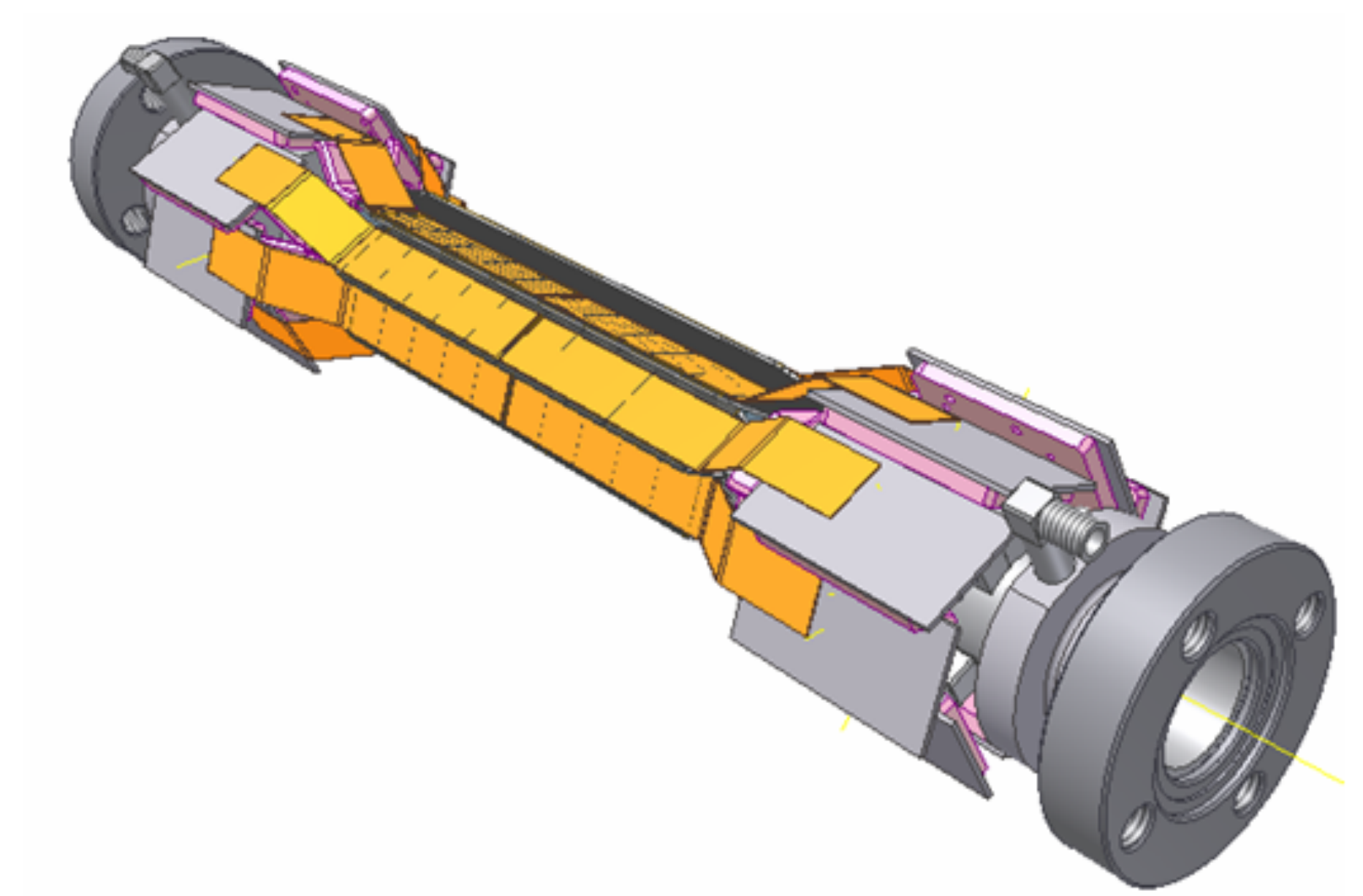}
\end{center}
\caption{Schematic drawing of the full Layer0 made of 8 pixel modules mounted 
around the beam pipe with a pinwheel arrangement.} 
\label{fig:layer0_beampipe}
\end{figure}

In all the pixel options under evaluation,  sharing the same 
multichip module structure, the material budget of 
all the components must be kept under control to minimize the 
detrimental effect of multiple 
scattering.

The main contributions to the material budget of pixel modules in different 
technologies are discussed in this section. See Table~\ref{table:material_budget}
for a summary and a comparison with the striplets option.

In the hybrid pixel solution, the contribution of the silicon of the sensor 
(100-200 $\mu m$) and of the front-end chip (100-150 $\mu m$) can 
be in the range of 0.25-0.4\%  $X_0$. 
In the CMOS MAPS option, the sensor and the front-end electronics are 
integrated in the same CMOS chip, 
that could be thinned down to 50 $\mu m$ reducing this contribution down to 
0.05\%  $X_0$.

Another important contribution to the material comes from the pixel bus, for
the connection of front-end chips to the periphery of the module. 
This will be 
realized with an Al/kapton multilayer bus (now under development) and should satisfy 
the present requirements on speed (high bandwidth due to a hit 
rate of 100 MHz/cm$^2$) and power consumption 
(about 1.5W/cm$^2$). The estimated material budget for the pixel bus is about 
0.15-0.3\%  $X_0$, depending on 
the achievements of present R\&D on this item. 

The pixel module support structure needs to include a cooling 
system in the active area to evacuate the power dissipated by the front-end electronics, 
a total of about 1.5W/cm$^2$. 
In order to minimize the material budget, a light carbon fiber support 
structure with integrated active cooling, based 
on microchannel technology~\cite{bosi:pixel2010} and forced liquid 
convection,
has been developed.
The support with integrated cooling is built with carbon fiber 
micro-tubes, with a hydraulic diameter of about $200 \mu$m, obtained by 
a pultrusion process. 
Measurements on the support prototypes, with a total material budget 
as low as 0.11\% $X_0$, indicate that such approach is a viable solution to
the thermal and structural problem of Layer0~\cite{bosi:elba2012}.
An innovative idea is also under development 
to integrate the cooling system into the silicon itself and is
based on microchannels made by DRIE technology. 
The embedded microchannels have diameters even below $100 \mu$m, and
feature a peculiar geometry.
In the final step a thin oxide layer is deposited to seal the channels.  Results
showed this to be reliable for operating in high-pressure conditions.
This novel technique permits the integration of the cooling system within 
the detector with  advantages for the efficiency  of thermal bridges 
and reduction of material budget~\cite{boscardin:silicon_microchannels}.


\tdrsec{Backgrounds} 

\label{svt_background}

A detailed analysis of background effects is fundamental to 
having a reliable estimation of performance and expected lifetime of the tracker.

The different sources of background have been simulated with a detailed Geant4-based 
detector model and beamline description (see Sec.~\ref{sec:Full_Simulation}).
The detailed simulation is needed because not only the detectors themselves, but also the support structure 
(``dead material'') play an important role in stopping or creating background particles.
The raw output of the simulation has then to be processed to obtain information useful  for the detector design.

As described in Sec.~\ref{chap:mdi}, in addition to well-known background 
sources, such as Touschek and beam-gas scattering, we have a significant contribution from physics
processes that occur in the interaction region.
The very high luminosity of the machine produces an unprecedented rate for 
additional pairs and for radiative Bhabha processes.
The effect of those physics processes cannot be mitigated by
optimizing the machine optics, because they scale with the luminosity and the machine goal
is always to maximize the luminosity.
In the SVT the contribution from the radiative Bhabha source is 
greatly reduced by the tungsten shields that surround the final focus regions.
On the contrary, the pair production source of background, with the typical low momentum 
electrons and positrons coming directly from the IP, cannot be shielded; 
this represents the main background contribution in the SVT.
The relative contribution of the various sources to the total background in the SVT layers is shown 
in Figure~\ref{fig:background_fractions}.
\begin{figure}[!h]
\begin{center}
\includegraphics[width=0.5\textwidth]{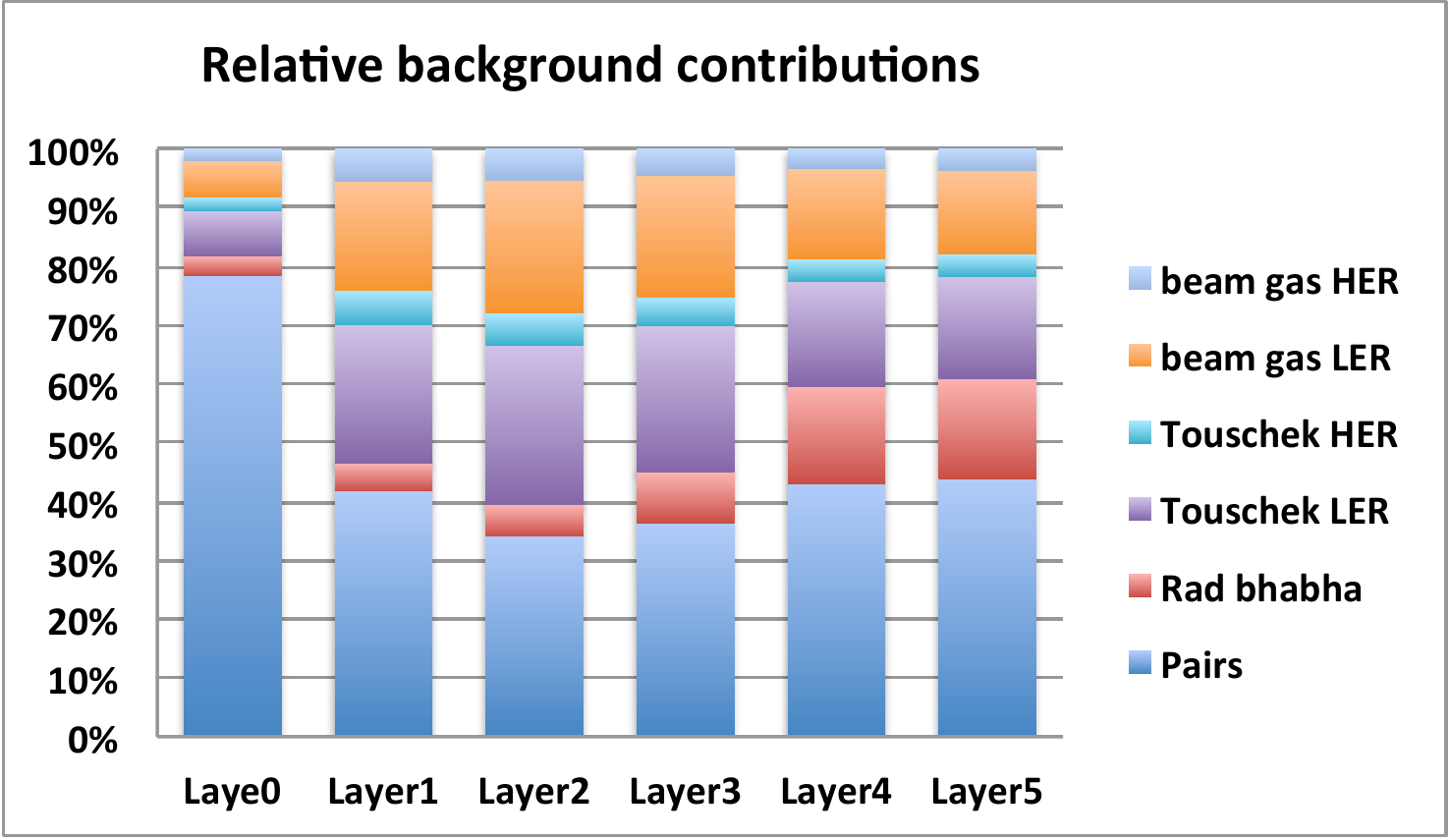}
\end{center}
\caption{Relative contribution of the various background sources to the total track rate in the SVT layers.}
\label{fig:background_fractions}
\end{figure}
%



The summary of the relevant variables extracted from background simulation is shown in 
table~\ref{tab:background_summary} and defined in the rest of this section.
\begin{table*}[!htb]\smaller
\caption{Summary of  expected nominal backgrounds in the sensor area. The SVT has been designed to withstand $\times$5 the nominal background. Numbers shown do not include the $\times$5 safety factor.
Simulation results for both Layer0 options, 
striplets and pixels, the latter being at a slightly lower average radius, are reported. The definition of the various quantities shown (track, cluster and strip/pixel rates) is given in section~\ref{svt_background}. 
}
\begin{center}
    \begin{tabular}{|c|c|c|c|c|c|c|c|c|} 
\hline
& & & \multicolumn{3}{|c|}{Total Rate/Area} & Total  & & \\
 Layer & Radius & Pitch {\small ($\phi$--$z$)}   & Track & Cluster 
& Strip {\small ($\phi$--$z$)} & Strip Rate & TID  & NIEL \\ 
  & {\small (mm)}  & {\small ($\mu$m)}  &  \multicolumn{3}{|c|}{({\small MHz/cm$^2$)}} 
& {\small (kHz)}  & {\small (Mrad/yr)} & {\small (n/cm$^2$/yr)}  \\ 
\hline
\hline 
 0 (pixel) &  13.9   &  50--50  & 2.32     & 5.86      &  30   & 0.8 (pixel)   & 3.3                &  5.2$\times 10^{12}$         \\
 0 (striplets) &  15.6   &  54--54 ($u$,$v$) & 1.62     & 4.10      &  20--20  ($u$,$v$)  & 187--187    & 2.3                &  3.6$\times 10^{12}$         \\
 1  &  33   &  50--100          & 0.217     & 0.540      &  3.2--2.7   &  170--134         & 0.3           &  7.9$\times 10^{11}$         \\
 2  &  40   &  55--100          & 0.163     & 0.393      &  2.0--1.9  & 134--134        & 0.2           &  5.1$\times 10^{11}$         \\
 3  &  59   &  100-110          & 0.079     & 0.208      &  0.60--0.83 & 116--79             & 0.1           &  3.0$\times 10^{11}$         \\
 4  &  120   & 100--210          & 0.022     & 0.037      &  0.08--0.06 & 25--13         & 0.01          &  2.0$\times 10^{11}$         \\
 5  &  140   & 100--210          & 0.014     & 0.022      &  0.04--0.03 & 16--9             & 0.01           &  1.8$\times 10^{11}$         \\

\hline

\end{tabular}
\end{center}
\label{tab:background_summary}
\end{table*}

From simulation we calculate the track rate per unit area, obtained by
counting particles that cross the sensor, normalized to the sensor area.
Since a single particle can cross the sensor 
several times, generating multiple clusters ({\it i.e.}, low-momentum charged particles that 
spiral in the detector magnetic field --- sometimes referred to as `loopers'), 
we also evaluate the cluster rate per unit area, 
taking into account multiple crossings of the sensor area by the same particle.
The simulated cluster rate and the cluster occupancy 
are the relevant variables used in section~\ref{sec:performance_and_background}
to evaluate the impact of background on tracking performance. 
For each cluster several strips/pixels can be fired, 
depending on the incident angle of the particle with the sensor, therefore
the strip/pixel rate can be significantly higher than 
the cluster rate.
The strip/pixel rate, which is particularly relevant for the design of the 
readout electronics, is also evaluated from background simulation
and shown in Table~\ref{tab:background_summary}.

The integrated radiation dose, and the 1 \mev neutron equivalent fluence, according to NIEL (non-ionizing energy loss) scaling, are 
extracted from simulation and are used to determine the radiation damage 
in the SVT layers over the entire lifetime of the experiment, including 
a $\times$5 safety factor on nominal simulated background.

\tdrsec{Detector Performance Studies}

\label{sec:svt_performance}

\tdrsubsec{Introduction}
\label{sec:SVT:performance:intro}

The \superb vertex detector is an evolution of the \babar\ SVT.
It is capable of maintaining adequate performance for time-dependent measurements 
 in the presence of a lower boost of the center-of-mass (CM) frame
($\beta\gamma = 0.24$ compared to $\beta\gamma = 0.55$ of \babar) and much higher background, 
mainly related to the increased instantaneous luminosity of about a factor of 100 larger than \babar.

The beam pipe features a reduced radius of about $1.0~\cm$  which allows the positioning of the Layer0 at 
an average radius of about $1.5~\cm$. The additional Layer0 measurement, 
along with the low radial material budget of the beam pipe ($0.42\%~X_0$) 
and of Layer0 ($0.45\%~X_0$ with the striplet option), 
is crucial for improving the decay vertex reconstruction of the \B mesons and 
 obtaining adequate proper-time resolution for time-dependent \CP violation measurements.
In addition, the small size of the luminous region, about $(1\times1)\mum^2$ in the transverse plane, 
also contributes to the improvement of the decay vertex reconstruction when imposing the constraint 
that the particles originate from the interaction point.
The baseline solution for the Layer0 uses short strip (stirplet) technology; an upgrade to a pixel design 
is foreseen once the machine starts operating at nominal luminosity.

In the following we discuss the baseline layout of the SVT and how the design has been optimized 
(Section~\ref{sec:SVT:performance:layout}), 
 the impact of the Layer0 on detector performance (Section~\ref{sec:SVT:performance:ImpactL0}), and 
the tracking performance (Section~\ref{sec:SVT:performance:tracking}). 
The impact of the machine background on SVT performance is discussed 
(Sections~\ref{sec:SVT:performance:impact_of_bkg} and \ref{sec:SVT:performance:striplets_sensitivity_studies_background})
and the performance with a Layer0 pixel detector is presented 
(Section~\ref{sec:SVT:performance:Layer0_pixel_performance}). 
In Section~\ref{sec:SVT:performance:PID_with_dedx} the SVT sensor performance
for particle identification based on ionization \dedx is described.

\tdrsubsec{The SVT layout}
\label{sec:SVT:performance:layout}

The baseline SVT is composed by 6 layers of double-sided silicon strip detectors 
and has a symmetric coverage in the laboratory frame 
down to $300 \mrad$ ($17.2 ^\circ$) with respect to the forward 
and backward directions, corresponding to asolid angle coverage of 95\% in the CM. 
The inner three layers perform track impact parameter measurements, while the outer layers
are required for pattern recognition and for tracking of low transverse momentum ($p_t$) particles.

The Layer0 strips are short (`striplets') and are oriented at $\pm 45^\circ$ with respect to the beam direction.
The Layer1 to Layer5 silicon strip detectors are very similar to those in \babar\ in terms of radial position and 
strip pitches. 
The optimization of the strip $z$ and $\phi$ pitches for 
 the strip detectors is discussed in Section~\ref{par:SVT:SensorDesign}.
 A dedicated study to optimize the SVT layout as a function of number of silicon sensors and 
radial positions was performed~\cite{walsh:frascati09}. 
Several figures of merit were studied: track parameter resolution, 
reconstruction efficiency, and kinematic variable 
 resolutions of \B decays with low momentum tracks using the benchmark channel $\Bz\to D^{*-}K^+$. 
Since low momentum tracks do not reach the DCH, they are reconstructed only using SVT information.
The \babar\ experiment has shown that at least 4 hits in the $\phi$ view and 3 hits in the $z$ view
 are necessary for robust track reconstruction~\cite{Aubert:2001tu,long:standalone_patrec}.
The main result obtained for \superb is that the 6-layer design is 
superior and more robust compared to the alternatives 
investigated, \ie~4- and 5-layer layouts where intermediate layers are removed.  
Indeed, when accounting for possible 
 inefficiencies in hit reconstruction (by removing intermediate layers), 
due to damaged modules or high background, 
 the 6-layer design ensures a higher reconstruction efficiency for low momentum tracks 
 compared to the other solutions investigated.

Table~\ref{table:SVT:EfficiencyLayout}  shows the 
 reconstruction efficiencies for the decay $\Bz\to D^{*-}K^+$ 
for the 4-, 5- and 6-layer configuration in different running conditions: 
ideal conditions (A), with a damaged module in Layer3 (B) 
 and with additional hit inefficiency in Layer0 with respect to case B (C).
\begin{table}[tbp]
 \caption{
Reconstruction efficiencies for $\Bz\to D^{*-}K^+$ decays for different SVT layout (4, 5 ,6 layers) and 
 running conditions (A, B, C). Case A correspond to ideal running conditions, 
 B represents SVT with a damaged module in Layer3 with $z$ hit efficiency of 70\%.  
Case C introduces additional inefficiency with respect to case B in Layer0: 
60\% hit efficiency for $z$ and $\phi$ views.}
\begin{center}
\begin{tabular}{|l|c|c|c|} 
\hline
         &    A          & B          & C           \\ 
         &    eff. (\%)     & eff. (\%)     & eff.  (\%)     \\
\hline
6 layers &    $66.0\pm0.3$     & $65.0\pm0.3$    & $64.0\pm0.3$   \\
5 layers &    $64.0\pm0.3$     & $62.0\pm0.3$    & $60.0\pm0.3$   \\
4 layers &    $60.0\pm0.3$     & $56.0\pm0.3$    & $53.0\pm0.3$    \\
\hline
\end{tabular}
\end{center}
\label{table:SVT:EfficiencyLayout}
\end{table}
The outer radius of the SVT is ultimately constrained to about $20 \cm$ by the DCH inner radius.  
It has been demonstrated that there is no real advantage in increasing the outer layer of the SVT with 
respect to the \babar\ design ($14.4~\cm$)~\cite{neri:perugia09, rama:perugia09, simi:perugia09}. 
Moreover, construction cost and technical difficulties would increase if one attempted to increase 
this parameter. Radial positions of the Layer1 to Layer4 modules have very little impact on track resolution
 when comparing a layout with detectors spaced equally or with a \babar-like radial layout.
   
\tdrsubsec{Impact of Layer0 on detector performance}
\label{sec:SVT:performance:ImpactL0}

The additional Layer0 measurement is crucial for maintaining adequate resolution on the 
 \Bz meson proper-time difference $\Delta t \simeq \Delta z / (\beta\gamma c)$ with the relatively low CM boost value $\beta\gamma = 0.24$ of \superb.
 The average separation $\Delta z$ between the decay vertex positions of the two \B mesons along the $z$ axis
 is $\Delta z \simeq \beta \gamma c \tau_{B} = 110~\mum$, where $\tau_B$ is the \Bz lifetime, which is about $1.5~\ps$.
 Hence, in order to be able to separate the \B meson decay vertices in \superb, their decay positions
  have to be determined with a significantly better precision than the average separation $\Delta z$.
 In addition in \superb the \B vertex separation in the transverse plane, which is about $25 \mum$,
 is not completely negligible with respect to the average $\Delta z$ separation of about $110 \mum$,
and therefore contributes to the determination of $\Delta t$.  
The reference value for the $\Delta t$ resolution, $\sigma(\Delta t)$, was determined by the 
 resolution obtained in the \babar\ experiment according to the Fast Simulation (sec.~\ref{sec:fastsim}), 
see Table~\ref{table:SVT:vtx_results}. 
Figure~\ref{fig:SVT:SperEventErrorVsSigmaDeltaT}  shows the dependence of the 
  per-event error on the physics parameter $S$ as a function of $\sigma(\Delta t)$, with the sensitivity 
 obtained in \babar\ superimposed. In this simplified model $\sigma(\Delta t)$ corresponds to the width of the core Gaussian of 
 the $\Delta t$ resolution function.
 The $S$ per-event error is defined as the error on the parameter $S$ normalized to the number of signal events. 
 $S$ is measured in time-dependent analyses and corresponds to $\sin(2\beta)$ for $B^0\to J/\psi\KS$ decays. 
\begin{figure}[!htb]
\begin{center}
\includegraphics[width=0.5\textwidth]{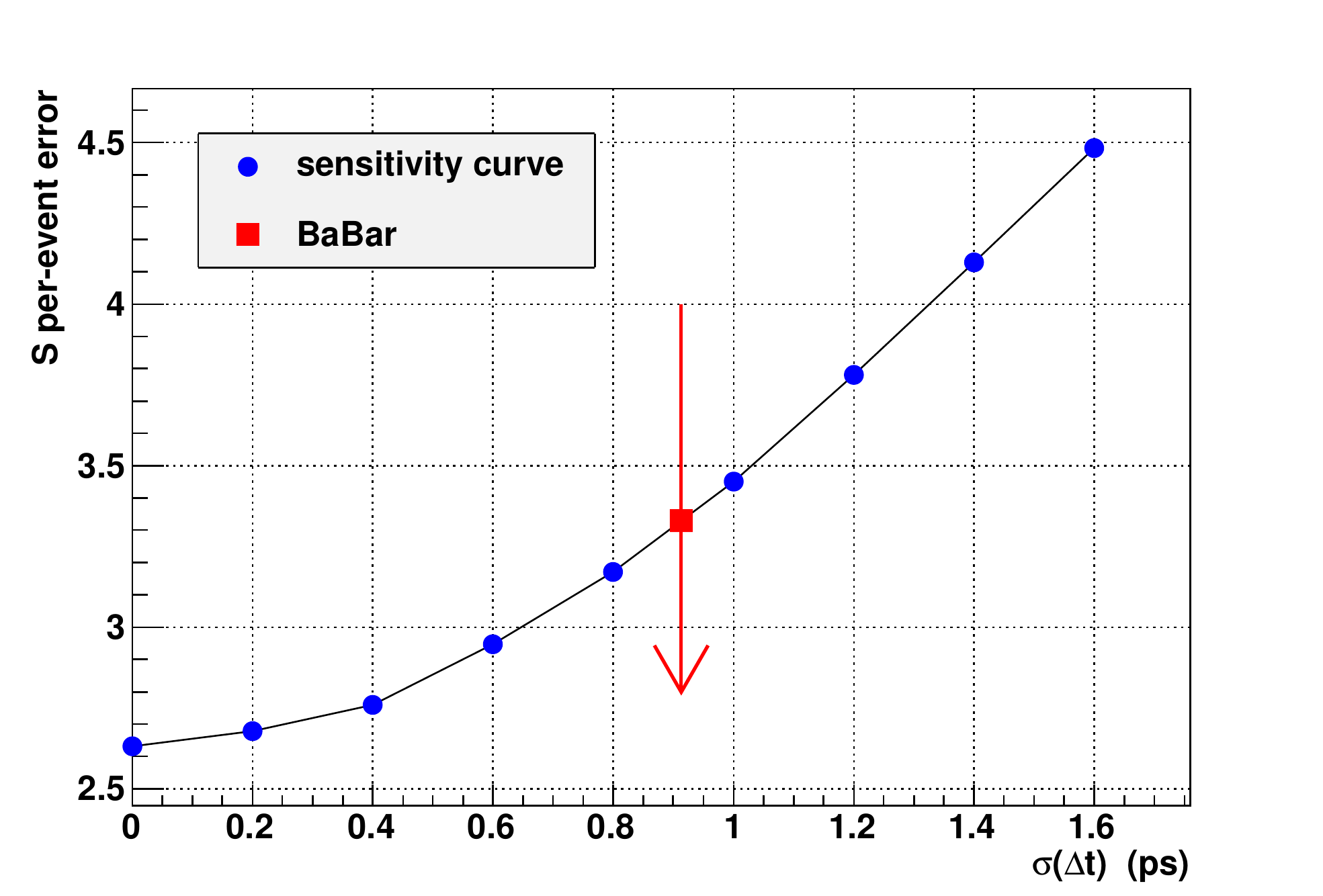}
\end{center}
\caption{The curve represents the dependence of the error on the physics parameter 
$S$ (\eg,\ $\sin(2\beta)$) as a function of $\sigma(\Delta t)$.  
The arrow indicates the $\sigma(\Delta t)$ value obtained in \babar\ according to the Fast Simulation 
and the square point is the relative value on the sensitivity curve.}
\label{fig:SVT:SperEventErrorVsSigmaDeltaT}
\end{figure}
The resolution $\sigma_z$ on the $z$ coordinate of the track depends on the geometry of the vertex detector and 
 the hit resolution. In a simplified model with two hits measured at radii $r_0$ and $r_1$ (where $r_1 > r_0$) 
with $z$ hit resolutions $\sigma_0$ and $\sigma_1$ respectively, $\sigma_z$ can be approximated as:
\begin{equation}
\sigma_z = \frac{\sigma_0^2+(\sigma_1r_0/r_1)^2}{1-(r_0/r_1)^2}. 
\label{eq:SVT:sigmaZ}
\end{equation}
In addition, the tracks undergo multiple scattering interactions with the
 material in the tracking volume. The scattering angle distribution can be approximated by 
 a Gaussian with a width given by~\cite{Nakamura:2010zzi}:
\begin{equation}
\theta_{\textrm{m.s.}}= \frac{13.6 \mevc}{p_t \beta}\sqrt{\frac{x}{X_0}}\left[ 1+ 0.0038 \ln\left(\frac{x}{X_0}\right)\right]
\label{eq:SVT:thetaMS}
\end{equation}
where $p_t$ is the transverse momentum, $x$ is the thickness of the material and 
$X_0$ is the radiation length. In order to minimize the uncertainty on $\sigma_z$ it is important to measure 
the first hit at an $r_0$ as small as possible with good hit resolution $\sigma_0$.  
Minimizing the material close to the interaction region, \eg,\ the beam pipe and Layer0 material, is also important.
\begin{figure}[!htb]
\begin{center}
\includegraphics[width=0.5\textwidth]{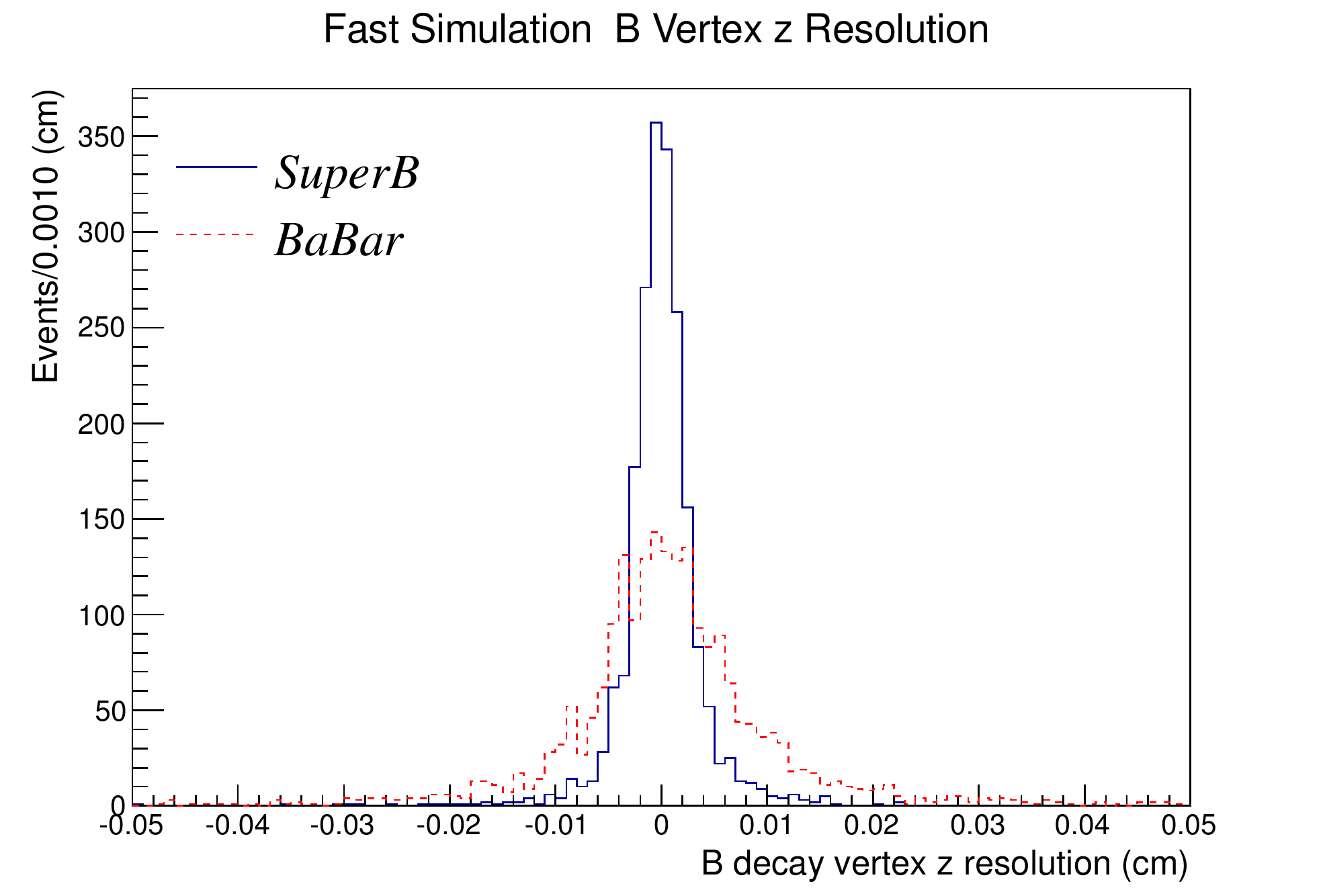}
\includegraphics[width=0.5\textwidth]{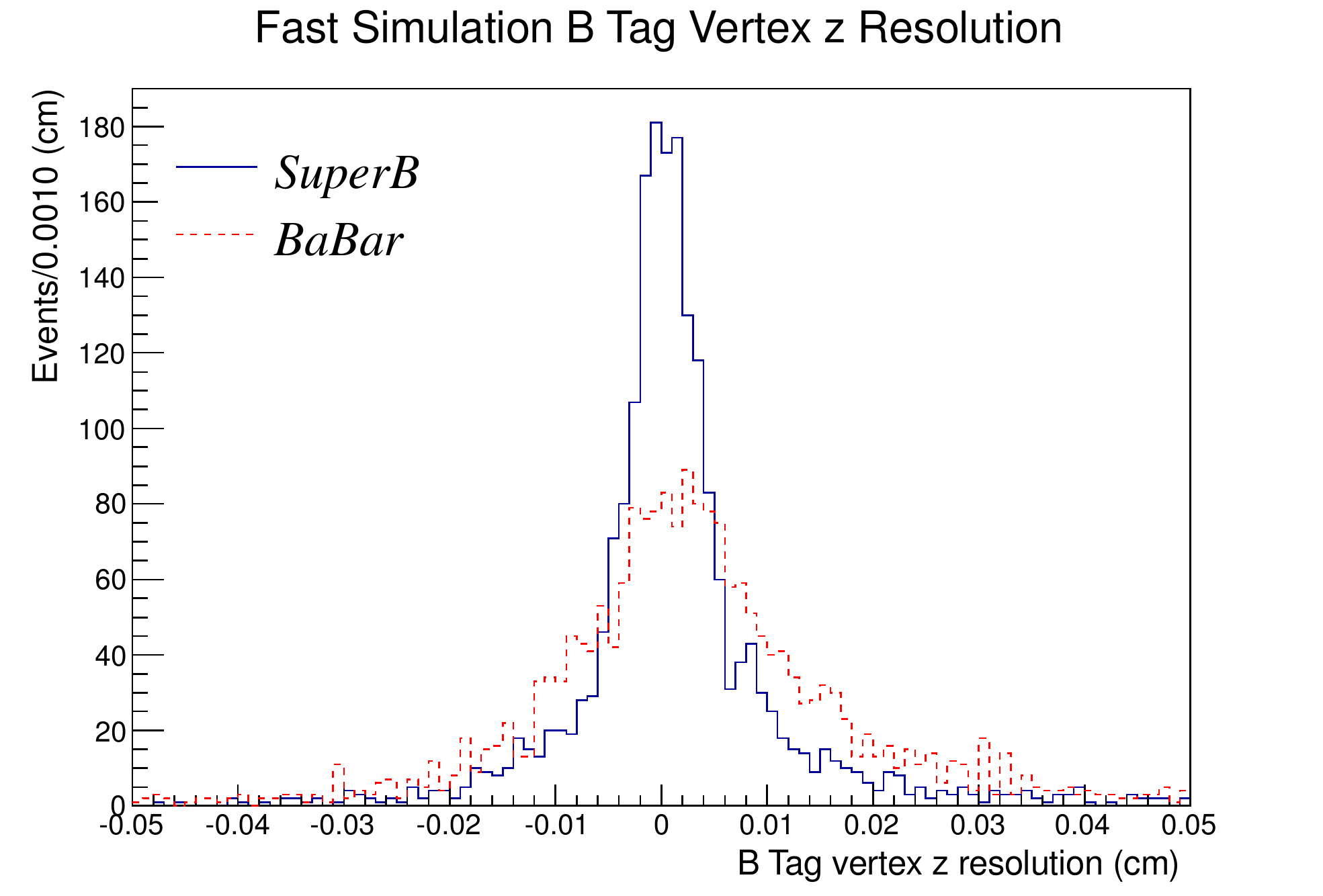}
\end{center}
\caption{\B decay vertex $z$ position (top) and $B$ tag $z$ position (bottom) 
residual distributions in \superb with Layer0 striplets 
 (continuous line) compared with \babar\ (dashed line) according to 
 Fast Simulation studies.}
\label{fig:SVT:VtxZ_Res}
\end{figure}
\begin{figure}[!htb]
\begin{center}
\includegraphics[width=0.5\textwidth]{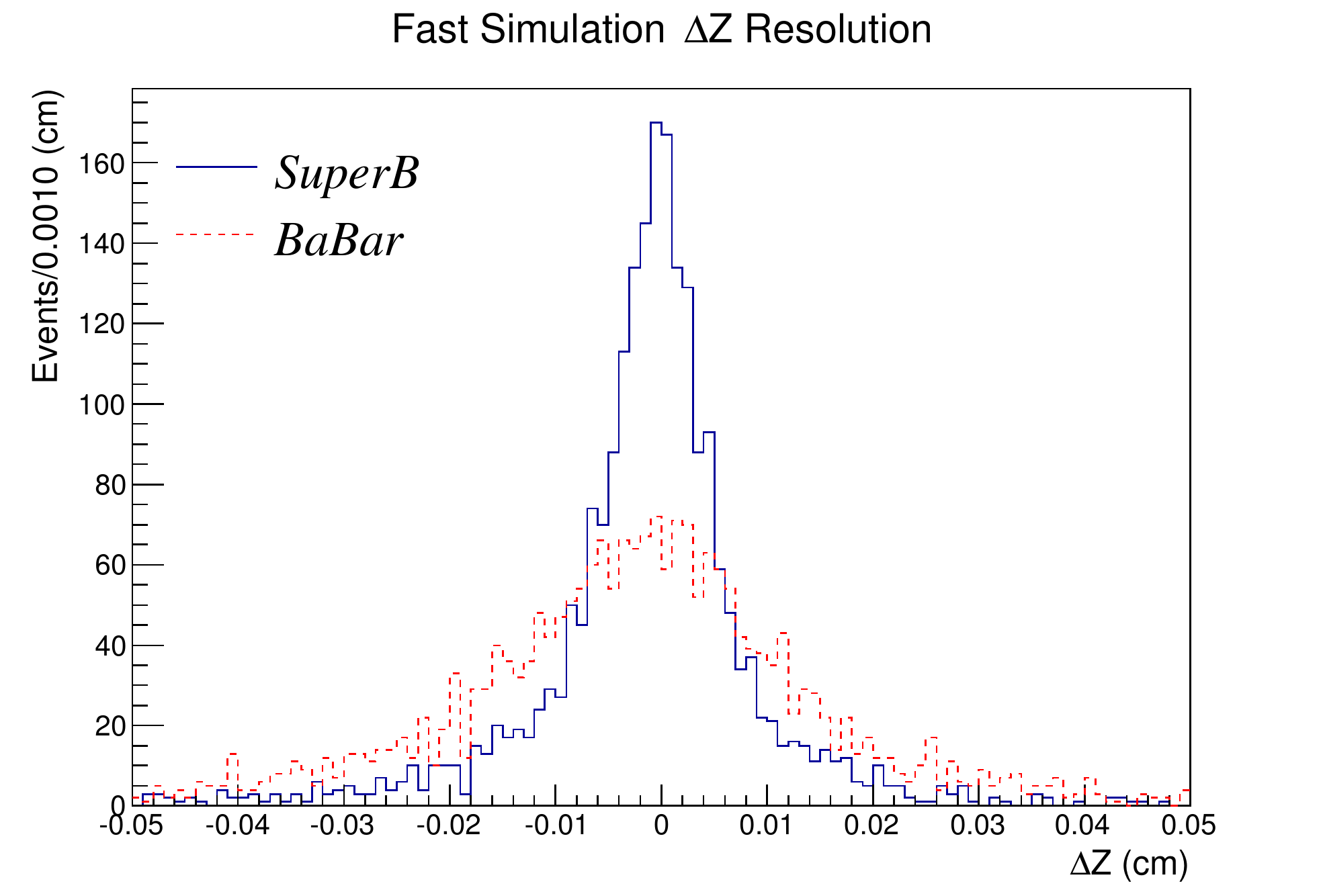}
\includegraphics[width=0.5\textwidth]{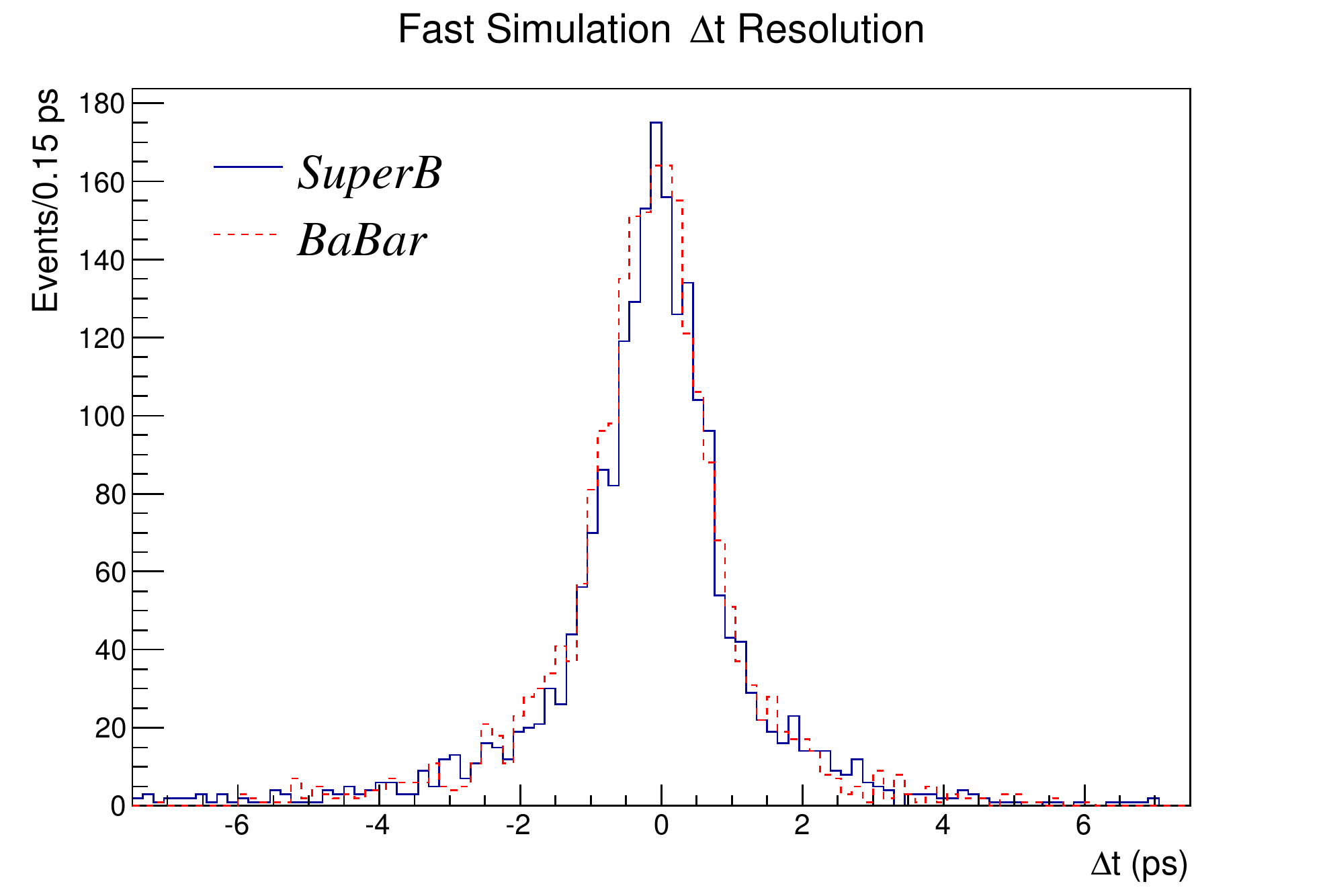}
\end{center}
\caption{$\Delta z$ (top) and $\Delta t$ (bottom) residual distributions in \superb
with Layer0 striplets (continuous line) compared with \babar\ (dashed line) according to Fast Simulation studies.}
\label{fig:SVT:DeltaZ_DeltaT_Res}
\end{figure}

Figure~\ref{fig:SVT:VtxZ_Res} shows the residual distributions of \B decay vertex $z$ positions
for exclusively (top) and inclusively  (bottom) reconstructed   \B decays.  
Figure~\ref{fig:SVT:DeltaZ_DeltaT_Res} shows the $\Delta z$ (top) and $\Delta t$ (bottom) 
residual distributions.  
One \B is exclusively reconstructed in the $\B^0\to\phi\KS$ mode ($B_\textrm{reco}$), while 
 the other \B is inclusively reconstructed using the remaining tracks of the event and is 
  also used for flavor tagging ($B_\textrm{tag}$).
The Fast Simulation results for \superb with Layer0 striplets are compared with the \babar\ results and 
 summarized in Table~\ref{table:SVT:vtx_results}.
\begin{table}[tbp]
 \caption{RMS of the residual distributions for decay vertex $z$ position for exclusively reconstructed 
$\B^0 \to \phi\KS$ decays ($B_\textrm{reco}$), inclusively reconstructed \B decays ($B_\textrm{tag}$), 
$\Delta z$ and $\Delta t$ at \superb and compared with \babar\ results, according to Fast Simulation studies.}
\begin{center}
{\small
\begin{tabular}{|l|c|c|} 
\hline
   &  \superb  & \babar  \\ 
\hline
$B_\textrm{reco}$ ($\mum$) &    $40\pm 1$    & $105\pm 1 $  \\
$B_\textrm{tag}$ ($\mum$)  &    $100\pm 2 $   & $145\pm 2$  \\
$\Delta z$ ($\mum$)   & $105\pm 2 $ & $165\pm 2$   \\
$\Delta t$ (\ps)   & $1.40 \pm 0.02 $ & $1.45\pm 0.02$    \\
\hline
\end{tabular}
}
\end{center}
\label{table:SVT:vtx_results}
\end{table}
In \superb we assume $\sigma_0=10\mum$ for both $u$ and $v$ hits, extrapolating from 
data from a recent beam test on a striplets module prototype.
The Layer0 has a pinwheel configuration with radii in the range 
$1.53-1.71\cm$; it is modeled in FastSim as a cylinder with average radius of about $r_0=1.6\cm$.
The total material budget for the beam pipe and the Layer0 striplets is about $x/X_0\simeq 0.9\%$.
For striplets detectors, $u$ and $v$ coordinates are oriented at $\pm45^\circ$ with respect to the $z$ axis
 and are perpendicular to each other.
In \babar\ we have $\sigma_0=14\mum$ for the $z$ hits and  $10 \mum$ resolution for the $\phi$ hits, 
 with $r_0=3.32\cm$ and $x/X_0\simeq 1.6\%$.
 In \superb, the improved  $\Delta z$ resolution is compensated by the reduced boost 
  value, yielding a  $\Delta t$ resolution very similar to \babar.
Figure~\ref{fig:SVT:SigmaDeltaT_X0} shows the $\Delta t$ resolution obtainable with different 
  Layer0 radii ($r_0 $= 1.4 and 1.6 $\cm$) and material budgets ($x/X_0$ ranging from 0.1 to 1.0\%). 
The dashed line represents the reference value of \babar. 
\begin{figure}[!htb]
\begin{center}
\includegraphics[clip=true,trim=0cm 1.6cm 1.3cm 0cm,width=0.5\textwidth]{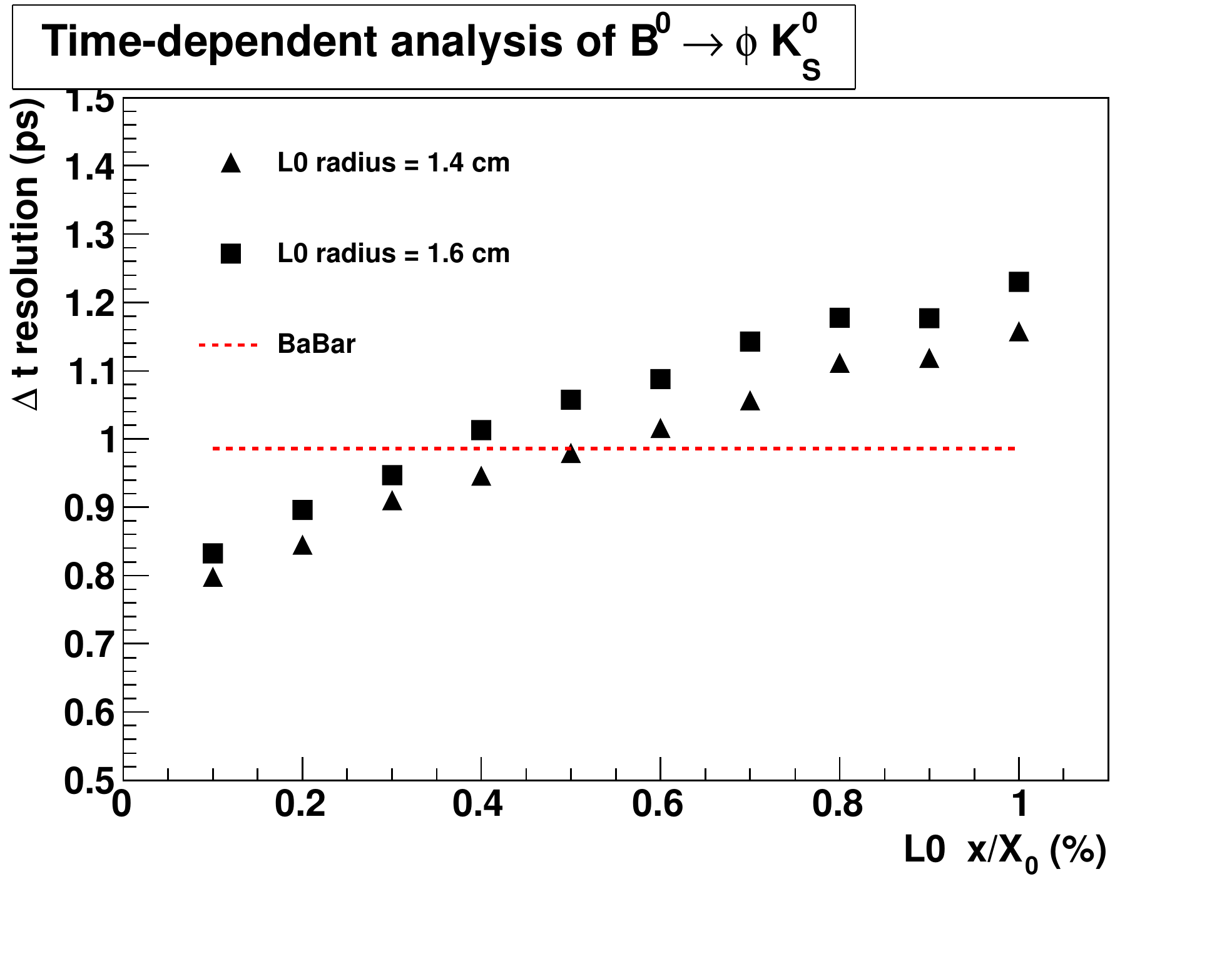}
\end{center}
\caption{Resolution on $\Delta t$ for different Layer0 configurations in terms of radius  ($r_0$ = 1.4 and 1.6 $\cm$)  
and material budget $(x/X_0=0.1-1.0\%)$ compared with the reference value of \babar\ (dashed line).}
\label{fig:SVT:SigmaDeltaT_X0}
\end{figure}
\par
The impact of the hit resolution on the decay vertex reconstruction has also been studied. 
With $10 \mum$ hit resolution in both views of Layer0, the error on the vertex position due to multiple scattering 
interactions with the material dominates the overall vertex uncertainty.
This is true even for high momentum tracks from $B^0\to\pip\pim$ decays. 
Hence,  without further reduction of the material budget, there is no real advantage in improving 
the hit resolution with respect to this value.
The hit resolution from Layer1 to Layer5 has been chosen according to the \babar\ SVT design which 
was optimized for low momentum track reconstruction. 
The intrinsic detector hit resolution and hit efficiency values used in the 
 Fast Simulation are reported in Table~\ref{table:SVT:hit_reso_eff} for the different SVT layers.
The efficiencies have been estimated from simulations taking into account possible hit 
 losses due to the overlap of pulses in the analog section of the FEE with nominal background conditions.
\begin{table}[tbp]
 \caption{Intrinsic detector hit resolution and hit efficiency for the $\phi$ and $z$ sides (Layer0 $u$ and $v$ sides) 
 for the different layers. }
\begin{center}
{\small
\begin{tabular}{|l|c|c|c|c|} 
\hline
         &  res.  & res.  & eff.  & eff.       \\
         &  $u$ ($\mum$) & $v$ ($\mum$) & $u$ (\%) & $v$ (\%)      \\
\hline
Layer0 &    10   & 10  & 99  &  99    \\
\hline
         &  res.  & res.  & eff.  & eff.       \\
         &  $z$ ($\mum$) & $\phi$ ($\mum$) & $z$ (\%) & $\phi$ (\%)      \\
\hline
Layer1 &    14   & 10  & 98 & 98   \\
Layer2 &    14   & 10  & 98 & 98    \\
Layer3 &    15   & 15  & 98 & 96   \\
Layer4 &    25   & 15  & 99 & 98   \\
Layer5 &    25   & 15  & 99 & 98   \\

\hline
\end{tabular}
}
\end{center}
\label{table:SVT:hit_reso_eff}
\end{table}
\tdrsubsec{Tracking performance}
\label{sec:SVT:performance:tracking}

The tracking performance at \superb has been studied considering alternative solutions for 
 the SVT and DCH layout~\cite{walsh:frascati09,neri:perugia09,rama:perugia09}. 
In particular, we have studied alternative 
 SVT configurations: with different values of the SVT outer radius (from about 14 to 22 \cm), 
 without Layer2 modules, different radial positions of the layers (\eg,\ uniform distance between layers), 
 different hit resolutions accounting for variations of about 50\% with respect
 to the nominal ones reported in 
 Table~\ref{table:SVT:hit_reso_eff}. The main result was that the \babar-like layout for Layer1-Layer5 
 was very close to be the optimal choice in terms of resolution for track parameters. 
 Small improvements in track parameter resolution would have been possible by removing Layer2. On the 
 other hand, the 6-layer layout has been proven to be more robust against possible problems that might cause
 loss of efficiency in some layers of the detector~\cite{walsh:frascati09} and was preferred for this reason.
Optimization of the strip pitches for the $z$ and $\phi$ sides of the different layers 
is discussed in Section~\ref{par:SVT:SensorDesign}.
Figure~\ref{fig:SVT:d0res} shows the resolution on the impact parameter $d_0$,
 as a function of $p_t$ with the \babar\ and \superb detectors.
The parameter $d_0$ is defined as the distance of the point of closest approach of the track to the $z$-axis 
from the origin of the coordinate system in the $x-y$ plane. 
Results for alternative configurations of the SVT 
 layout, with extended outer radius, with DCH lower radius, 
and without Layer2, are also shown.
A significant improvement in the $d_0$ resolution of about a factor 2 is achieved with the \superb detector 
 with respect to the \babar\ one. The alternative SVT layout options investigated give consistent results 
with the nominal \superb solution.
\begin{figure}[!htb]
\begin{center}
\includegraphics[width=0.5\textwidth]{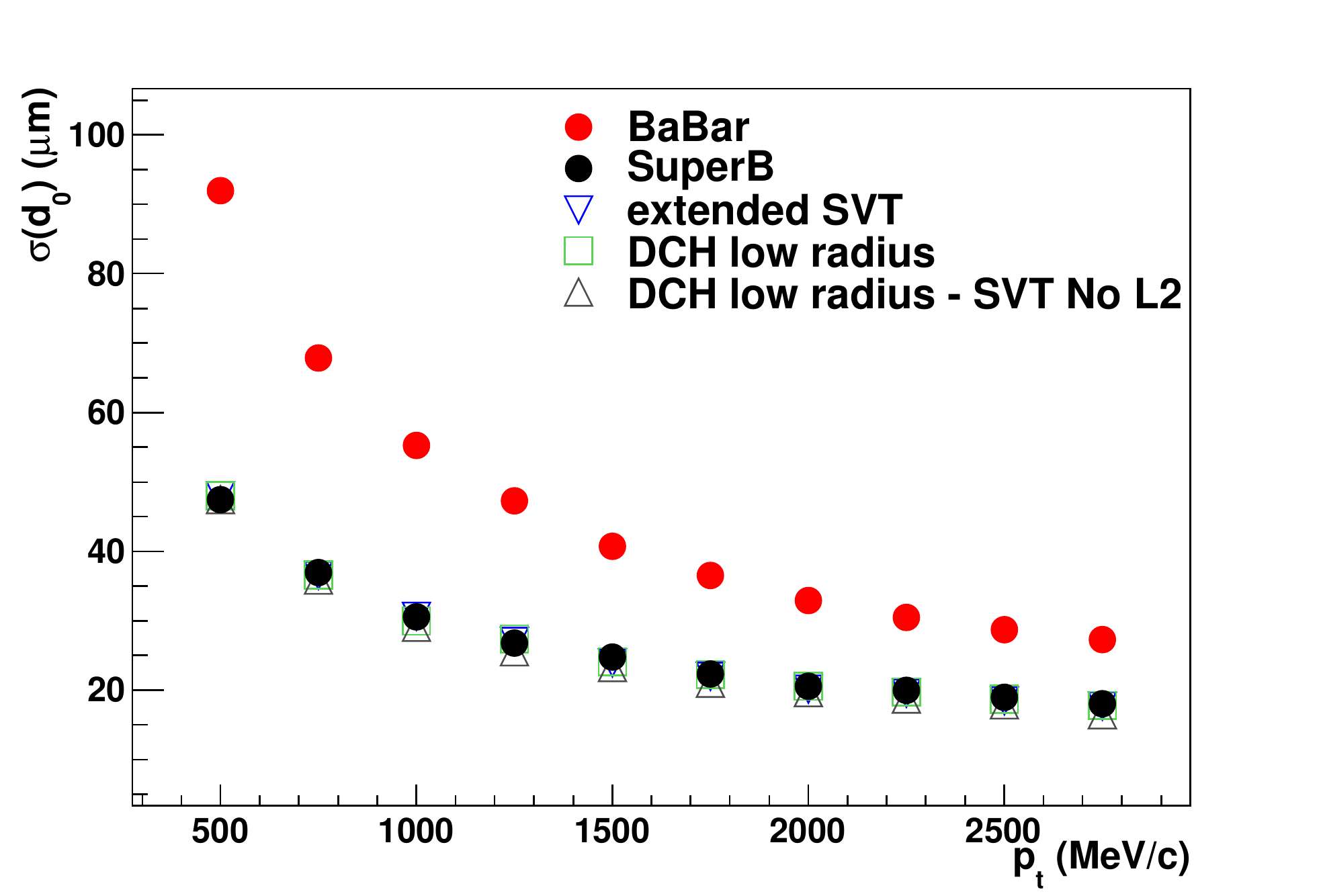}
\end{center}
\caption{Resolution $\sigma(d_0)$ on the impact parameter $d_0$ as a function of $p_t$
 in \babar\ and for various \superb\ tracking detector configurations.}
\label{fig:SVT:d0res}
\end{figure}
\tdrsubsec{Impact of machine background on tracking performance}
\label{sec:performance_and_background}
\label{sec:SVT:performance:impact_of_bkg}

As described in Section~\ref{svt_background}, the background conditions will be more severe in \superb than in \babar.
The fast front-end electronics of the SVT  provides 
 very good resolution on the time of passage of the particle or time of arrival of the hit. 
The Time over Threshold (ToT) of the shaper output is used to correct for the interval between the arrival
 of the hit and the time the shaper exceeds threshold.
The latter is referred to in the following as time stamp (TS) of the hit, and registered with a TS clock. 
 The resolution on the time of arrival of the hit
 depends on the SVT layer due to the different shaper peaking times of the front-end electronics, and 
  has been estimated using $ad hoc$ simulations~\cite{ratti:TimeRes12}.
Several peaking times  summarized in Table~\ref{tab:main_parameters}, will be available on the strip detector readout chips.
The resolution for all layers and sides is reported in Table~\ref{table:SVT:t0_res} for the 
 nominal peaking-time configuration and for the shortest peaking times in Layer4 and Layer5.
 It ranges from about 10 ns for Layer0 up to 50 ns for Layer5.
In our studies, hits outside a $\pm 5\sigma$ acceptance time window from the event time 
(determined by the DCH) are discarded. A similar procedure was used 
 in the reconstruction algorithm used by the \babar\ experiment.
\begin{table}[tbp]
 \caption{Resolution on the time of arrival of the hit for the $\phi$ and $z$ sides (Layer0 $u$ and $v$ sides) 
 for the different layers with selected peaking times. The ``old'' detector corresponds to the same detector after 7.5 years of running 
 and includes a safety factor of 5 times the level of radiation with respect to nominal.}
\begin{center}
{\small
\begin{tabular}{|l|c|c|c|} 
\hline
       & shaper      &  time res.          &   time res.       \\
       &peak. time   &  ``fresh'' det.     &   ``old'' det.      \\      
       & ($\ns$)     & ($\ns$,$\ns$)       &  ($\ns$,$\ns$)           \\
\hline
Layer0 &  25   &($\phantom{0}$9.7,$\phantom{0}$9.7)     & ($\phantom{0}$9.7,$\phantom{0}$9.7)      \\
Layer1 &  75   &(10.7,10.2)   & (11.0,10.8)     \\
Layer2 &  100  &(11.4,10.5)   & (12.0,11.5)      \\
Layer3 &  150  &(12.4,11.7)   & (15.1,14.1)     \\
Layer4 &  500  &(28.8,24.2)   & (41.6,49.4)     \\
Layer5 &  750  &(42.6,34.3)   & (54.6,46.7)     \\
\hline
Layer4 &  250  &(17.8,15.9)   & (21.1,19.3)     \\
Layer5 &  375  &(24.9,20.5)   & (27.6,23.8)     \\
\hline 
\end{tabular}
}
\end{center}
\label{table:SVT:t0_res}
\end{table}

The Fast Simulation tool does not apply a pattern recognition algorithm. 
Tracking performance is based on parameterizations tuned on
\babar\ measured performance. Hits from neighboring tracks may be merged or associated with
wrong tracks, but all generated tracks with a number of reconstructed hits above a given threshold are reconstructed; no fake tracks are added. 
This fast simulation tool thus allows one to study track parameter resolution, but not the tracking inefficiency arising from pattern recognition.
The impact of background on the resolution of the track impact parameters is shown in Figure~\ref{fig:SVT:d0res_bkg}.
\begin{figure}[!htb]
\begin{center}
\includegraphics[width=0.5\textwidth]{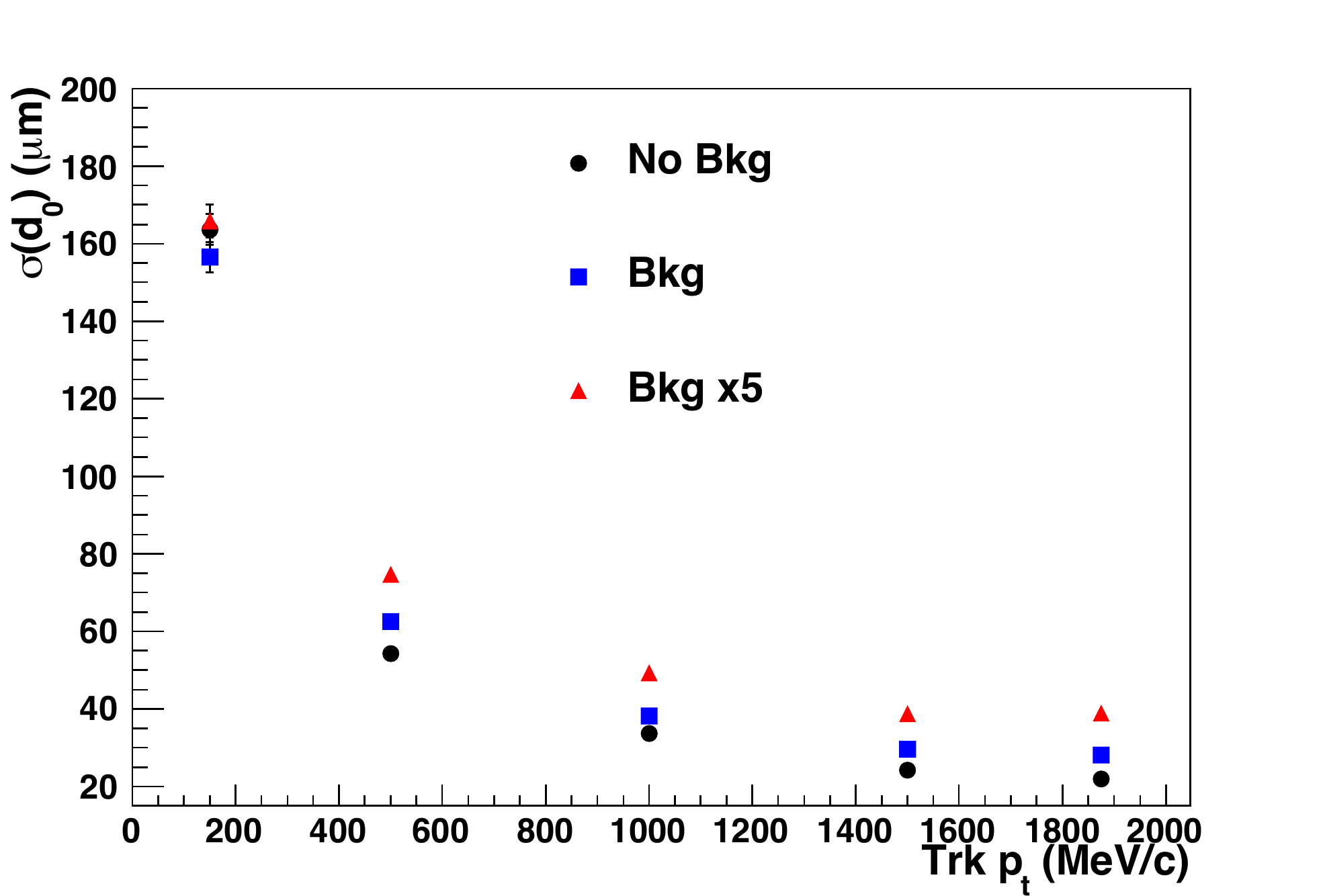}
\end{center}
\caption{Resolution on the impact parameter of the track $d_0$ as a function of $p_t$ 
for the \superb detector with Layer0 striplets. The results shown assume no background (points), 
nominal background (squares) and 5 times the nominal background (triangles).
}
\label{fig:SVT:d0res_bkg}
\end{figure}
In order to address the issue of the pattern recognition capability to reconstruct tracks in the high 
background environment of \superb, SVT detector occupancies estimated in \superb\ 
 have been compared to those observed in \babar.
In particular we compared the cluster occupancy, defined as the detector strip occupancy after applying the 
 time window cut, divided by the hit multiplicity in a cluster. 
The average cluster occupancy over all layers and sides 
is estimated to be about $0.4\%$ with nominal background in \superb.
This cluster occupancy in \superb\ is  smaller than the maximum value of about $0.7\%$ reached at high luminosity in \babar,
 thanks to  the improved hit time resolution.
When considering a scenario with an additional $\times5$ safety factor on background predictions for \superb, 
  the estimated average cluster occupancy is about $2\%$ with nominal peaking-time configuration and 
 about $1.5\%$ with the shortest peaking times in Layer4 and Layer5. 
\babar\ studies~\cite{svt:bad707} of SVT performance in high background conditions have been used to 
 estimate the efficiency to assign a hit to a track as a function of the cluster occupancy,
 that was found to be greater than 95\% up to a 3\% occupancy~\cite{ripp:elba12}.
These studies indicate that the pattern recognition should be able to function without major problems even
 in the presence of 5 times the nominal background.
Moreover, for low momentum tracks not reaching the DCH, the additional Layer0 measurements should help
 the pattern recognition when using SVT hits only. 
 Improvements in the pattern recognition algorithm may also be 
 foreseen with respect to that used in \babar.
%
%
\tdrsubsec{Sensitivity studies for time-dependent analyses}
\label{sec:SVT:performance:striplets_sensitivity_studies_background}

The sensitivity to the physics parameter $S$ has been considered as a figure of merit 
 for time-dependent analyses of \Bz decays. Several decay modes have been studied:
 $\Bz \to \phi\KS$, $\Bz \to \pip\pim$, $\Bz \to J/\psi\KS$, $\Bz \to D^+D^-$, and, in addition,
 decay modes such as $\Bz \to \KS\KS$, $\Bz \to \KS\piz$, in which the impact of the additional Layer0
 measurement is less effective due to the presence of neutral and long-lived particles in the
 final state. The per-event error on the $S$ parameter estimated in \superb\ using
 the Fast Simulation is consistent 
 with that observed with the \babar\ detector for all decay modes apart from
 $\Bz \to \KS\KS$ and $\Bz \to \KS\piz$ decays, 
 where a reduction in sensitivity of about 15\% is observed~\cite{neri:perugia09,simi:perugia09}.
 Only the impact of the $\Delta t$ resolution on the measurements has been included
 in the Fast Simulation studies.
In particular, possible improvements of the \superb reconstruction efficiency, due to the 95\% angular coverage in the CM frame
(with respect to 91\% in \babar), and better flavor tagging performance due to improved particle identification, 
have not been considered in these studies. 
\par
In the case of time-dependent analyses for mixing and \CP violation in the neutral $D$ meson system, 
the determination of the proper time \t relies on the measurement of the 3-dimensional flight length ($\vec{L}$) 
and the momentum $\vec{p}$ of the $D^0$ according to 
$$
t = \frac{\vec{L}\cdot\hat{p}~M}{|\vec{p}|}
$$
where $M$ is the \Dz nominal mass.
\Dz mesons produced in $\epem \to \ccbar$ events 
 gain a natural boost in the reaction. 
Even though the CM boost is reduced with respect to \babar, 
 the resolution on the \Dz proper time in \superb is about 2 times better than that for \babar~\cite{rolfa:slac09}. 
Figure~\ref{fig:SVT:D0_ProperTimeError} shows the distribution of the \Dz proper-time 
error in \superb, compared to that obtained from \babar. The average proper-time error is about 0.16 \ps in \superb
 and 0.30 \ps in \babar\ for $\Dz\to\KS\pip\pim$ decays, which should 
be compared with the \Dz lifetime of about 0.41 \ps.  Similar results have been obtained for $\Dz\to h^+h^-$ decays, 
where $h=\pi, K$.  It has been shown that the mean per-event error for $\Dz\to h^+h^-$ decays at charm threshold
is also about 0.17 \ps~\cite{svt:note:SB-PHY-2012-020}.
\begin{figure}[!h]
\begin{center}
\includegraphics[width=0.5\textwidth]{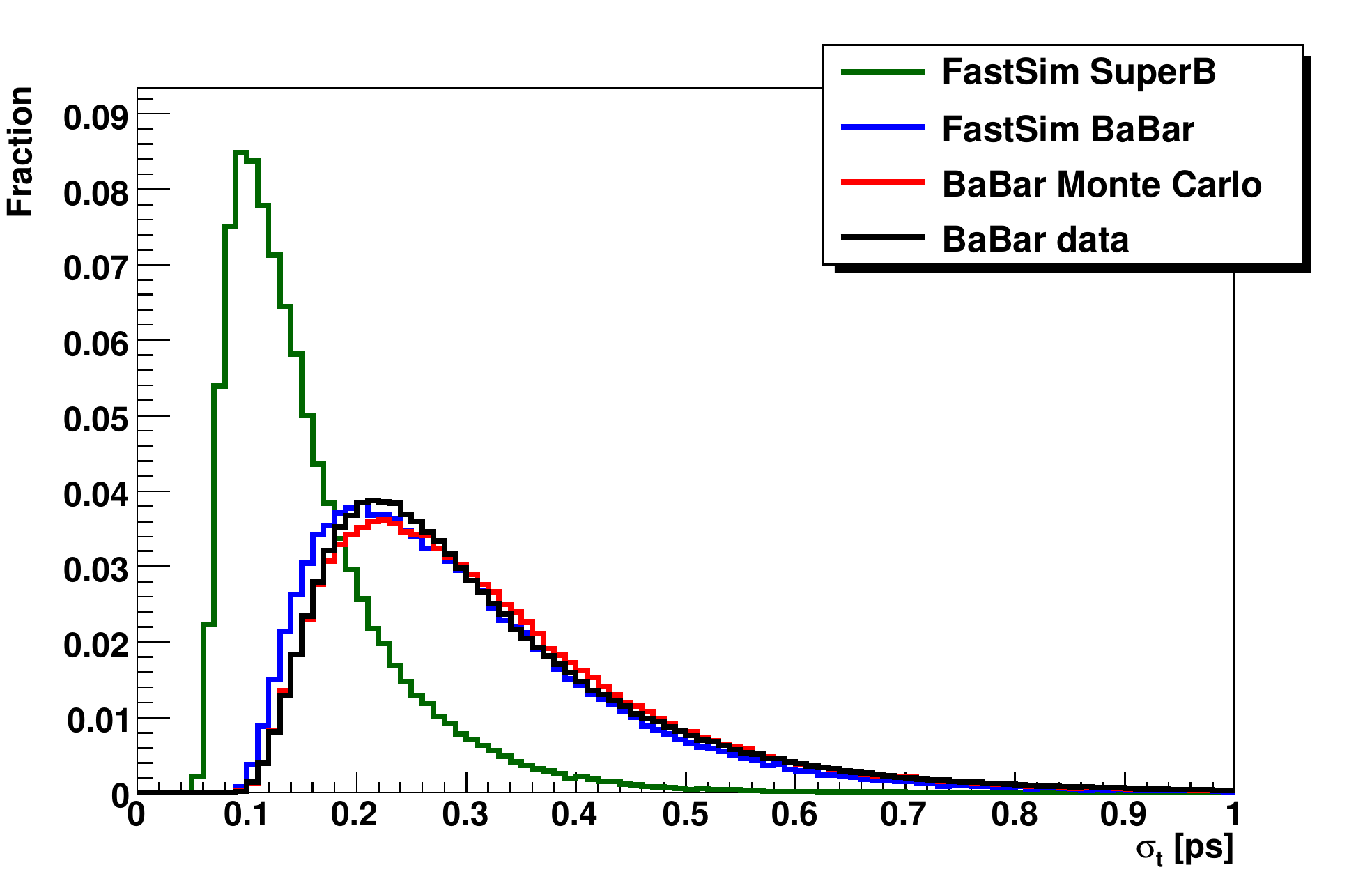}
\end{center}
\caption{\Dz proper-time error distributions obtained with the \superb (green line) and  
the \babar\ (blue line) detectors according to Fast Simulation studies. 
Distributions from \babar\ Monte Carlo (red line) and data (black line) are also shown in the plot.}
\label{fig:SVT:D0_ProperTimeError}
\end{figure}
\par

The impact of machine background events on the SVT performance has been studied by adding 
background hits to signal events according to the rates estimated using Full Simulation. Details on the 
 estimates of the machine background can be found in Section~\ref{svt_background}. 
 Background hits may reduce the hit reconstruction efficiency, increase the effective hit resolution, and
  reduce the efficiency of pattern recognition for charged tracks, along with the increase of fake tracks.
 Most of the  above effects have been included in our Fast Simulation, assuming that the charged track pattern 
 recognition algorithm will work with similar performance to that of \babar, but fake tracks are not simulated.
Hit efficiency of the readout chips used in the Fast Simulation studies 
 can be found in Table~\ref{tab:main_parameters} for the case of nominal background and with 
 5 times the nominal background.
Figure~\ref{fig:SVT:FOM_Bkg} shows the impact of the machine background events on the 
 physics parameter $S$ for the case of nominal background and with 5 times the nominal background rates.
 Background hits are rejected if they are not within a time window of $\pm 3 \sigma$ ($\pm 5 \sigma$)  
 with respect to the time of the event. 
The values of the hit time resolution ($\sigma$) are reported in Table~\ref{table:SVT:t0_res}.
 The reduction of the sensitivity to $S$ is quite limited with nominal background ($< 3\%$) and is about 9\% (14\%) 
 with 5 times the nominal background conditions when applying a $\pm 3 \sigma$ ($\pm 5 \sigma$) time window cut.
\begin{figure}[!h]
\begin{center}
\includegraphics[clip=true,trim=0cm 1.3cm 1.3cm 0cm,width=0.5\textwidth]{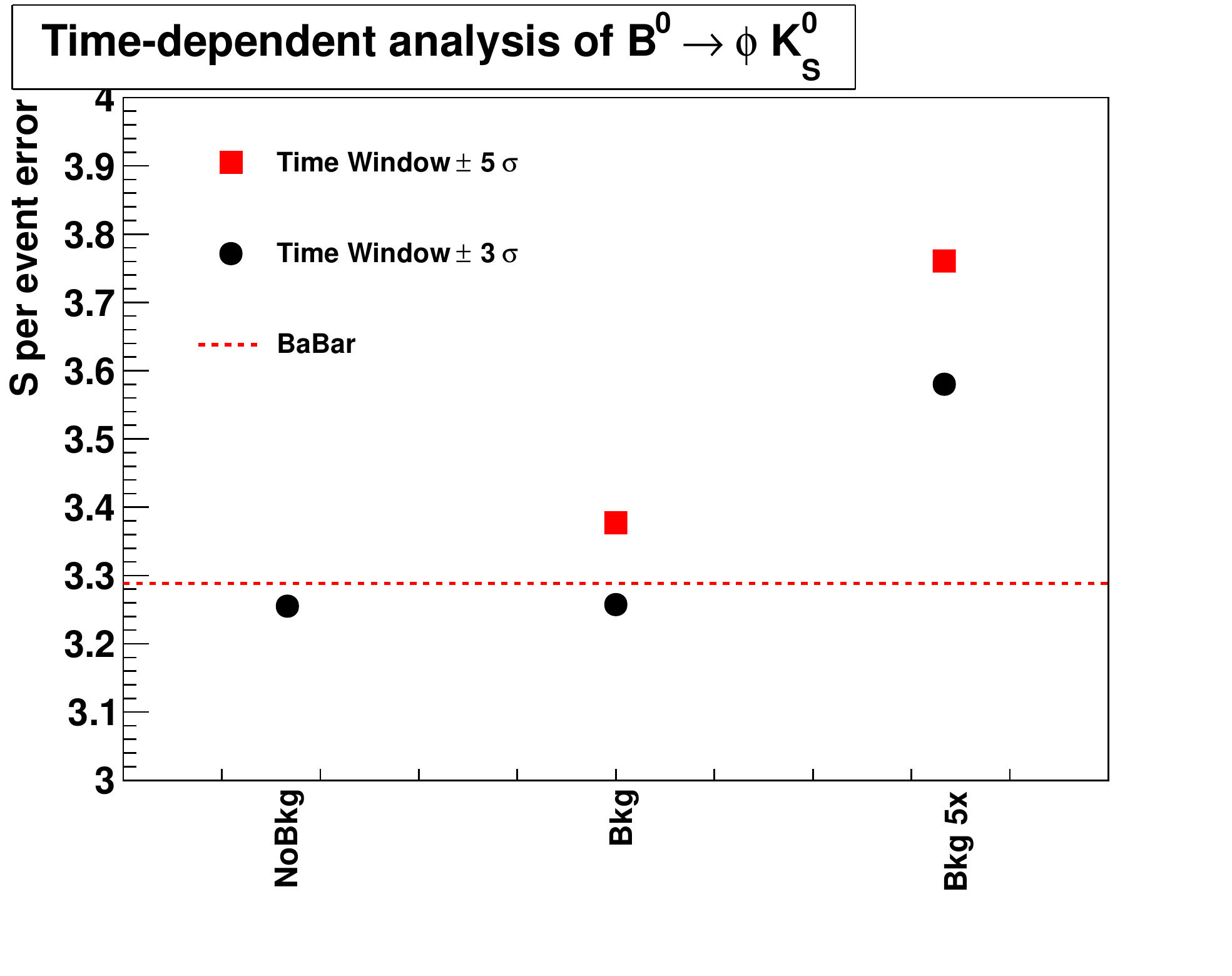}
\end{center}
\caption{Variation of the $S$ per-event error in $\Bz\to\phi\KS$ time-dependent analysis in presence of nominal 
 background events and with 5 times the nominal background. A cut on the time of arrival of the hits has been applied 
 at $\pm 3 \sigma$ and $\pm 5 \sigma$ with respect to the time of the event.}
\label{fig:SVT:FOM_Bkg}
\end{figure}
\tdrsubsec{Performance with Layer0 pixel detectors}
\label{sec:SVT:performance:Layer0_pixel_performance}

A Layer0 technology based on a high granularity silicon pixel sensor, \eg~with~$50 \times 50~\mum^2$ cell, is 
 considered for an upgrade of the baseline striplets solution. 
The different Layer0 technology options are described in Sec.~\ref{sec:pixel_upgrade} are based on either
hybrid pixels or thin CMOS MAPS. 
These solutions all adopt a digital sparsified readout with the area of the pixel cell of about
 $2500-3000 \mum^2$. The shape of the pixel can be optimized in such a way to reduce the sensor pitch in the $z$ 
 direction and to improve the relative hit resolution while keeping the pixel area constant. 

As already discussed in Sec.~\ref{sec:SVT:performance:ImpactL0}, the determination of the decay vertex position is driven by the 
 performance of Layer0. The advantage of the Layer0 pixel solution is to guarantee good detector performance 
 in presence of relatively high background. The detector occupancy, defined as the probability of 
 having a noise hit in the sensitive time window, is about two orders of magnitude lower than for the 
 striplet case, taking into account the different detector granularity and resolution on the time of 
 arrival of the hits. Occupancies at the level of $10^{-4}-10^{-3}$ in a Layer0 pixel detector would correspond to 
 occupancies of $10^{-2}-10^{-1}$ with the striplet solution, which are about the highest achievable values in the Layer0  striplets at \superb.
 Therefore, the impact of the background hits on the determination of the decay vertex and of the
track impact parameters is moderate at \superb with a Layer0 pixel solution.

 Figure~\ref{fig:SVT:FOM_PixelX0Radius}  shows the sensitivity to the $S$ parameter in time-dependent \CP 
 violation analysis of $B \to \phi\KS$ decays as a function of the Layer0 radius 
 ($r_0$ =1.4 and 1.6$\cm$) and of its material budget ($x/X_0 = 0.1-1.0\%$), in case of nominal background and $\times 5$ the nominal background.
The dashed line represents the reference value  obtained in \babar. 
A material budget in the range $x/X_0 = 0.35-0.50 \%$ 
 ($x/X_0 = 0.55-0.85 \%$) is achievable for a Layer0 pixel solution based on CMOS monolithic active pixel sensors 
 (hybrid pixels) depending on the results of the ongoing R\&D activities. 
 The $S$ sensitivity is very  similar to the one obtained in \babar. 
 The maximum difference is about 6\% (10\%) in the worst case considered (including 5 times the nominal background).

\begin{figure}[!htb]
\begin{center}
\includegraphics[clip=true,trim=0cm 1.6cm 1.3cm 0cm,width=0.5\textwidth]{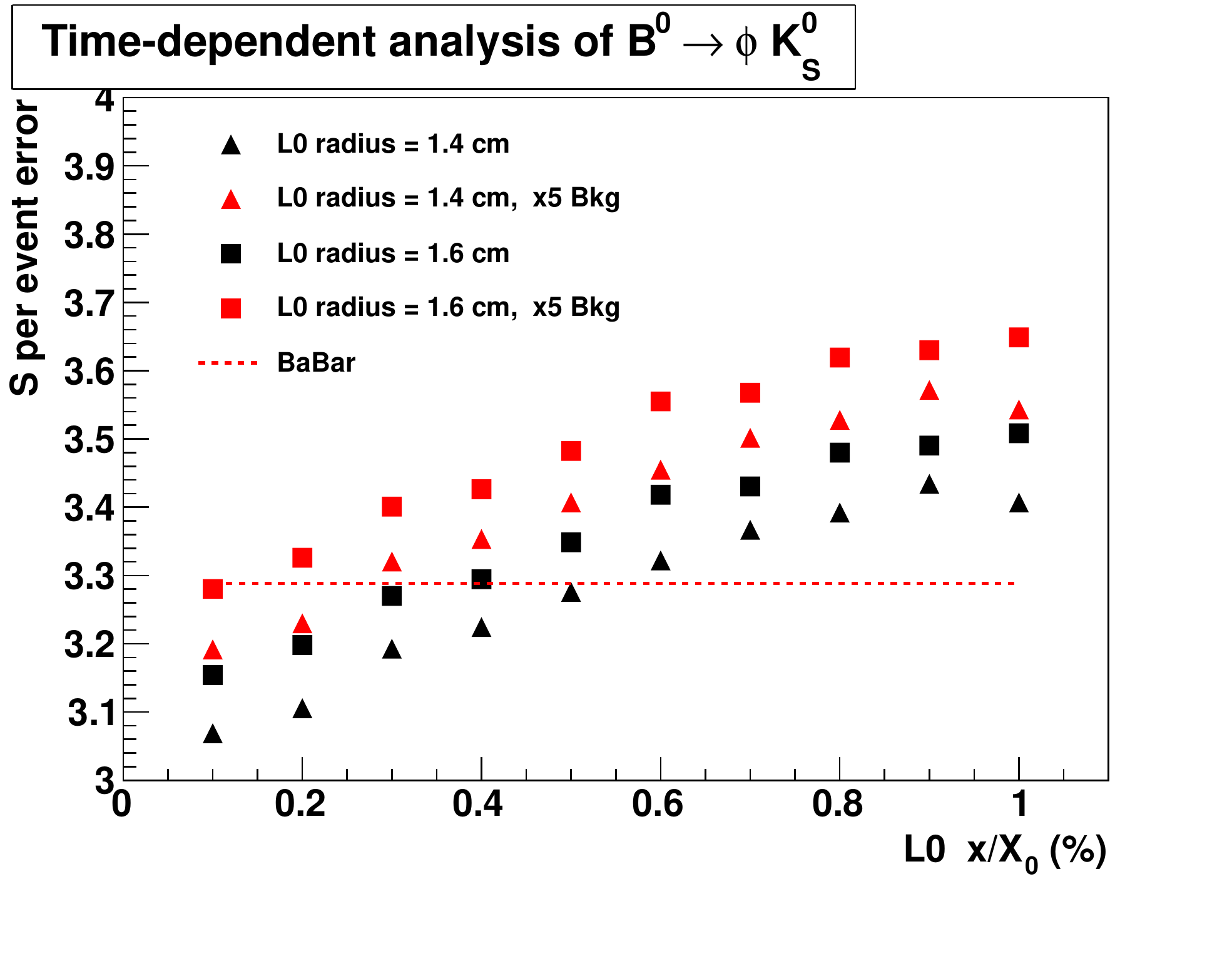}
\end{center}
\caption{$S$ per-event error in $\Bz\to\phi\KS$ time-dependent analysis for different Layer0 
 radii ($r_0$ = 1.4 and 1.6$\cm$) and material budget ($x/X_0=0.1-1.0\%$) 
compared with the reference value of \babar\ (dashed line). Results in presence of 5 times the nominal background 
 are also shown in the plot.}
\label{fig:SVT:FOM_PixelX0Radius}
\end{figure}

\tdrsubsec{Particle identification with \dedx}
\label{sec:SVT:performance:PID_with_dedx}

The measurement of the ToT value by the front-end electronics enables one to obtain the pulse height, and
hence the ionization energy loss \dedx in the SVT sensors. 
The dynamic range of the analog readout is about 
 10-15 times the value corresponding to minimum ionizing particles, which is sufficient to make the
 \dedx capability of the SVT useful for particle identification~\cite{svt:note126}.
 
Each sensor will provide 2 measurements of \dedx, one for each sensor side, 
 for a total of 12 \dedx measurements in the SVT. 
In \babar, where a total of 10 \dedx measurements (5 layers) were available, 
for every track with signals from at least four sensors in
the SVT, a 60\% truncated mean \dedx was calculated.
The cluster with the smallest \dedx
energy was also removed to reduce sensitivity to electronic noise.
For MIPs a resolution on the truncated mean \dedx of approximately 14\% was obtained.

The intrinsic smearing from the distribution of the
 energy deposition in the silicon sensors and from the atomic binding effects in the silicon 
will dominate the uncertainty 
 on the measured \dedx~\cite{svt:note126}. 
The contribution to the \dedx uncertainty from the electronic noise 
 should be relatively small. Therefore the resolution on \dedx for MIPs is expected to be similar 
to that achieved  in \babar.
However, the \dedx precision is approximately inversely proportional to the square root of the number 
of \dedx samples used for the truncated \dedx mean calculation~\cite{svt:bad1500}.  
Two additional measurements in the Layer0 should improve the average resolution of a factor  
 $\sqrt{5/6}=0.9$, where 5 is the average number of \dedx samples used in \babar and 6 is the expected 
average number in \superb.
 The $e/\pi$ separation is expected to be larger than $3\sigma$ for momenta lower than 150\mevc and 
 will be very useful for rejecting low momentum electrons from background QED processes.

\tdrsec{Silicon Sensors} 
\label{svt_sensors}

Layers 1 to 5 of the SVT will be based on 300~$\mu$m thick double-sided silicon strip sensors, with integrated AC-coupling capacitors and polysilicon bias resistors. These devices are a technically mature and conservative solution that matches the requirements to be met by the SVT i.e. to provide precise, highly segmented tracking near the interaction point.  For the new Layer0, the baseline option also foresees the use of double-sided silicon strip detectors, with short strips (`striplets'), 20~mm long, oriented at $\pm 45$ degrees from the beam direction, fabricated on 200~$\mu$m thick substrates. The requirements that the sensors must meet are discussed below.

\tdrsubsec{Requirements}

\paragraph{Material budget.}

To achieve good vertex resolution, it is especially important to minimize the material up to  and including the first measurement.  This requirement, and the need to provide precise vertexing in both $z$ and $r\phi$ views, leads to the choice of double-sided detectors.  For Layers 1 to 5 we plan to use 300~$\mu$m thick silicon wafers, which are a standard choice and present acceptable handling properties. For Layer0, given the very stringent limitations on the amount of material, we are forced to go to 200~$\mu$m thick substrates.  

\paragraph{Efficiency.}

The silicon detectors must maintain high single-point efficiency in order to meet the requirements given in Section~\ref{svt:overview} for high overall track reconstruction efficiency.  Loss of efficiency can occur from defective sensor strips, from bad interconnections, or from faulty electronics channels. Sensor related inefficiencies can be due to fabrication defects or handling damage, which can result in strips with high leakage currents, poor insulation or broken AC-coupling capacitor.  Our goal is to achieve an overall strip failure rate below 1\% for each sensor. The experience gained from a large production of double-sided AC-coupled detectors for the ALICE Inner Tracking System indicates that a total rate of defective strips below 1\% can be achieved with reasonable yield ($>$ 70\%).

\paragraph{Resolution.}

As described in Section~\ref{svt:overview}, we require the intrinsic point resolution to be 15~$\mu$m or better in both $z$ and $\phi$ for the inner layers (Table~6.2).
These are the point resolutions for tracks at near-normal incidence.  As the angle between the track and the plane normal to the strip increases, the resolution initially improves, then degrades.
We require the resolution to degrade by no more than a factor of $\sim3$ for angles up to 75$^\circ$ from normal incidence.

\begin{table*}[ht]
 \caption{Physical dimensions, number of strips and pitches for the nine different sensor models. Model VI has a trapezoidal shape.}
\begin{center}
\begin{tabular}{|l|c|c|c|c|c|c|c|c|c|} 
\hline
Sensor~Type       &   0 & I    & II   & III  & IVa   & IVb    & Va   & Vb   & VI \\ 
\hline\hline
Dimensions (mm)  &   &    &      &     &     &      &     &     &  \\
$z$ Length (L)   & 105.2 & 111.7  &  66.4   &  96.4   &  114.6   &  119.8   &  102.2  &  106.0  & 68.0 \\
$\phi$ Width (W)    & 15.1 &    41.3 & 49.4 & 71.5   & 52.8    &   52.8   &  52.8   &  52.8   & 52.8-43.3 \\
Thickness & 0.20  & 0.30  &   0.30   & 0.30    &  0.30   &   0.30   &  0.30   &  0.30   &  0.30 \\
\hline
PN junction side reads & $u$ & $z$ & $z$ & $\phi$ & $\phi$ & $\phi$ & $\phi$ & $\phi$ & $\phi$ \\ 
\hline
Strip Pitch ($\mu$m)  &  &      &     &     &      &     &    &     &  \\
$z$ ($u$ for Layer0)  & 54 &   50  &  50   &  55  &  105  &  105  &  105  &  105  & 105 \\
$\phi$ ($v$ for Layer0)  & 54 &  50   &  55  & 50  &  50  &   50  &  50  &  50  &  50 $\rightarrow$ 41\\
\hline
Readout Pitch ($\mu$m)  &  &      &     &     &      &     &    &     &  \\
$z$ ($u$ for Layer0)   & 54 & 100 & 100 & 110 & 210 & 210 & 210 & 210 &  \\
$\phi$ ($v$ for Layer0)  & 54 & 50 & 55 & 100 & 100 & 100 & 100 & 100 &  100 $\to$ 82\\
\hline
Number of Readout Strips  &  &      &     &     &      &     &    &     &  \\
$z$ ($u$ for Layer0)  & 1536 &  1104 &  651 & 865 &  540 &  565  &  481 & 499 &  318\\
$\phi$ ($v$ for Layer0) & 1536 &  799 &  874 & 701 &  512  &  512 &  512 &  512 &  512\\
\hline
\end{tabular}
\end{center}
\label{table:SVT:SensorGeometry}
\end{table*}

\begin{table*}[htb]
 \caption{
Values for the total 1~MeV neutron equivalent  fluence $\Phi_{tot}$, effective dopant concentration $N_\mathrm{eff}$ and total depletion voltage $V_{td}$ after 7.5 years at nominal radiation load, without and with the the 5$\times$ safety factor. 
For all Layers the initial (pre-irradiation) resistivity was assumed to be 7~k$\mathrm{\Omega}$~cm, corresponding to a donor concentration  
$N_d=6.3\times 10^{11}$~cm$^{-3}$.}
\begin{center}
\renewcommand{\arraystretch}{1.2}
\begin{tabular}{| c || c || c | r | c || c | r | c ||} 
\hline
 &  & \multicolumn{3}{| c ||}{ Nominal radiation load in 7.5 years }
                               &\multicolumn{3}{| c ||} {With 5$\times$ safety factor included} \\

 & Pre-Irrad.       & $\Phi_{tot}$ & Post-Irrad.   & Post-Irrad. & 5$\times$$\Phi_{tot}$     & Post-Irrad.                         & Post-Irrad. \\ 
 & $V_{td}$(V)   & (cm$^{-2}$)    & $N_\mathrm{eff}$ (cm$^{-3}$) & $V_{td}$(V)    &   (cm$^{-2}$) & $N_\mathrm{eff}$ (cm$^{-3}$) & $V_{td}$(V)\\
\hline\hline
Layer0  & 19 &  2.7~$10^{13}$  &  --1.3~$10^{12}$   &  40   &  1.4~$10^{14}$   &  --7.3~$10^{12}$  &   220   \\
Layer1  & 44 &  5.9~$10^{12}$  &  --6.4~$10^{10}$   &  4.4  &  3.0~$10^{13}$   &  --1.4~$10^{12}$  &   100   \\
Layer2  & 44 &  3.8~$10^{12}$  &  +1.2~$10^{11}$  &  8.1  &  1.9~$10^{13}$   &  --8.7~$10^{11}$ &    60    \\
Layer3  & 44 &  2.3~$10^{12}$  &  +2.9~$10^{11}$  &  20  &  1.1~$10^{13}$   &   --4.2~$10^{11}$ &    29    \\
Layer4  & 44 &  1.5~$10^{12}$  &  +3.9~$10^{11}$   & 27  &  7.5~$10^{12}$   &   --1.8~$10^{11}$ &    12    \\
Layer5  & 44 &  1.3~$10^{12}$  &  +4.1~$10^{11}$   & 29  &  6.6~$10^{12}$   &   --1.2~$10^{11}$ &     8.0  \\
\hline
\end{tabular}
\end{center}
\label{table:SVT:SensorDamage}
\end{table*}

\paragraph{Radiation hardness.} 

The sensors must hold up to integrated doses of ionizing radiation and to 1 MeV equivalent neutron fluences as high as reported in Table \ref{tab:background_summary}.
This requirement leads to the use of AC-coupled sensors in order to avoid the problems associated with direct coupling of the large leakage currents caused by the irradiation. It also has implications in the choice of the strip biasing scheme, making the punch-through technique unsuitable and limiting the maximum values of the polysilicon resistors. 

Another important effect of the irradiation is a change in the effective dopant concentration of the substrate and, as a consequence, on the minimum bias voltage necessary to deplete the sensor and to achieve full collection of the signal charge. An estimate of these effects is given in the next section. 

\tdrsubsec{Sensor design and technology}
\label{par:SVT:SensorDesign}

From the above requirements and from the discussion in Section~\ref{svt:overview}, we have arrived at the detector specifications and design parameters described in the following. 

\paragraph{Sensor models and sizes.}

Given the increased module length with respect to the \babar  SVT, in order to reduce the insensitive area between adjacent sensors and the complexity of the assembly operations -- and to simplify the detector alignment task -- we seek to minimize the number of sensors making up each SVT module. Taking into account the constraints on the module sizes coming from the overall SVT design (Section~\ref{svt:overview}), this goal can be met by designing a specific sensor model for each module type -- plus the wedge shaped  sensor -- and by having the sensors fabricated on 150~mm diameter wafers. 

We are therefore led to nine different sensor models, whose overall sizes are listed in Table~\ref{table:SVT:SensorGeometry} together with strip numbers and pitches, which will be discussed later.

Although 300~$\mu$m thick sensors fabricated on 150~mm substrates are by now an available option from several suppliers, processing the Layer0 double sided sensors on 200~$\mu$m thin, 150~mm diameter wafers is a significant challenge, which very few manufacturers are willing to tackle. Unfortunately, while Layers 1-5 could also be assembled from smaller sensors, fitting inside 100~mm wafers, Layer0 sensors require larger wafers. This is due to the requirement to have only one sensor per Layer0 module, which in turn is dictated by the need to minimize insensitive regions and mechanical support structures, and also by limitations on the available number of readout channels. These difficulties are mitigated by the very small number of Layer0 sensors required and the fact that five of them can comfortably fit into a single 150~mm wafer. Because of this, a lower fabrication and assembly yield can be tolerated for Layer0 sensors.

\paragraph{Fabrication technology.}

The microstrip sensors will be fabricated on $n$-type wafers, with $p^+$ strips on the junction side and $n^+$ strips on the ohmic side,  insulated by patterned $p^+$ implants ($p$-stops) in between, or by a uniform $p$-type implant ($p$-spray). 
The strips will be AC-coupled, with integrated capacitors and polysilicon bias resistors. The alternative biasing method, exploiting the punch-through effect, does not offer adequate radiation tolerance. \\
This has proven to be a mature, reliable technology, requiring no R\&D.

\begin{table*}[ht]
 \caption{Estimated strip capacitance and series resistance for the different sensor models.}
\renewcommand{\arraystretch}{1.2}
\begin{center}
\begin{tabular}{|c|c|c|c|c|c|c|} 
\hline
              & \multicolumn{3}{|c|}{$z$ readout side ($u$ for model 0)}
                               &\multicolumn{3}{|c|}{$\phi$ readout side ($v$ for model 0)} \\
\cline{2-7}
Detector Model & $C_{strip}/\ell$ 
                    & $C_{AC}/\ell$
                          & $R_{series}/\ell$
                               &$C_{strip}/\ell$
                                     & $C_{AC}/\ell$
                                          & $R_{series}/\ell$ \\
              & (pF/cm) 
                    & (pF/cm) 
                          & ($\mathrm{\Omega}$/cm)
                               & (pF/cm)
                                     & (pF/cm)
                                          & ($\mathrm{\Omega}$/cm) \\
\hline\hline
0             & 2.5 & 40  & 9  & 2.5 & 30 & 14 \\
I             & 1.7 & 40  & 5  & 2.5 & 30 & 9 \\
II            & 1.7 & 40  & 4  & 2.5 & 30 & 7 \\
III           & 1.7 & 30  & 7  & 1.7 & 40 & 4 \\
IVa,IVb,Va,Vb       & 1.7 & 60  & 3 & 1.7 & 40 & 4 \\
VI           & 1.7 & 60  & 3 & 1.7 & 30 & 4.5 \\
\hline

\end{tabular}
\end{center}
\label{table:SVT:MeasuredCR}
\end{table*}

\begin{table*}[htb]
 \caption{
Total 1~MeV neutron equivalent  fluence $\Phi_{tot}$ for the different layers and maximum leakage current $I_{leak}$ for  $\phi$ and $z$-side strips of a module, after 7.5 years at nominal radiation load, without and with the 5$\times$ safety factor. An operational temperature of 20~$^\circ$C is assumed.}
\begin{center}
\renewcommand{\arraystretch}{1.2}
\begin{tabular}{| c || c | r | c || c | r | c ||} 
\hline
 &  \multicolumn{3}{| c ||}{ Nominal radiation load in 7.5 years }
                               &\multicolumn{3}{| c ||} {With 5$\times$ safety factor included} \\

 & $\Phi_{tot}$ & $\phi$ strips               & $z$ strips                       & 5$\times\Phi_{tot}$ &  $\phi$ strips  &  $z$ strips  \\ 
 & (cm$^{-2}$) & $I_{leak}$ ($\mu$A) & $I_{leak}$  ($\mu$A) & (cm$^{-2}$)      & $I_{leak}$ ($\mu$A) & $I_{leak}$ ($\mu$A) \\
\hline\hline
Layer0  &  2.7~$10^{13}$  &  0.15   &  0.15  &  1.4~$10^{14}$   &    0.75  &    0.75   \\
Layer1  &  5.9~$10^{12}$  &  0.28   &  0.40  &  3.0~$10^{13}$   &   1.4  &   2.0   \\
Layer2  &  3.8~$10^{12}$  &  0.23  &   0.31  &  1.9~$10^{13}$   &   1.2 &    1.6    \\
Layer3  &  2.3~$10^{12}$  &  0.37  &   0.30  &  1.1~$10^{13}$   &   1.8 &    1.5    \\
Layer4  &  1.5~$10^{12}$  &  0.39  &   0.41  &  7.5~$10^{12}$   &   1.9 &    2.0    \\
Layer5  &  1.3~$10^{12}$  &  0.43  &   0.36  &  6.6~$10^{12}$   &   2.1 &    1.8  \\
\hline
\end{tabular}
\end{center}
\label{table:SVT:LeakageCurrent}
\end{table*}

\begin{table*}[ht]
 \caption{
Number of the different sensor types per module, area of the installed sensors, number of installed sensors  and number of sensors including spares. Spare sensors include one spare module per module type (two for Layer0), plus additional sensors accounting for possible losses during the whole SVT assembly process.}
\begin{center}
\begin{tabular}{|l|c|c|c|c|c|c|c|c|c|c|} 
\hline
Sensor Type  &   0 & I    & II   & III  & IVa   & IVb    & Va   & Vb   & VI & All \\ 
\hline\hline
Layer0  & 1 &  -  &  -   &  -   &   -   &  -   &   - &   -  & - & 1 \\
Layer1  & -  &  2  &  -   &  -  &    -  &  -    &  -   &   -  & - & 2 \\
Layer2  &  - & -  &  4   &  -  &   -   &   -   &  -   & -    & - & 4 \\
Layer3  & - &  -  &  -   &  4   &   -   &   -  &  -  &  -   &  & 4 \\
Layer4a  & -  &   -   &  -   &  -  &  4    &   -  &  -   &  -   &  2 & 6 \\
Layer4b &  - &  -   &  -   & -   &  -    &   4   &  -   &   -  &  2 & 6 \\
Layer5a & -  &   -   &   -  &  -   &   -   &  -   &  6  &   -  &  2 & 8 \\
Layer5b & -  &   -   &   -  & -   &   -   &   -   &   -  &   6  &  2 & 8\\
\hline
Silicon Area (m$^2$) & 0.013 &  0.055 &  0.079 &  0.167 &  0.194 &  0.203 &  0.291 &  0.302 &  0.222 & 1.52 \\
Nr. of Sensors &  8 & 12 & 24 & 24 & 32 & 32 & 54 & 54 & 68 & 308 \\
Nr. Including Spares & 20 & 20 & 40 & 35 & 44 & 44 & 72 & 72 & 92 & 439 \\
\hline
\end{tabular}
\end{center}
\label{table:SVT:SensorMultiplicity}
\end{table*}

\begin{table*}[ht]
 \caption{
List of different mask sets for 150~mm wafers, specifying the content of each wafer layout, the minimum value of the distance 
between the sensors and the wafer edge, the number of wafers required for each design and the total number of wafers. 
The numbers quoted include the spare sensors, but not the fabrication yield.}
\begin{center}
\begin{tabular}{|l|c|c|c|} 
\hline
Mask Design       & Wafer content & Min. Clearance to & Number  \\ 
                               &                           & Wafer Edge (mm) & of Wafers\\
\hline\hline
A &  5$\times$Mod 0 &  10.2  & 5  \\
B &  Mod I + Mod VI &    8.2  & 20  \\
C &  Mod III &  15.0  & 35  \\
D &  Mod IVa &  11.9  & 44  \\
E &  Mod IVb &    9.5  & 44  \\
F &  Mod Va + Mod VI  &    9.8  & 72  \\
G & Mod Vb + Mod II &   6.9  & 72  \\
\hline
 Total  &  &  & 287 \\
\hline
\end{tabular}
\end{center}
\label{table:SVT:MaskDesign}
\end{table*}

\paragraph{Substrate resistivity and depletion voltages.} 

The wafer resistivity is assumed to be in the range 6--8~k$\mathrm{\Omega}$\cm, corresponding to a depletion voltage of 50--38~V for 300~$\mu$m thick sensors. These values seem to be a reasonable compromise between the need to limit the initial depletion voltage and peak electric fields on one hand, and on the other hand the desire to delay the onset of type inversion due to radiation damage. \\
In fact, the relatively high radiation levels expected will cause a significant displacement damage in the substrate. The expected 1 MeV neutron equivalent fluences over 7.5 years of operation at nominal luminosity, both without and with an additional safety factor of 5, are reported in Table~\ref{table:SVT:SensorDamage} for the various layers, together with the effective dopant concentration $N_\mathrm{eff}$ and the total depletion voltage $V_{td}$ expected at the end of the irradiation period. For all Layers the initial (pre-irradiation) resistivity was assumed to be 7~k$\mathrm{\Omega}$~cm, corresponding to a donor concentration $N_d=6.3\times 10^{11}$~cm$^{-3}$ and to a total depletion voltage of 19~V for the 200~$\mu$m thick Layer0 sensors and 44~V for the 300~$\mu$m thick sensors of the other layers. 
The changes in effective dopant concentration have been predicted using the model and the data given in \cite{RadiationDamage1,RadiationDamage2,RadiationDamage3,Spieler}. Due to the long exposure time, the damage component undergoing short-therm anneal has been neglected; the reverse anneal component has been estimated taking into account, for each operation year, the remaining time until the end of the 7.5 year operation period. An operational temperature of 20~$^\circ$C has been assumed. Although the possibility of cooling the SVT sensors to lower temperatures (in the range 8-12~$^\circ$C) is being evaluated for the purpose of reducing the leakage current, the beneficial effect on the reverse annealing would not be very significant, considering also the fact that the detector would in any case stay at room temperature for extended periods. \\
With the 5$\times$ factor included, the resulting final depletion voltage is still comfortable (below 100~V), except for the Layer0 sensors, for which it would reach 220~V. While this is not an unbearable value, it must be kept in mind that the Layer0 is designed to be replaceable and that the striplet detectors are foreseen to be superseded by pixels in the later part of the \superb\ operation. 

\paragraph{Configuration of $z$ and $\phi$ readout strips.}

The choices regarding the strip readout pitch and which side of the detector (junction or ohmic) should read which coordinate ($z$ or $\phi$) largely follow those adopted for the \babar SVT.  

At equal pitch, the strip capacitance and, consequently, its noise contribution is somewhat smaller on the junction side than on the ohmic side.
Furthermore, the series resistance of the metal strip can be made lower, because the absence of $p$-stops leaves room for a wider metal readout strip.  For these reasons, and because the $z$ vertex measurement is more important from the point of view of physics, we use the junction side for the $z$ strips on the inner Layers 1 and 2 (in Layer0 the two sides read equivalent coordinates, being oriented at $\pm45^\circ$ from the beam direction). 

The choice of the strip pitch is influenced by several factors, such as the number of available readout channels, the strip capacitance, series resistance and leakage current, the limited area available for the bias resistors, the need to maintain a large enough signal from angled tracks. Profiting from the optimization work made for the \babar SVT, and the experience gained from its operation, the following choices have been made:
	\begin{itemize}
	\item both sides of Layer0 are read by strips with a 54~$\mu$m pitch, without intermediate floating strips;
	\item the $z$-side of Layers 1, 2 is read by $p$-type strips with a 100~$\mu$m pitch with one intermediate floating strip;
	\item the $z$-side of Layer3 is read by $n$-type strips with a 110~$\mu$m pitch with one intermediate floating strip;
	\item the $z$-side of Layers 4, 5 is read by $n$-type strips with a 210~$\mu$m pitch with one intermediate floating strip;
	\item the $\phi$-side of Layer1 is read by $n$-type strips with a 50~$\mu$m pitch without intermediate floating strips;
	\item the $\phi$-side of Layer2 is read by $n$-type strips with a 55~$\mu$m pitch without intermediate floating strips;
	\item the $\phi$-side of Layers 3, 4, 5 is read by $p$-type strips with a 100~$\mu$m pitch with one intermediate floating strip.
	\end{itemize}

The strip pitches and the resulting numbers of strips for the various sensor models are listed in Table~\ref{table:SVT:SensorGeometry}. 

\paragraph{Strip capacitance and resistance.}
Important strip parameters are the capacitance, the series resistance and the leakage current, all of which are proportional to the strip length $\ell$. For long strips, as will be those of the SVT, the noise contribution of the strip resistance becomes more important, because it is proportional to the strip capacitance multiplied by the square root of the resistance. This gives an $\ell ^{3/2}$ dependence on the strip length, to be compared with an $\ell$ dependence for the noise contribution of the strip capacitance and  an $\ell ^{1/2}$ dependence for that of the leakage current. In order to reduce the series resistance of the strips, we plan to increase the thickness of the aluminum metallization beyond the standard value of $\sim 1~\mu$m. 

Table~\ref{table:SVT:MeasuredCR} reports the expected values for the strip capacitance -- as deduced from measurements on  \babar sensors and other strip sensors -- and for the strip series resistance, as calculated from the design width of the strips and a metallization thickness of 2.5~$\mu$m.

\paragraph{Strip leakage current.}
After an initial period of operation, the strip leakage current will be dominated by carrier generation at radiation-induced defects. Table~\ref{table:SVT:LeakageCurrent} reports the total 1~MeV neutron equivalent  fluence for the different layers and the maximum leakage current expected for $\phi$ and $z$-side strips, after 7.5 years at nominal radiation load, without and with the 5$\times$ safety factor. Because of the long irradiation time, spanning 7.5 years, the currents are calculated taking into account the effect of the annealing, by assuming a current damage coefficient $\alpha = 2.8\times 10^{-17}$~A/cm. The values reported in the table refer to a module strip, composed of a few sensor strips daisy-chained together into a readout channel, by direct bonding on the $\phi$-side and by using the fanout circuits on $z$-side. For the $z$-side, different numbers of strips (up to three) can be connected together to a single readout channel within the same module; in this case the table reports are the maximum value of the current.

\paragraph{AC coupling capacitors.}
The strips are connected to the preamplifiers through a decoupling capacitor, integrated on the detector by interposing a dielectric layer between the $p$ or $n$-doped strip and the metal strip. AC coupling prevents the amplifier from integrating the leakage current with the signal; handling the high leakage currents due to radiation damage imposes an additional heavy burden on the preamplifier design. On each sensor, the value of the decoupling capacitance must be much larger than the total strip capacitance on the same sensor, a requirement which is easily met by the current fabrication technologies. 

\paragraph{Bias resistors.}  
The bias resistor values will range between 1 and 8~M$\mathrm{\Omega}$, depending on the layer and the detector side. The choice of the $R_B$ value is constrained by two requirements. A lower limit is determined by the need to limit the noise contribution, which has a $\sqrt{\tau /R_B}$ dependence, and if several strips are ganged together the effective resistance is correspondingly decreased. The requirement that, for floating strips, the product $R_B \cdot C_{TOT}$ must be much larger than the amplifier peaking time in order to allow for capacitive charge partition is fulfilled with ample margin for any reasonable values of $R_B$.
An upper limit to $R_B$ is dictated by the allowable potential drop due to the strip leakage current, which depends mainly on the irradiation level and decreases going from inner to outer layers. The maximum resistance value is also limited in practice by the need to limit the area occupied on the wafer. Values of 40~k$\mathrm{\Omega}$/square for the sheet resistance of polysilicon can be achieved with sufficient reproducibility. Thus, it is possible to fabricate a 10~M$\mathrm{\Omega}$ resistor with a 6~$\mu$m wide, 1500~$\mu$m long polysilicon resistor. With a suitable shaping of the polysilicon line, the space required by the resistor will be less than 200~$\mu$m at 100~$\mu$m pitch (corresponding to strips at 50~$\mu$m pitch with resistors placed at alternate ends).  A final requirement is that the bias resistance be  sufficiently stable against the expected radiation exposure, a condition that is easily satisfied by polysilicon resistors. 

\paragraph{Edge region and active area.}

Considering the space needed to accommodate the bias resistors and to prevent the depletion region from reaching the cut edge of the sensor, we specify the active region of the detectors to be 1.4~mm smaller than the physical dimensions, that is, the dead region along each edge has to be no more than 700~$\mu$m wide. This is the same specification chosen for the \babar  strip detectors and, although stricter than adopted by most silicon sensor designs, has proven to be feasible without difficulty, thanks to the choice of placing the polysilicon resistors in the edge region, outside the guard ring. For Layer0 sensors, which have a reduced thickness of 200~$\mu$m and smaller value, shorter bias resistors, we specify a 600~$\mu$m wide inactive edge region.

\tdrsubsec{$z$-side strip connection options}

On $z$-side, the readout pitch is set to $100~\mu$m in Layers 1 and 2, $110~\mu$m in Layer3 and $210~\mu$m in Layers 4 and 5, with a `floating' strip in between, to improve spatial resolution for particle tracks with close to normal incidence. Since the number of readout strips exceeds the number of available electronic channels, it is necessary to `gang' together up to three (depending on the SVT layer) strips. The `ganging' scheme -- adopted in the \babar SVT -- connects two or three far apart strips to the same readout channel (Figure~\ref{fig:SVT:ganging}), thus preserving the strip pitch at the expense of a higher capacitance and series resistance (both resulting in higher noise), in addition to introducing
ambiguities in the hit position. \\
For tracks at small $\theta$ angles with respect to the beam direction (that is, large incidence angles on the sensor), the signal-to-noise ratio is further degraded by the fact that a track traverses several $z$-strips (up to nine in the inner layers) and the signal becomes approximately proportional to the strip readout pitch (only 1/3 the wafer thickness in Layers 1 to 3). 
This suggests that one should adopt an alternative connection scheme, in which two (or more, at large incidence angles) \textit{adjacent} strips are bonded to a single fanout trace, effectively increasing the strip pitch and the signal into a readout channel, with a less than proportional increase in capacitance, and no increase in series resistance. We call this connection scheme `pairing'.\\
At small $\theta$ angles, this gives better S/N and, consequently, 
higher detection efficiency when compared to individually connected strips. The improvement is even more important in comparison to the `ganging' scheme, where the strip capacitance is proportional to the number of strips ganged together, but the signal remains that of a single strip. Moreover, for paired  strips also the fanout capacitance and resistance can be made lower, because of the larger trace pitch. \\
Due to the lower noise, at small $\theta$ angles pairing is also expected to give better spatial resolution with respect to ganging. 
In order to avoid a significant increase of the input capacitance, pairing will be made between the `readout'  strips (at $100~\mu$m pitch) so that a `floating' strip is always present \textit{between} two adjacent groups of paired strips.  However, we are evaluating the option of connecting also the intermediate (otherwise floating) strips \textit{within} a group of paired strips. 

Strip capacitance measurements performed on test sensors \cite{irina:elba2012} confirm that pairing yields significantly lower capacitance with respect to ganging the same number of strips; the advantage in capacitance of pairing with respect to ganging increases for higher pairing/ganging multiplicity.
The additional increase in total capacitance when connecting also the intermediate strips is  $4-5\%$ on the$p$-side,  
and $\sim 6\%$ on the $n$-side. In addition to this, a better charge collection efficiency is expected.

For Layers 1-3 the combination of track incidence angles and strip pitch is such that all strips can be connected to the available readout channels using a pairing-only scheme. For Layers 4-5, in part because of the `lampshade' SVT design (limiting the track incidence angles for large $|z|$ in Layer4 and Layer5)  and in part because of the larger ratio of strip to channel numbers, it is necessary to adopt a combined pairing/ganging scheme, as indicated in Table~\ref{tab:svt_layout}.

\tdrsubsec{Wafer layout and quantities}

Table~\ref{table:SVT:SensorMultiplicity} reports the sensor composition of the different detector modules, the number of installed sensors of each type, with the corresponding silicon areas, and the total numbers of sensors including spares. Spare sensors account for one spare module of each type (two for Layer0), plus an additional 20\% to compensate for possible losses during the assembly process. 
We see that the current design employs nine different types of sensors, for a total of 308 installed sensors covering 1.52~m$^2$.
Using 150~mm diameter wafers and a dedicated sensor model for each module type allows one to cover the $\sim1.5$ times larger area with a smaller number of sensors with respect to \babar , at the expense of having nine different models of sensors. However, through optimized usage of the wafer area it is possible to accommodate all nine sensor types in seven different wafer layouts, i.e. seven mask sets, and to fabricate all 439 sensors (spares included) on 287 wafers. This is illustrated in Table~\ref{table:SVT:MaskDesign}.

\tdrsubsec{Prototyping and tests}
Although both the design and the technology of double-sided microstrip sensors are well developed, and the \superb SVT is a direct evolution of the proven \babar  design, some specific aspects need to be tested on prototypes before starting the sensor production. 

Prototypes of the striplet sensors for Layer0, absent in \babar , have been thoroughly tested with beam at CERN in 2008 and 2011 \cite{bettarini:slim5_testbeam,fabbri:elba2012}, showing adequate performance for use in the \superb SVT. 

Additional tests will be required to check the performance of sensors irradiated up to the maximum levels expected at \superb which, although much lower than those relevant for LHC detectors, significantly exceed the radiation load experienced in \babar . We plan to irradiate existing prototype sensors with designs similar to that forseen for the \superb SVT with neutrons and/or charged particles, as well as some spare modules remaining from the \babar production.

At least one small batch of dedicated prototype sensors will be ordered and qualified before issuing the tender for the sensor supply. Given the quite different demands posed by processing double sided sensors on thin 150 mm diameter substrates, and the very small number of wafers required, fabrication of striplet sensors for Layer0 could likely be awarded to a different company than the one supplying the other sensors. Because of their special characteristics and their fragility, the striplet sensors will also follow a different, dedicated testing procedure.

A limited number of pre-series sensors will be qualified by a full electrical test using a probe station before releasing the series production. Testing double-sided sensors with automatic loading from a cassette, besides requiring a very specialized custom-made prober, would not allow sufficient flexibility for testing the many different sensor sizes equipping the SVT. The experience gained from testing \babar sensors and, more recently, those for the ALICE experiment \cite{irina:firenze2003} shows that the routine parametric test of a double-sided sensor can be performed in about two hours using a semiautomatic probe station, with manual mounting and positioning of the sensors on dedicated support jigs designed and made in house. The  time increases, of course, if some peculiar results of the test require additional dedicated measurements. 

The number and the detailed characteristics of the measurements that will be included in the routine acceptance test of production sensors will depend on the results of the tests themselves. Starting from a full test on the first batches, it is possible that the set of parameters tested could be reduced if the quality and consistency of the measured characteristics will indicate that one would be advised to do so. 

Overall, based on past experience, we estimate that the testing of all production sensors will take about one year if performed at a single location.

\tdrsec{Fanout Circuits}

The routing of the signals from the silicon sensors to the HDIs
is performed by the so-called \emph{fanout circuits}; they consist in flexible circuits that 
bring the signals from the end of the detector module (where they are wire bonded) to the 
front-end electronics IC (located a few tens of centimeters from the IP).

The fanout for the $\phi$ and $z$ coordinates are designed to minimize the material 
crossed by the particle within the angular acceptance. However, while the $\phi$ fanout 
circuits are just one-to-one connections, the $z$ ones are more complicated since they cover 
the full length of the detector modules (up to $\sim$40~cm for the outer layers) and have to 
provide the interconnection (ganging or pairing) where the number of readout strips exceeds 
the available readout channels.

In the following sections the details about the chosen technology and the geometrical and 
electrical properties of the fanout circuits are described; the layer 0 fanout will be 
analyzed in a dedicated section.

\tdrsubsec{Fanouts for Layer0}
\label{sec:L0fanout}
The fanout circuit for the innermost layer of the SVT vertex detector is part of the striplet module and is shown in 
Figure~\ref{fig:layer0_striplets_module}. 
The complex shape of this passive circuit is derived from the geometry of the vertex detector.
The Layer0 sensor is closely surrounded by the other layers of the vertex detector as shown in Figure~\ref{fig:superb_svt} and Figure~\ref{fig:svt_rphi}.
From the mechanical model of the vertex detector it can be seen that the clearance between 
two adjacent detector layers is not constant 
along the z direction. They generally have a minimum in the center of the detector and increase slightly moving away from the center.
Such geometries have been studied for each layer and the shape of the fanout maximizes the area available for the circuits, avoiding mechanical 
interference between the modules.
The resulting outline of the Layer0 fanout is shown in Figure~\ref{fig:SVT:fanoutL0}. 
The minimum width of the circuit is 29 mm while the overall length is approximately 300 mm. 
Each detector has one fanout. Half of the strips (768) are read from each end of the circuit. 

\begin{figure*}[tbh]
\begin{center}
\includegraphics[width=0.9\textwidth]{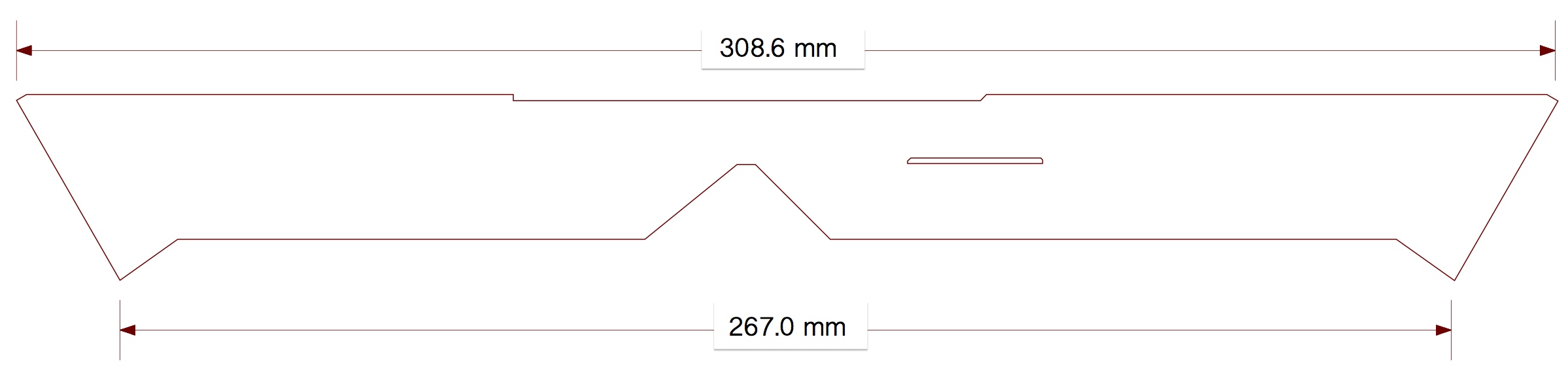}
\end{center}
\caption{Layer0 fanout outline.}
\label{fig:SVT:fanoutL0}
\end{figure*}

\tdrsubsubsec{Requirements and Technology}

Due to the size, the number of lines and the shape of the fanout, the requirement of reducing 
the radiation length of each component of the striplet module to a minimum affects 
the choice of the technology as well as the line pitch to be used in the design.
The technology proposed for the \superb\ fanout is similar to the one used in the past for the \babar\ fanouts. The base material is 
 $50\mum$ polyimide with metal directly deposited on the dielectric, $5\mum$ of copper or $10\mum$ of aluminum. 
The direct deposit of the metal on the polyimide improves the ease with which the circuit can be manufactured.
The maximum trace resolutions for copper or aluminum are different and depend also on the technique used to expose the photo-resist 
that defines the pattern to be etched.
Moreover, the maximum trace resolution depends also on the length of the lines to be etched. 
Typically, with the presently available best technology, the minimum line width/space is about $15/15~\mum$ for copper and 
approximately $65/65~\mum$ for aluminum on short traces (up to few centimeters), 
being on the contrary up to $20/20~\mum$ for copper and $70/70~\mum$ for aluminum on long traces (up to tens of centimeters).
The feasibility of the quoted pitch values has been verified on prototypes.
The use of aluminum instead of copper has the advantage of significantly reducing the radiation length of the fanout and represents 
a novelty to be pursued even though a larger fanout is needed to accommodate 
the same number of connections.

A two-layer fanout design has been preferred to a single-layer solution for the following reasons:

\begin{itemize}

\item the orientation of the sensor strips at $\pm 45$ degrees from the beam axis and the aspect ratio of the circuit, long and narrow, 
requires that the strips are bonded to the fanout along the two long sides of the sensor, in order to keep the width of the circuit 
within 29~mm;

\item the minimum line width/space is approximately $68/68 \mum$ for a two-layer solution, while is of the order of $20/20 \mum$ for 
the single-layer one. 
The former is a standard consolidated solution at least for a copper fanout, while the latter is at the present technology limit over the length
of the Layer0 fanout (300 mm);

\item the Layer0 fanout has to be partially bent to be installed in the detector, as shown in Figure~\ref{fig:layer0_module}. The bending takes place outside the sensor area and close to 
the region where the circuit width increases in a funnel like shape. Having $68\mum$ lines instead of $\sim 20 \mum$ lines increases reliability;


\item the two-layer fanout can be made either in copper or in aluminum. The aluminum solution has not been manufactured before and therefore 
has to be considered as an R\&D effort;

\item the length of the fanout does not present any particular technological challenge.

\end{itemize}

\tdrsubsubsec{Design}

The CAD drawing of the module and the layout of one of the two layers of the fanout are shown in Figure~\ref{fig:SVT:CADL0} and 
Figure~\ref{fig:SVT:LayoutFanoutL0}, respectively. The stack up of the fanout is reported in Table~\ref{tab:SVT:L0FanoutMaterial}.

\begin{table}[!tbp]
\caption{Material breakdown of Layer0 fanout, surface coverage and equivalent radiation length $x/X_0$. 
The coverage of the surface area is very small for gold since it is present only on the pads.}
\begin{center}
\scalebox{0.8}{
\renewcommand{\arraystretch}{1.2}
\begin{tabular}{|l|c|c|c|} 
\hline
 Material & Thickness ($\mum$) & Coverage & $x/X_0$ (\%)\\
\hline
Kapton & 50  &  1 & 0.017 \\
Cu (Al) & 5 (10)  & 0.5 & 0.018 (0.006)   \\
Glue  & 10 & 1 & 0.003   \\
Kapton & 50  & 1 & 0.017 \\
Cu (Al) & 5 (10) & 0.5 & 0.018 (0.006) \\
Au  & 1.5 & $\le 10^{-2}$ & $\le 0.001$ \\
\hline
Total   &  120 (130)   &   &  0.073 (0.049) \\
\hline
\end{tabular}
}
\end{center}
\label{tab:SVT:L0FanoutMaterial}
\end{table}
\begin{figure*}[!tbh]
\begin{center}
\includegraphics[width=0.9\textwidth]{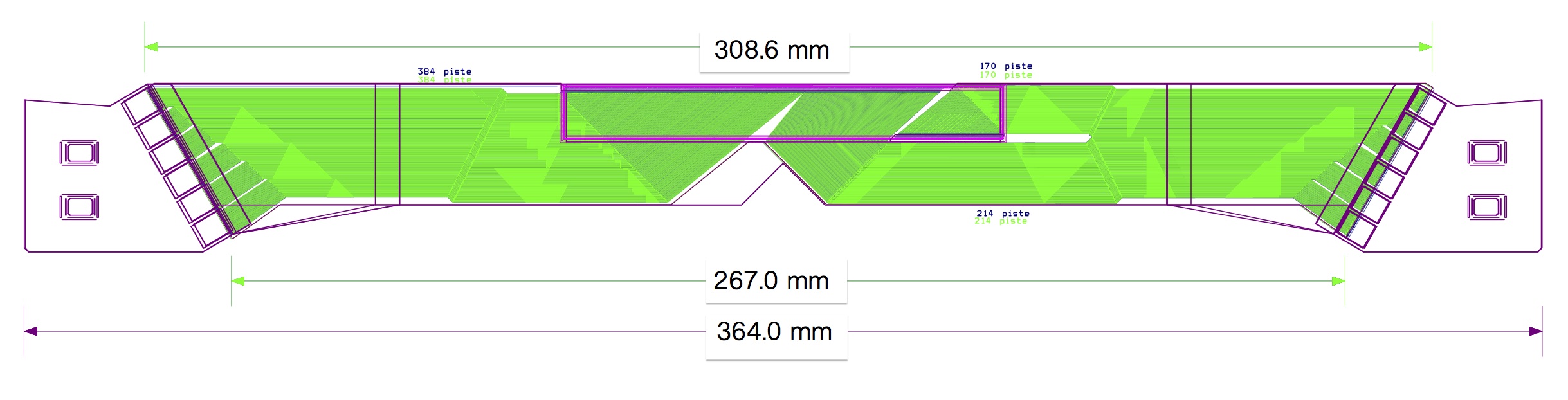}
\end{center}
\caption{CAD drawing of the Layer0 module.}
\label{fig:SVT:CADL0}
\end{figure*}
\begin{figure*}[!tbh]
\begin{center}
\includegraphics[width=0.9\textwidth]{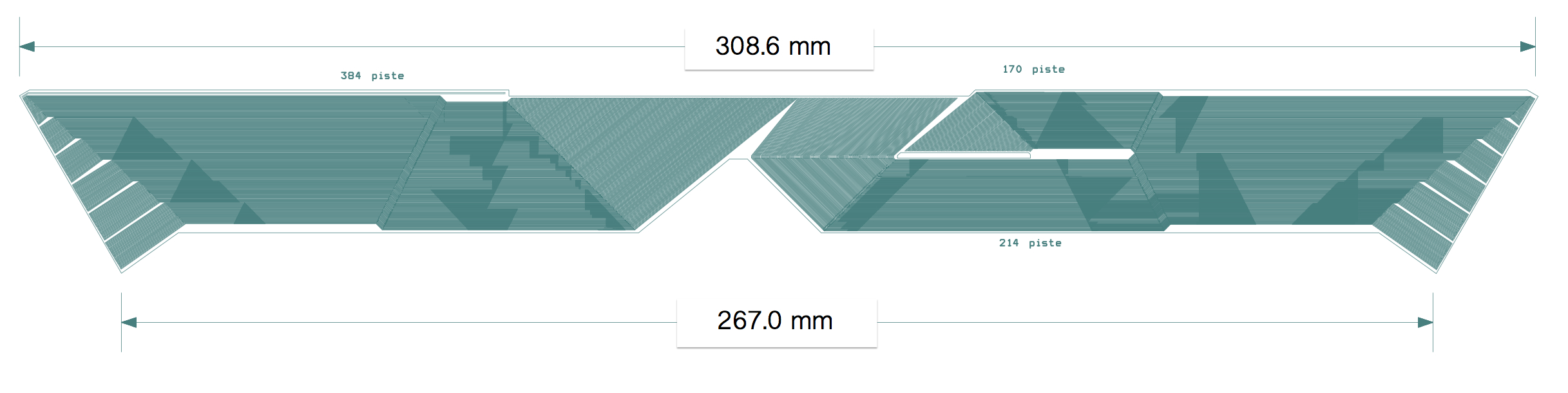}
\end{center}
\caption{Layout of the Layer0 fanout.}
\label{fig:SVT:LayoutFanoutL0}
\end{figure*}

From the figures it can be seen that a cut-out, 1 mm wide, is needed in the central area of each layer to bond some of the strips 
and at the same time to properly route the traces without violating the $68\mum$ minimum pitch rule.
The traces are also tapered and bent at the extremities of the circuit to match the pitch of the front end electronics pads, to which the 
fanout lines are wire bonded.
Only in the pad region the traces are gold plated (1.5 $\mum$) for bonding. This solution has been adopted to minimize the radiation length of the assembly.
The relaxed $68 \mum$ pitch should result in a simulated crosstalk of about 1\% on adjacent traces of the same layer.

Each layer can be individually manufactured, tested and qualified before assembly. After an initial optical inspection to spot defects, 
the electrical characteristics of each trace of the circuit  will be measured, using a "bed of nails" to contact a group of adjacent strips. 
Only defect-free layers will be glued together to build the two-layer fanout assembly.
Repairs of defective layers are not foreseen at the moment. The small number of fanouts to be produced, 16 pieces, justifies the request of 
having 100\% or more redundancy during production. Moreover the copper solution does not represent a real technological challenge and 
so it is considered the baseline solution until the aluminum solution is proved feasible and reliable.

\tdrsubsubsec{Prototyping and tests}

All fanout assemblies will be tested against shorts and opens. Each trace will be measured in terms of resistance and capacitance 
to the neighbor traces and the data saved for further analysis.  No repairs are foreseen on the fanout.
A first batch of prototypes is in production at the CERN PCB facility. They have the previously shown layout and will be used not only 
to validate the design and the test procedure but also for assembling `mechanical' versions of the striplet modules.

\tdrsubsec{Fanouts for outer layers}
\tdrsubsubsec{Requirements}
The geometrical requirements will be fixed by the detector designs. 
From the technological point of view, the design requires a typical line width/space 
of 45/45~$\mu$m and a small region of 15/15~$\mu$m. This 
region (corresponding to the bonding area) is 1.5~mm long and it extends to the total 
width of the fanout itself. No constraints are present on the fanout length given the same 
machines used for the micro-pattern gas detector production will be used. 

\tdrsubsubsec{Material and production technique}
The \babar fanouts were produced on 50~$\mu$m Upilex substrates (manufactured by UBE Industries, 
Ltd.\footnote{http://www.ube-ind.co.jp/english/about/index.htm}) with a deposit of 
150~nm of Cr, 4.5~$\mu$m of copper followed by a layer of 150~nm of Cr and 1.5~$\mu$m of 
amorphous gold. The \superb SVT fanouts will be produced on a similar material by 
UBE (50~$\mu$m of polyimide with 5~$\mu$m of copper directly deposited on the 
base material) which should ensure fewer defects and thus a better yield. This 
material will be tested in the prototype phase. The old Upilex is anyway still 
available if the new material would prove to be inadequate. 

A new fabrication technique will be implemented in order to reduce the 
production time. In the \babar production line, the photo-resist was exposed 
through a mask after being deposited on the Upilex, which requires working in a clean 
room. For \superb, the idea is to expose the photo-resist directly with a scanned laser beam, 
without a photo-mask; the photo-resist is solid and the whole procedure, performed 
in a clean micro-environment inside a machine, becomes much faster. This technique has already 
been tested on samples with the same pitches foreseen for the \superb SVT fanouts.

The increase in the production speed allows one to repeat the fabrication of pieces with 
defects without delaying the SVT assembly. All the fanouts will be gold plated 
with 1.5~$\mu$m of amorphous gold for bonding.

\tdrsubsubsec{Design}
The design will follow the same rules of the \babar fanouts adapting it 
to the different length of the modules. Unlike the \babar pieces, 
no test-tree is foreseen (see next section). To allow the gold plating, all the 
lines will be shorted together by a small extension of the circuit. A suitable cutting device will be developed to cut 
away the shorting extension after visual inspection.

Table~\ref{table:SVT:Fanout} summarizes the geometrical parameters\footnote{The fanouts dimensions 
have been taken from the SVT Mechanics talk presented at the $4^{th}$ \superb Collaboration 
Meeting (June 2012, La Biodola - Elba IT).} as well as the number of readout strips and channels, 
the typical pitch and the total number of required circuits per layer and type.

\begin{table*}[!htb]
\renewcommand{\arraystretch}{1.1}
\caption{Summary of fanout circuit characteristics.}
\begin{center}
    \begin{tabular}{|c|c|c|c|c|c|c|c|c|} 
\hline
Layer          &  Fanout  &\multicolumn{2}{|c|}{Length (mm)}  & \multicolumn{2}{|c|}{Number of Readout}  & \multicolumn{2}{|c|}{Typical Pitch at ($\mu$m)}  & Number\\ 
\cline{3-8}
               & Type     & Left        & Right               & Strips  & Channels                       & Input         & Output                           & of Circuits \\
\hline
\hline
 1             &   $z$    & 216.76      & 217.52              & 1104    & 896                            & 100           & 45                               &  12 \\
               &  $\phi$  & 105.60      & 105.82              & 799     & 896                            & 50            & 45                               &  12 \\
               &          &             &                     &         &                                &               &                                  &     \\
 2             &   $z$    & 209.00      & 208.80              & 1302    & 896                            & 100           & 45                               &  12 \\
               &  $\phi$  & 76.20       & 76.00               & 874     & 896                            & 55            & 45                               &  12 \\
               &          &             &                     &         &                                &               &                                  &     \\
 3             &   $z$    & 250.26      & 246.26              & 1730    & 1280                           & 110           & 45                               &  12 \\
               &  $\phi$  & 54.76       & 53.46               & 701     & 1280                           & 100           & 45                               &  12 \\
               &          &             &                     &         &                                &               &                                  &     \\
 4a            &   $z$    & 332.91      & 328.60              & 1398    & 640                            & 210           & 45                               &  16 \\
               &  $\phi$  & 35.71       & 31.40               & 512     & 640                            & 82            & 45                               &  16 \\
               &          &             &                     &         &                                &               &                                  &     \\
 4b            &   $z$    & 332.15      & 327.97              & 1448    & 640                            & 210           & 45                               &  16 \\
               &  $\phi$  & 24.55       & 20.37               & 512     & 640                            & 82            & 45                               &  16 \\
               &          &             &                     &         &                                &               &                                  &     \\
 5a            &   $z$    & 407.13      & 411.40              & 1761    & 640                            & 210           & 45                               &  18 \\
               &  $\phi$  & 32.53       & 36.80               & 512     & 640                            & 82            & 45                               &  18 \\
               &          &             &                     &         &                                &               &                                  &     \\
 5b            &   $z$    & 406.63      & 411.00              & 1815    & 640                            & 210           & 45                               &  18 \\
               &  $\phi$  & 20.63       & 25.00               & 512     & 640                            & 82            & 45                               &  18 \\
               &          &             &                     &         &                                &               &                                  &     \\

\hline
\end{tabular}
\end{center}
\label{table:SVT:Fanout}
\end{table*}

Figure~\ref{fig:SVT:ganging} presents a sketch of the ganging principle proposed for the design of layers $z$ 3-4-5.

\begin{figure}[tbh]
\begin{center}
\includegraphics[width=0.5\textwidth]{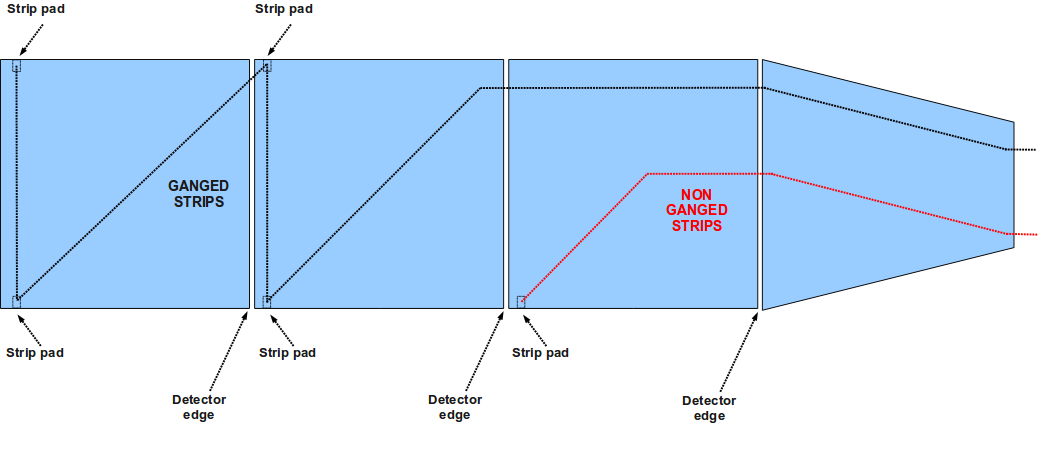}
\end{center}
\caption{Schematic view of two $z$ strips ganged through the fanout circuit.}
\label{fig:SVT:ganging}
\end{figure}

\tdrsubsubsec{Tests and prototyping}
All the fanouts will be automatically optically checked by a dedicated 
machine which will use the CAD layout files of the fanouts to find 
shorted or open lines. The machine can work with 25~$\mu$m lines; the region 
with smaller lines (15~$\mu$m with a 15~$\mu$m space) will have to be controlled 
manually.

Given the short time needed for the production, no correction 
is foreseen for shorted or open lines; the damaged pieces will 
be produced again. However, if a short is present in the larger 
pitch region, the same correction procedure used for \babar (involving the use 
of a micro-probe to remove the short) can be implemented.

As far as the tests are concerned, two batches of prototypes have been manufactured up to now.
The first batch (designed starting from the \babar layout) has been used to check the performance 
of the fabrication machine and the whole production chain. The second batch, conforming to the \superb 
Layer3 design (shown in Figure~\ref{fig:SVT:FanoutPrototype}), has been used to estimate the trace capacitance 
and resistance: values of C/l $\sim$0.6~pF/cm and of R/l $\sim$1.2~$\mathrm{\Omega}$/cm have been found. 
These pieces in principle can be used with working detectors to test also the assembly procedure.

\begin{figure}[!tbh]
\begin{center}
\includegraphics[width=0.5\textwidth]{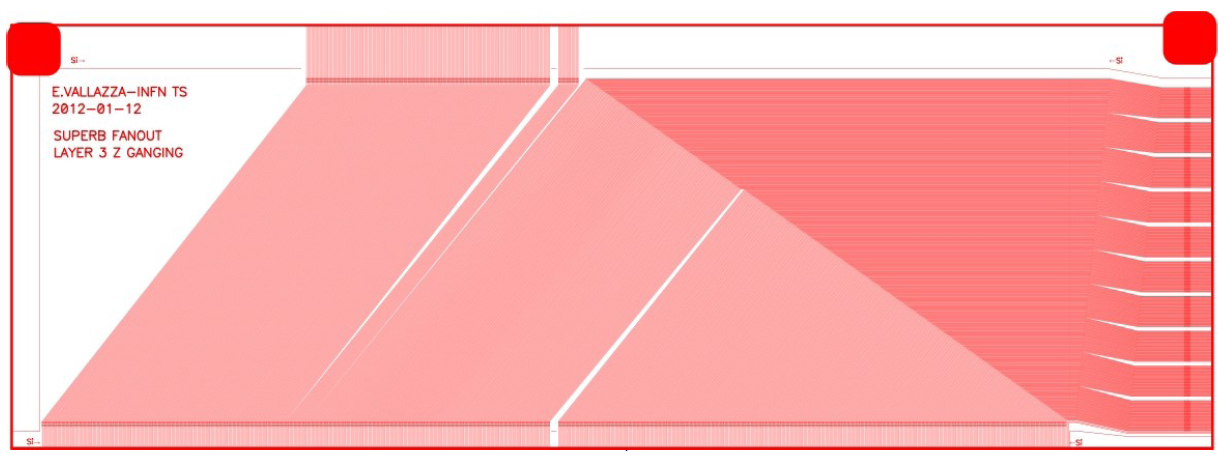}
\end{center}
\caption{Design of a $z$ fanout prototype of layer 3.}
\label{fig:SVT:FanoutPrototype}
\end{figure}

Figures~\ref{fig:SVT:FanoutPrototype2} and~\ref{fig:SVT:FanoutPrototype3} show a picture of the prototype of the layer 3 $z$ and a step of the cutting procedure.

\begin{figure}[!tbh]
\begin{center}
\includegraphics[width=0.5\textwidth]{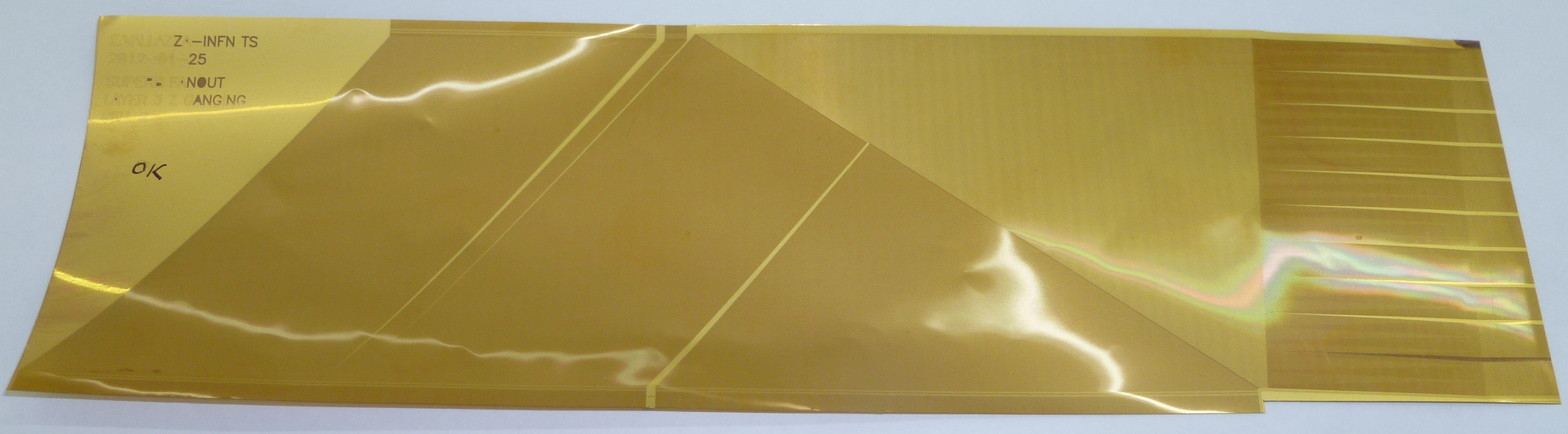}
\end{center}
\caption{Picture of the prototype of the layer 3 $z$.}
\label{fig:SVT:FanoutPrototype2}
\end{figure}

\begin{figure}[!tbh]
\begin{center}
\includegraphics[width=0.5\textwidth]{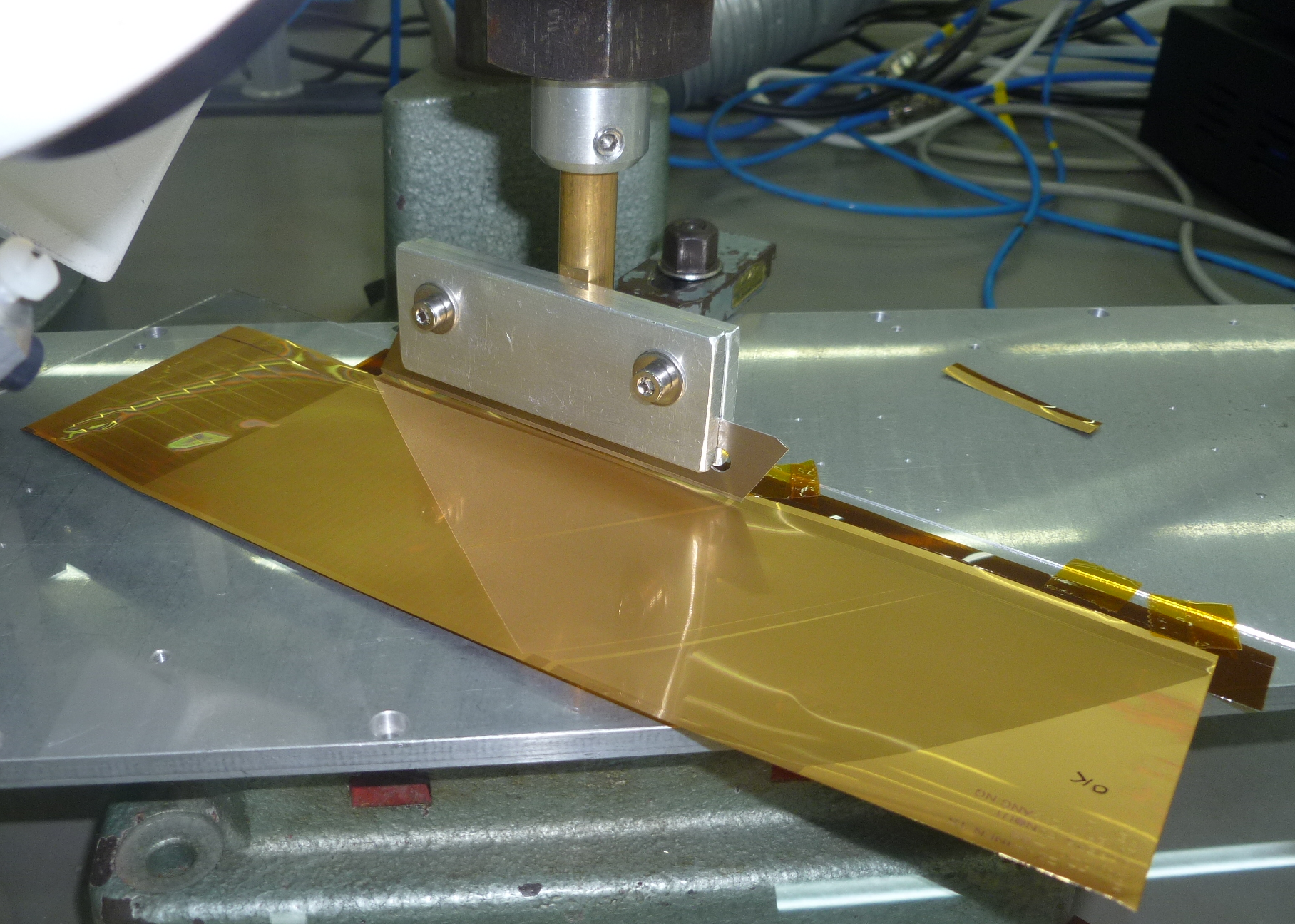}
\end{center}
\caption{Picture of the prototype of the layer 3 $z$ during the cutting procedure.}
\label{fig:SVT:FanoutPrototype3}
\end{figure}

\tdrsec{Readout Electronics}
\label{svt:electronicsreadout}
The SVT electronics chain consists of various separate hardware components.
Starting from the detector and going to the ROM boards, the chain contains: a) the readout chips 
mounted on an HDI placed at the end of the sensor modules; 
b) wire connections to a transition card (signals in both directions and power lines); 
c) a transition card, placed about 50-70 cm from the end of the sensors, 
hosting the wire-to-optical conversion; 
d) a bidirectional optical line running above 1 Gbit/s; 
e) a receiver programmable board (front-end board: FEB). 
A sketch of the full data chain is given in Figure~\ref{fig:svt_data_chain}

\begin{figure*}[tbp]
\begin{center}
\includegraphics[width=0.8\textwidth]{etd-svt.pdf}
\caption{Schematic drawing of the full SVT data chain.}
\label{fig:svt_data_chain}
\end{center}
\end{figure*}



\tdrsubsec{Readout chips for Strip and Striplet Detectors}

The front-end processing of the signals from the silicon strip detectors will be performed by custom-designed ICs mounted on hybrid circuits that distribute power and signals, and thermally interface the ICs to the cooling system. As discussed below, the very different features of the inner (Layer 0-3) and outer layers (4 and 5) of the SVT set different requirements on the readout chip, which makes it necessary to include programmable features in the readout ICs, in order to adjust operating parameters over a wide range. This obviously also holds in the case that a different technology (pixels) is adopted for Layer 0 instead of short strips (striplets). Generally speaking, the ICs will consist of 128~channels, each connected to a detector strip. The signals from the strips, after amplification and shaping will be compared to a preset threshold. If a signal exceeding the threshold is detected, a 4~bit analog information about the signal amplitude will be provided with the Time Over Threshold technique. The analog information is useful for position interpolation, time walk correction, $dE/dx$ measurements, as well as for calibration and monitoring purpose. The dimensions of the readout IC are expected to be about 6$\times$4~mm$^2$. As discussed in the SVT HDI subsection of this TDR, the dimensions of the HDI set a 6~mm upper limit on the side of the chip with the bonding pads for the interconnection with the strip sensors.  The power dissipation will be below 4~mW/channel including both analog and digital sections. For each channel with a signal above threshold, the strip number, the amplitude information, the chip identification number and the related time stamp will be stored inside the chip waiting for a trigger signal for a time corresponding to the trigger latency (about 6~$\mu$s, with 150~kHz trigger rate). When a trigger is received, data will be read out and transmitted off chip, otherwise they will be discarded. The data output from the microstrip detector will be sparsified, i.e. will consist only of those channels generating a hit. The readout integrated circuits must remain functional up to 5 times nominal background. 

The option of operating in a data push fashion could be preserved for the external layers, where this is feasible as a result of the low strip hit rate. This will provide the possibility to feed data from these layers to the trigger system.  

\subsection{Readout chips requirements}
\label{sec:chips_requirements}

The microstrip electronics must ensure that the detector system operates with adequate efficiency, but also must be robust and easy to test, and must facilitate testing and monitoring of the microstrip sensors. AC coupling is assumed between the strips and the readout electronics.  The following summarises the required characteristics:

\begin{itemize}
\item \textbf{Mechanical Requirements:} \\
Number of channels per chip: 128 \\
Chip size:  width $\leq$6~mm, length $\leq$4~mm \\
Pitch of input bonding pads: $<$45~$\mu$m\\

\item \textbf{Operational Requirements:}
Operating temperature: $<$40~$^{\circ}$C\\
Radiation tolerance: $>$3~Mrad/year, $>$5$\cdot$10$^{12}$ n$_{eq}$/cm$^2$/year (these are the expected values in Layer 0; in outer layers, radiation levels are at least one order of magnitude lower)\\
Power dissipation: $<$4~mW/channel\\
Detector and fanout capacitance: 10 pF~$\leq$$C_D$$\leq$~70 pF (the chip must be stable when sensor strips are disconnected from the input pads of the analog channels)\\

\item \textbf{Dynamic range:} The front-end chips must accept signals from either P or N-side of the strip detectors. A linear response of the analog processing section is required from a minimum input charge corresponding to 0.2~MIP up to a full dynamic range of 10-15~MIP charge for $dE/dx$ measurements.\\

\item \textbf{Analog Resolution:} The front-end chips have to provide analog information about the charge collected by the detector, which will be also used for calibrating and monitoring the system. A resolution of 0.2~MIP charge is required for $dE/dx$ measurements. In the case of a compression-type ADC, based on the time-over-threshold technique (ToT), this may translate into 4 bits of information. \\

\item \textbf{Efficiency:} At design luminosity, the microstrip readout must have a hit efficiency of at least 90\% during its entire operational lifetime.  This includes any loss of data by readout electronics or readout dead time.\\

\item \textbf{Readout bandwidth:} Data coming out of the chip will be substantially reduced by operating in a triggered mode. The chips can use up to 2 output LVDS lines with 180~MHz clock, as it is needed to handle the higher data throughput in inner SVT layers. \\

\item \textbf{Radiation Tolerance:} All the components of the microstrip readout system must remain operational over the entire lifetime of the experiment, including a safety factor of 5 on the nominal background expected, this corresponds to 7.5$\times$5 years at nominal peak luminosity of 10$^{36}$.

\item \textbf{Operational lifetime:}
Up to 10 years of \superb running at the nominal luminosity.\\

\item \textbf{Peaking Time:} The constraints for the peaking time of the signal at the shaper output are dictated by different needs in inner and outer layers. In Layer~0, the high occupancy due to background and the need to avoid pulse overlap and consequent hit inefficiencies set the required peaking time in the range of $t_p$=25-50~ns, which also allows for a high time resolution (see below). In the external layers 4 and 5, where background hit frequency is much smaller and where strips are longer and have a larger capacitance, the peaking time will be mostly determined by the need to reduce series noise contributions and will be in the range of 0.5-1.0~$\mu$s.\\

\item \textbf{Signal-to-Noise Ratio:} Concerning the signal, this requirement has to take into account the different thickness of silicon detectors in inner (200~$\mu$m) and outer (300~$\mu$m) layers, as well the signal spread among various strips that depends on the track angle inside detectors and that, again, may vary in different SVT layers. Noise-related parameters (strip capacitance and distributed resistance) also vary significantly across the SVT. A signal-to noise ratio of 20 has to be ensured across the whole SVT and should not decrease significantly after irradiation. Here are the two extreme cases (where the equivalent noise charge ENC includes the thermal noise contribution from the distributed resistance of the strips):\\

	\begin{itemize}
	\item Layer 0 striplets: ENC $\approx$900 e-  at $C_D$=14.9~pF and at $t_p$=25~ns\\
	\item Layer 5 strips: ENC $\approx$1100 e-  at $C_D$=66.2 pF and at $t_p$=750~ns \\
	\end{itemize}

\item \textbf{Threshold and Dispersion:}  Each microstrip channel will be read out by comparing its signal to a settable threshold around 0.2~MIP. Threshold dispersion must be low enough that the noise hit rate and the efficiency are degraded to a negligible extent. Typically, this should be 300~rms electrons at most and should be stable during its entire operational lifetime.\\

\item \textbf{Comparator Time Resolution:} The comparator must be fast enough to guarantee that the output can be latched in the right time stamp period.\\

\item \textbf{Time Stamp:} A 30 ns time stamp clock is required for inner layers to get a good hit time resolution in order to reduce the occupancy in the target offline time window (100-150~ns). In the outer layers the time stamp resolution is less critical since the hit time resolution will be dominated by the long pulse shaping time. A single 30~ns time stamp clock in all layers will be used.\\

\item \textbf{Chip clock frequency:} Two main clocks will be used inside the readout chip, the time stamp clock (about 30~MHz) and the readout clock (120 MHz or 180 MHz). These clocks will be synchronized with the 60~MHz \superb system clock. In the case that the analog-to-digital conversion is based on the Time-Over-Threshold method, a ToT clock has to be generated inside the chip. The ToT clock period should at least match the pulse shaping time to get a good analog resolution. A faster ToT clock could slightly improve the analog resolution but an upper limit ($\approx$3.5) on the ratio between ToT clock frequency and the shaping time frequency is imposed by the required dynamic range needed for low momentum particle $dE/dx$ measurements ($\approx$10-15 MIP) and the number of bits available for ToT. With the experience of the \babar Atom chip a ToT clock frequency 3 times higher than the pulse shaping frequency could be used.\\

\item \textbf{Mask, Kill and Inject:}  Each micro-strip channel must be testable by charge injection to the front-end amplifier. By digital control, it shall be possible to turn off any micro-strip element from the readout chain.\\

\item \textbf{Maximum data rate:} Simulations show that machine-related backgrounds dominate the overall rates. At nominal background levels (including a safety factor of~5), the maximum hit rate per strip goes from about 1~MHz/strip in Layer~0 to about 50~kHz/strip in Layer 5, z-side.\\

\item \textbf{Deadtime limits:} The maximum total deadtime of the system must not exceed 10~\% at a 150~kHz trigger rate and background 5 times the nominal expected rate.\\

\item \textbf{Trigger specifications:} The trigger has a nominal latency of about 6~$\mu$s, a maximum jitter of 0.1~$\mu$s, and the minimum time between triggers is 70~ns. The maximum Level~1 Trigger rate is 150~kHz.\\

\item \textbf{Cross-talk:} Must be less than 2~\%. \\

\item \textbf{Control of Analog Circuitry on Power-Up:}  Upon power-up, the readout chip shall be operational at default settings.\\

\item \textbf{Memory of Downloaded Control of Analog Circuitry:}  Changes to default settings shall be downloadable via the readout chip control circuitry, and stored by the readout chip until a new power-up cycle or additional change to default settings.\\

\item \textbf{Read-back of Downloadable Information:}  All the data that can be downloaded also shall be readable.  This includes data that has been modified from the default values and the default values as applied on each chip when not modified. \\

\item \textbf{Data Sparsification:}  The data output from the microstrip detector shall be only of those channels that are above the settable threshold. \\

\item \textbf{Microstrip output data content:} The microstrip hit data must include the time stamp and the microstrip hits (strip number and relevant signal amplitude) for that time stamp. The output data word for each strip hit should contain 16~bits (7 strip addresses, 4~ToT, 1~type (Hit or Time Stamp) 4~bits to be defined). A 10-bit time stamp (with 6 additional bits: 1~type, 5~bits to be defined) will be attached to each group of hits associated to a given time stamp (hit readout will be time-ordered).\\

\end{itemize}

\subsection{Readout Chip Implementation}

\begin{figure*}[tbp]
\begin{center}
\includegraphics[width=0.9\textwidth]{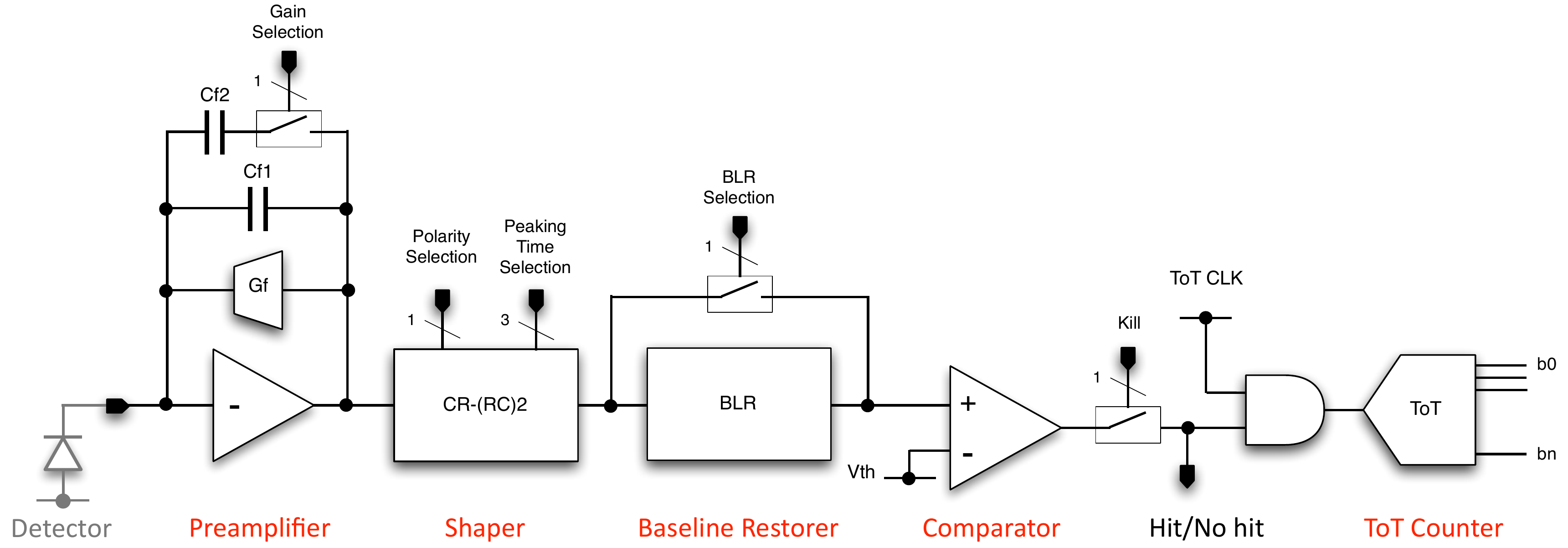}
\caption{Analog channel block diagram. }
\label{fig:channel}
\end{center}
\end{figure*}

The \superb SVT readout chips are mixed-signal integrated circuits in a 130~nm CMOS technology and are being designed to comply with the requirements discussed above. Each chip comprises 128 analog channels, each consisting of a charge-sensitive preamplifier, a unipolar semi-Gaussian shaper and a hit discriminator. A polarity selection stage will allow the chip to operate with signals delivered both from n- and p-sides of the SVT double-sided strip detectors. A symmetric baseline restorer may be included to achieve baseline shift suppression. When a hit is detected, a 4~bit analog-to-digital conversion will be performed by means of a Time-Over-Threshold (ToT) detection. The hit information will be buffered until a trigger is received; together with the hit time stamp, it will then be transferred to an output interface, where data will be serialized and transmitted off chip on  LVDS output lines. An n-bit data output word will be generated for each hit on a strip. A programming interface accepts commands and data from a serial input bus and programmable registers are used to hold input values for DACs that provide currents and voltages required by the analog section. These registers have other functions, such as controlling data output speed and selecting the pattern for charge injection tests.

Given the very different requirements of inner and outer layers, in terms both of detector parameters and hit frequency, several programmable features will be included in the chips such as the peaking time, the gain and the size of the input device.
The block diagram of the analog channel is shown in Figure~\ref{fig:channel}.

\begin{figure*}[tbp]
\begin{center}
\includegraphics[width=0.7\textwidth]{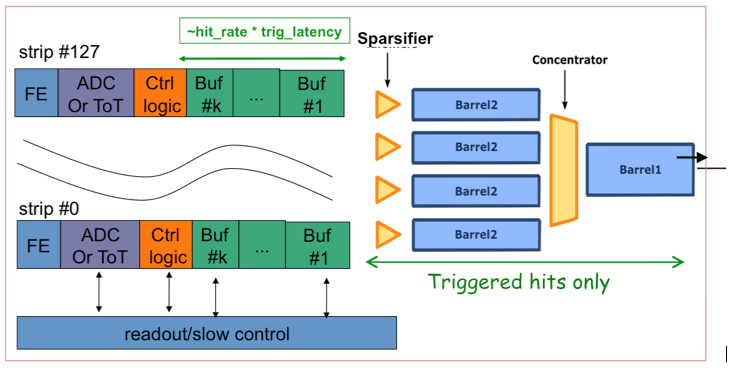}
\caption{Readout architecture of the SVT strip readout chips.}
\label{fig:readout_architecture_strips}
\end{center}
\end{figure*}

The digital readout of the matrix will exploit the architecture that was originally devised for a high-rate, high-efficiency readout of a large CMOS pixel sensor matrix. A schematic concept of the FE chip readout
is shown in Figure~\ref{fig:readout_architecture_strips}. 
Each strip has a dedicated array of pre-trigger buffers, which can be filled by hits with different time stamps. The size of this buffer array is determined by the maximum strip hit rate (inner layers) and by the trigger latency. After arrival of a trigger, only hits with the same time stamp as the one provided by the triggering system send their information to the back-end. The array of 128 strips is divided in four sections, each with a dedicated sparsifier encoding the hits in a single clock cycle. The storage element next to each sparsifier (barrel level-2) acts like a FIFO memory conveying data to a barrel-L1 by a concentrator which merges the flux of data and preserves the time order of the hits. This barrel-L1 will drive the output data bus which will use up to 2 output lines depending on the data throughput and will be synchronous to a 180~MHz clock.

\subsection{R\&D for strip readout chips}
\label{sec:svt_rd_strip_readout_chips}

\begin{figure*}[tbp]
\begin{center}
\includegraphics[width=\textwidth]{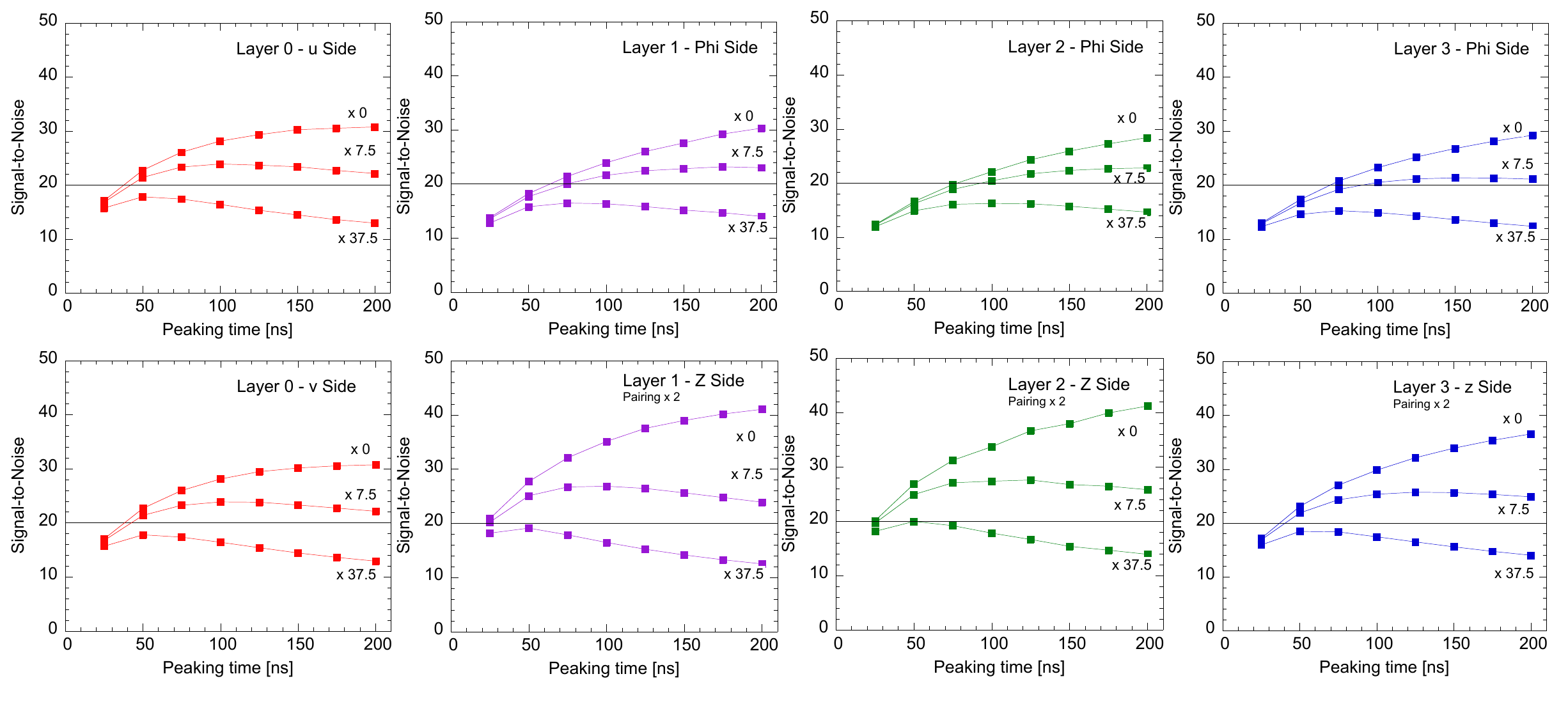}
\caption{Simulated S/N for Layer0-3, as a function of the shaping time. Results are shown 
for un-irradiated detectors, with negligible noise  contribution from leakage current (x0), and  
for irradiated detectors (higher leakage current noise contribution) 
with nominal background (x7.5, after 7.5 years) and with a safety factor x5 on nominal background applied (x37.5).}
\label{fig:layer0_3_noise}
\end{center}
\end{figure*}

The R\&D to support the development of the \superb strip readout chips began in 2011. The chosen technology for integration is a 130 nm CMOS process: this has an intrinsically high degree of radiation resistance, which can be enhanced with some proper layout prescriptions such as enclosed NMOS transistors and guard rings. There is a large degree of experience with mixed-signal design in this CMOS node that was gained in the last few years inside the HEP community.

The readout architecture is being tested with inputs generated by the detailed Monte Carlo simulation of 
the expected backgrounds in the SVT.
Verilog simulations demonstrate that the chip will be able to operate with a 99~\% digital readout efficiency in the worst case scenario, which includes a safety factor of 5 in the background levels.  

\begin{figure}[tbp]
\begin{center}
\includegraphics[width=0.5\textwidth]{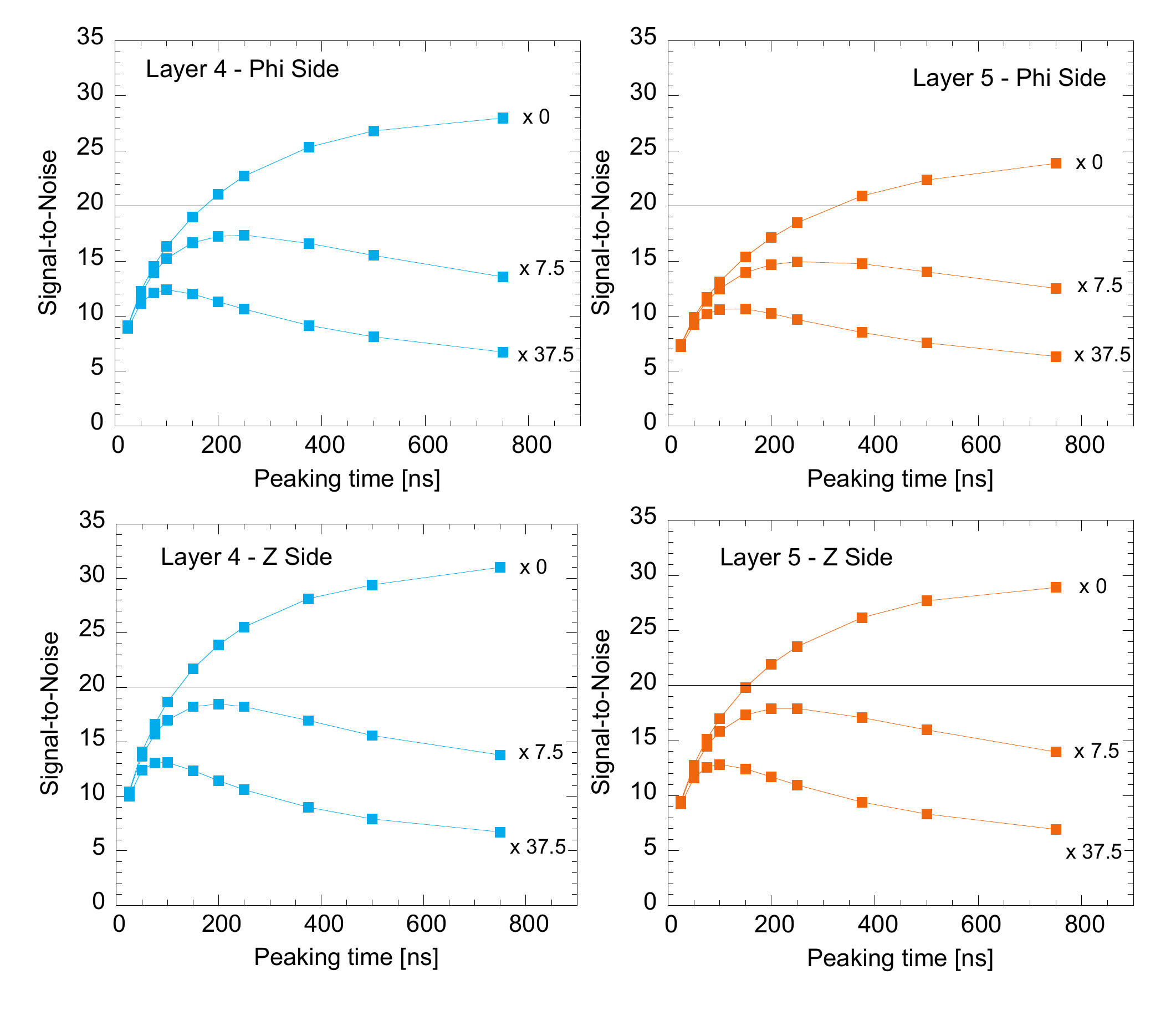}
\caption{Simulated S/N for Layer4-5, as a function of the shaping time, for un-irradiated detectors,
with negligible noise  contribution from leakage current (x0), and  
for irradiated detectors (higher leakage current noise contribution) 
with nominal background (x7.5, after 7.5 years) and with a safety factor x5 on nominal background applied (x37.5).}
\label{fig:layer4_5_noise}
\end{center}
\end{figure}
The analog section of the chip is being optimized from the standpoint of noise, comparator threshold 
dispersion and sensitivity to variations of process parameters. 
It will be possible to select the peaking time of the signal at the shaper output in a wide range 
(25-200~ns typical values for inner layers, 350-750~ns typical values for outer layers) 
by changing the value of capacitors in the shaper. 
In this way the noise performance of the chip can be optimized according to the background occupancy, in 
order to preserve the required efficiency. The shaping times can be varied with time to take into account 
the radiation damage and the related increase in the sensor leakage current.
Figures~\ref{fig:layer0_3_noise} and~\ref{fig:layer4_5_noise} show the simulated S/N for all the layers as a 
function of the shaping time with different background conditions and related sensor 
radiation damage.
Table~\ref{tab:main_parameters} shows the main parameters of the analog section, for some reference shaping times,
according to simulation 
estimates for realistic values of detector parameters and strip hit rates.  
The loss in efficiency is determined by the limits in the double pulse resolution of the analog section, 
which depends on the signal peaking time. 
An acceptable compromise will be found here with the noise performance. 
Although the safety factor of 5 used in noise estimation  after 7.5 years represents a really worst case, 
different strategies will be pursued to mitigate the noise increase after irradiation. In particular, 
especially for Layer4 and 5, S/N may benefit from moving to shorter peaking times after irradiation, 
as shown in Figure~\ref{fig:layer4_5_noise}, and from reducing the temperature of a few degrees with respect to the ambient 
temperature. As a further possibility, the replacement of the irradiated detectors with un-irradiated ones can also be considered.

\begin{sidewaystable*}[tbp]
\caption{Main parameters of the analog section of the SVT strip readout chips. The total ENC and the efficiency of the analog section has been 
evaluated with different background conditions: nominal and with the safety factor $\times$5 applied ($\times$5 bckg.).}
\begin{center}
{\small 
{\renewcommand{\arraystretch}{1.4}
\begin{tabular}{lccccccccc}
\hline
 Layer	& $C_D$	 	& $t_p$  		& $t_p$ 		& Total ENC 	& Total ENC 	& Total ENC 	& Hit rate/strip	& Efficiency	& Efficiency		\\
 		& [pF] 		& [ns]		& [ns]		& [e rms]		& [e rms]		& [e rms]		& [kHz]		& 1-N		& 1-N			\\
 \hline
  		& with fanout 	& Available	& Selected 	&  			& after	7.5 years	& after 7.5 years		& nominal		& nominal		& $\times$5 bckg. 	\\
    		& \& ganging		& 			& 			& 			&                       & \&  $\times$5 bckg.	&  			& 			& 	\\

 \hline
   0-side u	& 14.9		& 			& 25			& 936		& 952		& 1016		& 187		& 0.99		& 095	\\
   0-side v	& 14.9		& 			& 25			& 939		& 956		& 1019		& 187		& 0.99		& 0.97	\\   
 \cline{1-2}\cline{4-10}
   1 phi	& 33.4		& 			& 75			& 1122		& 1197		& 1457		& 170		& 0.98		& 0.92	\\
   1 z		& 16.2		& 25-200		& 75			& 748		& 899		& 1342		& 134		& 0.98		& 0.91	\\   
\cline{1-2}\cline{4-10}
   2 phi	& 37.2		& 	 	 	& 100		& 1085		& 1174		& 1476		& 134		& 0.98		& 0.90	\\
   2 z		& 18.0		& 			& 100		&  711		&  876		&  1346		& 134		& 0.98		& 0.88 	\\
\cline{1-2}\cline{4-10}
   3 phi	& 35.7		& 			& 150		&  897		&  1125		&  1763		& 116		& 0.96		& 0.82	\\
   3 z		& 24.6		& 			& 150		&  707		&  935		&  1540		& 79			& 0.98		& 0.90	\\
\hline
   4 phi	& 53.1		& 			& 500		& 1086		& 1709		& 3090		& 25			& 0.98		& 0.92	\\
   4 z		& 47.2		& 375, 500	& 500		& 819		& 1555		& 3041		& 13.4		& 0.99		& 0.95	\\
\cline{1-2}\cline{4-10}
   5 phi	& 66.2		& 750		& 750		& 1106		& 2099		& 3859		& 16.2		& 0.98		& 0.93	 \\
   5 z		& 52.2		& 			& 750		& 808		& 1727		& 3375		& 8.8		& 0.99		& 0.95 	\\
\hline
\end{tabular}
}
}
\end{center}
\label{default}
\label{tab:main_parameters}
\end{sidewaystable*}%

In 2012, the submission of a chip prototype including 64 analog channels and a reduced-scale version of the readout architecture is foreseen.

The submission of the full-scale, 128-channel chip prototypes is scheduled for late 2013. This version will have the full functionality of the final production chip.





\tdrsubsec{Hybrid Design}
\label{sec:SVT:HDI}

The \superb\ SVT hybrid is conceptually similar to the High Density Interconnect (HDI) hybrid of the \babar\ SVT.
It is a multipurpose structure that has to satisfy essentially three different types of requirements: mechanical, thermal and electrical.
Each of the six detector layers of the SVT is readout by the readout chips described in the previous section. 
The readout chips will be mounted on a custom designed ceramic circuit called HDI hybrid. 
The HDI is a thick-film ceramic circuit fabricated from aluminum nitride (AlN). 
Since the strip sensors are double sided,  the readout chips are mounted on both sides of the 
HDI together with external passive components. 
The HDIs are mounted through ``two bushings'' on metal cooling rings which are fixed on the carbon fiber support cones.
The HDI is connected to the detector by the flexible fanout circuits, described earlier, and to the so called tail, 
which will be described in the following section. The HDI provides the physical support for the readout chip, 
the thermal interface between the chips and the cooling system, the electrical interconnects among the chips 
and the electrical connections with other components. It also represents 
the mechanical interface between modules and the support cone.

\tdrsubsubsec{Hybrid mechanical requirements}

The HDI is located outside the active region in a very limited space (approximately $1\cm$ thick) between the tracking acceptance cone 
and the accelerator stay-clear volume as shown in Figure~\ref{fig:superb_svt}.
The limited space forces the need for different types of HDI. Currently 4 different models of hybrids are foreseen and they substantially 
differ in the number of front-end chip (4 to 10) mounted on the hybrid as well as in shape and dimension. 
An additional reduction in number of types will be studied only when final readout electronics will be available.
The approximate dimensions of the HDIs and the quantity of each HDI type to be installed are summarized in the Table~\ref{table:SVT:HDI}.
\begin{table*}[tbp]
\caption{HDI dimensions and quantities for each SVT layer. The dimensions of the width ($w$) and the length ($l$) are in $\mm$.}
\label{table:SVT:HDI}
\begin{center}
\begin{tabular}{|l|c|c|c|c|c|c|} 
\hline
Layer & View & Module & HDI & Chip & $w$, $l$ ($\mm$) & type \\
\hline
\multirow{2}{*}{L0} & $u$     &  \multirow{2}{*}{8} &  \multirow{2}{*}{16}  &  6  &  $w$ = 40 & \multirow{2}{*}{Type 0}  \\ 
                    & $v$  &                     &                       &  6  &  $l$ = 55 &                       \\
\hline
\multirow{2}{*}{L1} & $z$     &  \multirow{2}{*}{6} &  \multirow{2}{*}{12}  &  7  &  $w$ = 47 & \multirow{2}{*}{Type I}  \\ 
                    & $\phi$  &                     &                       &  7  &  $l$ = 41 &                       \\
\hline
\multirow{2}{*}{L2} & $z$     &  \multirow{2}{*}{6} &  \multirow{2}{*}{12}  &  7  &  $w$ = 47 & \multirow{2}{*}{Type I}  \\ 
                    & $\phi$  &                     &                       &  7  &  $l$ = 41 &                       \\
\hline
\multirow{2}{*}{L3} & $z$     &  \multirow{2}{*}{6} &  \multirow{2}{*}{12}  &  10 &  $w$ = 66 & \multirow{2}{*}{Type II}  \\ 
                    & $\phi$  &                     &                       &  6  &  $l$ = 38 &                       \\
\hline
\multirow{2}{*}{L4} & $z$     &  \multirow{2}{*}{16} &  \multirow{2}{*}{32} &  5  &  $w$ = 34 & \multirow{2}{*}{Type III}  \\ 
                    & $\phi$  &                     &                       &  4  &  $l$ = 44 &                       \\
\hline
\multirow{2}{*}{L5} & $z$     &  \multirow{2}{*}{18} &  \multirow{2}{*}{36} &  5  &  $w$ = 34 & \multirow{2}{*}{Type III}  \\ 
                    & $\phi$  &                     &                       &  4  &  $l$ = 44 &                       \\
\hline
\end{tabular}
\end{center}
\end{table*}

Chips are mounted on both sides of the HDI in order to read the $\phi$ and $z$ strip of half of the detector module. 
The number of chips and the association with the layer is summarized in Table~\ref{table:SVT:HDI}.
An important parameter in the design is the total thickness of the HDI including the external mounted components. 
Based on the \babar\ experience the thickness should be no more than 8 mm.
The HDI is the mechanical interface between the detector module and the support cone. 
Special care must be taken for the mechanical requirements of the AlN substrate and of the technical realization of the circuit:

\begin{itemize}

\item accuracy in chip positioning: $\pm 50 \mum$;
\item planarity tolerance in the stay-clear regions for the contact with the ``buttons'': $\pm 10 \mum$;
\item accuracy of the positions of holes for ``buttons'' and vias: $\pm 50 \mum$;
\item accuracy in the cut of the substrate: $\pm 100 \mum$;
\item stay-clear region on the front part of the HDI for gluing the fanout (minimum value): $5 \mm$;
\item stay-clear region on the four corners of the HDIs for mechanical tools access:   $\sim 2.5\times1 \mm^2$.

\end{itemize}

On the opposite side to the fanout, a low profile multi-pin Panasonics connector is mounted in
order to contact the HDI to the tail.  As a result 
the mounting requirements of these connectors also have to be respected.

A picture of the layer 0 HDI is shown in Figure~\ref{fig:SVT:HDIL0}.

\begin{figure}[tbp]
\begin{center}
\includegraphics[width=0.5\textwidth]{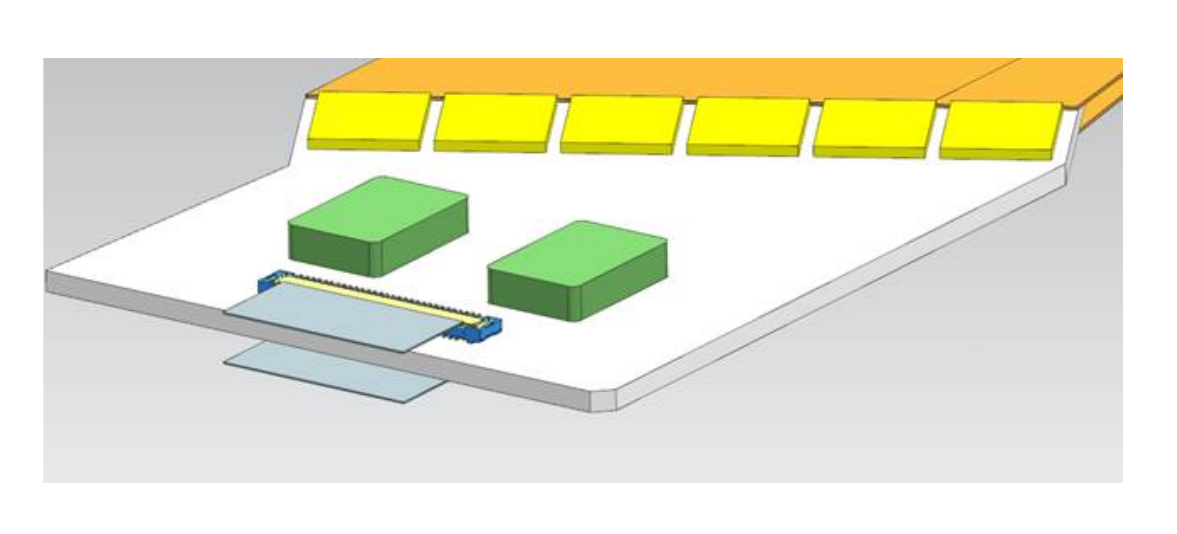}
\end{center}
\caption{In the picture the yellow rectangles represent readout chips, the green boxes are 
the thermal contacts (``bushing'') and the Panasonics connectors are in blue. 
Two connectors per HDI are needed.}
\label{fig:SVT:HDIL0}
\end{figure}

\tdrsubsubsec{Hybrid Electrical requirements}

The HDI must allow all functionality of the readout chip. In particular:

\begin{itemize}

\item provide separate analog and digital power through low impedance planes;
\item two different current returns, one for the digital current and one for the analog current;
\item each power line must be locally filtered;
\item differential command, control and data lines have to be distributed from each chip to the Panasonics connector;
\item impedance of differential lines have to be controlled at 5\% to guarantee LVDS communication between front-end chip and transition card;
\item whenever possible control and command lines need to be redundant;
\item the detector bias voltage must be capacitively coupled to the analog power (representing the analog reference voltage of the readout chip);
\item the two readout sections ($\phi$ and $z$) of the HDI must be capacitively coupled;
\item each HDI must host and provide connection to one resistive temperature monitor;
\item each HDI must provide connections for remote sensing lines for all the supply voltages;
\item detector fan-outs are glued on the hybrid edge and chip inputs are wire bonded to the fan-outs.

\end{itemize}

\tdrsubsubsec{Layout requirements and implementation}

The layout has to fulfill the electrical requirements. 
The preamplifier, which is the first step in the signal processing, is particularly sensitive to noise. 
For this reason as a general rule, the layout must be developed to minimize the coupling of the analog and digital sections. 
Crosstalk and noise coming from the power planes must also be minimized. 
Furthermore power supplies have to be distributed in wide planes to reduce trace induction as much as possible and achieve 
good coupling with current return. 
Each power line is filtered locally with capacitors to the common return. 
The capacitors have to be reliable for high frequency behavior, aging effects, temperature coefficients, dimension and values. 
Following the \babar\ experience, SMD capacitors with X5R dielectric are proposed.
To suppress common mode noise coming from the detector, the two HDI readout sections ($\phi$ and $z$) need to be coupled. 
The connection between the two HDI sides will be realized with a via close to the Panasonics connector, i.e. in an area where space is not critical.
Four or more layers (depending on the specific electric layout) on each side will allow electrical connections between the detector and chips 
from one side and the chips and the transition cards and power supplies on the other side.
The thick film technology proposed for the realization of the HDI is a well known industrial solution so that no R\&D is required.
All fabrication processes are completely under control and there is also possibility of some rework. 
Such a kind of hybrid is a robust solution and has proven long-term reliability. 
Based on the \babar\ HDI, shown in Figure~\ref{fig:SVT:BaBarHDI}, it is reasonable to assume that each layer has a thickness of 
about $65\mum$ ($15\mum$ conductor and $50\mum$ dielectric), traces are $15\mum$ thick, $250\mum$ wide, traces pitch is $400\mum$ and 
pad dimensions are $250 \times 400 \mum^2$. The minimum distance between two vias is $400\mum$. 
It is likely that some of these technological parameters will need to be tuned during the final layout implementation either 
to take advantage of the latest available technology or to improve some of the electrical parameters, for example the impedance of the differential lines.

\begin{figure}[tbp]
\begin{center}
\includegraphics[width=0.5\textwidth]{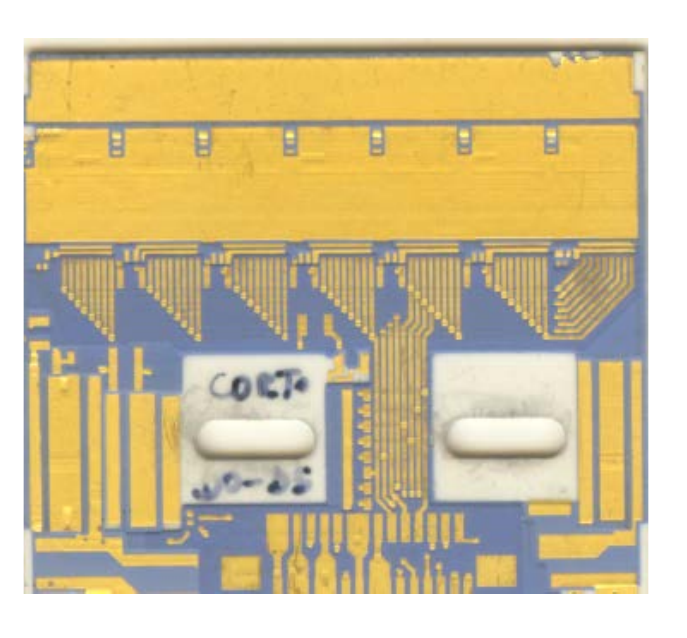}
\end{center}
\caption{\babar\ HDI, shown as an example of the technology to be used.}
\label{fig:SVT:BaBarHDI}
\end{figure}

Because multiple layers must be screened on the HDI high quality workmanship has to be followed and layer per layer full inspection will be implemented 
in close collaboration with the manufacturer.
For a good isolation of two conductive layers the dielectric layer must have a minimum thickness of $45-50\mum$. 
This will be realized with three different dielectric depositions (printing and thermal processes) and two via fillings:

\begin{itemize}

\item deposition of a conductive layer ($10-15 \mum$ thick);
\item first dielectric deposition ($15 \mum$ thick);
\item first via filling;
\item second dielectric deposition ($15 \mum$ thick);
\item last dielectric deposition ($15 \mum$ thick);
\item second via filling;
\item deposition of another conductive layer.

\end{itemize}

Moreover component placement and electrical association of layers will play an important role in design. 
Again using the experience gained in \babar, the following criteria will be followed:

\begin{itemize}

\item separation of components connected to the analog and digital part: the components ``linked'' to the analog part 
are mostly placed close to the border of the HDI. The other components are placed in between the mechanical supports;

\item dedicated layer/s for clock, control lines and data lines;

\item dedicated layers for power/return lines;

\item dedicated layer for component mounting. It will contain most of the traces and pads for soldering components. 
Power sense lines and temperature monitoring will also be on this plane;

\item shielding layer will be added if necessary.

\end{itemize}

Final layout of the HDI will need detail knowledge of IC pinout which is not yet available.

\tdrsubsec{Data Transmission 
}
\label{sec:SVT:DataTransmission}

The SVT data transmission consists of various separate hardware components: 
the HDIs, the tails, the transition cards and the Front end Boards (FEBs).
The characteristics and functionality of the FEBs are described in Sec.~\ref{sec:ELEX:svt}.

A simplified drawing of the data transmission chain with the locations of the various components up to the transition card is shown in 
Figure~\ref{fig:SVT:DataChain}.

\begin{figure*}[tbp]
\begin{center}
\includegraphics[width=0.9\textwidth]{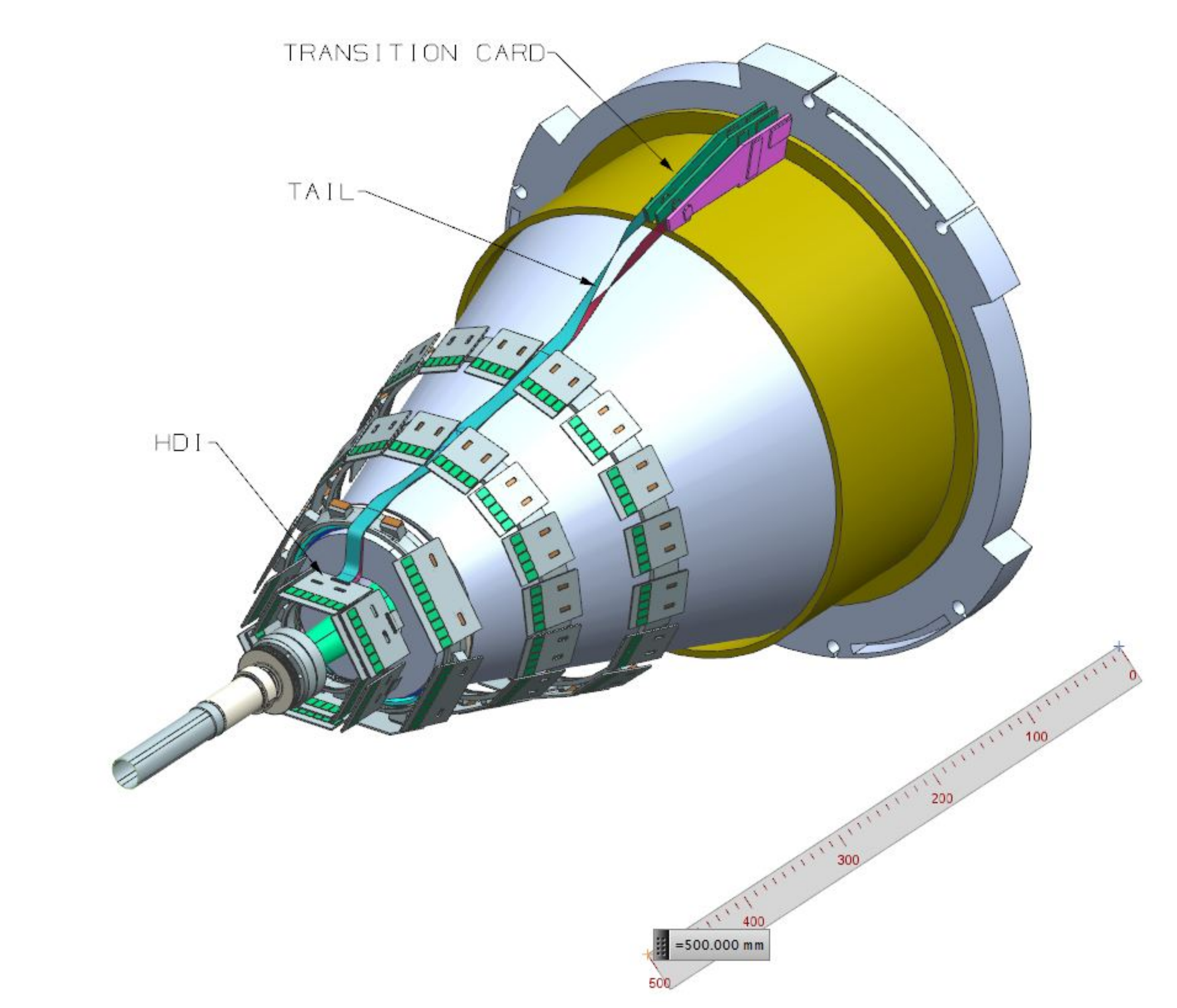}
\end{center}
\caption{Axonometric view of the combination HDI, tail and transition card.}
\label{fig:SVT:DataChain}
\end{figure*}

The FEBs are connected to the transition card output via optical fibers and are located after 
the radiation wall.
As explained in the previous section in the HDI, the signals generated in the detector are processed 
and translated in digital data, properly formatted, by the readout ICs.
The output data from the ICs are transferred from the HDI to the next component of the data transmission, 
the tail, without further processing. 
The limited space available for the HDI prevents mounting additional electronics on the hybrid.
The specification/characteristics of the data lines can be summarized as follow:

\begin{itemize}

\item the data signals follow a LVDS standard, consequently they are differential signals;
\item the number of data lines, per IC, varies from:
\begin{itemize}
\item 1 to 3 if synchronized to a 120 MHz clock
\item 1 to 2 if synchronized to a 180 MHz clock
\end{itemize}
depending from the data throughput;
\item the data are serialized and encoded using a 8b/10b protocol in the readout chip.

\end{itemize}

The data lines needed for the different HDIs (reference clock of 120 and 180 MHz) are summarized in 
Table~\ref{tab:SVT:DataTransmission}.

\begin{table*}[tbp]
\caption{Number of data lines, tails, transition cards (TCs) and optical fibers for the six layers of the SVT detector.}
\label{tab:SVT:DataTransmission}
\begin{center}
\scalebox{0.8}{ 
\begin{tabular}{|l|c|c|c|c|c|c|c|c|c|c|} 
\hline
Layer & View & Modules & HDIs & HDI  &  Chips & Data lines    & Data lines   & Tails    & TCs & Optical \\
      &      &         &      & Type &        & per HDI side  & per HDI side & (Cu Bus) &     & Fibers  \\
      &      &         &      &      &        & (120 MHz)     & (180 MHz)    &          &     &         \\
\hline
\multirow{2}{*}{L0} & top     &  \multirow{2}{*}{8} &  \multirow{2}{*}{16}  &  \multirow{2}{*}{Type 0}  &   6  &   18 & 12 &  \multirow{2}{*}{32} &  \multirow{2}{*}{32} &  \multirow{2}{*}{32} \\
                    & bottom  &                     &                       &                           &   6  &   18 & 12 &                      &                      &                      \\
\hline
\multirow{2}{*}{L1} & $z$     &  \multirow{2}{*}{6} &  \multirow{2}{*}{12}  &  \multirow{2}{*}{Type I}  &   7  &   14 & 7 &  \multirow{2}{*}{24} &  \multirow{2}{*}{24} &  \multirow{2}{*}{24} \\
                    & $\phi$  &                     &                       &                           &   7  &   21 & 14 &                      &                      &                      \\
\hline
\multirow{2}{*}{L2} & $z$     &  \multirow{2}{*}{6} &  \multirow{2}{*}{12}  &  \multirow{2}{*}{Type I}  &   7  &   14 & 7 &  \multirow{2}{*}{24} &  \multirow{2}{*}{24} &  \multirow{2}{*}{24} \\
                    & $\phi$  &                     &                       &                           &   7  &   14 & 14 &                      &                      &                       \\
\hline
\multirow{2}{*}{L3} & $z$     &  \multirow{2}{*}{6} &  \multirow{2}{*}{12}  &  \multirow{2}{*}{Type II}  &  10  &   20 & 10 &  \multirow{2}{*}{24} &  \multirow{2}{*}{24} &  \multirow{2}{*}{24}\\
                    & $\phi$  &                     &                       &                           &   6  &    12 &  6 &                      &                      &                       \\
\hline
\multirow{2}{*}{L4} & $z$     &  \multirow{2}{*}{16} &  \multirow{2}{*}{32}  &  \multirow{2}{*}{Type III}  &  5  &   10 & 5 &  \multirow{2}{*}{64} &  \multirow{2}{*}{32} &  \multirow{2}{*}{32} \\
                    & $\phi$  &                     &                       &                              &   4  &   8 & 4 &                      &                      &                       \\
\hline
\multirow{2}{*}{L5} & $z$     &  \multirow{2}{*}{18} &  \multirow{2}{*}{36}  &  \multirow{2}{*}{Type III}  &  5  &   5 & 5 &  \multirow{2}{*}{72} &  \multirow{2}{*}{36} &  \multirow{2}{*}{36} \\
                    & $\phi$  &                     &                       &                              &  4  &   4 & 4 &                      &                      &                        \\
\hline
\end{tabular}
}
\end{center}
\end{table*}

The large number of data lines suggests that the reference clock should be the 180 MHz one, so that the maximum number of lines is 14 for one side 
of the Type I HDI.
Together with the data lines (and always to LVDS standards) all the ICs of same HDI share 6 input lines 
(Reset, Clock, FastClock, Timestamp, Trigger, RegIn) and 1 output line (RegOut).
\tdrsubsubsec{The Tail}
\label{sec:SVT:Tail}
The total number of differential lines to be transferred from a HDI to the next processing block of the chain is 21 or less. 
As the data rate per line is not excessively high ($\sim$ 180 Mbps) and a custom designed shielded, multilayer copper bus could be used to connect 
the output of the HDI with the input of the transition card. 
The length of this bus, called the ``tail'', is between 50 to 70 $\cm$ and changes slightly from layer to layer. 
The tail should have a rectangular section  not exceeding $10\times 4 \mm^2$, because it has to be pulled along the supporting cone 
and below the HDI itself during detector integration. In many cases the tail has to pass between the HDI thermal standoff, the ``buttons''.
Similar dimensional constraints apply to the tail terminating connectors. A very low profile ($0.8 \mm$) narrow pitch  ($0.35 \mm$) 
Panasonics connector (series AXE, 70 pins) can be utilized for this purpose (see Figure~\ref{fig:SVT:Panasonic}). 
By using the repetitive pin sequence Gnd/D+/D-/Gnd it is possible to connect up to 22 LVDS lines (11 per side) to this connector.  
Each LVDS line is individually shielded if the lines are realized as strip-lines.
\begin{figure}[tbp]
\begin{center}
\includegraphics[width=0.5\textwidth]{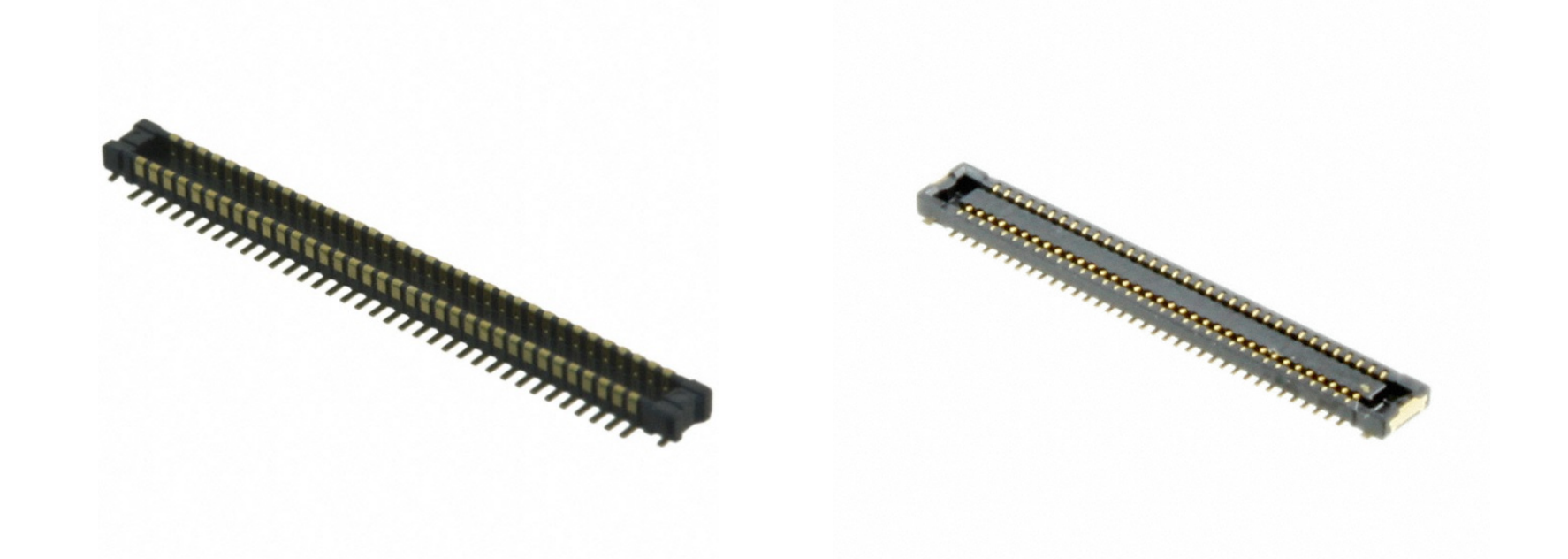}
\end{center}
\caption{Panasonics AXE series connector. Header (left) and socket (right).}
\label{fig:SVT:Panasonic}
\end{figure}

The socket ($2.5 \mm$ wide) is mounted on the HDI and the transition card, while the header ($2.0 \mm$ wide) is mounted on the tail. 
The dimensions of this surface mount connector are compatible with the space reserved on the HDI for the output connectors. 
The retention mechanism of this connector provides a holding force of at least 0.20 N/contact, 
which ensures a reliable lock between the elements of the data chains.
This connector is also adequate for the transition card as explained in the following section.
In the present design, the tail is a multilayer flexible katpon circuit, whose main electrical parameters are the following:

\begin{itemize}

\item five layer circuit. Stack-up plane/signal/plane/signal/plane;
\item LVDS lines: strip lines on the two signal layers;
\item LVDS line width/space: 100/100 $\mum$;
\item LVDS copper thickness  $ \sim 20\mum$;
\item dielectric thickness:  $\sim 2.5\mm$ ;
\item Zdiff: $\sim 100$ Ohm;
\item kapton thickness: $50\mum$;
\item via diameter: $250-300\mum$;
\item via clearance: $150\mum$;
\item width: $10 \mm$ min, $15.8 \mm$ max (at the connector).
\end{itemize}

The plane layers available in the stack-up can be used both as a shield for the LVDS traces as well as Power/Ground planes. 
The tail does not only carry the digital signals but also the power for the readout ICs.
The parameters for the plane segmentation of the prototype design are:

\begin{itemize}

\item power plane width:  $\sim 4.7 \mm$;
\item ground plane width: $\sim 9.4 \mm$;
\item metal thickness:  $> 50 \mum$;
\item current capability:  $\sim 3-3.5$ Amps;
\item voltage drop (both ways):  $\sim 250$ mV.

\end{itemize}

The minimum bending radius of the resulting multilayer circuit has to be carefully evaluated to verify that during 
installation the tail can be bent into the required shape. 

\tdrsubsubsec{The transition card}

The next element of the transmission chain is a printed circuit board made of halogen free material, 
called transition card (TC). The geometrical aspect, dimensions and shape are defined by the detector geometry 
as shown in Figure~\ref{fig:SVT:DataChain} and in more detail in Figure~\ref{fig:SVT:TransitionCard}.

\begin{figure*}[!htb]
\begin{center}
\includegraphics[width=0.9\textwidth]{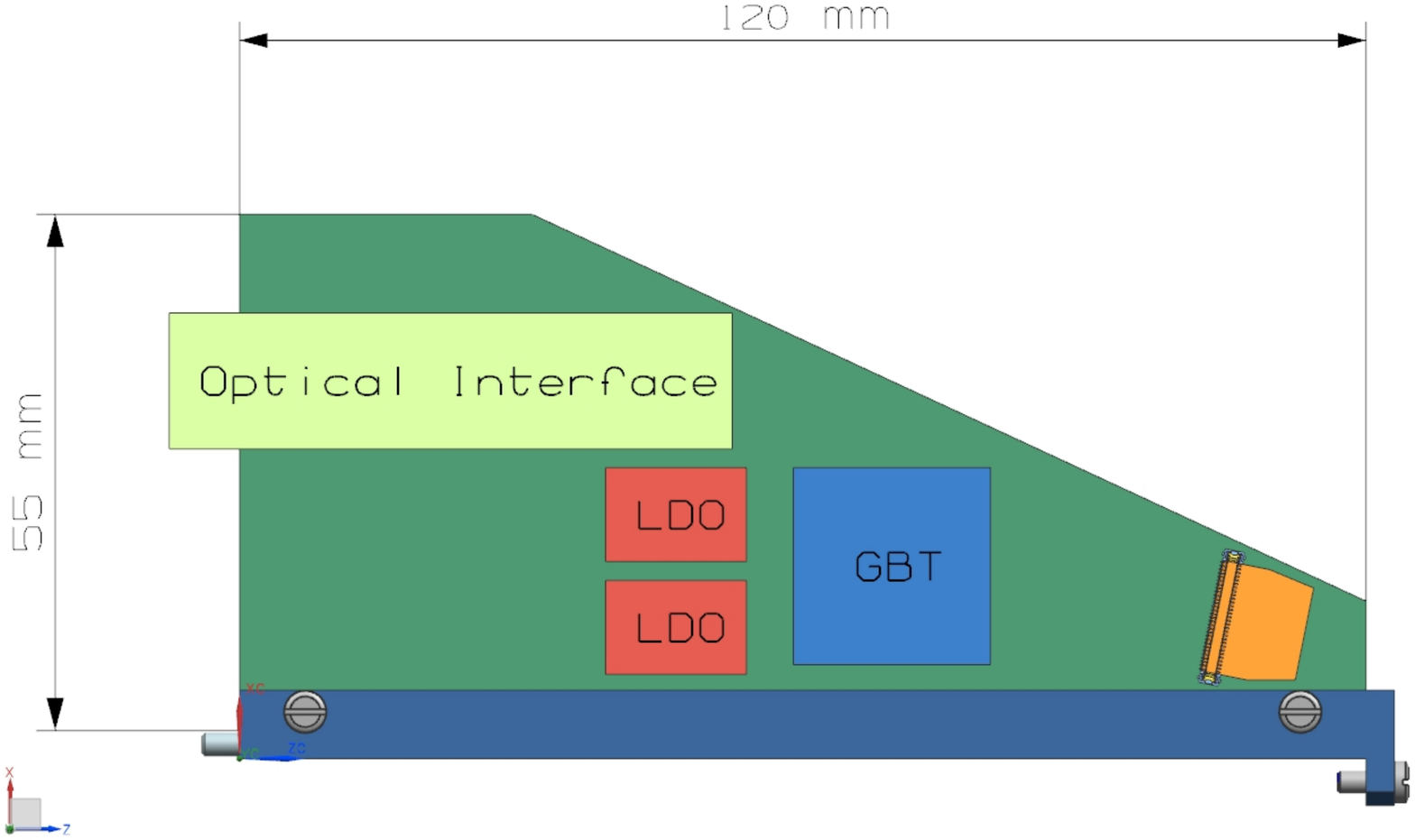}
\end{center}
\caption{Shape/dimensions of a transition card. The main active components to be mounted on the PCB are also shown  
(GBT, LDOs, optical transmitter) as well as the mechanical support required to mount the board (visible at the bottom)
of the figure.}
\label{fig:SVT:TransitionCard}
\end{figure*}

The transition cards are located approximately $50\cm$ from the detector sensors and they are electrically connected 
inbound to the HDI by means of the tail. 
They are also connected outbound by optical fibers to the FEBs and by cables to the remote power supplies 
(both low and high voltage).

A transition card has various functions:

\begin{itemize}

\item it will receive and distribute the input and control signals for the front-end chip through the optical fiber from the FEB;
\item it will receive, format and serialize, the data coming from the front-end chips;
\item it will perform the electrical-to-optical conversion of the data and it will transmit the data to the FEB to begin the data acquisition;
\item it will regulate and distribute the power to the front-end chip on the HDIs as well as the HV power for the microstrip detectors. 
       It also filters the supplies to have low ripple voltages sent to the HDIs.
\end{itemize}

Because the data volume decreases substantially from the innermost to the outermost detector layer (from $\sim$ 1 Gbits/trig, to $\sim$ 0.25 Gbits/trig per TC),
 one transition card will be used per HDI side for the layers 0 to 3. 
Instead one transition card will serve both sides of the same HDI for layers 4 and 5.

Such a reduction in total number of TC is advantageous both in having a uniform data link speed (<2Gbit/s) with reasonable link utilization 
for all TCs but also in maximizing the surface area available per TC and increasing the space between two adjacent TCs.

The typical geometrical disposition of the TCs is shown in Figure~\ref{fig:SVT:Turbine}. 
The minimum space between two cards is $3.6~\mm$, barely enough to mount the tail. 
The total number of TC is 86 per detector side (172 in total).

\begin{figure*}[tbp]
\begin{center}
\includegraphics[width=0.55\textwidth]{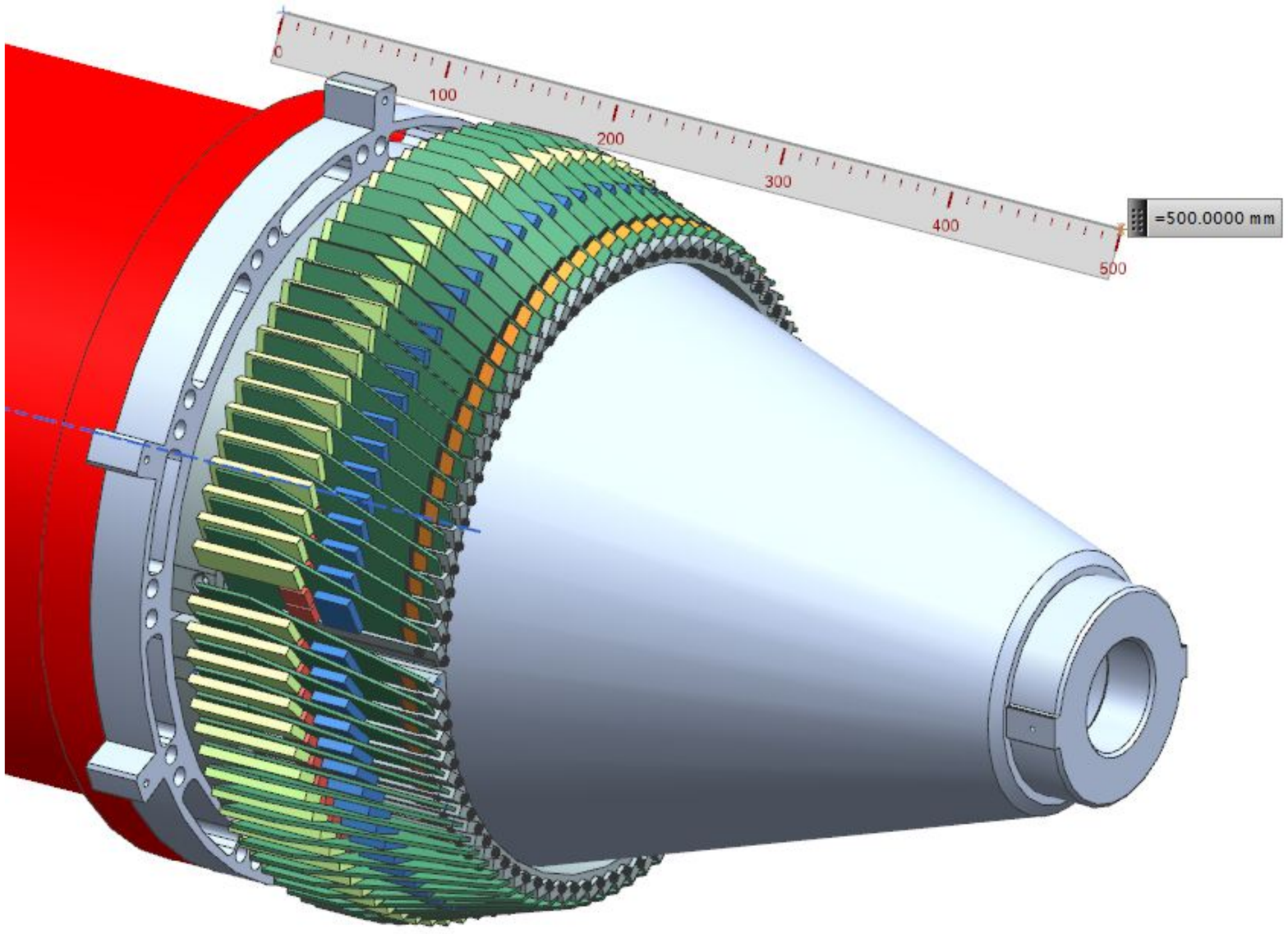}
\includegraphics[width=0.40\textwidth]{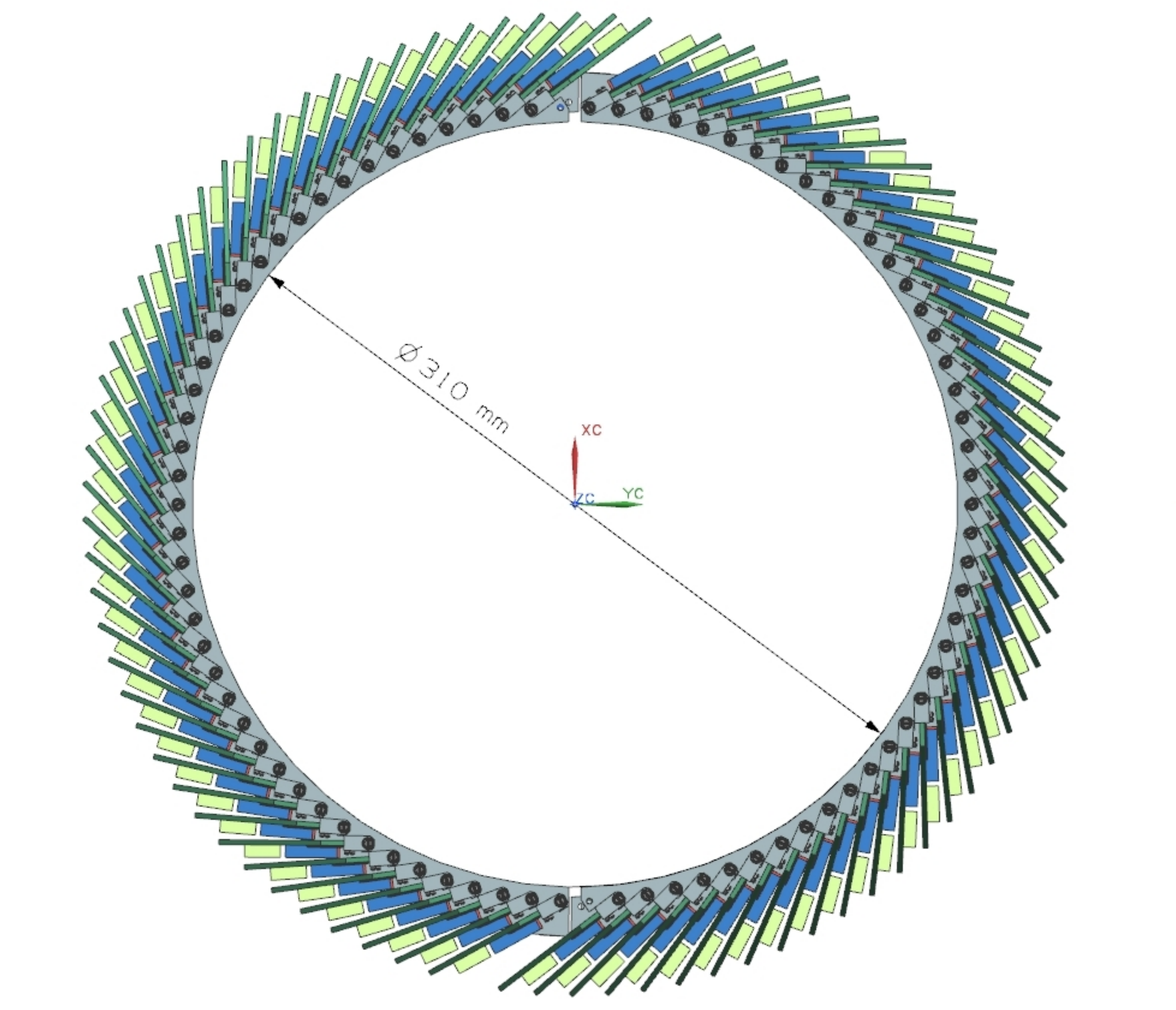}
\end{center}
\caption{Turbine arrangement of the TC on the support cone.}
\label{fig:SVT:Turbine}
\end{figure*}

The TC requirements can be fulfilled by having at least the following logic blocks:

\begin{itemize}

\item an interface circuit to group, before serialization, the LVDS data lines coming from the HDI;
\item a serializer/deserializer;
\item an optical transmitter/receiver;
\item slow control interface for handling the detector control signals.

\end{itemize}

This device has not been designed yet, however we are closely following various CERN designs to take advantage 
of the development already started. 
In particular the CERN GBT project which is well advanced and which is trying to address very similar needs for the LHC experiments~\cite{GBT}, 
is our baseline design for signal transmission.

The contact undertaken with CERN designers have indicated that the GBT IC characteristics match and even exceed SVT requirements in term of 
data link, being the GBT project intended to operate up to 4.8 Gbit/s. 

The physical dimension of the IC, less than $400 \mm^2$, the power ($\sim$ 2 W/chip) and the expected delivery time (not later than 2013) 
are compatible with the TC design, construction and testing.
Moreover the CERN design has been proved to be radiation tolerant up to radiation levels well exceeding the ones expected in \superb. 

The maximum reference clock of the GBT IC is related to the LHC machine clock (40 MHz). 
Such a frequency will need to be locally generated at the level of the transition card to properly operate the device. 
If this reference clock, as recently discussed could be an option at \superb level, the TC design will take advantage of this possibility.

Similarly, the design of the optical interface and the choice of the optical fibers will take advantage of the results obtained by the 
CERN VERSATILE LINK project~\cite{V-Link}, which again is well suited for the SVT application. 
In particular the new ``versatile small form factor transceiver'', the design of which is ongoing at CERN,
could fit dimensionally in the TC and 
has provision for connecting two optical fibers in the same package, one fiber for control signals, the other for data.

Finally, at the present stage, the regulation of the low voltages needed for the front-end IC will be implemented by using the same adjustable 
low dropout voltage regulators (LDOs) widely adopted at the LHC: LHC4913. 
This component, manufactured by ST, has been thoroughly tested against radiation and can generate a clean reference voltage down to 1.25 Volts.
It also has remote sense capability to compensate voltage drops between the point of regulation and the point of load due to cables. In the baseline design two LDOs, one for the analog and the other for the digital supply, will be used for each HDI.

A preliminary thermal analysis of the transition card has been performed based on the typical heat dissipation of the various electrical blocks 
mounted on it. A conservative 5 Watt per TC has been used.

A dedicated cooling system is being designed to effectively remove the heat without increasing the temperature of the surroundings. 
The thermal connection between the board and the cooling system can be achieved by means of the TC mounting support, 
the design of which is shown in Figure~\ref{fig:SVT:Turbine}.




\tdrsec{Mechanical Support and Assembly}
An overview of the mechanical support and assembly is provided in section~\ref{svt:overview}.
In this section we provide a more detailed account of the constraints of the mechanical design due to the accelerator components of the 
Interaction Region (IR) and describe the details of the detector assembly, installation, survey, and monitoring.

\tdrsubsec{IR Components \& Constraints}
The support structure design of the SVT is dictated by the configuration and 
assembly procedure of the IR machine components, as well as by the SVT geometry. 
A schematic drawing of the IR assembly with all the components described in the rest of this section is shown 
in Figure~\ref{fig:long_ir_section}.
The whole IR assembly consists of the Beryllium beam-pipe, the SVT, the forward and backward Final Focus (FF) permanent magnets, 
the conical Tungsten shields 
for background reduction and the cryostats of the FF superconducting magnet system.
\begin{figure*} [tbp]
\centerline{\includegraphics[clip=true,trim=6.5cm 2cm 6.5cm 2cm, angle=90, width=\textwidth]{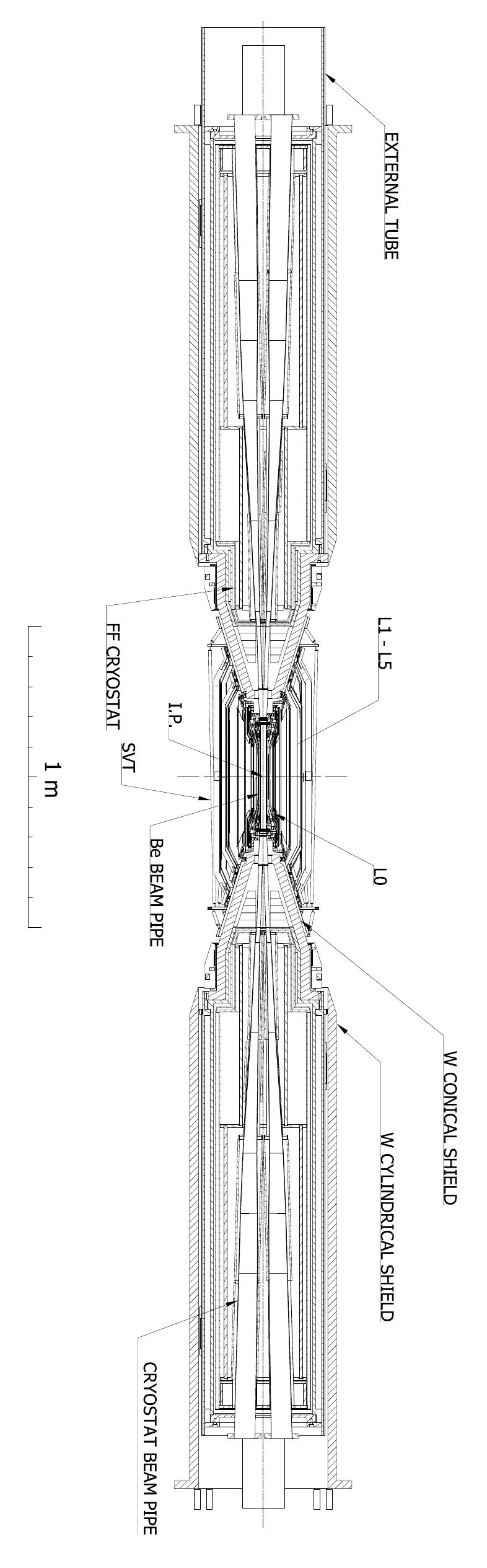}}
\centerline{\includegraphics[clip=true,trim=5cm 3cm 5.5cm 3.5cm, angle=90, width=0.9\textwidth]{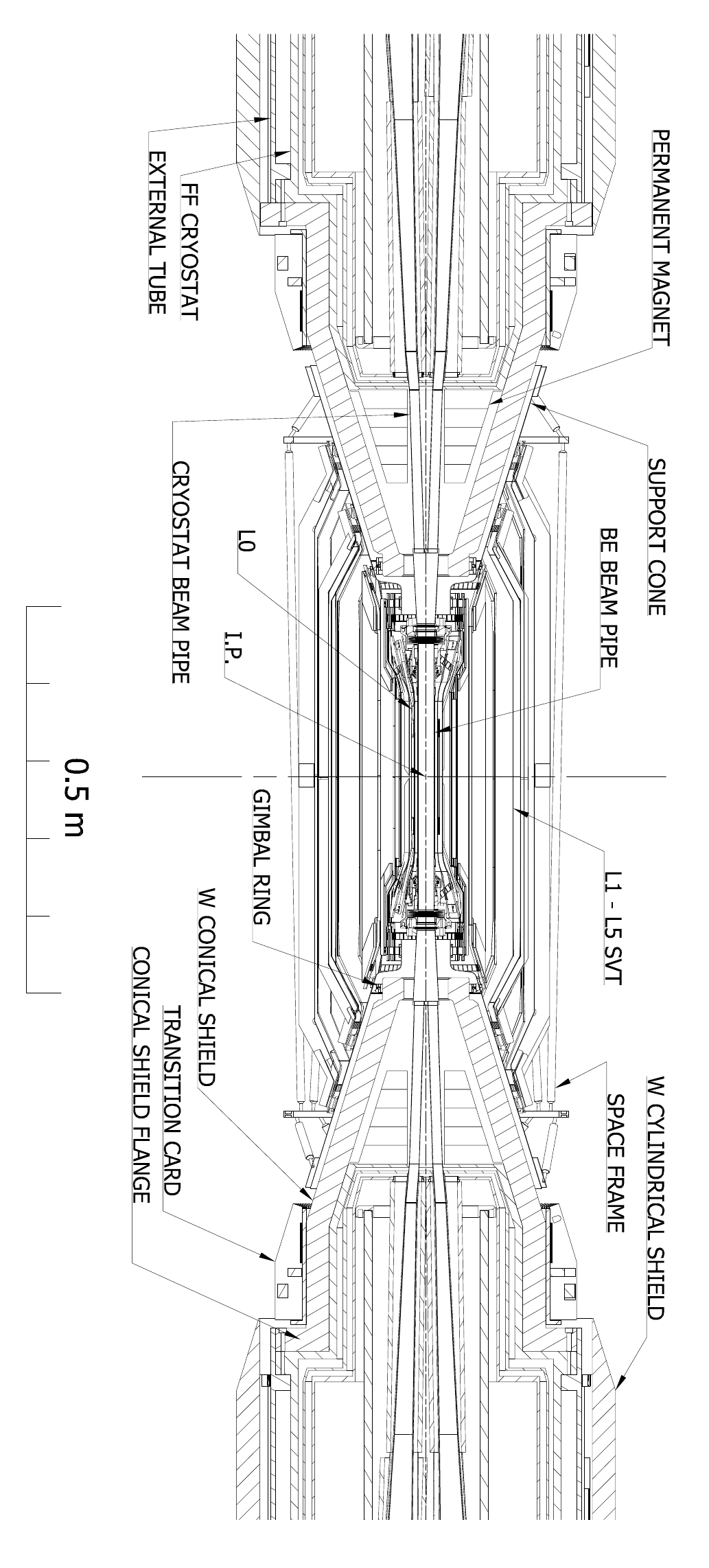}}
\caption{Longitudinal section of the IR in the \textit{x-z} plane (top) with a detailed view (bottom).}
\label{fig:long_ir_section}
\end{figure*}

%


The background conditions impose the need for a pair of conical Tungsten shields located about 25 cm from the IP on either side. 
There are two further Tungsten shields of cylindrical shape that are located symmetrically with respect to the IP, starting  
just at the end of the conical ones. The cylindrical shields are rigidly attached to the iron structure of the return yoke of the detector magnet in the forward and backward barrel face of the detector. 
The conical shields and the final focus permanent magnets occupy most of the region below $17.2^{\circ}$ (300 mrad) on both sides of the IP . 
In order to minimize the mass inside the active tracking volume,  all of the electronics are mounted below the 300 mrad cone. 
This requires that in backward and forward directions electronics, cooling, cabling and supports must be confined in a volume of about one centimeter thick around the conical shields. The use of this tight space below 300 mrad must be carefully arranged with the needs of the accelerator. The angular acceptance in the laboratory frame of the SVT is therefore restricted to the region 
$17.2^{\circ} < \theta < 162.8^{\circ}$, where $\theta$ is the polar angle. 
The radial position of the SVT fiducial volume is imposed by the inner radius of the DCH at about 27 cm.\\
The Be beam-pipe, cooled by liquid forced convection, is about 2 cm in diameter and 40 cm long. It is positioned symmetrically with respect to the IP and it directly supports the Layer0 modules. 
The eight striplet modules, arranged in a pin wheel geometry, are mounted on the two cooled flanges. 
The Layer0 is mechanically decoupled from the other SVT layers to allow its replacement without a complete disassembling of the other SVT modules.
The SVT layers from 1 to 5 are supported on the backward and forward sides on the conical shields by two gimbal rings, the kinematic mounts that allow
the necessary degree of freedom to prevent SVT over-constraints during the installation/removal operations. 
The central Be beam-pipe is connected at each end through a bellow and a flange to the beam-pipe coming out of the cryostat of the FF 
superconducting magnets. In this region the beam pipe is split in two arms: the LER and HER pipes represent the warm internal vessel of 
the FF cryostats. The forward and backward cryostats are symmetrically positioned around the IP and they are the extreme components of the IR, 
with the cryostat beam-pipe terminal flanges placed at about 2.2 m away from the IP, as shown in Figure~\ref{fig:long_ir_section}.
The cryostat is rigidly connected to the conical shield flanges and, through the terminal back flange, at a very rigid coaxial external tube, 
allowing a free space of about 2 cm in radius along its extension for routing the SVT cables out of the inner 
detector region.

Since the L1-L5 and the Layer0 must be installed with all the components of IR system in place, they must be split in two halves and then clam-shelled around the beam-pipe.
The whole IR system, just described, will be assembled and aligned in a clean area outside the experimental hall and then
transported and installed inside the \superb detector.
From the mechanical point of view, the IR system consists of two very rigid systems  (conical shield + cryostat + external tube) 
joined by a very weak 
system (the Be beam-pipe and the SVT) that could be damaged during the transportation and installation/removal operation.
On the other hand in order to reduce the passive material between the SVT and the DCH, and improve the detector performance, no permanent 
stiffening structure (i.e. a support tube as in the \babar detector) is foreseen in \superb to rigidly connect these systems. 
Therefore a removable structural support has been designed, in the following called temporary cage, to connect rigidly the backward and forward 
conical shields. The temporary cage will be inserted to absorb the mechanical stress present at the moment of the first transportation to the 
experimental hall and during all the installation/removal operations of the IR assembly/SVT/Layer0, and it will be  
removed for the data taking (see Sec.~\ref{sec:temporary_cage}).


\tdrsubsec{Module Layout \& Assembly}
The SVT is built with detector modules, each mechanically and electrically independent. Each module consists of silicon wafers glued to fiber composite beams, with a high density interconnect (HDI) hybrid circuit at each end. The HDIs are electrically connected to the silicon strips by means of flexible circuits and they are mechanically supported by the fiber composite beam. The entire module assembly is a rigid structure that can be tested and transported in its case. As an example a drawing of a detector module from layer 3 is shown in Figure~\ref{fig:l3_module}.\\
\begin{figure*}[tbp]
\begin{center}
\includegraphics[clip=true,trim=6cm 2.5cm 3.5cm 2.5cm, angle=90, width=\textwidth]{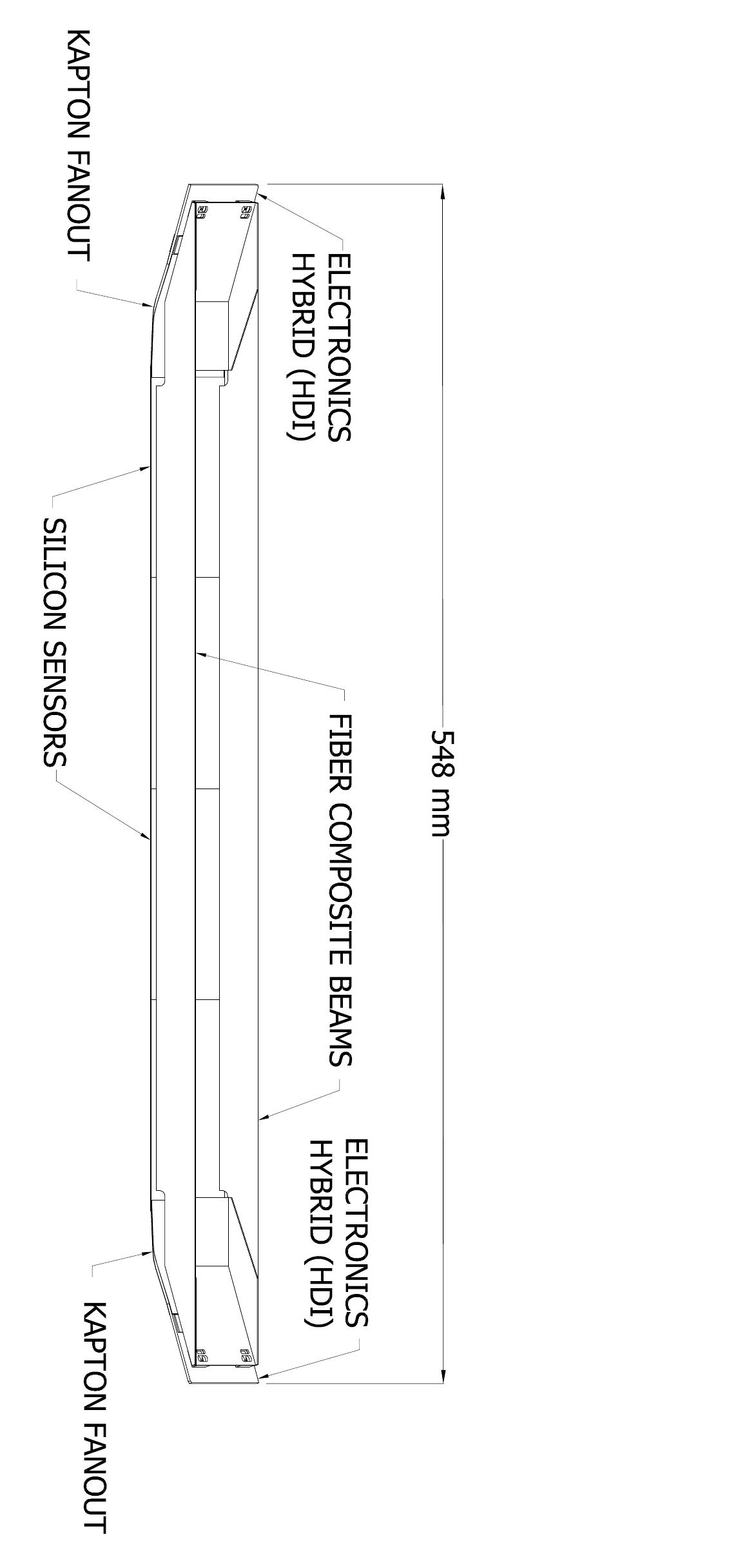}
\end{center}
\caption{Drawing of a layer 3 silicon strip module.}
\label{fig:l3_module}
\end{figure*}
The assembly of a detector module will follow the procedure used for the \babar\ SVT modules~\cite{Bozzi:2000} and 
begins after the preparation of all the necessary parts. The silicon detectors must be fully tested, including a long-term stability (burn-in) test under full bias voltage. The fanout circuits will be optically inspected and single trace tested for spotting shorts or opens. The readout hybrid must be assembled and tested, with the HDI supports, the front-end chips and additional passive components.
Finally, the completed beam structure, which provide mechanical stiffness, must be inspected to ensure it  meets the specifications. These individual parts will be fabricated by different 
laboratories and then shipped to the place where the module assembly is carried out. The hybrids will be tested again after shipment.\\
The assembly of the inner barrel-shaped modules and the outer arch-shaped modules is necessarily different. However, there are common steps. 
The procedure is the  following:
\begin{enumerate}
\item each silicon detector is precisely aligned using its reference crosses and then head to head glued to the adjacent to form the module;
\item the \textit{z} and $\phi$ fanouts are glued to the detectors and wire-bonded to the strips. The ganging bonds between $\phi$ strips are then performed. Electrical tests, including an infrared laser strip scan, are performed to assess the quality of  the detector-fanout assembly (DFAs);
\item the silicon detectors and the readout hybrid are held on a suitable fixture and aligned. The fanouts are glued to the hybrids and wire bonded to the input channels of the readout ICs;
\item the final assembly stage is different for different layers. The module of the layer 1 and 2 are then joined together by gluing the beams on the top of layer 1 to the bottom of layer 2 to give a combined structure called the "sextant". For the layer 3,  the DFA is bonded flat to the fiber composite beams. 
For the modules of layer 4 and 5, the flat module is held in a suitable fixture and bent on a very precise mask at the corners of the arch and at the connection to the HDI. The fiber composite beams are glued to the module with fixtures insuring alignment between the silicon detectors and the mounting surfaces on the HDI. This procedure has been already successfully adopted for the external modules of the SVT of the \babar experiment;
\item once completed, these detector modules are extremely rigid ladders that can be stored and submitted to the final electrical characterization. They are then stored in their shipping boxes for the final installation on the detector.
\end{enumerate}
A drawing of an arch-shaped module is shown in Figure~\ref{fig:layer5_module}.
\begin{figure}[tbp]
\begin{center}
\includegraphics[clip=true,trim=4cm 4cm 3cm 4cm, angle=90, width=0.5\textwidth]{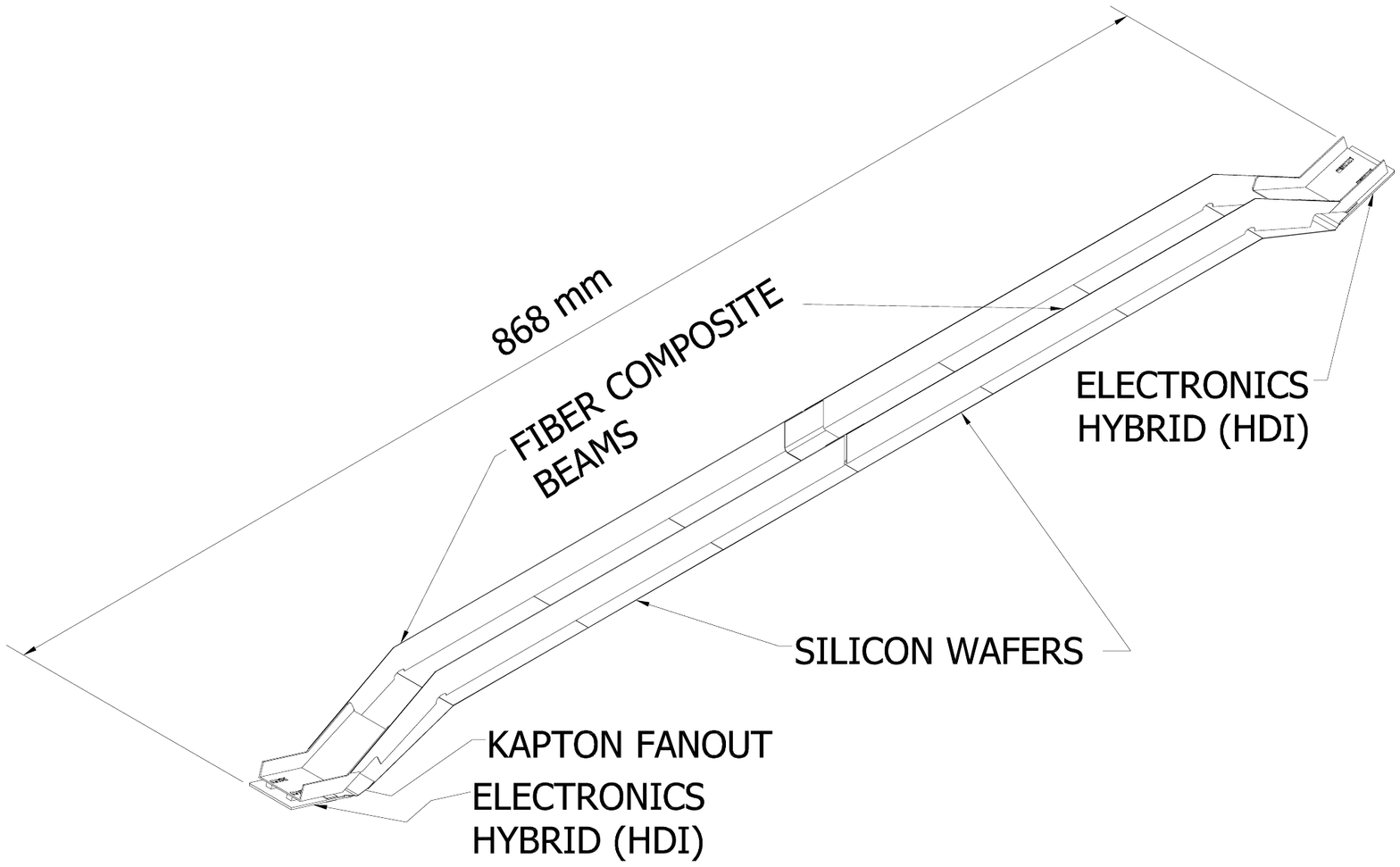}
\end{center}
\caption{Drawing of a layer5 silicon strip arch-module.}
\label{fig:layer5_module}
\end{figure}

The assembly procedures for a Layer0 module (shown in Figure~\ref{fig:layer0_module}) in general follow that reported for the other SVT layers, 
although the smaller sensor dimensions represent a greater difficulty in the handling. The procedure is the following:
\begin{enumerate}
\item no detector alignment is required because there is only one detector;
\item the \textit{u} and \textit{v} fanouts are glued on each face of the detector on both sides and wire-bonded to the strips. A two-layer fanout is necessary to read out each face;
\item the silicon detectors and the readout hybrid are held on a suitable fixture and aligned. The fanouts are glued to the hybrids and wire bonded to the input channels of the readout ICs. Electrical tests, including an infrared laser strip scan, are performed;
\item in the final assembly stage for Layer0 module there is the requirement for a mask that is able to position the HDI in a very precise position in a plane inclined at 10$^\circ$ in order to be positioned outside the active region. The fiber composite beams are glued to the module with fixtures assuring alignment between the silicon detectors and the mounting surfaces on the HDI.
\end{enumerate}
\begin{figure*}[tbp]
\begin{center}
\includegraphics[clip=true,trim=6.5cm 4.5cm 6cm 3.5cm, angle=90, width=\textwidth]{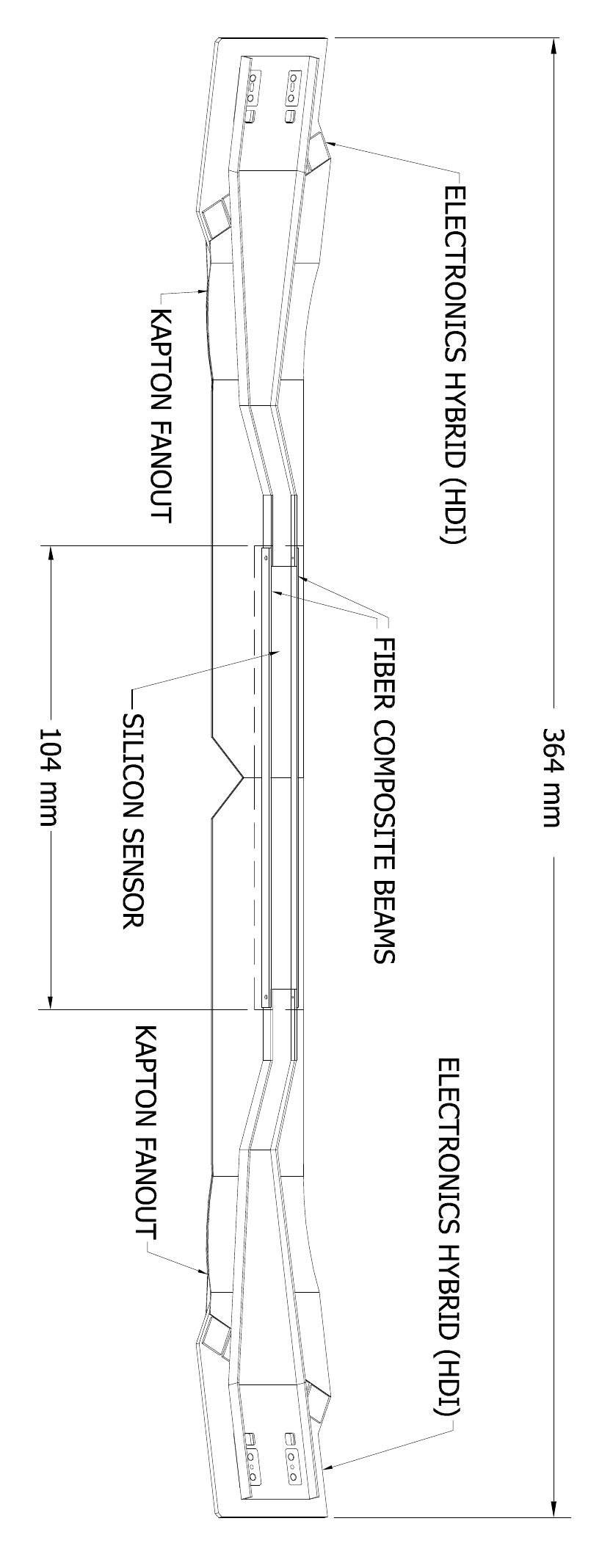}
\end{center}
\caption{Drawing of a Layer0 striplets module.}
\label{fig:layer0_module}
\end{figure*}

\tdrsubsec{L1-L5 Half detector assembly}
The L1-L5 detector is assembled in two halves in order to allow the device to be mounted around the beam pipe. The detector modules are supported at each end by cooling/support rings in brass that are attached on the support cone realized in laminated carbon-fiber. Water circulates in the cooling rings and cools the mounting protruded pieces (buttons) in thermal contact with the HDIs. The cones are split along a vertical  plane and have alignment pins and latches that allow them to be connected together around the conical shield. 
See Figure~\ref{fig:cooling_cone} for a drawing of the support cone.

\begin{figure*}[tbp]
\begin{center}
\includegraphics[clip=true,trim=0.5cm 1cm 0.5cm 2cm, width=0.7\textwidth]{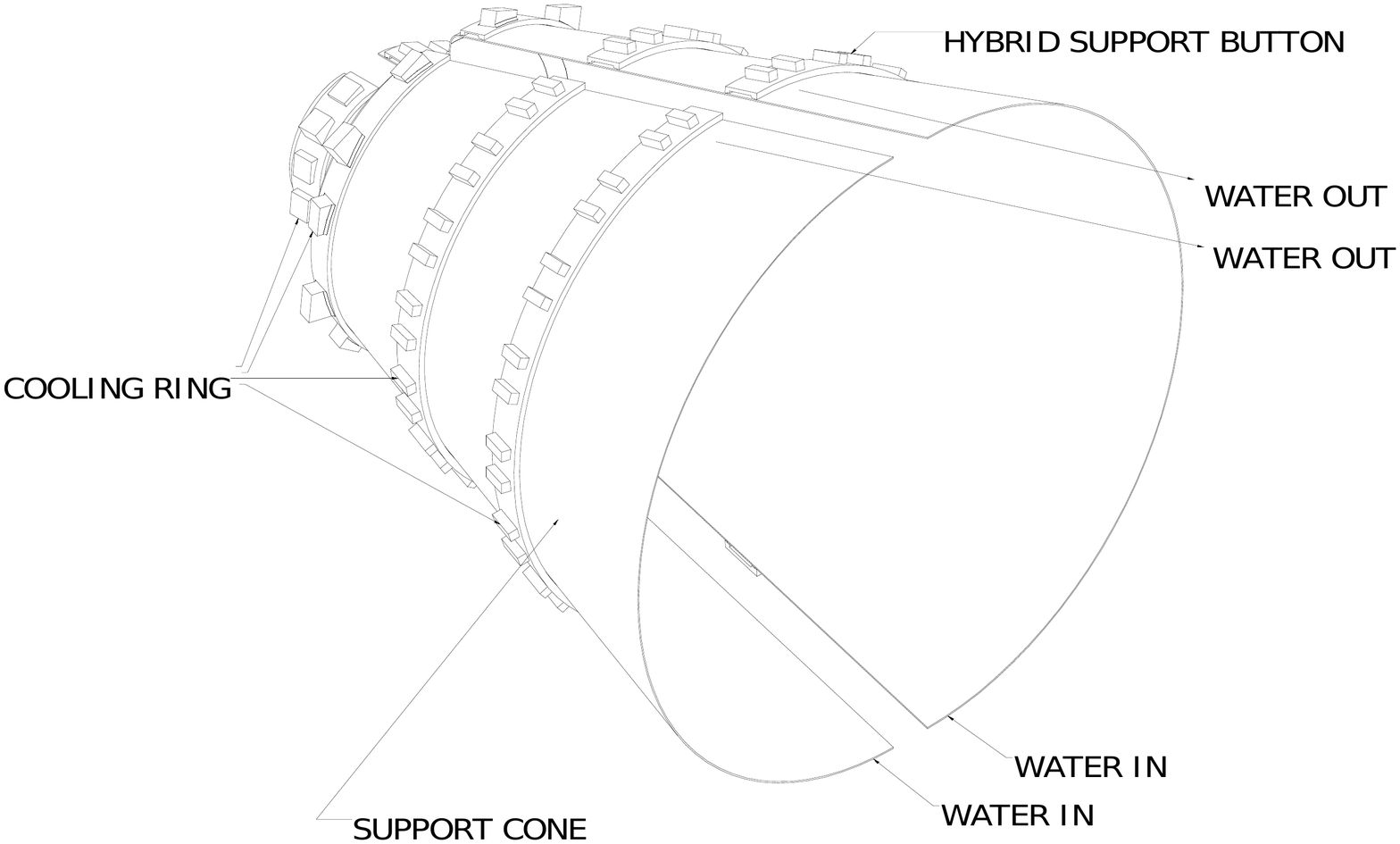}
\end{center}
\caption{The support cone with the four cooling rings for layers 5, 4, 3 and for the sextants of layers 1 and 2. }
\label{fig:cooling_cone}
\end{figure*}

During the half detector assembly, the two half-cones will be held in a fixture which holds them in a precise position relative to each other. The detector modules are then mounted on the half-cones to each end. A fixture holds the detector modules during this operation and allows for well-controlled positioning of the module on the half-cones. Pins located in the buttons provide precise positioning of the modules, which are then screwed down. Accurate alignment of the mounting with respect to the silicon wafers is achieved by a pair of mating fixtures. One is a dummy module (mistress mask) and the other simulates the mating surface of the cone (master mask). These fixtures are constructed together and mate perfectly. One is used to verify the correct machining of the mounting of the cooling ring on the cones. The other is used to position the mounting points on the HDIs during the assembly of the module.\\
The connection between the module and the cone (called a foot) provides both accurate alignment of the module and a thermal bridge from the HDI to the coolant.
A schematic of the forward foot region, which contains the readout electronics and the cooling ring with the mounting buttons is shown in Figure~\ref{fig:ffoot}.
\begin{figure}[tbp]
\begin{center}
\includegraphics[clip=true,trim=4cm 7cm 6cm 6cm, angle=90, width=0.5\textwidth]{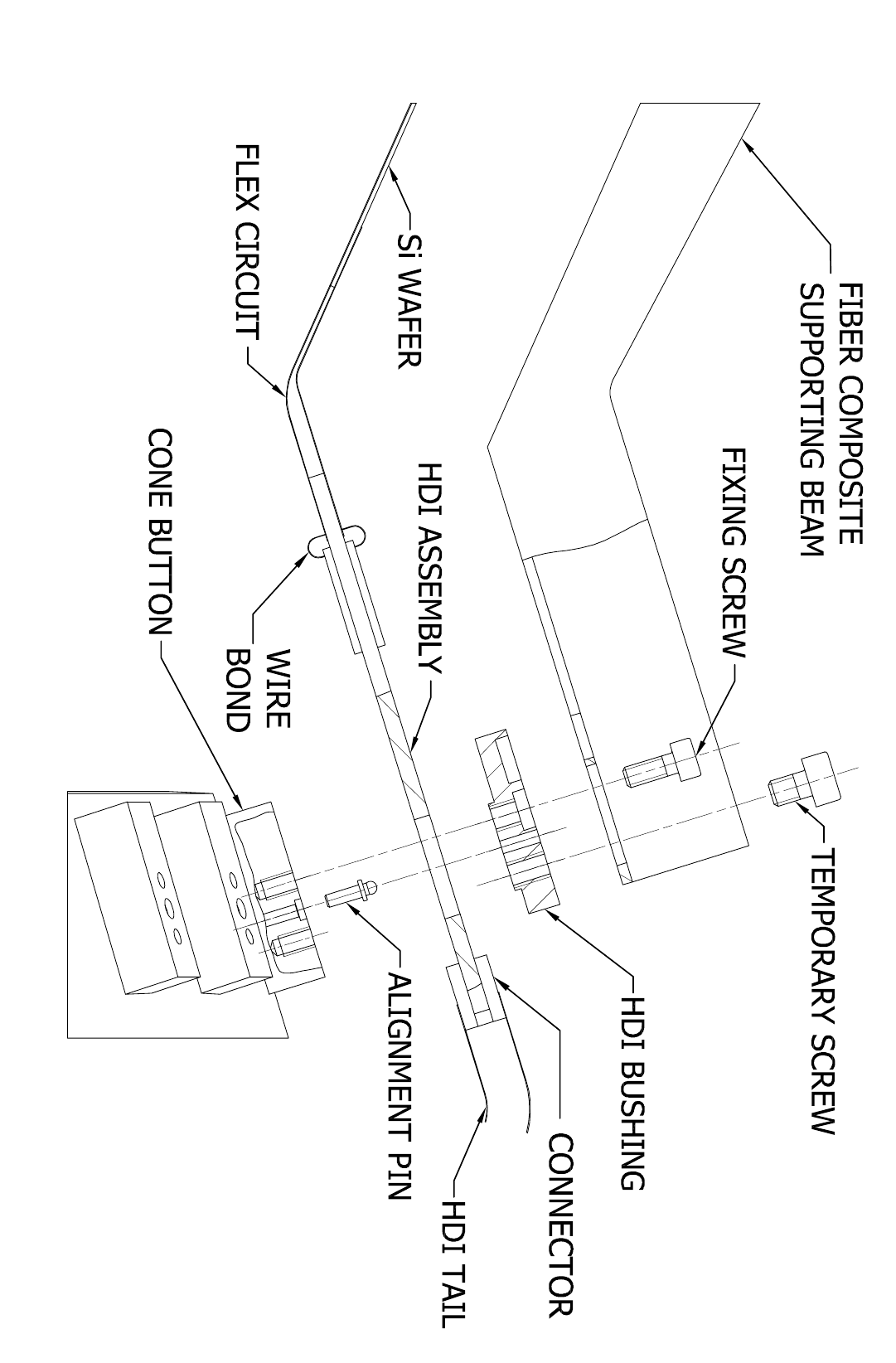}
\end{center}
\caption{The forward foot region.}
\label{fig:ffoot}
\end{figure}

On the backward side, the foot region is more complicated: in case of bad mounting or temperature variations the elongation of the module in the horizontal plane must be  allowed, avoiding over constraint due to the inclined geometry. 
A schematic of the backward foot region is shown in Figure~\ref{fig:bfoot}.
\begin{figure}[tbp]
\begin{center}
\includegraphics[clip=true,trim=4cm 7cm 4cm 7cm, angle=90, width=0.45\textwidth]{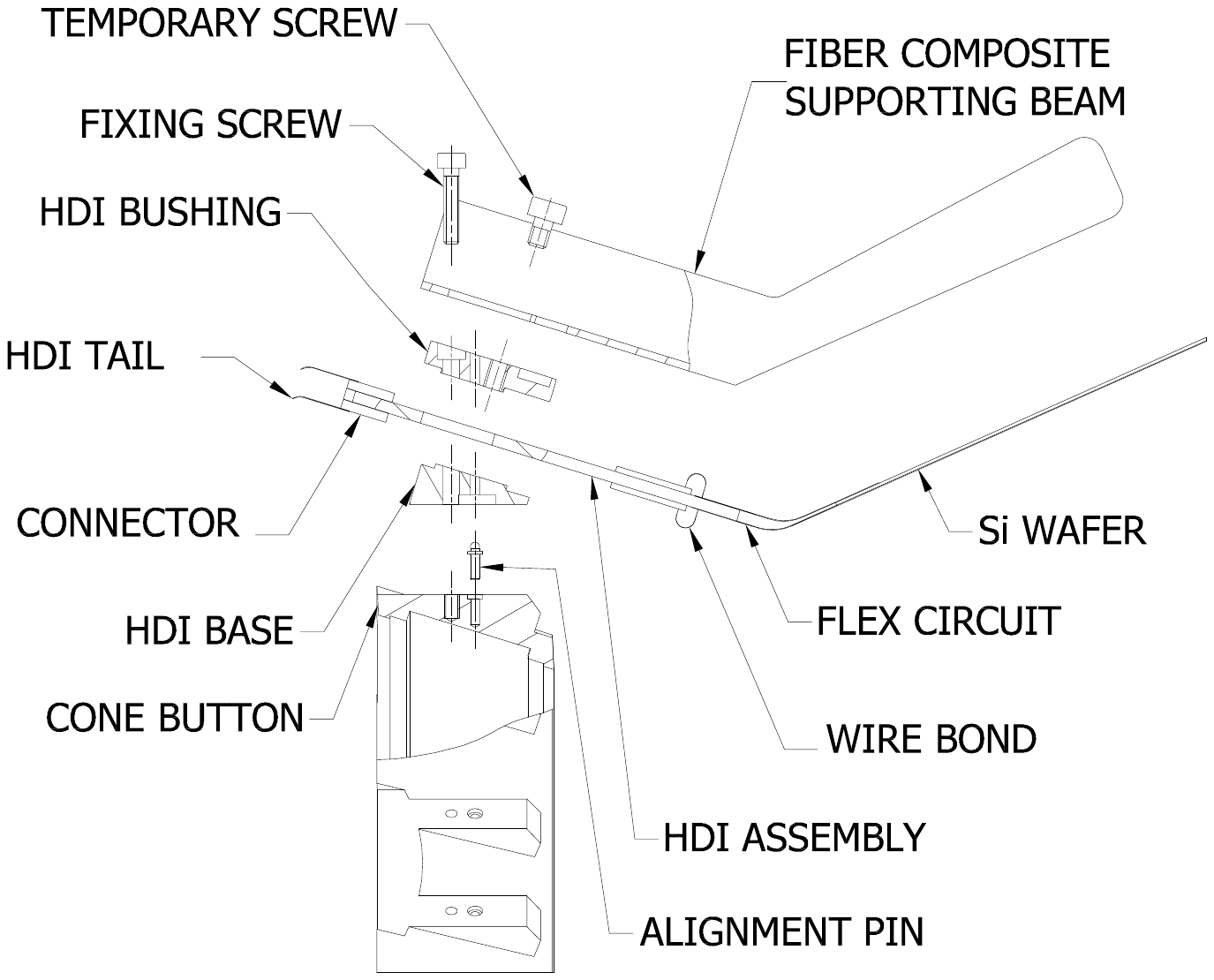}
\end{center}
\caption{The backward foot region.}
\label{fig:bfoot}
\end{figure}

After verification of the alignment, the connection between the two HDIs and the support beam is permanently glued. The glue joint allows for the correction of small errors in the construction of either the cones or the module. After the beam is glued, the module may be removed and re-mounted on the cone as necessary. The design of the foot allows this glue joint to be cleaved and remade should major repair of the module be required. After the detector module is mounted, it is electrically tested to verify its functionality. As each layer  is completed, it is optically surveyed and the data are entered in a database. Finally, the two cones are connected together with the space frame, resulting in a completed detector assembly ready to be opened and re-mounted around the layer0.

\paragraph{Space Frame}
The two carbon fiber support cones are mechanically connected by a low mass 
carbon fiber space frame (see Figure~\ref{fig:spaceframe}).\\
The struts of the frame are made of carbon fiber tubing and the rings are 
carbon fiber skins with foam core. The frame is constructed by gluing the 
pieces together while they are held rigidly in fixtures. The joints between the 
struts and the rings use end pieces of carbon fiber material machined to fit 
the inner diameter of the strut. The space frame must split into two halves 
just as the cones do. The rings at the ends of the space frame are sections of 
cones to match the attaching surfaces of the support cones.
\begin{figure}[tbp]
\begin{center}
\includegraphics[clip=true,trim=2cm 5cm 2cm 5cm, angle=90, width=0.5\textwidth]{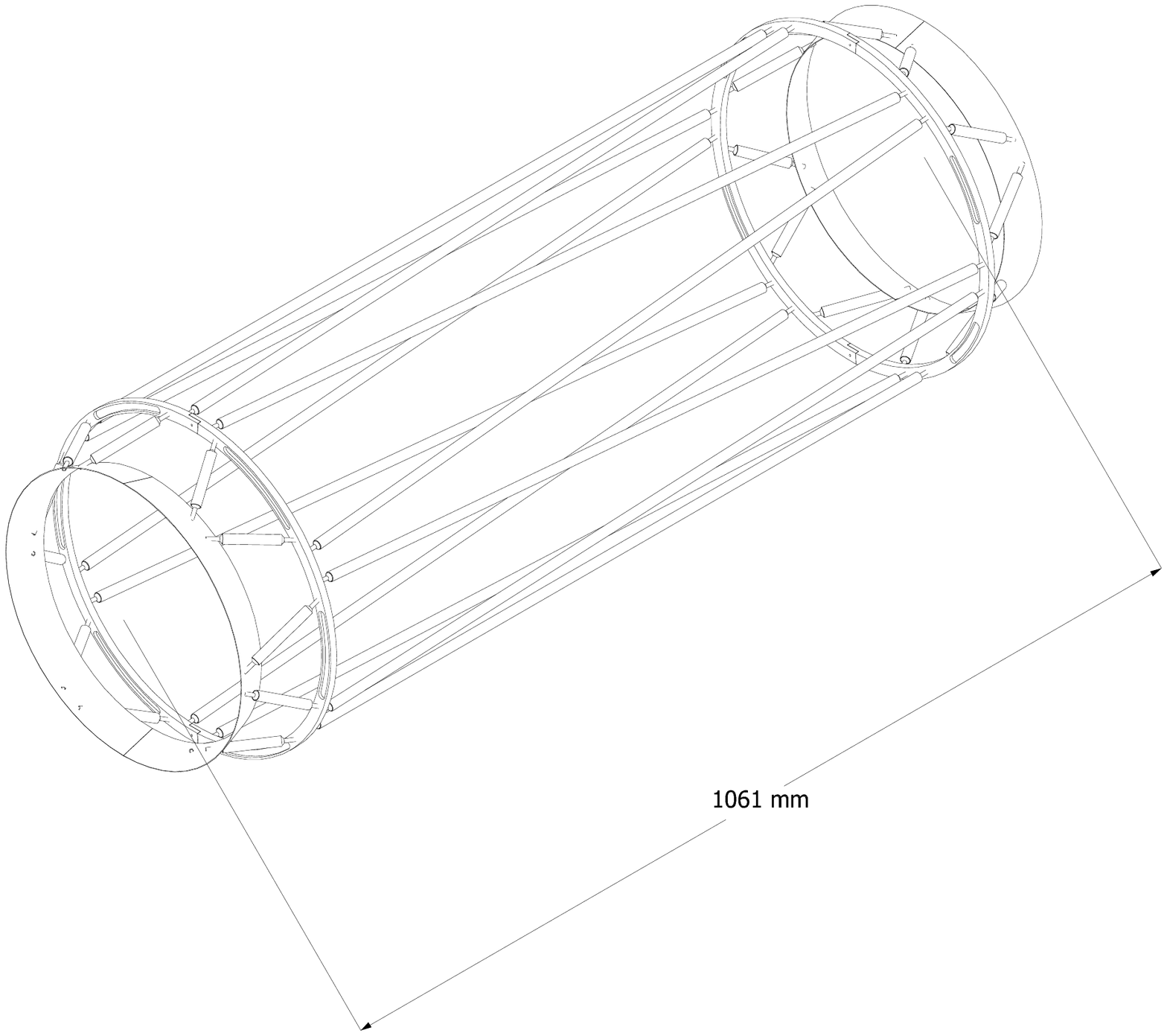}
\end{center}
\caption{The carbon fiber space frame.}
\label{fig:spaceframe}
\end{figure}

\tdrsubsec{Layer0 Half Detector Assembly}
Also the Layer0 detector is assembled in halves in order to be clam shelled around the Be beam-pipe. The detector modules are supported at each end by the cooled half flanges. Cooling water circulates inside each half-flange and cools the protruded mounting wing-pieces supporting the HDIs and comes in thermal contact with them. The Layer0 half-flanges are mated in the vertical plane and have alignment screw-pins to be wrapped in a precise and reproducible way on the robust stainless steel part of the Be beam-pipe.
See Figure~\ref{fig:L0flanges} for a drawing of the Layer0 half-flanges fixed on Be beam-pipe.
\begin{figure*}[tbp]
\begin{center}
\includegraphics[clip=true,trim=3cm 2cm 3cm 2cm, angle=90, width=0.9\textwidth]{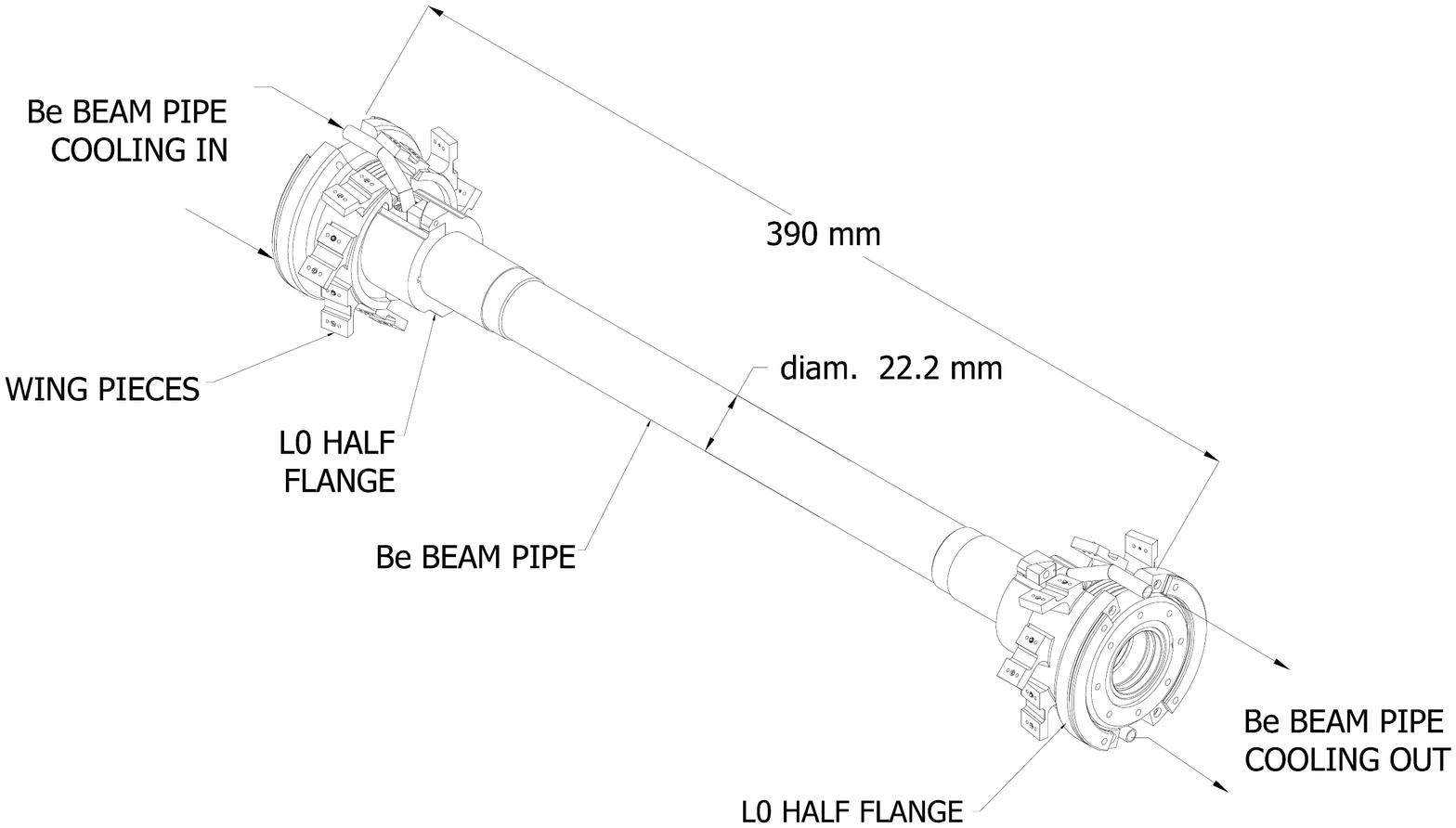}
\end{center}
\caption{The Layer0 half-flanges fixed on the Be beam-pipe.}
\label{fig:L0flanges}
\end{figure*}

During the half detector assembly, the two half-flanges will be held in a fixture in a precise relative position. The detector modules are then mounted to the half-flange wing pieces and positioned through precision pins and then screwed down.  The procedure to mount a Layer0 module is the same as that used for the L1-L5 modules, using similar fixtures that function in the same way as those for L1-L5.
The connection between the modules and the half-flange wing piece feet provides accurate and reproducible alignment of a module and conduction of heat from the HDI heat sink to the cooling water circulating in the half-flanges.  The technique to mount and connect the HDI to the support beam is the same as that described for L1-L5 modules. Also in this case, the glue joint allows for the correction of small errors in the construction of either the half-flange or the module. After each Layer0 detector module is mounted it is electrically tested  to verify its functionality. Once the Layer 0 is complete, it is optically surveyed and the data are entered in a database. Finally the two half-flanges are connected together with a proper fixture forming a rigid half shell and are then ready to be assembled around the Be beam pipe.
An illustration of the two Layer0 half-shells around the Be beam-pipe is shown Figure~\ref{fig:L0shells}.
\begin{figure*}[tbp]
\begin{center}
\includegraphics[clip=true,trim=1.0cm 1.0cm 1.5cm 4.0cm, width=0.9\textwidth]{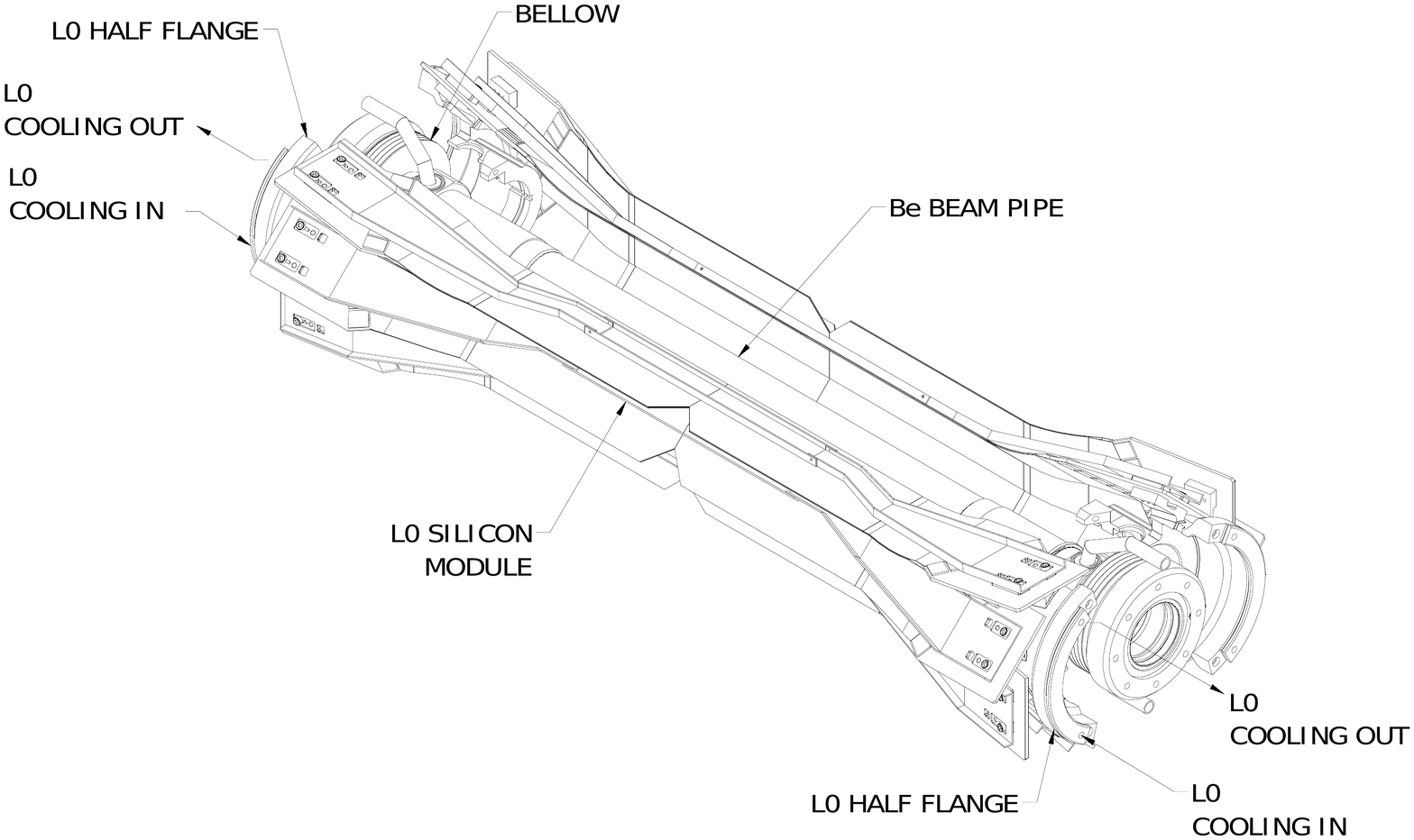}
\end{center}
\caption{The two Layer0 half-shells around the Be beam-pipe.}
\label{fig:L0shells}
\end{figure*}

\tdrsubsec{Mount Layer0 on the  Be-pipe and L1-L5 on the conical shield}
When the two Layer0 half detector assemblies are completed, they are brought to the staging area where the Be beam pipe, conical shields and cryostats are already mounted. The assembly fixture used to hold Layer0 half-shells,  is also used to bring together and clam-shell them around the Be beam-pipe. An isostatic mounting of the Layer0 with respect to the Be beam-pipe is equipped, to allow relative motion due to differential thermal expansion. Radial and circumferential keys ensure precise and reproducible positioning of the Layer0 flanges and thus of the detectors with respect to the Be beam-pipe. The two half detectors are mated and locked together with pin-screws. The cables from HDIs are attached to the conical shields and routed to the transition cards, which are mounted in cooling support flanges at the end of the conical shield. An optical survey is performed and the detector is tested.
In the same way, when the two L1-L5 half-detector assemblies are completed, they are brought to the staging area where both conical shield  and  Final Focus superconducting Magnet Cryostat  are mounted and Layer0 has been assembled on the Be beam-pipe. Fixtures are employed to hold the cones as they are brought together and clam-shelled around the Layer0 and Be beam-pipe. The two half detector assemblies are mated and the latches between them are closed. The cables from HDIs are attached on the cones and routed to the transition cards. The entire detector is then thoroughly tested. Afterwards the Temporary Cage Sectors are mounted on the Conical Shield flanges and rigidly fixed in place. At this point the assembly is relatively rigid and can be transported to the interaction hall and installed in the experiment.

\tdrsubsubsec{Gimbal Rings}
The detector assembly as described above forms a rigid structure as long as the cones and space frame are connected together.
This structure is supported on the conical shields. During the transportation of  the IR Assembly to the interaction hall, it is possible for the forward side of the IR Assembly to move by as much as 1 mm relative to the backward side. This motion is reversible, and they will return to their original alignment when installed under normal conditions. In addition, differential thermal expansion may affect the relative alignment of the magnets during periods in which the temperature is not controlled, and relative motion of the magnet and the beam-pipe may occur should there be seismic activity.  

The support of the detector on the conical shield must allow for this motion without placing stress on the silicon wafers. In addition, the position relative to the IR must be reproducible when installed. These constraints are met by mounting the support cones on a pairs of Gimbal Rings. One gimbal ring connects the forward cone to the Conical Shield  to constrain its center, while allowing rotation about the \textit{x} and \textit{y} axes. A second set of Gimbal Rings supports the  cone in the backward direction in a similar manner, with an additional sleeve that allows both for motion along \textit{z} and rotation about the \textit{z} axis.

\tdrsubsec{Installation \& Removal of the Complete IR Assembly in \superb}
The clearance between the SVT Layer0 and the beam pipe is of the order of 1 to 2 mm. During the transportation of the IR Assembly, the critical 
clearances must be monitored in real time to ensure that no accidental damage to the detectors occurs.
In its final position, the IR Assembly will be supported along the External Tube on the Cylindrical Shield. \\ 
One possible installation/removal scenario employs a crane that handles a long stiff beam that rigidly supports the entire IR Assembly 
through the forward and backward external tube and it is able to carefully lay this system on the external cradle support shown in 
Fig:~\ref{fig:qdbefore_after}. 
This support is aligned with respect to the Cylindrical Shield and positioned in front of the forward end of the detector. \\
With this procedure it will be possible to slide the IR Assembly (in or out) within the DCH using the translation 
system of the cylindrical 
shield,  described in Sec.~\ref{sec:temporary_cage}, which has been previously aligned with respect to the \superb detector. 
Once the IR assembly is in the final position the External Mechanical Cradle is removed.
			
\tdrsubsubsec{Quick Demounting of the SVT, Layer0 and Be beam-pipe} 
%
\begin{figure*} [tbp]
\centerline{\includegraphics[clip=true,trim=6cm 2cm 7cm 1cm, angle=90, width=\textwidth]{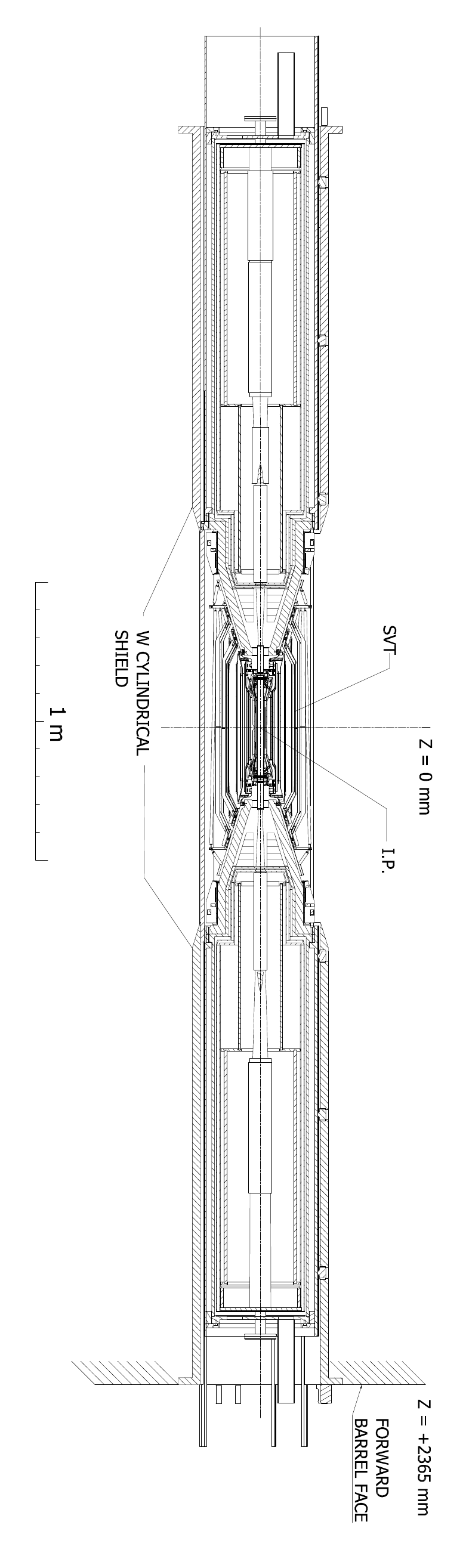}}
\centerline{\includegraphics[clip=true,trim=7cm 2cm 6.5cm 2cm, angle=90, width=\textwidth]{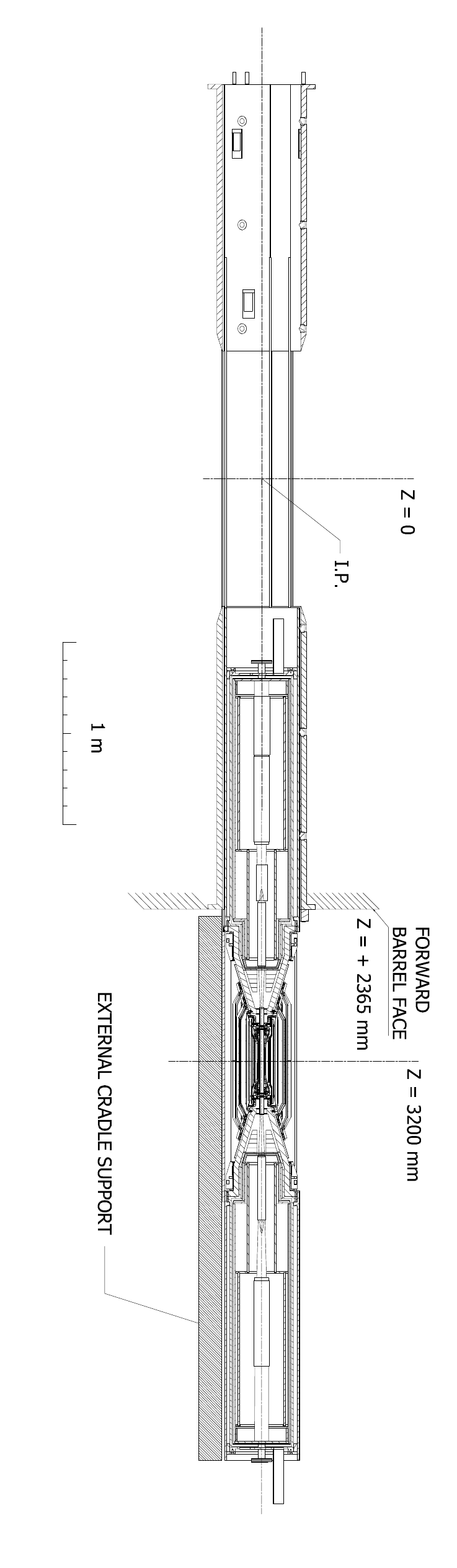}}
\caption{Quick demounting operation: a longitudinal section in the y-z plane of the IR system in the initial (top) and final (bottom) position.}
\label{fig:qdbefore_after}
\end{figure*}

For a rapid access to some of the IR components (SVT, Layer0 and the Be beam-pipe) the 
so called quick demounting procedure has been developed to allow experts to operate on these components inside 
the experimental hall, avoiding the complete removal of the IR assembly from the \superb detector, and then reducing the associated 
down time with respect to what was needed in \babar\ (several months).
This operation can be followed in case that Layer0 
needs to be repaired or replaced with a new detector in a short amount of time, as well as for a rapid access to the SVT or 
the Be beam pipe. 
As shown in Figure~\ref{fig:qdbefore_after} the procedure foresees that one slide the IR Assembly only partially outside the detector, 
leaving it close to the forward door, in order to access the central region of the IR and perform any necessary work, 
and then slide the entire IR Assembly back inside the detector. 
The quick demounting operation plans to rigidly move all the IR Assembly components along the \textit{z} axis in the forward direction up to a position that 
allows the SVT detector to be completely out of the forward side of the iron of the magnet return yoke (forward end plug open), at \textit{z}=+2650 mm.
In this hypothesis it is assumed that the cylindrical backward and forward shields are rigidly attached on a solid structure and perfectly aligned 
along the \textit{z} axis direction, having a supporting function in the IR translation.\\
The stroke necessary for the SVT demounting position is about 3200 mm.
The total weight of the IR assembly components is of about 1.65 t.



%
%
A beam profile box of about 3$\times$4$\times$3 $m^3$ volume will be mounted on the forward end of the detector and it will be equipped with the proper filters and fans to maintain ISO 8 cleanness conditions for a four-person team working on the SVT, Layer0, beam-pipe.  

\tdrsubsubsec{Temporary Cage and Translation System}
\label{sec:temporary_cage}

The SVT system represents the weak ring of the the IR assembly mechanical chain. The Be-beam pipe is joined to the Cryostat beam-pipe through a system of  flexible bellow flanges. In a different way from \babar\, where the SVT region was stiffened by the CF support tube, in \superb\ it was decided not to insert any stiffening structure that would add passive material in front of the DCH. Therefore it was necessary to design a temporary and removable structural support (the  temporary cage). This rigidly fixes the forward and backward conical shields around the SVT region; it is able to absorb all the mechanical stress that could be present during all the installation/removal 
operations; it can be removed by operating from the forward end region of the detector, once the SVT is in data-taking position.  On doing so one removes the passive material of the temporary cage from between the SVT and the DCH in 
preparation for data taking.\\
When the temporary cage is mounted and rigidly connects  the two opposite conical shields,  the whole IR assembly (forward side + backward sides + SVT) can be considered a rigid body supported by several recirculating spheres. These spheres are embedded in the cylindrical shields, acting on the three rails positioned at 120${^\circ}$ that are formed on the external tube profile. The translation system is able to guide the entire IR assembly and prevent any rotation during the installation/removal.  The cylindrical shields also include a mechanical system (the Radial Blocking Device) able to rigidly block the IR assembly at the correct position with respect to the IP. 
This blocking is enacted by longitudinal bars that push on a mechanical conical device embedded in the cylindrical shield, able to block radially the external tube with respect to the cylindrical shield. This blocking system is also useful at the moment of the temporary cage mounting/demounting on the conical shield flange operations, in order not to transfer any mechanical stress to the SVT detector and the Be beam-pipe.\\
Due to the presence of translation rails on the external tube, the temporary cage is cylindrically shaped in three independent separate sectors, confined in a radial space of about 2 cm between the external tube and the cylindrical shield.  The temporary cage sectors are  made from a metal sandwich structure with very high flexural resistance.  Each temporary cage sector is moved towards the IP position supported by two removable beam-rails mounted and embedded in the cylindrical shield. The beam-rails have sufficient length in order to be fixed on the opposite cylindrical shields in order to support the temporary cage sector. The temporary cage sector has a special mechanical connection on the front side in order to perform a coupling in a secure conical way on the backward conical shield flange. On the backward side it has a special radial bushing device able to be fixed to the forward conical flange avoiding any mechanical stress to the SVT detector. The temporary cage sector fixing screws are tightened with a long special screw driver by acting at the working area in front of the forward end of the detector and in front of the Horse Shoe region (backward side).
A mechanical Support Cradle Facility has the function of extending the rail and support for the IR Assembly 
when it is slid in to the final working position outside the forward end of the \superb\ detector (see Figure~\ref{fig:qdbefore_after}).
An illustration of the IR Assembly showing some components of the installation/removal operation is represented 
in Figure~\ref{fig:artistic_v}.
\begin{figure}[tbp]
\begin{center}
\includegraphics[clip=true,trim=4.5cm 10.5cm 3.5cm 7.5cm, angle=90, width=0.45\textwidth]{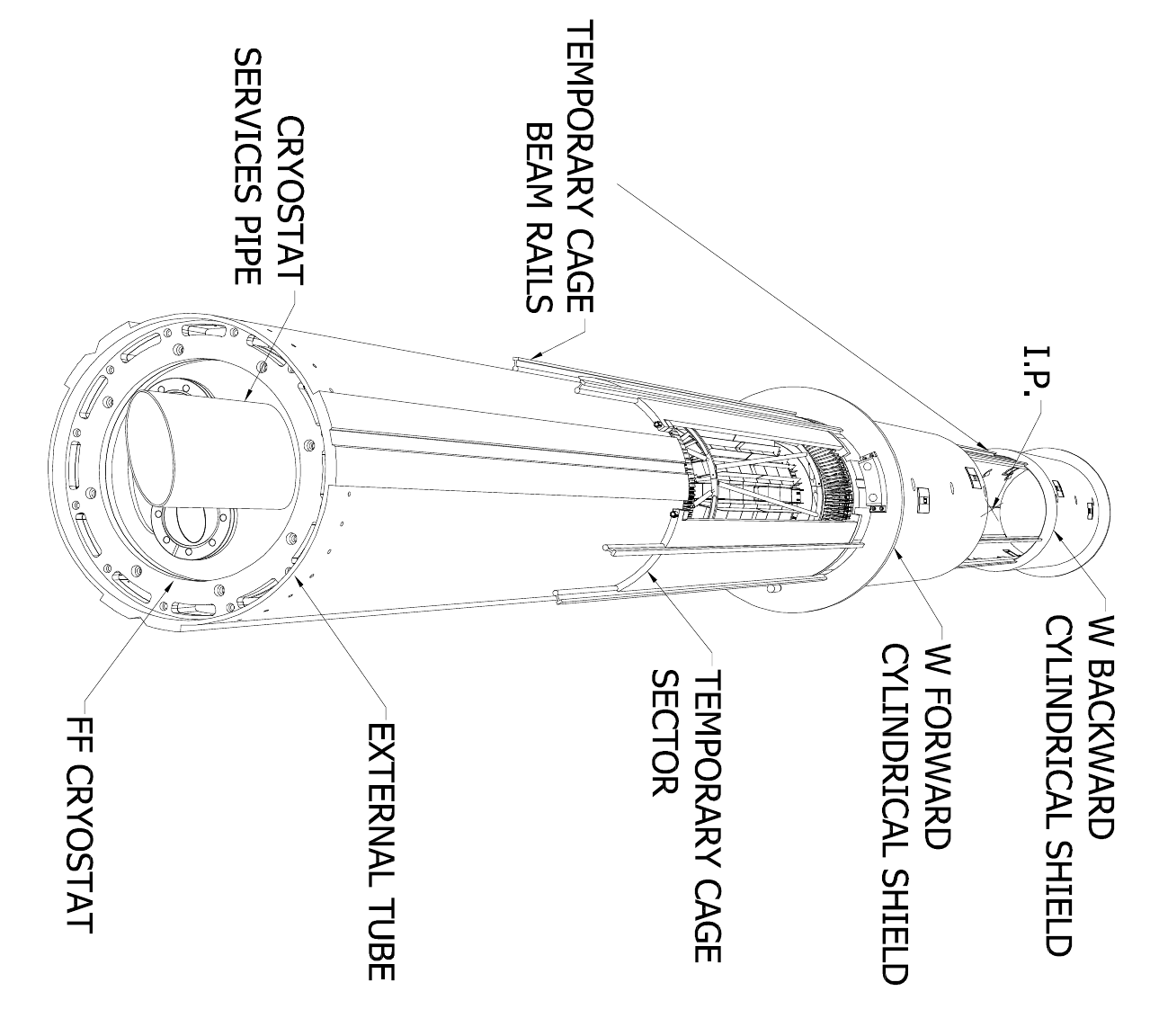}
\end{center}
\caption{An artistic view of the IR Assembly outside the forward end of the \superb\ detector,  showing some components 
used for the installation/removal operation.}
\label{fig:artistic_v}
\end{figure}
To enable the sliding of the IR Assembly an extension with flexible pipe connections has to be foreseen 
for the cryogenic service to the Cryostat Final Focus Superconductor Magnets.  
Monitoring position devices are planned to be installed on the cylindrical backward and forward Shields in order to control the distances of the various components during the mounting-demounting operations.
Also a strain-gauge set is planned to be mounted on the Temporary Cage Sectors to monitor the mechanical stress in the different positions during the translation process.

\tdrsubsec{Detector Placement and Survey}

The SVT must provide a spatial resolution of the order of 10 $\mu$m. The final location of each of the wafers relative to each other (local alignment) and to the DCH (global alignment) will be determined by alignment with charged-particle tracks. This requires a certain degree of overlap of the modules within a layer.  There must be overlap in z as well as $\phi$, so as to accurately locate the z position of the wafer in a single module with respect to each other. 

Mechanical tolerance and measurements must be such that the process of detector alignment with charged-particle tracks converges in a reasonable time. 
In \babar the optical survey precision was estimated to be 5 \mum in the wafer plane and 20 \mum out of the plane.

This leads to the requirement that the relative position of the various wafers be stable to the 5 $\mu$m level over long periods of time (months or more).
The position of the entire detector structure with respect to the DCH can be followed more easily, so that variations of the order of few hours can be tracked. 

The relative positions of silicon wafers should be stable in order to avoid the need for frequent local alignments with tracks.
Preliminary calculations of the thermal expansion of the entire structure predict of the order of 1 $\mu$m/$^\circ$C change over the length of the detector active region. If the temperature inside the SVT is maintained constant within 1 $^\circ$C, thermal expansion is not a problem.
	 
\tdrsubsec{Detector Monitoring}		

\tdrparagraph{Position Monitoring System}
Although the final placement of the silicon wafer will be measured and monitored with charged particle tracks which traverse the silicon detector and drift chamber, two displacement monitor systems will be designed to measure relative changes in the position of the silicon detector with respect to the machine elements and the cylindrical shield. One displacement monitoring system will be used to monitor the relative position during transportation of the IR Assembly with the silicon detector inside it and during data taking. This system consists of either a capacitive displacement monitor or a LED-photodiode reflection monitor which are sensitive to relative displacements between the silicon detector and the machine components such as the beam pipe, magnet and Cylindrical Shield. \\
In addition, a laser system will monitor the displacement of the outer layer of detectors with respect to the drift chamber during data taking. Given that the SVT layers are not mounted on the same support as the drift chamber, it is possible that motion between the two will occur. To monitor this motion, short infrared laser pulses are brought in with fiber optics (e.g., 50 $\mu$m core diameter) which are attached to the drift chamber. The laser light shines in the active region and reaches the outer layers of the silicon detector to calibrate the relative position of the DCH and SVT.

\tdrparagraph{Radiation Monitoring}
To protect the silicon detector system against potentially damaging beam losses and to monitor the total radiation dose that the detectors and electronics receive, diamond detectors will be installed on a crown around the beam-pipe, in the vicinity of Layer0. If the radiation dose exceeds a certain threshold, a beam dump signal will be sent to the accelerator control-room. This sort of radiation protection system has already been used successfully in the \babar experiment and it is described in detail in Sec.\ref{sec:svt_rad_mon}.

\tdrsubsec{R\&D Program}
The following R\&D projects are planned before the final design of the SVT mechanical configuration is finalized.		
\tdrparagraph{Cables}
Prototypes of the cable from the hybrid to the transition card will be constructed. This allows one to verify the detail of cable routing plans and mechanical robustness. It will also allow the electrical properties to be measured to verify simulations.
\tdrparagraph{Hybrid}	
Realistic mechanical models of the High Density Interconnect (HDI) are required. The HDI is a critical element both in the cooling of the electronics and the mounting of the detector modules.  Models will be tested for heat transfer capability and for the module mounting schemes.
\tdrparagraph{Be Beam pipe}
A full scale of Be beam pipe will be constructed with the cooling design actually available. The part in Be will be realized in a light Aluminum alloy and the cooling and structural tests performed will be renormalized towards Be material. 
\tdrparagraph{Layer0 Module}
A full-scale mock-up of the Layer0 module and its cooled supporting flanges will be constructed. This will be used to verify  thermal stability calculations and to investigate the effect of nonuniform beam pipe cooling. It will also be used to design Layer0 assembly fixtures and test and practice the assembly technique.
\tdrparagraph{Inner layer sextant}
A full-scale mock-up of the L1-L2 layer sextants will be constructed. This will also be used to test and practice assembly technique.
\tdrparagraph{Arch modules}
Full-scale mock-ups of the arch detector module will also be constructed and used with the prototype cones to verify mechanical stability and mounting techniques.
\tdrparagraph{Cones and space frame}
A set of prototype cones and space frame will be built to provide realistic test of cooling, mechanical rigidity and thermal stability. In addition they will be used to design L1-L5 assembly fixture and test the assembly technique.		
\tdrparagraph{Full-scale model of IR and Cylindrical Shield}
A model of the entire IR Assembly with the beam pipe near the IP and forward/backward tungsten cylindrical shield will be constructed. This will allow the identification of any interference problems and verify mounting schemes. It will also provide a test bench for the design of various installation fixtures.
\tdrparagraph{Quick Demounting test}
Tests of the Quick Demounting operation will be made using the full scale model of IR Assembly and 
cylindrical shield.  Sensor gauges will be installed on the SVT mock up module to measure stress and deformation eventually induced by relative movement of the forward/backward cryostat, although blocked  by temporary cage sectors. Also the translation mount/demount stages and blocking of the IR Assembly with respect to the Cylindrical shield will be tested.

\tdrsec{Layer0 Upgrade Options} 
\label{sec:pixel_upgrade}


With the machine operating at full luminosity, the SVT Layer0  may benefit from 
being upgraded to a pixellated detector that has a more stable performance in 
high background conditions than the baseline striplet option, as discussed in Sec.~\ref{sec:layer0_upgrade_overview}. 
In particular:

\begin{itemize}

\item the occupancy per detector element from machine background is expected to fall to a few kHz, with a major impact on the speed specifications for the front-end electronics, mainly set by the background hit rate in the case of the striplet readout chip;

\item better accuracy in vertex reconstruction can be achieved in presence of relatively high background; the shape of the pixel can be optimized in such a way to reduce the sensor pitch in the $z$ direction while keeping the area in the range of 2500-3000~$\mu$m$^2$, which guarantees enough room for sparse readout functionality.

\end{itemize}



\noindent Two main technologies are under evaluation for the upgrade of Layer0:
hybrid pixel and thinner CMOS Monolithic Active Pixel Sensor (MAPS).
The first option, mature and with adequate radiation hardness and readout speed capability for application 
in Layer0, tends to be too thick and has too large a pitch in the detectors presently operated at the LHC. 
On the contrary for CMOS MAPS, that can be realized with small pitch and material budget, the requirements on 
readout speed and radiation hardness are more challenging.

Specific R\&D activities are in progress to understand advantages and potential issues of the different options described in the  
following sections.

\tdrsubsec{Hybrid pixels}

Hybrid pixel technology has reached quite a mature stage of development. Hybrid pixel detectors are currently used in the LHC experiments~\cite{peric06,reiner07,bortoletto07,kluge07}, with a pitch in the range from 100~$\mu$m to a few hundred $\mu$m, and miniaturization is being further pushed forward in view of the upgrade of the same experiments at the High Luminosity LHC (HL-LHC)~\cite{garcia-sciveres11,hugging11,favaro11}. Hybrid pixel systems are based on the interconnection between a sensor matrix fabricated in a high resistivity substrate and a readout chip. Bump-bonding with indium or indium-tin or tin-lead alloys is the mainstream technology for readout chip-to-sensor interconnection. 

The design of a hybrid pixel detector for the SVT innermost layer has to meet some challenging specifications in terms of material budget and spatial resolution. 
Since the readout chip and the sensor are laid one upon the other, hybrid pixels are intrinsically thicker detectors than microstrips. Interconnect material may further degrade the performance, significantly increasing the radiation length equivalent thickness of the detector (see Table~\ref{table:material_budget}).
As far as the readout and sensor chips are concerned, substrate thinning to 100-150~$\mu$m and subsequent interconnection are within the present technological reach. Further thinning may pose some issues in terms of mechanical stability and, as the detector thickness is reduced, of signal-to-noise ratio and/or front-end chip power dissipation. Concerning interconnection, the vertical integration processes currently under investigation in the high energy physics community might help reduce the amount of material. Among the commercially available technologies, the ones provided by the Japanese T-Micro (formerly known as ZyCube), based on so called micro-bumps, and by the US based company Ziptronix, denoted as direct bonding technique, seem the most promising~\cite{garrou08}. The Fraunhofer EMFT has developed a bonding technique called SLID and based on a very thin eutectic Cu-Sn alloy to interconnect the chips~\cite{klumpp04}. 

The spatial resolution constraints set a limit to the area of the elementary readout cell and, as a consequence, to the amount of functionalities that can be included in the front-end electronics. 
A planar, 130~nm CMOS technology may guarantee the required density for data sparsification and in-pixel time stamping in a 50$\times$50~$\mu$m$^2$ pixel area (as already observed, a different aspect ratio might be preferred to improve the resolution performance in one particular direction). 
\begin{figure}[b!]
\begin{center}
\includegraphics[width=0.5\textwidth]{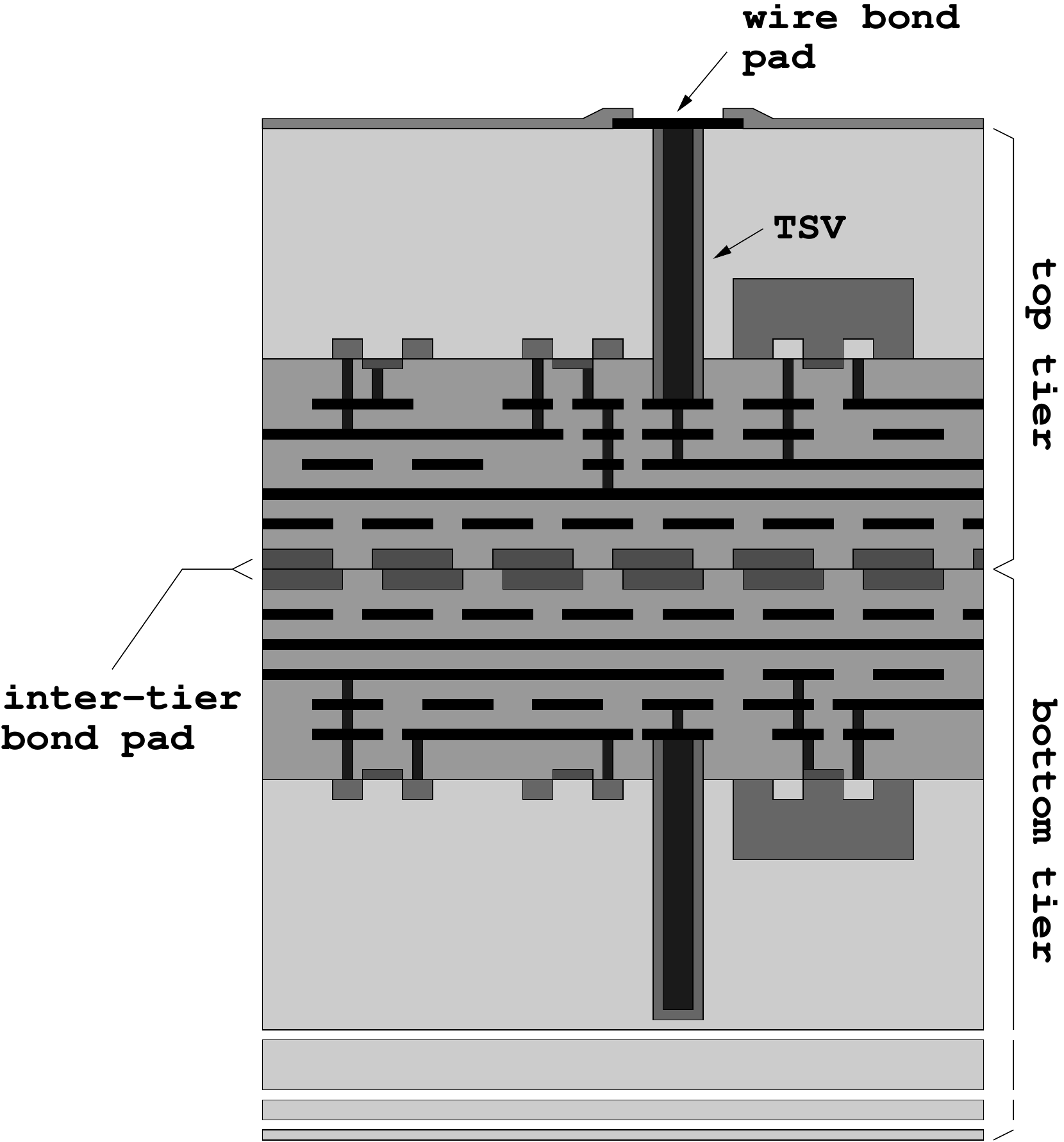}
\end{center}
\caption{Cross-sectional view of a double-layer 3D process.}
\label{fig:3D_tech}
\end{figure}
The above mentioned interconnection techniques can fully comply with the detector pitch requirements (in the case of the T-Micro technology, pitches as small as 8~$\mu$m can be achieved). A fine pitch (30~$\mu$m minimum), more standard bump-bonding technology is also provided by IZM. This technology has actually been successfully used to bond the SuperPIX0 front-end chip (to be described later on in this section) to a 200~$\mu$m thick pixel detector.\\
Denser CMOS technologies belonging to the 90 or 65~nm technology can be used to increase the functional density in the readout electronics and include such functions as gain calibration, local threshold adjustment and amplitude measurement and storage. In this case, costs for R\&D (and, eventually, production) would increase significantly. 

Vertical integration (or 3D) CMOS technologies may represent a lower cost alternative to sub-100~nm CMOS processes. The technology cross section shown in Figure~\ref{fig:3D_tech} in particular points to the main features of the extremely cost-effective process provided by Tezzaron Semiconductor~\cite{patti06}. 
The Tezzaron process can be used to vertically integrate two or more layers, specifically fabricated and processed for this purpose by Chartered Semiconductor (now Globalfoundry) in a 130 nm CMOS technology. In the Tezzaron/Chartered process, wafers are face-to-face bonded by means of thermo-compression techniques. Bond pads on each wafer are laid out on the copper top metal layer and provide the electrical contacts between devices integrated in the two layers. The top tier is thinned down to about 12 $\mu$m to expose the through silicon vias (TSV), therefore making possible the connection to the buried circuits. Among the options available in the Chartered technology, the low power (1.5 V supply voltage) transistor option is considered the most suitable for detector front-end applications. The technology also provides 6 metal layers (including two top, thick metals), dual gate option (3.3 V I/O transistors) and N- and P-channel devices with multiple threshold voltages. 

The main advantages deriving from a vertical integration approach to the design of a hybrid pixel front-end chip can be summarized as follows:

\begin{itemize}

\item since the effective area is twice the area of a planar technology from the same CMOS node, a better trade-off can be found between the amount of integrated functionalities and the detector pitch;

\item separating the digital from the analog section of the front-end electronics can effectively prevent digital blocks from interfering with the analog section and from capacitively coupling to the sensor through the bump bond pad.

\end{itemize}

\noindent The design of a 3D front-end chip for pixel detectors is in progress in the framework of the VIPIX experiment funded by INFN~\cite{giorgi11}. 


\tdrsubsec{CMOS monolithic sensor options}\label{sec:cmosmaps}

\tdrsubsubsec{Deep N-well CMOS MAPS}\label{sec:dnwmaps}

Deep N-well (DNW) CMOS monolithic active pixel sensors (MAPS) are based on an original design approach proposed a few years ago and developed in the framework of the SLIM5 INFN experiment~\cite{slim5}. The DNW MAPS approach takes advantage of the properties of triple well structures to lay out a sensor with relatively large area (as compared to standard three transistor MAPS~\cite{degerli06}) read out by a classical processing chain for capacitive detectors. 
This new design allowed for the first time the implementation of thin CMOS 
sensors with similar functionalities as in hybrid pixels, such as 
pixel-level sparsification and fast time stamping in order to match the readout speed requirements for application in \superb.

As shown by the technology cross section in Figure~\ref{fig:triple-well}, the sensor, featuring a buried N-type layer with N-wells (NW) on its contour according to a typical deep N-well scheme, collects the charge released by the impinging particle and diffusing through the substrate, whose active volume is limited to the uppermost 20-30 $\mu$m thick layer below the collecting electrode. Therefore, within this extent, substrate thinning is not expected to significantly affect charge collection efficiency, while improving 
resolution performance in charged particle tracking applications. As mentioned above, DNW MAPS have been proposed chiefly to comply with the intense data rates foreseen for tracking applications at the future HEP facilities. The area taken by the deep N-well collecting electrode can actually be exploited to integrate the NMOS parts of the analog front-end inside the internal P-well. A small amount of standard N-well area can be used for PMOS devices, instrumental to the design of high performance analog and digital blocks taking full advantage of CMOS technology properties. In this way, both analog functions, such as signal shaping, and digital functions, such as time stamping and data storing, buffering and sparsification, can be included in the pixel operation. Note that the presence of N-wells other than the sensor is instead strongly discouraged in standard MAPS design, where the operation of the tiny collecting electrode would be jeopardized by the presence of any N-type diffusion in the surrounding.
\begin{figure}[t!]
\begin{center}
\includegraphics[width=0.45\textwidth]{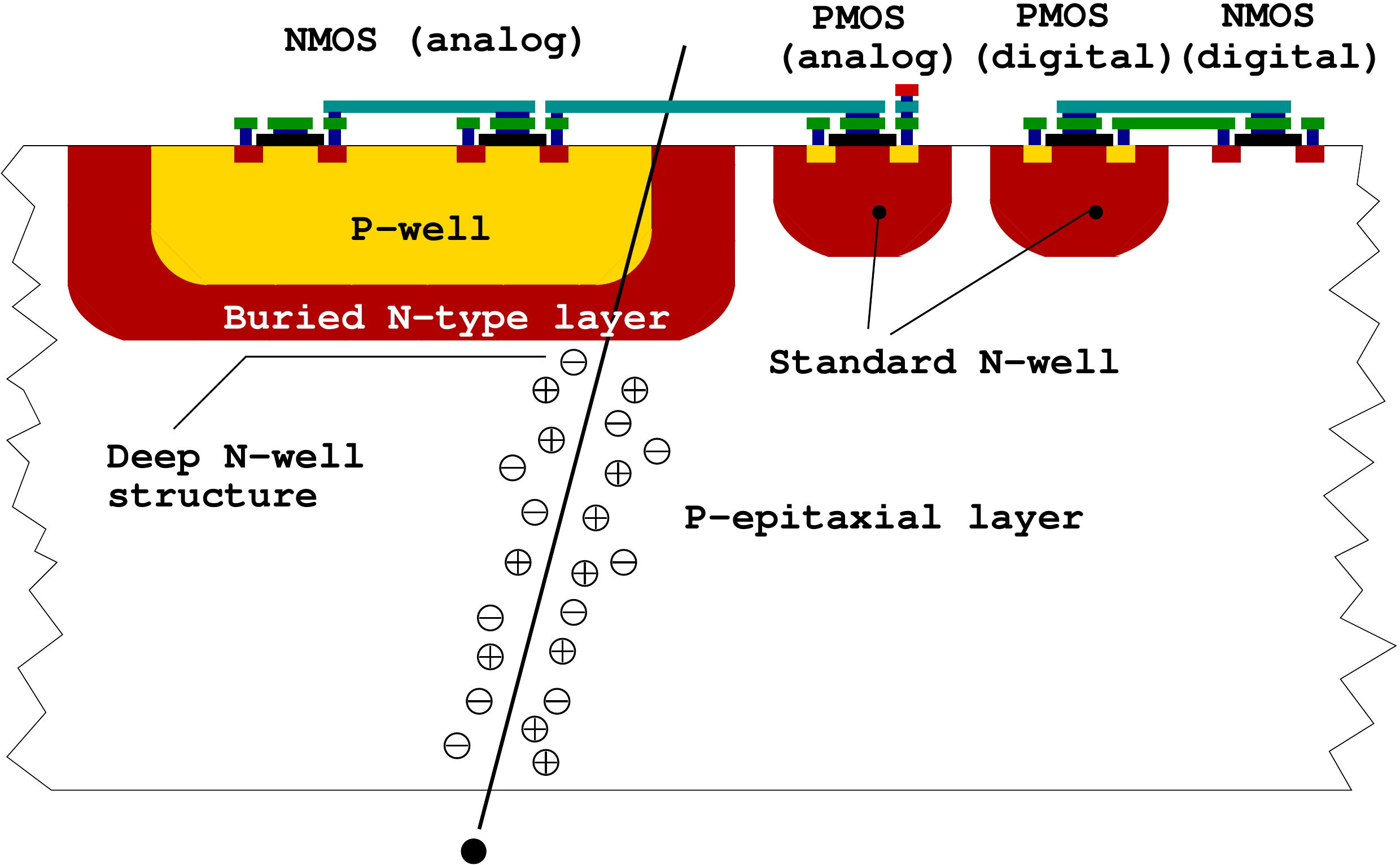}
\end{center}
\caption{Simplified cross-sectional view of a DNW MAPS. NMOS devices belonging to the analog section may be built inside the sensor, while the other transistors cover the remaining area of the elementary cell, with PMOSFETs integrated inside standard N-wells.}
\label{fig:triple-well}
\end{figure}
Based on the concept of the DNW monolithic sensor, the MAPS detectors of the Apsel series (see Section~\ref{sec:apsel}), which are among the first monolithic sensors with pixel-level data sparsification \cite{gabrielli07,stanitzki07}, have been developed in a planar, 130~nm CMOS technology. In 2008, the Apsel4D, a DNW MAPS with 128$\times$32 elements has been successfully tested at the Proton Synchrotron facility at CERN~\cite{bettarini10}. 

\tdrsubsubsec{DNW MAPS in vertical integration technologies}

More recently, vertical integration technologies, like the ones discussed in the previous section for hybrid pixels, have been considered for the design of 3D DNW monolithic sensors. Some specific advantages can derive from the vertical integration approach to DNW MAPS. In particular,  all the PMOS devices used in digital blocks can be integrated in a different substrate from the sensor, therefore significantly reducing the amount of N-well area (with its parasitic charge collection effects) in the surroundings of the collecting electrode and improving the detector charge collection efficiency. The first prototypes of 3D DNW MAPS~\cite{casarosa10,rattinima09} have been submitted in the framework of the 3D-IC collaboration~\cite{3dicsite}. Characterization has started in the last quarter of 2011. 

\tdrsubsubsec{Monolithic pixels in quadruple-well CMOS technology}\label{sec:inmaps}

\begin{figure}[b!]
\begin{center}
\includegraphics[width=0.45\textwidth]{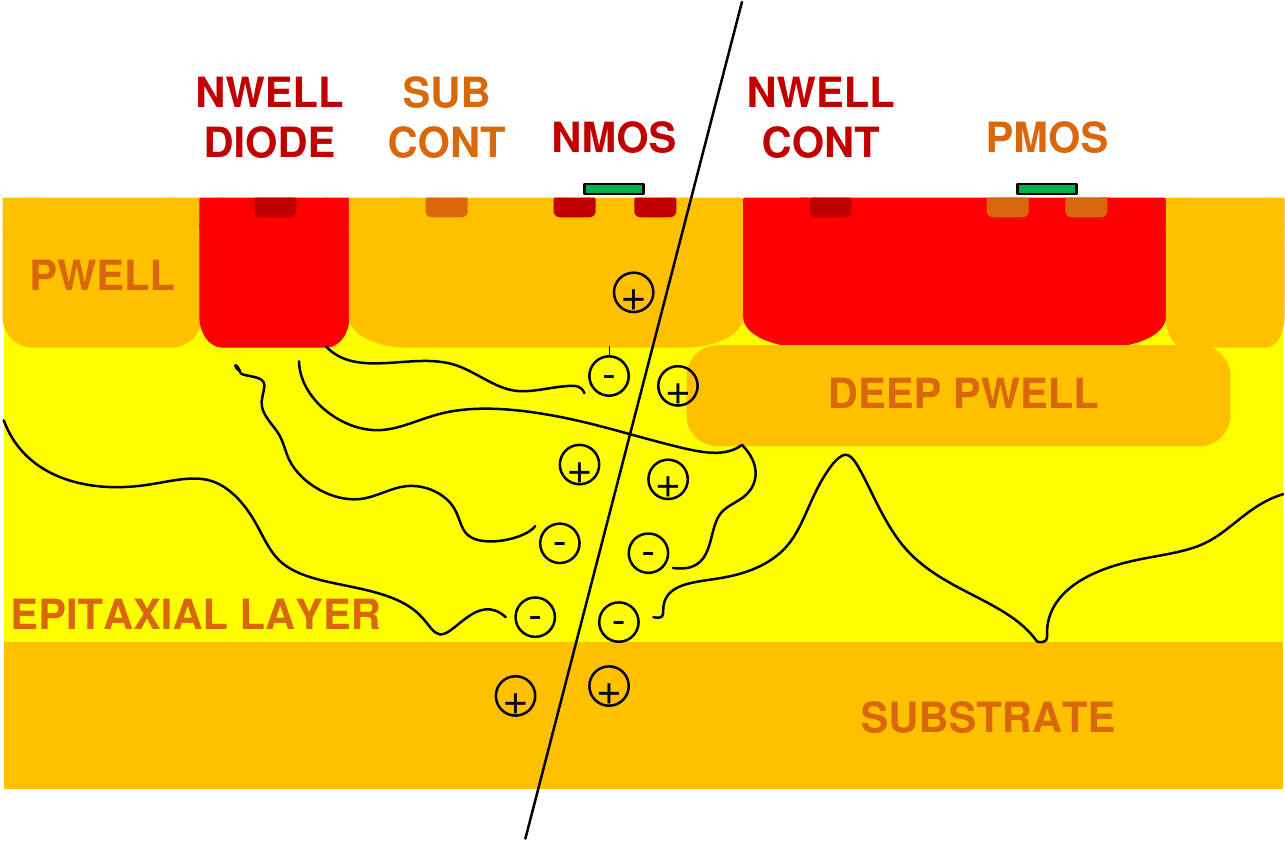}
\end{center}
\caption{Cross-sectional view of the INMAPS CMOS technology; emphasis is put on the deep P-well layer.}
\label{fig:inmaps-crossec}
\end{figure}

In DNW MAPS, charge collection efficiency can be negatively affected, although to a limited extent, by the presence of competitive N-wells including the PMOS transistors of the pixel readout chain, which may subtract charge from the collecting electrode. Inefficiency is related to the relative weight of N-well area with respect to the DNW collecting electrode area. An efficiency of about 92\% was demonstrated in the aforementioned test beam at CERN~\cite{bettarini10}. 

A novel approach for isolating PMOS N-wells has been made available with a planar 180 nm CMOS process called INMAPS, featuring a quadruple well structure~\cite{stanitzki07}. Figure~\ref{fig:inmaps-crossec} shows a simplified cross section of a pixel fabricated with the INMAPS process. By means of an additional processing step, a high energy deep P-well implant is deposited beneath the PMOS N-well (and not under the N-well diode acting as collecting electrode). This implant creates a barrier to charge diffusing in the epitaxial layer, preventing it from being collected by the positively biased N-wells of in-pixel circuits and enabling a theoretical charge collection efficiency of 100\%. The NMOS transistors are designed in heavily doped P-wells located in a P-doped epitaxial layer which has been grown upon the low resistivity substrate. 

The epitaxial layer is obviously expected to play an important role in improving charge collection performance. Actually, carriers released in the epitaxial layer are kept there by the potential barriers at the P-well/epi-layer and epi-layer/substrate junctions. 
The use of high resistivity epitaxial layer, also available in this process,
can further improve charge collection and radiation 
resistance of INMAPS sensors with respect to 
standard CMOS devices.

Epitaxial layers with different thickness (5, 12 or 18~$\mu$m) and resistivity (standard, about 10~$\Omega\cdot$cm, and high resistivity, 1~k$\Omega\cdot$cm) are available. 
A test chip, including several different test structures to characterize both the readout electronics and the collecting electrode performance has been submitted in the third quarter of 2011. Results from the preliminary characterization of the prototypes are discussed in Section~\ref{sec:apsel4well}.

\tdrsubsec{Overview of the R\&D activity}

\tdrsubsubsec{Front-end electronics for hybrid pixel detectors in 130 nm CMOS technology}\label{FEforHP}
A prototype hybrid pixel detector named SuperPIX0 has been designed as a first iteration step aimed at the development of a device to be used for the layer0 upgrade. The main novelties of this approach are the sensor pitch size (50$\times$50~\mum) and thickness (200~\mum) as well as the custom front-end chip architecture providing a sparsified and data-driven readout. 
The SuperPIX0  pixel sensor is made of n-type, Float Zone, high-resistivity silicon wafers, with a nominal resistivity larger than 10~k$\Omega\cdot$cm. 
The SuperPIX0 chip, fabricated in the STMicroelectronics 130nm CMOS technology, is composed of 4096 channels (50$\times$50~$\mu$m$^2$) arranged into 128 columns by 32 rows. Each cell contains an analog charge processor (shown in Figure~\ref{fig:schematic}) where the sensor charge signal is amplified and compared to a chip-wide preset threshold by a discriminator. The in-pixel digital logic, which follows the comparator, stores the hit in an edge-triggered set reset flip-flop and notifies the periphery of the hit. 
The charge sensitive amplifier uses a single-ended folded cascode topology, which is a common choice for low-voltage, high gain amplifiers. 
\begin{figure}[b!]
\begin{center} 
\includegraphics[width=0.48\textwidth]{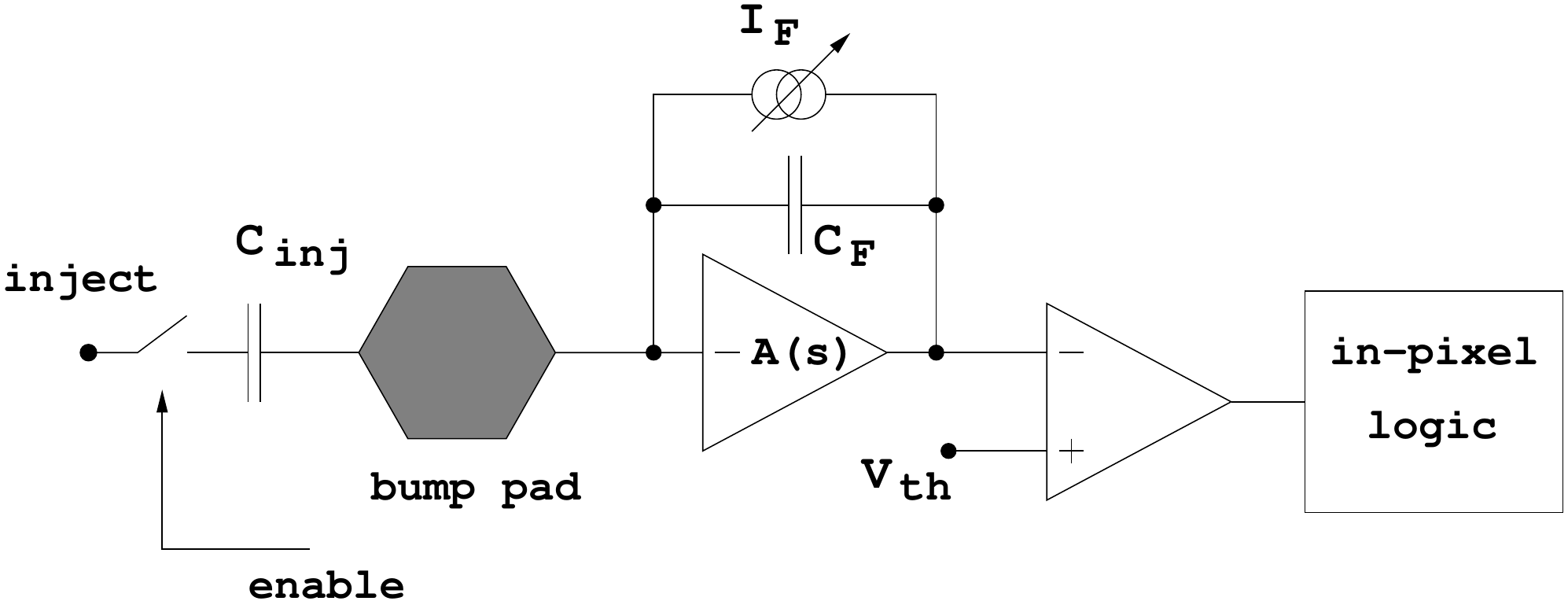}
\caption{Block diagram of the analog front-end electronics for the elementary cell of the SuperPIX0 readout chip.}  
\label{fig:schematic} 
\end{center}
\end{figure}  
The 20~fF MOS feedback capacitor is discharged by a constant current which can be externally adjusted, giving an output pulse shape that is dependent upon the input charge. The peaking time increases with the collected charge and is of the order of 100~ns for 16000~electrons injected. 
The charge collected in the detector pixel reaches the preamplifier input via the bump bond connection. Alternatively, a calibration charge can be injected at the preamplifier input through a 10~fF internal injection capacitance so that threshold, noise and crosstalk measurements can be performed. The calibration voltage step is provided externally by a dedicated line. Channel selection is performed by means of a control section implemented in each pixel. This control block, which is a cell of a shift register, enables the injection of the charge through the calibration capacitance.  Each pixel features a digital mask used to isolate single noisy channels. This mask is implemented in the readout logic. 
The input device (whose dimensions were chosen based on detection efficiency optimization criteria~\cite{ratti09}) featuring an aspect ratio  W/L=18/0.3 and a drain current of about 0.5~$\mu$A, is biased in the weak inversion region. A non-minimum length has been chosen to avoid short channel effects. The PMOS current source in the input branch has been chosen to have a smaller transconductance than the input transistor. The analog front-end cell uses two power supplies. The analog supply (AVDD) is referenced to AGND, while the digital supply is referenced to DGND. Both supplies have a nominal operating value of 1.2~V. Since single-ended amplifiers are sensitive to voltage fluctuations on the supply lines, the charge preamplifier is connected to the AVDD. The threshold discriminator and voltage references are connected to the AVDD and AGND as well. The in-pixel digital logic is connected to the digital supply. The substrate of the transistors is connected to a separate net and merged to the analog ground at the border of the matrix. The SuperPIX0 chip has been fabricated in a six metal level technology. Special attention has been paid to layout the channel with a proper shielding scheme. Two levels of metal have been used to route the analog signals, two for the digital ones and two for distributing the analog and digital supplies. The supply lines, at the same time, shield the analog signals from the digital activity.
\begin{figure}[t!]
\begin{center} 
\includegraphics[width=0.48\textwidth]{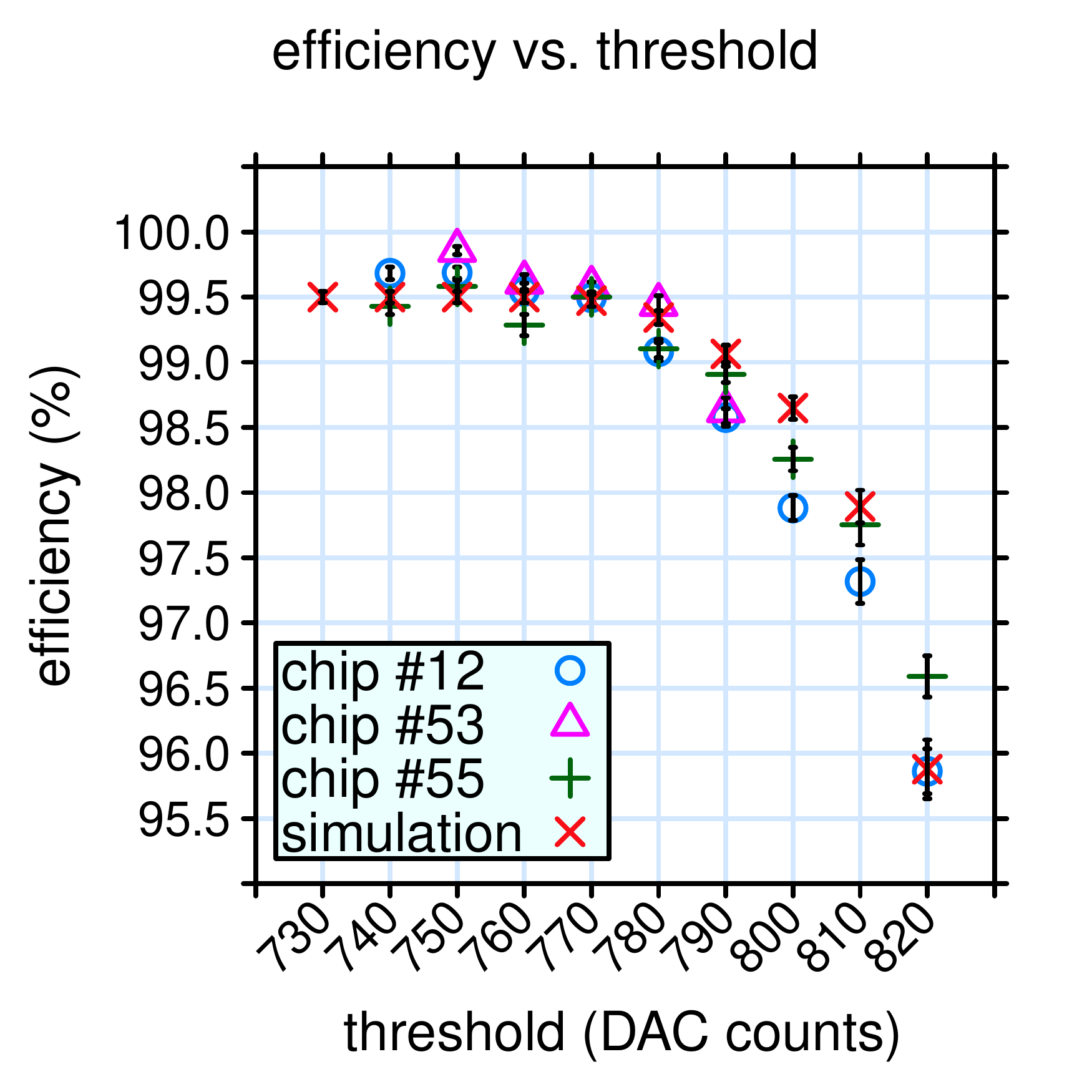}
\caption{Superpix0 efficiency as a function of the voltage discriminator threshold in the case of normal incidence angle.}  
\label{fig:spix0_eff} 
\end{center}
\end{figure}  
For nominal bias conditions the power consumption is about 1.5~$\mu$W per channel. 
More details on the design of the analog front-end can be found in the literature~\cite{traversi11}. 
The measured threshold dispersion in the chip is around 490 $e^-$ with an average pixel noise of about 60 $e^-$ (without the sensor connected). Since the threshold dispersion is a crucial characteristic to be considered in order to meet the required specifications in terms of noise occupancy and efficiency, circuits for threshold fine-adjusting have to be implemented in the next version of the chip. These results have been extracted using the gain measured with an internal calibration circuit, implemented in the pixel, injecting a charge from 0 to 12 fC in each channel preamplifier. An average gain of about 40 mV/fC with a dispersion at the level of 5$\%$ has been obtained. The front-end chip has been connected by bump-bonding to a high resistivity pixel sensor matrix of 200~$\mu$m thickness. The bump-bonding process has been performed by the Fraunhofer IZM with electroplating of SnAg solder bumps. Measurements on the bump-bonded chip show a working sensor and a good quality of the interconnection at 50~$\mu$m pitch. The measured gain and threshold dispersion are compatible with the ones extracted from the front-end chip only. We observe an increase of the noise of around 20$\%$, up to about 76 $e^-$, due to the added capacitive load of the sensor connected. The Superpix0 chip, bump bonded to a high resistivity silicon pixel detector, was also tested on the beam of the Proton Synchrotron (PS) at CERN. The measured efficiency is shown in Figure~\ref{fig:spix0_eff} as a function of the voltage threshold in the discriminator. Efficiencies larger than 99\% were obtained for thresholds up to 1/4 of a MIP, corresponding to more than 10 times the pixel noise.

\tdrsubsubsec{The Apsel DNW MAPS series}\label{sec:apsel}

\paragraph{DNW MAPS in planar CMOS technology}
Deep N-well MAPS were proposed a few years ago as possible candidates for charged-particle tracking applications. 
The Apsel4D chip is a 4096 element prototype MAPS detector with data-driven readout architecture, implementing twofold sparsification at the pixel level and at the chip periphery. In each elementary cell of the MAPS matrix integrated in the Apsel4D chip, a mixed signal circuit is used to read out and process the charge coming from a DNW detector. This design approach, relying upon the properties of the triple well structures included in modern CMOS processes, has been described in Section~\ref{sec:dnwmaps}. In the so called DNW MAPS is integrated with a relatively large (as compared to standard three transistor MAPS) collecting electrode, featuring a buried N-type layer, with a classical readout chain for time invariant charge amplification and shaping. In the Apsel4D prototype, the elementary MAPS cells feature a 50~$\mu$m pitch and a power dissipation of about 30~$\mu$W/channel. The block diagram of the pixel analog front-end electronics is shown in Figure~\ref{fig:analog_fe}. The first block of the processing chain, a charge preamplifier, uses a complementary cascode scheme as its forward gain stage, and is responsible for most of the power consumption in the analog section. The feedback capacitor $C_F$ is continuously reset by an NMOS transistor, biased in the deep sub-threshold region through the gate voltage $V_f$. The preamplifier input device, featuring an aspect ratio $W/L=14~\mu m/0.25\mu m$ and a drain current of 20~$\mu$A, was optimized for a DNW detector about 900~$\mu$m$^2$ in area and with a capacitance $C_D$ of about 300~fF. The charge preamplifier is followed by a CR-RC, bandpass filtering stage, with open loop gain $T(s)$, featuring a programmable peaking time which can be set to 200 or 400~ns. $C_1$ is a differentiating capacitor at the CR-RC shaper input, while $G_m$ and $C_2$ are the transconductance and the capacitance in its feedback network. A discriminator is used to compare the processed signal to a global voltage reference $V_t$, thereby providing hit/no-hit information to the cell digital section. More details on the design of the analog front-end can be found in the literature~\cite{ratti06}. 
\begin{figure}[t!]
\centering
\includegraphics[width=0.45\textwidth]{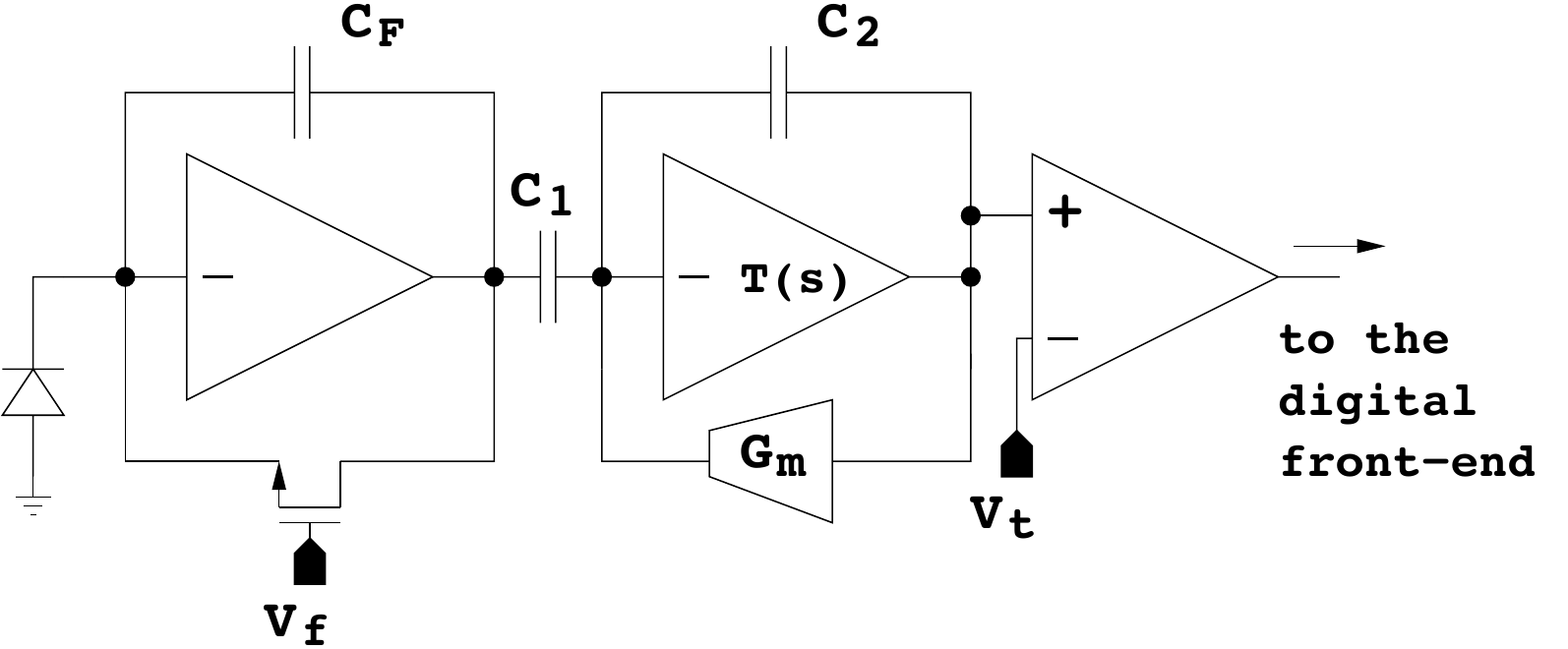}
\caption{Block diagram of the analog front-end electronics for the
elementary cell of the Apsel4D prototype.}
\label{fig:analog_fe}
\end{figure}

A dedicated readout architecture to perform on-chip data sparsification has been implemented in the Apsel4D prototype. The readout logic provides the timestamp information for the hits. 
The timestamp (TS), which is necessary to identify the event to which the hit belongs, is generated by a TS clock signal 
(also called Bunch-Crossing clock for historical reasons). 
\begin{figure}[t!]
\centering
\includegraphics[width=0.45\textwidth]{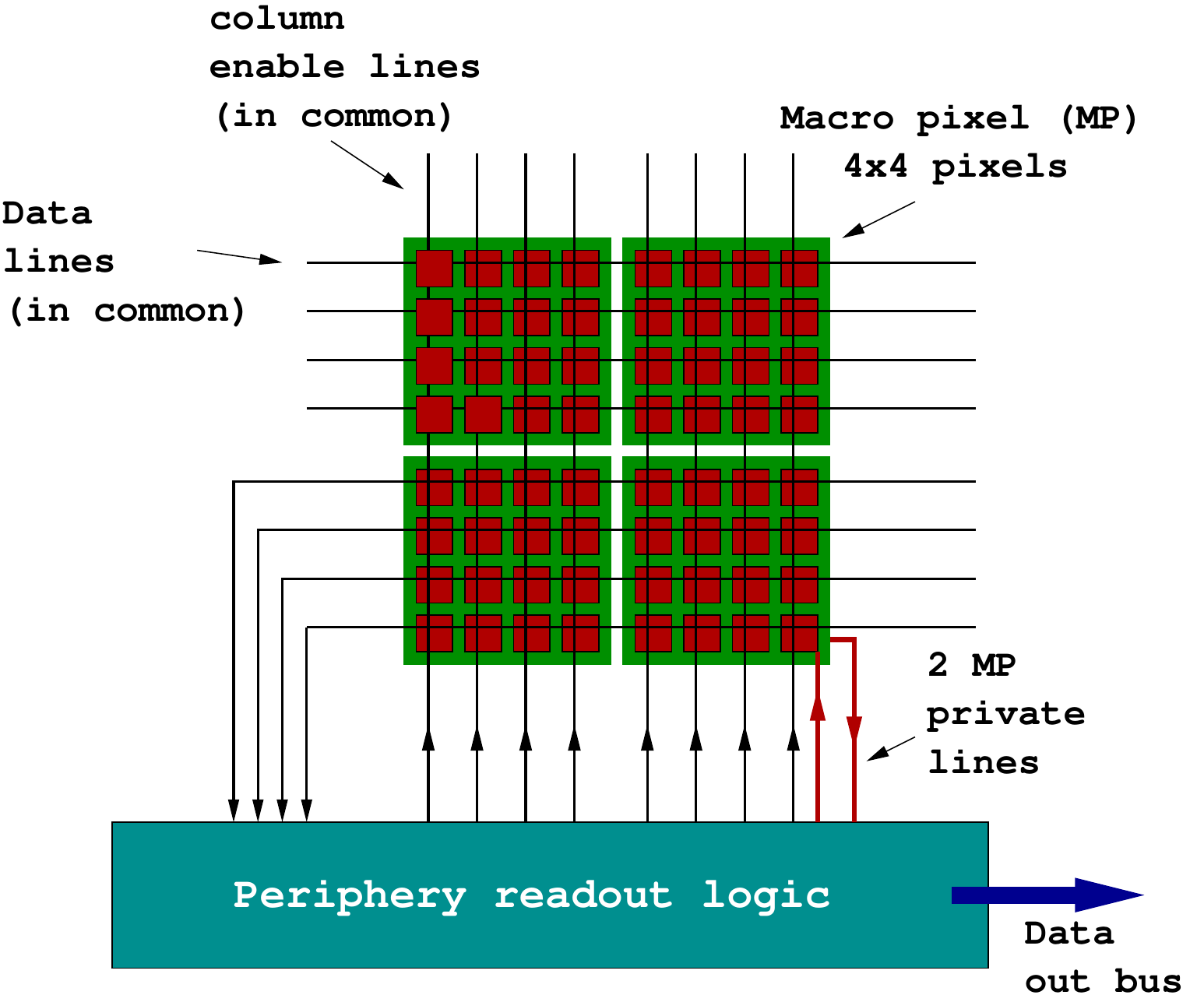}
\caption{Schematic concept of the architecture for MAPS matrix readout.} 
\label{fig:readout_architecture}
\end{figure}
The key requirements in this development are 1) to minimize logical blocks with PMOS inside the active area, thus preserving the collection efficiency, 2) to reduce to a minimum the number of digital lines crossing the sensor area, in particular its dependence on detector size to allow the readout scalability to larger matrices and to reduce the residual crosstalk effects, and 3) to minimize the pixel dead time by reading hit pixels out of the matrix as soon as possible. 

With these criteria a readout logic in the periphery of the matrix has been developed, as schematically shown in Fig~\ref{fig:readout_architecture}. To minimize the number of digital lines crossing the active area the matrix is organized in MacroPixels (MP) with 4x4 pixels.  Each MP has only  two private lines for point-to-point connection to the peripheral logic: one line is used to communicate that the MP has got hits, while the second private line is used to freeze the MP until it has been read out.  When the matrix has some hits, the columns containing fired MPs are enabled, one at a time, by vertical lines.
Common horizontal lines are shared among pixels in the same row to bring data from the pixels to the periphery, where the association with the proper timestamp is performed before sending the formatted data word to the output bus. The chip has been designed with a mixed mode design approach. While the pixel matrix has a full custom design and layout, the   periphery readout architecture has been synthesized in standard cell starting from a VHDL model; automatic place-and-route tools have been used for the layout of the readout logic \cite{gabrielli07}. The chip has been designed to run with a readout clock up to 100 MHz (20 MHz in test beam), a maximum matrix readout rate of 32 hit pixels/clock cycle and a local buffer of maximum 160 hits to minimize the matrix sweep time. 

Apsel4D, together with smaller test matrices 3x3 with full analog output, have been successfully tested with 12 GeV/$c$ protons at the PS-T9 beam line at CERN~\cite{bettarini:slim5_testbeam}. On the analog test structures the measured most probable value for the cluster signal
for MIPs was about 1000 electrons  with a Signal-to-Noise ratio of about 20.  
The efficiency of the Apsel4D DNW MAPS as a function of threshold for two devices with different silicon thickness (100~$\mu$m and 300~$\mu$m thick) has also been measured. Figure~\ref{fig:eff_v_thr} shows the measured hit efficiency, determined as described in a published work~\cite{bettarini:slim5_testbeam}.  
\begin{figure}[t!]
\centering
\includegraphics[width=.50\textwidth]{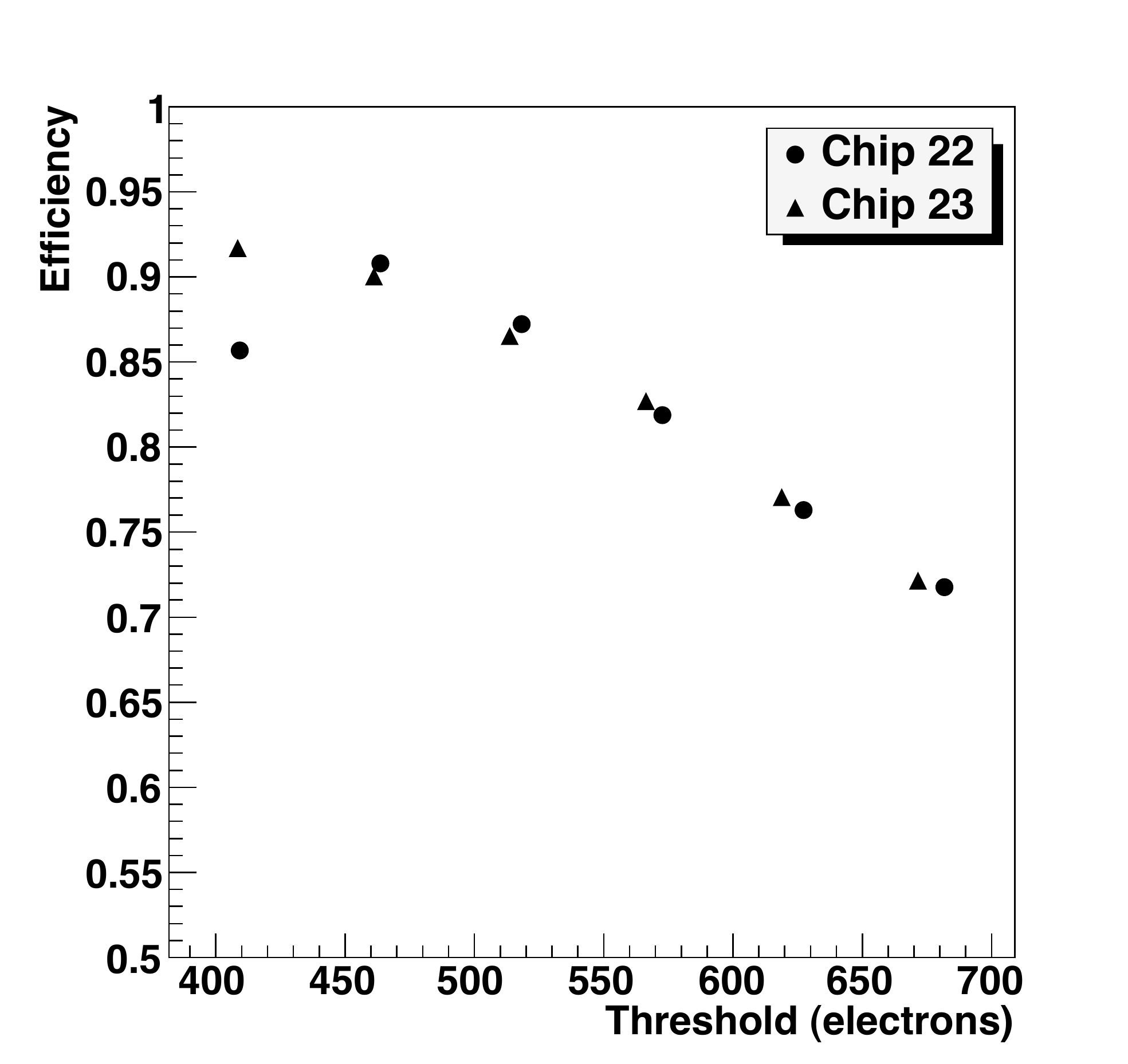}
\caption{Efficiency results for two MAPS detectors. Chip 22 is 300~$\mu$m thick while Chip 23 is 100~$\mu$m thick. The statistical uncertainty on each point is smaller than the size of the plotting symbol.}
\label{fig:eff_v_thr}
\end{figure}
\begin{figure}[t!]
\centering
\includegraphics[width=0.5\textwidth]{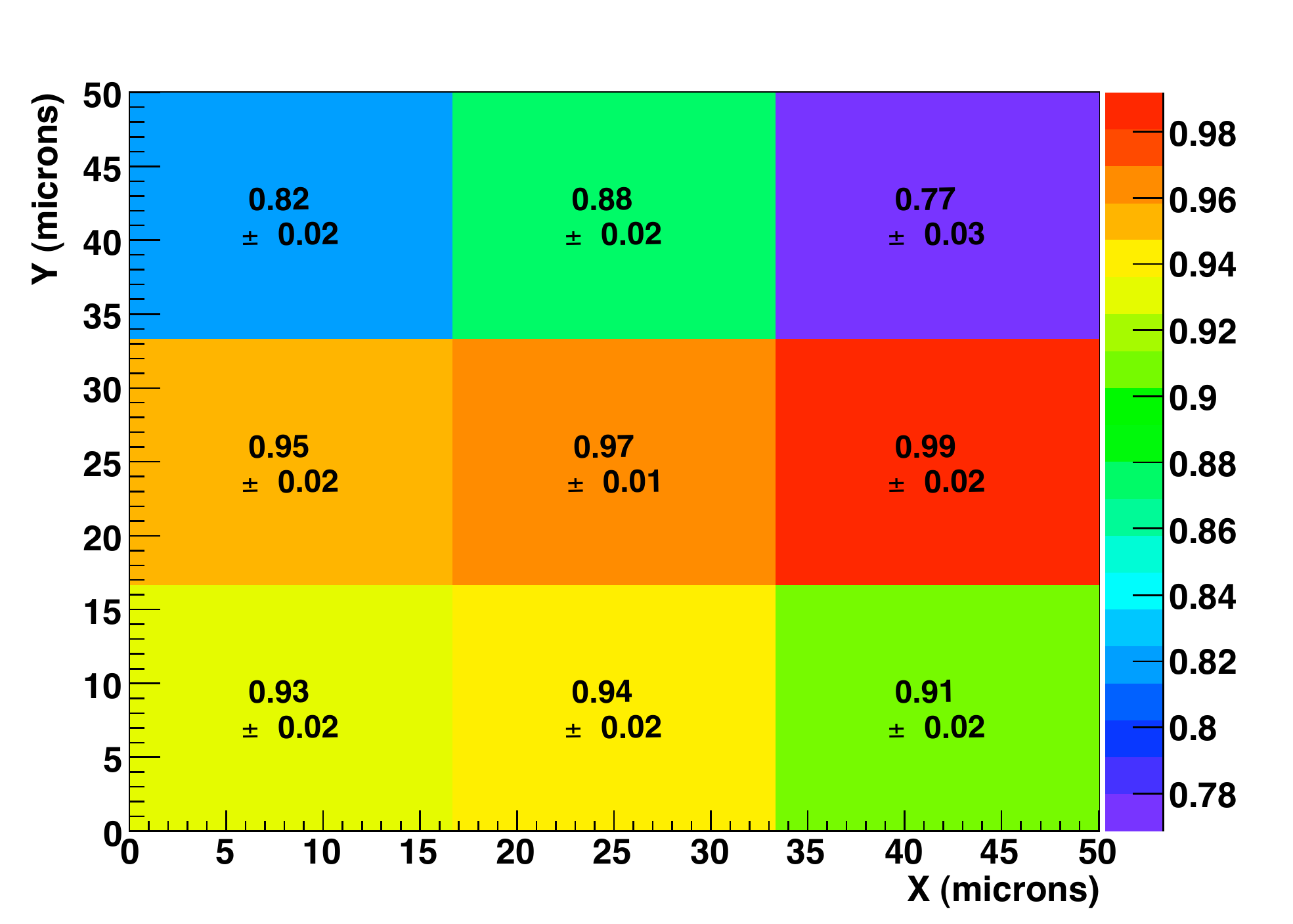}
\caption{Hit efficiencies measured as a function of position within the
  pixel (the picture, which is not to scale, represents a single pixel
  divided into nine sub-cells).}
\label{fig:eff_pxl_map}
\end{figure}
At the lowest thresholds a maximum efficiency of approximately 92\% and the expected general behavior of decreasing efficiency with increasing threshold can be observed. The noise occupancy for this range of thresholds was found to vary from $2.5 \times 10^{-3}$ to $1 \times 10^{-6}$.  The low efficiency observed for the 300~$\mu$m chip at the lowest threshold appears to have been caused by a readout malfunction. Investigations have shown that a small localized area on the detector had very low efficiency, while the rest of the detector behaved normally with good efficiency. Additionally, the efficiency for detecting hits as a function of the track extrapolation point within a pixel has been studied.  Since the pixel has internal structures, with some areas less sensitive than others, we expect the efficiency to vary as a function of position within the cell. The uncertainty on the track position including multiple scattering effects is roughly 10 microns, to be compared to the 50 $\mu$m pixel
dimension. The pixel has been divided into nine square sub-cells of equal area and the hit efficiency within each sub-cell has been measured. The efficiencies thus obtained are ``polluted'' in some sense due to the migration of tracks among cells. We obtain the true sub-cell efficiencies by unfolding the raw results, taking into account this migration, which we characterize using a simple simulation. The result can be seen in Figure~\ref{fig:eff_pxl_map}, where the efficiency measured in each sub-cell is shown. A significant variation in sensitivity within the pixel area can be observed, as expected.  In particular, the central region is seen to be virtually 100\% efficient, while the upper part of the pixel, especially the upper right-hand sub-cell, shows lower efficiency due to the presence of competitive n-wells.
The position of this pixel map relative to the physical pixel is not fixed.  This is a consequence of the alignment, which determines the absolute detector position by minimizing track-hit residuals, as described above. If the pixel area is not uniformly efficient, the pixel center as determined by the alignment will correspond to the barycenter of the pixel efficiency map. Thus, it is not possible to overlay Figure~\ref{fig:eff_pxl_map} on a drawing of the pixel layout, without adding additional information, for example a simulation of internal pixel efficiency.
The efficiency as a function of position on the MAPS matrix has also been investigated, since non-uniformity could indicate inefficiencies caused by the readout.  Generally, a uniform efficiency across the area of the MAPS matrix was observed. The intrinsic resolution $\sigma_{hit}$ for the MAPS devices was measured as already described in a published paper~\cite{bettarini:slim5_testbeam}. The expected resolution for cases where the hit consists of a single pixel is given by $50/\sqrt{12} = 14.4$ $\mu$m, where 50 microns is the pixel dimension.  

\paragraph{DNW MAPS in 3D CMOS technology}

\begin{figure}
\begin{center}
\includegraphics[width=0.4\textwidth]{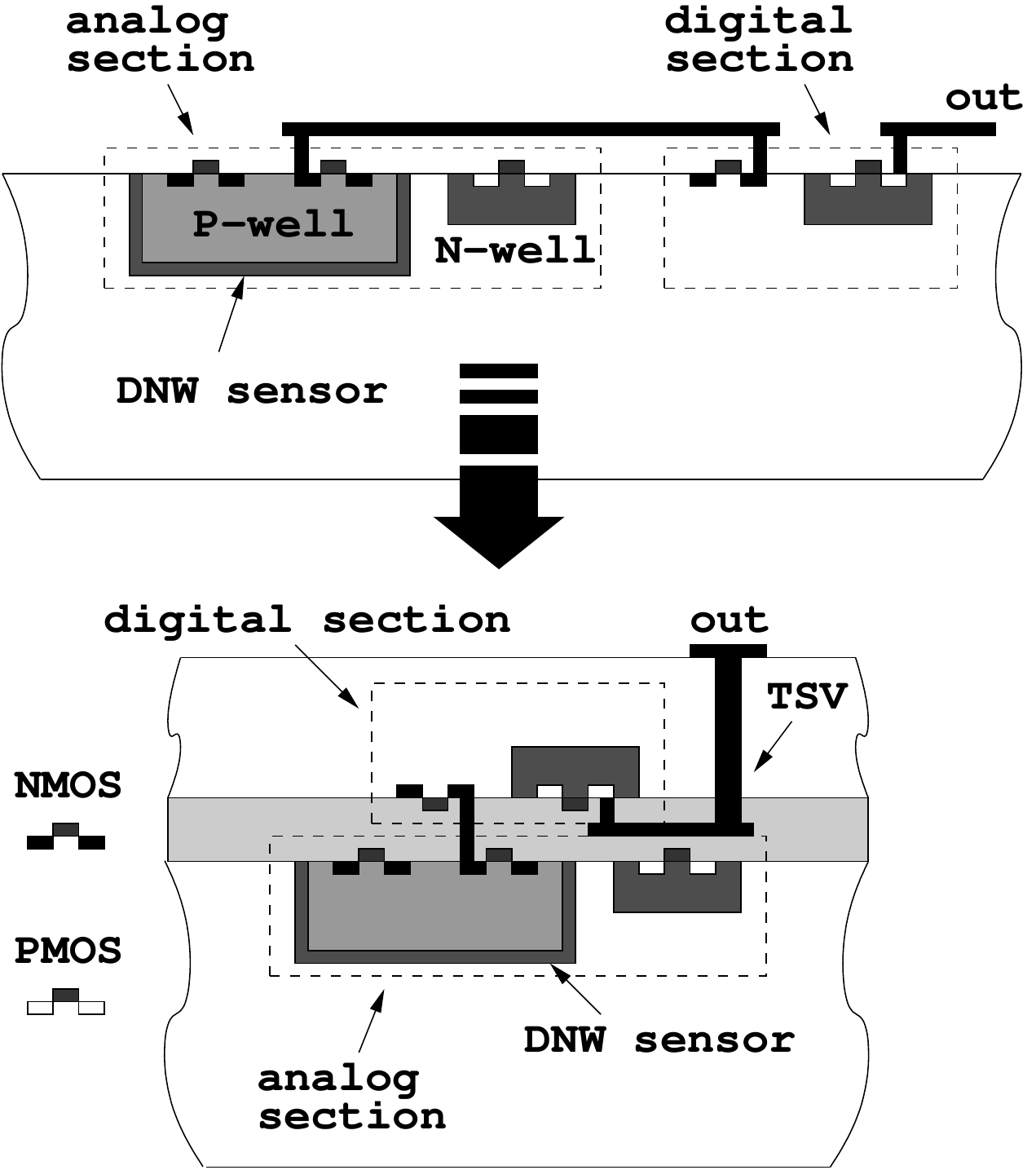}
\end{center}
\caption{Cross-sectional view of a DNW CMOS MAPS: from a planar CMOS technology to a 3D process.}
\label{fig:2Dto3D}
\end{figure}

As already mentioned in Section~\ref{sec:dnwmaps}, the DNW monolithic sensors have been designed and fabricated also in the Tezzaron/Globalfoundry technology, based on the vertical integration of two 130~nm CMOS layers. The conceptual step from the DNW MAPS in a planar CMOS technology to its vertically integrated version is illustrated in Fig~\ref{fig:2Dto3D}, showing a cross-sectional view of a 2D MAPS and of its 3D translation. 
The prototype includes two small 3$\times$3 matrices for analog readout and charge collection characterization and a larger one, 8$\times$32 in size, equipped with a digital readout circuit with data sparsification and time stamping features. 
In the 3D pixel cell, with 40~$\mu$m pitch, the fill factor 
has been increased up to 94\% (90\% in 2D DNW MAPS version) and the 
sensor layout has been optimized; according to device simulation, 
where a collection efficiency in excess of 98\% is expected.

A number of different problems were encountered during fabrication of the first device batch. Among them, the misalignment between the two tiers prevented the analog and digital sections in each pixel cell to communicate to each other~\cite{hiroshima11}. At the time of the TDR writing, other 3D wafers are being processed and devices from the first run are undergoing characterization. 

\begin{figure}
\begin{center}
\includegraphics[width=0.45\textwidth]{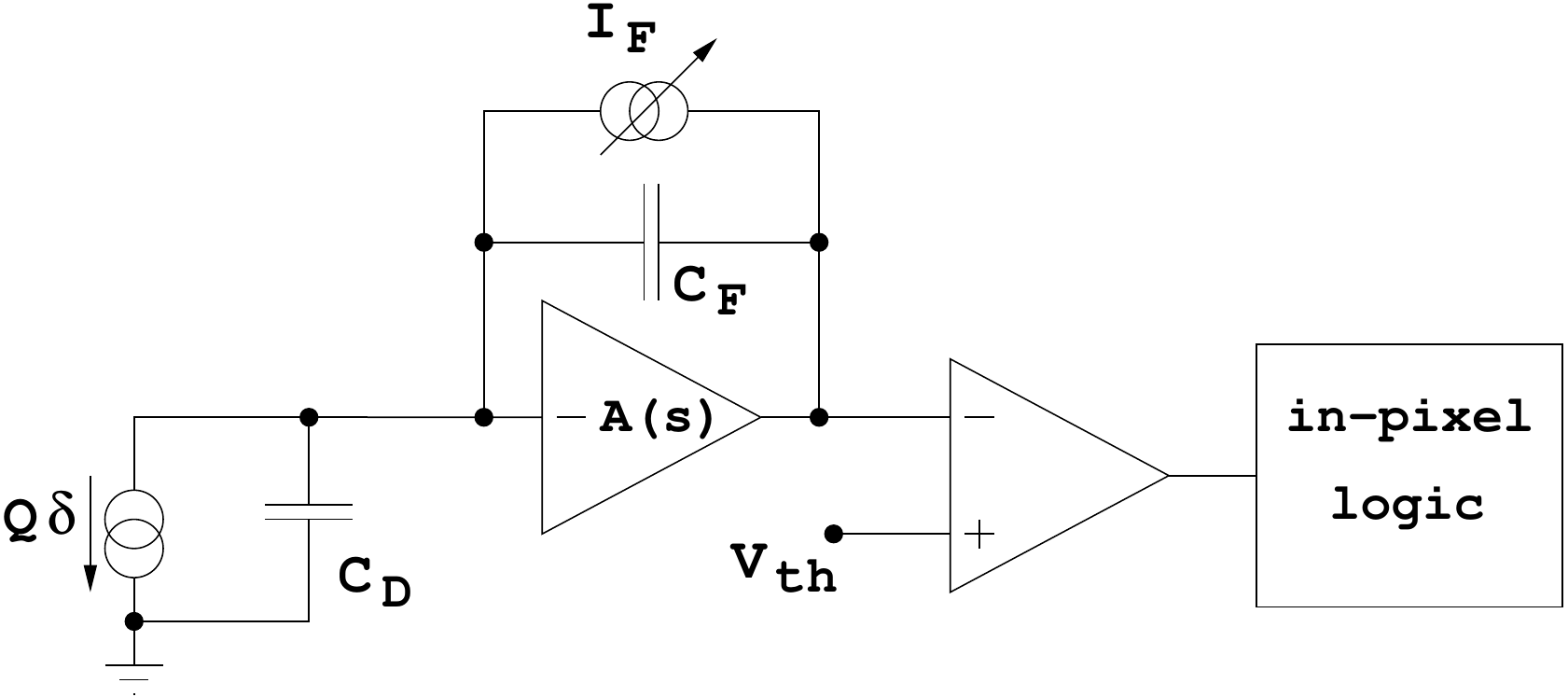}
\end{center}
\caption{Block diagram of the analog front-end electronics for the elementary cell of the 3D DNW MAPS of the apsel family.}
\label{fig:3Dmaps_FE}
\end{figure}

\begin{figure}[t!]
\begin{center}
\includegraphics[width=0.48\textwidth]{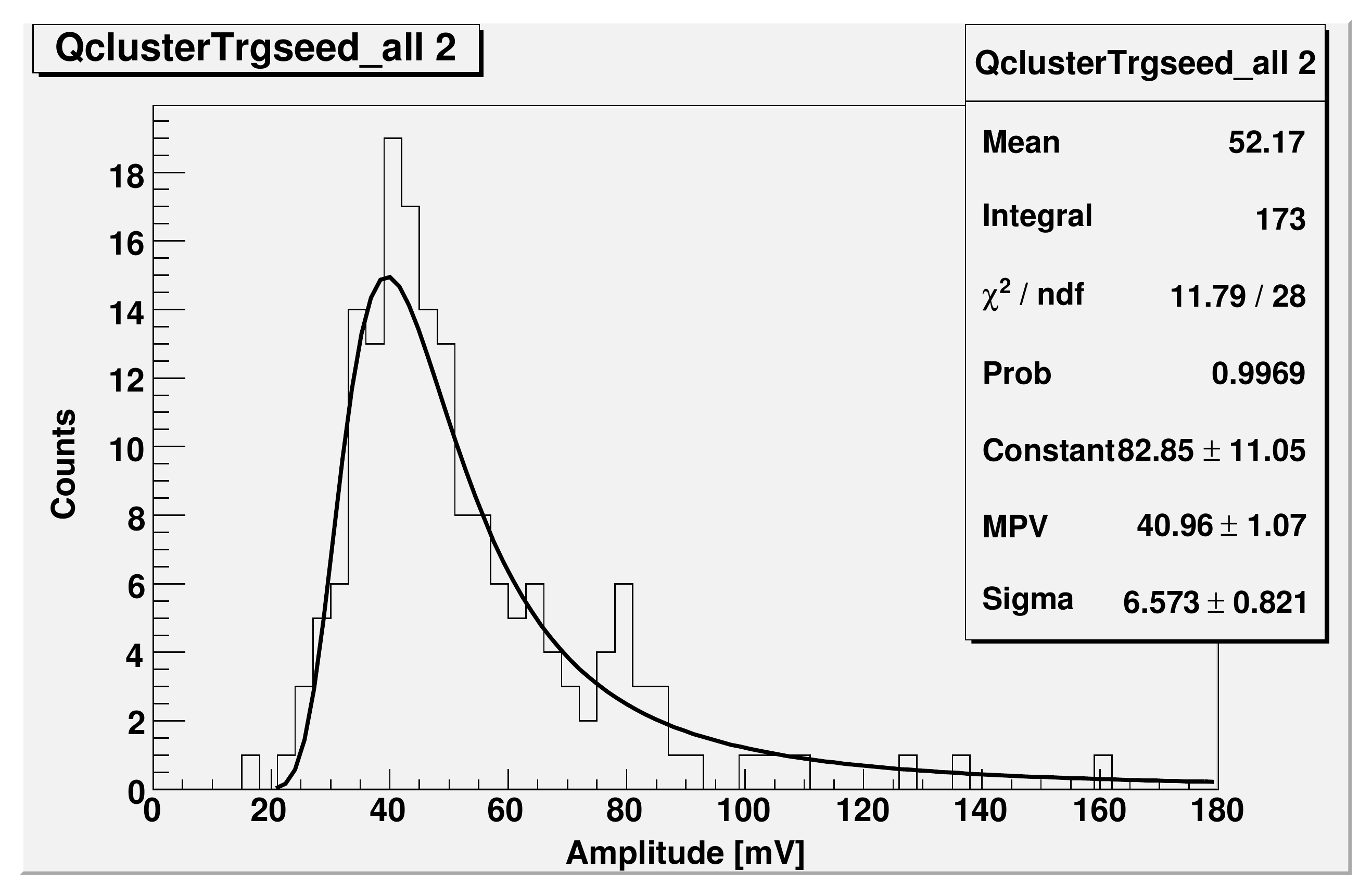}
\end{center}
\caption{3$\times$3 cluster signal for electrons from a $^{90}$Sr source for the 3D DNW MAPS.}
\label{fig:3Dsr90}
\end{figure}

\begin{figure}[t!]
\begin{center}
\includegraphics[width=0.45\textwidth]{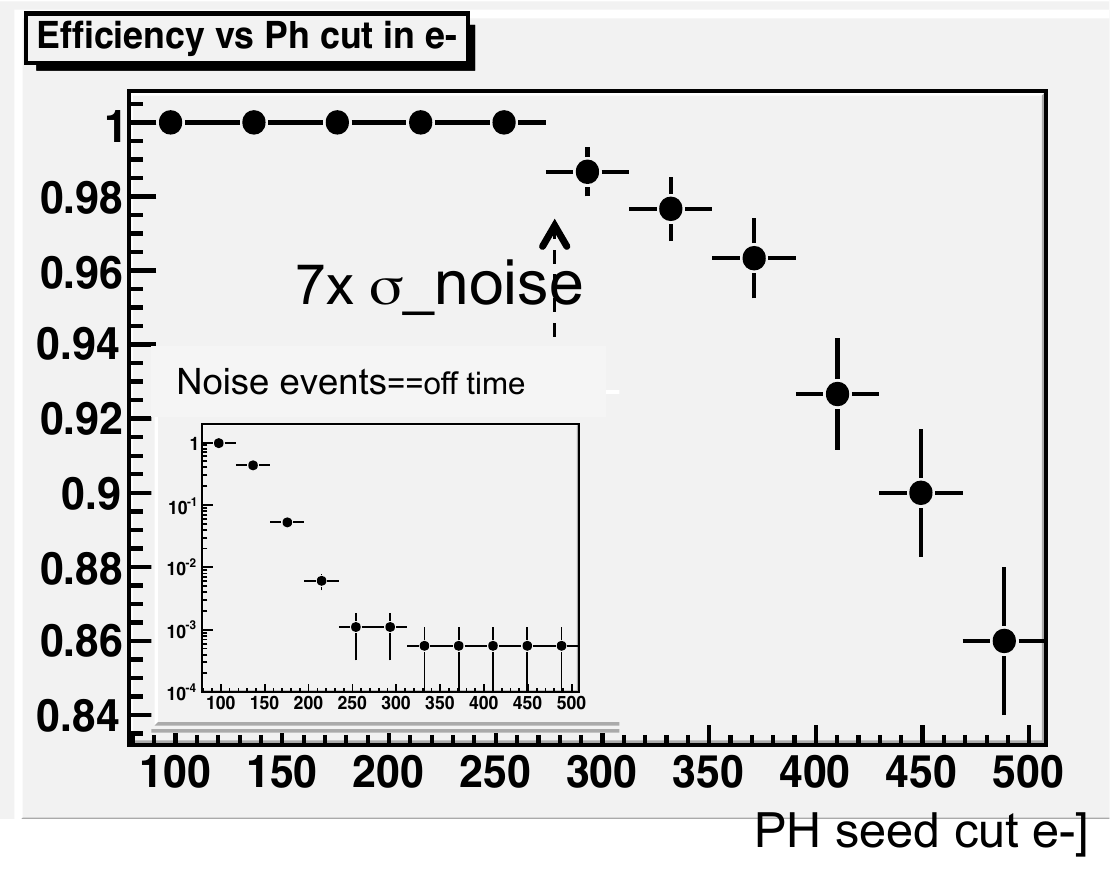}
\end{center}
\caption{Detection efficiency of a pseudo-3D DNW MAPS as a function of the cut on the pulse height of the detected events. The efficiency is 1 up to threshold values of 7 times the pixel noise.}
\label{fig:apsel5_tc_effi}
\end{figure}

Figure~\ref{fig:3Dmaps_FE} shows the analog front-end channel of the 3D DNW MAPS (quite similar to the analog processor of the SuperPIX0 chip, see Figure~\ref{fig:schematic}), simply consisting of a charge preamplifier, whose bandwidth was intentionally limited to improve the signal-to-noise ratio (the so called shaper-less configuration). 

Tests on the CMOS layer hosting the 
sensor and the analog front-end, already available, allowed a 
first characterization of the process.
Equivalent noise charge of between 30 and 40 electrons (in good agreement with circuit simulations) and a charge sensitivity of about 300~mV/fC (a factor of 2 smaller than in simulations) were obtained. 
Figure~\ref{fig:3Dsr90} shows the $^{90}$Sr spectrum detected by the cluster of 3$\times$3 pixels in a small matrix. 
The most probable value of the collected charge is about 1000 electrons. 
Pseudo-3D DNW MAPS have 
been tested also on the PS beam at CERN. Here, the term pseudo-3D refers to devices consisting of just one tier but suitable for 3D integration. Very promising results were obtained with hit detection efficiency above 98\%, as displayed in Figure~\ref{fig:apsel5_tc_effi}.

\tdrsubsubsec{The Apsel4well quadruple well monolithic sensor}\label{sec:apsel4well}


\begin{figure}
\begin{center}
\includegraphics[width=0.45\textwidth]{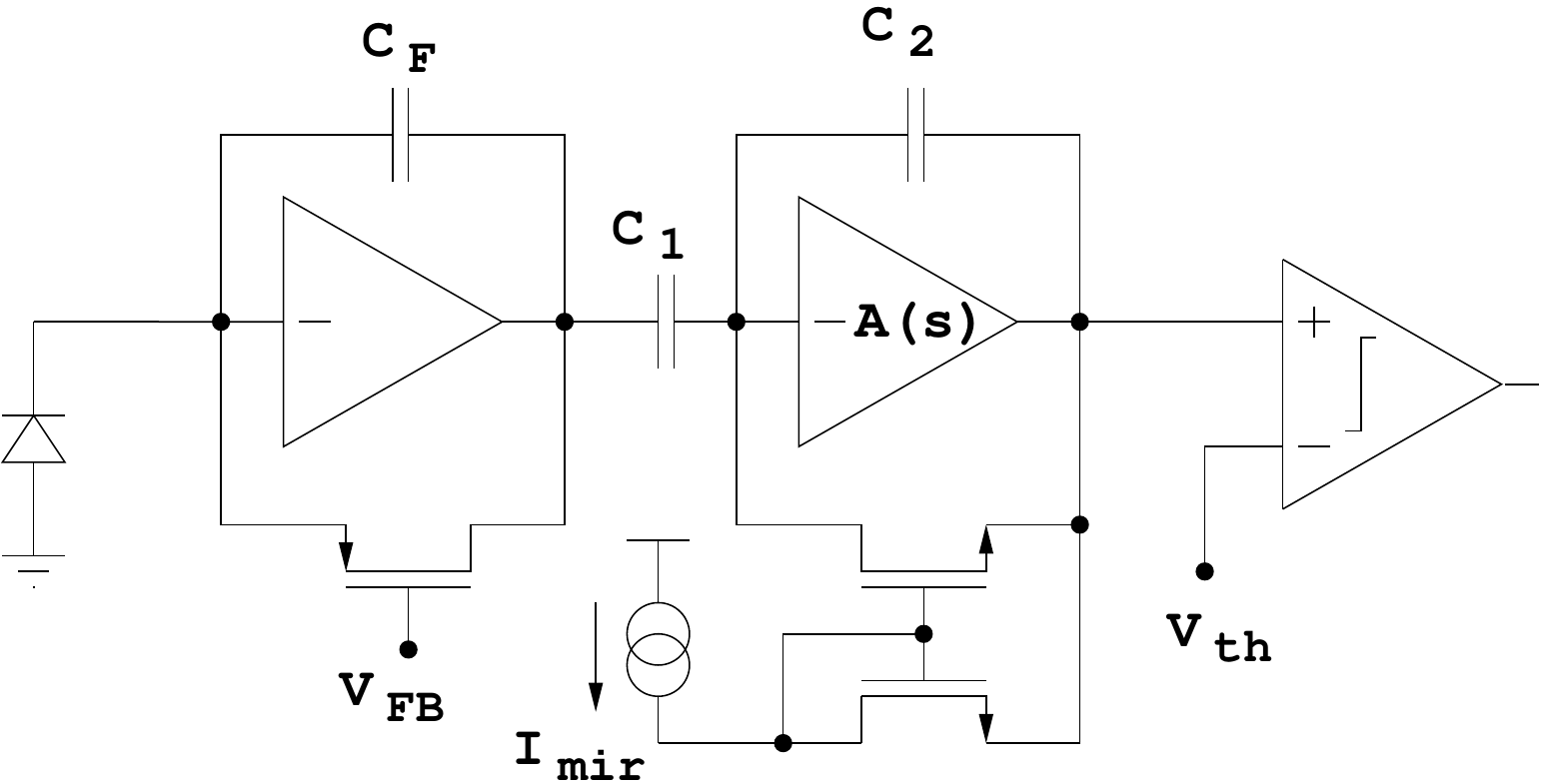}
\end{center}
\caption{Block diagram of the analog front-end electronics for the elementary cell of Apsel4well monolithic sensor.}
\label{fig:INMAPS_FE}
\end{figure}

As already mentioned in section~\ref{sec:inmaps}, a test chip in the INMAPS, 180 nm CMOS technology, called Apsel4well, has been submitted in August 2011. The chip includes four 3$\times$3 matrices with different number (2 or 4) of the collecting electrodes (each consisting of a 1.5~$\mu$m$\times$1.5~$\mu$m N-well diffusion), with or without the shielding deep P-well implant, with or without enclosed layout transistors as the input device of the charge preamplifier. The prototype also contains a 32$\times$32 matrix with sparsified digital readout, with an improved version of the architecture developed for the DNW MAPS and now optimized for the Layer0 target hit rate of 100~MHz/cm$^2$.

The first test chips have been realized with different 
thicknesses (5-12 $\mu$m) and
epitaxial layer resistivity (standard, about 10~$\Omega\cdot$cm, and high 
resistivity, 1~k$\Omega\cdot$cm).  

In the Apsel4well pixel ($50~\mu$m pitch) 
the collecting electrode  is readout by a channel 
similar to previous versions of DNW MAPS, including 
charge preamplifier, 
a shaping stage with a current mirror in the feedback network and a two-stage threshold discriminator (Figure~\ref{fig:INMAPS_FE}),   
followed by the in-pixel logic needed to implement the new readout architecture described in section~\ref{sec:improved_pixelreadout}.
With the new pixel design the total sensor capacitance is only about 40 fF,
an order of magnitude smaller than in previous DNW pixels,
allowing a reduction of almost a factor of 2 in power consumption (now  
18 $\mu$W/pixel). 


\begin{figure}
\begin{center}
\includegraphics[width=0.42\textwidth]{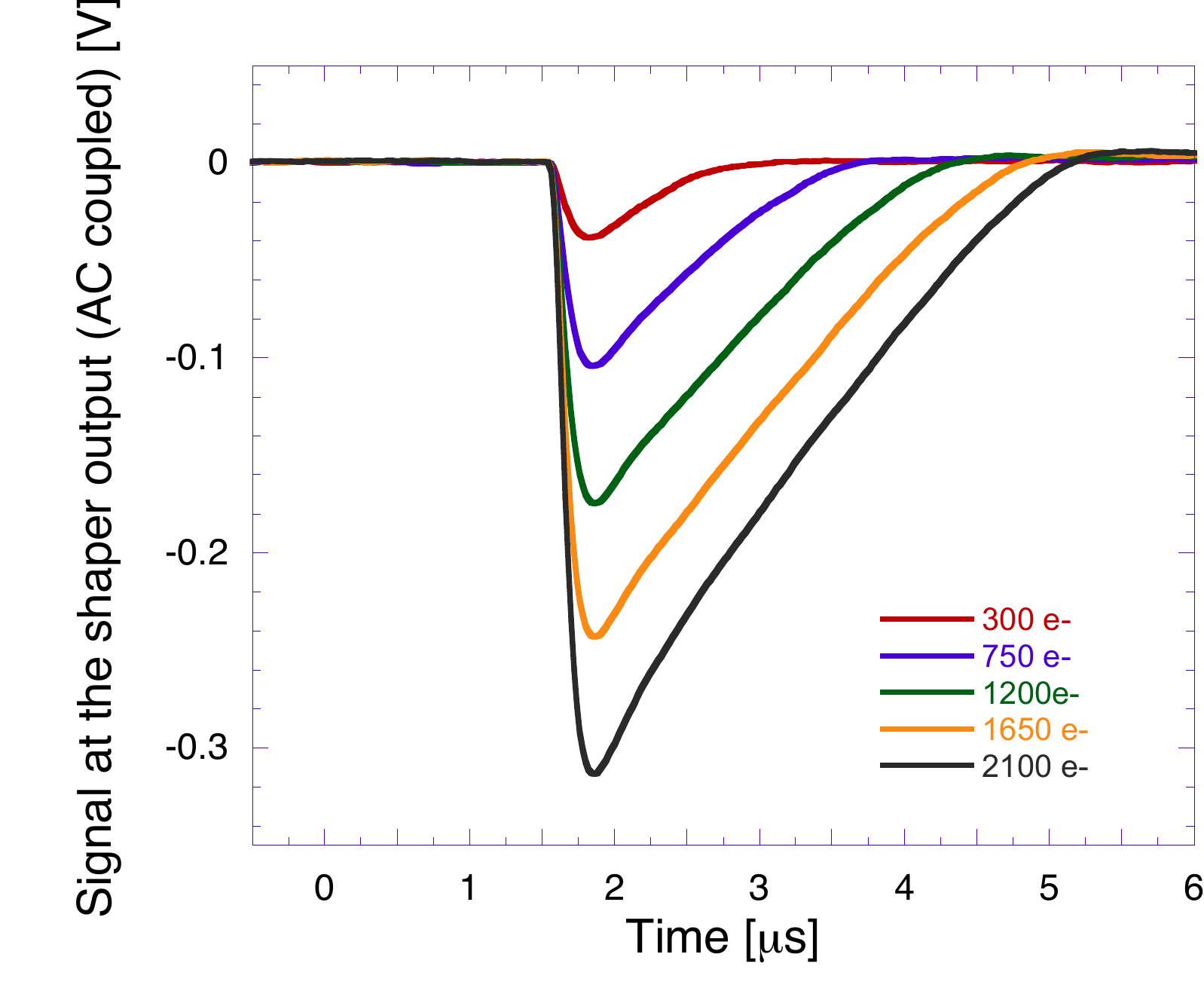}
\end{center}
\caption{Signal at the shaper output as a response to an input charge signal with varying amplitude in an Apsel4well sensor.}
\label{fig:shapout}
\end{figure}

Preliminary results on the different INMAPS test structures realized
showed very promising results.
Figure~\ref{fig:shapout} shows the signal at the shaper output as a response to an input charge signal with varying amplitude.
In the 3x3 analog matrix an average equivalent noise charge of 
38 e$^-$ and a gain of about 900~mV/fC were measured, 
both in good agreement with post layout simulation.


The characterization with laser and radioactive sources confirmed that the 
deep p-well is beneficial in preserving charge collection by the 
small Nwell diodes,
although a much larger area inside the pixel is now covered by competitive 
Nwells with the deep p-well implant beneath.
The plot in Figure~\ref{fig:laserplot} represents the collected charge in a Apsel4well pixel (12~$\mu$m epitaxial layer thickness, standard resistivity) illuminated with an infrared laser source. The layout of the N-wells (both the collecting electrodes and the N-wells for PMOS transistors) included in the elementary cell is superimposed onto the plot. The position of the collecting electrodes is easily detectable.
\begin{figure}
\begin{center}
\includegraphics[width=0.45\textwidth]{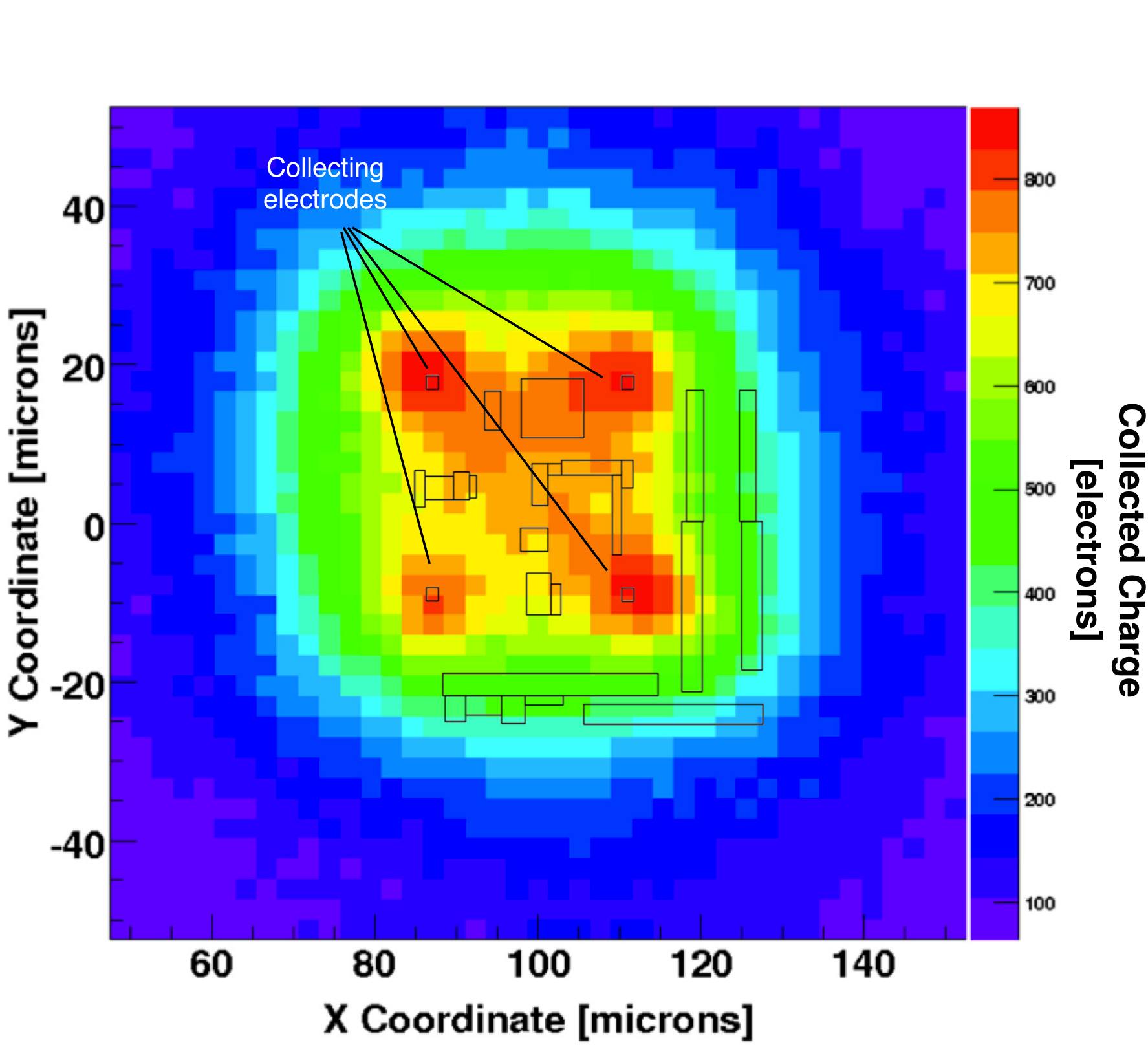}
\end{center}
\caption{Collected charge in a Apsel4well pixel illuminated with an infrared laser source.}
\label{fig:laserplot}
\end{figure}

\begin{figure}
\begin{center}
\includegraphics[clip=true,trim=0cm 0.cm 0cm 0.cm, width=0.43\textwidth]{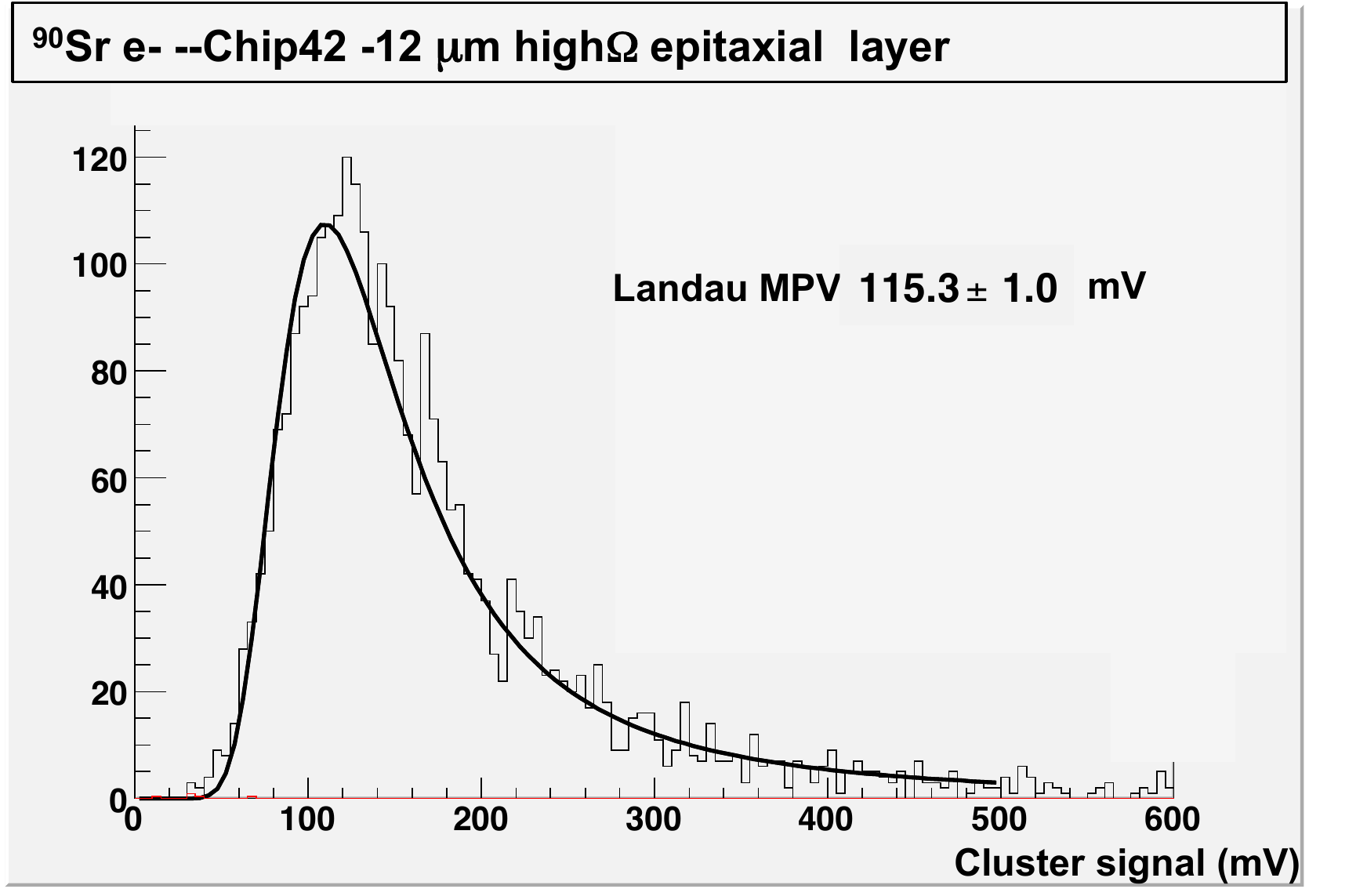}
\end{center}
\caption{Cluster signal for electrons from a $^{90}$Sr source detected by a 3$\times$3 matrix of the Apsel4well chip with high resistivity epitaxial layer.}
\label{fig:inmaps_sr90_highres}
\end{figure}

Measurements with electrons from a $^{90}$Sr source are shown in 
Figure~\ref{fig:inmaps_sr90_highres}, for a chip with 12$\mu$m high 
resistivity epitaxial layer.
The cluster signal has been fitted with a Landau distribution with a 
MPV corresponding to about 930 electrons. The Signal-to-Noise ratio of the 
samples featuring a high resistivity epitaxial layer ranges from 20 to 25.
 
Evaluation of the threshold dispersion and of the noise variation among pixels 
has been performed on the 32x32 digital matrix, measuring the hit rate as a 
function of the discriminator threshold. 
With a fit to the turn-on curve we report a threshold dispersion of about  
2 times the average pixel noise, and 
a noise variation of about 35-40\% inside the matrix. 
These figures were obtained with a readout clock of 10 MHz, with no 
significant change in the performance with measurements 
done up to 100 MHz readout clock.

Specific tests on the 32x32 digital matrix were also performed to verify 
the new readout architecture in both the 
operation modes available on chip: data push 
and triggered. 
Both functionalities are working as expected with 
very similar performance verified  with measurements of 
the pixel turn-on curve. 

Radiation hardness of these devices is being 
investigated for application in \superb.
Irradiation with neutrons up to $10^{14}$~n/cm$^2$ 
has been performed and devices are currently under measurement with laser
and radioactive sources (see section~\ref{sec:inmaps_radhard}).

\tdrsubsubsec{Fast readout architecture for MAPS pixel matrices}\label{sec:improved_pixelreadout}

An evolution of the readout architecture implemented in the 2D DNW MAPS 
was developed~\cite{gabrielli:nim2011} to benefit from the larger area available for the in-pixel logic 
in some  technologies: 3D MAPS with vertical integration and INMAPS realized with the deep p-well implant.
Having more room for the in-pixel logic makes it possible to 
latch the TS at the pixel level when a 
hit occurs. 
When a TS request is sent to the matrix, a pixel FastOR signal activates if the latched TS is the same as the requested one. 
The columns with an active FastOR signal are enabled and read out in a sequence; 1 clock cycle per column is needed. 
A conceptual view of the digital readout architecture is shown in Figure~\ref{fig:INMAPS_digro}. 
The matrix readout is timestamp sorted and 
can  be operated either
in data push mode (selecting all the TS) or in triggered mode, where 
only triggered TS are readout.
The readout periphery takes care of encoding, buffering and serializing/sorting the hits retrieved from the sensor matrix. 
In order to achieve the remarkably high readout frequency set by the \superb experiment, the architecture can be subdivided 
in a number of modules, each serving a submatrix. 
This choice improves the scalability features of the readout section and makes it suitable for experiment scale detectors.
 
\begin{figure*}[tbp]
\begin{center}
\includegraphics[width=0.85\textwidth]{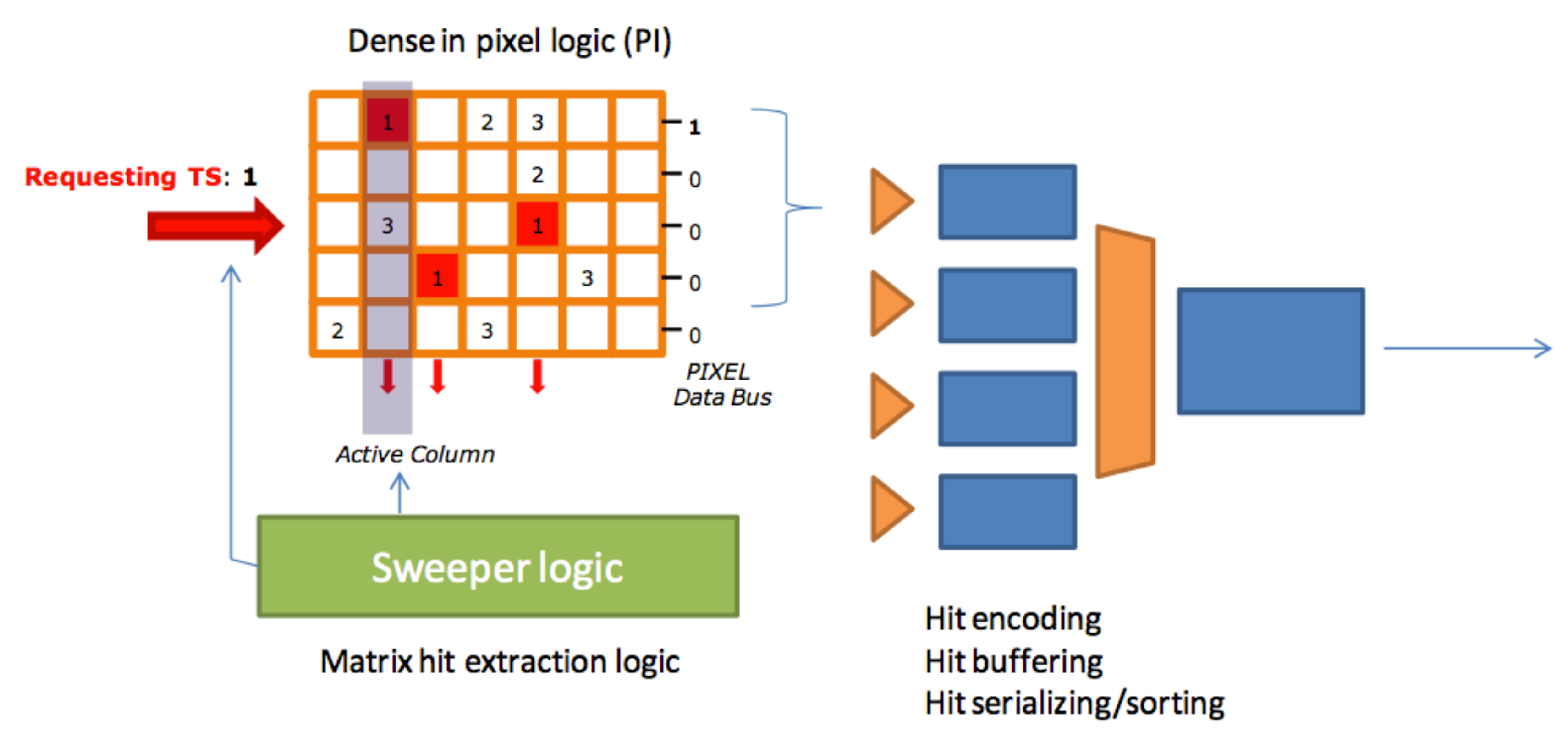}
\end{center}
\caption{Conceptual view of the new pixel readout architecture implemented in the Apsel4well chip.}
\label{fig:INMAPS_digro}
\end{figure*}

With this architecture, a TimeStamp of 100 ns can be 
used, greatly reducing the TS 
resolution with respect to standard CMOS sensor readout, where large 
frame readouts are needed. For application in Layer0, where high rates 
are expected, a fast TS ($< 1\mu$s) is needed to reduce the module 
readout bandwidth to an acceptable level ($<$5 Gbit/s).
Full VHDL simulation of the readout has been implemented to test the 
performance on the full size matrix 
(1.3 cm$^2$) with a target Layer0 hit rate of 100 MHz/cm$^2$. 
Results from simulation showed that with a TS clock of 100 ns and
a 50 MHz readout clock one can achieve 
efficiency above 99\% for the data push architecture.  
Studies on the triggered architecture showed an efficiency of 
about 98\%, for 6 $\mu$s trigger latency: in this configuration, 
where the pixel cell acts as a single memory during the trigger latency,  
the associated pixel dead time dominates the inefficiency.

This new readout has been implemented in the Apsel4well test chip as well as in two devices in preparation for 
next 3D submission: a second prototype of a front-end 
chip for hybrid pixels
(Superpix1, 32x128 pixels with $50 \times50$ $\mu$m$^2$ pitch)  
and a large DNW MAPS matrix with the cell optimized for the vertical 
integration process (APSELVI, 128x100 pixels with $50 \times50$ $\mu$m$^2$ 
pitch).


\tdrsubsec{Radiation tolerance}

\paragraph{Hybrid pixels.} The high degree of radiation tolerance of modern CMOS technologies, coming as a byproduct of the aggressive scaling down of device minimum feature size, is having a beneficial impact in high energy physics applications. Beginning with the 130 nm CMOS processes, which entered the sub-3 nm gate oxide thickness regime, direct tunneling contribution to the gate current has assumed a significant role as compared to trap assisted mechanisms~\cite{ghetti00}. This may account for the very high degree of radiation hardness featured by devices belonging to the most recent technology nodes, which might benefit from relatively fast annealing of holes trapped in the ultra-thin gate oxides. Tolerance to a few hundred of Mrad(SiO$_2$) has been recently proven in front-end circuits for hybrid pixel detectors~\cite{garcia-sciveres11}. Charge trapping in the thicker shallow trench isolation (STI) oxides is considered as the main residual damage mechanism in 130 nm N-channel MOSFETs exposed to ionizing radiation~\cite{re06,ratti08}, especially in narrow channel transistors~\cite{faccio05}. Ionizing radiation was found to affect also the 90~nm and 65~nm CMOS nodes, although to an ever slighter extent, this is likely to be due to a decrease in the substrate doping concentration and/or in the STI thickness. As far as analog front-end design is concerned, ionizing radiation damage mainly results in an increase in low frequency noise, which is more significant in multifinger devices operated at a small current density. This might be a concern in the case of the front-end electronics for hybrid pixel detectors, where the input device of the charge preamplifier is operated at drain currents in the few $\mu$A range owing to low power constraints. However, at short peaking times, typically below 100~ns, the effects of the increase in low frequency noise on the readout channel performance is negligible. Also, use of enclosed layout techniques for the design of the preamplifier input transistor (and of devices in other critical parts of the front-end) minimizes the device sensitivity to radiation~\cite{snoeys02}.
\begin{figure}
\begin{center}
\includegraphics[width=0.5\textwidth]{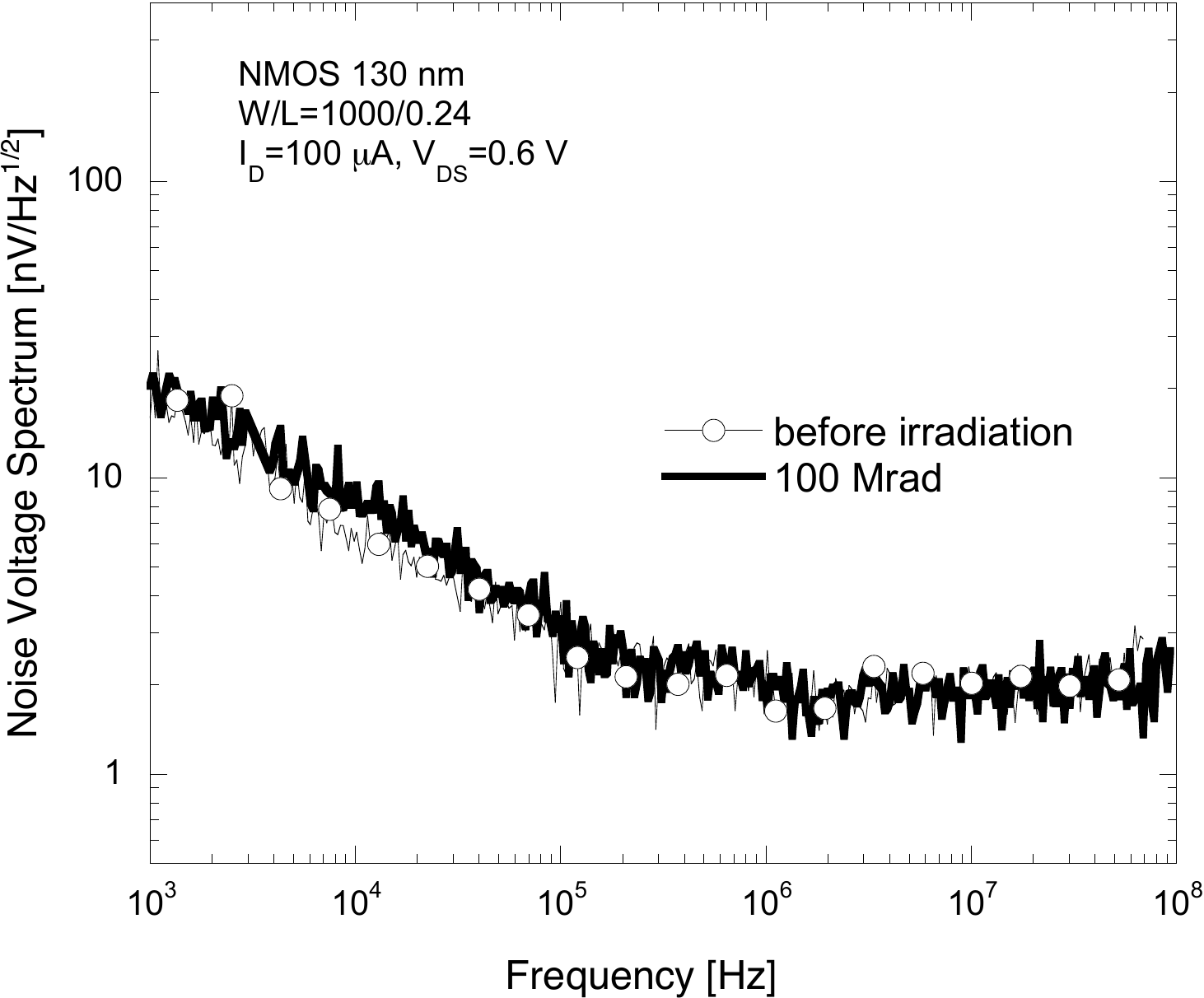}
\end{center}
\caption{Noise voltage spectrum for a 130~nm NMOS device with enclosed layout.}
\label{fig:spectrum_ibm}
\end{figure}
For this purpose, Figure~\ref{fig:spectrum_ibm} shows the noise voltage spectrum for a 130~nm NMOS transistor with enclosed layout, featuring no significant changes after irradiation with a 100~Mrad(SiO$_2$) total ionizing dose. On the other hand, CMOS technologies are virtually insensitive to bulk damage, since MOSFET transistor operation is based on the drift of majority carriers in a surface channel.

\paragraph{DNW CMOS MAPS.}
DNW MAPS have been thoroughly characterized from the standpoint of radiation hardness to evaluate their limitations in harsh radiation environments. In particular, the effects of ionizing radiation, with total doses of about 10 Mrad(SiO$_2$), have been investigated by exposing DNW MAPS sensors to a $^{60}$Co source~\cite{ratti10}. 
In that case, some performance degradation was detected in the noise and gain of the front-end electronics and in the sensor leakage current, while no significant change was observed as far as the charge collection properties are concerned. 
Figure~\ref{fig:enc_gamma} shows the equivalent noise charge as a function of the absorbed dose and after the annealing cycle for a DNW monolithic sensor. The significant change can be ascribed to the increase in the flicker noise of the preamplifier input device as a consequence of parasitic lateral transistors being turned on by positive charge buildup in the shallow trench isolation oxides and contributing to the overall noise. Use of an enclosed layout approach is expected to significantly reduce the effect of ionizing radiation.
\begin{figure}
\begin{center}
\includegraphics[width=0.5\textwidth]{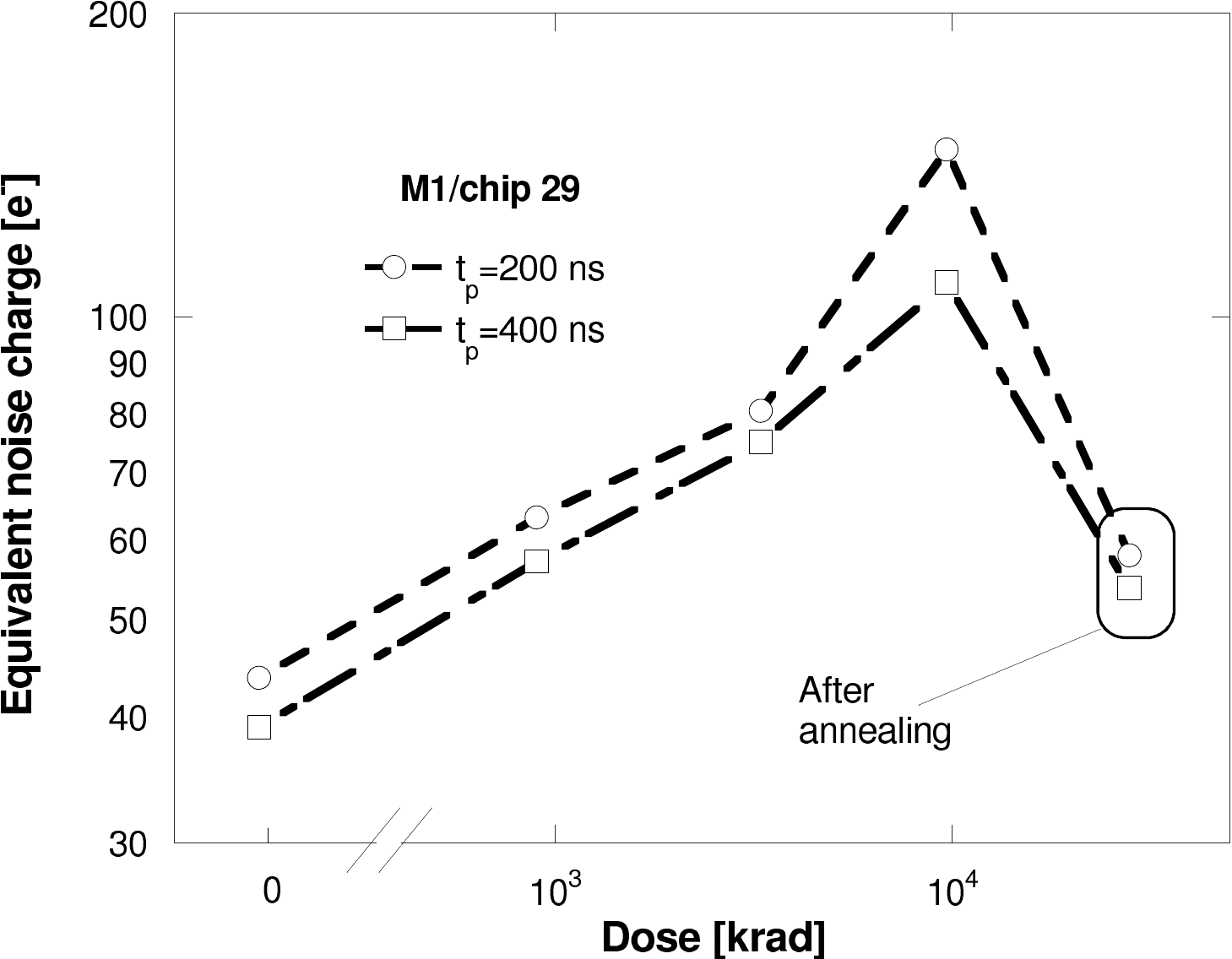}
\end{center}
\caption{Equivalent noise charge as a function of the absorbed dose and after the annealing cycle for DNW monolithic sensor. ENC is plotted for the two available peaking times.}
\label{fig:enc_gamma}
\end{figure}
\begin{figure}
\begin{center}
\includegraphics[width=0.5\textwidth]{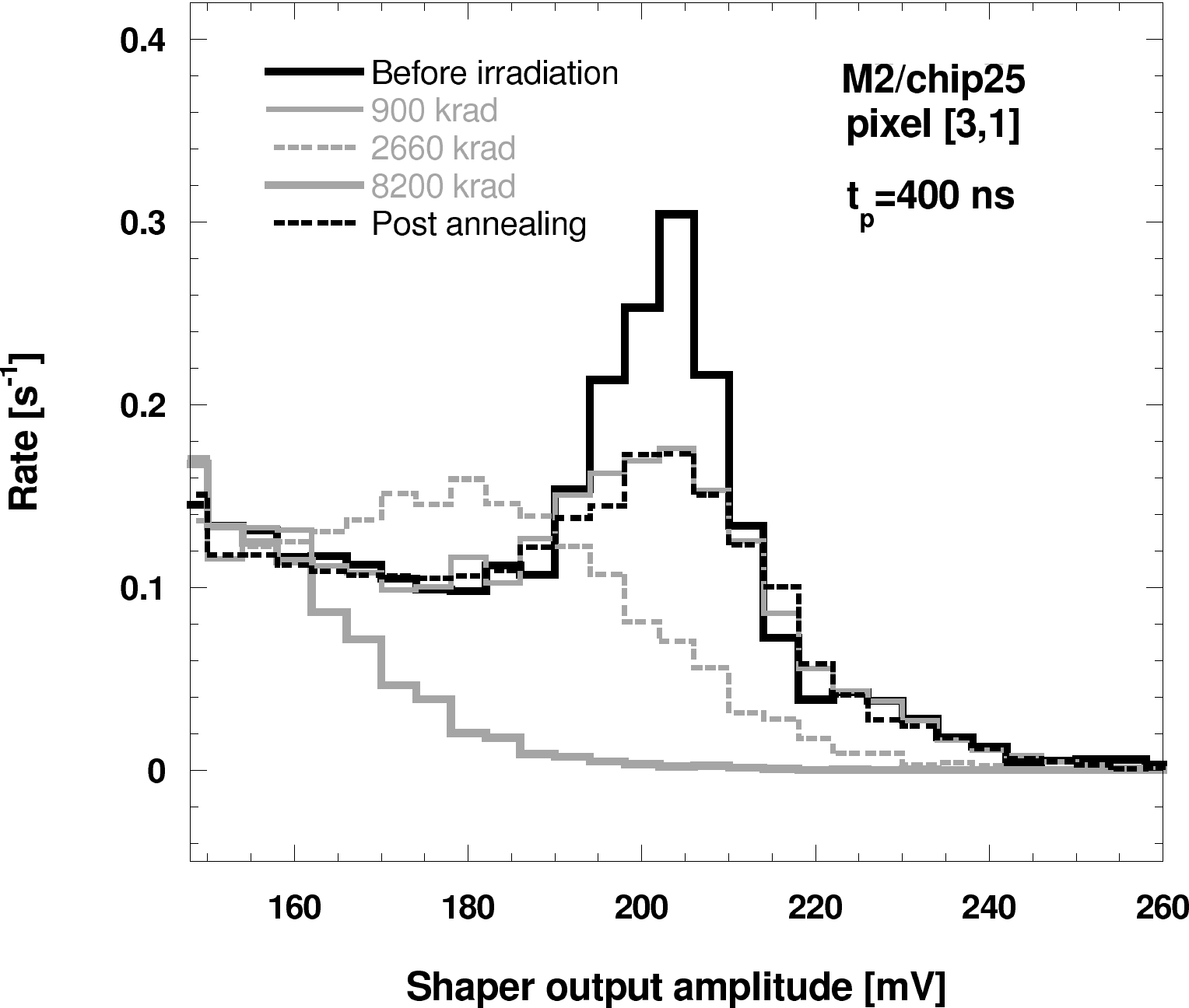}
\end{center}
\caption{Event count rate for a DNW monolithic sensor exposed to a $^{55}$Fe source before irradiation, after exposure to $\gamma$-rays and after the annealing cycle.}
\label{fig:fe_gamma}
\end{figure}

Figure~\ref{fig:fe_gamma} shows event count rate for a DNW monolithic sensor exposed to a $^{55}$Fe source before irradiation, after exposure to $\gamma$-rays and after the annealing cycle. As the absorbed dose increases, the 5.9 keV peak gets broader as a consequence of the noise increase (in fair agreement with data in Figure~\ref{fig:enc_gamma}. 
\begin{figure}
\begin{center}
\includegraphics[width=0.5\textwidth]{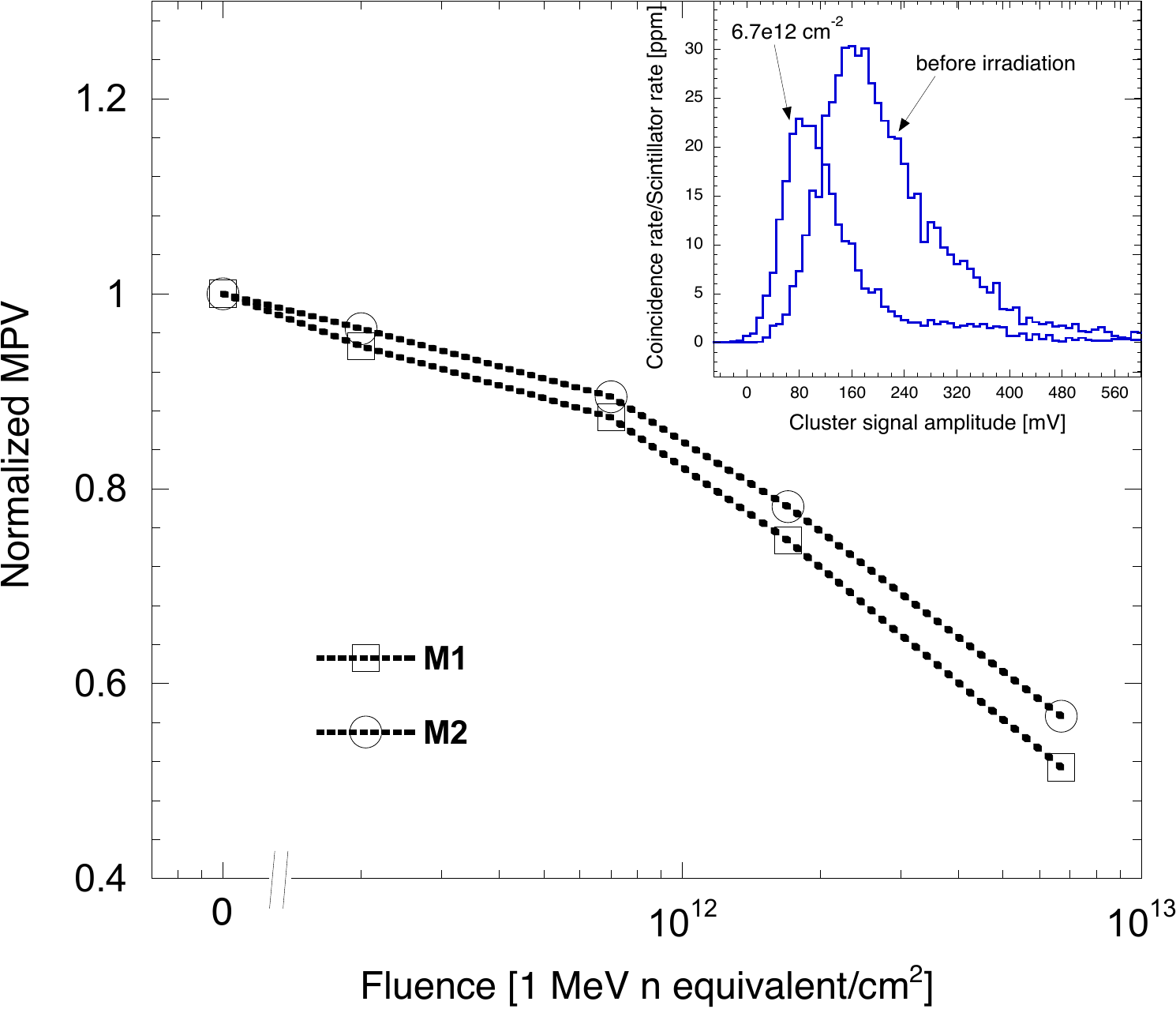}
\end{center}
\caption{Most probable value (MPV) of the $^{90}$Sr spectra (shown in the inset for one of the tested chips before irradiation and after exposure to a 6.7$\times$10$^{12}$~cm$^{-2}$ neutron fluence) normalized to the pre-irradiation value as a function of the fluence for DNW MAPS with different sensor layout.}
\label{fig:sr_neutron}
\end{figure}
At the same time, the peak is shifted towards lower amplitude values, as a result of a decrease in the front-end charge sensitivity also due to charge build up in the STI of some critical devices. 

DNW MAPS of the same kind have also been irradiated with neutrons from a Triga MARK II nuclear reactor to test bulk damage effects~\cite{zuccaradecs11}. The final fluence, 6.7$\times$10$^{12}$~1-MeV-neutron equivalent/cm$^2$, was reached after a few, intermediate steps. The devices under test (DUT) were characterized by means of several different techniques, including charge injection at the front-end input through an external pulser, sensor stimulation with an infrared laser and spectral measurements with $^{55}$Fe and $^{90}$Sr radioactive sources. Neutron irradiation was found to have no sizable effects on the front-end electronics performance. This can be reasonably expected from CMOS devices, whose operation is based on the drift of majority carriers in a surface channel, resulting in a high degree of tolerance to bulk damage. Exposure to neutrons was instead found to affect mainly the charge collection properties of the sensors with a reduction in the order of 50\% at the maximum integrated fluence. Figure~\ref{fig:sr_neutron} shows the most probable value (MPV) of the $^{90}$Sr spectra normalized to the pre-irradiation value as a function of the fluence for DNW MAPS with different sensor layout. A substantial decrease can be observed, to be ascribed to a degradation in the minority carrier lifetime.
A higher degree of tolerance was instead demonstrated in monolithic sensors with high resistivity (1~k$\Omega\cdot$cm) epitaxial layer~\cite{deveaux11}. Actually, doping concentration plays a role in determining the equilibrium Fermi level, which in turn influences the effectiveness of neutron-induced defects as recombination centers~\cite{messenger92}.

\paragraph{Quadruple-well CMOS MAPS.}\label{sec:inmaps_radhard}

Radiation hardness study of quadruple-well MAPS (see sections~\ref{sec:inmaps} and \ref{sec:apsel4well}) is in progress during the writing of this TDR. The preliminary results, shown in the following, are quite encouraging and suggest that the INMAPS 180~nm CMOS process might overcome the radiation tolerance issues found in DNW MAPS and become the candidate technology for the Layer0 pixel upgrade. A number of Apsel4well chips, featuring a high resistivity, 12~$\mu$m thick epitaxial layer, have been irradiated with  neutrons from the same source as the one used for DNW MAPS tests. Also samples with standard resistivity epitaxial layer (same epi-layer thickness) have been irradiated for the sake of comparison. 
\begin{figure}
\begin{center}
\includegraphics[width=0.5\textwidth]{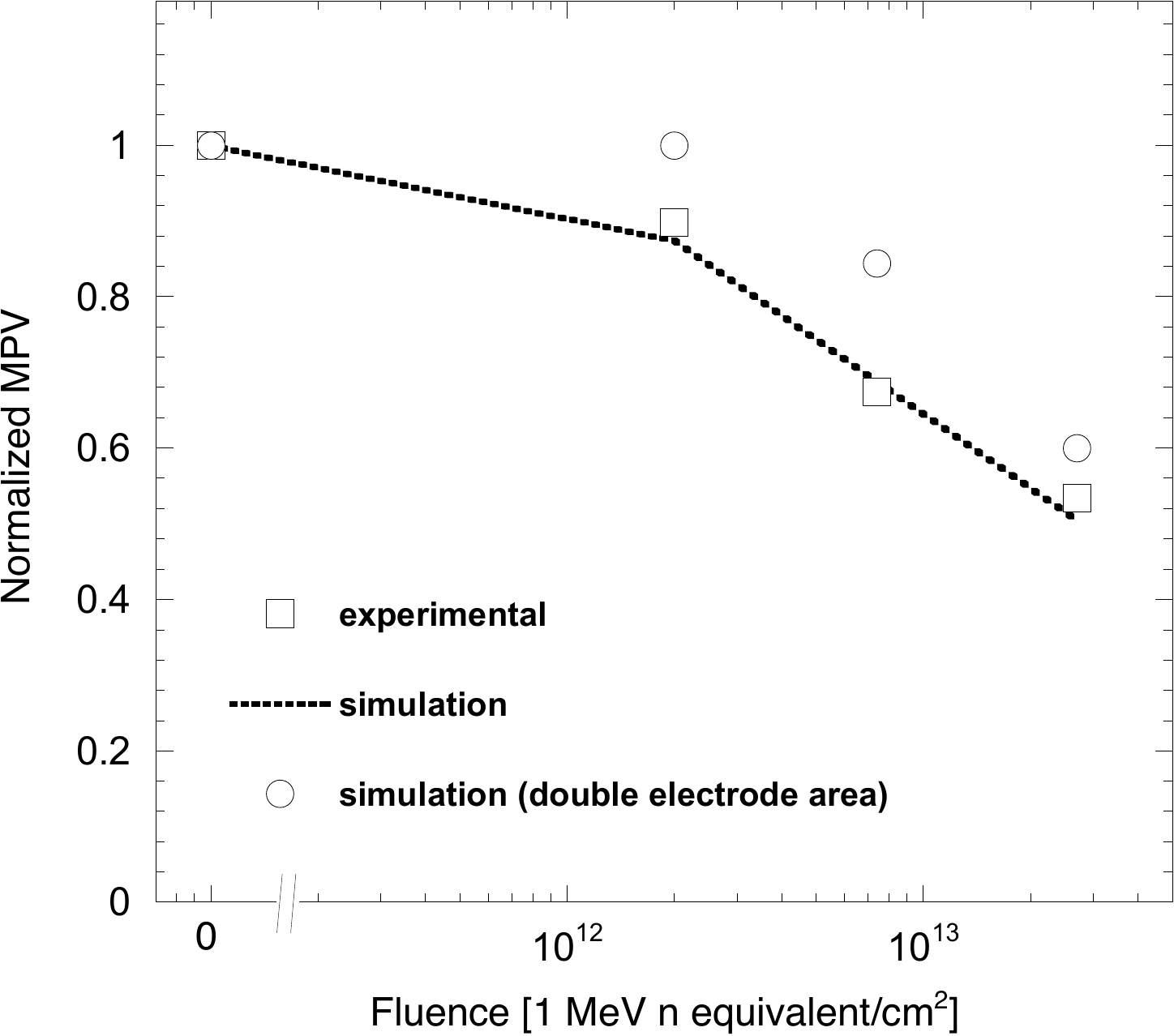}
\end{center}
\caption{Most probable value (MPV) of the $^{90}$Sr spectra as a function of the fluence for quadruple-well MAPS with high resistivity epitaxial layer. Data are normalized to the MPV measured in the non-irradiated device.}
\label{fig:inmaps_hr_90sr}
\end{figure}
The maximum fluence was 10$^{14}$~\mbox{1-MeV-equivalent neutron/cm$^2$}. The irradiated devices where tested with bench top instrumentation and with laser and radioactive sources. 
As already discussed in the case of DNW MAPS, irradiation with neutrons was found to affect mainly the collecting electrode performance. 
Figure~\ref{fig:inmaps_hr_90sr} shows the MPV (open square markers) of the measured $^{90}$Sr spectra, normalized to the MPV obtained in the non-irradiated device, as a function of the fluence, for quadruple-well MAPS with high resistivity epitaxial layer. Note that at a fluence of about $7\times 10^{12}$~cm$^{-2}$, the loss in collected charge is about 32\% of the pre-irradiation value. Data in Figure~\ref{fig:sr_neutron}, relevant to DNW MAPS, show a charge loss of about 50\% at a slightly smaller fluence. In Figure~\ref{fig:inmaps_hr_90sr}, results from device simulations, based on a purposely developed Monte Carlo model~\cite{radecs12}, are also shown. Simulations performed on devices with the same electrode geometry as in the test structures (dashed line in Figure~\ref{fig:inmaps_hr_90sr}) reproduce the experimental data with good accuracy. The same Monte Carlo model (see the plot with circle markers again in Figure~\ref{fig:inmaps_hr_90sr}) anticipates that, by doubling the collecting electrode area, the charge loss should be about 40\% at a fluence of $2.5\times 10^{13}$~cm$^{-2}$, which is the expected fluence, including a $\times5$ safety factor, for a 1 year operation in the Layer0. 
Optimization of the collecting diode dimensions is under evaluation to further improve S/N performance for irradiated devices.
%
In addition, the characterization of monolithic sensors developed in a very similar 180~nm CMOS technology (including deep P-well and high resistivity epitaxial layer options), but featuring a different readout approach, has recently provided some promising results in terms of radiation hardness \cite{alice_upgrade,rsmdd12}. In this case, less than 10\% loss in collected charge and a slight increase in the noise (about 20\%) were detected in MAPS samples irradiated with a neutron fluence of $3\times 10^{13}$~cm$^{-2}$.

\tdrsec{Services and  Utilities}

The vertex detector requires the following services, which must be brought inside the support tube to a location near the outboard of the W conical shield. 
	
\tdrparagraph{Power Supply}
The readout ICs require two low-voltage power supply lines 
(analog and digital); the silicon sensors need the bias voltage. 
Each power supply board must provide a unique low voltage level for each 
HDI side (referenced to the bias level of the silicon sensor's side). 
The further split between the analog and digital power takes place on the HDI. 
The bias voltage will be raised during the experiment life up to 200 V 
for the Layer0, to fully deplete the sensor in presence of radiation damage.  
The power supplies will be specially procured to match the vertex 
detector specifications to take under control the electronic noise.

\tdrparagraph{Cooling water}	
The readout electronics and transition card will be water cooled. Two sets of water connections for each cone (since each cone is constructed from two halves) and two set for each L0 cold flange (since each flange is constructed from two halves) will be required. In the same way, since each transition card support is constructed in two halves, two water connections for each transition card support is required. Also a connection for the cooled Be beam pipe is needed. The cooling water will be supplied by a special low volume chiller system dedicated to the vertex detector system.

\tdrparagraph{Dry air or nitrogen}
The vertex detector requires a dry, stable environment. Cooled and dry air or nitrogen from each side 
is planned to be passed through the internal volume of the SVT.  The airflow needed will be 
planned in relation to the silicon detector temperature required, as based on results of 
background simulation studies and thermal finite element analysis.

\bibliographystyle{\tdrbibstyle}
\balance
\bibliography{SVT/SVT-bib}


\renewcommand{\ChapLabel}{chap:DCH} 
\tdrchap{Drift Chamber}
\label{sec:DCHmain}  

\newcommand{\dch} {drift chamber\xspace}
\newcommand{\Dch}{Drift chamber\xspace}
\newcommand{\DCH}{Drift Chamber\xspace}
\newcommand{\pT}{\ensuremath{p_\perp}\xspace}
\newcommand{\HeIso}[2]{\ensuremath{{#1}\%\mathrm{He}-{#2}\%i\mathrm{C}_4\mathrm{H}_{10}}\xspace}
\def \epsstereo {\ensuremath{\varepsilon}\xspace}
\def \cc {cluster counting\xspace}
\def \CC {CC\xspace}
\def \SR {SR\xspace}
\def \Cc {Cluster counting\xspace} 
\def \Breco {\ensuremath{\B_\mathrm{reco}}\xspace} 
\def \fsr {\ensuremath{\mathcal{R}_{f:s}\xspace}}
\def \REP {\ensuremath{R^\mathrm{\scriptstyle EP}\xspace}}
\def \LDCH {\ensuremath{L^\mathrm{\scriptstyle DCH}\xspace}}
\def \RDCH {\ensuremath{R^\mathrm{\scriptstyle DCH}\xspace}}
\let\dEdx=\dedx
\newcommand\Tp{\rule{0pt}{2.6ex}}	
\newcommand\Bt{\rule[-1.2ex]{0pt}{0pt}}	

\setcapindent{0pt} 

\tdrsec{Introduction}
\label{sec:DCH_Overview}
The \superb \DCH (DCH) is the main tracking detector of the \superb
experiment. Immersed in a 1.5 T solenoidal magnetic field, the DCH
provides several position measurements along the track for a precise
reconstruction of the particle momentum.
It also measures the ionization energy loss used for
particle identification. In fact, the DCH is the primary device in
\superb to measure velocities  of particles having momenta below
approximately 700\mevc and, at least in the initial phase of the
experiment when no forward PID device is foreseen, it is the the only
particle identification device for tracks with $\vartheta\le25^\circ$.

Finally the \dch, together with the electromagnetic calorimeter, 
will provide the trigger of the experiment (Sec.~\ref{subsubsec:DCT}).

\tdrsubsec{Physics Requirements} 

Similarly to \babar\,\cite{bib:BaBarNIM}, one of the goals of the
\superb experiment is to reconstruct exclusive and inclusive final
states with high and well-controlled efficiency. This requirement
implies maximal solid angle coverage and highly efficient
reconstruction of tracks with \pT as low as $100\mevc$, the minimum
transverse momentum for a track produced at the interaction point to
leave enough hits to be reconstructed in the \dch.

Background rejection requires excellent momentum resolution
over the full momentum spectrum. In \babar the momentum resolution 
has been measured\,\cite{bib:BaBarNIM} on muon tracks to be
$\sigma_{\pT}/\pT=(0.13\pT+0.45)\%$, \pT in \gevc.  
In the \superb \dch we aim at obtaining a similar result, or better.

The average charged particle momentum in \superb is less than 1\,\gevc, 
smaller than in \babar due to the reduced boost. Multiple
scattering will therefore be the main limiting factor to track parameter
resolution for the majority of charged tracks in \superb. The effect
is controlled by keeping to a minimum the material budget in the DCH.
A thin inner wall helps matching 
tracks reconstructed inside the vertex detector with those in the DCH and
improves the \pT and the $\vartheta$ resolution by effectively
increasing the lever arm of the tracking system when adding high
precision measurement points from the vertex detector. It also
reduces photon conversion thus minimizing background in the chamber.
Low-mass endplates and outer wall limit photon conversions and
multiple scattering which could degrade the performance of the
detectors surrounding the DCH.

Finally, although the track dip angle is measured with high precision
by the vertex detector, the DCH must provide an independent
measurement of this angle to allow the reconstruction of secondary vertices of
long-lived particles decaying outside the vertex detector region, such as
\KS mesons or $\Lambda$ baryons. The DCH must also provide a measurement of the hit $z$
coordinate. A stand-alone full 3D tracking in the
DCH helps in extrapolating tracks to the calorimeter, to the FDIRC 
detector and  back to the vertex tracker, and allows their use in the L1 trigger.

\tdrsubsec{Design Overview}
\label{subsec:DCH_Design_Overview}
The physics requirements outlined in the previous section can be met
with a cylindrical \dch with performance similar to the \babar one.
We have therefore taken the \babar \dch\,\cite{bib:Babar_dch} as a starting point, seeking
improvements in two main areas: reduction of the detector material
and improvement of the \dEdx measurement.

We shall show that the first goal can be met with the adoption of modern 
construction techniques for the mechanical structure 
involving the use of composite materials (Sec.\,\ref{sec:DCH_Mechanics}), together with the use of lighter gas
mixtures and reduced wire material (Sec.\,\ref{sec:DCH_Optimization}).
Improvements in the energy loss measurements could be achieved
by adopting the cluster counting (\CC) technique\,\cite{bib:CC1,bib:CC2,bib:CC3,bib:CC4} for the DCH readout.
These two solutions constitute the baseline of the present
design and will be discussed in detail in the remaining sections of this
chapter.

We note that a fallback solution for the readout is represented
by the standard TDC/ADC readout chain discussed in
Sec.~\ref{subsec:DCT_Standard_RO}.

To reach its ambitious goal on the integrated luminosity, the \superb
experiment will work in ``factory-mode'', taking data continuously
over extended run periods.  This poses stringent requirements on the
reliability of all its subdetectors, including the DCH.  The \babar
\dch already met these requirements, and we expect that our choice of
separating the signal digitization electronics from the on-detector
preamplification boards will further improve the system reliability.

\tdrsubsec{Cluster Counting}
\label{subsec:clusterCounting}

Signals in drift chambers are
usually split to an analog chain which integrates the charge, and to a
digital chain recording the arrival time of the first electron,
discriminated with a given threshold. 
The \cc technique consists
instead in digitizing the full waveform to count and measure the time
of all individual peaks. On the assumption that these peaks can be
associated to the primary ionization acts along the track, the energy
loss and to some extent the spatial coordinate measurements can be
substantially improved.
In counting the individual clusters, one indeed removes the sensitivity
of the specific energy loss measurement to fluctuations in the
amplification gain and in the number of electrons produced in each
cluster, fluctuations which significantly limit the intrinsic
resolution of conventional \dEdx measurements.

The ability to count the individual ionization clusters and measure
their drift times strongly depends on the average time separation
between them, which is, in general, relatively large in He-based gas
mixtures thanks to their low primary yield and slow drift velocity.
Other requirements for efficient cluster counting include good
signal-to noise ratio with no or limited gas-gain saturation, high
preamplifier bandwidth, and digitization of the signal with a sampling
speed of the order of 1Gs/sec. Finally, it is necessary to extract
online the relevant signal features (\ie\ the cluster times), because
the DAQ system of the experiment would not be able to manage the
enormous amount of data from the digitized waveforms of the about
10\,000 drift chamber channels.

\tdrsubsec{Geometrical Constraints}
\label{subsec:DCH_geom_constraints}

A side view of the \dch is shown in Fig.~\ref{fig:DCH_SideView}.
The DCH inner radius is constrained by the final focus cooling system,
by the Tungsten shield surrounding it and by the SVT service space to
$R_{inner}=270\mm$. The outer radius is constrained to
$R_{outer}=809\mm$ by the DIRC quartz bars, as in \babar.
The total length available for the DCH is $L=3092\mm$,
including the space for front-end electronics, signal and high voltage
cables, and the needed services (\eg\ cooling, gas pipes). The space
necessary for possible future upgrades of the \superb detector such as
the backward electromagnetic calorimeter (\secref{EMCback}\ref{sec:EMCback}) and the
forward PID system (Sec.~\ref{section:forwardPID}) has already
been taken into account.

\begin{figure*}[tbp]
\begin{center}
\includegraphics[width=0.9\textwidth]{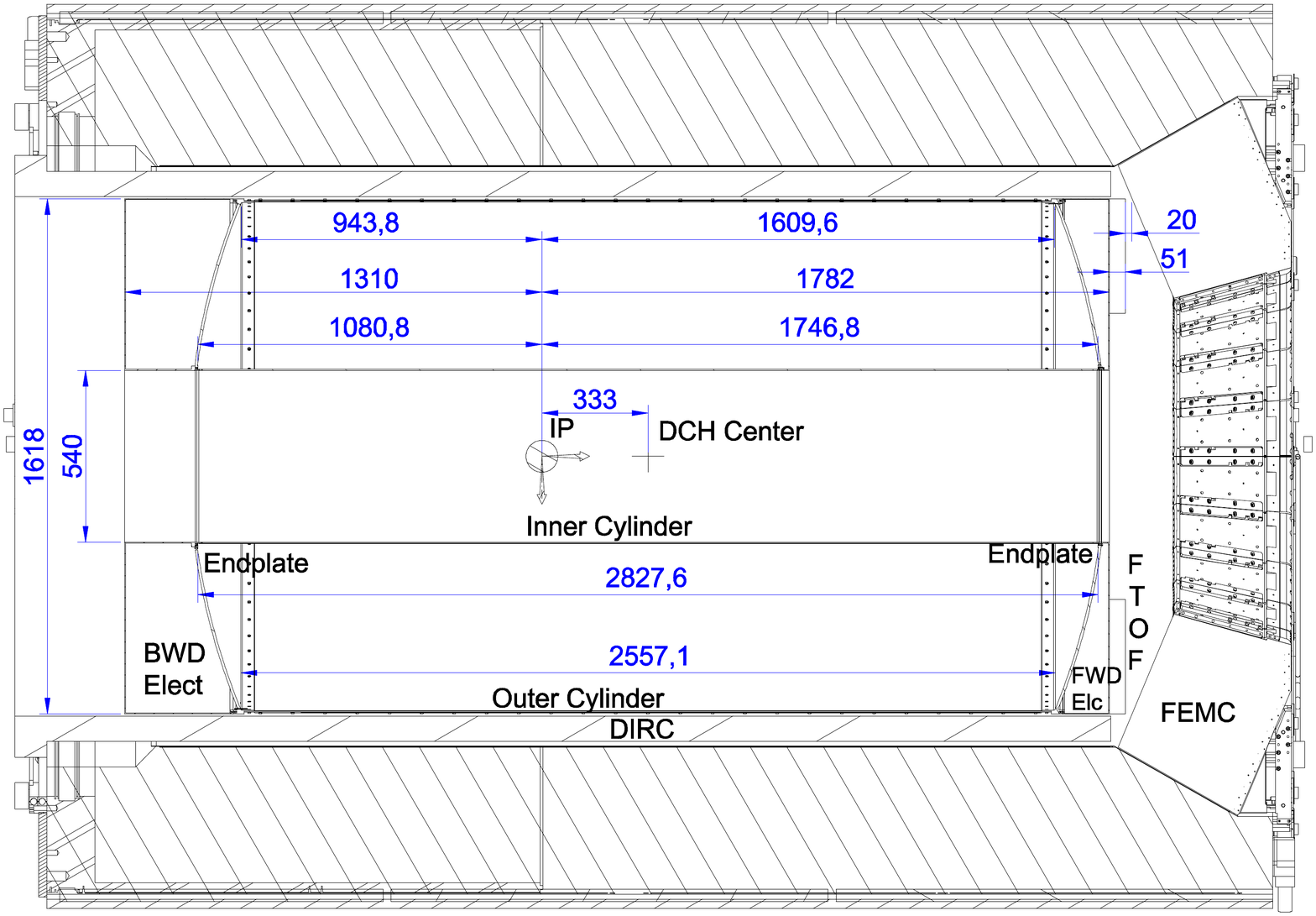}
\caption{Longitudinal section of the DCH with principal dimensions.}
\label{fig:DCH_SideView}
\end{center}
\end{figure*}

The drift chamber is shifted with respect
to the interaction point by an offset of 333\mm, resulting from the fact that we want to exploit all the available space in the backward and forward regions (Figure\,\ref{fig:DCH_SideView}).
 .

\tdrsubsec{Machine Background Considerations} 
\label{subsec:DCH_backgrounds}

Background processes are responsible for the largest part of the DCH rate. 
A rate higher than the design level would affect the charged track
reconstruction, due to spurious hits, and generally degrade tracking performance.
Additionally, the DAQ electronics can be progressively damaged when being exposed 
to a significant rate of particle for a long period of time. 
In both cases, an accurate simulation of the radiation rates is needed to estimate 
the real performances and the life-span of the electronics.

The main tool used to perform this estimate is 
the full simulation program described in detail in Sec.~\ref{sec:Full_Simulation}.
The chamber gas volume is described as a homogeneous
medium with a density corresponding to the weighed average of the gas and the wires densities. 
The chamber structure is instead accurately described including carbon fiber inner and outer walls
and endcaps, each with the appropriate density. 
Three silicon plates are included on the backwards side
to simulate the amount of material corresponding to the front-end electronics.
The basic information in the simulation output corresponds to a single interaction (or step)
of the particle simulated by GEANT4\,\cite{bib:G4}.
Each particle step inside the gas volume and silicon plates is recorded, together with additional information
about the type of particle, its position, incident energy, vector momentum, and deposited energy. 
The simulation is not aware of the internal wire structure: this allows us to simulate background events
faster, and rates can be computed in each layer for different wire configurations in a post-simulation analysis,
as described below.
The post-simulation analysis evaluates also the radiation dose, the equivalent fluence of 1 \mev neutrons,
and fluences of different kind of particles in the plates representing the DAQ electronics.

A dedicated application has been developed to analyze the simulation output 
and produce the relevant quantities and plots. 
A given wire structure is provided as input to the application, with all the information on the cell size
and stereo angles.
For each step, the helix parameters of the track trajectory are computed starting from the initial and final points, and the momentum. Assuming \dEdx constant along the step, the total deposit energy is divided between each cell crossed by the track proportionally to the path length inside the cell itself.
After having analyzed all the steps from charged tracks in the event, 
the table is used to fill a histogram with one entry for each layer: 
the number of cells with deposited energy 
greater than zero in each layer is added into the corresponding interval. 
A histogram is produced for all the background sources and then rescaled 
for the cross-section/frequency of each process.
For each given transverse radius of the DCH, the tracks produced by 
background processes are with good approximation
isotropic in azimuthal angle, so no information is lost 
when averaging the cells fired over the same layer. 

The largest contributions to the rate are from radiative Bhabha and Pair productions.
The simulation includes also contributions from Touschek scattering and beam scattering with residual gas.
Photons from synchrotron radiation give a negligible contribution to the DCH rate
and radiation levels, so they are omitted in the following. 
Figure \ref{fig:mdi_dch_rate_axial} shows the total rate and the separate contribution 
for each background source assuming a typical configuration with axial cells only.
\begin{figure}[tbh]
  \begin{center}
    \includegraphics[width=0.5\textwidth]{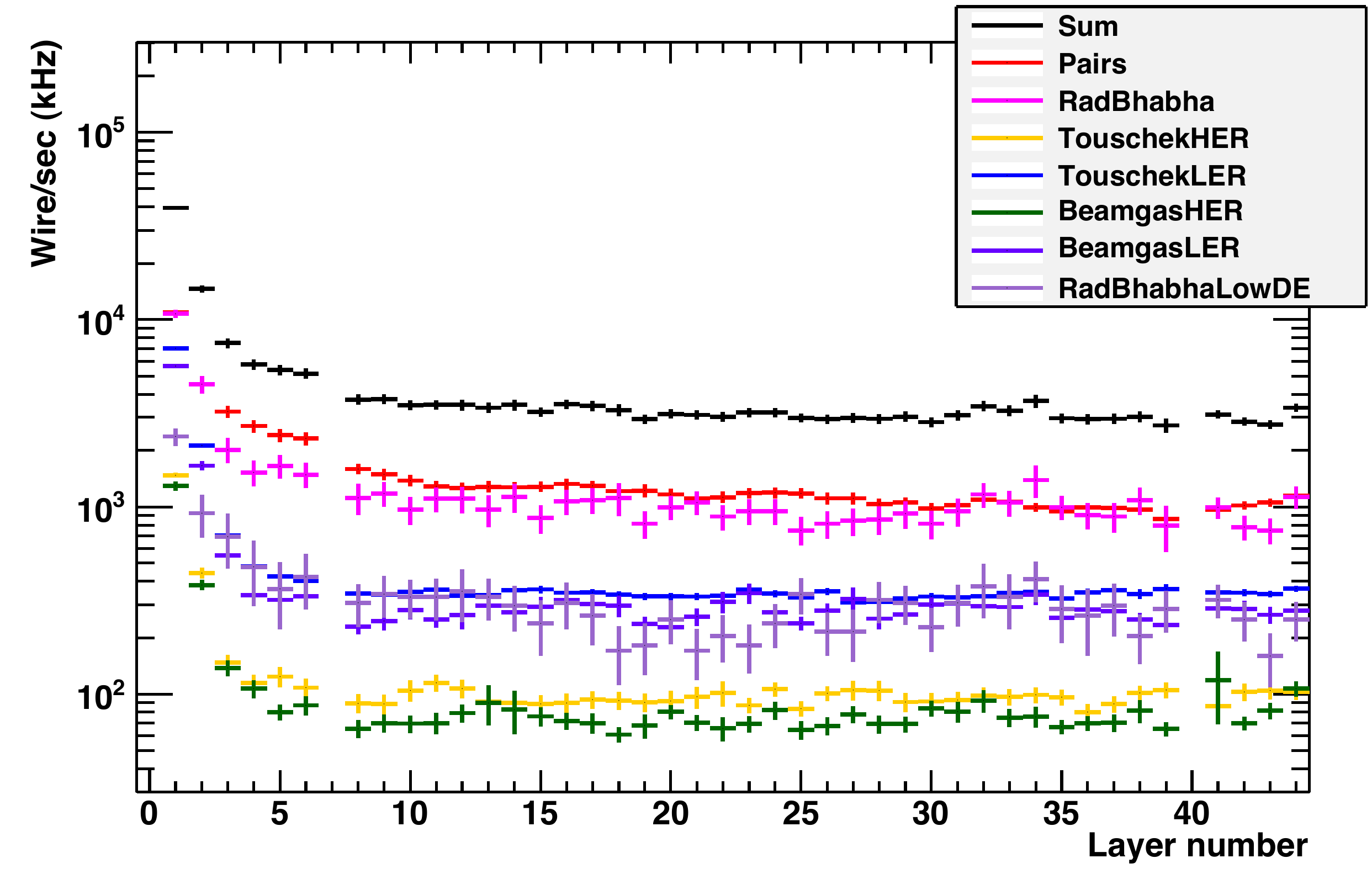}	
    \caption{Total rate for each DCH layer for a typical axial configuration.
      The empty intervals correspond to empty spaces between groups of layers.
      The separate contributions from each background source are shown in different colors
      (color online, see Table \ref{tab:fullsim_samples} for more details on the samples).}
    \label{fig:mdi_dch_rate_axial}
  \end{center}
\end{figure}
Most of the background activity is located around the interaction point, so
the rate decreases fast when moving to outer layers.
No significant difference in the shape of the rate is observed between different background sources, 
because most of the particles that hit the chamber are not primaries, 
but particles that exit the final focus shielding after many interactions, as explained below.

We found that the DCH rate had a significant dependence on the maximum step length allowed in the GEANT4 simulation.
A dedicated study showed that this is due to a poor simulation of the multiple scattering over long distances
in low density materials. For this reason the maximum step length in the
gas chamber has been fixed to 1 mm, reducing this dependence to a negligible level.

The studies showed that the largest part of the rate is not due to charged tracks coming directly from the interaction point, 
but mostly from photons coming out of the shielding and kicking out
electrons off the chamber walls when they enter the chamber.
Looking specifically at those photons coming out of the final focus elements, 
we were able to detect some areas where the number of particles was higher than usual. 
The shielding has been subsequently optimized to reduce the number particles
from those areas.

Photons impinging on the DCH wall produce low energy electrons that can
still travel long distances in the gas. If their transverse momentum is
small, they can even traverse the whole chamber spiraling along the $z$ coordinate. 
In the case of stereo layers, one such a particle travelling along the
longitudinal direction can thus cross multiple cells, 
explaining the higher rate observed in the stereo layers shown in Figure~\ref{fig:mdi_dch_rate_stereo}.
\begin{figure}[tbh]
  \begin{center}
    \includegraphics[width=0.5\textwidth]{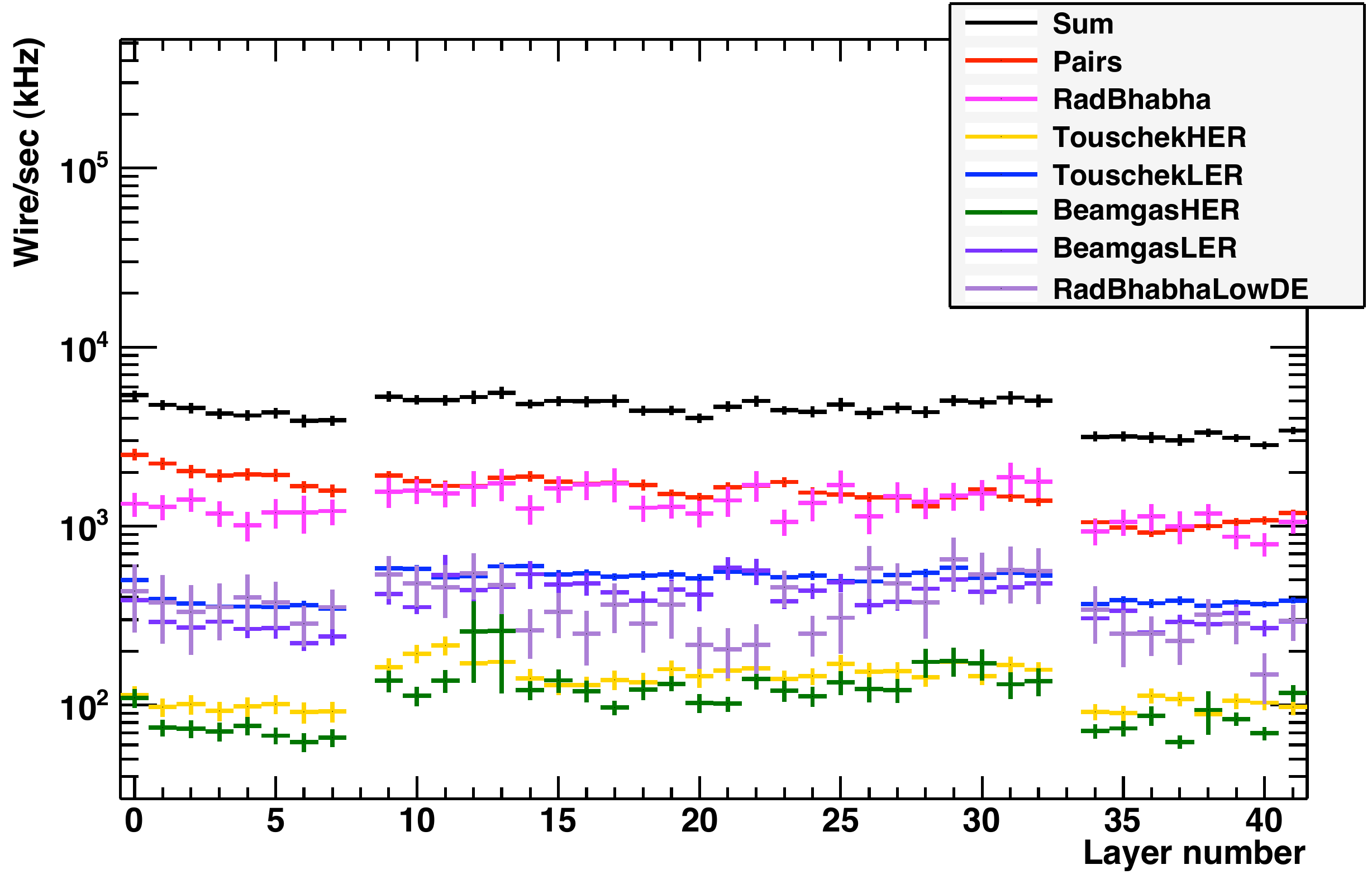}	
    \caption{Total rate for each DCH layer for a typical stereo configuration (layer 11--34, stereo angle of 70--90 mrad; the other layers are axial).
      The empty intervals correspond to empty spaces between groups of layers.
      The separate contributions from each background source are shown using different colors
      (color online, see Table \ref{tab:fullsim_samples} for more details on the samples).}
    \label{fig:mdi_dch_rate_stereo}
  \end{center}
\end{figure}
The wires in the central block from layer 11 to 34 have a stereo angle from 70 to 90 mrad,
and show clearly a rate higher than the neighbor axial wires.
The higher rate in the first layers observed in the plot of  Figure~\ref{fig:mdi_dch_rate_axial}, 
is not present here because, for this configuration, the first layer is 2 cm far from the inner wall.

With the geometry used for the simulation, a typical configuration with a central section of stereo layers, 
and an integration time of $1\mu \rm s$, the occupancy is estimated at $\sim5\%$. 

The radiation levels estimated in the region where the electronics will be located are generally low.
For a period corresponding to an integrated luminosity of $10~\rm ab^{-1}$
the estimated dose is below 8 Gy, the equivalent 1 MeV neutron fluence is below $3 \times 10^{11}~n_{eq}/cm^2/yr$,
and the fluence due to hadrons with kinetic energy larger than 20 MeV is below $7 \times 10^{10}~n_{eq}/cm^2/yr$.

\tdrsec{Design Optimization}
\label{sec:DCH_Optimization}
The \babar\ \dch operational record has been quite good, both for 
performance and reliability.
For this reason, several features of the \superb DCH, which will operate in
a similar environment, derive from the \babar \dch.

However, given that its design and construction dates back over
fifteen years, we have looked for possible technological advances and
design improvements to build a \dch with performances even better than
\babar.

\tdrsubsec{Performance Studies}
\label{subsec:DCH_performance_studies}

We have made extensive simulation studies to evaluate the impact on
track reconstruction of a number of design choices concerning both the
external structure (radius and material of the inner cylinder,
position and shape of the endplates) and the internal layout (cell
shape, wire orientation). The studies have been performed using the
\superb\ fast simulation tool {\it FastSim}, whose main features are
described in Chapter~\ref{sec:computing}. The main results of these
studies are summarized in the following sections.

\tdrsubsubsec{Radius and Thickness of the Inner Wall}

The simulation of several signal samples with both high (\eg\
$B\to\pipi$), and medium-low (\eg\ $\B^0\to D^{*-} K^+$) momentum
tracks indicates that, as expected, the momentum resolution improves
as the minimum \dch radius $R_{min}$ decreases, as long as the hit
rate per layer is low enough not to significantly impact the
reconstruction.  However, as discussed in
Sect.\,\ref{subsec:DCH_geom_constraints}, $R_{min}$ is actually
limited by mechanical constraints from the cryostats and radiation
shields, and is set to $R_{min}=270$~mm.

The thickness of the inner wall can affect significantly the
reconstruction of charged particles, especially those with lower
momentum where multiple scattering is larger.  For example, the width
of the $\Delta E$ distribution in $B^0\to D^{*-} K^+$ events gets
about 7\% narrower if the thickness of the carbon fiber inner wall is
decreased from 1~mm---equal to 0.4\%~\Xrad, comparable to the 
1~mm of Beryllium used 
in \babar---to about 0.2~mm (0.1\%~\Xrad). A carbon
fiber, 0.2~mm-thick inner cylinder is the current baseline, although
an even thinner solution is being considered as discussed in
Sec.~\ref{subsec:Endplates}.

\tdrsubsubsec{Endcap shape and position}
\label{subsubsec:DCH_Length_Studies}

\begin{figure}[!tbp]
\includegraphics[trim=0mm 0mm 20mm 0mm,clip,width=0.48\textwidth]{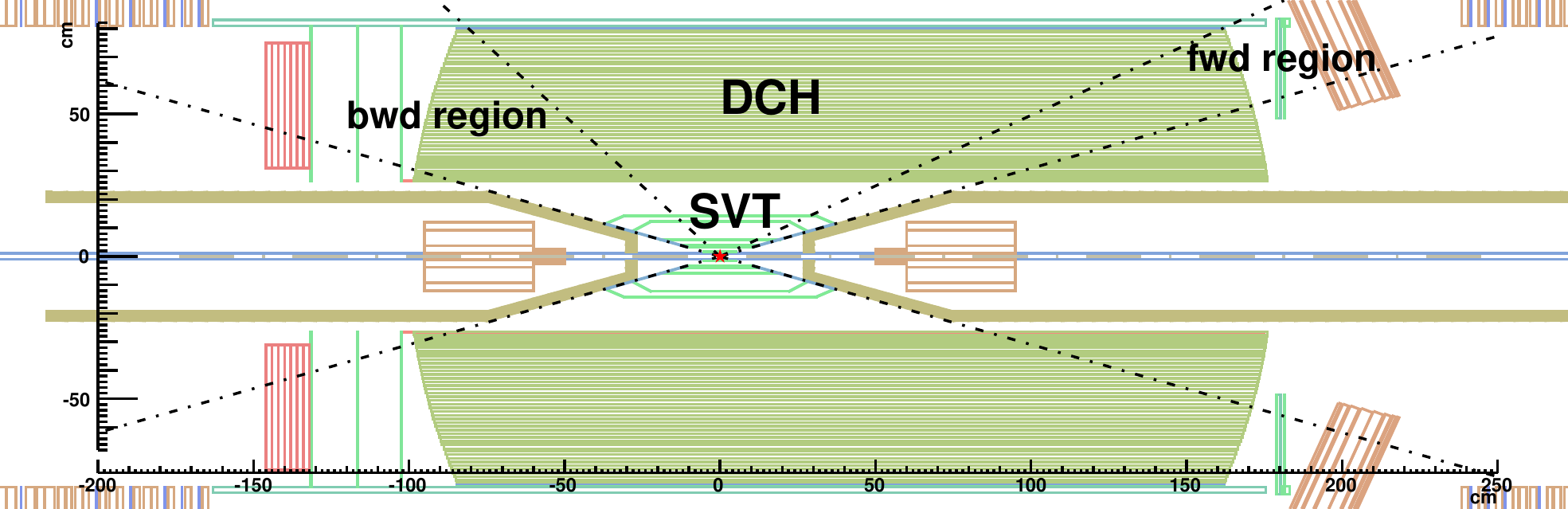}\\

\includegraphics[trim=0mm 0mm 20mm 0mm,clip,width=0.48\textwidth]{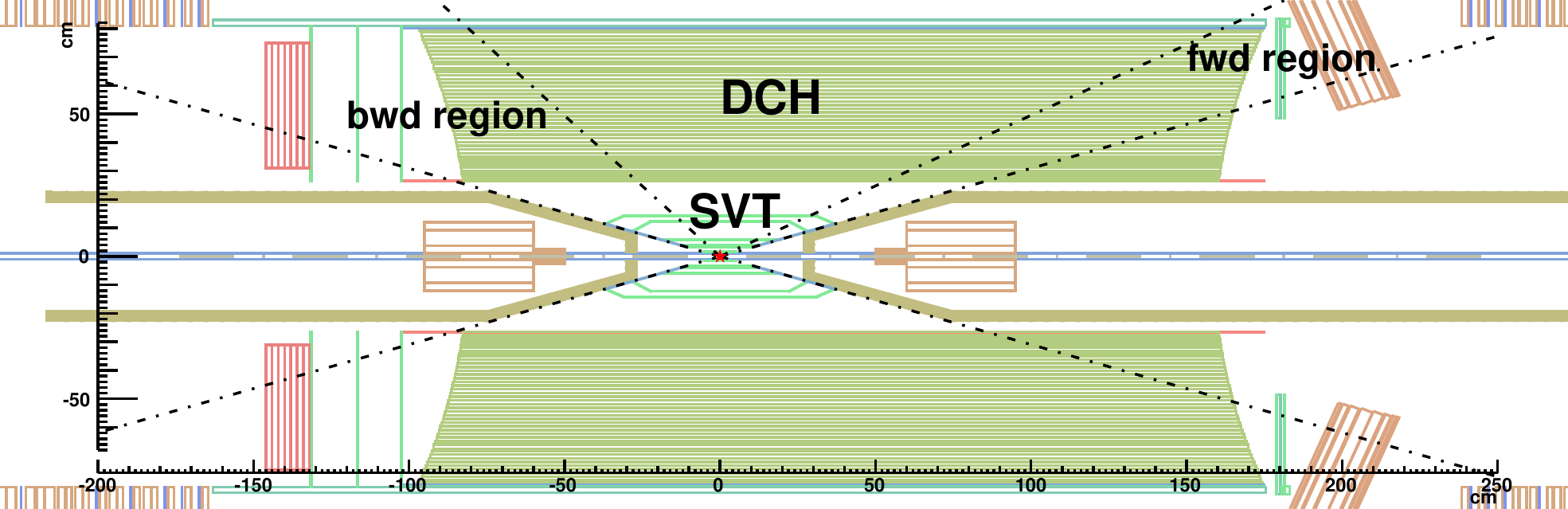}
\caption{Example of DCH configurations where the endcaps are placed in
  the same position (same $z_{min}$ and $z_{max}$) but their shape is
  different: convex (top) and concave (bottom). The forward and
  backward regions are indicated.}
\vskip -10pt
\label{fig:dch_endcap_shapes}
\end{figure}

A study was carried out to compare the momentum and \dEdx measurement
resolution of tracks reconstructed in the forward or backward
directions as a function of the endcap shape (convex or concave) and
its position along the beam ($z$) axis.
Figure~\ref{fig:dch_endcap_shapes} shows two of these configurations
together with the definition of the forward and backward regions.  We
simulated samples of single charged pions with flat $p$ and
$\cos\theta$ distributions and samples of $B^0\to D^{*-}K^+$ decays.
We compared the \pT resolution, the $K/\pi$ separation based on \dEdx
and the reconstruction efficiency of $B^0\to D^{*-}K^+$ for different
configurations. At fixed $z$ position, the average \pT resolutions for
the concave and convex shapes are consistent within a 3\% relative
uncertainty in both the forward and backward regions. The average
$K/\pi$ separations are also consistent within 1--2\%.  Differences of
this order of magnitude depend on the specific polar angle
distribution of the tracks.  The reconstruction efficiency of $B^0\to
D^{*-}K^+$ events is not affected by the endcap shape in the forward
region, while it is slightly higher ($\approx 1\%$) when the endcap in
the backward region is convex. The average variation of
$\sigma(\pT)/\pT$ as a function of the endcap position is roughly
consistent with 1\% (1.5--2.0\%) per additional cm of DCH length in the
forward (backward) direction. Similarly, the average $K/\pi$
separation based on \dEdx increases by about 0.5\% (1\%) per
additional cm of DCH length in the forward (backward) region. This
asymmetry between the forward and backward direction is ascribed to
the different average number of crossed layers.

To summarize, also as a result of weighing  the effects discussed
above for the fraction of tracks crossing these regions (typically
below 10\% of the total), little difference in terms of performance is observed
between the concave and convex endplate shapes \textit{per se},
while the length does have a significant impact. Due to the details of
mechanical design of the endplate outer flanges, convex endplates
allow for a longer DCH.

\tdrsubsubsec{Wire and Gas Material} The \babar \dch featured
concentric layers of hexagonal cells, organized in axial (A), positive
stereo (U) or negative stereo (V) super-layers. This choice was
motivated by the good symmetry of the hexagonal cells, and by the
small number of field wires per sense wire (field-to-sense wire ratio
$\fsr=2$) because many field wires are shared among adjacent cells.
However, such a small value \fsr\ required larger wire diameters---and
therefore more material in the tracking volume and more force on the
endplates---to avoid gas amplification at the field wires. In
addition, the advantage of stringing fewer field wires was outbalanced
by the guard wire layers needed at every super-layer transition to
keep the electric field homogeneous across layers (hexagonal cells
with different stereo angles cannot intersect radially and must be
separated).
\begin{table}[tbp]
\caption{
Combined gas+wires material budget for different cell geometries in a
\HeIso{80}{20} mixture. Values from the \babar DCH are shown in boldface.}
 \vskip10pt
 \begin{tabular}{|lccc|} \hline
Cell  & $\rho(g/cm^3)$ & \Xrad (m) &\# \\
layout & $\times10^4$ & & layers\\ \hline
                       & coat/nocoat & coat/nocoat &\\
Hex 2:1 & \textbf{10.0}/9.5 & \textbf{270}/350 & 40\\
Sqr 3:1  & 9.8/8.5 & 310/400 & 44\\
Sqr 4:1  & 8.5/8.1 & 320/430 & 44\\ \hline
\end{tabular}
\label{table:DCH_gas_and_cell_material}
\end{table}
In Table\,\ref{table:DCH_gas_and_cell_material} we report the material
budget for the hexagonal \babar-like layout and for two alternative
solutions using rectangular cell geometry with different values for
\fsr.  All the numbers shown in
Table\,\ref{table:DCH_gas_and_cell_material} and used in the
simulations below were obtained for an inner DCH radius of
$R_{min}=210$ mm and a cell height of $12$ mm (the final configuration
will have cells of slightly different heights, as discussed in
Sec.\,\ref{subsec:DCH_CellStructure}).  The case in which the field
Aluminum wires are coated with a 0.5\mum thick gold plating is
compared to the case where the coating is absent.  The hexagonal
configuration has 40 layers while the other two have 44 layers.  Since
in the square cell configurations there is no need to leave space
between superlayers with different stereo orientation, for a given
cell size it is possible to allocate more measurement layers compared
to the hexagonal configuration. The average single hit spatial
resolutions are estimated to be similar in all configurations.
Figure~\ref{fig:DCH_cell_study} compares $\sigma(\pT)/\pT$~vs~\pT for
the configurations in Tab.~\ref{table:DCH_gas_and_cell_material}. The
average \pT resolution in the \fsr=4 square cell configuration is
about 5\% better than in the hexagonal cell scenario.  Both the
material and the wire count is optimized with rectangular cells.
\begin{figure}[tbp]
\includegraphics[trim=0mm 0mm 20mm 0mm,
clip,width=0.48\textwidth]{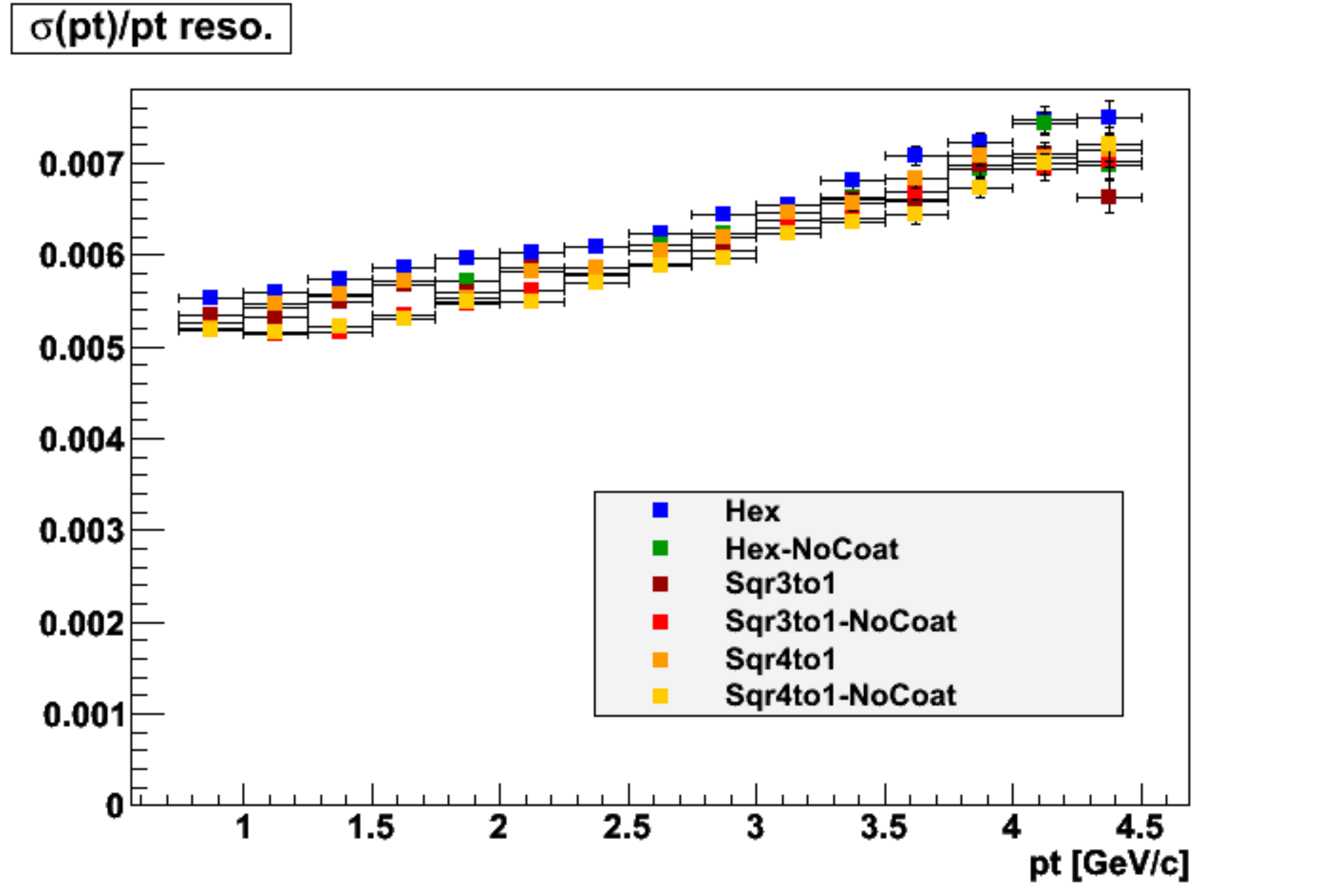}
\caption{Relative \pT resolution for different cell geometries.
  The color codes refer to the configurations in
  Table\,\ref{table:DCH_gas_and_cell_material}.}
\vskip -10pt
\label{fig:DCH_cell_study}
\end{figure}

The effect of the material budget of different gas mixtures was also
studied.  We considered square cell configurations with \fsr=3 and
three gas mixtures: He80-Ibu20, He90-Ibu10 and He80-Met20.
Figure~\ref{fig:DCH_gas_study} shows $\sigma(\pT)/\pT$~vs~\pT for the
three scenarios.  $\sigma(\pT)/\pT$ for the He80-Ibu20 configuration
is 10-15\% worse than in the He80-Met20 mixture.
\begin{figure}[tbp]
\includegraphics[trim=0mm 0mm 20mm 0mm,
clip,width=0.49\textwidth]{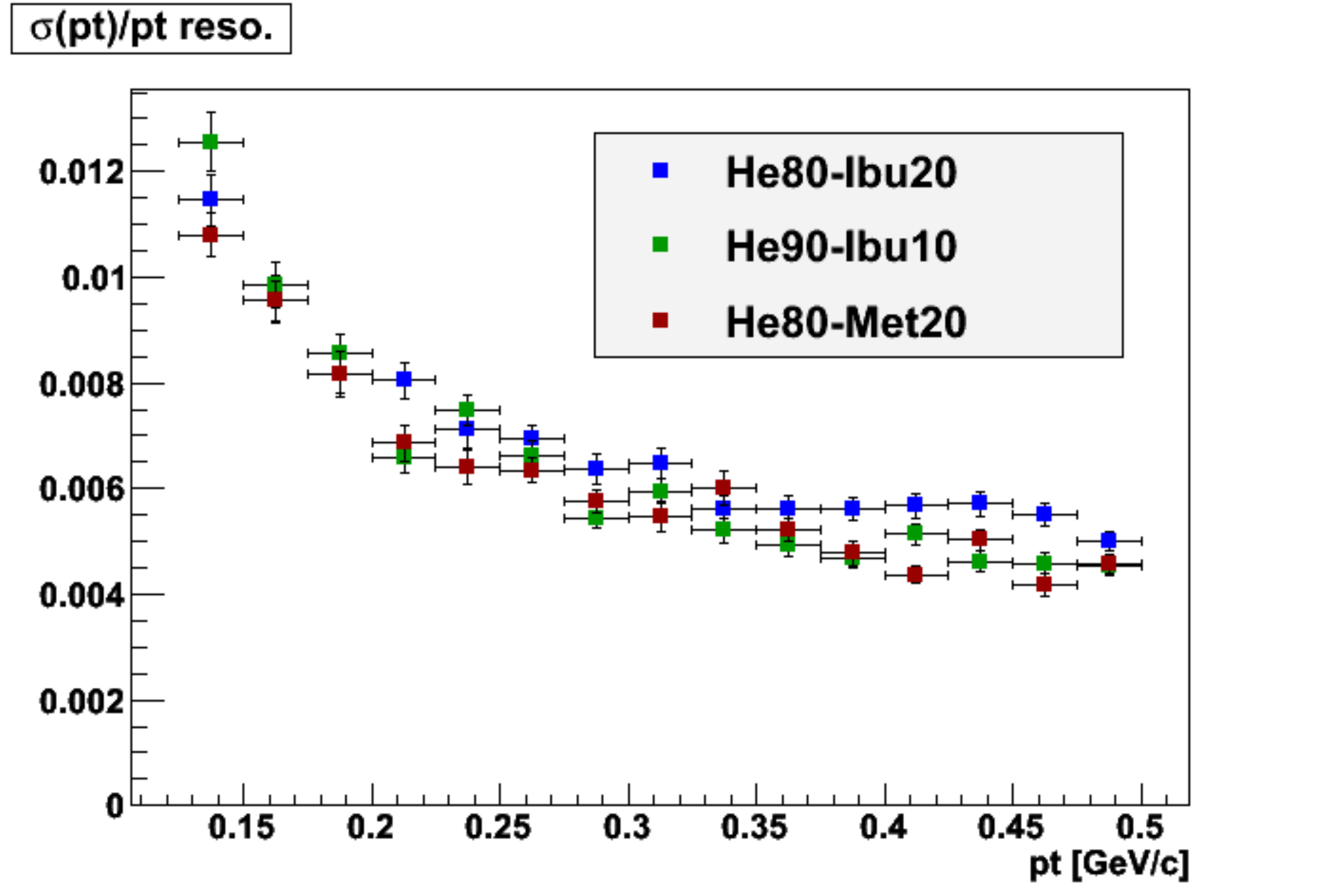}
\caption{Relative \pT resolution for different gas mixtures in the
  configuration with rectangular cells and $\fsr=3$.}
\vskip -10pt
\label{fig:DCH_gas_study}
\end{figure}

\tdrsubsubsec{Stereo Angle Layout}
The impact of the stereo angle layout on tracking performance was
studied by comparing the reconstruction of $\B\to\pipi$ and
$\B\to\Dstar K$ events in four
different configurations: \textit{i)} a \babar-like AUVAUVAUVAUVA;
\textit{ii)} a completely axial DCH; \textit{iii)} a ``mostly stereo
SL'' layout in which the  innermost and outermost superlayers are
axial, and the remaining ones alternate positive and negative stereo
orientations, AUVUVUVUVA; and \textit{iv)} a ``stereo layers''
configuration in which two axial superlayers placed in the innermost and 
outermost position surround a set of alternate stereo angle layers (alternate 
stereo angle layers were \eg\ used in the KLOE \dch).
The outcome of this study is that since the Silicon
Vertex Tracker largely dominates the polar angle measurement,
the arrangement of the stereo layers does not impose tight constraints
on tracking: for example both the \pT and the $\Delta E$ resolution
are almost identical in the four configurations (configuration
\textit{iii)} is slightly better for the case of relatively softer
tracks in $\B\to\Dstar K$ events). 
Compared to the completely axial configuration, however,
configurations with stereo layers are preferred because they can
provide -- in addition to 3D trigger information --  some constraints
on the polar direction of those few tracks that are not reconstructed
in the SVT.

\tdrsubsubsec{Cluster Counting}
\label{subsec:CCMcStudy}
We introduced in Sect.\,\ref{subsec:DCH_Design_Overview} the possibility
of enhancing the performance of particle identification (PID) in the
\DCH by counting the individual ionization clusters. In general, an
improved PID would have several advantages, including better
reconstruction efficiency in inclusive and exclusive modes, better
tagging efficiency, and reduced backgrounds.

In order to estimate the benefits of the \cc technique, we evaluated
the efficiency of the semi-exclusive hadronic \B reconstruction
(``\Breco''), a technique developed by \babar to obtain
high-statistics samples of \B candidates used in recoil analysis, \eg\
of rare decays.

We generated four different samples of 10 million generic \Bpm decays each in which at
least one $K$ is present in the final state.
In one sample the \dEdx performance is set equal to the \babar one and
used as a reference. 
To simulate the improvement due to \cc, in the other three samples $S_1,S_2,S_3$ the momentum- and $\vartheta$ angle-averaged 
$K/\pi$ separation is set to be higher than the
reference one by a factor of 1.3, 1.5 and 1.7, respectively. 
The results, given in terms of the ratios of their \Breco reconstruction 
efficiency with respect to the reference, are summarized in Figure\,\ref{fig:ElisaPlot}. 
It is observed that this ratio improves by 4--6\% when \cc\ is used.
The figure shows that the improvement is present also if kinematical cuts on $m_{ES}$ and $\Delta E$ are
used in conjunction with particle ID information. 

The actual improvement achievable with the use of \cc\ must be determined experimentally.
Preliminary results from test beam data are discussed in Section\,\ref{subsec:SingeCellProto}.  
\begin{figure}[tbp]
\begin{center}
{\includegraphics[width=0.48\textwidth, trim=0mm 5mm 20mm 0mm, clip]{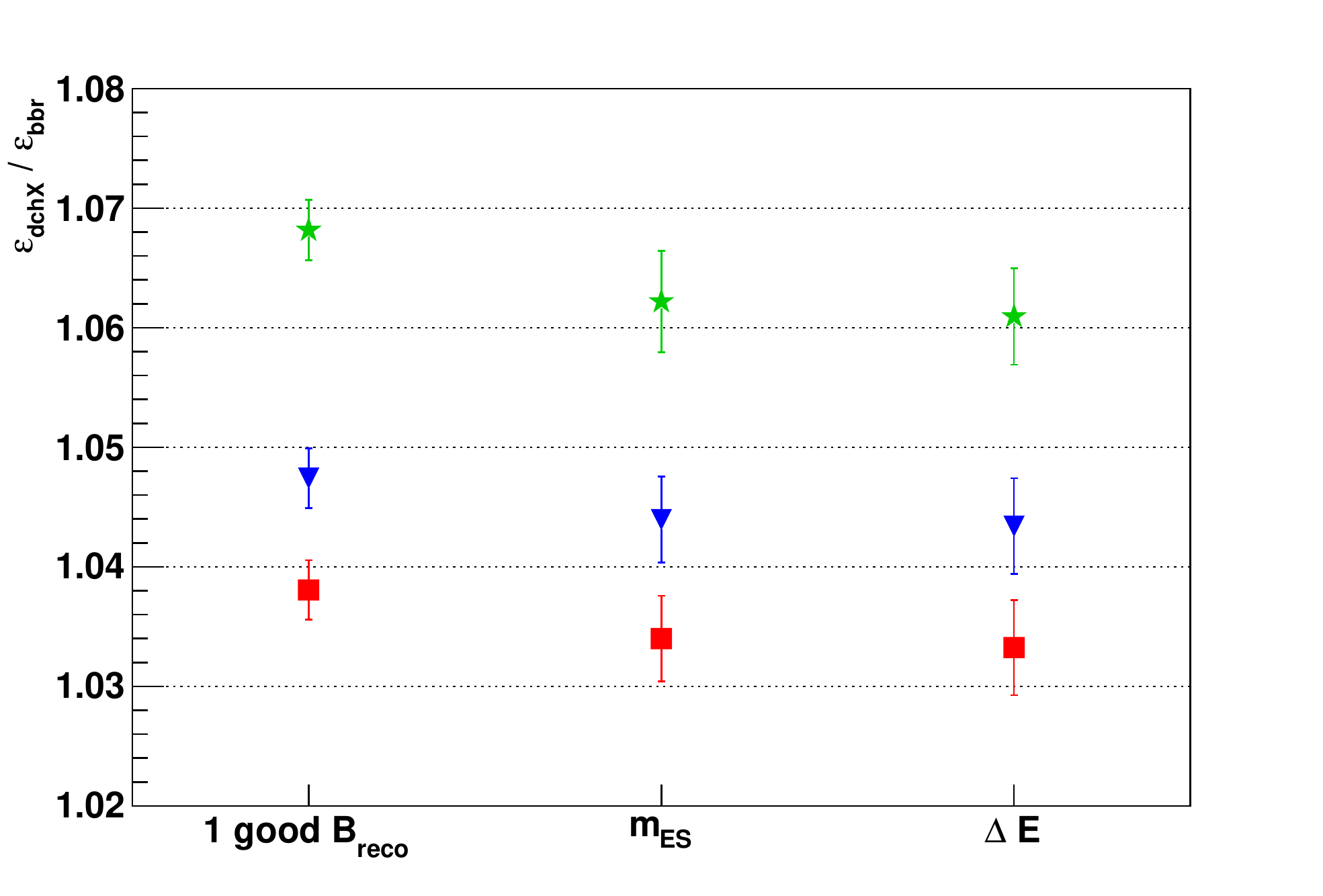}}
\end{center}
\caption{Ratio of the \Breco reconstruction efficiency with \cc to the
   reference sample for the $S_1$ (red squares), $S_2$ (blue triangles) and $S_3$ (green asterisks) 
samples described in the text. 
}
\label{fig:ElisaPlot}
\end{figure}

\tdrsubsec{Cell Design and Layer Arrangement}
\label{subsec:DCH_CellStructure}

The cell must be designed to optimize the homogeneity of the
electric field inside it; this is particularly relevant with the
non-saturated mixture we intend to use. 
Other critical parameters are the wire material and an
optimal use of the \dch volume for accommodating as many measurement
points as possible.

The design for the \superb drift chamber employs small rectangular
cells arranged in concentric layers about the axis of the chamber.
The $z$ coordinate of the track hits is measured by orienting a subset
of the wire layers at a small positive or negative stereo angle,
$\epsstereo$, relative to the chamber axis. 
Such a measurement is performed with precision
$\sigma_z\simeq\sigma_{R\phi}/\tan\epsstereo$.
As in \babar, four consecutive cell layers are grouped radially into a
superlayer (SL). This will allow to keep the same \babar\ algorithms
for track-segment finding, both in the track reconstruction and in the
formation of the \dch trigger. 

The rectangular cell layout ensures the most efficient filling of the drift 
chamber volume, because the transition between superlayers of opposite
stereo angles does not require to leave free radial space, nor layers
of field-shaping guard wires. Indeed, the latter are only used at a
radius inside the innermost SL and at a radius outside the
outermost SL. Such guard wires also serve the purpose to
electrostatically contain very low momentum electrons produced from
background particles showering in the DCH inner cylinder and in the
SVT, or background-related backsplash from detector material just
beyond the outer SL.

Simulations\,\cite{bib:Garfield} have shown that a field:sense wire
ratio of 3:1 ensures good homogeneity of the electric field inside
the cells. In this configuration, used successfully by previous
experiments \cite{bib:CLEODC,bib:KLOEDC},  each sense wire is
surrounded by 8 field wires.

The radial positions of the stereo wires in the $j$-th layer vary
with the $z$ coordinate, being larger at the endplates than at
the center of the chamber by the ``stereo drop'' $\delta_j\equiv\REP_j-R_j$.
The cell shapes are most uniform when $\delta_j=\delta$ is a constant
for all layers: this is obtained by changing the stereo angle with the
radius, by the relation $\tan\epsstereo_j =
2\delta/L_j\sqrt{2\REP_j/\delta-1}$ ($L_j$ is the 
chamber length at layer $j$).

Additional constraints used to determine the cell layout include:
\begin{enumerate}
\item[a)] the number of cells of width $w_j$ on the $j$-th sense wire
layer, $N_j=2\pi R_j/w_j$, must be an integer number; 
\item[b)] to keep a fixed periodicity in signal and high voltage
distribution, it is convenient that the number of cells per layer is
incremented of a fixed quantity $\Delta N$ when passing
from a SL to the next one.
\item[c)] since the density of both physical tracks and background hits
is higher at smaller radii, we choose to have smaller cells in the
innermost layers of the drift chamber.
\end{enumerate}
A possible choice for the drift chamber layout, obtained for
$\delta=8\mm$ and $\Delta N=16$ is shown in
Table\,\ref{table:DCH:superb_SL} and \ref{table:DCH:superb_layer}.
In this arrangement the two innermost SL's contain 1472 cells with
height $h=10\mm$ and widths $w=(10.2-11.7)\mm$. The cells in the
remaining superlayers have $h=12.85\mm$ and $w=(16.1-19.1)\mm$.
There are a total of 7872 cells in the \dch.
The first two superlayers have an axial orientation; this minimizes
the occupancy from background hits due to low-momentum
spiraling electrons which traverse the \dch along its axis (see
Sec.\ref{subsec:DCH_backgrounds}). The two outermost superlayers
are also axial. The fact that the innermost and outermost superlayers
do not exhibit the stereo drop deformation $\delta$ matches the axial
symmetry of the inner and outer \dch cylinders. The six internal SL's
have a stereo arrangement, with angles as shown in the Table.
\begin{table}[t] 
\caption{
A possible \dch SL structure, specifying the number of cells per
layer, the radius at the center of the chamber of the innermost sense
wire layer in the SL, the cell widths, and wire stereo angles, which
vary over the four layers in a SL as indicated.}
 \vskip10pt
 \begin{tabular}{|rcccc|} \hline\hline
SL & N$_{\rm cells}$ & $R\,$ & width & Angle \\
      &         & [mm] & [mm] & [mrad] \\ \hline
   1 &  176 & 292.5 &   $10.4-11.5$ &   0 \\ 
   2 &  192 & 332.5 &   $10.9-11.9$ &   0 \\ 
   3 &  144 & 375.4 &   $16.4-18.1$ & $+(60-63)$ \\ 
   4 &  160 & 426.8 &   $16.8-18.3$ & $-(63-66)$ \\ 
   5 &  176 & 478.2 &   $17.1-18.4$ & $+(67-69)$ \\ 
   6 &  192 & 529.6 &   $17.3-18.6$ & $-(69-71)$ \\ 
   7 &  208 & 581.0 &   $17.5-18.7$ & $+(72-73)$ \\ 
   8 &  224 & 632.4 &   $17.7-18.8$ & $-(74-75)$ \\ 
   9 &  240 & 691.8 &   $18.1-19.1$ &   0 \\ 
 10 &  256 & 743.2 &   $18.2-19.2$ &   0 \\ \hline
\end{tabular}
\label{table:DCH:superb_SL}
\end{table}

With this wire layout and using a \HeIso{90}{10} gas mixture (see
Sec.\ref{sec:gas_mixture}) , the total radiation length inside the
\dch is $35.1\,g/\cm^2$ (545\,m).

The corresponding wire map in the region with angle
$|\varphi|<10^\circ$ is shown in Figure\,\ref{fig:DCHwirelayout} at the
center of the chamber (a) and at the endplates (b).

It is seen that the axial-stereo transition between SL 2 and SL 3
creates some additional radial space close to the endplates, which
disappears at the DCH center. The opposite happens at the stereo-axial
transition between SL 8 and SL 9. It is clear that the electric field
should be as uniform as possible across layers to ease the \dch
calibration; however, simulation studies have shown that the field
distortion at the two transition radii is moderate and does not
require compensation by layers of guard wires, which would add
material and reduce the sensitive volume.

In Figure~\ref{fig:DCHfieldlines} we show the drift lines and the
isochrone curves for two sample rectangular cells of the proposed
\superb drift chamber with a \HeIso{90}{10} gas mixture, in  a 1.5\,T
magnetic field. The rectangular cells with field:sense wire ratio of 3:1
are indeed a satisfactory compromise, ensuring that the field lines
are sufficiently contained within the cell and the isochrone lines are
isotropic for most of the drift region, while at the same time the
number of field wires is not excessively large.

\begin{figure*}[tbp]
\begin{center}
\subfloat[Cell layout at the \dch center.]{\includegraphics[trim=7mm 0mm 18mm 5mm, clip, width=0.48\linewidth]
{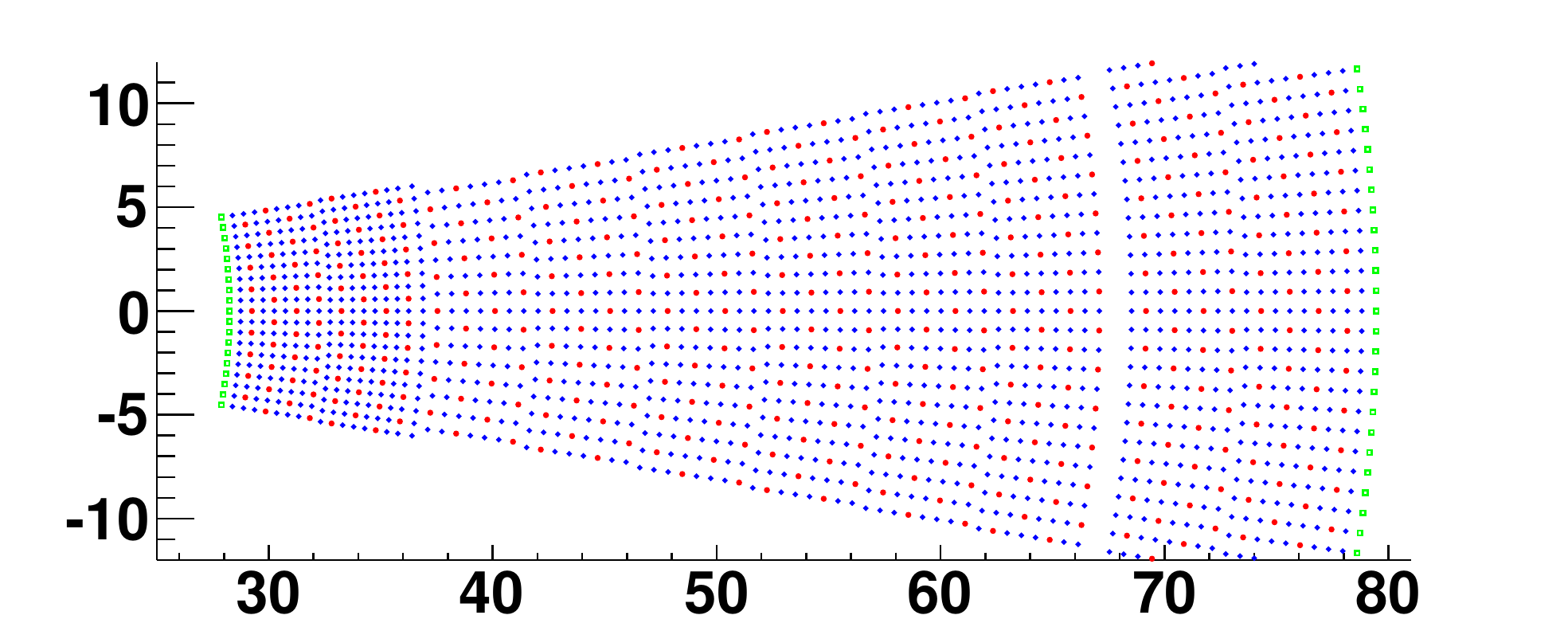}}
\hfill
\subfloat[Cell layout at the \dch endplates.]{\includegraphics[trim=7mm 0mm 18mm 5mm, clip, width=0.48\linewidth]
{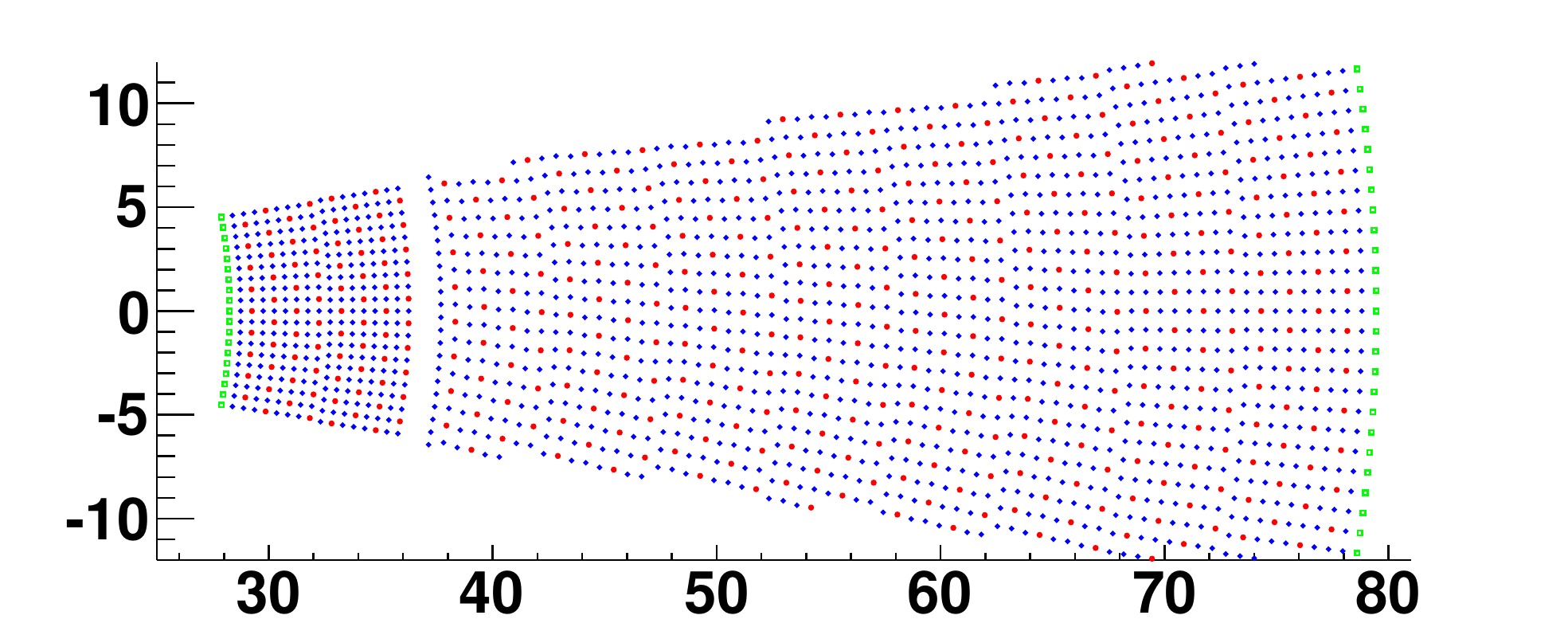}}
\end{center}
\caption{A possible cell layout of the \superb\ \dch with
  $h_\mathrm{in}=10\mm$, $h_\mathrm{out}=12.85\mm$. Open green squares:
  guard wires; open
  blue circles: field wires; full red circles: sense wires. Note how
  the boundary regions after the first 8 layers of axially strung
  wires in the inner part of the chamber and after the following 24
  layers of stereo layers map differently at the \dch center and at
  the endplates.}
\label{fig:DCHwirelayout}
\end{figure*}

\begin{table*}[t] 
\caption{
A possible \dch layer structure, specifying the number of cells per
layer, the wire layer radius at the center of the chamber, the cell
width and the wire stereo angle.}
\centering
 \vskip10pt
\begin{subfloat}\centering
 \begin{tabular}{|rcccr|} \hline\hline
layer & N$_{\rm cells}$ & $R\,$ & width & Angle \\
      &         & [mm] & [mm] & [mrad] \\ \hline
  1 &  176 & 292.5 &   10.4 &   0.0 \\ 
  2 &  176 & 302.5 &   10.8 &   0.0 \\ 
  3 &  176 & 312.5 &   11.2 &   0.0 \\ 
  4 &  176 & 322.5 &   11.5 &   0.0 \\ 
  5 &  192 & 332.5 &   10.9 &   0.0 \\ 
  6 &  192 & 342.5 &   11.2 &   0.0 \\ 
  7 &  192 & 352.5 &   11.5 &   0.0 \\ 
  8 &  192 & 362.5 &   11.9 &   0.0 \\ 
  9 &  144 & 375.4 &   16.4 &  55.7 \\ 
10 &  144 & 388.2 &   16.9 &  56.7 \\ 
11 &  144 & 401.1 &   17.5 &  57.7 \\ 
12 &  144 & 413.9 &   18.1 &  58.7 \\ 
13 &  160 & 426.8 &   16.8 & $-59.7$ \\ 
14 &  160 & 439.6 &   17.3 & $-60.7$ \\ 
15 &  160 & 452.5 &   17.8 & $-61.7$ \\ 
16 &  160 & 465.3 &   18.3 & $-62.7$ \\ 
17 &  176 & 478.2 &   17.1 &  63.7 \\ 
18 &  176 & 491.0 &   17.5 &  64.7 \\ 
19 &  176 & 503.9 &   18.0 &  65.7 \\ 
20 &  176 & 516.7 &   18.4 &  66.7 \\ \hline

\end{tabular}
\end{subfloat}
\begin{subfloat}\centering
 \begin{tabular}{|rcccr|} \hline\hline
layer & N$_{\rm cells}$ & $R\,$ & width & Angle \\
      &         & [mm] & [mm] & [mrad] \\ \hline
21 &  192 & 529.6 &   17.3 & -67.7 \\ 
22 &  192 & 542.4 &   17.7 & -68.6 \\ 
23 &  192 & 555.3 &   18.2 & -69.6 \\ 
24 &  192 & 568.1 &   18.6 & -70.6 \\ 
25 &  208 & 581.0 &   17.5 &  71.6 \\ 
26 &  208 & 593.8 &   17.9 &  72.6 \\ 
27 &  208 & 606.7 &   18.3 &  73.5 \\ 
28 &  208 & 619.5 &   18.7 &  74.5 \\ 
29 &  224 & 632.4 &   17.7 & -75.5 \\ 
30 &  224 & 645.2 &   18.1 & -76.5 \\ 
31 &  224 & 658.1 &   18.5 & -77.5 \\ 
32 &  224 & 670.9 &   18.8 & -78.5 \\ 
33 &  240 & 691.8 &   18.1 &   0.0 \\ 
34 &  240 & 704.6 &   18.4 &   0.0 \\ 
35 &  240 & 717.5 &   18.8 &   0.0 \\ 
36 &  240 & 730.3 &   19.1 &   0.0 \\ 
37 &  256 & 743.2 &   18.2 &   0.0 \\ 
38 &  256 & 756.0 &   18.6 &   0.0 \\ 
39 &  256 & 768.9 &   18.9 &   0.0 \\ 
40 &  256 & 781.7 &   19.2 &   0.0 \\ \hline
\end{tabular}
\end{subfloat}
\label{table:DCH:superb_layer}
\end{table*}

\begin{figure*}[tbp]
\begin{center}
\subfloat[Field and isochrone lines in a sample ``small'' cell, layer 6.
Horizontal and vertical dimensions are in cm.]
{\includegraphics[trim=20mm 70mm 30mm 60mm, clip,
  width=0.45\linewidth]{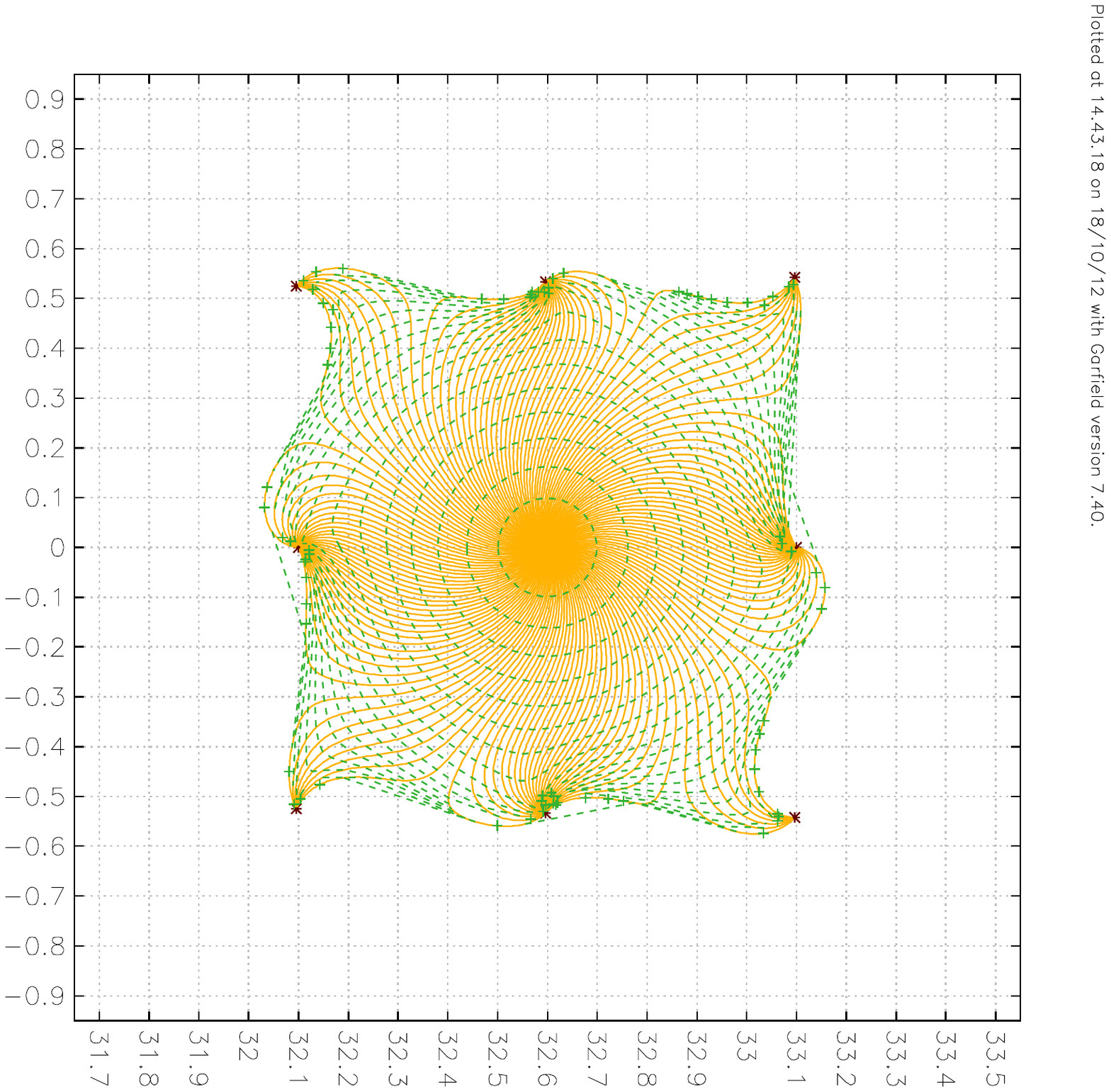}}
\hfill
\subfloat[Field and isochrone lines in a sample ``big'' cell, layer 22.
Horizontal and vertical dimensions are in cm.]
{\includegraphics[trim=20mm 70mm 30mm 60mm, clip,
  width=0.45\linewidth]{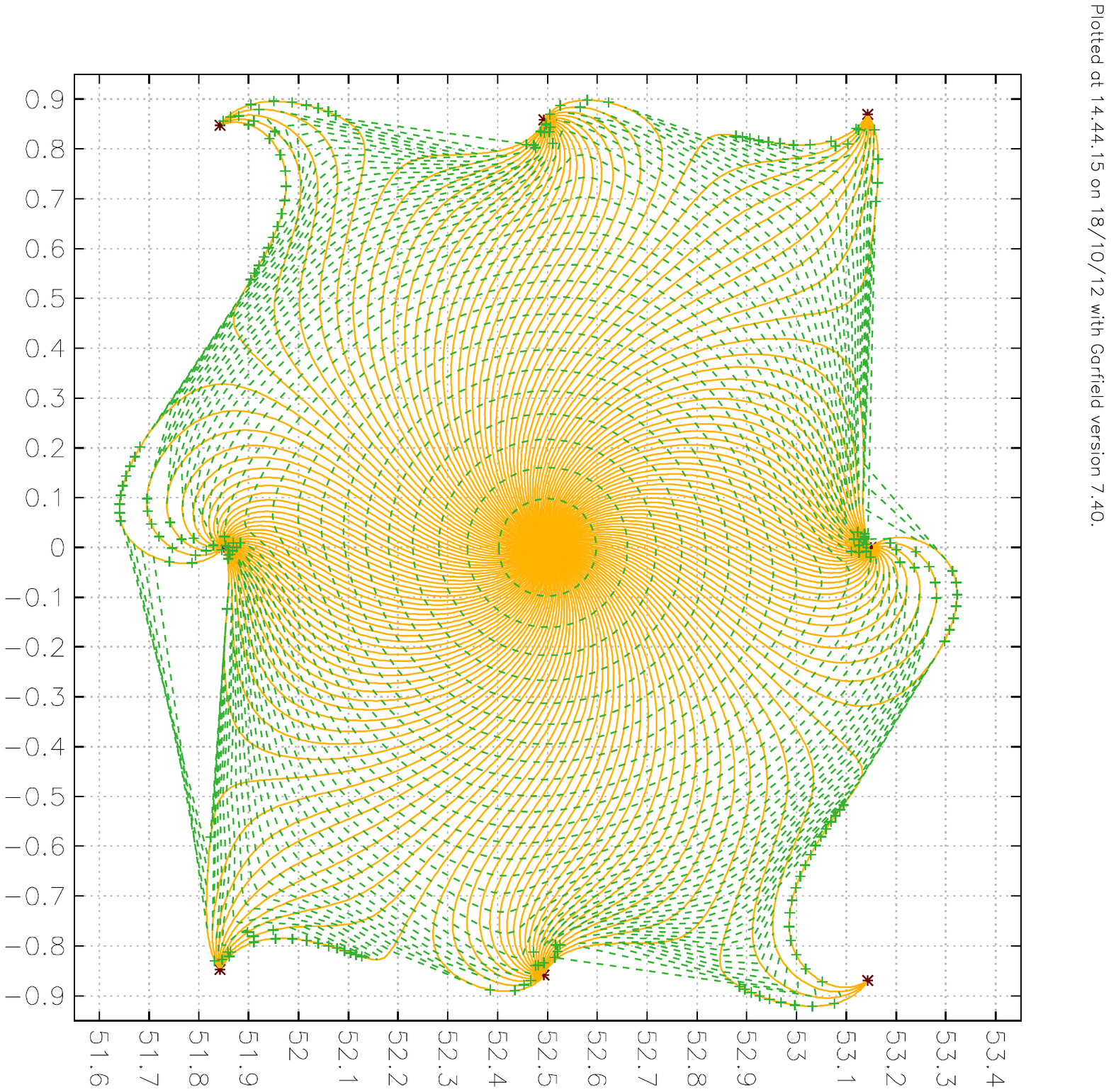}}
\end{center}
\caption{Field lines and isochrone curves (shown with a 25 ns step) in
  a cell belonging to the first 8 layers (left) and in a larger cell
  of the outermost 32 layers (right).}
\label{fig:DCHfieldlines}
\end{figure*}

\tdrsubsec{Gas Mixture}
\label{sec:gas_mixture}

The gas mixture for the \superb \dch is chosen to allow optimal
resolution in the measurement of both momentum and energy loss.
It must also be operationally stable (\eg, have a wide high voltage
plateau), and be little sensitive to photons with $E\leq10\kev$ to
help controlling the rate of background
hits (see Sec.\,\ref{subsec:DCH_backgrounds}).
Finally, aging in the chamber should be slow enough to match the
projected lifetime of a typical High Energy Physics experiment (about
15 years).
These requirements already concurred to the definition of the \babar\
\dch gas mixture (\HeIso{80}{20}).
Indeed, a high Helium content reduces the gas density and thus the
multiple scattering contribution to the momentum resolution.
Good spatial resolution 
calls for high single electron efficiency and for small diffusion
coefficient. The effective drift velocity in Helium-based gas mixtures
is typically non saturated. It thereofore depends on the local electric
field, and on the Lorentz angle. These dependences can be taken into
account by a proper calibration of the space-time relations and in
principle do not pose limits to attaining the required spatial
resolution.  In practice, a careful choice of the cell shape (see the
discussion in Sec.\,\ref{subsec:DCH_CellStructure}), and a small value
of the Lorentz angle are an advantage.

To match the more stringent requirements on occupancy
rates of \superb, it could be useful to select a gas mixture with a
higher drift velocity in order to reduce ion collection times and so
the probability of hits overlapping from unrelated events.  The \cc
option would instead call for a gas with low drift velocity and
primary ionization.
As detailed in Section~\ref{subsec:DCH_RandD}, R\&D work is ongoing to
optimize the gas mixture.

\tdrsubsec{R\&D and Prototype Studies}
\label{subsec:DCH_RandD}
In order to optimize the gas mixture for the \superb\ environment, and to
asses both the feasibility and the operational improvements for the
\cc technique  a complete R\&D program has been  proposed.
The program includes both beam tests and cosmic ray stands to
monitor performances of {\it ad hoc} built prototypes.
While the \dEdx resolution gain of the \cc method is in
principle quite sizable compared to the traditional total charge
collection, the actual capability of the measured number of clusters
might not retain  the same analyzing power, due to a pletora of
experimental effects that should be studied in detail so that  the
energy loss measurement derating  should be assessed and, if 
possible, cured.

A number of prototypes were built and operated to answer the above mentioned
questions.
\tdrsubsubsec{Prototype 1}
The first one is a small aluminum chamber, 40\cm long,  with a geometry
resembling  the original \babar \dch. It consists of 24 hexagonal
cells organized in six layers with four cells each. A frame of guard
wires with appropriate high voltage settings surrounds the cell
array to ensure uniformity of the electric field among the cells.
The device was operated in a cosmic ray test stand in conjunction with
an external telescope, used to extrapolate the track trajectories with
a precision of 80\mum or better.
Different gas mixtures have been tried in the prototype: starting with
the original \babar\ mixture (80\%He-20\%iC$_4$H$_{10}$) used as a
calibration point, both different quencher proportions and different
quenchers have been tested in order to assess the viability of lighter
and possibly faster operating gases.

As an example, the correlation between the extrapolated drift distance
and the measured drift time is shown in Figure\,\ref{fig:proto1_STRfit} for a
75\%He-25\%C$_2$H$_6$ gas mixture. 
The result of a fit to a
$5^{th}$-order Chebychev polynomial is superimposed to the
experimental points.
\begin{figure}[tbp]
\includegraphics[width=0.48\textwidth]{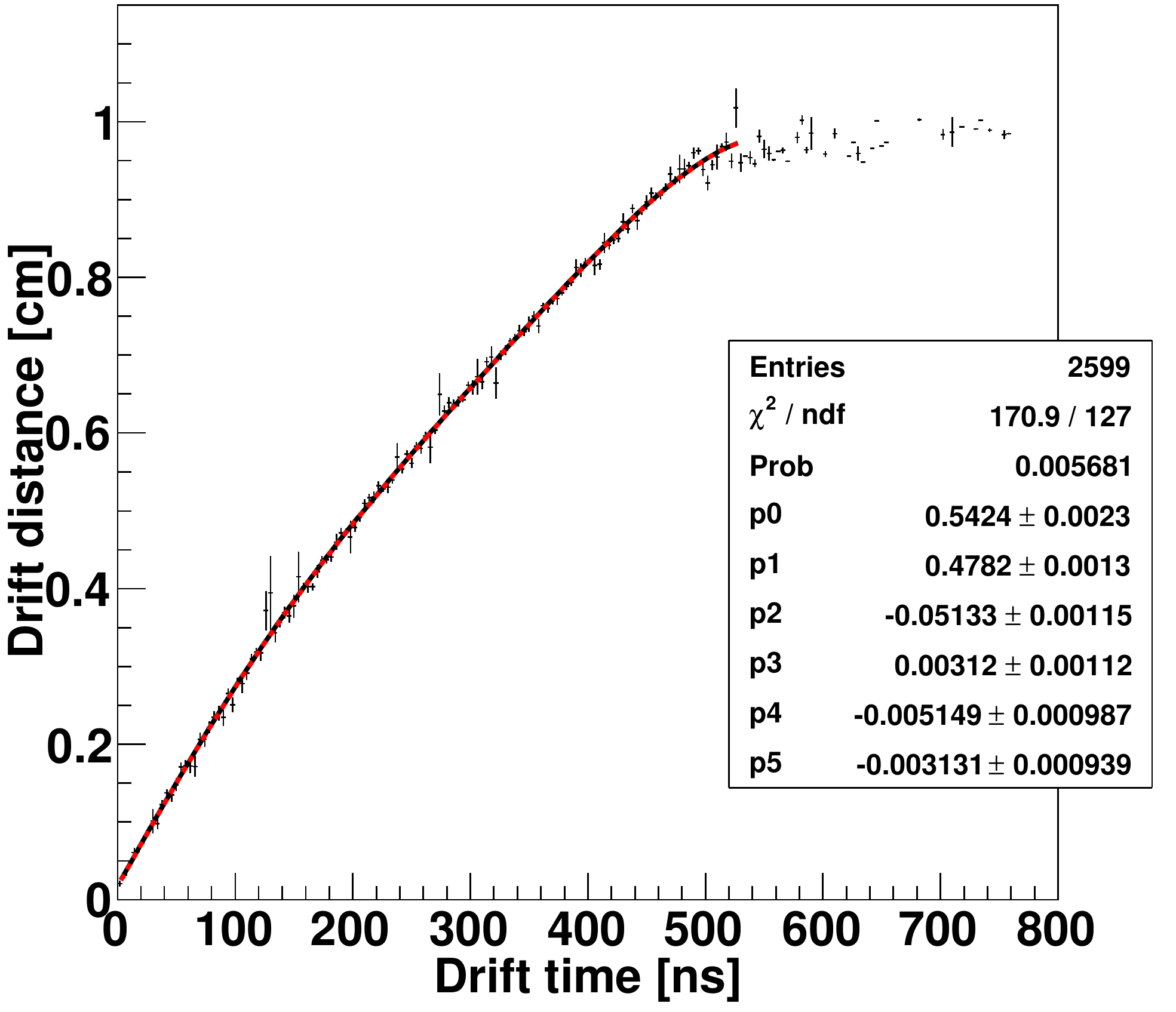}
\caption{Track distance vs. drift time in a cell of the prototype.
The line is the result of a fit with a $5^{th}$-order Chebychev polynomial.}
\label{fig:proto1_STRfit}
\end{figure}
Track-fit residuals and spatial resolution as a function of the drift
distance for the same gas mixture are show in
Figure\,\ref{fig:proto1Resolution}.
\begin{figure}[btp]
\begin{center}
\includegraphics[trim=1mm 0mm 20mm 0mm, clip,width=0.48\textwidth]{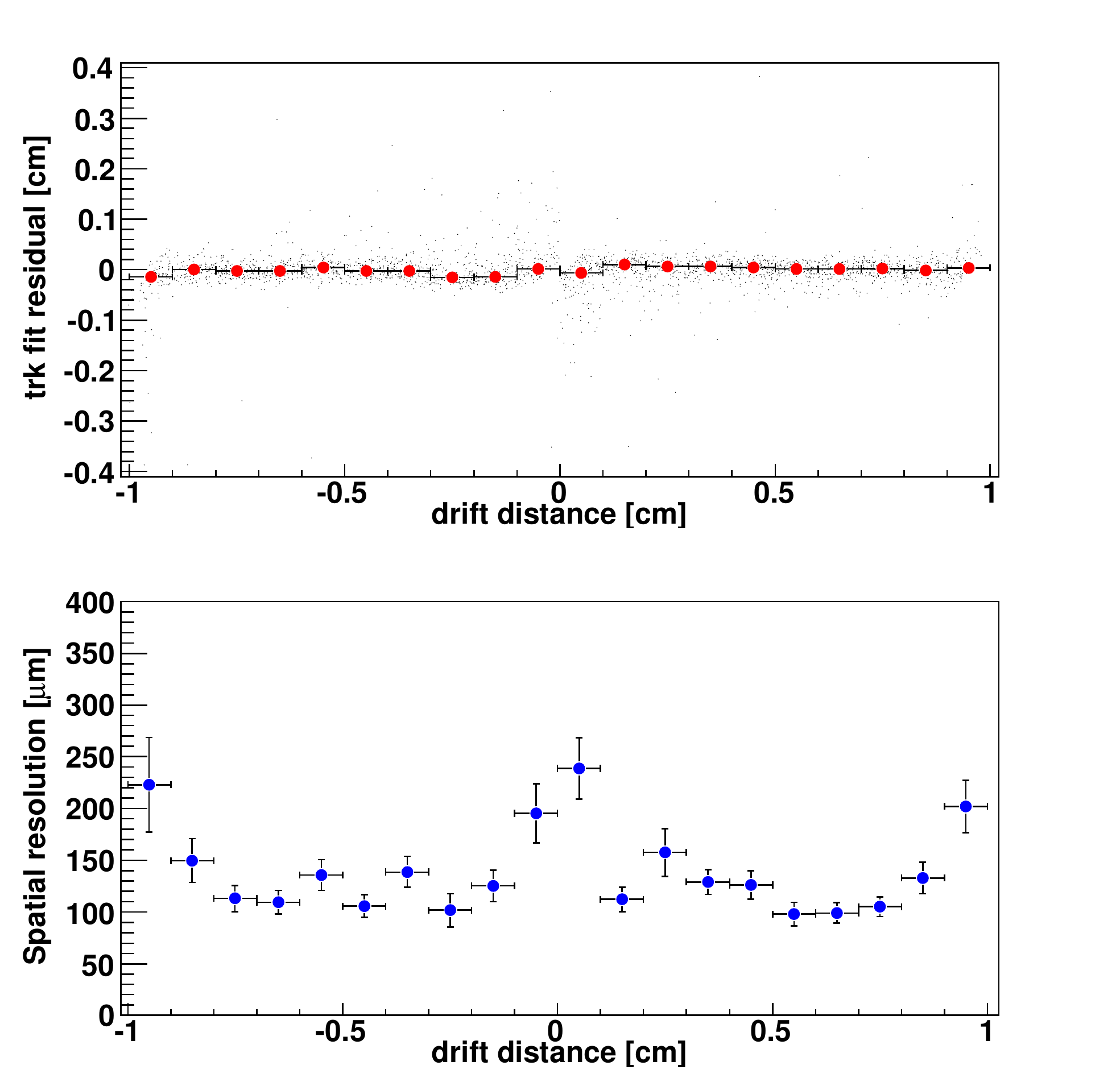}
\caption{Track fit residuals (top) and spatial resolution (bottom)
  as a function of the drift distance.
}
\label{fig:proto1Resolution}
\end{center}
\end{figure}

The results obtained with this prototype with the \babar gas mixture, agree well with the operating performances of the \babar \dch and constitute therefore
a reliable reference for our studies.

\tdrsubsubsec{Prototype 2}
A full-length drift chamber prototype was
designed, built and commissioned to study \cc in a
realistic environment, including signal distortion and attenuation along
2.5\,meter long wires. The prototype, which is also meant to serve as
a test bench for the final Front-End electronics and for the \dch
trigger, is composed by 28 square cells
with 1.4\cm side, arranged in eight layers and---as in the final
\superb drift chamber---with a  field-to-sense wire ratio of 3:1. The
eight layers have either 3 or 4 cells each, and are staggered by half
a cell side to help reduce the left-right ambiguity. Tracks with
angle $|\vartheta|\leq\pm20^\circ$ cross all the eight layers of the
chamber. A set of guard wires surrounds the matrix of 28 cells to
obtain a well-behaved field distribution at the boundary of the active
detector volume.
\begin{figure}[tbp]
\begin{center}
\includegraphics[width=\linewidth]{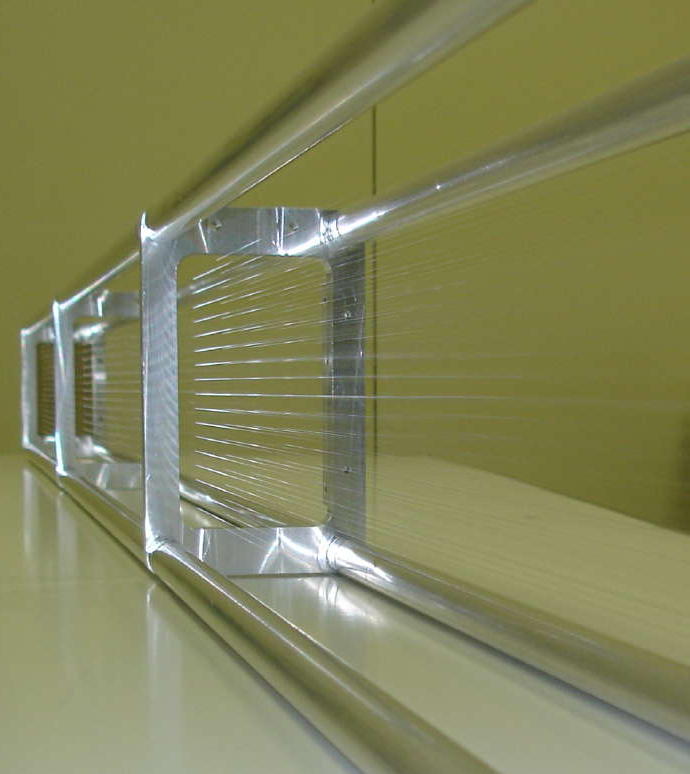}
\caption{Prototype 2: detail of the strung wires.}
\vskip -10pt
\label{fig:proto2_wires}
\end{center}
\end{figure}
 Most of the cells feature a 25\mum gold-plated Molybdenum sense wire,
while for reference  seven cells in two adjacent layers are strung
with a 25\mum gold-plated  Tungsten wire, traditionally used in drift
chambers. The reason for using the Molybdenum wire is its lower
resistivity, which produces smaller dispersion for pulses travelling along
the wires. A picture of the chamber after stringing completion is
shown in Figure\,\ref{fig:proto2_wires}.
The entire wire structure is enclosed in an Aluminum container 3\,mm
thick; three pairs of thin windows have been carved in the middle and
at the extremities in order to have smaller amount of material in the
path of low energy particles measured by the device.
Four preamplifier boards are used to extract the cell signals.
Each board serves seven channels, each with a transimpedence
preamplifier (rise time of about 2.4ns), at a nominal gain of 8mV/fC
and a noise of 2200\,$e$ rms. Each board also has a
test input, both unipolar and differential outputs (50\Ohm and
110\Ohm). The latter are used for a test implementation of the Drift
Chamber first-level trigger. 

\begin{figure}[tbp]
\begin{center}
\includegraphics[width=0.40\textwidth,clip=]{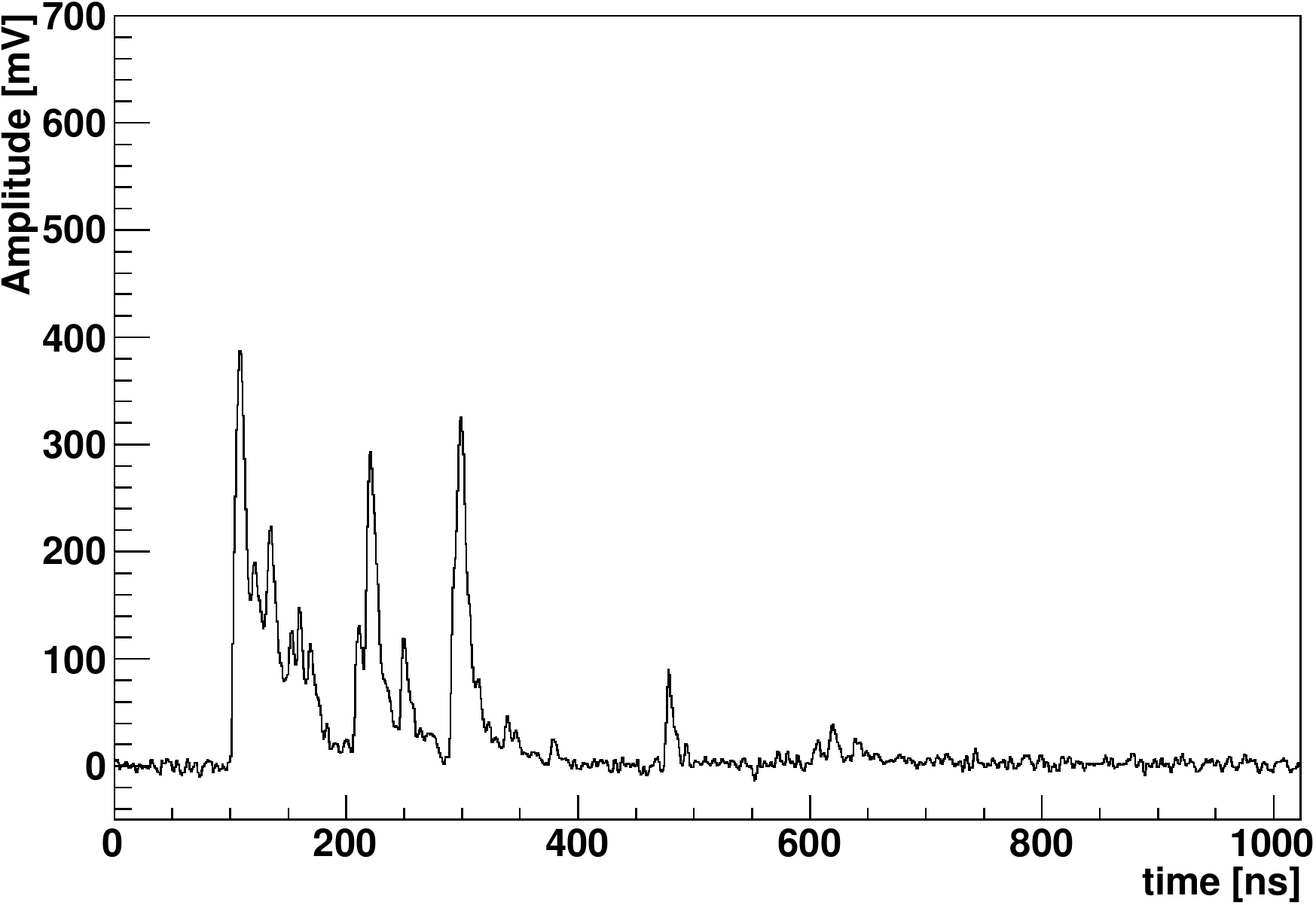}\\
\vskip 5pt
\includegraphics[width=0.40\textwidth,clip=]{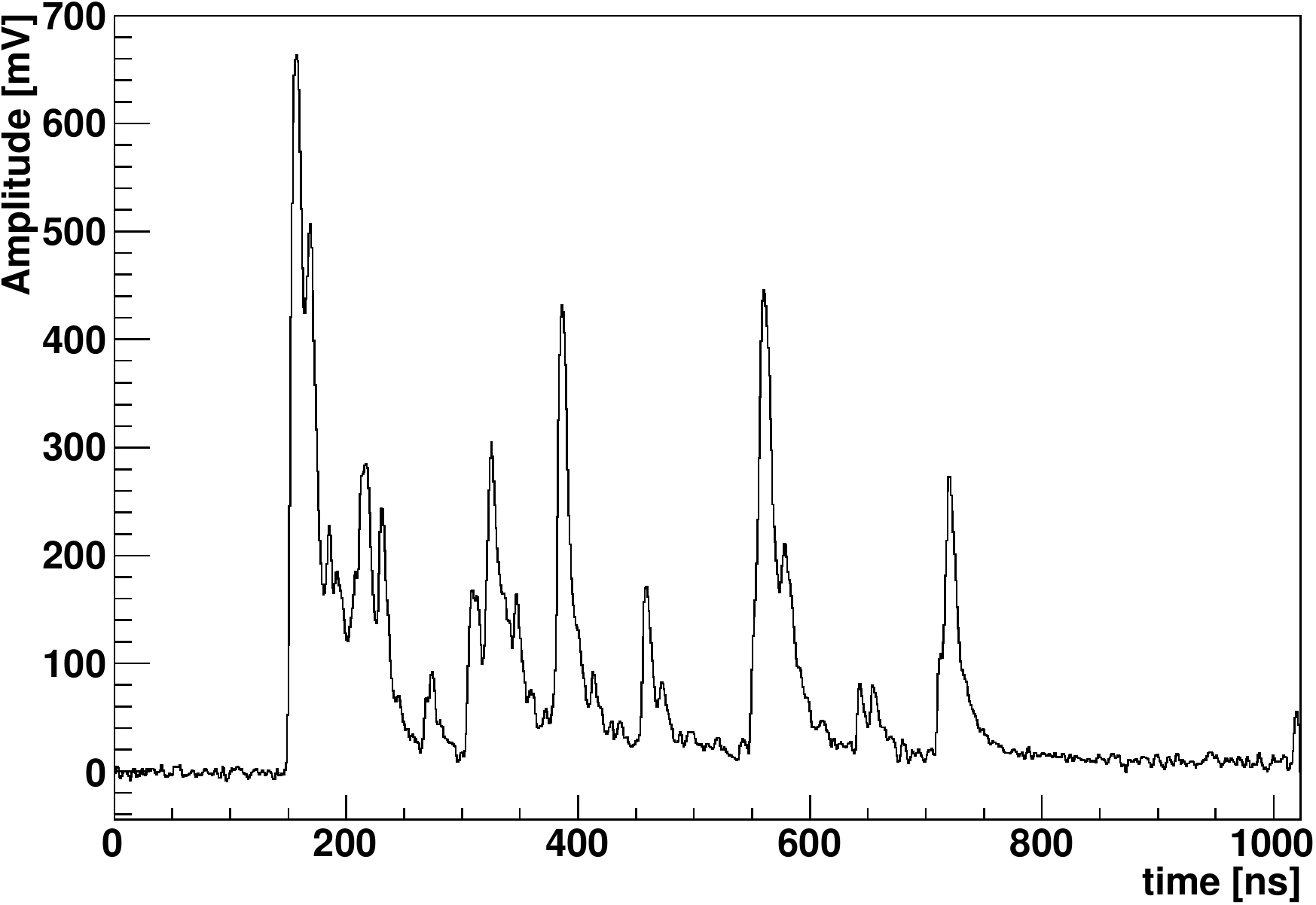}
\end{center}
\vskip -5pt
\caption{Sample waveforms from two different cells of the full-length drift chamber prototype.}
\vskip -10pt
\label{fig:proto2_sample_WFs}
\end{figure}
The data collected with this prototype are fed into a switch capacitor array digitizer\footnote{\texttt{CAEN V1742}: ~~
\url{http://www.caen.it/csite/CaenProfList.jsp?parent=13&Type=WOCateg}},
which samples the wire signals at 1\,Gs/sec with and input
BW$\geq500\,$MHz. The challenge of detecting the ionization
clusters in signals with a wide dynamic range and non-zero noise
levels
is  apparent from the two sample waveforms shown in
Fig.\,\ref{fig:proto2_sample_WFs}, recorded in the cosmic-ray setup.
\begin{figure}[tbp]
\begin{center}
      \includegraphics[width=0.40\textwidth,clip]{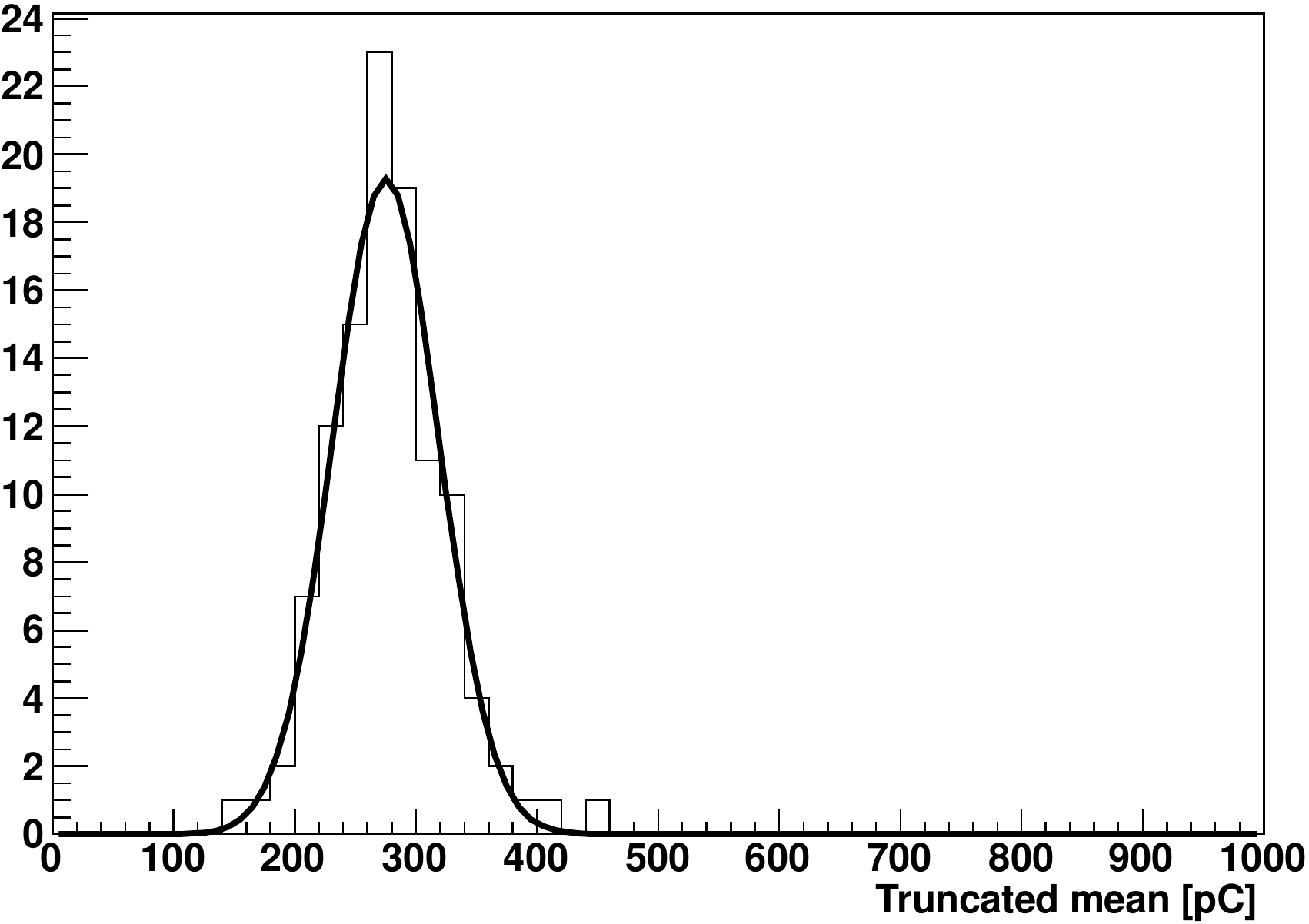} \\
\vskip 5pt
      \includegraphics[width=0.40\textwidth]{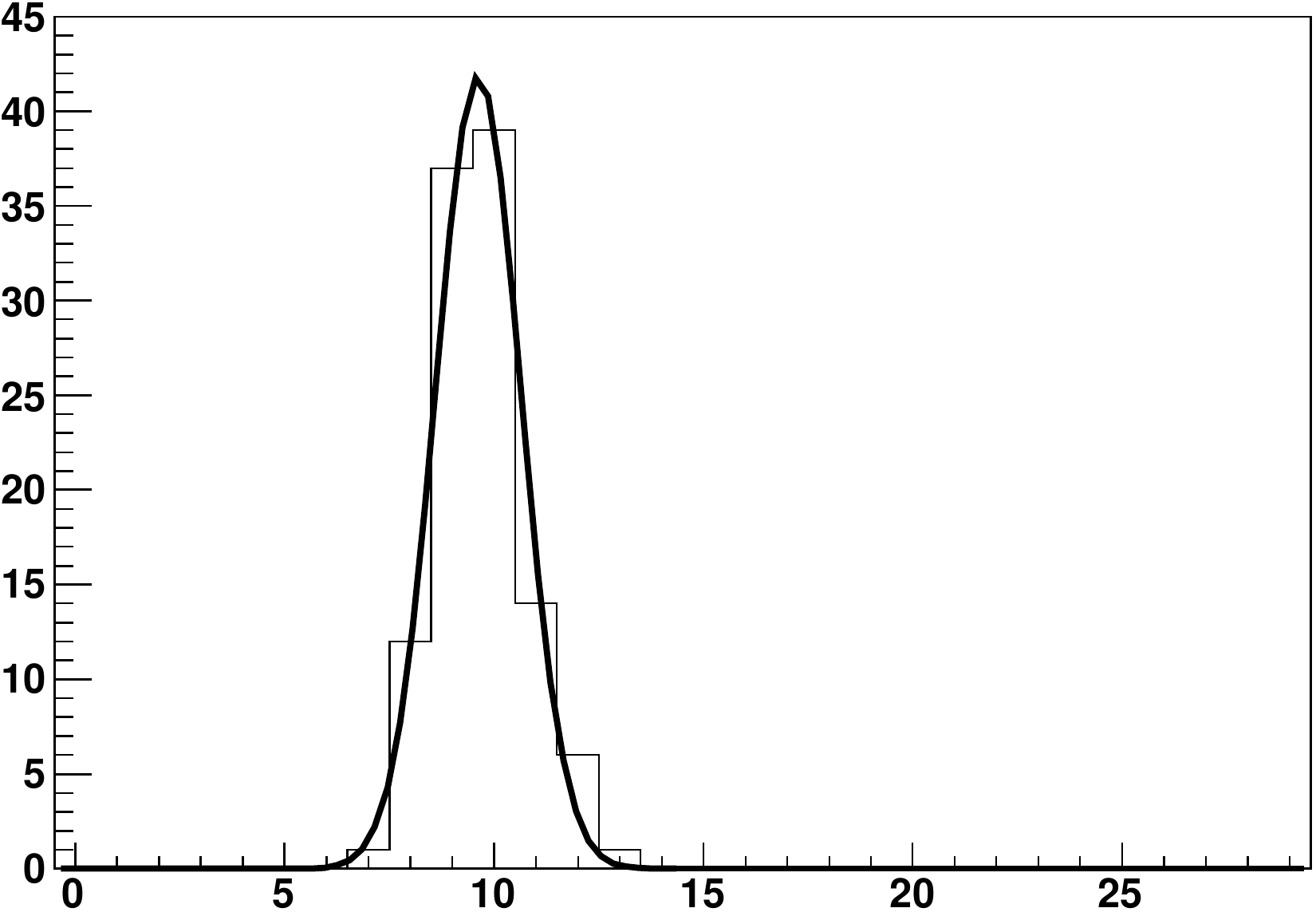} \\
\end{center}
\vskip -5pt
\caption{Average truncated mean \dEdx (top) and number of clusters (bottom) from 10 samples of a single cell belonging to a track
  reconstructed in the prototype.}
\vskip -10pt
\label{fig:proto2_dEdx_NClu_Cell}
\end{figure}
Hits associated to cosmic ray tracks reconstructed in the
drift chamber prototype are used to compare the performances in the
energy loss measurement of the traditional truncated mean
algorithm and of the \cc method.
Preliminary results when 10 samples from a single prototype cell are
used to form a 70\% truncated mean or to count the average number of
clusters are shown in Figure\,\ref{fig:proto2_dEdx_NClu_Cell}.
In the experimental conditions of our test, \cc yields a
40--50\% better relative resolution than the truncated mean
method. Additional R\&D efforts are ongoing to extend this
encouraging result to different momentum regions, and to study
how the $K-\pi$ resolving power in the range of interest of \superb
($|p|\leq5\gevc$) improves with the \cc technique.

The spatial resolution measured in the rectangular cells of the
prototype operated in a 90\%He-10\%iC$_4$H$_{10}$ gas mixture is shown
in Figure\,\ref{fig:resolP2_He90_Iso10} as a function of the drift distance.
\begin{figure}[!tbp]
\includegraphics[width=0.48\textwidth]{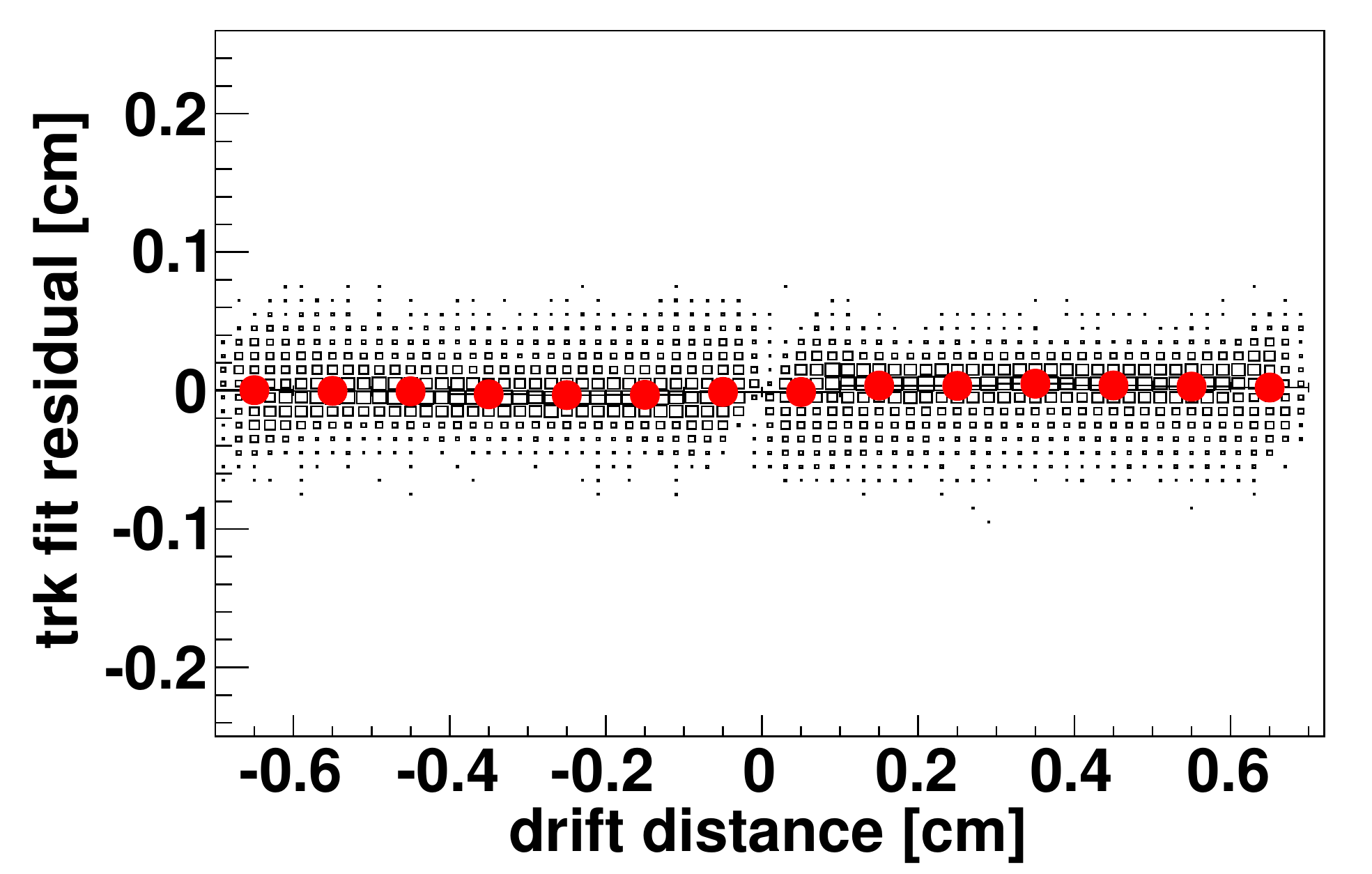}\\
\includegraphics[width=0.48\textwidth]{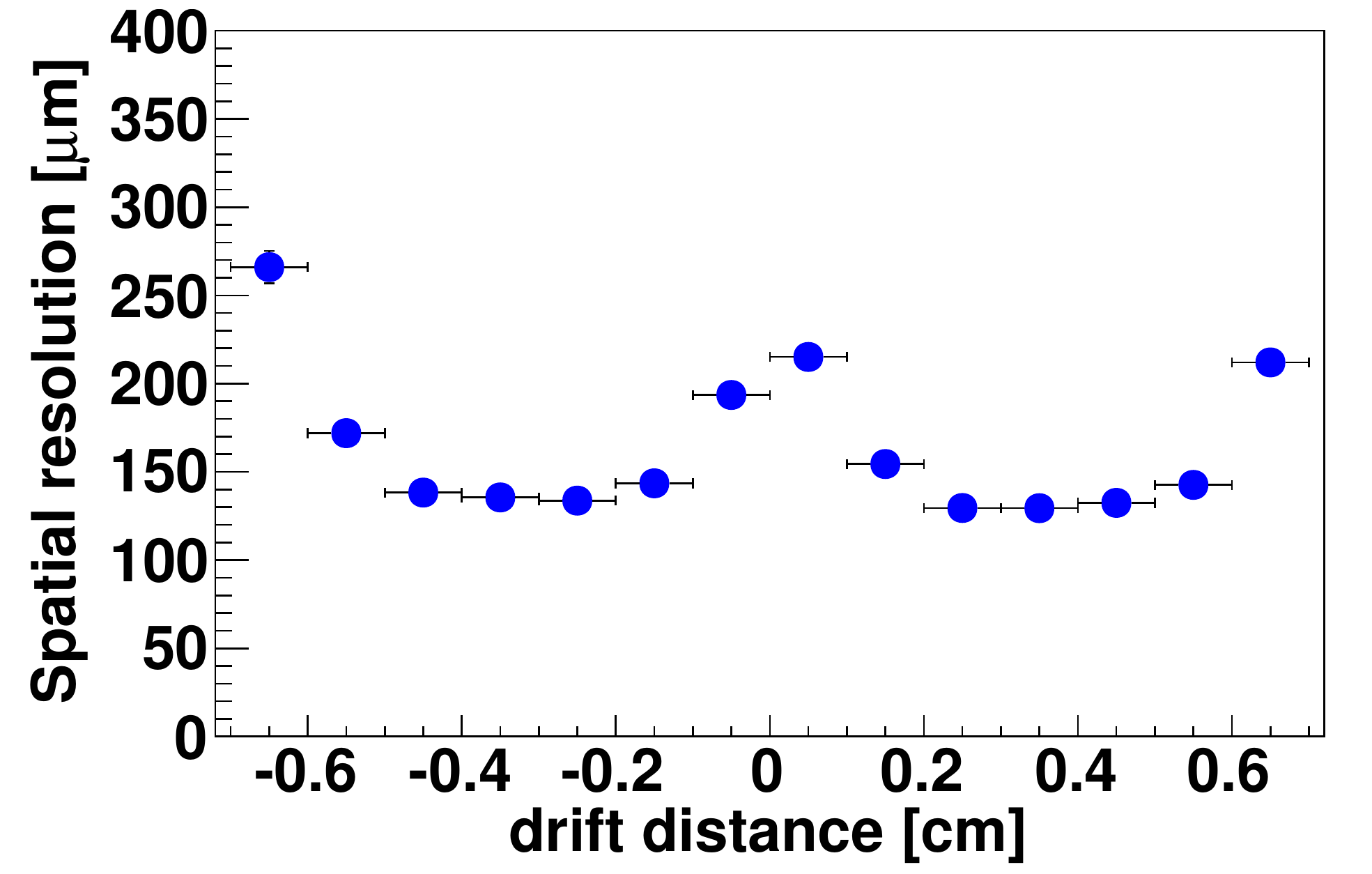}
\caption{Track fit residuals (top) and spatial resolution (bottom)
  as a function of the drift distance in the rectangular cells of
prototype 2 operated in a 90\%He-10\%iC$_4$H$_{10}$ gas mixture.}
\vskip -5pt
\label{fig:resolP2_He90_Iso10}
\end{figure}

\tdrsubsubsec{Single Cell Prototypes}
\label{subsec:SingeCellProto}
Two beam tests of single-cell drift chamber prototypes have been carried
out at the TRIUMF M11 beam line.  The goals of the tests were:
\begin{itemize}
\item to establish the benefits of clusters counting for particle
  identification;
\item to study the suitability for \cc of various
  amplifier prototypes provided by the University of Montreal
\item to quantify the impact on particle identification performance of
  various design choices, including sense wire diameter, cable for
  transmitting the analog signal, connectors, termination, and gas
  gain.
\end{itemize}
The beam test in November 2011 used a single prototype, with 25\mum\
diameter sense wire, while the test in summer 2012 used two
prototypes, one with 20\mum\ sense wire, and the other with either 25
or 30\mum.  The prototypes were 2.7m long, and consisted of a single
15~mm square
cell surrounded by an array of bias wires that adjusted the electric
field distribution within the cell to be that expected for a large
drift chamber [Figure~\ref{fig:singlecell}]. A 90\% helium / 10\% isobutane
(90:10) mixture was used for the summer 2012 tests, while the November
2011 test also tested 80:20 and 95:5.
\begin{figure}[!htbp]
\centering
\includegraphics[width=0.4\textwidth]{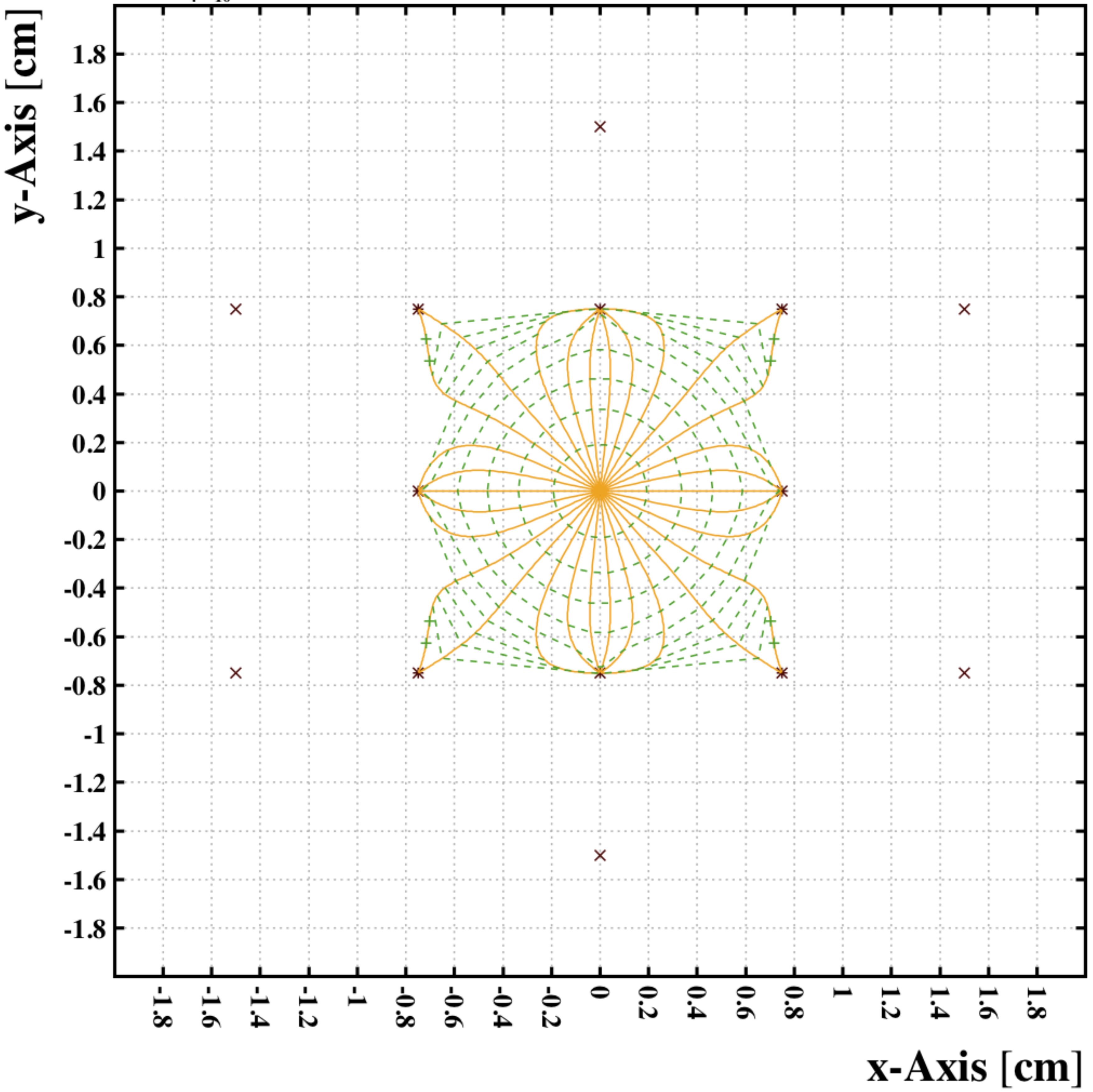}
\caption{Cell design of the single cell prototypes. Isochrones are in 
50~ns steps.}
\vskip -10pt
\label{fig:singlecell}
\end{figure}

Five different amplifier
prototypes were built for the test, with three different input impedances: 50, 170, and $380
\Omega$.  The impedance of the cell is $380 \Omega$, and for most
tests, the cell was terminated at this value on the non-readout end.
The 170 and $380 \Omega$ consisted of an impedance-matching front end,
followed by a $100 \times$ gain stage (of $50 \Omega$ input
impedance); the $50 \Omega$ amplifier consisted only of this gain
stage.

The TRIUMF M11 beam line delivers positively-charged electrons, muons, and pions, with
momenta ranging from 140 to 350\mevc.  In this momentum range, the
particle identification separation between muons and pions is
comparable to that between pions and kaons at the 2--3\gevc\ range
relevant for \superb. The trigger and time-of-flight (TOF) system
consisted of two scintillator counters, each with a pair of Burle
micro-channel plate phototubes. The trigger rate was typically tens of
Hertz, and the TOF resolution was 130~ps, providing clean separation 
between the particle species.

In the November 2011 test, the drift chamber waveforms were digitized
using a CAEN switched-capacitor array, read out using the MIDAS data
acquisition system. The bandwidth of this module is 300~MHz, which may
be less than required for \cc, so in the Summer of 2012,
a 4~GHz bandwidth LeCroy oscilloscope was instead used for
digitization. In both cases, the amplifier and digitizer were
connected by a 10~m long cable, the distance expected in the final
design. Events were written to disk at 10--15~Hz.  

The benefits of \cc on particle ID performance are
characterized by comparing the separation between muon and pion tracks
using \dEdx only to the combination of \dEdx and \cc.
Analysis is in progress, and results shown here are preliminary.

A track is formed by randomly selecting 40 samples of the same
particle species as determined by the TOF system. The track \dEdx is
obtained by discarding the largest 30\% of the samples.  Clusters are
identified by a simple threshold on a smoothed version of the
waveform.  This algorithm is not necessarily optimal, and other
methods are under study.  Conversely, adequate performance may be
achieved using a simpler algorithm that could be implemented in
hardware, as opposed to an FPGA.  The track is characterized by the
average of the number of clusters in the 40 samples. The track \dEdx
and \cc values are combined in a likelihood ratio that is
used to label the track as a muon or pion. 
Figure~\ref{fig:M11results} shows
examples of the results from the two beam tests. 
\begin{figure*}
\centering
\includegraphics[width=0.48\textwidth]{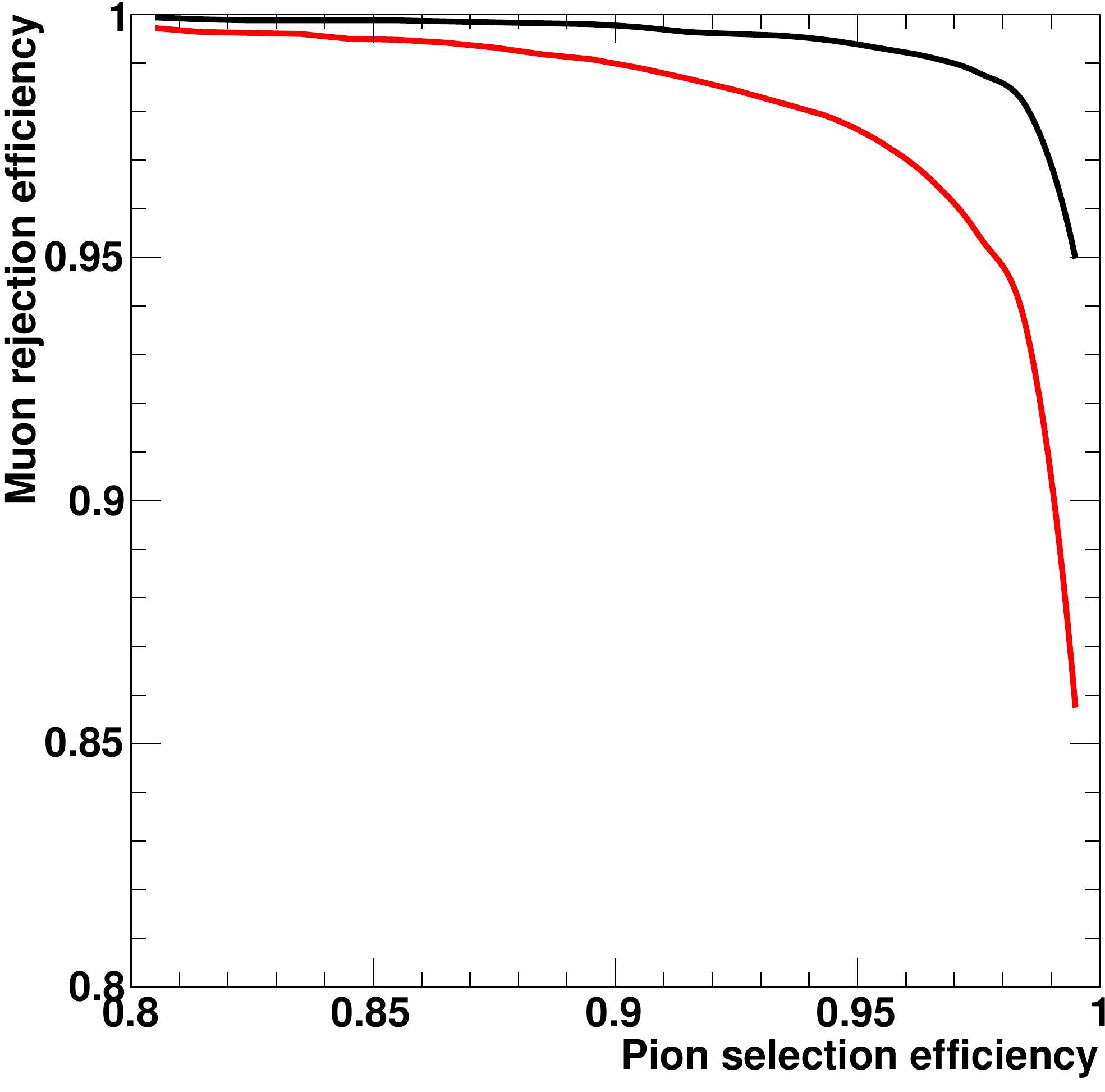}\hfill
\includegraphics[width=0.48\textwidth]{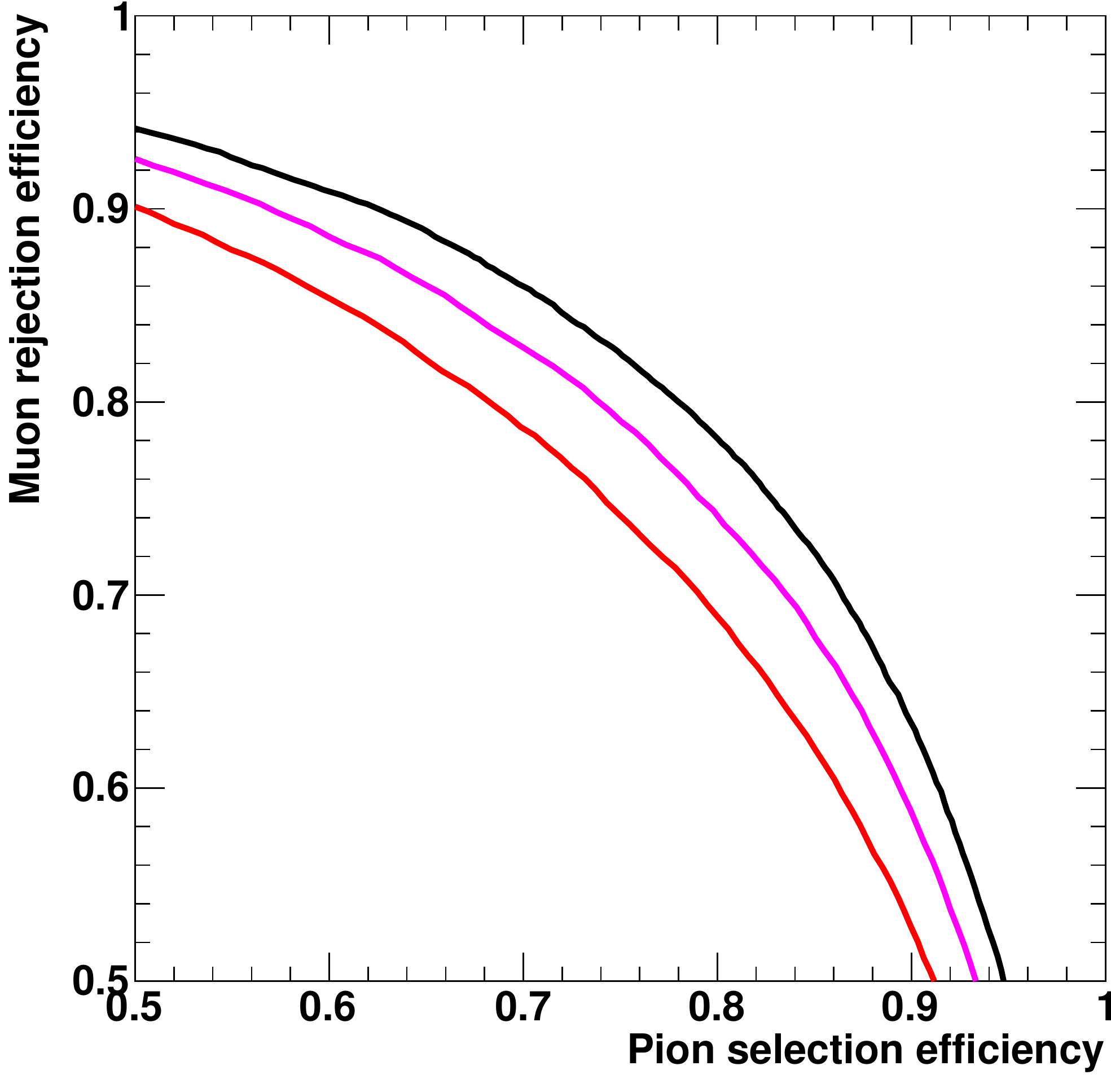}
\caption{Muon rejection efficiency versus the fraction of
  pions satisfying the selection criteria, for \dEdx only (red), 
  cluster counting only (magenta, not available in left plot), and for the
  combination of \dEdx and \cc\ (black). Left: 140\mevc\ data,
  November 2011 test. Right: 210\mevc\ data, summer 2012 test.}
\vskip -5pt
\label{fig:M11results}
\end{figure*}

The data seem to indicate that the use of \cc\ in combination with \dEdx\ leads to a significant
improvement in the pion efficiency. At 210\mevc, for example, for a misidentification probability of 10\% the
pion selection efficiency increases from 50\% to 64\%, an improvement of $\simeq 30\%$.
As the $\mu/\pi$ separation at 210\mevc is approximately the same as the $K/\pi$ at 2\gevc, a similar
improvement can thus be expected for real physics channels relevant at \superb.
The Monte Carlo study discussed in Section\,\ref{subsec:CCMcStudy} shows that a 30\% improvement in $K/\pi$ 
separation leads to about 4\% increase in the \Breco efficiency.
This result, albeit preliminary, is encouraging because it translates directly into an equivalent gain in the statistics collected
for a given integrated luminosity.

\tdrsubsubsec{Aging studies}
The goal of the aging studies is to establish that the proposed drift 
chamber can survive for a lifetime of at least 100~ab$^{-1}$. 

These studies use an $^{55}$Fe source both to age a
test chamber, and to characterize its performance. 
The initial studies are using a 30~cm long chamber containing a single
\babar-like hexagonal-shaped cell, and an 80\% helium 20\% isobutane 
gas mixture.
The sense wire is 20 micron gold-coated
tungsten, and the field and bias wires are 120 micron gold-coated aluminum. 
The chamber is exposed to a 100~mCi $^{55}$Fe source.  The resulting current
is monitored, along with temperature and atmospheric pressure, as a way 
to characterize gain as a function of accumulated charge. Once per week,
the hot source is replaced with a low-intensity one and the pulse-height
spectrum is recorded (Figure~\ref{fig:Fe55spectrum}).  The location of the 
$^{55}$Fe peak is an additional measurement of gain.  
The number of very 
small pulses is sensitive to the Malter effect\,\cite{bib:Malter}, a form of aging of the
field wires in which they accumulate an insulating coating. A second 
single-cell chamber, which is not exposed to the hot source, is used to 
calibrate out any possible gain effects due to gas variations, and
to verify the gas density corrections. The chamber is operated at a
voltage such that the electric field on the field wires is 
less than 20~kV/cm in order to minimize the Malter effect. 

\begin{figure}[tbh]
\centering
\includegraphics[width=0.45\textwidth]{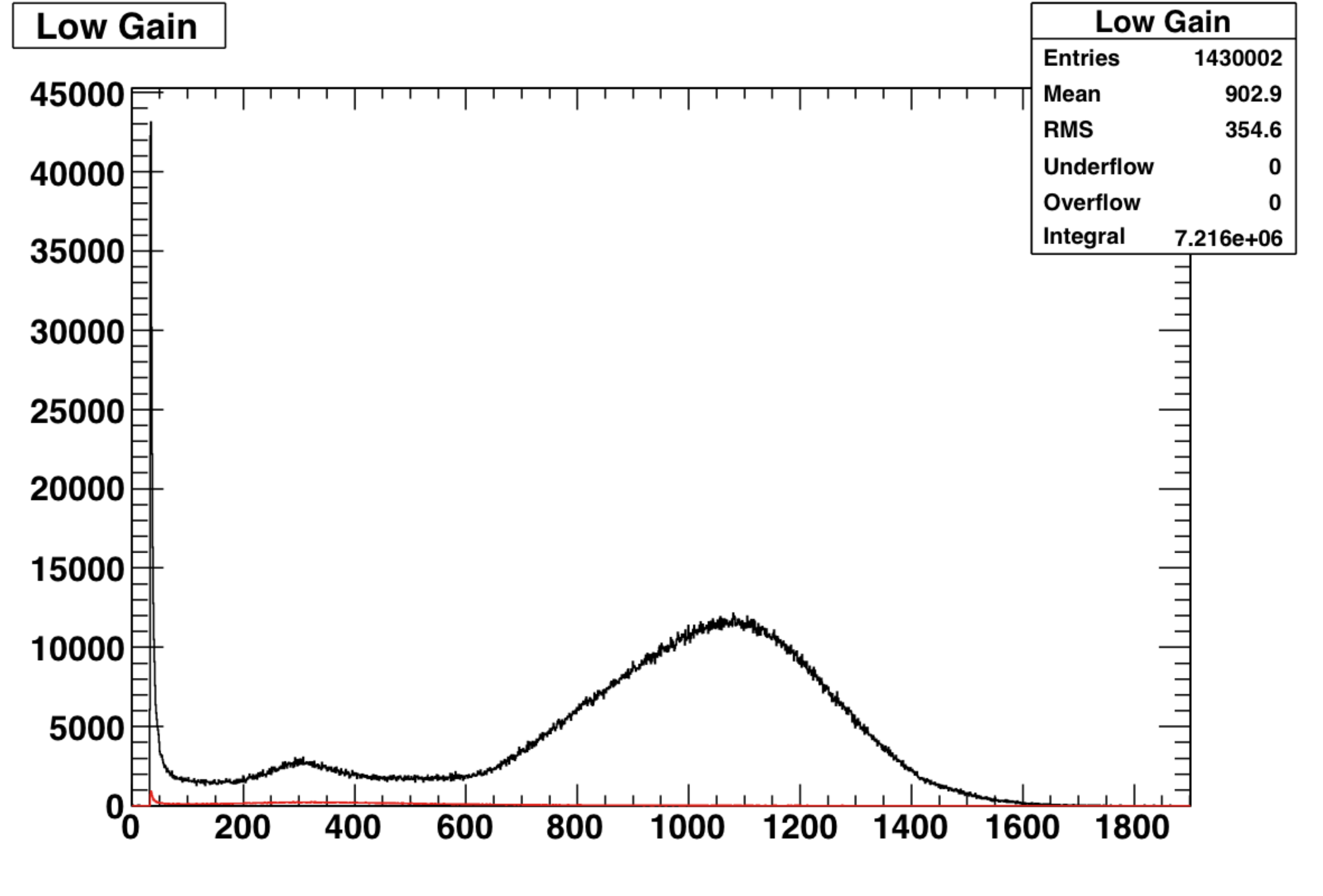}
\caption { Pulse height spectrum (arbitrary units) recorded from an
  $^{55}$Fe source by an aging chamber. The red curve is the
  underlying cosmic ray background.  }
\label{fig:Fe55spectrum}
\end{figure}

The aging chamber shows a gain drop of 25\% after accumulating
310~mC/cm over the last 20 months. 
This lifetime is significantly in excess
of the 34~mC/cm accumulated by the \babar\ drift chamber. The \babar\ chamber
saw a 10\% loss of gain over that time.

The next aging chamber is currently under construction.  It will
include seven square cells, so that the central field wires are
surrounded by sense wires, and will therefore accumulate the correct
amount of charge.
It will use
\superb\ materials, gold-coated 20 micron Molybdenum sense wires, and
80 or 90 micron bare aluminum field wires, and 90:10 helium-isobutane
gas.  The structure of the chamber will be aluminum, as for the
current aging chamber, but the walls will be covered by samples of the
carbon-fiber material that will be used in the actual chamber.

The amount of charge per cm expected for the \superb\ chamber is a function 
of the chamber occupancy, the gas gain, and the total running time. 
Current background calculations indicate occupancy levels comparable to 
\babar, including a five-times safety factor. The \superb\ running time will
also be comparable to \babar. The gas gain may be higher, due to the 
requirements of \cc. The gas gain required is a function of 
the amplifier; the 50 ohm input impedance amplifier prototypes
require four times the gas gain of the 370 ohm amplifiers, due to the 
impedance mismatch between the amplifier and the drift cell. 
The field wire diameter is, in turn, a function of the sense wire 
operating voltage, given the need to keep the electric field at the
surface of the field wire below 20~kV/cm.

\tdrsubsec{R\&D Future Developments}

At the time of this writing, a two week test beam data taking campaign has
been completed with the full scale prototype (Prototype 2) 
at the M11 beam line in the TRIUMF laboratory, Canada. Data were collected 
with a beam of $e$, $\mu$, and $\pi$ particles with momenta between 120
and 210\mevc. 

The goal of this test is to reproduce and study in further details,
with a more realistic full length, multiple cell prototype, the promising preliminary results
on cluster counting and \dEdx obtained with the single cell prototype described in Sec.\,\ref{subsec:SingeCellProto}.
The analysis of these data has just started and will be the object of a
future paper.

Another line of future development is centered in the realization of 
a prototype on-board feature extraction.

\tdrsec{Mechanical Design}
\label{sec:DCH_Mechanics}

The drift chamber mechanical structure must sustain the wire load with
small deformations, while at the same time 
be as light as possible to minimize the degradation of the performances of
the surrounding detectors. The structure is also required to ensure
tightness for the gas filling the drift volume.
We opted for a structure entirely in Carbon Fiber (CF) composite, with
an approximately cylindrical geometry. Given the studies shown in
Subsec.\,\ref{subsubsec:DCH_Length_Studies}, the active length of the
chamber has been maximized. In particular the length in the backward
direction is increased with respect to \babar despite the 160\mm
reserved in \superb for the possible backward EMC to be installed at a
later stage; this is possible thanks to the new design of the \dch
front-end electronics (Sect.\ref{sec:DCH_Electronics}), which require
substantially less space than \babar.

\tdrsubsec{Endplates}\label{subsec:Endplates}
The wires defining the cell layout are strung between the two
endplates, which are required to:
\begin{enumerate}
\item[a)] sustain the total wire load of about 2 tons (see
sec.\,\ref{subsec:DCH_CellStructure}) with minimal deformations;
\item[b)] be as transparent as possible to avoid degrading the
performances of the forward calorimeter.
\item[c)] have 33,000 precisely machined holes to allow positioning the
crimp feedthroughs with tolerances better than 50\mum;
\end{enumerate}

The endplates are two identical pieces of 8\mm thick CF composite with
inner radius of 270\mm and outer radius of 809\mm. Deformations under
load can be minimized using a shaped profile. An
optimization taking into account different constraints resulted in 
spherical convex endplates, with a radius of curvature of 2100\mm.
Two CF stiffening rings (the outer one shown in
Figure\,\ref{fig:dch_outercylinder_endplate}) 
on the inner and outer rims help minimize
radial (axial) deformations.
\begin{figure}[tbp]
\includegraphics[width=0.49\textwidth]{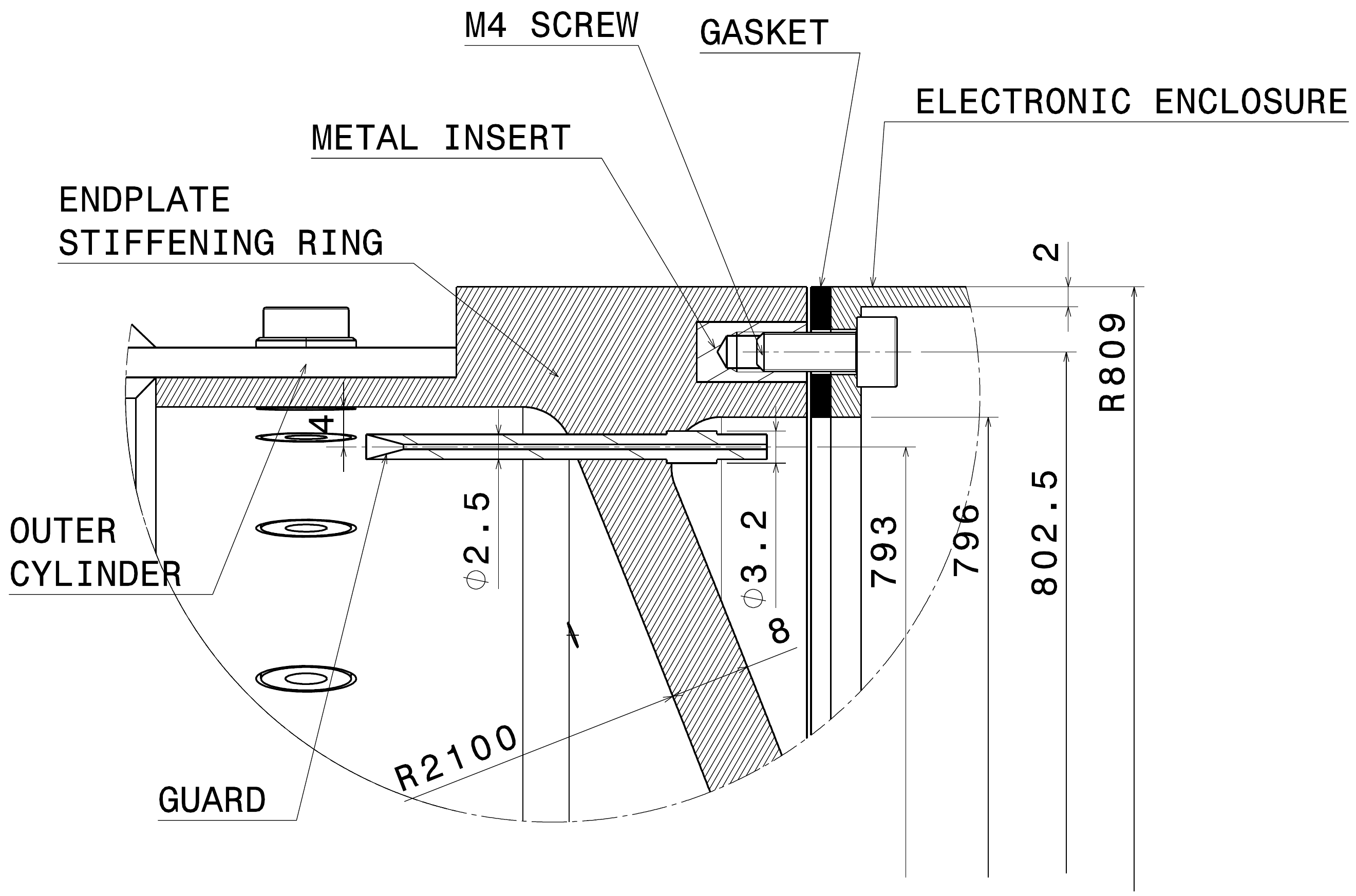}
\caption{Detail of the outer flange of the DCH endplates.}
\label{fig:dch_outercylinder_endplate}
\end{figure}
In the calculation an intermediate modulus carbon fiber 
(T300, with a Young modulus of 45 GPa) was used. The 8\mm CF
laminate is built using stacks of 250\mum plain weave plies with a
$[0/45/45/0]_8$ structure.
It was conservatively assumed that the average material characteristics
are degraded by about 30\,\% after drilling the 33,000 holes
on each endplate; detailed studies on this aspect will be performed on
custom samples.
The maximum displacements on the endplates is calculated to be less
than 600\mum (Figure\,\ref{fig:dch_endplate_Zdisplacement}).
\begin{figure}[tbp]
\includegraphics[width=0.49\textwidth]{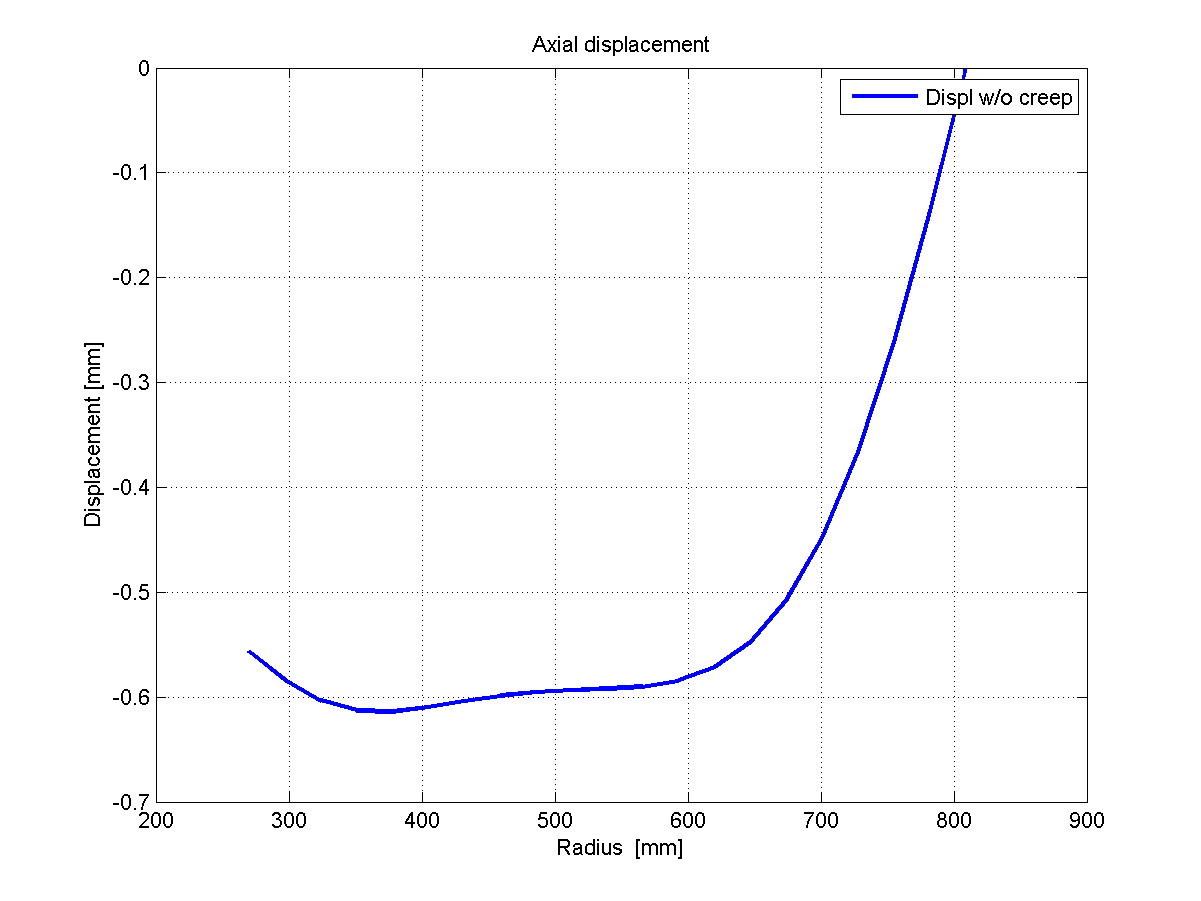}
\caption{Displacement of each endplate due to the wire load.}
\label{fig:dch_endplate_Zdisplacement}
\end{figure}
As a comparison, the thickness of flat endplates to have such small
deformations under the wire load would be 52\mm.

\tdrsubsec{Inner cylinder}
\label{subsec:Inner_Cylinder}
The \dch inner cylinder should be as transparent as possible to
minimize the multiple scattering degradation to the \pt measurement.
For this reason it was designed as a non-load-bearing structure: it
must only guarantee gas tightness, and sustain possible differential
pressures of about 20\,mbar\footnote{The nominal differential pressure
in normal operation of the \superb DCH will be 2\,mbar or less, but the
cylinder has been designed to sustain 20\,mbar.} between the inside
and outside of the chamber.
The 270\mm radius cylinder is composed by two 90\mum CF sheets
sandwiching a 3\mm thick honeycomb structure to increase the moment of
inertia and improve the buckling load. A 25\mum thick aluminum foil is
glued on each CF skin for RF shielding.
The possibility is being explored to build an even thinner inner
cylinder, replacing the two 90\mum CF plies with two 50\mum Kapton
foils.
During the stringing phase
the inner cylinder will be free to move longitudinally, being fixed
only to one endplate. Only after stringing, when all endplate
deformations are settled, will the cylinder be glued to the second
endplate. A detail of the interface between the inner cylinder and the
endplate is shown in Figure\,\ref{fig:dch_innercylinder_endplate}.
\begin{figure}[tbp]
\includegraphics[width=0.49\textwidth]{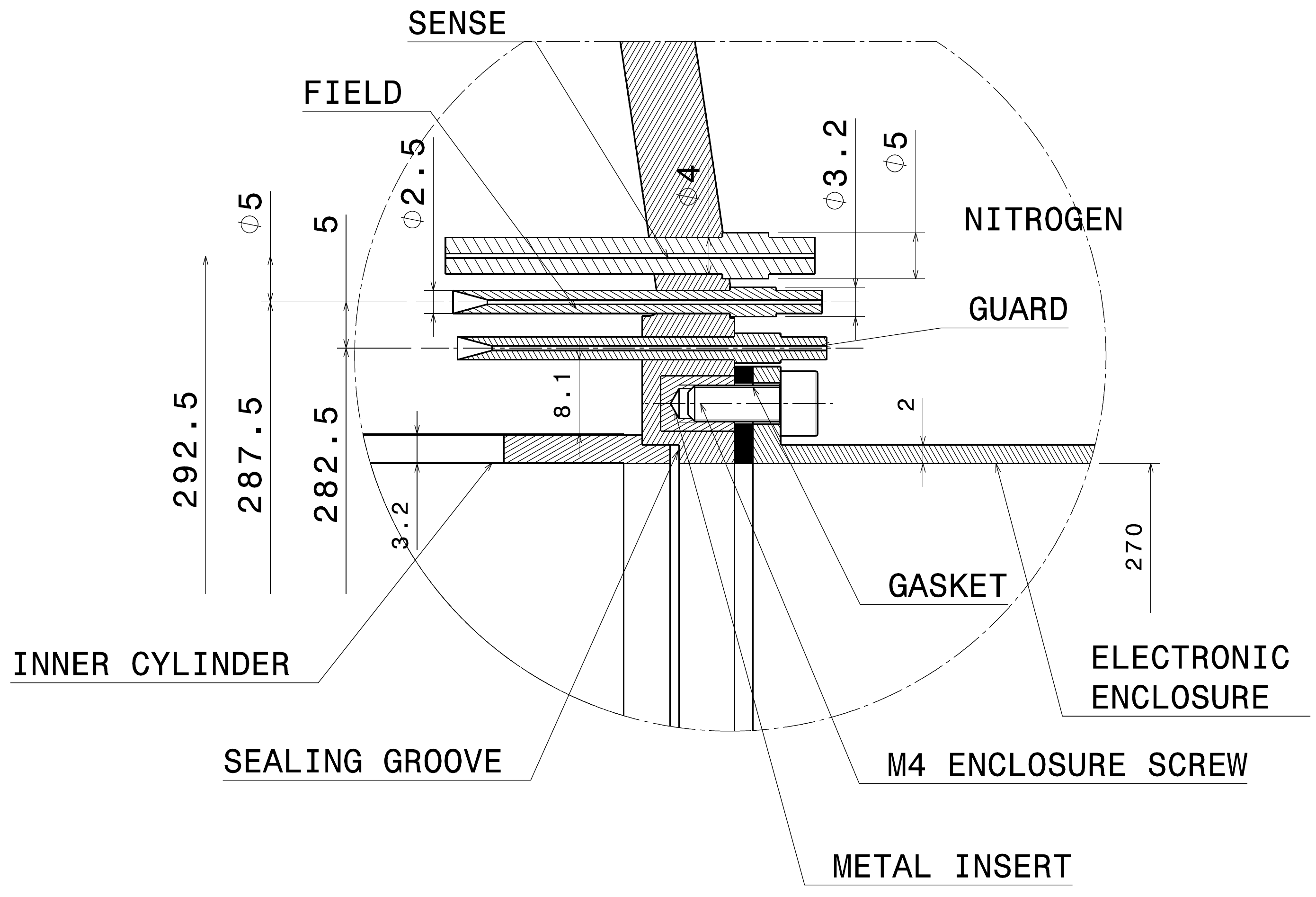}
\caption{Coupling of the DCH inner cylinder to the endplate.}
\label{fig:dch_innercylinder_endplate}
\end{figure}

\tdrsubsec{Outer Cylinder}
In addition to guaranteeing gas tightness and withstanding a
differential pressure as the inner cylinder, the outer cylinder
will also carry the entire wire load. It will be installed after
completion of the wire stringing. To ease the construction and the
mounting procedures, the cylinder is longitudinally divided in two
half shells.
Each shell consists of two 1\mm-thick CF skins laminated on a
6\mm-thick honeycomb core. A 25\mum thick aluminum foil
is glued to each CF skin to ensure the RF shield.
The sandwich structure guarantees a high bending stiffness and a high
safety factor for global buckling.

\tdrsubsec{Analysis of Buckling Instabilities}
The buckling stability of the system was studied for all elements
working under compression
with the ANSYS Finite Element Analysis Program.
Different methodologies were used for the main system elements, as
described below.
\tdrsubsubsec{Endplates}
The results of a buckling linear analysis performed varying the endplate
curvature radius are summarized in
Table\,\ref{tab:DCH_buckling_endplates}.
The first buckling mode for the configuration with the nominal
endplate curvature radius (2100\mm) is shown in
Figure\,\ref{fig:DCH_buckling_endplates}.
\begin{figure}[tbp]
\includegraphics[width=0.49\textwidth]{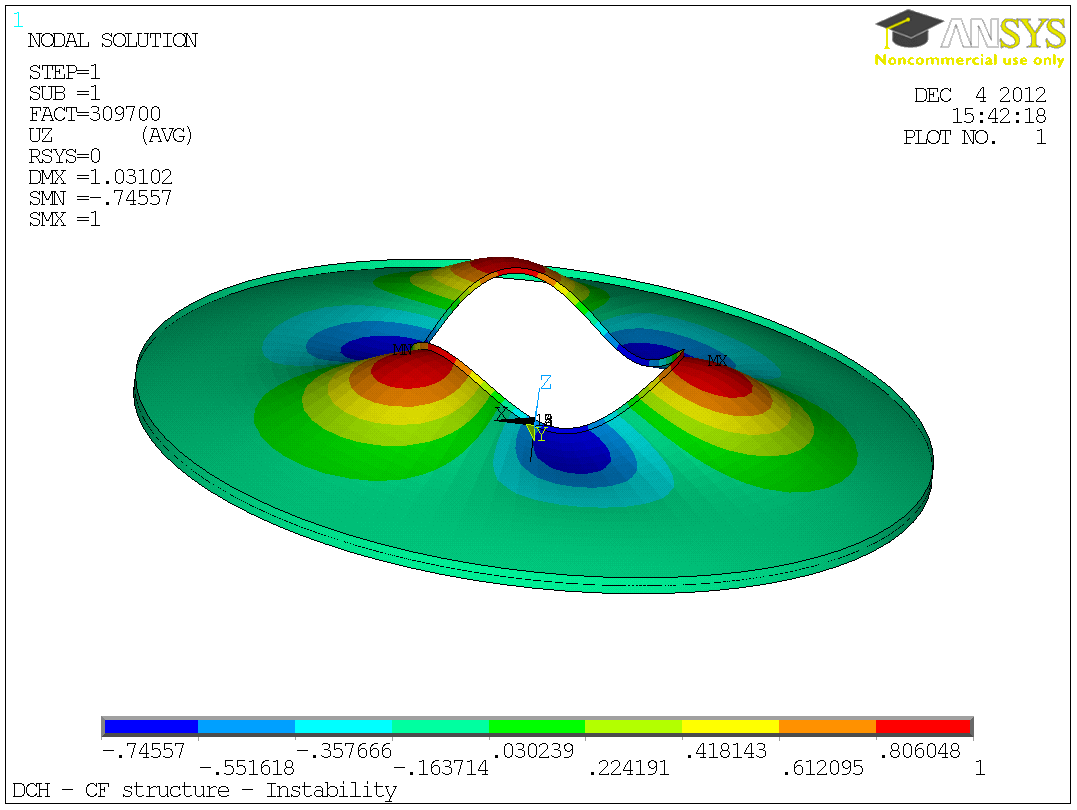}
\caption{Calculated first buckling shape of the endplates.}
\label{fig:DCH_buckling_endplates}
\end{figure}

\begin{table}[t]
\begin{center}
\caption{
Maximum endplate deformations $\delta_{max}$, static and buckling safety factors as
a function of the endplate curvature radius $R_{E.P.}$.}
 \vskip10pt
 \begin{tabular}{|ccccc|} \hline\hline
$R_{E.P.}$ & Thickness &   $\delta_{max}$ &  Static &      Buckling  \\
    mm  &     [mm]   & [mm]                 & S.F.    &        S.F. \\ \hline
  2100 &  8 & 0.6 &   16.0 &   16.0 \\ 
  3000 &  8 & 1.1 &   12.4 &    8.0 \\ 
  4000 &  8 & 1.8 &   10.0 &    4.8 \\ \hline 
\end{tabular}
\vskip -10pt
\label{tab:DCH_buckling_endplates}
\end{center}
\end{table}
The nominal configuration with $R_{E.P.}=2100\mm$ is observed not only
to minimize the displacement, but also to show the best buckling safety
factor. We consider a safety factor above 10 adequate even after the
uncertainties due to material imperfections and the effect of holes on
the endplates are taken into account.

\tdrsubsubsec{Inner Cylinder}
A linear buckling analysis was carried out on the inner cylinder
structure which, as discussed in Sect.\ref{subsec:Inner_Cylinder}, is
only expected to sustain a gas overpressure of a few\,mbar. The first
buckling modes of a perfect structure are shown in
Figure\,\ref{fig:DCH_buckling_inner}, for a very high load of 120~mbar.

\begin{figure}[tbp]
\includegraphics[width=0.49\textwidth]{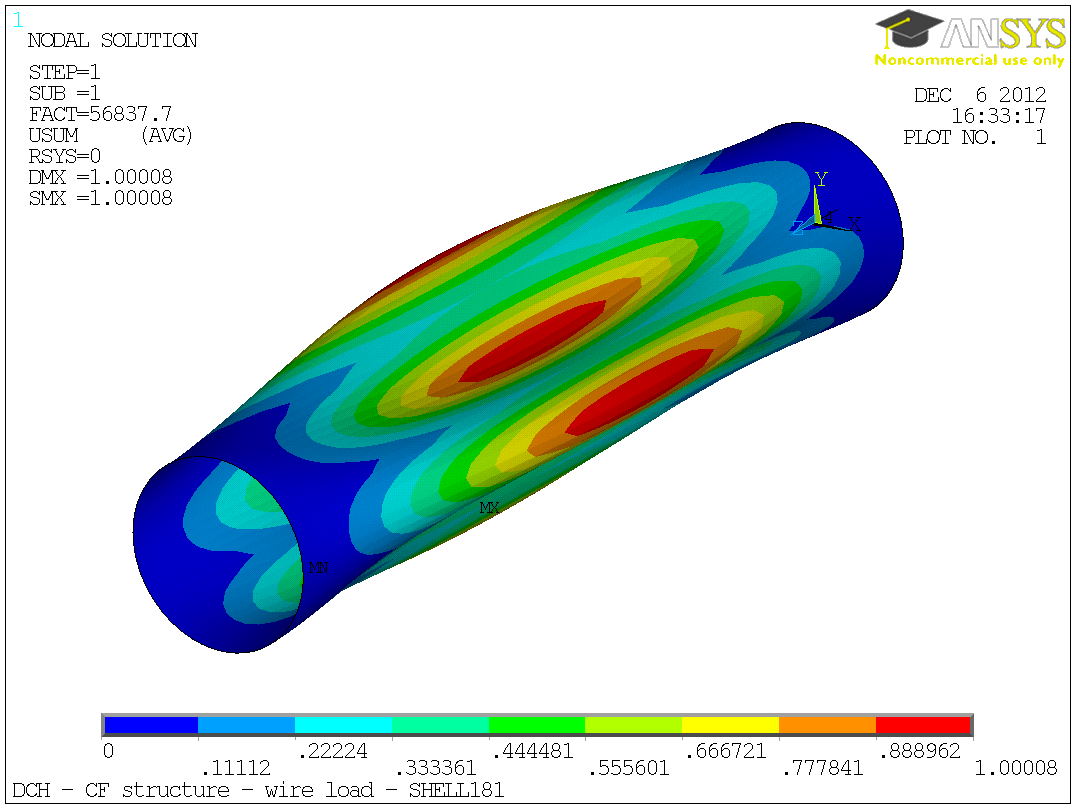}
\caption{Calculated first buckling shape of the inner cylinder.}
\vskip -5pt
\label{fig:DCH_buckling_inner}
\end{figure}

It is well known that
cylindrical thin shells are sensitive to geometrical imperfections. In
order to realistically estimate the buckling safety factor, we have
introduced such imperfections in our model, evaluating the critical
load in each case with a non linear analysis. The results are shown in
Figure\,\ref{fig:DCH_buckling_inner_imperfections} and in
Table\,\ref{tab:DCH_buckling_inner_imperfections}: as expected, the
larger the geometrical imperfections, the smaller the critical load.

\begin{figure}[tbp]
\includegraphics[width=0.49\textwidth]{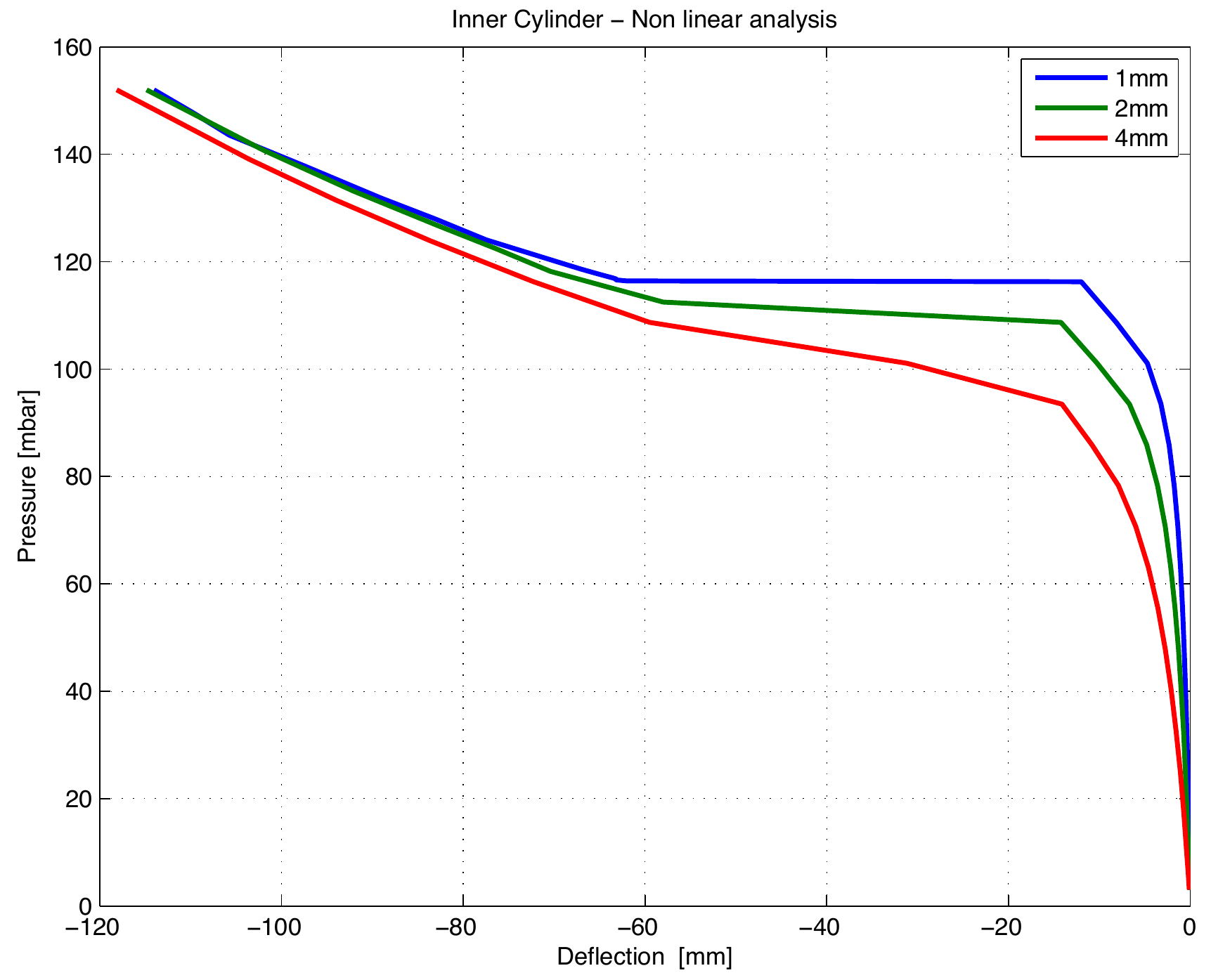}
\caption{Inner cylinder critical load for different values of the
  geometrical imperfections.}
\vskip -10pt
\label{fig:DCH_buckling_inner_imperfections}
\end{figure}

The table also shows that very small values are obtained when the
inner cylinder is only made of two (0.09 thick) CF skins.
The maximum load dramatically increases when a 3\mm thick honeycomb
structure is inserted between the CF skins.
An acceptance threshold on the geometrical imperfections at the level of
$2-4\mm$ can certainly be  imposed, which leaves a reassuring
factor 30 on the buckling load of the inner cylinder.

\begin{table}[t] 
\begin{center}
\caption{
Buckling load [mbar] as a function of the maximum size [mm] of
geometrical imperfections in the inner cylinder.
}
 \vskip10pt
 \begin{tabular}{|l|cccr|} \hline\hline
Max. size [mm] / &   0  &  1 &  2  &    4 \\
  Config   &  &   &   &   \\ \hline
  0.09/0.09 &  0.84 & 0.74 &  0.68 &   0.56 \\
  0.09/3/0.09 &  120 & 108 &   99 &   86 \\  \hline
\end{tabular}
\label{tab:DCH_buckling_inner_imperfections}
\end{center}
\end{table}

\tdrsubsubsec{Outer Cylinder}

A linear buckling analysis was performed on the outer load-bearing
structure similarly to what discussed in the previous subsection.
Also in this case geometrical imperfections were included,
evaluating the safety factor under an axial load of about $21$\,kN. 

\begin{figure}
\includegraphics[width=0.95\linewidth]{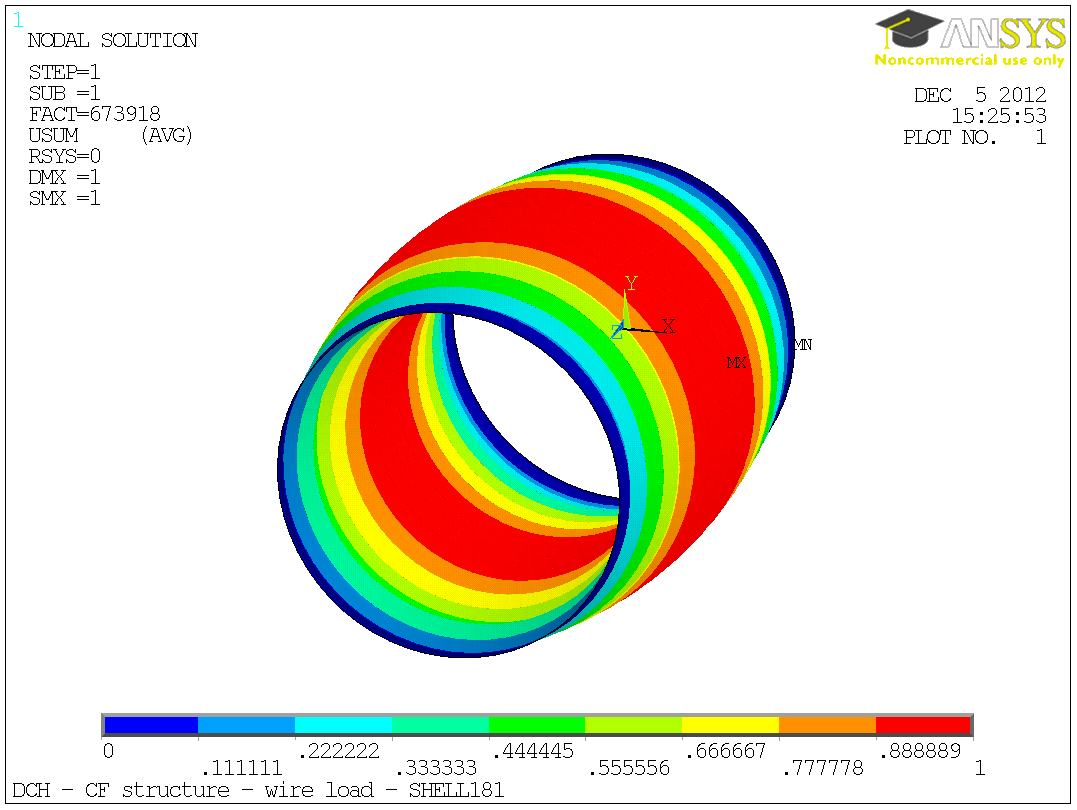}
\includegraphics[width=0.95\linewidth]{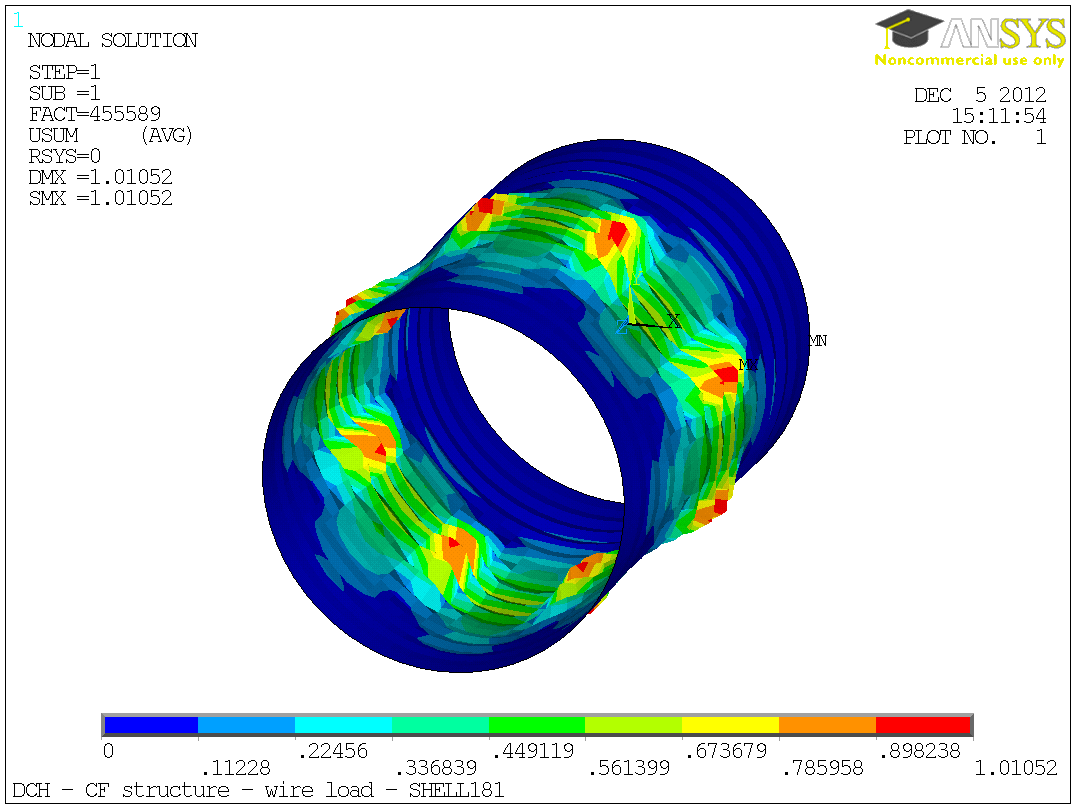}
\includegraphics[width=0.95\linewidth]{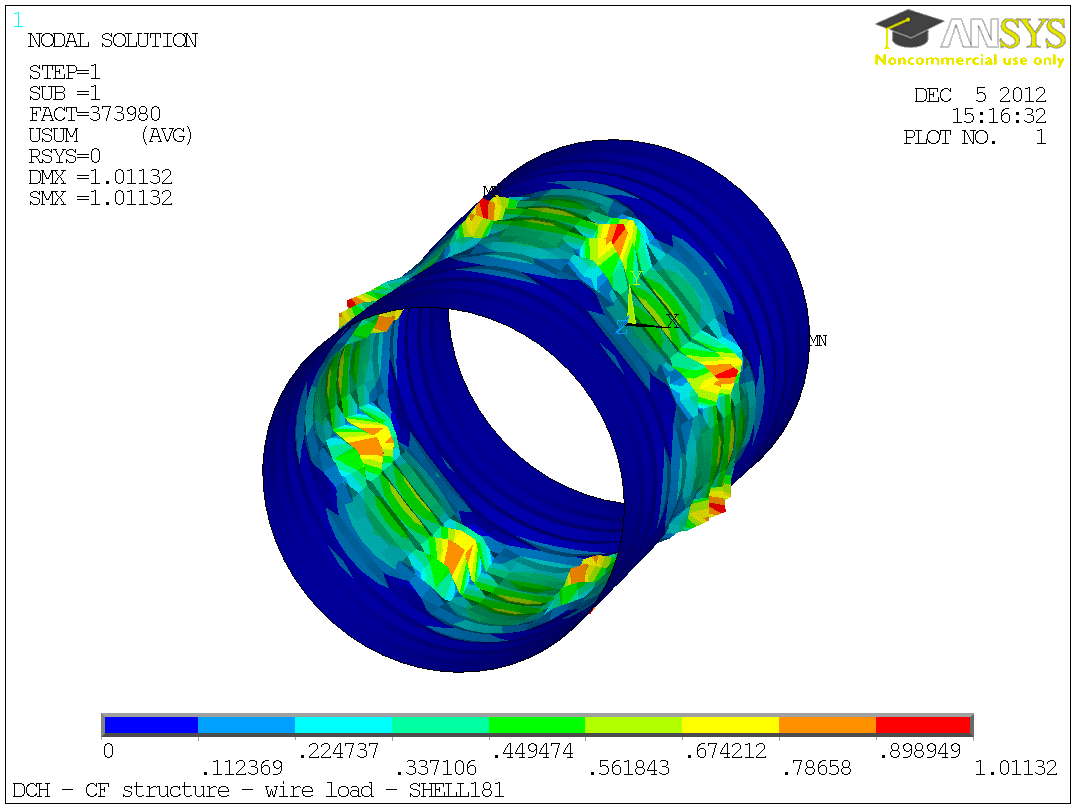}
\caption{Calculated outer cylinder 1$^{st}$ buckling shape for
(a) no imperfections, (b) 2~mm imperfections, or (c) 4~mm imperfections.}
\label{fig:DCH_buckling_outer_imperfections}
\end{figure}

These results are
shown in Figure\,\ref{fig:DCH_buckling_outer_imperfections} for the
final structure (1CF 6HC 1CF ) and summarized in
Table\,\ref{tab:DCH_buckling_outer} for 
different carbon fiber sandwiches.

\begin{table}[t]
\begin{center}
\caption{
Outer cylinder safety factors for different thickness in mm of the
Carbon Fiber (CF) and honeycomb (HC) components, and for different
sizes of the  geometrical imperfections.}
\vskip10pt
\begin{tabular}{|cccc|} \hline\hline
Config             & Nominal  &   2\mm &  4\mm  \\
                       &    S.F. & S.F. &   S.F. \\ \hline
 2CF                &  24.0   & 10.8 &   8.0  \\
 1CF 3HC 1CF &  22.0   & 12.8 &   8.3  \\
 1CF 4HC 1CF &  33.6   & 13.5 &   9.0  \\
 1CF 6HC 1CF &  33.6   & 22.8 &   18.6 \\ \hline
\end{tabular}
\label{tab:DCH_buckling_outer}
\end{center}
\end{table}

\tdrsubsec{Choice of wire}
As described in \subsecref{DCH_CellStructure}, each cell has one
sense wire surrounded  by a rectangular grid of eight field wires. 
A smaller value for the sense wire diameter is generally 
preferred for several reasons: any given gas amplification, therefore electric field,
can be reached with smaller applied high voltage; and the avalanche
starts closer to the wire, with smaller space-charge effects. In
practice the wire diameter is limited by ease of handling, and by the
mechanical resistance.
For sense wires, we intend to use 20\mum gold-plated Molybdenum
wires. Molybdenum has lower resistivity than the more
conventional W-Rh alloy, therefore smaller dispersion for pulses
traveling  along the wires. 
The density is also smaller than for Tungsten wires,
resulting in slightly less tension (160\,kg) on the endplates,
and more importantly in less material in the tracking volume.
The radiation length for all wires (field + sense + guard) increases from
4.25\cm with W-Rh wires to 6.95\cm with Mo wires, while \Xrad for
the wires and the \HeIso{90}{10} gas mixture increases from
480\,m to 545\,m.
A mechanical tension of approximately 23\,g will be applied to each
wire, needed for it to show a gravitational sag of 200\mum. Such a
tension is consistent with electrostatic stability and with the yield
strength of the wire.

The field wires are chosen to be 90\mum diameter Al-5056 wires with a
``stress-relieved'' temper. The selected diameter keeps the electric
field at the cathode surface below the limit of 20\,kV/cm, considered
a safe value in drift chambers not to trigger gas multiplication and
help suppressing the Malter effect. The
temper considerably increases the wire tensile and yield strength, and
reduces the loss of tension due to the creep effect. A
nominal tension of 73\,g will be needed to equalize the gravitational
sag of sense and field wires (in the stringing phase the latter will be
pulled at a slightly higher tension to compensate for the creep
effect). Again, this tension is well within the elastic limit of the wires.

\tdrsubsec{Feedthrough design}
The feedthroughs locate the wires to within the specified tolerances, 
hold the wire tension, and, in the case of the sense wires, insulate
against the high voltage. They must achieve these goals while 
maintaining a helium-tight gas seal.

A feedthrough is made from two components, a plastic outer insulator, 
and a conducting crimp pin (Figure~\ref{fig:dchfeed}). The insulators are
injection-molded parts formed from Celenex 3300-2, chosen for its
low shrinkage during molding, dimensional stability, and high 
dielectric strength. The crimp pins for the aluminum field wires are 
aluminum 6063.  Studies are planned to determine whether copper or
aluminum crimp pins are more suitable for the molybdenum sense wires.
The crimp pins will have a gold-flash coating with a nickel underlayment. 

\begin{figure}[tbh] 
\centering
\includegraphics[width=0.45\textwidth]{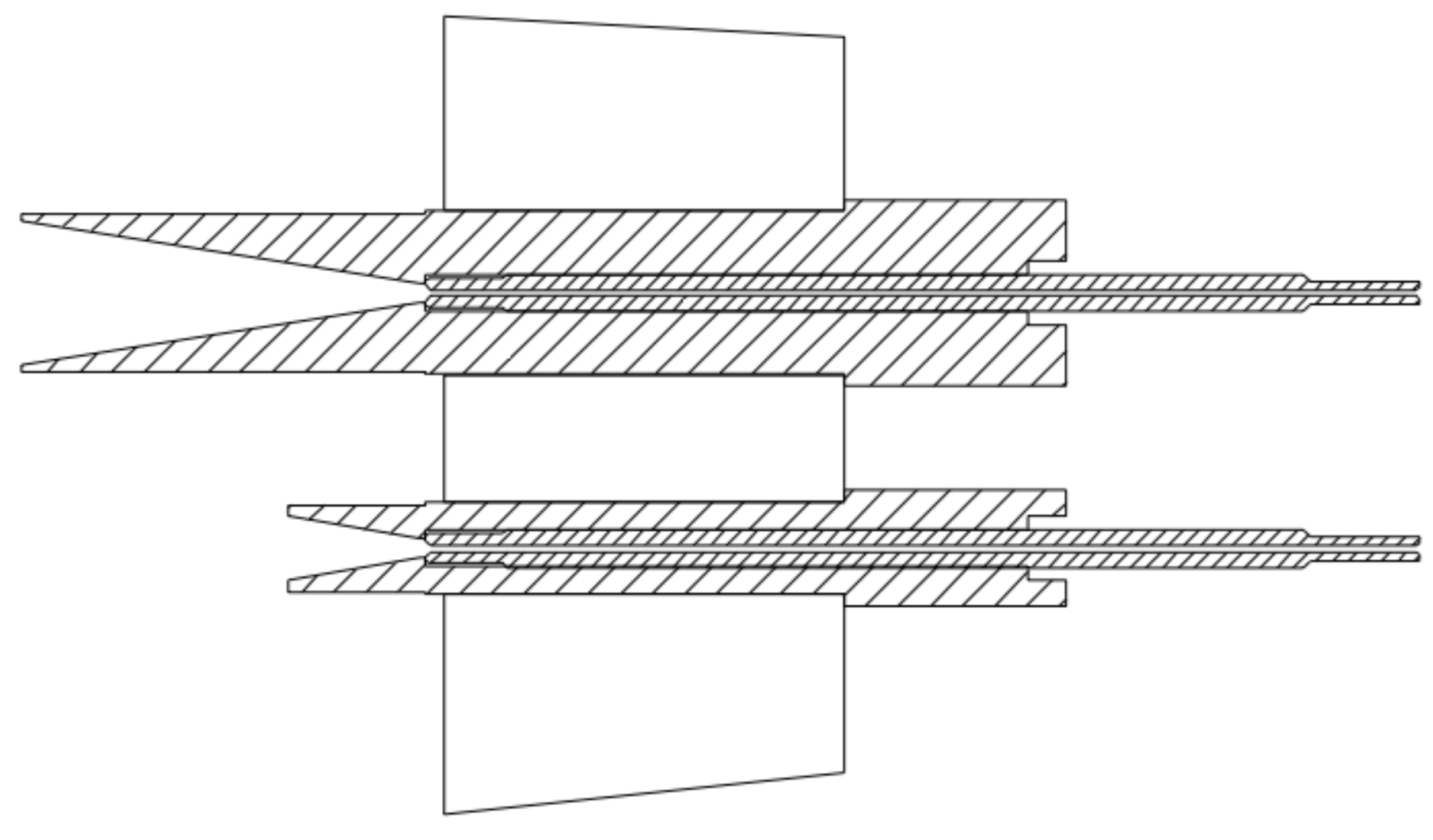}
\caption {
Sense wire (top) and field wire (bottom) feedthroughs from the 
\babar\ drift chamber.  The \superb parts will be similar. 
}
\label{fig:dchfeed}
\end{figure}

The two parts are glued together with an epoxy that is dyed so that 
extraneous epoxy on the crimp pin can be identified and removed. 

The tolerances are specified to ensure that the contribution to 
cell resolution is small, with tolerances on the sense wire parts 
significantly tighter than those on the field wires.  
Inner diameter of sense wire crimp pins at the wire release point will
be 100 microns, and 200 microns for field wires. Concentricity of the 
pin hole with respect to the shaft diameter will be less than 30 microns, 
and eccentricity of the shaft will be less than 25 microns. 

Each of the approximately 75000 feedthroughs will be individually 
tested against the specifications. 
Sense wire feedthroughs will have an additional test
to verify HV performance.

\tdrsubsec{Endplate systems}
\label{sub:DCHendplate}

\tdrsubsubsec{Electronics enclosures}

The amplifiers mounted on the backward endplate of the drift chamber,
and the high-voltage components mounted on the forward endplate are
covered by the electronics enclosures. The volumes are filled with
nitrogen to ensure that a leak of the flammable drift gas through the
endplate cannot form an explosive mixture. The enclosures also protect
the components inside, and provide the mounting points for the chamber
as a whole. 

The
enclosures are light aluminum structures (Figure~\ref{fig:enclosures}). 
The main features are
feedthroughs for the various signals and services used by the chamber:
chamber gas, nitrogen, cooling water, amplifier power and signal, and
high voltage. Each $1/16^{\rm th}$ sector corresponds to approximately 
500 signal cables, 
arranged into 8-channel ribbon cables and feedthroughs.
These feedthroughs may also carry the power and control lines for each 
amplifier. 
The feedthroughs are mounted on removable panels, while the cooling 
lines are on the fixed ribs. The panels allow access for installation and
repairs, although such accesses are expected to be rare. Although the 
endplates are curved, the panels can be flat, reflecting the geometries of
the backward calorimeter and the forward time-of-flight system on either
side of the drift chamber. This will greatly simplify the necessary gas 
seals. 

\begin{figure}[tbh]
\centering
\includegraphics[clip=true, width=0.50\textwidth, trim=0 3.1inch 0 3.1inch]{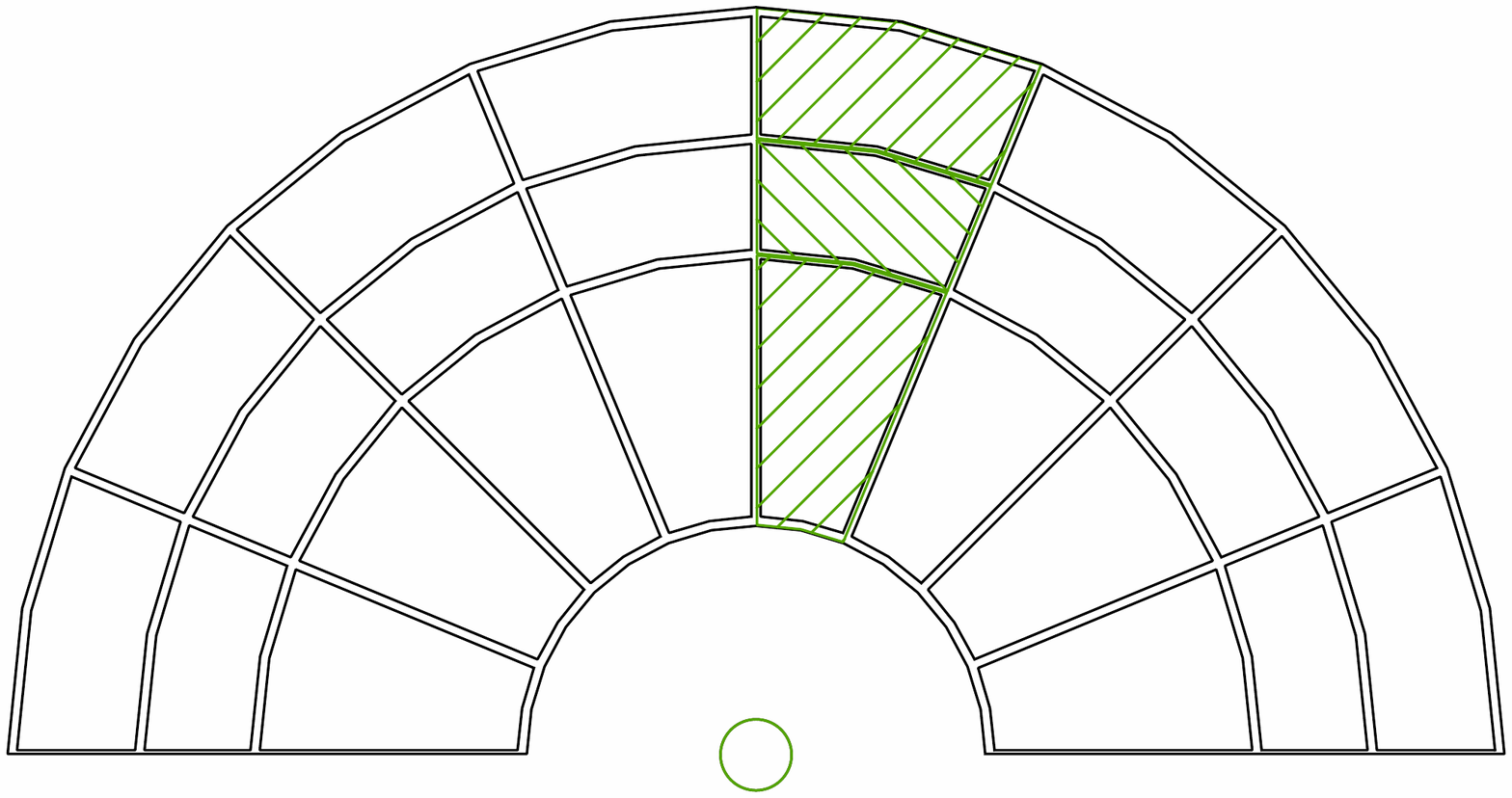}
\includegraphics[clip=true, width=0.50\textwidth, trim=0 2.9inch 0 3.4inch]{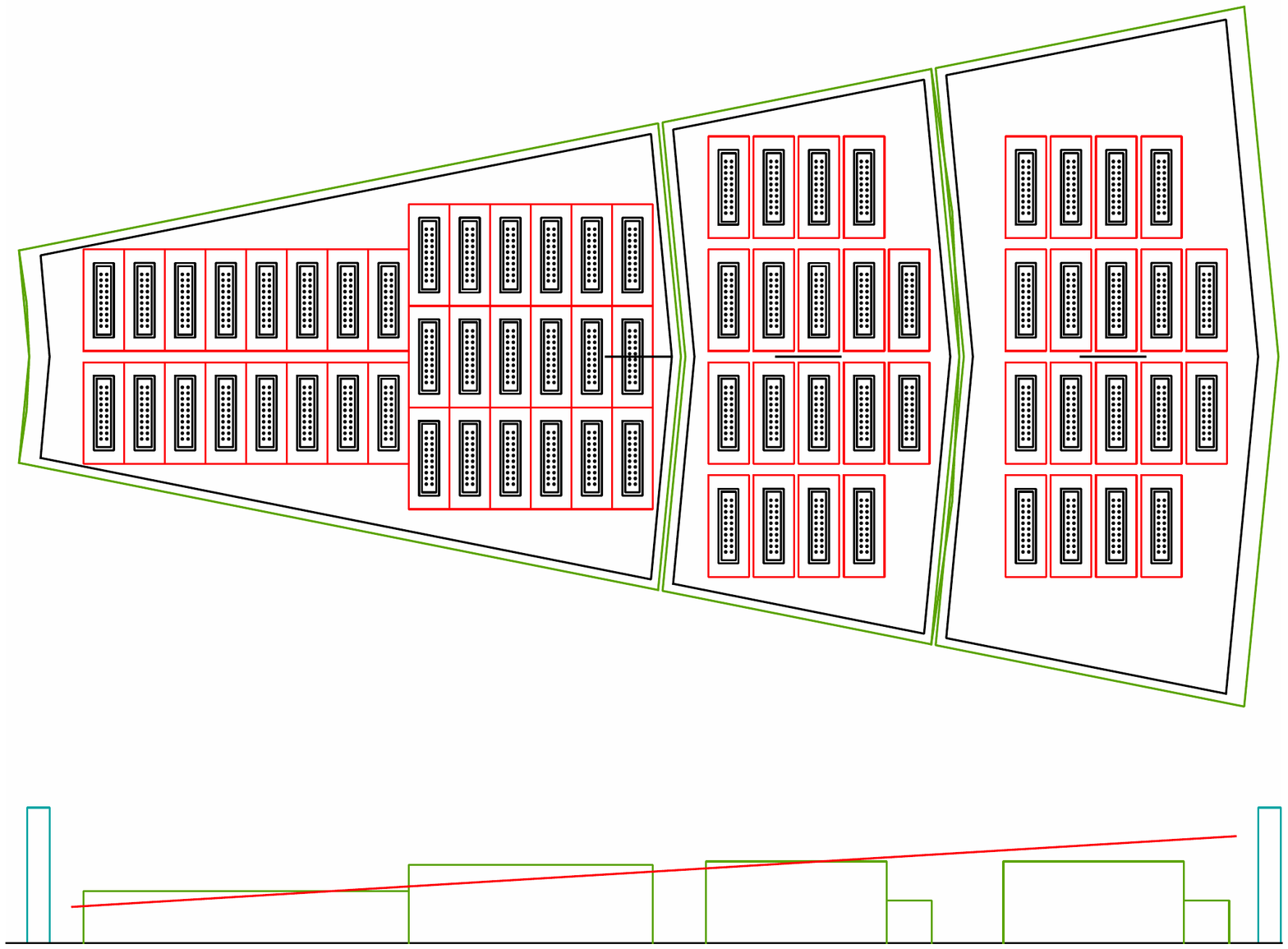}
\caption {
(a) The rib structure of each enclosure is machined from a solid piece of 
aluminum.
(b) The feedthroughs are on panels mounted to the ribs.  
}
\label{fig:enclosures}
\end{figure}

\tdrsubsubsec{Cooling}

The fast amplifiers mounted on the drift chamber endplates produce
approximately 1200~W of heat. This heat must be removed to keep the
temperature of the \dch\ and nearby detectors stable and uniform.  The
cooling system to accomplish this will be water based, operating at a
pressure below atmospheric pressure. Small leaks will therefore cause
air to leak into the water, rather than leading to water leaking into
the electronics enclosure. The system is quite similar to one 
recently built for the near detector of the T2K experiment\,\cite{bib:t2k}. 

The major components of the system include two water reservoirs, one
at atmospheric pressure, and the other maintained at an absolute
pressure of 0.3 atmospheres; a pump to move water between the systems;
cooling lines to the \dch\ and inside the electronics enclosures; valves
and gauges; a heat exchanger that connects to the laboratory chilled
water system; and a control system to maintain the desired water
temperature. Appropriate corrosion inhibitors and microbiocides will be
added to the water. 
The lines within the enclosures will be mounted to the fixed
ribs, and may include fins or other features to increase the surface area. 
These features, combined with turbulent nitrogen flow in the enclosure, may
generate sufficient thermal flow from the amplifiers to keep the temperatures
at an acceptable level.   
Mockups and thermal calculations will be undertaken to test this concept. 
The alternative will be cooling straps. 

Although there is no heat generated on the forward endplate, cooling lines
will be run to ensure a uniform temperature across the drift chamber. 

\tdrsubsubsec{Shielding}

The aluminum structure of the electronics enclosures, together with the
aluminum skins on the outer and inner cylinders, form a Faraday cage 
that encloses the amplifiers and the chamber wires. A 25 micron thick
aluminum skin bonded to the endplates provides additional shielding 
between the chamber wires and the enclosure volumes. 

\tdrsubsubsec{Electromechanical boards}

Electrical connections to the crimp pins are required at both ends of 
every wire. At the forward end, the HV distribution boards
ground the 
field wires and provide HV to the sense wires via the circuit  
described in Sec.~\ref{sec:DCH_HV}, and terminate the sense wire
at the characteristic impedance of the cell. 
At the backwards end, 
service boards ground the field wires, and gather the signals from 
eight sense wires onto a single multi-conductor connector. The only 
active components on the service boards are the HV blocking capacitors. 

Each of the ten superlayers requires a different size of HV
distribution board and service board.  Only a single style of
8-channel preamp card is required.  Each board will service eight
cells (four in radius by two in azimuth), 
typically corresponding to 32 crimp pins.  A number of crimp pin
connections much larger than this would make the board difficult
to insert and remove. 

The connections to the crimp pins are made using low-insertion-force
connectors, such as the Hypertronics connectors used for \babar. 
The details of how the curvature of the endplate is handled in the 
design of the cards will require prototypes and mockups. 

Jumpers between adjacent boards will connect the ground planes. 
At the HV end, jumpers will distribute the HV among the typically 
three boards serviced by a single HV channel. 

The endplate will include two blind holes per board that will be used
with dowel pins to align the boards during insertion. 
We do not anticipate using
pull-down screws, which would require a large number of tapped holes 
in the endplates.
Each board will have two threaded holes that will be used to push 
against the endplate if it is necessary to remove the board. 
The boards will be removed rarely, if 
ever. One of the two holes will be used during normal operations to 
provide an electrical connection to the aluminum RF shield on the 
endplate via a low-force spring connection.

\tdrsubsec{Stringing}

The chamber will be strung horizontally, with the inner cylinder in place
(but not under load), and the outer cylinder absent. The wire load is held
by a mechanical structure that attaches to 
the outer radius of each endplate and to a central shaft inserted through
the inner cylinder. The chamber can be rotated about this shaft.

There will be two stringing shifts per day, five days per week. Each
shift will have two teams consisting of two stringers plus a
wire-transport robot.  The robot is used to carry the wire between
the endplates, reducing the chance of errors or damage to the
previously strung wires, and keeping the interior of the chamber
clean by eliminating the need for people to work in that area.
The work will be done in a class 10000 or better clean room, with 
the interior of the chamber (including the robots and the previously 
strung wires) curtained off for greater cleanliness.

An abbreviated sequence of steps involved in stringing a wire is:
\begin{itemize}
\item A stringer attaches the wire to a steel needle and inserts it
  through a plastic sleeve into the appropriate feedthrough hole,
  where it is captured by the magnetic head of the robot.

\item The robot carries the wire to the other endplate, where the needle is 
taken by the second stringer, who pulls the additional wire 
needed.  The robot moves into position for the next wire when the
second stringer indicates that they have finished pulling the wire. 
Both ends of the wire are then placed in wire holders before being cut. 

\item Each stringer feeds the wire through a feedthrough, places a 
ring of glue around the shaft of the feedthrough, and inserts it in the
hole. The glue will be thixotropic gel cyanoacrylate adhesive such as 
Loctite 454 or 455, dyed red for ease of inspection. 

\item The second stringer crimps her crimp pin using a foot-operated
  pneumatic tool, such as the Simonds MSP-1 Squeeze Pliers. 
The first stringer then applies the appropriate
  tension using a pulley and weight, and crimps her crimp pin.
  Because the deflection of the endplate is negligible under the wire
  load (Figure~\ref{fig:dch_endplate_Zdisplacement}), wires can be
  tensioned at their nominal values.  Excess wire is cut at both ends,
  leaving 1~mm protruding for inspection purposes.

\item Once both teams have finished ten wires, the chamber is rotated 
into position for the next set of ten. 
\end{itemize}

The third shift each day is used for quality control. 
The steps include
measuring the tension, visually inspecting the glue seals around the
feedthroughs, looking for excess glue on crimp pins, and straightening
bent crimp pins.
Wire that have tension out of tolerance (a gravitational sag more than 
15~microns from nominal) are removed, and the holes cleaned. These 
wires are restrung at the start of the next stringing shift. 
The third shift also makes the gas seal between the wire and crimp pin
using a low-viscosity cyanoacrylate adhesive such as Loctite 493, dyed
red for easier inspection. 

One stringer starts work an hour before the others to prepare the glues, 
lay out the parts needed that day, 
and to verify that all crimp tools are producing crimps of the correct size.

Each team is expected to complete 80--100 wires per shift, for a total of
approximately 350 per day. 

\tdrsec{Electronics} 
\label{sec:DCH_Electronics}

\tdrsubsec{Design Goals} 

The \superb Drift Chamber front-end electronics is designed to
extract and process approximately 8000 sense wire signals to:

\begin{itemize}
\item measure the electrons drift times to characterize a track (momentum of charged particles)
\item measure the energy loss of particles per unit length, $dE/dx$
  (particle identification) 
\item provide hit information to the trigger system (trigger primitives)
\end{itemize}

Two methods will be discussed for the energy loss measurements.
We first present a solution already widely used in existing drift chambers, based on
the measurement of the integrated charge on each sense wire (standard
readout, \SR).
Then we present the baseline of our choice, which is based on 
the counting of the primary ionization clusters (\CC), which has never been used in a large scale \dch before.
Because the front-end requirements for the two options are quite different, we shall discuss them in separate sections.
 
\tdrsubsec{Standard Readout}
\label{subsec:DCT_Standard_RO}
\tdrsubsubsec{Charge measurement} 
The method is based on an integrated charge measurement, which allows the
use of relatively low bandwidth preamplifiers. This makes the
front-end chain less sensitive to noise pickup and instabilities, a
condition quite useful in a system with a large number of channels.
The three main specifications for a charge measurement are: resolution,
dynamic range and linearity.
 
\tdrparagraph{Resolution} The goal of a charge measurement to be used
for particle identification is to determine the most probable energy
loss per unit length to a precision of the order of 7.5\%.  This can
be achieved, despite the large fluctuations present the ionization
process, by sampling the collected charge many times (once per DCH hit
cell) and by applying the ``truncated mean'' method to obtain the peak
value of the resulting distribution to a precision of several percent.
 
The \superb \dch design parameters and expected working
conditions aim at obtaining an overall single cell energy loss resolution ($\sigma_E$) of
about $25-30\%$, mainly driven by the detector
contribution. To make the
contribution of the front-end electronics $\sigma_{EL}$ negligible, we set a limit at $\sigma_{EL} \le \sim 5\%$.
 
Finally, if we assume that the charge collection due to a
minimum ionizing particle crossing orthogonally the cell is about 50 $fC$  ($\sim 2 fC/e$ @
$10^5$ nominal gas gain),  we can infer a limit to the Equivalent Noise
Charge (ENC) for a single front-end channel of about  $50  \; fC
\cdot 0.05 \simeq 2.5\;  fC$. 
 
\tdrparagraph{Dynamic range}
With 8 bits ADCs the dynamic range is
$2.5-500\;fC$. This is more than enough to satisfy the system requirements.
 
\tdrparagraph{Linearity}
As stated above, a single cell energy
resolution is about $25-30\%$. Therefore, a linearity of the order of $2\%$
largely satisfies the system requirements.
  
\tdrsubsubsec{Time measurement} 
As for the charge measurement, we have three main specifications: resolution,
dynamic range and linearity.
 
\tdrparagraph{Resolution} One of the \superb \dch requirements is the
reconstruction of charged particle tracks. The measurement consists of
recording the arrival time at the sense wire of the first ionization
electron.  The spatial resolution ($\sigma_S$) is strictly related to
the time resolution and gets contributions from primary ionization
statistics, electrons diffusion time and time measurement accuracy.
To achieve the required point resolution of $\sigma_S\sim110\mu m$,
and assuming that the first two effects account for about $100\;\mu
m$, the upper limit for the electronic contribution can be deduced to
be $\sigma_{EL} \le \sim50\; \mu m$.  Because helium based gas mixtures
are characterized by a non saturated drift velocity up to high
fields\,\cite{bib:Sharma}, this velocity rapidly increases as the
electrons approach the sense wire. A value of 2.5 $cm/\mu s$ (25 $\mu
m/ns)$\,\cite{bib:LaCava} has been used to evaluate the maximum
acceptable error in a time measurement, that is $\sigma_{t} \le 50[\mu
m]/25[\mu m/ns] \simeq$ 2 $ns$.
 
Discarding the bunch length contribution (tenths of ps), there are two
main error sources in time measurements: the discriminator jitter and the
TDC resolution (digitization noise). The discriminator jitter, in turn, has two
main contributions: signal noise and time-walk.
 
The signal noise contribution is generally small and can be evaluated
according to $\Delta t = \sigma_{noise}/(dV/dt)\simeq
\sigma_{noise}\cdot\tau/V_{max}$ where $\tau$ is the
preamplifier-shaper peaking time.
Assuming that a single electron cluster generates a signal of amplitude $\sim$
20 $mV$, and that the noise and the peaking time associated with the signal
are $\sigma_{noise}\sim 3 \;mV(rms)$ and $\tau\sim
5\;ns$, we get a noise contribution to the time resolution of about 
  0.8 $ns$.
 
The time-walk effect is caused by the signal amplitude variation. With a
peaking time of about 5 ns, a time-walk contribution for a low-threshold
leading-edge discriminator can be estimated to be about 1.5 $ns$.
 
Finally, the digitization noise depends on the digitization unit $\Delta$
according to the formula $\sigma=\Delta/\sqrt(12)$.
Using  $\Delta \; \simeq \; 1.5 \; ns$ a digitizing
noise of about 0.45 $ns$ is obtained.
 
In summary, without corrections, the time resolution is dominated by
time-slewing effects. It can thus be estimated to be about 1.8 $ns$
(including all contributions). Nevertheless corrections can be applied
using digitized signals to minimize time-slewing effects thereby reducing
the time walk contribution (Figure~\ref{fig:ReadoutBoardBlockDiagram}).
 
\tdrparagraph{Dynamic range}
The TDC range depends on the drift velocity and on the cell size. A
maximum drift time of about 600 $ns$ has been estimated for \superb
\dch cells. Providing some safety factor,  a TDC range of about 1
$\mu s$ is enough to include any jitter in trigger generation and distribution.
 
\tdrparagraph{Linearity}
A linearity of the order of 1\% fully satisfies time measurement
requirements.
 
\tdrsubsubsec{Front End Electronics}
 
The \dch FEE chain (Figure\,\ref{fig:FeBlockDiaTM}) is split in two blocks: on-detector and off-detector electronics.
 
In the following paragraphs we will give a description of the on-detector 
electronics while the description of the Off Detector Electronics can be found in Sec.\,\ref{sec:ETD_DCH}.

\tdrparagraph{On Detector Electronics}
Preamplifier boards will contain HV blocking capacitors, protection
networks, preamplifiers and (possibly) shapers-amplifiers. Because of
the small cell dimensions, many cells can be grouped in a single,
multi-channel preamplifier-shaper board. Signals and power supply
cables will be connected to the boards by means of suitable connectors.
 
In addition to the requirements on the Signal to Noise Ratio (SNR), each 
preamplifier should be characterized
by enough bandwidth to preserve signal time information meeting the power
requirement of not more than  20--30\;mW per channel, to limit the total
power dissipation on the backward end-plate to 160--240\;W. This will
allow the use of a simpler and safer forced-air based cooling system (no risk of leak).
 
\begin{table}[h!]
\begin{center}
\smaller
\caption{Preamplifier main specifications}
\begin{tabular}{ll} \hline\hline
\Tp Non-linearity & $<2\%~(1--100 fC)$ \\ 
Output Signal Umbalance & $ < 2\%~(1--100 fC$) \\ 
Gain (Differential)  & $\sim$ 5.2 mV/fC\\
$Z_{IN}$ & 110 $\Omega$\\
$Z_{OUT}$ & 50 $\Omega$\\
Rise time & $\sim$  2 ns ($C_D = 24 pF$) \\
Fall time & $\sim$  13 ns ($C_D = 24 pF$) \\
Noise & 1350 erms ($C_D = 24 pF$) \\
$V_{SUPPLY}$ & 4V \\
$P_{D}$ \Bt & $\sim$ 30mW \Bt \\\hline\hline
\end{tabular}
\label{tab:PreSpecs}
\end{center}
\vskip -15pt
\end{table}
 
Concerning the circuit implementation, since the channel density is
low and a simple circuit topology can be used, an approach based
on SMT technology can be adopted thus avoiding a specially designed (and
expensive)  ASIC development.
 
\begin{figure}[tbp]
\begin{center}
\includegraphics[angle=0, scale = 0.40]{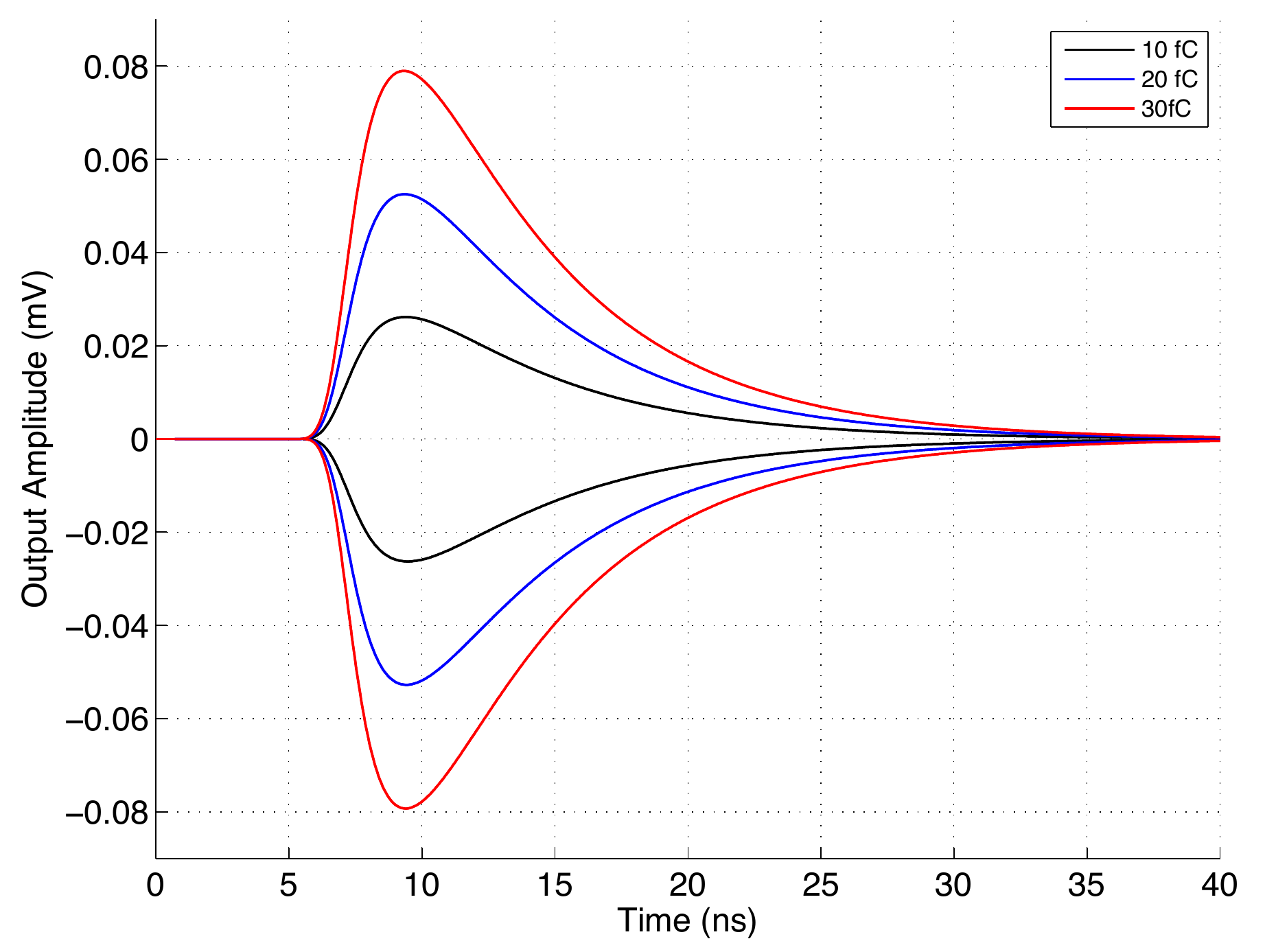}
\caption{Preamplifier output for a 10, 20 and 30 $fC$ test pulse $(C_{DET}$ = 24 $pF$) }
\label{fig:DiffOutputAmplitude}
\end{center}
\end{figure} 
 
As an example, the simulation of a three stages transimpedance
preamplifier based on SiGe transistors has been carried out. The first
stage dominant pole is around 26 $MHz$ while other stages have been
designed with wider bandwidth thus resulting in a good separation in
terms of cutoff frequencies. 
The simulation results are given in Table\,\ref{tab:PreSpecs} 
while Figure \ref{fig:DiffOutputAmplitude} shows the (simulated) output waveforms
for 3 different input charges (10, 20 and 30 $fC$) injected through the
test input.
 
\tdrsubsec{Cluster Counting}
\label{subsec:DCH_Waveform_RO}
 
The \cc technique is very powerful as it leads to an improved 
particle identification. 
To fully exploit the technique, individual ionization clusters
must be identified.

In our system we use slow drift gas mixtures ($\sim$1 $\mu s/cm$),
state of the art high sampling frequency digitizers (at least 1 GSPS)
and fast processing (data throughput must sustain the \superb expected
$150 \; kHz$ average trigger rate). These modules require a large
amount of power, forcing us to limit the number of channels to 8--16
channels per Amplifier Digitizing Board (ADB).
 
Moreover, fast amplifiers must be used on the preamplifier boards,
which results in a larger power requirement than that of the \SR. As a
consequence, we are considering the use of a local liquid cooling
system. The wide bandwidth requirement also impacts on the type of
cables used to interconnect the preamplifiers with the ADBs and on the
noise pick-up sensitivity of the full system .  In addition, the use
of \cc requires that the signal reflection in the sense wires must be
eliminated. This is done by mean of termination resistors ($R_T$),
which results in a lower limit on the system intrinsic noise.
 
Provided the cluster detection efficiency is sufficiently high, information from the \cc measurements 
can be used for tracking purposes. This requires storing the time of arrival at the sense wires of all the clusters, 
instead of simply counting them.
 
Specifications for the \CC measurements, as for the \SR, must be given on resolution, dynamic range
and linearity. 
 
\tdrparagraph{Resolution} The resolution of the digitizers depends on
the lowest signal amplitude to be sampled and the system baseline
noise. Assuming an average input signal of $\sim 6 fC/e$ @
$3\cdot10^5$ gas gain, a preamplifier-shaper gain of 10~$mV/fC$ and a
safety factor of 2 to account for gas gain fluctuations, the average
cluster signal is about 30~$mV$ for a single electron.  We can
estimate the preamplifier ENC from that of the dominant noise source,
which is the termination resistor. Assuming a $CR-RC$ shaping circuit
and a 3~$ns$ peaking time, we get an ENC of about 0.2~$fC$, that is
about 2~$mV\; rms$ for a preamplifier gain of 10~$mV/fC$. Thus a
voltage resolution of about 2~$mV$ allows good control of the system
noise and of the cluster signals reconstruction.
 
\tdrparagraph{Dynamic range} The \cc method requires the observation
of peaks (corresponding to the clusters) in the digitized signals. An
upper limit of the signal dynamic range (discarding gas fluctuations)
is then given by the expected total ionization.  Helium based gas
mixture have already been well characterized \cite{bib:Burchat}.  We
can assume that a minimum ionizing particle (m.i.p.)  crossing
orthogonally a $1.2$ cm square cell filled with a \HeIso{90}{10} gas
mixture, will generate on average about 22 electrons. Thus the dynamic
range of an 8 bits ADC is fully adequate for \cc measurements.
 
\tdrparagraph{Linearity}
As we are interested in finding (and tagging) signal peaks, a
linearity of 2\% fully satisfies the requirements.
 
\tdrsubsubsec{Front End Electronics}
 
The \dch FEE chain for \cc is similar to that for the \SR. In this
scenario, the electronics modules will be connected with mini coaxial
cables. Because of the smaller number of channels per ADB, both the
number of crates and ADBs will increases significantly (Tables
\ref{tab:LinksBoardsCrates} and \ref{tab:PowerRequirement}).
 
\tdrparagraph{On Detector Electronics} Because preamplifier boards
will host high bandwidth ($\sim$ 350 MHz) amplifiers, the layout and
assembly are more difficult compared to the \SR scenario.
Particularly, special attention must be provided to avoid ground loops
to minimize instabilities and external noise pickup.  The baseline
solution is the use of commercially available fast
transimpedance amplifiers, but the design of a preamplifier based
on high bandwidth and low-noise SiGe SMT transistors will be
investigated as well.

\tdrsec{High Voltage system}

\label{sec:DCH_HV}


\tdrsubsec{Main System and HV Distribution Boards}
The high voltage distribution network will be
located on the forward end-plate. The distribution board modularity
will match the preamplifier modularity while the number of
distribution boards connected to a single HV channel will depend on
the layer (example: inner layers = 2
boards, outer layers = 5 boards).

The HV distribution system consists of power supplies and distribution
boards, along with associated cables. 
The voltage will be supplied by a CAEN SY4527 Universal Multichannel
System supply with 16 A1535N distribution boards, giving a total of
384 channels. This is sufficient granularity that a single channel
failure will have a very small impact on detector performance.  
Spares of both the SY4527 and the A1535N will be available. 
The A1535N permits individual channels to operate in current-generator mode in
case of over-currents. This feature was found to be extremely useful
for the \babar\ drift chamber as it permitted the chamber to handle 
locally high
background rates without ramping down the chamber HV.

Individual HV channels are brought to the drift chamber from the
A1535N boards via multistrand cables with 
A996 52 pin Radiall connectors at both ends.

The multiconductor cable connects to a 
filter box containing a low-pass filter, located at the inner radius of the 
forward enclosure. The individual channels are fanned out within the 
enclosure to the HV distribution boards (Sec.~\ref{sub:DCHendplate}).
Each HV channel supplies two or three 8-channel distribution boards, 
depending on the superlayer. 

The HV distribution and sense wire termination circuit is shown 
in Figure~\ref{fig:CcHvDistib}. 
If termination is not used, the termination resistor ($R_T$) and the 500~pF capacitor per sense
wire are removed. 

\begin{figure}[tbh]
\begin{center}
\includegraphics[width=0.5\textwidth]{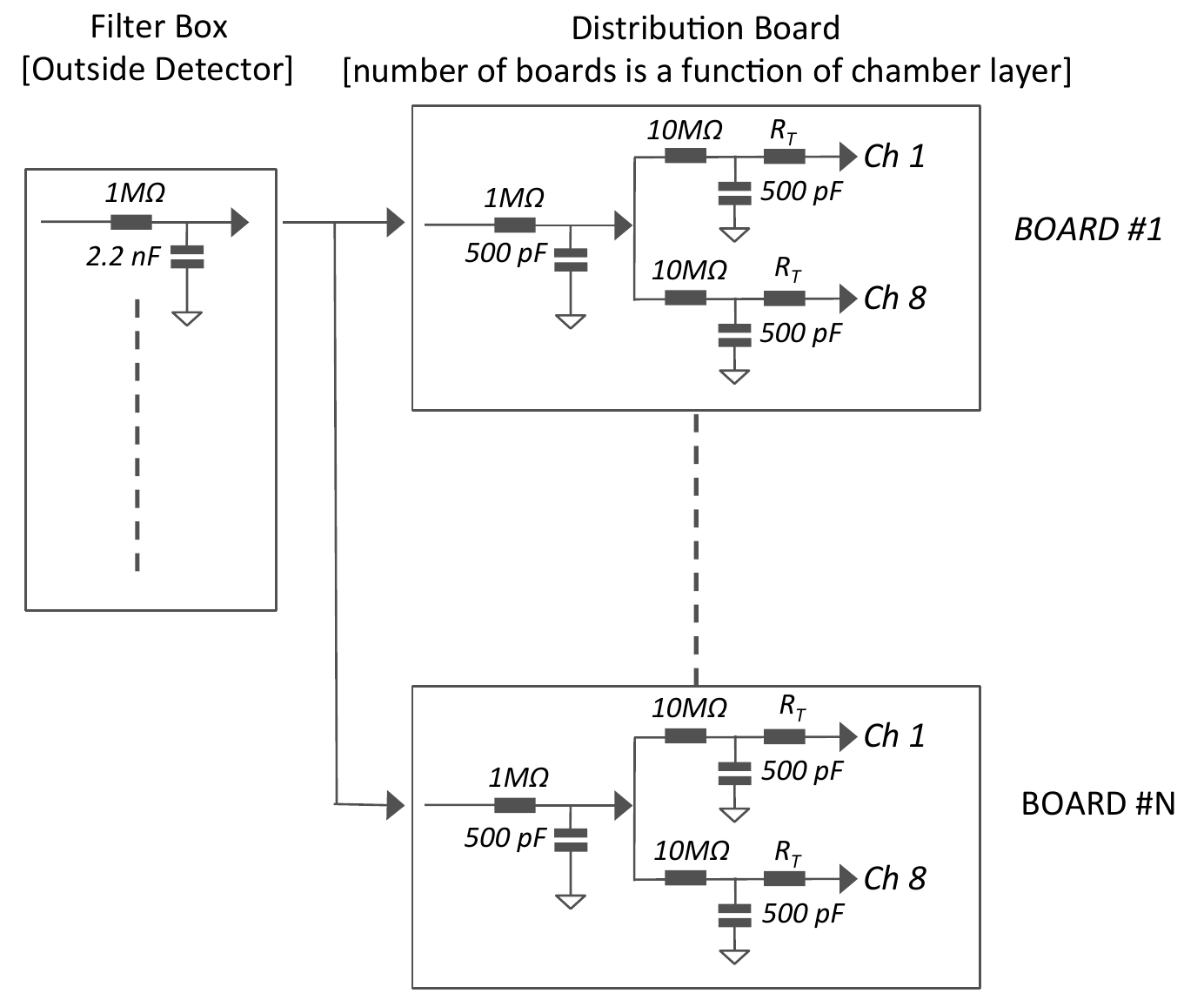}
\caption{HV distribution network.}
\vskip -10pt
\label{fig:CcHvDistib}
\end{center}
\end{figure}

\tdrsec{Gas system}
\label{sec:DCH_GasSystem}
The drift chamber is filled with a gas mixture of 90\% helium and 10\%
isobutane. The gas system supplies the appropriate gas mixture to the
chamber, while maintaining the required flow rate, pressure, purity,
and composition stability. It includes  safety items such as
flammable gas sensors and release valves to protect the gas system,
detector, and personnel from dangerous conditions caused by component
failures or operator errors. To reduce the operating costs, 85\% of
the gas will be purified and recirculated.

The mixing will be done using mass flow controllers that will maintain
the isobutane fraction at $(10.0\pm0.1)$\%. A parallel set of
rotameters may be included to allow for high flow rate flushing. The
system will allow a fraction of the flow to pass through a
temperature-controlled water bath, which will be used if we decide to
add water to help control the symptoms of aging.

The gas composition is verified using a set of analyzers to measure
isobutane, oxygen, and water. The analyzer set will be able to sample
gas at a variety of points in the gas system, such as before or after
the gas enters the chamber or the filters. 

This will be a recirculating gas system, which reduces operating costs and
air pollution from the isobutane. The total flow will be 15 liters per minute, 
of which 2.5 liters per minute will be fresh gas. For a chamber volume of 
5000 liters, this corresponds to 
four volume changes per day, or one volume of fresh gas every 1.5 days. 
The flow is controlled by an explosion-proof compressor, which is
regulated to maintain a chamber pressure of $4.00\pm0.05$~mbar (0.4\% of
an atmosphere) above atmospheric 
pressure. 

The primary gas lines between the gas mixing station and the detector
will be welded and pressure-tested stainless pipe, 1.5 inches
diameter. This will be reduced to 0.75 inch diameter in the cable
trays through the detector.  The input line is fanned out to 8 lines
of 5~mm diameter on the rear endplate, while the output line is fanned
out to 16 lines of 5~mm diameter on the forward endplate.

The gas returning from the detector passes through a palladium catalytic
filter which removes oxygen by the reaction
$13 {\rm O}_2 + 2 {\rm C}_4 {\rm H}_{10} \to 
8 {\rm CO}_2 + 10 {\rm H}_2{\rm O}$. The resulting water is 
removed by an alumino-silicate molecular sieve. The system contains 
two such sieves, so that one can be regenerated (i.e., have the 
absorbed water removed) by flushing with helium at elevated temperature
without stopping operations. This filter system was originally 
built for the \babar\ drift 
chamber and will be reused for \superb. 

The gas temperature and pressure will be monitored at various points in the 
system, along with atmospheric pressure. These quantities will be used to 
calculate gain correction due to gas density. We will also monitor gas gain
using a small, single-cell chamber that will be mounted on the return line
from the chamber. Any variations in the current induced
by an $^{55}$Fe source after applying the density correction would indicate
gain variations due to gas composition or chamber aging. 

The majority of the gas system components will be in the gas hut (or
room), which will be located at an exterior wall of the interaction
hall.  Two additional racks close to the detector will contain
bubblers, pressure sensors, and valves.  The gas storage areas will be
outside, under cover, immediately adjacent to the hut. The isobutane,
since it is flammable, will be stored in a physically separate area
from the other gases. The isobutane tanks and lines will be heated and
insulated.

The gas system includes an extensive safety system to protect personnel and 
equipment. This system will be reviewed and approved by the laboratory. 
Aspects of the safety system include:
\begin{itemize}
\item ventilation in the gas hut, which, when combined with flow
  restrictors on the lines into the hut, ensure that a leak cannot
  create an asphyxiation hazard.

\item nitrogen flows in the exhaust lines and into the electronics
enclosures on both endplates.

\item flammable gas sensors in the gas hut and the bubbler rack. 

\item an oxygen sensor on the return line.

\item bubblers and redundant pressure sensors to protect the chamber against
over pressure.

\item an independent helium line and regulator to protect the chamber against
sub-atmospheric pressures.

\item administrative controls on changes to the gas system. 

\end{itemize}
The system is designed such that it will remain safe even during extended 
power outages. We will undertake regular maintenance and keep sufficient
spare parts to ensure reliable operations. 

\tdrsec{Monitoring and Calibration}
\label{sec:DCH_Calibration}

\tdrsubsec{Monitoring}

Environmental and systems-related parameters that are potentially time dependent
 will be continuously monitored while \SuperB\ is taking data.
These will be collected asynchronously from the e$^+$e$^-$ collision data stream
within  the experiment's ``slow controls''  system. The parameters will be redundantly monitored and include:
\begin{itemize}
\item temperature: four in the experimental hall; four on the outside of each gas enclosure bulkhead; four inside the gas enclosure at each end
\item pressure: two in the experimental hall; two in each gas enclosure volume 
\item high voltage and current (read for each HV card) 
\item high voltage crate voltages, currents and temperatures (read for each HV card) 
\item  preamplifier low voltage and current (read for each read-out board)
\item  digital electronics voltages, currents and temperature (read for each crate)
\item discriminator levels set for cluster counting either in the FPGA or local derivative discriminator (one per channel)
\item oxygen, helium, isobutane and H$_2$O concentrations in the gas system at the input and output lines of the drift chamber
\item average $^{55}$Fe pulse heights from two gas quality monitoring chambers, one on the input the other on the output gas line of the drift chamber
\item cooling system monitors: temperature and pressure on the input and output lines and immediately before and after the chamber
\end{itemize}

 The quantities will be averaged over each minute and made available to the data-taking shift crew
 and drift chamber experts in the form of digital ``strip charts''. These will be actively monitored by 
specialized software and stored in a database for future examination. The annual data volume will, at most,  be a few Gigabytes.
The active monitoring will read the quantities and determine if a quantity is outside  tolerance. If a monitored quantity falls outside
its tolerance band, the shift crew will be alerted by an alarm alerting them to the need for action.

The gas quality will be monitored using the pulse height information from the $^{55}$Fe source in the two monitoring chambers.
Problems with the gas mixture will be revealed by changes in the gas gain after correcting for expected changes in gas density
associated with atmospheric pressure, or perhaps experimental hall temperature, changes.

In addition to these environmental and systems-related parameters, the drift chamber will be actively monitored to ensure that
 it is operating properly.
 Histograms of the number of hits per cell integrated over 10 minutes for  High Level Trigger events
  will provide on the order of 10$^3$ hits per cell for an event rate of 25~kHz
(based on the High Level Trigger accepting a cross-section of about 25~nb for the 10$^{36}$cm$^{-2}$s$^{-1}$ design luminosity).
 This will provide the ability to monitor if any given cell's occupancy deviates by more than about 10\% from nominal every ten minutes.
 In addition, histograms of the number of hits per layer can be used to monitor if a drift chamber layer's occupancy deviates by more than a percent 
every minute. Specialized software will alert the shift crew if cell and layer occupancies fall outside tolerances.
Trigger rates for triggers using  the drift chamber will also be monitored and used to establish that the drift chamber is operating as expected.
 
The monitoring of some quantities made available by the online reconstruction, such as D$^0$ and K$^0_S$ mass peaks, 
m$_{ES}$, and $\Delta$E will also be used to validate drift chamber operational integrity.

\tdrsubsec{Calibration}

The time-to-distance relations will be calibrated initially with cosmic ray muons. Each cell will have its own calibration, although
cells with a common electrostatic configuration are expected to have the same time-to-distance parameters. Comparisons of 
parameters of those families of cells will provide a measure of quality assurance of the calibration process.
 
When beams are colliding, low multiplicity physics events, such as
e$^+$e$^-\rightarrow \mu^+\mu^-$ and Bhabha events, will be used to
calibrate the time-to-distance relations and the calibrations will be
continuously updated at a rate determined by the luminosity.  The
\dEdx calibrations will use physics events in which kinematics provide
a cleanly identified particle. For example, $\phi\rightarrow K^+K^-$,
$K^0_S \rightarrow \pi^+\pi^-$, and $D^{*+}\rightarrow \pi^+
D^0$, $D^0\rightarrow K^-\pi^+$ can be used to select kaons and pions;
and $\Lambda \rightarrow p \pi^-$ can be used to select protons.
Control samples of leptons for particle identification calibration are
provided by e$^+$e$^-\rightarrow \mu^+\mu^-\gamma$ and
e$^+$e$^-\rightarrow$ e$^+$e$^-\gamma$ events.  These same control
samples will be used to calibrate the optimal parameters in the
cluster counting algorithm used for particle identification.


\tdrsec{Integration}
\label{sec:DCH_Integration}

\tdrsubsec{Overall geometry and mechanical support}

The envelope of the \dch\ is determined by the tungsten shield and the 
DIRC at the inner and outer radii, and by the backward calorimeter 
and the FTOF in the negative and positive $z$ directions. 
There are 5~mm radial clearances between the \dch\ and the 
surrounding components, and 5~mm clearance between the
\dch\ envelope and the backward calorimeter. The FTOF is 
directly mounted onto the \dch. The envelope in the backward direction
includes the space occupied by the signal cables after they exit the 
enclosure. 

In \babar, the chamber was supported at the backward end by 
turnbuckles connected to the inner surface of the 
DIRC central support tube (CST). 
We envision using a similar system, although the actual mounting
points used by \babar\ will be obscured by the backward calorimeter. 

In the forward direction, the \dch\ and FTOF form an integrated
mechanical package supported by the CST. 
Figure~\ref{fig:DCHmount} shows the mounting components 
used by the \babar\ drift chamber. Note that the support point
on the DIRC is on the $z$ end surface of the CST, not the inner radius. 
Because the \superb\ chamber is shorter in the forward direction 
than \babar, the corresponding support tabs will be on the FTOF,
not the chamber. 

\begin{figure}[tbh] 
\centering
\includegraphics[width=0.45\textwidth]{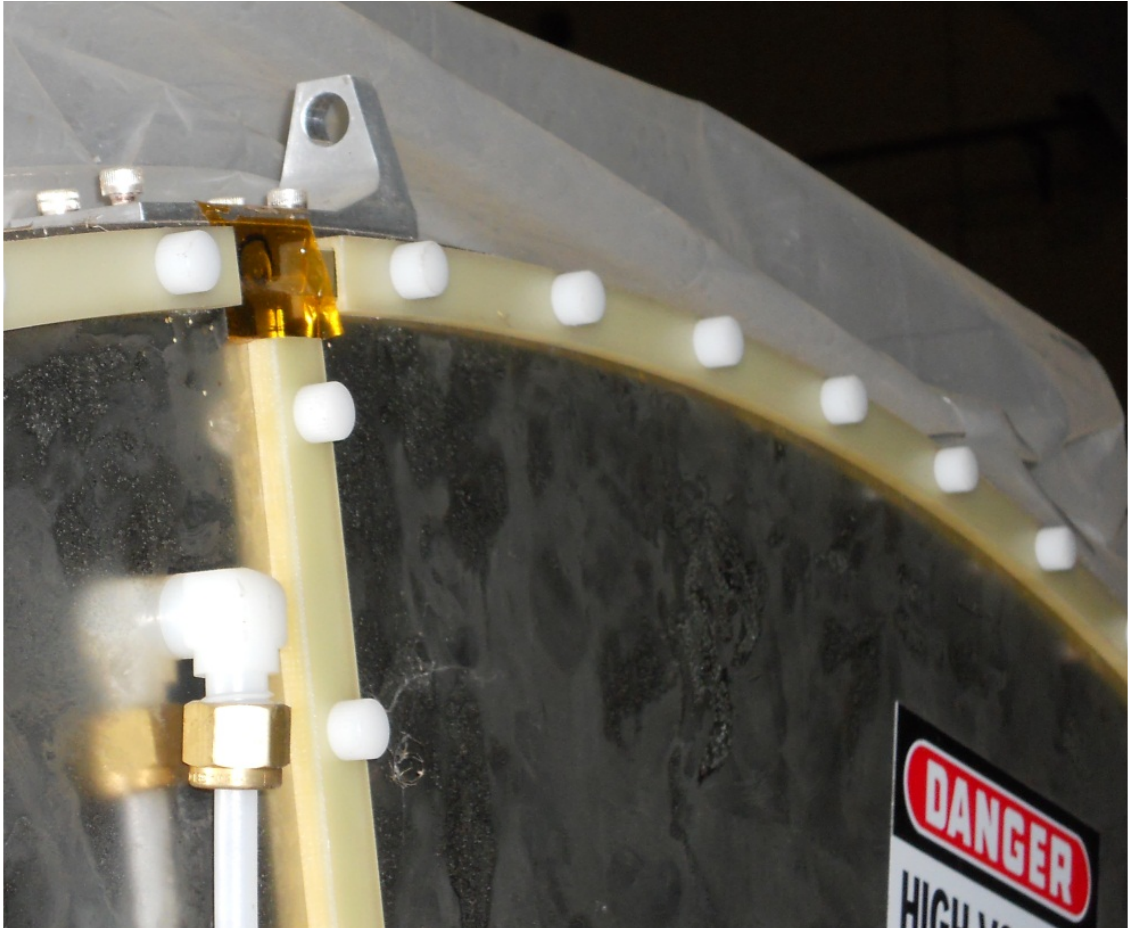}
\includegraphics[width=0.45\textwidth]{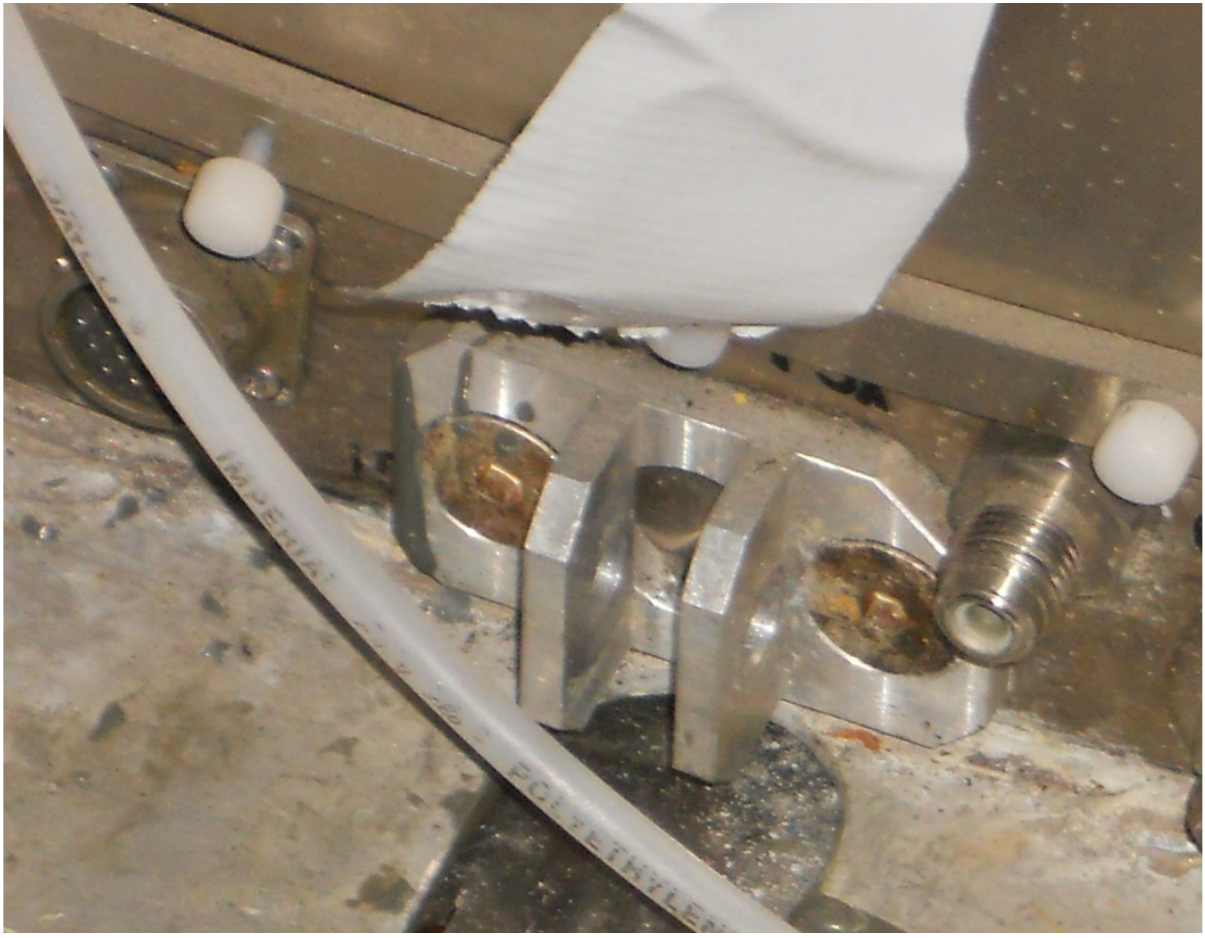}
\caption {
Mechanical mounting points for the \babar \dch on the forward (Top)
and backward ends (Bottom). Both ends are attached to the DIRC central
support tube.
}
\label{fig:DCHmount}
\end{figure}

\tdrsubsec{Installation and alignment}

The chamber and FTOF will be installed prior to the forward and
backward calorimeters and the tungsten shielding. The installation
will reuse the existing \babar\ equipment, which is currently stored
at SLAC.  The chamber is supported at the inner radius and slid along
a supporting beam that passes through the inner cylinder
(Figure~\ref{fig:dchinstall}).

\begin{figure}[tbh] 
\centering
\includegraphics[width=0.45\textwidth]{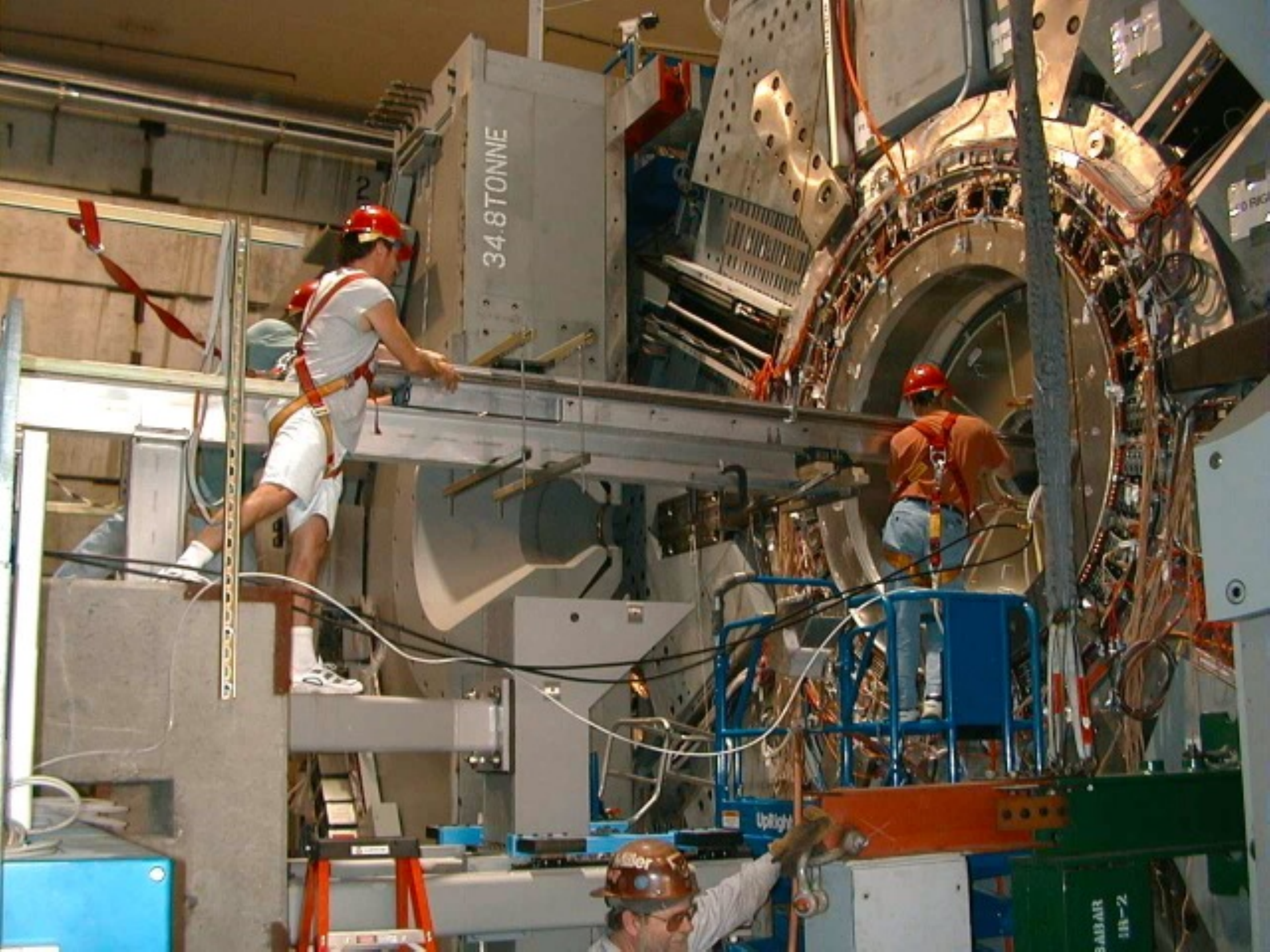}
\caption {
\babar\ drift chamber during installation. The same tooling is available
for use by \superb.
}
\label{fig:dchinstall}
\end{figure}

Both forward and backward enclosures will contain a number of 
precision 6~mm dowel holes into which target holders for
corner-cube reflectors can be mounted. The enclosures in turn are 
doweled to precise reference holes on the endplates, referencing
the target locations to the sense wire locations. 

The mounting systems at both ends allow for 
several mm of adjustment in $x$ and $y$. The chamber location will
be adjusted to center the chamber in $x$ and $y$ on the interaction 
point and to align the sense wire direction with the 
magnetic field.  The tolerances on these alignments have not yet been 
specified. 
The mounting system fixes the chamber location in $z$ at the forward end, 
but permits a small amount of motion at the backwards end in case of 
relative thermal expansion.  
The tolerance on the location in $z$ will be significantly 
looser than those on $x$ and $y$.

\tdrsubsec{Services}

The services required for the backward end are listed below. These will 
reach the backward enclosure via 16 slots in the outer radius of the 
steel plug located within the DIRC strong tube. Each slot will be 
approximately 50~mm in radius by 250~mm wide. Cables continuing to the
digitizing crates, located on the top of the detector, will be routed
to wireways at the end of the IFR iron immediately after exiting
the DIRC strong tube. 

Note that within the radial extent of the backward calorimeter, 
the signal cables (and all other services) are routed to stay within 
the \dch\ envelope. 

\begin{itemize}

\item Signal cables: approximately 8000 coaxial cables, RG-179, 2.54~mm in
diameter, organized into 8-cable ribbons. These are 10~m in length, 
and travel to the digitizing electronics crates.

\item Calibration cables: RG-179, one for every eight signal channels.
Also originate at the digitizer crates.

\item Low-voltage power: Four 1/0 welding cables, 14.7~mm diameter. 
Originate in the electronics hut. 

\item Cooling lines: 16 lines (8 separate circuits, each with a supply
  and a return line), 26.2~mm reinforced PVC. The sub-atmospheric
  water-based cooling system (Sec.~\ref{sub:DCHendplate}) will be
  close to the detector, but outside of the radiation area.

\item Drift gas: one line, 19~mm diameter stainless steel. Originates in 
the gas hut. This line increases to 38~mm diameter after exiting the 
detector.

\item Nitrogen flush gas: two lines, 19~mm diameter stainless steel. Also
from the gas hut. 

\end{itemize}

The services for the forward end will exit the detector in the 
radial space between the tungsten shield and the FCAL. The services
for the forward region are:

\begin{itemize}

\item High voltage: 16 56-conductor cables, 14~mm diameter. Originate
at the HV supplies in the electronics hut. 

\item Drift gas: one line, 19~mm diameter stainless steel. Originates in 
the gas hut.  This line increases to 38~mm diameter after exiting the 
detector.

\item Nitrogen flush gas: two lines, 19~mm diameter stainless steel. 
From the gas hut. 

\item Cooling lines: 2 lines, 19~mm reinforced PVC. From the cooling
system. 

\end{itemize}

\bibliographystyle{\tdrbibstyle}
\balance
\bibliography{DCH/DCH-bib}


\renewcommand{\ChapLabel}{chap:PID} 
\tdrchap{Particle Identification}




\tdrsec{ Summary of Detector Performance goals }

\tdrsubsec{ DIRC Detector concept }
  
    The DIRC (Detector of Internally Reflected Cherenkov light)~\cite{Ratcliff:1992hd,blair_2008} is an innovative detector technology
that has been crucial to the performance of the \babar\ science program. 
The main DIRC performance parameters are the following: (a) measured time resolution of ${\sim} 1.7$~ns per photon, 
driven by
the photomultiplier (PMT) transit time spread (TTS) of 1.5~ns; (b) single photon Cherenkov angle resolution of 9.6\mrad 
for tracks from di-muon events; (c) Cherenkov angle resolution of 2.5\mrad per track in di-muon events; 
and (d) $\pi/K$ separation greater than $2.5 \sigma$ over the entire track momentum range, from the pion Cherenkov 
threshold (${\sim} 130$\mevc for fused silica) up to 4.2\gevc. The \babar\ DIRC performed reliably and efficiently over the whole
\babar\ data taking period (1999-2008). Its physics performance remained
consistent throughout the whole period, although some upgrades, such as the addition of
shielding and replacement of electronics, were necessary to cope with evolving machine conditions. 

This excellent DIRC performance did not come without substantial effort, and this experience is useful when 
designing the next DIRC-like detector.
To maintain the background at an acceptable level in \babar, it was necessary to: 
(a) apply a tight timing cut of $\pm 8$~ns around the expected arrival time of each Cherenkov photon;
(b) add substantial shielding around the beam pipe under the DIRC, which 
reduced the PMT background by at least a factor of ${\sim} 6$; (c) install many background detectors (here we found, for example, that 
the rate in neutron detectors correlates with the DIRC background very well, indicating a common origin); and (d) improve operation 
of the machine.

After ${\sim} 10$ years of operation, the PMT gain was reduced by ${\sim} 45\%$ in total, but the tubes operated well 
until the end, requiring only a few voltage adjustments to correct for the gain loss~\cite{adam_2012}. The electronics 
dead time was limited to ${\sim} 1\%$ at a rate of a few \MHz per PMT~\cite{Adam:2004fq}. 

Excellent flavor tagging will continue to be
essential for the program of physics anticipated at \superb, and the
gold standard of particle identification in this energy region is that provided by internally reflecting ring-imaging
devices (the DIRC class of ring imaging detectors). The challenge for
\superb\ is to retain (or even improve) the outstanding performance
attained by the \babar\ DIRC~\cite{Adam:2004fq}, while also gaining an
essential factor of at least 100 in background rejection to deal with the much higher luminosity. 

A new Cherenkov ring imaging detector is being planned for the
\superb\ barrel; it is called the Focusing DIRC, or FDIRC. It will reuse the
existing \babar\ bar boxes and mechanical support structure. Each bar box
will be attached to a new photon 'camera', which will be
optically coupled to the bar box exit window. The new camera design
combines a small modular focusing structure that images the photons
onto a focal plane instrumented with very fast, highly pixelated
PMTs~\cite{Va'vra:2010zz}.

To cope with the higher luminosity, the sensitivity of the FDIRC photon camera to background will be reduced by an overall factor of 250 
compared to the \babar\ DIRC. Indeed,
\begin{itemize}
\item it is $25\times$ smaller in volume than the DIRC water-based camera;
\item its new, highly pixelated photon detectors will be about $10\times$ faster;
\item the photon camera is built using radiation-resistant fused silica material~-- instead of water or oil which is more 
sensitive to neutron background.
\end{itemize}

Other important characteristics of the FDIRC are the following: 
(a) the entire system will have ${\sim} 18,000$~pixels (each photon detector includes 32 pixels with
48 detectors per photon camera and 12 sectors total); (b) it will reconstruct photons in 3D (two space coordinates and time); 
(c) although the Cherenkov angle can be reconstructed using the pixel information only, time will be included in the final PID likelihood hypothesis, which incorporates both particle time of flight and photon propagation time in the FDRC system;
(d) time also plays an important role in FDIRC to reduce the background, and 
(e) the expected timing resolution of $\sigma {\sim} 200$~ps makes it possible to reduce the chromatic 
broadening of the Cherenkov angle resolution by 0.5-1\mrad, depending on the photon propagation path angle and length.

Although \babar had no PID in the forward direction, several options have been considered for a possible PID detector for \superb\ to cover the acceptance hole in
the forward direction: (a) ``DIRC-like TOF'' \tof (FTOF)~\cite{jjv_toflike_1,jjv_toflike_2}, (b) 
pixelated TOF~\cite{Va'Vra:2009zza} and (c) Aerogel RICH~\cite{farich}. As discussed later in this chapter, the \superb\ collaboration has determined there is physics merit in reserving a longitudinal gap of $\sim 5-10 $ cm for a forward PID device and tentatively identified the Dirc-like FTOF, which is based upon the TOF technique, as the leading candidate technology. 
Tests of a full-scale prototype of one sector of the FTOF detector are 
foreseen for the near future; if they are successful, the FTOF will be included in the \superb\ baseline.
See section~\ref{section:forwardPID} for more information on this topic.

\tdrsubsec{ Charged Particle Identification }

Charged particle identification at \superb\ relies on the same 
framework as the \babar\ experiment, utilizing several detectors in addition to the specific PID systems being described in this chapter. Electrons and muons are
identified by the EMC and the IFR respectively, aided by \dedx\ 
measurements in the inner trackers (SVT and DCH).  Separation 
for low-momentum hadrons is primarily provided by \dedx. 
At higher momenta (above 0.7\gevc for pions
and kaons, above 1.3\gevc for protons) the dedicated systems being described here, the
FDIRC~-- inspired by the successful \babar\ DIRC~-- will perform most of the
$\pi/K$ separation in the barrel region, aided by \dedx\  information from the tracking devices. The forward system, if included, would utilize similar combinations of detectors. The FDIRC and possible FTOF are described in the remainder
of this chapter. More details about the properties of other \superb\ detectors
can be found in the corresponding TDR chapters.

\tdrsec{ Particle Identification Overview }

\tdrsubsec{ Experience with the \babar\ DIRC }

The \babar\ DIRC~-- see Figure~\ref{fig:DIRC}~-- is a novel ring-imaging Cherenkov detector. The
Cherenkov light angular information, produced in ultra-pure synthetic fused silica bars 
(average refraction index 1.473), is preserved while propagating along the bar via
internal reflections to the camera (called the `StandOff Box' in the figure, SOB for short) where an image is produced and detected.

\begin{figure}[tbp]
\begin{center}
\includegraphics[width=\linewidth]{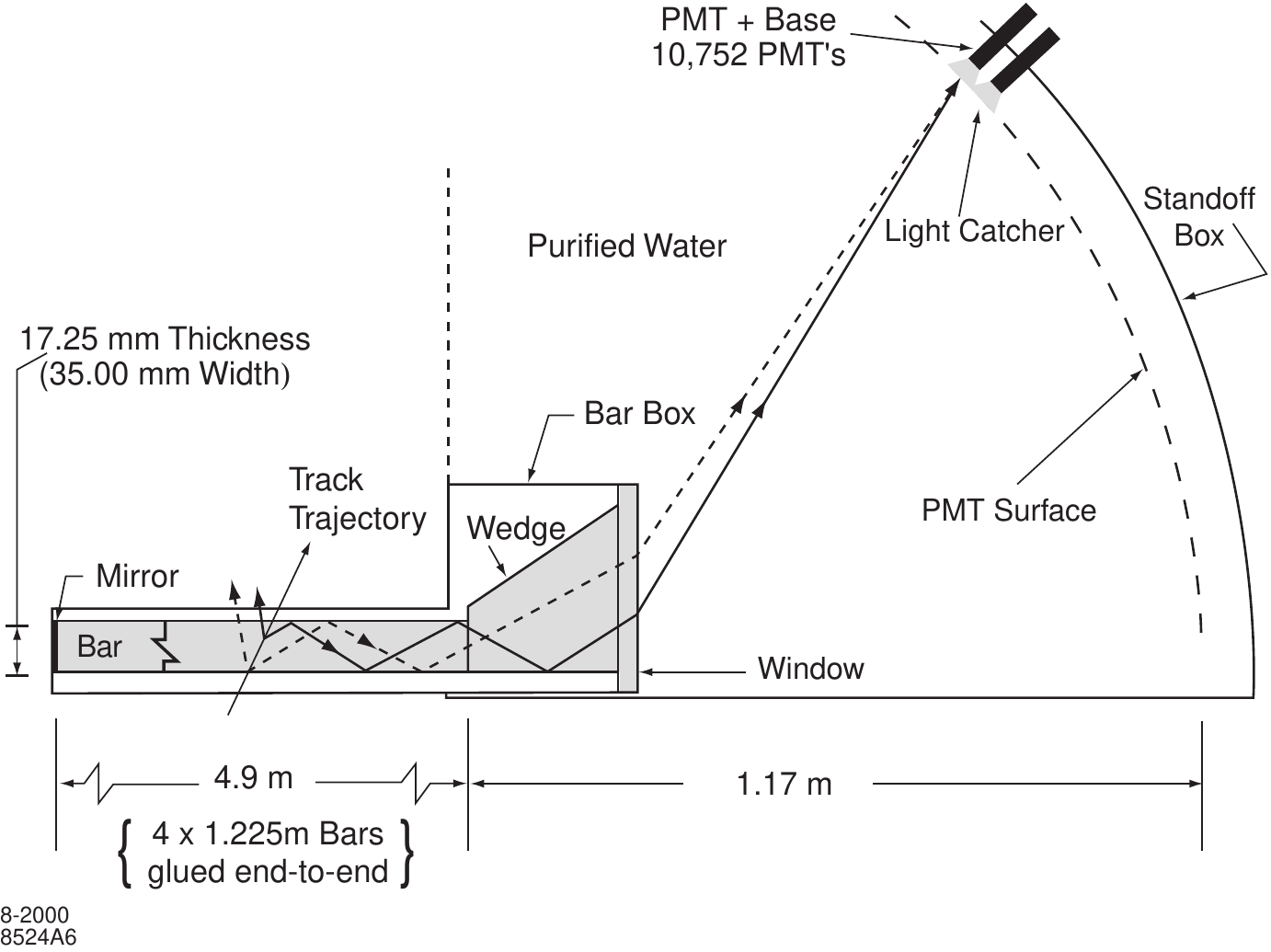}
\caption{Schematic of the \babar\ DIRC.}
\label{fig:DIRC}
\end{center}
\end{figure}

The entire DIRC contains 144 quartz bars, each 4.9\m long. They are aligned 
along the beam line and cover the whole azimuthal range. 
Thanks to an internal reflection coefficient of ${\sim} 0.9997$, a small absorption coefficient in the bulk material, and orthogonal bar faces,
Cherenkov photons
are transported to the back end of the bars with the magnitude of
their angles conserved with only a modest loss of photons. They exit into a  pinhole camera consisting of a large volume of purified water
(a medium chosen because it is inexpensive, transparent, and easy to clean, with average index of refraction and relative 
chromatic dispersion reasonably close to those of the fused silica). The photon detector PMTs are located at the rear of the SOB, about
1.2\m away from the quartz bar exit window. 

The reconstruction of the Cherenkov angle uses information from the
tracking system together with the positions of the PMT hits in the
DIRC.  Information on the time of arrival of hits is
used to reject background; to resolve ambiguities
due to the unknown path of a given photon in the quartz; and also provides some limited direct PID
information.

\tdrsubsec{ Barrel PID: Focusing DIRC (FDIRC) }

   As discussed above, the PID system in \superb\ must cope
with much higher luminosity-related background rates than in \babar~-- current estimates are 
on the order of 100 times higher. The basic strategy is to make the camera much smaller and faster.  
A new photon camera imaging concept, based on focusing optics, is therefore envisioned.  
The focusing blocks (FBLOCK), responsible for imaging the Cherenkov photons onto the PMT cathode surfaces, will be
machined from radiation-hard pieces of fused silica. The major design constraints for the new camera are
the following: (a) it must be consistent with the existing \babar\ bar box design, as these invaluable components will
be reused in \superb; (b) it must coexist with the \babar\ mechanical support and the \superb\ magnetic field constraints; 
(c) it requires very fine photon detector pixelation and fast photon detectors.

Imaging is provided by a mirror structure focusing onto an image
plane containing highly pixelated PMTs. The reduced
volume of the new camera and the use of fused silica for coupling
to the bar boxes (in place of water as 
in the \babar\ SOB),  is expected to 
reduce the sensitivity to background by about a factor of ${\sim} 25$
compared to \babar\ DIRC. The very fast timing of the new PMTs and of the
associated readout chain is expected
to provide many additional advantages: (a) improvement of the
Cherenkov resolution using the combination of photon propagation time and particle time of flight; (b) a measure of the chromatic dispersion term
in the radiator~\cite{Benitez:2006ei,Va'vra:2007zz,Benitez:2008zz} which leads to improved performance; (c) 
better separation of ambiguous solutions in the folded optical system; and
(d), an additional factor of 10 in background rejection.

\begin{figure*}[tbp]
\begin{center}
\subfloat[FDIRC optical design (dimensions in \cm).]{%
    \includegraphics[width=0.65\textwidth]{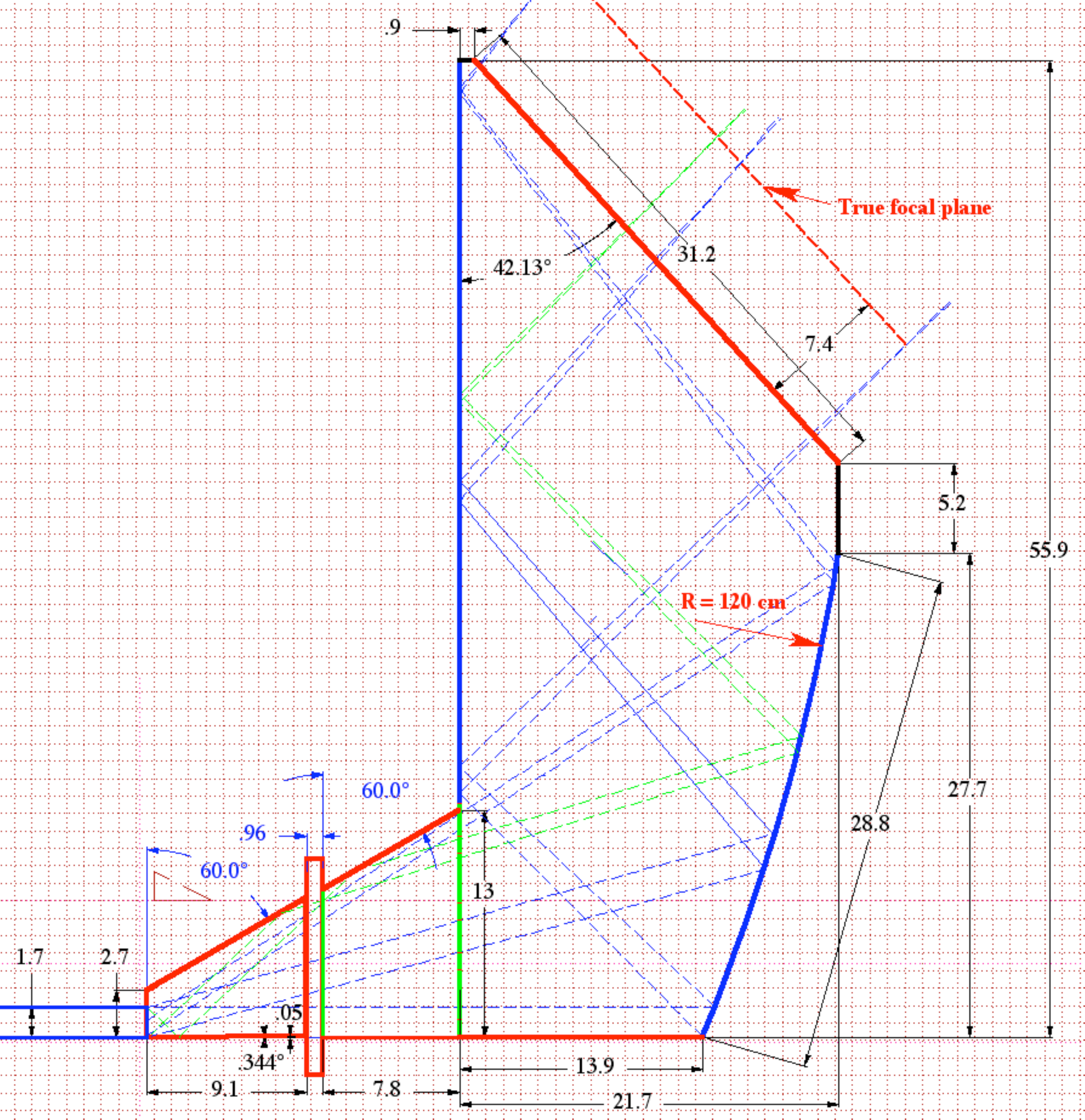}}
\hspace{5mm}
\subfloat[Its equivalent in the \geantFour\ MC model.]{%
  \includegraphics[trim=0 70 0 0,width=0.3\textwidth,clip]{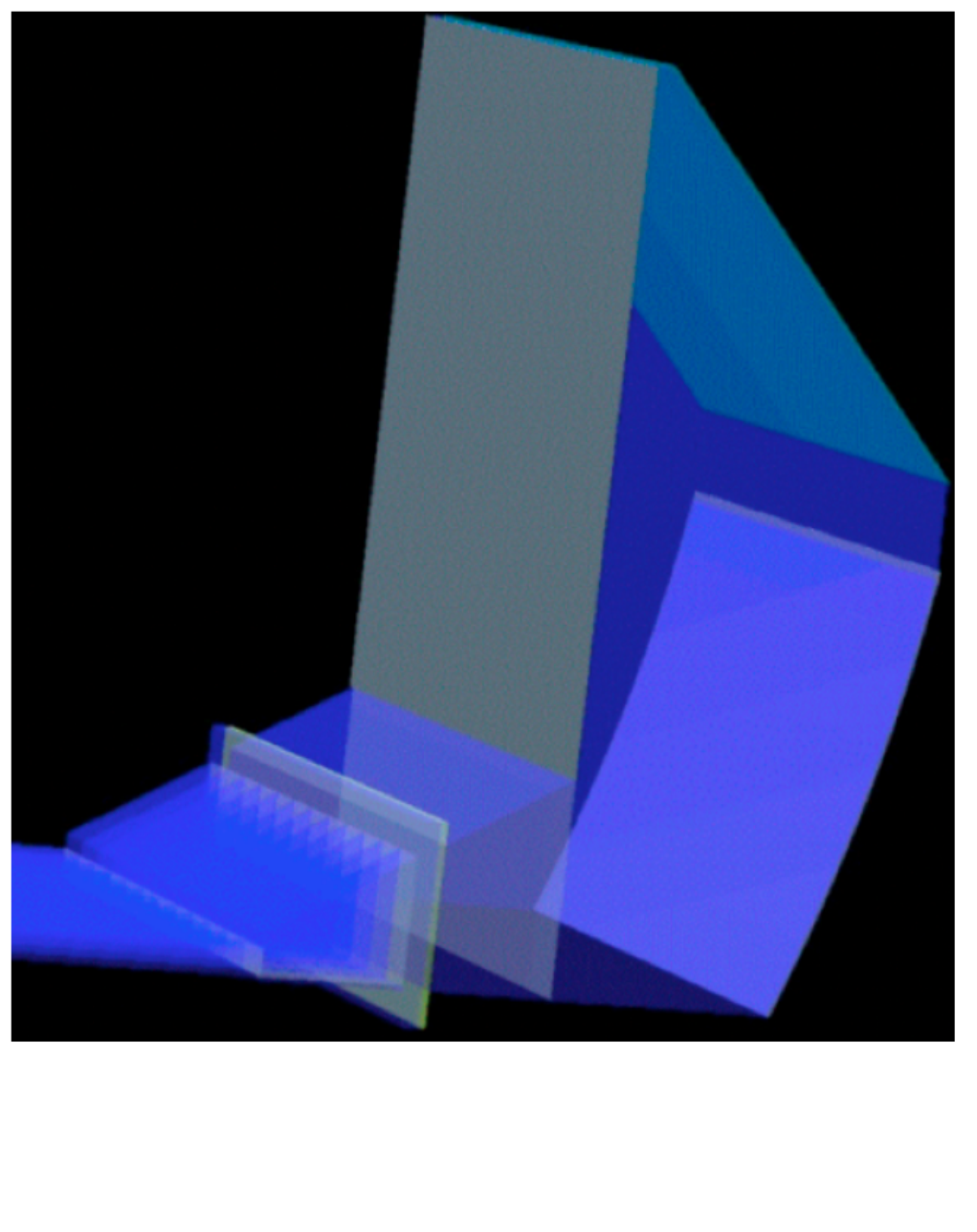}}
\end{center}
\caption{Barrel FDIRC Design.}
\label{fig:FDIRC_design}
\end{figure*}

Figure~\ref{fig:FDIRC_design} shows the new FDIRC photon camera design (see
Ref.~\cite{Va'vra:2010zz,Va'vra:2008zz,Va'vra:2009zzb,mathematica} for more details). It consists of two parts:
(a) a focusing block (FBLOCK) with cylindrical and  flat mirror surfaces, and (b) a new wedge. 
The wedge at the end of the bar rotates rays with large transverse angles in the focusing plane 
before they emerge into the focusing structure. The old \babar\ wedge is too short so that an additional 
wedge element must be added to insure that all rays strike the cylindrical
mirror. The cylindrical mirror reflects all rays onto the FBLOCK flat mirror, preventing reflections
back into the bar box itself; the flat mirror then reflects rays onto
the detector focal plane with an incidence angle of almost 90\degrees,
thus avoiding reflections. The photon detector plane is located in a slightly
under-focused position to reduce the FBLOCK size and therefore its
weight.  Precise focusing is unnecessary,  as the finite pixel size
would not take advantage of it. The total weight of a bare solid fused
silica FBLOCK is about 80 kg. This substantial weight requires good mechanical support.

\begin{figure}[tbp]
\begin{center}
\includegraphics[width=\linewidth]{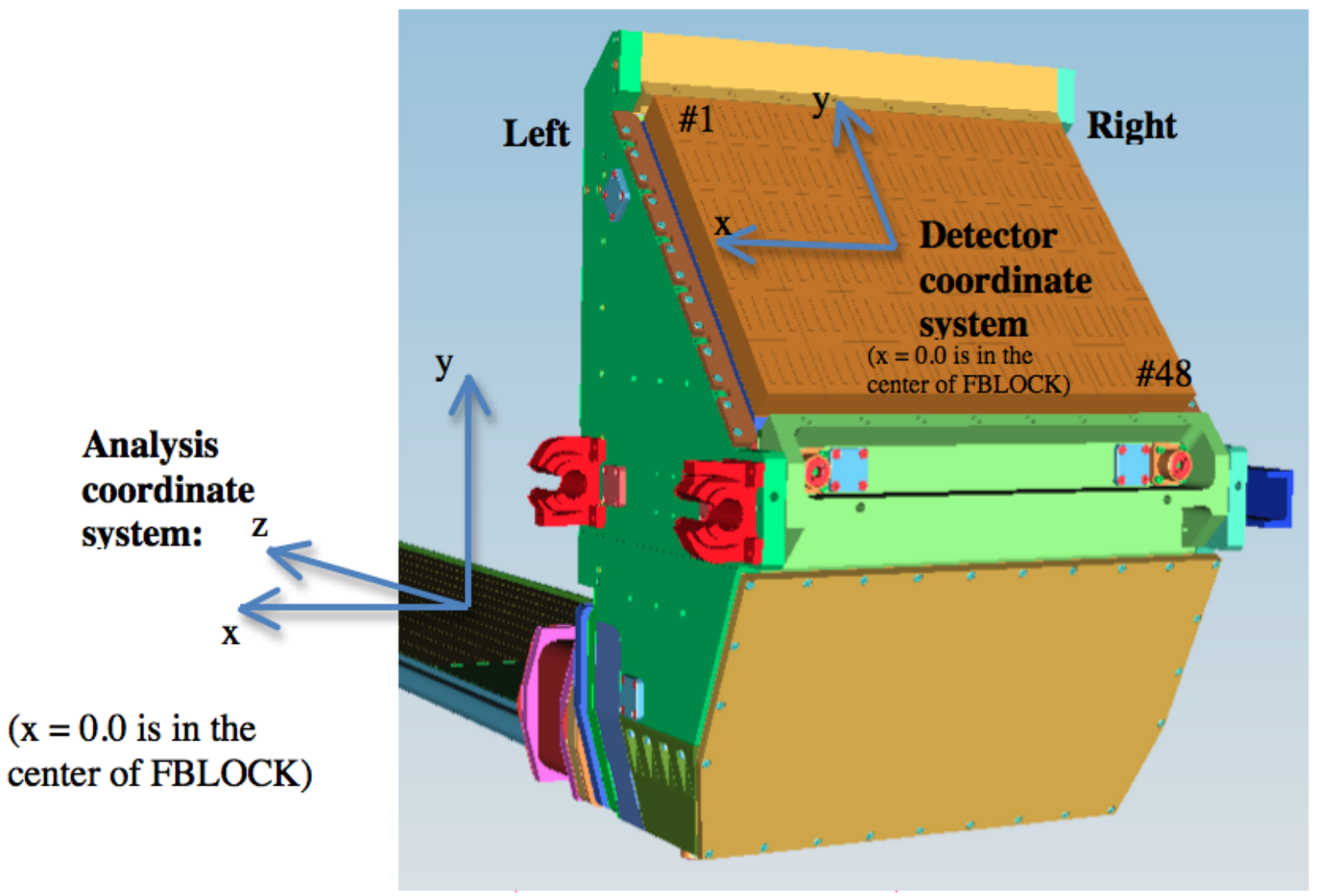}
\caption{FDIRC coordinate systems used in this chapter.}
\label{fig:FDIRC_coord_syst}
\end{center}
\end{figure}

There are several important advantages gained in moving from the \babar\
pinhole focused design with water coupling to a focused optical design made of solid fused
silica: (a) the design is modular; (b) as explained above, the sensitivity to background,
especially to neutrons, is significantly reduced; (c) the pinhole-size
component of the angular resolution in the focusing plane is
removed by focusing with a cylindrical mirror; (d) the total number of photomultipliers is reduced by about one half compared to a
non-focusing design with equivalent performance; (e) there is no risk
of water leaks into the \superb\ detector, and no time-consuming
maintenance of a water system, as was required to operate \babar\ safely.

Figure~\ref{fig:FDIRC_coord_syst} shows the FDIRC coordinate system as used in this document.

\begin{figure}[tbp]
\begin{center}
\includegraphics[width=\linewidth]{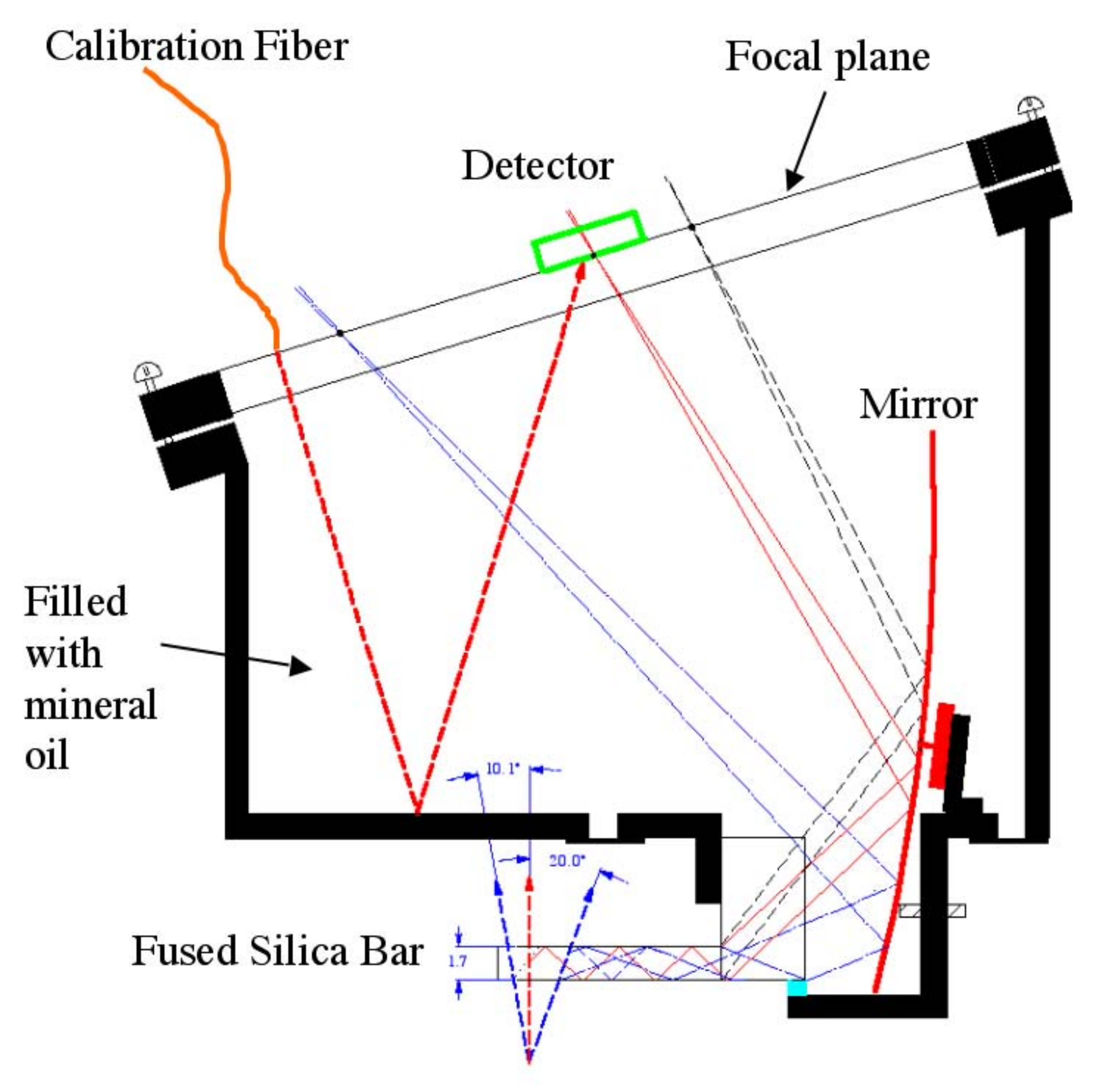}
\caption{The first FDIRC single-bar prototype employing a spherical mirror, an oil-filled photon camera, and highly-pixelated 
photon detectors~\cite{Benitez:2006ei,Va'vra:2007zz,Benitez:2008zz}.}
\label{fig:1st_prototype}
\end{center}
\end{figure}

Each new camera will be attached to its \babar\ bar box with optical RTV glue, which will be 
injected in a liquid form between the bar box window and the new camera, and cured in place. 
As Figure~\ref{fig:FDIRC_design} shows, the cylindrical mirror focuses in the radial $y$-direction, 
while pinhole focusing is used in the direction out of the plane of the schematic (the $x$-direction). 
Photons that enter the FBLOCK at large $x$-angles reflect from the parallel sides, leading to additional 
ambiguities. However, the folded design makes the optical piece small, and places the photon detectors in an 
accessible location, improving both the mechanics and the background sensitivity. Since the optical mapping 
is 1 to 1 in the $y$-direction, this ``folding" reflection does not create an additional ambiguity. Since a 
given photon bounces inside the FBLOCK only 2-4 times, the requirements on surface quality and polishing 
for these optical pieces are much less stringent than those required for the DIRC bar box radiator bars.

\begin{figure}[tbp]
\begin{center}
\includegraphics[width=\linewidth]{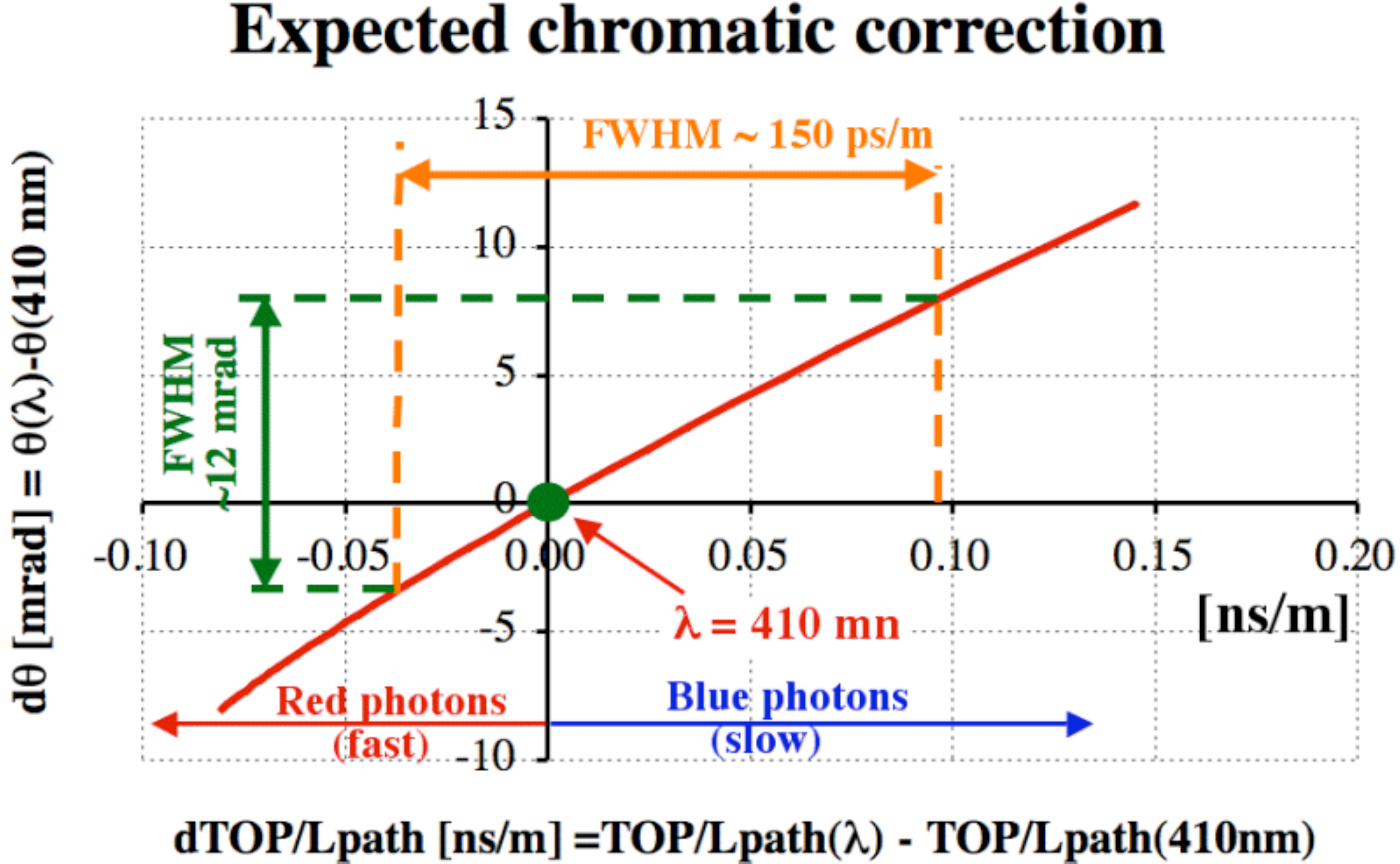}
\caption{Analytical correlation between the variation of the Cherenkov angle and the change in 
time-of-propagation (TOP), relative to 
the mean wavelength of 410~nm~\cite{Benitez:2006ei,Benitez:2008zz}.}
\label{fig:Chrom_corr_a}
\end{center}
\end{figure}

\begin{figure}[tbp]
\begin{center}
\includegraphics[width=\linewidth]{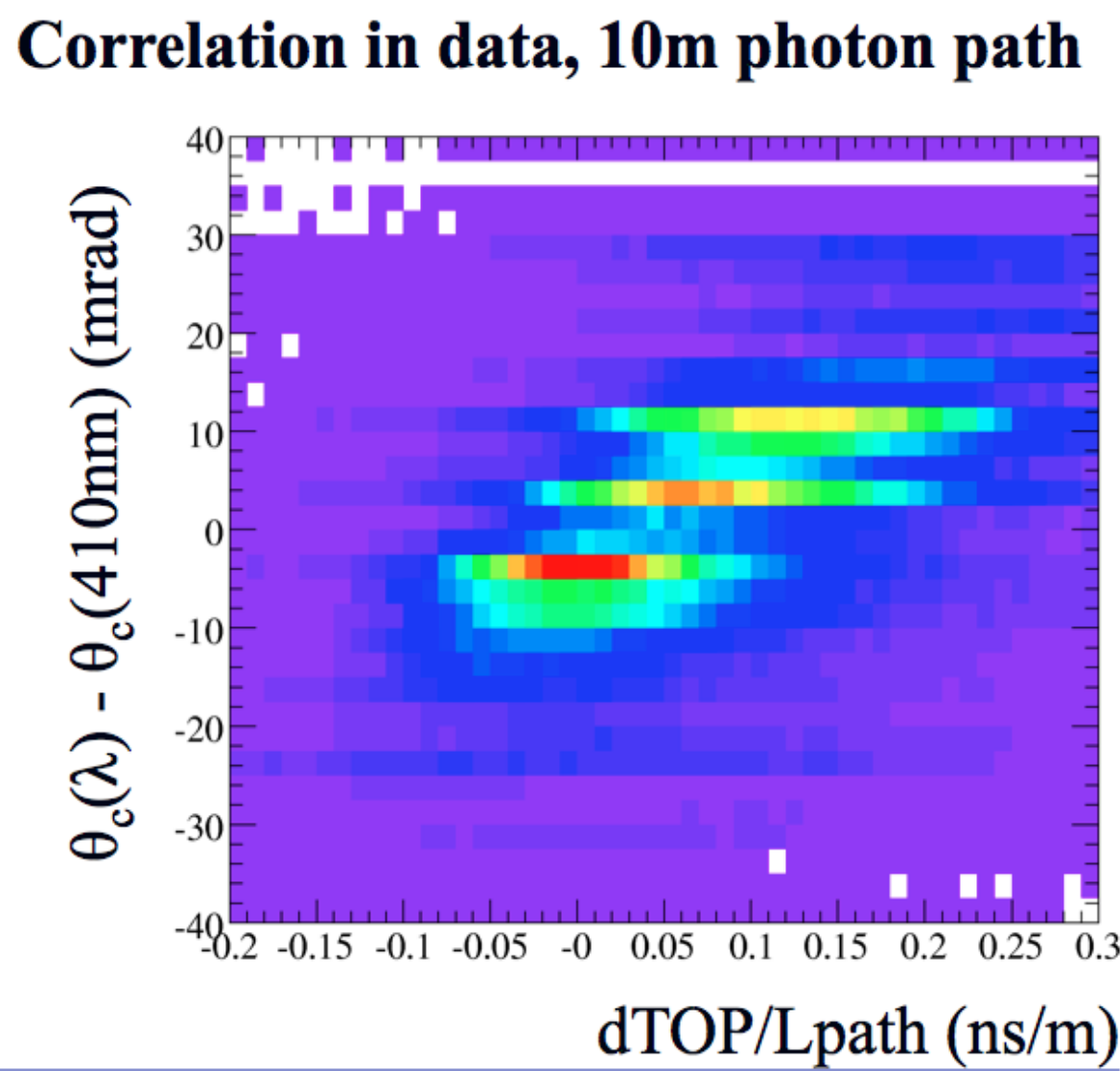}
\caption{The same correlation between the change of the Cherenkov angle and the change in TOP for photons
propagating 10 meters in the bar, as seen in the beam test data in the first FDIRC prototype~\cite{Benitez:2006ei,Benitez:2008zz}.}
\label{fig:Chrom_corr_b}
\end{center}
\end{figure}

As an intermediate step towards an upgrade of DIRC for the \superb\ detector, we have built
and tested a first FDIRC prototype (see Figure~\ref{fig:1st_prototype}), 
in order to learn how to design a more compact `DIRC' with new highly-pixelated fast PMTs~\cite{Benitez:2006ei,Benitez:2008zz}. 
This prototype demonstrated for the first time that chromatic correction could be done with timing. The principle 
is displayed in Figure~\ref{fig:Chrom_corr_a}, showing the analytical correlation between the change of the Cherenkov angle 
and the change in time-of-propagation (TOP), relative to the mean wavelength of 410~nm. Figure~\ref{fig:Chrom_corr_b} 
shows the same plot using data from photons propagating 10 meters in the bar. Figure~\ref{fig:Chrom_corr_c} shows 
the final result of the chromatic correction in the beam test data. One can see that the chromatic
correction improves the Cherenkov angle resolution by 0.5-2\mrad depending on the photon propagation path length:
the longer the path, the more efficient the correction. To achieve such result, a single 
photoelectron timing resolution of ${\sim} 200$~ps (provided by H-8500 or H-9500 multi-anode PMTs, MaPMTs) is required.

\begin{figure}[tbp]
\begin{center}
\includegraphics[width=\linewidth]{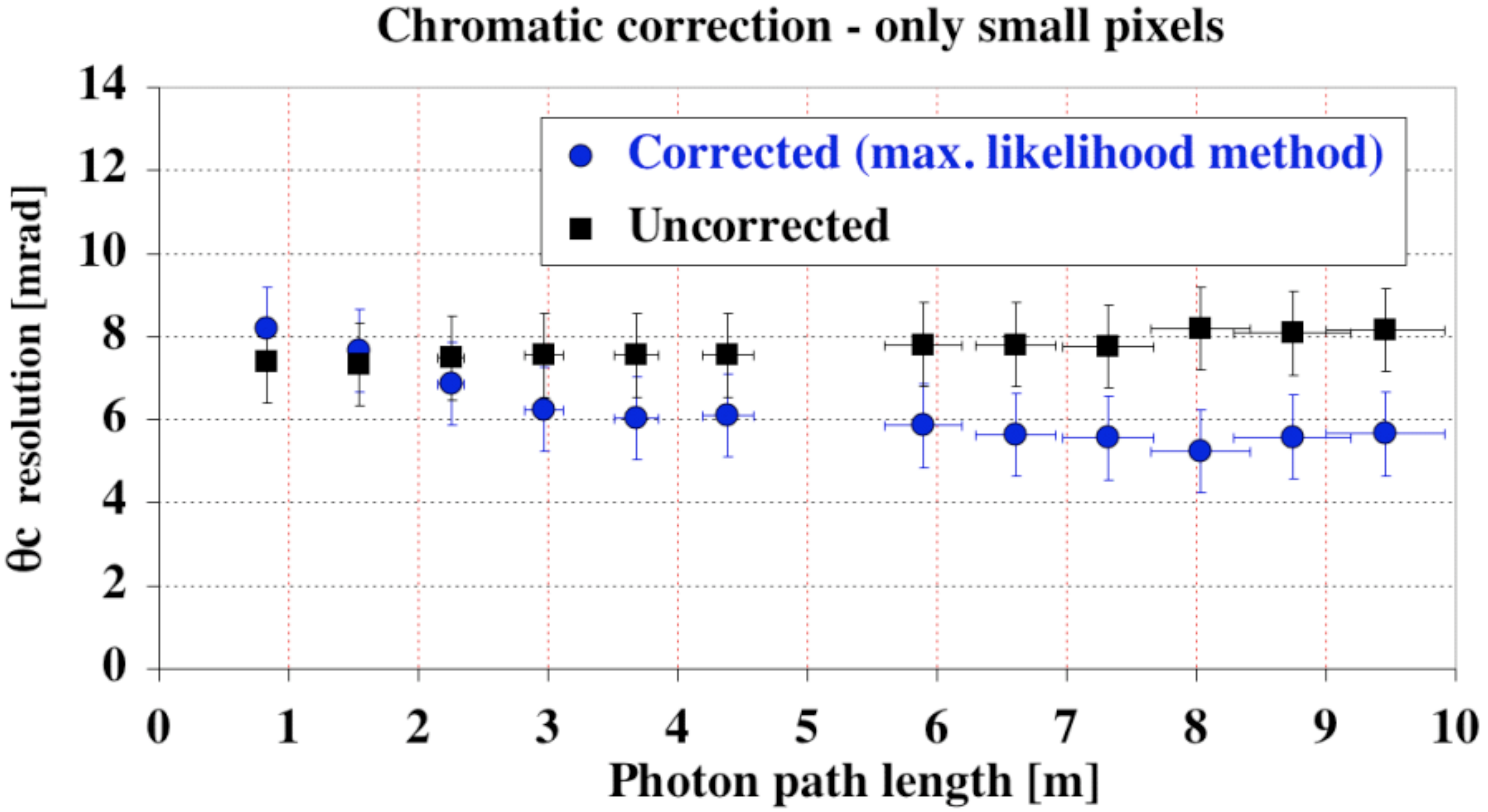}
\caption{Beam test data showing the effect of the chromatic correction for $3 \mm \times 12 \mm$ pixels obtained with H-9500 MaPMTs
in the first FDIRC prototype. Note that the \superb\ active region starts over 2 meters away from the camera detector end. Results with the 
H-8500 MaPMTs are similar, but with slightly worse resolution~\cite{Benitez:2006ei,Benitez:2008zz}.}
\label{fig:Chrom_corr_c}
\end{center}
\end{figure}

We also demonstrated from the first prototype that the Cherenkov ring has worse resolution in the wings than in the 
center, due to the optical aberration caused by the bar, which is amplified by the mirror. Figures~\ref{fig:abberation_a} 
and ~\ref{fig:abberation_b} show that this simulated error contribution goes from 0\mrad (at ring center) to ${\sim} 9\mrad$ 
(near the ring wings) and that it is $z$-dependent. This aberration is present in the non-focusing \babar\ DIRC as well, but it is smaller, 
\ie, the mirror amplifies this effect. The effect is similar for spherical, parabolic and spherical mirrors. 
For more details see Ref.~\cite{Va'vra:2008zz,Va'vra:2009zzb,mathematica}.

\begin{figure}[tbp]
\begin{center}
\includegraphics[width=\linewidth]{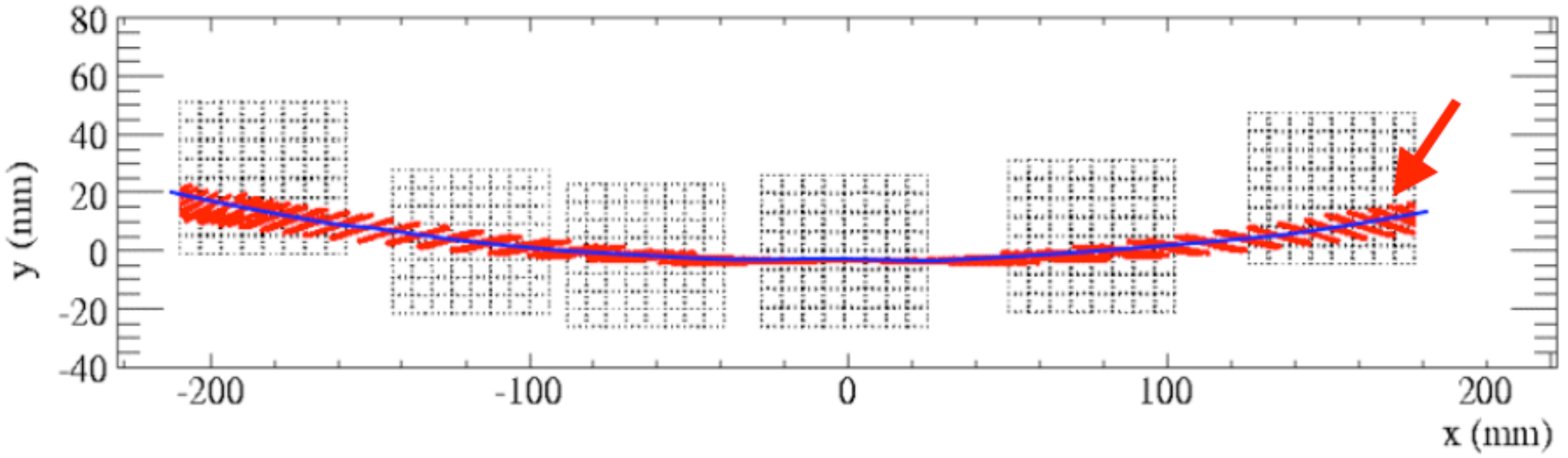}
\caption{Simulated optical ring aberration near the ring wings for the first FDIRC prototype with overlaid detectors and their pixels showing that it is a substantial effect~\cite{Benitez:2006ei}.}
\label{fig:abberation_a}
\end{center}
\end{figure}

\begin{figure*}[tbp]
\begin{center}
\includegraphics[width=0.7\textwidth]{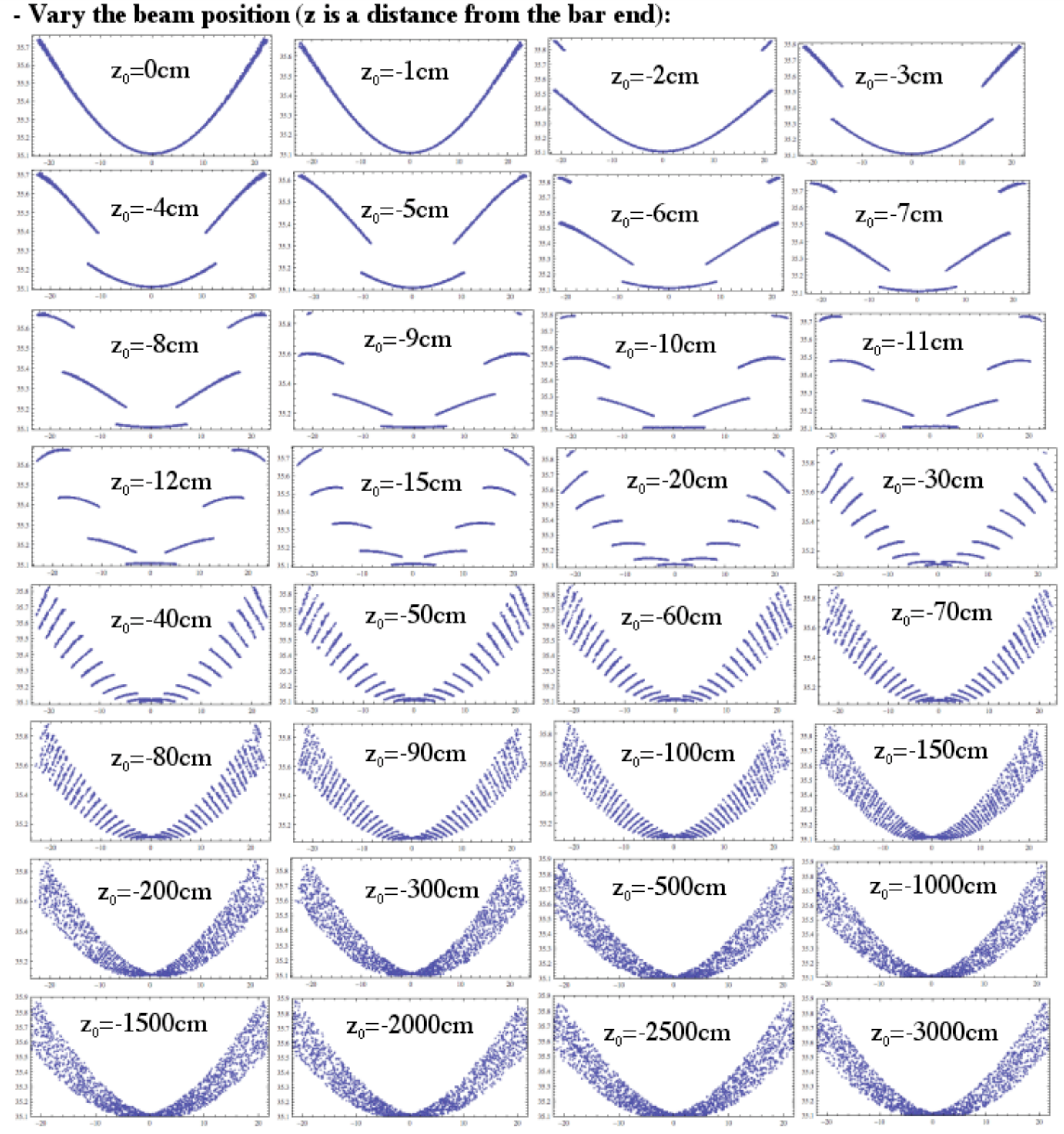}
\caption{Simulated optical ring aberration near the ring wings for the first FDIRC prototype as a function of the beam position
along the bar ($z$ coordinate)~\cite{Va'vra:2008zz,Va'vra:2009zzb,mathematica}.}
\label{fig:abberation_b}
\end{center}
\end{figure*}

Each DIRC wedge inside an existing bar box has a 6\mrad angle at the bottom. This was done intentionally 
in \babar\ to provide simple step-wise ``focusing" of rays leaving the bar towards negative $y$ to reduce the 
effect of  bar thickness. However, in the new optical system, having this angle on the inner wedge somewhat 
worsens the  design FDIRC optics resolution. Figure~\ref{fig:micro-wedge} shows the result of a simulation with 
Mathematica ~\cite{Va'vra:2008zz,Va'vra:2009zzb,mathematica}; it shows three images, one from the bottom of bar, one from the bottom of the old
wedge, which has inclined surface, and one from top of the old wedge. There are two 
choices: (a) either leave it as it is, or (b) glue a micro-wedge at the bottom of the old wedge, inside 
the bar box, to correct for this angle. Though (b) is possible in principle, it is far from trivial, as the 
bar box must be opened. Because of this difficulty we have decided to exercise option (a).

\begin{figure}[tbp]
\begin{center}
\includegraphics[width=\linewidth]{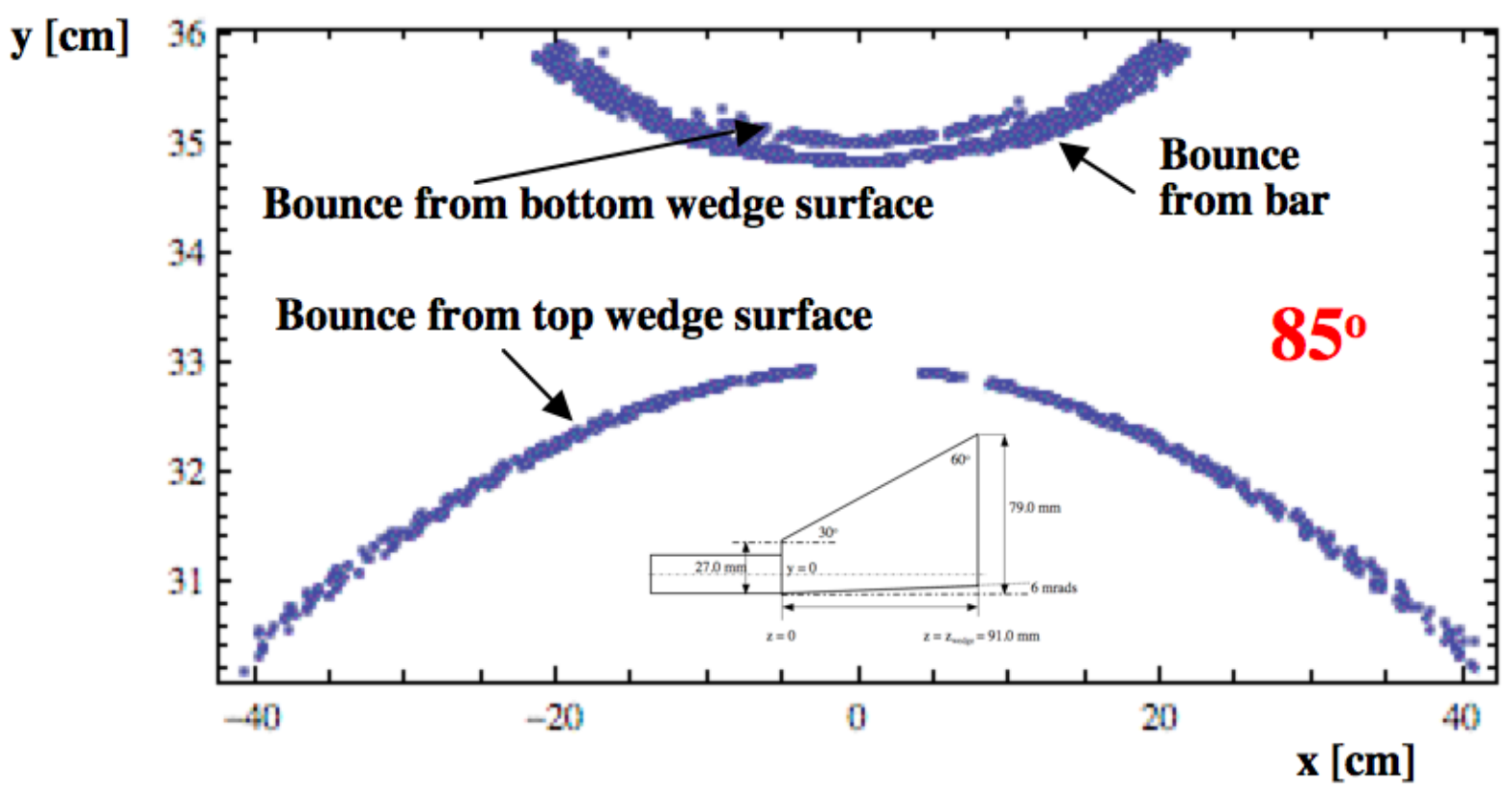}
\caption{Split of Cherenkov ring caused by the inclined bottom surface of the old (\babar) wedge~\cite{Va'vra:2008zz,Va'vra:2009zzb,mathematica}.}
\label{fig:micro-wedge}
\end{center}
\end{figure}

Figure~\ref{fig:ring_image} shows a Cherenkov ring image for one of the central bars in the new FDIRC. The ring image is more 
complicated than those from the \babar\ DIRC or from the first FDIRC prototype. This is due to reflections from the sides of the 
FBLOCK, and the pattern is different for each bar in a given bar box. The ring radius is not used in the analysis; instead, we use a dictionary 
of MC assignments for each pixel. These assignments are generated by placing a random photon source in
one bar at a time and checking which pixels were hit by these photons. In this way, we create pixel assignments of
$\hat{k}_{\mathrm {pixel}} = (k_x, k_y, k_z)$. One can also generate time-of-propagation for
direct and indirect photons TOP$_{\mathrm {direct}}$ and TOP$_{\mathrm {indirect}}$ for
tracks with $\theta_{dip}$ = $90^\circ$ and crossing each bar at $z = z_{\mathrm {middle}}$. For
any track direction $\hat{k}_{\mathrm {track}}$, one can then calculate the Cherenkov angle simply as a dot product
of two vectors:  $\cos\theta_C = \hat{k}_{\mathrm {track}} \cdot \hat{k}_{\mathrm {pixel}}$. This procedure
has been used successfully with the first FDIRC prototype in the cosmic ray telescope with 3D
tracking~\cite{Nishimura:2012zz}. In the final physics analysis, the measured photons for each track
will be tested against probability distribution functions (PDFs) for each particle hypothesis.

\begin{figure}[tbp]
\begin{center}
\includegraphics[width=\linewidth]{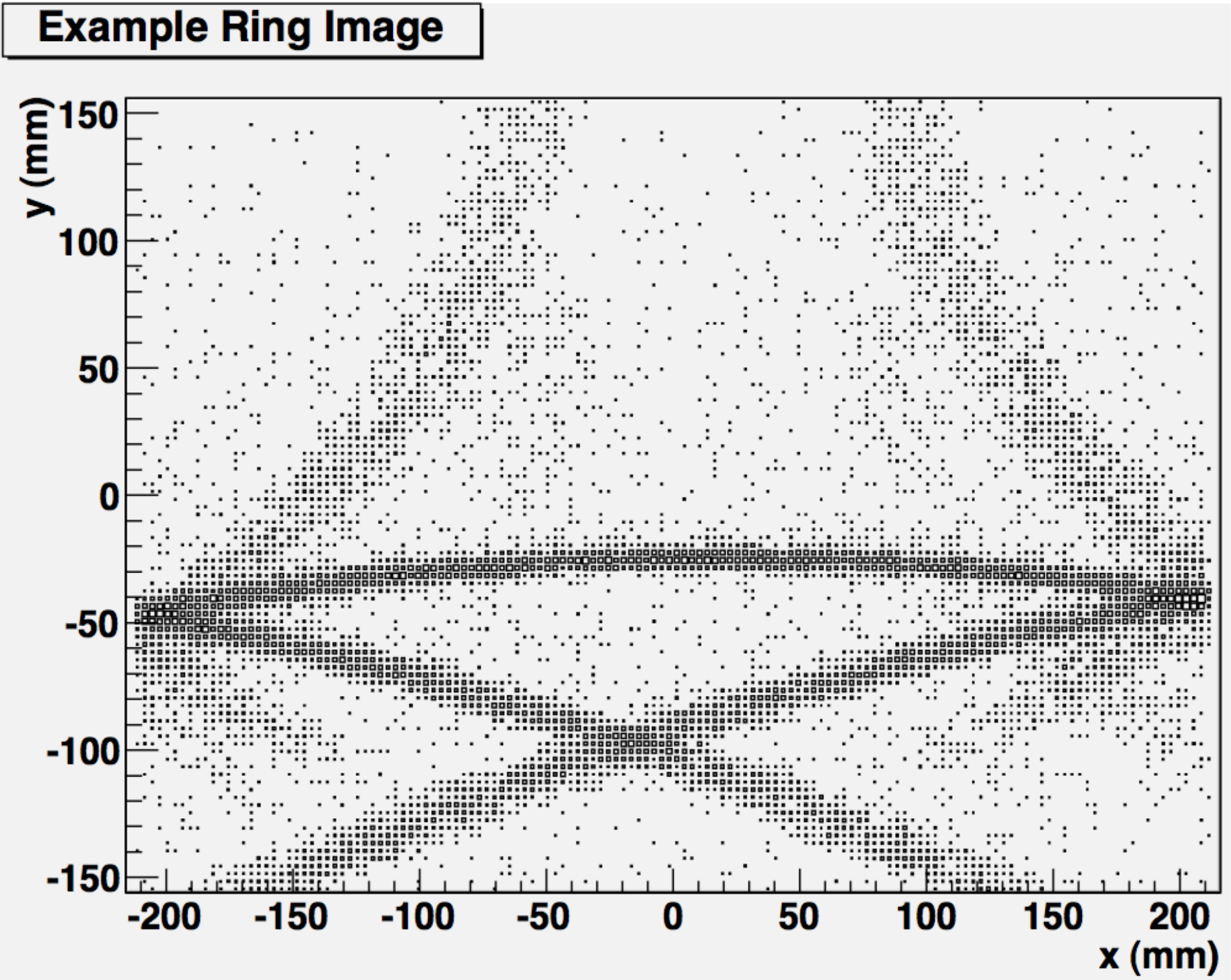}
\caption{Cherenkov ring image from \geantFour\ for tracks with $\theta_{\mathrm {dip}} = 90^\circ$ in the central bar at 4\gevc ~\cite{doug}. }
\label{fig:ring_image}
\end{center}
\end{figure}

\tdrsec{ The Barrel FDIRC Detector Overview }

\tdrsubsec{ Photodetectors }

\tdrparagraph{ Photon Detector choice  }
  There were three photon detectors under consideration: the H-8500 (64 pixels), the H-9500 (256 pixels) and the
R-11265-00-M64 (64 pixels) multi-anode PMTs (MaPMT) by Hamamatsu. At present, we have selected the 12-dynode H-8500 tube from 
several reasons: a) it is the tube preferred by the medical community and is therefore produced in a larger quantity, b) it has a much 
smaller unit price than the H-9500 MaPMT, c) it has a smaller single electron timing spread 
($\sigma_{\mathrm {TTS}}$ ${\sim} 140-160$\ps for H-8500 vs. $\sigma_{\mathrm {TTS}}$ ${\sim} 220$\ps for H-9500), d) it can be obtained 
with somewhat ~enhanced~ quantum efficiency (QE${\sim} 24\%$ for H-8500 vs. ${\sim} 20\%$ for H-9500), e) it has more uniform gain response across 
its face (2:1 for H-8500 vs. 5:1 for H-9500), and f) Hamamatsu ~strongly~ recommends not to consider H-9500 tube to keep a reasonable delivery 
schedule of large quantities. On the other hand, the H-9500 MaPMT can provide finer sampling in the $y$-direction and thus 
a significantly better Cherenkov angle resolution. We should keep it on the list of possible tube choices.

The Hamamatsu tube, R-11265-00-M64, came up recently for a consideration~\cite{maino_2012}. Its main attractions 
are (a) Super-bialkali QE of 36\%, (b) small 2.8\mm pixels, which would allow a thiner binning in $y$-direction, 
and therefore a better Cherenkov angle resolution, and (c) small dead space around tube boundaries.
We would combine 8 small pixels horizontally to create wide pixels in the $x$-direction, where we do not have focusing; at the
same time we would keep the same total number of electronics channels in the system.
We will test this tube and decide later. It would require 2304 tubes (12 photon cameras and $48\times 4$ tubes per camera) of this type in 
the FDIRC system. One should add that Hamamatsu also makes R-11265-00-M16 tube, which has the same pixel size as H-8500 tube.
It would still be useful to consider this tube as it has a Super Bialkali QE, and we would benefit from using it.
But smaller pixel size tubes are preferred at given QE.

The performance of the new FDIRC is simulated with a \geantFour\ based program ~\cite{doug}. Preliminary results for 
the expected Cherenkov angle resolution (for single photons) are shown in Table~\ref{table:FDIRC_resolution} for different layouts ~\cite{Va'vra:2010zz}. 
Design \#1 (a $3\mm \times 12\mm$ pixel size with the micro-wedge glued in) gives a resolution 
of $\sigma {\sim} 8.1\mrad$ per photon for 4\gevc pions at 90\degrees dip angle. As explained earlier, the micro-wedge option was supposed to remove 
a ${\sim} 6\mrad$ inclined surface on the old wedge. Since the micro-wedge will not be used, options \#1 and \#3 are excluded. Option \#3 is still 
considered and possibly chosen if a suitable PMT candidate is found in future. Presently, the selected 
option is \#4, which would give $\sigma {\sim} 9.6\mrad$ per photon. This would be a performance about the same as in \babar\ DIRC~\cite{Adam:2004fq}. 
However, if the chromatic correction would be implemented successfully, one could reduce the error by 0.5-2\mrad depending 
on the photon path length~\cite{Benitez:2008zz}. 

\begin{table}
\begin{center}
\caption{\geantFour\ MC simulation of the FDIRC performance (single photon resolution)~\cite{doug}.}
\begin{tabular}{|c|c|c|}
\hline \parbox[t][1.5cm][t]{1.4cm}{\centering FDIRC Design} & Option & \parbox[t][1.5cm][t]{1.7cm}{\centering $\theta_C$ \\ resolution [\mrad]} \\
\hline 1 & \parbox[t][1.4cm][t]{2.3cm}{$3\mm \times 12\mm$ pixels with a micro-wedge} & 8.1 \\
\hline 2 & \parbox[t][1.4cm][t]{2.3cm}{$3\mm \times 12\mm$ pixels and no micro-wedge} & 8.8 \\
\hline 3 & \parbox[t][1.4cm][t]{2.3cm}{$6\mm \times 12\mm$ pixels with a micro-wedge} & 9.0 \\
\hline 4 & \parbox[t][1.4cm][t]{2.3cm}{$6\mm \times 12\mm$ pixels and no micro-wedge} & 9.6 \\
\hline
\end{tabular}	
\label{table:FDIRC_resolution}
\end{center}
\end{table}

\begin{figure*}[tbp]
\begin{center}
\subfloat[H-8500 MaPMT single electron pulse without an amplifier (response to a 30\ps laser pulse at 407~nm).]{\includegraphics[height=0.23\textheight]{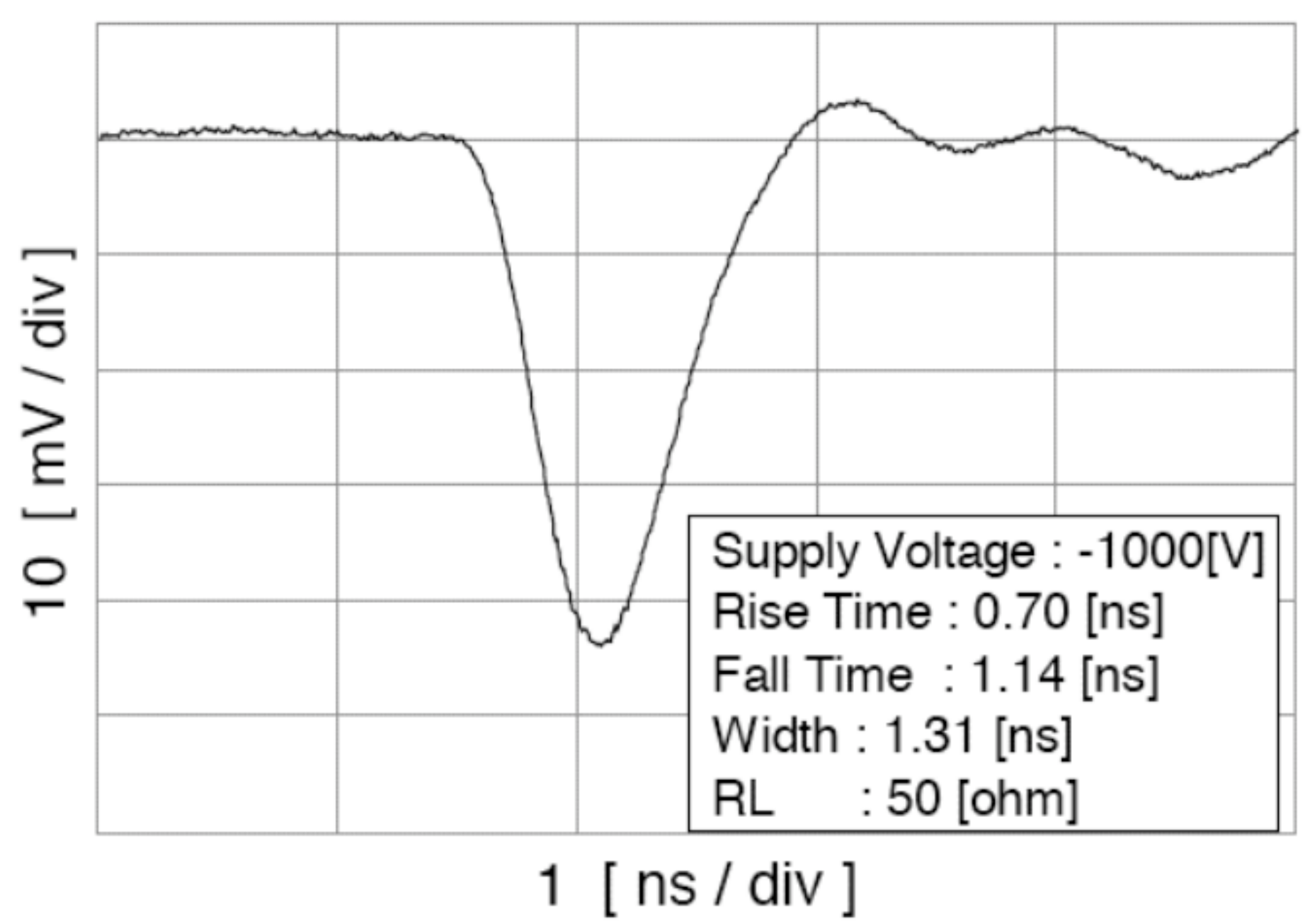} }
\hspace{5mm}
\subfloat[H-8500 MaPMT single electron pulse height spectrum.]{\includegraphics[height=0.23\textheight]{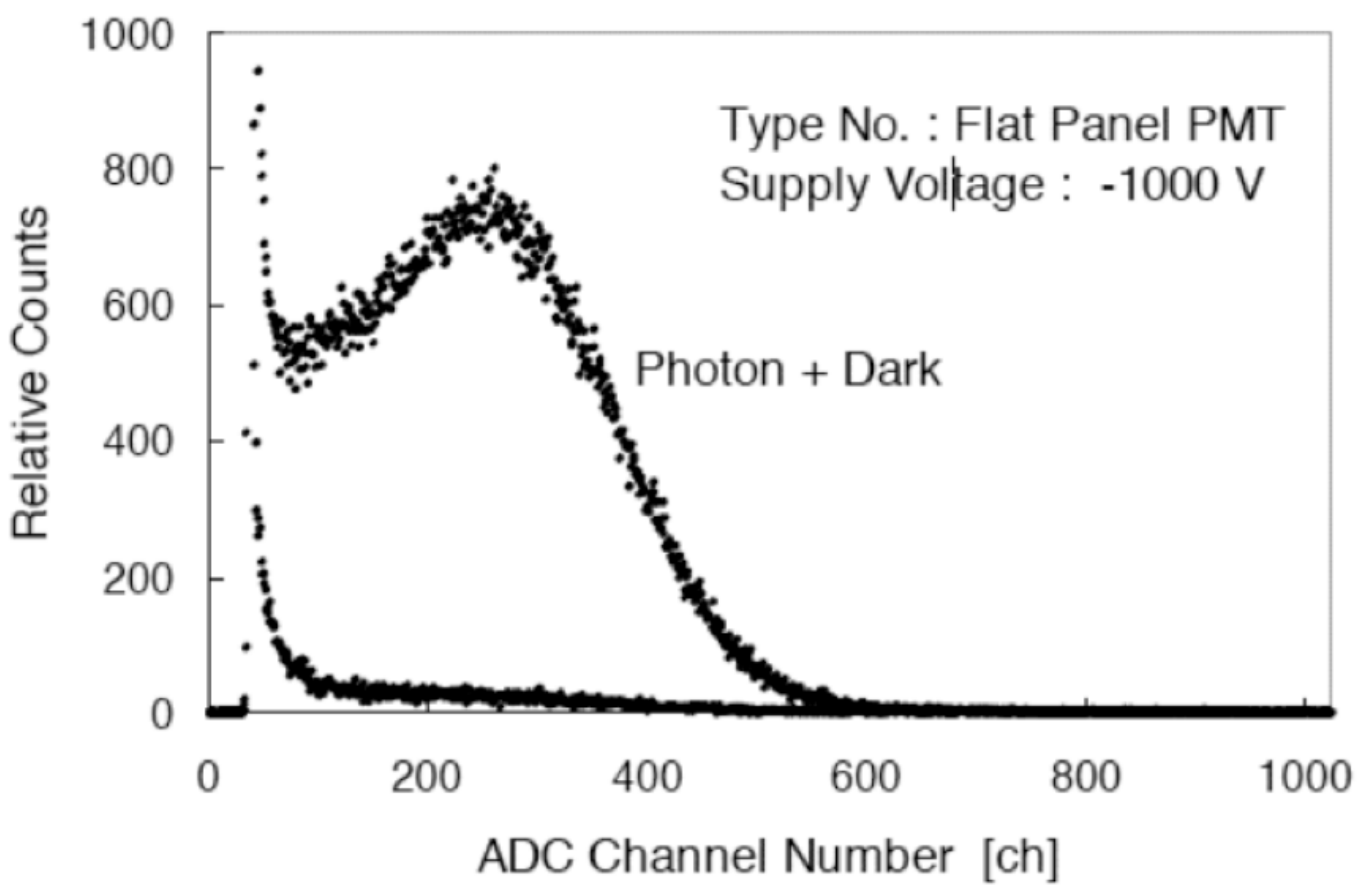}}
\end{center}
\caption{H-8500 MaPMT single electron pulse, noise and single electron pulse height distribution, Hamamatsu data~\cite{kyushima_2000}.}
\label{fig:H-8500_performance}
\end{figure*}

During the prototyping stage we used the H-8500C  version of the H-8500 tube, which has an internal resistor chain and a HV cable. 
In the final application, we plan to use the H-8500D version of this tube, which receives HV via local pins, which allows an easy 
distribution of HV among 48 tubes in the photon camera. This tube comes with a 1.5\mm-thick Borosilicate glass window with 
a spectral sensitivity between 300 and 650\nm. We would ask Hamamatsu to select a minimum QE of ${\sim} 24\%$. The dark anode 
current of this tube is very low (0.1~nA per pixel and 6~nA total), and the after-pulse rate is also 
almost negligible. Given the design of the dynode structure of H-8500 MaPMT preventing direct ion back-flow to the 
photocathode, we expect an usual cathode PMT aging behavior. It is not as good as in a classical PMT dynode structure, but 
expected to be close to it. We should confirm this by doing aging tests in the future.

\begin{figure}[tbp]
\begin{center}
\includegraphics[width=\linewidth]{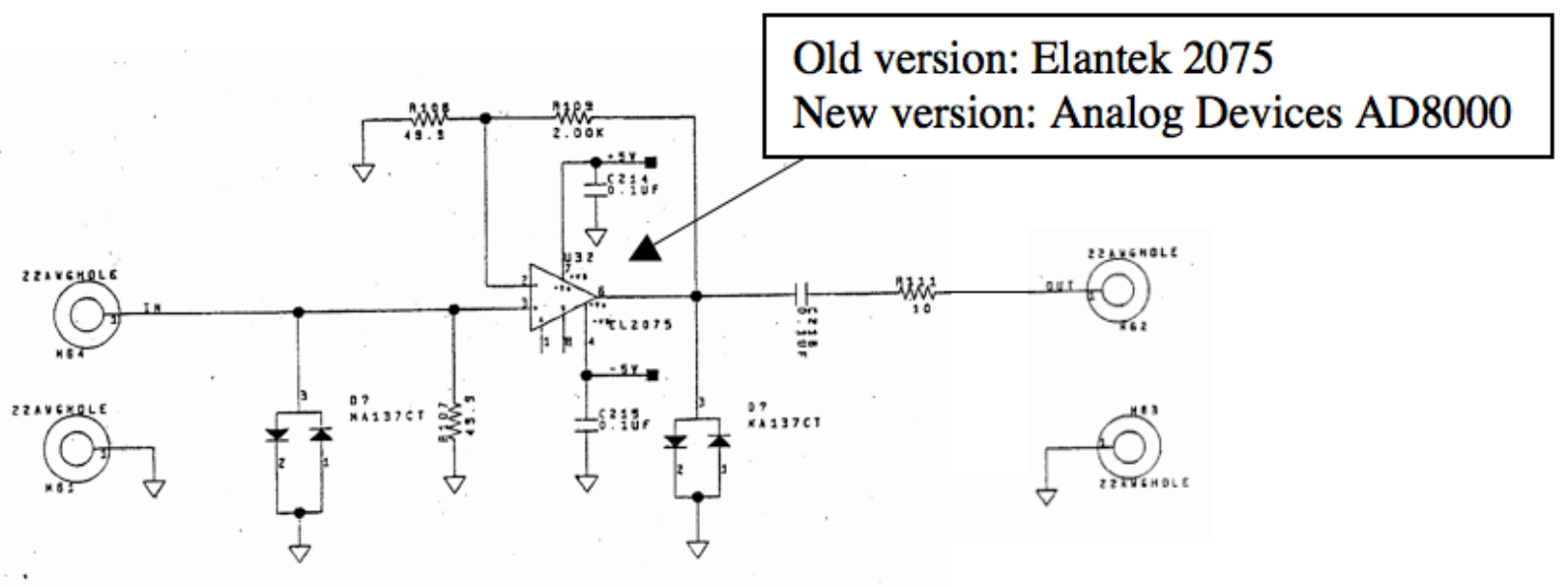}
\caption{SLAC amplifier used in all FDIRC prototypes, both in the SLAC test beam or for the CRT tests. The amplifier has a rise time of about 1.5~ns 
and a gain of ${\sim} 40\times$.}
\label{fig:slac_amplifier}
\end{center}
\end{figure}

It is important to state that all measurements presented in the following section were done with various electronics setups, which are not
the \superb\ final electronics, \ie\ many conclusions are somewhat preliminary.

Figure~\ref{fig:H-8500_performance} shows (a) a single photoelectron pulse from H-8500 before the amplification at ${\sim} 1.0\kV$,
(b) a single electron pulse height spectrum. As one can see on the plot, the rise time of H-8500 tube 
is about 0.7~ns. Figure~\ref{fig:slac_amplifier} shows the SLAC amplifier used for these tests. Hamamatsu points out that the pulse height 
spectra are not uniform across all pixels in the H-8500 tube. How this effect translates 
into the detection efficiency depends on the type of electronics, noise level and threshold; it will be 
studied in detail in the FDIRC prototype, first using the SLAC amplifier with the IRS-2 waveform digitizing 
electronics~\cite{varner_2011}, and then be compared to the \superb\ CFD electronics~\cite{Beigbeder_2011}. One should pay attention to areas 
in between pixels where pulse height is smaller due to charge sharing. One needs a sufficient electronics gain to get full efficiency.

\begin{figure}[tbp]
\begin{center}
\includegraphics[width=\linewidth]{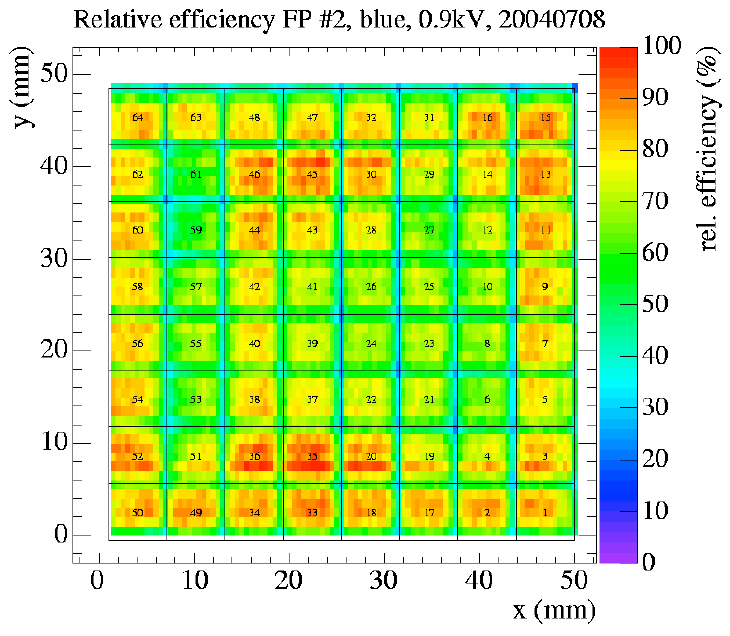}
\caption{Single photoelectron efficiency of an entire H-8500 tube, normalized to the highest efficiency spot within this particular tube~\cite{Field:2005wd}.}
\label{fig:2D_scan_a}
\end{center}
\end{figure}

\begin{figure}[tbp]
\begin{center}
\includegraphics[width=\linewidth]{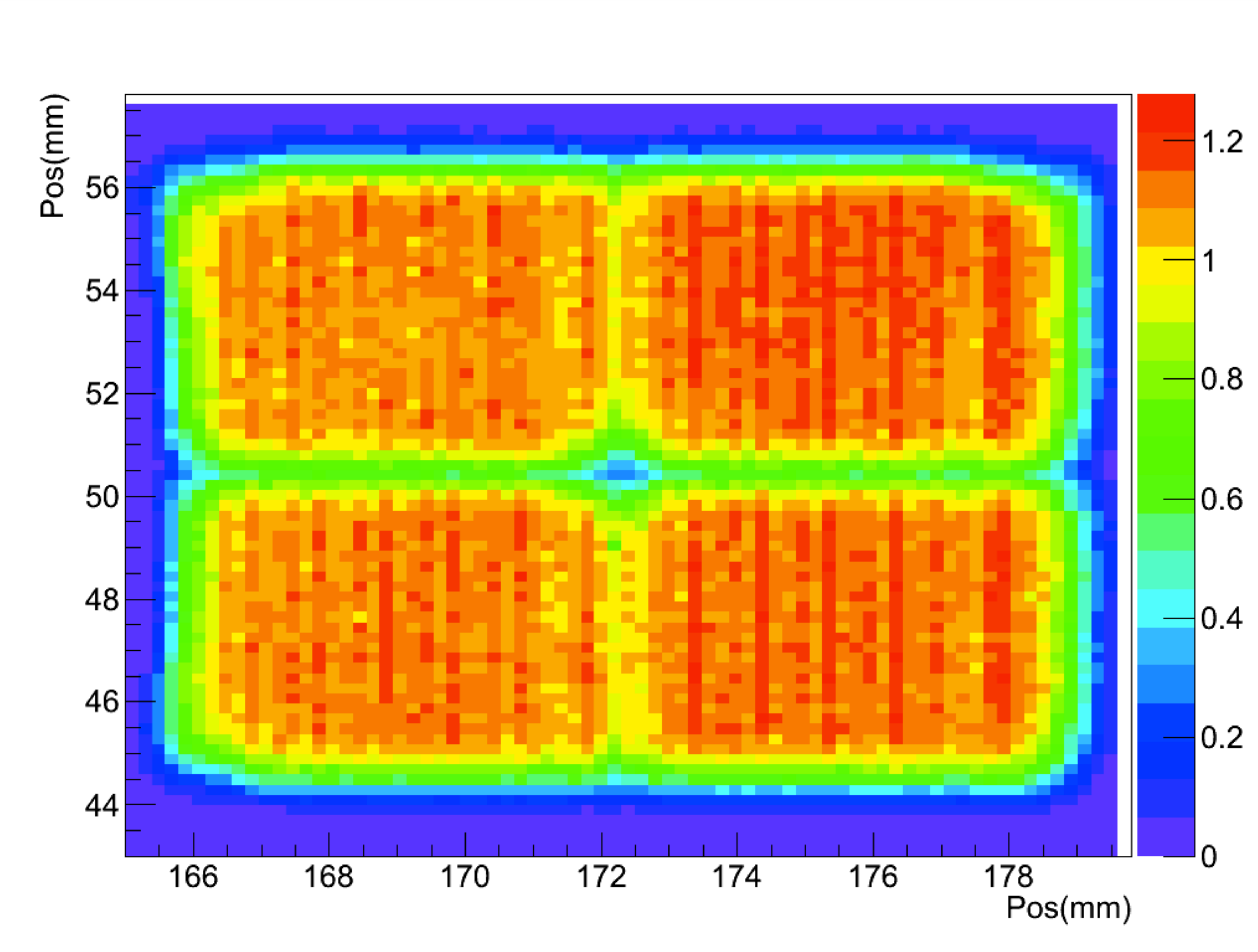}
\caption{Detailed efficiency scan of four pixels in H-8500 tube, normalized to the XP2020 tube~\cite{fabio_2012}.}
\label{fig:2D_scan_b}
\end{center}
\end{figure}

Figure~\ref{fig:2D_scan_a} shows the single photoelectron efficiency of a H-8500 tube, operating at -0.9~kV and normalized to the highest 
efficiency spot within this particular tube~\cite{Field:2005wd}. This plot was obtained with a SLAC amplifier, a version with two Elantek 2075 chips 
(a voltage gain of $130\times$); notice that there is no significant drop in the detection efficiency around pixel edges. There is the efficiency drop in between
pixels. This has been recently confirmed with another detailed scan~\cite{fabio_2012} shown on Figure~\ref{fig:2D_scan_b}. The data analysis and 
scope measurements indicate that some photoelectrons are lost due to the focusing electrode. Figure~\ref{fig:electrode_structure} shows the focusing
electrode geometry; it can steer electrons one way or another relative to the pixel boundary, and somewhat diminishes the effect of charge sharing.  
All these issues need to be evaluated carefully with the final electronics and for final tube deliveries.

\begin{figure}[tbp]
\begin{center}
\includegraphics[width=\linewidth]{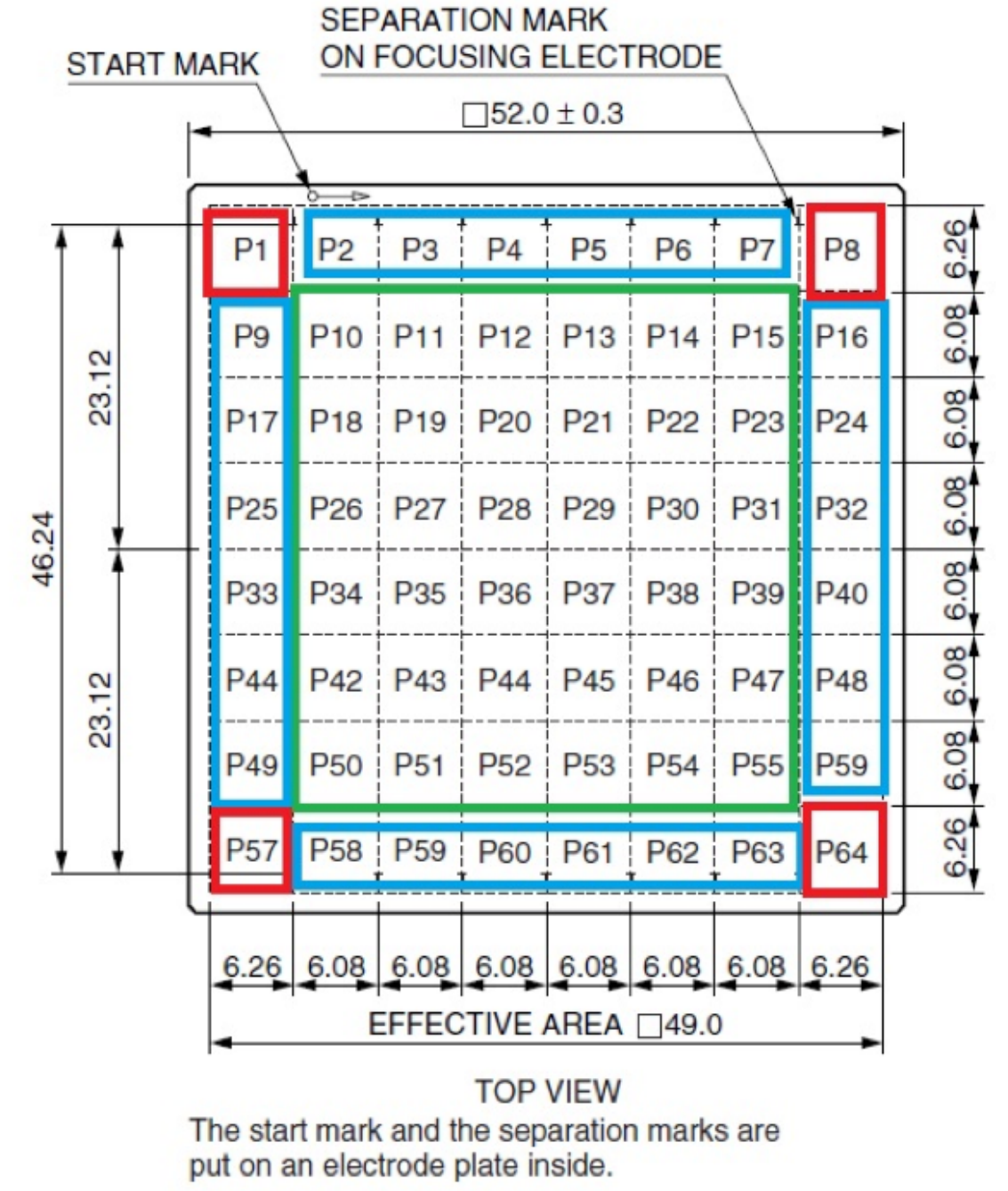}
\caption{The nominal H-8500 pixel geometry.}
\label{fig:H-8500_pixels}
\end{center}
\end{figure}

We plan to short two neighboring pixels in the $x$-direction, as there is only pinhole focusing available, and thus 
create $6\mm \times 12\mm$ pixels (H-8500), providing 32 readout channels per tube. Each photon camera would have 48 H-8500 MaPMT detectors, 
which corresponds to a total of 576 tubes for the entire \superb\ FDIRC, resulting in 18432 pixels total. 
Figure~\ref{fig:H-8500_pixels} shows the H-8500 tube pixel geometrical layout.

\begin{figure}[tbp]
\begin{center}
\includegraphics[width=\linewidth]{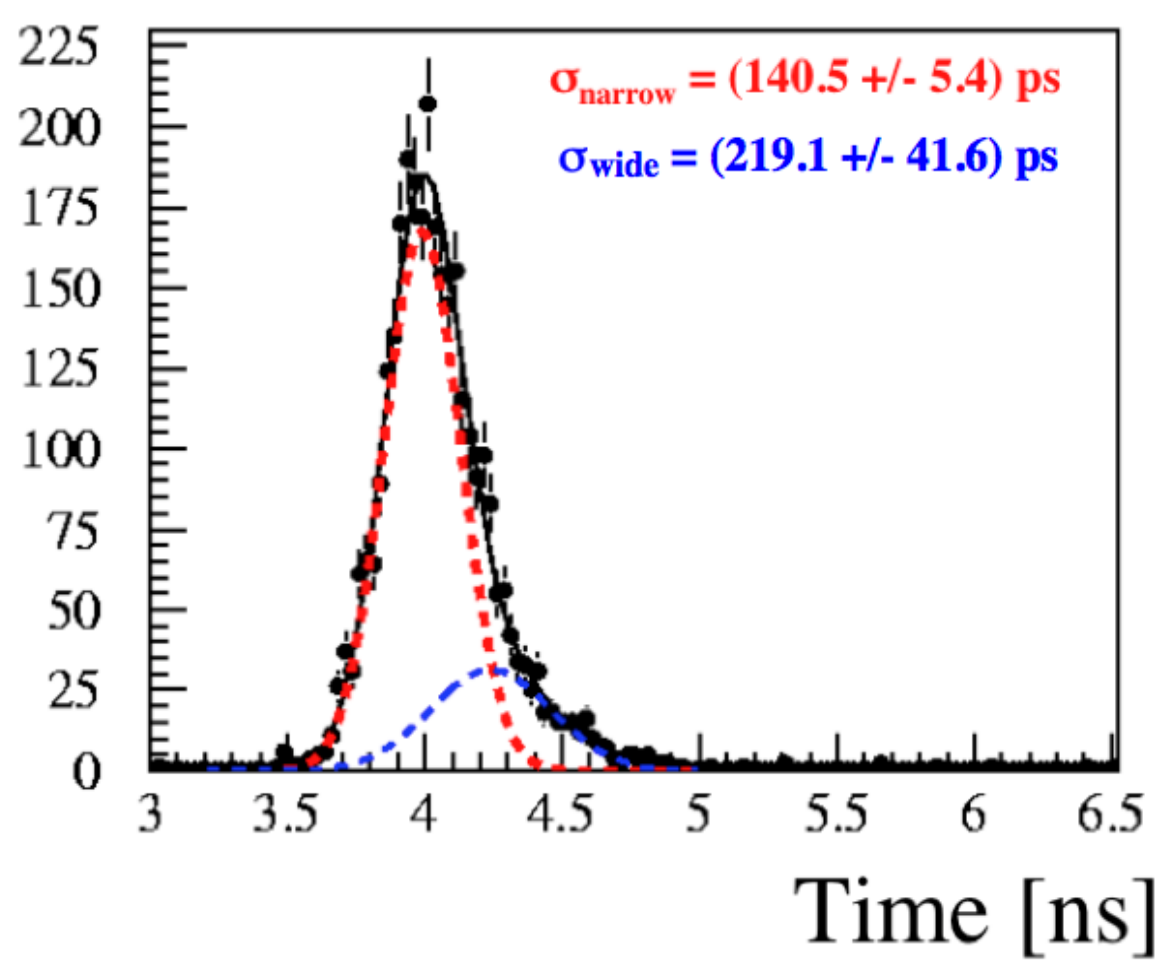}
\caption{The H-8500 single photoelectron transit time resolution is $\sigma_{\mathrm {TTS}} {\sim} 140\ps$~\cite{Field:2005wd}, if one 
fits the distribution with a double-Gaussian function.}
\label{fig:TTS_dist}
\end{center}
\end{figure}

\begin{figure}[tbp]
\begin{center}
\includegraphics[width=0.5\linewidth]{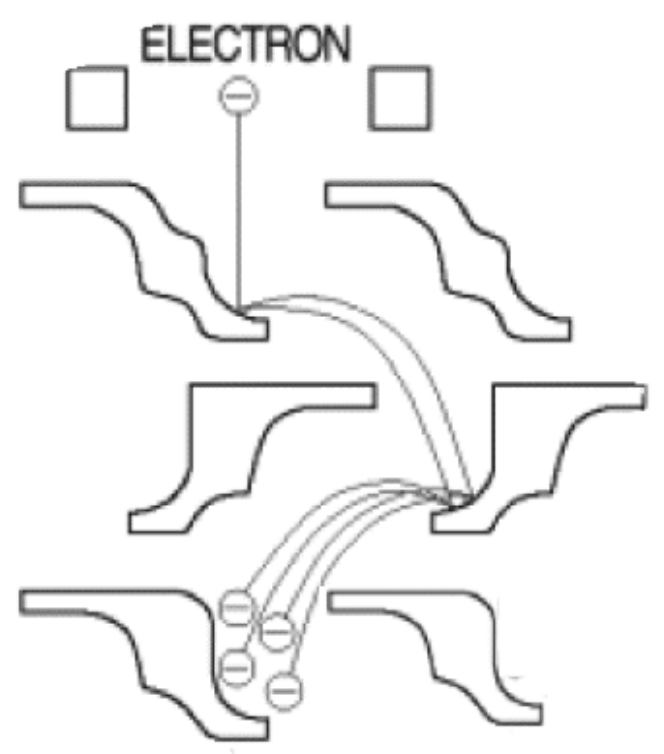}
\caption{The H-8500 electrode structure including focusing electrodes.}
\label{fig:electrode_structure}
\end{center}
\end{figure}

Figure~\ref{fig:TTS_dist} shows the H-8500C tube TTS timing resolution with single photoelectrons, indicating 
$\sigma_{TTS}~{\sim} 140\ps$~\cite{Field:2005wd}, obtained with a laser pointing to the center of a pixel. Hamamatsu 
data sheets for H-8500 tube indicates a value for the full width at half maximum (FWHM) of ${\sim} 400\ps$,
which gives $\sigma_{\mathrm {TTS}}~{\sim} 170\ps$, using a single Gaussian fit. 
Another measurement comes from R. Montgomery showing an average H-8500 tube TTS resolution of 
$\sigma_{\mathrm {TTS}}~{\sim} 154\ps$~\cite{montgomery_2011}. We measured $\sigma_{narrow}~{\sim} 140\ps$ and
$\sigma_{wide}~{\sim} 270\ps$~\cite{fabio_2012}. They also show that the TTS resolution depends on the position within a 
given pixel~\cite{gargano_2012}, which is driven by the PMT electrode structure, shown on Figure~\ref{fig:electrode_structure}.

The electrode structure and PMT edge effects and gain variation generally degrade 
the overall TTS resolution, so one should probably assume that the single photoelectron timing resolution is more like 200-250\ps. 
This agrees with Figure~\ref{fig:TTS_dist_in_1st_FDIRC_prot} where we plot TTS timing resolution in the first FDIRC 
prototype~\cite{Field:2005wd}, where laser photons populate the entire H-8500 face, \ie, pixels were hit uniformly, including their edges. 
Figure~\ref{fig:1st_prototype} shows how the laser calibration was done in the first FDIRC prototype.
This plot probably represents what will be a real TTS performance in practice. Notice also that edge pixels tend to 
have worse resolution.   

This timing performance, coupled to the electronics timing resolution contribution of $\sigma_{\mathrm {Electronics}}~{\sim} 100\ps$, allows 
corrections of the chromatic error for photon path length of more than 2 meters~\cite{Benitez:2008zz}.

\begin{figure}[tbp]
\begin{center}
\includegraphics[width=\linewidth]{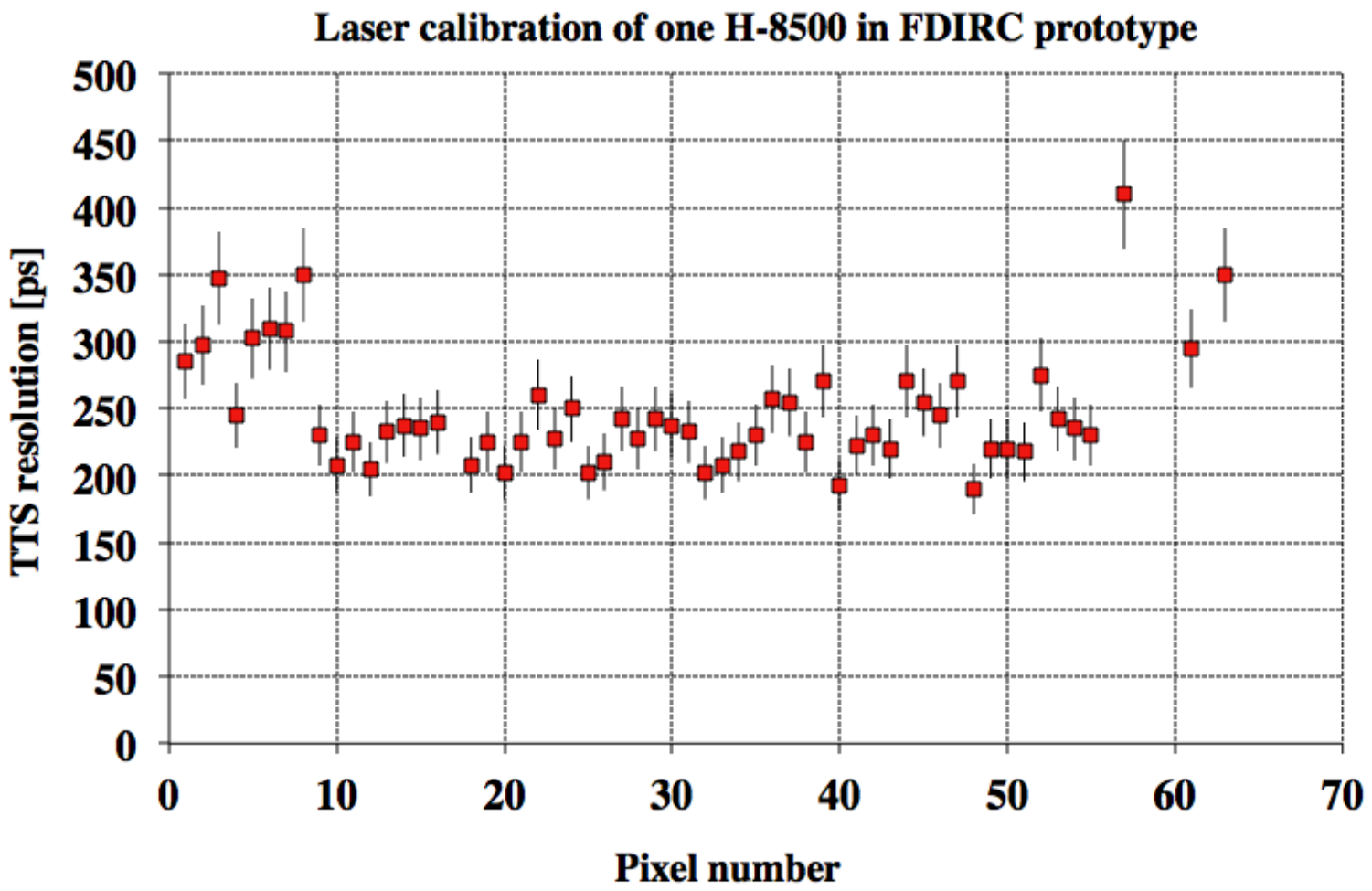}
\caption{The H-8500 single photoelectron transit time resolution across all pixels from a particular tube~\cite{Field:2005wd}.}
\label{fig:TTS_dist_in_1st_FDIRC_prot}
\end{center}
\end{figure}

There are two effects to take into account when considering interaction between two neighboring pixels: the pixel-to-pixel 
cross-talk, and the charge sharing avalanche between two pixels. The neighbor pixel-to-pixel single photoelectron cross-talk 
was measured to be ${\sim} 3$\% of the primary pixel signal, when a laser light was placed on the center of a pixel while 
looking at its neighbor~-- see Figure~\ref{fig:Cross-talk_vavra}. This test used a newer SLAC amplifier with AD8000 
chip~-- see Figure~\ref{fig:slac_amplifier}. 
However, the pixel-to-pixel cross-talk is even more complex in multi-anode tubes~\cite{montgomery_2011}. 
Figure ~\ref{fig:Cross-talk_montgomery} shows that the cross-talk depends on the position within a pixel. This 
will clearly require more study with the FDIRC final electronics. We had hoped to use the charge sharing, which is 
related to the avalanche size, to reduce the size of pixels in the $y$-direction by charge interpolation. 
However, the attempt to use charge sharing was not successful for this particular tube as it has 
entrance focusing electrodes defining pixel boundaries, which sweep electrons away from pixel 
boundaries, \ie\ Hamamatsu has designed the MaPMT electrode structure in such a way to suppress 
the charge sharing in these tubes. Both H-8500 and H-9500 have this charge sharing-suppressing feature. 
Such feature does not exist in MCP-PMT detectors.

\begin{figure}[!h]
\begin{center}
\includegraphics[width=\linewidth]{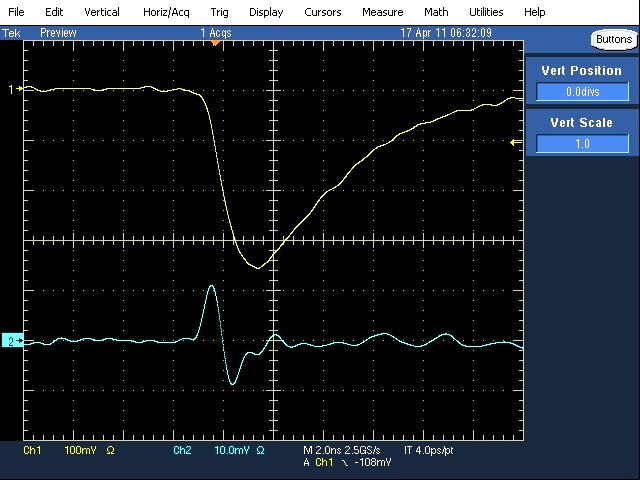}
\caption{The H-8500 tube pixel-to-pixel single electron cross-talk was measured to be ${\sim} 3\%$, when the laser 
light was placed on the center of a pixel while looking at its neighbor~\cite{jjv_elba_2011}.}
\label{fig:Cross-talk_vavra}
\end{center}
\end{figure}

\begin{figure*}[tbp]
\begin{center}
\includegraphics[width=0.95\textwidth]{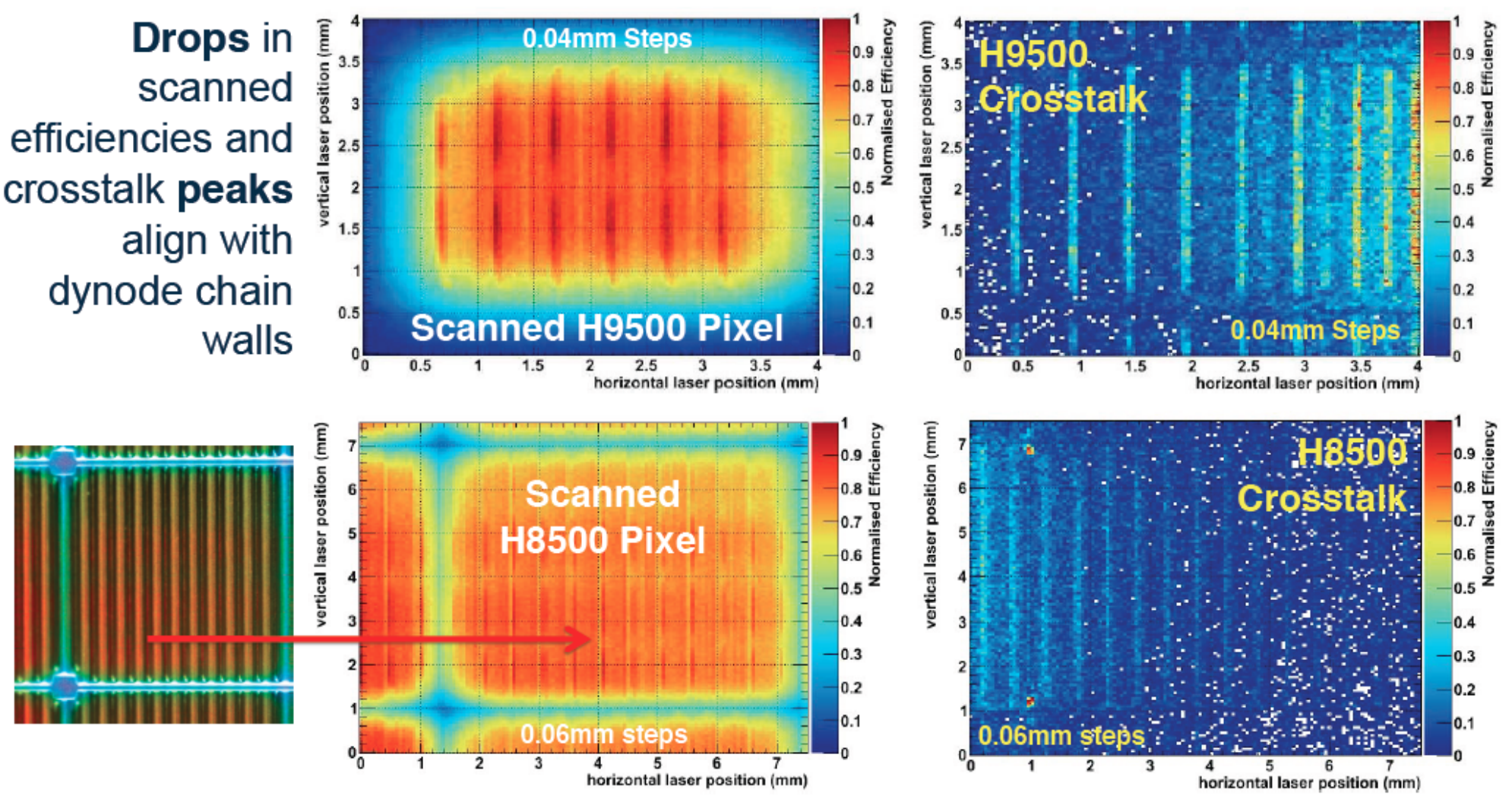}
\caption{The observed periodicities in single electron efficiencies and cross-talk are aligned with the dynode 
electrode structure~\cite{montgomery_2011}.}
\label{fig:Cross-talk_montgomery}
\end{center}
\end{figure*}

\begin{figure}[tbp]
\begin{center}
\subfloat[Double-Polya fit to single electron distribution as observed in R7600-03-M16 MaPMT. The lower peak
originates from photoelectrons which are missing one amplification stage in the MaPMT~\cite{Abbon:2007aa}.]{\includegraphics[width=\linewidth]{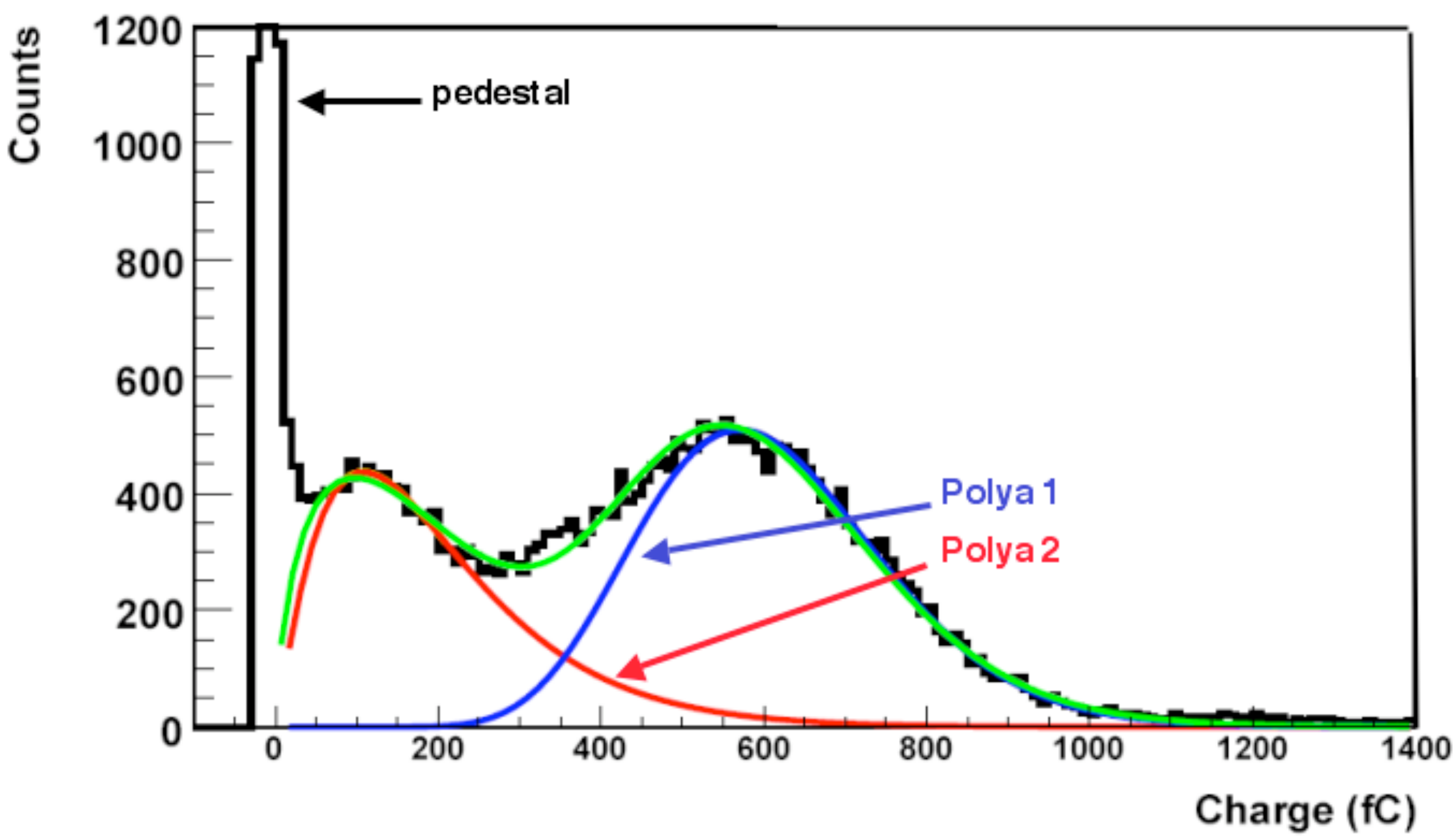} }
\hspace{5mm}
\subfloat[Single electron distributions in H-8500 MaPMT. The lower peak shows up only as a shoulder~\cite{Pauly_2011}.]{\includegraphics[width=\linewidth]{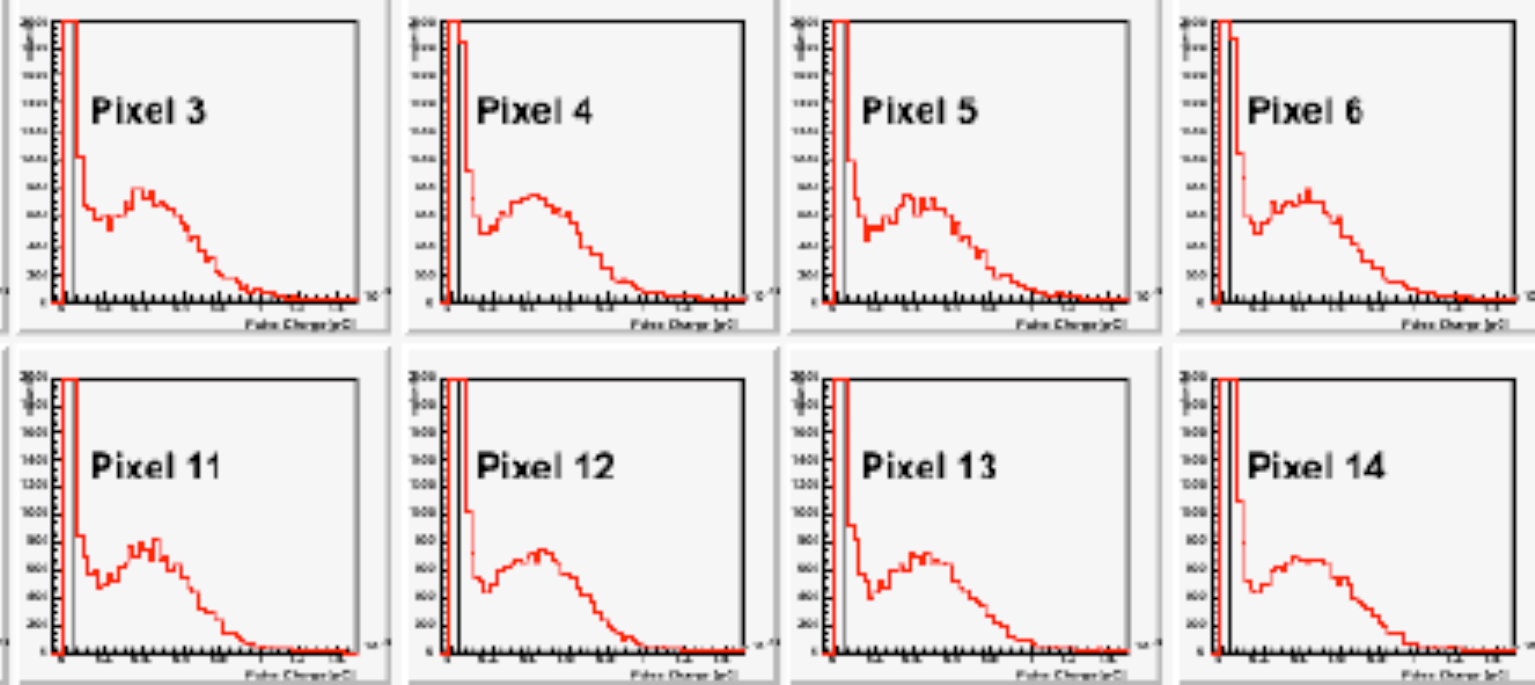}}
\end{center}
\caption{Single electron pulse height distribution in MaPMTs.}
\label{fig:MaPMT_PH}
\end{figure}

Another special feature of all MaPMT detectors is a double Polya distribution, one corresponding to a photoelectron 
produced at the cathode and the amplification using all 12 dynodes (this is the nominal distribution), and another 
one corresponding to a case that a photon produces a photoelectron striking the very first dynode rather than at the 
photocathode, and the amplification is only using 11 dynodes instead of 12. Missing one amplification stage produces 
a gain 2-3 smaller than the nominal amplification process, while the pulses arrive 2-2.5~ns earlier
(see Figure~\ref{fig:Pre_pulses}~\cite{jjv_elba_2011}). Figure~\ref{fig:TTS_time_spectrum_a}
shows a time spectrum of normal photoelectrons produced at the cathode, 
pre-pulse spectrum produced at the first dynode arriving ${\sim} 2.5$ ns earlier, and back-scattered
photoelectrons arriving ${\sim} 6$~ns later~\cite{simi_Frascati_2011}. Similar measurements indicate that this 
timing spectrum depends on the position within the pixel~\cite{fabio_2012}, \ie\ whether the photoelectron hits the focusing electrode or moves
in between two~-- see Figure~\ref{fig:TTS_time_spectrum_b}. However, one should point out that we are dealing with logarithmic scale
and that the total rate of off-time pulses is less than a few percents. Figures~\ref{fig:MaPMT_PH}
show the resulting single electron pulse height spectra which exhibit either a small shoulder 
near the pedestal at lower gain~\cite{Pauly_2011}, or a clear double-Polya distribution at higher gain~\cite{Abbon:2007aa}. 
Although the pre-pulses are a nuisance, they can be used as normal photoelectrons in the Cherenkov ring analysis and their time 
can be calibrated out. 

\begin{figure}[tbp]
\begin{center}
\includegraphics[width=\linewidth]{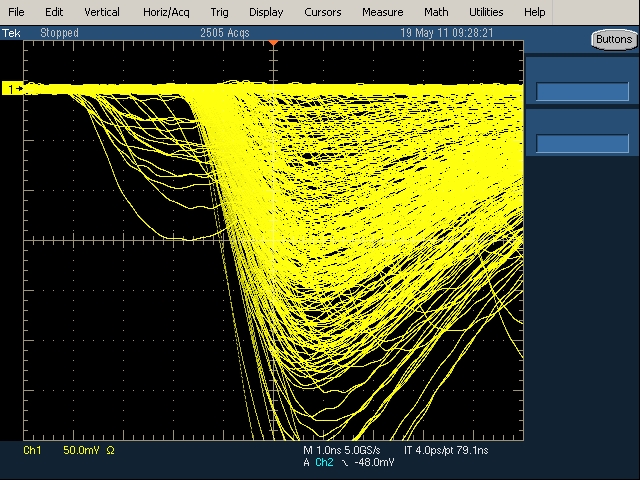}
\caption{The H-8500 single electron pre-pulses corresponding to a case when a photon produces a photoelectron at the very first 
dynode rather than at the photocathode, and the amplification is using only 11 dynodes instead of 12~\cite{jjv_elba_2011}.}
\label{fig:Pre_pulses}
\end{center}
\end{figure}

\begin{figure}[tbp]
\begin{center}
\includegraphics[width=\linewidth]{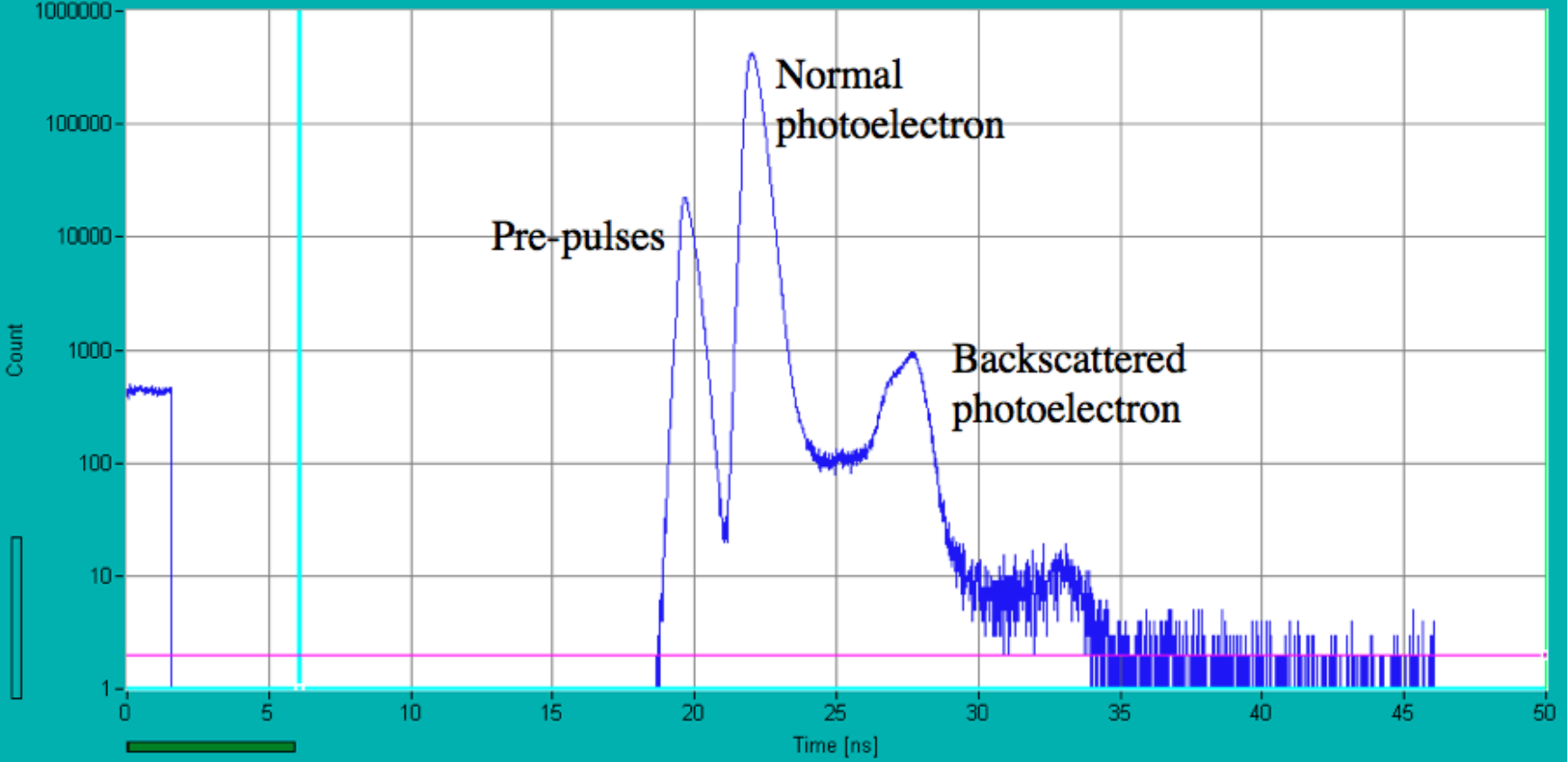}
\caption{The H-8500 single electron time spectrum showing normal photoelectrons, pre-pulses and the back-scattered photoelectrons~\cite{simi_Frascati_2011}.}
\label{fig:TTS_time_spectrum_a}
\end{center}
\end{figure}

\begin{figure}[tbp]
\begin{center}
\includegraphics[width=\linewidth]{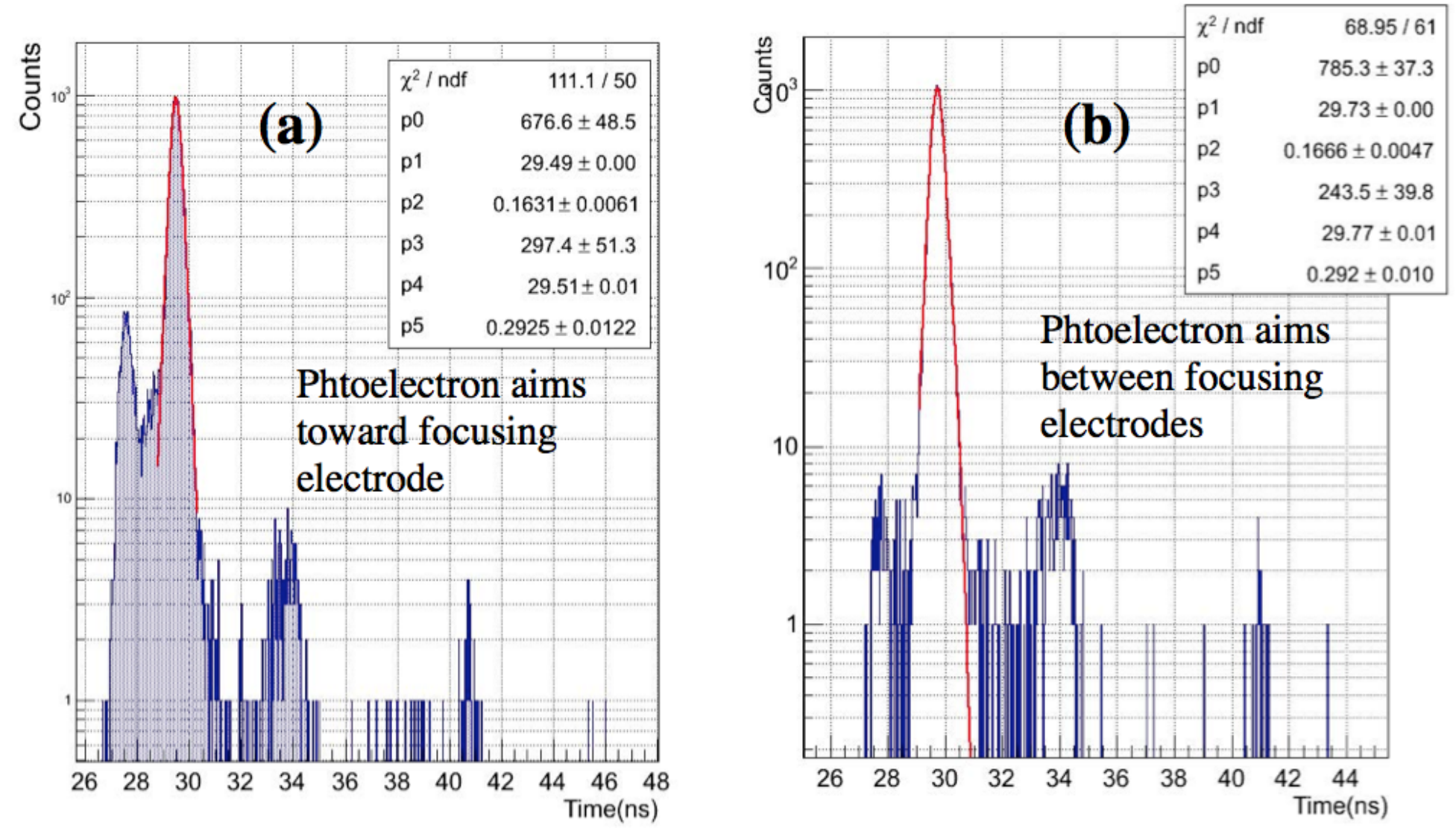}
\caption{The H-8500 single electron time spectrum depends on the photoelectron position relative to the focusing electrodes~\cite{fabio_2012}.}
\label{fig:TTS_time_spectrum_b}
\end{center}
\end{figure}

\begin{figure}[tbp]
\begin{center}
\includegraphics[width=\linewidth]{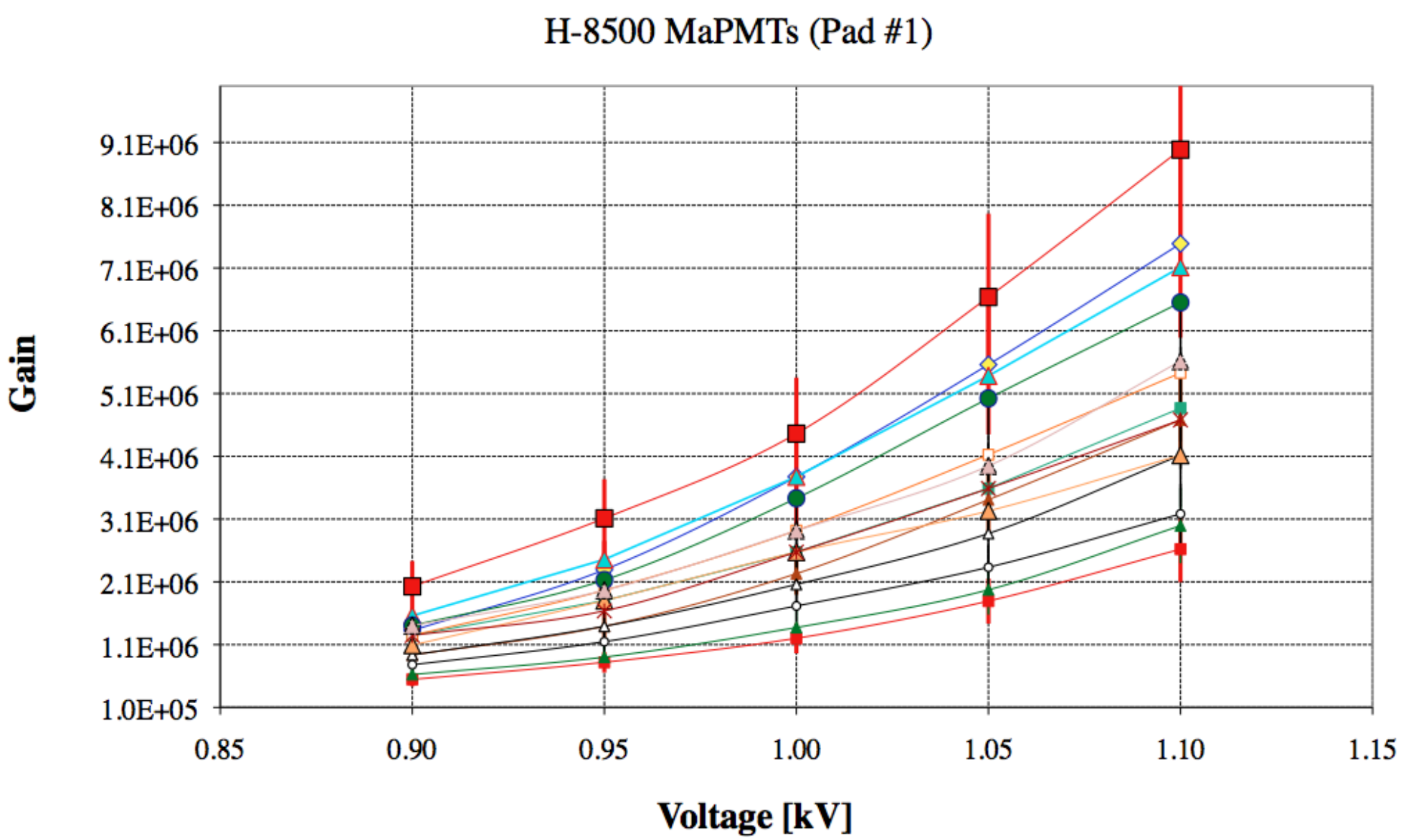}
\caption{The H-8500 tube gain range and dependence on voltage for 14 tubes using pixel 1 (see Figure~\ref{fig:H-8500_pixels}
for its location) in each tube~\cite{jjv_elba_2011}.}
\label{fig:Gain_dependence}
\end{center}
\end{figure}

Figure~\ref{fig:Gain_dependence} shows that the H-8500 tube gain range is $1-3 \times 10^6$ for nominal
operating voltage of $-1.0\kV$~\cite{jjv_elba_2011}. There is a variation of gain from pixel-to-pixel due
to non-uniformities in the multi-anode structure. 

\begin{figure*}[tbp]
\begin{center}
\subfloat[Max-min efficiency uniformity across pixels of 15 PMTs. This is compared to anode current max-min response across 64 pixels to a fixed photon flux.]{\includegraphics[height=0.18\textheight]{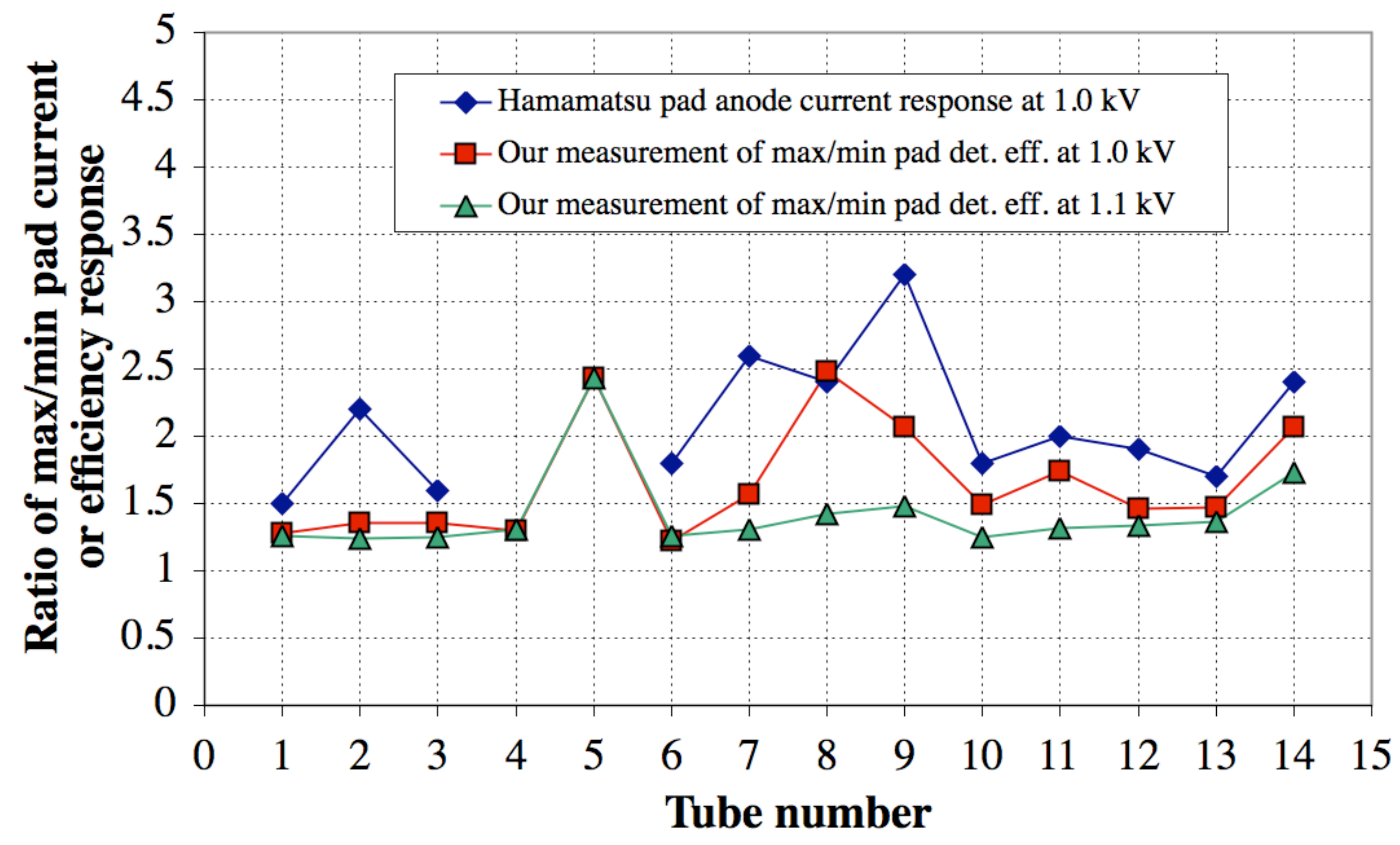} }
\hspace{5mm}
\subfloat[Relative efficiency scan of H-8500 tubes operating at $-1.0\kV$, normalized to XP2262/B PMT.]{\includegraphics[height=0.18\textheight]{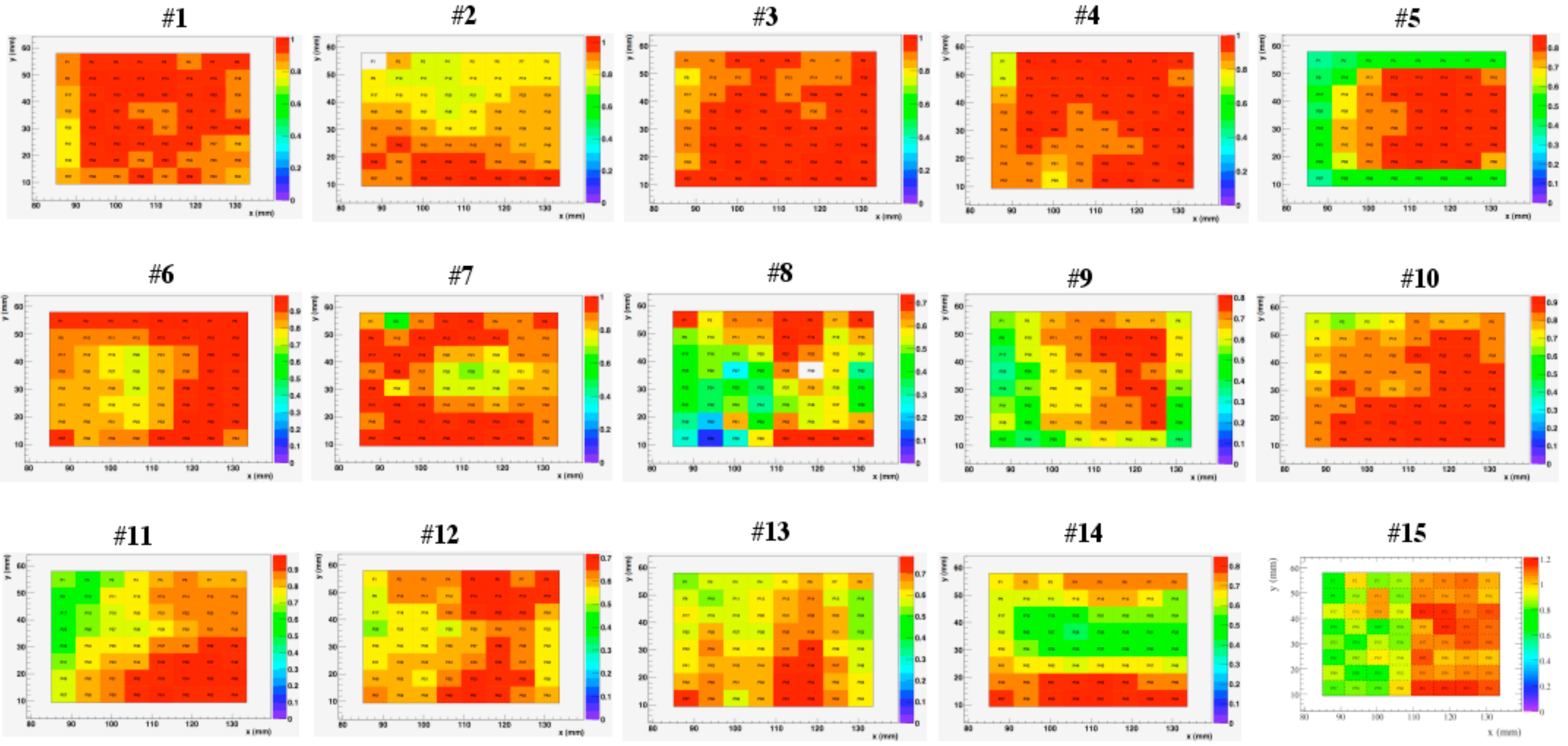}}
\end{center}
\caption{2D single electron detection efficiency across pixels of 15 H-8500 tubes~\cite{jjv_elba_2011}.}
\label{fig:2D_eff_uniformity}
\end{figure*}

Figure~\ref{fig:2D_eff_uniformity} shows scans of 15 tubes~\cite{jjv_elba_2011}, operating at -1.0\kV and -1.1\kV, 
amplifier gain of ${\sim} 40\times$, a threshold electronics of -25 mV, and indicates that typically 
the best-to-worst single electron detection efficiency might vary as much as 2:1 across the H-8500 PMT face. This is compared to 
the anode current response across all pixels to a fixed high photon flux (Hamamatsu data).
This figure also shows the efficiency maps of 15 tubes~\cite{jjv_elba_2011}, all operating at -1.0\kV, 
with an amplifier gain of $40\times$, with a simple threshold electronics, and normalized to the Photonis Quantacon XP2262/B PMT. 
It indicates that the best-to-worst detection efficiency variation is as much as 2:1 across the H-8500 PMT face. 
One should stress that the detection efficiency relative to the Quantacon XP2262/B PMT is typically at a level of 40-50\% for the worst pixels, 
and 80-100\% for the best pixels.  

\begin{figure}[tbp]
\begin{center}
\includegraphics[width=\linewidth]{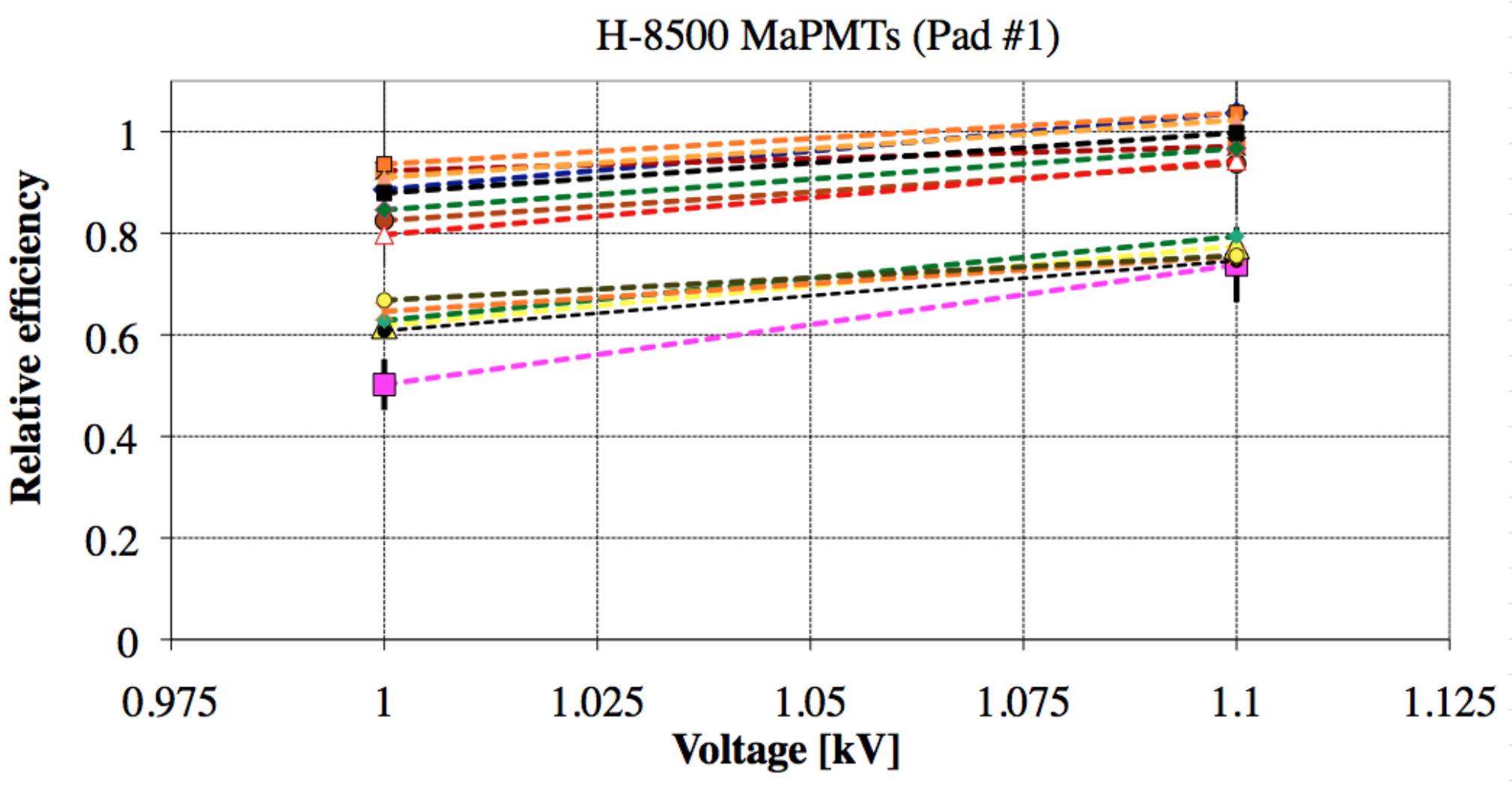}
\caption{The H-8500 single electron detection efficiency dependence on voltage~\cite{jjv_elba_2011}.}
\label{fig:Eff_increase_per_100V}
\end{center}
\end{figure}
 
Figure~\ref{fig:Eff_increase_per_100V} shows how the H-8500 tube single electron detection efficiency depends on 
voltage~{\cite{jjv_elba_2011}. One can see that the detection
  efficiency can be improved by 10-20\% per 100~V increase.

There are three possible ways to deal with the pixel-based gain non-uniformity: 
(a) process each tube, equipped with the final electronics, in a scanning setup; record the individual 
relative efficiency values and store them into an analysis database, or (b) adjust a discriminator threshold on 
each pixel, or (c) adjust an amplifier gain on each pixel. This concept has yet to be worked out in detail as this 
effect depends on details of the final electronics. When doing the overall gain adjustment, one should remember that 
the absolute maximum voltage on H-8500 is -1.1\kV (Hamamatsu recommendation). We want to set the initial voltage low enough to 
have some headroom for later period, when the detector will loose gain due to aging.  

\begin{figure}[!h]
\begin{center}
\includegraphics[width=\linewidth]{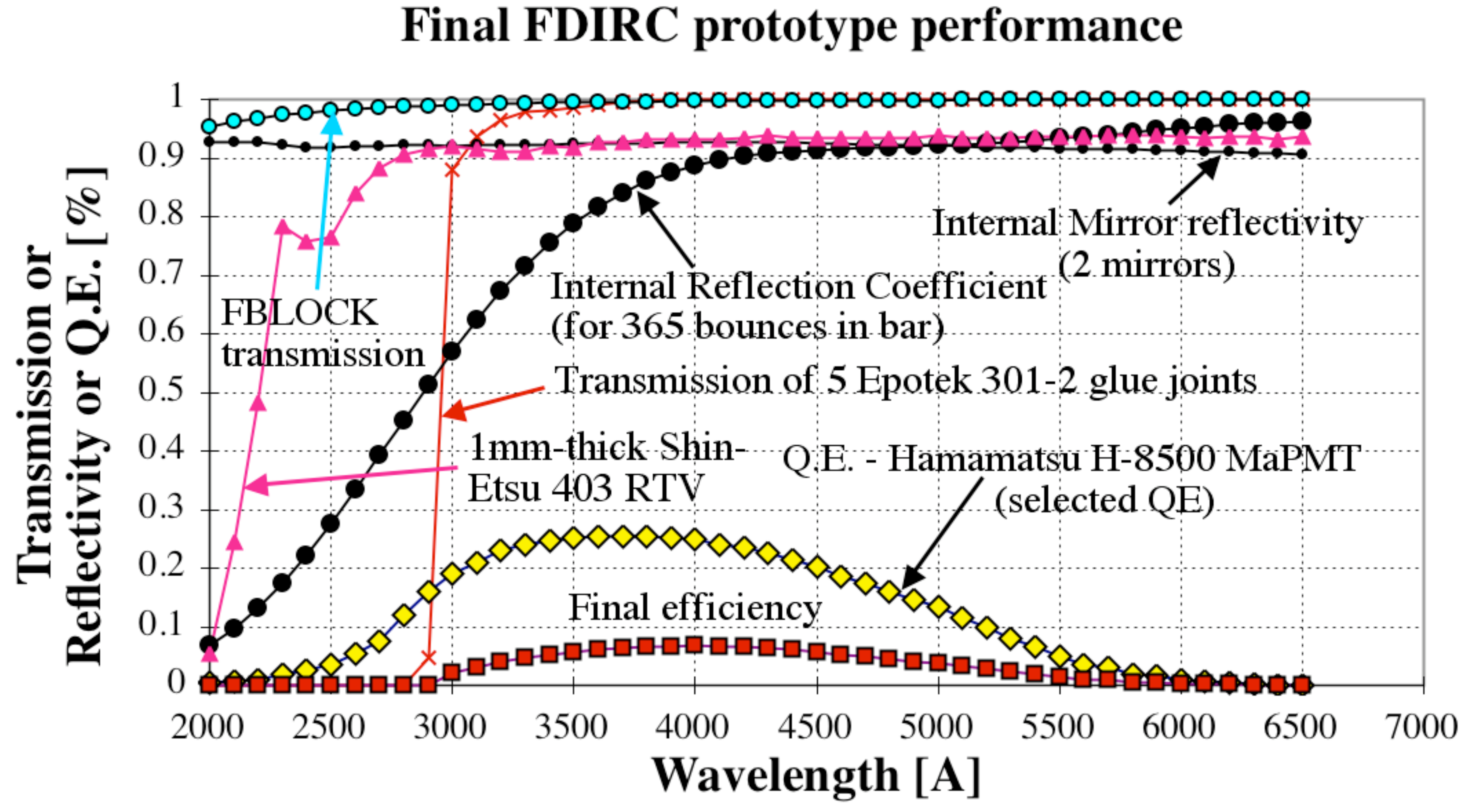}
\caption{The FDIRC wavelength response is limited on low wavelength side by the Epotek 301-2 glue used to glue bars together~\cite{Va'vra:2010zz}.}
\label{fig:wavelength_response}
\end{center}
\end{figure}

\tdrparagraph{ Prediction of number of photoelectrons per ring }

Figure~\ref{fig:wavelength_response} shows the FDIRC wavelength bandwidth~\cite{Va'vra:2010zz}. One can see that we operate 
in the visible wavelength region and that the effective filter is the Epotek 301-2 epoxy. Assuming a peak QE of 24$\%$, and 
no optical grease coupling between PMTs and the FBLOCK, one obtains ${\sim} 32$~photoelectrons for di-muon tracks with $\theta_{\mathrm {dip}} = 90^\circ$ 
using a simple spreadsheet calculation. Figure~\ref{fig:Optical_coupling} shows the MC simulation of the number of photoelectrons as a function 
of the dip angle~\cite{doug}. At $\theta_{\mathrm{dip}}$ = $90^\circ$ it predicts ${\sim} 27$~photoelectrons. 

\tdrparagraph{ Modularity: photodetector mechanical packing fraction }

\begin{figure*}[tbp]
\begin{center}
\includegraphics[width=\linewidth]{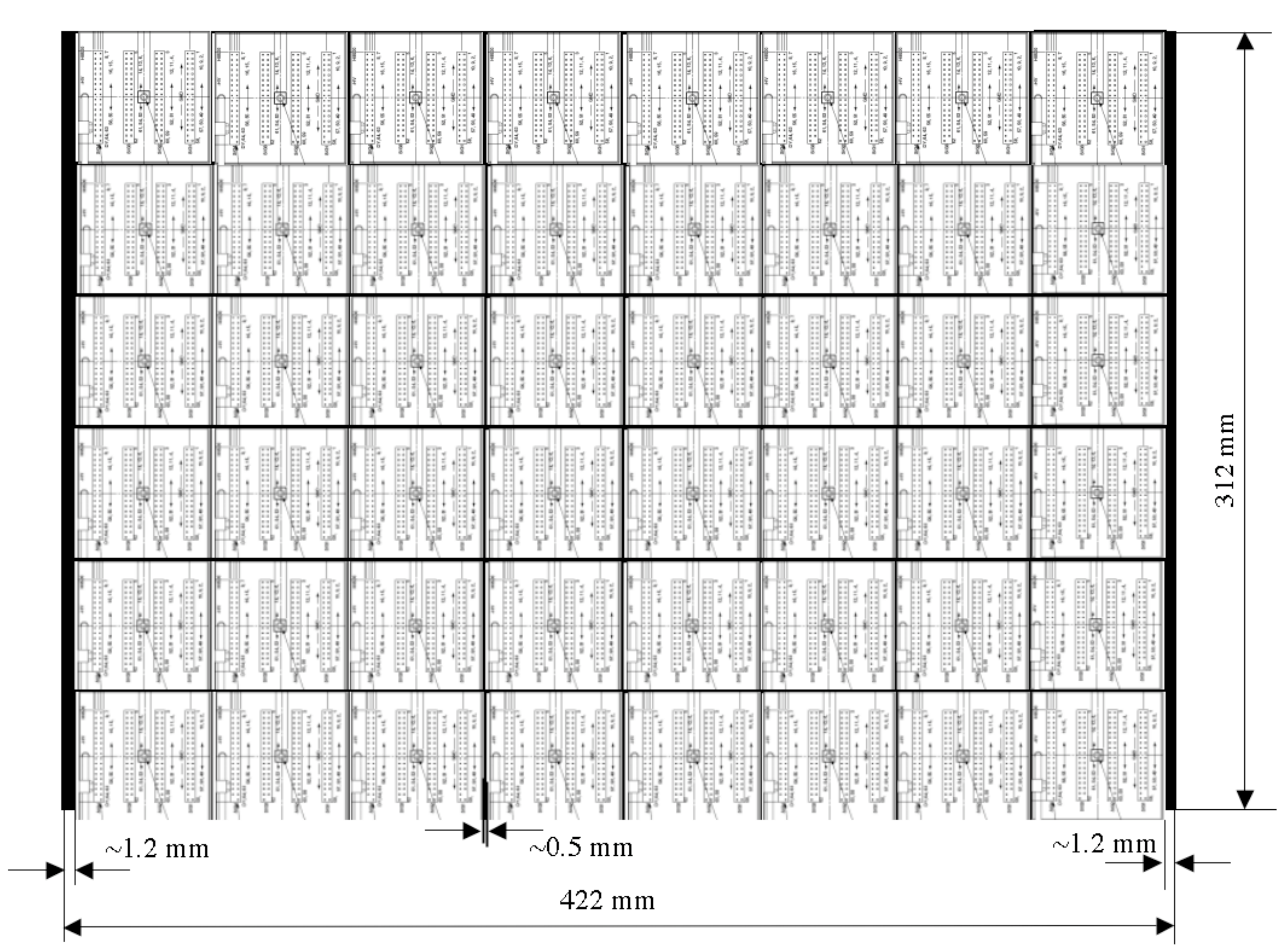}
\caption{Detector matrix on one FDIRC detector camera with 48 H-8500 MaPMTs. The entire FDIRC system needs 576 such tubes 
for a total of 18,432 pixels~\cite{Va'vra:2010zz}.}
\label{fig:Packing_fraction}
\end{center}
\end{figure*}

  There are two factors to consider when determining the photon coverage: (a) detection coverage in the focal 
plane of the photon camera, and (b) coverage within each tube (we will consider detection losses within the dynode structure 
later). Figure~\ref{fig:Packing_fraction} shows the H-8500 matrix of 48 tubes in one photon camera. 
The size of each H-8500 tube is $52.0 \pm 0.3\mm$, and the gap between each tube is ${\sim} 0.5\mm$; this gives a contribution 
to the packing fraction of ${\sim} 98.6\%$. The photon packing density (effective area/external size) within each tube 
is ${\sim} 89\%$. These factors give the overall photon packing efficiency of ${\sim} 88\%$ for the photon 
camera based on 48 H-8500 tubes.

\tdrparagraph{ Optical coupling of detectors to FBLOCK }
    MC simulation shows (see Figure~\ref{fig:Optical_coupling}) that we loose 8-25\% of photons if we do not optically 
couple PMTs to the FBLOCK~\cite{doug}. The event Cherenkov angle resolution 
($\sigma {\sim} 2.94\mrad$) improves by ${\sim} 10\%$ with optical coupling~\cite{doug}. 
On the other hand, an access to a single failed detector would become complicated. If we use optical coupling, we are 
considering eight vertical segments, each handling six detectors; each segment could be 
removable by sliding it vertically off the FBLOCK.  
In addition, we have not yet selected the optical coupling grease.
One should realize that any leak of the grease on the FBLOCK side optical surfaces, 
would result in a serious loss of photoelectrons. The optical coupling concept has yet to be tested 
to investigate its practicality, reliability and radiation hardness, and therefore this remains an open issue.

\begin{figure}[tbp]
\begin{center}
\includegraphics[width=\linewidth]{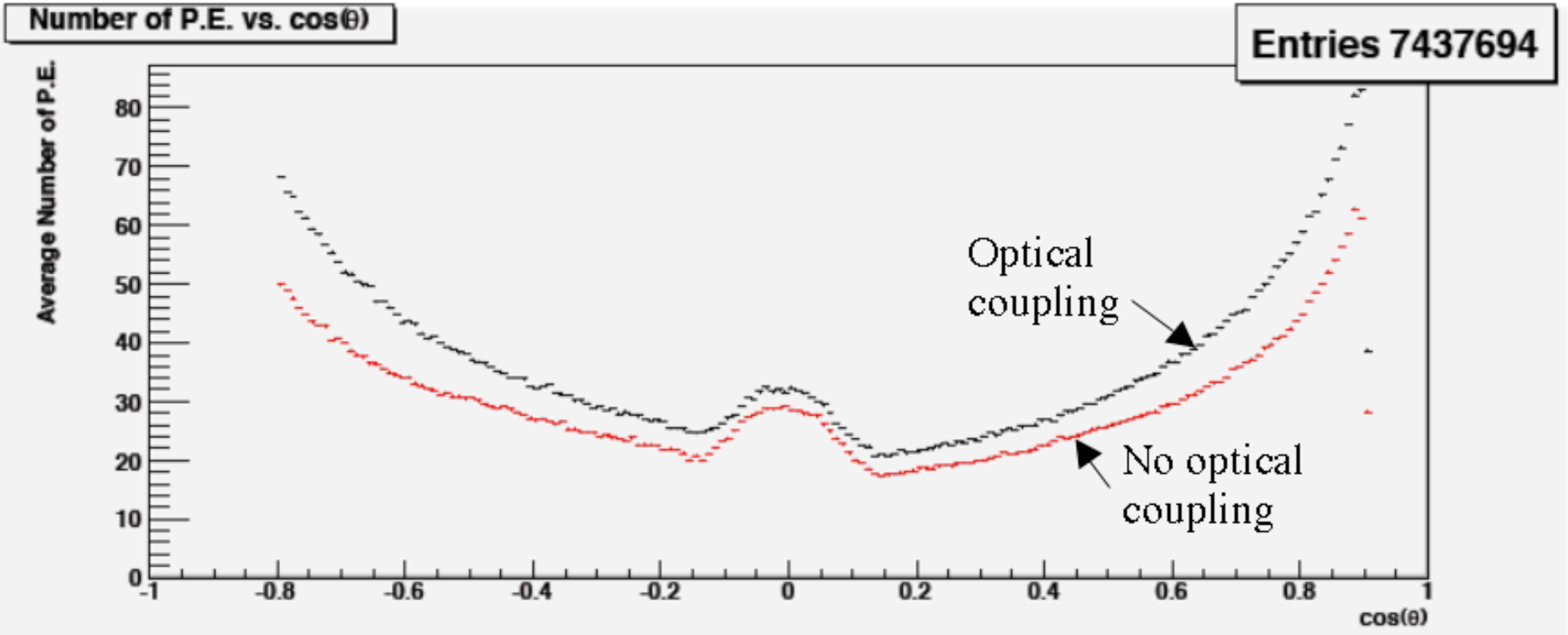}
\caption{Simulated (\geantFour\ MC) number of detected photoelectrons as a function of the polar angle for two cases: with 
and without optical coupling between detector face and the FBLOCK~\cite{doug}.}
\label{fig:Optical_coupling}
\end{center}
\end{figure}

\tdrparagraph{ Temperature requirements }
   There are two major sources of heat in the detector enclosure: (a) HV resistor dividers, and (b) electronics. 
Each tube has a HV divider. All dividers together draw ${\sim} 9$~W per 48 tubes, which is a small amount of heat. 
Assuming for now that the electronics will dissipate ${\sim} 500$~W per 48 detectors. We will need a water-cooled heat exchanger. 

  Another worry is what happens if we loose cooling. Based on tests with the FDIRC prototype detector enclosure,
the temperature would rapidly climb beyond ${\sim} 80^\circ$C. That would be dangerous for tubes, optical grease coupling, 
glues and that could also create mechanical stresses. Therefore, we need an automatic power shutoff system.

\tdrparagraph{ Rates and aging issues in H-8500 PMTs }
    One strong point of our design is that we share a total photon background load from 
a single bar box among 48 H-8500 detectors, and this results in acceptable rates, even 
at the highest luminosity, and an acceptable total charge load after 10 years of operation.

   We use two methods to estimate the FDIRC rates: (a) an empirical scaling (ES) from Belle-I Aerogel counter 
rates, assuming that the background rate scales as the luminosity. (b) We use the \superb\ MC simulation, which simulates 
all physics background processes involved in the background production and includes the precise modeling of beam line 
magnet components all the way up to 16 meters from the interaction point (IP) in either direction. It uses the correct FDIRC geometry with a proper 
handling of optical photons and includes all background shielding of the photon camera. The ES method is rather close to the 
MC prediction: 75~kHz (ES) vs. 60~kHz (MC) per double-pixel, or 2.4~MHz (ES) vs. 1.9~MHz (MC) per tube, which would correspond 
to the total accumulated charge of 1.3~C/cm$^2$ (ES) vs. 1.2~C/cm$^2$ (MC) for a total integrated luminosity 
of L$_{int}$ ${\sim} 50$~\invab. Table~\ref{table:Rates} summarizes rates under various conditions. 
Figure~\ref{fig:Det_layout_f} shows the FDIRC shielding design. 

As far as the neutron background on the FDIRC front end electronics (FEE), the MC predicts a rate of 
$3.3 \times 10^{10}$~n/cm$^2$/year of 1~MeV-equivalent neutrons with our up-to-date shielding.
This makes the use of SiPMTs detectors for the
photon camera impossible. However, this rate is tolerable for the FDIRC electronics.

One should point out that it is very crucial to shield the FDIRC photon camera, located outside of the magnet. Its contribution 
would be 550~kHz (MC) per double-pixel without shielding. Before the final shielding was put into the MC simulation, the dominating 
background was due to the Bhabha scattering, and the contribution of the Touschek effect was one order of magnitude smaller. After 
the shielding was added, all background contributions are at similar level. 

Given the design of the H-8500 dynode structure, which aims at preventing a direct ion backflow to the 
photocathode, we expect that MaPMT tube cathode-aging rate to be similar to usual PMT aging 
behavior, which means that the above numbers appear to be safe. For example, \babar\ DIRC PMTs 
accumulated at least ${\sim} 150$~C per tube during ${\sim} 10$~years of \babar\ operation, the PMT gain was reduced 
by ${\sim} 45\%$ in total, but tubes operated well until the end, with a few voltage adjustments to correct for the gain loss~\cite{adam_2012}. 
The above estimates mean that each H-8500 tube would accumulate about 1.3~C of charge during the \superb\ data taking, 
which is considered a relatively small amount. However, one should point out that aging tests are yet to be done for the H-8500 tube. 
Figure~\ref{fig:pmt_aging} shows the Hamamatsu aging data for R8400-00-M64 MaPMT running at $100~\muA$ for 10,000~hours and 
operating at $-1.0$~kV. There is no obvious large effect which could not be corrected by a voltage adjustment. 

One should also worry about unusual background conditions caused by the machine misbehavior, changes in tune, beam losses, etc., 
especially in the early periods before reaching the full luminosity. Hamamatsu recommends that the absolute 
maximum current be ${\sim} 100\muA$ per tube or ${\sim} 2\muA$ per pixel. Another constraint is the capability of the 
electronics to cope with high rates. The \superb\ FDIRC electronics can handle rates up to 
${\sim} 20\MHz$ per pixel, if one pixel is firing, and up to ${\sim} 2.5\MHz$ per tube if all pixels are firing.

\begin{table*} [tbp]
\begin{center}
\caption{MC prediction of FDIRC pixel rates, integrated charge dose and tube current, with and without shielding at tube gain of 10$^6$.}
\begin{tabular}{|c|c|c|c|c|c|}
\hline \parbox[t][1.0cm][t]{2.5cm}{Shielding} & Double-pixel rate & Tube rate  & Total charge dose & Current \\
\hline Yes & 60~kHz & 1.9~MHz & 1.3~C/cm$^2$/50$~\invab$  & 0.3~$\muA$/tube \\
\hline No & 550~kHz & 17.8~MHz & 12.7~C/cm$^2$/50$~\invab$ & 2.8~$\muA$/tube \\
\hline
\end{tabular}	
\label{table:Rates}
\end{center}
\end{table*}

\begin{figure}[tbp]
\includegraphics[width=\linewidth]{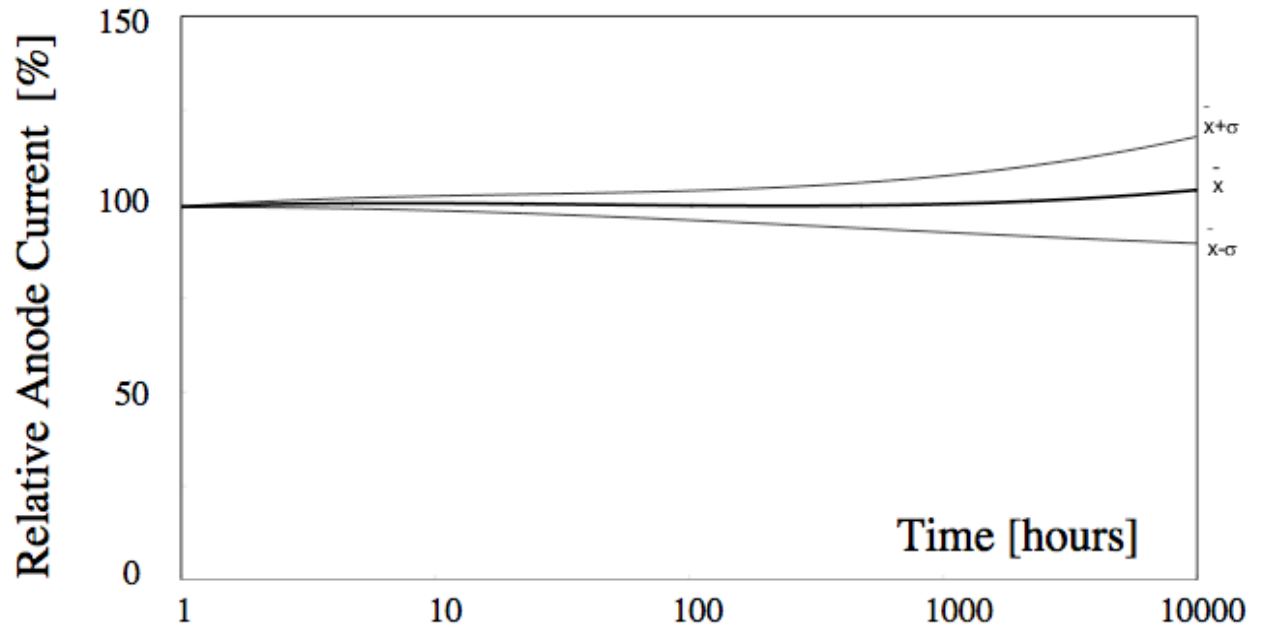}
\caption{Hamamatsu data for R8400-00-M64 MaPMT showing that there is no large drop in photo-current when running $100\muA$ 
for 10,000 hours (${\sim} 400$~days).}
\label{fig:pmt_aging}
\end{figure}

\tdrparagraph{ Background shielding to protect FDIRC electronics \& photon detectors}
The aim of this shield is to reduce the background contribution from the FBLOCK, located outside of the magnet. To design the shielding, 
two main constraints have to be taken into account: first, to allow an easy access to detectors and electronics, and then to minimize 
the overall weight of the shielding. Figure~\ref{fig:Det_layout_f} shows the FBLOCK shielding design, which was also modeled in the MC simulation. 
It consists of 10 cm of Boron-loaded polyethylene layer sitting on 10-15~cm lead-steel sandwich, both located on inner radius and front 
side of the FBLOCK with its detectors and electronics. The front section of the shielding is moving on the magnetic door allowing a quick access to electronics and detectors~-- see Figure~\ref{fig:Det_layout_f}.  

\begin{figure}[tbp]
\begin{center}
\includegraphics[width=\linewidth]{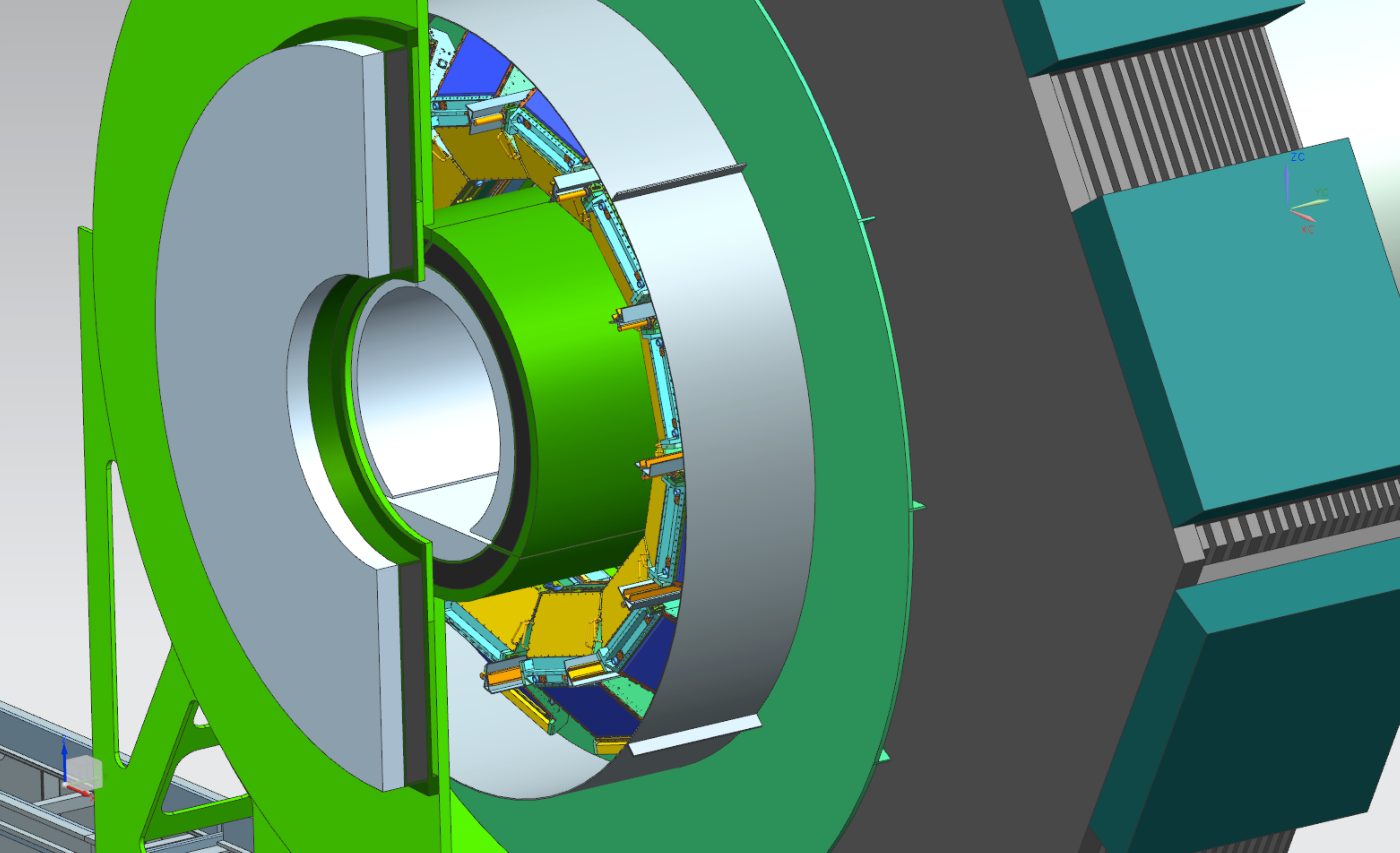}
\caption{Details of local shielding around the FDIRC photon camera (a layer of 10 cm of Boron loaded polyethylene 
followed by 10-15 cm of lead-steel sandwich, located both on inner and front sides of the FBLOCK with its detectors and electronics).}
\label{fig:Det_layout_f}
\end{center}
\end{figure}

\tdrparagraph{ Magnetic shield of H-8500 PMTs }

\begin{figure*}[tbp]
\begin{center}
\subfloat[The effect on boundary pixels.]{\includegraphics[width=0.465\linewidth]{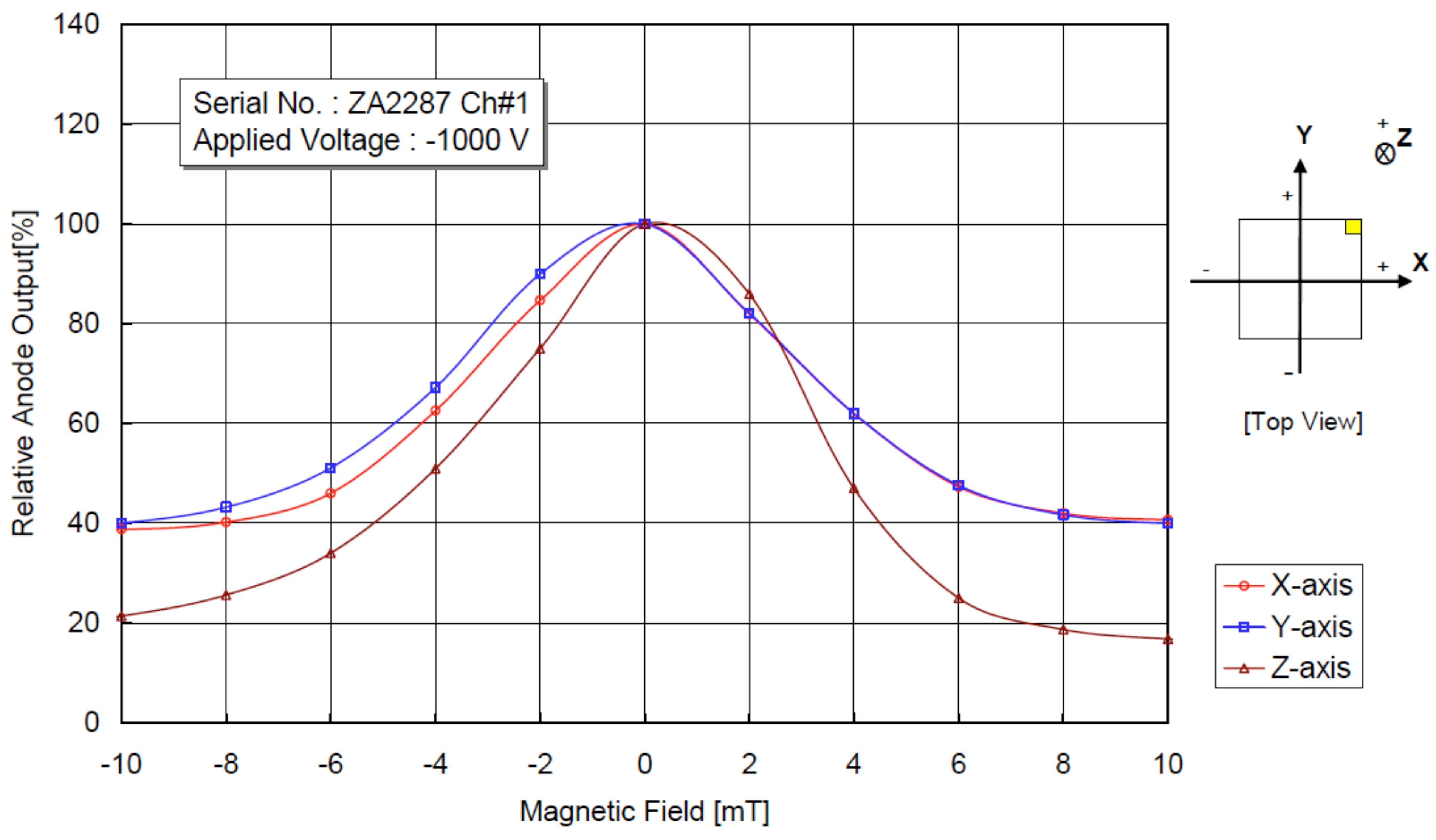}}
\hspace{5mm}
\subfloat[The effect on pixels near center.]{\includegraphics[width=0.465\linewidth]{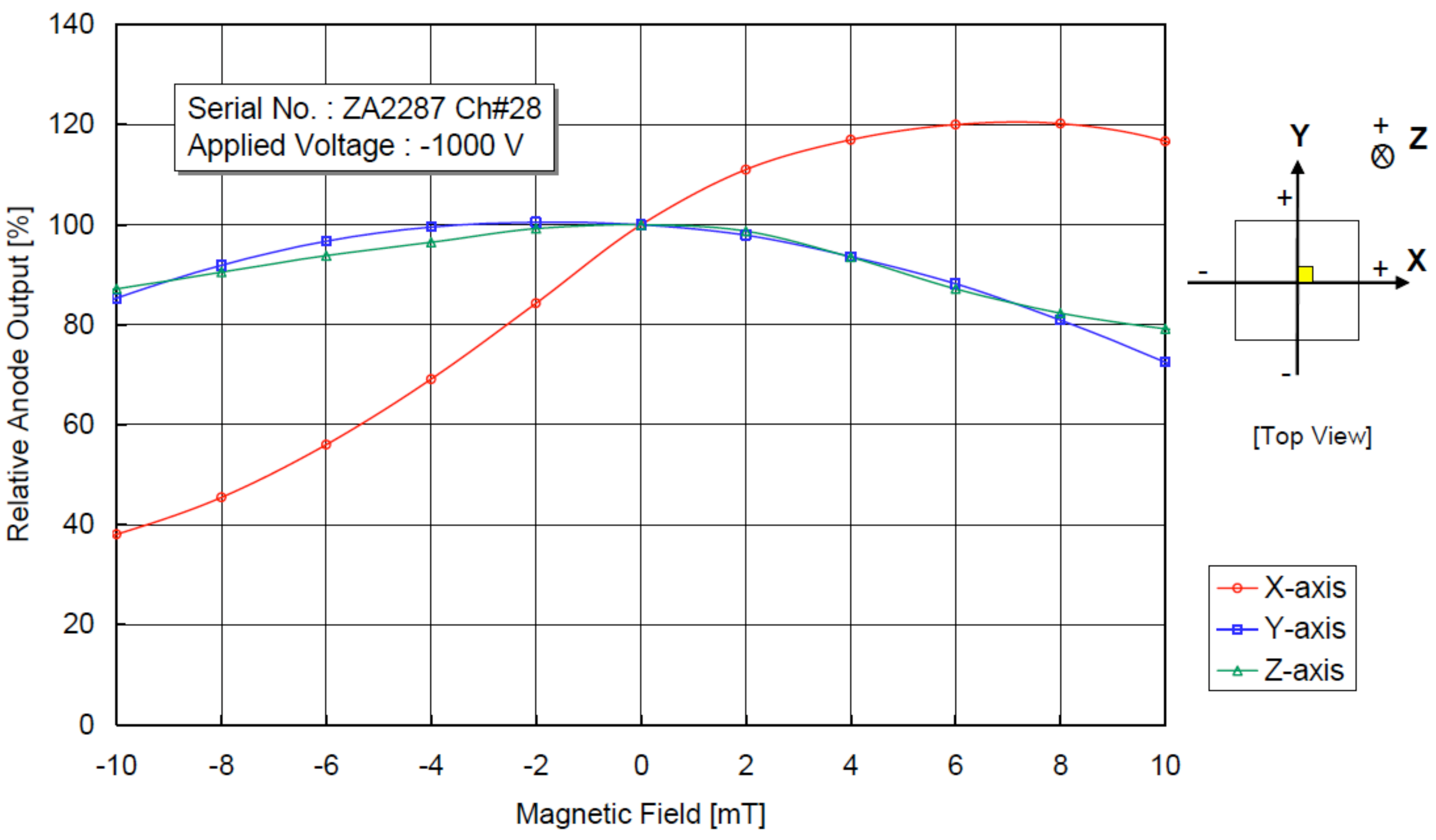}}
\end{center}
\caption{Magnetic field effect on the H-8500 MaPMT pulse height (Hamamatsu data).}
\label{fig:Mag_field}
\end{figure*}

For \babar\ DIRC PMTs, which have a classical PMT dynode design, it was necessary to keep the magnetic 
field below ${\sim} 1$~gauss in the SOB to prevent a serious degradation of pulse height spectra~\cite{Adam:2004fq}. 
To do that, it was necessary to enclose the entire photon camera into a large magnetic shield. 
Figure~\ref{fig:Mag_field} shows the effect of the magnetic field on the H-8500 tube pulse height. 
One can notice that the effect is different near the tube boundary compared to its central region. 
We conclude that we can tolerate a residual magnetic field up to a level of a few gauss with no effect 
on the pulse height. We plan to use a magnetic shield similar to that of \babar.

\tdrparagraph{ Radiation damage of optical components }

\begin{figure*}[tbp]
\begin{center}
\subfloat[Transmission of Epotek 301-2 epoxy (50-75 microns thick).]{{\includegraphics[height=0.195\textheight]{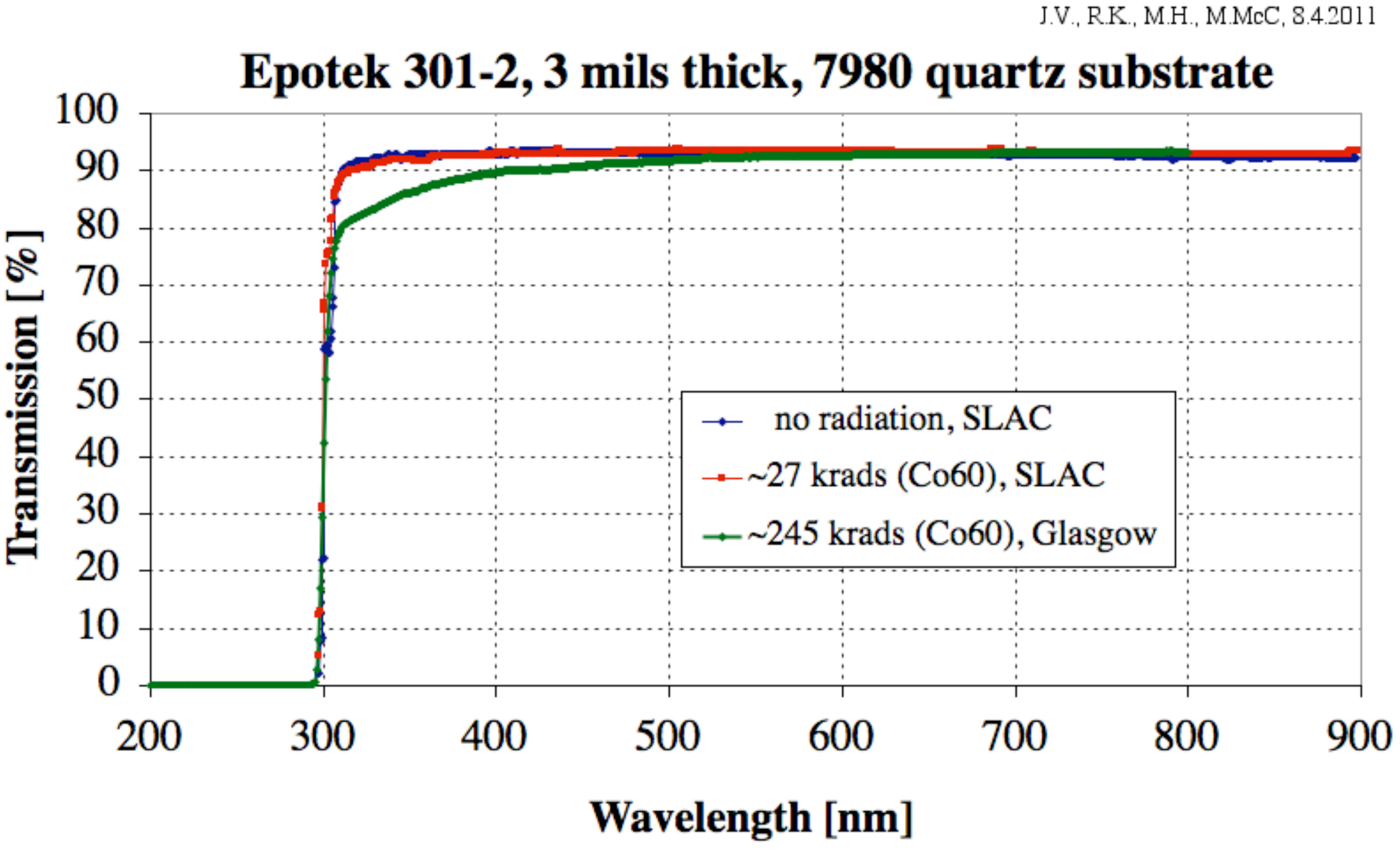}}}
\hspace{5mm}
\subfloat[Transmission of Shin-Etsu 403 RTV (1\mm thick).]{{\includegraphics[height=0.205\textheight]{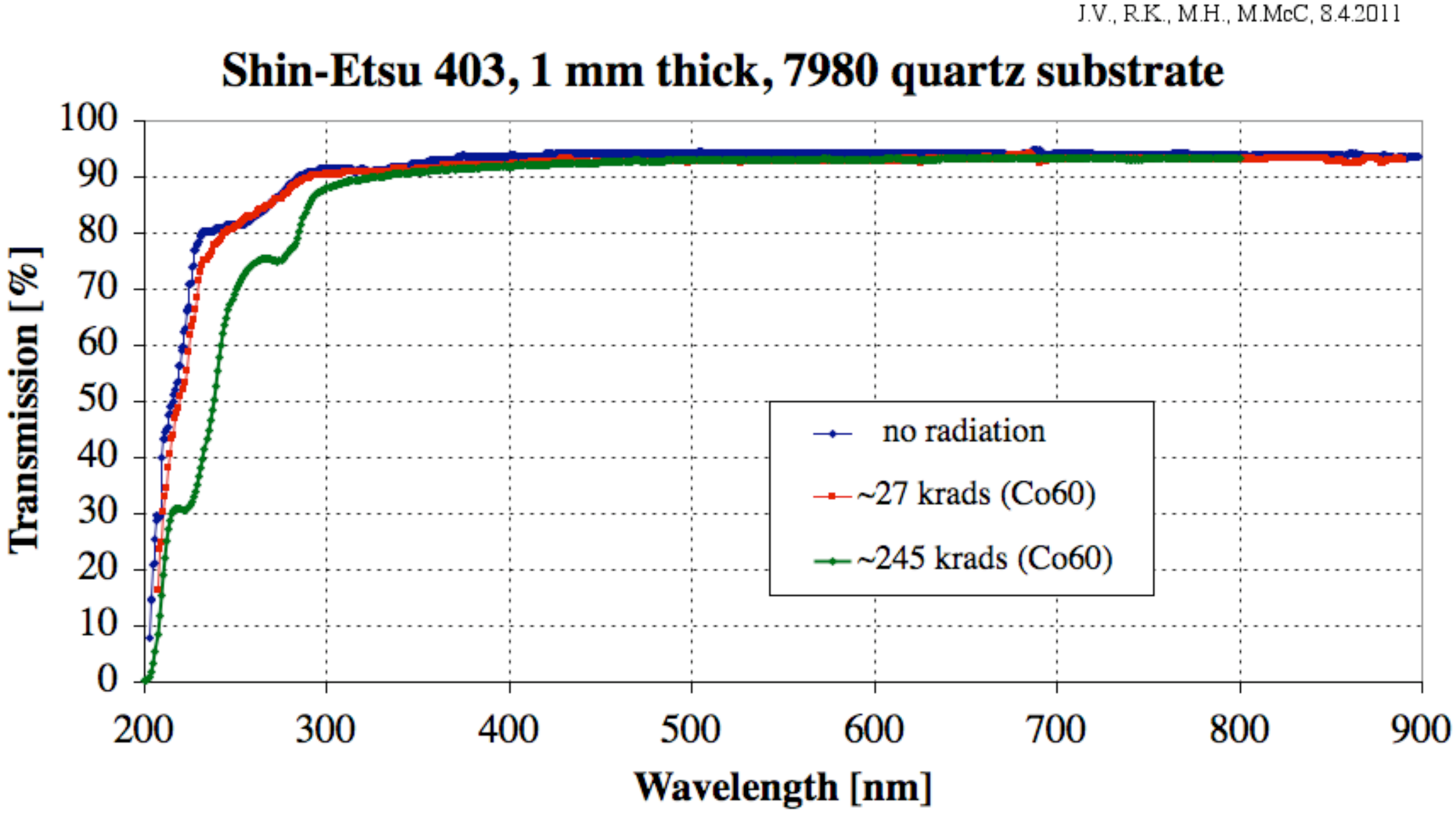}}}
\end{center}
\caption{Radiation damage by a $^{60}$Co source on glues used in the construction of the photon camera~\cite{jjv_london_2011}.}
\label{fig:Rad_damage}
\end{figure*}

 We used a $^{60}$Co source for the irradiation of the glue samples. First, we have investigated the radiation damage
of Corning 7980 Fused Silica 3\mm-thick coupons used for support of glue samples and, as expected, found no loss of
transmission up to 250\krad. Figure~\ref{fig:Rad_damage} shows the irradiation of the Epotek 301-2 epoxy, used for
coupling of the new Wedge to the bar box window, and the Shin-Etsu 403 RTV, used for coupling of the FBLOCK and the
new Wedge. We show that these glues are acceptable, although the Epotek 301-2 shows a loss of transmission 
at $\sim$~245~krad~\cite{jjv_london_2011}.

\tdrsubsec{ Laser calibration system  }

\begin{figure*}[tbp]
\begin{center}
\subfloat[Optical details of laser entry~\cite{jjv_elba_2011}.]{\includegraphics[height=0.23\textheight]{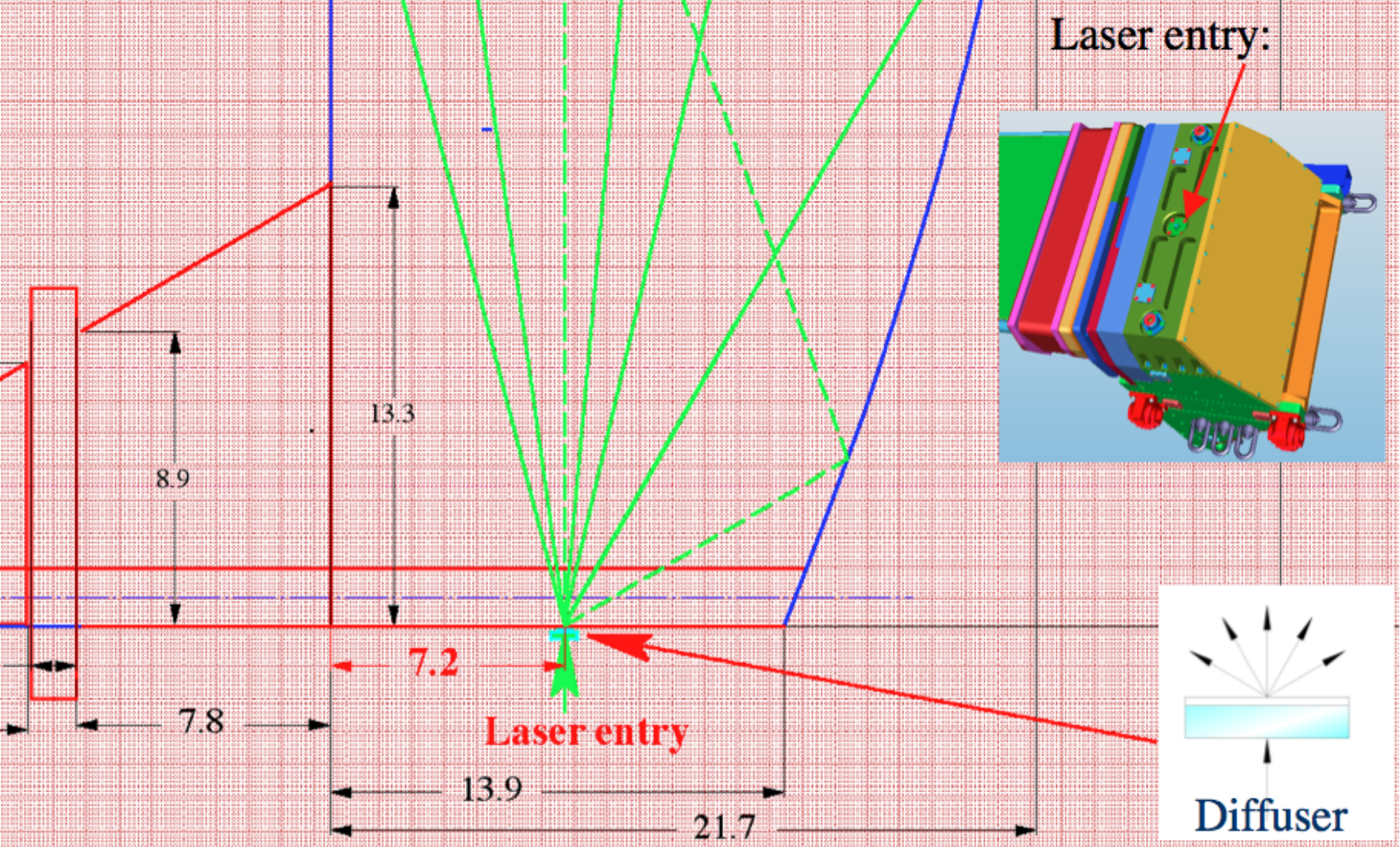} }
\hspace{5mm}
\subfloat[OPAL diffuser used to spray laser photons into the FBLOCK.]{\includegraphics[height=0.23\textheight]{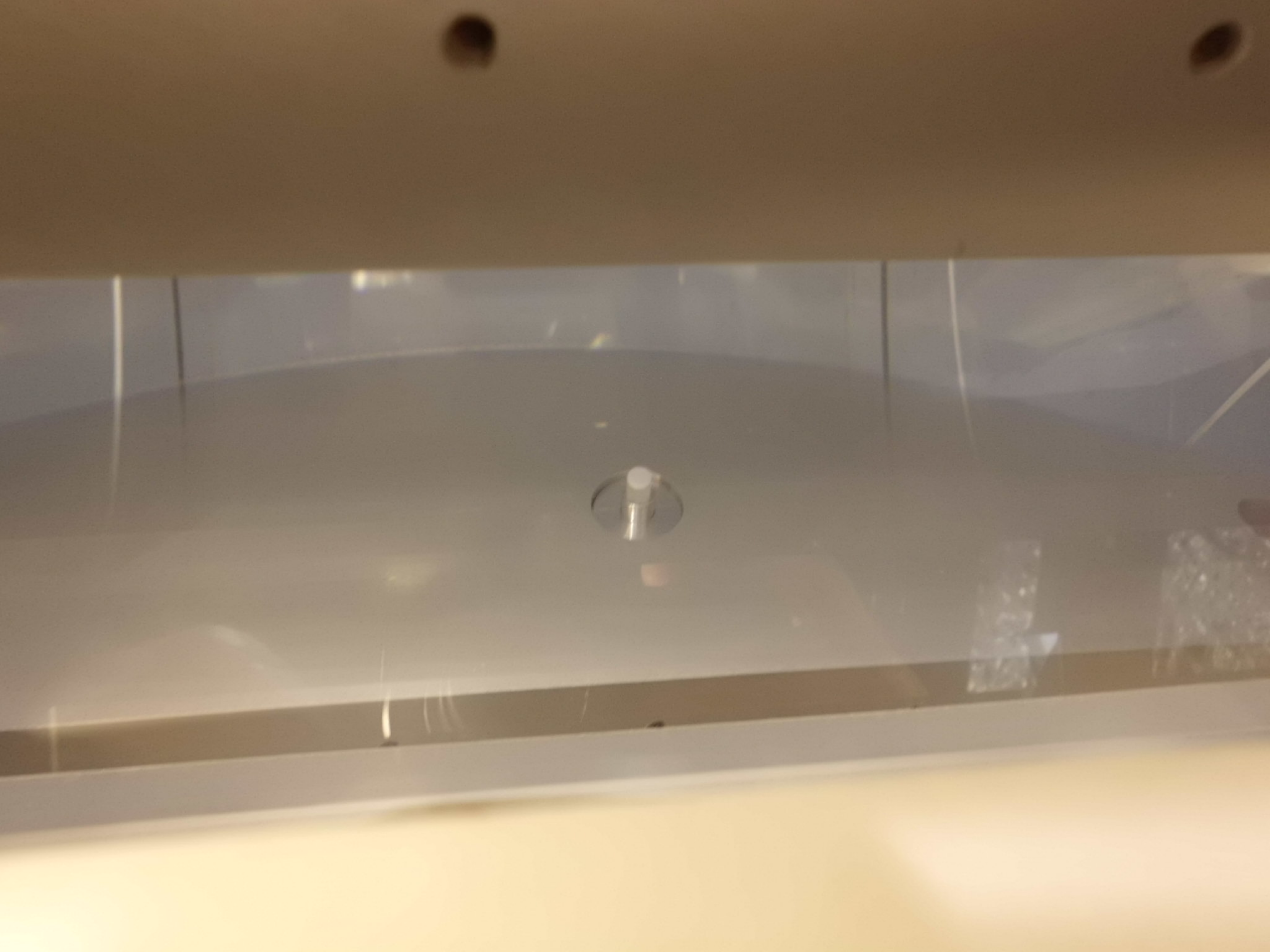}}
\end{center}
\caption{Laser entry into the FBLOCK.}
\label{fig:laser_entry}
\end{figure*}

\begin{figure}[!h]
\begin{center}
\includegraphics[width=\linewidth]{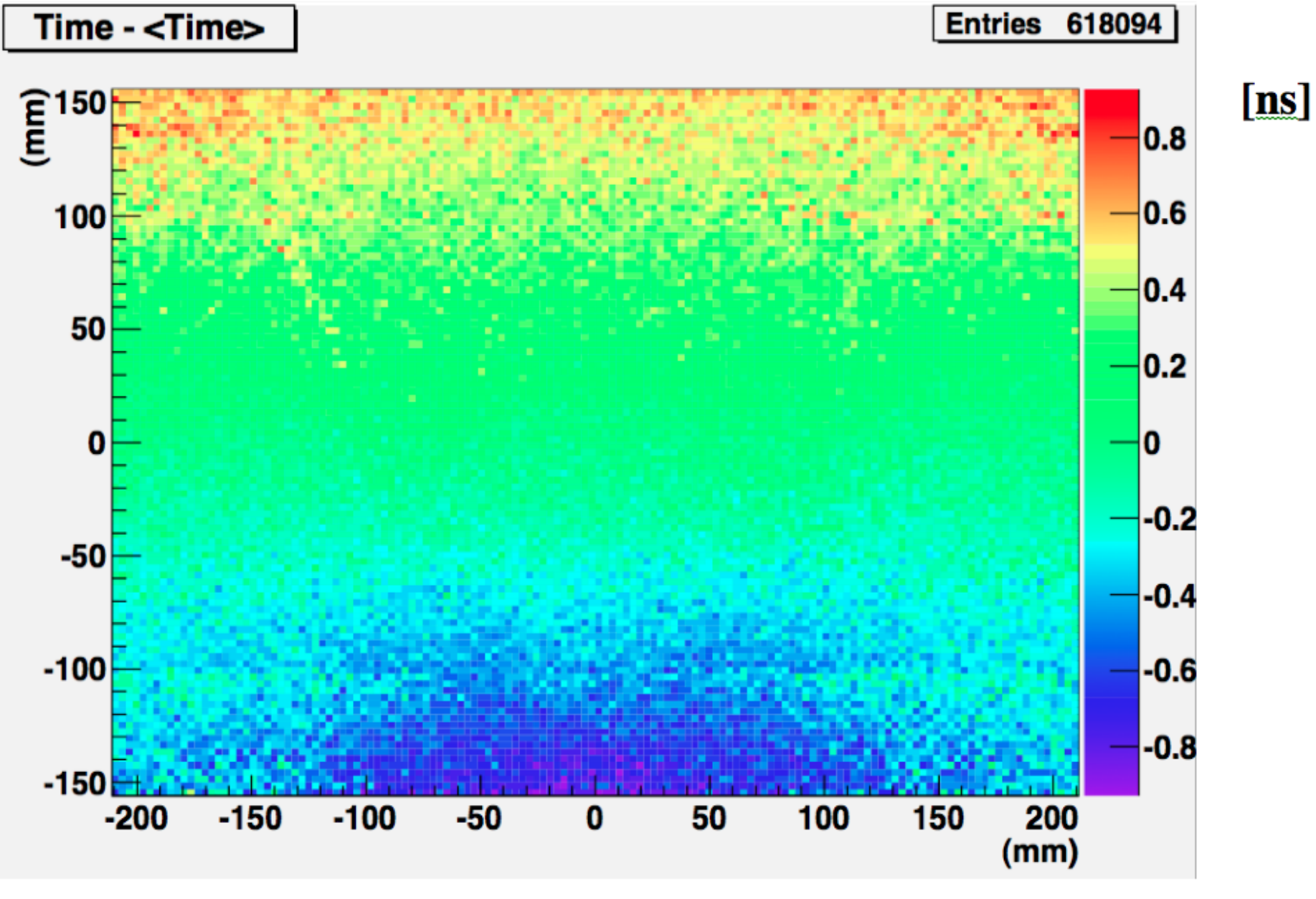}
\caption{The laser time spread across the focal plane is at a few ns level~\cite{doug}.}
\label{fig:laser_timing}
\end{center}
\end{figure}

\tdrparagraph{ Optics of calibration }
  The aim of this calibration is twofold: (a) to check the operation of tubes and electronics, (b) to provide pixel offset constants for 
FDIRC timing calibration, which was found to be useful in the first FDIRC prototype when doing the chromatic corrections~\cite{jjv_elba_2011}. 
Figure~\ref{fig:laser_entry} shows the laser entry into the FBLOCK as implemented in the final FDIRC prototype. The fiber plugs 
into a connector with a lens (F230FC-A), which makes a parallel laser beam, which then strikes a 5\mm diameter Opal diffuser, which 
was selected out of several choices for its uniform light diffusing effect. The small diameter diffuser is necessary to limit losses 
of real Cherenkov photons. We found experimentally that the best arrangement is if the diffuser is pressing against the bottom surface 
of the FBLOCK with the help of a spring (no gluing as it affects the uniformity of the scattered light). There is 
one fiber entry per photon camera serving one bar box. Figure~\ref{fig:laser_timing} shows a MC simulation indicating that 
there is a time spread across the focal plane of about 2~ns. The aim is to determine this offset for each pixel and correct any 
deviation from the expected time. One this correction is determined, we align all pixels to a single reference time, called $t_0$. 

\tdrparagraph{ Laser and fiber optics choice } 
   We will use PiLas laser diodes providing a light of 407\nm wavelength. We would like to split the light from one PiLas source into 2 branches. 
Each branch has to be adjusted to provide the same intensity, which should be low enough to generate single photoelectrons in each pixels. 
If this works, we will need six PiLas control units serving the entire system. 

\tdrsubsec{ FDIRC Mechanical Design }

\tdrparagraph{ Description of \babar\ bars, bar boxes }
   We will reuse the DIRC bar boxes. They will not be modified as it is considered too difficult to do, as discussed before. 
One potential problem is that the Epotek~301-2 glue has seen ${\sim} 10$ years of radiation during the \babar\ experiment. 
Extensive studies were performed with the \babar\ di-muon data and no detrimental effect was found on the glue 
transmission~\cite{nicolas_2009}. However, we do need to be extra careful when transporting bar boxes, as it is not known 
if the Epotek~301-2 glue strength was not affected. Similarly we will have to verify that 
Hexel panels used to build bar boxes can be transported by air (their Hexel-cells are sealed by glue and therefore will be affected by the pressure variation).

\begin{figure}[tbp]
\begin{center}
\subfloat[\babar\ DIRC bar box.]{\includegraphics[width=\linewidth]{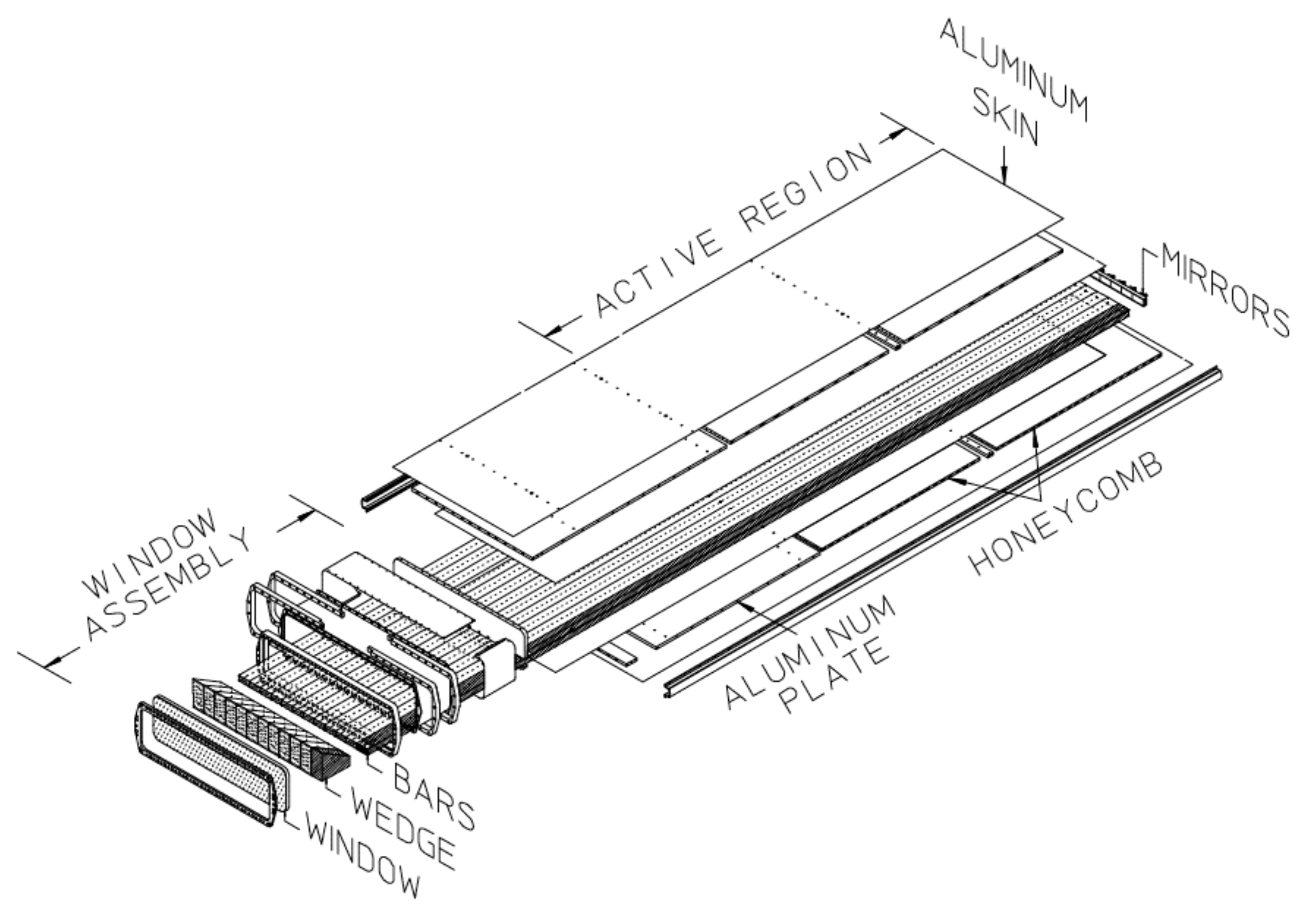} }
\hspace{5mm}
\subfloat[One bar segment with nominal dimensions.]{\includegraphics[width=\linewidth]{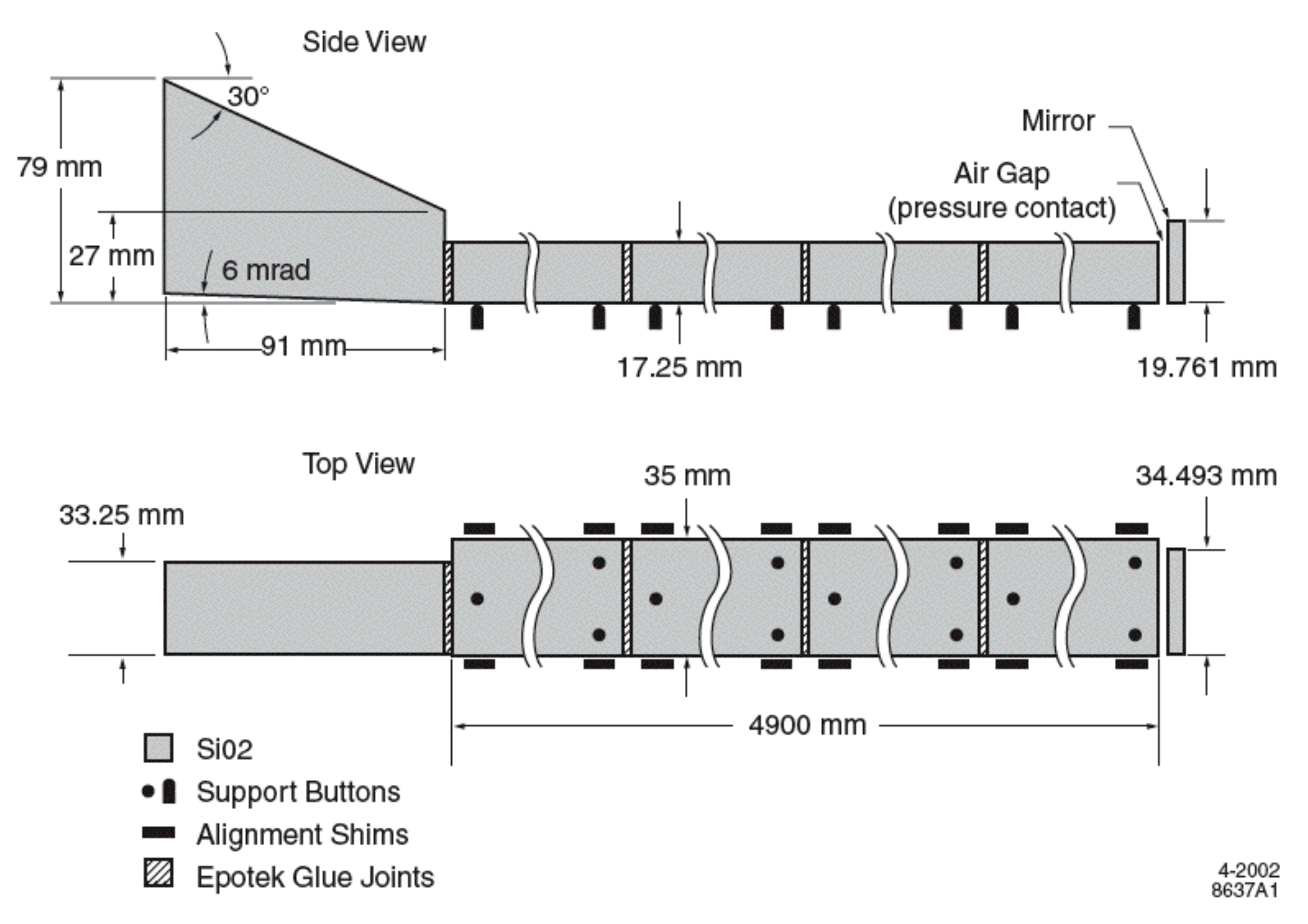}}
\end{center}
\caption{Bar box and bar dimensions~\cite{Adam:2004fq}.}
\label{fig:BaBar_bar_box}
\end{figure}

Figure~\ref{fig:BaBar_bar_box} shows the \babar\ DIRC bar box with its 12 Fused silica bars~-- each glued out of four 122~cm-long bar 
segments~-- and the nominal dimensions of each bar including  the wedge~\cite{Adam:2004fq}. In the real life, it is 
somewhat more complicated, as bar dimensions vary and each bar box is slightly different. 
This has been recorded in spreadsheets~\cite{jjv_1999}. Figure~\ref{fig:DIRC_bar_box_bars} shows the cross-section 
of a bar box containing 12 fused silica bars. Figure~\ref{fig:DIRC_bar_mirror} shows the bar forward end with a mirror. 
There are altogether 12 bar boxes and 144 full-length bars in the entire system. 

\begin{figure}[tbp]
\begin{center}
\includegraphics[width=\linewidth]{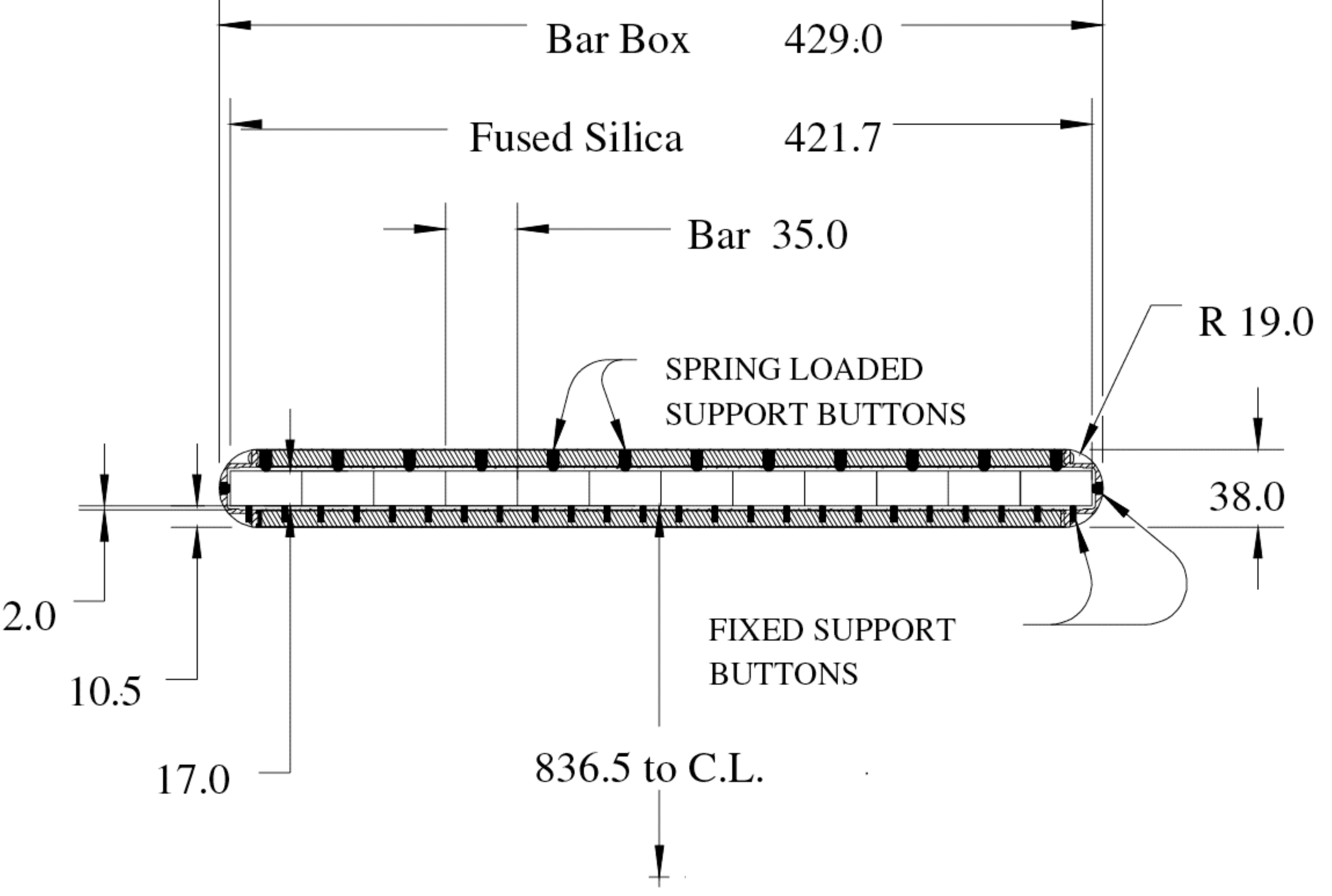}
\caption{Cross-section of a bar box with its 12 fused silica bars~\cite{Adam:2004fq}.}
\label{fig:DIRC_bar_box_bars}
\end{center}
\end{figure}

\begin{figure}[tbp]
\begin{center}
\includegraphics[width=\linewidth]{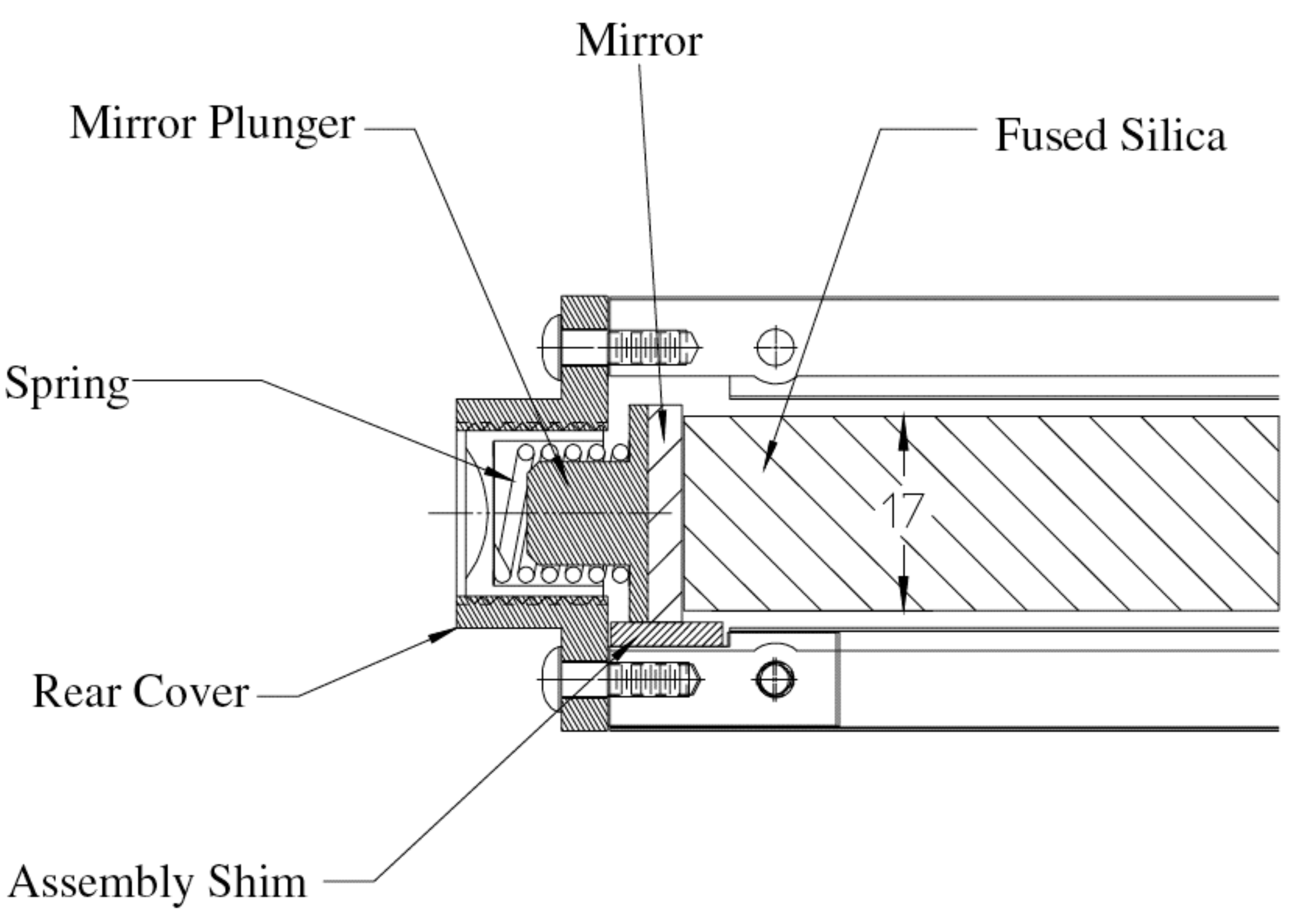}
\caption{Bar end with a mirror (forward side of \superb)~\cite{Adam:2004fq}.}
\label{fig:DIRC_bar_mirror}
\end{center}
\end{figure}

\tdrparagraph{ Fused silica optics: New Wedge and FBLOCK }

The new Wedge and the FBLOCK will most probaby be made of radiation hard Fused silica Corning 7980~--
this is the material used for the new FDIRC prototype which tests are starting at SLAC.
Corning Co. makes fused silica 7980 material in
a form of boules of up to 60" diameter~-- see Figure~\ref{fig:Corning_material_a}. The striae are running 
typically perpendicular to axis of a 60 inch dia. boule. The best homogeneity of the refraction index $dn/n$ is along the axis of the boule. 
There are two types of 7980 material: (a) standard (much less characterized and therefore requiring more checks), 
and (b) the so-called KrF (very characterized material; Corning qualifies striae with an optical interferometer). 
We chose the "standard" material, as the cost of the KrF material is about 2-3 times higher. For the standard
fused silica material these are typical specifications: (a) $dn/n$ is less than 1~ppm over the scale of a mm; 
(b) the bottom-to-top of the boule along the axis: the refraction index uniformity $dn/n$ is less than 5~ppm at 
200~nm and better at longer wavelengths; c) $dn/n$ is about 5-7~ppm in the direction perpendicular to the boule axis. 
Part-to-part variation is expected at a level of $dn/n {\sim} 20$~ppm in the visible wavelength range.

One should avoid the very bottom and top of the boule as there could be larger stria. Therefore, one should buy
thicker boule, and this issue should be remembered for the final production. We visited this company and tested
the material for stria with a laser. None was detected.
Out of one boule, one expects to make three blocks such as the one shown in Figure~\ref{fig:Corning_material_b}. One has to pay attention to 
orientation of the FBLOCK within the raw block. The back side of the FBLOCK needs to be at the bottom of
the boule as contamination from sand used as a seed of fused silica material deposition is more likely in this area.

\begin{figure}[!h]
\begin{center}
\includegraphics[width=\linewidth]{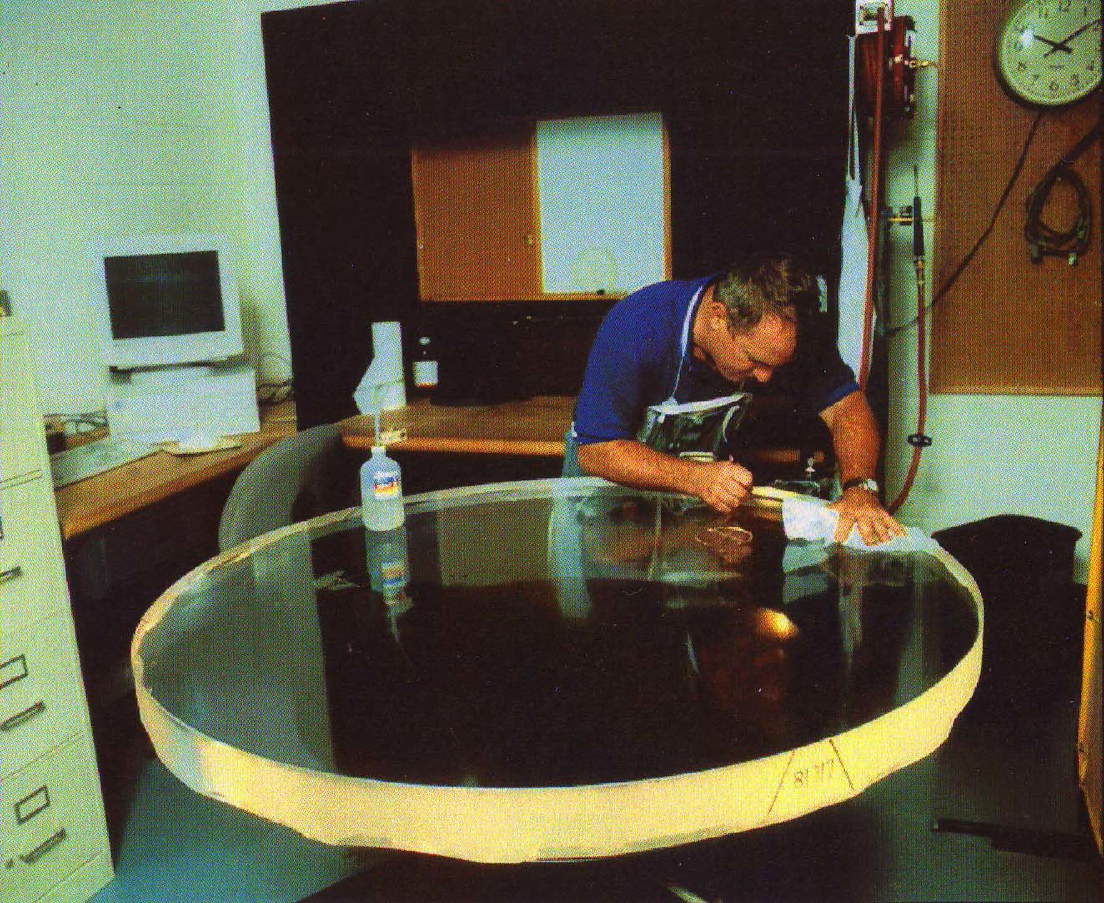}
\caption{An example of the fused silica 7980 material in the form of a 60 inch dia. boule made by Corning.}
\label{fig:Corning_material_a}
\end{center}
\end{figure}

\begin{figure}[tbp]
\begin{center}
\includegraphics[width=\linewidth]{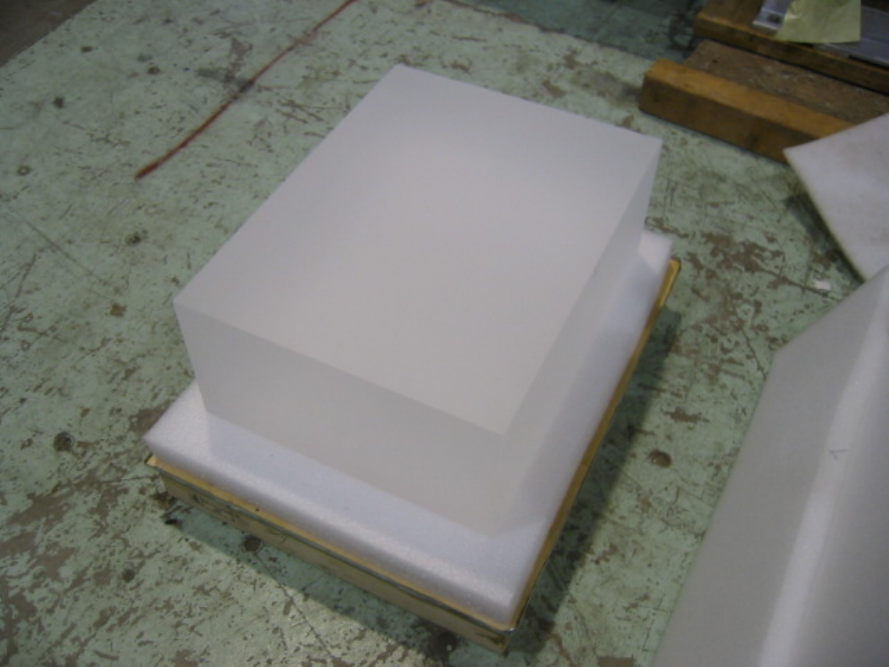}
\caption{Fused silica 7980 material in a form of block ready for machining; the FLOCK used for the FDIRC prototype has been machined from this piece of quartz.}
\label{fig:Corning_material_b}
\end{center}
\end{figure}

The manufacturing was split into four steps done in three different companies: (a) grinding final shapes about 1-2\mm oversized, 
(b) polishing to final size and surface polish of better than 30~\AA\ rms, (c) coating two FBLOCK reflecting surfaces with 
aluminum with SiO$_2$ overcoat to protect it, and (d) the final quality control (QC) of the completed pieces. Figure~\ref{fig:New Camera} shows the finished 
new Wedge and FBLOCK (before the two mirror plating step). These optical pieces were successfully produced, which demonstrates that the
new camera optics is doable. However, there is a number of critical steps where error can be made, for example: (a) damages when handling,
(b) surface pollution either before plating or in a final assembly, (c) stria problem needs to be checked, (d) accidental swaps of correct and wrong
materials, etc.

\begin{figure*}[tbp]
\begin{center}
\subfloat[FBLOCK after polishing but before plating.]{\includegraphics[height=0.35\textheight]{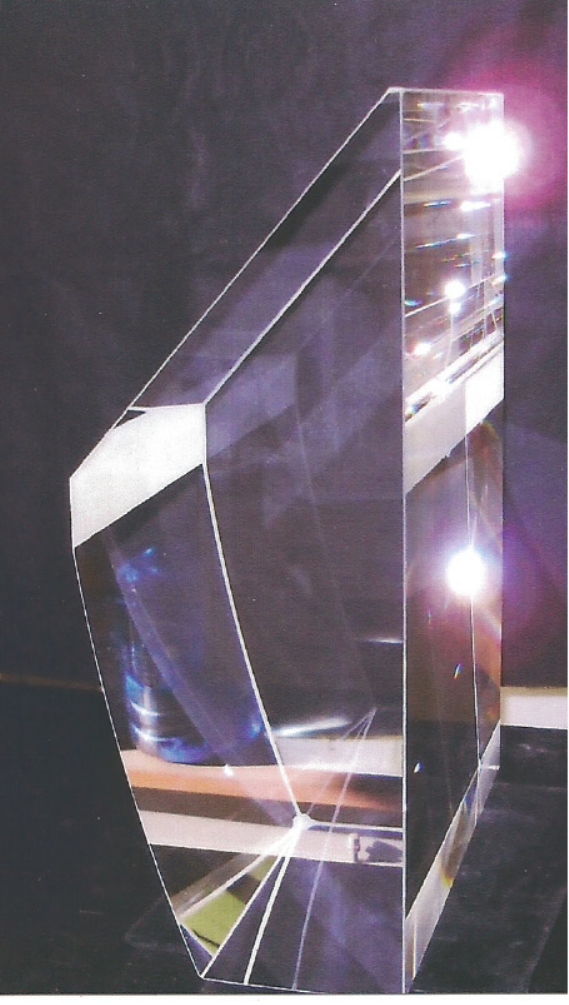} }
\hspace{5mm}
\subfloat[New Wedge after polishing.]{\includegraphics[height=0.35\textheight]{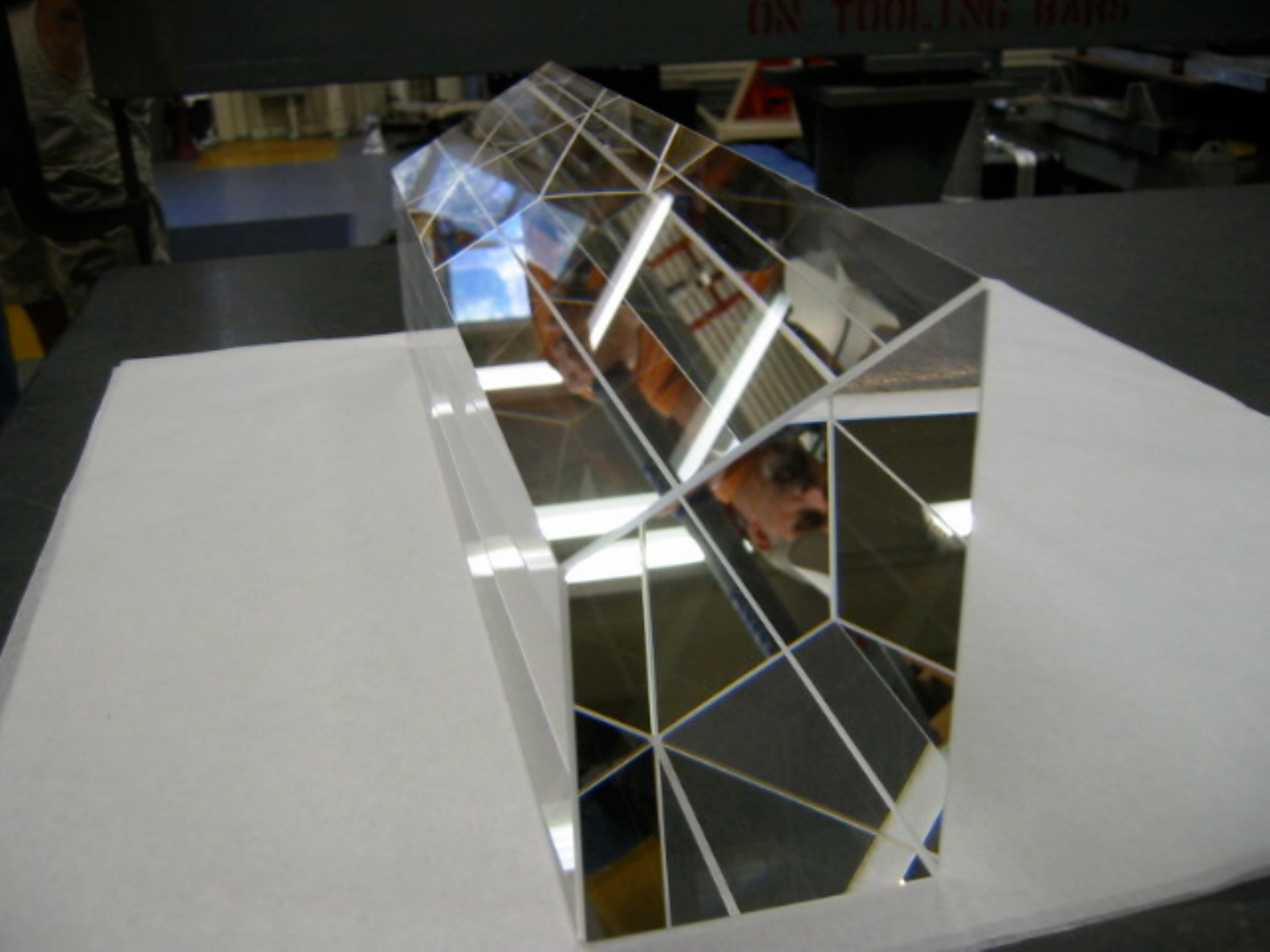}}
\end{center}
\caption{New photon camera parts: New Wedge and FBLOCK.}
\label{fig:New Camera}
\end{figure*}

\tdrparagraph{ FBLOCK mirror surfaces }
   It is absolute mandatory to have a very good surface cleanliness before the aluminum plating is attempted. Any contamination
will result in peeling problems. The two FBLOCK aluminum-plated mirror surfaces are protected by a SiO$_2$ layer. 
Even though there is this protection layer, mirror surfaces are still fragile, especially if the surface is polluted during handling. 

Another complicated issue is the FBLOCK shipping from the polishing company to the laboratory where it will be used.
The FBLOCK is very heavy; its polished 
surfaces and the two mirrored sides can be easily damaged by a rubbing motion created by shipping. Surfaces have to be protected 
by a plastic film during the shipment, but the film must not stick to mirror surface to cause peeling problems. Based on our tests, 
we have decided to use the Grafix plastic vinyl film in future, which adheres to surfaces via electrostatic forces, does 
not remove plated layer, and does not leave a surface pollution, which would be difficult to clean.

\tdrparagraph{ Gluing the new Wedge to the Bar Box Window }

\begin{figure*}[tbp]
\begin{center}
\subfloat[The bar box with the new Wedge in the SLAC clean room.]{\includegraphics[width=0.465\linewidth]{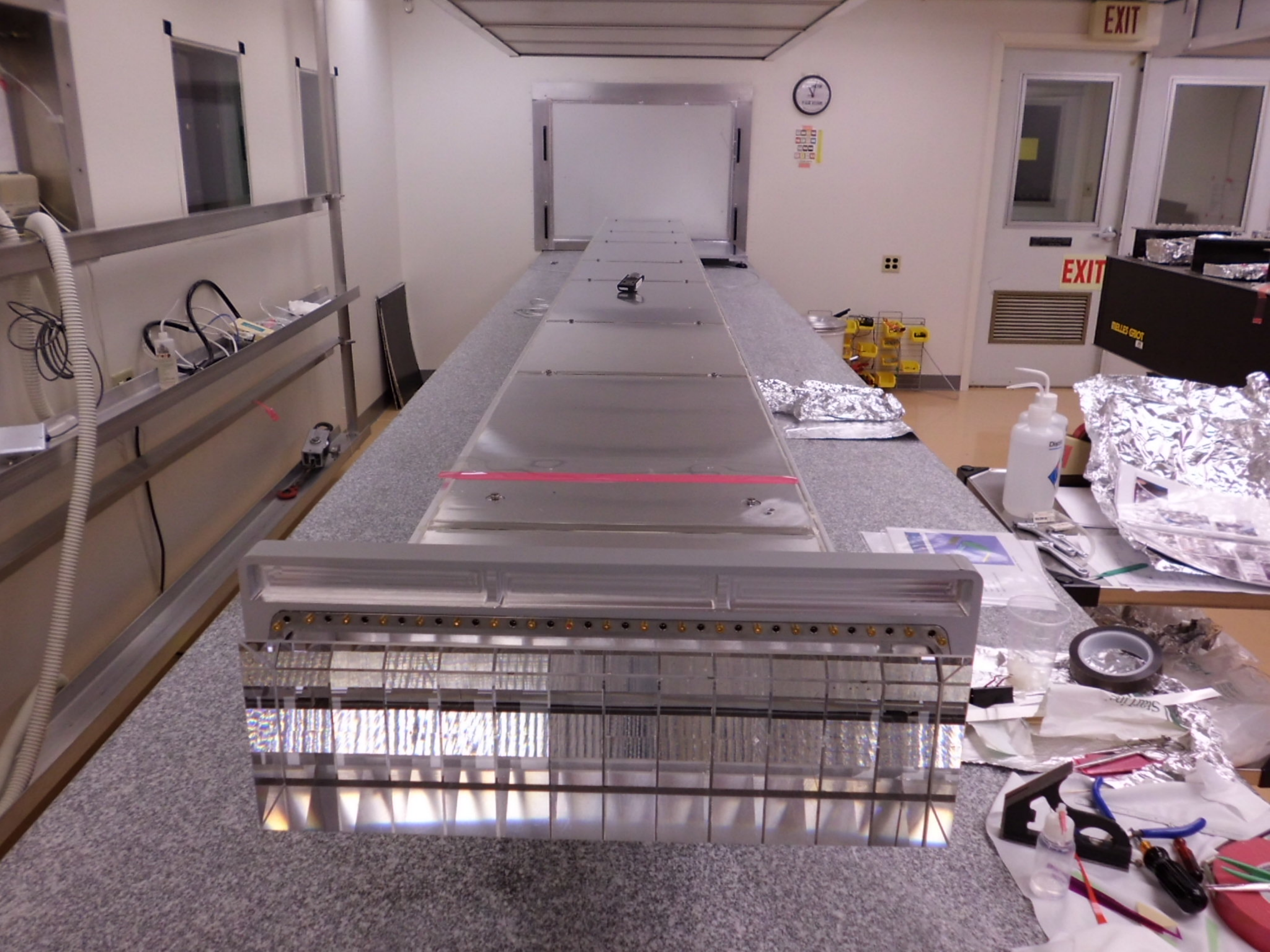} }
\hspace{5mm}
\subfloat[Detailed view of the new Wedge and the bar box window.]{\includegraphics[width=0.465\linewidth]{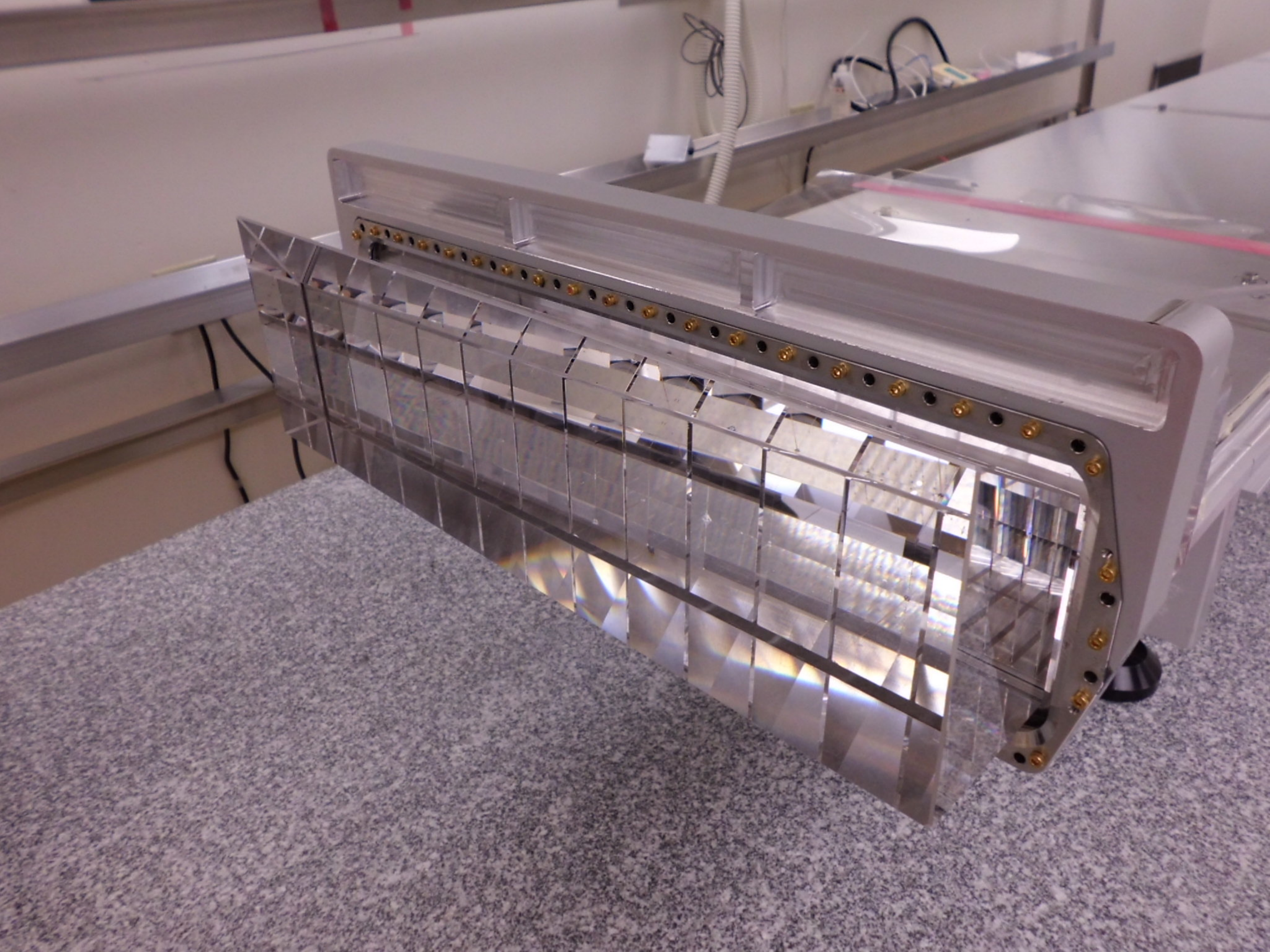}}
\end{center}
\caption{Coupling of the new Wedge to the bar box.}
\label{fig:Wedge_support}
\end{figure*}

 This optical coupling of Wedge to bar box window is done in the clean room. Figure~\ref{fig:Wedge_support} shows a detail of 
coupling of the new Wedge to the bar box. The coupling is done with the Epotek 301-2 optical epoxy of 25-50 micron thickness. 
The bottom of the new Wedge is aligned to the bottom bar surface, \ie, not to the old wedge as it has a ${\sim} 6\mrad$ angle. 
The new Wedge is centered left-right in the bar box window. This coupling is not possible to remove in the future, as one
would risk damaging the bar box. 

\tdrparagraph{ FBLOCK mechanical enclosure: the Fbox }

\begin{figure*}[tbp]
\begin{center}
\subfloat[Various components for the optics enclosure.]{\includegraphics[width=0.465\linewidth]{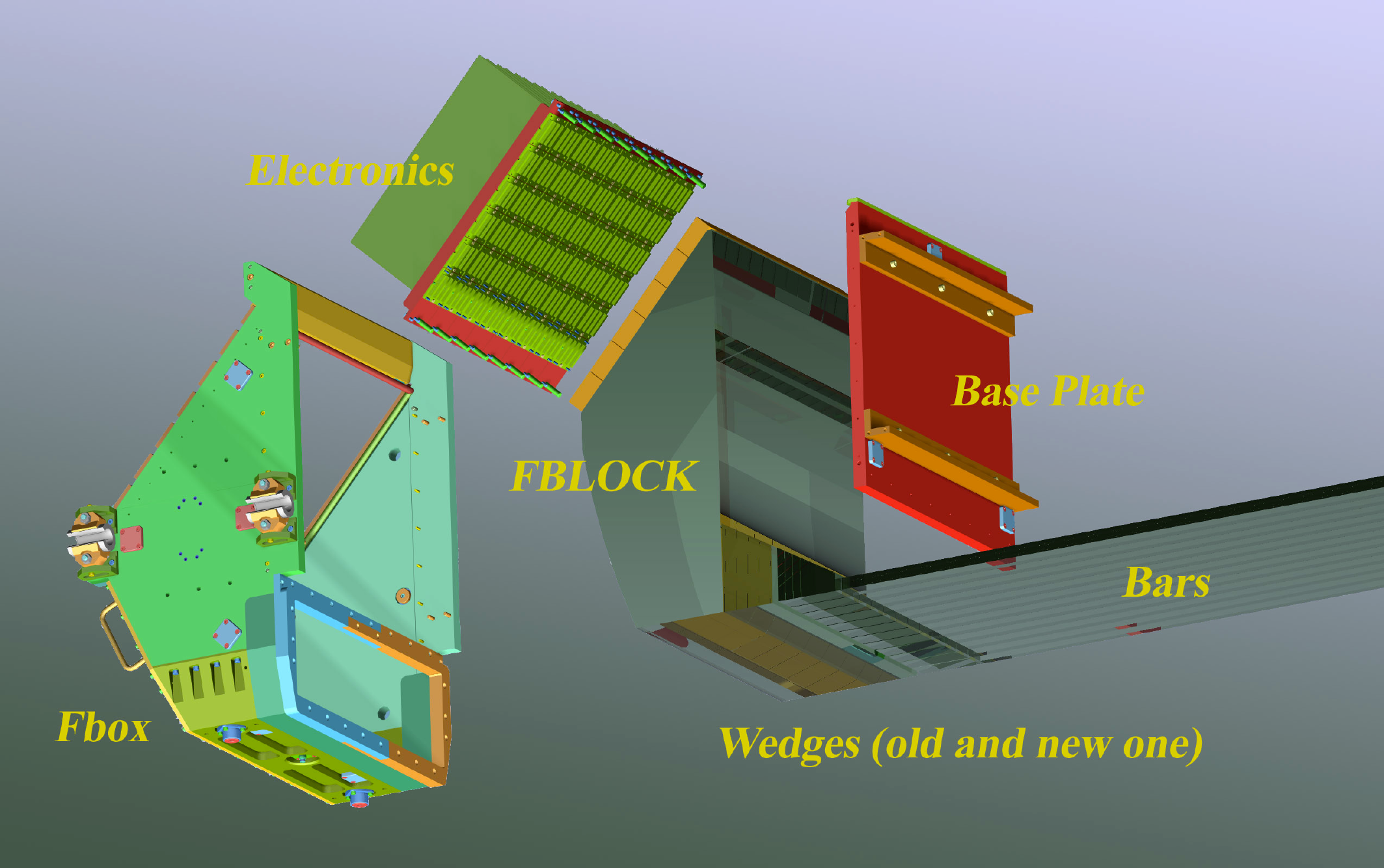} }
\hspace{5mm}
\subfloat[Complete Fbox enclosure, including bar box.]{\includegraphics[width=0.465\linewidth]{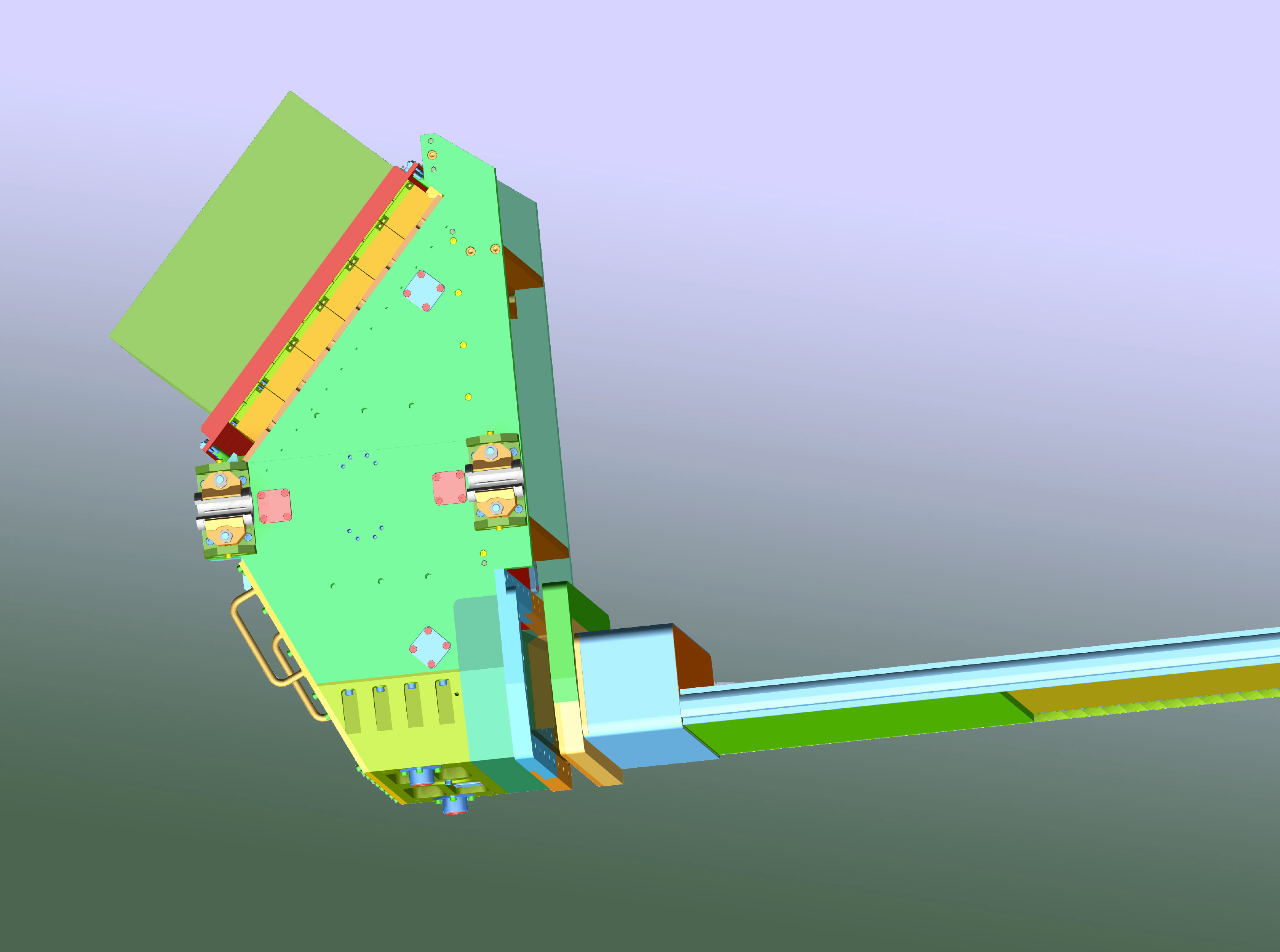}}
\end{center}
\caption{Fbox enclosure of FBLOCK optics, including wedges, bars, detectors and electronics.}
\label{fig:Fbox_enclosure}
\end{figure*}

\begin{figure*}[tbp]
\begin{center}
\subfloat[Top view.]{\includegraphics[width=0.465\linewidth]{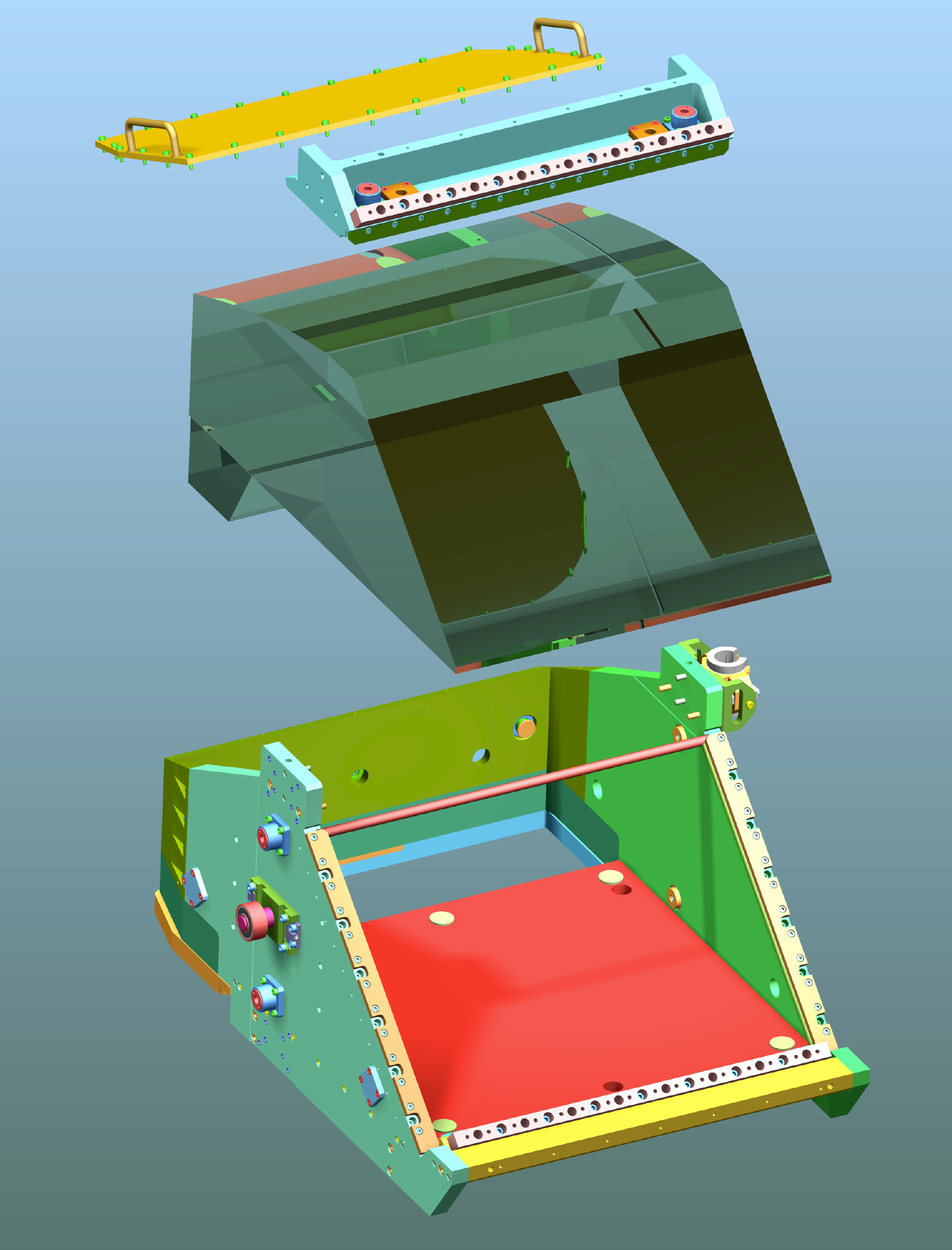} }
\hspace{5mm}
\subfloat[Bottoms view.]{\includegraphics[width=0.465\linewidth]{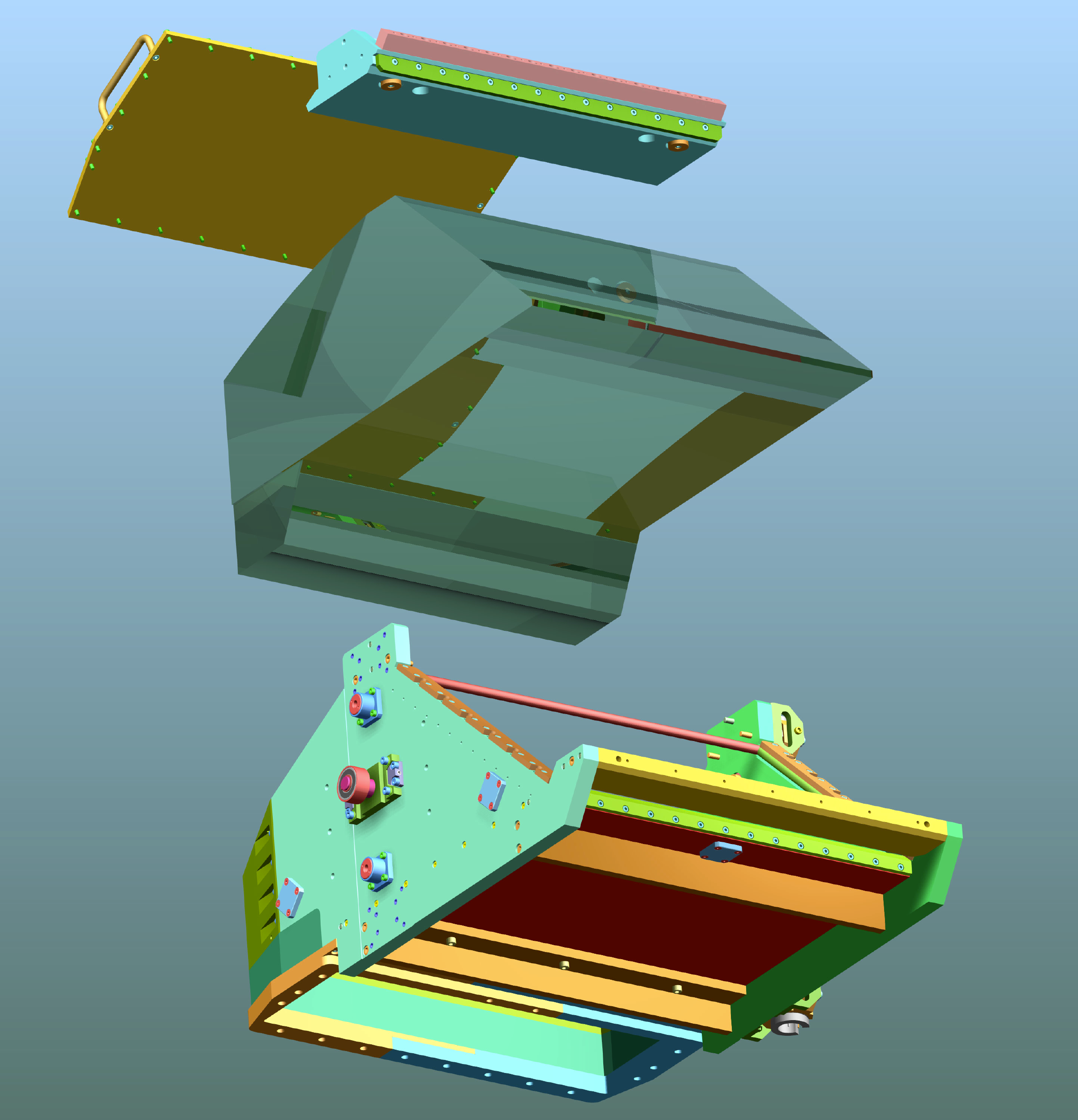}}
\end{center}
\caption{Button support of FBLOCK optics in Fbox.}
\label{fig:FBLOCK_support_a}
\end{figure*}

\begin{figure}[!h]
\begin{center}
\includegraphics[width=\linewidth]{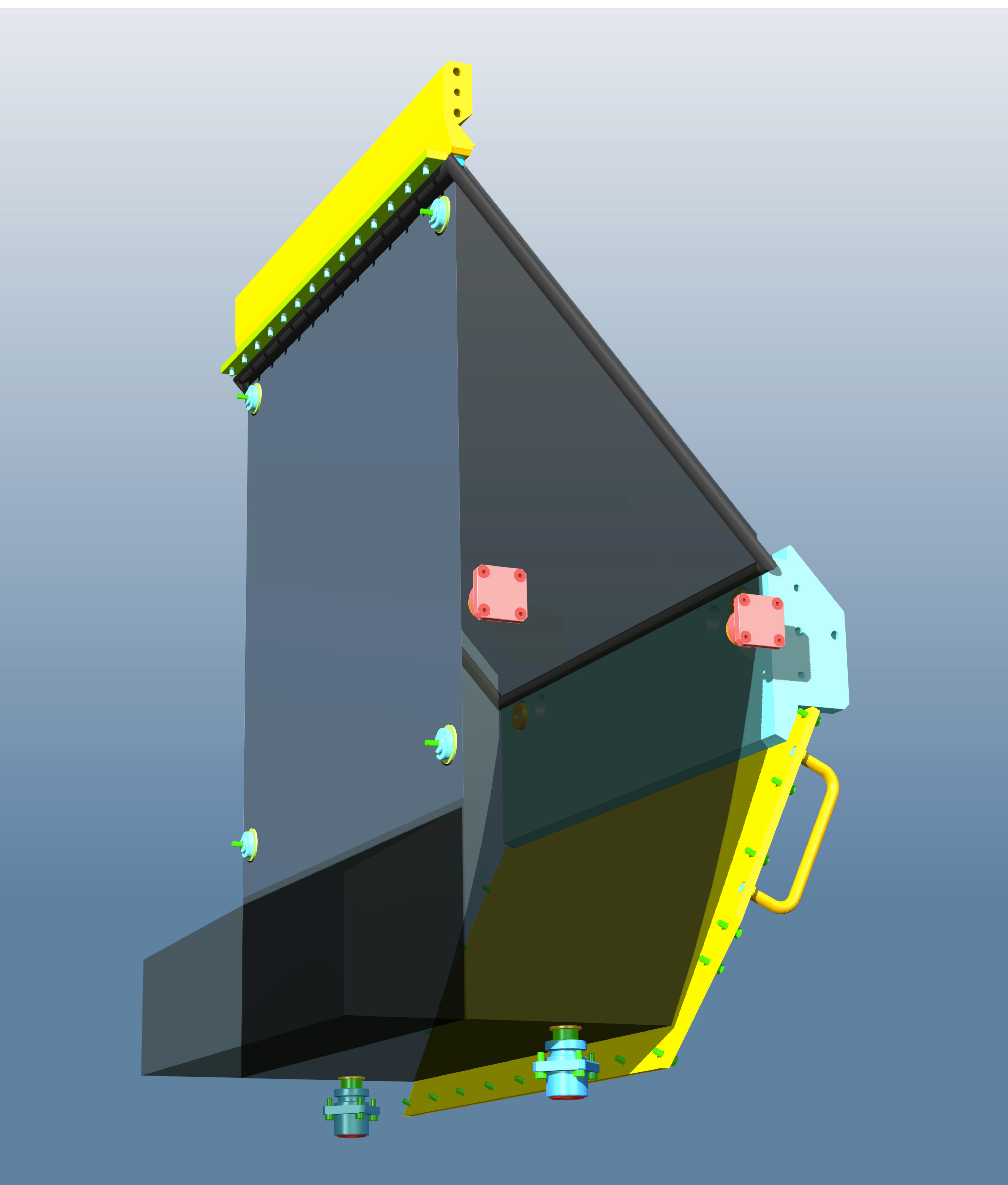}
\caption{FBLOCK support buttons: the small buttons are fixed while the larger ones (two from the bottom and two from one side) are adjustable 
to keep the FBLOCK stable even if the Fbox, made of aluminum, expands due to thermal effects.}
\label{fig:FBLOCK_support_b}
\end{center}
\end{figure}

 Figure~\ref{fig:Fbox_enclosure} shows the Fbox enclosure of the optics. Figures~\ref{fig:FBLOCK_support_a} 
and ~\ref{fig:FBLOCK_support_b} show details of how the FBLOCK is supported by plastic buttons. Plastic buttons, 
made of PET (polyethylene terephthalate), prevent FBLOCK optical surfaces from touching the aluminum surface 
of Fbox. Some buttons are fixed and some are spring-loaded. The spring loading is made using a stack of 8 belleville washers, 
which is the most compact way to produce predictable force. They are set to offset the total weight of FBLOCK and to take 
into account thermal effects. Placing the FBLOCK into the Fbox requires a very careful procedure as it is very heavy (${\sim} 80$~kg) 
and easy to be damaged. It was very useful to work out a step-by-step procedure~\cite{massimo_elba_2011} with a dummy plastic 
FBLOCK~\cite{mazziotta_london_2011}.

We already have the experience of putting together the real 
photon camera with the Fused Silica FBLOCK. Figure~\ref{fig:FBLOCK_assembly_a}(a) shows the first step of Fbox assembly where we 
placed the FBLOCK on four plastic support buttons. We have chosen a four-point support rather than a three-point one because it 
was judged to be easier to place the FBLOCK on the Fbox base plate, which is actually a very tricky operation. The front 
mirror surface is protected by four quartz coupons about 1.5\mm-thick, glued to the flat mirror surface by Epotek 301-2 epoxy. 
These four coupons are then touching plastic buttons located in the Fbox. The idea is that a rubbing motion due to thermal 
effects will be better dealt with if plastic buttons slide on quartz coupons rather than on the mirror plating directly. 
Figure~\ref{fig:FBLOCK_assembly_a}~(b) shows a fully assembled Fbox with the real Fused Silica FBLOCK.  

  The Fbox and its bar box have to be optically coupled. Figure~\ref{fig:FBLOCK_assembly_b} shows an example of how this is done in 
the CRT setup.  

\begin{figure*}[tbp]
\begin{center}
\subfloat[FBLOCK placed on the base plate of Fbox.]{\includegraphics[width=0.465\linewidth]{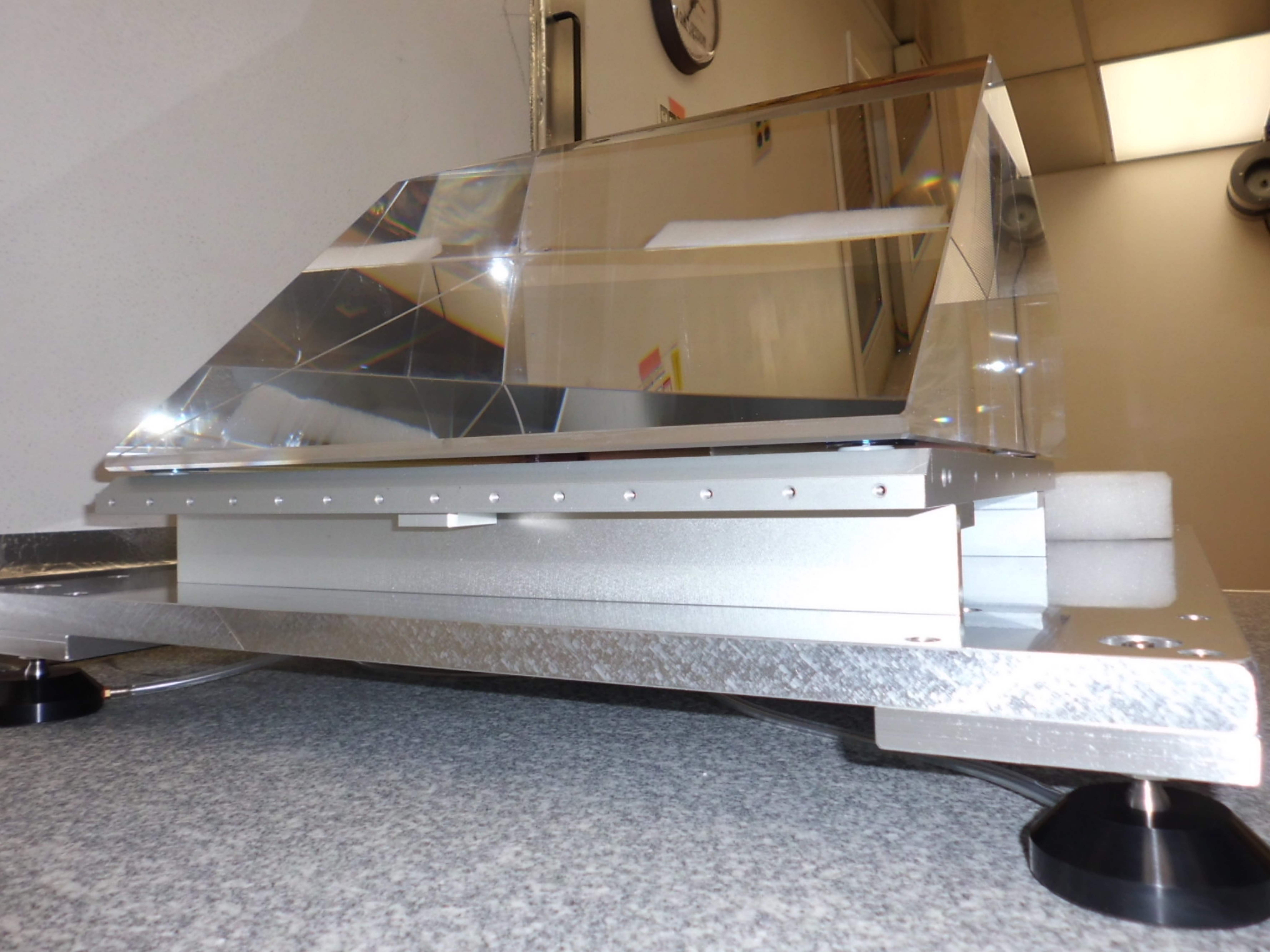} }
\hspace{5mm}
\subfloat[Assembled Fbox in front of bar box.]{\includegraphics[width=0.465\linewidth]{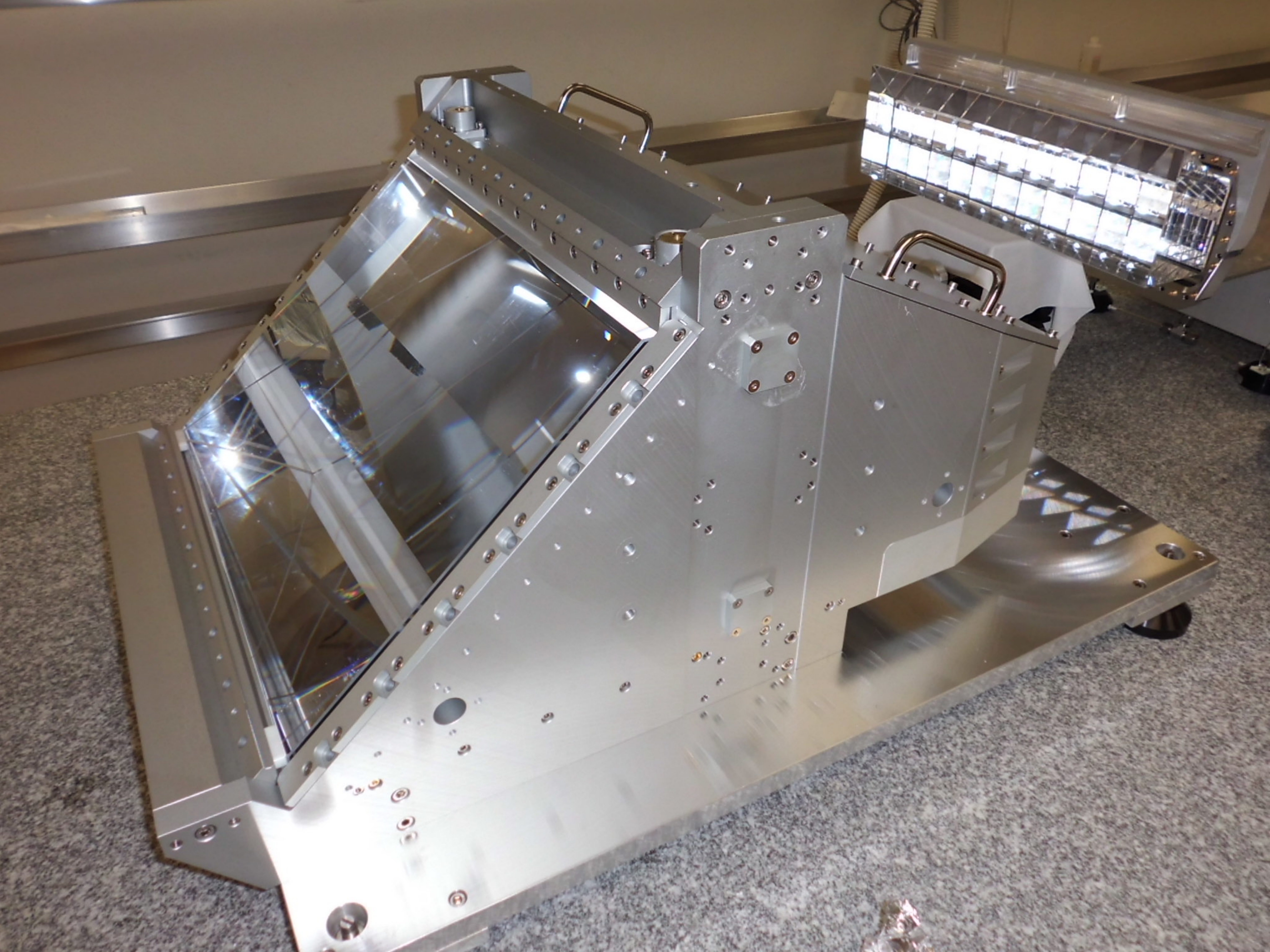}}
\end{center}
\caption{Fbox assembly around the real Fused Silica FBLOCK.}
\label{fig:FBLOCK_assembly_a}
\end{figure*}

\begin{figure*}[tbp]
\begin{center}
\subfloat[Fbox and bar box in the CRT setup.]{\includegraphics[width=0.465\linewidth]{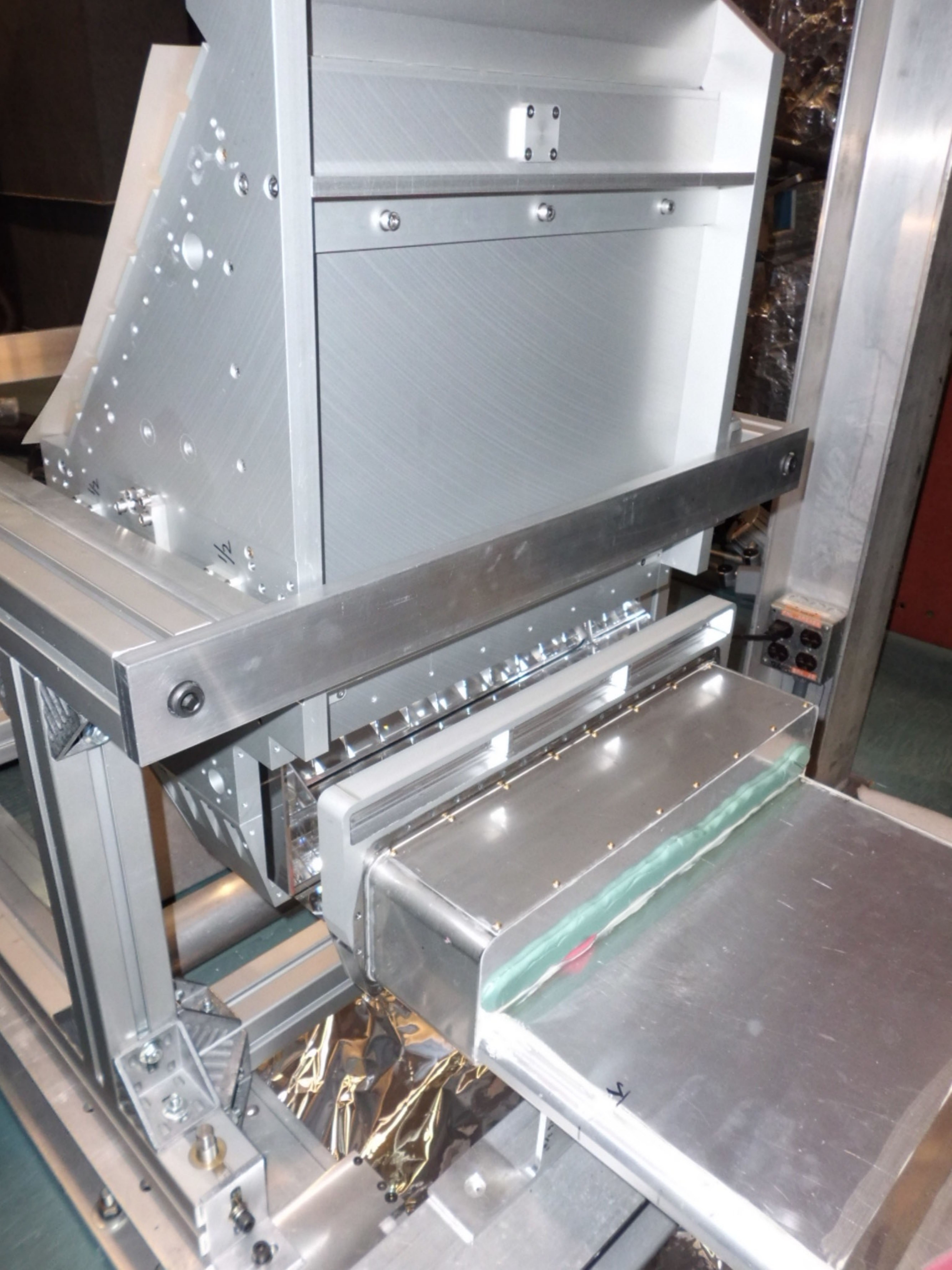} }
\hspace{5mm}
\subfloat[Details of coupling of new Wedge to FBLOCK.]{\includegraphics[width=0.465\linewidth]{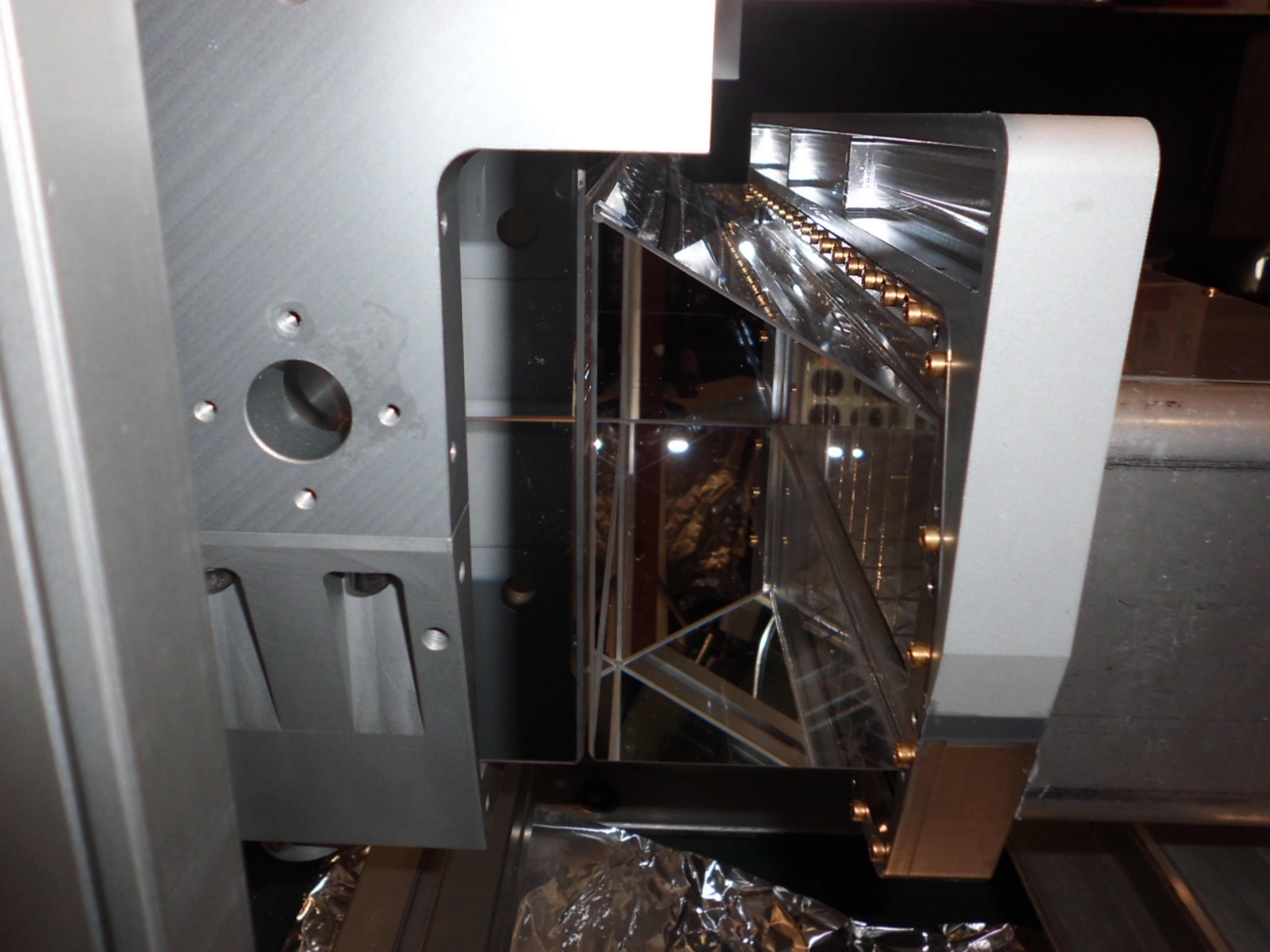}}
\end{center}
\caption{Fbox coupling to bar box in the CRT setup.}
\label{fig:FBLOCK_assembly_b}
\end{figure*}

\tdrparagraph{ Protection of optical surfaces }
   As one deals with internal reflections, all optical surfaces have to be very clean, and therefore every part of 
the Fbox was very carefully cleaned before final assembly to prevent outgassing. In addition, optical surfaces are protected 
against environmental pollution and moisture condensation by flowing a boil-off N$_2$ through the sealed Fbox. 
Fbox is sealed with a combination of Viton flat gaskets, Viton O-ring and Gore gasket tape (near the detector area), and 
in some difficult sections simply with DP-190 glue. Based on experience in \babar, each bar box requires a flow of about 100~cc/min. 

\tdrparagraph{ Bar box storage at SLAC }
  Figure~\ref{fig:Storage} shows the present storage of the 12 DIRC bar boxes.  They are supported on pre-aligned shelves to prevent 
mechanical stresses due to support distortions. They are under a constant flow of 
boil-off N$_2$ and thermally insulated. The storage box is kept at a nominal temperature of $18^\circ$C. 
In addition, there is no light to prevent yellowing of the Epotek 301-2 glue, an important issue to consider in future.  

\begin{figure}[tbp]
\begin{center}
\includegraphics[width=\linewidth]{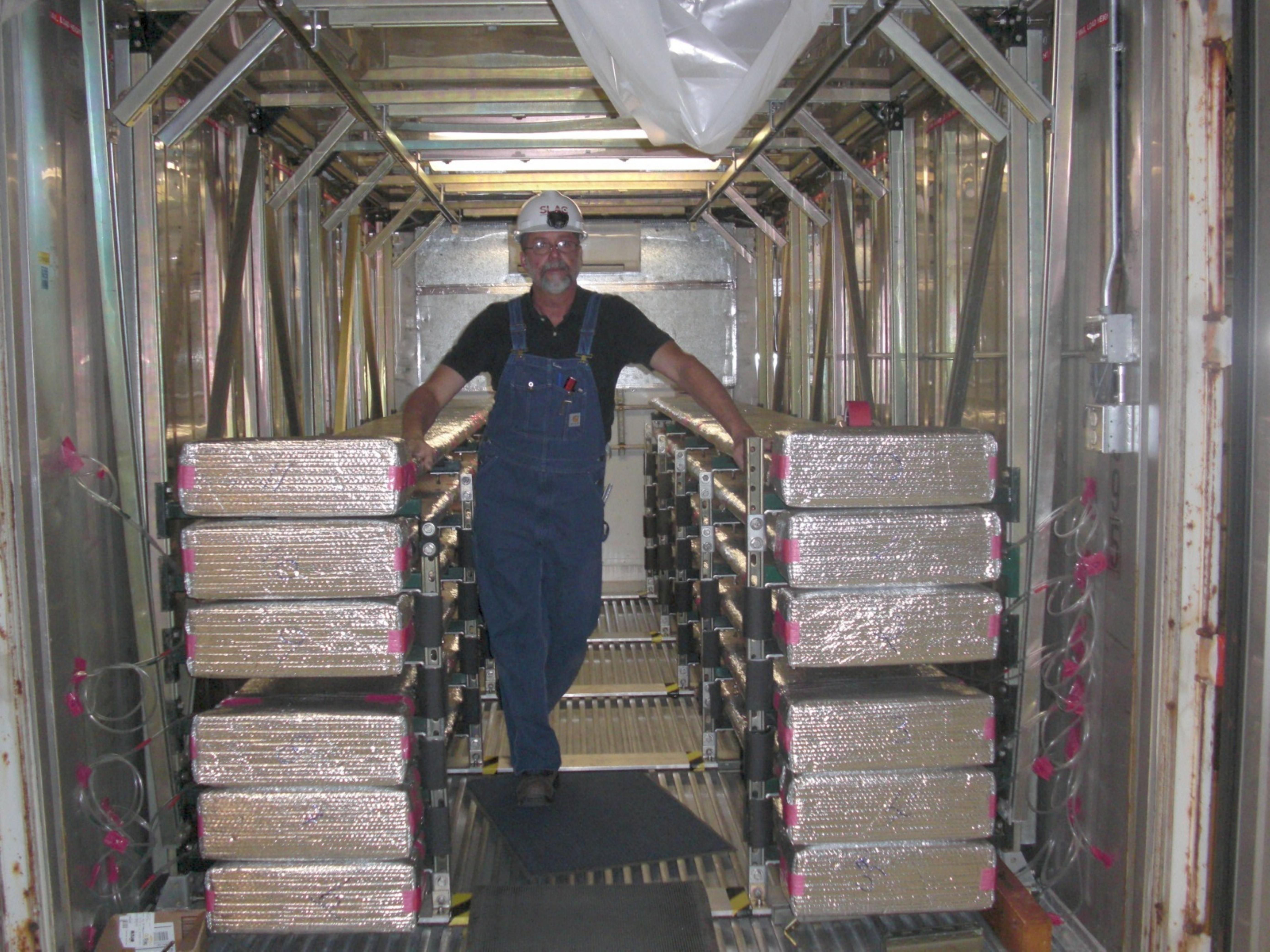}
\caption{Bar box storage at SLAC.}
\label{fig:Storage}
\end{center}
\end{figure}

\tdrparagraph{ Gluing FBLOCK to Wedge }
   The optical coupling between the new Wedge and the FBLOCK is done in situ, and is in principle removable. We set the gap between 
the new Wedge and the FBLOCK to 1\mm, and fill it with Shin-Etsu 403 RTV. In case of problems with the Fbox, one can first separate 
the two pieces using a thin razor wire, then clean the surfaces and finally couple them again. The penalty for this option is that an 
RTV joint is not as strong as an epoxy joint. The breaking force of this RTV coupling was measured to be ${\sim} 520$~kg using 
glass windows of the correct size; it was found that this value depends strongly on the glass cleaning procedure. 
See Figure~\ref{fig:Det_install_a} and chapter "Support of Fbox in the \superb\ magnet" below for more details on the installation procedure.

\tdrparagraph{ Support of Fbox in the \superb\ magnet }
  The plan is to install bar boxes with the new Wedge already glued to the bar box windows. The Fbox will be installed in situ.
Figure~\ref{fig:Det_install_a} shows the procedure. The space between bar boxes is very tight. Therefore, the bar box has to be 
moved beyond the neighboring already installed Fboxes, so one has enough room for gluing. With a temporary rail support 
it is possible to bring the Fbox close to the bar box; in that way the Wedge and the FBLOCK surfaces are parallel 
and the gap is set to 1\mm, bottom surfaces of FBLOCK and the new Wedge are aligned, and both are centered left-right. 
The gap between FBLOCK and the new Wedge is then filled with the Shin-Etsu 403 RTV. Once the RTV is cured, the gas sealing is made, 
the Fbox is pushed on the rail to its final position, and the earthquake bracing is installed. The temporary rail support is then removed. 
Figure~\ref{fig:Det_install_b} shows the camera after the installation procedure has been completed.

\begin{figure*}[tbp]
\begin{center}
\subfloat[Fbox installation fixture for position pointing up.]{\includegraphics[width=0.465\linewidth]{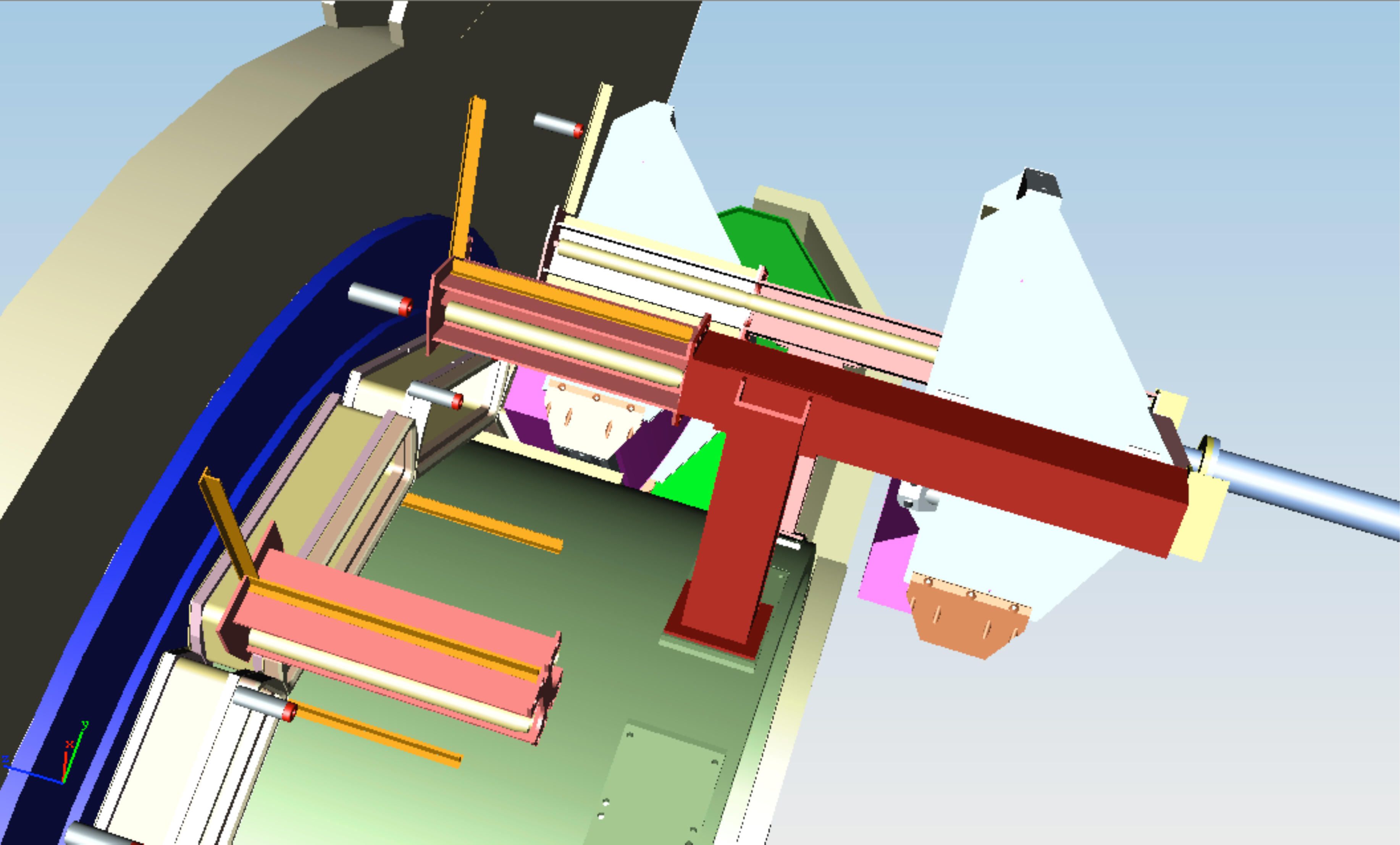} }
\hspace{5mm}
\subfloat[Fbox installation fixture for position pointing down.]{\includegraphics[width=0.465\linewidth]{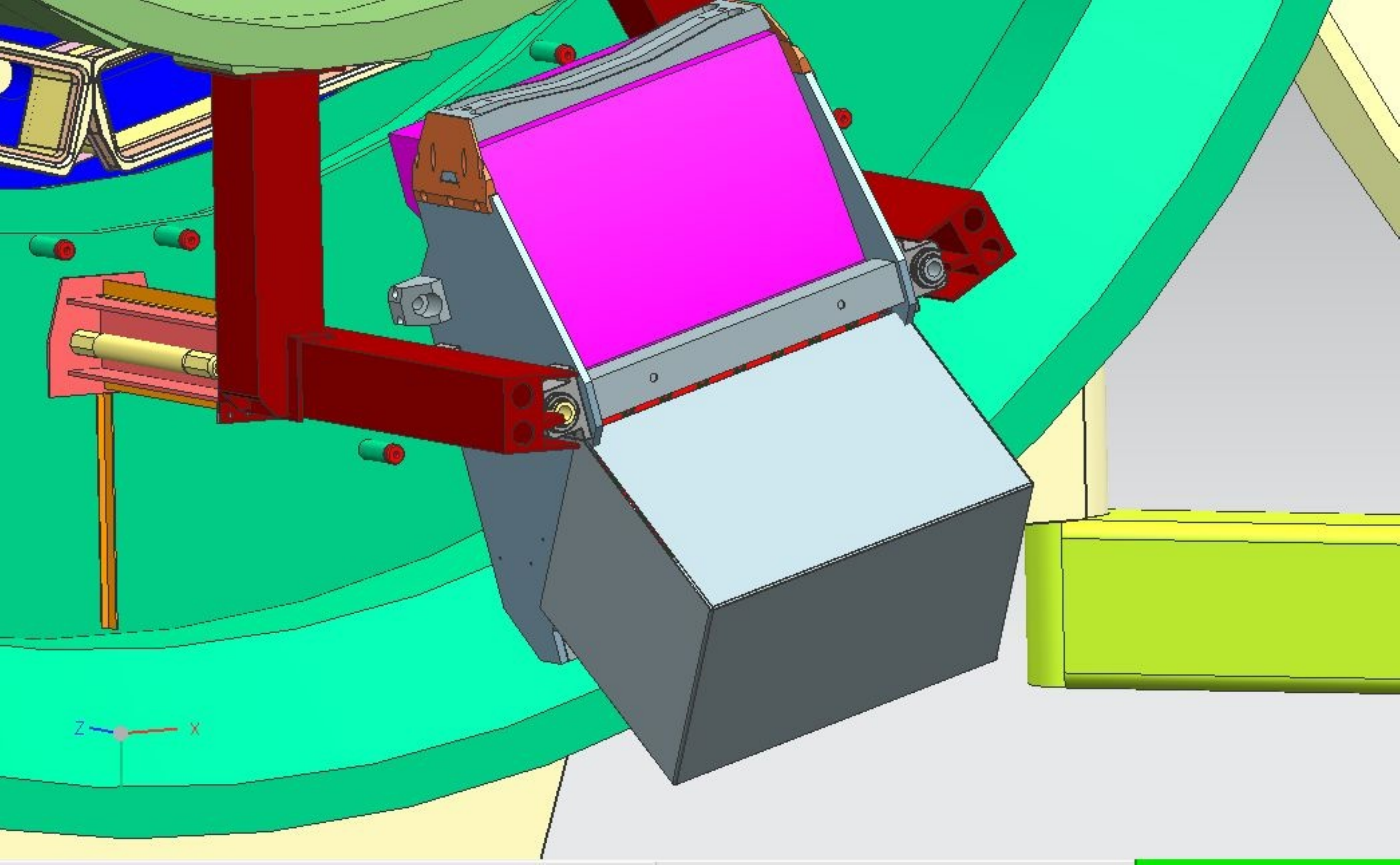}}
\end{center}
\caption{Fbox installation in the \superb\ magnet.}
\label{fig:Det_install_a}
\end{figure*}

\begin{figure}[tbp]
\begin{center}
\includegraphics[width=\linewidth]{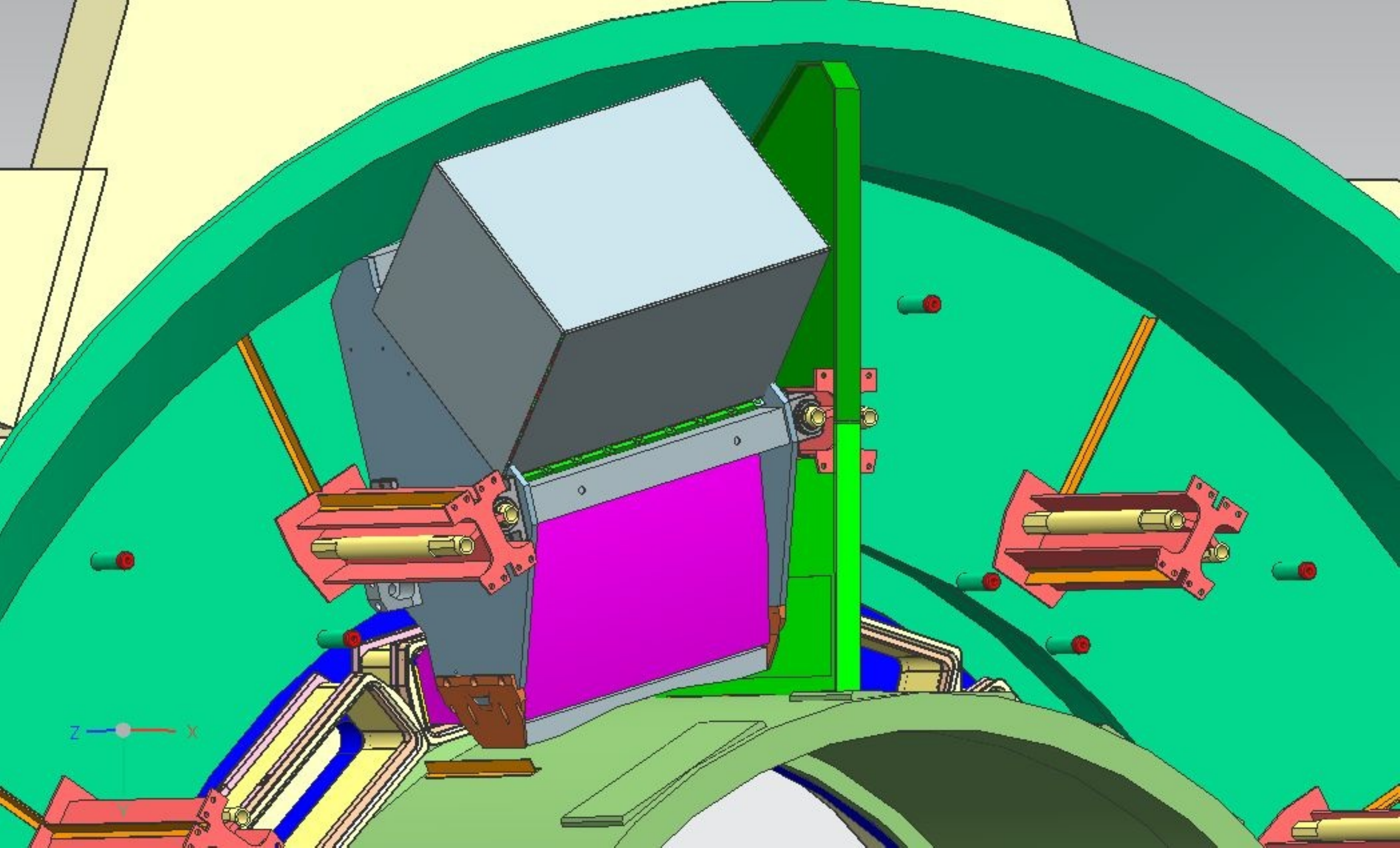}
\caption{Camera after completion of the installation procedure.}
\label{fig:Det_install_b}
\end{center}
\end{figure}

\begin{figure}[tbp]
\begin{center}
\includegraphics[width=\linewidth]{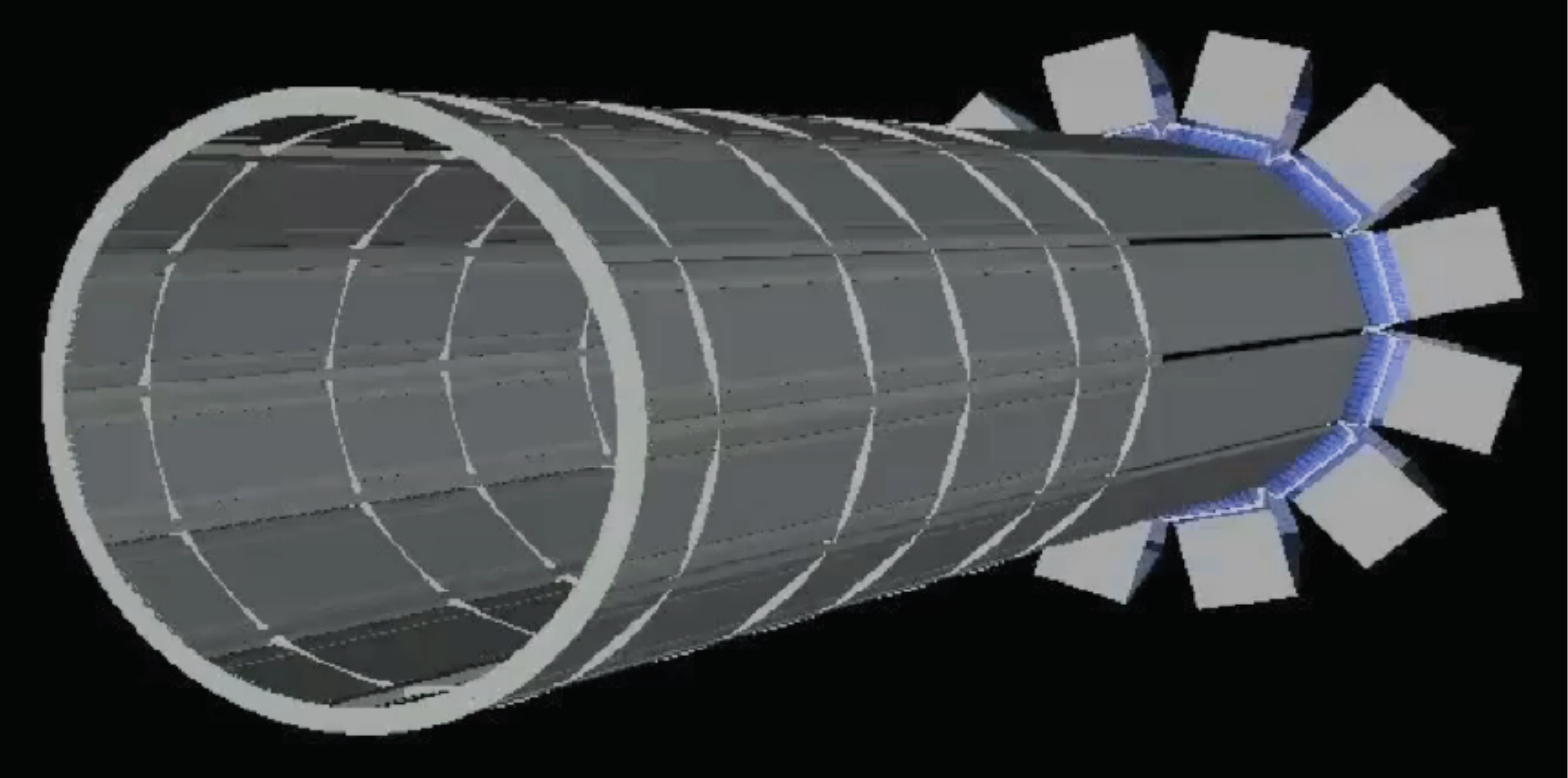}
\caption{Overall view of the FDIRC layout with 12 bar boxes and 12 photon cameras.}
\label{fig:Det_layout_a}
\end{center}
\end{figure}

Figure~\ref{fig:Det_layout_a} shows the overall FDIRC detector schematic layout with its 12 bar boxes, and the 12 corresponding photon cameras. 
Figures~\ref{fig:Det_layout_b} and~\ref{fig:Det_layout_c} show overall mechanical views of the FDIRC in the 
\superb\ experiment. Details of the shielding are shown on Figure~\ref{fig:Det_layout_f}. 

\begin{figure*}[tbp]
\begin{center}
  \subfloat[A 3D view showing the new magnetic shield and background
  shields, and
  Fboxes.]{\includegraphics[width=0.5\linewidth]{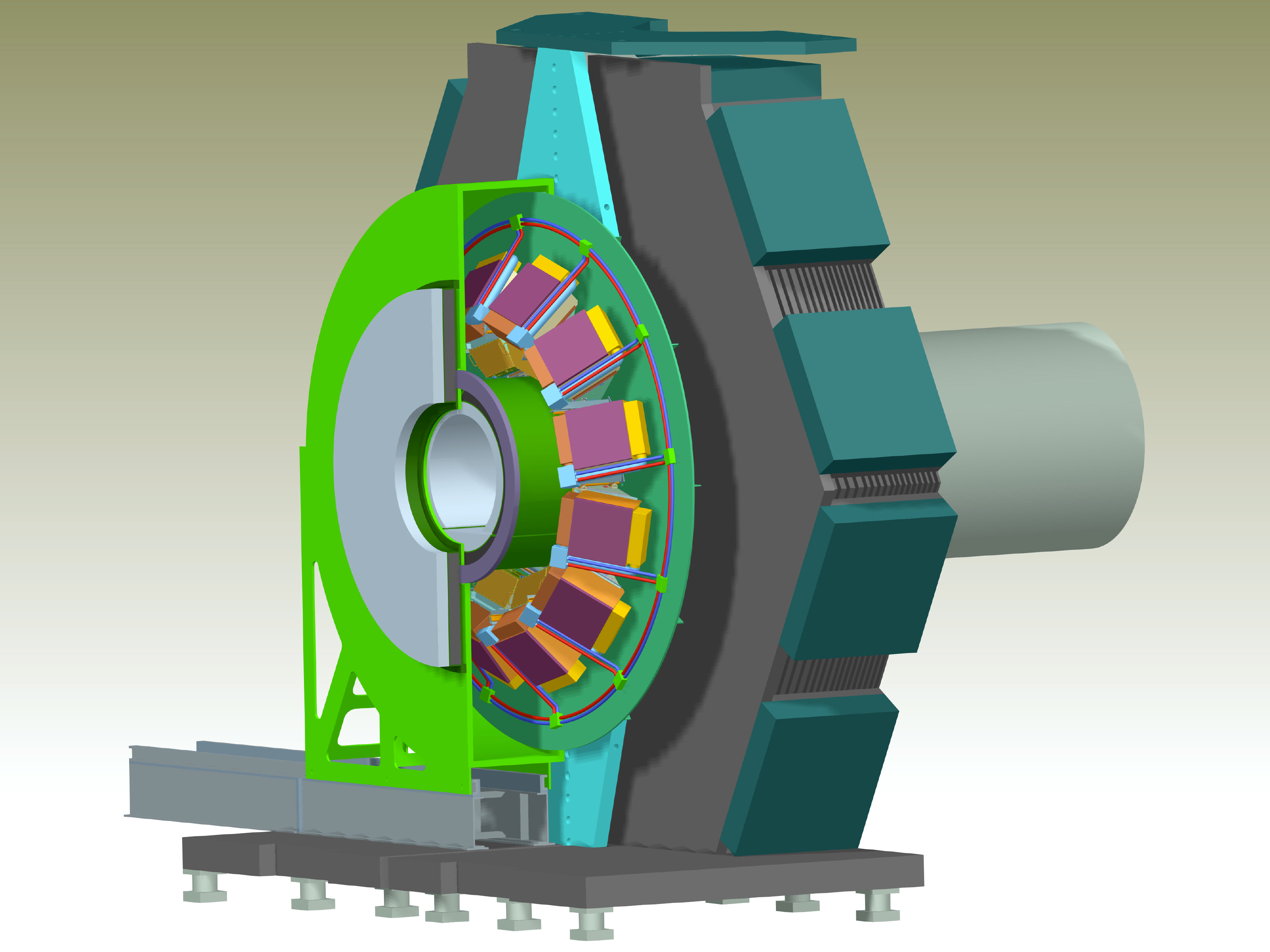}}
  \subfloat[Front view showing six Fboxes, the rest is hidden behind
  the magnetic and background
  shields.]{\includegraphics[width=0.5\linewidth]{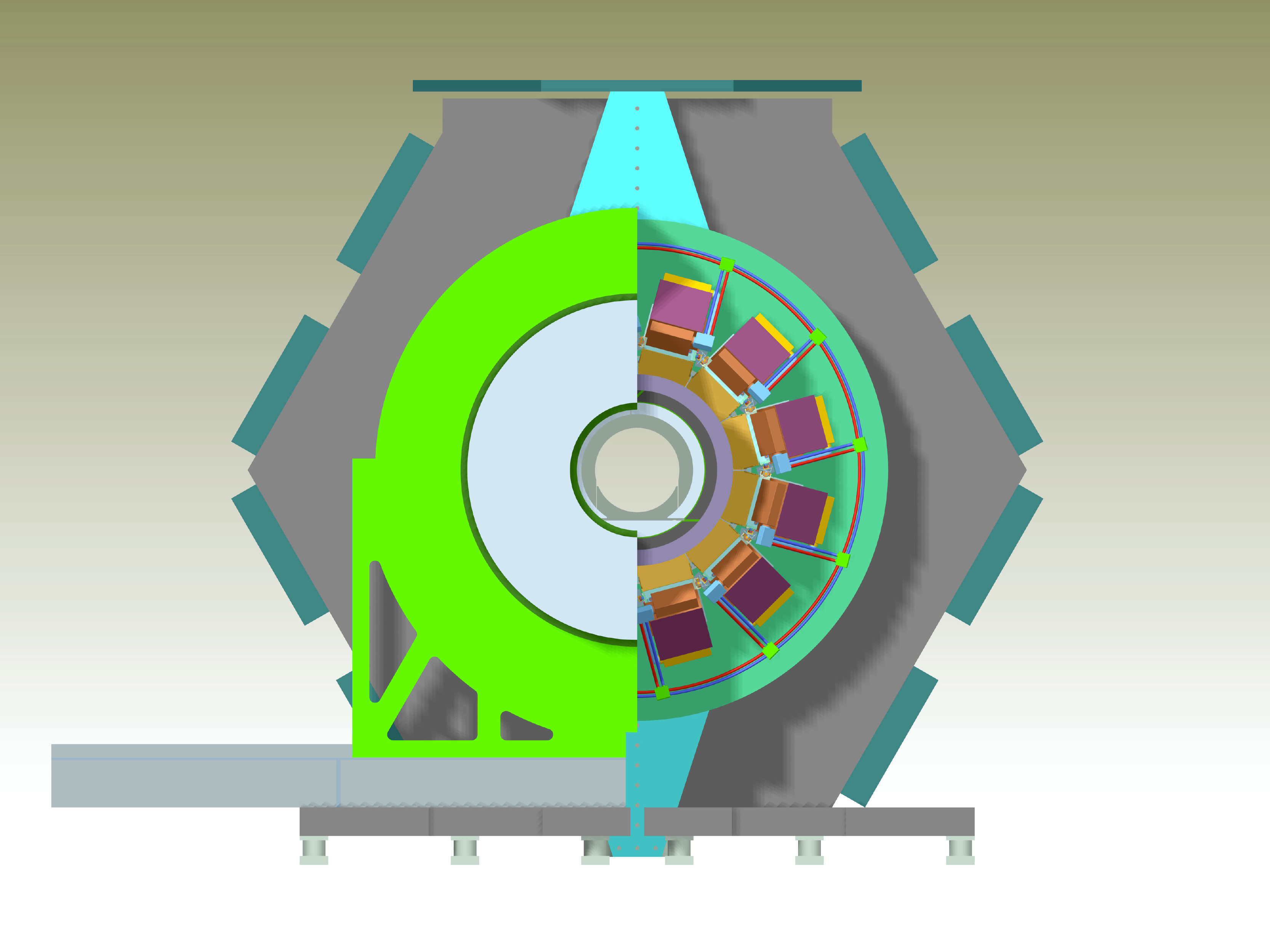}}
\end{center}
\caption{FDIRC in the magnet.}
\label{fig:Det_layout_b}
\end{figure*}

\begin{figure}[tbp]
\begin{center}
\includegraphics[width=\linewidth]{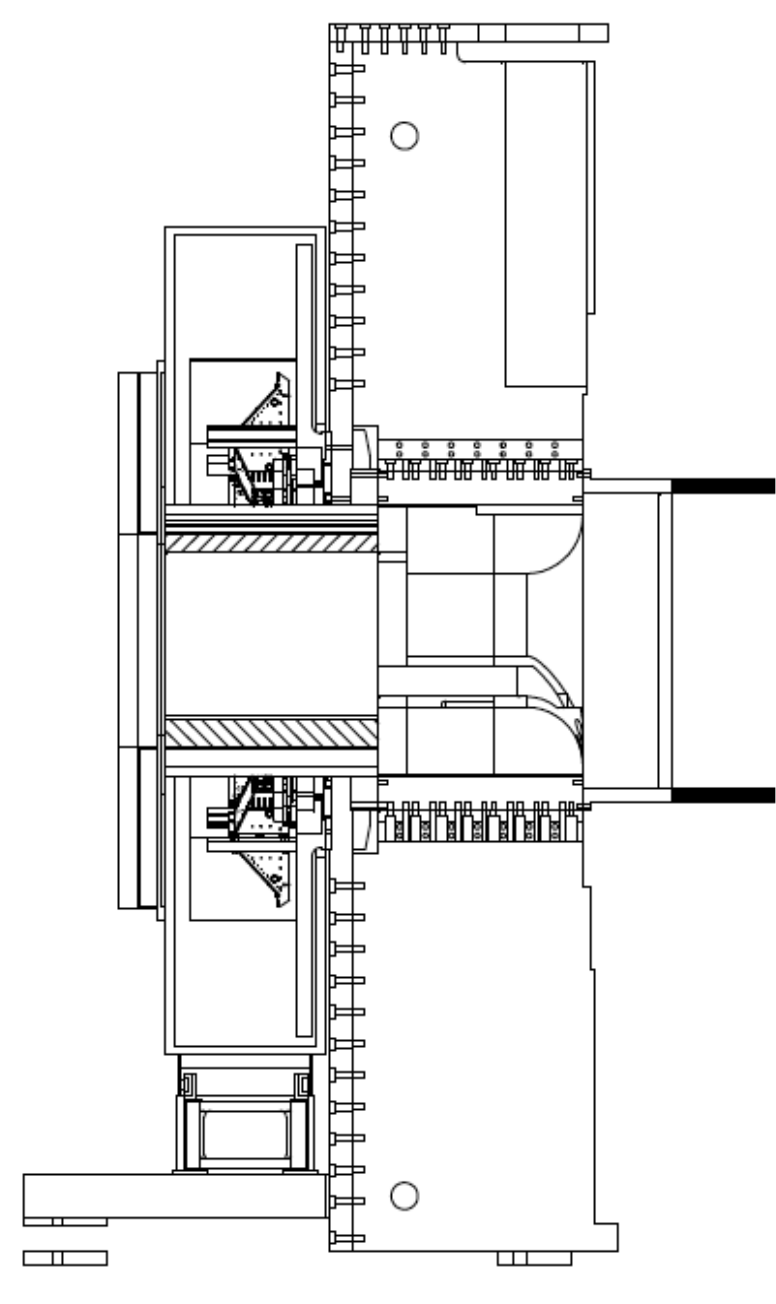}
\caption{Side view of the FDIRC showing magnetic and background shields, and rails on which they move.}
\label{fig:Det_layout_c}
\end{center}
\end{figure}

\tdrparagraph{ Experience with shipping FBLOCK }
   It is useful to share experience with shipping of finished FBLOCK. The polishing company worried about this step because the FBLOCK is very heavy
and any rubbing even against a foam padding could cause problems. They decided to cover all surfaces with a protective tape resembling a masking tape. 
We found that the tape adhesion was too strong and a small section of mirror plating was removed when the tape was peeled off. This triggered a small
R\&D effort to find out what type of protection tape would not cause this problem. We found that the best result was achieved with the 
electrostatically-adhered vinyl tape (Grafix Plasics, S-C2436, Cling vinyl). We also found that if one such problem occurs, the best way to fix 
it is to use a small section of aluminized Mylar sheet pressed uniformly by a foam on its entire surface. Any other method of 
stretching the aluminized Mylar sheet created wrinkles.

\tdrparagraph{ Bar box shipment to Italy }
   There are several issues to consider: (a) vibration and mechanical shocks, (b) thermal effects, (c) pressure changes, and 
(d) light exposure. Each bar box will have a container providing mechanical support and constant thermal environment. 
The vibrations and mechanical shocks will be mitigated by placing bar boxes on a precisely leveled support with a foam on top of it. 
The support structure will need telescopic mount to suppress large shocks. They will be thermally isolated and equipped with 
active thermal blankets to keep temperature constant. We will also provide a $N_{2}$ boil-off gas flow. Another important 
issue is pressure changes if air transport is used. The Hexel panels, used to construct bar boxes, do not 
have perforated walls, and therefore some stresses will be created, if one wants to ship bar boxes by air. 
The present thinking is to construct a container to ship two bar boxes at a time. This container would be air-shipped by a commercial
company Fedex or DHL. The shipment has to be insured for full value of bar box replacement (presently we think that the cost of such 
insurance is about 15\% of the total cost of shipment, judging from experience of Nagoya group).
Finally, bar boxes must not be left exposed to a strong light as it could yellow the Epotek 301-2 epoxy. 


\tdrsubsec{ Electronics readout, High and Low voltage }
   The electronics for the FDIRC can be seen as an upgrade of the electronics of the \babar\ DIRC. The new requirements of the 
experiment (trigger rate, background, radiation environment) and FDIRC specific requirements (resolution, number of channels 
and topology) have led to a similar but new design of the electronics chain.

The FDIRC electronics will handle 18,432 channels in total. The electronics chain is based on a 
high resolution and high count rate TDC, a time-associated charge measurement with 12-bit resolution, and an event data 
packing, sending data frames to the data acquisition system (DAQ). The target timing performance of the overall electronics 
chain is a time resolution of 100\ps rms. It has to deal with hit rates of 100 kHz per channel, a level-1 (L1) trigger rate up to 
150 kHz, and a minimum spacing between triggers of about 50 ns. 

The radiation level is expected to be about 1~gray per year, thus the use of radiation tolerant or off-the-shelf 
radiation-qualified components is necessary. However, the expected energy of the particles may make the latch-up 
effect almost impossible. Thus, the design has only to take into account Single-Event-Upsets (SEUs). We selected the Actel family 
FPGA components for their non-volatile flash technology configuration memories, which are well-adapted to low level radiation environment.

Several architectures have been considered: (a) all electronics directly mounted on the FBLOCK, (b) all electronics mounted next 
to the detector and linked to PMTs by cables, and (c) a part of it on the detector (the Front-end boards) and the other part, 
called 'crate concentrator', is located close to the detector (this board is in charge of interfacing with the Front-end, reading out 
event data, packing and sending it to the DAQ). 

The first solution has been chosen as baseline for the TDR for two main reasons. (a) The cost of cables (PMT to Front-end 
boards) is estimated to be close to 200 kEuros (1/3 of the price of the overall electronics cost), making this design too 
expensive. Moreover, the possible option to have pre-amps on the PMT bases does not prevent from having electronics and power 
supplies on the detector. (b) The large amount of data per channel leads to have the L1 trigger derandomizer and buffer on the 
Front-end boards. The Fast Control and Timing System (FCTS) receiver could be individually located on each Front end board but the number of cables needed 
pushes to distribute all the control signals on a backplane. Consequently, the board dedicated to receiving and transmitting 
FCTS signals on the backplane naturally tends to also become the event data concentrator and the link to the DAQ.

The baseline design assumes a 16-channel TDC ASIC, offering the required precision of ${\sim} 100$\ps rms, embedding an analog 
pipeline in order to provide an amplitude measurement transmitted with the hit time. Thanks to a 12-bit ADC, the charge 
measurement will be used for electronics calibration, monitoring and survey purposes. The Front-end board FPGA synchronizes 
the process, associates the time and charge information, and finally packs them into a data frame which is sent via the 
backplane to the FBLOCK control board (FBC). The FBC is in charge of distributing signals coming from the FCTS and the 
Experiment Control System (ECS), 
packing the data received from the front-end (FE) boards to a n-event frame including control bits and transferring it to the DAQ.

\tdrparagraph{ The FDIRC electronics }
The FDIRC electronics consists of a front-end analog part and a TDC both integrated in a common ASIC called CATS, and of an external ADC. The front-end analog part will soon be
validated in a prototype chip called PIF (PId Front end chip), which will contain an amplifier and a peak detector for the amplitude measurement (digitization will be performed by the external ADC), and a low walk
discriminator (called "Discri") to drive the TDC inputs.  This TDC has already been developed and tested in a prototype called SCATS. Figure~\ref{fig:PIF_concept} shows
the concept of one channel of the PIF chip and the principle of the "Discri". The latter is mainly based on an high gain amplifier and a discriminator with a programmable threshold .
The overall aim is to measure single electron pulses with a resolution of  ${\sim} 100$~ps rms. 
The TDC chip is derived from a design realized for the SuperNemo experiment~\cite{Pahlka:2008dw}.
It provides a time measurement with steps of 200\ps offering both a high resolution (70\ps rms) and a large dynamic range (53 bits). The architecture of this chip is based on the association of Delay Locked
Loops (DLLs) and of a digital counter, all of these components being synchronized to a 178.5~MHz external clock, multiple of the global system clock (59.5 MHz). 

The \superb\ TDC chip (CATS) will keep the same philosophy, but the
high input rate requirement leads to a complete re-design of the digital part, in order to minimize the dead time per channel and to
increase the output data rate. The solution chosen consists in transforming the former architecture, where data was read out by the FPGA, into one where data goes to FPGA via TDC. Moreover, instead of using registers for memorizing the hits, the new circuit makes use of individual FIFO memories in each channel in order to internally de-randomize the potential high frequency bursts on the chip inputs. With this architecture, data from the DLLs and the coarse counters is now written into this FIFO memory. 
When the writing is complete, the channel is automatically reset and ready for the next hit, within a delay lower than 50 ns. Simulations
of the readout state machine showed an output FIFO data rate capability of up to 100 MHz, thus permitting a safe readout at a frequency of either 59.5 or 89.25 MHz.
Time ranges for the DLLs and the coarse counter can be easily customized by adjusting the output data format (16, 32, 48 or 64 bits). Therefore, this chip is actually
suitable for various applications with either high count rate and short integration time, or low count rate and long
integration time. Figure~\ref{fig:TDC:chip} shows the block diagram of the \superb\ FDIRC TDC chip (CATS). 

\begin{figure*}[tbp]
\begin{center}
\subfloat[One-channel concept of PIF chip.]{\includegraphics[width=0.5\linewidth]{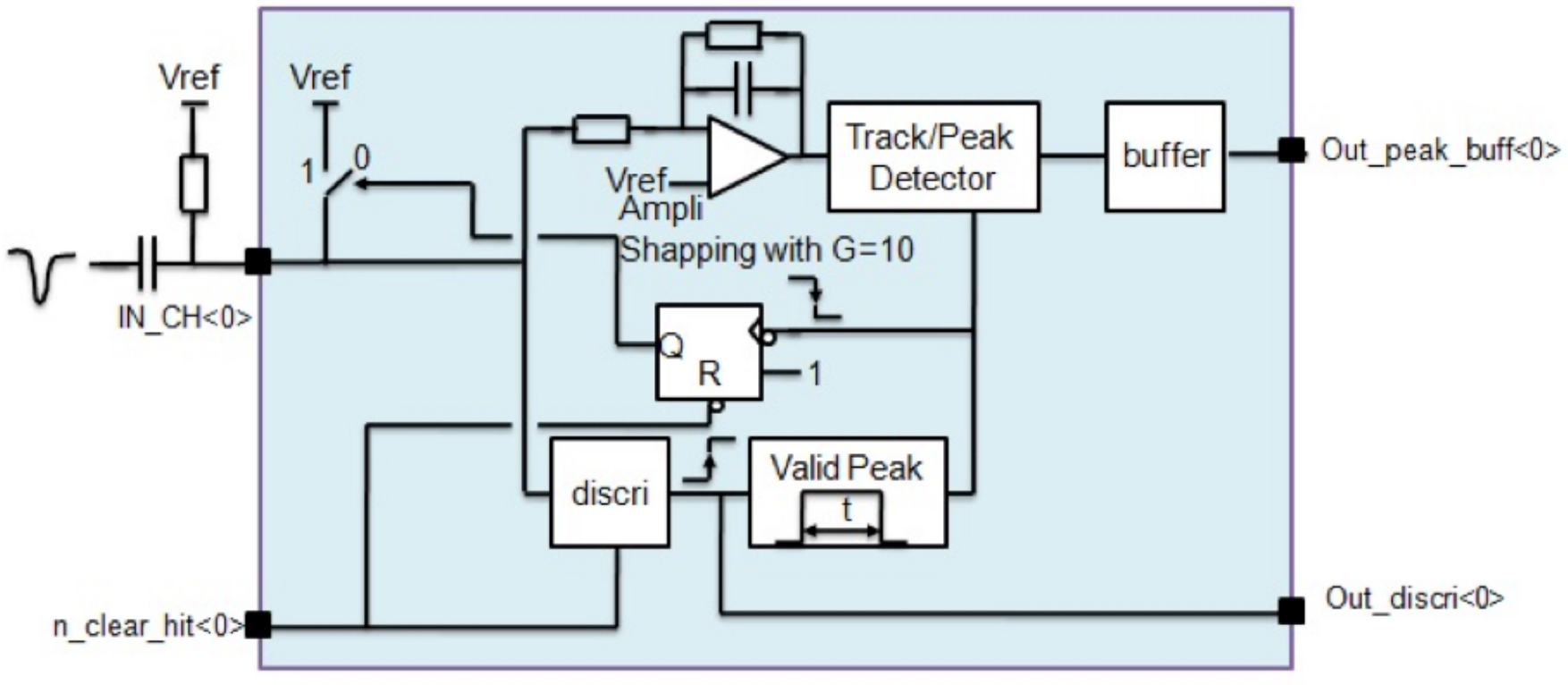} }
\hspace{5mm}
\subfloat[The principle of the "discri".]{\includegraphics[width=0.4\linewidth]{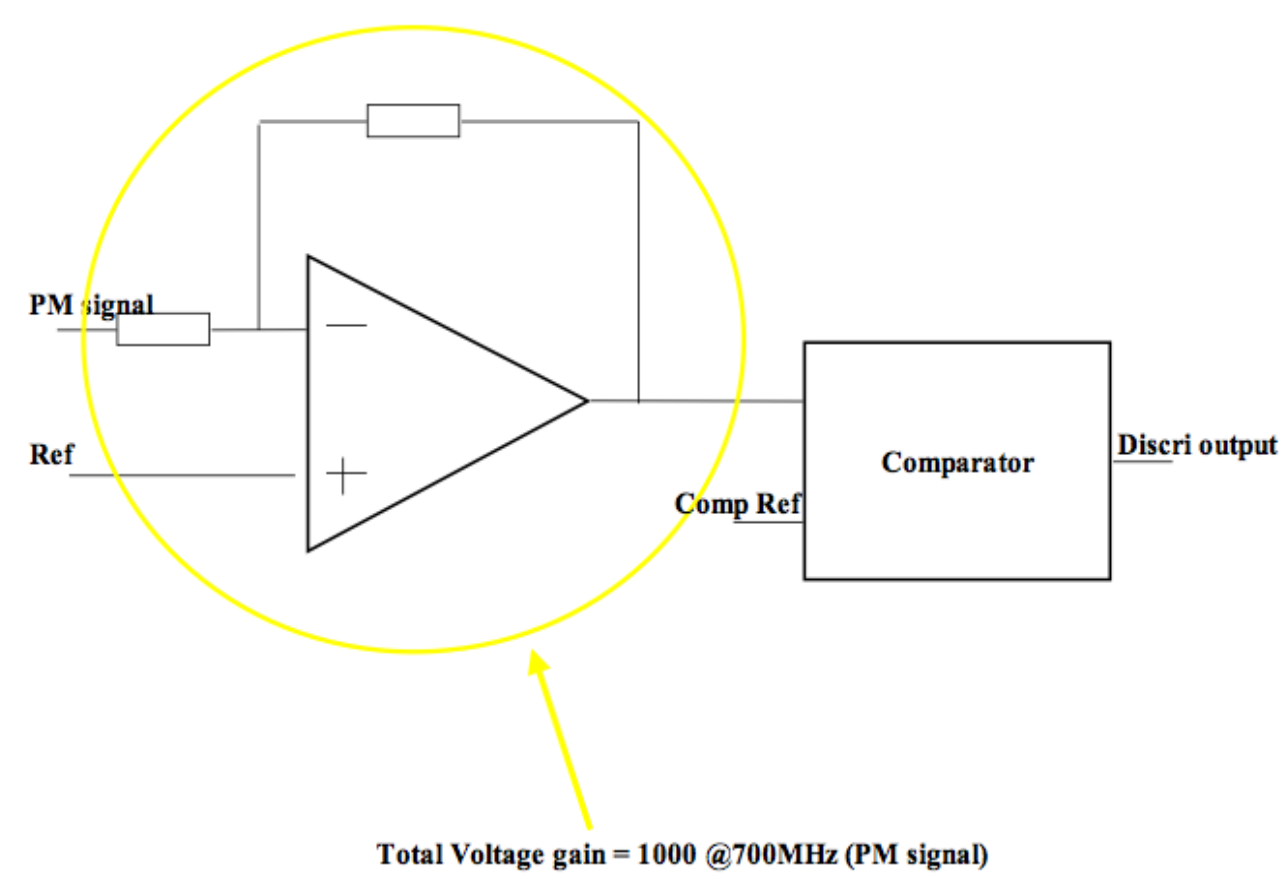}}
\end{center}
\caption{FDIRC front-end analog electronics}
\label{fig:PIF_concept}
\end{figure*}

\begin{figure*}[tbp]
\begin{center}
\includegraphics[width=\linewidth]{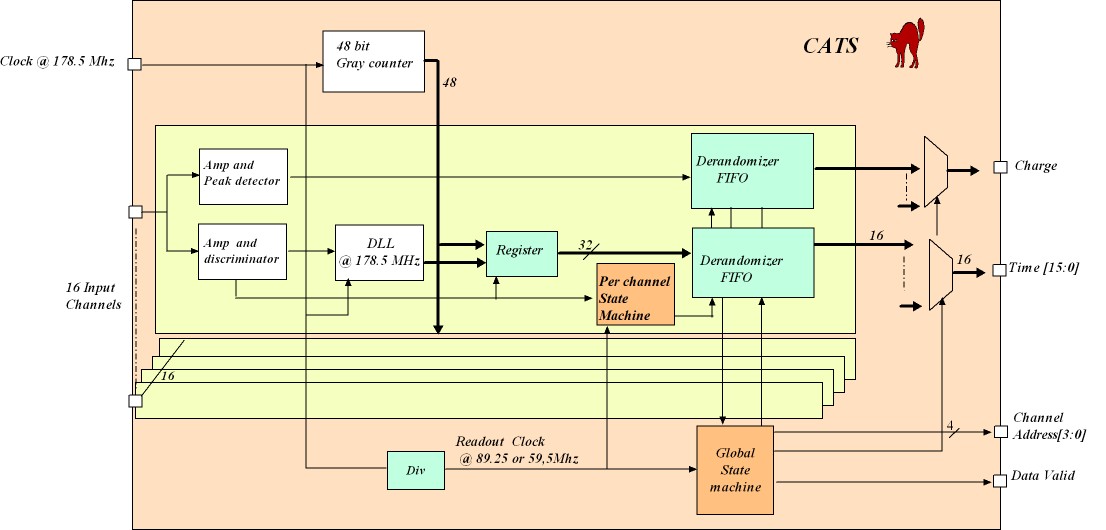}
\caption{\superb\ FDIRC TDC chip block diagram.}
\label{fig:TDC:chip}
\end{center}
\end{figure*}

A FIFO depth of 8 words (32-bit each) has been selected after simulation
with a exponential distribution model of delta time between
hits (mean rate ${\sim} 1$~MHz) applied to inputs. To design this FIFO, a
full custom RAM has been developed. In addition to increasing its speed, it permits reducing
the size of the chip and consequently its cost. The chip is designed using
known and proved mitigation techniques to
face SEU issues due to the low-level radiation environment. The prototype
version of the chip (called SCATS), which does not include
the analog FIFO and the discriminator, has been submitted in November 2011.
We plan to submit by end 2012 another prototype chip (PIF)~-- see Figure~\ref{fig:PIF_concept}~-- dedicated to the currently main missing parts: (a) a low walk discriminator
receiving the PM outputs and able to send a logic signal to the TDC part of the chip, and (b) an amplifier followed by a peak detector and an output buffer.
After testing and validation, a final version of the CATS chip including all these functions will then be assembled and submitted by end 2013. 

{\it Front-end Crate}\\ The board input will fit the topological distribution of the PMT on the FBLOCK~-- see Figure~\ref{fig:Backplane}. 
In each sector, the PMTs are arranged as a matrix of 6 in vertical direction by 8 in horizontal direction. Each column of 6 PMTs will be readout by two FE boards. One 
vertical motherboard will couple one front-end board to one half column of 6 PMTs. There will be a total of 8 motherboards. For each PMT, the motherboard will 
convert 4 H-8500 connectors into two other connectors, each connected to a FE board. Figure~\ref{fig:FBLOCK_el} shows the Fbox with 
the front-end electronics.
 
The motherboard will also distribute High Voltage to the PMT if we use H-8500D. However, in case that we choose the H-8500C PMT, 
each PMT will have its own HV cable and HV distribution will be separate. In addition to the 8 motherboards, the backplanes receives 
one communication board for distributing control signals and data between FE boards and the FBLOCK control board. The FB-crate will 
use many features of commercial crates, such as board guides, rails, etc.

\begin{figure*}[tbp]
\begin{center}
\includegraphics[width=\linewidth]{Fe_crate_HD.jpg}
\caption{Front-end crate: PMT backplane, communication backplane, FE-board, FBLOCK controller (FBC).}
\label{fig:Backplane}
\end{center}
\end{figure*}

\begin{figure}[tbp]
\begin{center}
\includegraphics[width=\linewidth]{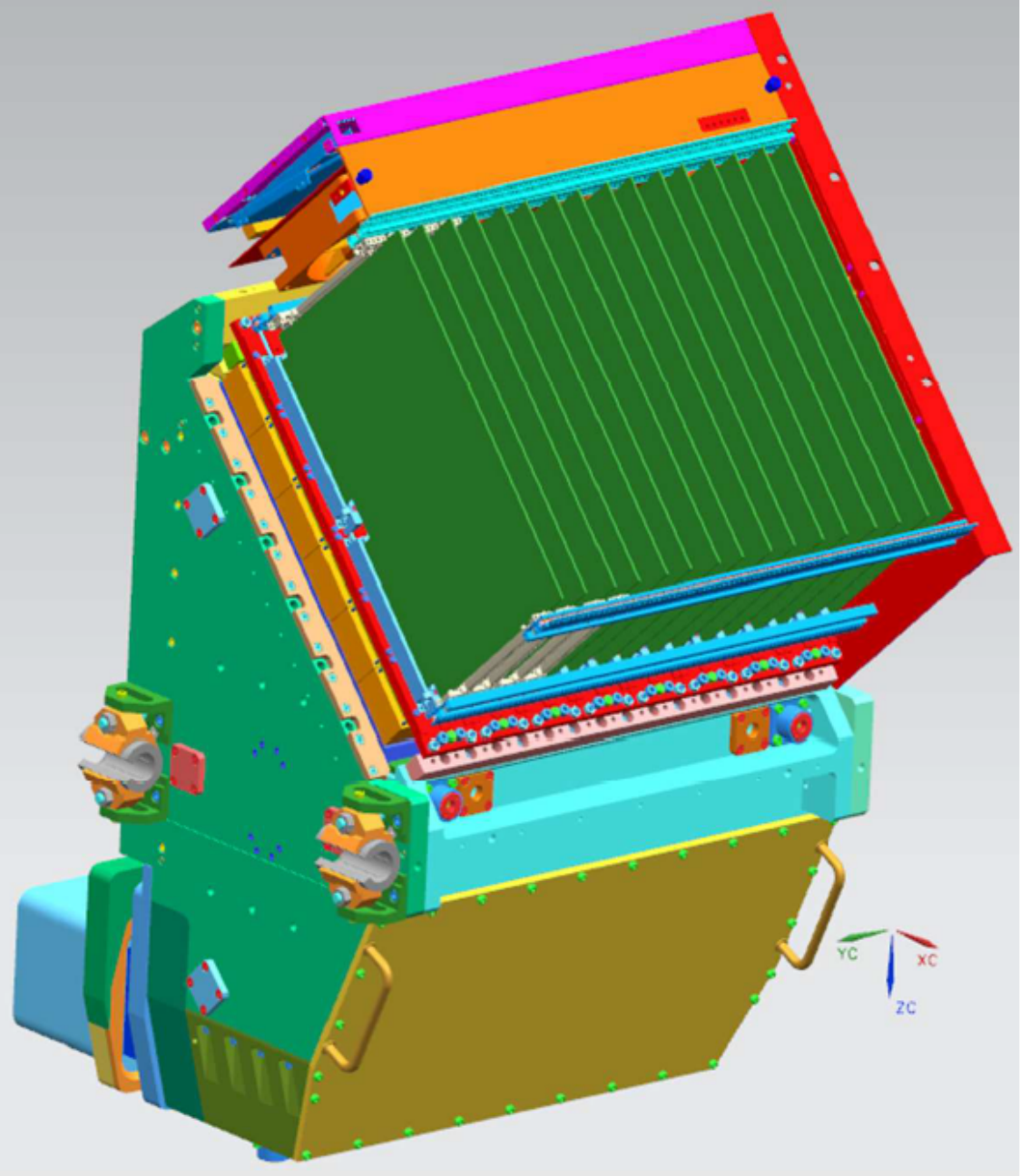}
\caption{Fbox equipped with electronics and its cooling.}
\label{fig:FBLOCK_el}
\end{center}
\end{figure}

\begin{figure*}[tbp]
\begin{center}
\includegraphics[width=\linewidth]{Fe_board_HD.jpg}
\caption{Front-end board connected to backplanes.}
\label{fig:Front-end_board}
\end{center}
\end{figure*}

{\it Communication Backplane}\\ The communication backplane distributes the ECS and FCTS signals from the FBC to the 8 FE boards thanks to point to point 
LVDS links, and connects each FE board to the FBC for data transfer. A serial protocol will be used between FE board and the 
FBC in order to reduce the number of wires and consequently improve the reliability. It will also distribute JTAG signals 
for FPGA board reprogramming, and all signals for the monitoring and control of the crate.

{\it PMT Backplane}\\ It is an assembly of 8 motherboards, each one corresponding to a column of 6 PMTs. One motherboard receives 2 FE boards. 
The 64 channels from 4 connectors per PMT are merged on the motherboard into two connectors to get into the Front end board to get 
16 channels per half PMT, \ie, 6 PMTs correspond to 96 channels per FE board. It also insures the ground continuity 
between the FE boards, the FE crate and the FBLOCK.

{\it The Front-end Board}\\ One FE board is constituted of many 6-channel processing blocks handling the board  96 channels. The 
channel-processing block is constituted of one TDC chip, one ADC, one small Actel FPGA and glue logics.
The FPGA receives event data from the TDC and the converted associated charge from the ADC. The 16-bit input bus carrying data from 
the 16 channels of the TDC is split into 16 individual trigger latency buffers where events are kept waiting for the level 1 trigger selection. Upon its reception, event data from the 6-channel processing blocks is sent to the master FPGA which packs the event and stores it into the readout derandomizer. The FE board then transfers
the event frame in differential LVDS to the FBC via the communication backplane.
Figure~\ref{fig:Front-end_board} shows the architecture of the FE-board connected to the backplanes.

{\it The crate controller board (FBC)}\\ The FBC board gathers front-end data and transfers it via optical fibers to the DAQ system. 
There will be one FBC board per crate. The board is separated in several functionalities:  (a) acquisition from the front-end boards and DAQ 
interface to ROM, (b) building of spy-data, (c) ECS interface, (d) deserialization of clock and control signal from FCTS, and (e) monitoring of the crate
temperature, power supplies, fans, etc.

{\it Cooling and Power Supply}\\ Electronics is located on the detector in a place enclosed by the magnetic doors. There are two major 
consequences: one is the problem of cooling, which must be carefully studied in terms of reliability and capability, and 
the second is that the location is ``naturally'' shielded against magnetic field. Consequently, the use of magnetic sensitive 
components as coils or fan trays is possible. An estimate of the overall electronics consumption lead to ${\sim} 6.1$~kW, not including 
the external power supplies. This can be broken down to individual contributions as follows: (a) electronics: 0.325~W/channel, hence
500~W/sector, 6~kW for the whole FDIRC; (b) HV resistor chain: 0.19~W/tube, hence 9.1~W/sector and 109~W for the whole FDIRC. The cooling system must be 
designed in order to maintain the electronics located inside at a constant temperature close to the optimum of 30 degrees. 
The air inside the volume must be extracted while dry, clean and temperature-controlled air will be flowing inside. 
Each FB crate will have its own fan tray like in a commercial crate. Targeting a difference of 10 degrees between inside and 
outside temperature drives to a rough estimate value of 300 m$^3$ per hour per crate. 4000 m$^3$ per hour can be considered 
as the baseline value for the whole detector.

\begin{figure}[tbp]
\begin{center}
{\includegraphics[trim=80 50 50 70,width=\linewidth,clip]{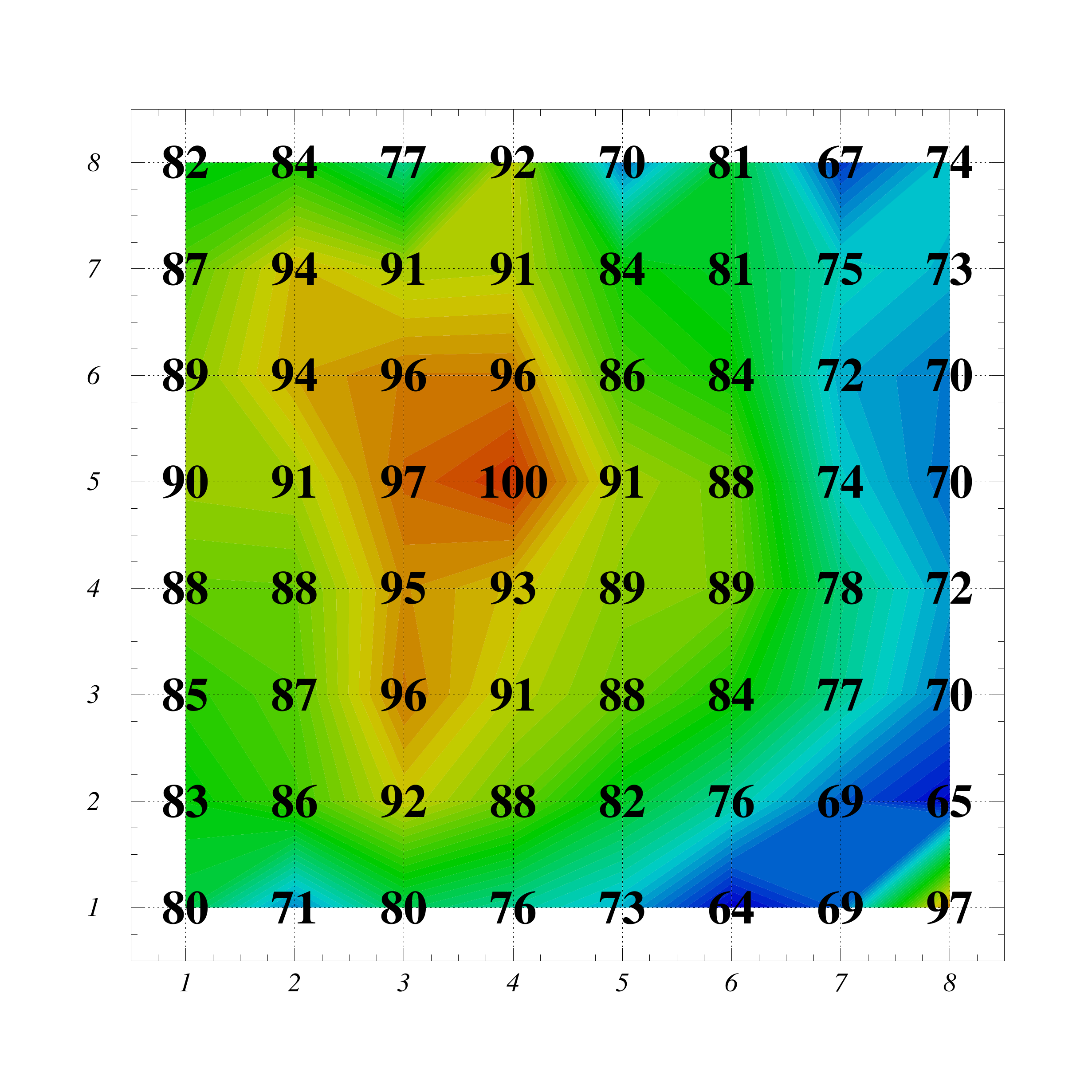}}
\caption{A relative pulse height response of H-8500 anodes to a pulse coupled to the last dynode~\cite{simi_Frascati_2011}.}
\label{fig:el_calibration}
\end{center}
\end{figure}

H-8500 PMT has an AC-connection to the last dynode, and this can be used either for triggering or calibration purposes. One
can do the calibration with HV off by injecting a pulse and looking at response of all anodes. Figure~\ref{fig:el_calibration} shows 
a relative pulse height response of 64 anodes to such pulse injection~\cite{simi_Frascati_2011}. It is not uniform, but it could be 
useful in order to identify possible electronics problems.  

\tdrparagraph{ Motherboard }

 We presently consider two choices for motherboard geometry: (a) a PC-board combining a group of 6 PMTs (see Figure~\ref{fig:Motherboard_d}), \ie, 
we need altogether 8 such motherboard per photon camera, or (b) a single PMT PC-board. The nominal choice is the 6-PMT 
motherboard. The total insertion force is 160N/FE board with ERNI connectors. To make sure that we do not bend pins in 
connectors, we will need guiding pins. We were considering also zero-insertion connectors (ZIF connectors), however, they are being discontinued and we were
advised by the TYCO company not to use them. There will be 16 FE boards per FBLOCK. These boards will either be inserted or extracted with a help of 
tools and rails, a similar procedure as in some commercial crates. Figure~\ref{fig:Motherboard_c} shows the complete photon camera with the electronics 
for 48 H-8500 PMTs and 1536 double-pixels.

\begin{figure}[tbp]
\begin{center}
\includegraphics[width=0.9\linewidth]{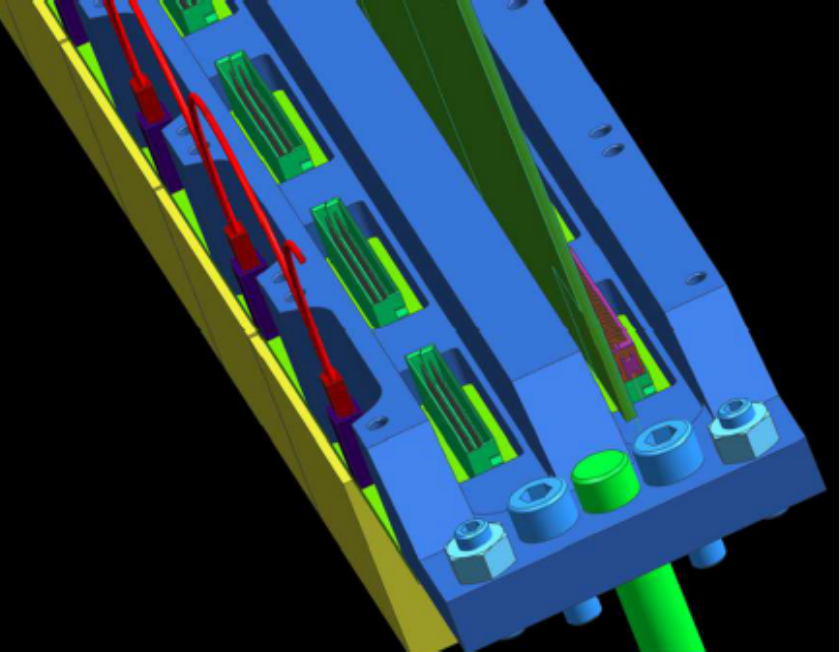}
\caption{A motherboard for six H-8500D tubes, which uses ERNI SMC-Q64004 press fit connectors to couple to FE boards. The total insertion force is 160N/FE board.}
\label{fig:Motherboard_d}
\end{center}
\end{figure}

\begin{figure*}[tbp]
\begin{center}
\subfloat[Photon camera with its electronics.]{\includegraphics[width=0.465\linewidth]{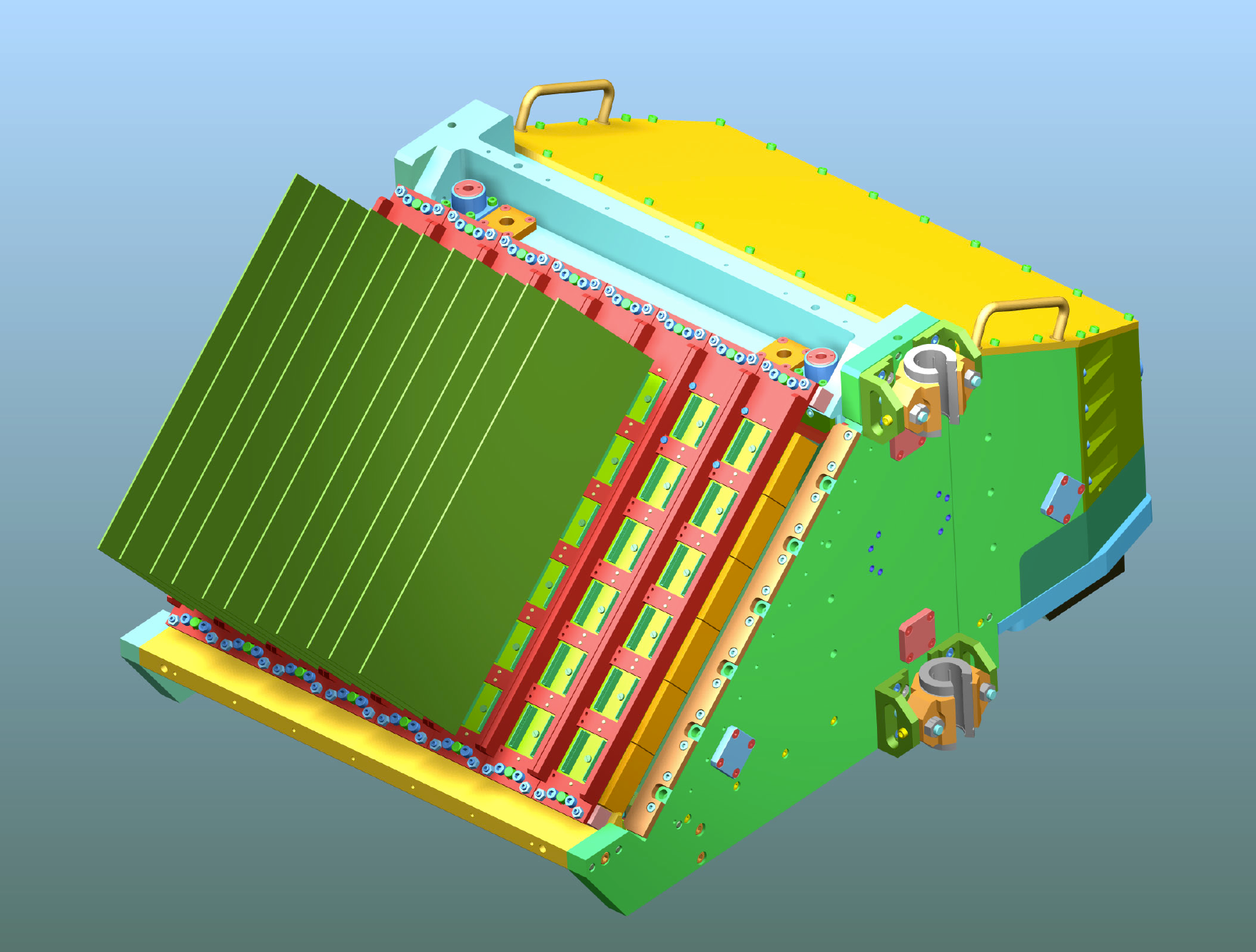} }
\hspace{5mm}
\subfloat[Photon camera with its electronics.]{\includegraphics[width=0.465\linewidth]{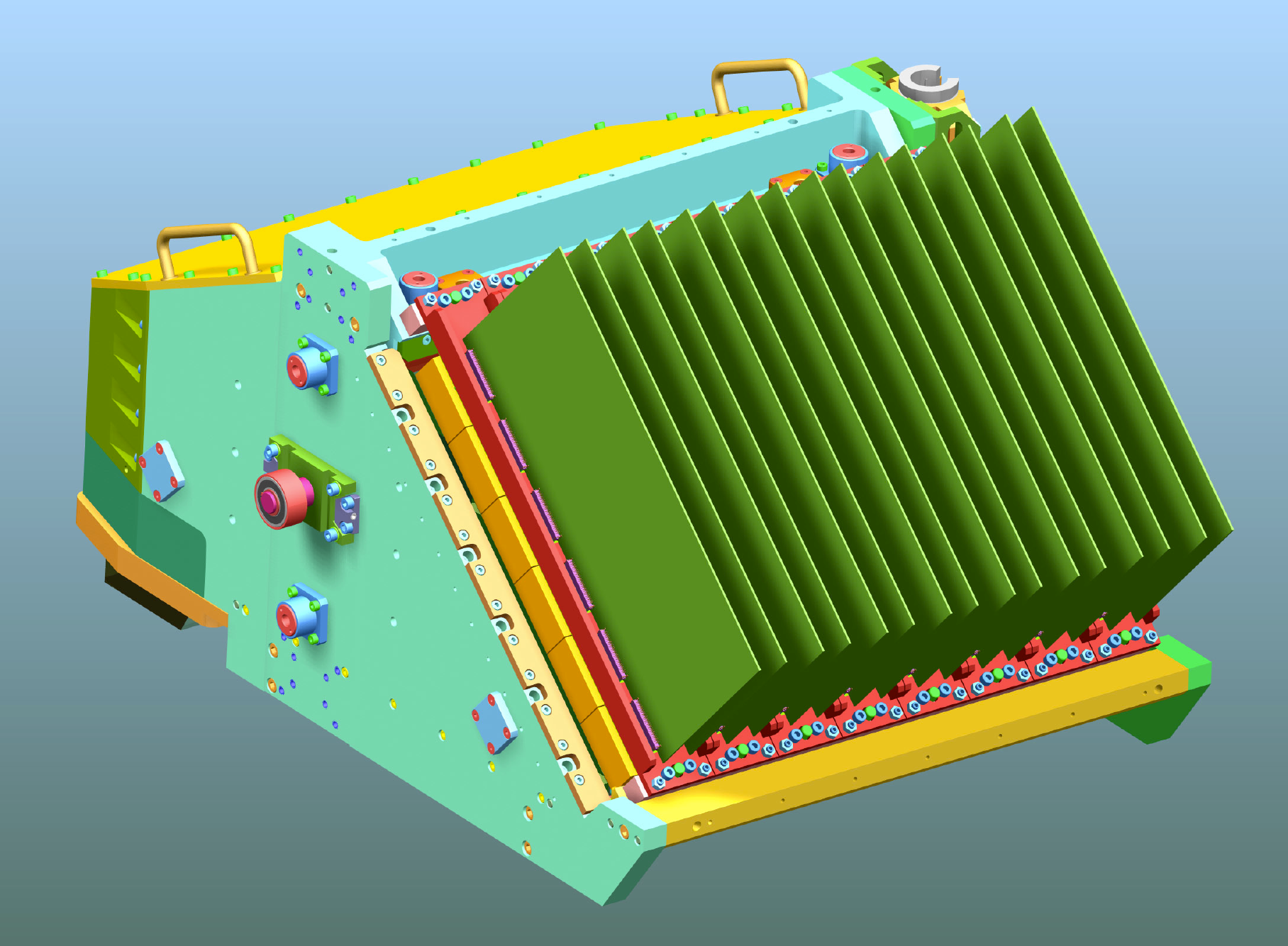}}
\end{center}
\caption{Photon camera with high density electronics reading out 48 H-8500 PMTs.}
\label{fig:Motherboard_c}
\end{figure*}

\tdrparagraph{ HV distribution and HV power supplies }

If the H-8500C tube is selected, HV cables will be routed under the motherboard as shown on Figure~\ref{fig:HV_distribution_b}. 
The drawback of this solution is that we will have 48 HV cables in a relatively small volume, and tubes will have to be rotated to fit HV 
cables in an efficient way.

The resistor chain of each H-8500 tube draws ${\sim} 150\muA$ at -1.0~kV. The HV power supply will be CAEN, Model A1835, or equivalent. It has 12 independent 
channels per module, each channel being able to provide up to 1.5~kV and either 7~mA or 0.2~mA (selectable by a jumper). The current monitor 
has 20~nA resolution. The entire FDIRC HV system would need 48 such HV power supplies, \ie, four per photon camera. They will be 
located behind the background shield in the non-radiation area. 

If the H-8500D tube is selected instead, HV will be distributed on the board as shown on Figure~\ref{fig:HV_distribution_a}.
The drawback of this solution
is that we would be grouping six PMTs on one HV power supply channel, which would have to supply 1~A. 

\begin{figure}[tbp]
\begin{center}
\includegraphics[width=\linewidth]{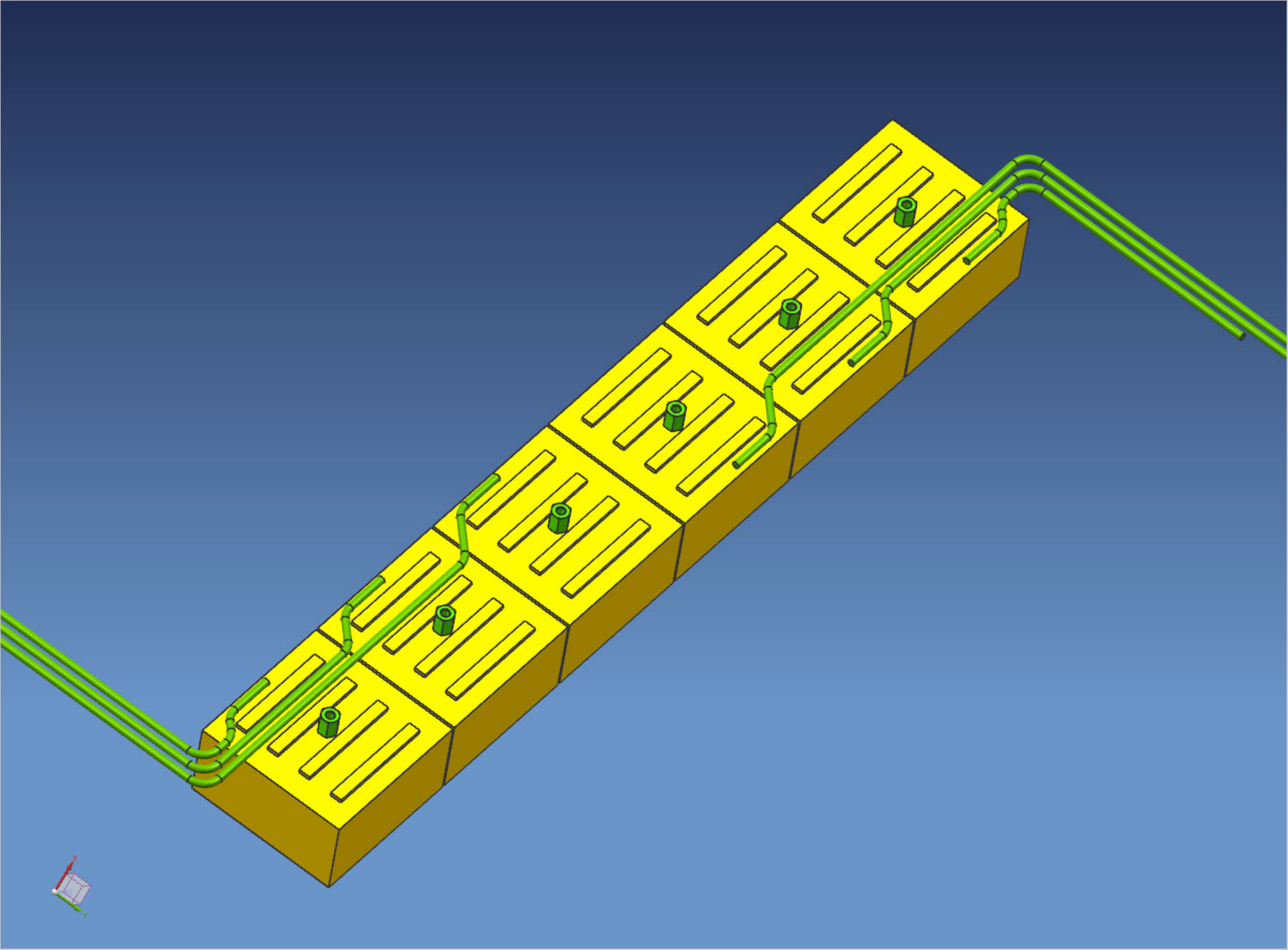}
\caption{HV distribution to H-8500C tubes along the vertical column. Half of tubes are rotated to pack cables efficiently.}
\label{fig:HV_distribution_b}
\end{center}
\end{figure}

\begin{figure}[tbp]
\begin{center}
\includegraphics[width=\linewidth]{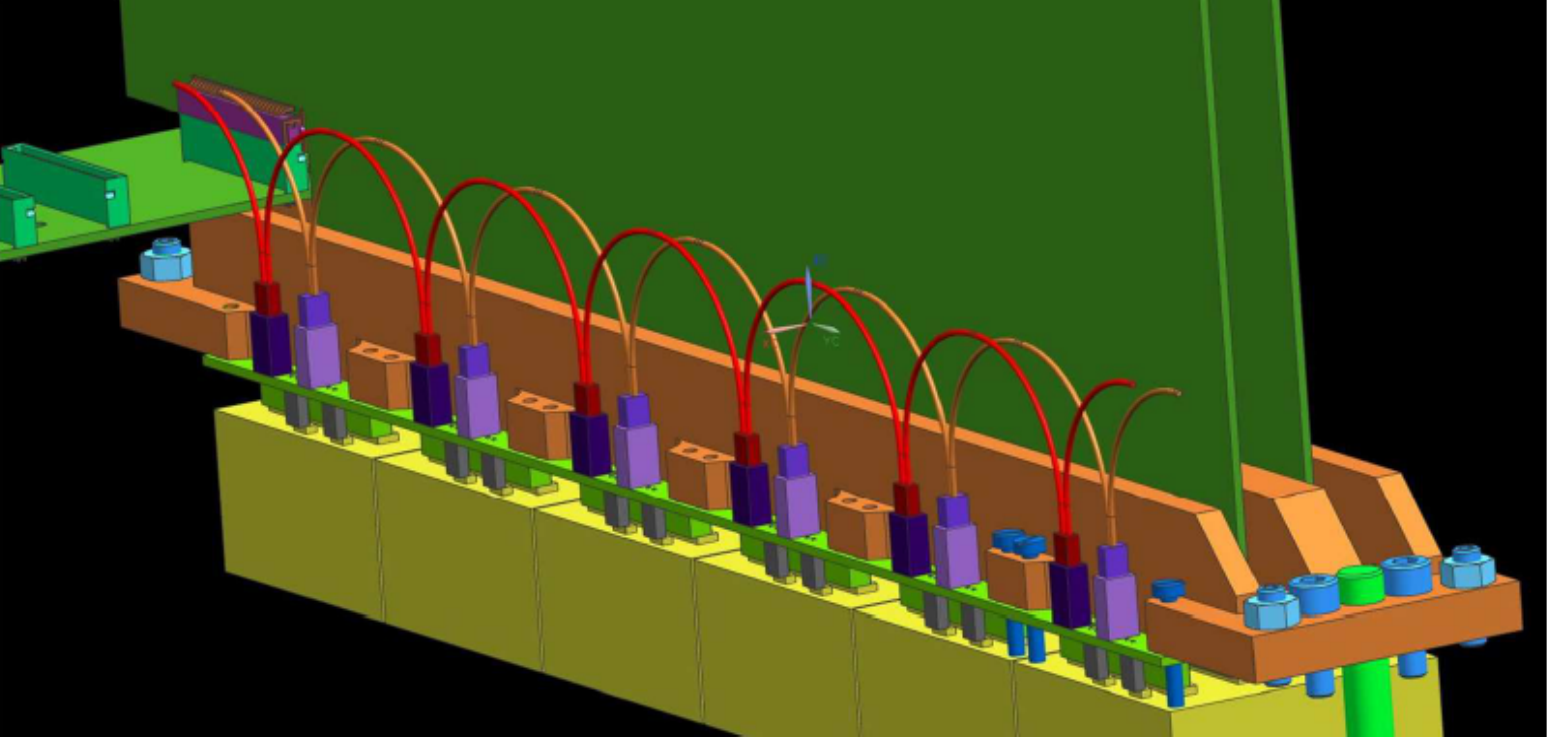}
\caption{HV distribution to H-8500D tubes along the vertical column.}
\label{fig:HV_distribution_a}
\end{center}
\end{figure}

\tdrparagraph{ Support services }
  FDIRC detector will need these services:
\begin{itemize}
\item A boil-off N$_2$ flow in each bar box up to 100~cc/min per bar box. The N$_2$ gas has to be distributed in stainless steel 
electropolished tubing. No oil bubbles to be used.
\item The total power dissipation in the entire system is about 6.1~kW. The cooling will be done with water-based heat exchangers and forced air.
\end{itemize}


\tdrsubsec{ Integration issues }

\tdrparagraph{ Background shield and access to detector maintenance }
  Because the front part of the FBLOCK background shield is mounted on the magnetic door, which is on rails, it will be easy 
to move it sideways to allow a quick access to the detectors and electronics~-- see Figure~\ref{fig:Det_layout_f}.

\tdrparagraph{ Earthquake analysis of FBLOCK \& bar box structure }
  Bar box axial and radial constraints will be equivalent to \babar\ DIRC setup. The Fbox system itself is not critical, being compact, 
rigid,  and with very limited lever arms. Of course  the support disk and the support rails structures of the Fbox must be
adequately stiff to avoid  resonance in the typical quake range. Axially, the Fbox must be constrained as the bar box. The increased risk 
relative to \babar\ DIRC consists in the coupling of Fbox and bar box. However, the presence of a RTV gluing layer instead of a rigid coupling 
and an adequately stiff support of the Fbox should prevent risks due to this coupling. Calculations are in progress.

\tdrparagraph{ PMT protection (helium, large backgrounds) }
   It is well-known that the PMT operation can be affected by a helium contamination, which can penetrate the PMT glass. These atoms 
convert to ions in the avalanche process, which can drift back to the photocathode creating secondary photoelectrons, often 
called 'after pulses'. Therefore, just like in case of the \babar\ DIRC (which had ${\sim} 10751$~PMTs) we should worry about any leak 
checking close to the \superb\ detector and using helium~-- for vacuum purpose. Even if it is done far away in the tunnel, air 
draft could bring it to the detector. We will need a helium detector to monitor this. 

Reference~\cite{mazziotta_Frascati_2011} summarizes the effect of helium contamination on a PMT. We should stress, however, that we did not 
do any experimental study with the H-8500 tube up to this point. 

The ion contamination in a PMT can be estimated by measuring the after-pulse rate. Figure~\ref{fig:after_pulses} shows how the 
$H^+$, $H_{2}^+$ and $He^+$ ion contamination~\cite{simi_Frascati_2011} affects a time spectrum of afterpulses in a H-8500 tube. 
The total measured rate of after pulses was less than ${\sim} 1\%$ rate for this tube. This measurement will have to be part of PMT QC 
procedure to weed out bad tubes. It will be useful to repeat it periodically on some tubes during the \superb\ data taking.

\begin{figure}[tbp]
\begin{center}
\includegraphics[width=\linewidth]{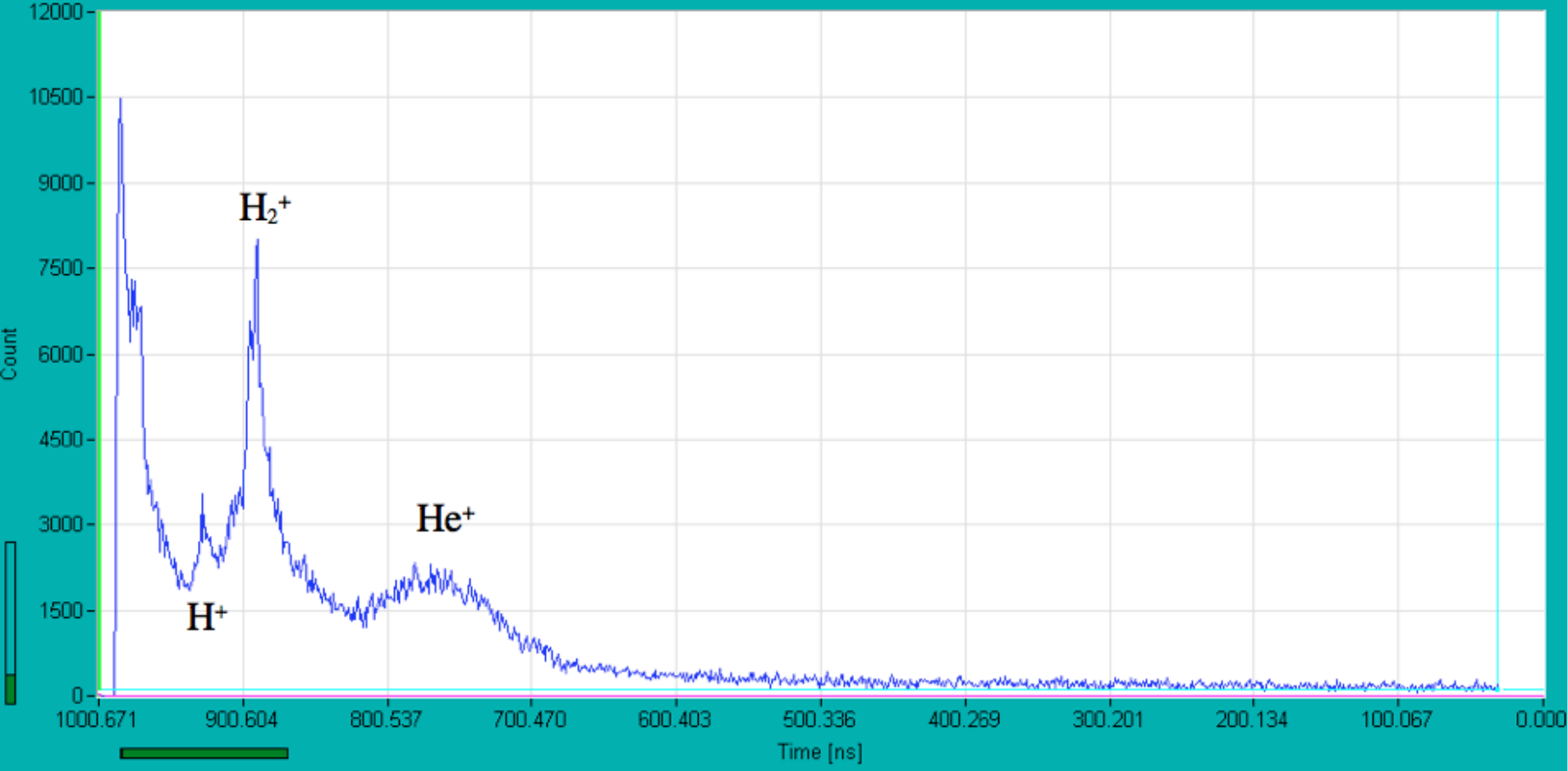}
\caption{A time spectrum of after pulses showing $H^+$, $H_{2}^+$ and $He^+$ contamination~\cite{simi_Frascati_2011}.}
\label{fig:after_pulses}
\end{center}
\end{figure}

\tdrsubsec{ FDIRC R\&D Results }

\tdrparagraph{ Test beam results from the first FDIRC prototype }

\begin{figure*}[tbp]
\begin{center}
\subfloat[Resolution with 3\mm pixels.]{\includegraphics[width=0.465\linewidth]{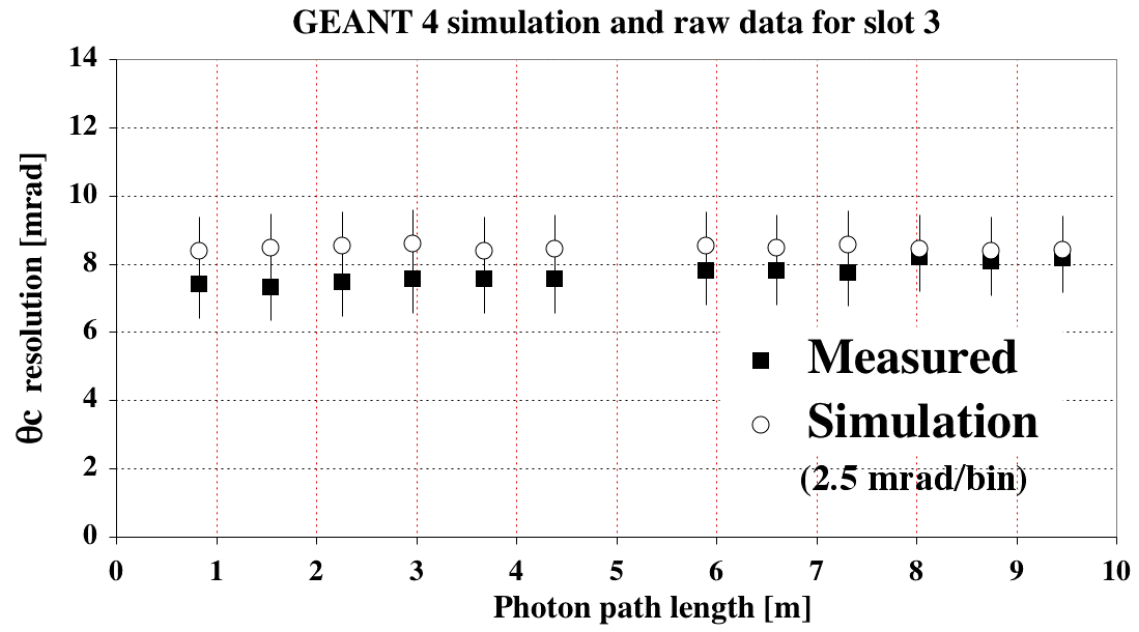}}
\hspace{5mm}
\subfloat[Resolution with 6\mm pixels.]{\includegraphics[width=0.465\linewidth]{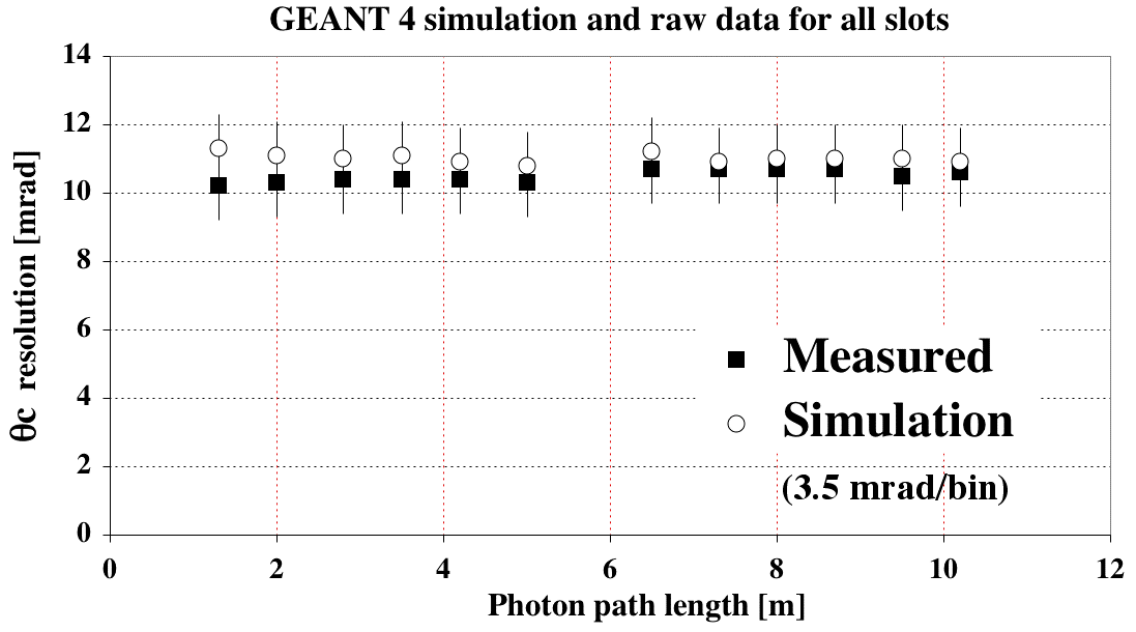}}
\end{center}
\caption{Measured and simulated Cherenkov angle resolution without chromatic correction. It is clear that 3~mm pixel choice would improve 
the FDIRC PID performance~\cite{Benitez:2008zz}}
\label{fig:1st_prototype_result_b}
\end{figure*}

Figure~\ref{fig:1st_prototype} shows the prototype. This prototype was tested in a 10~GeV electron 
test beam at SLAC. This beam entered the bar perpendicularly. It was a very successful R\&D program, resulting in a number of 
very useful results~\cite{Benitez:2006ei,Va'vra:2007zz,Benitez:2008zz}, which can be summarized as follows:

\begin{itemize}

\item We learned how to operate new fast highly pixelated detectors (Hamamatsu H-8500 and H-9500 MaPMTs; Burle MCP-PMTs). 
The H-9500 MaPMT was arranged to have $3\mm \times 12\mm$ pixel size, while the other two tubes had $6\mm \times 6\mm$ pixels.

\item Test achieved ${\sim} 10 \times$ better single-electron timing resolution than DIRC: $\sigma_{\mathrm{H-8500}} {\sim} 240\ps$, 
$\sigma_{\mathrm{H-9500}} {\sim} 235\ps$, and $\sigma_{\mathrm{MCP-PMT}} {\sim} 170\ps$.

\item We learned how to design a new optics, which is a combination of pin hole coupled to focusing optics, resulting in 
${\sim} 25 \times$ smaller photon camera that the \babar\ DIRC SOB.

\item This was the very first RICH detector establishing that the chromatic error can be corrected by timing~--
 see Figure~\ref{fig:Chrom_corr_c}. To be able to do such correction, one needs to achieve a timing resolution at a level 
of ${\sim} 200\ps$ per single photon, and the photon path length needs to be longer than 2-3 meters. The (F)DIRC bars 
are indeed longer~-- due to the bar extra length in the magnet iron~--, which helps to improve this correction.

\item With 6\mm size pixels we could reproduce \babar\ DIRC performance of Cherenkov angle resolution of ${\sim} 10\mrad$ per 
single photon if we do not perform the chromatic correction. With the chromatic correction, one could improve this
resolution by 0.5-2\mrad for photon path lengths longer than 2-3 meters~-- see Figure~\ref{fig:Chrom_corr_b}.

\item With 3\mm size pixels, we could substantially improve on the FDIRC performance~-- see Figure~\ref{fig:1st_prototype_result_b}. 
Figure ~\ref{fig:Final_prediction} shows the overall PID performance for various detector schemes. Clearly, smaller binning 
in the $y$-direction would be beneficial to improve the overall performance. Hamamatsu encouraged to use R11256 tube, which 
would give use 3~mm $\times$ 12~mm pad sizes, and possibly QE ${\sim} 36$\%. The comparison 
also includes the new Hamamatsu $8 \times 8$ SiPMT array, where we assumed PDE ${\sim} 52$\%; this is shown for comparative purposes only, as 
we do not assume to use it due to its large random noise rate at room temperature and further worsening by a possible neutron damage.

\item Discovered a new Cherenkov ring aberration, which worsens the resolution near the Cherenkov 
wings~-- see Figures~\ref{fig:abberation_a},~\ref{fig:abberation_b}. 

\end{itemize}

\begin{figure*}[tbp]
\begin{center}
\includegraphics[width=0.8\textwidth]{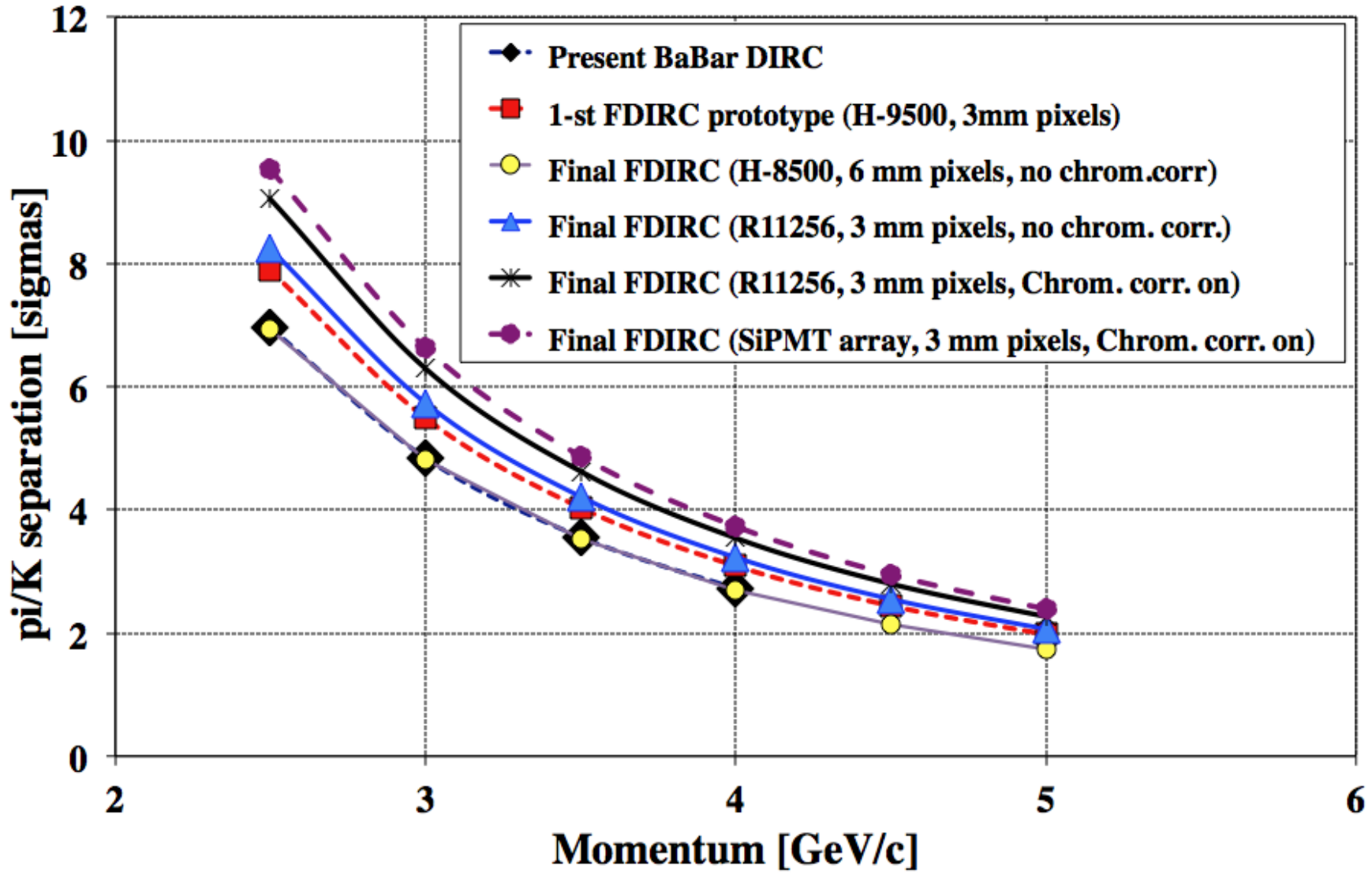}
\caption{Expected $K-\pi$ separation as a function of momentum for various detector schemes, including R11265 PMT and a
SiPMT $8 \times 8$ array (3~mm pixel sizes).}
\label{fig:Final_prediction}
\end{center}
\end{figure*}

\tdrparagraph{CRT test results from the first FDIRC prototype }

  We have built the cosmic ray telescope (CRT) to study the FDIRC prototype and various versions of TOF counters using 3D tracking~\cite{Va'vra:2009zz}. 
Figure~\ref{fig:CRT_setup} shows the CRT setup~\cite{Va'vra:2009zz} which provides a muon energy cutoff of 2~GeV in its latest version.

\begin{figure*}[tbp]
\begin{center}
\includegraphics[width=0.8\textwidth]{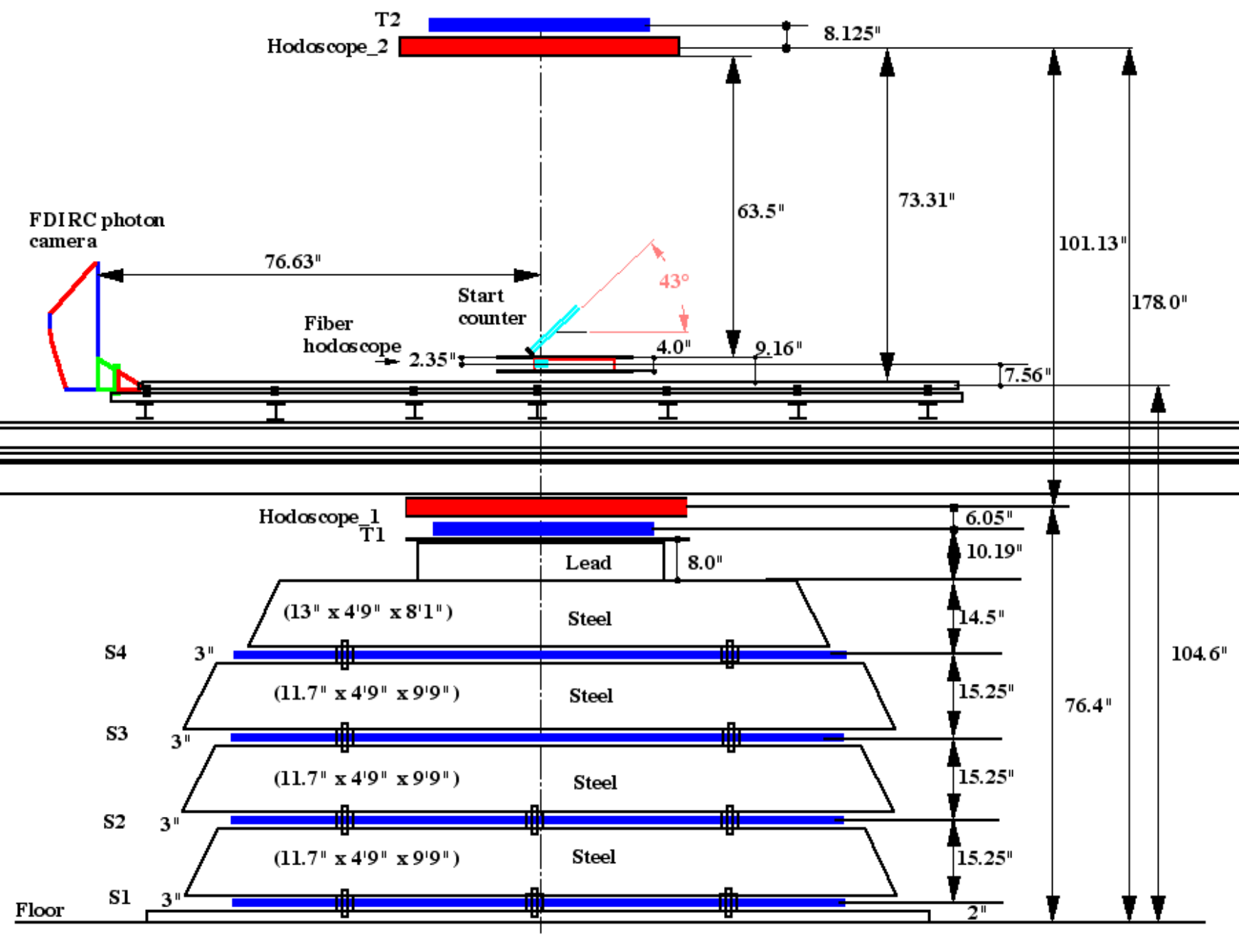}
\caption{The final FDIRC prototype in the CRT setup (shown is its latest version with addition of 8 inches of lead absorber).}
\label{fig:CRT_setup}
\end{center}
\end{figure*}

   The first prototype was also tested in the CRT setup. The SLAC CRT setup consists of energy absorber made of
4 feet-thick iron, which provides a muon energy lower cut-off of $\sim$~2~GeV. It also provides tracking with 1.5\mrad resolution over an angular range
of dip angles within 15 degrees from the vertical. This was also a significant test because it allowed to investigate the Cherenkov angle resolution 
with 3D tracks~\cite{Nishimura:2012zz}. Results can be summarized as follows:

\begin{itemize}
\item We learned how to handle 3D tracks in the Cherenkov angle analysis (during the beam test tracks entered 
perpendicularly~\cite{Benitez:2006ei,Va'vra:2007zz,Benitez:2008zz}). 
\item Tail in the Cherenkov angle distribution is related to the ambiguity treatment and it is more significant for 3D tracks. 
The first FDIRC prototype~\cite{{Nishimura:2012zz}} only had two ambiguities: we could not tell a sign of photon vector in the
$x$-direction for photons exiting the bar end, and therefore in the analysis we had to consider both signs. In the final FDIRC prototype, we will
have more ambiguities than in the first FDIRC prototype as there are more photon reflections in the new optics. 
This ambiguity effect enhances the tail as one cannot always reject wrong solution, 
and it is magnified by a presence of background such as knock-on electrons (delta-rays) or showers accompanying CRT muons. The CRT setup is very 
good to learn how to deal with these ambiguities. The main conclusion of the studies is that one has to use the quantity 
dTOP = TOP$_{\mathrm{measured}}$ - TOP$_{\mathrm{expected}}$,
where TOP is the photon time-of-propagation in the bar. 
One can make a tight cut on dTOP to help rejecting the background. 
\item Running CRT continuously allowed to test various versions of electronics very conveniently, and to produce
the Cherenkov angle resolution under real conditions.
\item The CRT trigger was also used to trigger a PiLas laser diode, which provided a single photoelectron monitoring of all pixels all
the time while we were running. The laser trigger did not overlap with the CRT data to avoid any confusion. This allowed to study 
the stability of FDIRC timing.
\item 
We plan to test the Final FDIRC prototype at high rate background 
in the CRT setup. This will be done by admixing an asynchronous random light source to the laser calibration signal, while taking regular
CRT data. This task will be accomplished with a fiber mixer, which will mix the laser signal with the random light source. 
In this way we can study the reconstructed Cherenkov resolution as a function of the random background 
in a controlled way, and see at what point the reconstruction algorithm breaks down. At the same time, we will 
be monitoring the timing resolution deterioration using the laser signal.
\end{itemize}

\tdrparagraph{ Scanning setups to test H-8500 PMTs and Electronics }
  We have several PMT scanning setups located at SLAC, Maryland, Bari, Padova and LAL-Orsay. These setups differ in their capabilities, 
designs and electronics. Although these setups do not use the final electronics yet, they were nevertheless already very useful to 
reveal many H-8500 detector details. So far, the following topics have been studied in some details: 
(a) efficiency uniformity across PMT for 15 tubes, (b) gain uniformity for 15 tubes, (c) cross-talk, (d) charge sharing, 
(e) after-pulses, (f) pre-pulses (amplification starts on first dynode), (g) anode response to pulses on the last 
dynode (for calibration purposes), etc. We used many results from these tests throughout this TDR chapter. 

   In these studies we learned that:
\begin{itemize}
\item Based on a study of 15 tubes (960 pixels), the gain uniformity among pixels is typically better than 2.5:1. 
\item The single photoelectron timing resolution (TTS) has a structure within each pad.
\item The charge sharing effect is very small in this particular tube due to its electrode structure, and it is not 
worthwhile to use it to reduce the effective pixel size, which would help to improve the Cherenkov angle resolution.
\item Although one could find a good spot in a H-8500 PMT giving a TTS resolution of ${\sim}140$~ps (see Figure~\ref{fig:TTS_dist}),
averaging over an entire pixel area gives a TTS resolution more like ${\sim} 200-250$~ps, with edge pixels being 
worse (see Figure~\ref{fig:TTS_dist_in_1st_FDIRC_prot}).
\item A typical pixel-to-pixel cross-talk in H-8500 tube is about 3\%, judging from the scope measurement of pulse amplitudes 
on neighboring pixels using the SLAC amplifier~-- see Figure~\ref{fig:Cross-talk_vavra}.
\end{itemize}

\tdrsubsec{Conclusion and ongoing FDIRC R\&D }

\tdrparagraph{ Experience with the final FDIRC prototype in CRT }
  During its construction we learned the following major points:
\begin{itemize}
\item It is possible to build this kind of optics, for affordable cost and within the required tolerances. 
\item It is possible to handle heavy FBLOCK fragile optics and to assemble Fbox around it.
\item We learned how to couple bar box window to the new wedge with Epotek-301 glue.
\item We learned how to couple optically the FBLOCK to the new Wedge. It is done with a 1\mm-thick RTV. It is a very large area optical coupling and we learned 
how to develop a bubble-free coupling. We have demonstrated that this RTV coupling can be cut by a razor wire, the surfaces cleaned and glued again.
\item The full size FDIRC prototype is now being studied in the SLAC cosmic ray telescope. The main aim is to verify the optical design and learn how
to perform the data analysis.
\end{itemize}

\tdrparagraph{ Detector studies in various scanning setups }
  The study of the H-8500 PMT continues in all scanning setups. So far all tests used various
kinds of electronics. It is essential to repeat some of these studies with the final electronics. This, however, cannot be done sooner 
than in 2013. We also want to make a decision of other possible tube choices, namely Hamamatsu H-9500 and R-11256.

\tdrparagraph{ Electronics R\&D }
  The final electronics is being developed at LAL. The detector motherboard is being designed at LAL, Padova and Bari. The final 
electronics will be then tested in the CRT setup at SLAC.

\tdrsubsec{ System Responsibilities and Management }

\tdrparagraph{ Management board structure }
  The PID group has a management board, where each institution has a representative. The role of this management board is to 
resolve monetary, manpower and other global issues within various institutions as they come during the construction stage.

\tdrsubsec{ Cost, Schedule and Funding Profile }

\tdrparagraph{ Cost }
The cost of the FDIRC construction, assembly and integration in the \superb\ detector
is given in Chapter~\ref{chap:Budget}.

\tdrparagraph{ Schedule and Milestones }
   The FBLOCK production determines the entire production time line, \ie, all other tasks take less time. 
Figure~\ref{fig:Schedule} shows the present time-flow estimate of the FBLOCK production (for clarity we show only two FBLOCK production cycles). 
The manufacturing speed is limited by FBLOCK machining and polishing. The total duration of FBLOCK production, as it stands now, 
is ${\sim} 28$~months. The next longest part of the project is the PMT production, which will take about 2 years, 
including procurement, delivery and testing. 

\begin{figure*}[tbp]
\begin{center}
\includegraphics[width=0.9\linewidth]{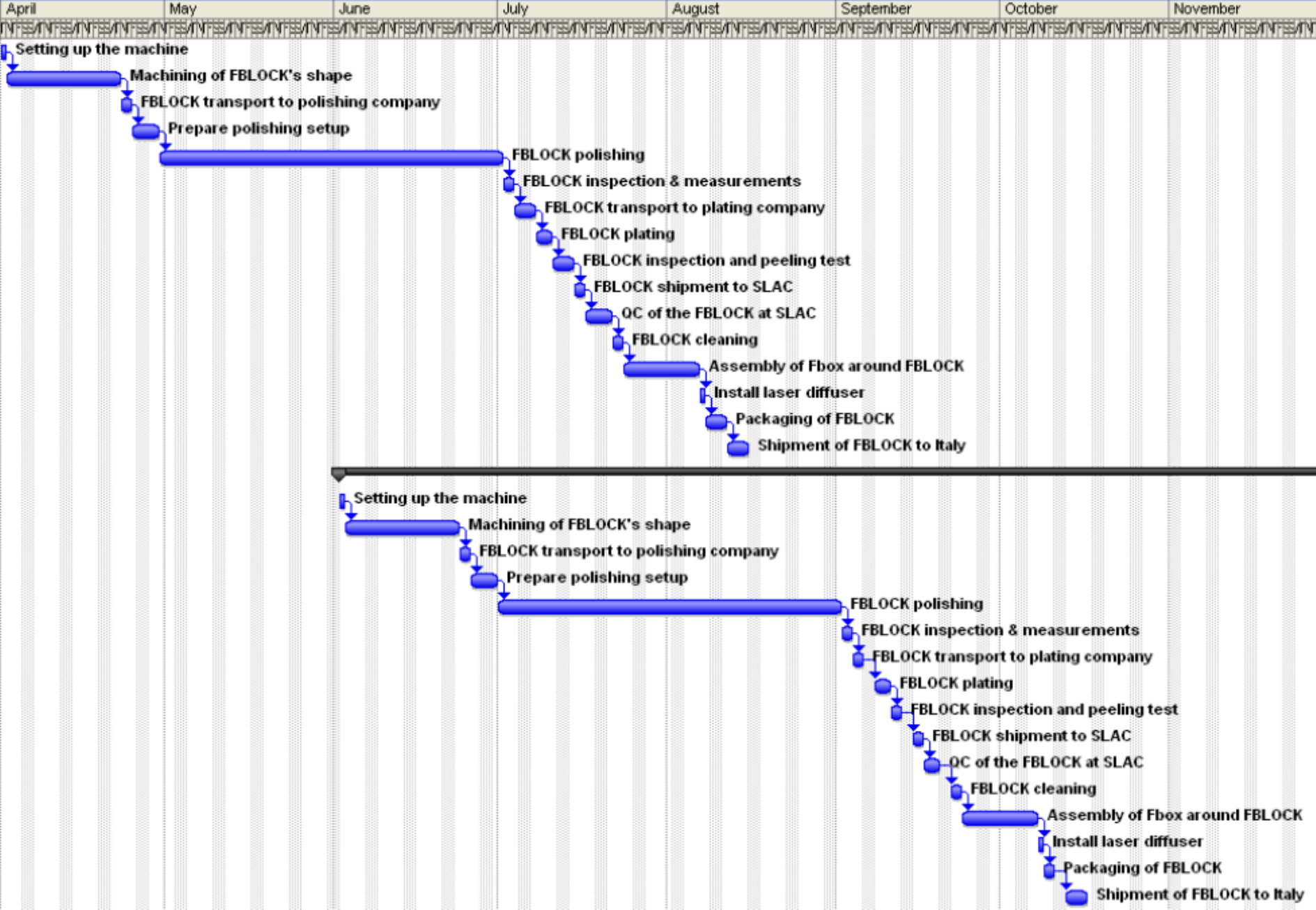}
\caption{FBLOCK production schedule (2 FBLOCK production cycles only).}
\label{fig:Schedule}
\end{center}
\end{figure*}

\tdrparagraph{ Critical path items }
  Clearly, the most critical path items are: (a) machining and polishing of the FBLOCK optics and (b) delivery of 600 Hamamatsu H-8500 
PMTs. The FBLOCK delivery is controlled by the production capacity, which limits deliveries to one FBLOCK every 6-8 weeks. We will try 
to find a way to speed it up, or find a parallel production possibility, but it is generally difficult to replicate relevant 
experience with different companies when one deals with a non-standard optics. Hamamatsu company told us that they can deliver 
600 H-8500 tubes over a period of 2 years, which is about 25 tubes per month. It is important that we check that delivered tubes 
have required performance. We may have to split testing into at least two different scanning setups to be able to 
cope with a delivery rate of 25 tubes per month.

\tdrsec{ A possible PID detector on the \superb\ forward side }
\label{section:forwardPID}

\tdrsubsec{ Physics motivation and detector requirements }

The \superb\ barrel region is covered by a dedicated PID detector: the FDIRC, described in the previous sections of this report.
The information from this detector, combined with the energy losses from the DCH, ensures a good $\kaon/\pi$ separation up to about
4~\gevc. On the other hand, PID in the \superb\ endcaps only relies on \dedx\ measurements from the tracking system
without the addition of a dedicated device. In the high
momentum region, pions and kaons are only separated at the ${\sim} 2 \sigma$ level~-- however, the use of the cluster counting
method in the DCH would increase this separation~\ref{subsec:clusterCounting}. Moreover, \dedx\ distributions
exhibit a 'cross-over' $\kaon/\pi$ ambiguity region around 1~\gevc, inside which charged hadrons cannot be properly identified as
the energy loss curves overlap~-- see Figure~\ref{fig:pi-K_separation}. One can see that a time-of-flight (TOF) resolution of only ${\sim} 100$~ps is 
good enough in the cross-over region. Though, a better timing accuracy is needed if one wants to be sensitive to higher momenta.

\begin{figure}[tbp]
\begin{center}
\includegraphics[width=\linewidth]{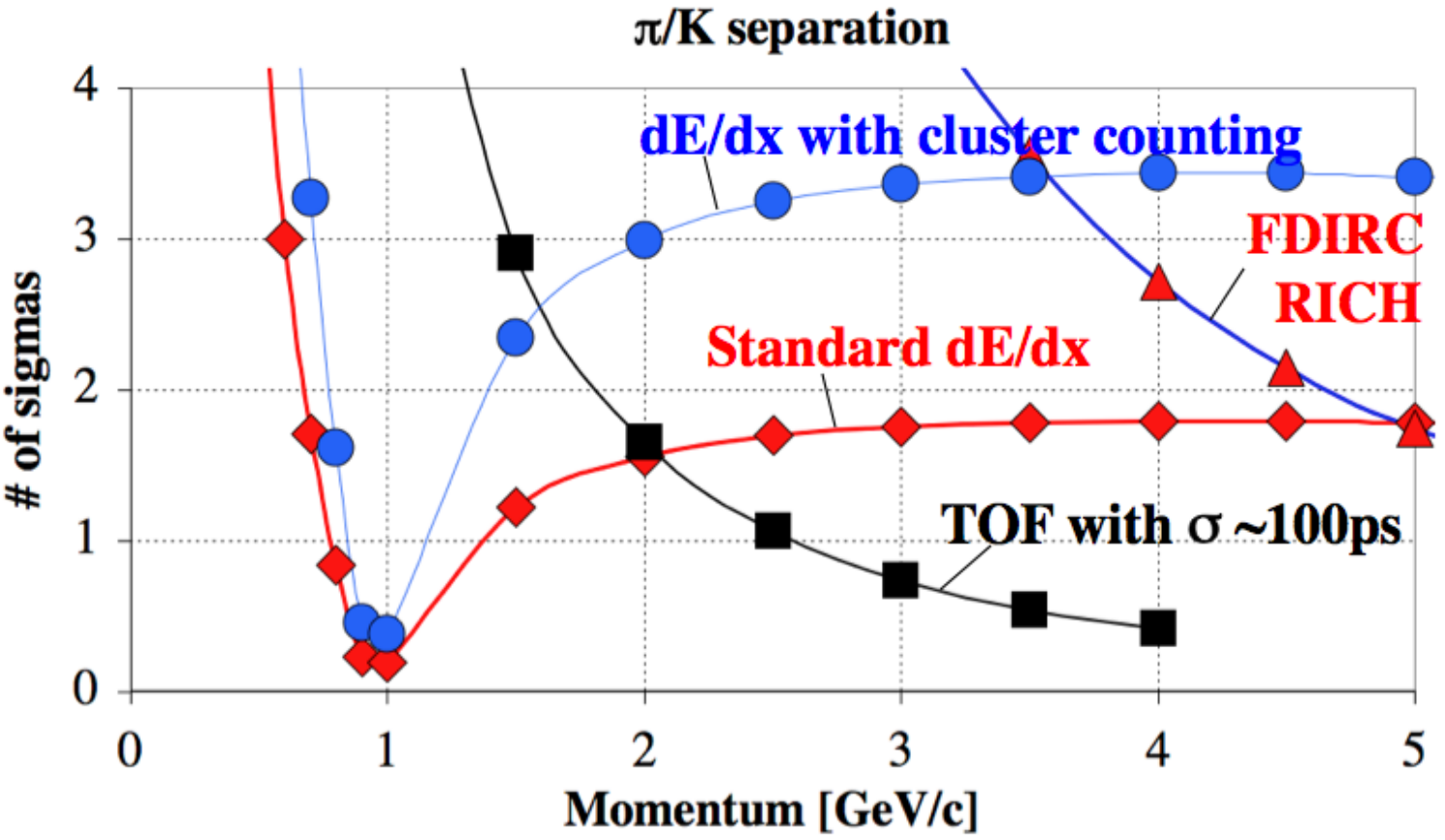}
\caption{Role of a TOF system with ${\sim} 100$~ps timing resolution for $\kaon/\pi$ separation at low momentum~\cite{jjv_toflyso_1,jjv_toflyso_2}.}
\label{fig:pi-K_separation}
\end{center}
\end{figure}

Improving PID in the two endcap regions requires thus new dedicated detectors which should be both powerful and relatively small.
The latter characteristic is needed in order to fit in the limited space available in the endcap regions.

In the backward side (where the particle momentum is quite low in average), the EMC group is proposing to install a veto
calorimeter to improve the \superb\ energy measurement~-- see section~\ref{sec:EMCback}. Should this device be
fast enough, it would allow one to separate kaons from pions using the particle TOF.

Due to the \superb\ boost, the forward region of the detector corresponds to a fraction of the geometrical acceptance larger than
its angular coverage in the laboratory frame~-- about $[17^\circ;25^\circ]$ in polar angle~-- while the particles have higher momentum in average.
Therefore, a forward PID detector has to be efficient from about 1~\gevc to 3~\gevc.

Physics-wise, the \superb\ performance would benefit from improved PID in many areas:
\begin{itemize}
\item larger efficiency in various rare and exclusive $B$ decays;
\item reduced background;
\item better \B\ tagging;
\item improved exclusive reconstruction of hadronic and semileptonic $B$ channels. This technique is used for recoil physics
analysis of rare and hard-to-reconstruct decays. Based on the coherent production of a $\BB$ pair at the \FourS, it first
identifies events for which one of the \B's, the $\B_{\mathrm {tag}}$, is fully reconstructed. Once this is done, the rest
of the event automatically corresponds to the (other) signal $B$: $\B_{\mathrm {sig}}$. This constraint provides additional
kinematical information on $\B_{\mathrm {sig}}$ decays which would otherwise be underconstrained~-- e.g. because of undetected
neutrinos in the final state. Each of the hundreds of exclusive modes used by this method to reconstruct the $\B_{\mathrm {tag}}$
meson is characterized by two numbers: its reconstruction efficiency and the purity of the selected sample. Both numbers increase with
an improved PID and the larger the number of particles in the final state, the faster the gain.
\end{itemize}

Given the integration constraints in the \superb\ forward region, a forward PID detector must be compact.
A total thickness around 10~cm would likely put some constraints on the position and dimensions of the DCH and of the forward EMC,
but studies~-- see section~\ref{subsec:FTF} below for a summary~-- have shown that such changes would only have a limited impact on the performances
of these two devices. Yet, the forward PID detector should add the smallest possible \Xrad\ fraction in front of the calorimeter
to limit preshowers.

Finally, the cost of the forward PID detector should be moderate in comparison to the one of the FDIRC as it covers a much smaller
fraction of the \superb\ acceptance.

\tdrsubsec{ The Forward task force and the selection of a forward PID technology }
\label{subsec:FTF}

From early 2009, several possible designs of forward PID detectors have been studied within the PID group. The proposals were
submitted to a task force representative of the whole collaboration. It was charged to review the different designs and to make
recommendations to the \superb\ Technical Board, which would then take the final decision. This step was completed in June 2011
and lead to the choice of a technology, as reported in the remainder of this section. 

Four main technologies were extensively studied:
\begin{itemize}
\item the focusing aerogel RICH (FARICH);
\item a pixelated \tof\ detector with Cherenkov radiators;
\item a scintillating detector coupled to G-APD arrays;
\item the \ftof\ (FTOF), finally selected~-- see section~\ref{subsec:FTOF} for the description of this detector.
\end{itemize}
More details about the first three forward PID concepts can be found in appendix~\ref{subsec:forwardPID-Appendix}.

\tdrparagraph{ Charge and activities of the Forward task force }

In July 2010, a taskforce on a possible forward PID system in the \superb\ detector was formed. Its primary charge was to provide an assessment of
\begin{itemize}
\item the physics impact of a PID system in the forward region of the \superb\ detector, roughly defined by the polar angle range $17^\circ$ to $25^\circ$;
\item the feasibility of the proposed detector technologies for the forward PID system.
\end{itemize}

The taskforce considered a list of questions and criteria for both physics and technology assessments, as discussed below.

\begin{itemize}
\item Physics evaluation
\begin{itemize}
\item A list of key benchmark physics channels that are affected by a forward PID system were evaluated for both
the gain and any potential negative impact due to the added material before the forward EMC or any required change to the DCH dimensions.
\item A preliminary performance evaluation was made of some of the proposed technologies, in the presence of background hits.
\item The impact of each proposed forward PID technology on \piz\ efficiency, and momentum and \dedx\ resolutions in the DCH was performed. 
\end{itemize}
\item Detector technology issues
\begin{itemize}
\item Estimates of cost, required manpower, and construction schedule for each of the proposed technologies were performed.
These include information on the availability of components on the time scale of the \superb\ construction schedule.
\item The need for a proof-of-principle was considered for each of the proposed technologies~-- at least with cosmic rays
and if possible with beam tests.  Issues common to nearly all devices are: performance in presence of background; the
effect of the \superb\ magnetic field on the photodetectors and on the overall performance of the device; photodetector aging.
\item Integration issues.
\end{itemize}
\end{itemize}

The taskforce met with forward PID proponents at all \superb\ workshops between September 2010 and May 2011 when the
committee recommendation was released. There were also phone meetings in between workshops. 

\tdrparagraph{ Summary and taskforce recommendation }	

~\\~\\ {\it Note: This report has been written in May 2011. Since then, both the \superb\ design as a whole and the design 
of the FTOF (the selected technology) have
evolved. For instance, the width of the Tungsten shield around the beam pipe has been increased from 3 to 4.5~cm,
leading to a dramatic reduction of the FTOF background. More information can be found in the following section
(\ref{subsec:FTOF}), where the current status of the FTOF R\&D activities is described.} \\

The physics gain from a forward PID device is around 4-5\% for the benchmark channel $B \to K^{(*)} \nu \bar{\nu}$:
roughly 2\%/kaon. No physics channel with higher gain has been identified. Results based on simulation and beam test
(1\gevc\ electrons) show no significant degradation of resolution \& efficiency for $\gamma$ and $\pi^0$ in the EMC. In addition,
the impact on tracking resolution due to a shortened DCH has been estimated to be ${\sim} 1\%$ degradation in momentum
resolution/cm cut. 

The overall assessments for the proposed detector technologies are the following.
\begin{itemize}
\item FARICH. On the whole, this technology is likely to yield the most powerful ~-- and robust~-- PID performance,
extending well above the nominal 4\gevc\ for $B$ decays. The expected performance is verified by impressive beam test
results. However, no physics channel that would significantly gain from the extended performance has been identified.
Moreover, the required cut of ${\sim} 17$~cm to DCH length significantly degrades momentum resolution in this angular
region. This is an unacceptably large negative impact on the detector performance and a too severe constraint on the
tracking system. Hence, the taskforce does not see this technology appropriate for forward PID in the \superb\ detector

\item Pixelated TOF (LYSO plus G-APD array option). This technique, due to its potential minimal disturbance on the rest of the 
detector and likely modest cost, was deemed very attractive. At the aimed resolution of ${\sim} 100$~ps, it would 
complement the \dedx\ measurements for $\kaon/\pi$ coverage below 2\gevc. However, with the obtained time resolution (${\sim} 230$~ps) 
for a full size LYSO crystal in cosmic ray tests, the proponent \& taskforce have concluded that this technique will not deliver
the required performance.

\item FTOF. Simulation studies and cosmic ray tests~-- see section~\ref{subsec:FTOF}~-- have demonstrated that key aspects of this technique
can be attained~-- including the goal of a time resolution of ${\sim} 90$~ps/hit. But there remains significant uncertainties
on the expected background level and its impact on PMT lifetime. The taskforce believes this technique could be appropriate
for the forward PID system provided that background (and radiation) issues are understood~-- which may require further studies
of the IR design and shielding~-- and that a full prototype of the system is developed and tested, to verify the expected
performance, in particular the pattern recognition in the presence of background hits.
\end{itemize}



The final recommendation for the forward \superb\ region is the following: the importance of hermeticity (and redundancy)
in PID coverage will increase as we approach the systematics-dominated era in the \superb\ physics program.  Hence, the
taskforce members believe~-- independently of the outcome of the current technology evaluation~-- that there is physics
merit to allowing a gap of ${\sim} 5-10$~cm in the forward region for a PID device. This would allow for introduction of
a system at a later stage of the experiment, in case the detector studies are not completed in time for the initial installation of the \superb\ detector.

\begin{figure}[tbp]
\begin{center}
\includegraphics[width=\linewidth,angle=90]{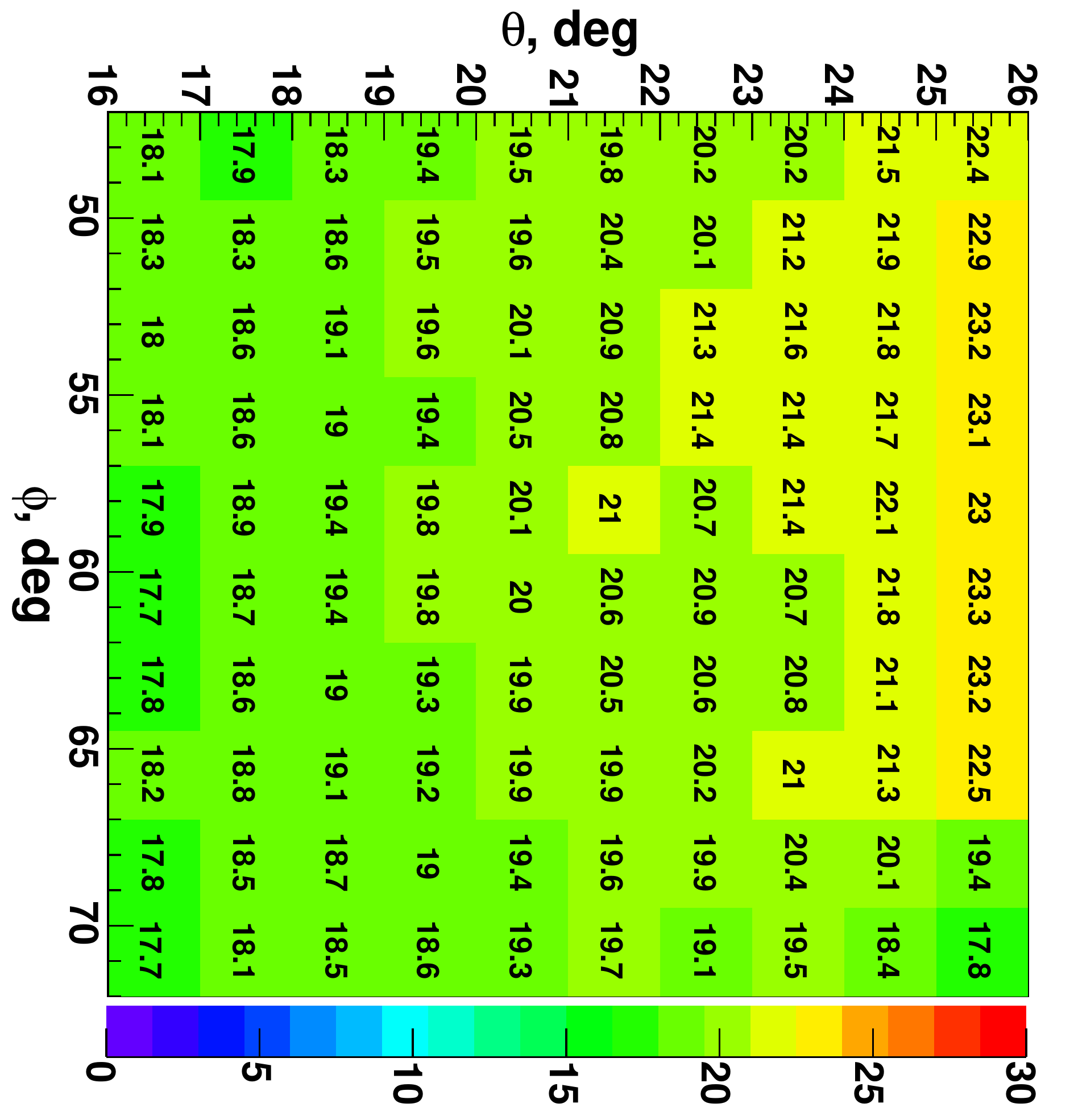}
\caption{Map giving the mean number of detected photoelectrons in a FTOF sector as a function of the track polar
($\theta$) and azimuthal ($\phi$) angles at origin. This map is drawn for 800~\mevc\ kaons shot from the IP in a simplified
detector model including the 1.5~tesla longitudinal solenoid field.}
\label{fig:PID-Forward:photonMap}
\end{center}
\end{figure}

\tdrsubsec{ The DIRC-like forward time-of-flight detector (FTOF) }
\label{subsec:FTOF}

As indicated by its name, the FTOF~\cite{jjv_toflike_1,jjv_toflike_2,Burmistrov:2011wka} belongs to the family of the DIRC detectors~\cite{Ratcliff:1992hd,blair_2008}.
One should also say that this particular geometry is close to the TOP concept, which is proposed for the Belle-II barrel detector~\cite{Iijima:2011zza}. Charged particles cross a thin layer of fused silica; provided that their momentum is high enough, they emit Cherenkov
light along their trajectories. Part of these photons are trapped by total internal reflection and propagate inside
the quartz until an array of MCP-PMT where they are detected. Unlike the (F)DIRC,
no attempt is made to reconstruct Cherenkov angles: $\kaon/\pi$ separation is provided by \tof~-- at given momentum, kaons
fly more slowly than pions as they are heavier.

Given the short flight distance between the IP and the FTOF (about
2~meters, hence a 90~ps difference for 3~\gevc particles), the whole detector chain must make measurements accurate
at the 30~ps level or better. This is one of the main challenges of this design, the others being the photon yield, the
sensitivity to the background (including ageing effects due to the charge integrated over time by the photon detectors)
and the event reconstruction. Regarding the latter point, one should emphasize that the FTOF is in fact a two-dimensional
device: the PID separation uses both the timing and spatial distributions of the photons detected in the MCP-PMT arrays.

\tdrparagraph{ Design optimization }

The performances of the FTOF depend on two main parameters which should be optimized simultaneously~\cite{Burmistrov:2011wka}. One is the photon
yield per track, which should be as high as possible; the other is the photon timing jitter, which should be minimized
in order to efficiently separate kaons from pions. Both are strongly related to the geometry of the detector.

The number of Cherenkov photons produced by a given track scales linearly with its path length in fused silica.
Simulations show that a 1.5~cm thickness for the quartz tile is a good compromise between light yield and detector
thickness in term of \Xrad. This finite size introduces an irreducible component of the photon time jitter
(about 50~ps), as Cherenkov photons are emitted uniformly along the track path inside the tile. Their initial
directions lie on a cone which opening angle depends on the track velocity (which is a function of the particle
measured momentum and of its unknown mass) and on the medium optical index ($n\approx 1.47$ in fused silica). The
main axis of the cone is aligned with the track direction. With the FTOF design extensively used to prepare this
report, tracks originating from the IP all give a large-enough number of photoelectrons, as shown on
Figure~\ref{fig:PID-Forward:photonMap} which is drawn using low-momentum kaon tracks~-- the most
unfavourable case as the Cherenkov photon yield increases with the particle velocity.

\begin{figure}[tbp]
\begin{center}
\includegraphics[width=\linewidth,angle=90]{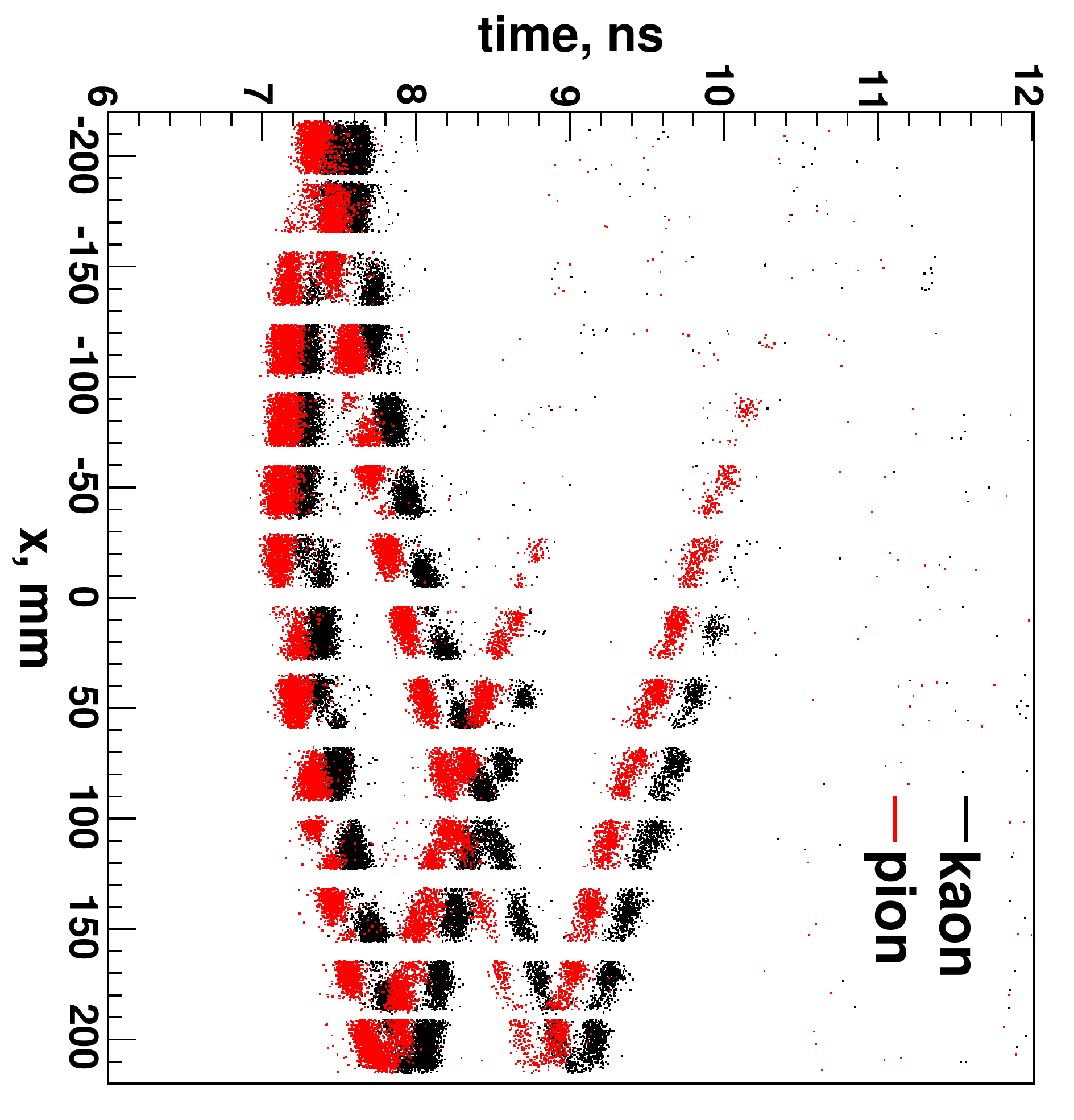}
\caption{The FTOF is a 2D-device: in addition to the photoelectron timing, the distribution of hits in the MCP-PMT
channels is also a discriminating variable to separate kaons from pions, as shown on the two plots superimposed in this
figure. Both show the time versus position of hits generated by 2~\gevc\ kaons (black) and pions (red).
$t=0$ corresponds to the particle generation at the IP while $x$ is an axis along the tile
width which shows the MCP-PMT position. The 14 photodetectors are clearly visible just as the small gaps between them.}
\label{fig:PID-Forward:2DMaps}
\end{center}
\end{figure}

Therefore, it is clear that photons from a given track can follow several different paths before being detected in a MCP-PMT channel.
This is another source of time jitter which has to be taken into account. One way to mitigate it is to select only the 'most-direct'
photons which reach the detectors after at most few reflections on the tile front and back faces. This can be done by covering the
tile sides by photon absorber, but the loss in photon yield is then too high. In the current design, only the tile inner radius
is covered by absorber to stop the 'downward-going' photons whose paths are too long and make their timing unusable. Indeed,
with tiles perpendicular to the beam axis, photons are mainly 'upward-going': they are travelling towards the tile outer radius,
which is why the MCP-PMTs are located in this area.

Given the accuracy of the timing measurements required by the FTOF, chromaticity is another effect which cannot be neglected:
the smaller the photon wavelength the larger its speed~-- 'red' photons are faster than 'blue' ones.

The FTOF photon yield also depends on the tile orientation as only Cherenkov photons trapped inside the quartz by total
internal reflection can be detected. Using vertical tiles as reference, simulations show that a $10^\circ$ tilt in the
forward (backward) direction increases (decreases) the yield by about 15\% in average. But configurations in which the
FTOF is bent forward are impossible in practice, due to integration constraints on the \superb\ forward side.

One point worth mentioning is that the FTOF is a 2D-device, as shown on Figure~\ref{fig:PID-Forward:2DMaps}. This means that
the $\kaon/\pi$ separation is not only coming from the photoelectron timing but also from the spatial distribution of the photon
hits on the MCP-PMTs.

\tdrparagraph{ FTOF design }


The \ftof\ is made of 12 thin fused silica tiles (1.5~cm thick each, which corresponds to 12\%~\Xrad) located perpendicularly
to the beam axis and covering $30^\circ$ in azimuth each~-- see conceptual drawings on Figure~\ref{fig:FTOF_Conceptual_Schema}.
The requirements for the tile dimension accuracy and the fused silica
polishing quality are less stringent than for the \babar\ DIRC bars, as the Cherenkov photons will bounce much less in the quartz.
Therefore, building the FTOF tiles should not be an issue given the knowledge accumulated in this area over the past two decades.


\begin{figure}
\begin{center}
\includegraphics[width=\linewidth]{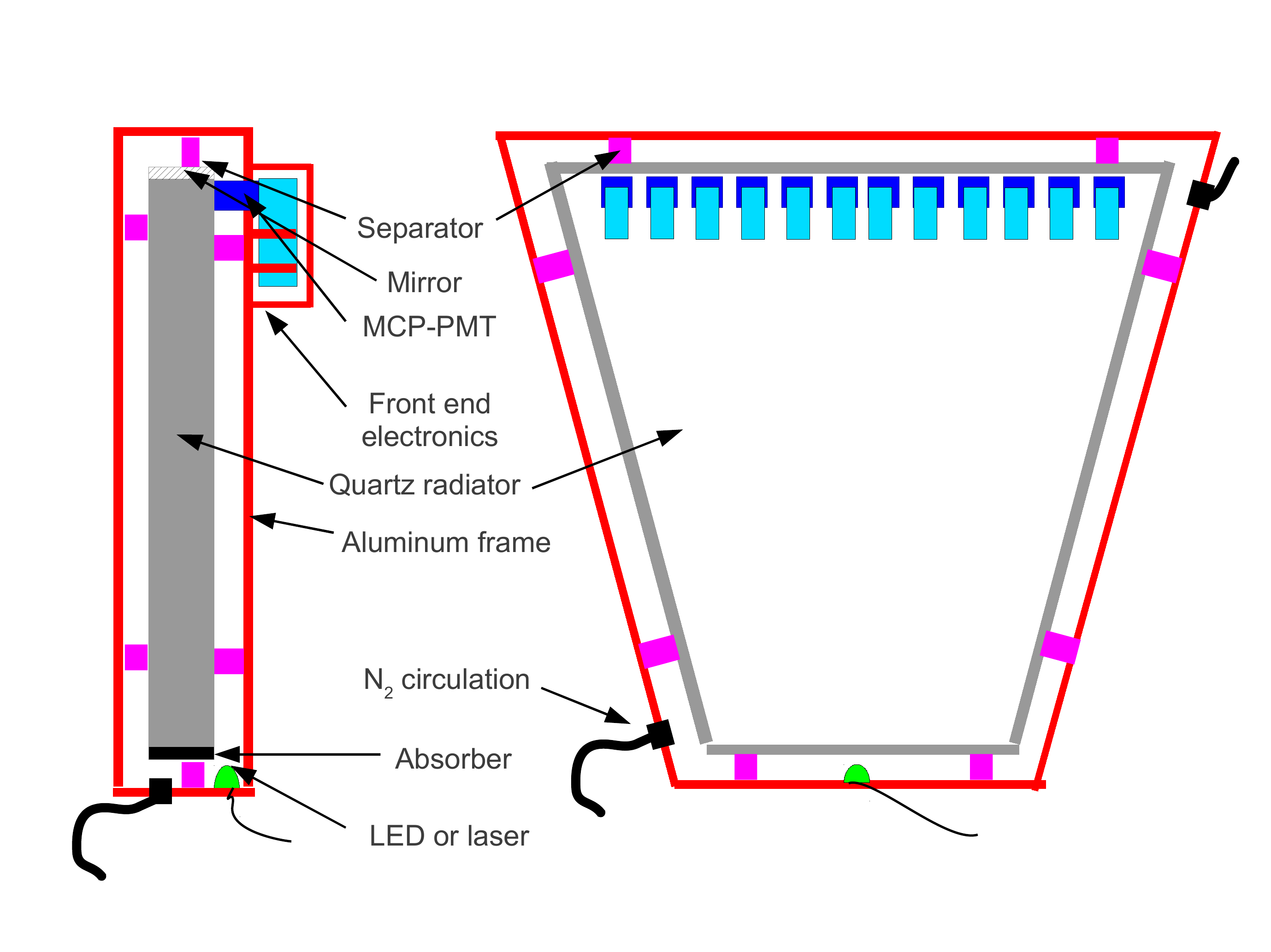}
\caption{Sketch of a FTOF sector: side (left) and back (right) views. 
These drawings are not to scale: the width of a FTOF sector is about
5~cm (frame included) while its height is a few tens of cm.}
\label{fig:FTOF_Conceptual_Schema}
\end{center}
\end{figure}


Each FTOF tile will be placed into a light aluminum box which is the equivalent of the DIRC bar boxes for the DIRC
quartz bars. To keep the fused silica tiles as clean as possible, the tile box should not contain any other detector component;
there will just be an optical contact between the quartz and the MCP-PMTs. Both the tile box and the box enclosing the
MCP-PMTs will be light-tight; in addition, the tile box will be leak-tight and N$_2$ will flow continuously inside it to
avoid any moisture.


\begin{figure}[tbp]
\begin{center}
\includegraphics[width=\linewidth]{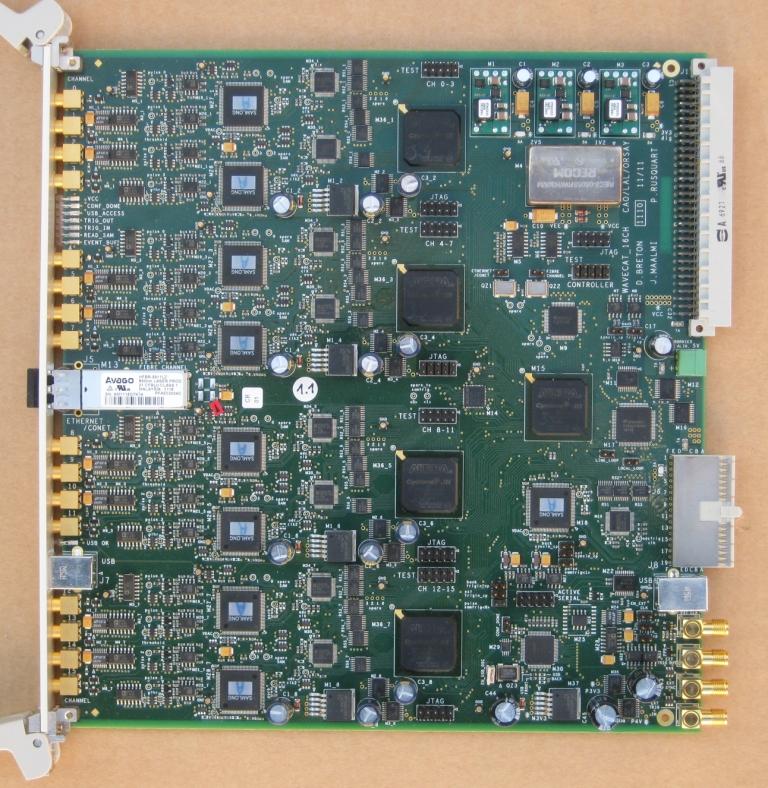}
\caption{The 16-channel Wavecatcher board.}
\label{fig:PID-Forward:wavecatcher16channels}
\end{center}
\end{figure}

The prototype FTOF front-end electronics is based on the new 'WaveCatcher' boards commonly developed by LAL Orsay
and CEA/Irfu Saclay~\cite{Breton:2010zza,Breton201261}. They are 12-bit 3.2 GS/s low power and low cost
waveform digitizers based on the SAMLONG ultra-fast
analog memory. The sampling time precision is as good as 10~ps rms, a value measured at the level of a crate hosting
eight 2-channel boards. The photon arrival time is extracted via digital Constant Fraction Discrimination (CFD). A new
16-channel board~-- see Figure~\ref{fig:PID-Forward:wavecatcher16channels}~--
recently designed has been characterized: it exhibits the same timing performances.

\begin{figure}[tbp]
\begin{center}
\includegraphics[width=\linewidth]{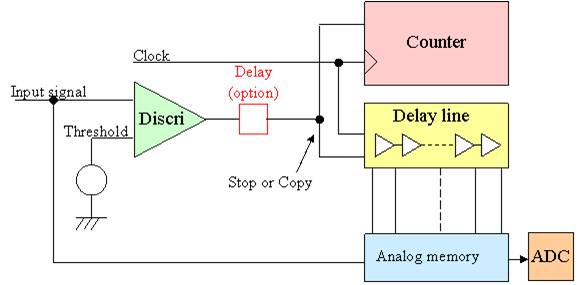}
\caption{Scheme of the new TDC for the FTOF electronics.}
\label{fig:PID-Forward:FTOF-TDC}
\end{center}
\end{figure}

The final electronics will be highly integrated and based on a new principle of TDC, called SAMPIC~-- see Figure~\ref{fig:PID-Forward:FTOF-TDC}.
The latter, designed in
AMS CMOS 0.18~\mum\ technology, will be able to tag the arrival time of 16 analog signals with a precision of a few ps thanks
to its embedded analog memory (running between 5 and 10 GS/s) and its embedded ADC. This will permit housing at least 64 channels
in a final front-end board. In the current FTOF design, a single 64-channel Wavecatcher
board will be enough to readout a sector (hence 12 boards will be needed in total for the whole FTOF).

\begin{figure*}[tbp]
\begin{center}
\includegraphics[width=0.7\linewidth]{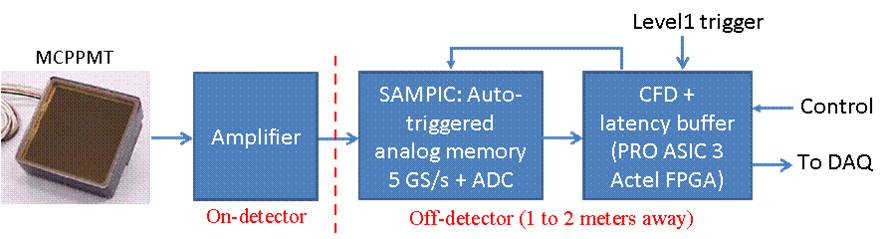}
\caption{Schematics of the FTOF whole electronics chain.}
\label{fig:PID-Forward:FTOF-Schematics}
\end{center}
\end{figure*}

The FTOF front-end boards will be located outside the detector, both for protection from background and radiation,
and to allow easy access for repair. The analog signal will be amplified right behind the MCP-PMTs and then transferred
to the front-end board via signal cables. Figure~\ref{fig:PID-Forward:FTOF-Schematics} shows the schematics of the
FTOF electronics chain.


\begin{figure}[tbp]
\begin{center}
\includegraphics[width=\linewidth]{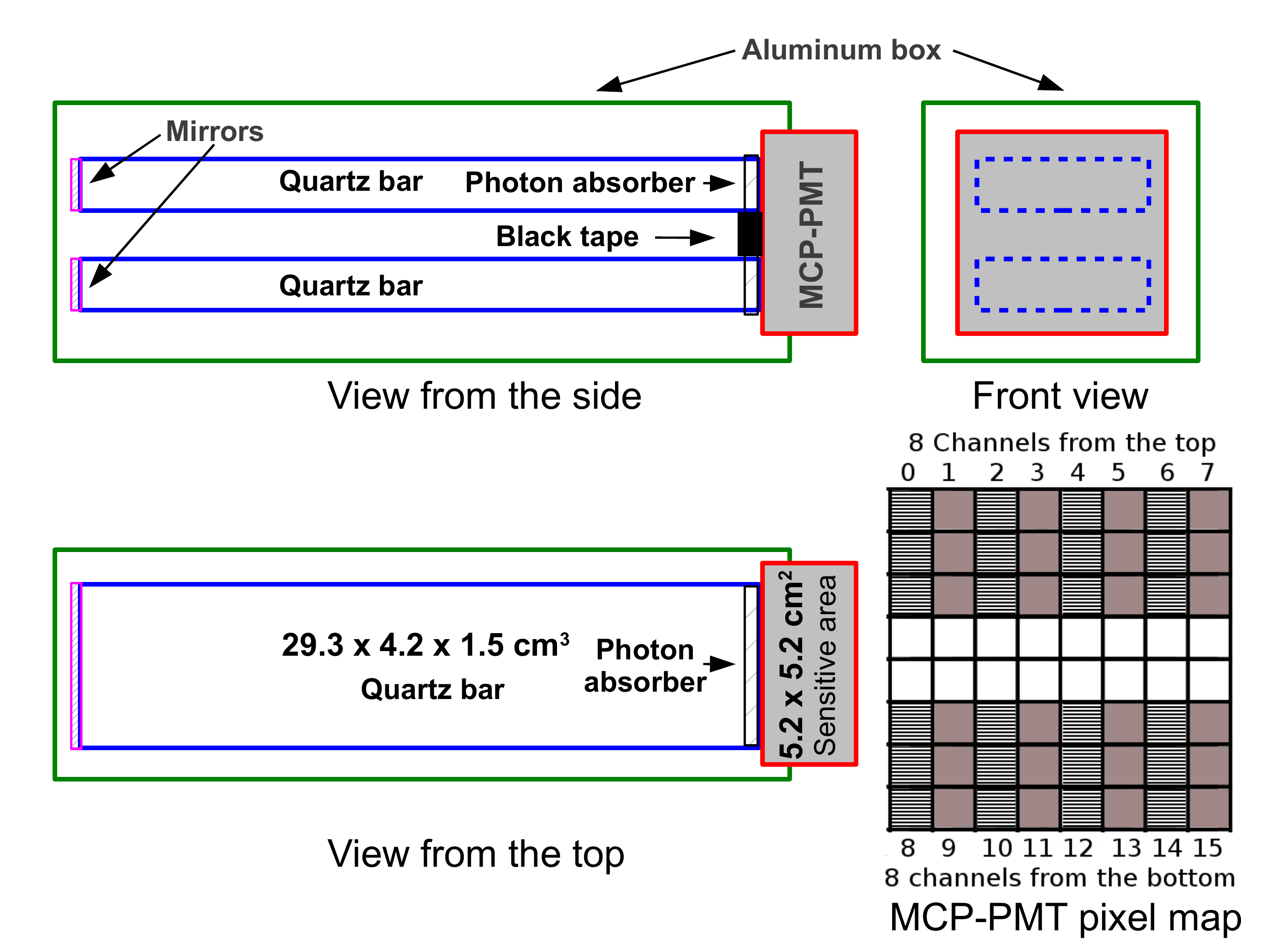}
\includegraphics[width=\linewidth]{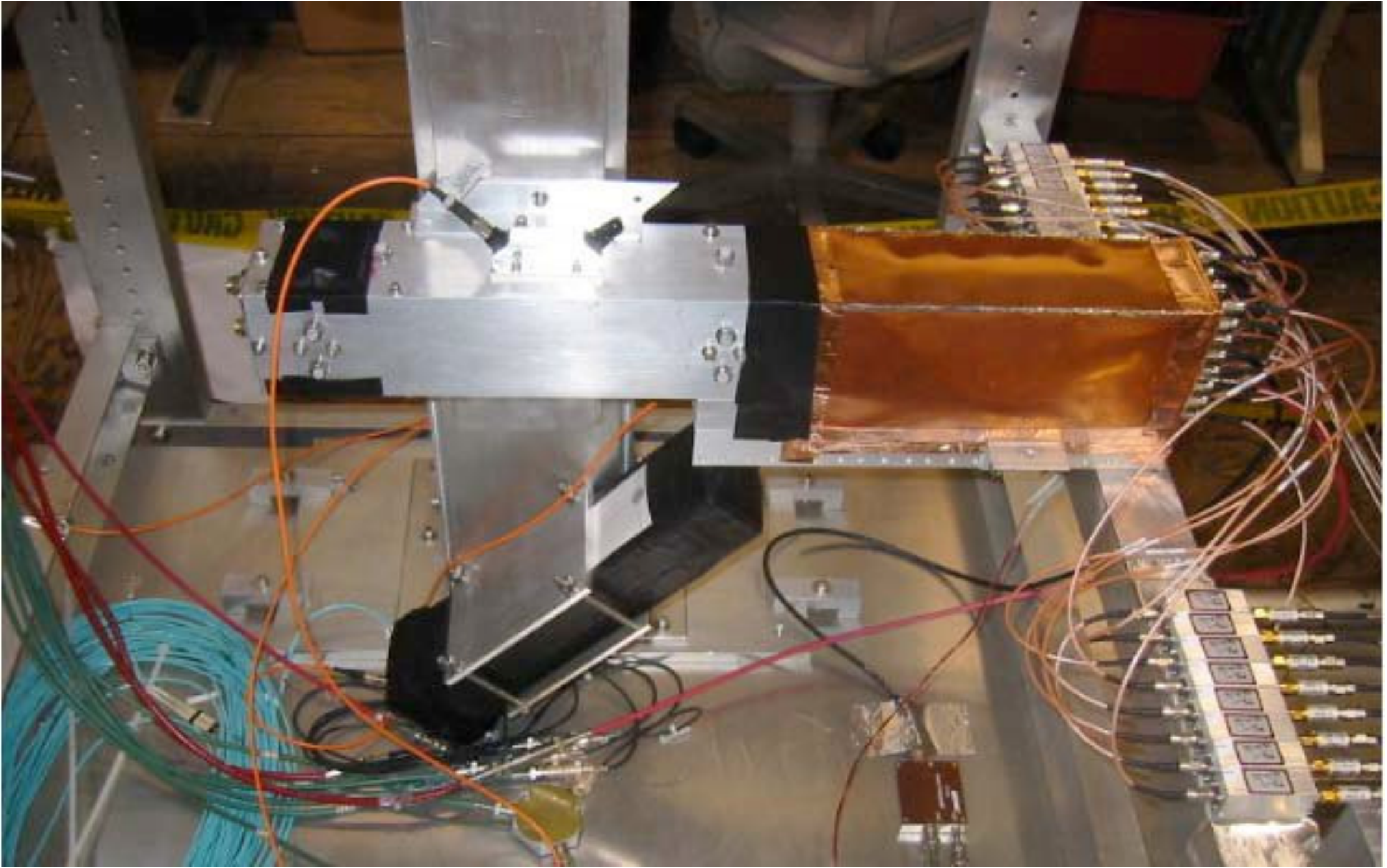}
\caption{Top: side and top views of the 2-bar setup; the 16 channels readout by the USBWC boards are shown on the
right-hand side. Bottom: picture of the apparatus in the SLAC CRT. The orange box is a Faraday cage containing the
USBWC electronics while the grey box contains the prototype. Under these items, one can see the enclosure of the sloping
fast quartz counter which is used to trigger the CRT in coincidence with the hodoscopes.}
\label{fig:PID-Forward:FTOFprototype}
\end{center}
\end{figure}

As explained above, the whole FTOF measurement \& processing chain must be ultra-fast. The time resolution per photon must be
at the level of 100~ps or better, which makes the use of MCP-PMTs mandatory. Among the products available on the market, the
Hamamatsu SL-10 is currently our baseline. With a time-transit-spread (TTS) of 70~ps (FWHM), an active area of
$22 \times 22$~mm$^2$, a quantum efficiency of 17\%~\cite{Inami_2010} and a lifetime of ${\sim} 2$~C/\cma~\cite{Inami_2010}, 
it offers a good compromise given the FTOF requirements. Two such photon detectors will be characterized at LAL-Orsay in 
the coming months. In the design used during the TDR preparation period,
one needs 14 SL-10 per sector, hence 168 in total; each MCP-PMT will host 4 different channels which will all have their own
HV channel. The HV power supplies will be located behind the detector shield wall and could then be accessed 24/7.

A fully instrumented FTOF sector is expected to be very light: about 12~kg total.

\tdrparagraph{ Tests with the first FTOF prototype }


A test of the FTOF detection concept was run at SLAC in the Cosmic Ray Telescope (CRT) between Fall 2010 and Spring 2011~\cite{Burmistrov:2011wka}.
The SLAC group provided the detector hardware (two short rectangular fused silica bars readout by a Photonis MCP-PMT~\cite{Vavra_2010}) 
and the CRT test facility; the LAL team brought new 2-channel USBWC boards~\cite{Maalmi_2010}. The test setup is shown in 
Figure~\ref{fig:PID-Forward:FTOFprototype}. Cosmic muons cross two bars located on top one another. The Cherenkov light propagates 
to the instrumented end of the bars. By shorting and grounding pixels, one defines 16 MCP-PMT vertical pads~-- 8 for each bar~-- which are connected to the USBWC electronics.

\begin{figure}[tbp]
\begin{center}
\includegraphics[width=\linewidth]{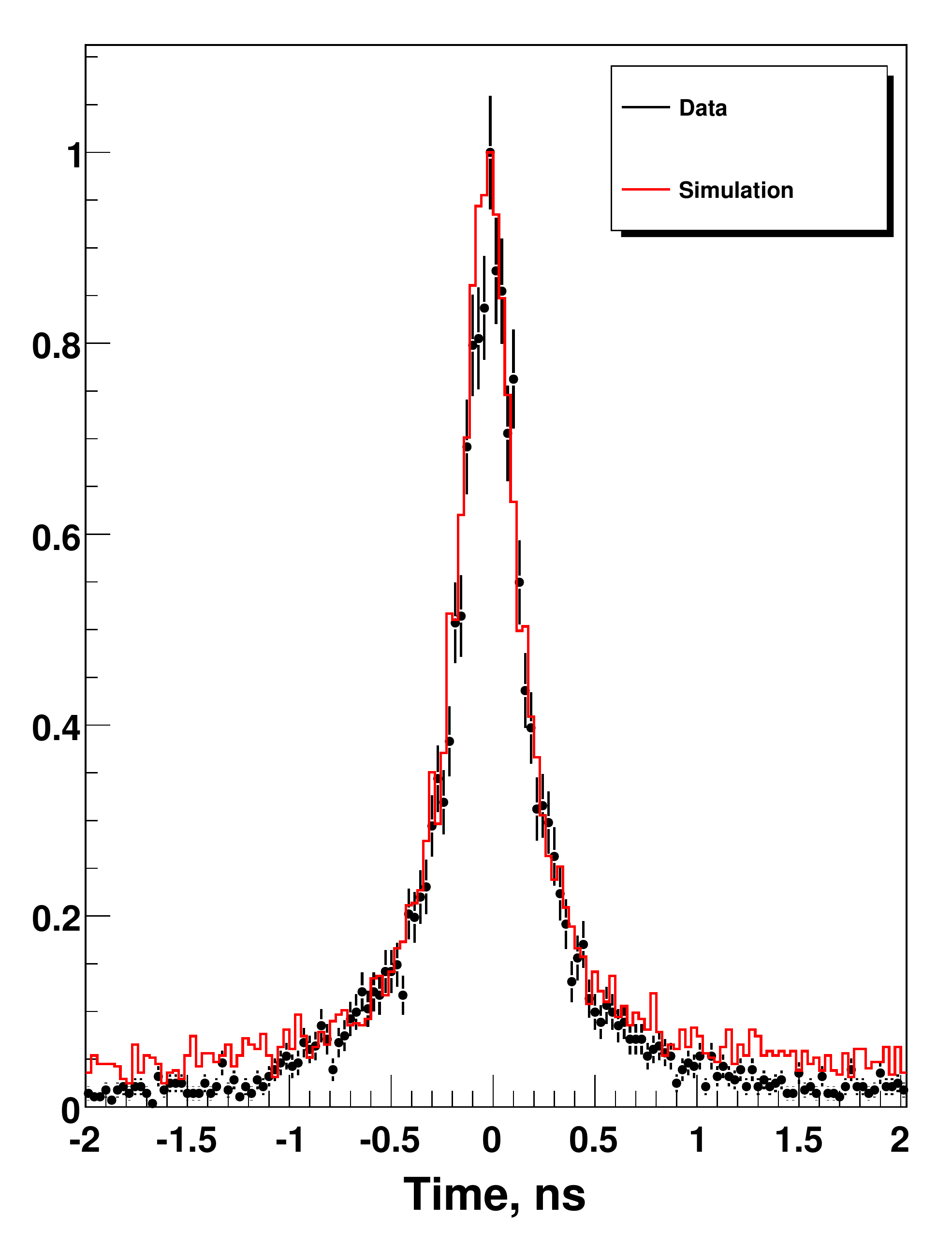}
\caption{Comparison between data (bullets) and simulation (solid-line histogram) for the time difference
between two MCP-PMT channels among the 16 readout by the USBWC boards. All distributions are scaled so that their peak
is equal to 1 in arbitrary units. The data-MC agreement is impressive, both for the core and the tails of the distributions.
Similar results are achieved for all pairs of channels studied.}
\label{fig:PID-Forward:data-MC}
\end{center}
\end{figure}

The CRT trigger start time is defined by a fast Cherenkov counter located under the two bars. For each trigger, the 16 USBWC analog 
buffers are readout to see if photons have been detected by one or more pads. If this is the case, the USBWC event is written to disk.
To handle overlapping waveforms in a given pixel, a photon absorber was placed in front of the MCP-PMT. The waveform time was 
measured by a CFD-based software algorithm.

\begin{figure}[tbp]
\begin{center}
\includegraphics[width=\linewidth]{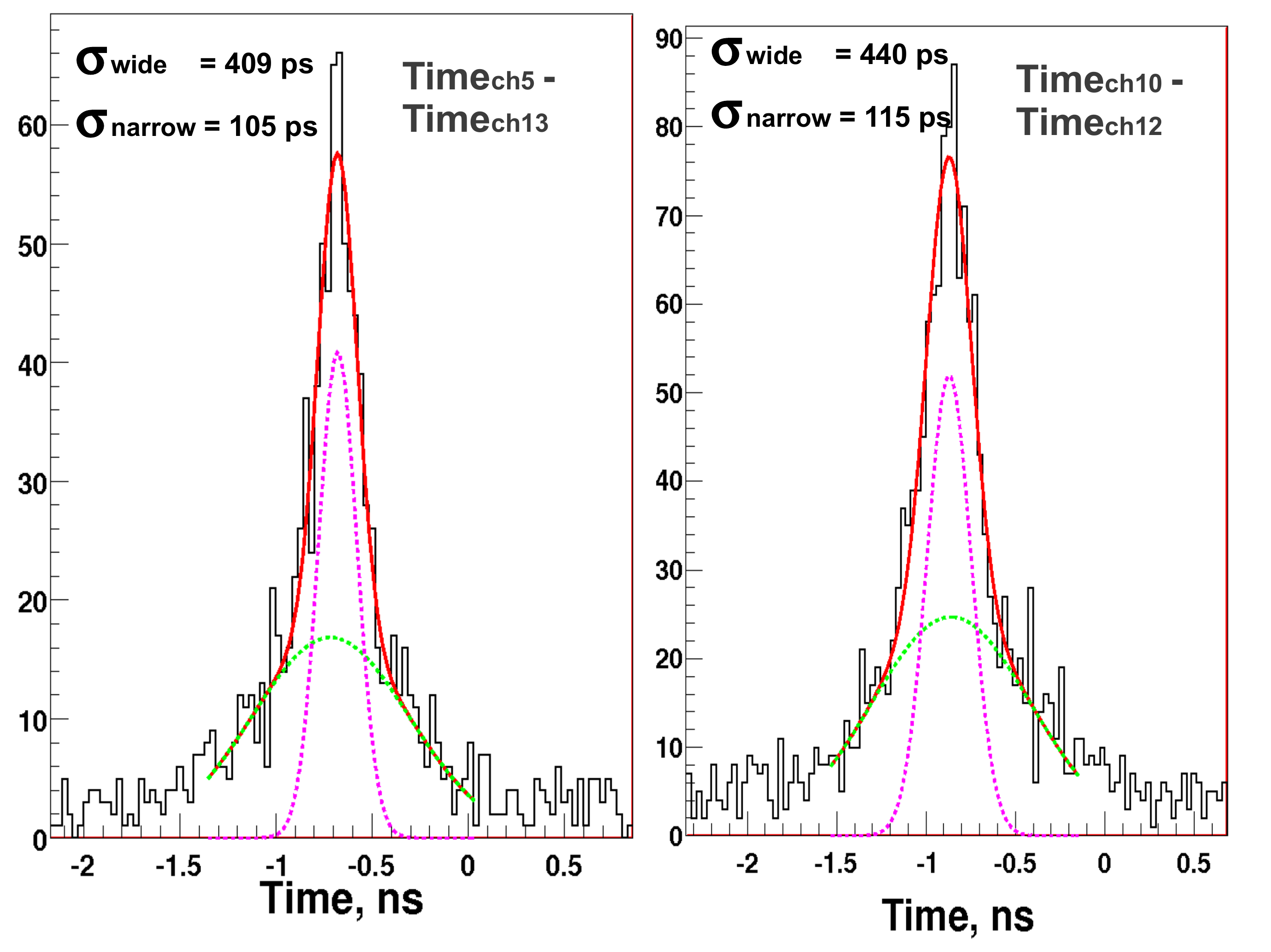}
\caption{Double Gaussian fits to two representative distributions of the time difference between two channels.
In both cases, the width of the narrow component is interpreted as the prototype device while the wide component accounts
for multiple photon paths (and hence multiple timings) between pads.}
\label{fig:PID-Forward:resolution}
\end{center}
\end{figure}

The CRT setup does not provide a sufficiently good start time (unless one would severely restrict tracking phase space). 
Therefore, the current determination of the prototype time resolution is based on time difference between channels, one 
chosen in the top bar and the other in the bottom bar. All pairs of channels have been considered but they are studied 
separately as the resulting timing distributions depend on the photon paths compatible with the particular two MCP-PMT 
pads considered. The interpretation of these data is not straightforward as photons emitted by a muon track
can follow several different paths to reach a particular pad. At present, a 3D-tracking treatment is not included in the current
analysis. Ideally, one should use a variable dTOP = TOP$_{{\mathrm measured}}$ - TOP$_{{\mathrm expected}}$, to determine the ultimate timing resolution; 
this is not easy in CRT as one deals with 3D tracks; the absence of the right treatment of the variable dTOP broadens the 
timing distribution and increases the size of tails. Yet, a good data-Monte-Carlo agreement~-- see Figure~\ref{fig:PID-Forward:data-MC}~-- 
makes us confident that this analysis is valid. As shown in Figure~\ref{fig:PID-Forward:resolution}, the measured single 
photon timing accuracy is in the range 80-100~ps, dominated by detector effects (spread of the Cherenkov photon emission 
time due to the finite quartz tile width; multiple photon paths inside the fused silica to the MCP-PMTs; variation of 
the cosmic muon track parameters, chromatic effects, etc.). This preliminary result should be confirmed by a 
reanalysis of the data based on 3D-tracking and aiming at comparing the photon measured and expected times.

\tdrparagraph{ Background and aging studies }

The FTOF sensitivity to background can be seen in two ways. First, background hits could  affect the CFD timing algorithm 
and add to confusion in the pattern reconstruction algorithms. In addition, the background rate will increase the MCP-PMT total charge.
The consequence of the latter will be a slow decrease of the MCP-PMT quantum efficiency; at an integrated charge
around 2~C/\cma~\cite{Inami_2010}, the photon detectors will cease working properly. Therefore, it is crucial to estimate the FTOF
background accurately and to find ways to decrease it as much as possible. In addition, the MCP-PMT gain should be
high enough to keep the TTS low, but not too high as it limits the MCP-PMT rate.

\begin{figure}[tbp]
\begin{center}
\includegraphics[width=\linewidth]{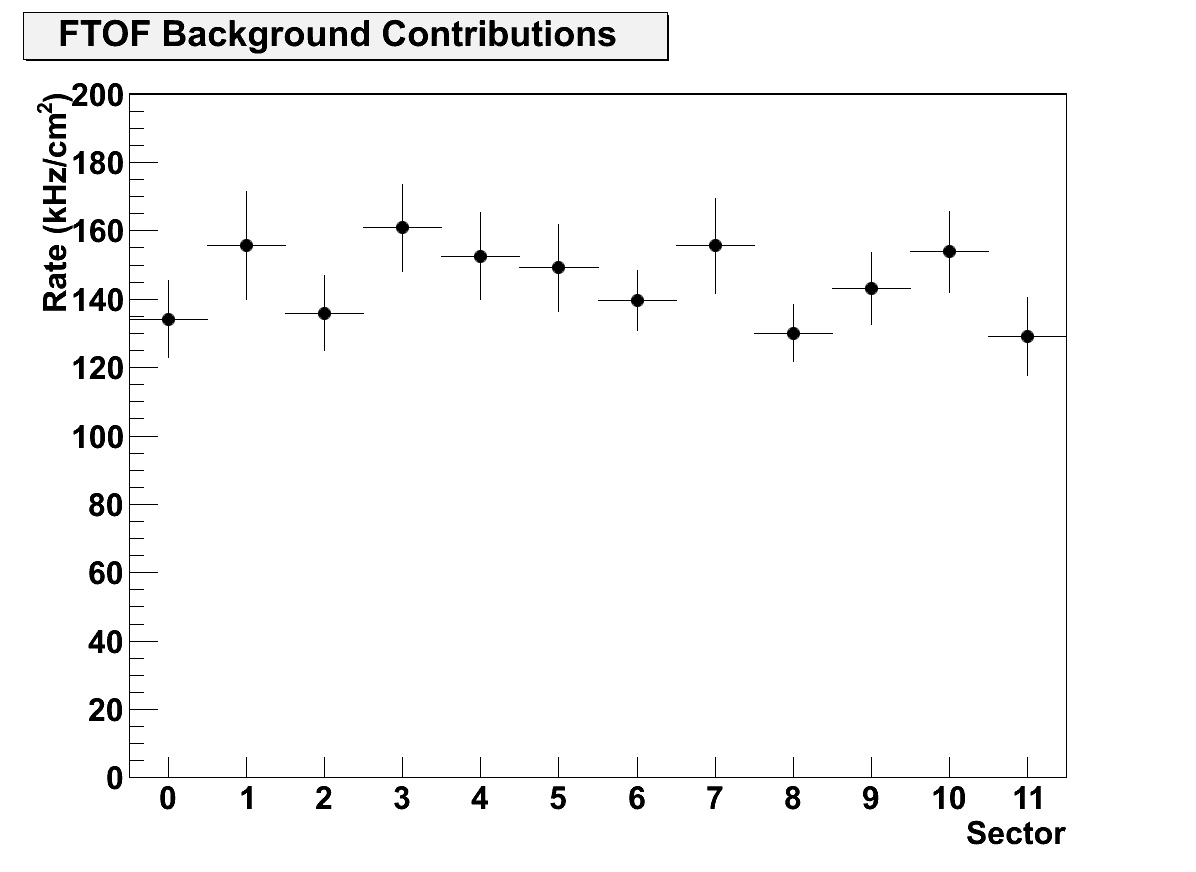}
\caption{Sum of all background contributions in the 12 FTOF sectors~-- no safety factor included.}
\label{fig:FTOF-sectorRates_3}
\end{center}
\end{figure}

Full \geantFour\ simulations of the \superb\ interaction region (detector included) simulating the main backgrounds~--
radiative Bhabha, Touschek particles, etc.~-- have shown that the dominant FTOF background is due to off-energy
positrons which hit the beam pipe about one meter away from the IP on the forward side. This localized background
source can be mitigated by increasing the thickness of the tungsten shield which protects the detector. Initial
studies have shown that going from 3~cm to 4.5~cm (currently the nominal value)
decreases the FTOF rate by a factor 3 to about 115~kHz/\cma. As shown in Eq.~\ref{eq:FTOF-gain}, the background rate can 
be directly related to the gain at which MCP-PMTs
can be operated, assuming 15\invab collected per year during five years. Therefore, work is ongoing to 
refine the FTOF background estimate. 

\begin{equation}
\begin{split}
&\text{Maximum gain} = \; 2.8 \, 10^{5} \\
&\times \left( \frac{ 2.5 \, \mathrm{C}/\cma }{ \text{ Maximum integrated charge } } \right) \\
&\times \left( \frac{ 150\kHz/\cma }{ \text{ Background rate } } \right) \\
&\times \left( \frac{ 5 }{ \text{Safety factor} } \right) \\
&\times \left( \frac{ 75\invab }{ \text{Integrated luminosity} } \right)
\end{split}
\label{eq:FTOF-gain}
\end{equation}

Figure~\ref{fig:FTOF-sectorRates_3} shows the simulated
background rates (summing up the contributions from all sources) in all 12 FTOF sectors, labelled clockwise as
seen from the IP~-- sector 0 is at the top of the FTOF.

\tdrparagraph{ Ongoing activities and future plans } 

As explained above, the FTOF technology has been chosen by the \superb\ collaboration for the forward PID
detector. Therefore, if one such device is to be built, that will be the FTOF. Yet, to be included in the
detector baseline, a full-size prototype of a FTOF sector must be built and validated, both in cosmics and test beam. 

\begin{figure*}[tbp]
\begin{center}
\includegraphics[width=0.95\textwidth]{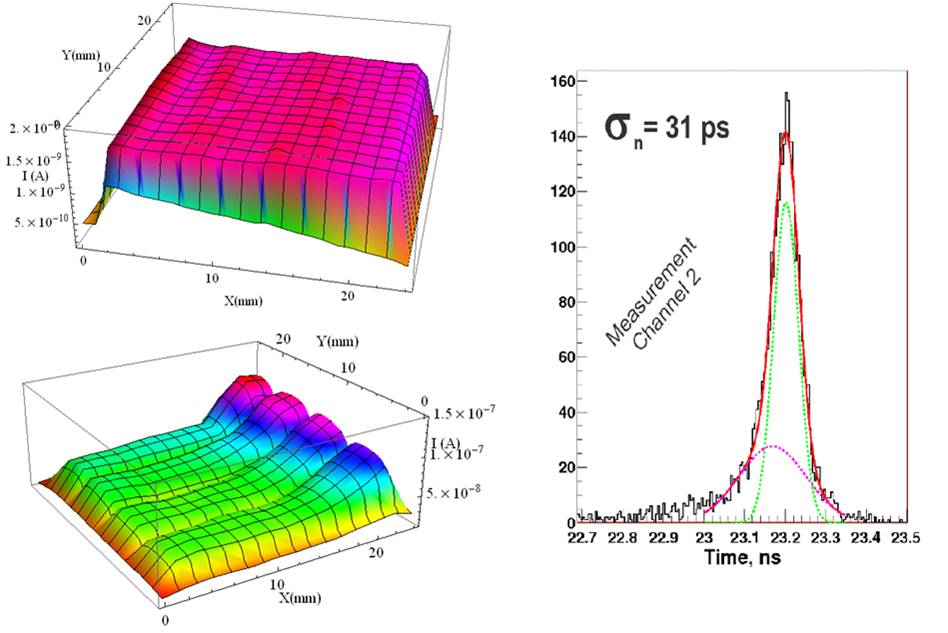}
\caption{Top left plot: SL-10 photocathode response mapping; bottom left plot: 4 anode response mapping; right plot:
single photoelectron timing resolution (FWHM).}
\label{fig:PID-Forward:mapping_TTS_SL10}
\end{center}
\end{figure*}

As a first step, a fused silica tile (Spectrosil 2000 from Heraeus) has been purchased by the LPSC Grenoble
group (which joined the FTOF development in 2011). It will be placed in a local cosmic ray telescope specifically
designed for the FTOF geometry with an active area of 0.23~m$^2$ and good resolution on position
(${\sim}0.3-0.6$~cm) and angle (${\sim}0.2$~deg). Different absorber thicknesses are used to select 7
momentum ranges of the muons from 300 MeV/c, the thicker absorber giving a lower cut at 1.7~GeV/c. The
yield of this telescope is 1~Hz for the full momentum spectra and only 0.2~Hz for the higher energy bin.
The main purpose of this setup is to study the photon yield versus the muon impact position and incident
angle, using several coatings on edges and faces.
The goals of these studies are to measure the photon collection efficiency and timing in various configurations.
These data will allow one to validate the full \geantFour-based optical simulations.


\begin{table*}[tpb]
\begin{center}
\caption{Comparison of two MCP-PMTs candidate for the FTOF~\cite{vero}.}
\begin{tabular}{|c|c|c|c|c|c|}
\hline MCP-PMT & Effective & Quantum & Typical & Transit & QE reduction \\
& area & efficiency & gain & time spread & after 2~C/cm$^2$ \\
& (mm$^2$) & (QE) @ 400~nm & & (FWHM) & accumulated charge \\
\hline R10754X-01-L4 & $22 \times 22$ & 17\% & $10^6$ & 70~ps & 20\% \\
\hline XP85112/A1 & $53 \times 53$ & 22\% & $10^5$ & 82~ps & Unaffected \\
\hline
\end{tabular}
\label{table:PID-Forward:MCP-PMT}
\end{center}
\end{table*}

In addition, the FTOF effort will benefit from another cosmic ray telescope: CORTO. 
This R\&D facility, which is being built at LAL, will be used by different labs of the Orsay campus to test detectors.
At CORTO, the FTOF-like fused silica tile will be tested with different MCP-PMTs: R10754X (SL-10) from Hamamatsu and XP85112
from Photonis. Table~\ref{table:PID-Forward:MCP-PMT} shows a comparison of these two devices.
Measurements of the characteristics of these photon detectors are in progress,
as shown in Fig~\ref{fig:PID-Forward:mapping_TTS_SL10}.
In addition, we may test in the future MCP-PMTs developed jointly by the Ekran company and BINP~\cite{Barnyakov:2011zz}. They are much
cheaper than the two models mentioned above. But they also need to be studied in details to see if they would match the
\superb\ requirements.

\begin{figure*}[htb]
\begin{center}
\includegraphics[width=0.8\linewidth,angle=90]{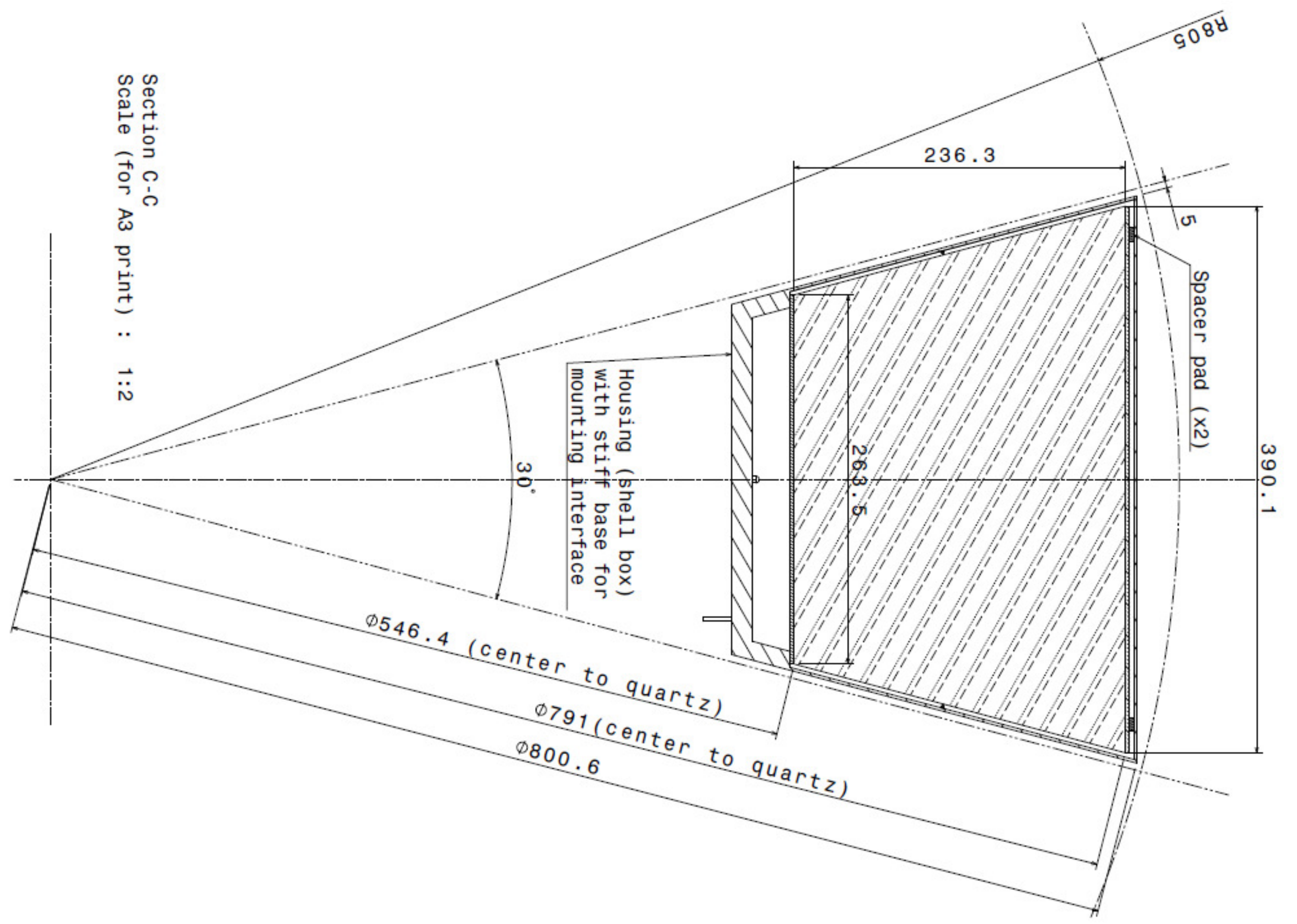}
\caption{Front view of a sector of the FTOF detector inside its 'tile box'.
The quoted dimensions are in millimeters and correspond to diameters.}
\label{fig:FTOF_engineers}
\end{center}
\end{figure*}

In the meantime, the FTOF design will go on, benefiting from improved simulations and progress on the
geometry of the crowded \superb\ forward region. Discussions are also ongoing about the best location
of the MCP-PMTs: at the outermost radius they are subject to less background but accessing them for
repair would be extremely difficult. On the other hand, putting the MCP-PMTs at the innermost radius
would require tilting the FTOF significantly (to have a larger fraction of Cherenkov photons
downward-oriented), which is not possible given the integration constraints on the \superb\ forward side.

Following a meeting at SLAC in July which was dedicated to detector integration issues, the geometry of the \superb\
forward region has been updated~\cite{integration_update} at the September 2012 \superb\ meeting in Pisa. Space constraints
at large radii make impossible the extension of the FTOF beyond the DCH outer diameter. Therefore, its maximum
diameter has to be reduced by about 10~cm with respect to the design which was used for all the studies reported in this
document. The new design is being propagated to all \superb\ software and will be used as baseline from now on.

Mechanical drawings have then been produced to present the new shape of the FTOF.
Figure~\ref{fig:FTOF_engineers} shows the front view of a FTOF sector embedded inside its 'tile
box'. Then, Figure~\ref{fig:FTOF_engineers_front} shows the dimensions of one such sector. Finally,
Figure~\ref{fig:FTOF_engineers_side} shows the corresponding side view of the sector.

\begin{figure*}[htb]
\begin{center}
\includegraphics[width=0.8\linewidth,angle=90]{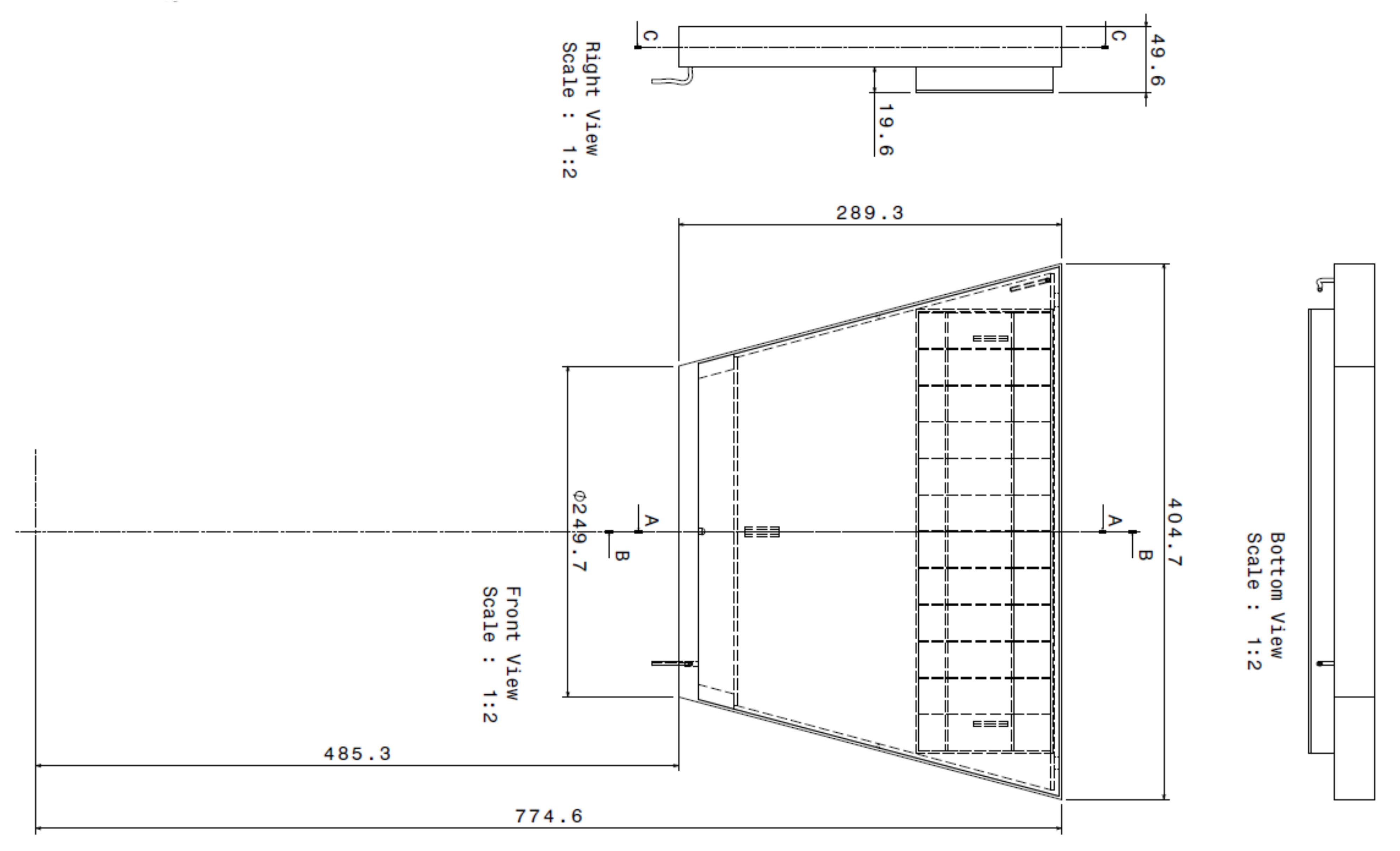}
\caption{Front view of a single FTOF sector showing its dimensions.}
\label{fig:FTOF_engineers_front}
\end{center}
\end{figure*}

The main parameters of the FTOF structure are the following.
\begin{itemize}
\item Aluminium side frame thickness: 3~mm
\item Aluminium front frame thickness: 1~mm
\item Absorber thickness: 3~mm
\item Mirror thickness: 3~mm~-- this mirror is used to reflect back photons which
miss the MCP-PMT line initially. It increases significantly the photon yield per track.
\item Mass of the quartz tile is around 2.5~kg
\end{itemize}

\begin{figure}[htb]
\begin{center}
\includegraphics[width=\linewidth]{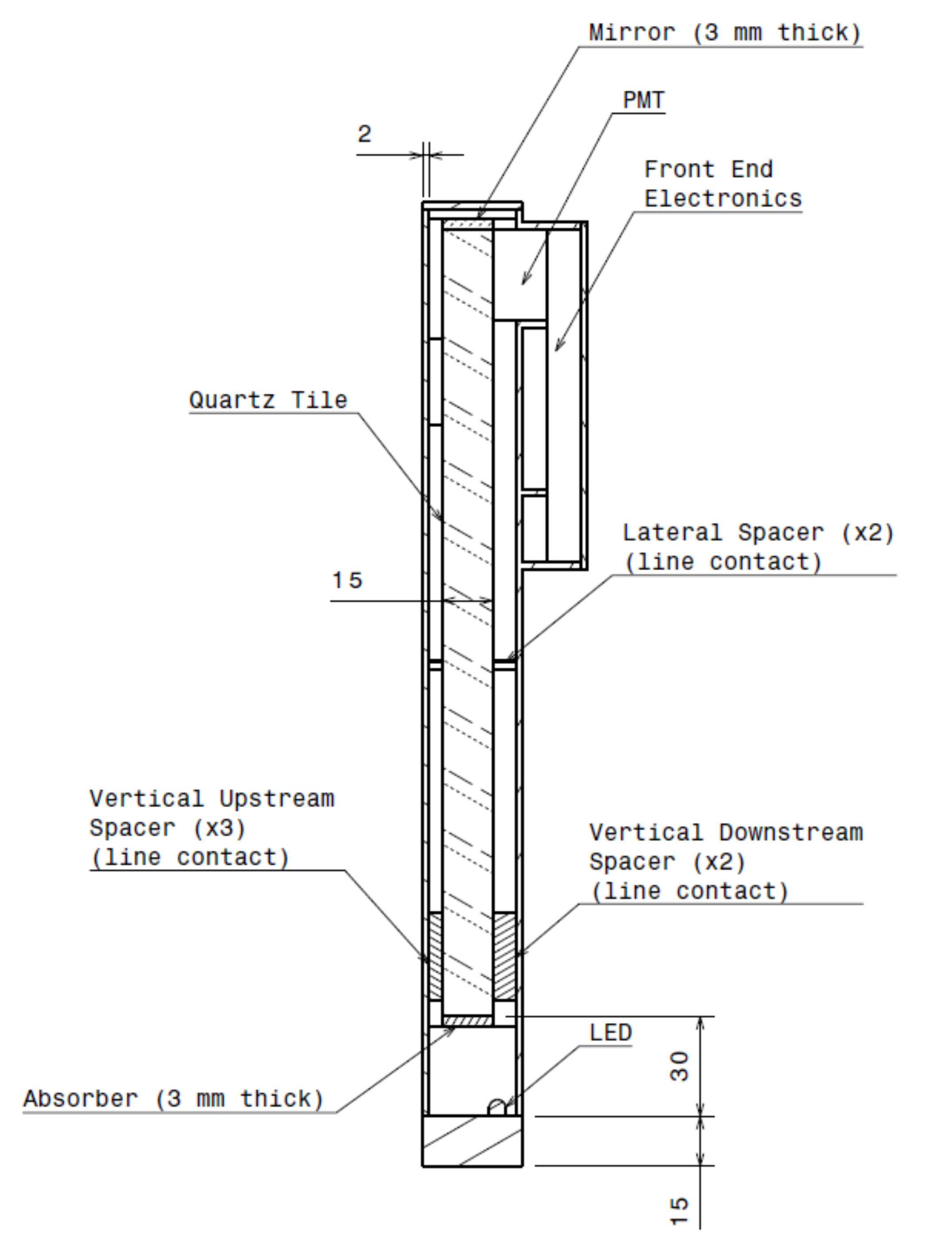}
\caption{Side view of a FTOF sector. On this drawing, the rear of the tile is on the right.}
\label{fig:FTOF_engineers_side}
\end{center}
\end{figure}

Once the FTOF design is frozen, the building of the prototype will start~-- its cost is strongly dominated
by the MCP-PMT price. Assuming that the whole project proceeds smoothly and that the FTOF is accepted
by \superb, a detailed appendix of the present detector TDR will be published to provide all the needed
information to support this innovative detector. 

\tdrsubsec{ Appendix: forward PID R\&D activities }
\label{subsec:forwardPID-Appendix}

In the following paragraphs, the main characteristics of the three
proposals which have not been selected by the Forward task force
are briefly summarized, such as their advantages and drawbacks.

\tdrparagraph{Focusing aerogel RICH (FARICH)}

Ring Imaging Cherenkov detectors are the most powerful instruments for charged hadron identification in a wide momentum region,
ranging from ${\sim} 0.5$ to ${\sim} 100$\gevc. The use of multilayer aerogel radiators in proximity focusing RICH detectors
significantly improves PID performances with respect to a single layer aerogel RICH device. To provide PID at momenta below
0.6~GeV/$c$, an additional radiator with high refractive index is added.  
 
From the beginning, it was known that the main drawbacks of the FARICH were integration issues (the free space on the forward side of the detector 
is limited) and cost. One of the factors which drives the latter is the number of channels. Therefore, the design was optimized
to minimize the number of channels, such as the total thickness of the system, while keeping the PID performance high.
The FARICH design is presented in Figure~\ref{fig:SuperBFARICH}; its main parameters are given below.
\begin{itemize}
    \item Expansion gap: 65~mm (total thickness of the system ${\sim} 150$~mm).
    \item Photon detector: the Photonis Multi-Channel Plate Photomultiplier (MCP-PMT) with $6 \times 6$~mm$^2$ anodes ($8 \times 8$ matrix);
          photoelectron collection efficiency: 70\%; packing efficiency: 80\%.
    \item A 2-layer 'focusing' aerogel (n$_1$=1.039, n$_2$=1.050) of total thickness 30~mm. 
    \item One additional layer of NaF radiator: n=1.33, 5~mm thickness.
    \item Number of PMTs: 312.
    \item Number of channels: 20000.
    \item Amount of material (X/X$_0$): 25\%~ = 2.4\% (aerogel) + 4.3\% (NaF) + 10\% (MCP~PMT) + ${\sim} 8$\% (support, electronics, cables). 
\end{itemize}

\begin{figure}[htbp]
\begin{center}
\includegraphics[width=\linewidth]{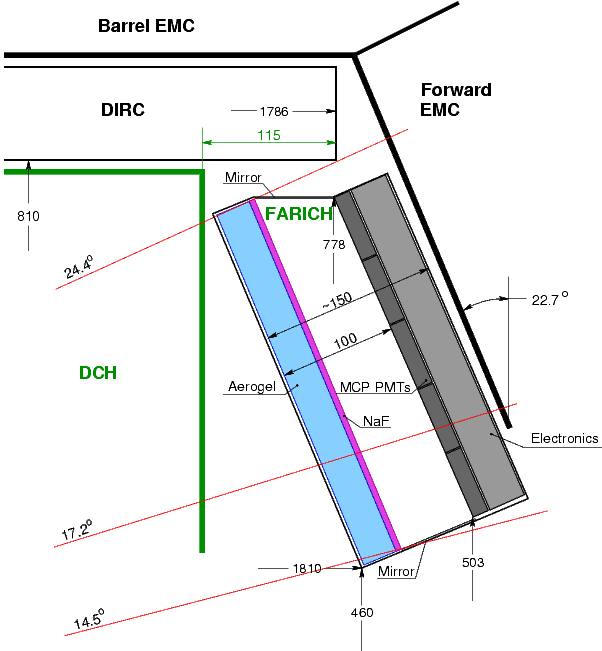}
\caption{Possible FARICH detector layout in the \superb\ detector.}
\label{fig:SuperBFARICH}
\end{center}
\end{figure}

A Monte Carlo simulation was developed for the proposed configuration. The number of detected photons for $\beta=1$ particles
are 16 from the aerogel radiator and 10 from the NaF one. The FARICH detector is able to perform $\kaon/\pi$ separation at the
$3 \sigma$ level and better from 0.2 to 5\gevc, $\mu/\pi$ separation from 0.13 to 1\gevc and $\pi$/p separation from 1 to 8\gevc. 

In addition, a prototype was tested in a beam. The main goal of this test was to demonstrate the 'focusing' capabilities of a
real multilayer aerogel radiator at short
expansion gap and to measure the contribution from aerogel radiator to the resolution on the Cherenkov angle. The measured aerogel
tile had 4 layers, with a maximum layer index of refraction of 1.05 and a total thickness of 30~mm.
The FARICH prototype used 32 MRS-APDs (SiPMs) from the CPTA company (Moscow, Russia) as photon detectors. Their active area
was $2.1 \times 2.1$\mma and they were selected as photodetectors for the prototype because of their good spatial resolution
($\sigma_{\mathrm{pixel}} = 2.1/\sqrt{12}=0.6$~mm). But such APDs could not be used for the \superb\ FARICH because of fast damage by neutrons.
Custom-made discriminator boards and the CAEN V1190B multi-hit TDC were used for the signal readout.

The test beam facility was constructed at the VEPP-4M collider at the Budker INP in Novosibirsk. 
This apparatus also included a trigger, some veto scintillation counters, coordinate drift chambers and
a NaI calorimeter. 

During this experiment, we measured simultaneously the parameters of 1\gevc electron tracks and the coordinates of the associated
Cherenkov photons detected in the multilayer aerogel radiator. The density of photoelectrons vs. radius for one of the
SiPMs used by the prototype is
presented in Figure~\ref{fig:1peRes}. As one can see, the distribution of Cherenkov photons vs. radius is well-described
by the sum of a signal Gaussian plus a linear background due to random coincidences with SiPM noise, at least within $\pm 5\sigma$
of the signal peak.
The measured radial resolution $\sigma_r$ for the Cherenkov photons is equal to 1.1~mm. 

\begin{figure}[htbp]
\begin{center}
\includegraphics[width=\linewidth]{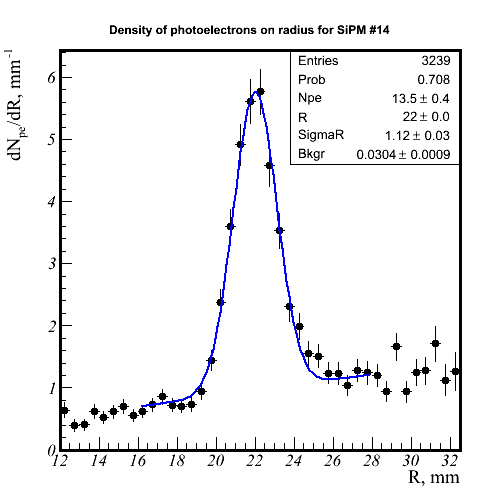}
\caption{Density of photoelectrons vs. radius for SiPM \#14.}
\label{fig:1peRes}
\end{center}
\end{figure}

With this input, we estimate the contribution to the single photon resolution from an aerogel radiator to be 
$\sigma_{\mathrm{Aerogel}}$ = $\sqrt{\sigma^2_r-\sigma^2_{\mathrm{pixel}}} = 0.91$~mm.
The expected $\kaon/\pi$ separation based on $\sigma_{\mathrm{Aerogel}}$ (taken from the test beam data) and data
provided by Photonis (MCP-PMT provider) is presented in Figure~\ref{fig:FARICHPID}.

\begin{figure}[htbp]
\begin{center}
\includegraphics[width=\linewidth]{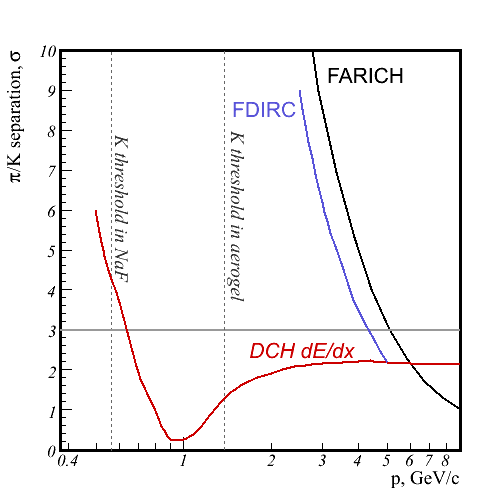}
\caption{The expected FARICH $\kaon/\pi$ separation based on $\sigma_{\mathrm{Aerogel}}$ from test beam data together
with the expected \superb\ FDIRC and \dedx\ $\kaon/\pi$ separations.}
\label{fig:FARICHPID}
\end{center}
\end{figure}

Our study has shown that a FARICH-based PID system could be used by an experiment like \superb.
First, multilayer 'focusing' aerogel radiators are now available on the market; moreover, the estimated backgrounds
are on the low-level side for this detection technology. Finally, the expected life time of the PMTs during the experiment is about
10 years or more.

After detailed investigations,
it was concluded that the gain in detection efficiency was not significant enough, in comparison with the cost of the system
and the necessity to cut the forward end of the \superb\ drift chamber in order to give the FARICH enough space.

\tdrparagraph{Pixelated \tof\ detector with Cherenkov radiators}


Figure~\ref{fig:Pixilated_TOF} shows a possible concept of a TOF detector. It uses polished and side-coated fused silica radiator 
cubes, which are optically isolated from each other. The radiator cubes are coupled to a MCP-PMT detector with 10 micron holes. This
concept is the most simple of all TOF detector designs, as it avoids complicated 3D data analysis and minimizes chromatic effects. 
On the other hand, it requires a large number of MCP-PMT detectors, which increases the cost prohibitively at present. However, if such 
detectors would become cheap at some point in future, one could revive this concept again.

\begin{figure}[tbp]
\begin{center}
\includegraphics[width=\linewidth]{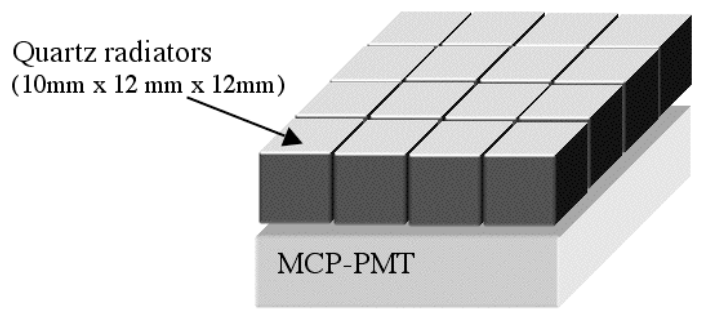}
\caption{A possible geometry for a pixelated TOF detector: polished and side-coated fused silica cubes coupled to a MCP-PMT detector.}
\label{fig:Pixilated_TOF}
\end{center}
\end{figure}

Figure~\ref{fig:Pixilated_prototype_TOF} shows a prototype of the pixelated TOF concept, with a coated fused silica radiator cylinder. 
The detector had a fiber allowing calibration and laser-based bench-top tests. The radiator was coupled to a Photonis MCP-PMT
with 10 micron holes. There were two identical detectors placed in the test beam, allowing to use one as a start and the other 
as a stop in the timing measurement, and therefore the quoted resolution is the measured resolution 
divided by $\sqrt{2}$. Reference~\cite{Va'vra:2008zza,Va'Vra:2009zza,Breton:2010zza,vavra_2012} summarizes all work towards this concept.

\begin{figure}[tbp]
\begin{center}
\includegraphics[width=\linewidth]{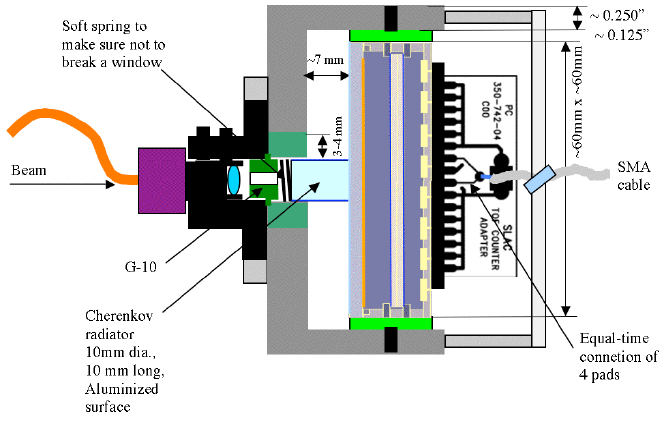}
\caption{The geometry of a pixelated TOF prototype detector used in our tests (two such detectors were actually used for the measurements).}
\label{fig:Pixilated_prototype_TOF}
\end{center}
\end{figure}

Figure~\ref{fig:Pixilated_prototype_TOF_best} shows the best resolution obtained with the TOF prototype in the Fermilab test beam. 
The detector operated at low gain, which means it did not achieve the ultimate resolution. The low gain operation was intentional
to avoid aging in the \superb\ environment.

\begin{figure}[tbp]
\begin{center}
\includegraphics[width=\linewidth]{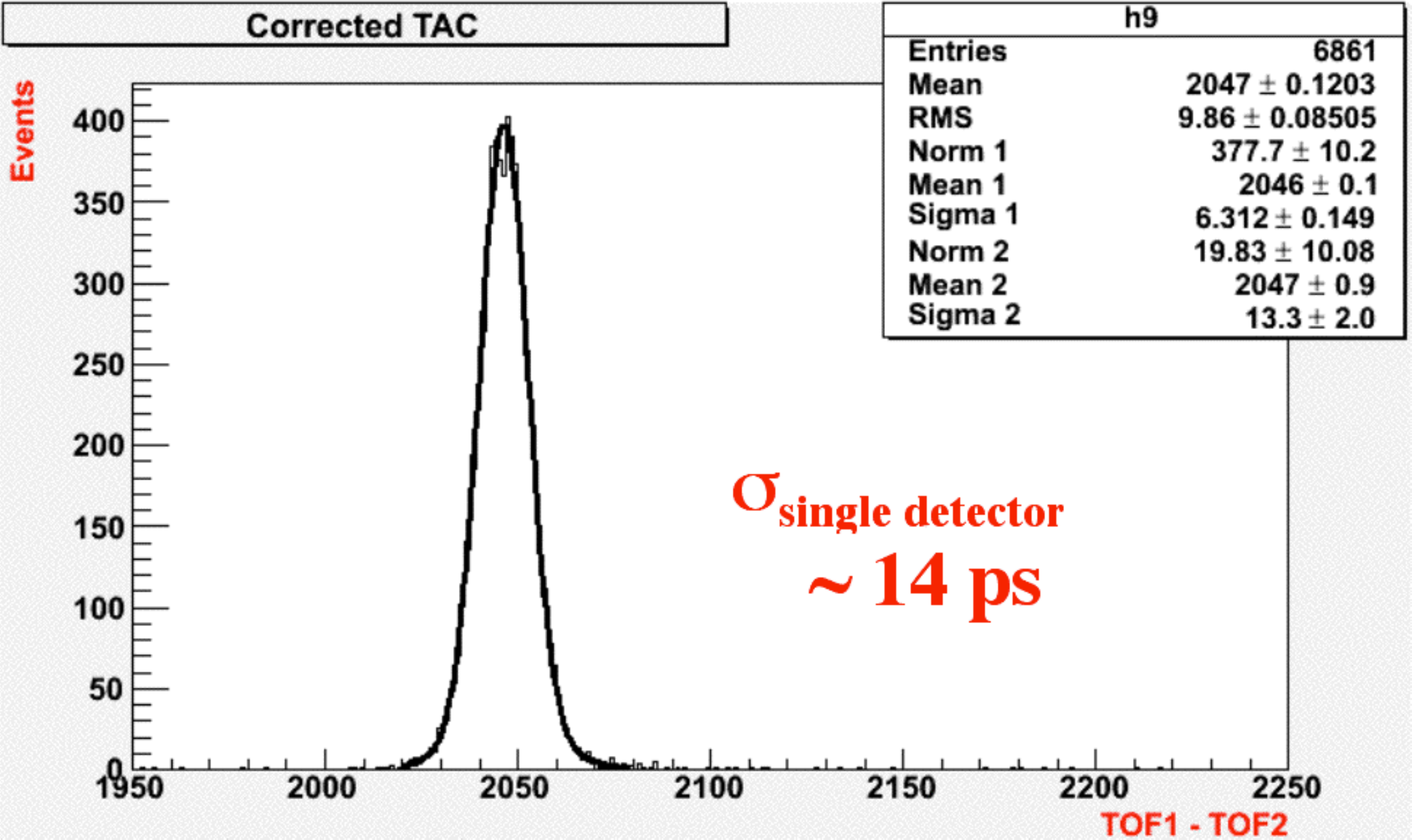}
\caption{The best result obtained in the Fermilab test beam with the Ortec 9327 electronics.}
\label{fig:Pixilated_prototype_TOF_best}
\end{center}
\end{figure}

Figure~\ref{fig:Pixilated_prototype_TOF_results_a} shows all our test results both in the test beams and laser-based bench-top
 experiments. These results were obtained with Ortec 9327 CFD electronics, WaveCatcher~\cite{breton_2009}~-- see section~\ref{subsec:FTOF}~--, DRS4~\cite{ritt_2009} 
and Target~\cite{Varner:2007zz} waveform digitizers. One can see that waveform digitizers "almost" reach the resolution of 
the classical CFD electronics, but not quite.

\begin{figure}[tbp]
\begin{center}
\includegraphics[width=\linewidth]{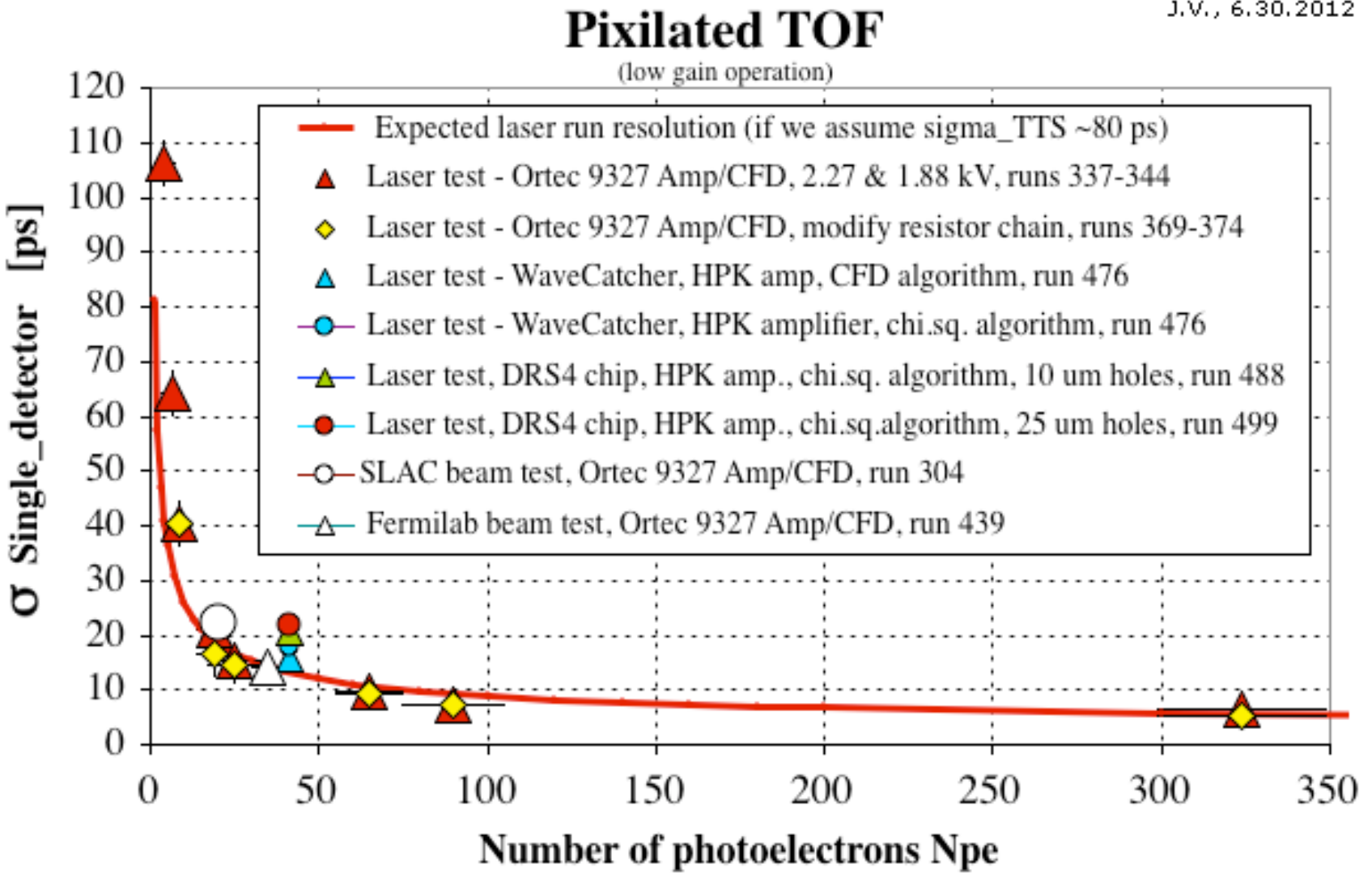}
\caption{A summary of all results with the pixelated TOF prototype, including beam and laser tests.}
\label{fig:Pixilated_prototype_TOF_results_a}
\end{center}
\end{figure}

Figure~\ref{fig:Pixilated_prototype_TOF_results_b} shows the laser-based test results as a function of the signal-to-noise ratio (S/N). 
A large value of S/N ratio is essential to achieve a good timing resolution. In these tests we achieved
$\sigma_{\mathrm{electronics}} {\sim} 2.42\ps$ with Ortec 9327 CFD electronics. Therefore the detector contribution to the final ultimate 
timing resolution was $\sigma_{\mathrm{detector}} {\sim} 3.6\ps$ for S/N of ${\sim} 1400$. To achieve such a high value of S/N ratio in practice 
is difficult, as there are many limitations. The most important one is due to MCP detector aging, which requires a low gain
operation. One could also have to avoid using an amplifier to improve the S/N ratio, as the amplifier is contributing significantly to 
the noise (Ortec 9327 CFD electronics has internal $10\times$ amplifier). It is probably more realistic in practice to target S/N ${\sim} 200$, 
\ie, a 10-12~ps resolution.

\begin{figure}[tbp]
\begin{center}
\includegraphics[width=\linewidth]{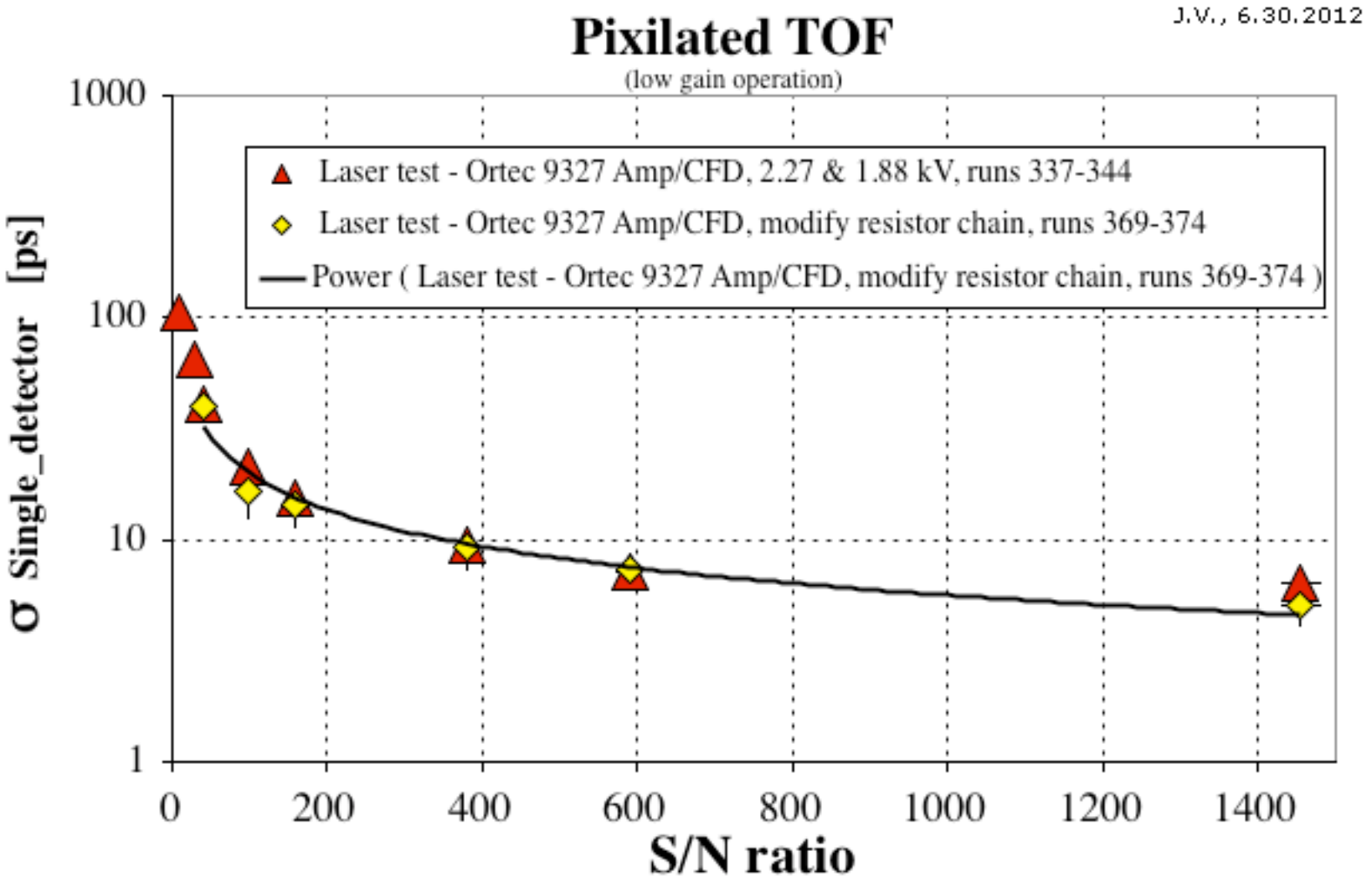}
\caption{A summary of our laser-based results with the pixelated TOF prototype with the Ortec 9327 CFD electronics as a function of signal to noise ratio (S/N).}
\label{fig:Pixilated_prototype_TOF_results_b}
\end{center}
\end{figure}

\tdrparagraph{Scintillating detector coupled to G-APD arrays}

 Because of the prohibitive cost of MCP-PMT detectors for the entire forward pixelated TOF detector design at present, we considered 
other possible "pixelated" schemes based on scintillators and G-APDs. We used the SLAC CRT to test various 
options using 3D muon tracks. The 3D tracks with dip angle up to 20 degrees will somewhat worsen the timing resolution, however, 
they provide more realistic results compared to test beam. We used various types of scintillators (BC-404, BC-420, a small LYSO crystal and a 
full size one) as radiator. The logic for the full size LYSO radiator was to see if one could "parasite" the endcap 
calorimeter by adding a $4 \times 4$ G-APD array on its front face. The photon detection in these tests was provided by either 
MCP-PMT (for a reference run only), or Hamamatsu $4 \times 4$ G-APD array (each pixel size is 3~mm~$\times$~3~mm), or simply a pair of single 
3~mm~$\times$~3~mm G-APDs (either MEHTI or Hamamatsu MCCP) coupled to the side of the small scintillator. The overall goal was to 
reach a resolution of ${\sim} 100$~ps only, in order to provide a PID identification in the dE/dx cross-over region. Table~\ref{table:scint_tof} shows a 
summary of all results~\cite{jjv_toflyso_1,jjv_toflyso_2}. One can see that we could reach resolutions between 110 and 180 ps for several 
small scintillators. However, the full size LYSO crystal results were considerably worse. Although one could search for some further 
improvements in future, it was felt that this type of TOF counter cannot compete with the FTOF concept at present.

\begin{table*}[tbp]
\begin{center}
\caption{Results with TOF pixelated detectors using scintillator radiators and 3D tracks in the SLAC CRT.}
\begin{tabular}{|c|c|c|}
\hline \parbox[t][1.0cm][t]{2.5cm}{Radiator} & Detector & \parbox[t][1.0cm][t]{2.0cm}{Measured resolution} \\
\hline Small LYSO ($17 \times 17 \times 17$~mm$^3$) & MCP-PMT & 109 \& 159 ps \\
\hline Small LYSO ($17 \times 17 \times 17$~mm$^3$) & $4 \times 4$ G-APD array (pixel 3 mm$^2$) & 140 ps  \\
\hline Large LYSO ($25 \times 25 \times 200$~mm$^3$) & $4 \times 4$ G-APD array (pixel 3 mm$^2$) & 220 ps \\
\hline Scintillator ($17 \times 17 \times 17$~mm$^3$) & $4 \times 4$ G-APD array (pixel 3 mm$^2$) & 136 ps \\
\hline BC-404 scintillator ($38 \times 38 \times 25$~mm$^3$) & Two single 3 mm$^2$ G-APDs & 156 ps \\
\hline BC-420 scintillator ($38 \times 38 \times 10$~mm$^3$) & Two single 3 mm$^2$ G-APDs & 177 ps \\
\hline
\end{tabular}	
\label{table:scint_tof}
\end{center}
\end{table*}

\bibliographystyle{\tdrbibstyle}
\balance
\bibliography{PID/PID-bib}


\renewcommand{\ChapLabel}{chap:EMC} 

\tdrchap{Electromagnetic Calorimeter}
\label{chap:Calorimeter}

As was the case with \babar, the \superb\ electromagnetic calorimeter
(EMC) will play an essential role in the study of flavor physics,
especially in the study of $B$ meson decays involving neutral
particles. The calorimeter provides energy and direction measurement
of photons and electrons, reconstruction of neutral hadrons such as
$\pi^0$'s, and discrimination between electrons and charged hadrons.
Channels containing missing energy due to the presence of neutrinos
will rely on information from the EMC to discriminate against
backgrounds.

\tdrsec{Overview}

The electromagnetic calorimeter has three major components: a CsI(Tl)
``barrel'' calorimeter covering the central region, a ``hybrid''
CsI(Tl)/LYSO(Ce) ``forward'' calorimeter covering the small angle
region in the direction of the high energy beam, and a
lead-scintillator ``backward'' calorimeter covering the small angle
region in the direction of the low energy beam.
Table~\ref{table:EMC:superb} shows the solid angle coverage for the
three components of the \superb\ EMC.


\begin{table*}[tbh]
\caption{Solid angle coverage of the electromagnetic calorimeters. Due the lower boost,
the barrel calorimeter is displaced $10$ cm in the backward direction with respect to the  position in \babar. The CM angular coverage and solid angle are for massless particles and nominal 4 on 7\gev beam energies.
}
\vskip10pt
\begin{center}
\begin{tabular}{|lccccc|}
\hline
Calorimeter & \multicolumn{2}{c}{$\cos\theta$ (lab)} & \multicolumn{2}{c}{$\cos\theta$ (CM)} & $\Omega$ (CM)(\%)\\
            & minimum & maximum     & minimum & maximum & \\
\hline
\hline
Backward & $-0.98$ & $-0.88$&$-0.99$&$-0.93$&2.9\\
Barrel  &$-0.79$&0.89&$-0.87$&0.81&84\\
Forward &0.89&0.97&0.81&0.94&6.8\\
\hline
\end{tabular}
\end{center}
\label{table:EMC:superb}
\end{table*}

The \superb\ EMC reuses the barrel part of the \babar\ EMC detector
consisting of 5760 CsI(Tl) crystals (Fig.~\ref{EMC:fig:xsecview}).
Due to the increased rates and radiation dose at \superb\ compared
with PEP-II, the \babar\ forward calorimeter must be modified: the
innermost CsI(Tl) rings of the forward endcap will be replaced by new
LYSO crystals better-matched to the \hbox{\superb\ }environment.

Reuse of the \babar\ barrel calorimeter still requires a substantial
effort. There is an ongoing engineering investigation of whether the
barrel structure can be shipped as a unit, or whether it is necessary
to disassemble it into its 280 individual modules for safe shipment.
In either eventuality, due to the substantial increase in background
and signal rates at \superb\ compared to PEP-II, it will be necessary
to replace the crystal-mounted preamplifier-shapers with versions
having a shorter integration time.


\begin{figure*}
\begin{center}
\includegraphics[width=13cm]{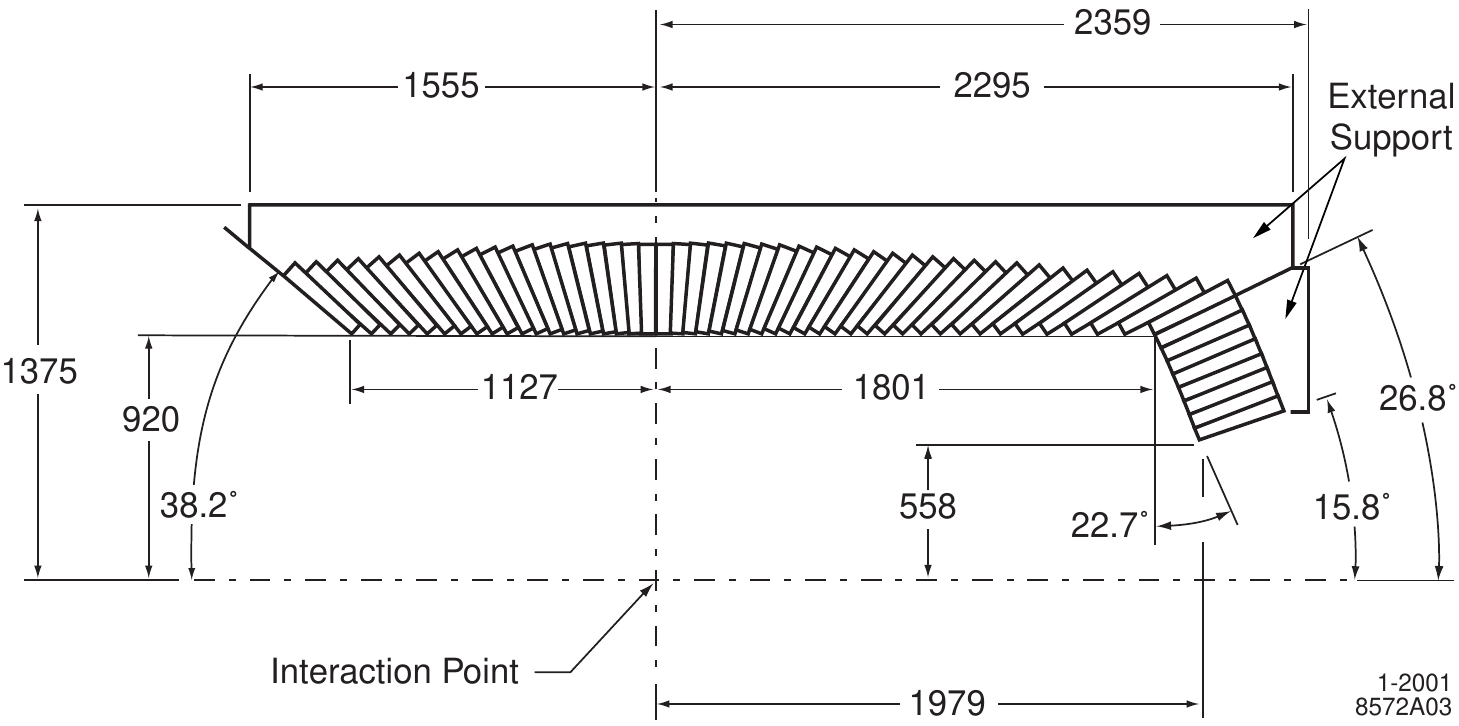}
\caption{Longitudinal \babar\ EMC cross-section (top half)
showing the arrangement of the 48 barrel and 8 endcap crystal rings.
The detector is axially symmetric around the $z$-axis. Dimensions are
in mm.}
\label{EMC:fig:xsecview}
\end{center}
\end{figure*}

After an intensive R$\&$D program, the baseline choice for the forward
endcap is to replace the inner rings of the \babar\ forward
calorimeter with faster and more radiation resistant LYSO crystals. As
will be discussed below, this is the clear favorite in terms of
performance and radiation hardness over the alternatives we have
considered. The faster response time and smaller Moli\`ere radius
serve together to address the higher event and background rates at
\superb.  LYSO is a fast scintillator heretofore used principally in
small sizes in medical applications. Our R\&D program has concentrated
on the optimization of the performance of large crystals (2 cm
$\times$ 2 cm $\times$ 20 cm).  This entails having high light yield,
optimized cerium doping to ensure the greatest possible uniformity of
light output along the crystal, and surface treatment to refine the
uniformity of response to meet a stringent specification.  Thanks to
this effort, there are now at least four producers able to grow LYSO
crystals for high energy physics applications. Table
~\ref{tab:crystals} is a comparison of LYSO and other crystals used in
high resolution electromagnetic calorimeters. The main issue with LYSO
is cost. The hybrid design is a compromise, aimed at balancing the
performance and cost of the forward endcap. We have studied several
additional alternatives, described below.

A lead-scintillator-sandwich backward endcap calorimeter improves the
hermeticity of the detector.  Its main purpose is to act as a veto for
events having ``extra'' energy in the backward region, which is
particularly important for studying decay channels with neutrinos in
the final state.  The backward EMC has excellent time resolution, and
may thus also have a role in particle identification by providing
time-of-flight for the relatively slow backward-going charged
particles.

The EMC calorimeter plays an important role in the experiment's
trigger. which is discussed in Chapter~\ref{chap:ETD}.

\tdrsubsec{Background and radiation issues}
\label{sec:EMC:OVERbkg}

One of the major concerns for the electromagnetic calorimeter is its
capability to sustain the radiation dose, which is larger than in
\babar\ due to the increased luminosity. The dominant contribution to
the radiation dose in \superb\ comes from radiative Bhabha events that
produce a large number of low energy photons at an extremely high
rate. The high photon rate has two main effects on the performance of
the detector: the integrated radiation dose can reduce the
transmittance of the crystals and therefore alter the calibration of
the detector over time, and the large number of background photons can
produce pile-up, degrading the energy resolution.

The impact of these effects have been simulated in detail; this is
discussed in Sec.~\ref{sec:MDI:EMCbkg}.  The energy deposited by
background events in each crystal is simulated at each beam crossing.
The resulting rates of energy deposit are shown in
Fig.~\ref{fig:emc_edep_ring_stack}.
The simulations show that there is no significant difference in the
radiation dose to individual crystals between the barrel and the
forward endcap. This is by construction. Due to the different Moli\`ere
radii of CsI(Tl) and LYSO, the transverse dimensions of the LYSO
crystals are chosen to be about half the size of the CsI(Tl) crystals
and the overall volume of a LYSO crystal is 6.7 times smaller (120
cm$^3$ vs 800 cm$^3$) than a typical CsI(Tl) crystal. Thus the
background rates and radiation dose per crystal are compensated by the
change in volume.

The radiation dose to which the crystals are subjected is defined as
the total energy deposited in a crystal per unit mass. The dose
expected per year, conventionally taken as a ``New Snowmass Year''
having $1.5\times 10^7$ s, was estimated to range between 0.15 and 0.3
krad.  Conservatively, assuming 10 years of operations, the crystals
must be radiation resistant up to at least 4.5 krad. The impact on the
energy resolution of a $\sim 0.045$ rad/hour dose rate needs to be
considered. In contrast to PbWO$_4$, the light output of neither
CsI(Tl) or LYSO is rate-dependent.


\tdrsubsec{Simulation tools}
\label{sec:EMC:simulation}

\tdrsubsubsec{FastSim}

A fast simulation (FastSim) tool based on the \babar\ software
framework has been developed to evaluate the detector performance,
optimize the geometry and study the physics reach. The detector
geometry is modeled as two-dimensional ``shells'' of basic topologies
such as cylinders, disks, cones, and rectangles. The event generators
are identical to those in \babar.  Each particle is tracked in the
detector volume. When it crosses a detector element, the interaction
type and the energy loss are determined according to the particle
type, detector material and thickness. Secondary particles are created
if necessary. Particle showers are represented by a quasi-particle
instead of a (large) number of individual shower particles.

Charged track hits are fitted with a Kalman filter. EMC clusters are
created based on the Moli\`ere radius, interaction type (EM shower,
hadron shower or minimum ionizing), detector segmentation, and
parameterized shower shape.  Geometry and model parameters can be
easily set up and modified.  The high level data structure of an event
is the same as that in \babar, including tracks, clusters, and
particle identification objects, so we were able to directly apply
available \babar\ analysis tools.

Background cannot be simulated with FastSim because it is highly
sensitive to the details of the beam line and interaction region
geometries, magnetic fields, and beam trajectories. The
secondary particles from beam and material interactions must be
simulated precisely, which can only be done with a complete GEANT4
simulation. Thus in order for the FastSim to include background
effects, periodic ``background frame'' productions are carried out
with GEANT4, including beamlines up to $\pm 5$ meters from the
interaction point.
For each bunch crossing, background particles that cross the boundary
between the beamline elements and the detector volume are recorded and
stored in a background frame file. In FastSim, for each signal event,
background particles of a number of bunch crossings are read in from
the background frame file and are appended to the list of signal
particles. FastSim then proceeds as usual. The number of bunch
crossings per signal event is determined by the crossing frequency and
an overall sensitive time window. Each sub-detector has its own
response timing structure to model the appropriate signal pulse.

\tdrsubsubsec{FullSim}

The calorimeter full simulation tool (FullSim) is based on GEANT4 and
the Bruno tool (Sec.~\ref{sec:Full_Simulation}), which is used
to simulate particle interactions and energy deposits. The geometry
description includes all of the \superb\ sub-detectors. In the case of
the calorimeter, it contains all the crystals and support structures.
The Bruno output information for the calorimeter is the deposited energy
per crystal.  Each single source of machine background is simulated
separately for the entire \superb\ detector and, if needed, these are
combined in a second step of the analysis.

At the present stage of development the FullSim does not perform the
digitization of signals and the reconstruction of clusters, which are
instead obtained with a two-step procedure through stand-alone
programs.  During the digitization stage, which is overlaid on the
GEANT4 simulation output, the electronic signal shape is generated for
each crystal with an energy deposit. Photostatistics, electronic noise
and time resolution effects are added using extrapolations from
\babar, or using actual measurements when possible.  To study the
calorimeter performance, signal particles events are overlapped with
machine background events, following the time structure of the bunch
crossing rate in a time window centered on the signal particle arrival
time. This allows the emulation of effects of the background,
including electronic signal pile-up.  The different sources of
background are mixed according to their probability and time behavior.

Starting from the outcome of the digitization step, energy clusters
are created following the \babar\ algorithm
\cite{bib:babar_clustering}; a random 1\% intercalibration error is
applied when the cluster energy is computed (see for example
\cite{bib:Bauer:2006bj}).  To be accepted for clustering, the signal
from a crystal is required to peak within a time window centered on
the expected signal time.


\tdrsec{Barrel Calorimeter}

We propose to reuse the barrel portion of 
the \babar\ EMC~\cite{bib:BaBarNIM}, 
retaining the mechanical structures and the
5760 CsI(Tl) crystals and the pair of photodiodes mounted on each
crystal, after carrying out modifications of the electronics required
for optimal performance at \superb.

We first describe below the condition, as of the end of \babar\
running, of the CsI(Tl) crystals.  We then discuss the \superb\
environment, which produces high pile-up and backgrounds. These must
be addressed to optimize the energy resolution.  Next, we describe the
mechanical design and calibration systems for the barrel, each of
which will be minimally changed from their current configurations, as
well as the detector readout system, which will be upgraded.  We then
discuss the resulting barrel energy and position resolution, as well
as the $\gamma \gamma$ mass resolution, and the effects of long-term
radiation exposure at \babar. These results are extrapolated to
estimate the effect on performance to be expected over the lifetime of
\superb.

We then describe the existing barrel crystal photodiodes, which will
not be replaced, and the changes proposed for the on-crystal
pre-amplifier package and the front-end electronics. Finally we
present the proposed plan for disassembly of the barrel EMC down to
its component structures, their packaging and shipping and local
storage before and reassembly at a facility at or near Tor Vergata.
\bigskip \tdrsubsec{Requirements for the \superb\ Environment}

\tdrsubsubsec{Crystal Aging in \babar at PEP-II}



Over the span of \babar running at PEP-II, the EMC crystal light yield
has decreased due to exposure to high levels of
radiation, which is monitored by 116~RadFETs distributed
on the front face of the calorimeter.
The most common form of radiation damage~\cite{Zhu:1998eh} comes
from the development of absorption bands that reduce
the crystal's light yield by reducing the transmittance. Although crystals in
the EMC endcap experienced higher levels of radiation
than those in the barrel, all EMC crystals from the
furthest backward to those more forward received
non-negligible integrated doses. The resulting changes in crystal
light yields were monitored and corrected for during
\babar\ operation using calibrations performed at both
ends of the dynamic range of the detector: a low-energy
calibration using a $6.13\mev$ radioactive photon source,
discussed in Sec.~\ref{EMC:subsec:calibration},
and a high-energy calibration with Bhabha events.

\begin{figure}[H]
\centering
\subfloat[Change in light yield versus dose (Rad.), for barrel backward]{
  \includegraphics[clip=true,trim = 0 0 45 0,width=\linewidth]{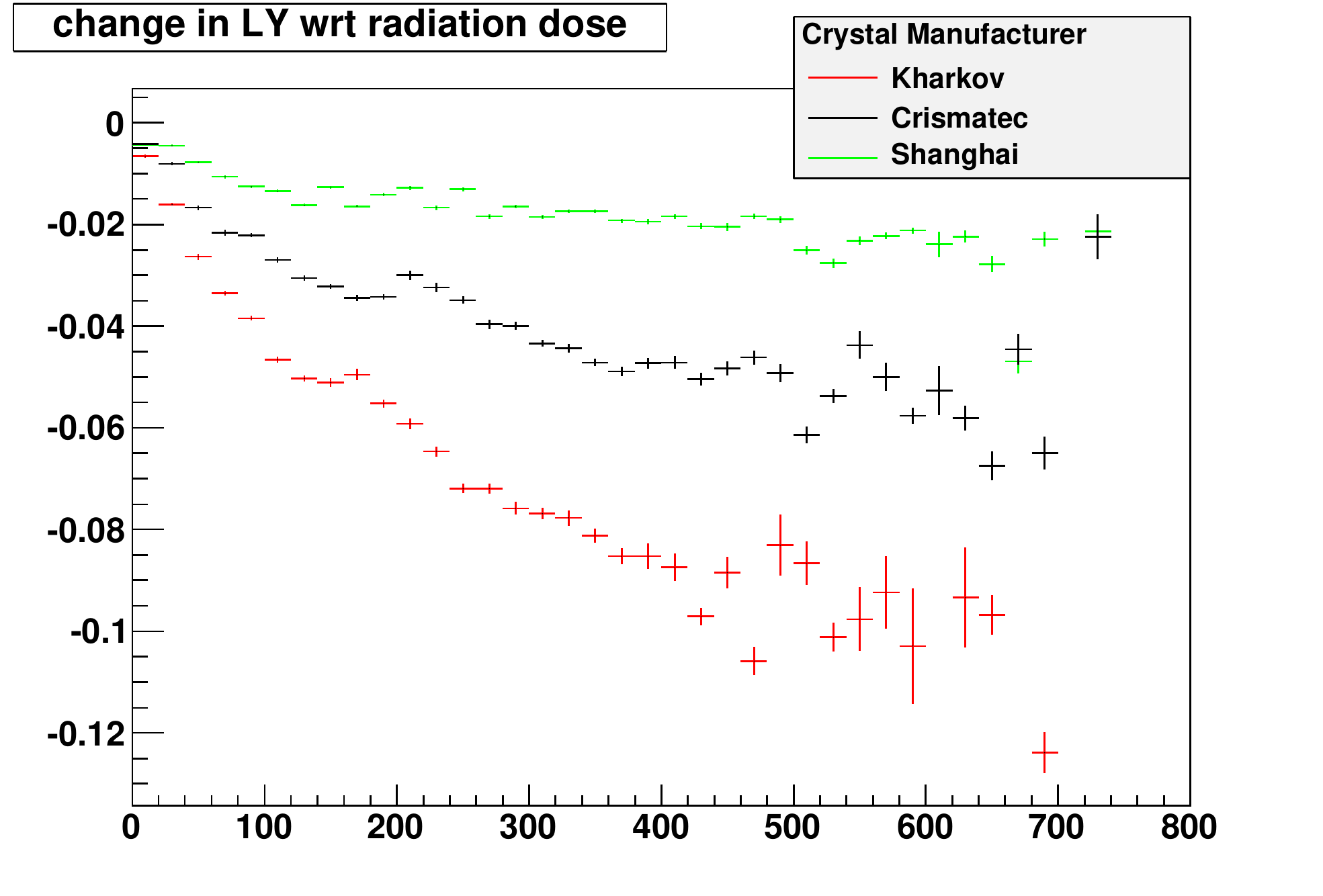}
}\\
\subfloat[Change in light yield versus dose (Rad.), for barrel forward]{
  \includegraphics[clip=true,trim = 0 0 45 0,width=\linewidth]{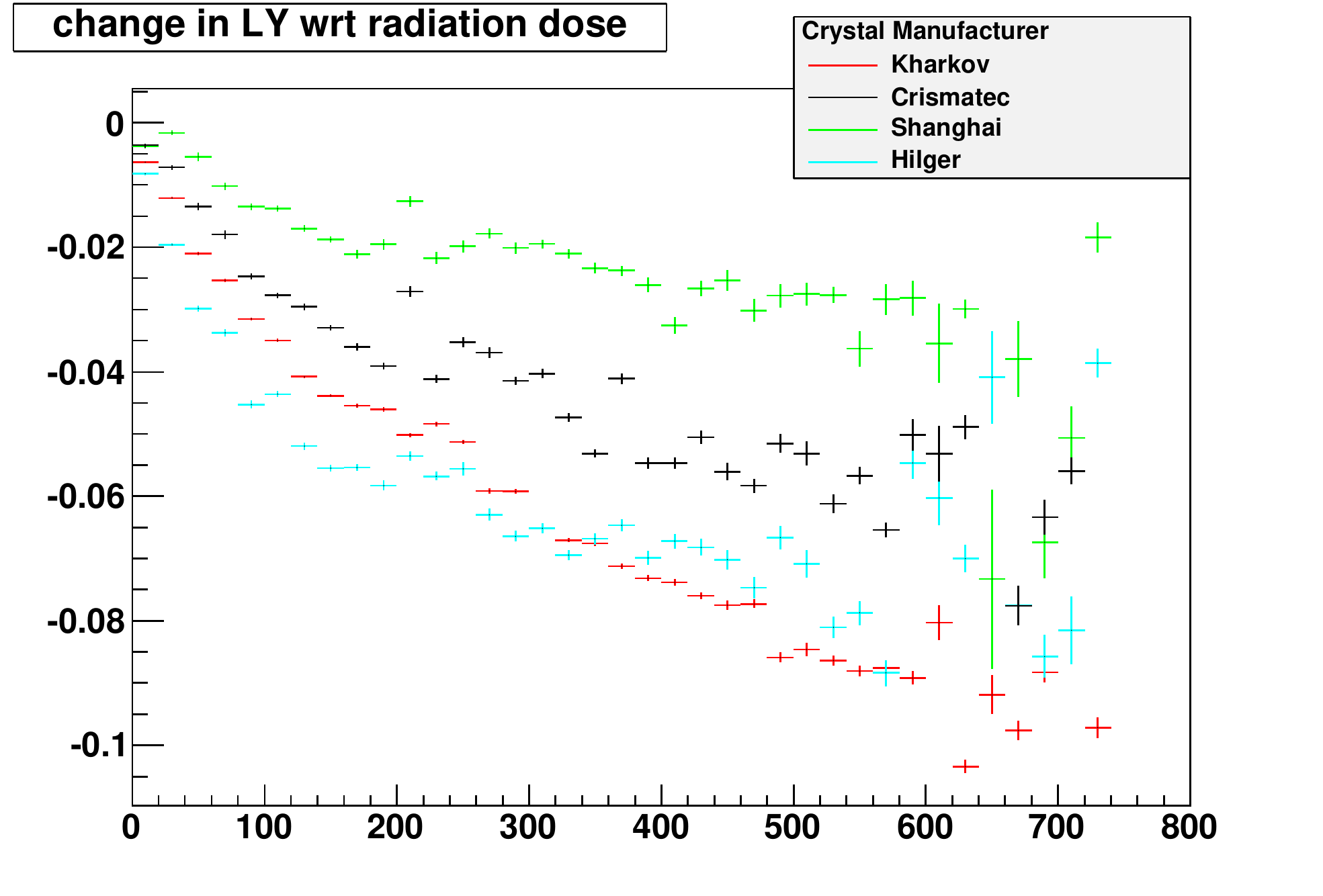}
}\\
\subfloat[$\Delta$ LY vs theta. The forward endcap is 6--9; the barrel starts
at 10, with the backward barrel at large values of the index.]{
  \includegraphics[width=\linewidth]{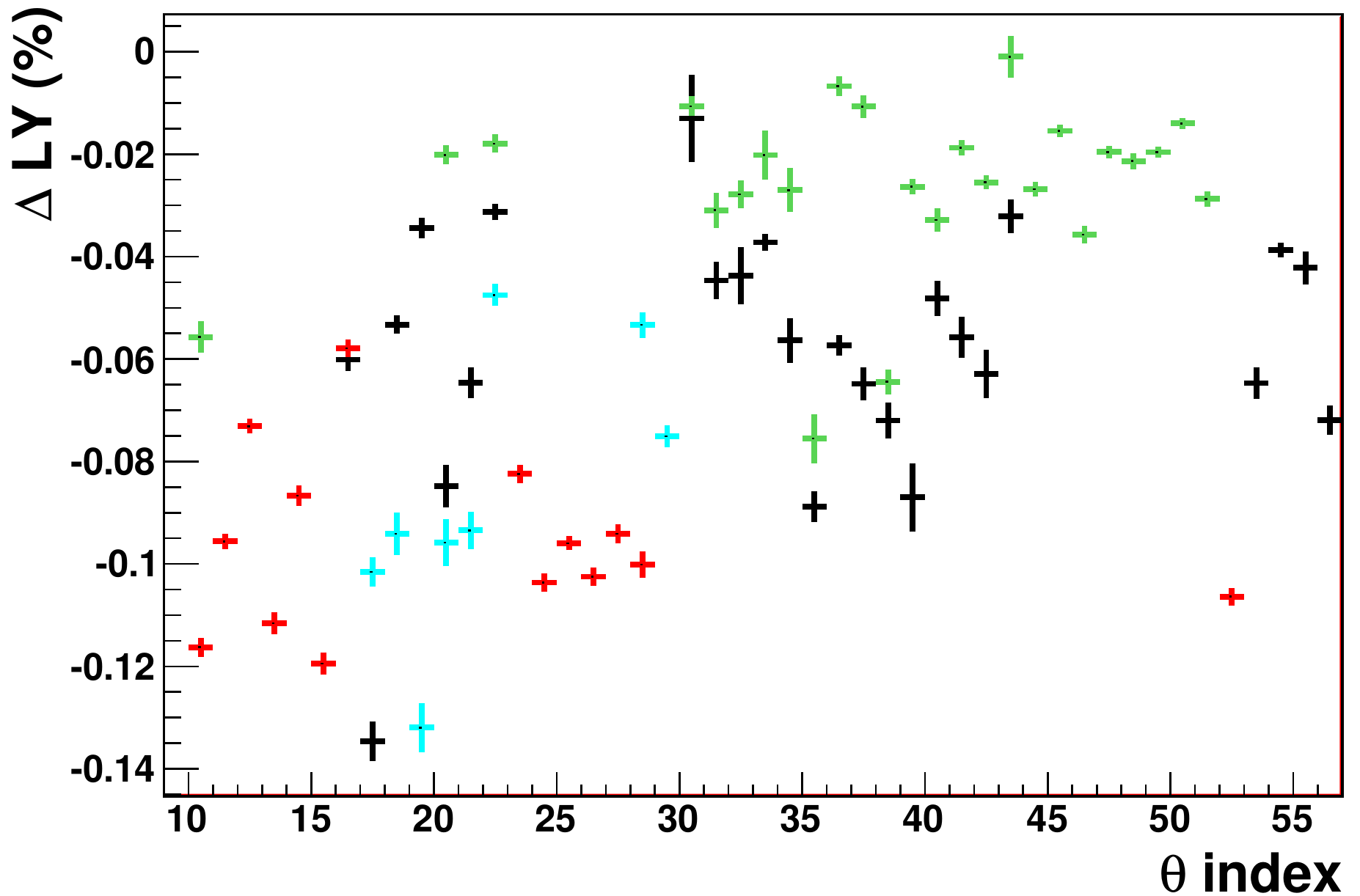}
}
\caption[Barrel crystal light yield decreases] {Backward and forward
  barrel crystal light yield decreases with absorbed radiation dose,
  with the different crystal providers indicated.  The bottom plot
  shows the light yield decrease as a function of polar angle index
  for each manufacturer using the same color legend.}
\label{fig:EMC:emc_ly_change}
\end{figure}

The change in light yield for barrel crystals as a function of
absorbed radiation dose, based on the low-energy calibration data, is
shown categorized by crystal manufacturer in
Fig.~\ref{fig:EMC:emc_ly_change}.  Though the crystals were initially
set up to have a relatively uniform response before they were
integrated into the detector, there have been varying degrees of
degradation in performance over time.  Depending on the manufacturer,
the total decrease in light yield can be up to $\sim\!\! 10\%$. Much
of the decrease occurred during the initial years of PEP-II running.
As the integrated dose increased, there was less light loss per unit
dose received.  However, as can be seen in
Fig.~\ref{fig:EMC:emc_ly_change}, in the \superb\ environment, the
eventual loss in light yield for the worst-performing barrel
crystals---generally those provided by Kharkov and Hilger---is
substantial.  The relatively poor performance of the crystals provided
by these manufacturers was known at the time of barrel construction.
To the extent possible, these manufacturers' crystals were placed as
far backwards in polar angle as possible
(Fig.~\ref{fig:EMC:emc_ly_change}).


\tdrsubsubsec{Backgrounds}
\label{sec:EMC:BARbkg}

In addition to crystal aging, backgrounds can degrade energy resolution
by causing electronic signal pile-up. The dominant source in \superb\
is expected to be photons and neutrons from radiative Bhabha
events. This effect was negligible in \babar, but could be substantial
in \superb, especially in the low energy range.

The pile-up effect is a function of signal pulse shape. Since \superb\
is reusing the \babar\ barrel, the limitation of the long decay time
of CsI(Tl) scintillation light must be accepted.  Readout and
electronics can, nonetheless, be optimized to minimize the impact of
the pile-up effect.  To ensure similar physics sensitivity as \babar,
the requirement is that background pile-up should have a negligible
effect on the energy resolution of high energy photons, and should not
dominate the energy resolution at low energy, which is approximately
4.5\% at 100~\mev in \babar.

\tdrsubsec{Description of \babar\ Barrel Calorimeter}

\tdrsubsubsec{Mechanical design}


\ignore{
\begin{figure*}
\begin{center}
\includegraphics[width=13cm]{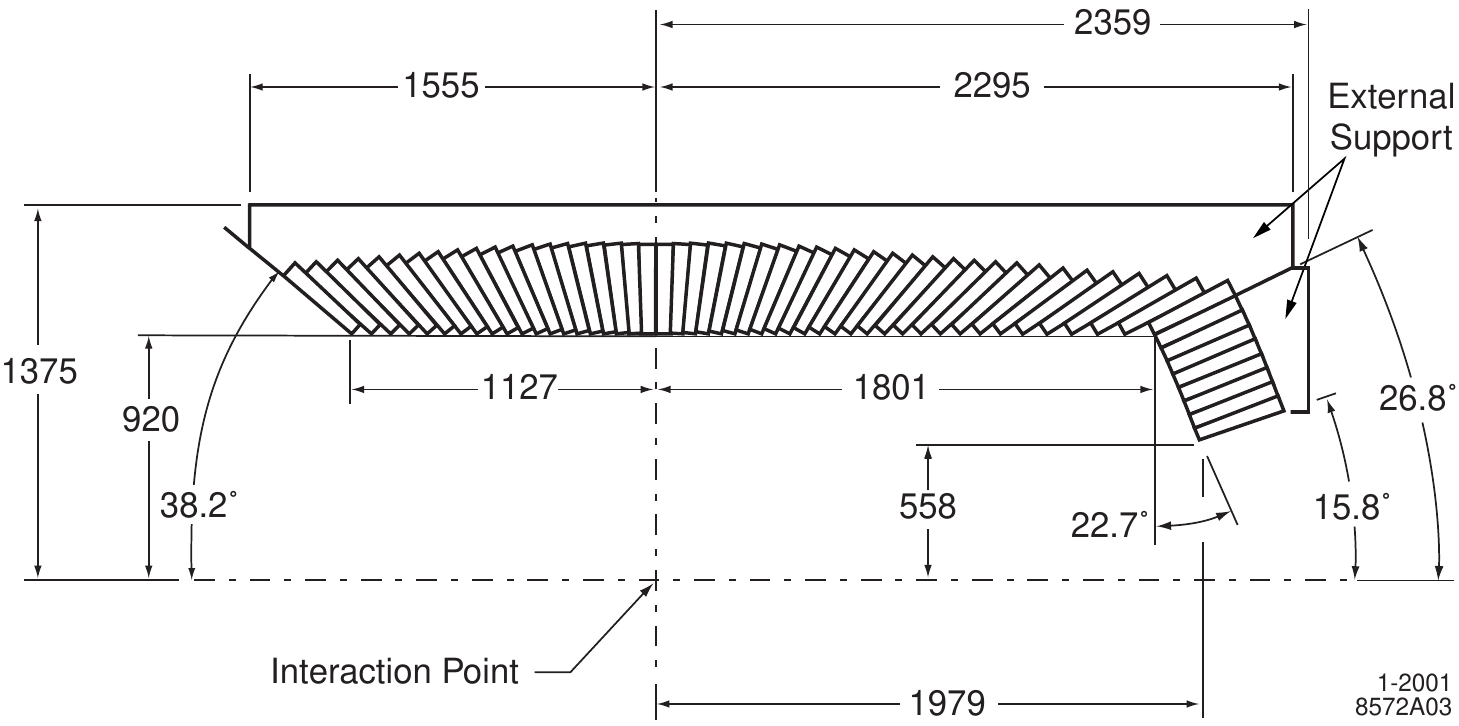}
\caption{Longitudinal \babar\ EMC cross-section (top half)
showing the arrangement of the 48 barrel and 8 endcap crystal rings.
The detector is axially symmetric around the $z$-axis. Dimensions are
in mm.}
\label{EMC:fig:xsecview}
\end{center}
\end{figure*}
}

The \babar\ barrel EMC consists of a cylindrical shell with full
azimuthal coverage, extending in polar angle from $26.8 \degrees$ to
$141.8 \degrees$. A longitudinal cross-section, including the forward
endcap, is shown in Fig.~\ref{EMC:fig:xsecview}.  The barrel EMC
contains 5,760~crystals arranged in 48 azimuthal rings, with
120~identically dimensioned crystals in each ring.  Each crystal has a
tapered trapezoidal cross section, with length increasing from $29.6
\cm$ furthest backward to $32.4 \cm$ furthest forward in order to
minimize the effects of shower leakage from increasingly higher energy
particles.  To minimize the probability of pre-showers, the crystals
are supported entirely at the outer radius, with only a thin gas seal
at the front. The amount of material in front of the crystal faces is
$0.3-0.6\Xrad$, mostly from material in the beampipe, SVT, DCH, and
DIRC.

Figure~\ref{EMC:fig:xtalpkg} is a schematic of a single
crystal, showing the layered crystal wrappings, silicon photodiodes,
diode carrier plate, preamplifier, and the aluminum housing.

The barrel crystals are inserted into carbon fiber modules that are supported individually
from an external cylindrical support structure that
carries the load
of the barrel modules and the forward endcap to the magnet steel
through four flexible supports, which decouple and dampen
any motion relative to the magnet steel during an
earthquake.

\begin{figure}[hb!]
\begin{center}
\includegraphics[width=5.8cm]{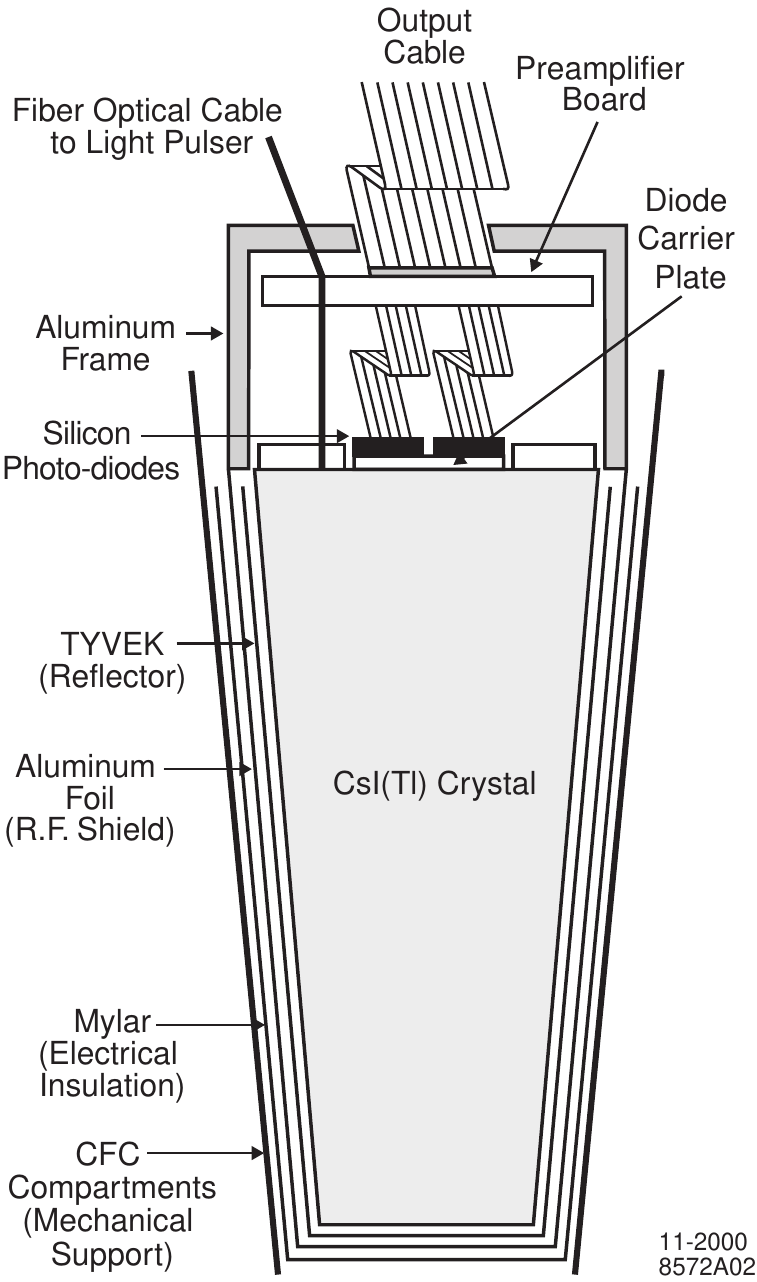}
\vspace{0.1in}
       \caption{A schematic (not to scale) of a wrapped barrel crystal and the
       front-end readout package mounted on the rear face. Also
       indicated is the tapered, trapezoidal CFC compartment, which is
       open at the front.}
\label{EMC:fig:xtalpkg}
\end{center}
\end{figure}

The crystal modules are tapered, trapezoidal compartments made from
carbon-fiber-epoxy composite (CFC) with 300\mum-thick walls.
Each compartment loosely holds a
single wrapped and instrumented crystal, assuring that the
forces on the crystal never exceed its own weight. Each
module is surrounded by a second layer of 300\mum\ CFC to
provide additional strength.  The modules are bonded to an aluminum
strong-back that is mounted on the external support structure.
Figure~\ref{EMC:fig:construction} shows some details of a module
and its mounting to the support cylinder.
This scheme minimizes inter-crystal material while exerting minimal force on the
crystal surfaces, preventing geometric deformations and surface degradation
that could compromise performance.

\begin{figure*}[htpb!]
\begin{center}
\includegraphics[width=15.0cm]{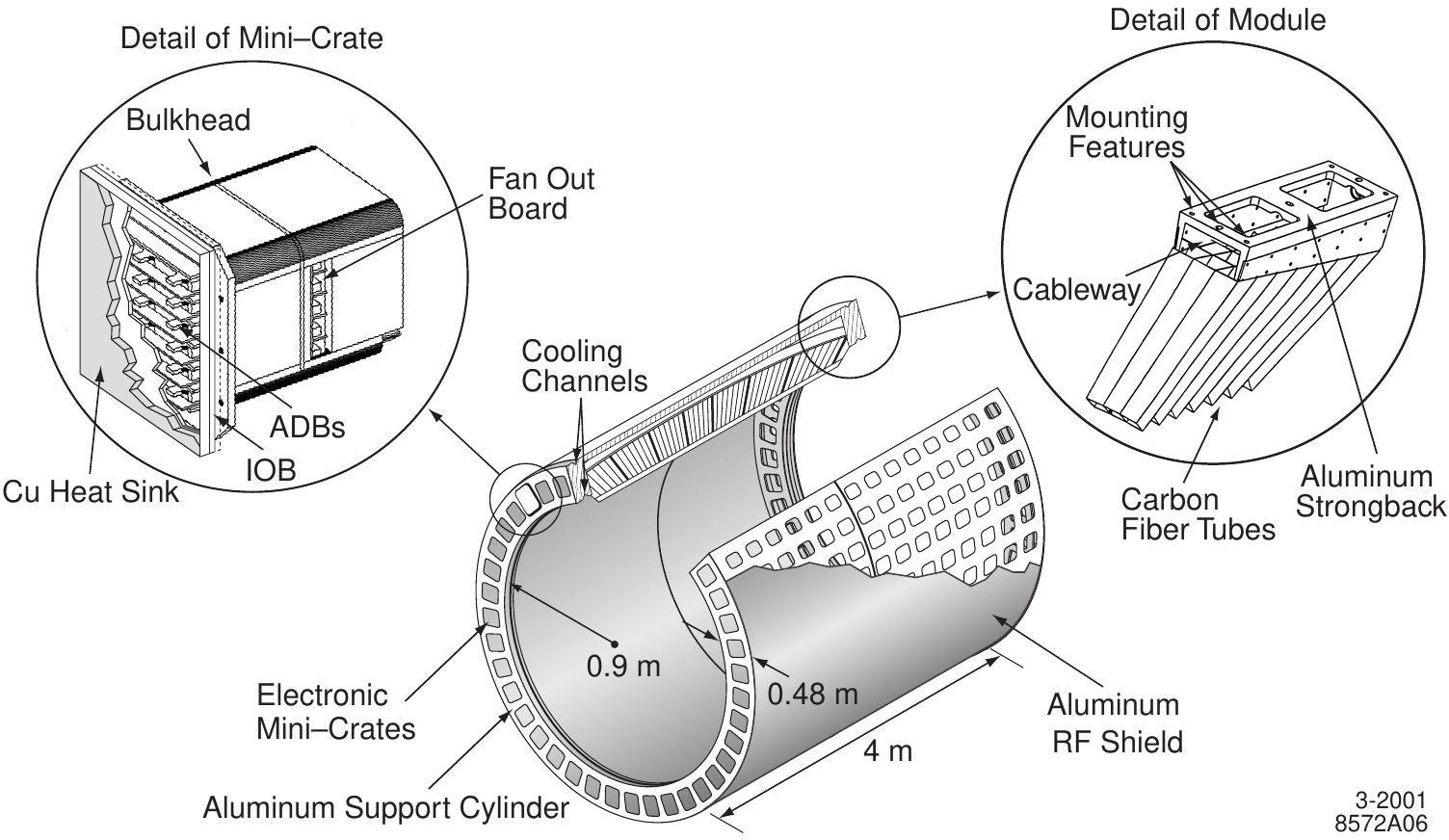}
\vspace{0.1in}
       \caption{The EMC barrel support structure, with details on the
       modules and electronics crates (not to scale).}
\label{EMC:fig:construction}
\end{center}
\end{figure*}

The barrel is composed of 280 separate modules, each
holding 21 crystals ($7 \times 3$ in $\theta \times \phi$),
except for the furthest backward modules which contain only $6 \times 3$ crystals.
After insertion of the crystals, the aluminum readout frames,
which also stiffen the module, are attached with thermally-conducting
epoxy to each of the CFC compartments.  The entire $\sim 100$~kg module is
then bolted and glued with conducting epoxy 
to an aluminum strong-back (Fig.~\ref{EMC:fig:construction}).
The strong-back contains alignment features, as well as channels that
couple it to the cooling system. Each module is installed into the
2.5\cm-thick, 4\m-long aluminum support cylinder, and subsequently
aligned.  On each of the thick annular end-flanges, the support cylinder
contains access ports for digitizing electronics crates with
associated cooling channels, as well as mounting features and
alignment dowels for the forward endcap. Figure~\ref{EMC:fig:construction}
shows details of an electronics mini-crate situated within the support cylinder.

The primary heat sources internal to the calorimeter are the
preamplifiers ($2 \times 50$~mW/crystal) and the digitizing
electronics (3~kW per end-flange).  In the barrel, the heat generated
by the preamplifiers is removed by conduction to the module strong
backs, which are directly cooled by Fluorinert
(polychlorotrifluoroethylene).  The digitizing electronics are housed
in 80 mini-crates mounted on the end-flanges of the cylindrical
support structure.  These crates are indirectly cooled by chilled
water pumped through channels milled into the end-flanges close to the
inner and outer radii.

The barrel is surrounded by a double Faraday shield
composed of two 1\mm-thick aluminum sheets, further shielding the diodes and
preamplifiers from external noise.  This cage
also serves as the environmental barrier, keeping the slightly
hygroscopic crystals in a dry, temperature-controlled,
nitrogen atmosphere.

The stability of the photodiode leakage current, which rises
exponentially with temperature, and the crystal light yield, which is
weakly temperature-dependent, were of particular concern during \babar\
data-taking.  The most important issue for the upgrade to \superb\ is
that the diode-crystal
epoxy joints experience as little stress as possible due to
differential thermal expansion during disassembly, transport and reassembly.
Studies performed prior to the
construction of the EMC showed that preserving the integrity of these
joints requires that crystals be kept at a temperature of $(20 \pm
5)$~C. The barrel and endcap are currently maintained in this state,
with temperatures monitored in real-time and recorded in the
conditions database.  Figure~\ref{EMC:fig:emc_temp_history2} shows the
temperature history for one year for the four quadrants of the barrel
and the two halves of the endcap.

\begin{figure}[hb!]
\centering
\includegraphics[width=0.9\linewidth]{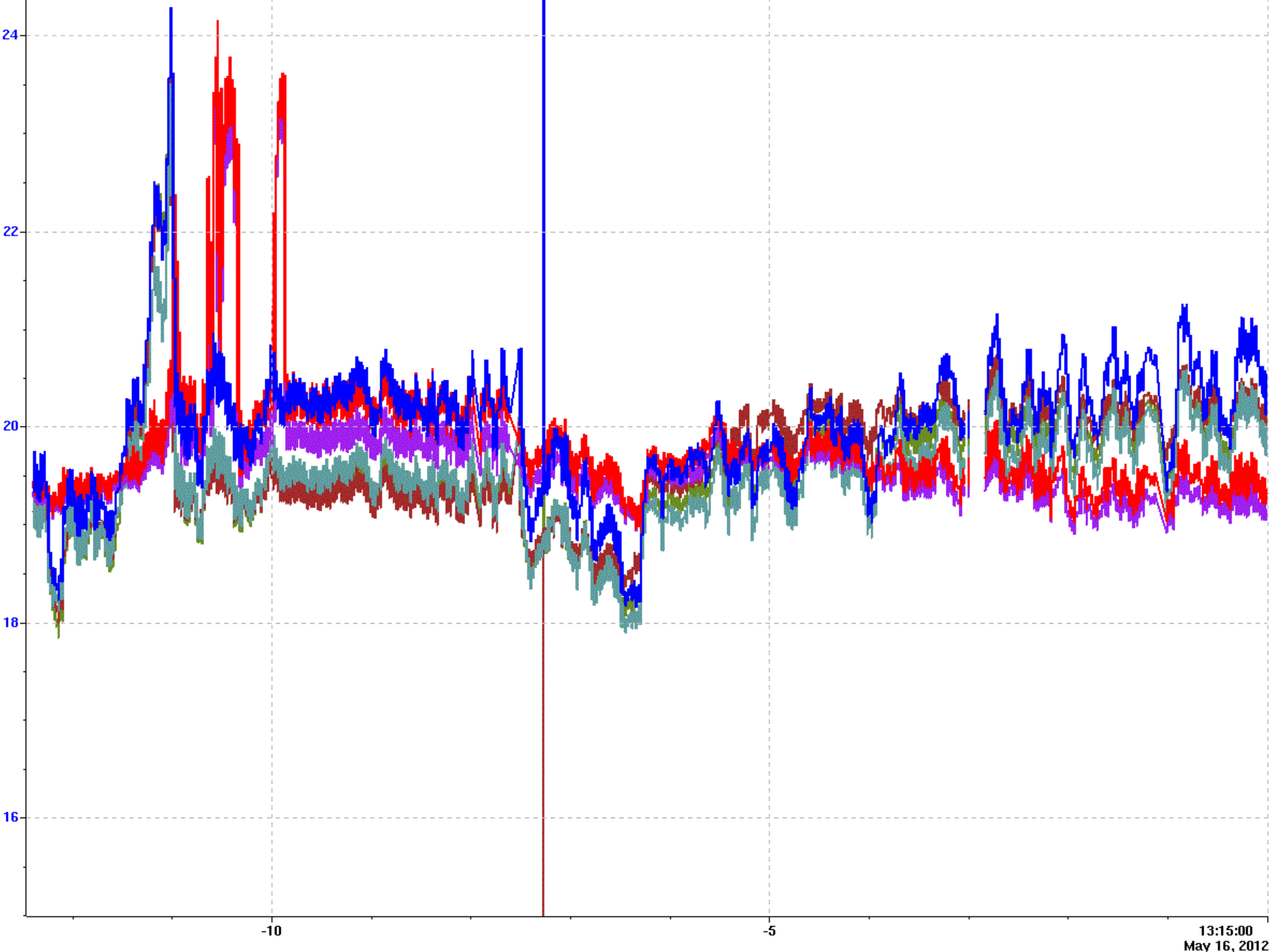}
\caption[Barrel and endcap thermal history]
{Barrel and endcap thermal history for one year.
The four barrel quadrants are blue, brown, and dark
and light green; the endcap halves are purple and red.
The horizontal axis shows months prior to 16 May 2012.
The vertical line extending the full range of the plot is
an artifact due to a bad database record.}
\label{EMC:fig:emc_temp_history2}
\end{figure}

\tdrsubsubsec{Readout}


Each CsI(Tl) barrel and endcap crystal is read out using
two $2 \times 1\cma$ silicon PIN diodes glued to a transparent
1.2~\mm-thick polystyrene substrate that is in turn glued
to the center of the rear face of the crystal with an optical
epoxy to maximize light transmission.
The surrounding area of the crystal back face
is covered by a plastic plate coated with white reflective
paint, which has two 3~\mm-diameter penetrations for
the fibers of the light pulser monitoring system.

The signal from each of the diodes is transmitted to a low-noise
preamplifier through a cable terminated in a connector that allows the
preamplifier to be detached from the photodiodes. The preamplifier
characteristics are discussed below in Sec.~\ref{sec:EMCbarrelElec}.
The entire diode/electronics assembly is enclosed in an aluminum
fixture (Fig.~\ref{EMC:fig:xtalpkg}). This fixture is electrically
coupled to the aluminum foil wrapped around the crystal 
silver-conductive epoxy and thermally coupled to the support frame to
dissipate the heat load from the preamplifiers.

\tdrsubsubsec{Low-energy Source Calibration}
\label{sec:EMC:source_calibration}
\label{sec:EMC:BARcalib}
\label{EMC:subsec:calibration}

Achieving the best possible performance of the EMC requires careful calibration and monitoring during data-taking.
The low-energy calibration system devised for \babar\ produces a 6.13~MeV photon line from a short-lived $^{16}$O transition that can be activated as required. This system was successfully used for routine biweekly calibrations of the \babar\ calorimeter, and is an ideal match to the \superb\ requirements. We will therefore refurbish the \babar\ system for use in \superb.
The final \babar\ calibration run was taken on 3 July 2008, and records of this run, as well as a few earlier ones,
will be maintained for comparison with the refurbished \superb\ EMC.


In this system, $^{19}$F, a component of
Fluorinert$^{\tiny\texttrademark}$ (polychlorotrifluoroethylene)
coolant liquid, is activated with a neutron source to produce
$^{16}$N, which $\beta$-decays to an excited state of $^{16}$O. This
state in turn emits a 6.13~MeV photon as it cascades to its ground
state.  The source spectrum, as seen with a CsI(Tl) crystal with the
PIN diode readout used in \babar, is shown in
Fig.~\ref{fig:EMC:calib}. There are three principal contributions to
the overall peak, one at 6.13~MeV, another at 5.62~MeV, and the third
at 5.11~MeV. The latter two represent \epem 0.511~\mev
annihilation-photon escape peaks. Since all three peaks have
well-defined energies, they simultaneously provide an absolute
calibration and an energy scale at the low end of the energy spectrum.

\begin{figure}[!h]
\centerline{ \includegraphics[width=60mm]{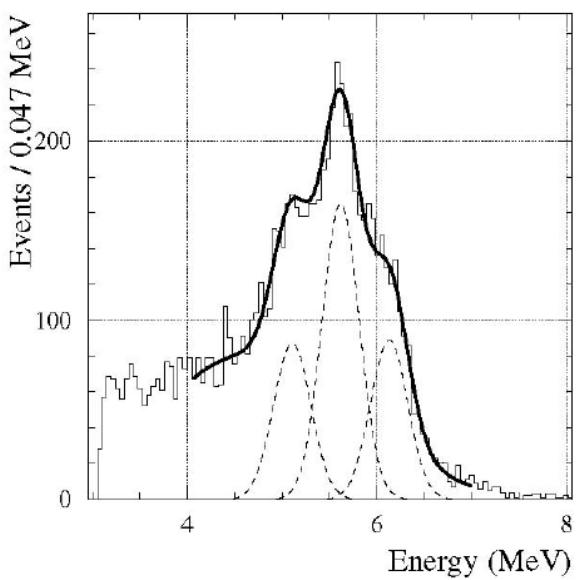}}
\caption {Energy spectrum in a single  \babar\ CsI(Tl) crystal irradiated with 6.13 MeV photons from an $^{16}$O* source, read out with a PIN diode. The solid curve is a fit to the spectrum, including Gaussian contributions at 6.13 MeV, 5.62 MeV, and 5.11 MeV, indicated by the dashed curves.  \protect\cite{bib:BaBarNIM}}
\label{fig:EMC:calib}
\end{figure}

The fluorine is activated by neutrons produced by a commercial
deuterium-tritium (DT) generator that produces 14.2~MeV neutrons by
accelerating deuterons onto a tritium target. Typical rates are
several times $10^8$ neutrons/second. The DT generator is surrounded
with a bath of the fluorine-containing liquid, which is then
circulated in a manifold to the barrel and endcap crystals. There are
many suitable fluorine-containing liquids available; Fluorinert
FC77~\cite{bib:FC77}, was used in \babar. The half-life of the
activated liquid is 7 s, hence residual radioactivity is not a
substantial concern when the DT generator is not operating.

The Fluorinert is stored in a reservoir near the DT generator. When a
calibration run is started, the generator and a circulating pump are
turned on.  Fluid is pumped from the reservoir through the DT bath to
be activated, and then to the calorimeter. The system is closed, with
fluid returning from the calorimeter to the reservoir
(Fig.~\ref{fig:EMC:calibschematic}). In \babar, the fluid is pumped at
3.5~l/s, producing a rate of approximately 40~Hz in each of 6500
crystals. This produces a calibration with a statistical uncertainty
of 0.35\% on peak positions in a single crystal in a 10-minute
calibration run~\cite{bib:BaBarNIM,bib:Weaver}.  The crystals are
approximately 12~m from the DT generator. Although Fluorinert FC-77 is
no longer manufactured, an adequate supply is on hand.  Should it be
necessary to substitute a commercially available coolant liquid (all
of which have slightly reduced fractions of fluorine), the pumping
rate or neutron source intensity would have to be adjusted
accordingly.

\begin{figure*}
\begin{center}
\includegraphics[width=10cm, angle=90, trim= 0cm 0cm 0cm 0cm, clip=true]{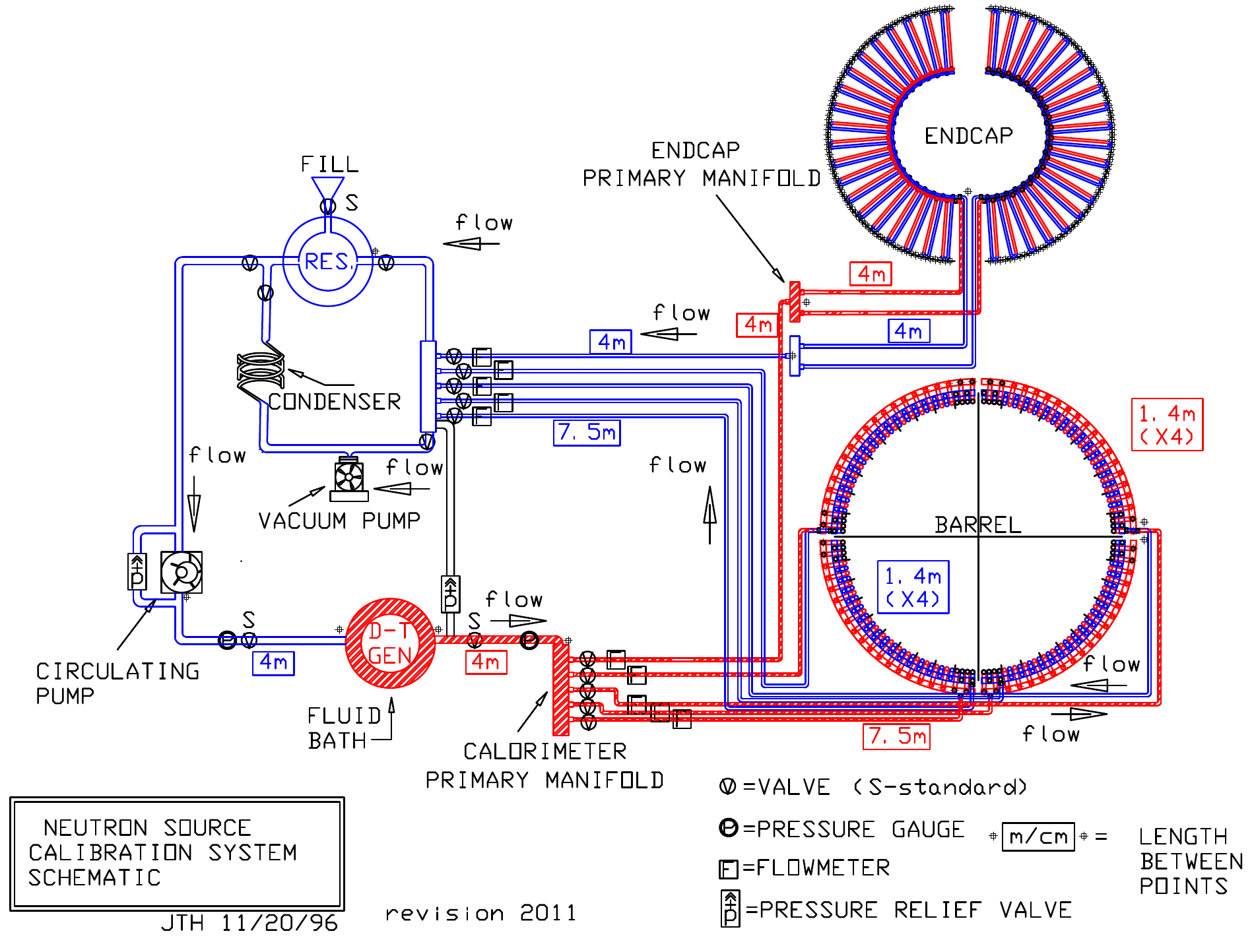}
\caption{Schematic of the \babar\ calibration system, which produces $^{19}$F 6.13 MeV photons from an $^{16}$O* source, as updated for \superb. The system is used for both the barrel and forward endcap calorimeters.}
\label{fig:EMC:calibschematic}
\end{center}
\end{figure*}


In \babar, the fluid transport manifold consisted of thin-wall (0.5
mm) aluminum tubing (3/8~inch diameter), flattened to meet space
constraints.  One millimeter of Al represents 1.2\% of a radiation
length. This material is located in front of the \babar\ crystals, as
is an additional 2~mm of Al in the structural support of the tube
assemblies, which are deployed as a set of curved panels. This system
can be reused for the barrel in \superb, or it could be rebuilt to
better integrate into the barrel structure, reducing the amount of
material and providing additional radial space.


The DT generator is a small accelerator; radiation safety protocols
factor into the mechanical design of the system. For example,
operation of the source will be done remotely, in a no-access
condition. It will be shielded according to INFN radiation safety
regulations. The shielding will be interlocked such that the DT
generator cannot be operated if the shielding is not in place. The
reservoir is capable of holding the entire volume of Fluorinert fluid.
The system is anticipated to be operated approximately weekly. In the
event of a fluid leak, the maximum exposure for the similar \babar\
system was calculated to result in an integrated dose of less than 1
mrem. A detailed hazard analysis will be performed in collaboration
with INFN radiation safety experts.

\tdrsubsubsec{Light Pulser}
\label{sec:EMC:light_pulser}

The light response of the individual crystals was measured daily
during \babar\ running using a light-pulser system, which transmitted
spectrally filtered light from a xenon flash lamp through optical
fibers to the rear of each crystal. We propose to refurbish the
\babar\ system for use in \superb.  The light pulse is similar in
spectrum, rise-time and shape to the scintillation light in the
CsI(Tl) crystals. The pulses were varied in intensity by
neutral-density filters, allowing a precise measurement of the
linearity of light collection, conversion to charge, amplification,
and digitization. The intensity was monitored pulse-to-pulse by
comparison to a reference system with two radioactive sources, 
${\rm^{241}Am}$ and ${\rm ^{148}Gd}$. Each of these was attached to a
small CsI(Tl) crystal read out by both a photodiode and a
photomultiplier tube. The response was stable to 0.15\% over a period
of one week.




\tdrsubsec{Performance of the \babar\ barrel}

\tdrsubsubsec{Energy and position resolution}

The energy dependence of the energy resolution of a homogeneous
crystal calorimeter is generally empirically described by two terms
summed in quadrature:
\begin{equation}
\frac{\sigma_E}{E}= \frac{a}{\sqrt[4]{E({\rm GeV})} }\oplus  b,
\end{equation}
where $E$ and $\sigma_E$ refer to the energy of a photon and
its {\it rms} error in $\gev$. The energy-dependent term arises from
fluctuations in photon statistics, electronics noise, and beam
background-generated noise.  The constant term arises from
non-uniformity in light collection, shower leakage, calibration
uncertainties, and energy deposited in the non-sensitive material
between and in front of the crystals.  The angular resolution is
primarily determined by the crystals' transverse dimension and is also
described empirically with energy-dependent and energy-independent
contributions.  Figure~\ref{fig:EMC:photon-resolution-Babar} shows the
energy and angular resolution of the \babar\ calorimeter derived from
various data control samples. The
performance may be parameterized as~\cite{bib:BaBarNIM}:
\begin{equation}
 \label{eq:EMC:babarNIMEres}
  \frac{\sigma_E}{E}=\frac{(2.32\pm  0.3)\%}
                          {\sqrt[4]{E({\rm GeV})}}
  \oplus (1.85\pm 0.12)\% 
\end{equation}
\vskip -12pt
\begin{equation}
  \sigma_{\theta} = \sigma_{\phi} =
   \frac{(3.87\pm 0.07)\hbox{ mrad} }  {\sqrt{E( \hbox{GeV})} }
   \oplus   (0.0\pm 0.04)\hbox{ mrad}
\end{equation}

\begin{figure}[hbtp]
  \begin{center}
      \includegraphics[width=6.1cm]{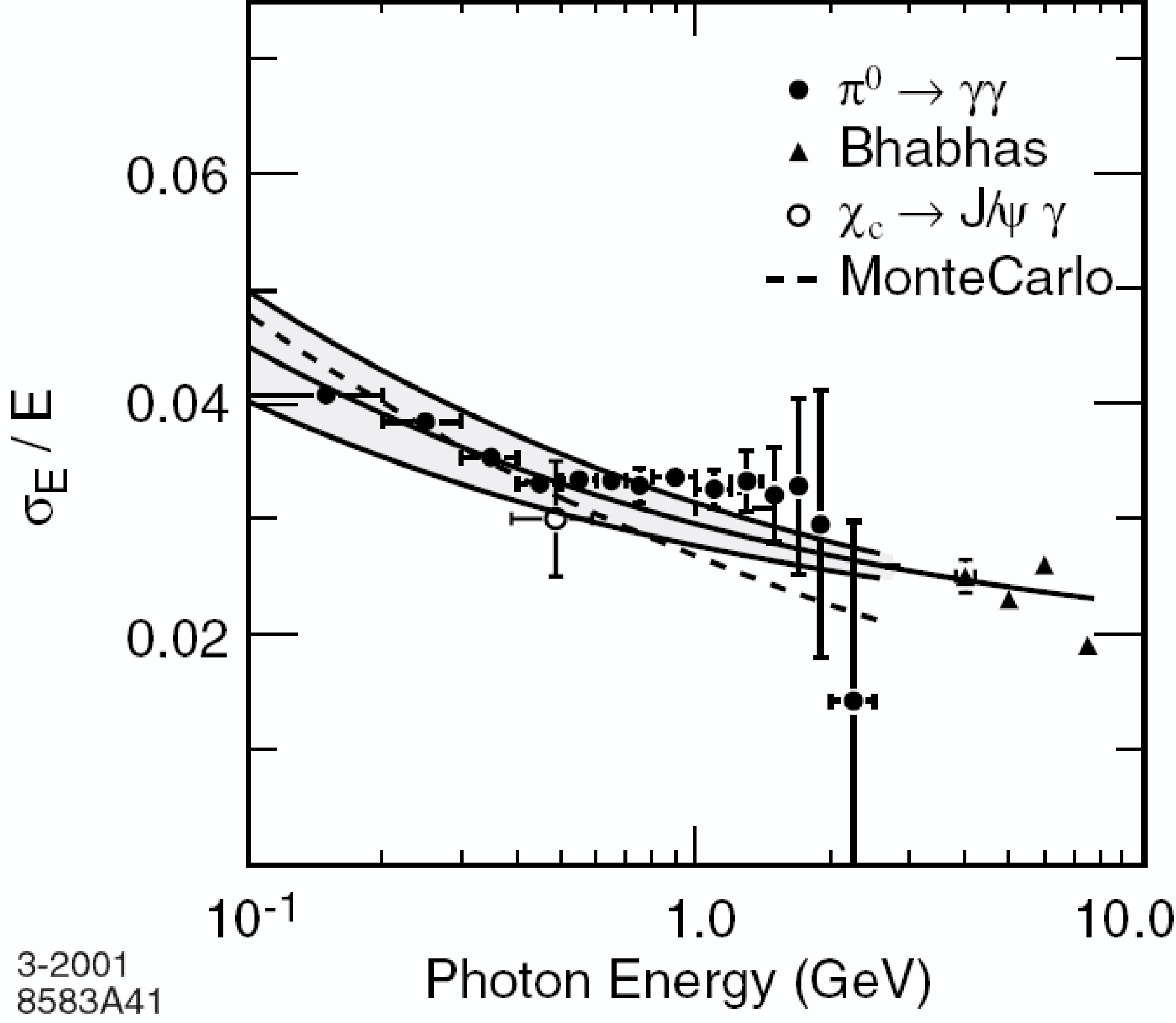}
      \includegraphics[width=5.8cm]{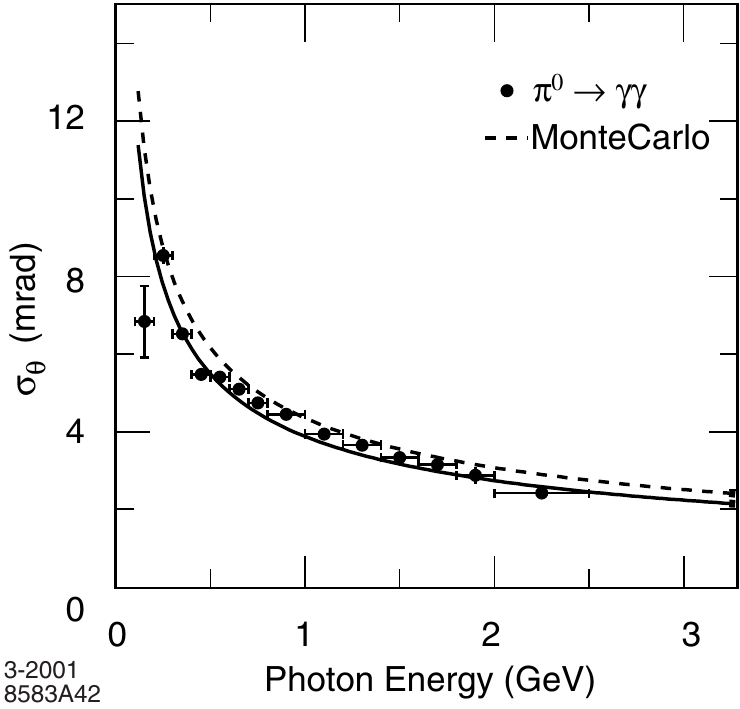}
      \caption{Photon energy (top) and angular (bottom) \babar\ EMC resolutions~\protect{\cite{bib:BaBarNIM}}. Solid curves are fits to the data 
(eq.~\ref{eq:EMC:babarNIMEres}); dashed lines are simulation.}
      \label{fig:EMC:photon-resolution-Babar}
  \end{center}
\end{figure}


\tdrsubsubsec{$\gamma\gamma$ mass resolution}

Figure~\ref{fig:EMC:gamgam-resolution-Babar} shows a two-photon
invariant mass distribution in the $\piz$ and $\eta$ mass ranges for
selected \babar\ data multi-hadron events from 2001.  Photons are
required to have energy $> 30 \mev$, with $\pi^{0}$
candidate energy $> 300 \mev$, and $\eta$ candidate
energy $> 1 \gev$.  The reconstructed $\piz$ mass is
found to be 134.9~\mevcc, with a width of 6.5~\mevcc. The fitted
$\eta$ mass is 547~\mevcc, with a width of 15.5~\mevcc.

\begin{figure}[hbt]
  \begin{center}
      \includegraphics[clip=true,trim = 5 5 50 5,width=0.45\textwidth]{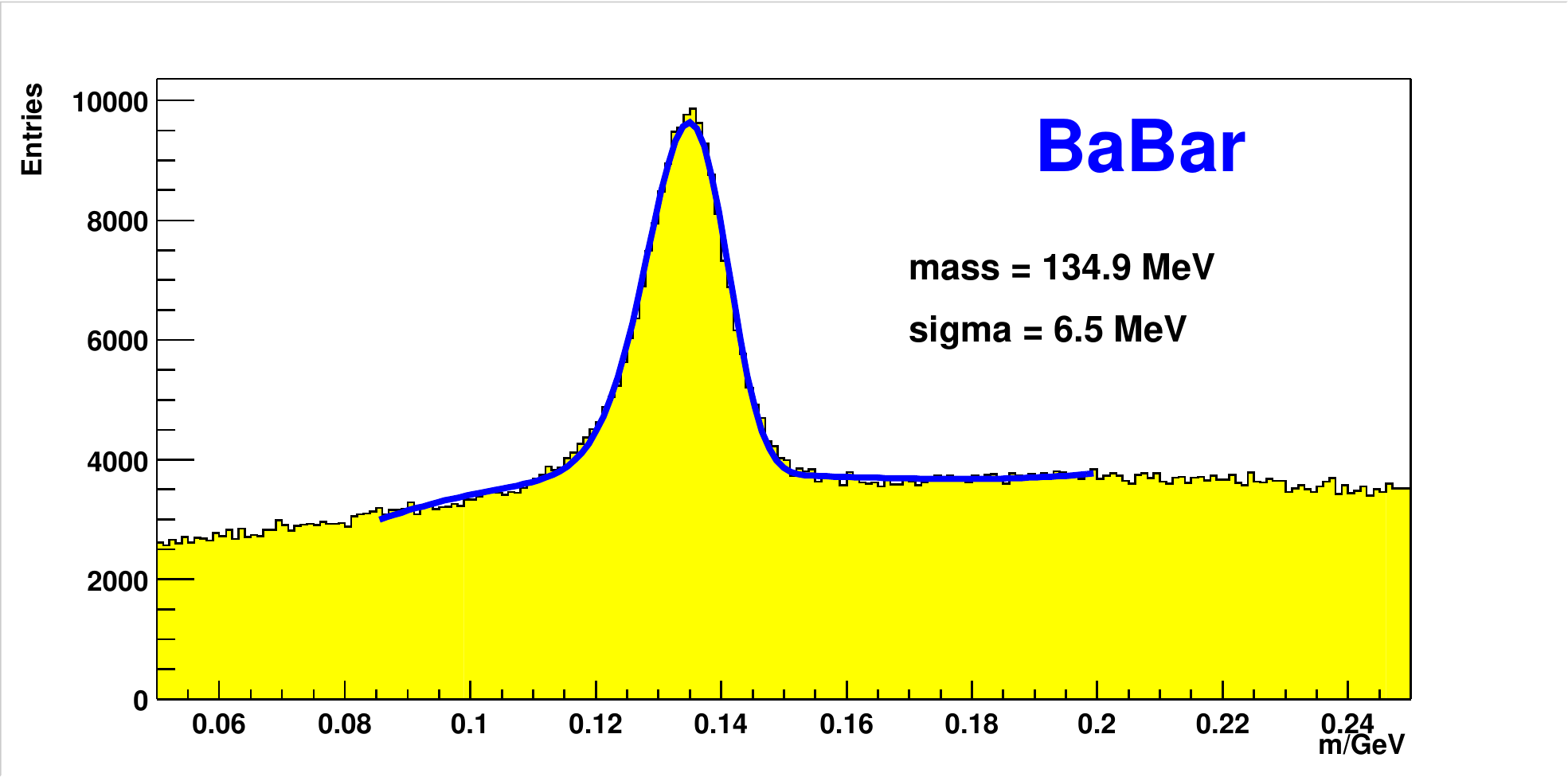}
      \includegraphics[clip=true,trim = 0 0 50 0,width=0.45\textwidth]{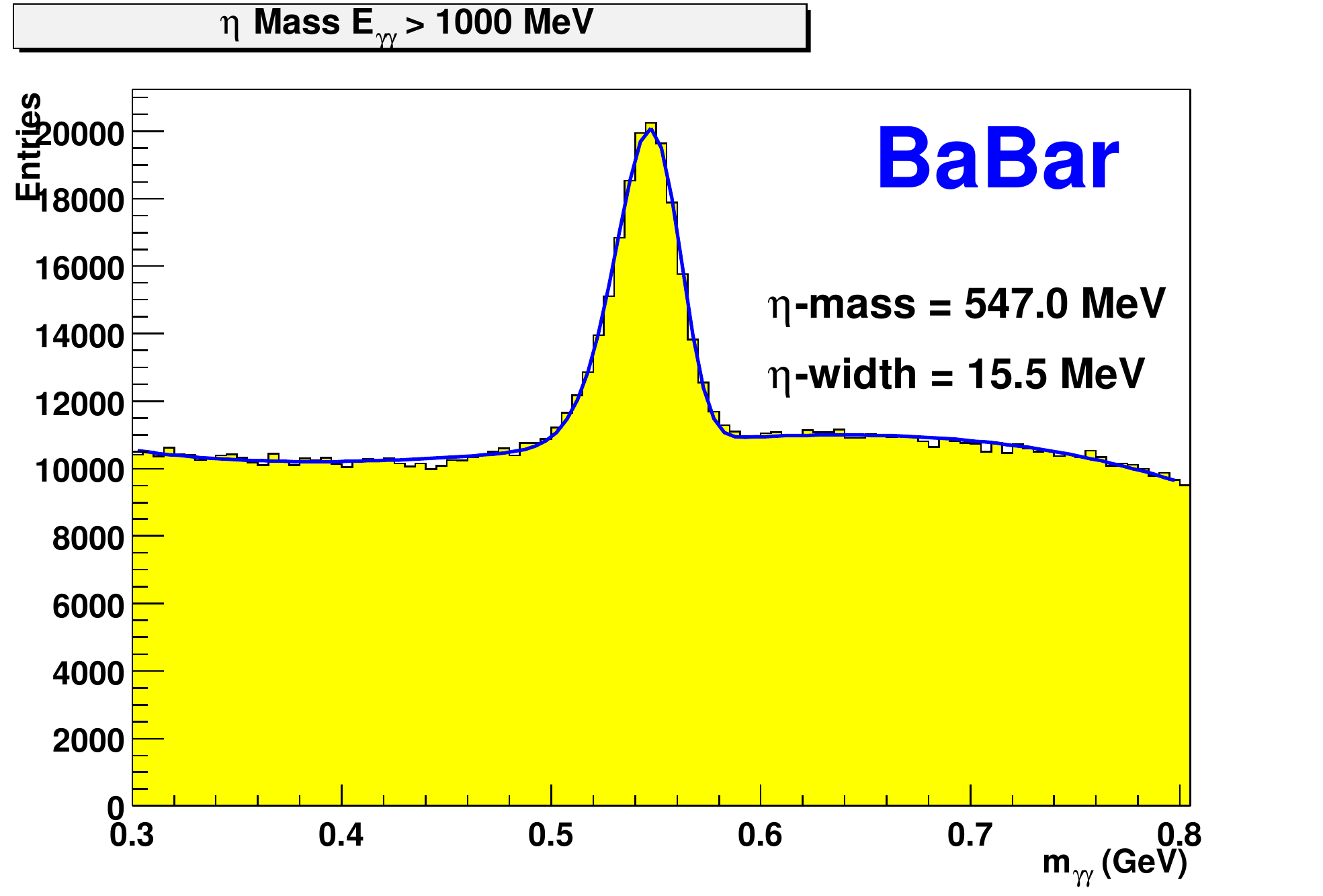}
      \caption{Two-photon invariant mass distribution in $\pi^0$ (top)
      and $\eta$ (bottom) mass ranges.}
      \label{fig:EMC:gamgam-resolution-Babar}
  \end{center}
\end{figure}

%



\tdrsubsubsec{Radiation Damage Effects on Resolution}

Beam-generated backgrounds are the major cause of reduction in the
light yield of the crystals over time. In order to monitor this source
of background, 116 RadFETs are placed in front of the calorimeter
barrel and endcap crystals. These RadFETs, real-time integrating
dosimeters based on solid-state MOS technology, are integrated into
the EPICS monitoring system. As can be seen in
Fig.~\ref{fig:emc_raddam}, the integrated dose is largest in the
endcap, which is closer the the beam line, as well as more forward in
polar angle, making it more susceptible to beam-generated background
photons from small angle radiative Bhabha events in which an $e^\pm$
strikes a machine element.


Radiation damage impacts CsI(Tl) through the creation of color centers
in the crystals, inducing absorption bands, resulting in a degradation
of response uniformity and light yield.  The nominal dose budget for
the \babar\ CsI(Tl) calorimeter is 10 krad over the lifetime of the
detector.
The dominant contribution to the dose arises from luminosity and
single-beam background sources, and hence is due to \mev photons and
presumably neutrons. The integrated dose does not scale with
integrated beam current, but does scale approximately linearly with
integrated luminosity.

A total dose of about 1.2~krad was received in the most heavily
irradiated regions for an integrated luminosity of 400 fb$^{-1}$,
resulting in a loss of about $\sim 15\%$ of the total light yield, but
with no measurable impact on physics performance.  It is notable that
most of the observed light loss occurred relatively early in \babar\
running, although the radiation dose accumulated relatively steadily,
and that crystals from different manufacturers have responded somewhat
differently to irradiation.


In order for the barrel calorimeter to function in the \superb\
environment, beam background rates must be maintained at a level of
approximately 1~MeV$/\mu$s or less per CsI(Tl) crystal.  If this
condition is achieved, then radiation dose rates are anticipated to be
roughly comparable to current \babar\ levels.  A dose budget of well
under 1 krad/year is expected to be achievable. At this a level, the
CsI(Tl) barrel would survive for the duration of \superb\ operations.
This assumption will, however, need to be verified by detailed
simulation.

\begin{figure}[htbp]
  \begin{center}
  \includegraphics[width=0.5\textwidth]{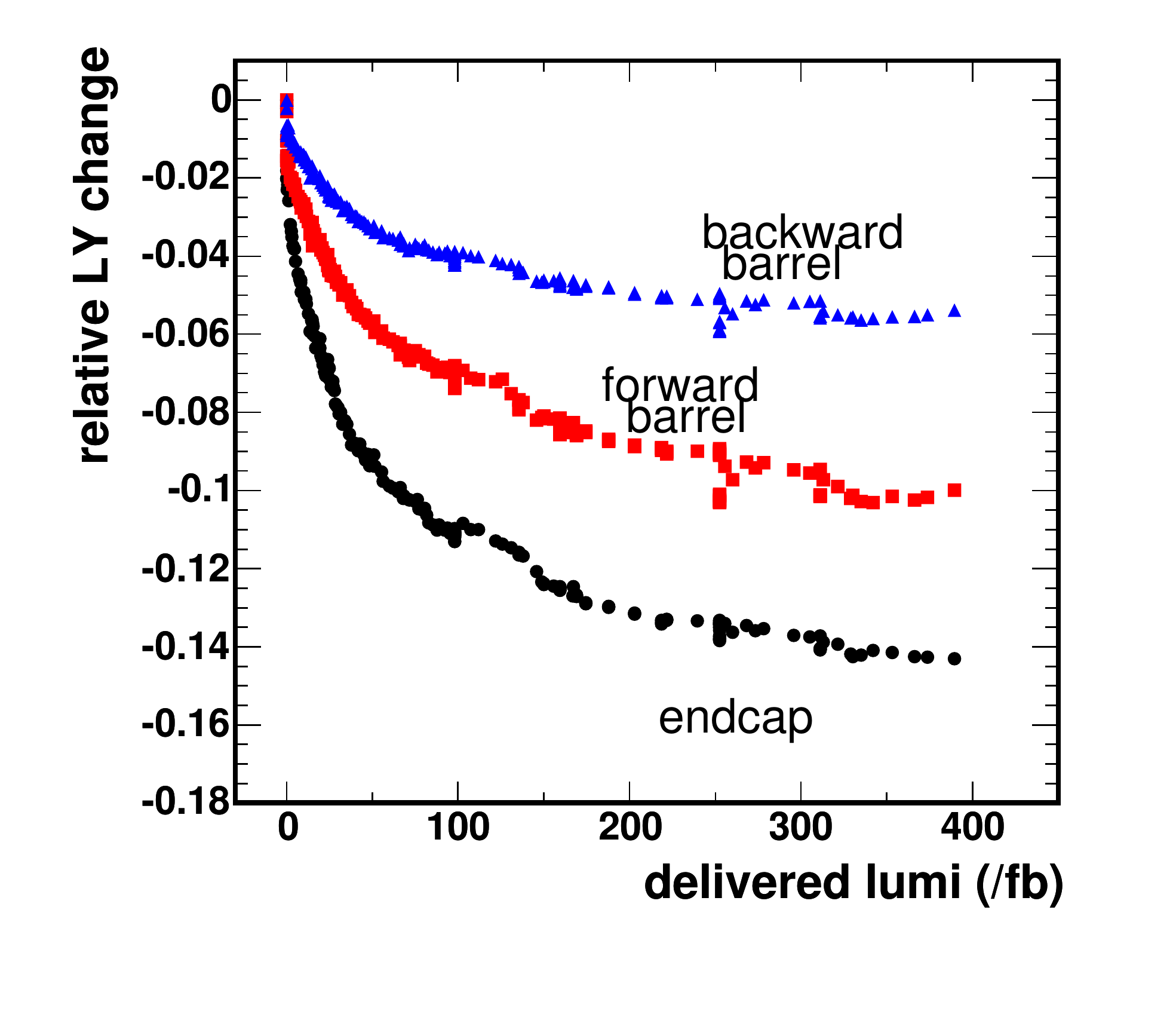}
  \caption{The light yield loss in the \babar\ CsI(Tl) crystals due to
    radiation damage as a function of integrated luminosity. The total
    dose received after 300\invfb is 1.2 krad in the endcap and 750
    rad in the barrel.}
  \label{fig:emc_raddam}
  \end{center}
\end{figure}

\tdrsubsubsec{Expected Changes in Performance at \superb}


The CsI(Tl) crystals used in the barrel calorimeters of both \babar\
and Belle are the most expensive elements of the two detectors. Based
on the performance that has been achieved, and the radiation damage
that has been observed so far, both collaborations have concluded that
the re-use of the barrel crystals is possible at a Super $B$ Factory.

The baseline assumption is that the geometry of the crystals is
unchanged from that of the current \babar\ detector. The one change
that should be made is to move the position of the interaction point
from $-5$ cm to $+5$ cm relative to the position of the crystal gap
normal to the beam axis. This adjustment retains the current
non-pointing geometry, but, in view of the reduced energy asymmetry,
moves the barrel to a slightly more symmetric position.  The effect of
the change in boost from $\gamma\beta=0.56$ to 0.28 and the shift of
the IP is to increase the angular coverage of the barrel from 80\%
($\cos\theta = -0.93$ to $+0.66$) to 84\% ($\cos\theta = -0.87$ to
$+0.81$).

The higher background rates will impact the performance of the barrel
calorimeter. This effect will be mitigated with the electronics
changes described below in Section~\ref{sec:EMCbarrelElec}. FastSim
studies modeling the new electronics and background rates indicate
that the barrel performance will continue to be acceptable.
Figure~\ref{fig:EMC:photon-resolution-SuperB-Fastsim} shows the
expected single-photon energy resolution
in the barrel at \superb, obtained with FastSim.  Little change from
the \babar experience is expected, except at the lowest photon
energies.
Considering generic $B^0\bar B^0$ events at predicted backgrounds, we
obtain the $\gamma\gamma$ mass plots in the $\pi^0$ and $\eta$ regions
in Fig.~\ref{fig:EMC:barrelGG}.  There is a slight shift in the mass
from the expected peak positions; the FastSim absolute calibration is
undergoing further tuning. The $\gamma\gamma$ mass resolution in the
MC could be somewhat underestimated, partly due to inaccurate modeling
of the angular resolution compared with data, which slightly affects
the $\gamma\gamma$ mass resolution here.  It is clear, however, that
the effect of the changes at \superb is not large.

\begin{figure}[hbtp]
  \begin{center}
      \includegraphics[width=7.0cm]{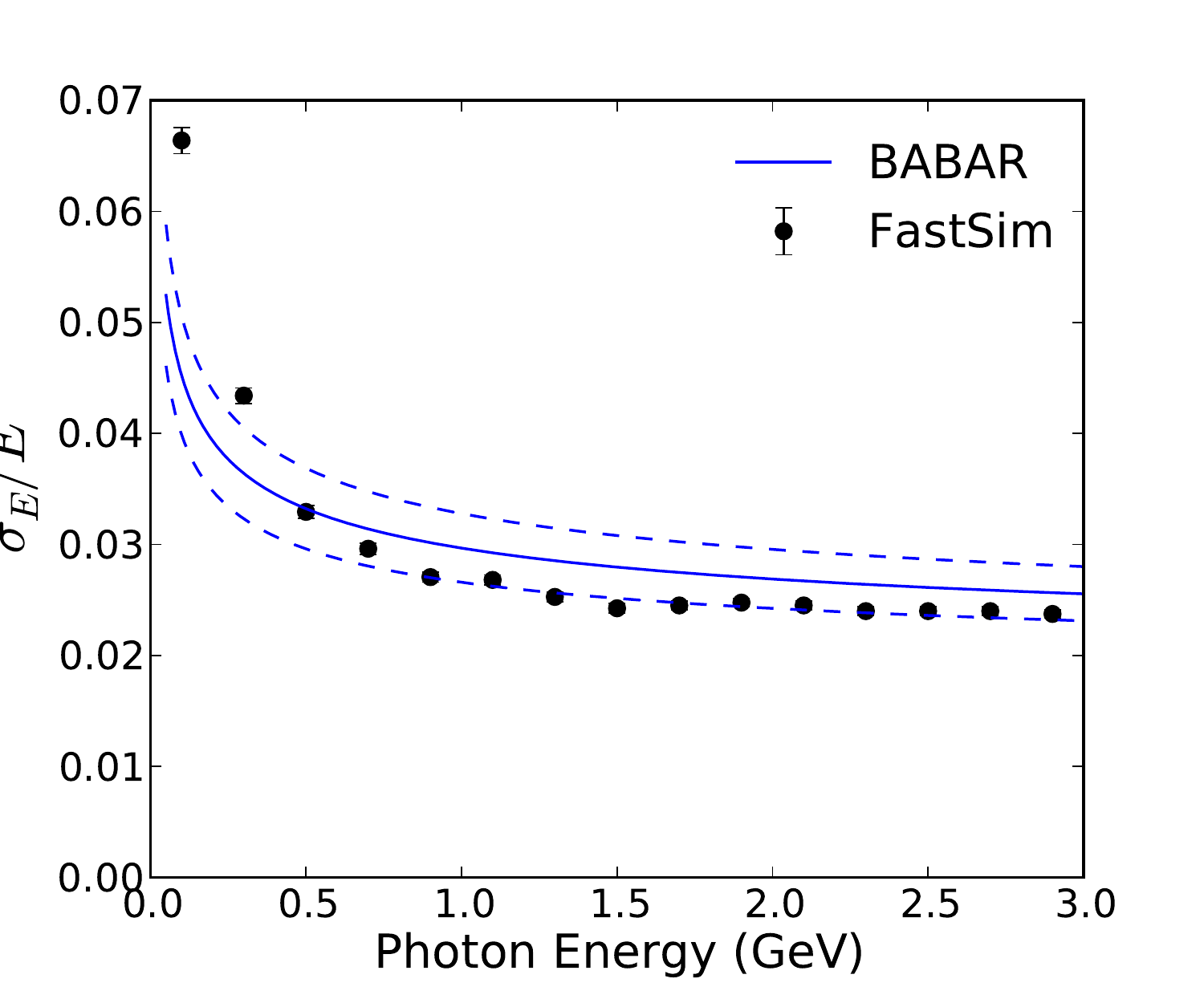}
     \caption{FastSim \superb single photon energy resolution in the barrel at expected background. The curves
show the \babar resolution, with error bands, according to Eq.~\protect{\ref{eq:EMC:babarNIMEres}}}
     \label{fig:EMC:photon-resolution-SuperB-Fastsim}
  \end{center}
\end{figure}

\begin{figure}[hbtp]
  \begin{center}
      \includegraphics[width=0.5\textwidth]{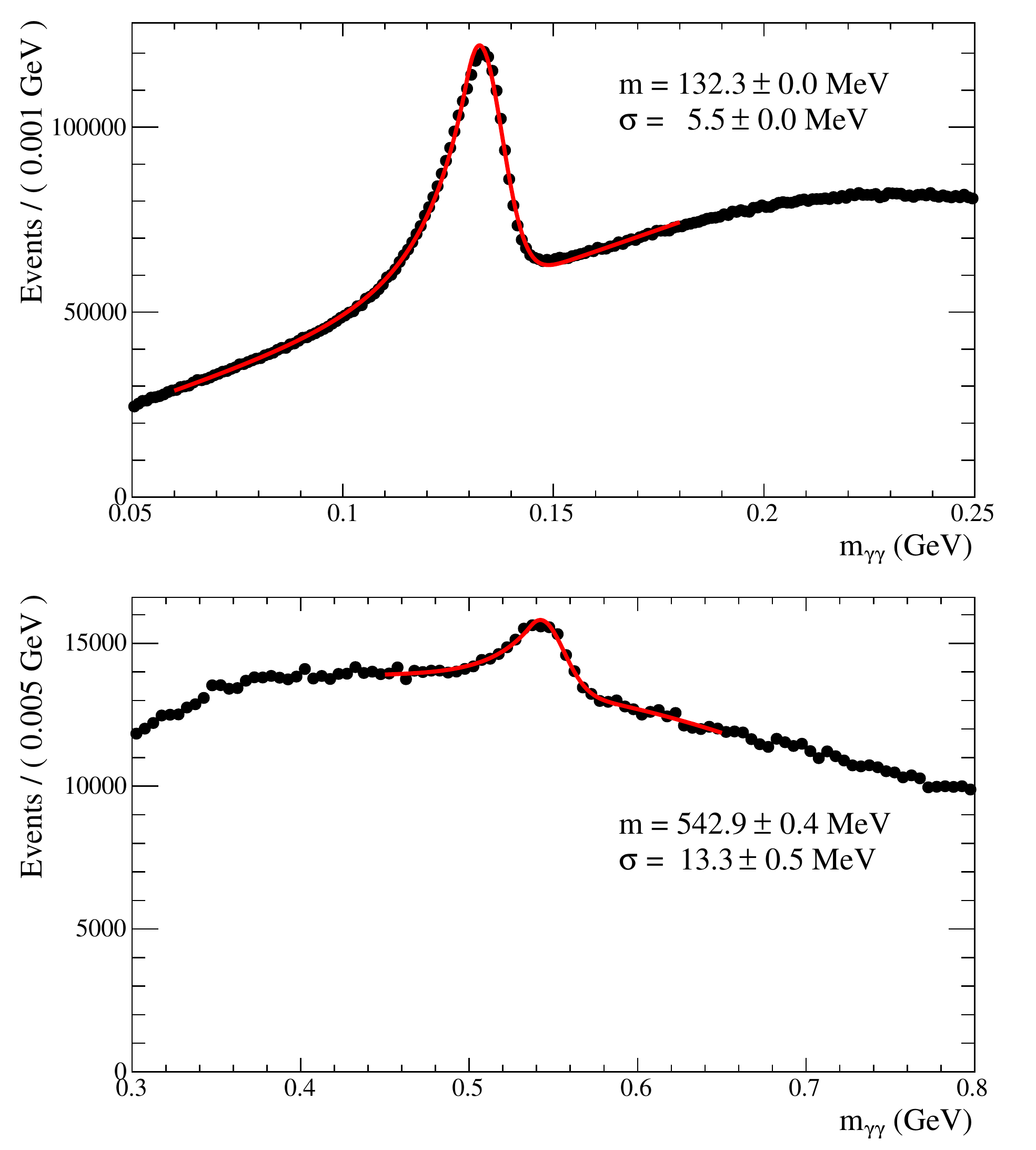}
     \caption{FastSim \superb $\gamma\gamma$ mass spectra for generic $\B^0\bar B^0$ events in the $\piz$ and $\eta$ meson mass regions,  with both photons in the barrel, at expected background rates.}
     \label{fig:EMC:barrelGG}
  \end{center}
\end{figure}


A possible change to improve the coverage in the backward region would
be to add one or two additional rings of crystals to the last module
ring in $\theta$, which currently only contains 6 rings of crystals.
This would require, however, major changes in the mechanical support
structure and a redesign of the electronics readout, so it will not be
undertaken unless there is a significant gain from the extra ring(s).
Changes to the rear section of the barrel also interact strongly with
the possible addition of a backward endcap calorimeter
(Sec.~\ref{sec:EMCback}).


\tdrsubsec{Electronics changes}
\label{sec:EMCbarrelElec}

The \babar\  EMC barrel crystals are coupled to two PIN  diodes (Hamamatsu S2744-08) glued to the rear face of each crystal.
The possibility of replacing these photo-detectors with new devices was investigated but discarded, due to the strong mechanical constraints and the difficulty of removing the PIN photodiodes from the crystals without the risk of damage.
Each PIN diode is read with a separate electronic chain composed of a charge-sensitive preamplifier and a CR-RC-RC shaper, with \hbox{800 ns$-250$ ns$-250$ ns} shaping times. To achieve the required dynamic range, the output of each channel is amplified in two chains, having gains of 1 and 32, respectively.

\tdrsubsubsec{Requirements}

The increased background radiation in the \hbox{\superb} environment calls for a shorter integration time, at the price of a reduction of the integrated light, and a corresponding  worsening of the electronics signal-to-noise ratio. The signal shaping time must also be reduced to a few hundred nanoseconds; this increases the bandwidth of the shaper and consequently the electronic noise. Finally, the need to have good timing response for use in the EMC trigger led us to replace the charge sensitive preamplifier with a transimpedence amplifier with a low feedback capacitance.

Achieving the desired signal-to-noise performance with this design requires the use of low noise components.
The new Front End Boards will also have increased power consumption, due to the larger bandwidth operational amplifiers and the use of off-the-shelf components. 

The final design choices will result from a compromise between rate capability and electronic noise performance.

\tdrsubsubsec{Electronic noise measurements}

A test stand has been setup in the Electronics Laboratory (LABE) of
the INFN in Rome in order to study the electronic noise of the FEE of
the Barrel EMC.  The aim of this test is to study the effect of short
integration times, of the order of 100 ns, by measuring the noise
level for different FEE configurations.  To this end, a CsI(Tl)
crystal from the \babar\ calorimeter, with dimensions of $\sim 6\times
6\times 30$ cm$^3$, has been equipped with the photosensor under test
on one side, and with an EMI9814B PMT on the opposite side to provide
a trigger.  Two photosensors have been tested: standard PIN diodes
from \babar\ with sensitive area $1\times 2$ cm$^2$ (only one of the
pair is read out), operated at 60 V, and the avalanche photodiodes
(APD) used in the CMS calorimeter (Hamamatsu S8664), with a sensitive
area $0.5\times 0.5$ cm$^2$, operated at two different voltages, 340
and 380~V.

The signal is provided by a $^{60}$Co radioactive source
having two well-defined $\gamma$ lines of 1.17 and 1.33 \mev.
The photosensor is read out with a charge-sensitive preamplifier
(CSP) followed by a Gaussian shaper.
Both the photosensor and the PMT signals are viewed with a LeCroy 12 bit
digital oscilloscope.
The waveforms have been recorded at a sampling rate of 20 GSamples/s; 
50000 events were collected for each configuration studied.
For each event, the maximum amplitude of the photosensor signal after the
trigger time is found; the region before the trigger is used
to evaluate the baseline.

The CSP is the Hamamatsu H4083 with 100~ns integration time, 
in combination with a Gaussian shaper with shaping times
ranging from 100 to 500~ns.

A prototype amplifier (LNA02V0) developed at LABE has also been
tested.  This prototype is very similar to the Hamamatsu CSP, but has
slower output signal rise time (400 {\it vs.} 300 ns), and narrower
bandwidth.

As a benchmark, a CSP with very large integration time, 140 $\mu$s, similar
to that used in the \babar\ readout chain, has also been used.   
Finally a test of a PIN diode with the original \babar\ FEE has also been
performed, to check the performance of our experimental setup.

\begin{figure}[htbp]
\begin{center}
\includegraphics[clip=true,trim = 35 50 55 90,width=0.5\textwidth]{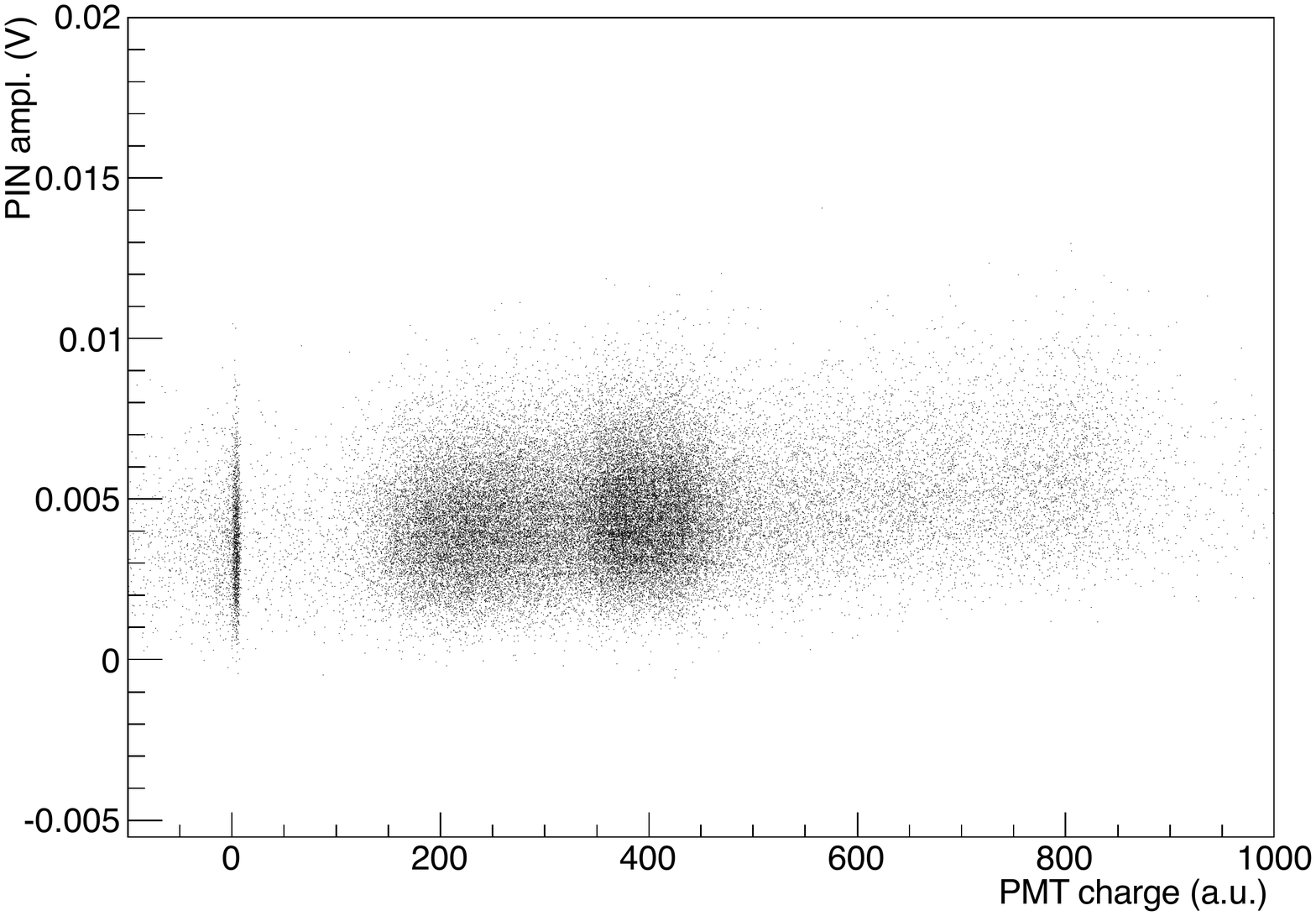}
\caption{APD maximum amplitude vs PMT response.\label{fig:EMC:BarElecScatter}
}
\end{center}
\end{figure}

The two $\gamma$ lines of $^{60}$Co are clearly visible in the charge spectrum of the PMT, as
is the small sum peak at 2.5\mev.
The basic idea is to use the two $\gamma$ lines and their sum to calibrate
the energy response of the photosensor, and to then use this calibration to
evaluate the equivalent electronic noise (in \mev) starting from the
absolute noise (expressed in mV).

The scatter plot of Fig.~\ref{fig:EMC:BarElecScatter}, shows an
example of the APD maximum amplitude as a function of PMT charge.
This scatter plot is divided into six regions: the pedestal around
$Q_{PMT} = 0$, the crystal background, the two $\gamma$ lines, the
tail of the crystal background and the region of the sum of the two
$\gamma$s.  The calibration of the energy scale is obtained from a
linear fit of the peaks of the distributions as function of the photon
energy (Fig.~\ref{fig:EMC:BarElecCalib}).  The dependence of the
widths of the same distributions on photon energy is also shown in
Fig.~\ref{fig:EMC:BarElecCalib}; as an estimate of the electronic
noise, the average of the first four points is taken, and the
equivalent noise is obtained by dividing by the calibration factor.
For the example shown:
\begin{equation}
 \frac{1.5 {\ \rm mV}}{0.8 {\ \rm mV/MeV}} = 1.9 {\ \rm MeV}
\end{equation}

The same analysis has been repeated for all the configurations studied .
Table~\ref{tab:EMC:noise} shows the absolute noise level.
The shortening of the shaping time increases the noise level; therefore in
the study of the PIN, only 500 ns shaping is considered.

\begin{table*}[tbh]
\caption{Absolute noise (mV) and equivalent noise (\mev) for several electronic solutions}
\vskip10pt
\begin{center}
\begin{tabular}{|lccc|}
\hline
CSP - integration time - shaping time & $V_{APD} = 340$ V & $V_{APD} = 380$ V & PIN diode \\
\hline\hline
Cremat - 140 $\mu$s - 500 ns & 2.4/1.3 & 3.3/0.62 & 1.5/0.56 \\
Hamamatsu - 100 ns - 500 ns & 2.9/1.3 & 4.5/0.85 & 2.5/2.5 \\
Hamamatsu - 100 ns - 250 ns & 5.1/3.4 & 6.0/1.3 & - \\
Hamamatsu - 100 ns - 100 ns & 5.3/4.4 & 6.0/1.5 & - \\
LNA02V0 - 100 ns - 500 ns & - & 3.0/0.57 & 1.5/1.9 \\
\hline
\babar\ FEE & - & - & 0.7/0.78 \\
\hline
\end{tabular}
\end{center}
\label{tab:EMC:noise}
\end{table*}

The energy calibration factors are related to the overall gain of the sensor/FEE chain, and 
the shortening of the shaping time corresponds
to a decrease of the effective gain.
Note also that the gain of the APDs at 380 V is about
three times the gain at 340 V, while the absolute noise
is only slightly higher.    


Table~\ref{tab:EMC:noise} shows the equivalent noise for each chain.
In this test, only one PIN of the two paired together is read out. If
both PINs were used, the signal would double, while the electronic
noise would increase by a factor of $\sqrt{2}$,improving the
equivalent noise performance by the same factor.  In this case, the
\babar\ FEE noise would go from 0.78 to 0.55 \mev; the Hamamatsu noise
from 2.5 to 1.8 \mev; and the prototype LNA02V0 noise from 1.9 to 1.3
\mev.

The impact of the reduction in integration and shaping time on noise
performance is significant. The search for further options and
optimization of the system including all effects is therefore still
ongoing.

\begin{figure}[tbh]
\begin{center}
\includegraphics[clip=true,trim = 5 5 10 5,width=0.5\textwidth]{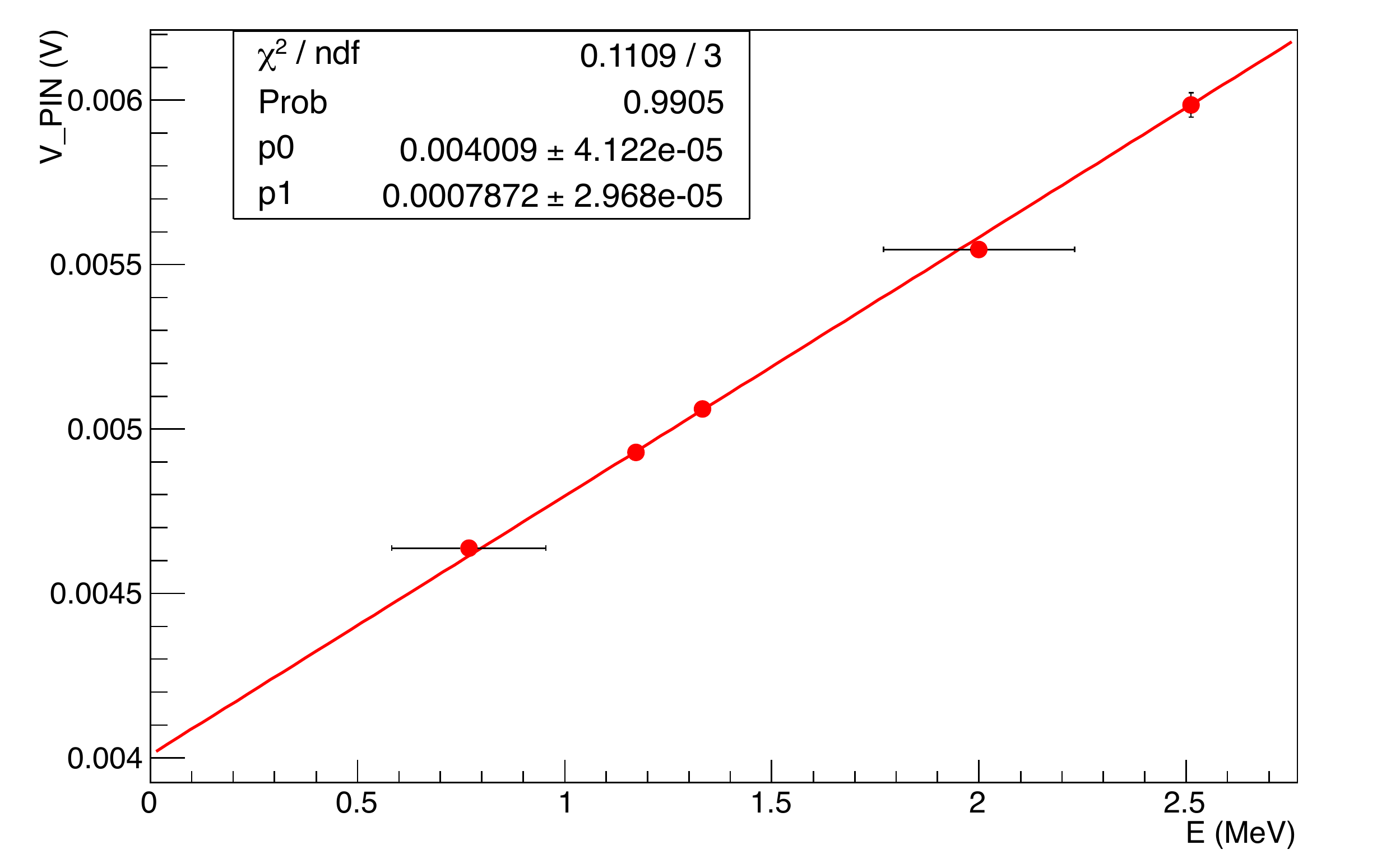}
\includegraphics[clip=true,trim = 20 50 70 140,width=0.5\textwidth]{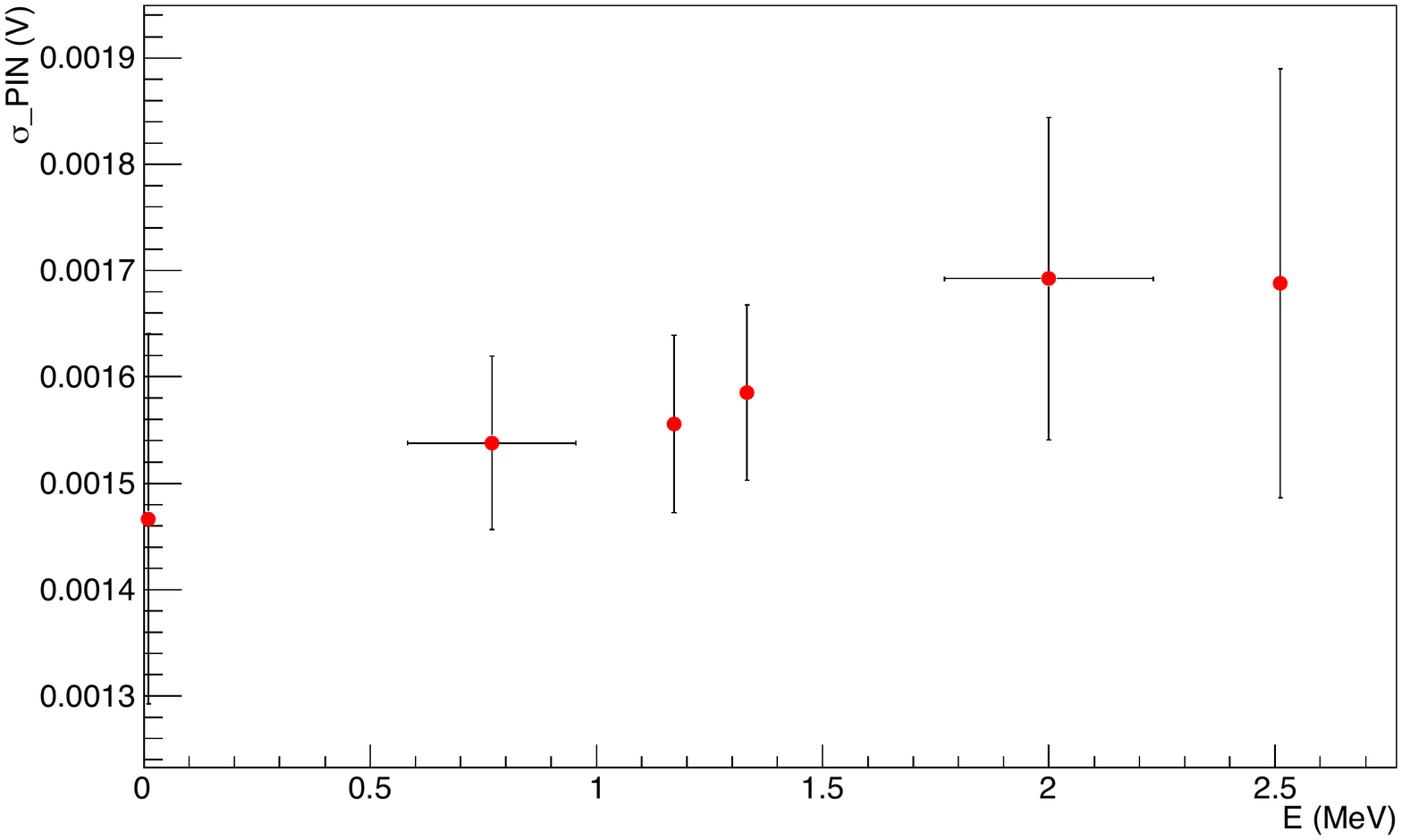}
\caption{(a) Calibration plot for the electronic noise measurements: 
average APD response vs energy measured by the PMT. 
(b) Resolution as a function of the deposited energy.
\label{fig:EMC:BarElecCalib}
}
\end{center}
\end{figure}

\tdrsubsubsec{Readout design} The readout of each PIN diode is done
using two separate channels. This choice is motivated by the need for
redundancy and to minimize the possibility of dead crystal channels in
the event of a PIN diode failure.  In order to minimize the noise of
the Front End Boards, we provide two outputs for each channel, one for
the low energy range and the other for the high energy range, with a
gain of 1 and 32 as in \babar.
The four signals for each crystal (two channels with two different gains) are combined in the digitizer board; normally the mean signal from two PIN diodes is used, but in case of failure of one PIN diode, the signal from the remaining diode can be selected.

\tdrsubsec{Disassembly at SLAC and Transport to Italy}

The \babar\ barrel calorimeter was assembled at SLAC and then inserted
and supported in the magnet flux return structure.  The EMC design did
not, therefore take into consideration the requirements of the
long-haul transportation that will be required to move it to Cabibbo
Lab.  We have studied two scenarios for shipment,
electronics refurbishment and final installation:
\begin{enumerate}
\item[a)] Ship the barrel calorimeter as a unit, in one piece, and then do the refurbishment of the front end electronics 
 in a suitable environment as close as possible to the
 \superb\ interaction area.
\item[b)] Disassemble the calorimeter at SLAC, shipping the 280 individual modules and the supporting 
structures for refurbishment in Naples, followed by transport of modules and
structures to the \superb\ assembly hall at Cabibbo Lab.
\end{enumerate}

Each option has pros and cons, which have been carefully evaluated,
leading to the decision to adopt scenario b) as the baseline.  To
evaluate the risks associated with the transport of the complete
barrel EMC, the original structural analysis calculations would have
to be updated, and some critical information, such as the current
condition of glued parts and internal structures, is difficult, and in
some cases impossible, to ascertain.  Moreover, design and
construction of a complex and expensive support structure able to damp
motions occasioned by an intercontinental air shipment, with
subsequent ground transportation, would be required. Air shipment of
the 30 ton barrel structure, with a volume of the order of $4 \times 4
\time 5$ cubic meters, requires the use of large cargo planes. The
cost of such a shipment has not been evaluated.

Shipping is simpler and safer in scenario b), at the cost of complex
disassembly and reassembly efforts.  Some disassembly is still
necessary in scenario a) in order to reach the front end electronics.
In scenario b), refurbishing work is much easier, allowing repairs
such as re-gluing of loose PIN diodes, that are difficult if the
modules are mounted in the cylindrical support structure. Last, but not
least, costs and availability of space resources and manpower have
been compared, tending in favor of the chosen baseline.

A well-coordinated effort to disassemble the barrel EMC is underway at
SLAC.  Most of of the tooling, from jigs and supporting structure to
critical tools like the robotic arm for module handling
(Fig.~\ref{fig:EMC:Barrelphotos}), has been located, and parts are
being tested and refurbished.  Procedures and schedules are
well-advanced, with the goal of readiness for shipment in summer 2014.

\begin{figure*}[tbh]
\begin{center}
\includegraphics[clip=true,width=0.45\textwidth]{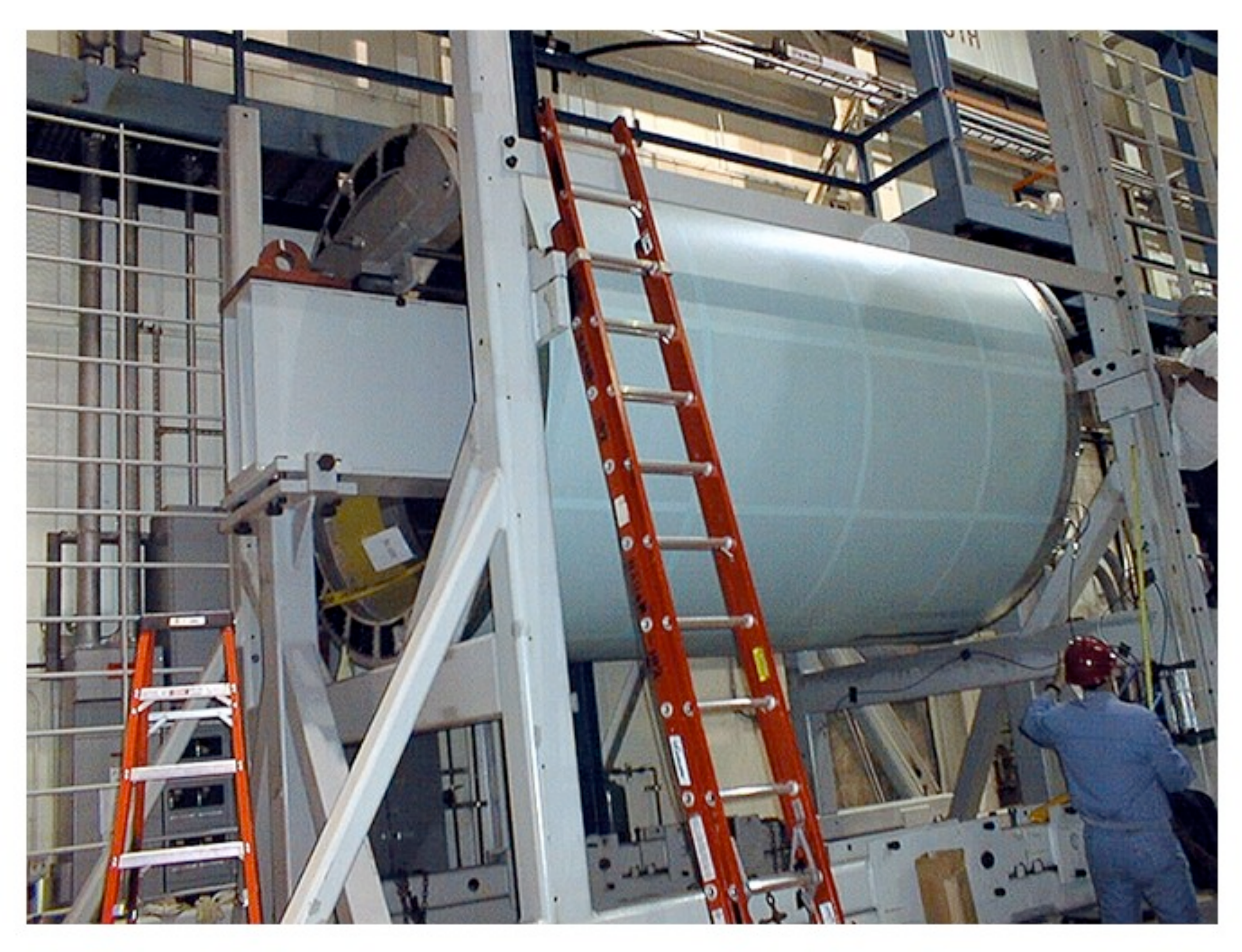}
\includegraphics[clip=true,width=0.45\textwidth]{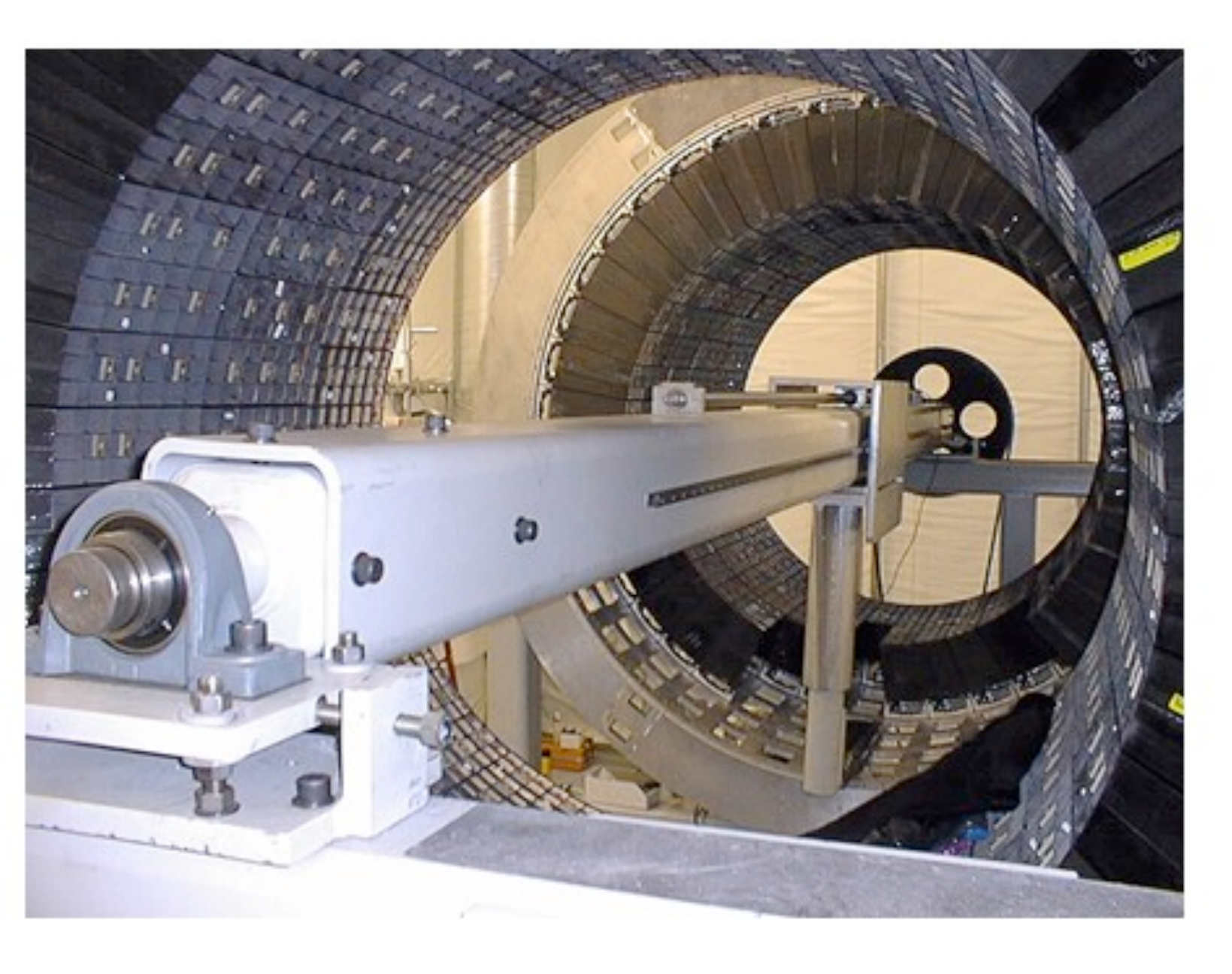}
\caption{Examples of tooling equipment for the EMC Barrel.}
\label{fig:EMC:Barrelphotos}
\end{center}
\end{figure*}

\tdrsubsec{Electronics refurbishment}

The guiding principle in the upgrade of the barrel electronics is to
retain the mechanics of the electronics components of \babar, using
the same form factor for the revised electronics boards so that they
fit into the existing mechanical structures.

The front end preamplifiers are located at the rear of each crystal,
inside a brass shield. The shield cages are connected to the crystal
photodiode assembly, which is very difficult to unglue and
disassemble. The brass cages themselves are robust, and can be
dismounted and reused without risk of damage.

With the intent to reuse this brass housing, we will design the
preamplifier form factor to fit, retaining as well the form factor of
the connector and cable.

The barrel digital electronics in \babar\ is housed inside 48 small
custom-built crates (minicrates) located on either side of the ends of
the cylindrical structure. Each minicrate contains a backplane that
divides the connection board from the digitizer boards and a fiber
optic signal transmission board.

We plan to reuse the mechanical structure of the minicrates, retaining
the same form factors for the four replacement electronics boards
(connection boards, backplane, digitizer board, and fiber optic
transmission board).


\tdrsubsec{Shipment, Refurbishing and Reassembly}

The CsI(Tl) crystals are sensitive to humidity, and the PIN sensors
glue joint can be broken by temperature excursions.  Tests in \babar\
have shown that temperature variations must be limited to $\pm
5^\circ$ C.  Accordingly, the module shipping containers must be
provide controlled temperature, a dry atmosphere, and protection
against mechanical and thermal shocks.  Shipment by air is foreseen in
order to shorten travel times and consequently decrease risks.

In due time, decisions on shipment and storage location of the rest of
the EMC mechanical components (such as the main support cylinders,
jigs, and tooling)  will be taken.  To ensure controlled
ambient conditions during module refurbishment in Naples, at the end
of 2013 part of the dedicated hall will be enclosed in a protective
tent ($10 \times 10$ m$^2$ area), whose temperature and humidity will
be actively controlled by a new air conditioning system that will work
in parallel with the already existing plant, as insurance against
failures. In this tent, the front end electronics will be refurbished
by teams working in parallel. Completed modules will be individually
tested with radioactive sources, light pulsers through optical fibers
reaching each crystal, and cosmic ray data taken inside an existing
telescope with tracking resolution in the millimeter range.  Queued
and completed modules will be stored in the same tent on layered
scaffolding.

Estimated refurbishing and test time, based on \babar\ experience, is
of the order of one year, placing the end of refurbishing work in
Naples approximately a year after the new preamps are available.
Depending on the status of the Tor Vergata site, shipment and the
remounting phase will start in 2016, or perhaps in 2015, using modules
already completed.  At Tor Vergata, in the construction hall close to
the interaction region, again in a air conditioned tent, modules will
be inserted into the steel support cylinder, cables and readout
electronics will be added, and calibration systems (light pulses on an
optical fiber distribution system, fluorine activation and
distribution system) will be implemented and tested during the
commissioning phase before detector assembly.  



\tdrsec{Forward Calorimeter}

The Forward Calorimeter extends the coverage of the electromagnetic
calorimeter to low forward polar angles, as detailed in
Tab.~\ref{table:EMC:superb}. To be effective, its performance must
therefore be comparable with the barrel calorimeter, despite the
higher rates seen in the forward direction. Hence the design choice is
a crystal calorimeter read out by compact photodetectors capable of
operating in magnetic field.

Taking as a performance objective the \babar\ calorimeter, the new
endcap should have a relative energy resolution of at most 4.3\% at
100~\mev and 2.7\% at 1~\gev. In order to retain appropriate
resolution on the $\pi^0$ invariant mass, and to allow 
$\pi^0\to\gamma\gamma$ reconstruction up to sufficiently high
energies, segmentation at least comparable to that of \babar\ is
needed. Since the transverse crystal size is dictated by the Moli\`ere
radius of the material, only crystals with a Moli\`ere radius no
larger than CsI(Tl) can be considered.  Finally hermeticity is also
important, so the requirement on mechanics is that the fraction of
particles originating from the interaction point passing through the
cracks between the crystals be minimal, comparable with the barrel.

As already described for the barrel calorimeter, the most stringent
constraints arise from the presence of large backgrounds due to the
extremely high luminosity.  As shown in Fig.~\ref{fig:EMC:dose}, the
expected radiation dose integrated in a year ranges from $\sim 250$
rad for the outermost rings to $\sim 320$ rad for the innermost ones.
Consequently, the dose rate the crystals must tolerate is about $\sim
0.1$ rad/h.
\begin{figure*}
\begin{center}
\includegraphics[clip=true,width=0.7\textwidth]{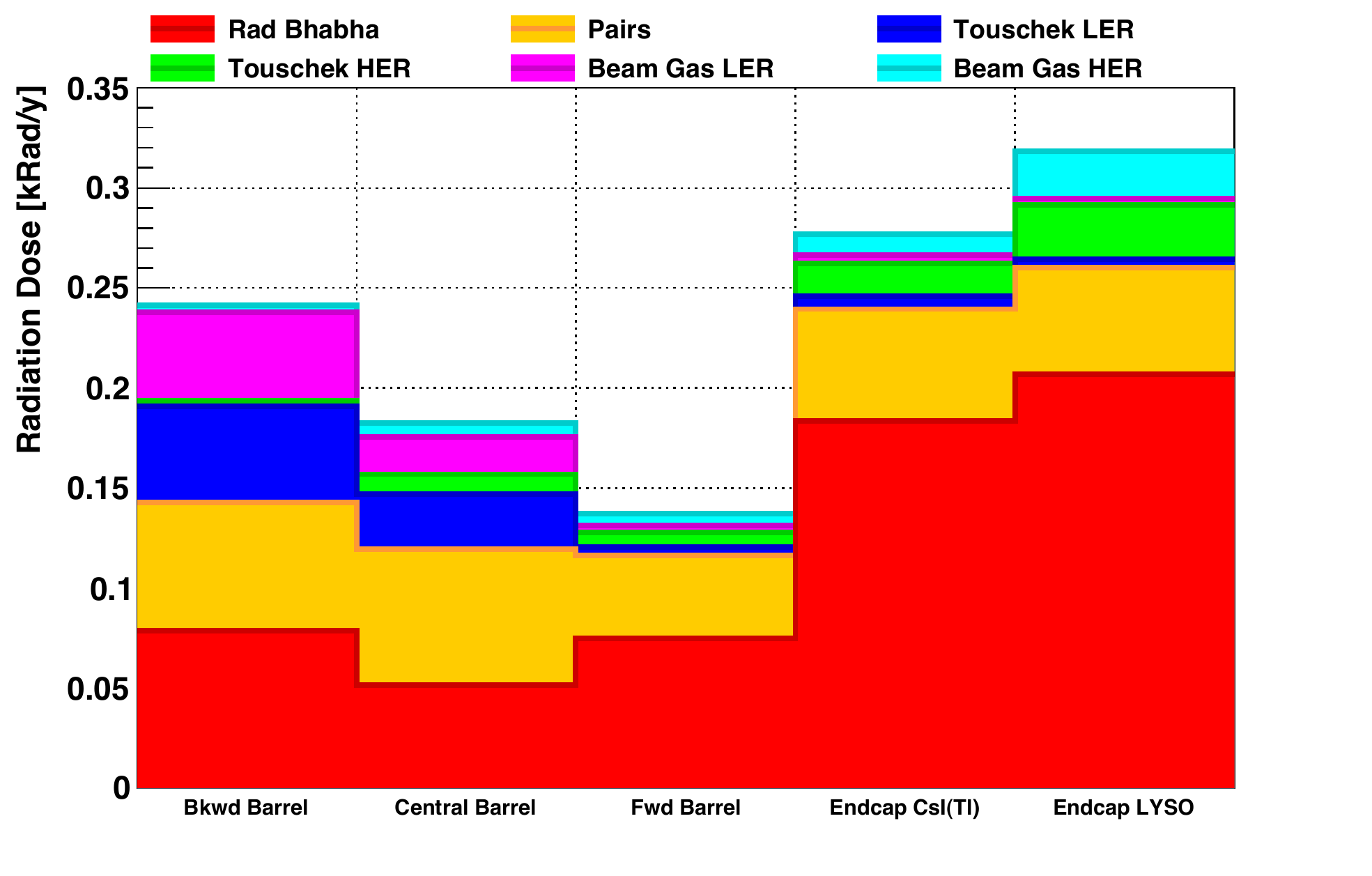}
\end{center}
\caption{Expected radiation dose per year as a function of the calorimeter region.}
\label{fig:EMC:dose}
\end{figure*}

As described in Sec.~\ref{sec:EMC:OVERbkg}, the large rate of low
energy photons can also damage the crystals themselves, reducing the
longitudinal transmission, and thus the light yield. The \babar\ experience is
that this is worse in the endcap than in the barrel (Fig.~\ref{fig:emc_raddam}), and this is 
expected to be true in \superb\ as well.  The large rate in the forward direction can degrade
the energy resolution by causing signal pile-up. This is a serious concern, as shown in 
Fig.~\ref{fig:EMC:EresCsI5x}, and in particular, the performance of the
\babar\ CsI(Tl) endcap is not adequate. The chosen crystal
must therefore have a stable light yield in the expected radiation
environment and the signal shape produced by the readout must be
compatible with the expected rates.

\begin{figure*}
\begin{center}
\includegraphics[clip=true,width=0.8\textwidth]{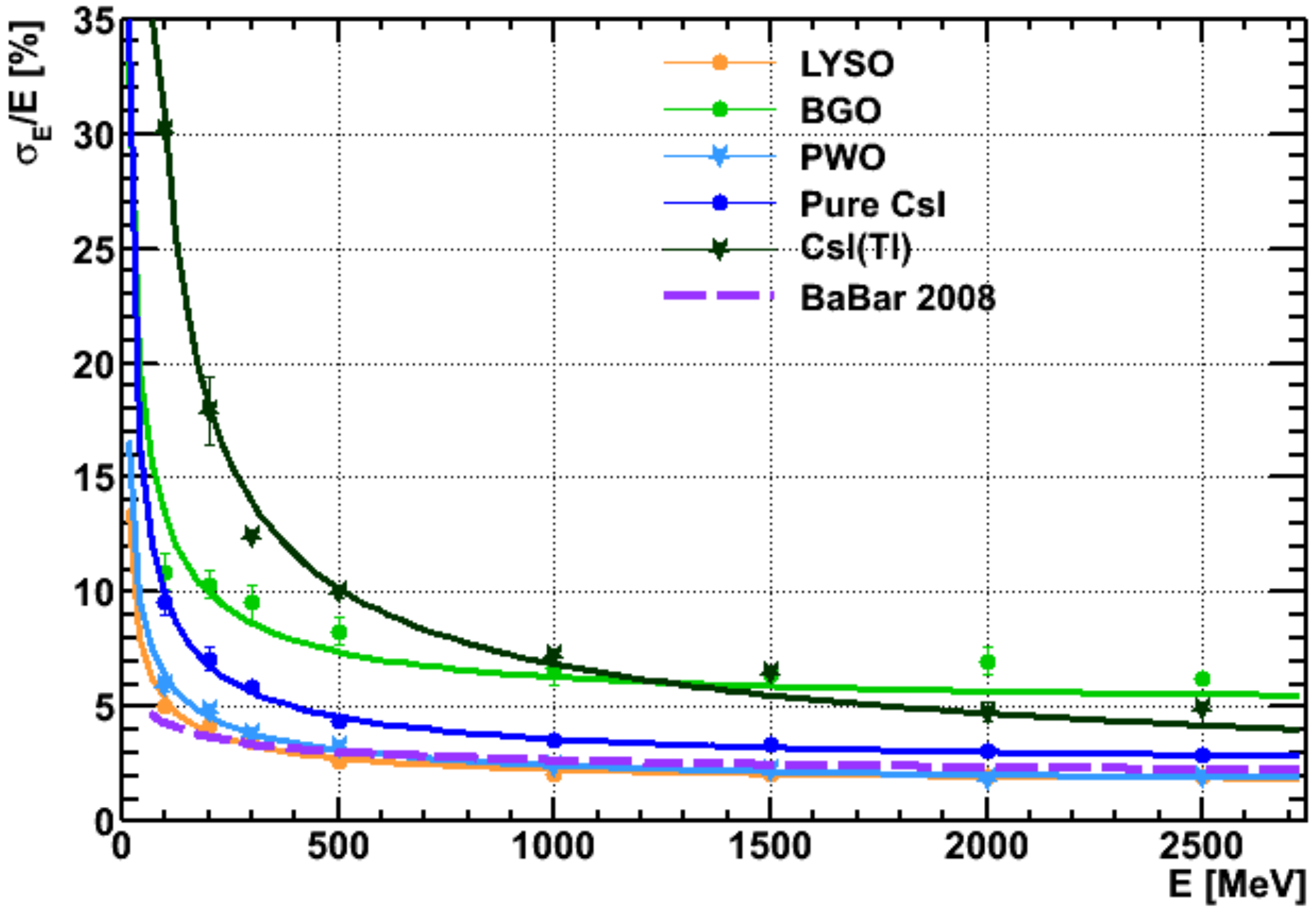}
\end{center}
\caption{Comparison of resolution in the forward calorimeter for several different 
crystals, at 5 times nominal background.}
\label{fig:EMC:EresCsI5x}
\end{figure*}

We have investigated several combinations of crystals and electronics,
as described below.
The option yielding the best performance is crystals made of LYSO,
with readout by avalanche photodiodes (APD), in a new low-mass
mounting structure. This configuration has been studied in detail.
However, budget restrictions force us to adopt as our baseline a
``hybrid'' scheme, making use of LYSO at the smallest radii, while
retaining the \babar\ CsI(Tl) crystals in the three outermost endcap
layers. This results in a significant reduction in cost, while
maintaining the benefits of LYSO where it is needed most, where the
rate and radiation issues are most severe. As part of this cost
compromise, the structural support of the \babar\ endcap will also be
reused, rather than adopting a lower mass structure that would improve
performance.

\def\Cherenkov{\v{C}herenkov }
\def\Cherenkovn{\v{C}herenkov}
\def\deg{$^\circ$C}                   %
\def\ea{{\it et al.}}
\def\BAF{BaF$_2$ }
\def\BAFN{BaF$_2$}
\def\CAF{CaF$_2$ }
\def\CEF{CeF$_3$ }
\def\CEFN{CeF$_3$}
\def\CSI{CsI}
\def\PBF{PbF$_2$ }
\def\PBFN{PbF$_2$}
\def\PBW{PbWO$_4$ }
\def\PBWN{PbWO$_4$}
\def\LSO{Lu$_2$SiO$_5$}
\def\LYSO{Lu$_{2(1-x)}$Y$_{2x}$SiO$_5$}
\def\GSO{Gd$_2$SiO$_5$}
\renewcommand{\arraystretch}{1.1}
\renewcommand{\topfraction}{1.}
\renewcommand{\bottomfraction}{1.}
\renewcommand{\textfraction}{0.}
\renewcommand{\floatpagefraction}{1.}
\renewcommand{\textfloatsep}{12pt}
\renewcommand{\floatsep}{12pt}
\renewcommand{\intextsep}{12pt}
\doublerulesep -1pt

\subsection{LYSO Crystals}
\label{sec:lyso_intro}

Over the last two decades, cerium-doped lutetium oxyorthosilicate
(\LSO\ or LSO)~\cite{bib:lso} and lutetium yttrium oxyorthosilicate
(Lu$_{2(1-x)}$Y$_{2x}$SiO$_5$ or LYSO)~\cite{bib:lyso1,bib:lyso2} have
been developed for the petroleum and medical industries; mass
production capabilities are firmly established. This section addresses
the crystal properties, specifications, production and testing.


\begin{table*}[tbp]
\renewcommand{\arraystretch}{1.2}
\caption{Properties of Heavy Crystal with Mass Production Capability}
\label{tab:crystals}
\begin{center}
\begin{tabular}{cccccccc}
\hline 
Crystal                 &NaI(Tl)&CsI(Tl)& CsI &BGO  &\PBWN  & LSO/LYSO(Ce)\\
\hline 
Density (g/cm$^{3}$)    &3.67   &4.51   &4.51   &7.13 &8.3    &7.40 \\
Melting Point (\deg C)  &651    &621    &621   &1050 & 1123  &2050 \\
Radiation Length (cm)   &2.59   &1.86   &1.86   &1.12 & 0.89  &1.14 \\
Moli\`{e}re Radius (cm) &4.13   &3.57   &3.57   &2.23 &2.00   &2.07 \\
Interaction Length (cm) &42.9   &39.3   &39.3   &22.7 &20.7   &20.9 \\
Refractive Index$^a$    &1.85   &1.79   &1.95   &2.15 &2.20   &1.82 \\
Hygroscopicity          &Yes    &Slight &Slight &No   &No     & No  \\
Luminescence$^b$ (nm)   &410    &560    &420    &480  &425    & 420 \\[-3pt]
(at Peak)               &       &       &310    &     &420    &     \\
Decay Time$^b$ (ns)     &245    &1220   &30     &300  &30     & 40  \\[-3pt]
                        &       &       &6      &     &10     &     \\
Light Yield$^{b,c}$     &100    &165    &3.6    &21   &0.30   & 85  \\[-3pt]
                        &       &       &1.1    &     &0.077  &     \\
d(LY)/dT$^{b,d}$ (\%/\deg C)&$-0.2$ &0.4  &$-1.4$   &$-0.9$ &$-2.5$   &$-0.2$ \\
                        &       &       &       &     &       &     \\
Experiment              &Crystal& CLEO  & KTeV & L3  & CMS   &{\bf Mu2e} \\
                        & Ball  & \babar & E787  &BELLE& ALICE &{\bf{Super}$\boldsymbol{B}$}\\
                        &       &BELLE  &      &     &PrimEx &{\bf HL-LHC?}\\
                        &       &BES III&       &     &{\bf Panda}  &  \\  
\hline
\multicolumn{8}{l}{a At the wavelength of the emission maximum.}\\[-3pt]
\multicolumn{8}{l}{b Top line: slow component, bottom line: fast component.}\\[-3pt]
\multicolumn{8}{l}{c Relative light yield of samples of 1.5 X$_0$ and with the PMT quantum efficiency taken out.}\\[-3pt]
\multicolumn{8}{l}{d At room temperature.}\\
\end{tabular}
\end{center}
\end{table*}
 
Table~\ref{tab:crystals}~\cite{bib:zhu_crystals} lists basic
properties of several heavy scintillating crystals: NaI(Tl), CsI(Tl),
pure CsI, bismuth germanate (Bi$_4$Ge$_3$O$_{12}$\ or BGO), lead
tungstate (\PBW or PWO) and LSO/LYSO. All have either been used in, or
are actively being pursued for, high energy and nuclear physics
experiments, which are also listed in the table. The experiment names
in bold refer to future crystal calorimeters. NaI(Tl), CsI(Tl), BGO,
LSO and LYSO crystals are also widely used in the medical and oil
industries. Mass production capabilities exist for all these crystals.

Because of their high stopping power, high light yield, fast decay
time, small temperature coefficient and excellent radiation hardness,
LSO and LYSO crystals have attracted broad interest for new intensity
frontier projects in high energy physics
(HEP)~\cite{bib:superB1,bib:superB2,bib:kloe,bib:mu2e,bib:zhu_cms}.
They are the best choice for the \superb\ forward calorimeter, either
the full uncompromised version or the adopted hybrid design.  Full
size LSO and LYSO crystals from the following vendors have been tested
during the R\&D phase of \superb: CTI Molecular Imaging (CTI), Crystal
Photonics, Inc. (CPI), Saint-Gobain (SG), Sichuan Institute of
Piezoelectric and Acousto-optic Technology (SIPAT) and Shanghai
Institute of Ceramics (SIC). Our development efforts over the past
several years have served to effect a substantial reduction in the
price of large LYSO crystals.

\label{sec:o&s}

\begin{figure*}[htbp]
\centering
\includegraphics[width=0.44\textwidth]{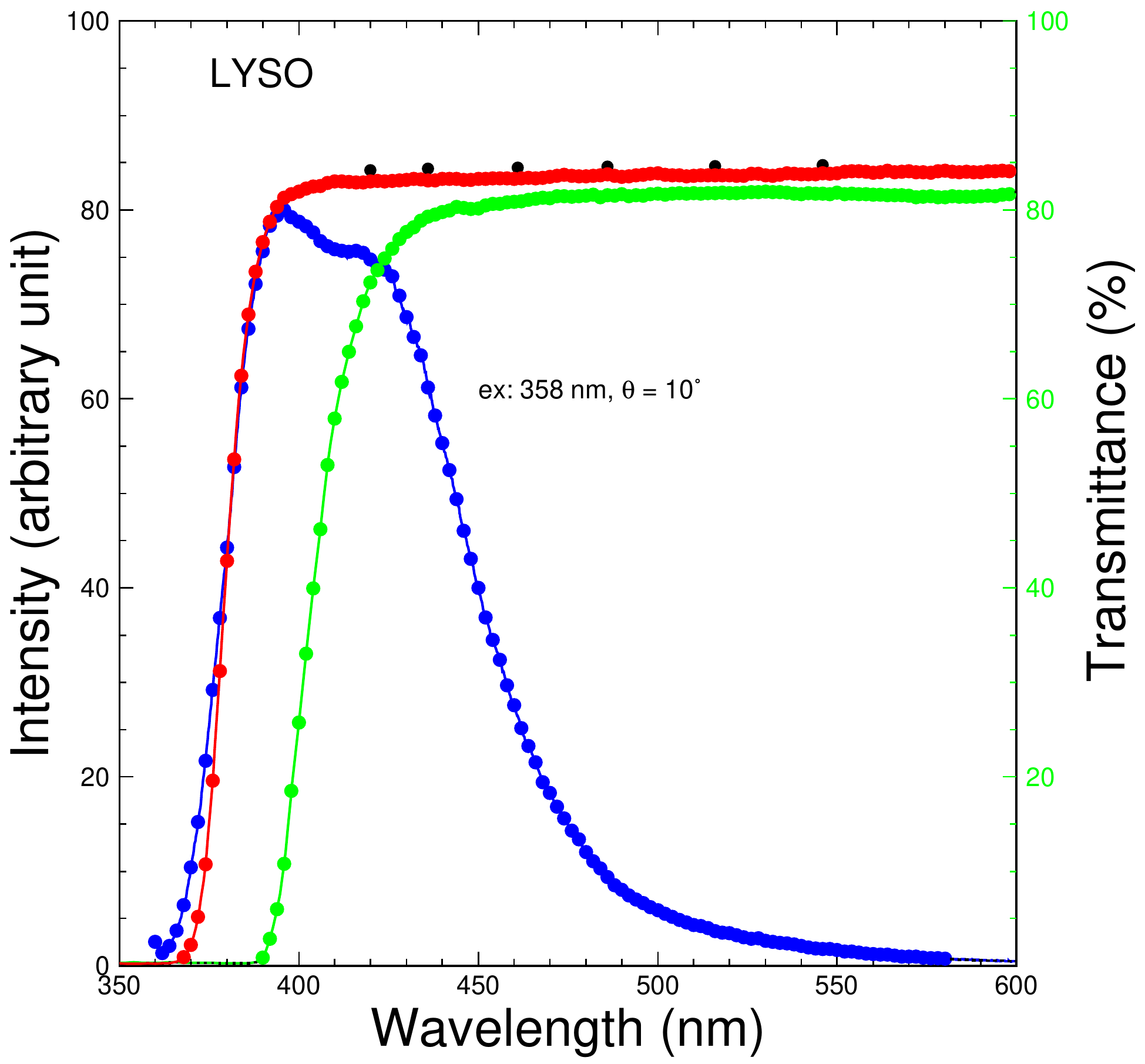}
\hspace{0.05\textwidth}
\includegraphics[width=0.41\textwidth]{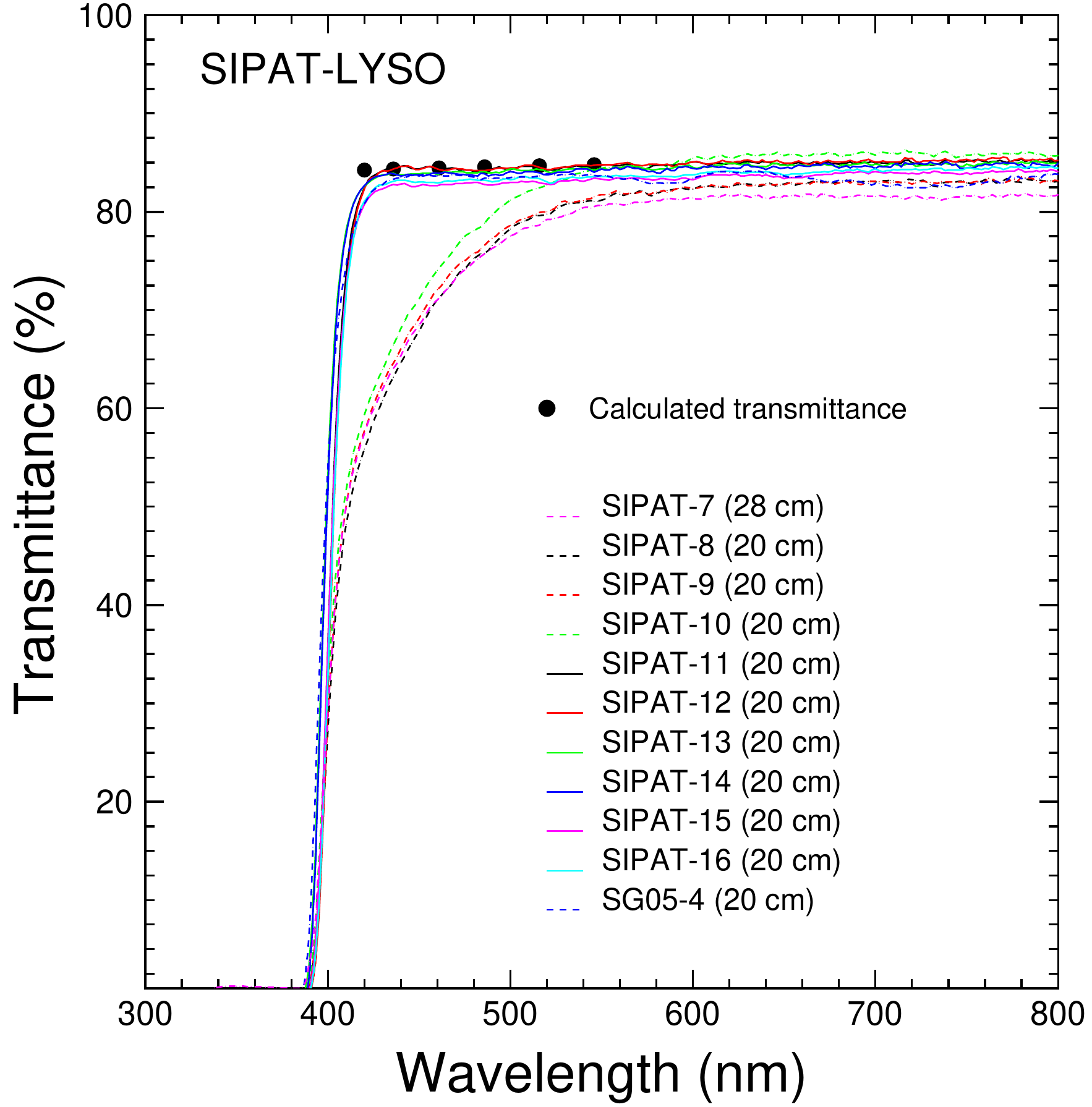}
\caption{Left: The longitudinal (green) and transverse (red) transmittance spectra and the photo-luminescence (blue) spectrum as a function of wavelength for a rectangular LYSO sample with dimensions 2.5 $\times$ 2.5 $\times$ 20 cm.  Right: Longitudinal transmittance spectra as a function of wavelength for eleven LYSO crystals: ten from SIPAT and one from Saint-Gobain. All except SIPAT-7 are 20 cm long.}
\label{fig:tt}
\end{figure*}

\subsubsection{{Transmittance and Emission}}
\label{sec:t&e}

Large LYSO crystals with good transmittance characteristics are now
routinely produced in industry. The left plot of Fig.~\ref{fig:tt}
shows the longitudinal (green) and transverse (red) transmittance
spectra measured for a rectangular LYSO sample with dimensions 2.5
$\times$ 2.5 $\times$ 20 cm (18 X$_0$). Significant red shift, caused
by internal absorption, is observed in the absorption edge of the
longitudinal transmittance, as compared to the transverse
transmittance. The black point are a fit to the theoretical
limit of transmittance calculated by using the refractive index,
assuming multiple bounces between two end surfaces and no internal
absorption~\cite{bib:zhu_lal}. It overlaps with the transverse
transmittance spectrum at wavelengths longer than 420 nm, indicating
the excellent optical quality of the crystal.  Also shown is the
photo-luminescence spectrum (blue)~\cite{bib:zhu_lyso08}. The fact
that a part of the emission spectrum is at wavelengths below the
absorption edge indicates that this part of the scintillation light is
absorbed internally in the crystal bulk; this is usually referred to
as the self-absorption effect. There is no such self-absorption effect
in other scintillation crystals commonly used for HEP calorimeters,
such as BGO, CsI(Tl) and PWO~\cite{bib:zhu_crystals}. While this
self-absorption has little consequence in the 6 mm long pixels used in
medical instruments, it can affect the light response uniformity for
the 20 cm long crystals used in the \superb\ calorimeter. This will be
discussed in section~\ref{sec:uni}.
%
\begin{table*}[htbp]%
\renewcommand{\arraystretch}{1.5}
\caption{Photo Luminescence-Weighted Quantum Efficiencies (\%)}
\label{tab:ewqe}
\begin{center}
\begin{tabular*}{\textwidth}{@{}l@{\extracolsep{\fill}}cccc}
\hline\hline 
Photo Luminescence   & LSO/LYSO & BGO & CsI(Tl)\\
\hline
Hamamatsu R1306 PMT  &  12.9$\pm$0.6 &8.0$\pm$0.4 &5.0$\pm$0.3 \\
Hamamatsu R2059 PMT  &  13.6$\pm$0.7 &8.0$\pm$0.4 &5.0$\pm$0.3 \\
Photonis XP2254b     &  7.2$\pm$0.4  &4.7$\pm$0.2 &3.5$\pm$0.2 \\
\hline
Hamamatsu S2744 PD  &  59$\pm$4 &75$\pm$4 &80$\pm$4 \\
Hamamatsu S8664 APD &  75$\pm$4 &82$\pm$4 &84$\pm$4 \\
\hline\hline 
\end{tabular*}
\end{center}
\end{table*}%

During the R\&D phase for crystal development poor longitudinal
transmittance was observed in some samples~\cite{bib:zhu_lyso10}. The
right plot of Fig.~\ref{fig:tt} shows that four samples (SIPAT-7 to
SIPAT-10) have poor longitudinal transmittance between 380 nm and 500
nm, showing an absorption band. Further investigation revealed that
this absorption band is located at the seed end, and is caused by
point defects~\cite{bib:zhu_lyso08} associated with a poor quality
seed crystal used in their growth. With rigorous quality control, LYSO
crystals grown later at SIPAT (SIPAT-11 to SIPAT-16) show no
absorption band at the seed end
(Fig.~\ref{fig:tt}). An increase of light output of about 30\% was found
after this problem was resolved. It is thus important to include in
crystal specifications a requirement on the crystal's longitudinal
transmittance.

The left plot of Fig.~\ref{fig:qe&decay} shows typical quantum
efficiencies of a PMT with multi-alkali cathode (Photonis XP2254b) and
an APD (Hamamatsu S8664)~\cite{bib:zhu_lyso071}. The photoluminescence
spectra of LSO/LYSO, BGO and CsI(Tl) crystals are also shown; the area
under the luminescence spectra is roughly proportional to the
corresponding absolute light output. Table~\ref{tab:ewqe} summarizes
the numerical values of the photoluminescence-weighted average quantum
efficiencies for various readout devices. These numbers can be used to
convert the measured photo-electron numbers to the absolute light
output in photon numbers.

\begin{figure*}[htbp]
\centering
\includegraphics[width=0.47\textwidth]{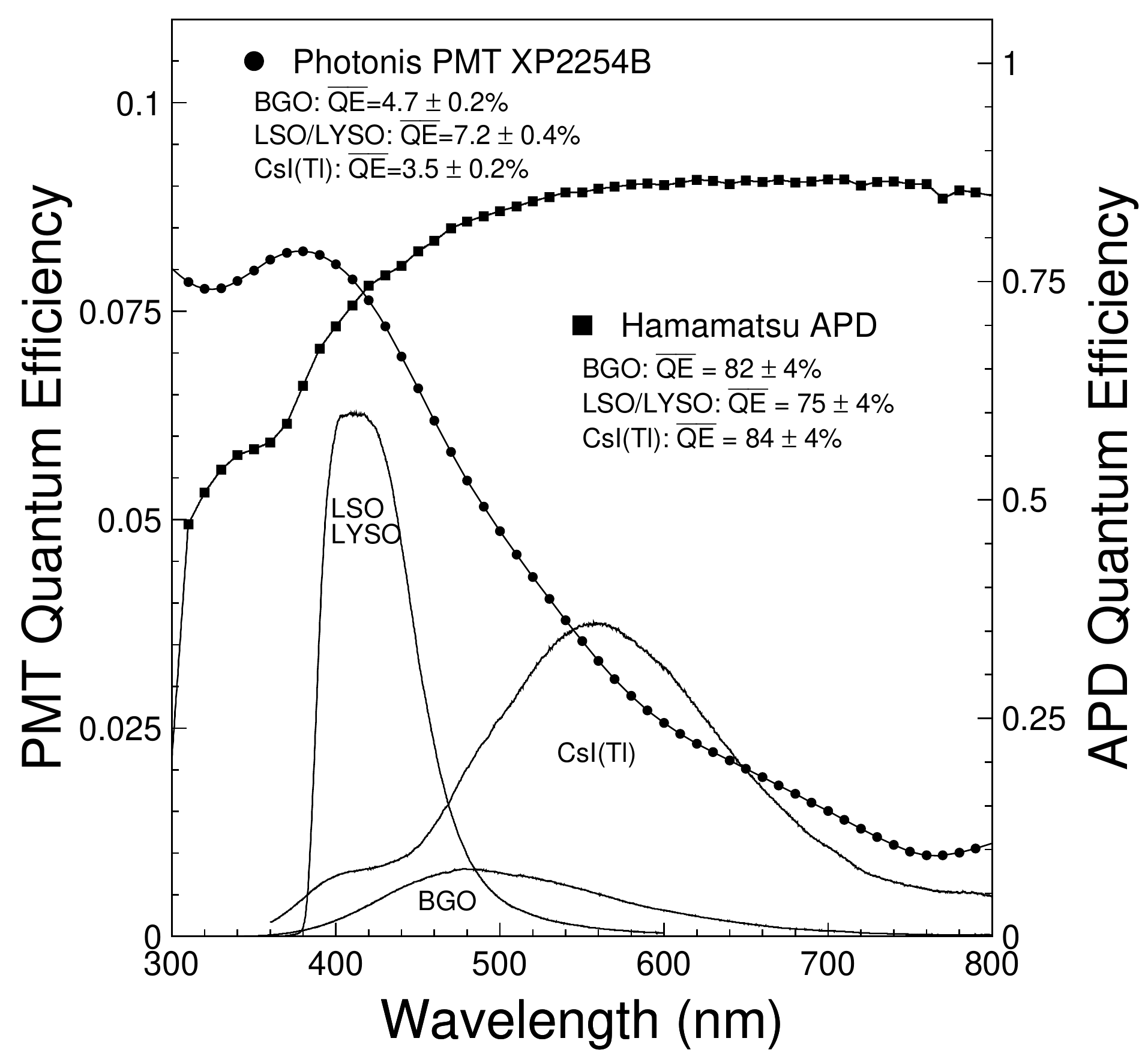}
\hspace{0.05\textwidth}
\includegraphics[width=0.45\textwidth]{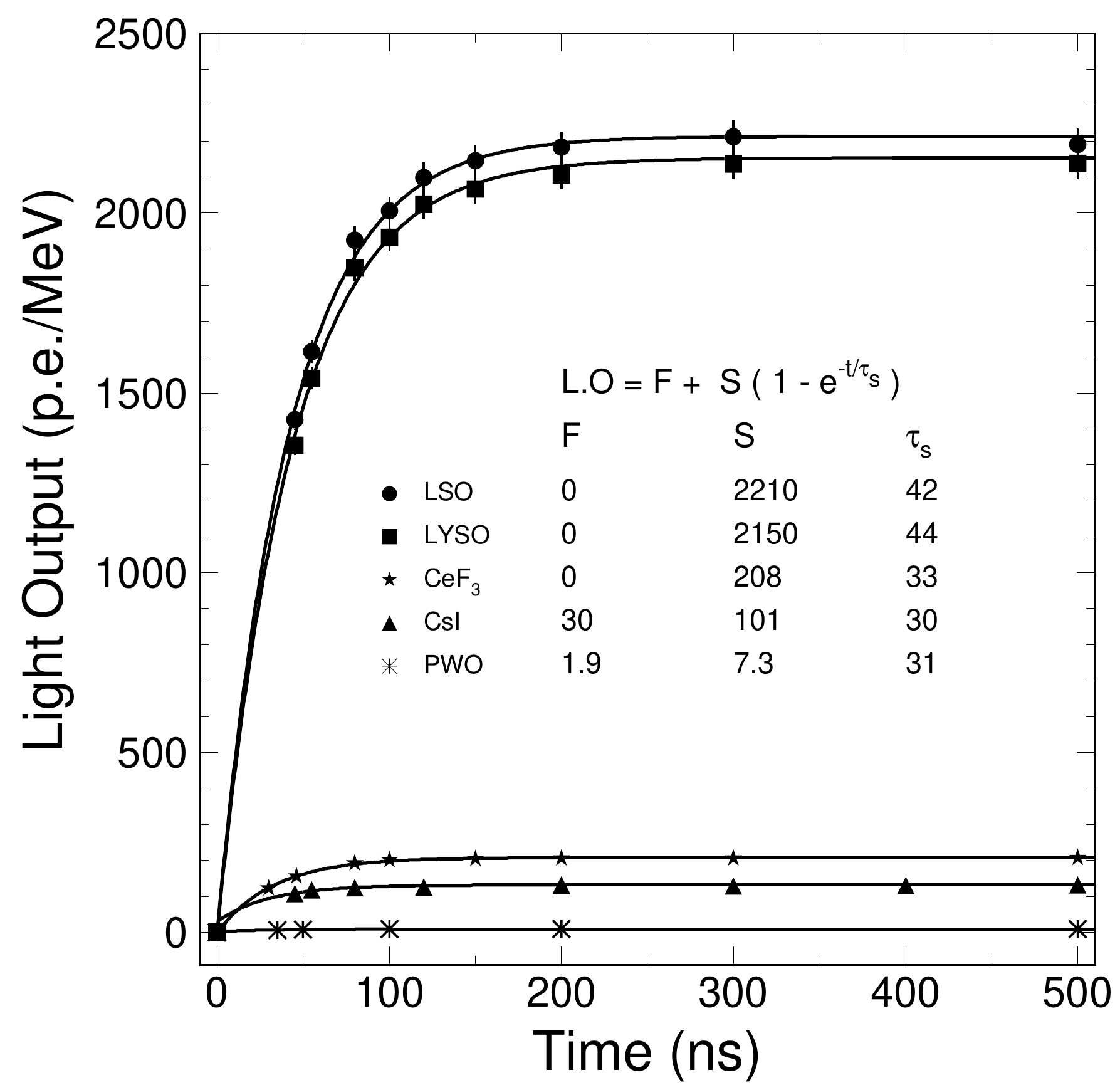}
\caption {Left: The quantum efficiencies of a Hamamatsu R2059 PMT (solid dots) and a Hamamatsu S8664 APD (solid squares) are shown as a function of wavelength together with photoluminescence spectra of the LSO/LYSO, BGO and CsI(Tl) samples, where the area under the luminescence spectra is roughly propotional to the corresponding absolute light output. 
Right: Light output measured by using a Photonis XP2254 PMT is shown as a function of integration time for six crystal scintillators.}
\label{fig:qe&decay}
\end{figure*}

A significant red component was observed in the $\gamma$-ray induced
luminescence spectra in the CTI LSO samples; this was not seen in the
LYSO samples from the other suppliers~\cite{bib:zhu_lyso08}. This red
component disappeared after a $\gamma$-ray irradiation with an
accumulated dose of 5 $\times$ 10$^3$ rad. This is the only
significant difference observed between the large size LSO and LYSO
samples~\cite{bib:zhu_lyso08}, indicating that LYSO is the preferred
choice.

\smallskip

\subsubsection{Decay time and Light Output}

The right plot of Fig.~\ref{fig:qe&decay} shows the light output for
six crystal scintillators in units of photoelectrons/\mev, measured
using a Photonis XP2254 PMT as a function of integration
time~\cite{bib:zhu_crystals}. The light output can be fit to the
following function to determine the fast and slow components and the
decay kinetics:
\begin{equation}
\label{eq:decay}
 LO(t) = F + S (1 - e^{-t/\tau_s}),
\end{equation}
where {\it F} is the fast component of the scintillation light with a
decay time of less than 10 ns, and {\it S} is the slow component with
a decay time of $\tau_s$ longer than 10 ns. The decay time of both LSO
and LYSO crystals is about 40 ns.

\begin{figure*}[!htb]
\centering
\includegraphics[width=0.47\textwidth]{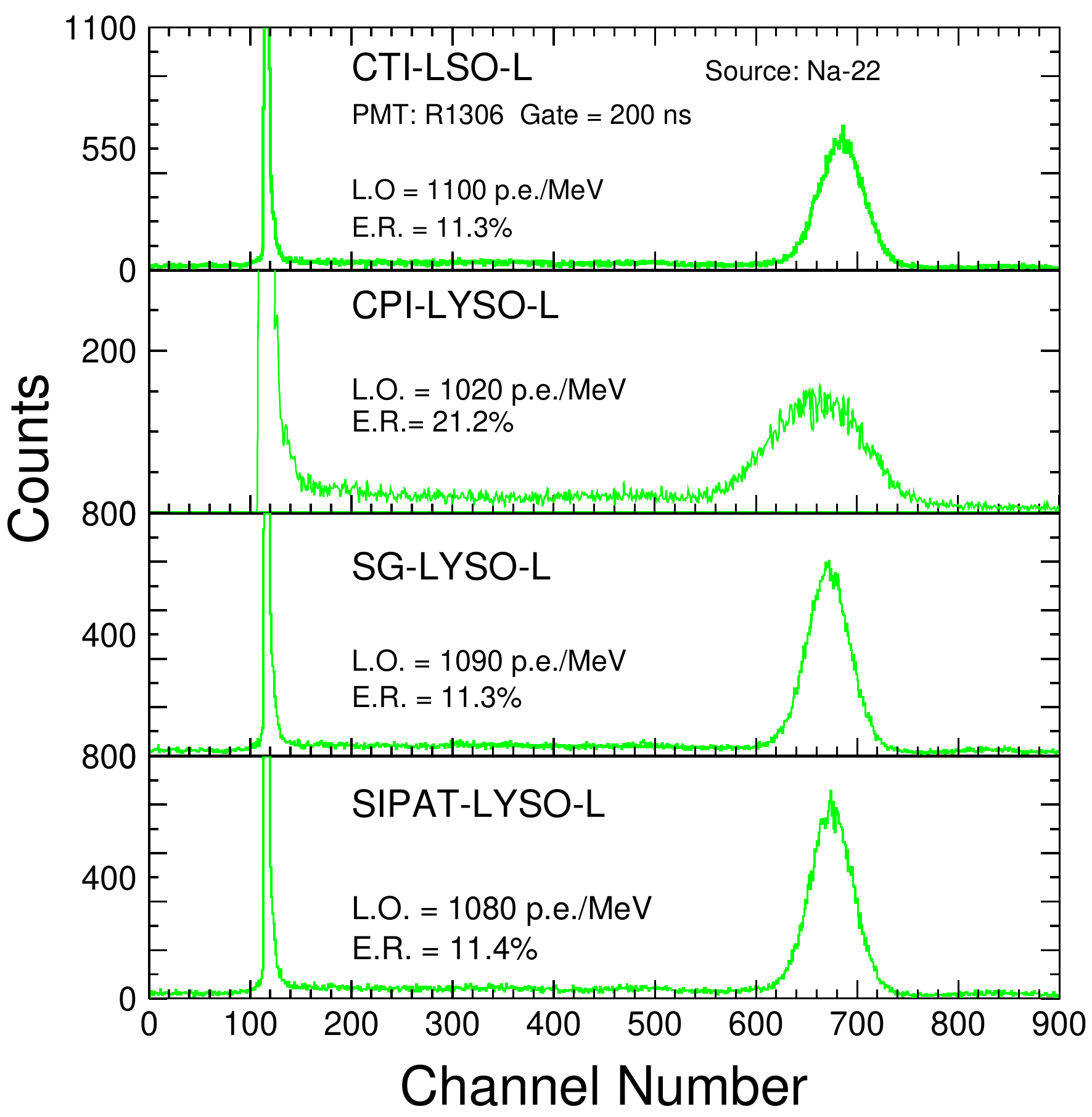}
\hspace{0.05\textwidth}
\includegraphics[width=0.45\textwidth]{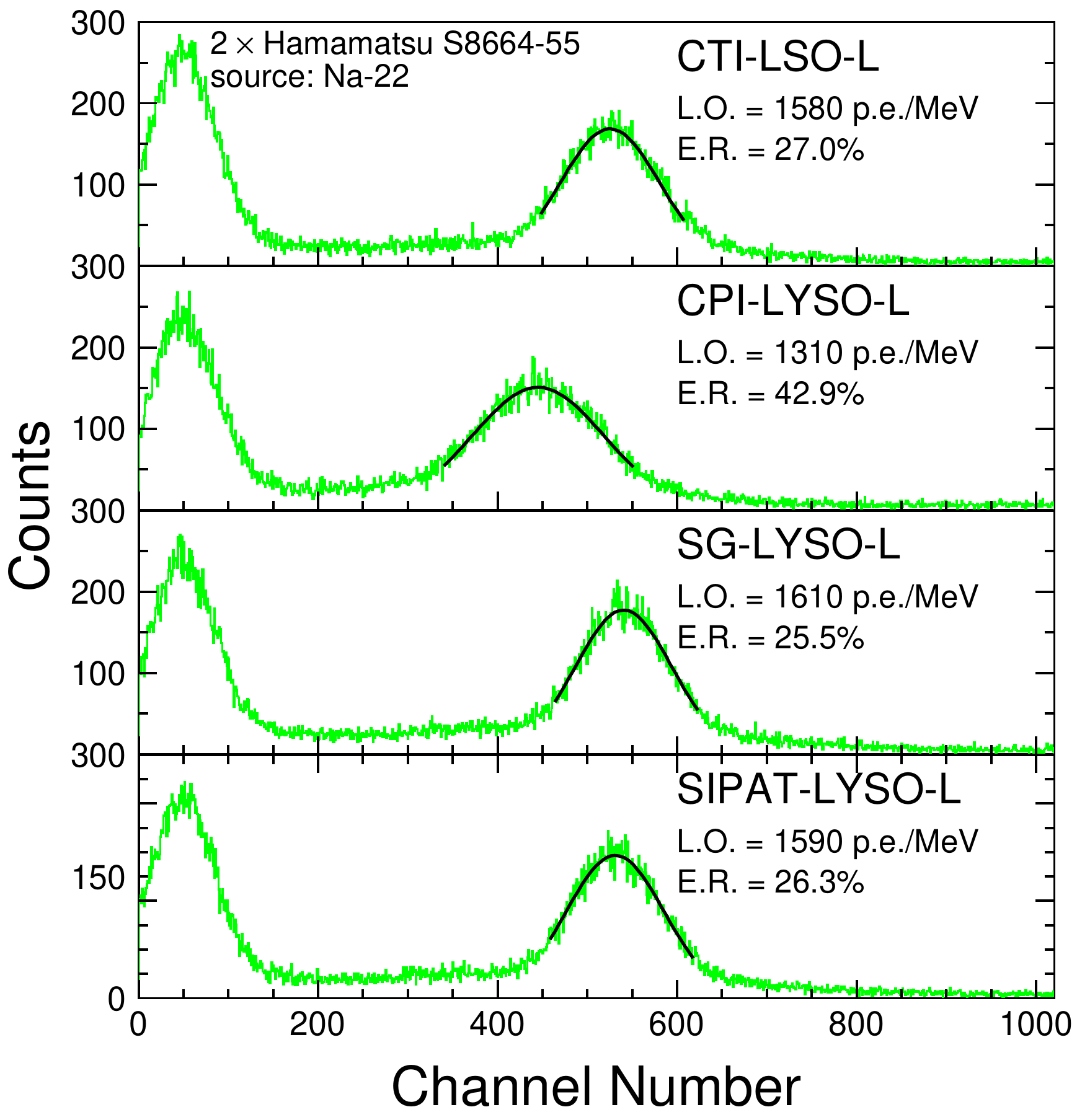}
\caption {0.511 MeV $\gamma$-ray spectra from a $^{22}$Na source, measured by a Hamamatsu R1306 PMT (Left)  and two Hamamatsu S8664-55 APDs (Right), with a coincidence trigger for four long LSO and LYSO samples of $2.5 \times 2.5 \times 20~{\rm cm}^3$.}
\label{fig:lyso}
\end{figure*}
As shown in Table~\ref{tab:crystals} LSO and LYSO crystals have high
light output, about 85\% and 50\% of NaI(Tl) and CsI(Tl) respectively,
and about 18, 4, and more than 200 times that of pure CsI, BGO, and
PWO, respectively.  Figure~\ref{fig:lyso} shows 0.511 $\gamma$-ray
pulse height spectra measured by a Hamamatsu R1306 PMT (left) and two
Hamamatsu S8664-55 APDs (right) for four LSO and LYSO samples of $2.5
\times 2.5 \times 20~{\rm cm}^3$ from CTI, CPI, SG, and SIPAT.  The
equivalent noise energy for the APD readout is less than
40~keV~\cite{bib:zhu_lyso071}. The CPI LYSO sample showed poor energy
resolution; the other samples did not. According to the grower this
was caused by intrinsic non-uniformity, which may be improved by
appropriate thermal annealing. It thus is important to include in
crystal specifications a requirement on the crystal's energy
resolution.

Because of their fast decay time and high light output, LSO and LYSO
crystals have also been used in time of flight (TOF) measurements for
medical applications, such as TOF PET (positron emission tomography).
Better than 500 ps FWHM was achieved for the time difference between
two photons. In HEP experiments, an {\it rms} time resolution of
better than 150 ps may be achieved for TOF measurements for single
particles. Since the intrinsic rise time of scintillation light is
about 30 ps for LSO and LYSO
crystals~\cite{bib:steve_time_resolution}, the measured time
resolution for LSO and LYSO is affected mainly by the response speed
of the readout device and the choice of time
pick-off~\cite{bib:zhu_time_resolution}. Calcium doping of LSO and
LYSO has been reported to reduce the decay time to about 20
ns~\cite{bib:chuck_ca}, which would help to improve the time
resolution.

\subsubsection{Response Uniformity}
\label{sec:uni}
\smallskip

Adequate light response uniformity along the crystal length is key to
maintaining the precision offered by a total absorption crystal
calorimeter at high energies~\cite{bib:zhu_RD}. The light response
uniformity of a long crystal is
parameterized as a linear function,
\begin{equation}
 \frac {LY}{LY_{\rm mid}} = 1 + \delta (x/x_{\rm mid} - 1 ),
 \label{eq_uni}%
\end{equation}
where ($LY_{\rm mid}$) represents the light output measured at the
middle point of the crystal, $\delta$ represents the deviation from
the flat response and $x$ is the distance from the photo-detector. To
achieve good energy resolution, the corresponding $|\delta|$ value for
\superb\ LYSO crystals of 18 X$_0$ length must be kept to less than
3\%~\cite{bib:cms_ecal}.
\begin{figure*}[!htb]

\centering
\includegraphics[width=0.47\textwidth]{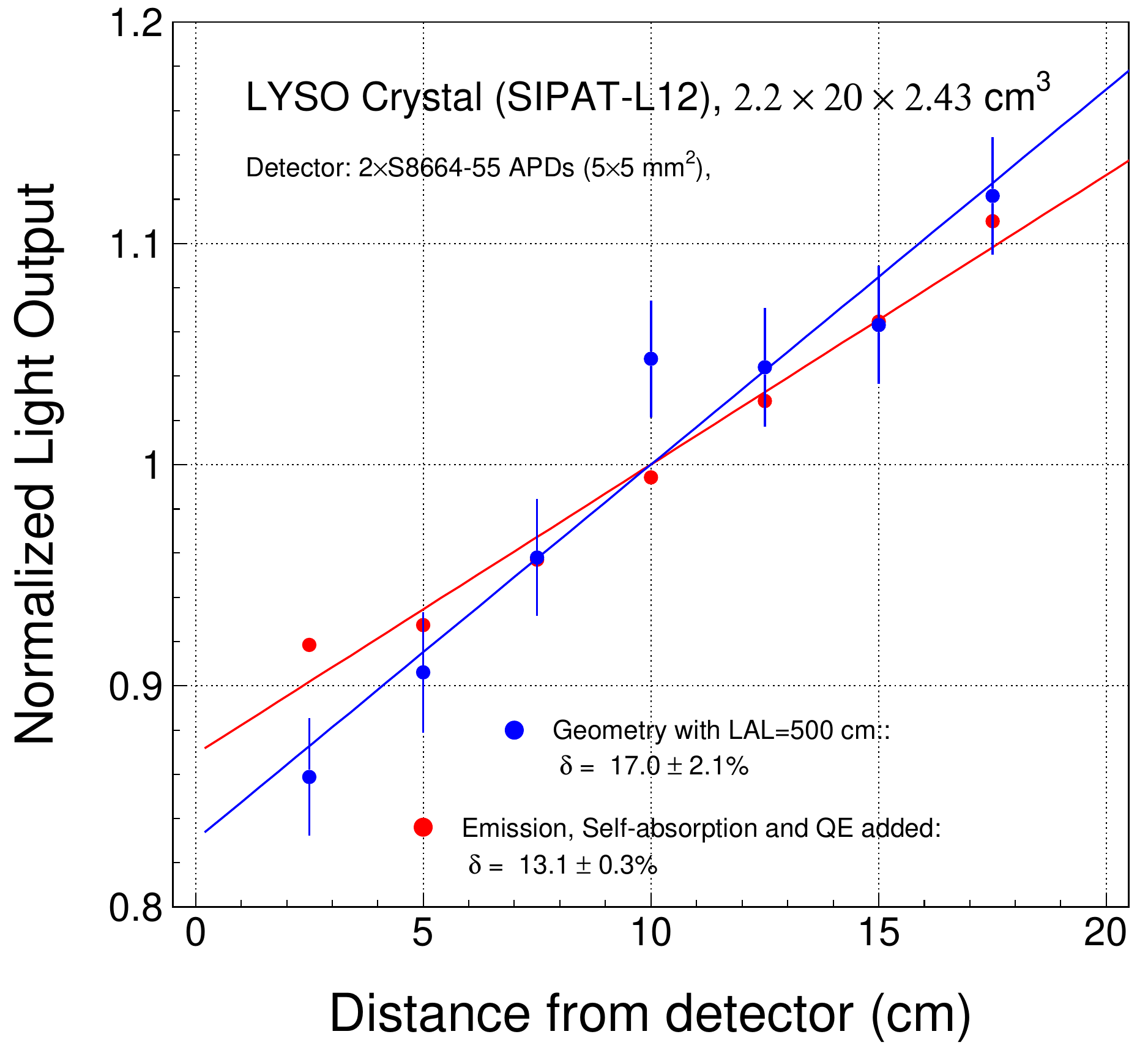}
\hspace{0.05\textwidth}
\includegraphics[width=0.45\textwidth]{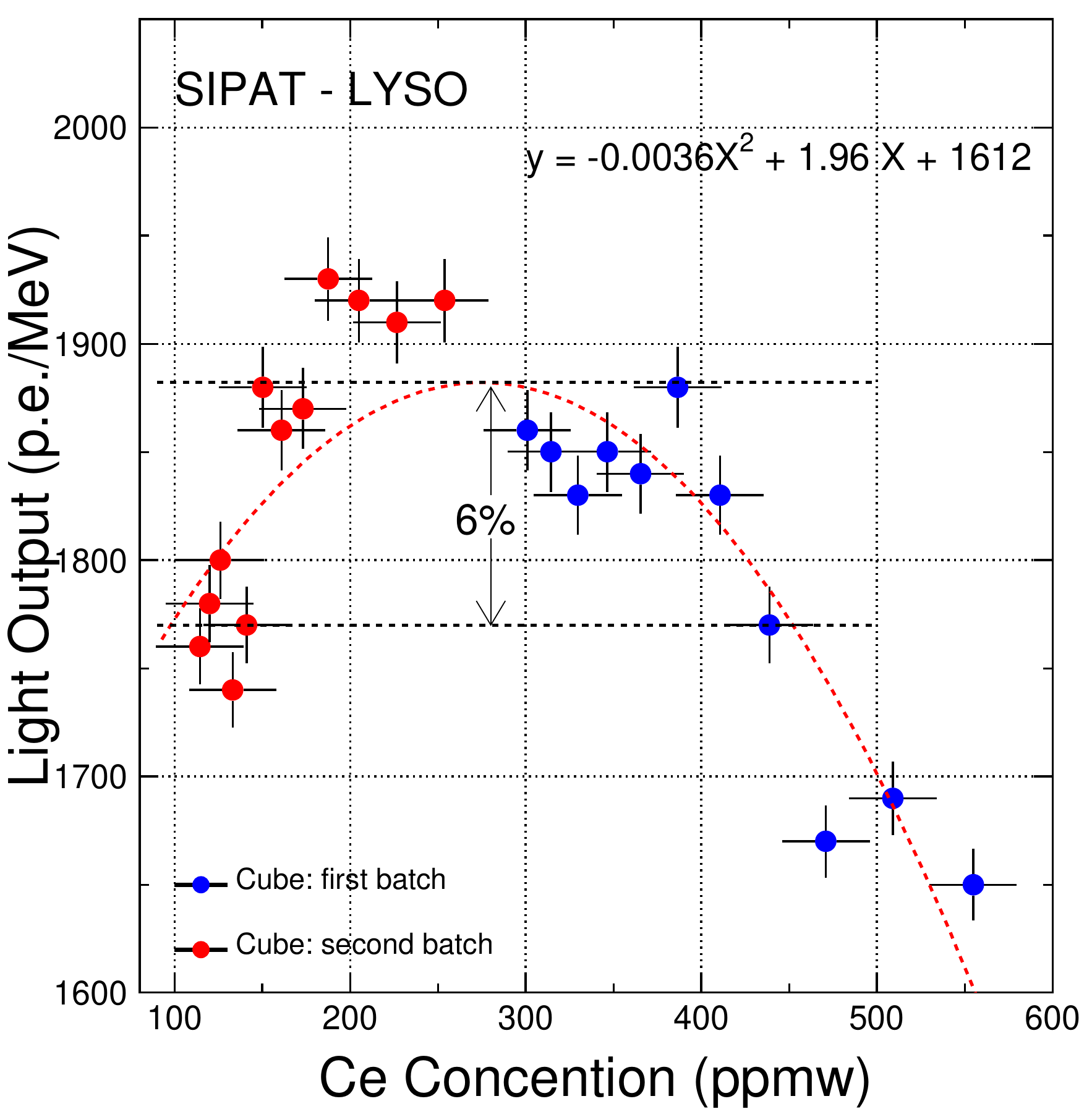}

\caption {Left: Light response uniformities without (blue) and with (red) self-absorption effects, calculated by a ray-tracing program, shown for a 20 cm long crystal with tapered geometry and two Hamamatsu S8664-55 APD readout. Right: Light output measured for 17 mm LSO/LYSO crystal cubes shown as a function of the cerium concentration.}
\label{fig:uni1}
\end{figure*}

Effective light collection requires good light reflection. The glass
fiber-based supporting structure designed for the originally
envisioned \superb\ full LYSO forward calorimeter was coated with a
thin layer of aluminum as reflector. All measurements and simulations
discussed in this section were therefore carried out with an
aluminum-coated glass fiber supporting structure cell, referred to as
the RIBA Cell, around the crystal.  In the baseline hybrid design, the
\babar\ endcap mechanical structure is retained; an equivalent
reflective wrapping to the four LYSO crystals that will occupy each
existing cell must be developed.

The light response uniformity of a long tapered LYSO crystal is
affected by three factors. First, the tapered crystal geometry leads
to an optical focusing effect, {\it i.e.,} the response for
scintillation light originating at the small end far away from the
photodetector would be higher than that at the large end, which is
coupled to the photodetector. This is caused by the light propagation
inside the crystal, and is common to any optical object with this
tapered geometry. Second, there is a self-absorption effect in LYSO
crystals, as discussed in section~\ref{sec:t&e}, since a part of the
emission spectrum is self-absorbed in the crystal bulk (left plot of
Fig.~\ref{fig:tt}). This effect is specific to LYSO crystals. Last,
there is a non-uniform light yield along the longitudinal axis of the
crystal, caused by segregation of the cerium activator in LYSO
crystals during growth. Because of the small segregation coefficient
(about 0.2) the cerium concentration increases from the seed end to
the tail end of the crystal. This effect is common to all Czochralski
method crystals doped with an activator, {\it e.g.}, CsI(Tl).
\begin{figure*}[htbp]

\centering
\includegraphics[width=0.47\textwidth]{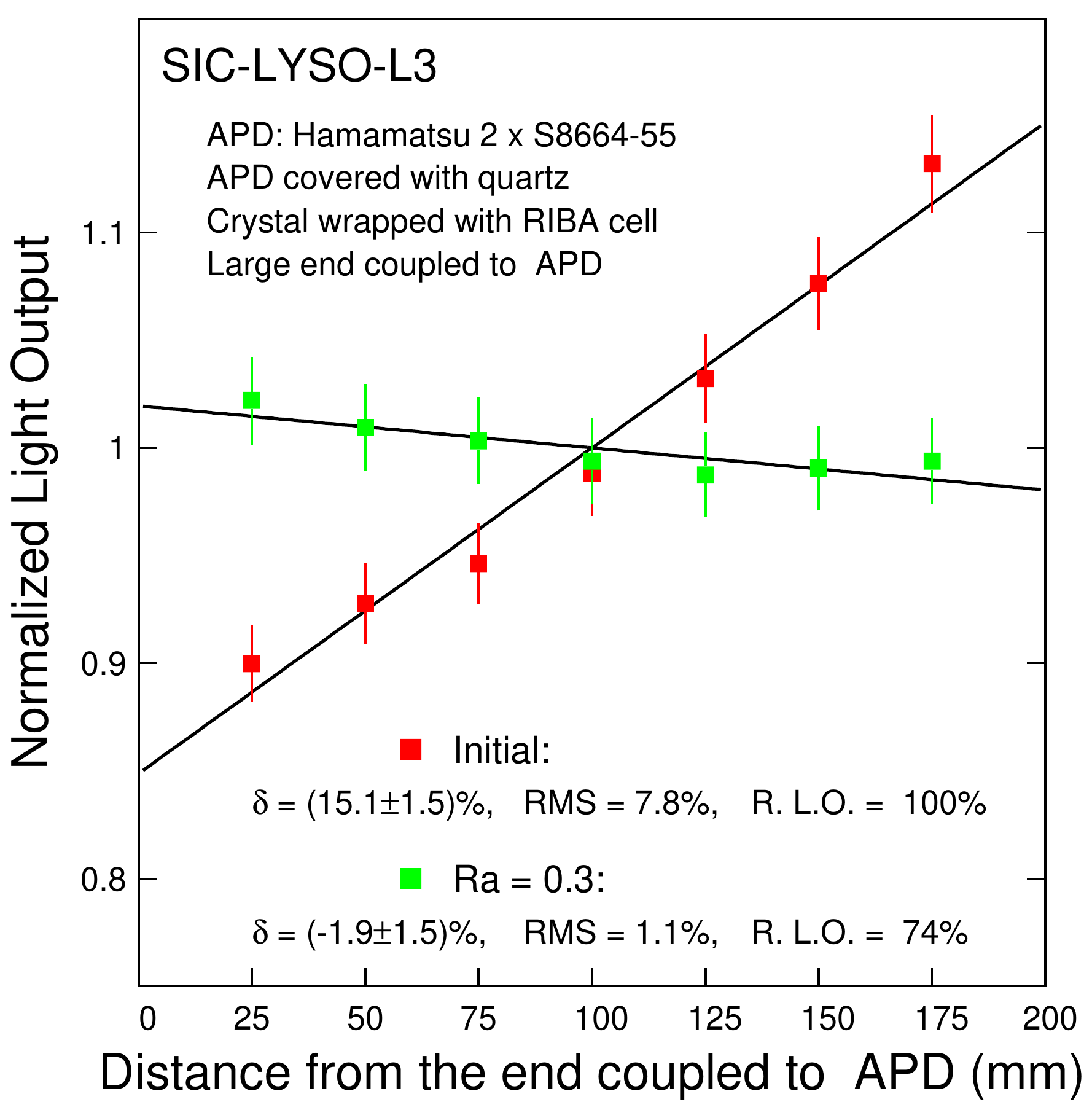}
\hspace{0.05\textwidth}
\includegraphics[width=0.42\textwidth]{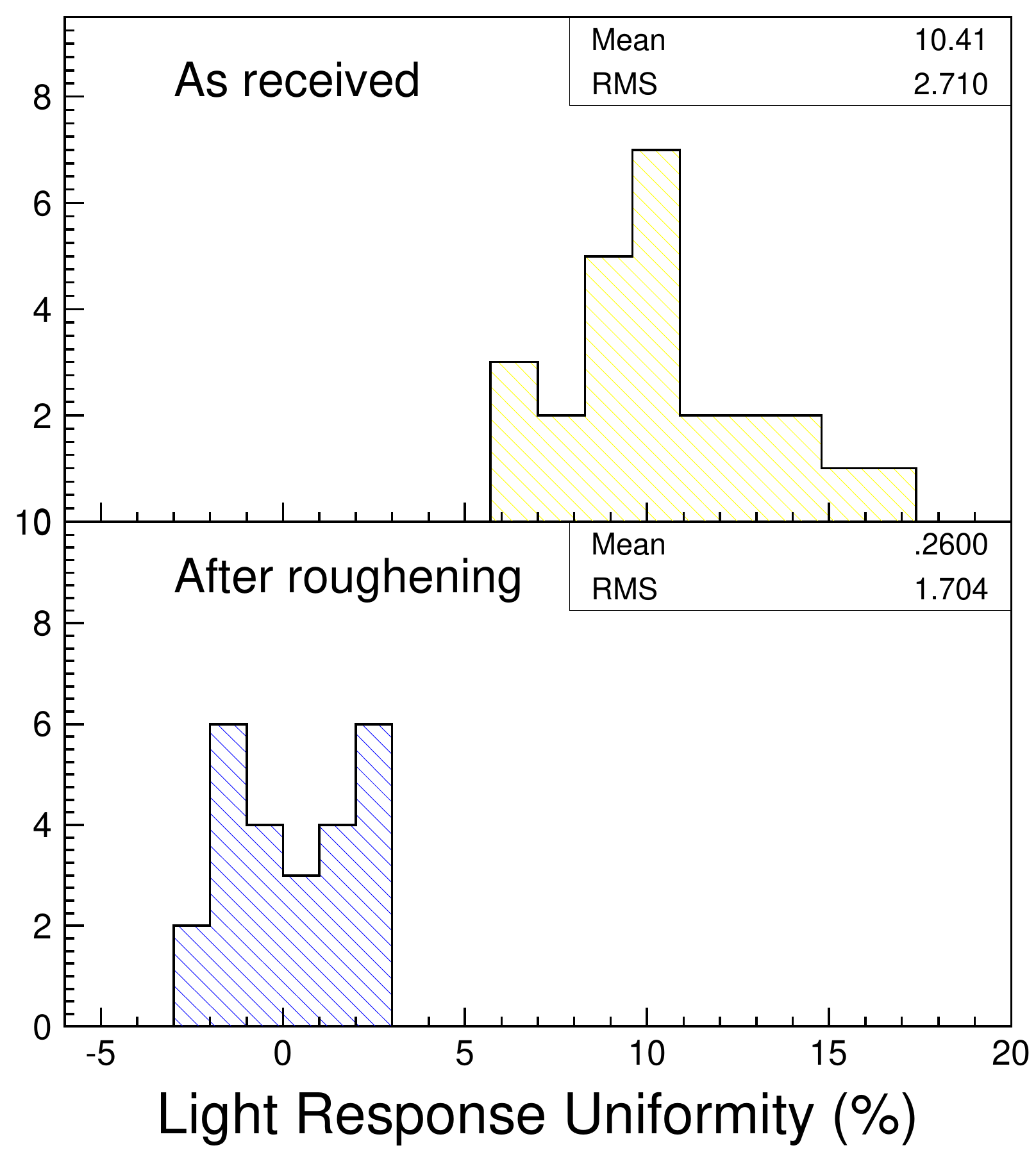}
\caption {Left: Light response uniformity measured with two Hamamatsu S8664-55 APD readout for a tapered \superb\ LYSO crystal SIC-L3 before (red dots) and after (green dots) uniformization by roughening the smallest side surface to Ra = 0.3. Right: The distributions of light response uniformity ($\delta$ values) are shown for 25 \superb\ test beam crystals before (top) and after (bottom) uniformization by roughening the surface of the smallest side.}
\label{fig:uni2}
\end{figure*}

The left plot of Fig.~\ref{fig:uni1} shows the light response
uniformities calculated using a ray-tracing program~\cite{bib:zhu_RD}
for a \superb\ LYSO crystal with tapered geometry and dual Hamamatsu
S8664-55 APD readout. The blue dots show the uniformity with only the
optical focusing effect; the red dots include, in addition, the
self-absorption. Numerically, the optical focusing effect alone causes
a $\delta$ value of 17\%, which is reduced to 13\% with
self-absorption included. This indicates that the self-absorption
effect can provide partial compensation for the optical focusing
effect, provided that knowledge of the crystal growth orientation is
retained. 

The right plot of Fig.~\ref{fig:uni1} shows the light output
measured for two batches of 17~mm LSO/LYSO crystal cubes (red and
blue) as a function of the cerium concentrations determined by Glow
Discharge Mass Spectroscopy (GDMS) analysis. It shows that the
optimized cerium doping level is between 150 and 450~ppmw because of
the interplay between the cerium activator density and the self
absorption caused by overdoping. Also shown in the plot is a second
order polynomial fit. By adjusting the cerium doping concentration,
the light yield difference along the crystal can be minimized. A
difference at the level of 10\% is more or less the maximum, which may
provide a variation of the $\delta$ value up to 5\%. Taking this into
account, the initial $\delta$ value of the \superb\ LYSO crystals may
vary between 8\% and 18\%.

Following the experiences of previous crystal calorimeters, such as L3
BGO and CMS PWO, a $|\delta|$ value of less than 3\% may be achieved
by roughening one side surface of the crystal to an appropriate
roughness~\cite{bib:zhu_lyso_uni}. The left plot of
Fig.~\ref{fig:uni2} shows the light response uniformities measured
with dual Hamamatsu S8664-55 APD readout for a tapered \superb\ LYSO
crystal, SIC-L3.  The $\delta$ value is reduced from 15\% before (red)
to $-1.9$\% after (blue) roughening the smallest side surface to Ra =
0.3. The right plot of Fig.~\ref{fig:uni2} shows a comparison of the
$\delta$ values before (top) and after (bottom) roughening for all 25
\superb\ test beam crystals, showing a reduction of the average
$\delta$ value from 10\% to 0.26\%. All 25 $|\delta|$ values after
uniformization are within 3\%. The reduction of light collection
efficiency caused by this uniformization is about 17\%. It is expected
that one or two Ra values would be sufficient to uniformize the full
cohort of mass produced LYSO crystals to achieve $|\delta|$ values of
less than 3\%.

\subsubsection{Radiation Hardness}

The radiation hardness of long LSO and LYSO samples has been
investigated with
$\gamma$-rays~\cite{bib:zhu_lyso072,bib:zhu_lyso09_gamma} and
neutrons~\cite{bib:zhu_lyso09_neutron}. It is found that the
scintillation mechanism of this material is not damaged. The damage
to light transmission does not recover at room temperature, indicating
no dose rate dependence~\cite{bib:zhu_RD}, but can be completely
eliminated by thermally annealing at 300\deg.  Studies also show that
LYSO is also more radiation-hard against charged
hadrons~\cite{bib:francesca_lyso09} than other crystals.
\begin{figure*}[htbp]

\centering
\includegraphics[width=0.47\textwidth]{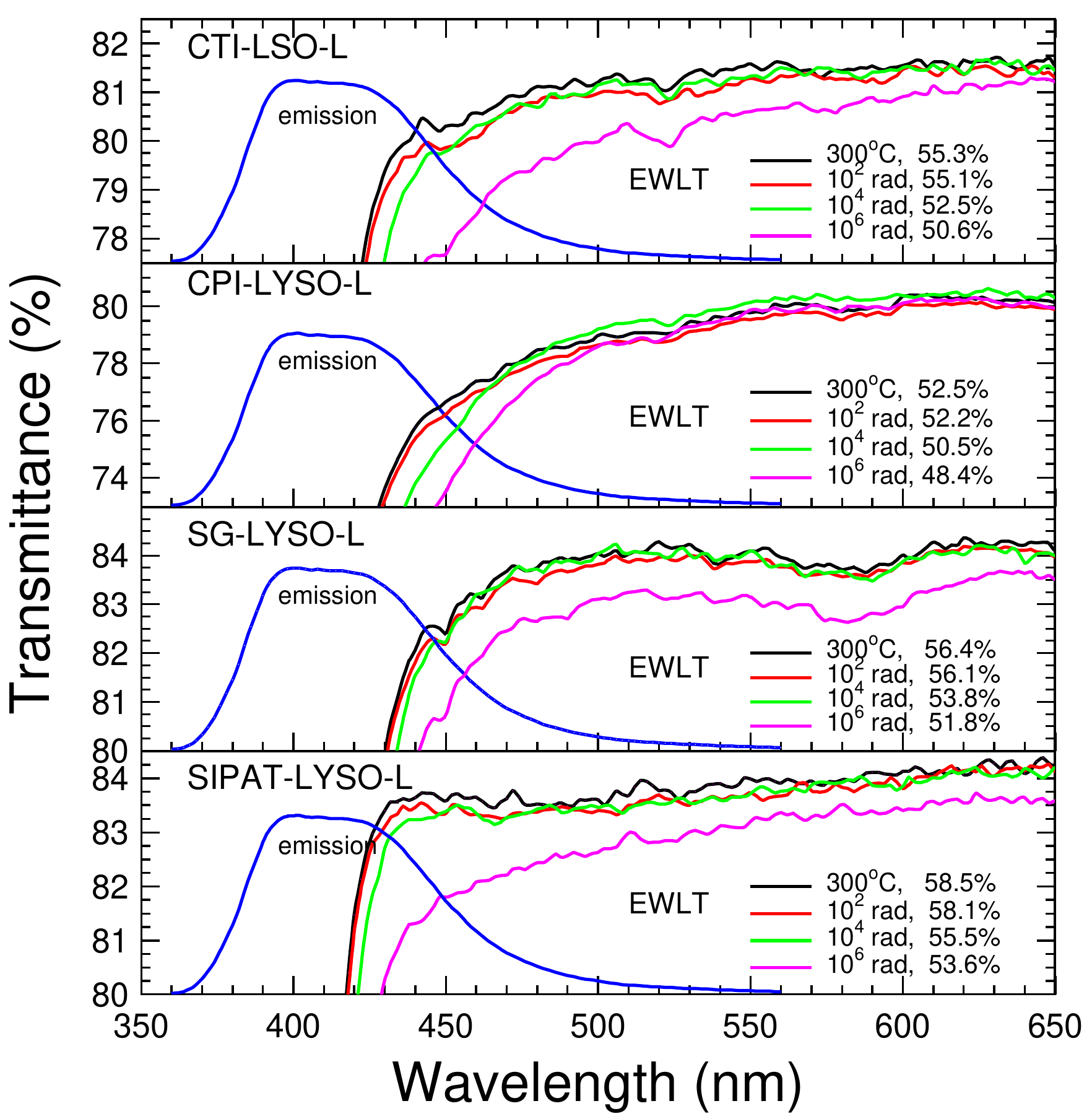}
\hspace{0.05\textwidth}
\includegraphics[width=0.45\textwidth]{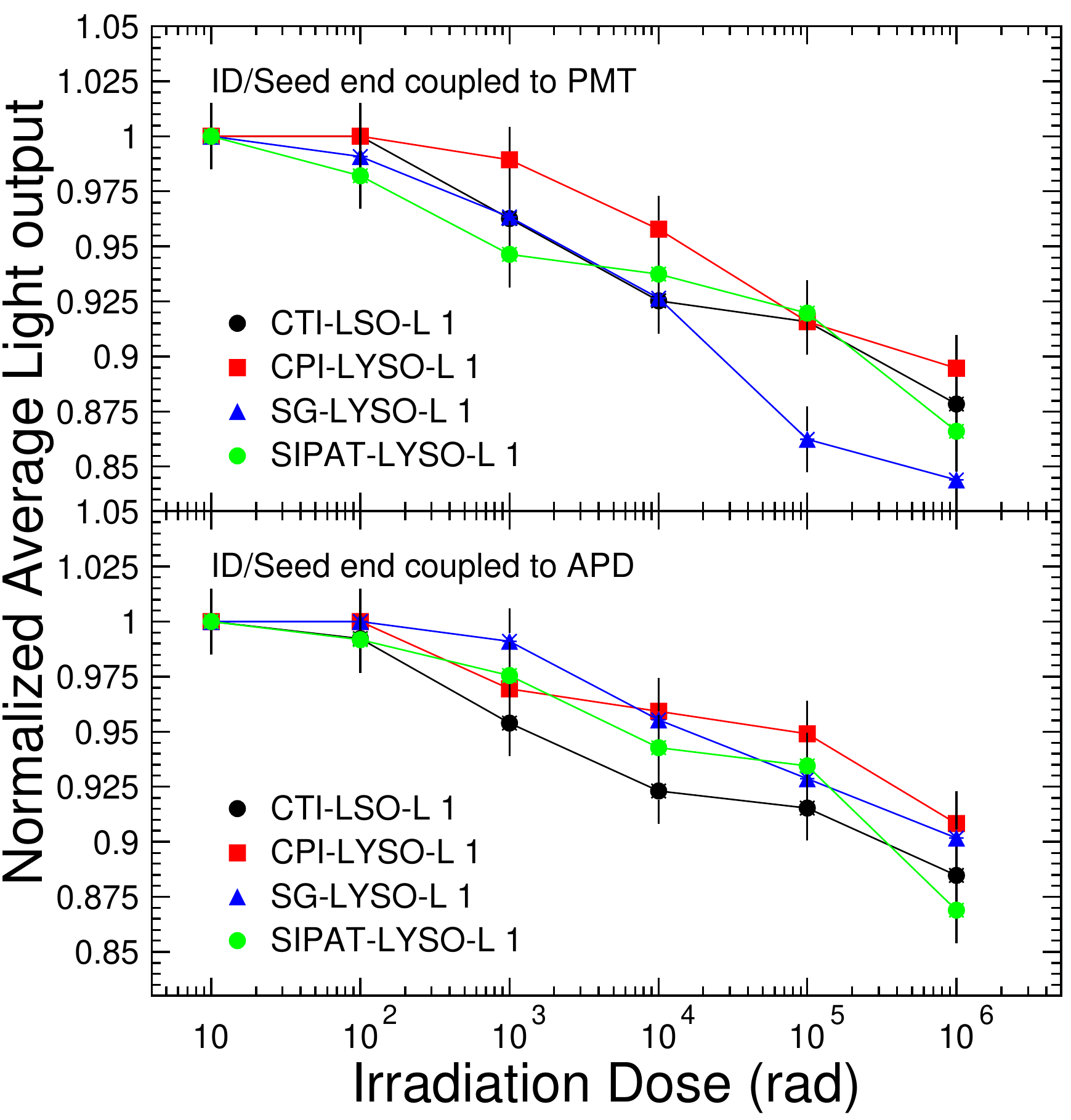}

\caption {The longitudinal transmittance spectra in an expanded view
  (left) and the normalized light output (right) are shown as a
  function of the integrated dose up to 1 Mrad for four LSO and LYSO
  samples of $2.5 \times 2.5 \times 20~{\rm cm}^3$. Also shown in the
  left plot is the photoluminescence spectrum (blue) in arbitrary
  units. The EWLT is listed in the legend.}
\label{fig:lyso_rad}
\end{figure*}

Figure~\ref{fig:lyso_rad} shows the longitudinal transmittance (left)
and normalized average light output (right) for four 20~cm LSO and
LYSO samples from CTI, CPI, SG and SIPAT. The light output was
measured by using a XP2254 PMT (top) and two S8664-55 APDs (bottom).
All samples tested have a consistent radiation resistance, with
degradations of the emission-weighted longitudinal transmittance
(EWLT) and the light output of approximately 12\% for a $\gamma$-ray
dose of 1~MRad. This radiation hardness is much better than other
scintillation crystals, such as BGO, CsI(Tl) and PWO.

\begin{figure*}[htbp]
\centering
\includegraphics[width=0.44\textwidth]{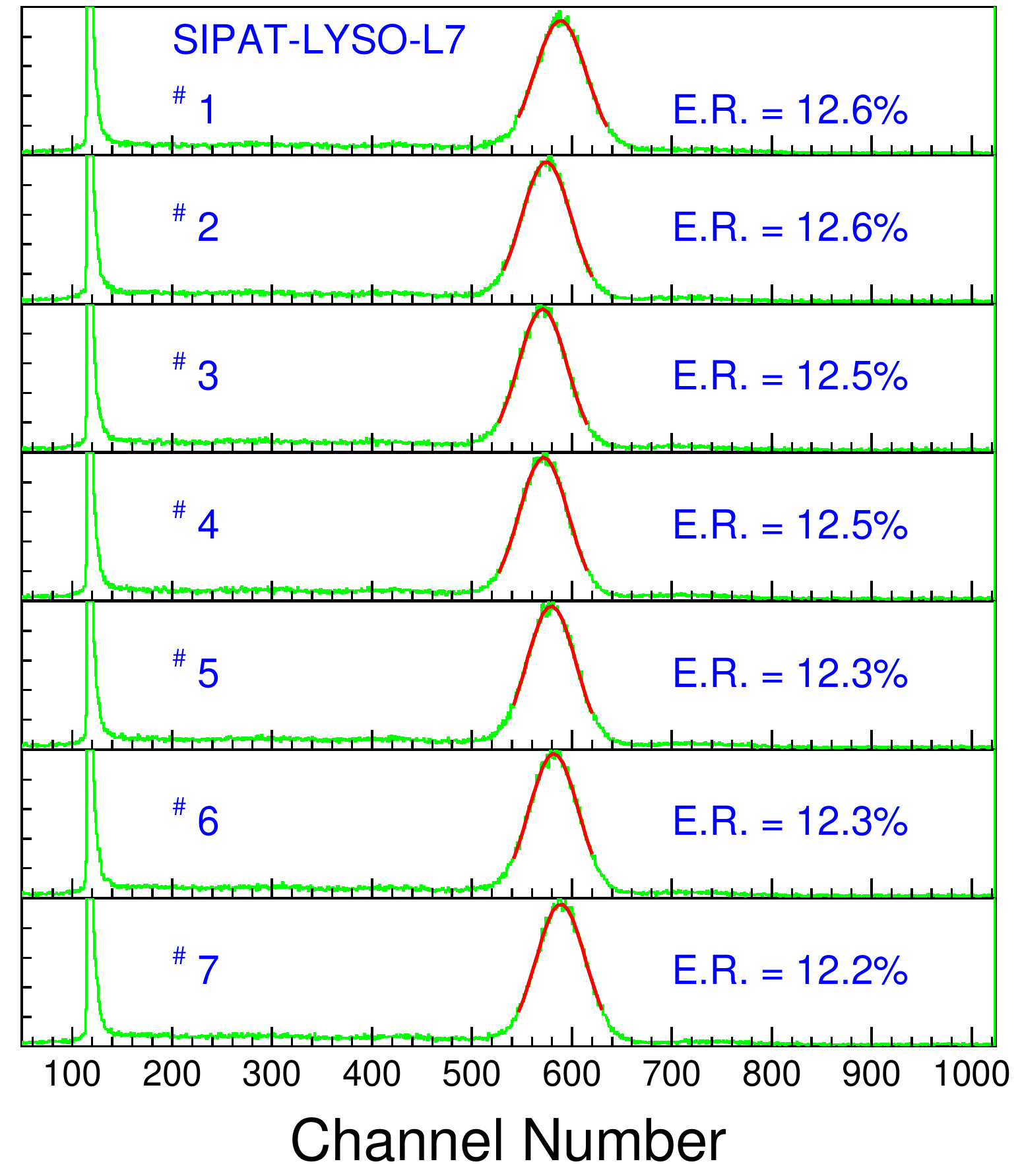}
\hspace{0.05\textwidth}
\includegraphics[width=0.49\textwidth]{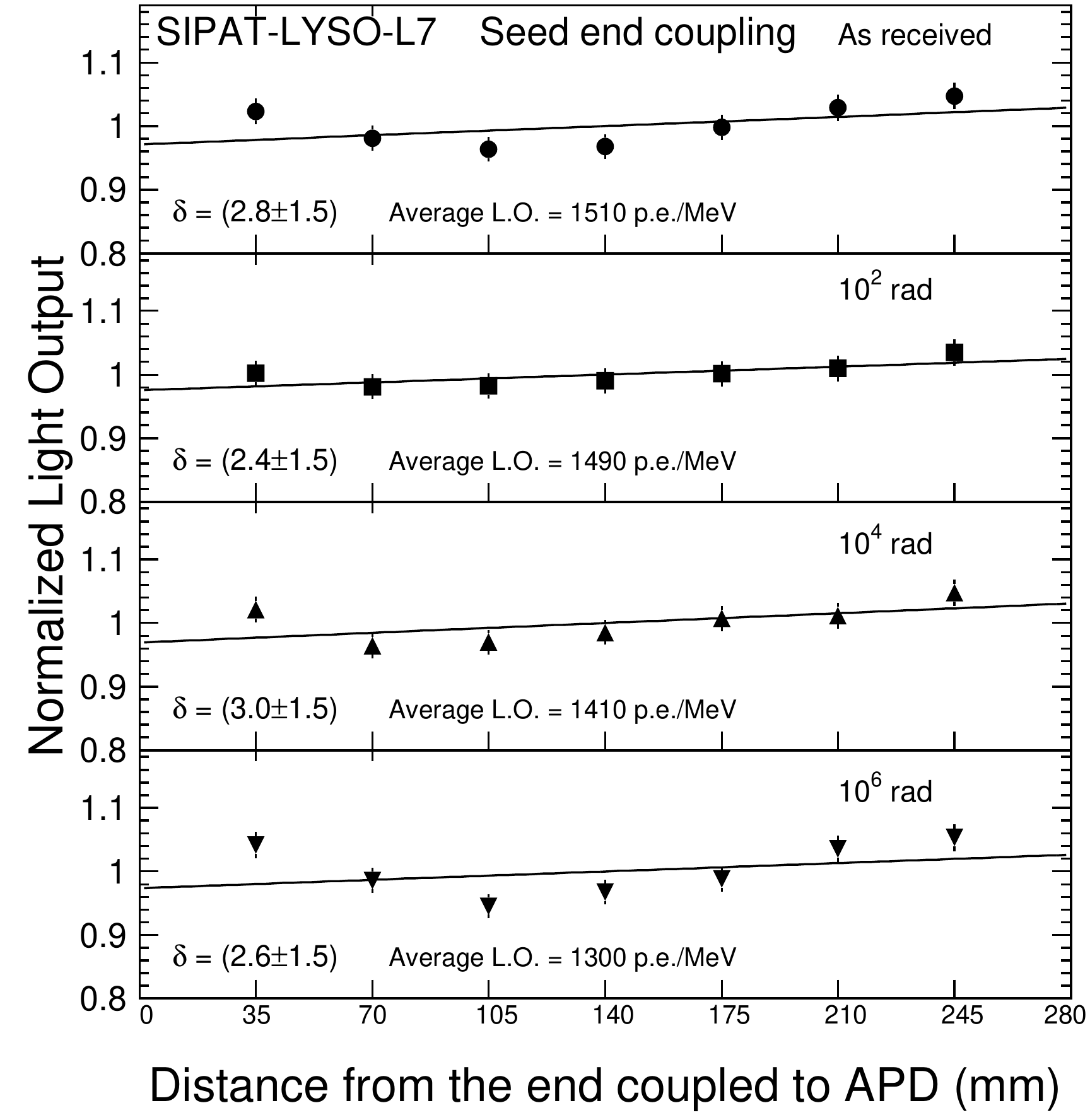}
\caption{Left: The pulse height spectra of 0.511 MeV $\gamma$-ray peaks (green) and corresponding Gaussian fits (red) measured by a Hamamatsu R1306 PMT are shown at seven points evenly distributed along SIPAT-LYSO-L7. Also  shown are the numerical values of the FWHM energy resolutions (E.R.). Right: Normalized light output and light response uniformity measured by two Hamamatsu S8664-1010 APDs, before and after $\gamma$-ray irradiations in several steps up to 1 Mrad are shown for SIPAT-LYSO-L7.}
\label{fig:sipat7}
\end{figure*}
Recently, a 28 cm (25 X$_0$) LYSO crystal (SIPAT-LYSO-L7) was grown at
SIPAT. 
This LYSO sample has consistent emission, adequate light response
uniformity and good radiation hardness against $\gamma$-rays up to 1
Mrad~\cite{bib:zhu_lyso10}. The left plot of Fig.~\ref{fig:sipat7}
shows the pulse height spectra measured by a Hamamatsu R1306 PMT at
seven points evenly distributed along SIPAT-LYSO-L7. The FWHM
resolutions obtained for 0.511 MeV $\gamma$-rays from the $^{22}$Na
source are about 12.5\%. This is quite good for crystals of such
length. The right plot of Fig.~\ref{fig:sipat7} shows normalized light
output and response uniformity measured by two Hamamatsu S8664-55 APD
before and after $\gamma$-ray irradiations with an integrated dose of
10$^2$, 10$^4$ and 10$^6$ rad. The degradation of the light output was
found to be about 13\% after 1 Mrad dose.
The light response uniformity of SIPAT-LYSO-L7 does not change even
after a 1 Mrad dose, indicating that its energy resolution will be
maintained~\cite{bib:zhu_RD}.

In summary, LSO and LYSO crystals are radiation hard crystal
scintillators, well-suited to the \superb\ rate and radiation
environment at forward polar angles.  These crystals are also likely
to find application in other severe radiation environments.

\subsubsection{Specifications, Production and Testing}

Following our extensive R\&D on LYSO crystals, the following
specifications have been defined for the procurement of high quality
LYSO crystals from various vendors for the \superb\ forward
calorimeter.

\begin{itemize}
  \item Dimension: +0.0/$-0.1$ mm.
  \item Longitudinal transmission at 420 nm: greater than $75\%$.
  \item FWHM energy resolution: $< 12.5\%$ for 0.511 MeV $\gamma$-rays
    measured by a Hamamatsu R1306 with Dow-Corning 200 fluid coupling
    at 7 points along the crystal.
  \item Light output will be required to be more than a defined percentage of a small crystal standard candle with air-gap coupling to the PMT.
  \item Light Response uniformity ($|\delta|$): $< 3\%$ measured with dual Hamamatsu S8864-55 APDs.
\end{itemize}

Crystals will be produced by various vendors. The total LYSO crystal
volume for the hybrid LYSO \superb\ forward calorimeter, which retains
the outer three CsI(Tl) rings) of the \babar\ endcap, is 0.24 m$^3$,
which is small compared to the volume of LYSO crystals grown for the
medical industry.

The following instruments are needed at each of the crystal vendors as
well as in the \superb\ crystal laboratory.

\begin{itemize}
  \item A station to measure crystal dimensions.  
  \item A photospectrometer with a large sample compartment to measure the longitudinal transmission along a 20~cm path.
  \item A PMT-based pulse height spectrometer to measure light output and FWHM energy resolution with 0.511 MeV $\gamma$-rays from a $^{22}$Na source.
  \item An APD-based pulse height spectrometer to measure light response uniformity with 0.511 MeV $\gamma$-rays from a $^{22}$Na source.
\end{itemize}

\tdrsubsec{Readout and Electronics}
\label{sec:EMCfwdReadout}

\tdrsubsubsec{APD Readout} The photosensors chosen for readout of the
LYSO crystals of the forward endcap are an independent pair of
10$\times$10 mm avalanche photodiodes (APDs). The APDs have several
advantages over photodiodes in this application: they are a better
match to the emission spectrum of LYSO, providing a quantum efficiency
integrated over the spectrum of 75\% (Fig.~\ref{fig:APD}); they
provide useful gain (of the order of 75) with low noise; and, as they
have a thinner sensitive region, they suffer less from the nuclear
counter effect, in which a hadron interaction in the sensitive region
produces a large anomalous signal. 

\begin{figure}[!htb]
\centerline{ \includegraphics[width=0.5\textwidth]{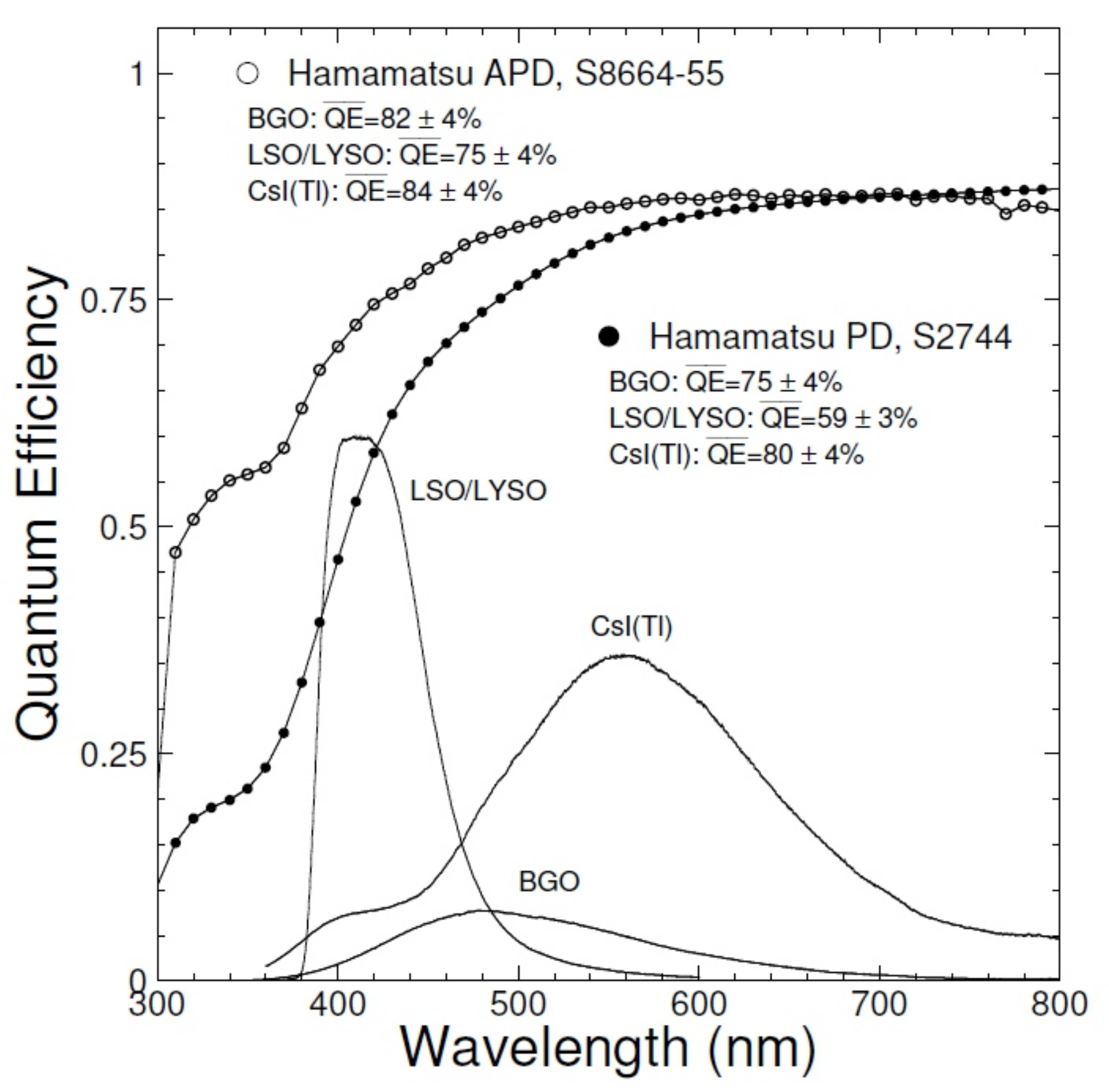}}
\caption {Quantum efficiency of a Hamamatsu APD and photodiode, together with the emission spectra of LYSO, BGO and CsI(Tl) crystals.}
\label{fig:APD}
\end{figure}

The gain with low noise of the APDs presents two additional
advantages: it allows a reduction of the shaping and integration time
constants, variables that
can be used optimize the signal-to-noise ratio in the presence of machine background; it also improves the signal-to-noise ratio for the signals used for calibration, allowing a crystal-by-crystal calibration (see Sec.~\ref{sec:EMC:FWDcalib}).

\tdrsubsubsec{Electronics Block diagram} 
The electronics of the
forward calorimeter uses the same components as the barrel
calorimeter.  In the hybrid design the 360 retained CsI(Tl) crystals
use exactly the same electronics as the barrel. The different
characteristics of LYSO crystals require a different preamplifier and
shaper, and a faster sampling time.

\tdrsubsubsec{Preamplifier and shaper} 
LYSO has a faster scintillation
decay time than CsI(Tl), typically 40 ns. The design of the
preamplifier uses the same active components as the CsI(Tl) case, but
with a different time constants in the passive network. Instead of a
transimpedance amplifier configuration as for the slower CsI(Tl),
in the LYSO case we use a CSP (charge sensitive preamplifier) with a
integration constant of 1~$\mu$s, which guarantees complete charge
collection.  The charge sensitive amplifier noise performance is
dominated by the low noise field effect transistor used as the first
stage, and by the APD capacitance of about 100~pf. The preamplifier
directly drives the shaper, which is also used as a buffer.  The
shaper circuit is the same as that of the CsI(Tl) circuit, with a
different passive time network. The shaper uses a 4 pole circuit
CR-RC$^{4}$ with a constant time of 100~ns, that produces a
near-Gaussian output shape for digital sampling. The shorter shaping
time avoids pile up and better uses the decay time characteristics of
the LYSO crystal. After the shaping, as with the CsI(Tl), there are
two separate drivers, one with unitary gain and the other one with a
gain of 32. Four signals, two for each independent photodiode, go to
the digitizer board.

\tdrsubsubsec{Digitization} The digitizer board is the same as the one
used for the CsI(Tl) crystal readout, but with a higher sampling rate
due to the shorter signal shape.  The sampling frequency chosen is
39.66 MHz (1/12 of the accelerator RF) using 32 samples in a total
time window of 806 ns.

\tdrsubsubsec{Requirements on mechanics} There are two major
differences with respect to the \babar\ setup as far as mechanics is
concerned.

On one side the power consumption of the electronics of each
individual crystal increases from 100~mW to approximately 180~mW per
crystal. This should not be a problem for the alveola where the
crystals is not changed.  In the 540 alveola where the single CsI(Tl)
crystal is replaced by 4 LYSO crystals there are additional mechanical
and cooling issues to be addressed. On one side, four channels instead
of one need to be lodged on the back of the crystals. To this aim one
can exploit the volume left free by the fact that the LYSO crystals
are shorter by about 10 cm. Nonetheless, dissipating 720~mW from cards
placed within the last 10cm of the alveola is a problem that still
needs to be addressed.

\tdrsubsec{Calibration}
\label{sec:EMC:FWDcalib}
\tdrsubsubsec{Initial LYSO calibration with sources} A goal of the
initial source calibration procedure is that the signal rate and the
signal-to-noise ratio with a typical radioactive source such as
$^{137}$Cs be sufficient to allow individual calibration of each
crystal with the readout device with which it will actually be
paired. As was the case with the \babar\ CsI(Tl) crystals, photodiode
readout of large crystals does not allow the use of sources for
calibration and setup. This is typically done with a reference
photomultiplier, with the results then convoluted with measurements of
the individually-calibrated photosensors. This procedure does not, of
course, fully account for the effects of surface oxidation of the
crystal or glue joint losses.  With APD readout of the LYSO crystals,
however, the response of the entire chain can be measured.

The full setup of each crystal assembly requires each crystal/readout
package to be individually adjusted to meet the uniformity
requirements {\it in situ} and the characteristics of each object to
be entered into a reference database. This involves appropriate
roughening of, typically, one crystal surface to conform to a light
collection uniformity specification, as discussed
in~\ref{sec:lyso_intro}. The output of this setup/calibration
procedure is then entered into a reference database, which serves as
the initial set of calibration constants for the calorimeter system.

The fully-assembled calorimeter is then calibrated with the circulated
Fluorinert system retained from \babar\ (see
Sec.~\ref{sec:EMC:BARcalib}) at appropriate intervals (one to four
weeks in the case of \babar). A substantial advantage of this approach
is that one obtains an individual pedestal and gain constant for each
crystal. A limitation is that the source is at a relatively low
energy, although it is at a higher energy than that of long-lived
radioactive sources.  Calibration with radiative Bhabhas can provide
calibration over the full energy range, but it requires development of
a complex matrix unfolding procedure, since high energy electrons
deposit shower energy in many crystals, not in a single crystal as in
the case of source calibration.  
if using crystals with intrinsic radioactivity, such as for instance
the LYSO. For such a crystal the 
properly designed to achieve the required accuracy in a sustainable
time.  Since LYSO is intrinsically radioactive, some of the endcap
crystals will have an irreducible background, and a configurable
trigger threshold is planned.  Estimates imply that the issue is
small. However a test with LYSO crystals using 6.13~MeV photons from
an $^{16}$O* state produced with neutrons from a $^{252}$Cf source
will be made.

\tdrsubsubsec{Temperature monitoring and correction} 
The characteristics of APDs place fairly stringent requirements on the
temperature control of the system, greater than those imposed by the
temperature variation of light output of the crystals, as well as on
the stability of the APD power supply voltage.

The Hamamatsu S-8664 APDs specified for the crystal readout have a
temperature coefficient of gain of $\Delta G/\Delta T$ of
2.5\%/$^\circ$C, while the LYSO light output varies
$-0.2\%$/$^\circ$C. A specification of an APD gain stability of $\pm
0.5\% $ requires knowledge of the temperature to $\pm 0.2^\circ$C.  
which is more stringent than that required by light output variation.
The CERN beam test demonstrated that a measurement of the calorimeter
temperature to 0.2$^\circ$C can be easily achieved. Note that this
specification is not on the stability of the temperature with time,
but rather on the knowledge of the temperature, as long as temporal
variations are slow.

The overall mechanical structure of the \babar\ calorimeter will be
retained. The entire calorimeter is surrounded by a double Faraday
shield and environmental enclosure composed of two 1mm-thick aluminum
walls. The diodes and preamplifiers are thus shielded from external
noise, and the slightly hygroscopic CsI(Tl) crystals reside in a dry,
temperature controlled nitrogen atmosphere.  The preamplifiers (2
$\times$ 50 mW/crystal) and the digitizing electronics ($\sim$3 kW per
end-flange) are the primary internal heat sources. The temperature was
monitored by 256 thermal sensors distributed throughout the
calorimeter. This system maintains the crystal environment at a
temperature of $20 \pm 0.5^\circ$C. Dry nitrogen circulation
stabilizes the relative humidity at $1 \pm 0.5\%$.

The gain stability of the APDs, for a gain of $\sim$75, with a reverse
bias voltage of $\sim$375~V, requires a voltage stability of better
than 1 volt. This requirement can be met by commercially available
computer-controlled high voltage supplies, such as those used for the
CMS calorimeter.

\tdrsubsec{Mechanical Structure}
\label{sec:EMC:FWDmec}

To reduce cost, the mounting structure of the \babar\ forward endcap
will be reused (Fig.~\ref{fig:EMC:FWDstruct}).  The \babar\ structure
was designed to minimize inactive materials, as was done for the
barrel, and to provide hermeticity and precision alignment at the
barrel/endcap interface.  The design is conceptually different from
that of the barrel. While the barrel remained unopened for the
lifetime of \babar, the endcap had to be designed to be capable of
rapid mounting and dismounting while retaining the precision of the
mating to the barrel itself. The forward endcap is a conic section,
with front and back surfaces tilted at 22.7$^\circ$ to the vertical,
matching the drift chamber endplate.  The total weight is
approximately four tons.  It is built in two monolithic D-shaped
halves, each containing ten carbon fiber crystal mounting structures,
to allow quick dismounting for access to the inner components of the
detector; It is supported from and precisely aligned with the
calorimeter barrel structure.

\begin{figure*}
\begin{center}
\includegraphics[clip=true,width=\textwidth]{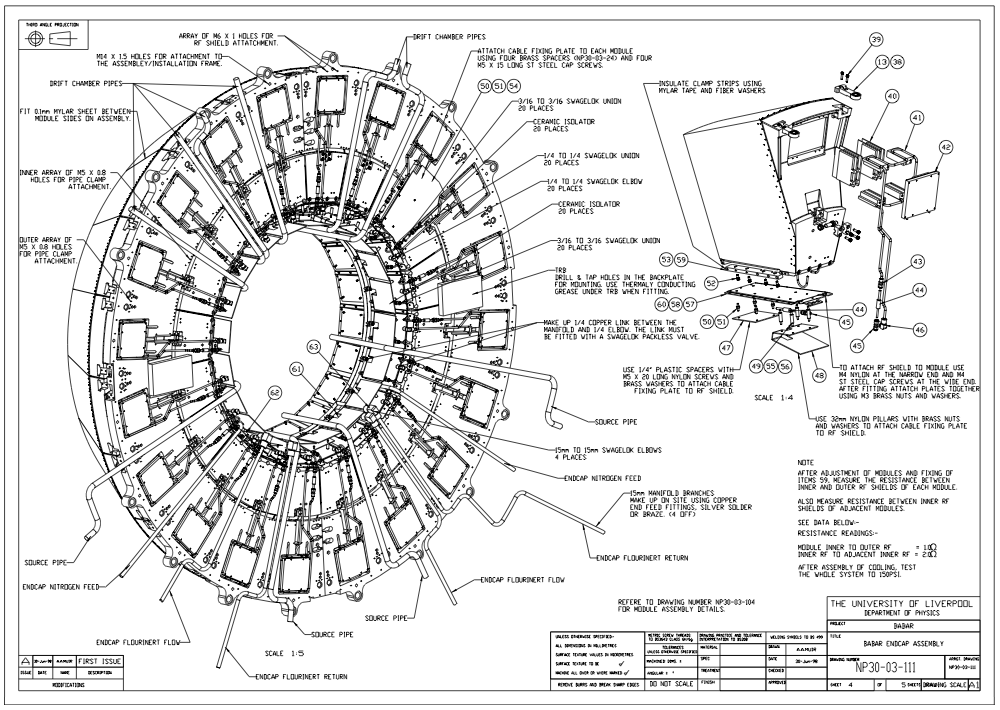}
\caption{Forward assembly drawing}
\label{fig:EMC:FWDstruct}
\end{center}
\end{figure*}

\tdrsubsubsec{Crystals} 

The \babar\ forward endcap structure contains
cells for a total of 900 CsI(Tl) crystals, arranged nine distinct
radial rings (3$\times$120, 3$\times$100, 3$\times$80), with
approximately the same crystal dimensions everywhere; total crystal
volume for the 900 compartments is 0.7 m$^3$.  In \babar, the inner
ring of 80 cells was used to hold lead shielding.

The outer eight rings hold the CsI(Tl) crystals. CsI(Tl) is a soft
material that cannot support substantial loads; for this reason the
CFC structure is designed to support the individual crystals without
transferring the load to neighboring cells; the CFC wall thickness of
250 $\mu$m wall thickness is chosen for adequate strength with minimum
radiation length.  The design and dimensions of these crystals are
described in Fig.~\ref{fig:EMC:FWDCsI}.

\begin{figure}[htbp!]
\begin{center}
\includegraphics[clip=true,width=0.5\textwidth]{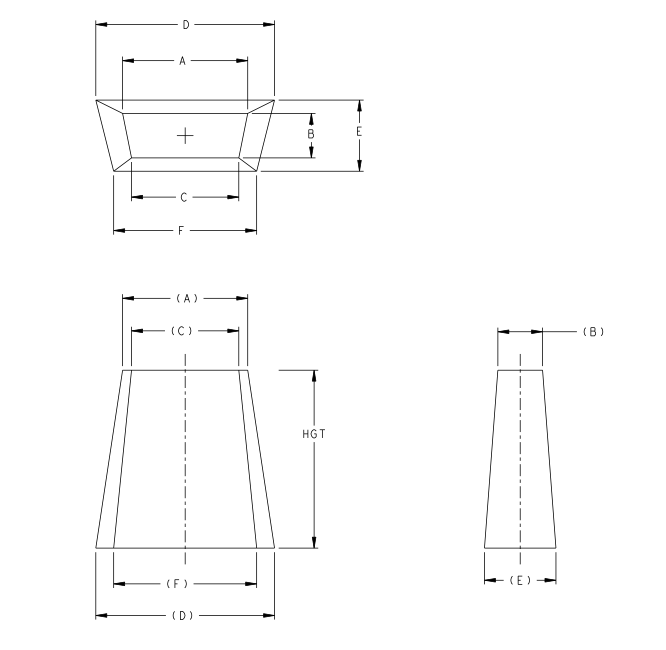}
\includegraphics[clip=true,width=0.48\textwidth]{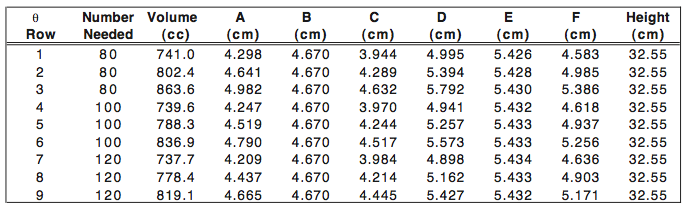}
\caption{Geometry of the projective crystals and their dimensions 
in the forward calorimeter of \babar.}
\label{fig:EMC:FWDCsI}
\end{center}
\end{figure}

The CsI(Tl) crystals, produced with a tolerance on transverse
dimensions of $\sim$225 $\mu$m, fit loosely into their compartments.
Each crystal is read out with 2 large areas photodiodes (Hamamatsu
S2744-08) glued onto a 1 mm polystyrene coupling plate, that itself is
glued to the rear crystal surface.  The crystals
(Fig.~\ref{EMC:fig:xtalpkg}) are wrapped with a double layer of Tyvek
(150 $\mu$m each) to improve the light yield, a layer of aluminum foil
for electrical shielding (75 $\mu$m) and a layer of mylar (12 $\mu$m)
for insulation and mechanical protection of the Al foil. A small
aluminum box covering the photodiodes at the back of crystals
contains the preamplifiers.

In the hybrid scheme, we will populate the inner six rings of the
endcap structure with LYSO crystals. As the Moli\`ere radius of LYSO
is approximately half that of CsI(Tl), the replacement will be four
LYSO crystals for each compartment. In this way, the same structure is
used, with four LYSO crystals per CFC cell. LYSO is
a very mechanically strong crystal; no additional support is required
within the compartment. Were further cost savings in the initial
implementation to be necessary, additional rings of CsI)Tl) may be
retained from the \babar\ endcap. A future upgrade of the detector
could include replacing more rings of CsI(Tl) with LYSO.

\tdrsubsubsec{Modules} 

The endcap consists of twenty individual modules
(Fig.~\ref{fig:EMC:FWDmodule}). The geometry is essentially
projective, with the compartments aligned so that the crystal axes
point to an axial position 5 cm from the interaction point. In order
to preserve optimum light collection and both spatial and energy
resolution, similar-sized crystals have been used, arranged in a
layout of nine rings of trapezoidal shaped compartments grouped in
three super-rings.

\begin{figure*} [hpt!]
\begin{center}
\includegraphics[clip=true,width=\textwidth]{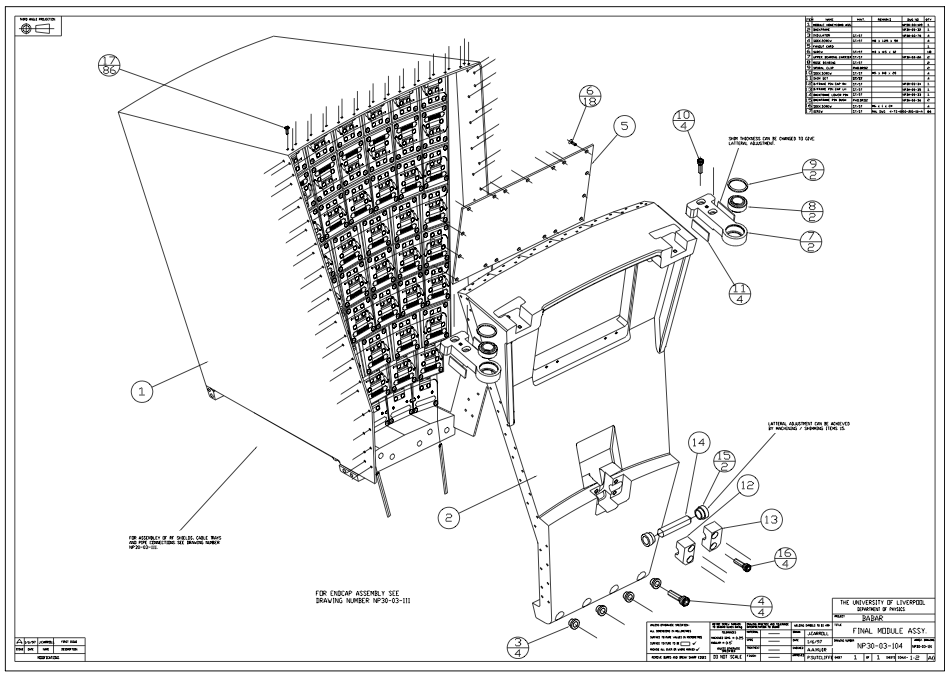}
\caption{Forward module assembly drawing}
\label{fig:EMC:FWDmodule}
\end{center}
\end{figure*}

\begin{figure}[bp]
\begin{center}
\includegraphics[clip=true,width=0.49\textwidth]{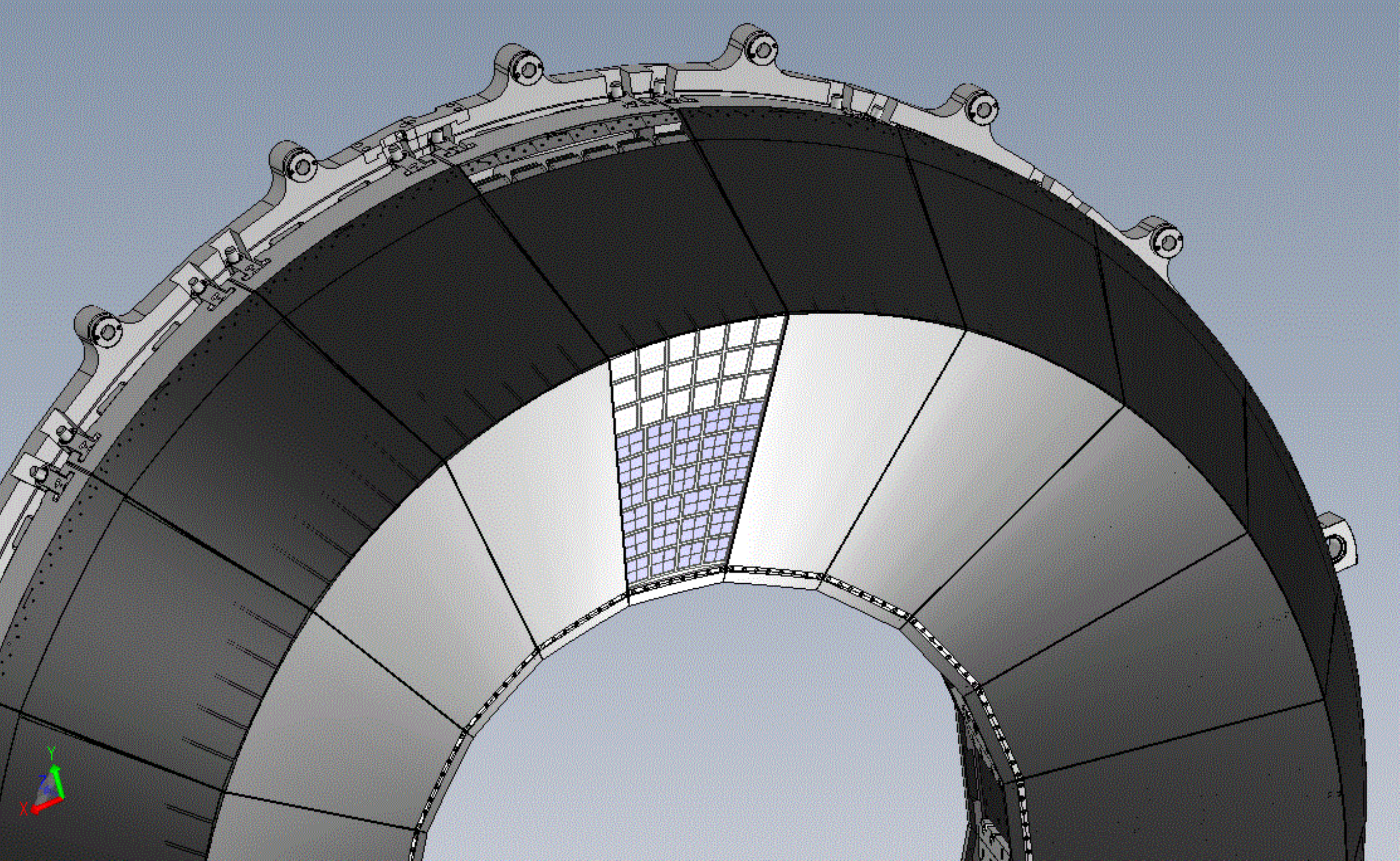}
\caption{Layout of a module in the forward endcap in the hybrid plan.}
\label{fig:EMC:FWDhybrid}
\end{center}
\end{figure}

As described above, for \superb\ we populate the six inner rings of
the endcap with LYSO, retaining the existing outer three rings of
CsI(Tl) (Fig.~\ref{fig:EMC:FWDhybrid}).

\tdrsubsubsec{Installation}

The installation tools and instruments used at \babar\ to install the
twenty modules and two D structures still exist at SLAC. Detailed
procedures for moving and installing the forward calorimeter are
derived from \babar\ are being prepared.
Fig.~\ref{fig:EMC:FWDinstallation} shows the existing Installation
Bridge, able to move individually each one of the two monolithic
endcap halves.

\begin{figure*} [htbp!]
\begin{center}
\includegraphics[clip=true,width=\textwidth]{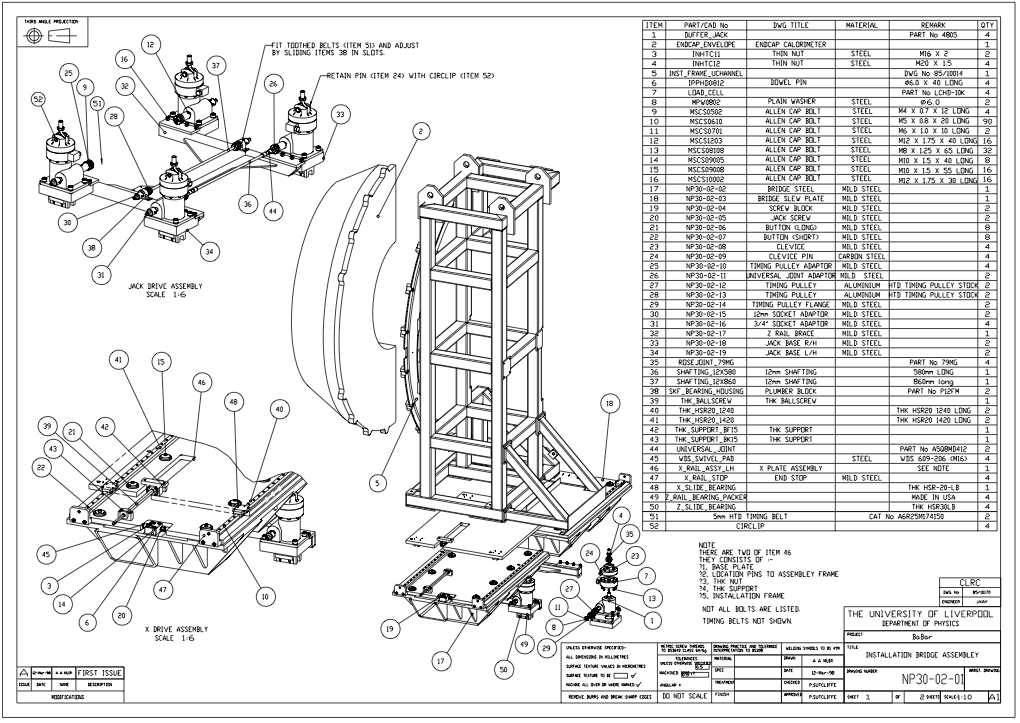}
\caption{The forward endcap installation tooling. }
\label{fig:EMC:FWDinstallation}
\end{center}
\end{figure*}

\tdrsubsubsec{Refurbishment of the \babar\ structure}

As with the \babar\ barrel, the endcap will be disassembled to the
module level for transport. The alternative of transporting each half
with modules installed carries mechanical risks, but will be
investigated further.

For the CsI(Tl) rings that are retained, each crystal diode pair and
preamps will be tested and measured for signal strength, before and
after shipment; repairs will be made as necessary. The testing in
Italy may be done at a remote site at the module level, for both the
CsI(Tl) and LYSO crystals. Final assembly of the endcap and
installation into the detector will be done at Tor Vergata.

The electronics for the retained Cs(Tl) rings will be refurbished as
for the barrel (Sec.~\ref{sec:EMCbarrelElec}.  The LYSO readout with
APDs requires new electronics (\ref{sec:EMCfwdReadout}).

\tdrsubsubsec{Tests of the Spare FWD modules}

In order to study the changes required to accommodate the new LYSO
crystals, two spare \babar\ endcap structural modules have been
shipped to Italy. Tests and measurements on them will be necessary to
define a procedure for the insertion of the new LYSO crystals, as well
as to understand any problems (in mounting the electronics, for
example), and to validate the existing technical information,
particular regarding the strength of the CFC compartments.


\tdrsubsec{Beam Tests}

Two beam tests have been performed with a prototype LYSO matrix,
one at CERN in October 2010 and one at the Beam Test Facility (BTF)
in Frascati in May 2011. 

\tdrsubsubsec{Description of the apparatus}

In both cases, the prototype matrix was composed of 25 LYSO crystals
of pyramidal shape with dimensions 2.3 cm$\times$2.3 cm$\times$22 cm
inserted into a support structure, described in detail in
Sec.~\ref{sec:EMC:FWDmec}, that was fabricated at the RIBA company
(Faenza, Italy).  To improve light output uniformity along the length
of each crystal, a 15~mm wide black band was painted at the end of its
smallest face. The area of the face not covered by the APD (or PIN)
was painted with a reflective white paint.  The mounting structure is
made of glass fiber, where each cell consists of the wrapping of
several layers of pre-impregnated glass fiber epoxy composite, a
single inner layer of a reflecting aluminum foil and a composite
bottom insert. Electrical connection is made to copper foils 35~$\mu
m$ thick at the back each compartment.  Between one cell and the other
there is a nominal thickness of 200~$\mu m$, while the external walls
are 135~$\mu m$ thick. Figure~\ref{fig:fwdEMCmatrix} shows a picture
of the test beam structure with one row of crystals inserted.

\begin{figure}[htbp]
\begin{center}
\includegraphics[clip=true,width=0.5\textwidth]{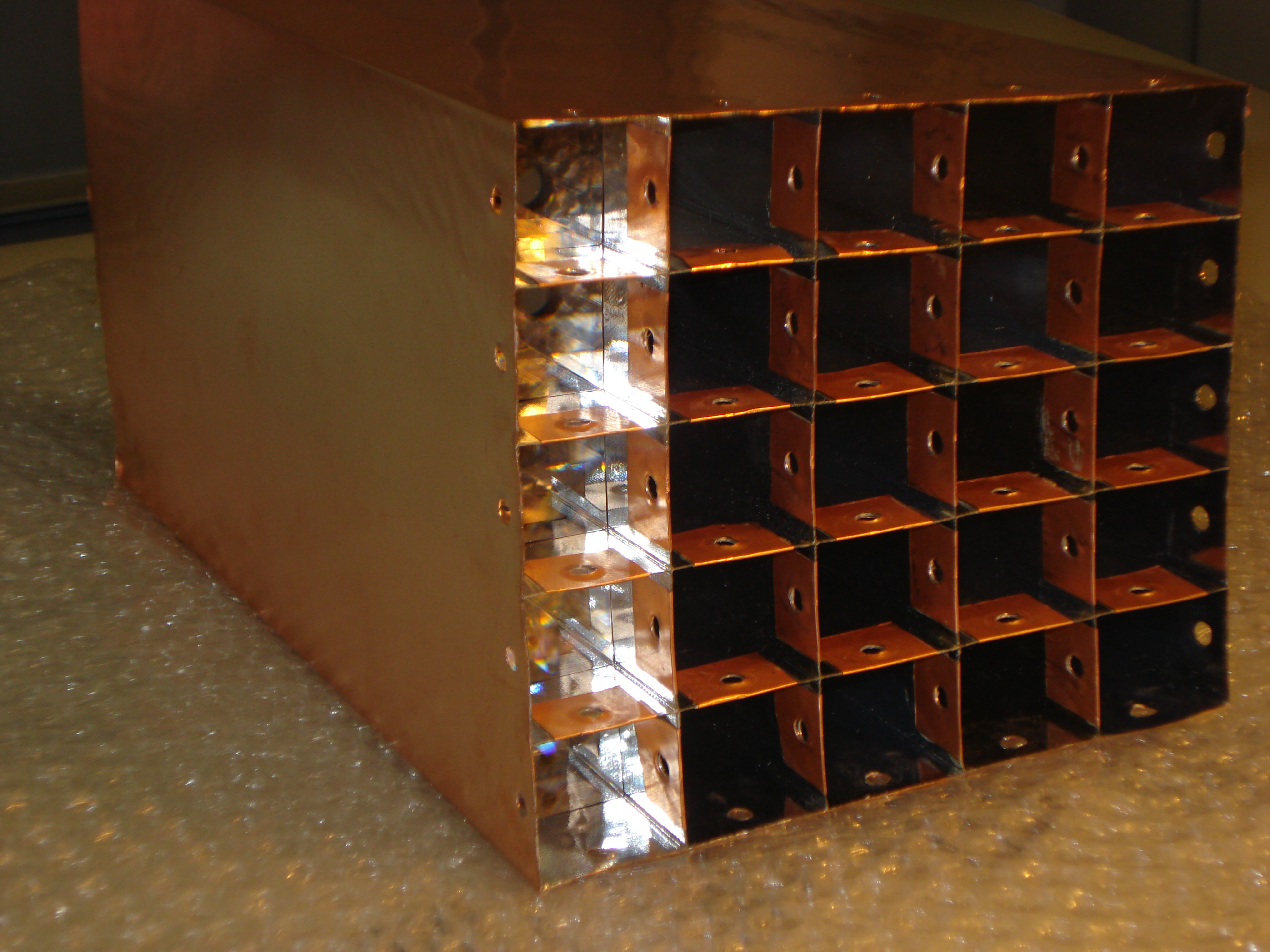}
\caption{The test beam mechanical structure with one row of
  LYSO crystals installed.\label{fig:fwdEMCmatrix} }
\end{center}
\end{figure}

Twenty of the crystals were read out with an Avalanche Photodiode
(APD) in both tests; the remaining five were read out with PIN Diodes
at CERN and with APDs at the BTF.  The readout chain was composed of a
front end board (VFE) containing a charge shaper preamplifier (CSP); a
shaper range board that further shapes the signal, and then divides
the signals according to the energy
(Fig.~\ref{fig:fwdEMCelectronics}). Two different ranges are foreseen
in the treatment of the signals, for energies lower than 200 \mev and
for energies greater than 200 \mev, although in the beam tests the
amplification was adjusted to use only a single range. A 12 bit CAEN
ADC digitized the analog outputs.
The APDs were biased at 308~V to avoid saturation effects.

\begin{figure*}[tbp]
\begin{center}
\includegraphics[clip=true,trim = 5 15 0 0,width=0.7\textwidth]{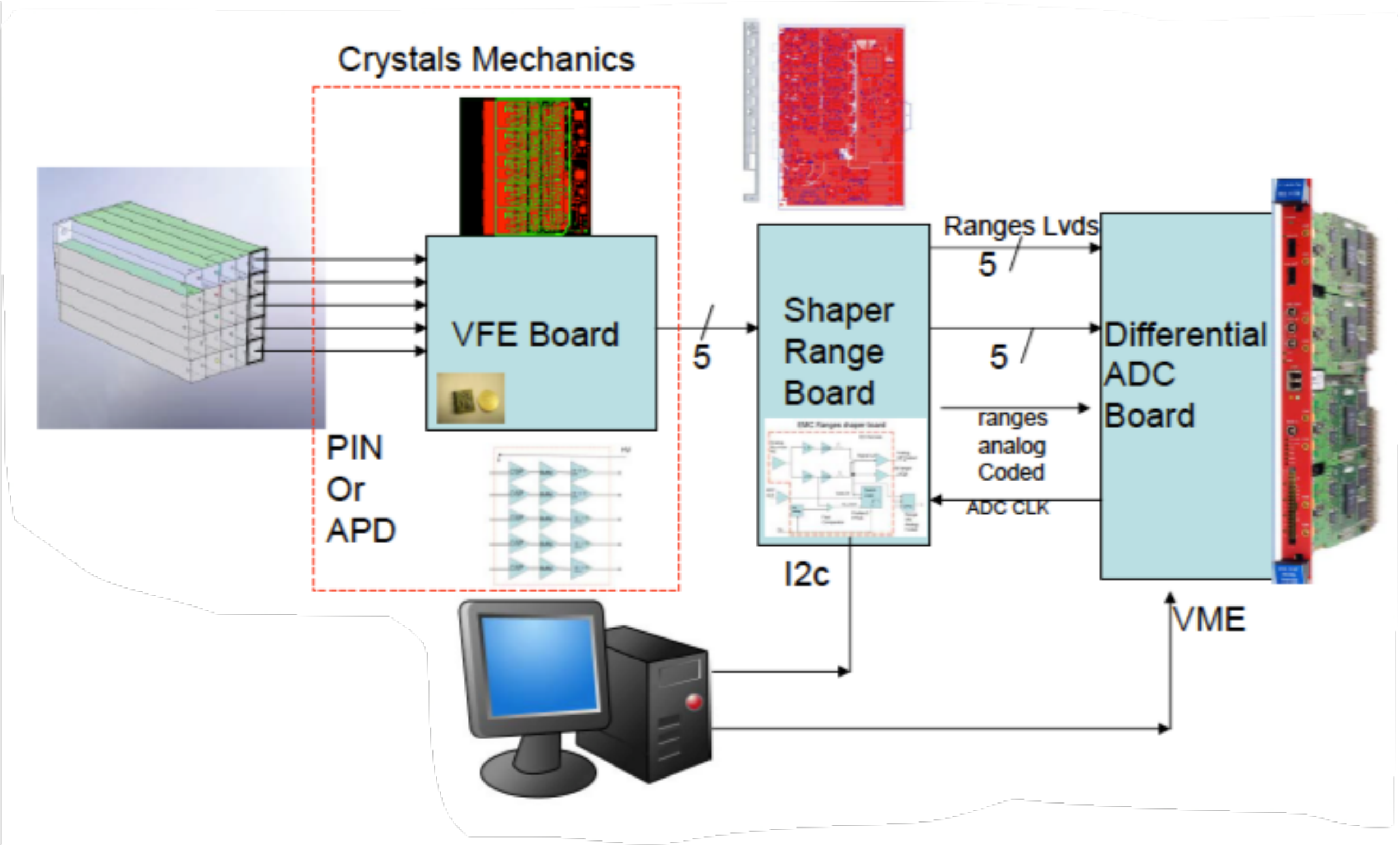}
\end{center}
\caption{Schematic view of the electronics chain for the forward EMC
  beam tests.}
\label{fig:fwdEMCelectronics}
\end{figure*}

\tdrsubsubsec{Description of the beams}

The beam test at CERN took place at the T10 beam line in the East
Area. The beam was composed mainly of electrons, muons and pions
produced by protons on aluminum and tungsten targets. The composition
of the beam is highly dependent on the energy: for electrons it ranges
from 60$\%$ at 1\gev to 1$\%$ at 6\gev. The maximum energy reachable
at this beam line is 7\gev with a nominal momentum spread $\Delta p/p
\simeq 1\%$. The distance between the end of the beam line and the
crystal matrix was about 15~m.  The event rate was of the order of
1~Hz.  Figure~\ref{fig:fwdEMCsetup} shows the experimental setup used
at CERN, composed of a Cherenkov detector for electron-pion
discrimination, two scintillators (finger counters) 2$\times$2~cm$^2$,
the box containing the matrix and the VFE boards.

\begin{figure}[htb]
\begin{center}
\includegraphics[clip=true,width=\linewidth]{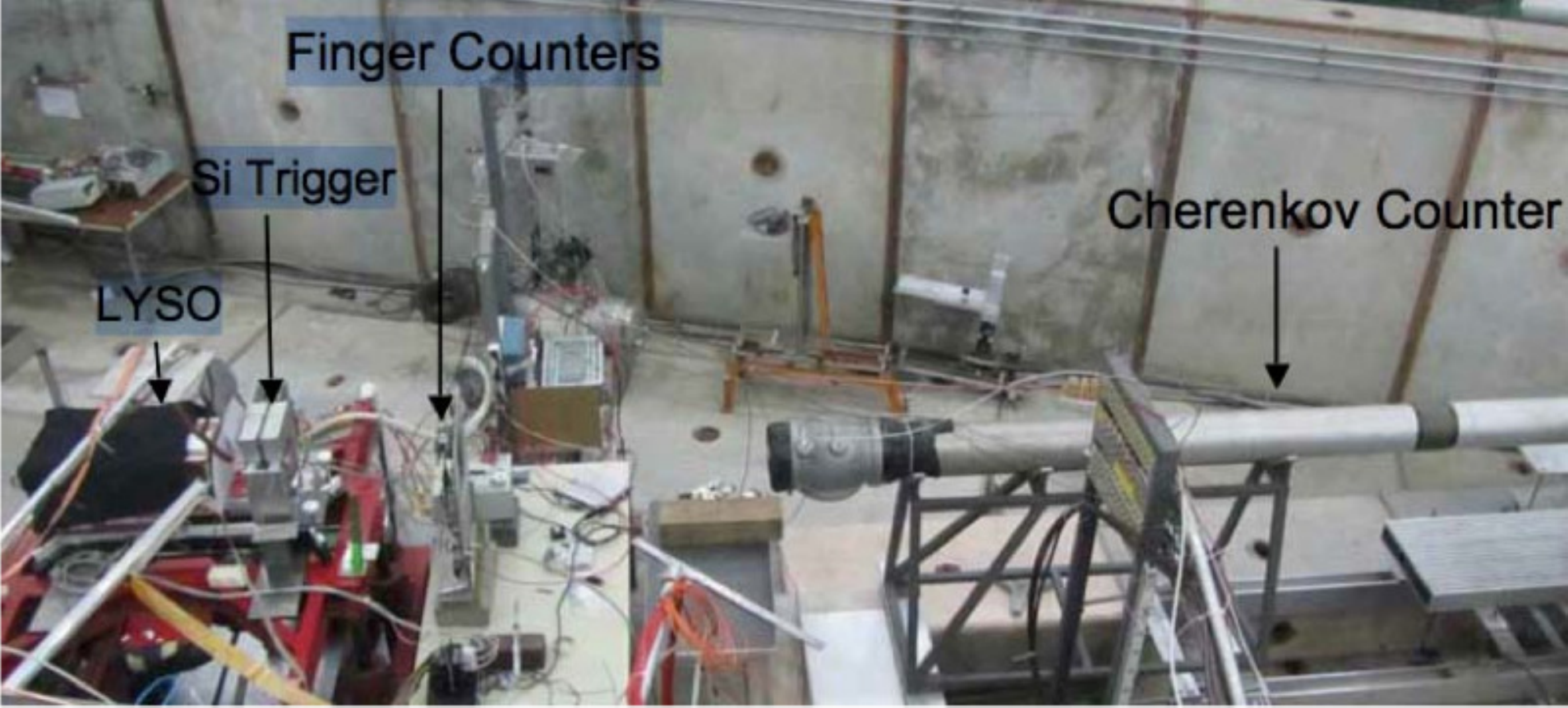}
\end{center}
\caption{Schematic view of the test beam setup used at CERN.}
\label{fig:fwdEMCsetup}
\end{figure}

Signals from the two finger counters formed the trigger. The Cherenkov
detector information was stored in order to select Minimum Ionizing
Particles (MIP) offline, and was also used for calibration and to
prescale the triggers.

The Beam Test Facility at the INFN Frascati Laboratory (LNF) is
part of the $\Phi$Factory, Da$\Phi$ne, composed of a linear
accelerator LINAC, a spectrometer, and two circular accelerators of
electrons and positrons at 510~\mev. The LINAC that fills the rings
also supplies the test beam line at the BTF. The pulsed beam of the
LINAC circulates electrons up to 800~\mev at a maximum current of 
550~mA/pulse and positrons at a maximum energy of 550~\mev with a current
of 100~mA/pulse. The typical pulse duration is 10~ns, with a
repetition rate of 50~Hz. A bending magnet selects the electron
momentum, and a system of slits controls the particle flux. The beam
energy spread is 1$\%$ at 500~\mev. The setup of the beam test of the
crystal matrix at the BTF is shown in
Fig.~\ref{fig:fwdEMCsetup-BTF}.

\begin{figure}[htb]
\begin{center}
\includegraphics[trim=70 70 85 50,clip=true,width=\linewidth]{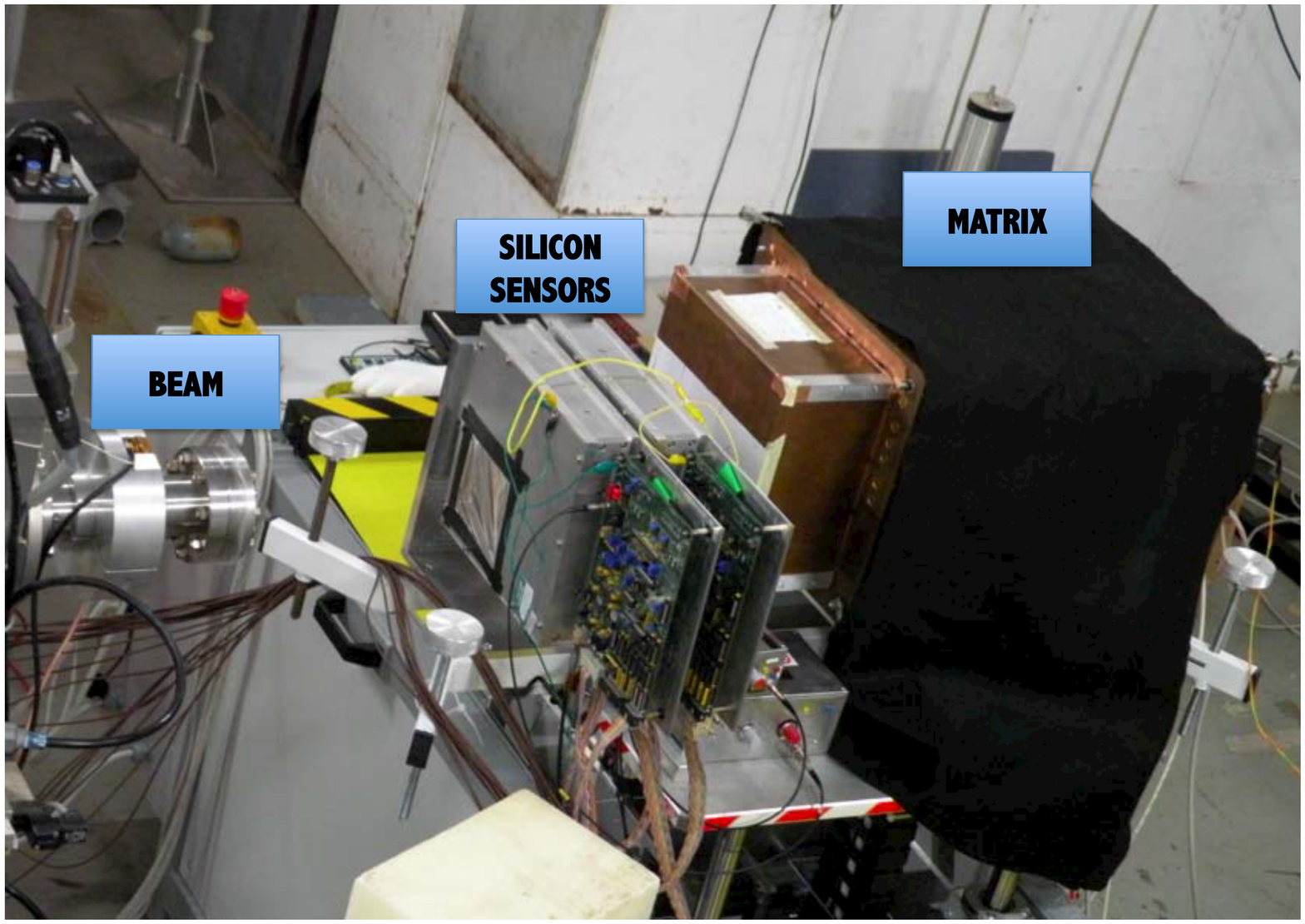}
\end{center}
\caption{Schematic view of the test beam setup used at the BTF LNF.}
\label{fig:fwdEMCsetup-BTF}
\end{figure}

The setup drawing shows the end of the electron beam line, four planes
of silicon strip detector (two measurements in $x$ and two
measurements in $y$) and the box containing the crystal matrix and the
VFE boards. As mentioned above, at the BTF all the crystals are
equipped with APDs. The gain of the VFE was changed from 0.5 to 1, and
an amplification factor was also introduced.  To control the position
of the beam with respect to the matrix, a detector composed of 
$16 \times 16$
3~mm scintillating fibers was used. The LINAC RF pulse signal (25~Hz)
was included in the trigger.

Based on the performance at BTF, we suspect that the beam energy spread
at the CERN facility was larger than the specification.  Thus, we use
the CERN TB data only to study the linearity at high energy;
resolution studies are performed exclusively on the BTF data.

\tdrsubsubsec{Description of data and calibration}

For each triggered event, the output of the readout is 384 samples of
the waveforms of the 25 channels.  The signal amplitude in each
channel was defined as the maximum of the waveform, extracted from a
Gaussian fit to the sampling distribution, with pedestal subtracted.
The pedestal is calculated for each crystal by averaging the first 60
samples on a reference run.  The pedestal-subtracted amplitude is
taken as the measurement of the energy deposit, provided it is above a
threshold, chosen to be three times the {\it rms} noise, whose value
is determined from a run taken with random triggers where no signal is
present.  After calibration, the energy of the electron that initiated
the shower was determined by summing all the energy deposits in all
crystals (``cluster energy'').

\begin{figure}[htbp]
\includegraphics[trim=30 0 45 40,clip=true,width=\linewidth]{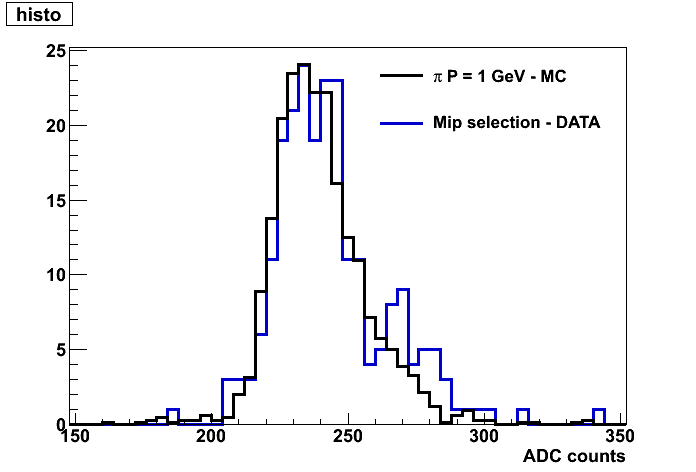}
\caption{Comparison between data and MC of the energy deposited in the
  MIPs sample. The hypothesis that after the selection, the beam is
  dominated by pions is made. }
\label{fig:EMC:FWDMIPS_MCcomp}
\end{figure}

At the CERN test beam hadrons traversing the crystals horizontally
were selected as Minimum Ionizing Particles (MIPs) by requiring no
significant signal in the other crystals and a signal consistent with
a hadron in the Cherenkov detector. Profiting from the fact that MIPs
deposit a constant energy in the crystal regardless of their energy,
the amplitude spectra of each crystal was fitted to extract the most
likely value. We determined by GEANT4 simulation the expected energy
deposit in each crystal (Fig~\ref{fig:EMC:FWDMIPS_MCcomp}); the
corresponding calibration constants could then be extracted.

At the BTF test beam, where no hadrons were available, the relative
intercalibration constants were obtained from the electron sample
itself. The relative cluster energy resolution was minimized by
floating a constant for each crystal except the central one. The
overall energy scale was then determined from the knowledge of the
beam energy. This procedure was applied on a small fraction of the
runs where the electrons were approximately 500\mev (the highest
energy reached in the tests) and the corresponding constants used in
all other runs. This intercalibration was also cross-checked using
cosmic-ray data obtained with an {\it ad hoc} trigger derived from two
plastic scintillator pads positioned above and below the crystal
matrix. The channels where there is a significant difference are those
where the electron data shows very little energy, because they are far
from the center of the matrix. In such cases, which have little impact
on the resolution studies since they contribute little to the total
energy measurement, the calibration constants estimated from the MIPs
are used.


\tdrsubsubsec{Electronics noise measurements}

The first information we could extract from the data are the
characteristics of the electronic noise.  From the signal distribution
in a random trigger run at the BTF we estimated
(Fig.~\ref{fig:EMC:FWDnoise}) that apart from two channels, the
noise is on average 2~ADC counts.  After applying the calibration, this
noise corresponds to approximately 0.4~MeV. To understand if there were
resonant components, we analyzed the power spectrum (PS) of the noise of
each crystal $i$ using waveforms acquired with a random trigger:
\begin{equation}
{\rm PS}_i(\omega_k) = \langle n_i(\omega_k)n_i^*(\omega_k) \rangle ,
\label{eq:noiseps}
\end{equation}
where $n(\omega_k)$ is the k-th component of the discrete Fourier transform (DFT) of a noise waveform,
and $<>$ denotes the average taken over a large number of waveforms.

The estimated power spectrum of a representative channel is shown in
Fig.~\ref{fig:EMC:FWDnoise}, where it can be seen that the dominant
source of noise is in the range 0--8~MHz, which corresponds to the
bandwidth of the shaper. Sources of noise occurring after the shaper
give a negligible contribution; the dominant contributions occur
before the shaper transfer function filter. Correlations among the
noise in the crystals was also estimated and found to be negligible.

\begin{figure}[htbp]
\begin{center}
    \includegraphics[clip=true,width=\linewidth]{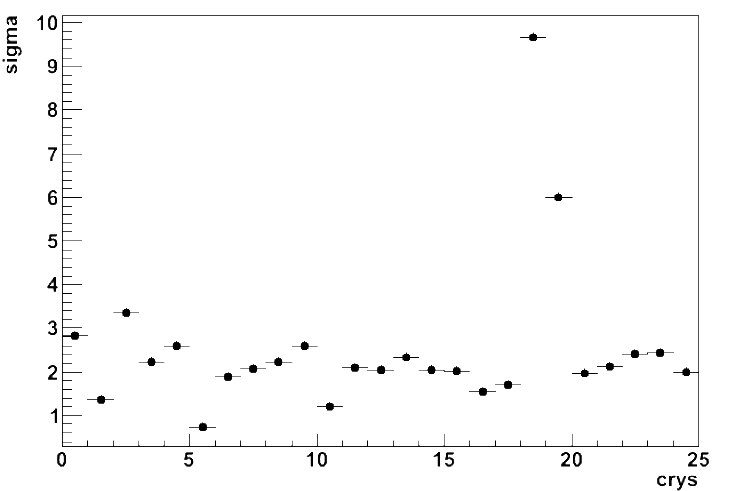}  
    \includegraphics[clip=true,trim = 0 0 40 30, width=\linewidth]{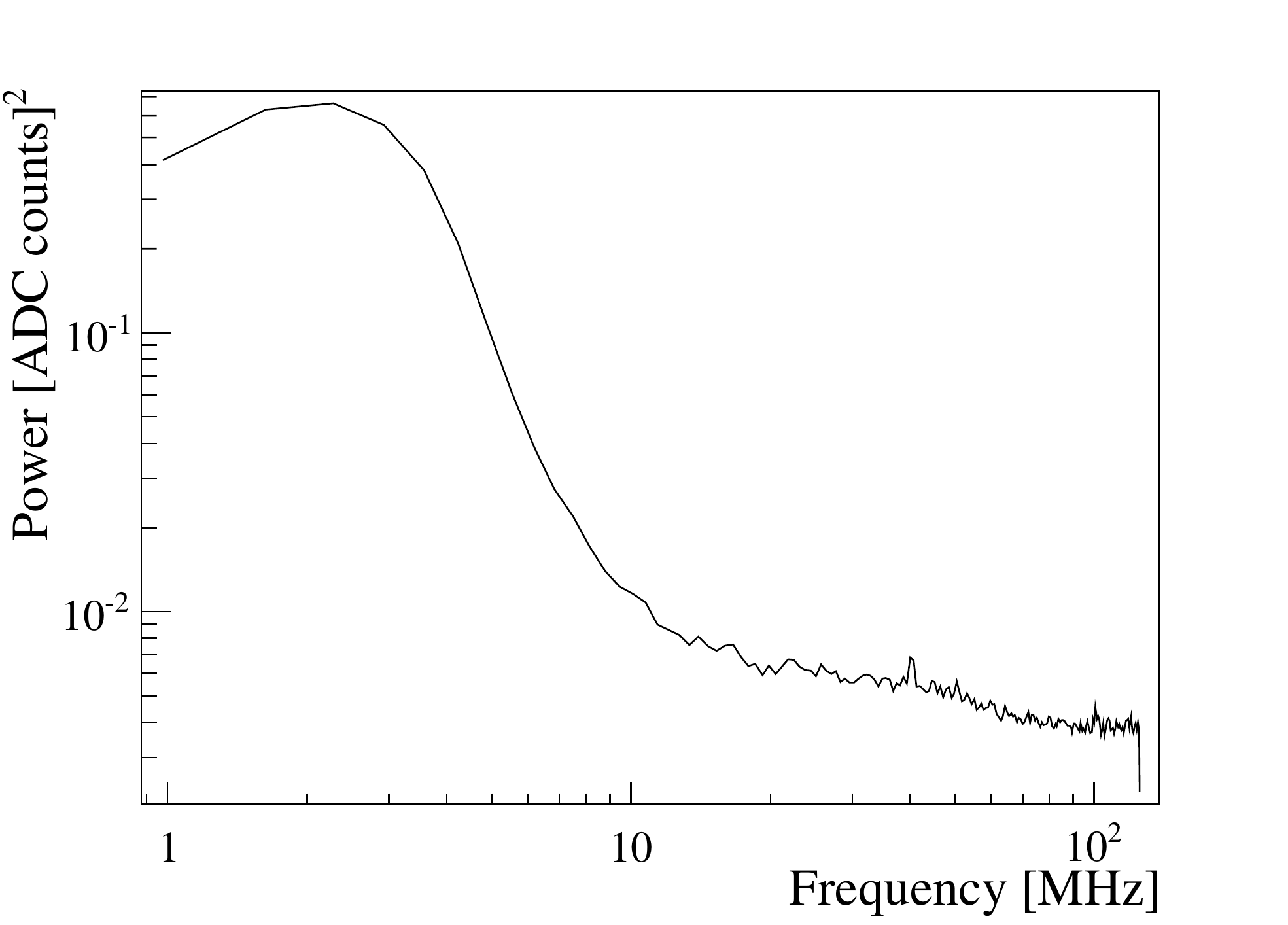}  
\caption{Top: Noise ({\it rms}) for each channel of the BTF test beam calorimeter. Bottom:  power spectrum of a representative channel.}
\label{fig:EMC:FWDnoise}
\end{center}
\end{figure}

\tdrsubsubsec{Temperature corrections}

A temperature dependence of several percent per degree is expected in
the light yield of the LYSO crystals and in the gain of the APDs. At
the CERN test beam the position of the MIP peak as a function of the
temperature measured by sensors placed on the rear of the crystals has
been used to extract the temperature correction: $E_{\rm corr}=E_{\rm raw}/(1-p_0 (T-T_0))$
where 
$E_{\rm corr}$ and $E_{\rm raw}$ are the corrected and uncorrected energies, 
$p_0=(2.8\pm 0.2)\times 10^{-3}$, and $T_0=34$~K. 


This correction proved irrelevant at the BTF test beam where the room
was climate-controlled and the temperature variation was negligible.

\tdrsubsubsec{Results}

In order to determine the linearity of the crystal matrix response and
the energy resolution, the distribution of the cluster energy was
fitted for each beam configuration with a Crystal Ball function. The
mean value of the Gaussian component is taken as the estimate of the
energy, while the resolution is evaluated from the FWHM of the entire
distribution.

Figure~\ref{fig:linearity-CERN} shows the mean value as a function of
the energy for the data collected during the CERN beam test.  The
linearity of the response is thus clear.

\begin{figure*}[htbp]
  \begin{center}
    \includegraphics[width=0.49\linewidth] {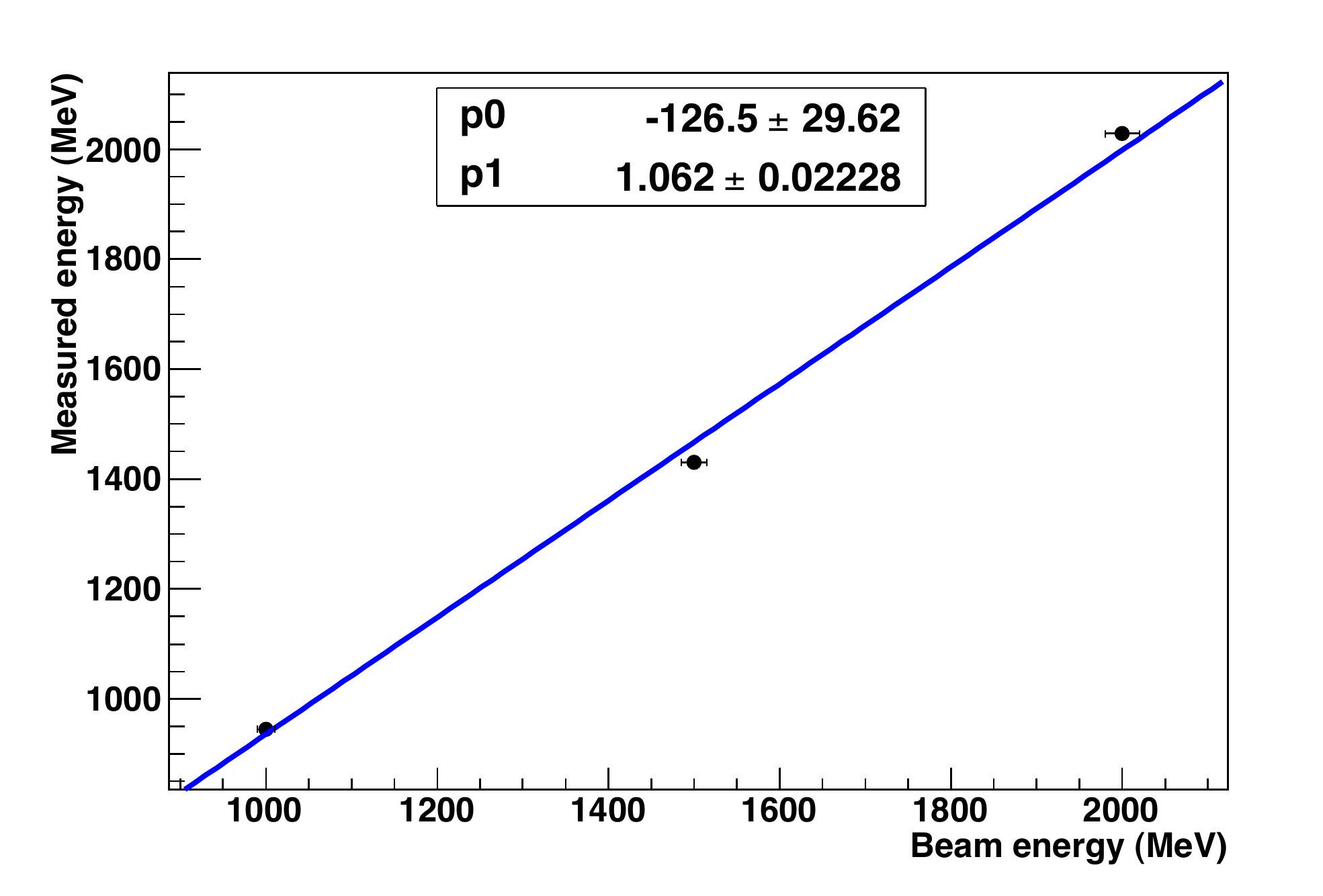}
    \includegraphics[width=0.49\linewidth] {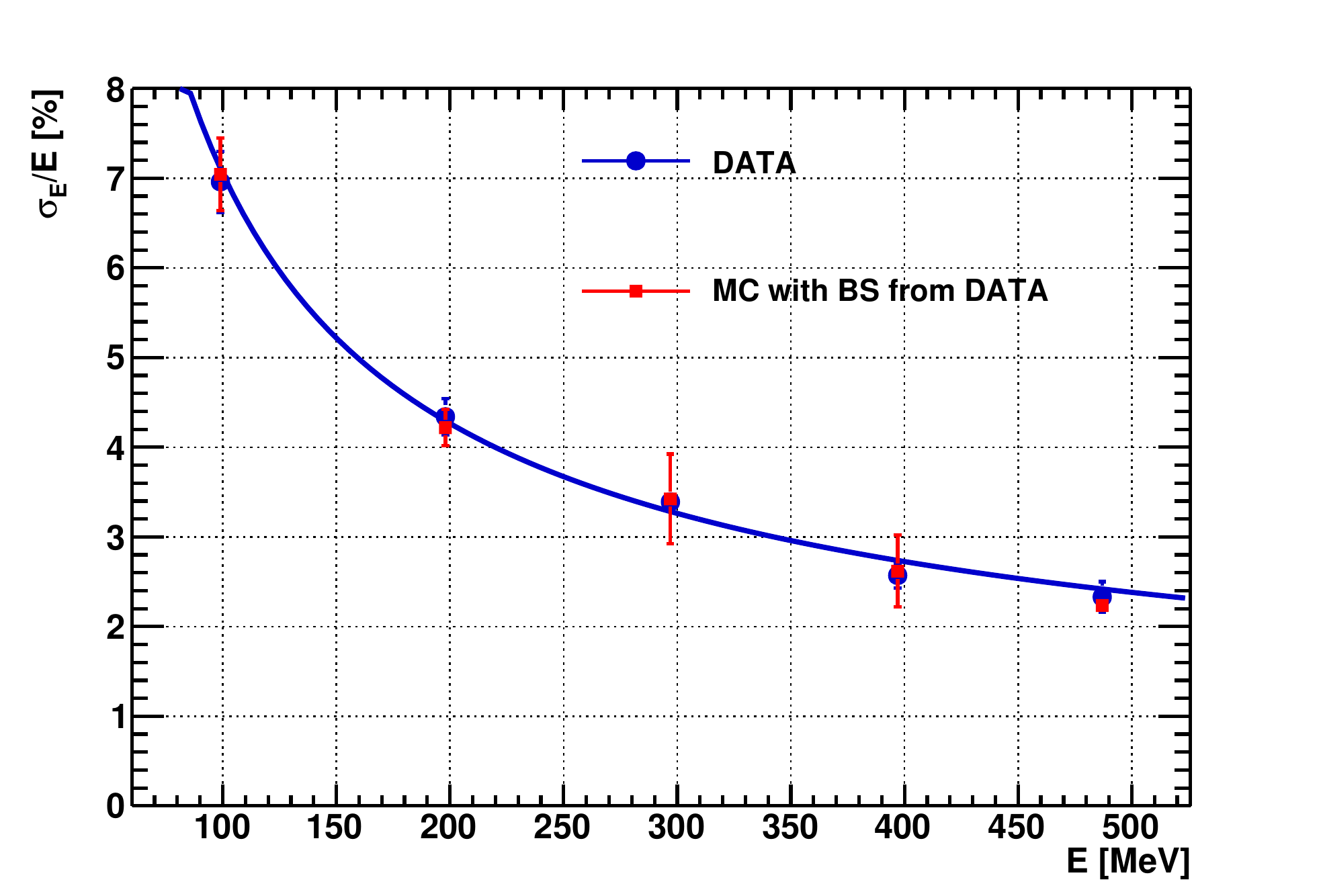}
  \end{center}
  \caption{Left: Measured mean value of the energy as a function of
    the beam energy at the CERN test beam. Right: relative resolution
    as a function of the deposited energy on data and MC simulation.
    Error bars in MC contains the contribution of the beam energy
    spread evaluated from data.}
\label{fig:linearity-CERN}
\end{figure*}

At the BTF, the beam was configured to allow as many as three
simultaneous electrons impacting the crystal matrix within the DAQ
dynamic range. This allowed us to have more than one resolution
measurement for each beam energy configuration
(Fig.~\ref{fig:linearity-CERN}). This in turn allowed us to estimate
the dependence of the beam spread on the energy by comparing
measurements of the same amount of deposited energy, but obtained with
different beam energies.  The intrinsic resolution obtained by
subtracting the measured beam spreads in quadrature from the average
resolutions is also shown in Fig.~\ref{fig:linearity-CERN}.

These intrinsic resolutions were then compared with the results of a
full simulation of the experimental setup
(Sec.~\ref{sec:EMC:simulation}). There is good agreement, with a
resolution smaller than 2\% at 1~GeV.

\tdrsubsec{Alternatives}
\label{sec:EMC:FWDalt}
We have described above the baseline design adopted for the Forward
EMC: a hybrid design that retains the existing \babar\ mounting
structure and the outer three rings of CsI(Tl) crystals, and populates
the six inner rings of the structure with LYSO crystals.

The best technical choice for the forward calorimeter would have been
to construct a totally new device using LYSO crystals throughout, with
a new thinner mechanical support structure to reduce non-sensitive
material. We have pursued extensive R\&D on this option. Cost
considerations do not, however, permit us to choose this device as the
baseline; it remains an alternative should adequate funding
materialize.  This section describes the original full LYSO design as
well as other alternatives that have been considered.

\tdrsubsubsec{Full LYSO calorimeter}

The driving design principles are the same as were used for the
\babar\ Electromagnetic Calorimeter (see Sec.~\ref{sec:EMC:FWDmec}),
the biggest changes resulting from the use of higher density LYSO
crystals that have a Moli\`ere radius (2.07~cm) nearly a factor two
smaller than that of \hbox{CsI (Tl)} (3.57~cm). The radiation length
is also significantly smaller (1.14~cm vs 1.86~cm). This
changes both the optimal transverse dimensions of the cells and the
position of the center of gravity of the modules, thus significantly
altering the original \babar\ design.  A detailed report of the
studies that have been carried out can be found in
Ref.~\cite{SuperB:EMC:FWD:Mech}. We stress herein only the relevant
points.

The new support structure would be subdivided into four rings in theta
(Fig.~\ref{fig:EMC:FWDrings}) each composed of 36, 42, 48, and 54
modules respectively. The cells are designed in order to keep the
dimensions of the front of each crystal as close to identical as
possible.

\begin{figure}[htbp]
\begin{center}
\includegraphics[clip=true,width=0.4\textwidth]{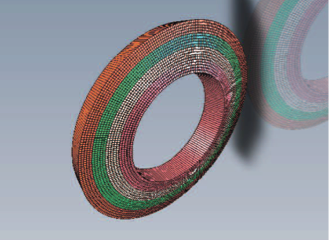}
\caption{Overview of the structure of the FWD EMC. }
\label{fig:EMC:FWDrings}
\end{center}
\end{figure}

The detailed placement of electrical, calibration and cooling services
was not studied in detail, but the \babar\ solutions would be
well-suited to the new geometry; the reduced crystal lengths leaves
further room for them.

The basic modular unit is a $5 \times 5$ alveolar matrix (\ie, a sort
of egg-crate structure, with a cell for each crystal) of crystals with
dimensions of approximately $110 \times 110 \times 230$~mm$^3$.  The
total weight of the crystals is about 25~kg.  The requirements on the
thickness and the material of the walls constrain the module to be
held in a very light container of 220~g, thus making the mechanical
requirements challenging.

The $\phi$ and $\theta$ symmetries of the design require only twenty
distinct types of crystals.  To achieve the required energy
resolution, crystal-to-crystal separation must be less than or equal
to 500 microns. The design guarantees a maximum distance between
crystal faces of 400 microns within a module and of 600 microns across
two modules,in both the $\phi$ and $\theta$ directions. The tolerance
on the transverse dimensions of the crystals is specified as
$^{+0}_{-100}$ microns.

Alveolar modules are assembled into a shell support structure, with
the front face positioned by 5 tubular carbon-fiber reinforced plastic
(CFRP) setpins and then glued to the structures front plate
(Fig.~\ref{fig:EMC:FWDalveola}).
             
 \begin{figure}[htbp]
\begin{center}
\includegraphics[clip=true,trim = 60 0 35 0,width=0.8\linewidth]{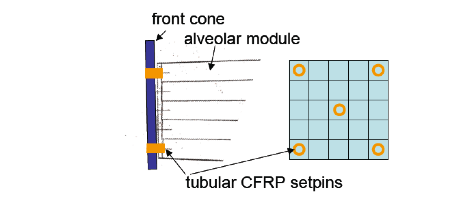}
\caption{Alveolar supports.\label{fig:EMC:FWDalveola}}
\end{center}
\end{figure}

To validate the submodule design, two prototypes of the alveolar
module were constructed (see a photo in Fig.~\ref{fig:fwdEMCmatrix}).
The first prototype (Proto1) was produced to validate the cell
structure concept and understand possible economies in the construction.
It was then used to hold the 25 LYSO crystals in the beam tests for
physics validation. Proto1 provided validation of the whole production
process. A 3D dimensional inspection performed showed that dimensional
tolerances could be met. Wall thickness measurements at the cell open
edge produced the
following values. For internal walls: nominal 200 microns, actual
150--220 microns; for external walls: nominal 135 microns, actual
130--170 microns.

A second prototype, Proto2, identical to Proto 1, was produced in
September 2011 to confirm the process repeatability and to evaluate the
global mechanical properties of the structure.  Global deformations of
the alveolar array are significant, and a load test is essential to
check for lack of interference between modules and the absence of
stress on the crystals.  As shown in
Fig.~\ref{fig:EMC:mectest}, the cells were loaded with dummy crystals
to simulate the mass, and different gravity vectors were investigated.
The mechanical tests performed on the modular structure provided basic
input data to a finite element analysis of the complete support
structure.

 \begin{figure}[htbp]
\begin{center}
\includegraphics[clip=true,width=\linewidth]{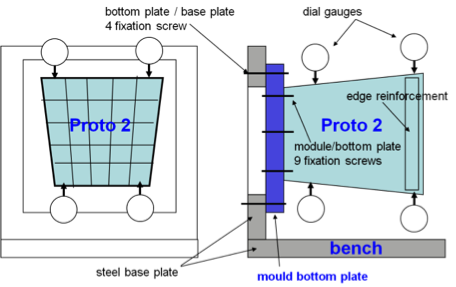}
\caption{Alveolar module test setup}
\label{fig:EMC:mectest}
\end{center}
\end{figure}

A detailed finite element analysis was performed on the alveolar
structure. An approximate module based on surfaces was used as a
reference to validate the Global Finite element model of the whole EMC
(Fig.~\ref{fig:EMC:FWDfinite}).  On the external wall two laminates of
total thickness 200 microns were used, together with four internal
walls laminates at $0^\circ$, $90^\circ$, $0^\circ$, and $90^\circ$.
The base was composed of 600 microns aluminium, equivalent to the
CFRP.  Five constraint points simulated the setpin constraints that
connect the module to the mechanical structure.

The load is the 1~kg weight of each crystal, simulated by a ``rigid
bar'' with five nodes along the center alveolar cell, and a ``lumped
mass'' having 1/5 of the crystal weight at each of the nodes. These
nodes are then rigidly connected with certain others near the alveolar
walls, and from there to the alveolar wall nodes with ``gap and
spring'' elements.
     
 \begin{figure}[tbph]
\begin{center}
\includegraphics[clip=true,width=\linewidth]{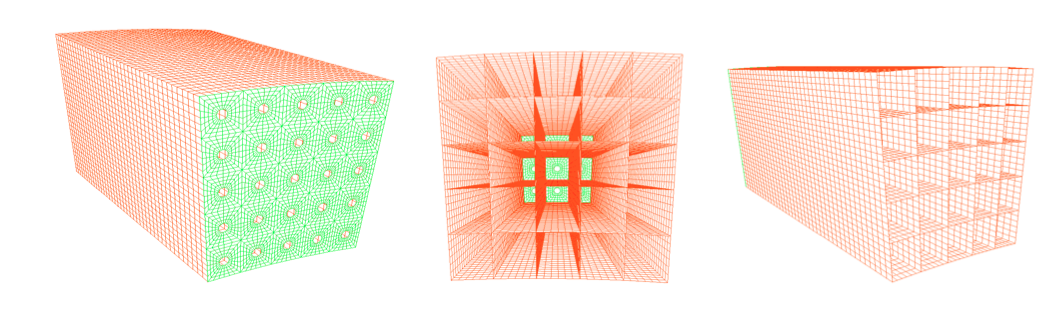}
\caption{Alveolar module finite element model}
\label{fig:EMC:FWDfinite}
\end{center}
\end{figure}

The resulting stresses predicted on the structures are depicted for
the $0^\circ$ case in Fig.~\ref{fig:EMC:FWDstress}. The corresponding
values for the ply stresses and for the index failure ({\it i.e.}, the
ratio between the predicted stress and the breakdown) are summarized
in Table~\ref{tab:EMC:FWDstress}.

\begin{figure}[htbp]
\begin{center}
\includegraphics[clip=true,trim = 800 150 50 200, width=\linewidth]{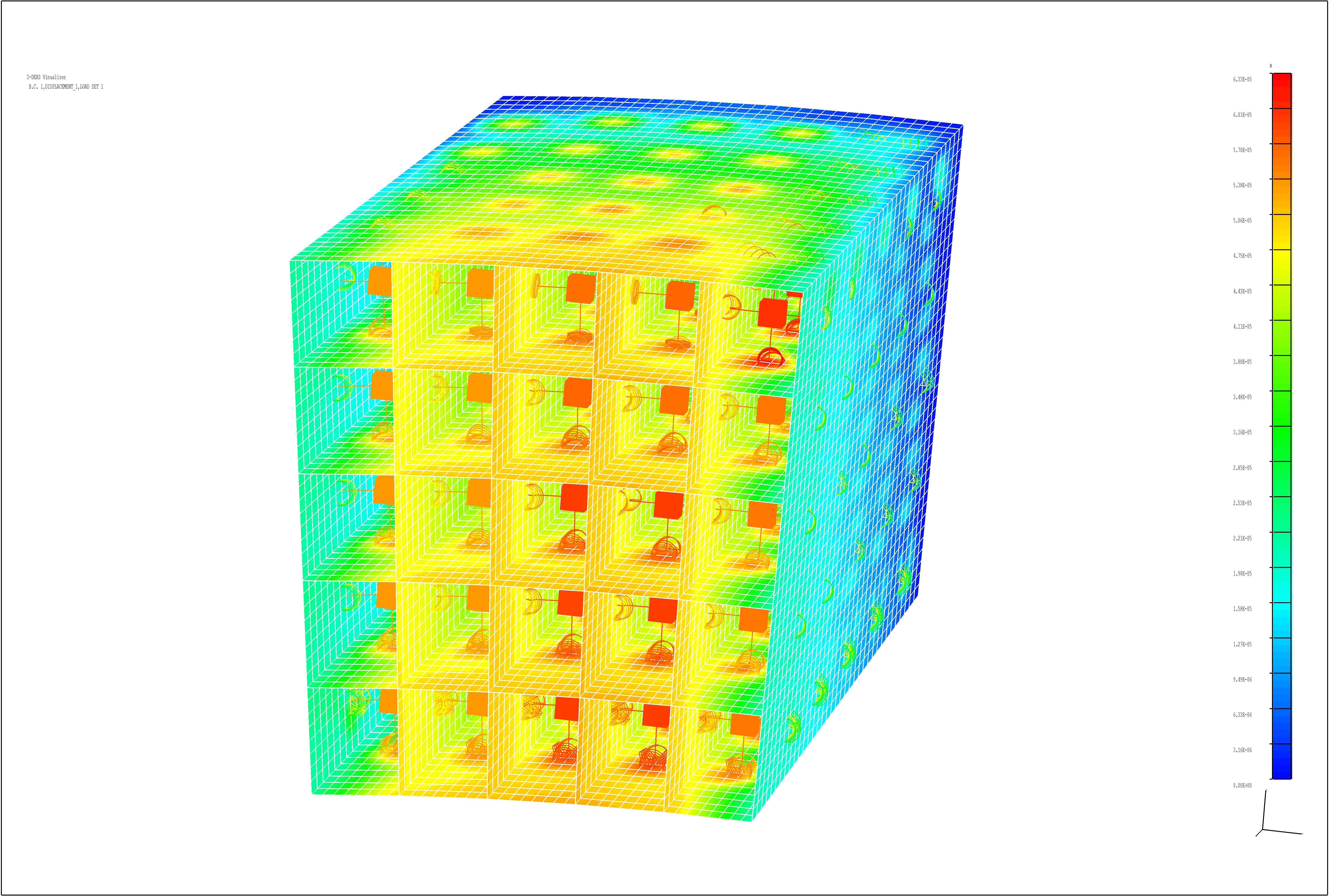}
\caption{ Results of the module finite element model at $0^\circ$. Scale
ranges from 0 (dark blue) to $6.3 \times 10^{-5}$ (Red).}
\label{fig:EMC:FWDstress}
\end{center}
\end{figure}

\begin{table}[tbp]
\caption{Expected ply stresses and index failures.}
\label{tab:EMC:FWDstress}
\begin{center}
\begin{tabular}{|c|c|c|}\hline
case ($^\circ$)/ & Ply Stress(MPa) & index failure \\
coordinate & & \\
 \hline\hline
0/x & 35 & 1/48 \\
0/y & 2.6 & 1/10\\ 
0/z & 1.4  & 1/16\\ \hline
90/x & 5 & 1/342\\
90/y&10 &1/2.6\\
90/z&1.7 & 1/13\\ \hline
180/x&30 & 1/57\\
180/y&7 &1/4\\
180/z&1.5 &1/15\\ \hline
\end{tabular}
\end{center}
\end{table}

The shell, shown in
Fig.~\ref{fig:EMC:FWDsupport}, consists of the outer cone and front
cone as a single unit of CFRP.  The inner cone is a 20~mm section of
7075 Al. The back plate is retained from \babar.  The volume is
defined by the lines AB, AD, and CD.

\begin{figure}
\begin{center}
\includegraphics[clip=true,width=\linewidth]{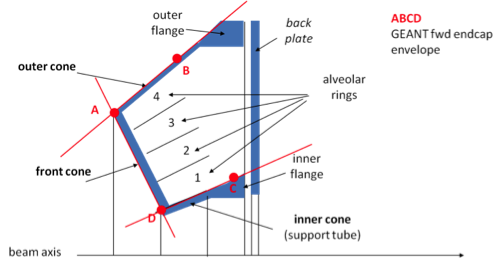}
\caption{Support shell structure}
\label{fig:EMC:FWDsupport}
\end{center}
\end{figure}
 
The outer cone is composed of 6 to 10~mm of CFRP, while the front cone
is either CFRP or a sandwich plate 20~mm thick.  The back plate
provides the EMC interface with the bearing points for position
reference and transmission of loads.  The alveolar array is
cantilevered from the shell front cone as detailed in
Fig.~\ref{fig:EMC:FWDsassebly}.

 \begin{figure}[htbp]
\begin{center}
\includegraphics[clip=true,width=\linewidth]{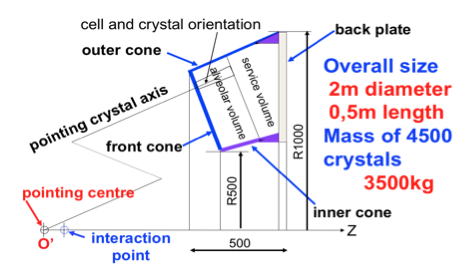}
\includegraphics[clip=true,width=0.4\linewidth]{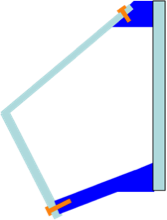}
\caption{Support shell assembly}
\label{fig:EMC:FWDsassebly}
\end{center}
\end{figure}

\tdrsubsubsec{Pure CsI}

Crystals of pure CsI have been used as fast scintillators in
electromagnetic calorimeters at high counting rates.  The
scintillation spectrum has two components: the fast component, at
310~nm has a decay time of 6~ns, with a light output 1.1\% that of
NaI(Tl), while the slow component at 420~nm has a decay time of 30~ns
and a 3.6\% relative light output.  The temperature dependence of the
CsI light yield is $-1.4 \%/^\circ$C. CsI has a relatively high
resistance to radiation damage.
Table~\ref{tab:crystals}~\cite{bib:zhu_crystals} compares the main
properties of pure CsI with other scintillators. The scintillation
spectrum matches the sensitivity of typical PMTs better than it
matches solid state readout devices.

Tests of pure full size ($5\times 5\times 30$
cm$^3$) CsI crystals have been carried out to measure the light yield
and signal-to-noise ratio.  Due to the low light yield of pure CsI
crystals, it is important to achieve the best possible electronic
noise performance.  The crystals were read out with a PMT on one end
and one or two APDs on the other, using the electronics developed for
the LYSO readout.  The measurement of the light output has been done
with a Photonis XP2266b PMT, integrating the signal in a 500 ns window, with
the crystal wrapped with Tyvek.  A $^{137}$Cs source (0.667 \mev
photons) was used. A light output of 21 photoelectrons/MeV has been
measured, in agreement with the PDG-quoted value.
Figure~\ref{fig:CsI-LY} shows the distribution of the measured charge.

\begin{figure}[!hb]
\centering
\includegraphics[width=0.5 \textwidth]{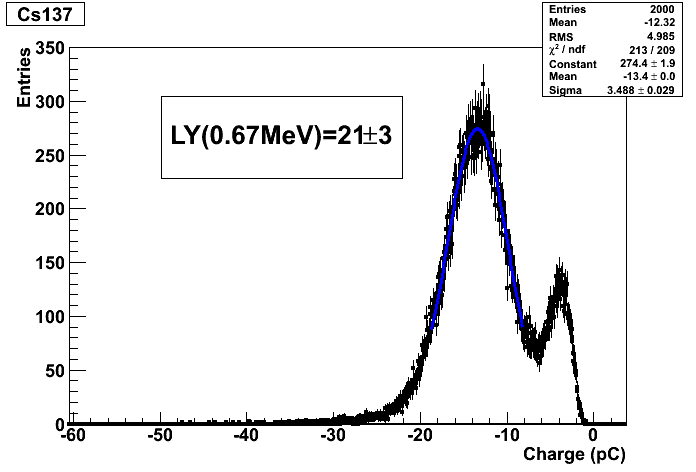}
\caption{Distribution of the measured charge.}
\label{fig:CsI-LY}
\end{figure}

The noise was measured using either one or two Large Area APDs (LAAPD)
(10mm $\times$ 10 mm) to read the crystal.  The signal was sent to a
charge shaper preamplifier and a shaper. A PMT on the opposite side of
the crystal provided the trigger. A calibration of the system has been
performed using cosmics collected with a PMT, selecting events which
passed completely through the crystal. Three measurements were made
with APD voltages at 365, 380, and 390~V; the noise was measured to be
11, 12 and 13~\mev, respectively. In the same configuration, but with
two LAAPD's, a noise of about 6 \mev has been measured. A second setup
used the front end developed for the LYSO measurement and beam test.
In this configuration, the trigger was provided by the coincidence of
two scintillator bars. In this case the best signal-to-noise ratio of
8.2 was obtained at 390~V (Fig.~\ref{fig:Signal-CsI-2LAAPD}).

\begin{figure}[!hb]
\centering
\includegraphics[width=0.5 \textwidth]{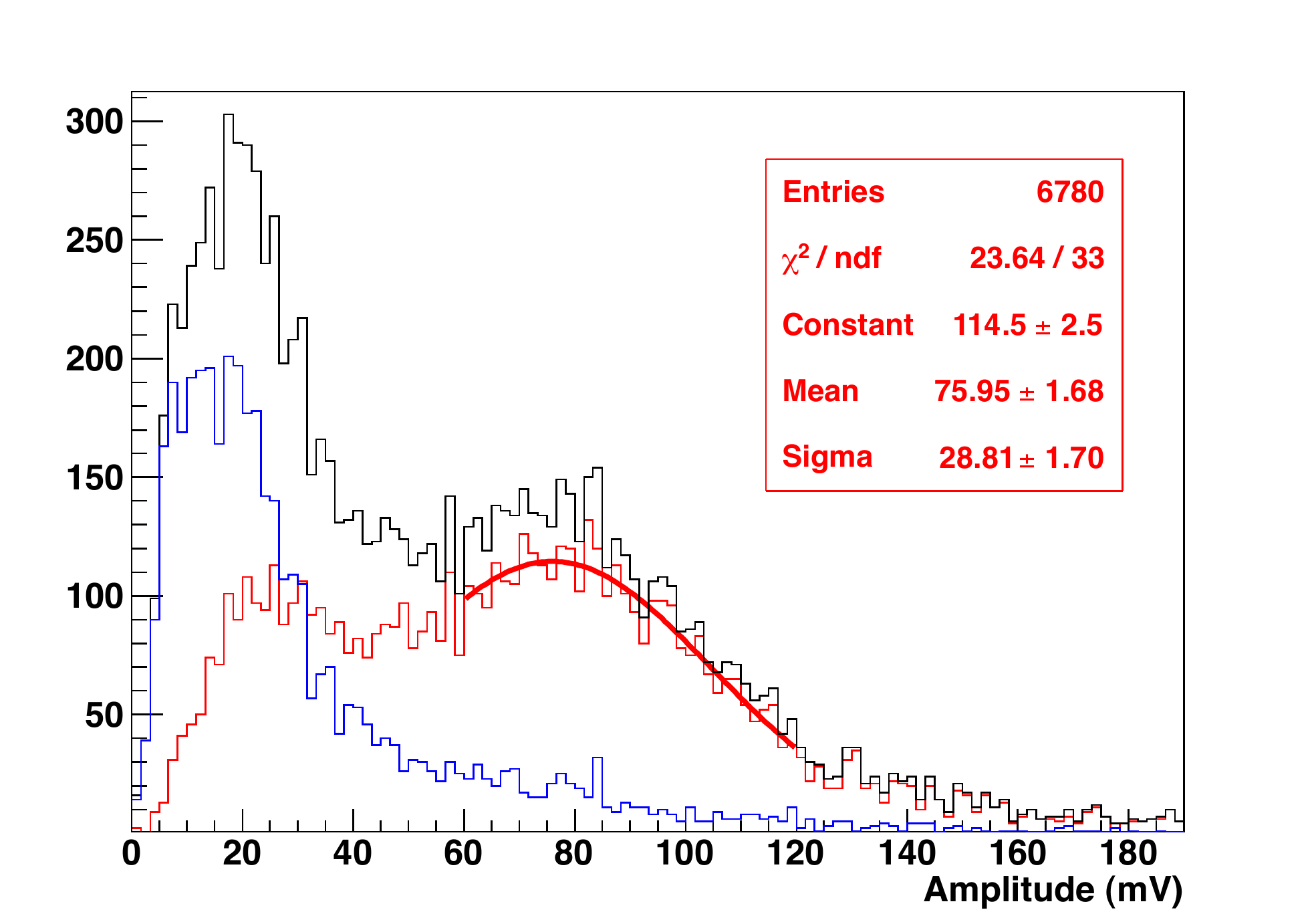}
\caption{Distribution of the cosmic ray signal measured with two
  LAAPDs (red curve). The blue curve is the measured noise and the
  black is the total signal.}
\label{fig:Signal-CsI-2LAAPD}
\end{figure}

The energy deposited by 1~\gev muons through the 5~cm thickness of
pure CsI has been calculated with simulations and is $\sim$30 \mev,
leading to a total measured noise of 3.7~\mev. This is a substantial
improvement with respect to the previous setup. It has, however, been
shown with simulations, that the performance of the calorimeter would
be adversely affected if the noise is higher than 1 \mev. To reach
this level of noise, further improvements in the electronics are being
developed, as well as a better selection of the events. A new front
end will also be developed to try a readout with a photopentode
instead of the LAAPD. In this case the gain of the photodetector would
be higher (120 to be compared with 50 for the LAAPD), but the noise
level has to be controlled a dedicated readout electronics. R$\&$D is
continuing. Radiation hardness measurements are foreseen before the
end of the year and a beam test with a pure CsI matrix of 25 crystals
is in preparation for 2013.

The option of the forward calorimeter using pure CsI crystals does not
require any particular change in the mechanical structure supporting
the detector. Due to the fact that the material is the same except for
the doping, pure CsI crystals will be of the same dimensions and of
the same shape of the original forward \babar\ CsI(Tl) calorimeter.

\tdrsubsubsec{BGO} Prior experience with calorimeters made of BGO
crystals in high energy physics come from L3~\cite{L3Det} and
Belle~\cite{BelleDet}. In the first case BGO was used for both barrel
and endcaps, while Belle used BGO to build the Extreme Forward
Calorimeter for the purpose of improving hermeticity close to the beam
line and to monitor beam background.  The information gained will be
summarized here, supplemented by tests performed to account for the
particular operating conditions expected at \superb: short integration
time, needed to reduce pile-up, and large radiation doses.

The properties of the BGO crystals are summarized in
Tab.~\ref{tab:crystals}: the mechanical properties are similar to
LYSO, but with a light yield four times smaller. The scintillation
decay time is intermediate between that of LYSO and CsI(Tl);
acceptable performance in a high rate environment is therefore
possible.

The performance of the L3 calorimeter, read by PIN diodes, was
$\sigma_E/E=(1.6/\sqrt(E)\oplus 0.35)\%$ and
$\sigma_\theta=((6/\sqrt(E)\oplus 0.3) {\ \rm mrad})$, i.e., the
statistical term of the resolution at 100\mev was 4\%.  Later studies
with APDs~\cite{Zhu2007} show that the system produces $\sim 420$
p.e./\mev, i.e., the statistical term at 100 MeV would be 0.5\%, well
within the requirements.

Nonetheless, due to the requirements dictated by machine background
(Sec.~\ref{sec:EMC:OVERbkg}) we would not be able to operate the
detector with an integrate that captures the entire light output of
the crystal. The degradation in resolution caused by the shorter
integration time was tested specifically for this application in the
INFN-Roma1 electronics laboratory with a single $2 \times 2\times 18$
cm$^3$ crystal. 0.667 \mev photons from a $^{137}$Cs radioactive
source were detected with a crystal read at both ends by two EMI-9814B
PMTs. The waveforms of the signals provided by the PMTs were acquired
by a LeCroy digital oscilloscope having a bandwidth of 300 MHz and
used with a sampling rate of 250 Ms/s. One of the two signals was used
to trigger the oscilloscope.  The other signal was processed with an
Ortec-474 and acquired by the scope.  The Ortec-474 module has an
integration time (RC) that can be chosen between 20, 50, 100, 200, and
500~ns.

For each value of RC and for each pulse shape both the total charge,
independent of the total time, and the maximum amplitude, the quantity
that will eventually be used in the experiment, were recorded. The
resolution on the charge was found to be 9.5\%, consistent with 130
p.e./\mev after detection and independent of the integration time of
the ORTEC shaper.  The resolution on the amplitude depends instead on
the latter, as shown in Fig.~\ref{fig:EMC:FWDampl} and compared with
the expectations from a simple Monte Carlo simulation in which only
the effect of the photo-electron statistics is
included~\cite{Bocci2012}. The results show that shortening the
integration time worsens the resolution, but only moderately. For
instance, even integrating only for 20 ns, compared with the 300 ns
scintillation time of the crystal, the contribution to resolution due
to the photo-statistics worsens only by 60\%.

\begin{figure}[tbp]
\centering
\includegraphics[width=0.5 \textwidth]{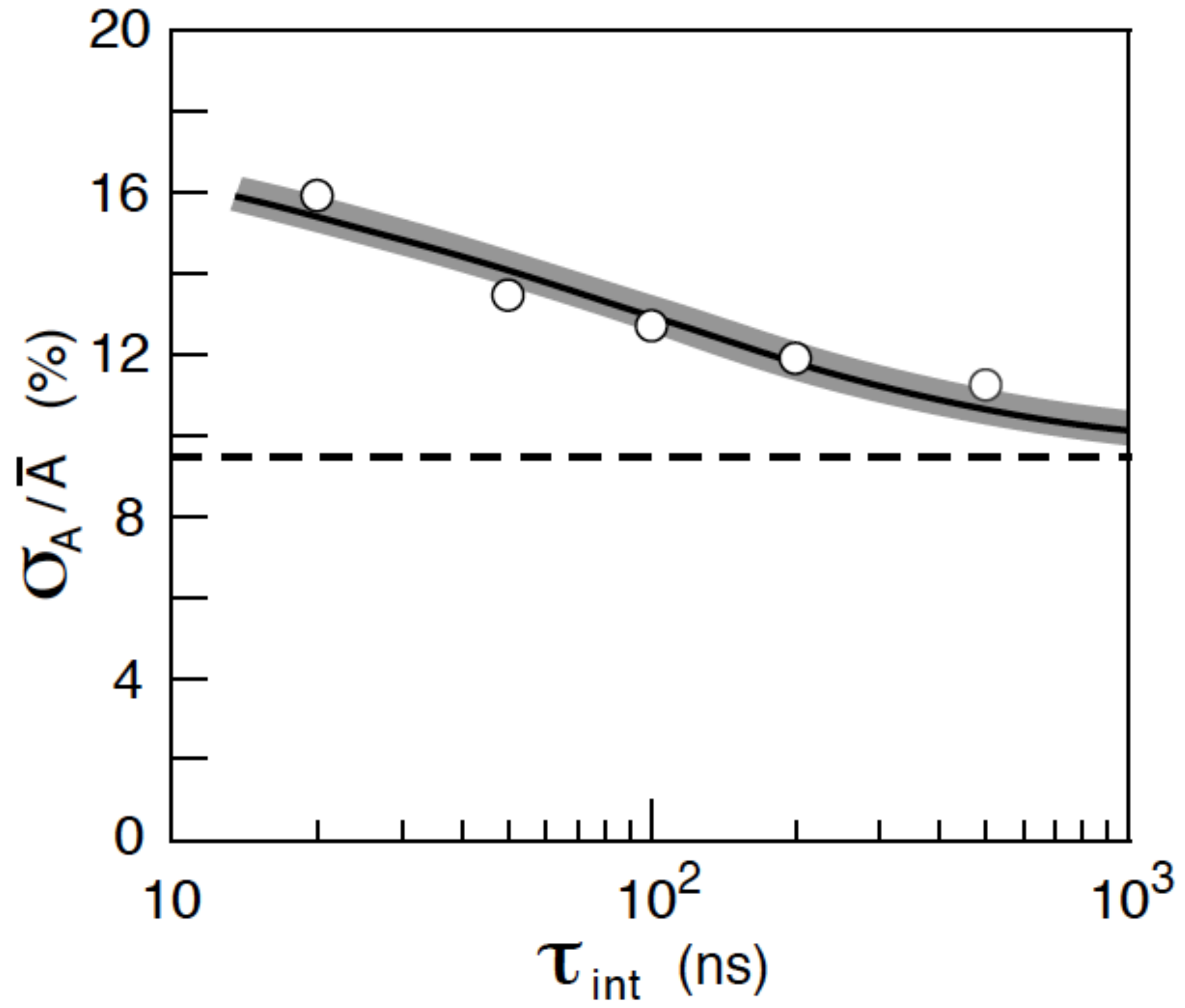}
\caption{Resolution on the measurement of the total charge (dashed line) and the amplitude of the pulse shapes (dots)as a function of the integration time. The results of a Monte Carlo simulation are superimposed (shaded line).}
\label{fig:EMC:FWDampl}
\end{figure}

A test to measure the electronic noise of the read out for BGO
crystals has been performed with the same experimental setup described
in Sec.~\ref{sec:EMCbarrelElec}: a $2\times 2\times 18$ cm$^3$ BGO
crystal read out with an $0.5\times 0.5$ cm$^2$ APD followed by a CSP
and a Gaussian shaper.  The test procedure is slightly different from
the one previously described: instead of using a $^{60}$Co source to
excite the crystal scintillation, the intrinsic background of the BGO,
due to the presence in the crystal of $^{207}$Bi, which decays into
$^{207}$Pb with the emission photons of different
energies~\cite{BGOrad} has been exploited.  Since the PMT resolution
is much better than the APD, the PMT signal, after the
calibration of the energy scale, is used as the ``true'' energy to
calibrate the APD response.  The scatter plot of the distribution of
the APD amplitude as a function of the PMT charge has been divided
into slices and for each slice the peak of the distribution is
obtained.  These peaks show a linear behavior with energy that has
been fitted to obtain the calibration of the APD response.  An
estimate of the equivalent noise has been obtained from the widths of
the distributions of the difference of the APD and PMT signals both
measured in \mev.  In the case of the Hamamatsu H4083 CSP with 100~ns
integration time, with a shaping time of 500~ns, the equivalent noise
is measured to be about 2.2~\mev.  To compare this result with a CSP
with a longer integration time, a measurement with a Cremat CSP with
140~$\mu$s integration time was also performed and the equivalent
noise in this case was about 1.7~\mev.

The BGO radiation hardness was tested up to 90Mrad~\cite{BelleDet}.
After a drop of about 30\% in the first 20 Mrad of integrated dose the
crystal light yield plateaus. It will nonetheless recover up to 90\%
of the original light yield in approximately 10 hours; it is not
documented what will happen if after this pre-irradiation the crystal
receives a small additional dose. For non irradiated crystals, a dose
of 115~krad implies a light yield loss of 30\% that fully recovers
with a lifetime of $\sim$1 hour.  These results refer to undoped,
recently produced crystals; they also depend strongly on the level of
doping and on the manufacturer.

\begin{figure}[tbp]
\centering
\includegraphics[width=0.5 \textwidth]{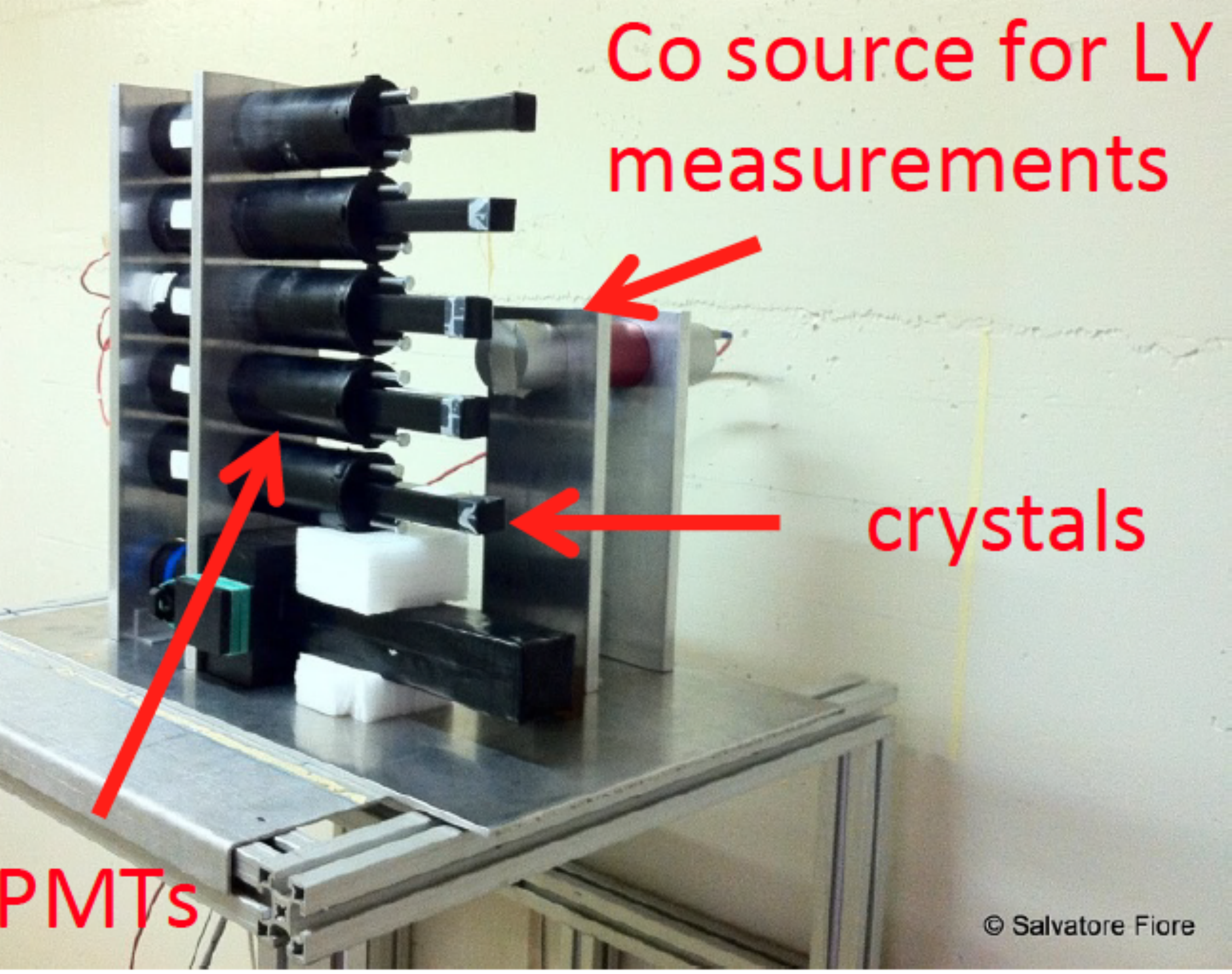}
\caption{Experimental set-up for light yield measurements. Crystal
  samples are shown while mounted on the PMTs. Crystals were coupled
  to the PMTs during low-dose irradiation, and data were acquired by
  shutting down the source for few minutes.}
\label{fig:IrradSetup}
\end{figure}

To customize the radiation hardness study to our case, we exposed four
samples of BGO crystals to $\gamma$-rays from a high-activity
$^{60}$Co radioactive source.  Two crystals of $2.2\times 2.2\times
18$ cm$^3$ have been previously used in the L3 experiment at CERN,
other two crystals of $2.5\times 2.5\times$20 cm$^3$ were recently
supplied by the Shanghai Institute of Ceramics (SIC).

Irradiations and measurements took place at the Calliope Gamma
Irradiation Facility at ENEA-Casaccia center (Rome)
(Fig.~\ref{fig:IrradSetup}).  The irradiation source is a cylindrical
array of $^{60}$Co source rods emitting $\gamma$-rays of 1.17 and 1.33
\mev, in an irradiation cell of $6\times 7\times 3.9$ m$^3$ plus an
attached gangway. Depending on the placement of the samples, dose
rates from few rad/h up to 230~krad/h are available. The source can be
moved outside or inside its shielding pool in less than two minutes.

Plastic mechanical supports for the crystals, not shielding them from
the radiation, allowed us to expose the samples to different dose
rates, and to couple them to EMI 9814B Photomultiplier Tubes (PMT) to
perform the light yield (LY) measurements using a low-activity
$^{60}$Co calibration source. PMTs were read out by a CAEN VME ADC.

We started our irradiation campaign by exposing the crystals to 5--10
rad/h dose rate for a few hours. We were able to measure the LY once
every 20 minutes of irradiation, with only two minutes delay after the
source was shut down. Using this approach we measured the progressive
LY reduction at different dose rates.  We measured the LY recovery
after 15--30 rad and 170 rad doses, and evaluated the recovery time
constant for the different samples by fitting the time evolution of
the LY after irradiation with the sum of two exponential functions.
The results summarized in Tab.~\ref{tab:EMC:FWDBGOhard} show that the
LY loss are important, and that there are several components in the
recovery time. The most relevant one, due to 5--10 rad/h dose rate is
about four hours.

\begin{table}[ht]
  \caption{Effect of irradiation on a BGO crystal: relative LY loss after irradiation and recovery times as measured with a fit to the time evolution with two exponential components.}
\small{
\begin{center}
\begin{tabular}{| c|c ||c | c|| c |  }
\hline 
dose  & rate & LY  & $\tau_1$  & $\tau_2$ \\
(rad) & (rad/h) & $drop $ & hr & hr \\
\hline 
15 &5 & 25\% & $2.7\pm 0.4$& -- \\
170 &10 & 67\% & $4.7\pm 0.2$& 380 (fixed)  \\
12$\times10^6$ & 23$\times10^4$ & 90\% & $5.0\pm 2.4$& $381.9\pm 3.5$  \\
\hline
\end{tabular}
\end{center}
}

\label{tab:EMC:FWDBGOhard}
\end{table}

We then exposed some crystals to a massive dose rate of 230~krad/h for
a total dose of 12~Mrad, and then to small doses with 1.5~rad/h rate,
to measure the sensitivity to small dose rates after strong delivered
doses. We measured LY reductions up to 1/10 of the pre-irradiation LY
value. A long-term recovery was then analyzed and exploited to measure
the recovery time constant. We fit the LY recovery data with a
double-exponential function, yielding a short time constant of
(5.0$\pm$2.4)~h and a long time constant of (381.9$\pm$3.5)~h.  BGO
crystals from L3 experiment showed a larger damage after each
irradiation, but also a faster recovery capability, while crystals
from SIC hardly recovered from radiation damage.

Light transmission spectra have been acquired before and after
irradiation, in order to identify the nature of the LY reduction. As
previously stated, the radiation-induced LY reduction could be due
either to a decrease of the light transmittance, or to a damage of the
scintillation mechanism itself. By comparing the LY and transmittance
measurements, we observed that the main effect of the radiation damage
was a reduction of the light transmission due to color center
formation.

\tdrsubsubsec{Comparison among options}
\label{sec:compAlt}

The decision on the technology to be used for the forward calorimeter
must weigh physics performance, technical feasibility and cost. On
this basis the baseline was chosen to be a hybrid design, with three
rings of CsI(Tl) and six rings of LYSO hybrid. Other options
considered were the use of a full LYSO design, retaining the existing
structure or constructing an improved one, pure CsI, or BGO crystals,
as described above, and the use of lead tungstate (PbWO$_4$) crystals,
for which the study was based on the existing literature.

In order to compare the physics performance of the various options, we
set up a full GEANT4 simulation that included the correct detector
geometry, the electronic noise obtained from beam or lab tests, when
available, and extrapolations from them when needed. The signal shape
was obtained from detailed simulation of the expected electronic
chain, using signal extraction algorithms that reproduced the actual
electronics and a clustering algorithm based on the one used by
\babar.

\begin{figure}[htbp]
\centering
\includegraphics[clip=true,trim = 0 0 50 30,width=\linewidth]
{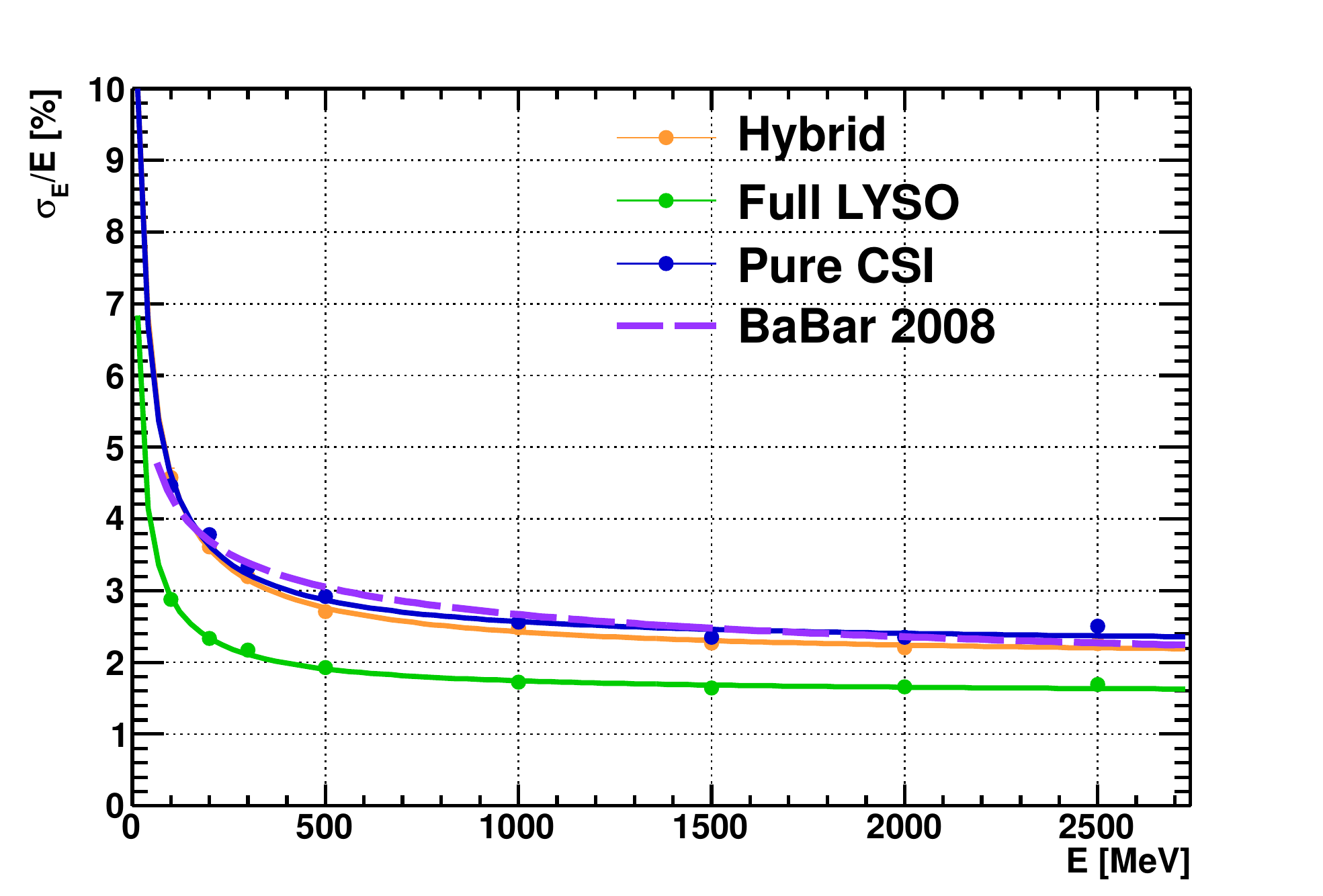}
\includegraphics[clip=true,trim = 0 0 50 30,width=\linewidth]
{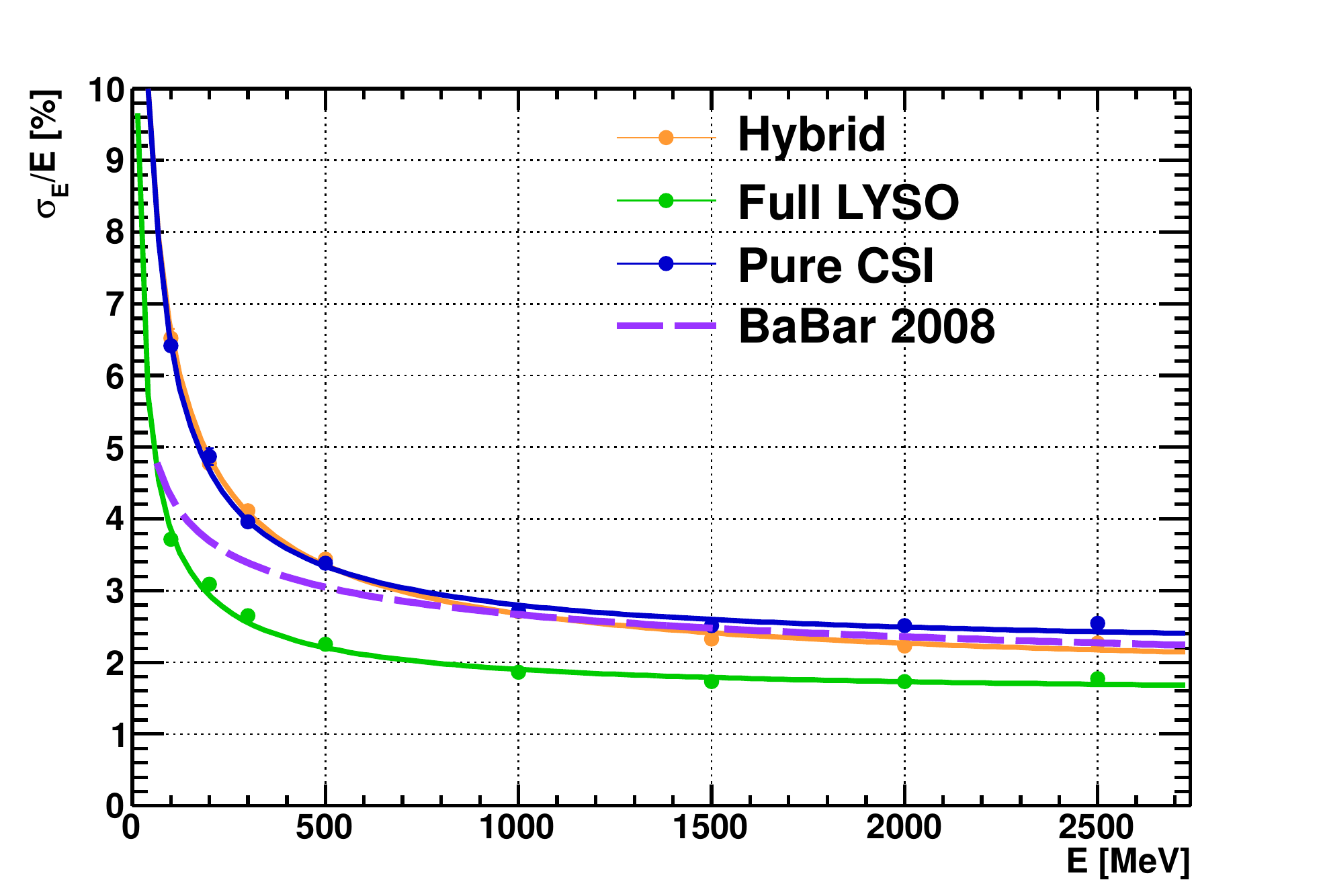}
\caption{Comparison of the relative energy resolution between several
  options of the forward EMC. Top: nominal machine background.
  Bottom: machine background increased by a safety factor of five.}
\label{fig:EMC:FWDCompReso}
\end{figure}

\begin{table*}[!htp]
\renewcommand\T{\rule{0pt}{2.6ex}}
\renewcommand\B{\rule[-1.2ex]{0pt}{0pt}}

\caption{Comparison of crystal volume and calorimeter structural and calibration costs for the baseline forward endcap configuration and several options. \label{tab:EMC:FWDAltComp}}
\vspace{-12pt}
{\scriptsize
  \begin{center}
\begin{tabular}{lcccccccc}
      \hline \hline
    \T Option  &  Number  & New crystal & Crystal & Crystal & Photo-  & Calibration & Total & Total \\
      &  of new & volume & cost/cm$^3$   & cost  & detectors & system  & structure & Cost  \\
        \B
              &   crystals  & (cm$^3$) & (k\euro)  & (M\euro) & (M\euro) & (M\euro)  & (M\euro)&  (M\euro) \\
      \hline\hline
       \T Baseline - Hybrid &&&&&&&\\
3 CsI(Tl) + 6 LYSO & 2160 &	245	& 19.23 &	4.76 & 0.38 & - & 0.19 & 5.33 \B  \\ \hline\hline
\T LYSO new structure &	4500 &	402 & 19.23 & 7.72 & 0.44& - & 1.75 & 9.91 \\ 
\T LYSO old structure  &	3600 &	402 & 19.23 & 7.72 & 0.44& - & 0.19 & 8.35 \\ \hline
 \T Reduced Hybrid &&&&&&&\\
4 CsI(Tl) + 5 LYSO & 1760 &	198	& 19.23	& 3.81 & 0.31 & - & 0.19 & 4.31 \\
5 CsI(Tl) + 4 LYSO & 1360 &	154  & 19.23	& 2.93& 0.24 & - & 0.19 & 3.38 \\ \hline
 \T Alternative crystals &&&&&&&\\
Pure CsI &	900	& 692 &	3.92 &	2.71 & 0.43 & - & 0.19 & 3.33 \\
BGO	& 4500	& 392 &	6.92 &	2.72 & 0.44 & 0.92-2.31 & 1.75 & 5.82-7.21 \\
\B PbWO$_4$	& 4500	& 306 &	2.96 &	1.18 & 0.44 & 0.92-2.31 & 1.75 & 4.28-5.67 \\
      \hline\hline
    \end{tabular}
\end{center}
}
\end{table*}

For each option, and for several energies, the response of the
detector to monochromatic photons was fitted with a Gaussian with
exponential tails. The relative energy resolution was computed as the
FWHM of the fitted shape for several values of photon energy. The
results for different options are compared in
Fig.~\ref{fig:EMC:FWDCompReso} for the current estimate of
the machine background and for a scenario in which the
background is increased by a safety factor of five
(5$\times$ background).

The best forward endcap option is clearly to use LYSO crystals
everywhere; the option suffering for the most problems is BGO. The
latter is very sensitive to the 5$\times$ background, most likely due
to the non-optimal geometry of the simulation, which was done {\it
  ad-hoc}. The hybrid solution also suffers from background; the
effect is comparable with the expected degradation in resolution of
the barrel. All other options have comparable performance,
intermediate between the full LYSO and the hybrid options.

To complete the picture, Tab.~\ref{tab:EMC:FWDAltComp} shows a
comparison of the volume and total cost of the scintillating crystals,
the cost of the photodetectors, the calibration system, and the
mounting structure, all of which must be included in evaluating
options.

Photo-detectors have roughly similar costs for all options. Some
saving can be achieved with the hybrid option by reusing existing
readout channels.  A more precise (and more expensive) calibration
system is needed only for those crystals whose light output degrades
under irradiation with a short recovery time, namely BGO and PbWO$_4$.
Finally, the carbon fiber mechanical support needs to be redone only
for the LYSO, BGO and PWO options; in the other options, the only
costs are the shipping and the refurbishment of the present structure.

The table also lists three hybrid options, in which differing numbers
of the outer CsI(Tl) rings of the endcap are retained, the inner rings
being replaced by LYSO crystals.

The two least expensive options are the CsI(Tl)/LYSO hybrid and the
pure CsI solutions. Simulations show that the hybrid option has a
somewhat worse resolution. It is, however, technically the
best-studied, has a proven readout scheme and is ready for
construction. The properties of pure CsI crystals and their readout
are still under study, and therefore the conservative TDR baseline
choice was the hybrid design. The budget has been based on
this option.

\tdrsec{Backward Calorimeter}
\label{sec:EMCback}

The backward electromagnetic calorimeter for \superb\ is a new device
designed to improve the hermeticity of the detector at modest cost.
Excellent energy resolution is not a requirement, since there is
significant material from the drift chamber in front of it. Thus, a
high-quality crystal calorimeter is not planned for the backward
region.  The proposed device is based on a multi-layer
lead-scintillator sampling calorimeter with longitudinal segmentation
providing capability for $\pi/e$ and $K/\pi$ separation at low
momenta. The design is derived from the analog hadron calorimeter for
the ILC~\cite{bib:ILC}.

The active region of the backward calorimeter is located behind the
drift chamber starting at $z=-1320$~\mm (Figure~\ref{fig:det:superb})
allowing room for the drift chamber front end electronics. The inner
radius is 310~\mm, the outer radius is $r_0 = 750$~\mm and its total
thickness is less than 180~\mm covering 12\Xrad. It is constructed
from a sandwich of 2.8~\mm Pb plates (0.5\Xrad) alternating with 3~\mm
plastic scintillator strips (\eg, BC-404 or BC-408). The scintillation
light of each strip is collected by a wavelength-shifting fiber (WLS)
coupled to a photodetector located at the outer radius. The
scintillator strips come in three different geometries, right-handed
logarithmic spirals, left-handed logarithmic spirals and radial wedges
(Figure~\ref{fig:backEMCgeom}). This pattern alternates eight times.
Each layer contains 48 strips producing a total of 1152 readout
channels.

The WLS fibers, Y11 fibers from Kuraray, are embedded in grooves
milled into the center of the scintillator strips.  Each fiber is read
out at the outer radius with a $1\times 1\mma$ multi-pixel photon
counter (SiPM/MPPC)~\cite{bib:MPPC}. A mirror is glued to each fiber
at the inner radius to maximize light collection.  
A preamplifier is coupled to the SPM/MPPC, then the signal is transmitted to readout boards mounted in a convenient location. The PID electronics is used, providing
12 bit ADCs and 200\ps TDCs.

\tdrsubsec{Requirements}

The main goal of the backward EMC is to increase the calorimeter
coverage by recording any charged or neutral particle in the backward
region. This information is important in particular for analyses that
utilize the recoil method with hadronic and semileptonic tags to
select $B$ meson decays with neutrinos in the final state. The
backward EMC helps to increase the selection efficiency and to improve
background rejection. For this task, excellent energy resolution is
not necessary. It is more important to keep the costs moderate. With
moderate energy resolution and good angular resolution, it will also
be possible to use the backward EMC in \piz
reconstruction.  Furthermore, the backward EMC has the
capability to measure time-of-flight and the energy loss via
ionization of charged particle well.  This information is useful
for particle identification, in particular $\pi/e$ and $K/\pi$
separation at low momenta.

\begin{figure}[tbp]
\begin{center}
\includegraphics[width=\linewidth]{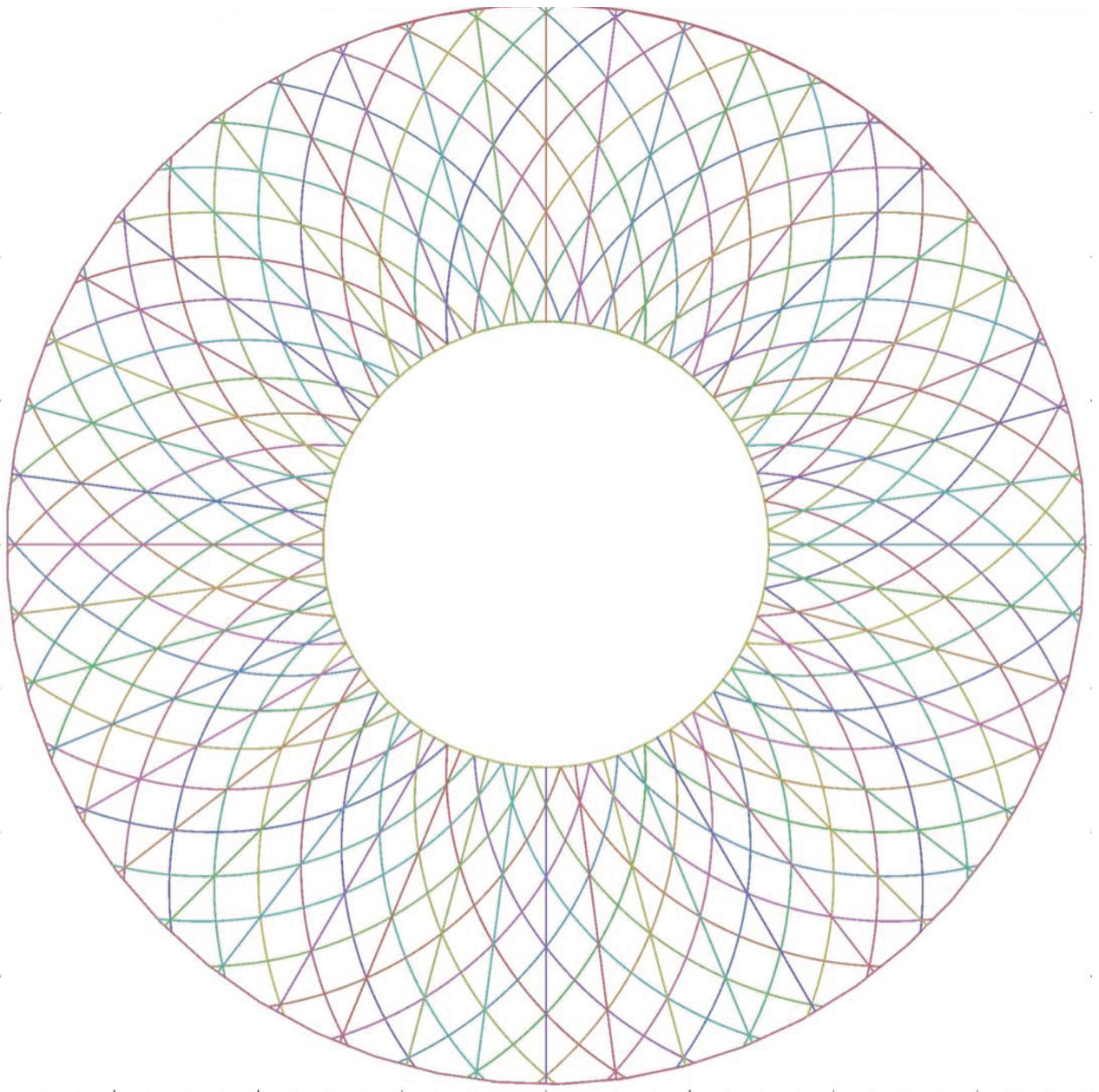}
\end{center}
\caption{The backward EMC, showing the scintillator strip
geometry for pattern recognition.}
\label{fig:backEMCgeom}
\end{figure}

\tdrsubsubsec{Energy and angular resolution}

Since the backward EMC prototype is still in the construction phase,
we do not yet have results on energy resolution and angular
resolution. However, electromagnetic sampling calorimeter prototypes
with plastic scintillator strips and tiles have been tested in test
beams within the CALICE collaboration~\cite{bib:calice-emc}.  The
energy resolution for the stochastic term is $15\%\sqrt{E[\gev]}$ and
for the constant term is around $1\%$. For the CALICE analog hadron
calorimeter which has a non-optimized geometry for electromagnetic
showers, the stochastic term was measured to be around
$20\%/\sqrt{E[\gev]}$. For low photon energies, an additional noise
term of $\sim 130\mev/E$ contributes.  Thus, the backward endcap EMC
is expected to have a similar performance with a stochastic term of
$15-20\%/\sqrt{E[\gev]}$.

A left-handed logarithmic spiral is defined by
\begin{eqnarray}
x(t)=r e^{b \cdot t} \cos{t}-r \\
y(t)=r e^{b \cdot t} \sin{t}, 
\end{eqnarray}
where $t$ is a arbitrary parameter, $r=r_o/2=37.5$~cm, and $b=0.2$.
Eight left-handed spiral strips overlap with eight right-handed spiral
strips defining a tile-shaped region. The radial strips
overlap with five left-handed and right-handed spiral strips.  In the
worst case, the resolution is estimated to be
$\sigma_r =\sigma_\phi \simeq 29\mm$ for
a single tile in the outer region. This is improved to $\sigma_r
=\sigma_\phi \simeq 12\mm$ in the inner region. If the shower is
distributed over several adjacent tiles, 
the center-of-gravity method will improve the position
resolution significantly.

\tdrsubsubsec{Background rates}

Present background simulations indicate that the worst rates for all
energies of $3~{\rm kHz}/\cms$ occur in the inner most region. In ten
years of running this amounts to $ 6.1\times 10^9~ n/\mma$ in the
region near the inner radius. The background rates drop significantly
toward the outer radius. At the location of the photodetector, the
rate is reduced by more than a factor of 10. Further simulation
studies are needed to study the impact on occupancy.  To deal with
this issue we may either subtract a higher average background energy
from each strip or divide the strips into two segments at the cost of
doubling the number of photodetectors.  The former solution has an
effect on the energy resolution since the background energy deposit
has a wide distribution. The latter solution is preferable, but is
about 75k Euros more expensive.

\tdrsubsubsec{Radiation hardness}

Irradiation of Si detectors causes the dark current to increase linearly with
flux neutron flux $\Phi$:

\begin{equation}
\Delta I=\alpha \Phi V_{\rm eff} G,
\end{equation}
where $\alpha= 6 \times 10^{-17} ~ {\rm A}/\cm~ V_{\rm eff}\sim
0.004\mm^3$, $\Phi$ is the flux, $V_{\rm eff}$ is the bias voltage and
$G$ is the gain. Since the initial resolution of MPPCs/SiPMs of $ \sim
0.15$ photoelectrons (pe) is much better than that of other Si
detectors, radiation effects start at lower fluxes. For example, at a
flux $\Phi=10^{10}~n /\cms$ the individual single pe signals are
smeared out. The MIP peak is still visible at $\Phi\sim 10^{11}
~n/\cms$. The number of observed hot spots and the noise rate
increases after irradiation of $3\times 10^9 ~n /\cms$. No significant
changes are observed on the cross talk probability as well as no
significant change on the saturation curves. The main effect is an
increase in noise after exposure to high $n$ dose. Hamamatsu has
produced new SiPM/MPPCs with $20\mum \times 20\mum$ and $15\mum \times
15\mum$, which have lower detection efficiency than the $25\mum \times
25\mum$ version due to additional boundaries and thus need a higher bias
voltage to compensate for losses.  Figure~\ref{fig:efficiency} shows
the detection efficiency as a function of bias voltage for $15\mum
\times 15 \mum$ detectors before and after irradiation with $10^{13}~
n/\cms$. For the new detectors, signal/noise and the equivalent noise
charge look fine after irradiation. We expect the backward endcap EMC
will record $10^{11}~ n/\cms$ after 10 years operation. If the $25\mum
\times 25 \mum$ pixel SiPM/MPPC is problematic we may switch to one of
the new SiPM/MPPCs with smaller pixel size.

\begin{figure}
\begin{center}
\includegraphics[width=\linewidth]{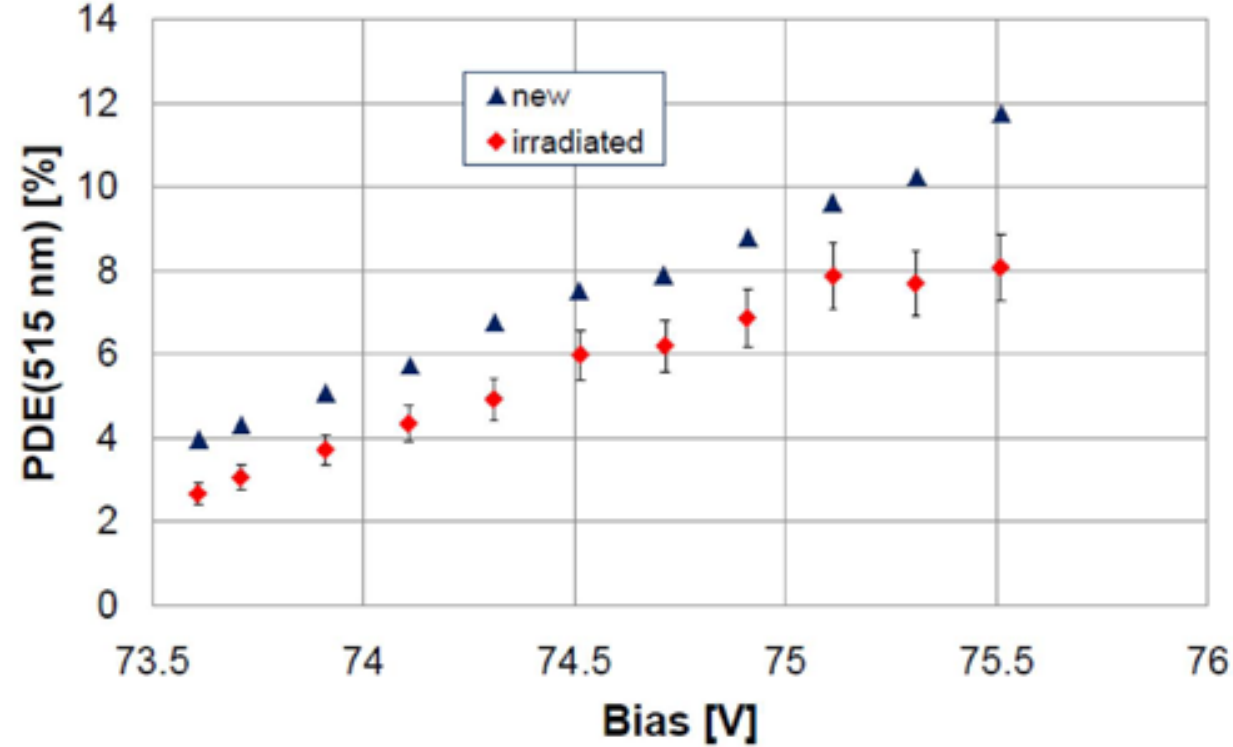}
\end{center}
\caption{The photon detection efficiency (PDE) of $15\mum \times
  15\mum$ pixel MPPCs as a function of the bias voltage before and
  after irradiation with $10^{13}\ n/\cms$.}
\label{fig:efficiency}
\end{figure}

\tdrsubsubsec{Solid angle, transition to barrel}

In the laboratory frame, the backward EMC covers a full polar angle
region from $231\mrad $ to $473\mrad $. Partial coverage extends the
polar angle region from $209\mrad$ to $517\mrad$. In the present
design, there is a gap between the backward EMC and the barrel
covering the region $517\mrad$ to $694\mrad$ in the laboratory frame.
In the center-of mass frame, full coverage of the backward EMC is in
the region $215\mrad$ to $442\mrad$, while partial coverage exists in
the region from $194\mrad$ to $482\mrad$. If the backward EMC could
move closer to the IR, the gap to the barrel calorimeter would be
reduced.

\tdrsubsec{Mechanical design}

The 3\mm thick scintillator strips are cut individually from a
scintillator plate.  Thus, the plate size and the cutting procedure
need to be carefully thought through to minimize the amount of waste.
For the spiral strips the least waste and fastest production is
obtained by fabricating a mould. However, this approach may be too
expensive, since the total number of spiral strips is rather small.
The preferred scintillator material is BC-404 from St.~Gobain, since it
has the smallest decay time for TOF capability and its emission
spectrum is reasonably matched to the Y11 absorption spectrum. The
strip width is 38\mm at the inner edge increasing to 98\mm at the
outer edge. The strip sides are painted with a white diffuse
reflector. Front and back faces are covered with reflectors sheets
(3M, Tyvek). Test bench measurements have shown that the yield along
the strips varies by more than a factor of two. To restore uniformity,
a pattern of black dots is printed onto the white reflector sheets.

In the center of each strip, a 1.1~\mm deep groove is milled into
which the 1~\mm thick Y11 WLS fiber is inserted. At the outer edge of
the strip, the groove is cut 0.4~\mm deeper so that the active area of
the SiPM/MPPC fully covers the fiber. The SiPM/MPPC is housed in a
small precisely cut pocket.  Especially fabricated fixtures out of
Teflon or Nylon will hold a strip. The fiber groove at the outer edge
is closed with a plug at the position of the photodetector. The Y11
fiber is pressed against the plug and held with a drop of glue.  After
removing the plug the SiPM/MPPC is inserted and is glued onto the Y11
fiber to match refractive indices. A mirror is placed at the other end
of the fiber to detect the light that moves away from the
photodetector. So tolerances in the length of the Y11 fiber are picked
up at the mirror end. The strip layout is shown in
Figure~\ref{fig:strip}.

\begin{figure}
\begin{center}
\includegraphics[width=\linewidth]{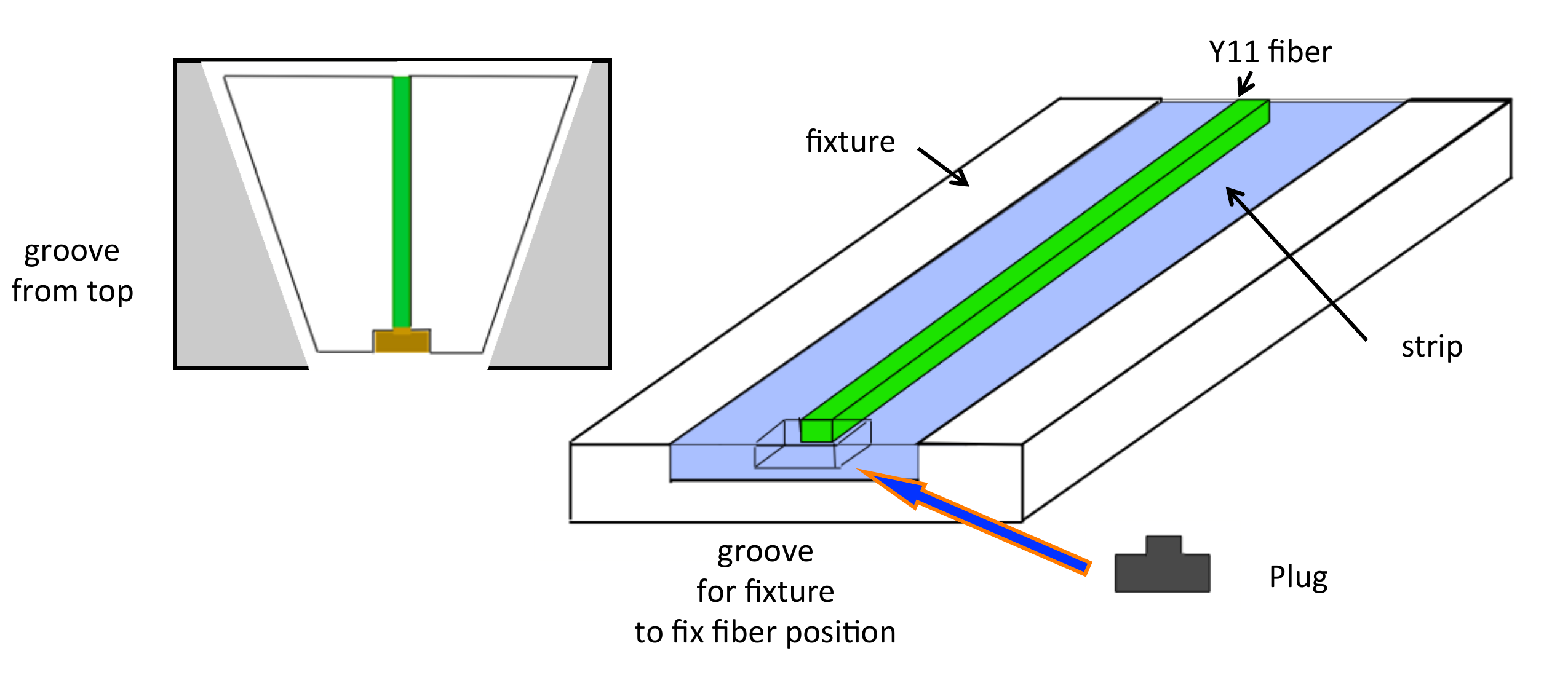}
\end{center}
\caption{Schematic of holding the scintillator strip in a fixture for mounting the MPPC to the Y11 fiber.}
\label{fig:strip}
\end{figure}

To hold the strips in each layer in place 1.5~\mm deep and 1~\mm wide
grooves are cut into the lead plates. The shape of groove matches that
of the strip.  A 3~\mm thick and 1\m wide and 550~\mm long plastic strip
is inserted into the groove and is glued.  This structure is strong
enough to hold the scintillator strips in place. The calorimeter can
be rotated by $90^\circ$. This is needed for operation with cosmic
muons that yield a MIP calibration and allows for testing the
calorimeter before installing it into the \superb\ detector.

The entire calorimeter weighs about 1300~{\rm kg}. An aluminum frame
with a strong rear plate and a strong inner cylinder will hold the
backward EMC in the \superb\ detector.  The front plate and outer
cylinder needed for closure and shielding can be thin.  Considering
cost and performance it is advantageous to build the backward EMC as a
single unit. This requires the calorimeter to slide back on the beam
pipe supported on the tunnel walls. It needs to be fixed at the tunnel
and rolled in. Since the inner radius is 31~cm, there is sufficient
clearance for pumps and other beam elements. The design of this
capability requires a detailed drawing of the beam pipe and the
position and size of machine elements.
 
It is possible, however, to build the backward EMC in two halves with
a vertical split. The impact of such a design is that ten strips per
layer have to be cut into two segments. The inner segments of these
strips have to be read out at the inner radius. This increases the
number of channels by 240, requiring 240 additional SiPM/MPPCs, additional readout boards
and four extra calibration boards. This layout will deteriorate the
performance near the vertical boundary. The effect needs to be studied
in simulations.  This also adds approximately 20\% to the cost.

\tdrsubsubsec{Calorimeter construction}

Each completed strip with Y11 fiber and MPPC mounted is tested in the
lab with a $^{106}$Ru source. All important properties, such as bias
voltage, gain, noise, and MIP position are recorded in a data base.
Stacking will start from the rear to the front. The rear Al plate
rotated by $90^\circ$ is placed on the mounting table and the inner
cylinder is bolted on to it. The back plate requires 48 radially
milled grooves into which the plastic strips are inserted that will
hold the scintillator strips. Next, all scintillator strips are
mounted, the MPPCs are connected to preamplifiers and a Pb plate with
right-handed logarithmic spiral grooves is placed on top completing
layer 24. This procedure is repeated 24 times. Finally, another
scintillator layer with radial strips and the front plate are stacked.
The scintillator layer 0 yields information on the shower origin.
 
Before the outer cylinder is bolted on, temperature sensors and clear
fibers need to be installed at the outer ring that transport light
from a UV LED to the strips. Each fiber illuminates 13 strips via a
notch spaced equidistantly every 12~\mm,requiring a total of 96
fibers. The fibers run through small holes in the back plate from the
rear to the front. Attaching four fibers per LED, 24 LEDs housed
outside the rear plate are needed.

\tdrsubsubsec{Support and services}

The preferred option is to build the backward EMC as one unit. In
order to avoid breaking the beam pipe when access is required to the
drift chamber endplate, the backward EMC must be be able to slide back
far enough. The explicit design requires a detailed design of the IR
with all machine elements in place. In the detector position, the
weight is supported by two brackets that are fixed at the rear
endplate and on the inner wall of the IFR backward endcap. When
rolling back the weight could be supported by the tunnel wall. To
facilitate sliding back, rollers are mounted on the inner support
cylinder.

The readout and calibration boards (CMBs) are mounted on the
rear support plate of the EMC. Each readout board has a multiplexed
output cable to the DAQ, a cable providing low voltage inputs, a cable for the MPPC bias voltage of
70~{\rm V}, an electronic calibration input and an analog output.

The four CMBs are also mounted on the rear support plate. Each CMB
holds six LEDs and six PIN photodiodes and needs to read out four
thermocouples. Including the low voltage supplies for the CMB
electronics, 80 cables are needed for the CMBs, yielding 272 cables in
total.

\tdrsubsec{SiPM/MPPC readout}

The photodetectors are SiPM/MPPCs from Hamamatsu (type S10362-11-025P)
with a sensitive area of $1\mm \times 1\mm$ holding 1600 pixels with a
size of $ 25\mum \times 25\mum $. These detectors are avalanche
photodiodes operated in the Geiger mode with a bias voltage slightly
above the breakdown, $50-70~{\rm V}$, providing a gain of a few times
$10^5$. They are insensitive to magnetic fields. Each pixel typically
has a quenching resistor of a few ${\rm M \Omega}$ so it recovers
within $100$~ns. The efficiency is of the order of $10-15\%$. Since
the SiPM/MPPCs record single photoelectrons (Figure~\ref{fig:pes}), they
are auto-calibrating.  SiPM/MPPCs are non-linear requiring
non-linearity corrections for higher energies. As an example,
Fig.~\ref{fig:saturation} shows the response curves of 10,000 SiPMs
measured at ITEP, most of which were installed in the CALICE analog
hadron calorimeter~\cite{bib:ILC}. The dynamic range is determined by
the number of pixels. Properties of several SiPM/MPPCs are listed in
Table ~\ref{tab:mppc}.

\begin{figure}
\begin{center}
\includegraphics[clip,trim = 15 20 20 35,width=\linewidth]{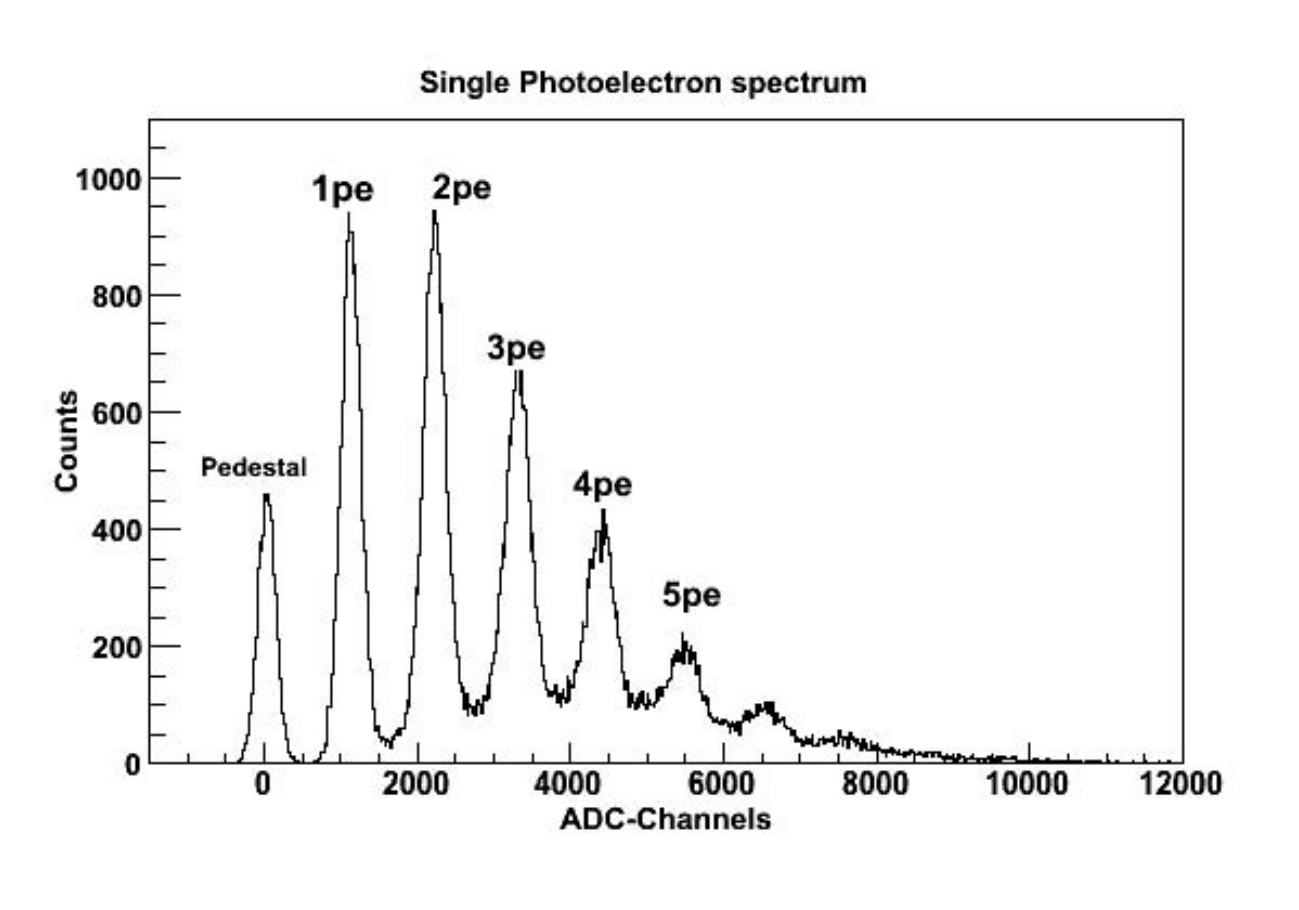}
\end{center}
\caption{Single photoelectron spectrum measured with a Hamamatsu S10362-11-025P MPPC.}
\label{fig:pes}
\end{figure}

\begin{figure}
\begin{center}
\includegraphics[width=\linewidth]{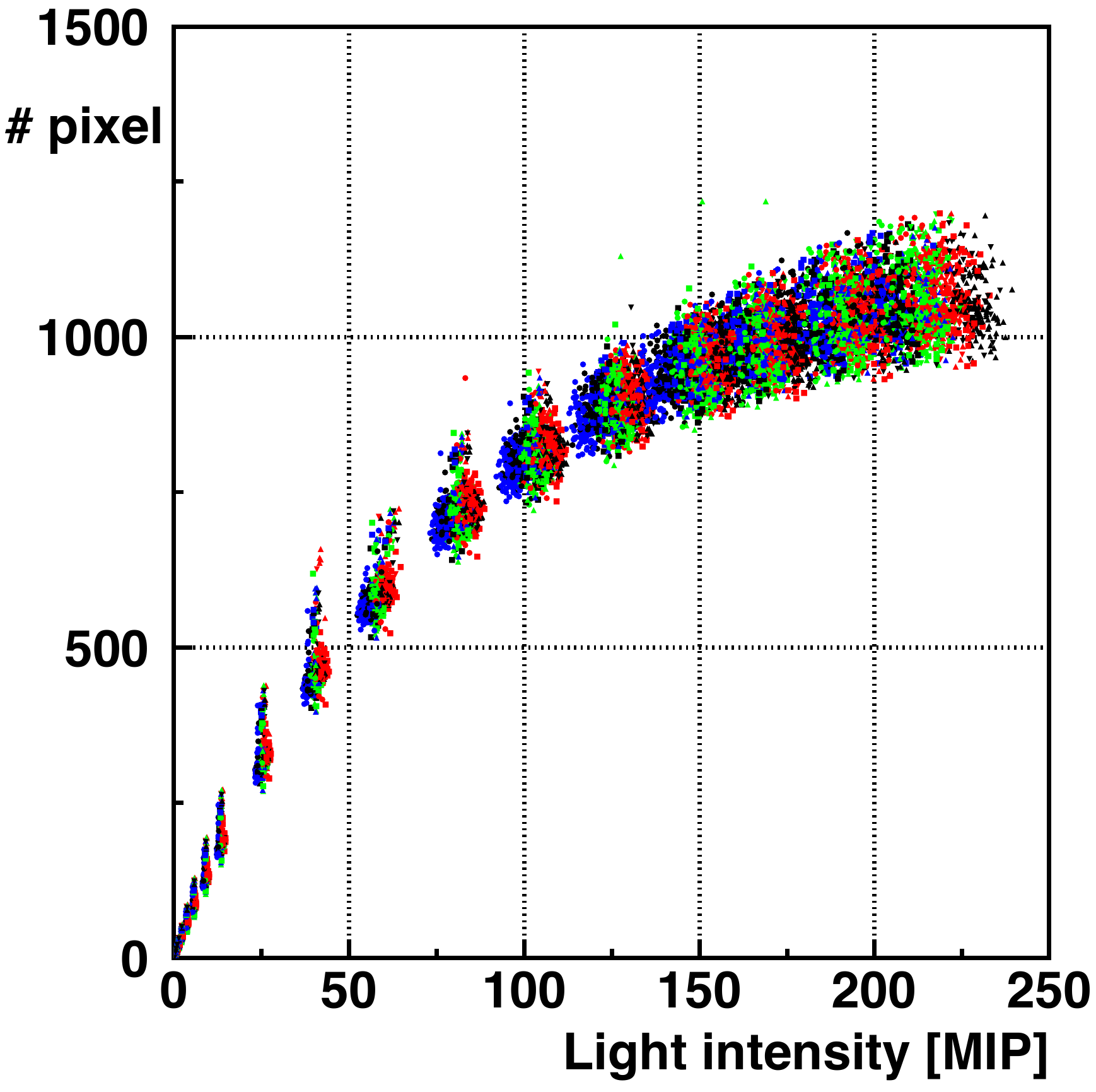}
\end{center}
\caption{Saturation curves of 10,000 SiPMs measured in the CALICE analog hadron calorimeter.}
\label{fig:saturation}
\end{figure}

\begin{table*}
\centering
\caption{Properties of Hamamatsu MPPCs}
\label{tab:mppc}
\begin{tabular}{lcccccccc}\hline \hline
MPPC  &  \# cells  & $C$ &  $R_{\rm cell}$& $C_{\rm cell}$ & $\tau=R_c \times C_c$ & $V_{\rm break~down}$ & $V_{\rm op}$ & Gain \\ 
type & $1/\mma$ & [pF]& ${\rm k}\Omega]$ & [fF]& [ns] & [V] at T=23~C &  [V] at T=23~C & $[10^5]$ \\ \hline
$15~\mum$ & 4489 & 30 & 1690 &6.75&11.4 &72.75 & 76.4&2.0 \\
$20~\mum$&2500&31& 305&12.4&3.8 &73.05&75&2.0 \\
$25~\mum$&1600&32&301&20&6.0&72.95& 74.75&2.75 \\
$50~\mum$& 400&36&141&90&12.7 &69.6&70.75& 7.5 \\
\hline
\hline
\end{tabular}
\end{table*}

A concern with SiPM/MPPCs is radiation hardness. Degradation in
performance is observed in studies performed for the \superb\ IFR,
beginning at integrated doses of order $10^8$ 1-\mev-equivalent
neutrons/\cma~\cite{bib:Faccini}.  This needs to be studied further,
and possibly mitigated with shielding. Another alternative is to
select a different photodetector.  Recently, Hamamatsu has produced
SiPM/MPPCs with pixel sizes of $ 20\mum \times 20\mum $ and $ 15\mum
\times 15\mum $ (see Table~\ref{tab:mppc}).  A study by CMS shows that
the performance of these new photodetectors deteriorates only slightly
after an irradiation of $10^{13} n/\cms$.

\tdrsubsec{Electronics}

For tests so far, the signal of the SiPM/MPPC is first amplified with
a charge-sensitive preamplifier then fed into an auto-triggered,
bi-gain SPIROC ASIC~\cite{bib:SPIROC}.  The SPIROC board has 36 channels. Each channel
has a variable-gain charge preamplifier, variable shaper and a 12-bit
Wilkinson ADC. It can measure the charge $Q$ from one photoelectron
(pe) to 2000 pe and the time $t$ with 100\ps accuracy.  A high-level
state machine is integrated to manage all these tasks automatically
and to control the data transfer to the DAQ. The SPIROC ASIC was designed
to supply the high voltage for the SiPM/MPPC.  Using a DAC, individual
high voltages can be supplied to each
photodetector within a range of $\pm 5$~V.

The SPIROC ASIC gives Gaussian signals with no tails, shows excellent
linearity and low noise. The endcap requires 32 ASIC readout boards.
The boards are mounted in two layers behind the endcap. The first
layer holds 20 boards and the second layer the remaining 12 boards.
Each board connects to 36 SiPM/MPPCs via a ribbon cable that has been
designed for the ILC.

It is unclear whether a version of the SPIROC board can be designed to
meet the data acquisition requirements at \superb.  Thus, we have
adopted the electronics of the PID system as our baseline.  This
provides 12 bit ADCs and 200\ps TDCs. Further details are in
section~\ref{subsec:PIDElectronics}.  To match the PID electronics,
the preamplifier will produce a negative pulse. Seventy-two of the
16-channel PID ASICs are required for the backward EMC. The circuit
boards are placed behind the backward EMC.

\tdrsubsec{Calibration}

An LED-based calibration system with fixed LED intensities is used to
monitor the stability of the strip-fiber-SiPM/MPPC system between MIP
calibrations, to perform gain calibrations, determine intercalibration
constants, and to measure the SiPM/MPPC response functions.  This is
necessary since the SiPM/MPPCs have a temperature and voltage
dependence of the gain of

\begin{eqnarray}
\frac{dG}{dT} \sim  -1.7\% /{^\circ\rm K},\\
\frac{dG}{dV} \sim 2.5\% / 0.1~{\rm V}.
\end{eqnarray}

The temperature and voltage dependence of the scintillation signal is 

\begin{eqnarray}
\frac{dQ}{dT} \sim  -4.5\% / {^\circ\rm K} \\
\frac{dQ}{dV} \sim 7\% / 0.1~{\rm V}.
\end{eqnarray}

The calibration system is based on a new design for the analog hadron
calorimeter. Light from a UV LED is coupled to clear fibers, notched
at equidistant positions. The shape of each notch changes such that it
emits about the same intensity of light.  For the backward EMC, 96
fibers with 12 notches each are sufficient to illuminate all strips.
If four notched fibers and one untouched fiber are coupled to one LED,
24 LEDs are needed whose light is monitored by a PIN photodiode via
the untouched fiber. A few thermocouples distributed throughout the
outer edge of the EMC near MPPCs measure temperature. Both temperature
and bias voltage are recorded regularly by the slow control system.
After temperature and PIN diode correction the stability of LED system
is better than $1\%$.

The calibration boards are $40\mm \times 140\mm$ in dimension and
house one LED, one PIN diode, the electronics to read out the PIN
diode, and monitor temperature and MPPC voltage. The boards are
sufficiently small to be mounted at the backplate near the outer
radius. They can be arranged such that the fibers are nearly straight.

\tdrsubsec{Backward simulation}

A backward EMC model exists in the GEANT4 detector simulation,
modeling 24 layers of scintillator and lead.  Each layer is modeled
by a complete disc without physical strip segmentations in r-$\phi$.
Support structure, fibers, electronics, and cables are presently
ignored.

In the fast simulation, the model does not separate lead from
scintillator.  It uses instead an artificial material that
approximates the overall density, radiation length, interaction length
and Moli\`ere radius of the mixture of lead and plastic. The volume is
divided into eight rings, each of which is divided into 60 segments.
The logarithmic spirals and the lead-scintillator layers, however, are
not modeled explicitly to avoid a complicated shower reconstruction
and the modeling of longitudinal shower energy distribution. The
energy resolution is set to 
$\sigma_{E}/E = 14\%/\sqrt{E(\gev)} \oplus 3\%$.

\tdrsubsec{Performance in simulations}

A GEANT4 simulation is performed to study the energy resolution under
the simplified conditions, ignoring the rest of the detector and
shooting mono-energetic photons perpendicular to the face of the disc.
All energy deposited in the scintillator is collected. No clustering
algorithm is performed. Figure~\ref{fig:BwdG4Edep} shows the energy
deposition for five different photons energies, 0.1~\gev, 0.2~\gev,
0.5~\gev, 1~\gev, and 2~\gev. For 0.1~\gev photons, approximately
9.5\% of the photon energy is deposited in the scintillator on
average.  This percentage drops to 9.0\% for 2~\gev photons.

The energy dependence of the energy resolutions on generated energy is
shown in Figure~\ref{fig:BwdG4Edep2}. The distribution can be fitted
with the function 
$\sigma_{E}/E = 10\%/E(\gev)^{0.485} \oplus 6\%$.

\begin{figure}
\begin{center}
\includegraphics[clip=true,trim = 20 15 5 5,width=\linewidth]{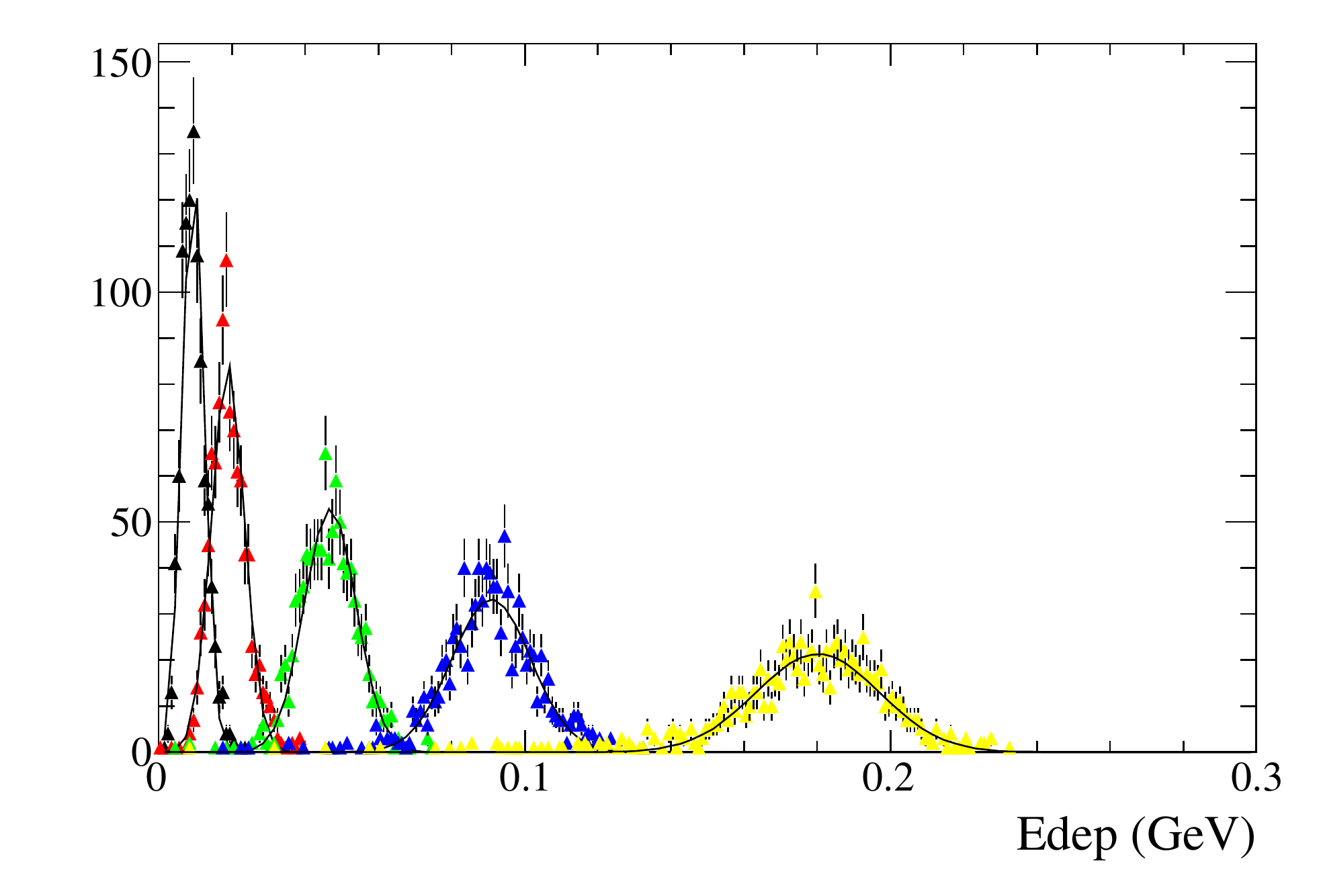}
\end{center}
\caption{Simulated energy depositions from mono-energetic photons  (0.1~\gev, 0.2~\gev, 0.5~\gev, 1~\gev, and 2~\gev) 
observed in the scintillators of the backward EMC. 
\label{fig:BwdG4Edep}
}
\end{figure}

\begin{figure}
\begin{center}
\includegraphics[width=\linewidth]{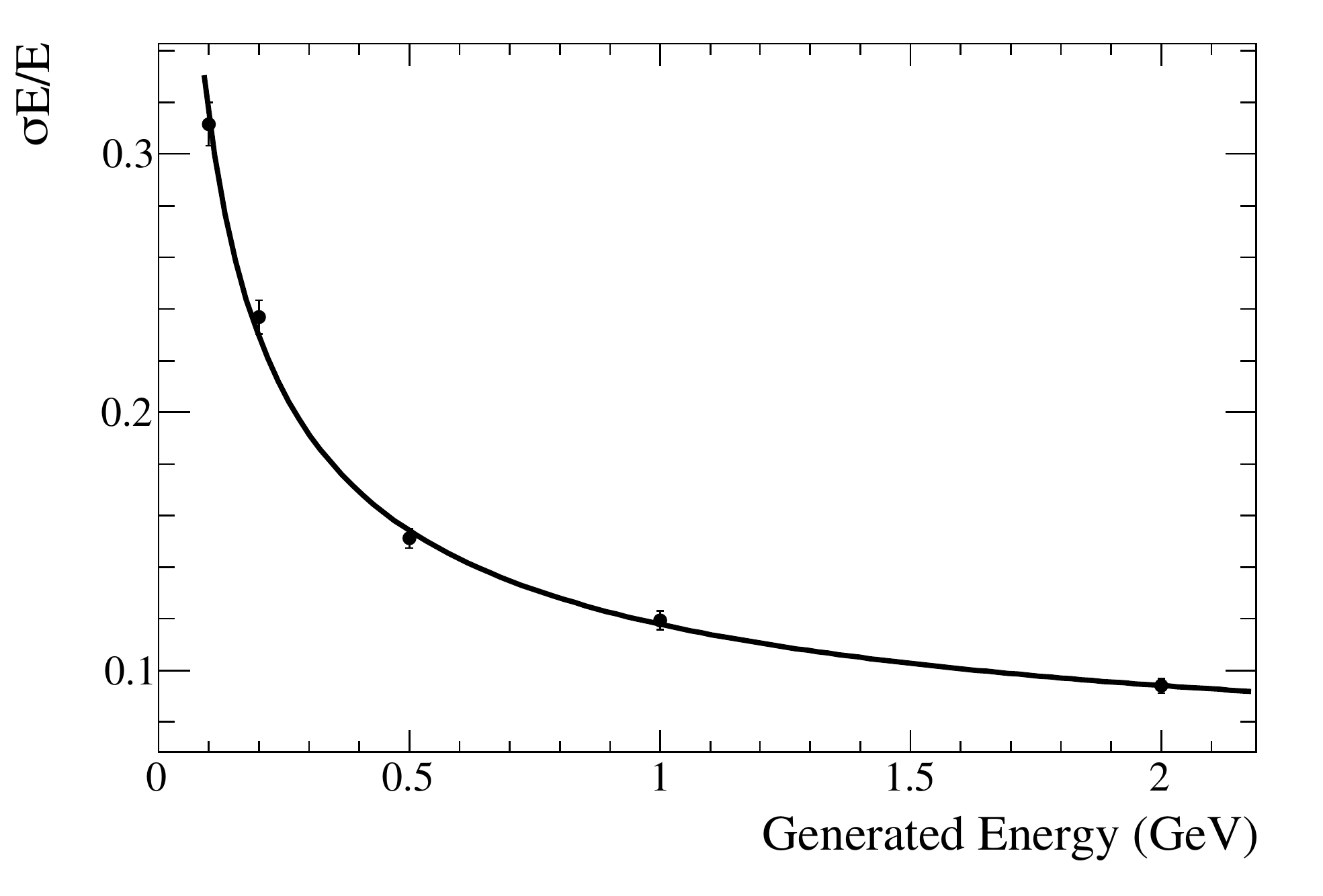}
\end{center}
\caption{The backward EMC energy resolutions, $\sigma_E/E$, where $\sigma_E$ 
and $E$ are the Gaussian width and mean in Figure~\ref{fig:BwdG4Edep}, as a 
function of generated photon energy. 
\label{fig:BwdG4Edep2}
}
\end{figure}

\tdrsubsec{Impact on physics results}

Fast simulation studies have been conducted to investigate the
performance gain achieved by the addition of the backward calorimeter.
The channels chosen to evaluate the impact of this detector are $B
\rightarrow \tau \nu$ and $B \rightarrow K^{(*)}\nu \bar\nu$, since
both are benchmark channels for the \superb\ physics program and
greater detector hermeticity improves signal reconstruction
efficiencies and background rejection.

The study of the leptonic decay $B \rightarrow \tau \nu$ is of
particular interest as a test of the Standard Model (SM) and a probe
for New Physics~\cite{bib:btn_teo}. The presence of a charged Higgs
boson (in {\it e.g.}, a Two Higgs Doublet Model) as a decay mediator
could significantly modify the branching ratio with respect to the SM
value, depending on the Higgs mass and couplings.  A detailed analysis
of this channel is therefore quite important in searches for physics
beyond the SM.  A complementary search for new physics can be
performed using $B \rightarrow K^{(*)}\nu \bar\nu$
decays~\cite{bib:bsnn_teo},~\cite{bib:bsnn_h2dm}. Being mediated by a
flavor-changing neutral current, these processes are prohibited at
tree level in the SM and the higher order diagrams may receive
contribution from a non-standard mechanism. Moreover, new sources of
missing energy may replace the neutrinos in the final state.

The reconstruction of both decay modes is challenging, since the final
state contains more than one neutrino and thus is only partially
reconstructible. Signal events are selected using a recoil method
analysis, in which the signal $B$-meson ($B_{\rm sig}$) is identified
as the system recoiling against the other tag $B$ meson ($B_{\rm
  reco}$). The tag $B$ meson is either reconstructed fully via its
hadronic decays or partially reconstructed from its semileptonic final
states~\cite{bib:bkstnn_exp}.  The rest of the event is assigned to
the $B \rightarrow \tau \nu$ candidate, if it is compatible with one
of the following decay modes of the tau lepton: $\mu\nu_\mu\nu_\tau$,
$e\nu_e\nu_\tau$, $\pi\nu_\tau$, $\pi\pi^0\nu_\tau$,
$\pi\pi^0\pi^0\nu_\tau$, $\pi\pi\pi\nu_\tau$, or
$\pi\pi\pi\pi^0\nu_\tau$.  These final states cover about $95\%$ of
all $\tau$ decays, and have one charged particle (1-prong) or three
charged particles (3-prong), with the possible addition of one or two
$\pi^0$s. Since final states containing one or more $\pi^0$ cover
about 40\% of the tau decay modes, an increase in the EMC coverage
improves substantially the efficiency of tau identification.

Candidates for the $B \rightarrow K^{(*)}\nu \bar\nu$ sample must be
compatible with one of the following final states: 
$K^{*+}\rightarrow K_S(\pi^+\pi^-) \pi^+,~ K^+ \pi^0$; 
$K^{*0}\rightarrow K^+ \pi^-$; $K^+$; or $K_S \rightarrow \pi^+\pi^-$ 
(semileptonic analysis only).  In
the analyses with a $K^{(*)}$ in the final state, further selection
criteria are applied using kinematic quantities related to the
goodness of the $B_{\rm reco}$ and $K^{(*)}$ reconstruction and event
shape variables that test the energy balancing in the event and the
presence of missing energy due to the neutrinos in the final state.
The present level of the analyses is very similar to that in
Refs.~\cite{bib:bkstnn_exp} and~\cite{bib:bknnSL_exp}.

Backgrounds are further rejected using $E_{\rm extra}$, the total
neutral energy in the calorimeter of particles not associated with
either $B$ meson. Signal events peak at low values of $E_{\rm extra}$,
while background events which contain additional sources of neutral
showers tend to have higher values of $E_{\rm extra}$. The
discriminating power of this observable obviously increases with the
calorimeter coverage.

\begin{table*}
\centering
\caption{Relative gain in  $S/\sqrt{S+B}$ by including the backward EMC in the event selection for $B \rightarrow K^{(*)}\nu \bar\nu$ and $B \rightarrow \tau \nu$ decay channels  reconstructed in the recoil of
  $B$ semileptonic and hadronic tags. The first uncertainty is statistical and the second is systematic.}
\label{tab:phys_bwdemc}
\begin{tabular}{lcc}\hline \hline
channel & Semileptonic  & Hadronic \\
\hline
$B \rightarrow \tau \nu$ & $(6.1 \pm 0.1 \pm 0.7)\%$ & $\simeq 3--5\%$\\
\hline
$B \rightarrow K^+ \nu \bar \nu$ & $(5.8 \pm 1.0 \pm 0.6)\%$ & -\\
\hline
$B \rightarrow K_S \nu \bar \nu$ & $(6.0 \pm 0.4 \pm 0.6)\%$ & -\\
\hline
$B \rightarrow K^{*+}(K^+\pi^0) \nu \bar \nu$ & \multirow{2}{*}{$(7.0 \pm 0.2 \pm 0.7)\%$}& $(5.9 \pm 2.5 \pm 0.5)\%$\\
$B \rightarrow K^{*+} (K_S \pi^+)\nu \bar \nu$ & & $(6.2 \pm 2.1 -0.5)\%$\\
\hline
$B \rightarrow K^{*0} \nu \bar \nu$ & $(9.1 \pm 0.4 \pm 0.7)\%$ & $(7.3 \pm 1.8 \pm 0.5)\%$\\
\hline
\end{tabular}
\end{table*}


\begin{figure*}[tbp]
\begin{center}
\includegraphics[width=0.49\linewidth]{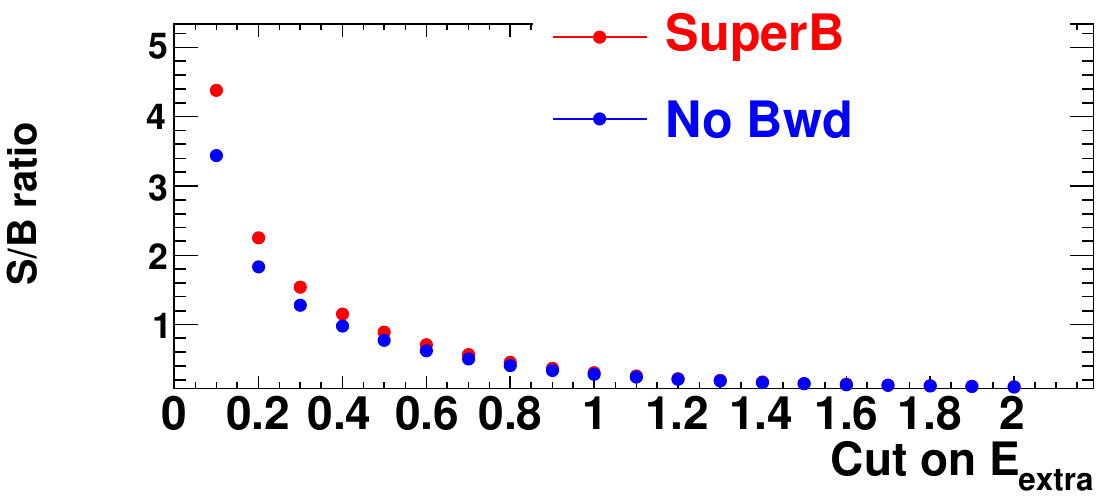}
\includegraphics[width=0.49\linewidth]{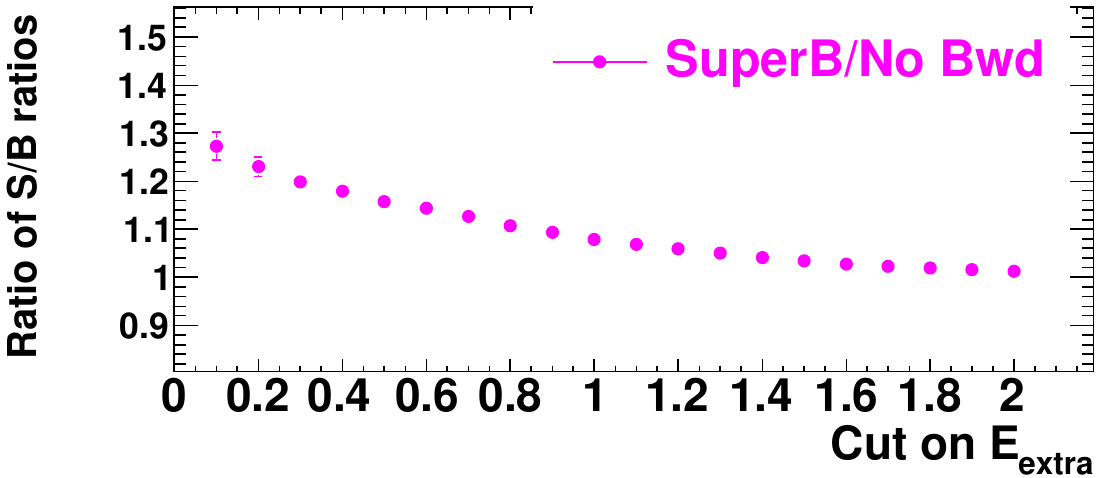}
\end{center}
\caption{Left: Signal-to-background ratio with and without
a backward EMC, as a function of the $E_{\rm extra}$ selection.
Right: Ratio of the S/B ratio with
a backward EMC to the S/B ratio without a
backward calorimeter, as a function of the $E_{\rm extra}$ selection.
}
\label{fig:BwdSB}
\end{figure*}

The performance of the backward EMC is assessed by comparing the
signal-to-background ratio, $S/B$, and the statistical precision,
$S/\sqrt{S+B}$, of the expected signal with and without the backward
EMC used as a veto device for extra-neutral energy. For this task, the
extra neutral energy reconstruction is split into two disjoint
calorimeter regions: $E^{\rm extra}_{\rm brrfwd}$ covering the barrel
and the forward region and $E^{\rm extra}_{\rm bwd}$ covering the
backward region only.  Furthermore, the effects of machine backgrounds
(Sec.~\ref{sec:machine_bkg}) superimposed on physics events are taken
into account.

The results for both rare decay channels are summarized in
Table~\ref{tab:phys_bwdemc}.  For $B \rightarrow \tau \nu$
reconstructed in hadronic tags, the improvements in $S/\sqrt{S+B}$ are
shown as a range sampling over the individual $\tau$ final states. For
$B \rightarrow \tau \nu$ reconstructed in semileptonic tags, the
improvement in $S/\sqrt{S+B}$ is presented as an average over all
$\tau$ final states.  Combining $S/\sqrt{S+B}$ of both tags, the net
gain is of the order of $10\%$. For $B \rightarrow K^*\nu \bar\nu$,
the net gain for hadronic and semileptonic tags combined ranges
between $8\%$ and $16\%$ depending on the final state. For $B
\rightarrow K\nu \bar\nu$ the net gain is about $6\%$.

These improvements in turn yield improvements in the precision of
measured physical observables
For
${\cal B}(B \rightarrow K^+\nu \bar\nu)$ and ${\cal B}(B \rightarrow
K^{*}\nu \bar\nu)$ the $3\sigma$ significance for evidence without the
backward EMC is reached at $5~{\rm ab}^{-1}$ and $51~{\rm ab}^{-1}$,
respectively. When adding the backward EMC, the $3~\sigma$
significance is already reached at $4.5~{\rm ab}^{-1}$ and $42~{\rm
  ab}^{-1}$, respectively.

The signal-to-background ratio for reconstruction of the rare decay
$B\to\tau\nu_{\tau}$ with and without a backward endcap in shown in
Figure~\ref{fig:BwdSB} (left). The ratio of S/B with and without a
backward endcap is shown on the right.

In addition to the measurement of neutral clusters, the backward EMC
can be used to improve the $\pi^0$ reconstruction efficiency. If one
photon is reconstructed in the backward EMC, the $\gamma\gamma$
invariant-mass resolution ranges from $\sim24\mev$ for $200\mevc$
$\pi^0$s to $\sim13\mev$ for $1\gevc$ $\pi^0$s.


It is worth mentioning that the inclusion of the backward EMC not only
improves the background rejection, but also the $B$-tagging
efficiency. The impact on the \B reconstruction efficiency is
determined by using the decays $\Bm\to\Dz\pim, \Dz\to\Km\pip\piz$.
Events are separated into two groups: the first uses only photons from
the barrel and forward endcap, while the second includes photons from
the backward EMC with polar angle between $-0.96<\cos\theta <-0.89$ as
well. The $\piz$ mass window is defined as 120--145\mev (100--180\mev)
for the first (second) group.  For $B$ candidates, \Dz s are selected
by $1.830< m_{K\pi\piz} < 1.880\gev$ and $-80 < \Delta E < 50\gev$.
The \mes distribution is fitted to determine the \B reconstruction
efficiency. Including the backward EMC, the signal efficiency
increases by nearly 4\% in this particular channel (see
Figure~\ref{fig:bwdemc_breco}).

\tdrsubsec{Use for particle identification}

Charged particles moving in the backward direction typically have
lower momentum. Thus, ionization loss and time-of-flight measurements
may provide useful information for particle identification, $e.g.$ for
$e/\pi$ and $K/\pi$ separation.

A preliminary time-of-flight study is performed with fast simulation.
Single kaons or pions are generated that move towards the backward
EMC. The true arrival time calculated at the first layer is smeared
with a Gaussian resolution to simulate the measured time distribution.
Velocity distributions are obtained for kaons and pions at a given
momentum from the measured time and reconstructed track path length.
Both mean and RMS values are extracted from these distributions to
determine $K/\pi$ separation in standard deviation units ($\sigma$) as
a function of momentum for different time resolutions, in which
uncertainties in the momentum measurement and path length
reconstruction are included. Figure~\ref{fig:BwdPID} shows $K/\pi$
separation in units of standard deviations as a function of momentum
for time resolutions of $100\ps,~50\ps,~20\ps,~10\ps$, and perfect
timing. For example, for a 100\ps time resolution, a $K/\pi$
separation of more than three $\sigma$ can be achieved for momenta up
to 1\gevc and approximately 1.5$\sigma$ up to 1.5\gevc.

Since each layer measures the time distribution, 24 measurements will
be averaged. In addition to timing, the ionization is measured in each
layer. For MIP-like particles, the average energy loss per layer is
$dE_{Pb}=4.3~\mev$ and $dE_{\rm scintillator} =0.6~\mev$.  A 0.5~\gevc
$\pi$ is at the ionization minimum, a 0.5~\gevc $K$ is below the
minimum and a 0.5~\gevc e is at the relativistic plateau. For MIP
particles, the ionization loss in the 24 layers is $\Delta E
=117~\mev$. Since the energy loss below the minimum increases with
decreasing momenta as $1/\beta^2$, $dE/dx$ measurements in the endcap
can be combined with the $dE/dx$ information from the SVT and DCH.
Figure~\ref{fig:dedx} shows the ionization curves for $e, \mu, \pi, K$
and $p$ as a function of momentum.  A $>3\sigma~K/\pi$ separation is
achievable for momenta up to $0.6-0.7\gev$.

\begin{figure*}[thb]
  \begin{center}
    \includegraphics[width=0.49\linewidth]{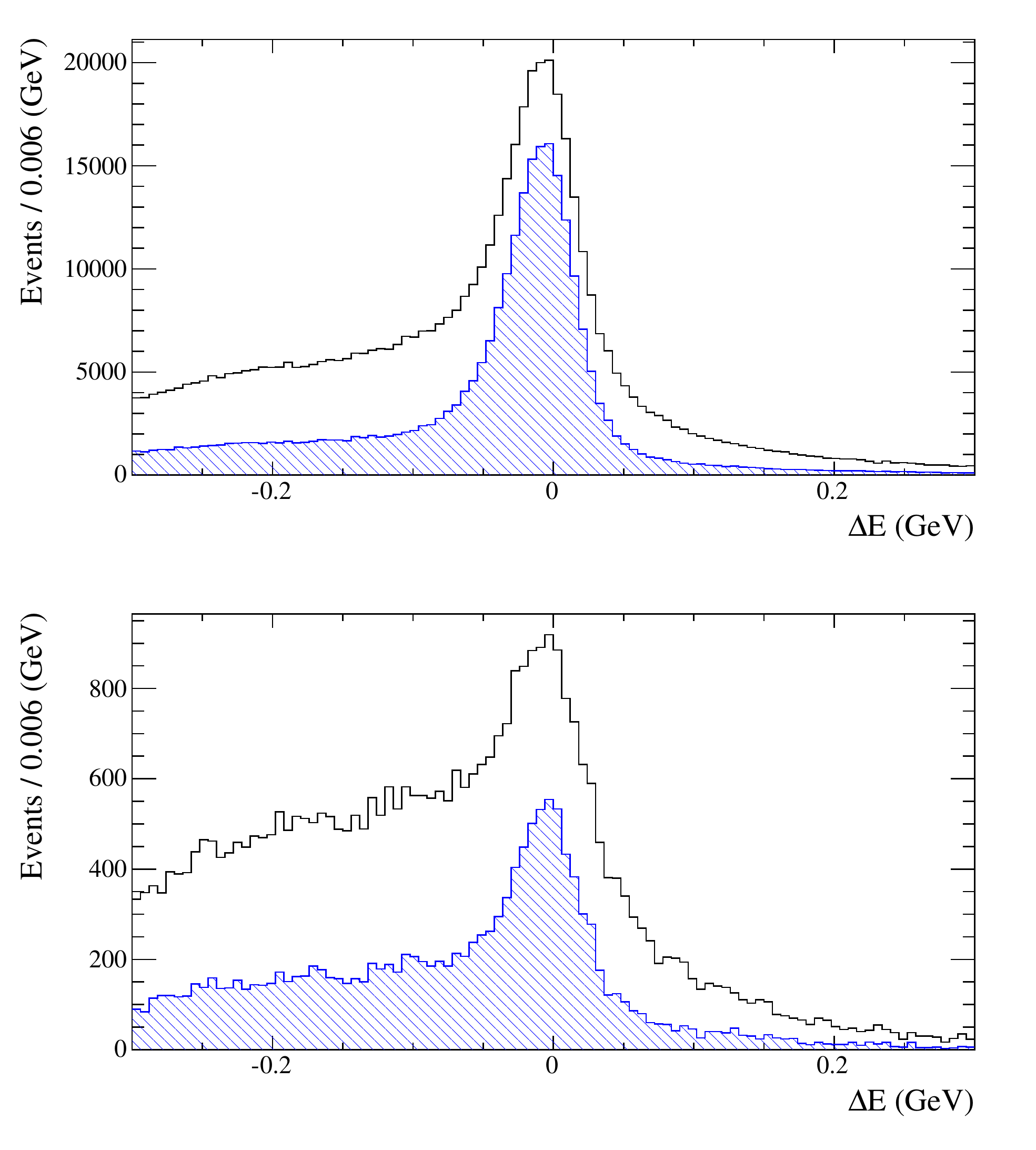}
    \includegraphics[width=0.49\linewidth]{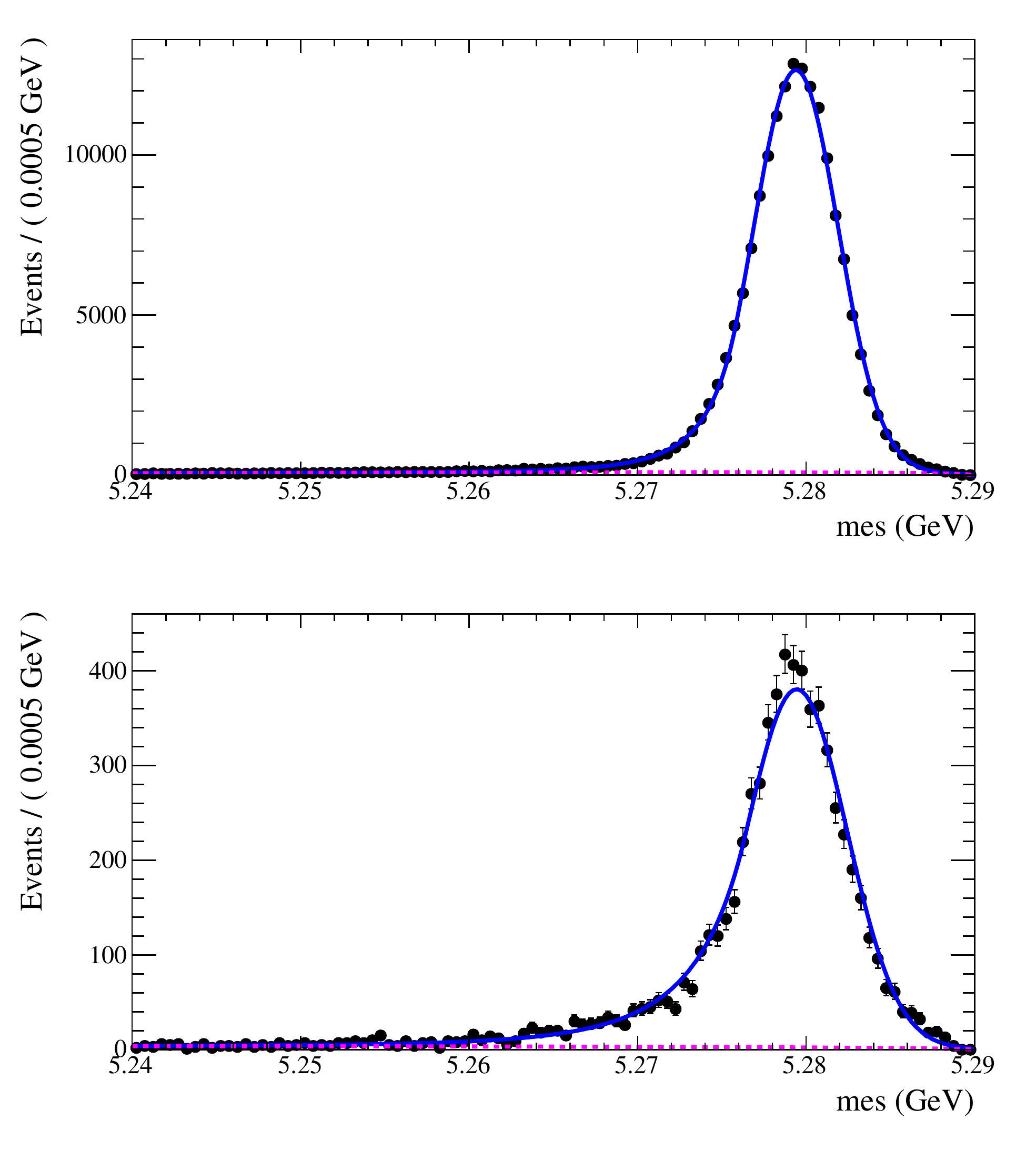}
    \caption{(left) $\Delta E$  and (right) $\mes$ distributions for the decay $\Bm\to\Dz\pim$ with 
      $\Dz\to\Km\pip\piz$ reconstruction for (top) both $\gamma$'s in barrel and forward 
      endcap and (bottom)  one $\gamma$ in the backward EMC. The two histograms in each $\Delta E$ distribution are
      before and after a requirement on the $\Dz$ mass.}
    \label{fig:bwdemc_breco}
\end{center}
\end{figure*}

\begin{figure}[tbp]
\begin{center}
\includegraphics[width=\linewidth]{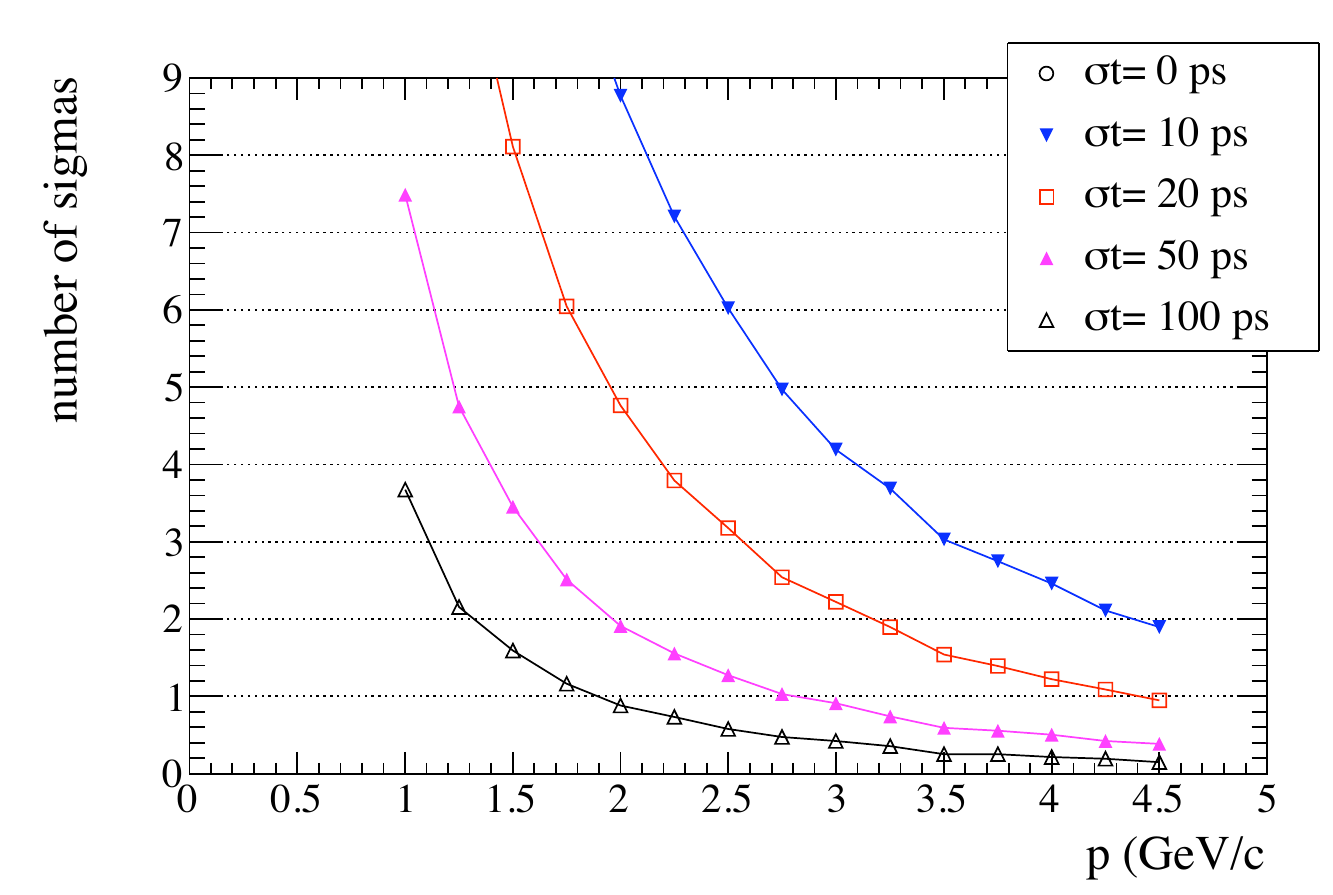}
\end{center}
\caption{$K/\pi$ separation versus measured momentum for different timing resolutions in the backward EMC. 
The $\sigma_t=0$ curve is offscale. The finite separation
in this case is caused by uncertainties in the momentum measurement.}
\label{fig:BwdPID}
\end{figure}

\begin{figure}[tbp]
\begin{center}
\includegraphics[width=\linewidth]{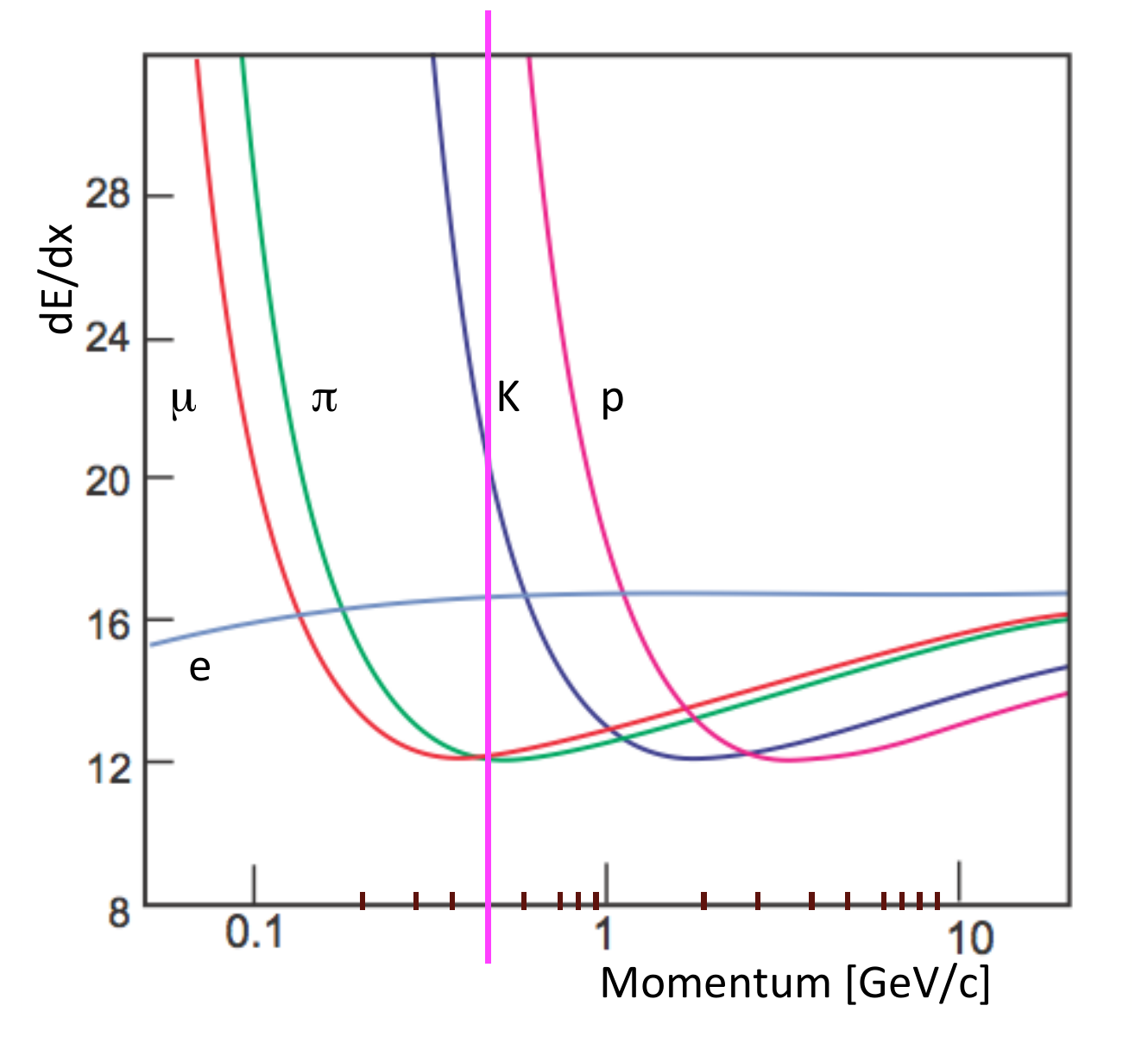}
\end{center}
\caption{Calculated energy loss curves versus momentum for $e,~\mu,~\pi,~K$, and $p$ in one layer of the backward EMC.}
\label{fig:dedx}
\end{figure}

\tdrsubsec{Discussion of task force conclusions}

\begin{itemize}

\item The device adds to the physics program in important channels
  through the increased hermeticity.

\item The design is technically plausible and cost-effective. However,
  operation of the readout in the radiation environment should be
  studied further, and a prototype should be demonstrated in a beam test.

\item The possibility of using the backward EMC for PID through
  time-of-flight is attractive and should be pursued with further
  R\&D.

\item The capability to install this device should be preserved, as
  opposed to, for example, extending the drift chamber.

\end{itemize}

\tdrsec{Environmental monitoring}

The calorimeter must be protected from various hazards.
We summarize these hazards in this section.

The principal heat loads in the EMC are from the electronics. The main
barrel readout electronics is water-cooled. Fluorinert is the coolant
for the barrel and endcap preamps and the endcap digital electronics.
Thermal stability is important: the CsI(Tl) light output, the PIN
diode leakage current and the APD gain are all temperature-dependent.
In addition, the PIN diode and APD glue joints can be broken due to
differential expansion. Temperature excursions of greater than $\pm
10^\circ$ are known to be capable of breaking the PIN diode glue
joint. A specification that variations be less than $\pm 5^\circ$ has
a comfortable safety margin. In \babar, the temperature was monitored
with 256 sensors on the barrel, and 100 sensors on the forward endcap.
Figure~\ref{EMC:fig:emc_temp_history2} provides an example of the data
on the thermal stability. This system will be retained in \superb. The
backward endcap requires additional temperature monitoring and
control.

The CsI crystals are slightly hygroscopic, so they must be kept dry.
This is achieved by flowing nitrogen through the crystal volume.  In
addition, both oxygen level and humidity are monitored.  The readout
for these sensors, as well as the temperature sensors is via a CANbus
General Monitoring Board~\cite{Anthony:1999xt}.

{\centering
\begin{figure}[htb]
\includegraphics[width=\linewidth]{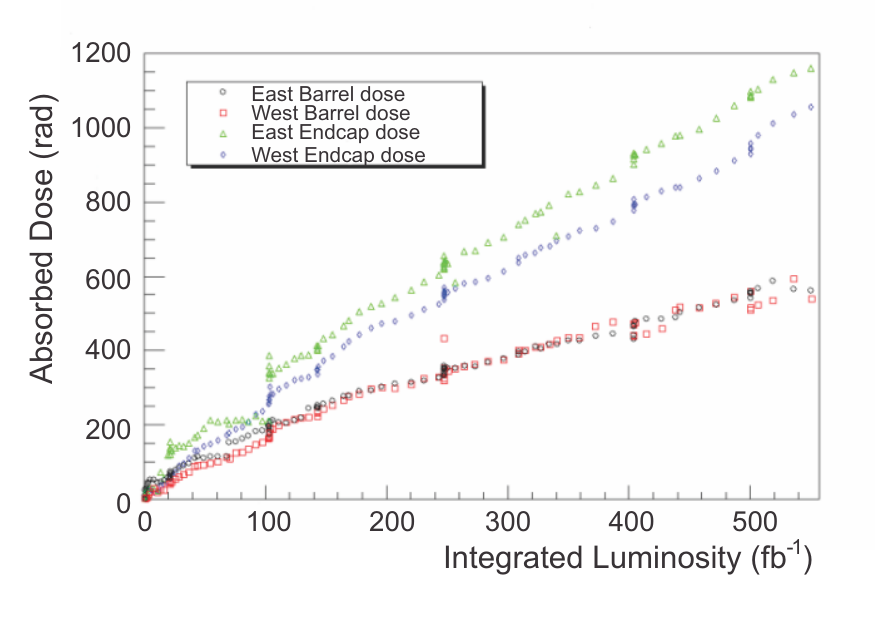}
\caption { Radiation dose in rads as measured by radFETs in the
  \babar\ barrel and endcap. The horizontal axis is the integrated
  luminosity in fb$^{-1}$. The curves show the barrel and endcap
  doses, divided into east and west sides of the detector.}
\label{fig:rad_dose}
\end{figure}
}

The CsI(Tl) crystal light output and uniformity also degrade with
radiation so the radiation exposure must be monitored.  Two radiation
monitoring systems for the EMC have been in use in \babar, and will be
reused in \superb.  First is a set of small CsI(Tl) crystals with PIN
diode readout for prompt monitoring and protection, and second is a
set of RadFET dosimeters to track long term exposure.

There are 16 CsI(Tl) crystal/PIN diode packages installed at both
forward and backward locations on the detector (4 backward and 12
forward; the 12 forward being in the inner and outer faces and the
inner radius of the forward EMC). The crystals are approximately
$4\times 4\times 1$ cm$^3$ with Hamamatsu S3590-01 PIN diodes.  The
diodes are run unbiased, in order to have a more stable leakage
current pedestal. In the current configuration at \babar, the signals
from the four packages at the forward EMC inner radius are summed and
used to provide an injection inhibit signal when the signal exceeds
0.7 V. This corresponds to a dose rate around 0.2 mRad/s.

The radiation dose over the long term is monitored by a set of 56
real-time integrating dosimeters (radFETs) placed in front of the
barrel and 60 radFETs in front of the endcap.  Readout of the radFETs
is performed with a RadFET Monitoring Board (RMB). The RMB ties the
radFETs to ground except during a reading, at which time a constant
current source injects a current and the resulting radFET voltage is
digitized~\cite{Khan:2006pj}.  The accumulated dose measured by these
radFETs over the life of the \babar\ detector is shown in
Fig.~\ref{fig:rad_dose}, separately for the endcap, the forward, and
the backward barrel of the calorimeter. These dose measurements may be
used to understand the changing light yields shown in
Fig.~\ref{fig:EMC:emc_ly_change}.

\bibliographystyle{\tdrbibstyle}
\balance
\bibliography{EMC/EMC-bib}


\renewcommand{\ChapLabel}{chap:IFR} 
\tdrchap{Instrumented Flux Return}


\tdrsec{Physics Goals and Performance Requirements}
       
The principal task of the Instrumental Flux Return (IFR) detector is the 
detection of muons and neutral hadrons. The detector is integrated
into the large iron structure that forms the magnet return yoke. 

Reliable muon detection is vital for searches of New Physics in
processes like $b\to s \mu^+ \mu^-$, $\tau\to\mu\gamma $, $\tau^+\to
\mu^+\mu^-\mu^+$, $\tau\to\mu X$, (with $X = \rho,\eta^{(\prime)},
f_0, ...$), $B_{d,s}\to \mu^+\mu^-$, $D^0\to\mu^+\mu^-$, to name but a
few~\cite{WP}~\cite{ADRIAN1}.  Moreover, efficient muon identification
identification is crucial for a substantial part of studies related to
the determination of sides and angles of the unitarity triangle, which
are based on semileptonic decays of $B$ mesons. In these decays, the
sign of the lepton charge determines the flavour of the parent
heavy meson, thus providing one of the tags for the CP-asymmetry measurements.\\

The asymmetry of the machine results in a strong correlation between
momenta and production angles of the final state particles. The
absolute population and the shape of the momentum spectra are quite
different in the backward, central and forward regions of the
detector. In this regard, the different \superb\ boost with respect to
\babar\ (see Chapter~\ref{chap:detoverview}) plays an important role
and must be taken into account for the detector optimization.
Figure~\ref{ifr:fig:babarsuperbmom} shows the momentum distribution
versus the polar angle for muons coming from $B$ semileptonic decays
in \babar\ and \superb: with the lower \superb\ boost the momentum
distribution is flatter and the barrel and backward regions gain
importance.
\begin{figure}[b]
\begin{center}
\includegraphics[width=\linewidth]{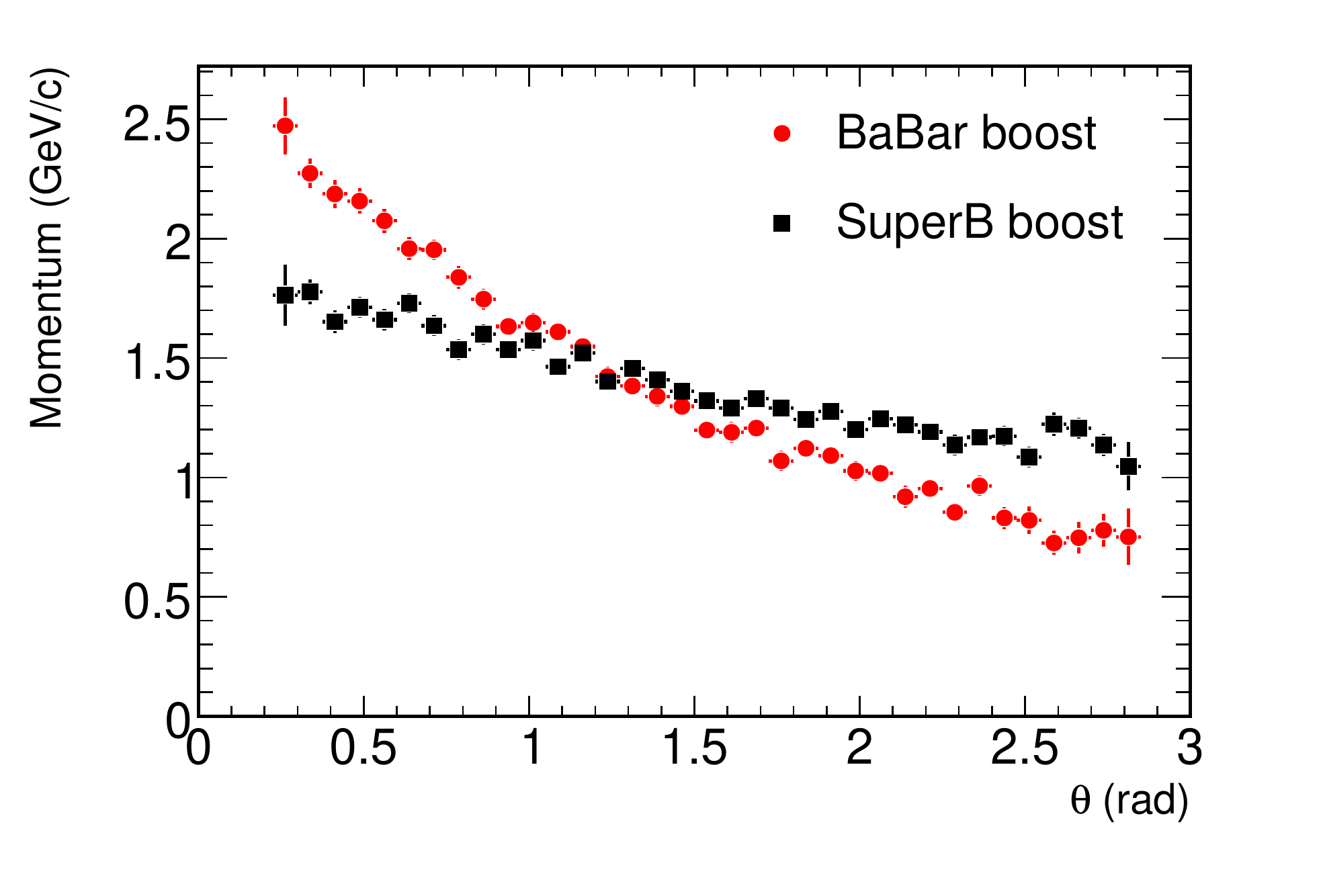}
\caption{Average momentum of muons coming from the semileptonic decay $B\to D^{(*)}\ell\nu$ as s function of the polar angle
with the \babar\ and \superb boost.}
\label{ifr:fig:babarsuperbmom}
\end{center}
\end{figure}
Almost one fifth of all $B$ decays contain at least one muon in the
\superb detector acceptance region.  Most of them are either produced
directly from the $B$ meson or from the cascade $D$s. $\tau$ decays
and $e^+e^- \to \mu^+\mu^-$ processes are an addtional small source of
muons. High momentum ($p>1\gevc$) muons come mostly from direct
decays, while muons from $D$ decays peak at lower momenta.
Figure~\ref{ifr:fig:mu-mom} presents the momentum and angular
distributions for muons coming from the decays $B\to K^{(*)}
\mu^+\mu^-$, $B\to \mu \nu$ and from $D\to K\mu\nu$, where $D$ mesons
originate from $B\to D\ell \nu$ events.  Pions are very abundant in
$B$ decays and their momenta peak below 1\gevc, as shown in
Figure~\ref{ifr:fig:semilepmu-mom}.

\begin{figure}[tbp]
\begin{center}
\includegraphics[width=\linewidth]{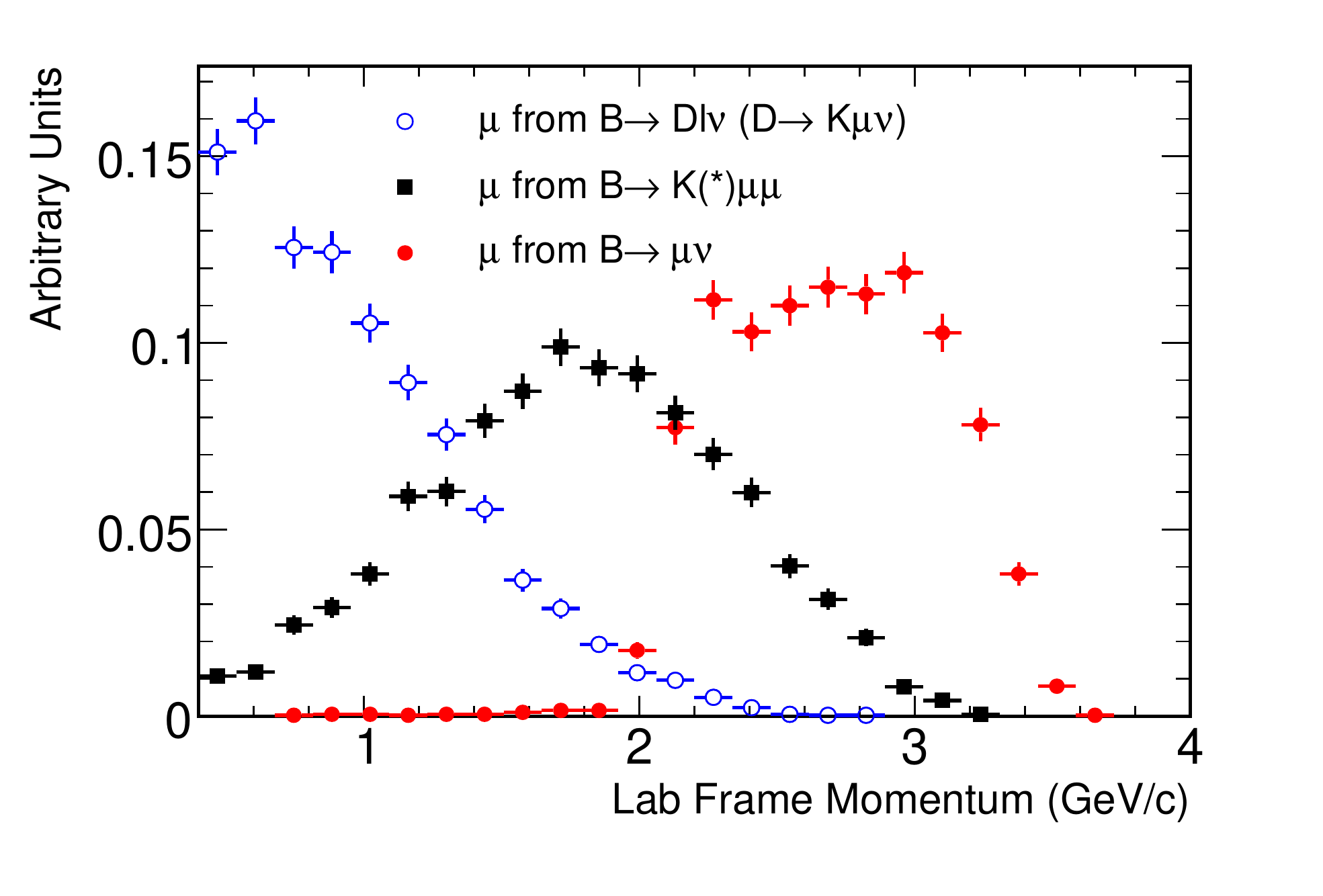}\\
\includegraphics[width=\linewidth]{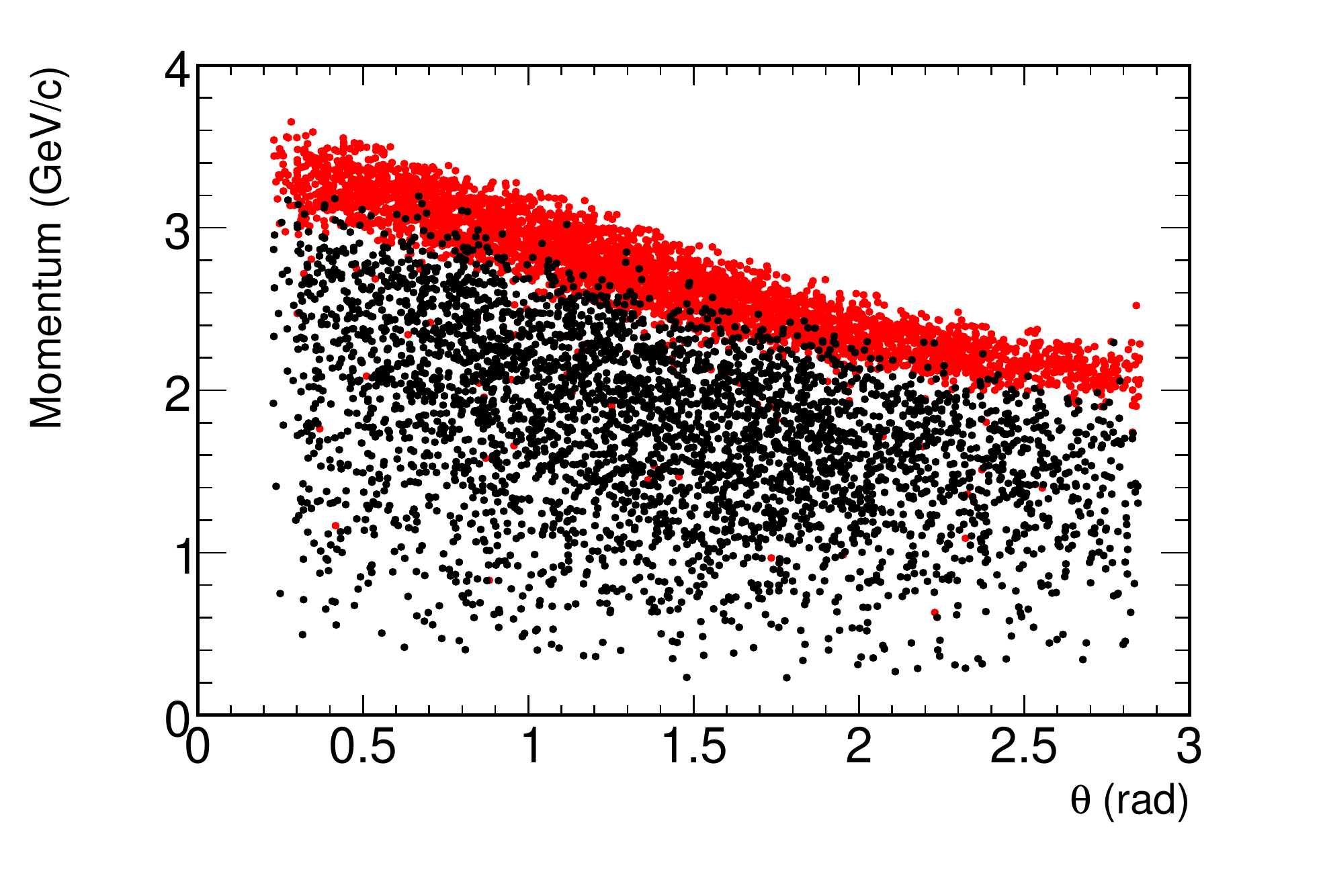}
\caption{Top: momentum distributions for muons coming from $B\to K^{(*)}\mu\mu$, from $B\to \mu\nu$
and from the $D$ decays $D\to K\mu\nu$ (where the $D$ meson originates from $B\to D\ell \nu$). Bottom: momentum versus the polar angle
for muons coming from $B\to K^{(*)}\mu\mu$ (black) and from $B\to \mu\nu$ (red).}
\label{ifr:fig:mu-mom}
\end{center}
\end{figure}

\begin{figure}[tbp]
\begin{center}
\includegraphics[width=\linewidth]{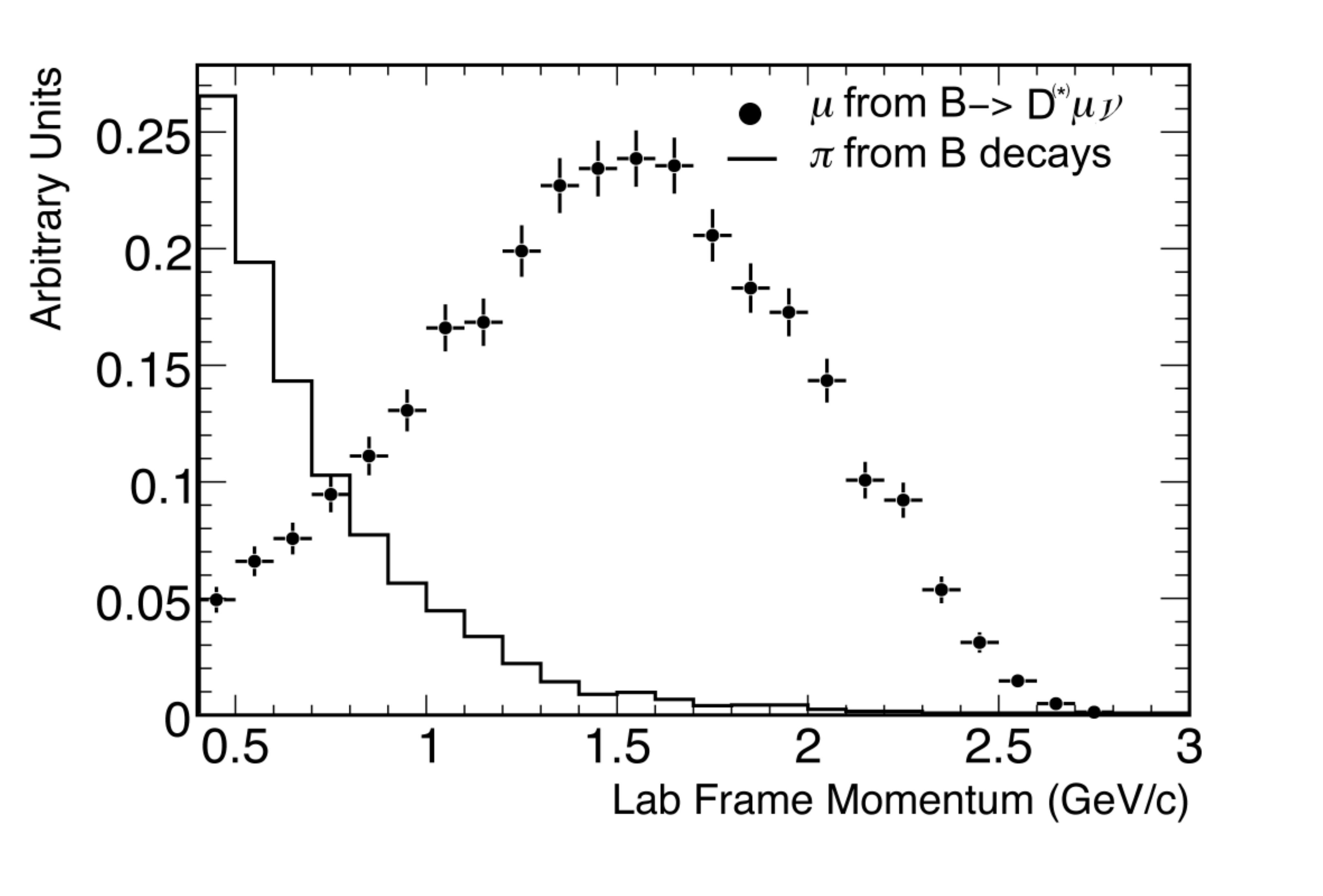}
\caption{Momentum distributions for muons coming from $B\to D^{(*)}\ell\nu$ (data points) decay and for pions coming
from generic B decays (black line).}
\label{ifr:fig:semilepmu-mom}
\end{center}
\end{figure}

The muon momenta to be covered by the IFR range from a few hundred
\mevc up to a few \gevc. The lower limit of this range is set by the
magnetic bending in the barrel region and by energy losses in the
inner detectors of the two endcap regions.

Charged tracks found with the internal tracking system will be matched
to the tracks in the IFR. Their identification as muons or hadrons
will result from detailed analysis of the hit patterns in the active
detector. The maximum tolerable
level of hadron misidentification will depend on the physics topic under study. \\

The IFR will also act as a hadronic calorimeter for the detection of
neutral hadrons. Special attention is paid to the reconstruction of
$K^0_L$ mesons which is performed by using IFR information in
conjunction with information provided by the electromagnetic
calorimeter. The ability to reconstruct the $K^0_L$ mesons allows, in
particular, to compare $CP$-violation effects in the decay channel
$B\to J/\psi K^0_L$ with those in $B\to J/\psi K^0_S$.  In other
analyses, the IFR will be used to veto events where a $K^0_L$ is
present.  These events are an important source of background for the
decays whose signature is characterized by a large missing-momentum,
like $B \to \tau \nu$ or $B \to \pi \tau \nu$.  About 60\pct of the
$K^0_L$ interact before reaching the IFR and can be detected with the
EMC, the remaining 40\pct interact only within the IFR.
Figure~\ref{ifr:fig:kl-mom} shows the momentum and angular
distributions for $K^0_L$ coming from $B$ generic decays and from
$B\to J/\psi K^0_L$.

\begin{figure}[h]
\begin{center}
\includegraphics[width=\linewidth]{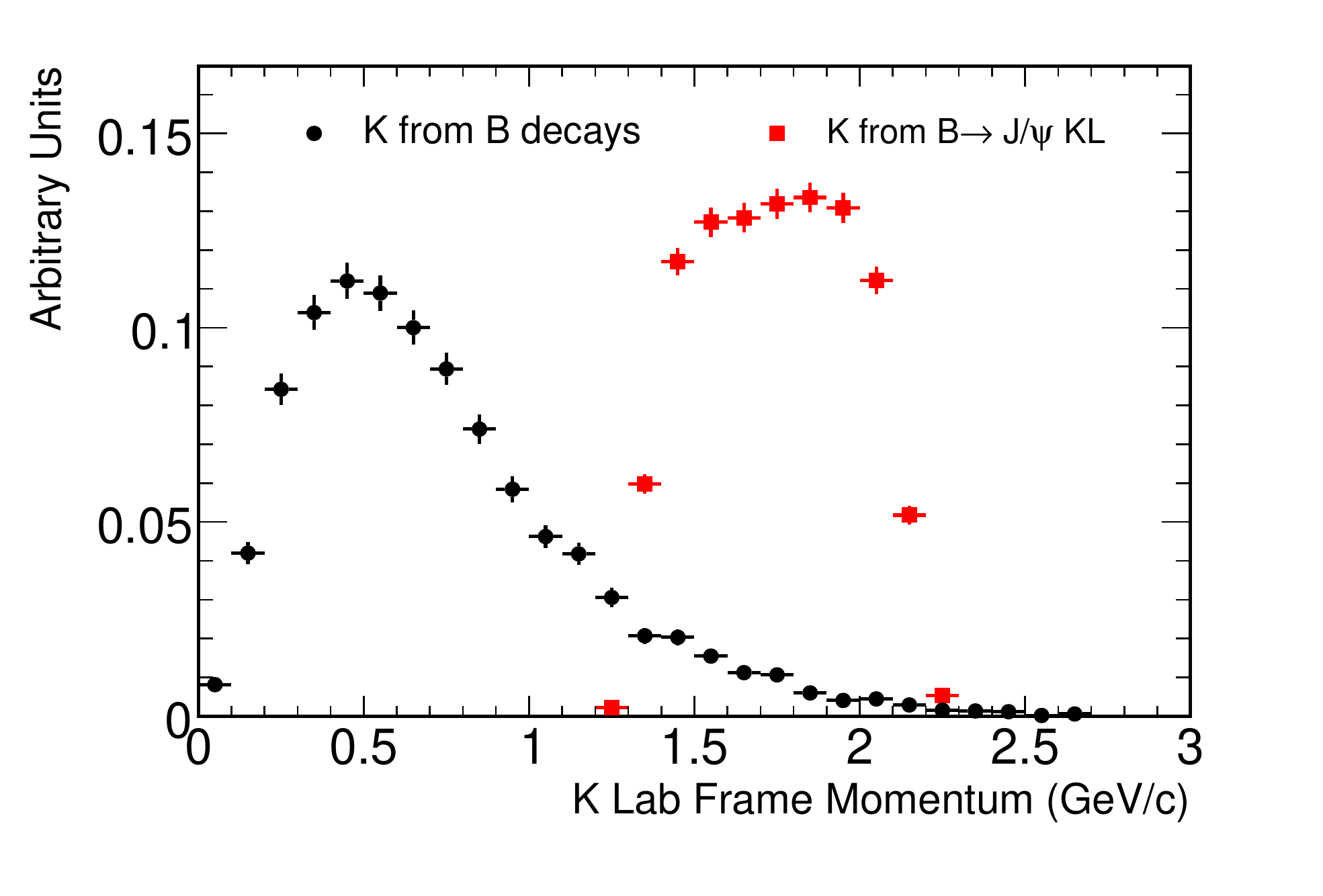}\\
\includegraphics[width=\linewidth]{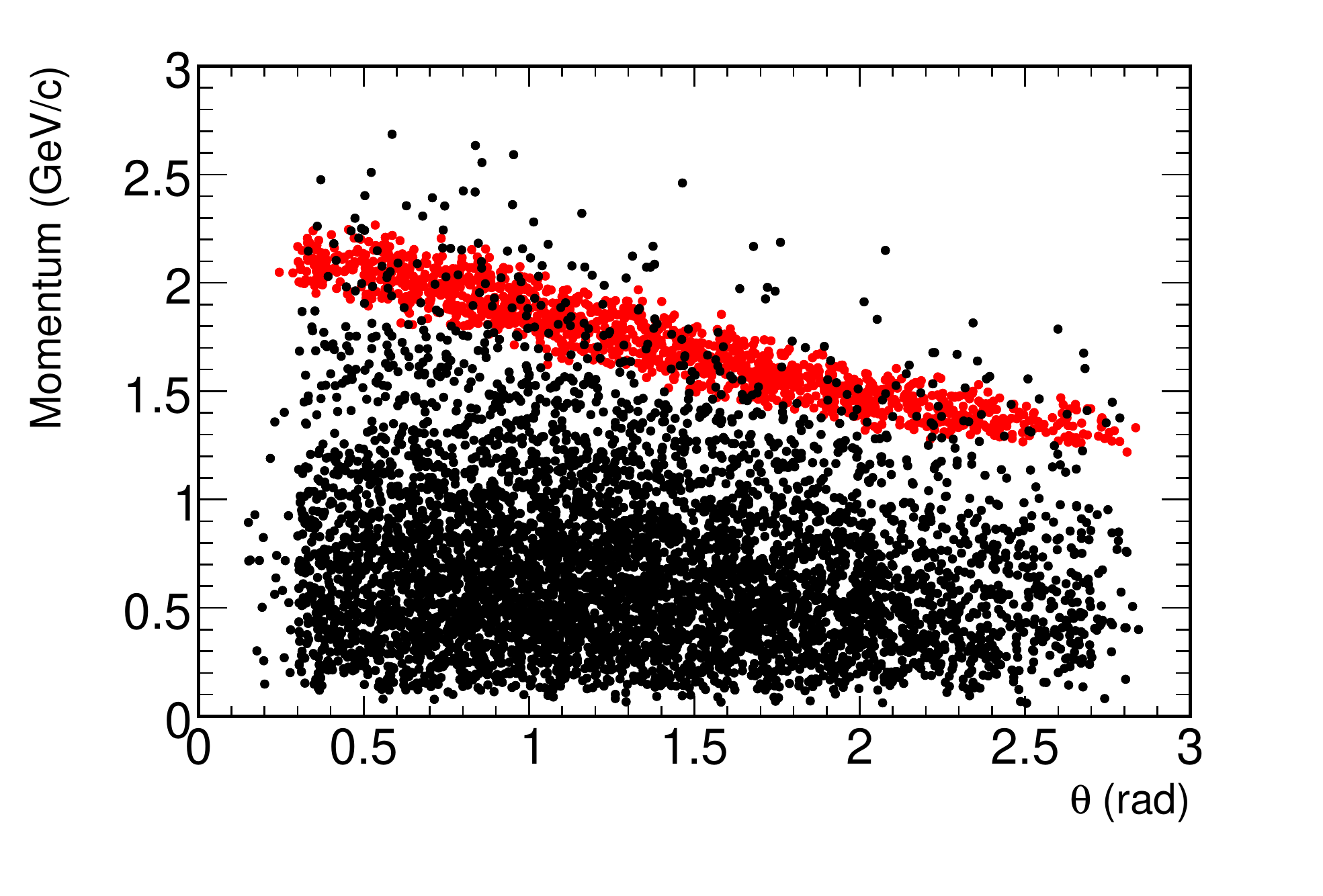}
\caption{Top: momentum distributions for $K^0_L$ coming from $B$ generic decays (black) and from $B\to J/\psi K^0_L$ (red).
Bottom: momentum versus the polar angle for the $K^0_L$ coming from the same events.}
\label{ifr:fig:kl-mom}
\end{center}
\end{figure}


\tdrsec{Detector Overview}

	\tdrsubsec{The Absorber Structure}
	
The \superb\ flux return structure, used as absorber material to
identify muons, will be composed of an hexagonal iron yoke with a
barrel and two endcaps (see Figure~\ref{IFR-flux-return}).  The barrel
is the assembly of six sextants, each one composed of an inner and an
outer wedge, while each of the two endcaps comprises two doors.  All
these components are made of welded steel plates with a thickness
ranging from 20\mm to 100\mm.  The geometry of the absorber will be
almost identical to the \babar\ IFR, except for the overall thickness of
the steel layers, which will be increased in \superb.

The thickness of the \babar\ IFR amounted to 650\mm (600\mm) for the
barrel (endcaps), respectively. The overall thickness of the steel in
\superb\ should be upgraded to 920\mm in order to increase the
particle identification capability of the flux return.

In \superb\ 9 layers of scintillator (instead of the 17 detector
layers in \babar) will be installed into slots of the flux return.  In
the barrel, the first layer of detectors will be mounted in front of
the first layer of steel and the last layer will be outside the outer
wedge plates, therefore only 7 slots will be needed for the detector
layers inside the steel. This arrangement is shown in Figure~\ref{sextant}.

In the doors 7 or 8 gaps will be used for detectors for the forward or
backward endcaps, respectively. On the backward side the available
space is limited by the horsecollar structure.

In order to minimize the cost, the baseline design foresees the reuse
of all the \babar\ structure, and the increase of the steel thickness
by inserting metal plates in those of the 17 slots that are left
empty. The nominal thickness of each detector plane is of the order of
25\mm, while the gap width of the \babar\ wedges and doors have
nominal dimension between 30\mm and 35\mm.

\begin{figure}[tbp]
\begin{center}
\includegraphics[width=\linewidth]{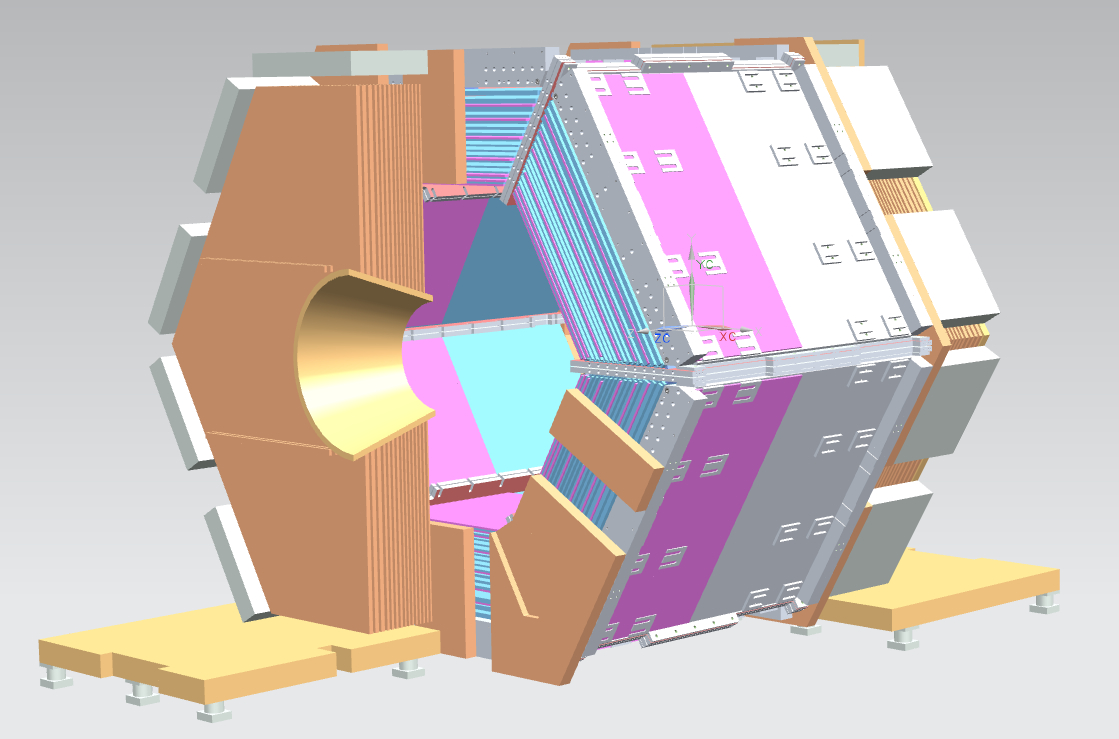}
\caption{The IFR flux return structure. }
\label{IFR-flux-return}
\end{center}
\end{figure}

The gaps not used by scintillators will be filled with metal plates.
To reach the 920\mm of the designed thickness there are 270\mm missing
from the \babar\ parts in the barrel, thus the gaps should be filled
with plates of 27\mm.  According to preliminary measurements of the
\babar\ barrel gaps, 25\mm plates should fit in all the barrel, except
for maybe one or two gaps of the lower sextant. The feasibility of the
insertion of 27\mm thick plates should be checked by measuring the
gaps with adequate resolution and proper gauges. 

The baseline design of the barrel foresees the reuse of the brass
plates of \babar\ and the insertion of 25\mm thick metal plates in the
remaining unused gaps; this allows to reach 882\mm of overall metal
thickness. The reuse of the six \babar\ brass layers inside the barrel
implies a reduction of the overall thickness by 18\mm compared to the
possible use of 25\mm plates. To still reach the design thickness, the
insertion of thin (3--5\mm) plates into the gaps where the brass is
already located, could be investigated.

\begin{figure}[tbp]
\begin{center}
\includegraphics[width=\linewidth]{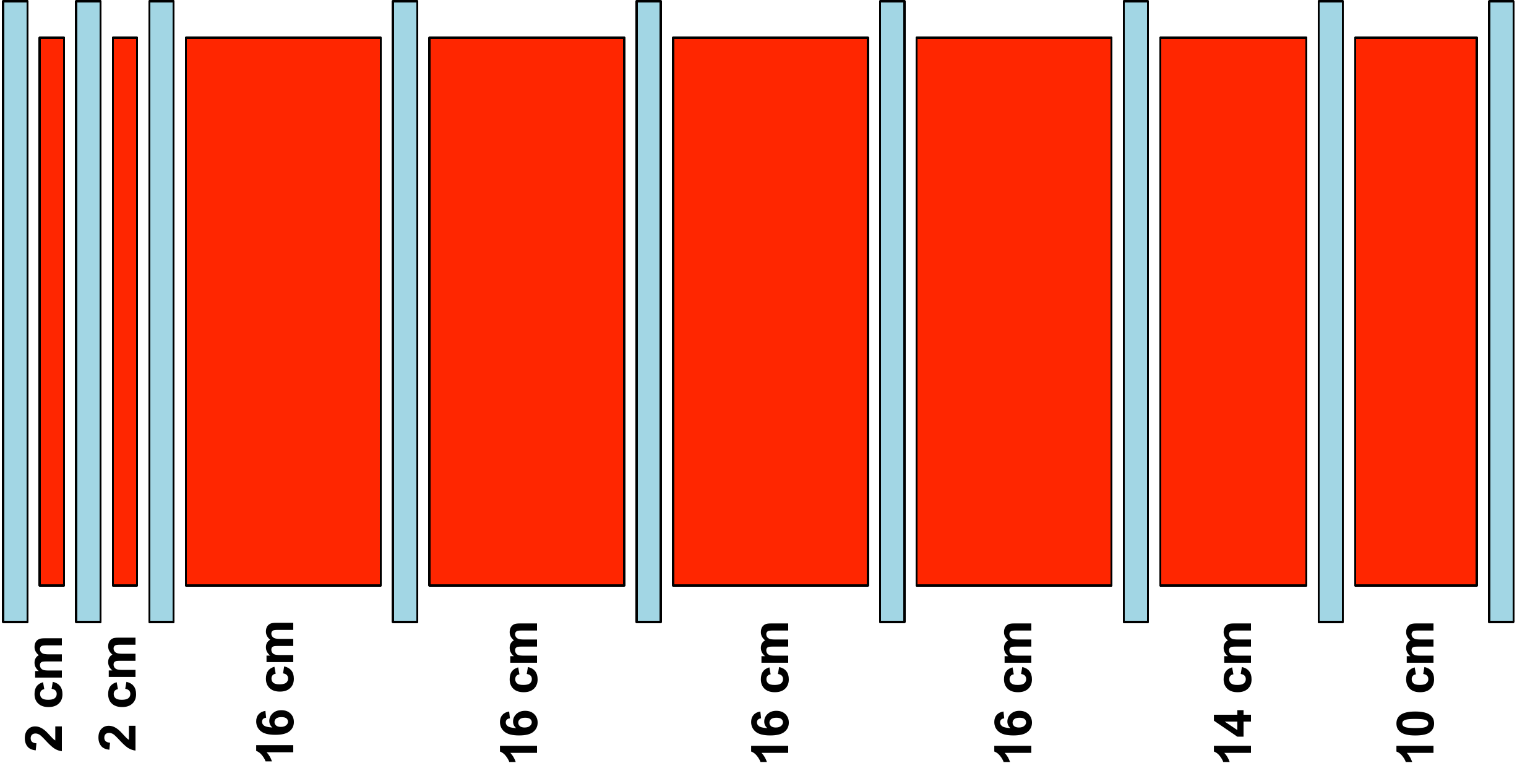}
\caption{Schematic of a possible IFR stratigraphy with 8 active layers; incoming particles travel from left to right.}
\label{sextant}
\end{center}
\end{figure}

For the doors the overall amount of steel can reach the thickness of
850\mm by filling the unused gaps.  The geometry of the doors allows
for the increase of their overall thickness by adding metal plates on
the outer face, as it was done in the forward doors of \babar, where a
plate of 100\mm (4 inches) was added on the outer face to increase the
shielding with respect to background. This additional slab can also be
reused in \superb to reach the nominal thickness of 900\mm. In the
backward door the addition of an external plate 100\mm thick could be
incompatible with the presence of the horsecollar, as it would limit
the movements needed for the opening and closing of the doors. The
thickness that can be added could be of the order of 50\mm, thus the
overall thickness can be 900\mm.  In order to define the maximum
thickness that can be added on the outer face of the backward door,
the opening movements of the doors need to be detailed and actual
dimensions of the horsecollar must be measured.

The metal composing the plates to be inserted in the unused gaps should match the following criteria: 
\begin{itemize}
\item non-magnetic material in order to avoid the increase of magnetic forces on the coil;
\item small interaction length;
\item acceptable flatness;
\item reasonable cost.
\end{itemize}

In \babar\ a few layers were filled with brass plates, but nowadays
this choice would be very expensive due to the cost of copper, which
suffered a strong increase in the last years. A viable alternative
choice could be a stainless steel 304L with certified low magnetic
permeability. While a reasonable value for the magnetic permeability
would be less than 1.04, it might be possible to reach a certified
permeability smaller than 1.02, depending on the production process
and eventual machining. The flatness of the plates is also important
and shall be checked and, if needed, corrected in order to easily fit
the plates in the gaps. The weight of the 25\mm stainless steel plates
to be inserted is about 110 ton, in addition to the brass already
available from \babar. A large part of this additional mass is due to
the filling of the backward doors and the barrel.
Another difference of the \superb\ IFR with respect to \babar\ will be
in the connection between the barrel wedges and the outer frame
(cradle and arches).  As the last detector layer will be located
outside the wedges, a useful gap for detectors will be realized in
between the structure and the wedges, reducing as much as possible the
connection between the two parts.  The connections will be made only
at the corners of the wedges leaving available the substantial part of
the top surface of outer wedges to lay down scintillators. As a
result, all the parts involved (cradle, arches, outer wedges) might
require some workshop intervention in order to modify the relative
coupling. The increasing of thickness also results in an increased
weight of the structure and the reduction of connections between the
outer structure and the wedges reduces the strength and stiffness of
the overall steel body. Therefore, the overall deformation of the
structure will be verified in detail to ensure that stresses are not
critical and deformations are compatible with the required overall
precision. The installation procedure is well defined: It will be
similar to the \babar\ installation, but in addition it can also rely
on the procedures that have been documented during the dismantling of
\babar.

During disassembly the \babar\ barrel has been measured with a laser
tracker. As a result, it will be possible to reference the \superb\
assembly to what was done in the past. The filling of the gaps in the
doors and wedges with plates could be done before the assembly of
these parts in the detector, allowing to save some time since those
processes can be parallelized. Once inserted, the plates will be fixed
in position by welding or by provisional brackets in order to prevent
their displacements during handling. Plate insertion could be assigned
to an external company and be performed either on the \superb\ site or
in a company workshop, depending on the incidence of costs of the
special transportation. The tools like spread beams, scaffolding,
platforms, etc., used to assemble and dismount \babar\ could also be
reused for \superb, provided that they are reviewed and certified
according to Italian law. In this case the cost of shipping, modifying
and certifying the tools should be evaluated and compared with the
cost for the production of brand new items. A preliminary offer for
filling the gaps of \babar\ with 110 tons of stainless steel plates is
around 600\,kEuro. The shipping cost of large pieces (wedges, doors,
cradle, arches) from the US to Italy is estimated to be about
500\,kEuro due to the non-standard dimensions, while import taxes and
duty are neglected for the purpose of this document. Other costs
encompass the shipping of the various IFR parts that can fit in
containers or open-rack platforms and the modifications to cradle,
arches and outer wedges that would be required to allow the addition
of the outmost layer of detectors.

According to these estimates, the overall cost of reusing the flux
return from \babar\ for \superb\ will be of the order of 1.5\,MEuro.

	\tdrsubsec{The Active Detector Choice}
	
The IFR active detectors will cover a surface of approximately
1200\msquare. They will be inserted in the gaps between the iron
plates, where access and replacement will be very difficult and in
some cases impossible. It is, therefore, necessary that the chosen
detector technology is simple and reliable as well as low in cost.

In the \superb\ environment, the critical regions for backgrounds are
the small polar angle sections of the endcaps and the edges of the
barrel internal layers, where we estimate that in the hottest regions
the rates are of the order of a few 100\Hzpcma. These rates are too
high for gaseous detectors. While the \babar\ experience with both
RPCs and LSTs has been, in the end, positive, detectors with high rate
characteristics are required for the high background regions of
\superb. A scintillator-based system provides much higher rate
capability than the gaseous detectors. For this reason, the baseline
technology choice for the \superb\ detector is extruded plastic
scintillator read out through WLS fiber and avalanche photodiode
pixels operated in Geiger mode.

Several detection modules will be joined together to fill each IFR
layer. Each module will be composed of orthogonal scintillator strips,
so that by combining their information each layer can measure space
points. A baseline design that guarantees sufficient performance to
match the physics requirements has already been established, however
the actual design of the modules is not completed yet. The
optimization of the number of WLS fibers for each bar and their
position as well as some mechanical details will be finalized in the
near future based on current R\&D studies. The total number of
channels in the current design is about 22,000.

For each layer, the modules will be inserted into the iron gaps using
appropriate tooling. The front end stage of the readout system will be
positioned in the IFR cable conduits near the detector, where they
will be readily accessible for testing and maintenance.


\tdrsec{Backgrounds}
	\label{ifr:sec:bkg}

Machine-related backgrounds are one of the challenges for the \superb\
detector and background considerations influence several aspects of
its design. The IFR detector has been simulated using \superb\ FullSim
based on \geantFour (see Section~\ref{sec:Full_Simulation}).  The
results of these simulations show that the primary source of
background are: radiative Bhabha events, Touschek scattering, pair
production, beam-gas scattering, and photons from synchrotron
radiation. For a detailed description of the background sources see
Chapter~\ref{chap:mdi}. For the IFR detector the dominant background
source is radiative Bhabha scattering as will be explained later in
this section.


\tdrsubsec{Neutron Background}

In this context there is a non-negligible production of neutrons via
giant resonances formation~\cite{giantresonance}.  The neutrons
produced with this mechanism have energies of some MeV but are
moderated when interacting with the detector material.  For this
reason the neutron energy spectrum in the \superb\ environment has a
very wide range as shown in Figure~\ref{fig:nspectrum} of
Sec.~\ref{neutronsection} where the neutrons are classified according
to their kinetic energy.

\begin{figure}[tbp]
  \begin{center}
   \includegraphics[clip=true,trim = 0 0 60 0,width=\linewidth]{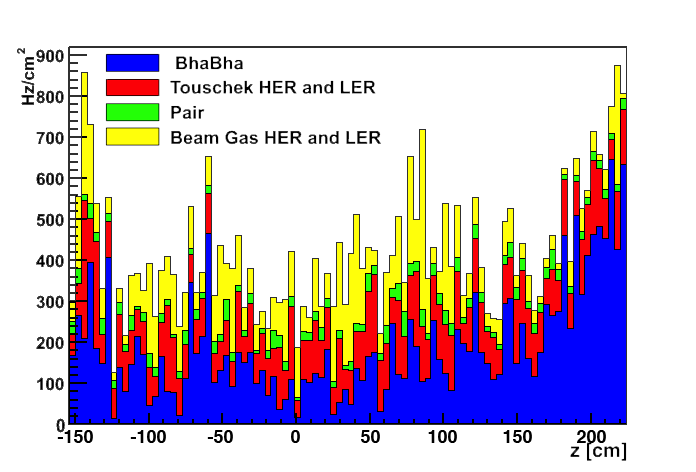}
   \includegraphics[clip=true,trim = 0 0 100 0,width=\linewidth]{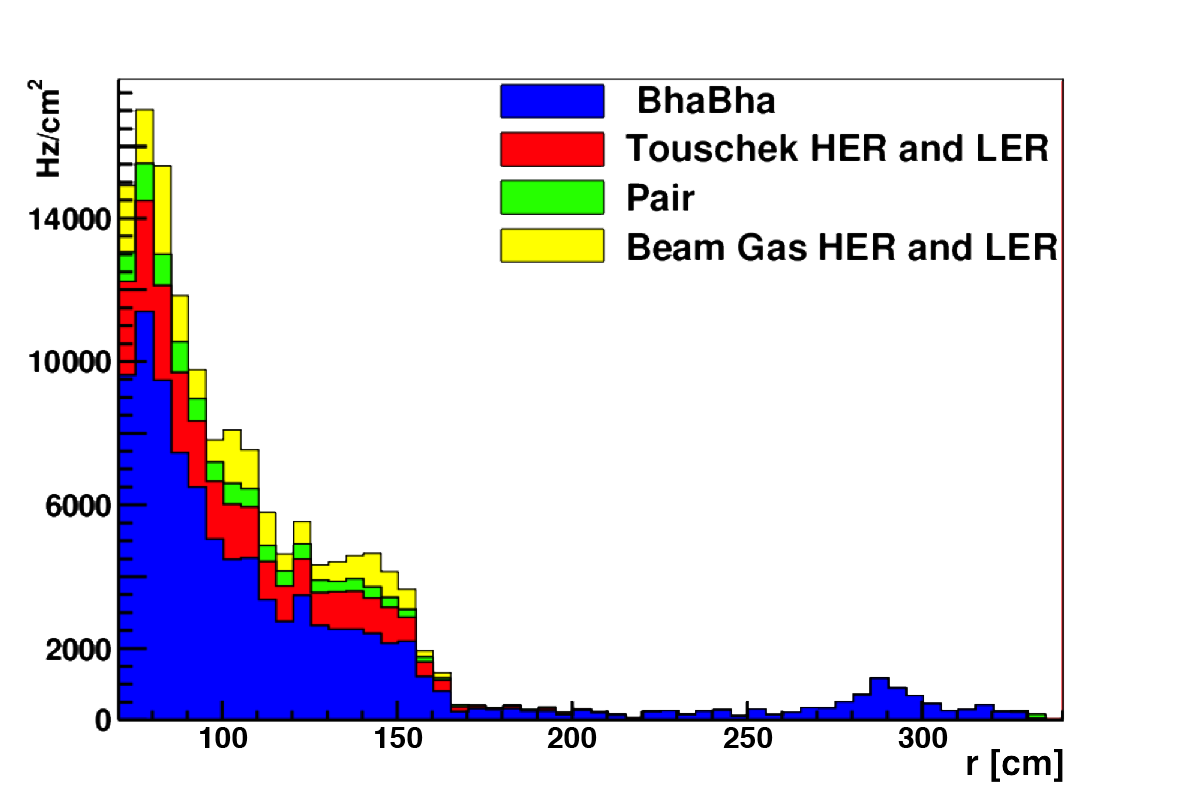}
   \caption{Neutron rate on IFR Barrel (top) and Forward Endcap
     (bottom) innermost layer for different background sources. The
     rates are normalized to 1\mev equivalent \cite{NeutronEq} .}
\label{fig:neutronrateL0FW}
\end{center}
\end{figure}

Since the neutron spectrum has such a large range, and the interaction
with matter strongly depends on it, it is standard to characterize the
neutron fluence from a source in terms of an equivalent mono-energetic
neutron (1\mev) fluence~\cite{NeutronEq,Lindstrom:2000gd}.
For this reason all neutron rates in this section are normalized to
the equivalent of a 1\mev neutron (${\mathrm {n}}_{eq}$).

Figure~\ref{fig:neutronrateL0FW} shows the neutron rate on the
innermost layer (layer 0) for the Barrel and Forward Endcap for the
different background sources. The rates are normalized to 1\mev as
explained above, and they include a safety factor ($\times 5$) that
takes into account the fact that the simulation could underestimate
the real background contribution.  The rate on the Forward Endcap
layer 0 is higher compared to that of the Barrel layer 0; this is due
to the fact that this layer cannot be shielded because of mechanical
constraints. The rate corresponding to small radii amounts to 2.5
$\times~10^{11}$ ${\mathrm {n}}_{eq}/{\mathrm {cm^{2}}}$ per year,
while for larger radii it is $6\times 10^{9}$
${\mathrm{n}}_{eq}/{\mathrm {cm^{2}}}$.

 \begin{figure}[t]
  \begin{center}
   \includegraphics[clip=true,trim = 0 0 60 0,width=\linewidth]{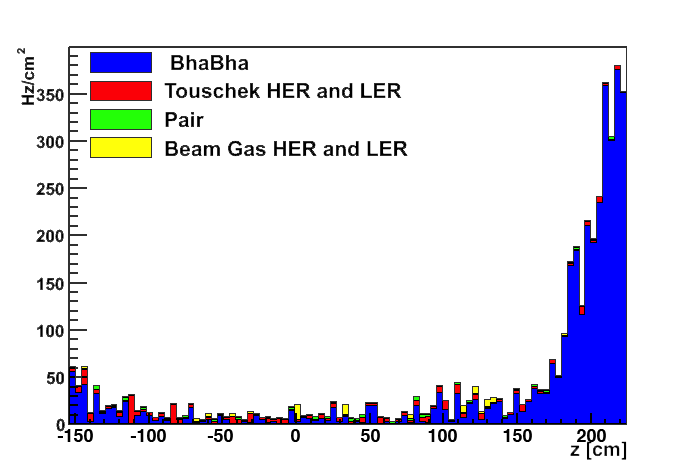}
   \includegraphics[clip=true,trim = 0 0 60 0,width=\linewidth]{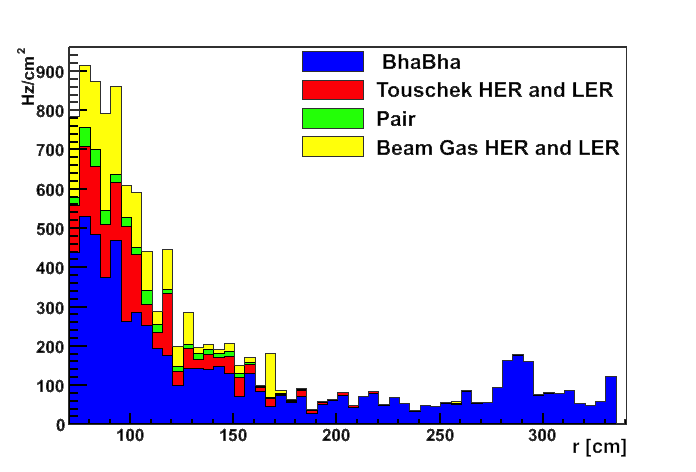}
   \caption{Electron/Positron rate for the IFR Barrel (top) and Forward Endcap (bottom) innermost layer due to different background sources. The threshold for the deposited energy is 150\kev.}
\label{fig:electronrateL0}
\end{center}
\end{figure}

\tdrsubsec{Charged Particles}

Charged-particle backgrounds can directly affect detector performance,
since they can produce additional signals in the detector, thus
modifying the track reconstruction and consequently the muon
identification.  Figure~\ref{fig:electronrateL0} shows the rate for
electrons and positrons coming from different background sources: as
for the neutrons, the dominant source of background is radiative
Bhabha scattering.  The rates shown in Figure~\ref{fig:electronrateL0}
are for electrons/positrons that deposit more than 150\kev, the
nominal energy threshold we have chosen for the detector.
 
\tdrsubsec{Photon Background}

Since all background sources come from QED processes, we also studied
the background contribution coming from photons. The photon energy
spectrum has a very broad energy range as shown in
Figure~\ref{fig:photonspectrumL0}. The lines from neutron capture on
Hydrogen (2.223\mev), from annihilation radiation (0.512\mev), and
from neutron capture on $^{10}B$ (0.48\mev) are clearly visible (the
$^{10}B$ is used for radiation shielding, see Sec.~\ref{sec:ifrshield}).

\begin{figure}[tbp]
  \begin{center}
   \includegraphics[width=\linewidth]{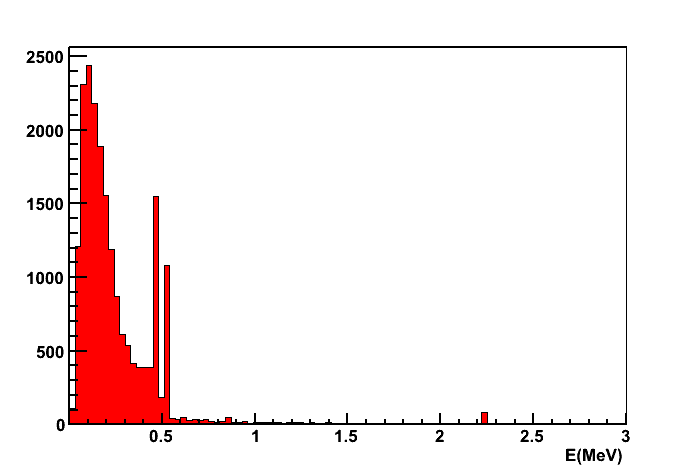}
   \caption{Photon energy spectrum for IFR Barrel Layer 0.}
\label{fig:photonspectrumL0}
\end{center}
\end{figure}

For the IFR, the photon rate contributes only to the radiation dose on
the scintillator, since the contribution from high energy photons that
convert is already taken into account in the rate of charged
particles.  The photon rates in the innermost layer of the barrel and
forward endcap are shown in Fig.~\ref{fig:photonrateL0} and, even
though high, are not expected to affect detector performance.

\begin{figure}[tbp]
  \begin{center}
   \includegraphics[clip=true,trim = 0 0 60 0,width=\linewidth]{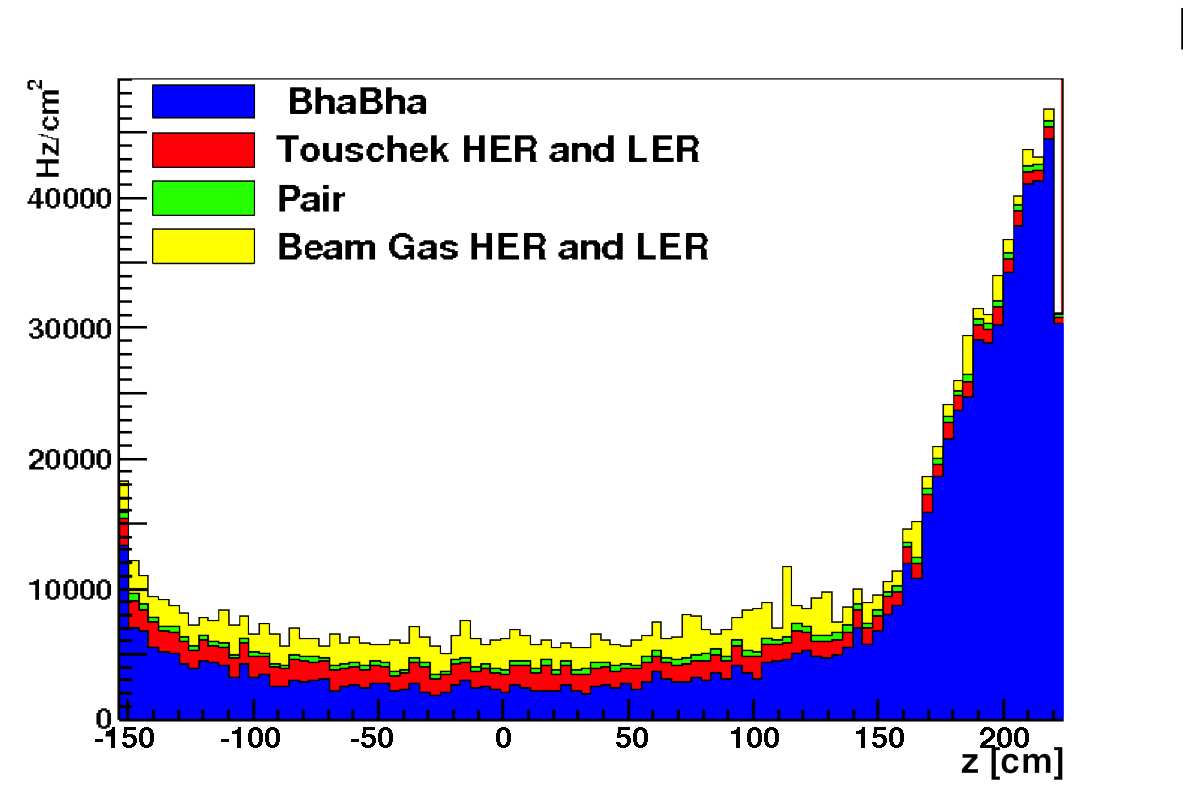}
   \includegraphics[clip=true,trim = 0 0 60 0,width=\linewidth]{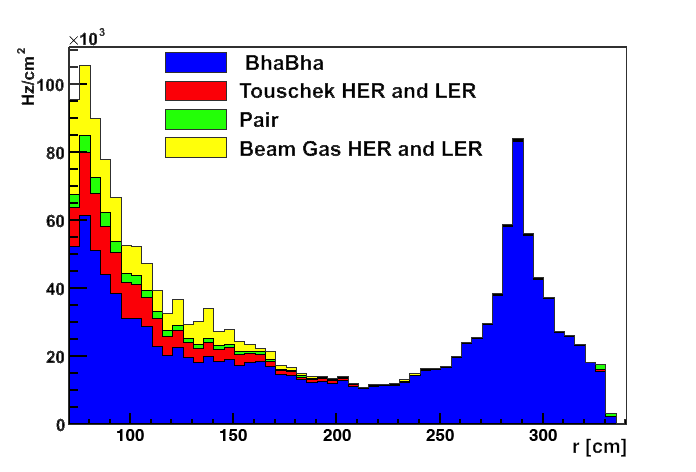}
   \caption{Photon rate on IFR Barrel (top) and Forward Endcap (bottom) innermost layer due to different background sources.}
\label{fig:photonrateL0}
\end{center}
\end{figure}



\tdrsec{Identification Performance}
	\tdrsubsec{Muon Detection}
\label{sec:muonid}

Muons are identified by their penetration range in the iron. Above
1.5--2.0\gevc, depending on the incident angle, muons penetrate all
layers.  Non-penetrating muons can be identified from the measured
range. More generally, separation of pions from muons is achieved by
using a combination of range and hit pattern
information. 

In the \superb\ baseline design the iron from \babar will be reused,
so many design parameters are fixed. The main questions that should be
addressed (and were already outlined in the {\it CDR}
\cite{ifr:superbcdr}) are:
\begin{itemize}
\item total amount of iron;
\item number of active layers;
\item size of the scintillation bars. 
\end{itemize}
 
The \superb\ full simulation (section \ref{sec:Full_Simulation}) based
on \geantFour\ is used to properly simulate the detector geometry and
the interaction of the particles with the elements of the detector.

The data taken with the IFR prototype have been used to validate the
simulation, in particular the algorithm of digitization needed to
generate actual hits from the hits generated by \geantFour (see the
prototype data analysis Section \ref{sec:protoanalysis}).

We focus our studies on the forward detector (FWD), but the
performance is the same in the other part of the setup.  We simulate
the binary readout (see section \ref{ifr:sec:systlay} for details)
assuming that each layer is constructed from a set of orthogonal bars
of 5\cm size.  For each layer the information we have are the
coordinate of the scintillator bars traversed by the charged track,
for the two orthogonal views, called {\it X-view} and {\it Y-view} in
the following.  Figure~\ref{fig:yz-view} shows the Y-view hit
positions for typical muons or pions fired into the FWD.

\begin{figure}[!htb]
\begin{tabular}{c}
\includegraphics[width=\linewidth]{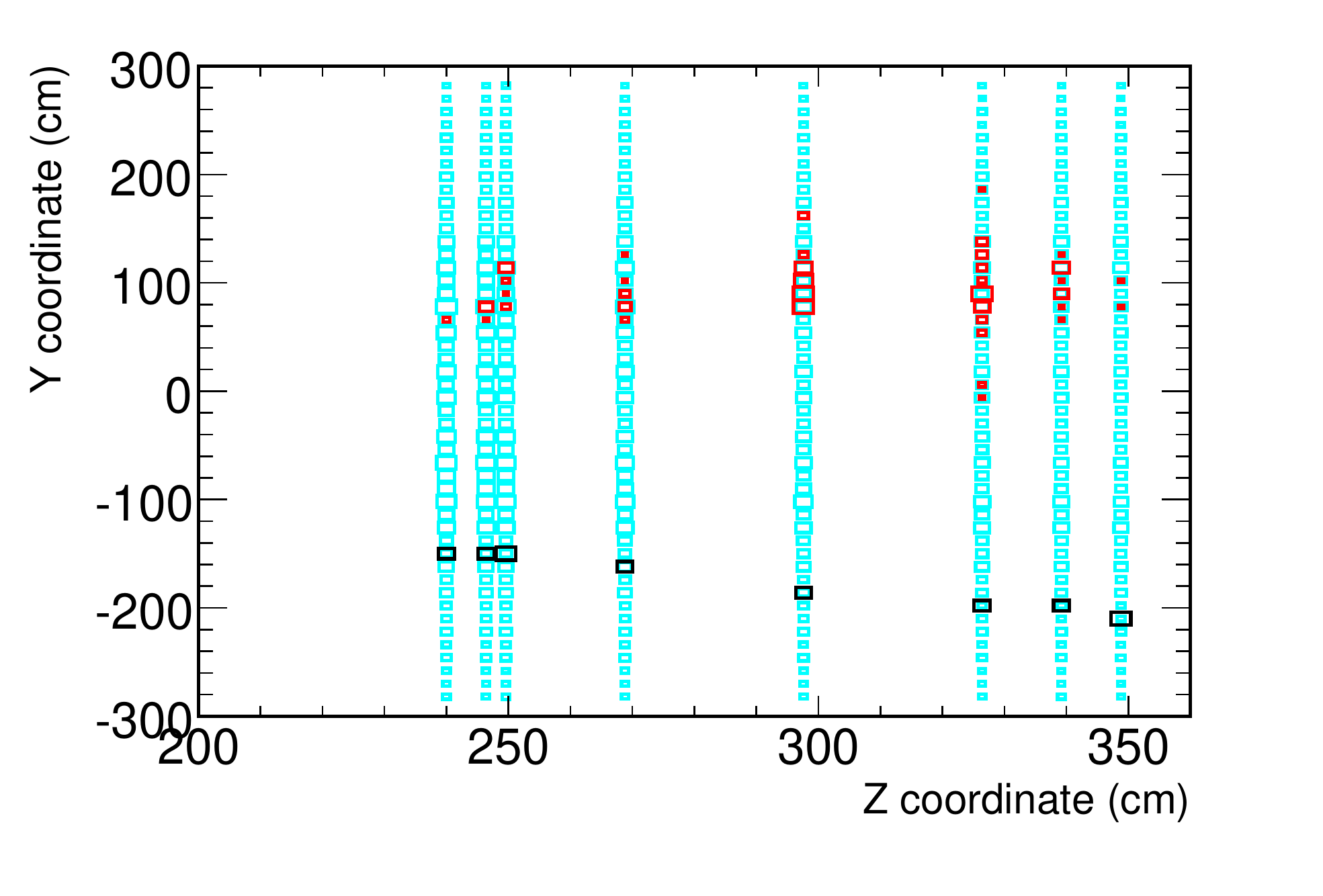}\\
\end{tabular}
\caption{
Hit position in the Y-view of the FWD detector. 
The particles come from the nominal interaction point located at (0,0), 
and are directed to the right. The hit patterns due to a muon 
(bottom hits) and a pion (top hits) are very different.}
\label{fig:yz-view}
\end{figure}

As already outlined before, the criteria for distinguishing muons from
hadrons are based on the differences in their interaction with matter.
A muon track typically generates just one hit in each layer, while
hadrons can interact strongly in the iron, resulting in a hadron shower
with several lower momentum particles and yieldig multiple hits per
layer. It is also possible that few neutrals (neutrons) are produced
in the hadronic interaction, travelling long distance before generating
a hit.  In general, for pion tracks, the number of consecutive active
layers is smaller than for muon tracks, and the average number of hits
per each hit layer is higher.

The hit positions associated with a track are fitted separately for
the X- and Y-view, with the functional form $y=Y(z)$ and $x=X(z)$,
where $z$ is the longitudinal coordinate (essentially the layer
number) and $x$ and $y$ are the two transverse coordinates given by
the bar position for the X- and Y-view, respectively.  The functions
$Y(z)$ and $X(z)$ are parameterized as second order polynomials.

The sum of the hit residuals squared, called $\chi^2$, and the fitted
parameters are used for the muon selection.  In particular, the
difference between the fit direction and a prediction extrapolated
from the inner tracking subdetectors is a very useful variable to
suppress part of the contamination from the pion and kaon decays in
fly before the first layer of the IFR.  So far we take the truth
direction information from the MC but these information will also be
available when the full track reconstruction will be available.


The amount of the energy released in the EMC, which has a slightly
different distribution for muons and pions, can also be used to reduce
the pion contamination. The DIRC response could in principle also be
used to improve muon detection at lower momenta---so far we have not
exploited this option.

\begin{figure*}[tbp]
\begin{tabular}{@{}cc@{}}
\includegraphics[width=0.49\linewidth]{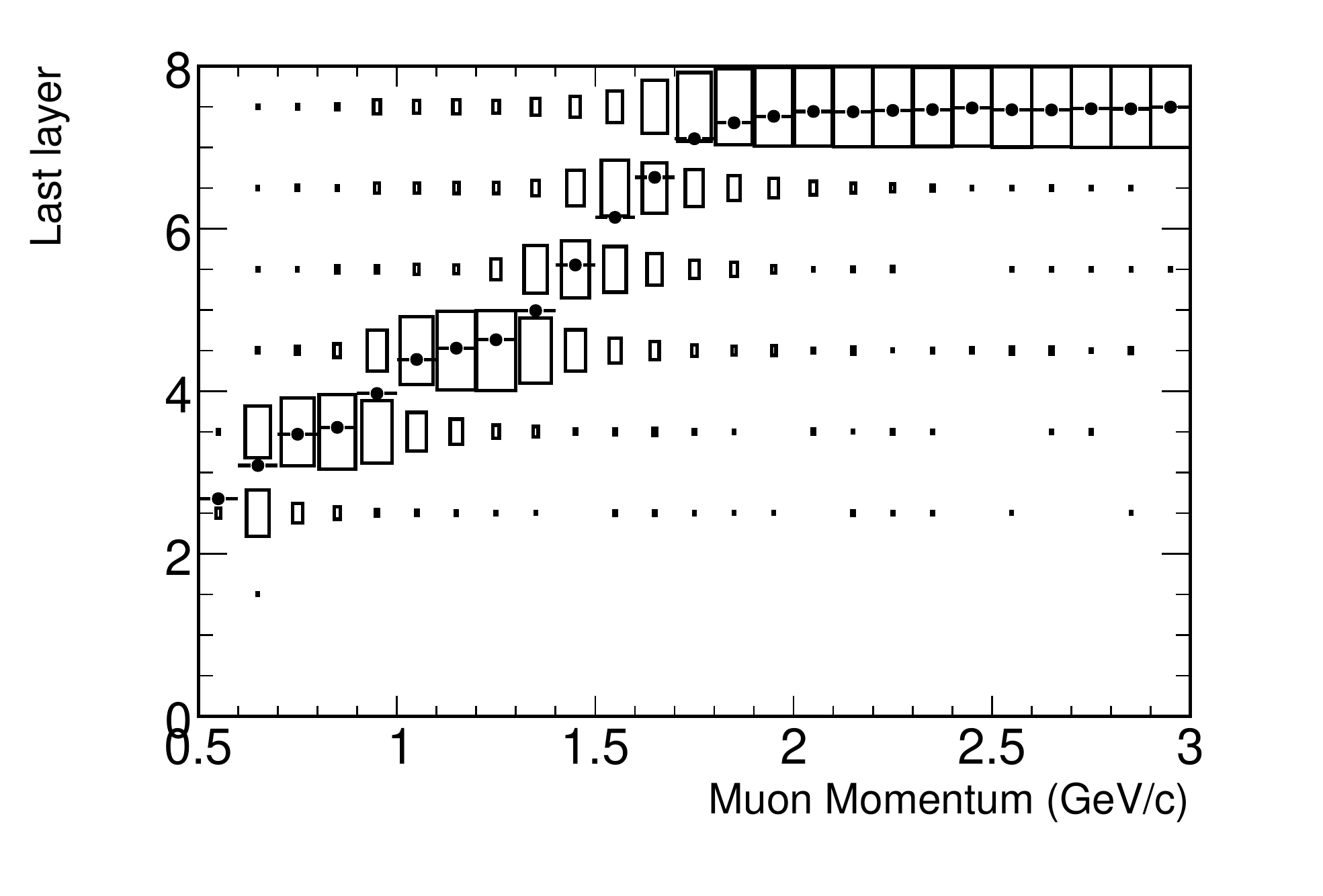}&\includegraphics[width=0.49\linewidth]{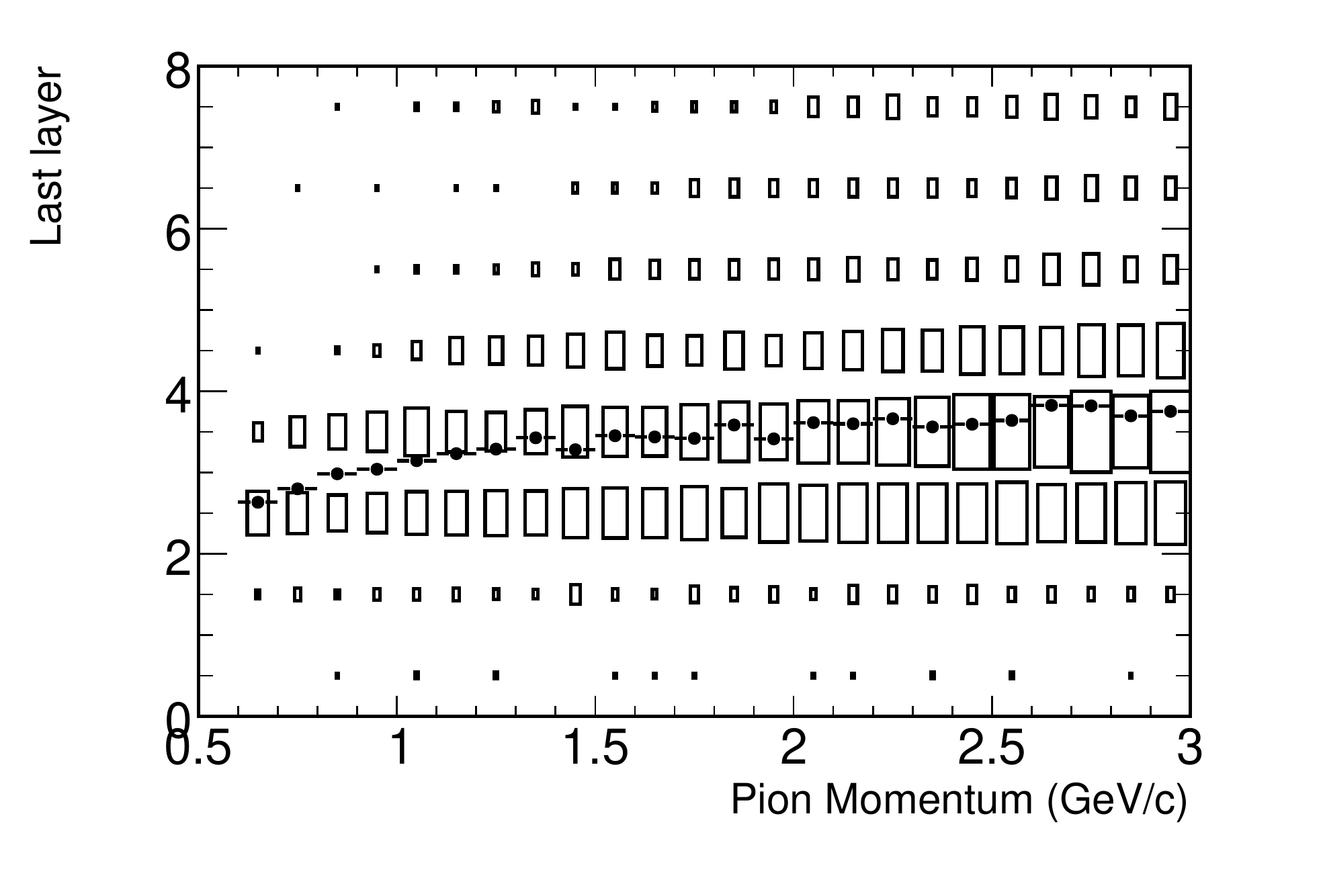}\\
\includegraphics[width=0.49\linewidth]{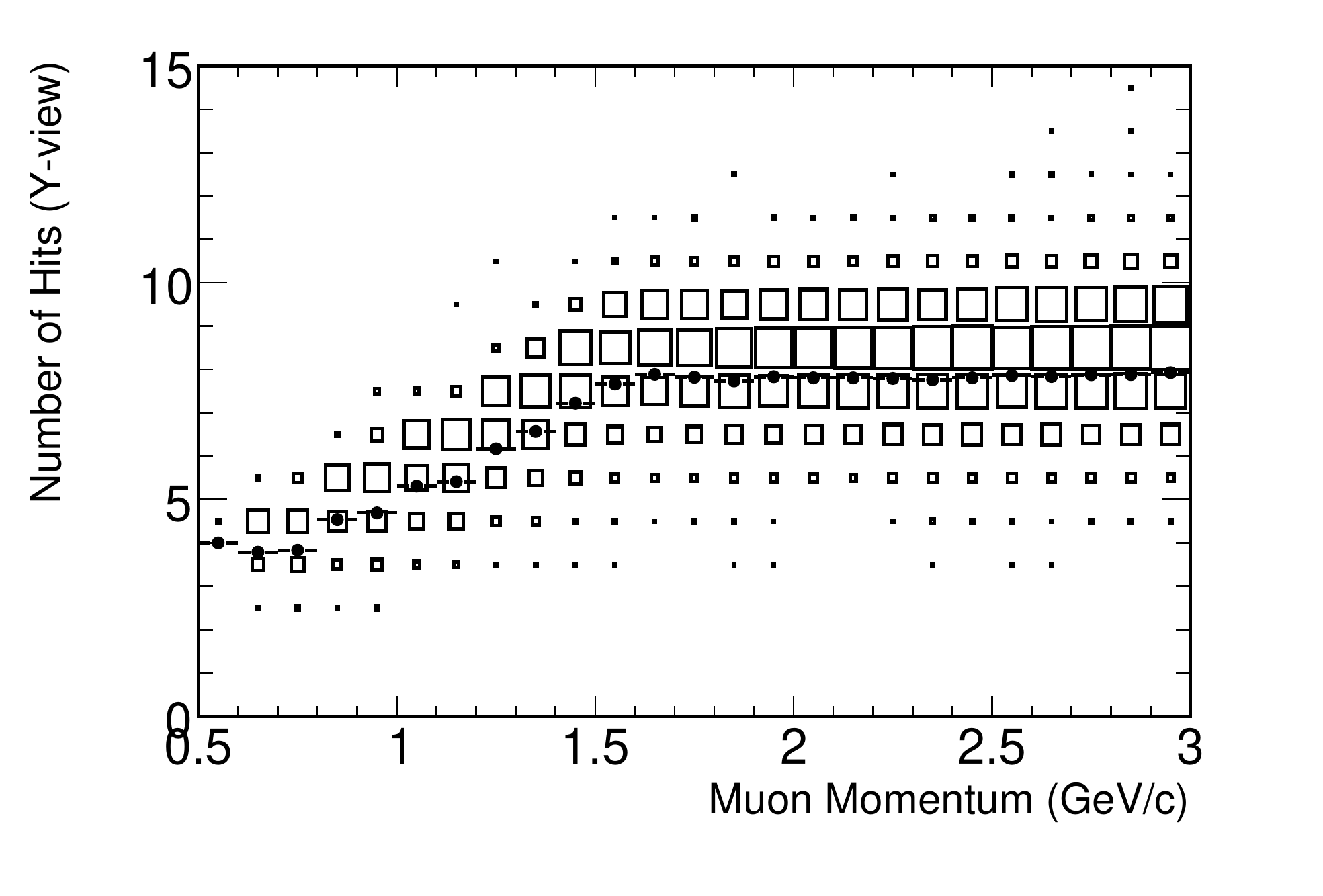}&\includegraphics[width=0.49\linewidth]{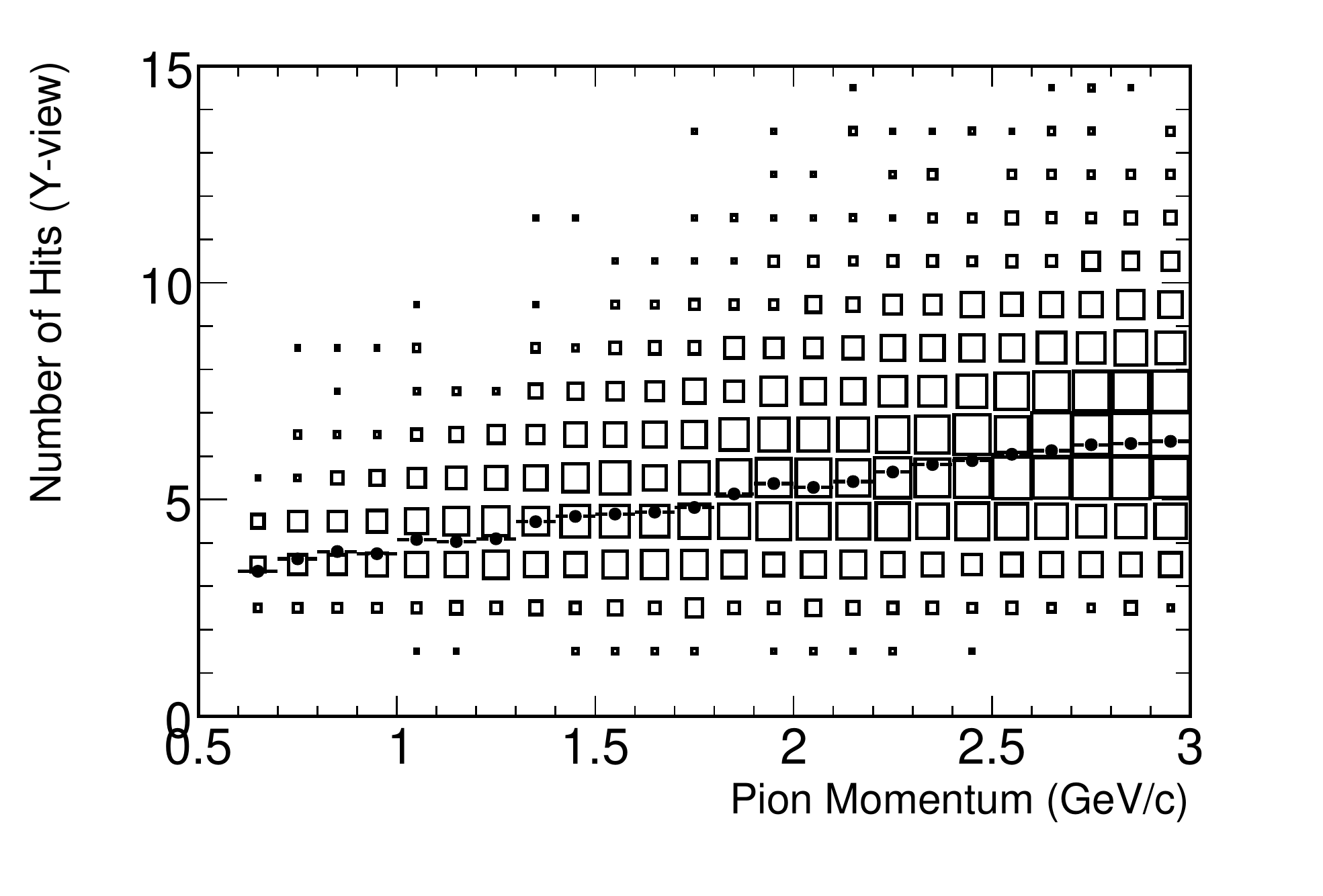}\\
\end{tabular}
\caption{
Distribution of the last detected layer reached (top), and the number of 
hits (bottom), as a function of the $\mu$ (left) and $\pi$ (right) momentum.
The dots represent the average value of the distribution.
}
\label{muid:variables}
\end{figure*}

To summarize, we study the muon identification performance, using a
consistent set of the following variables:

\begin{enumerate}
\item number of hits in the X- and Y-view;
\item number of active layers for the X- and Y-view;
\item the last layer containing hits, which is translated in the 
number of interaction lengths, $\lambda_{int}$;
\item track continuity;
\item $\chi^2$ of the hits to the tracks for the X- and Y-view;
\item parameters of the fit track;
\item the total energy released in the EMC.
\end{enumerate}

The average distribution of some of the above variables, as a function
of the track momentum, are shown in Figure~\ref{muid:variables}.  Even
simple sequential cuts on these variables, provide a good pion
rejection. The experience gained in \babar\ shows that a multivariate
approach gives the best performance for muon identification, even at
lower energy when the muons stop inside the IFR volume.  Because of
the non-trivial correlations between the reconstructed quantities, we
used a non-linear multivariate approach, in particular a boosted
decision tree (BDT), 
which turned out to be very powerful and robust with respect to
changes in the input parameters.

The BDTs are trained in 7 bins of momentum, ranging from 0.5 to
4\gevc.  $\pi$ rejection as functions of $\mu$ efficiency for the
0.5--1.0\gevc and 1.5--2.0\gevc momentum ranges obtained after
training are shown in Figure~\ref{fig:roc}. The performance can be
judged either by comparing these curves or by fixing the level of
$\pi$ rejection and comparing the $\mu$ efficiency.

\begin{figure}[t]
\begin{tabular}{c}
\includegraphics[width=\linewidth]{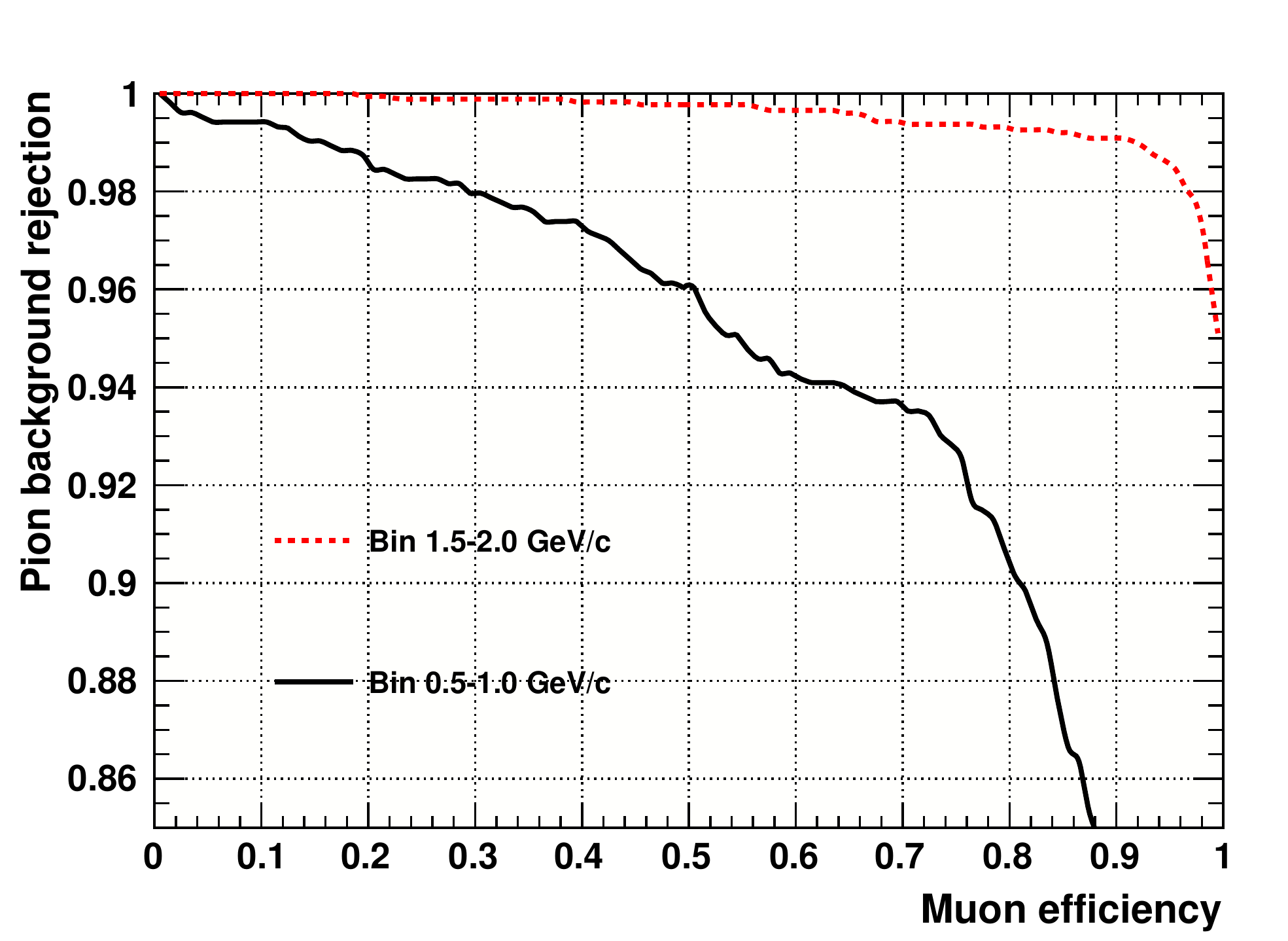}\\
\end{tabular}
\caption{ Plot of the pion rejection as a function of the muon
  efficiency.  The curves are reported for tracks in the 0.5--1.0\gevc
  (solid line) and 1.5--2.0\gevc range (dashed line).  }
\label{fig:roc}
\end{figure}

Plots of the $\mu$ efficiency as a function of the momentum, with the
$\pi$-misidentification probability, fixed at 2\pct and 5\pct, are
shown in Figure~\ref{fig:muon-efficiency}.

\begin{figure}[tbp]
\begin{tabular}{c}
\includegraphics[width=\linewidth]{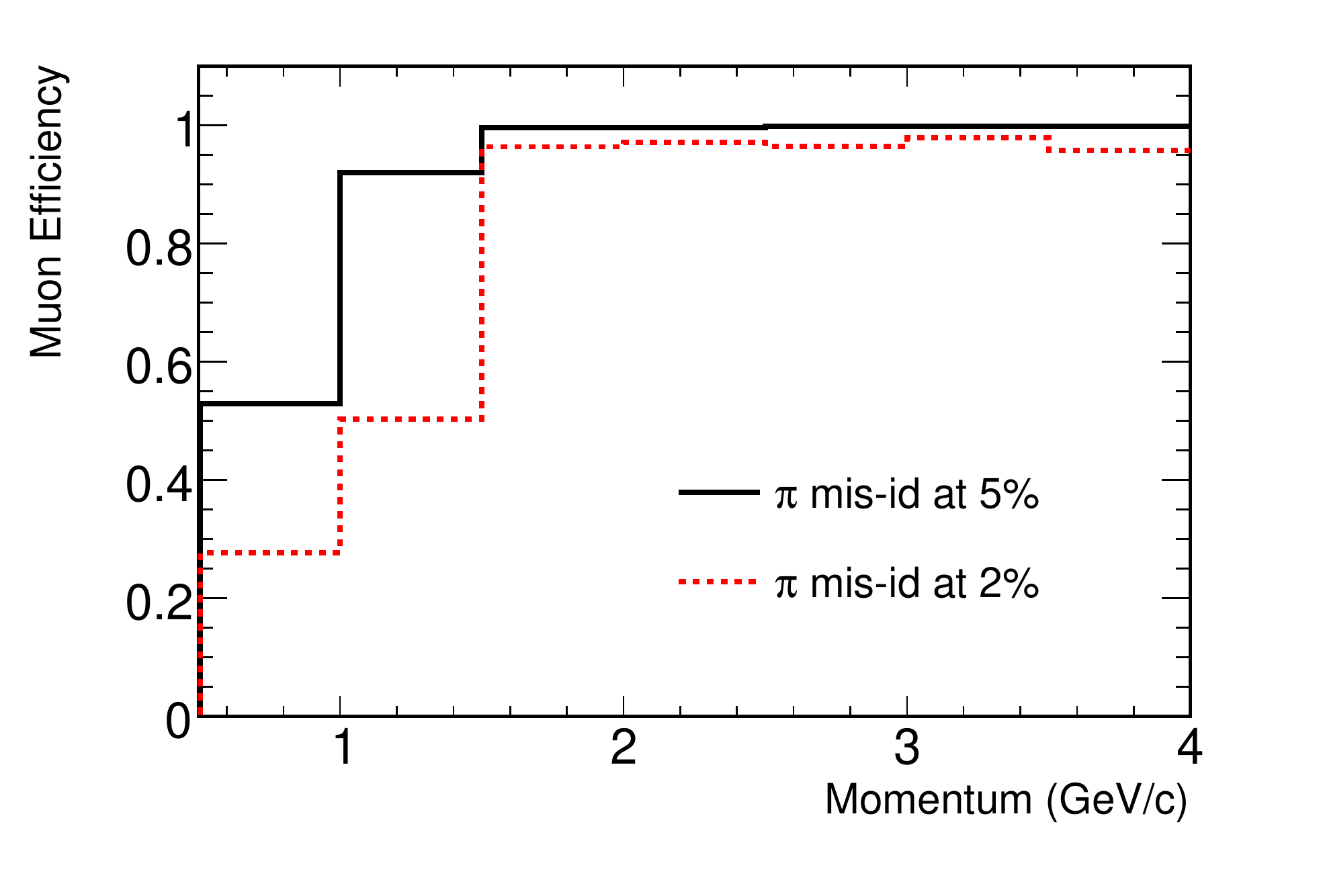}\\
\end{tabular}
\caption{
Muon efficiency as a function of the momentum, with $\pi$ mis-identification rate fixed in each momentum
bin at 2\pct (dashed line) and 5\pct (solid line).}
\label{fig:muon-efficiency}
\end{figure}

\begin{table}  
\centering
\caption{The efficiency for muon identification (in \%) in 7 bins of momentum between 
0.5 and 4\gevc and for two different value of the pion efficiency.}
\label{tab:eff}
\begin{tabular}{l c c }
\hline\hline
   $P (\gevc)$ & $\epsilon_\mu$ ($\epsilon_\pi=2\pct$) &  $\epsilon_\mu$ ($\epsilon_\pi=5\pct$) \\
   0.5-1.0   & 27.6$\pm$0.6 &  52.8$\pm$0.8\\
   1.0-1.5   & 50.2$\pm$0.4 &  91.9$\pm$0.5\\
   1.5-2.0   & 96.3$\pm$0.3 &  99.6$\pm$0.5\\
   2.0-2.5   & 97.1$\pm$0.3 &  99.9$\pm$0.5\\
   2.5-3.0   & 96.4$\pm$0.3 &  99.8$\pm$0.5\\
   3.0-3.5   & 97.9$\pm$0.3 &  99.9$\pm$0.5\\
   3.5-4.0   & 95.7$\pm$0.3 &  99.9$\pm$0.5\\ 
\hline\hline
\end{tabular}
\end{table} 

Muon efficiencies in bins of lepton momentum are also shown in
Table~\ref{tab:eff}.  The performance matches the expectations. Above
1.5\gevc, the efficiency is close to 100\pct for muons that hit the
detector. As expected, there is a drop in the signal efficiency below
1.5\gevc, where the muons stop within the IFR volume.  With the
segmentation and granularity studied so far, about half of the pion
contamination is due to the irreducible background from pions decaying
in flight.

We also studied different configurations and compared the response
with the default configuration, evaluating the impact of the following
factors:
 
\begin{itemize}
\item{efficiency lost}: an overall drop in the single layer efficiency ($\epsilon_{layer}$) 
to 50\pct and 80\pct;
\item{occurrence of a dead layer}: $\epsilon_{layer}=80\pct$ and absence of the last layer 
(total amount of the absorber equivalent to 820\mm of iron);
\item{background}: the effect of the background from radiative Bhabhas;
\item{background and efficiency lost}: the effect of the machine background and $\epsilon_{layer}=80\pct$.
\end{itemize}

\begin{figure}[!htb]
\begin{tabular}{c}
\includegraphics[width=\linewidth]{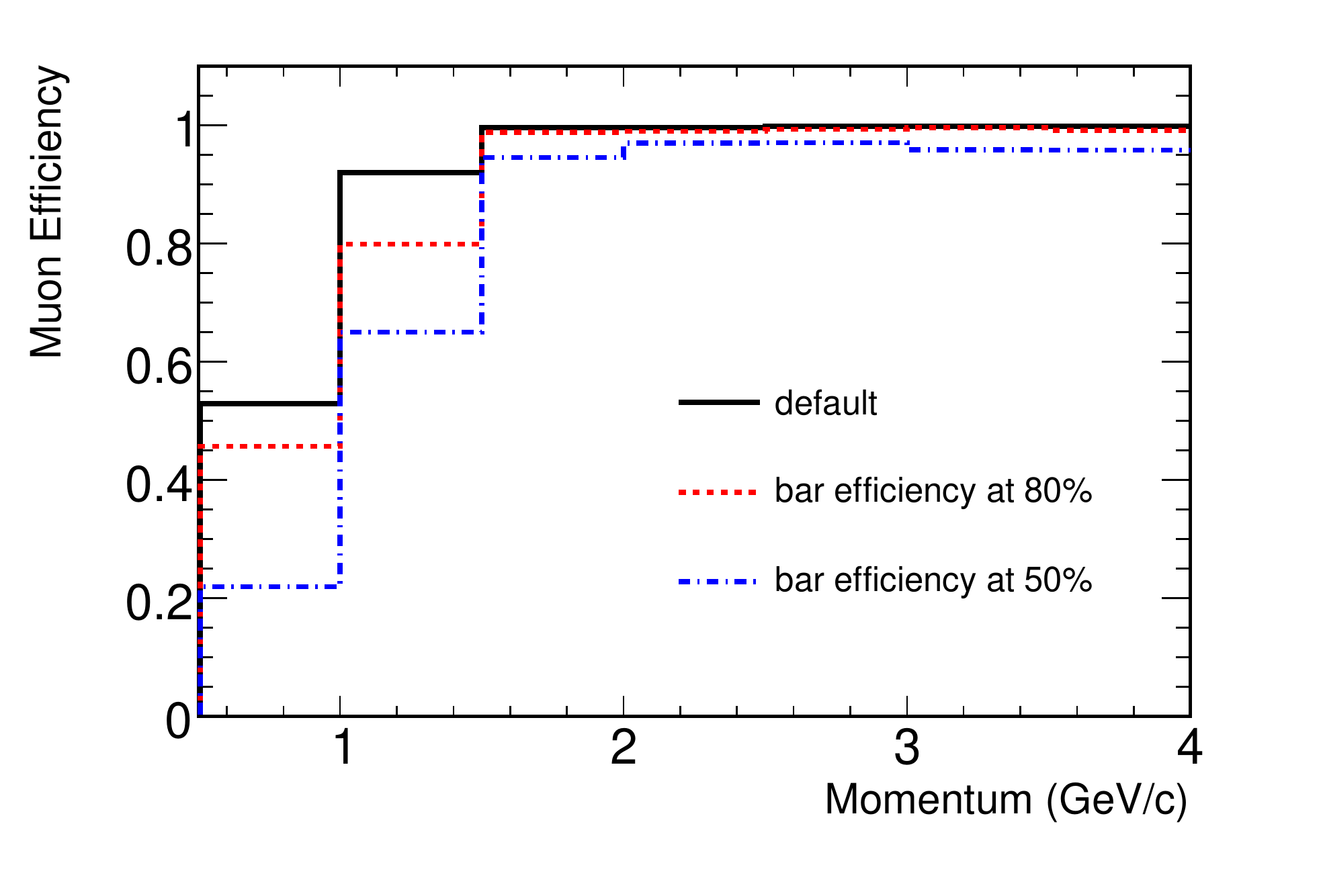}\\
\end{tabular}
\caption{ Muon efficiency as a function of the momentum, with $\pi$
  misidentification rate fixed at $5\pct$ for the default
  configuration (solid), compared with the assumpition of a layer
  efficiency at 80\pct (dashed) and 50\pct (dashed-dot).}
\label{fig:configurations}
\end{figure}

The resulting efficiencies are compared with the above default values
in Tab.~\ref{tab:configurations} and Figure~\ref{fig:configurations},
where we require a $\pi$ misidentification rate of $5\pct$. It is
evident that for momenta above 1.5\gevc, the impact of an average drop
in the single layer efficiency at $\epsilon_{layer}=80\pct$ is
negligible.  Moreover, the additional occurrence of the dead last layer
does not lead to a significant deterioration of the performance If the
average efficiency drop is reduced to 50\pct, we observe a large
impact at lower momenta.  This drop in the $\mu$ efficieny can be
compensated only by requiring a larger $\pi$ mis-identification rate.

We studied the impact of the background from Bhabha events by adding
samples of background hit events corresponding to five times the
expected Bhabha rate to events with single pions and muons. We repeat
the optimization of the $\mu$ selector obtaining the efficiencies
reported in the last two rows of Tab.~\ref{tab:configurations}.  This
background contribution reduces the detector performance mainly at
lower momenta.  Because we limit our study to the forward region,
where the background from Bhabha events is more important, the drop in
efficiency of about 10\pct can be considered the worst case.

\begin{table*}  
\centering
\caption{Muon efficiency (in \%) in 6 different bins of momentum between 
0.5 and 3.5\gevc for some configurations studied in the simulations. 
The muon efficiencies are obtained assuming $\epsilon_\pi=5\pct$. The statistical uncertainty
is 2-5\pct in the 0.5-1.0\gevc bin, and is less than 1\pct for the other bins.} 
\label{tab:configurations}
\begin{tabular}{l c c c c c c}
\hline\hline
   Momentum [\gevc]        &  0.5-1.0 & 1.0-1.5 & 1.5-2.0 & 2.0-2.5 & 2.5-3.0 &3.0-3.5\\
   Default & 52.5$\pm$0.7 & 91.9$\pm$0.5 & 99.6$\pm$0.5 & 99.6$\pm$0.5 & 99.6$\pm$0.5&99.8$\pm$0.06 \\
   Eff. at 80\pct   & 45.7&79.9 &98.8&98.9&99.3&99.6\\
   Eff. at 50\pct   & 21.9&65.0 &94.5&97.0&97.1&95.8\\
   Eff. at 80\pct and layer 8 dead & 44.9& 81.9& 98.6& 98.7& 99.2&99.4 \\
   Background                          & 53.5&73.5 &97.0 &98.5 &98.8 & 98.6\\
   Background and eff. at 80\pct   & 36.3& 71.3& 95.0& 96.7& 96.2& 97.5\\  
\hline\hline
\end{tabular}
\end{table*} 

\tdrsubsec{$K_L$ Detection}

The identification of the $K_L$ mesons in exclusive $B$ decays
requires the selection of the IFR hits belonging to the $K_L$ and the
determination of the $K_L$ direction from these hits. In the real
case, all the clusters near the extrapolation of charged tracks from
the inner detectors are eliminated.  The remaining clusters are due to
neutral particles and mis-associated charged hadrons.  A good angular
resolution is important to reduce the backgrounds.  To study our
capability to identify the $K_L$, and the $K_L$ angular resolution, we
fired single $K_L$ with momenta between 0 and 4\gevc towards the
forward endcap.  In Figure~\ref{fig:klong-lastl} we show the
distribution of the last detector layer reached.  About 65\pct of the
$K_L$ interact in the EMC, and the debris of the interaction generates
hits in the IFR. The hit pattern can be used to disentangle the $K_L$
from background, mainly $\gamma$ from $\pi^0$ and merged $\pi^0$ in
the EMC and charged hadrons.

\begin{figure}[!htb]
\begin{tabular}{c}
\includegraphics[width=\linewidth]{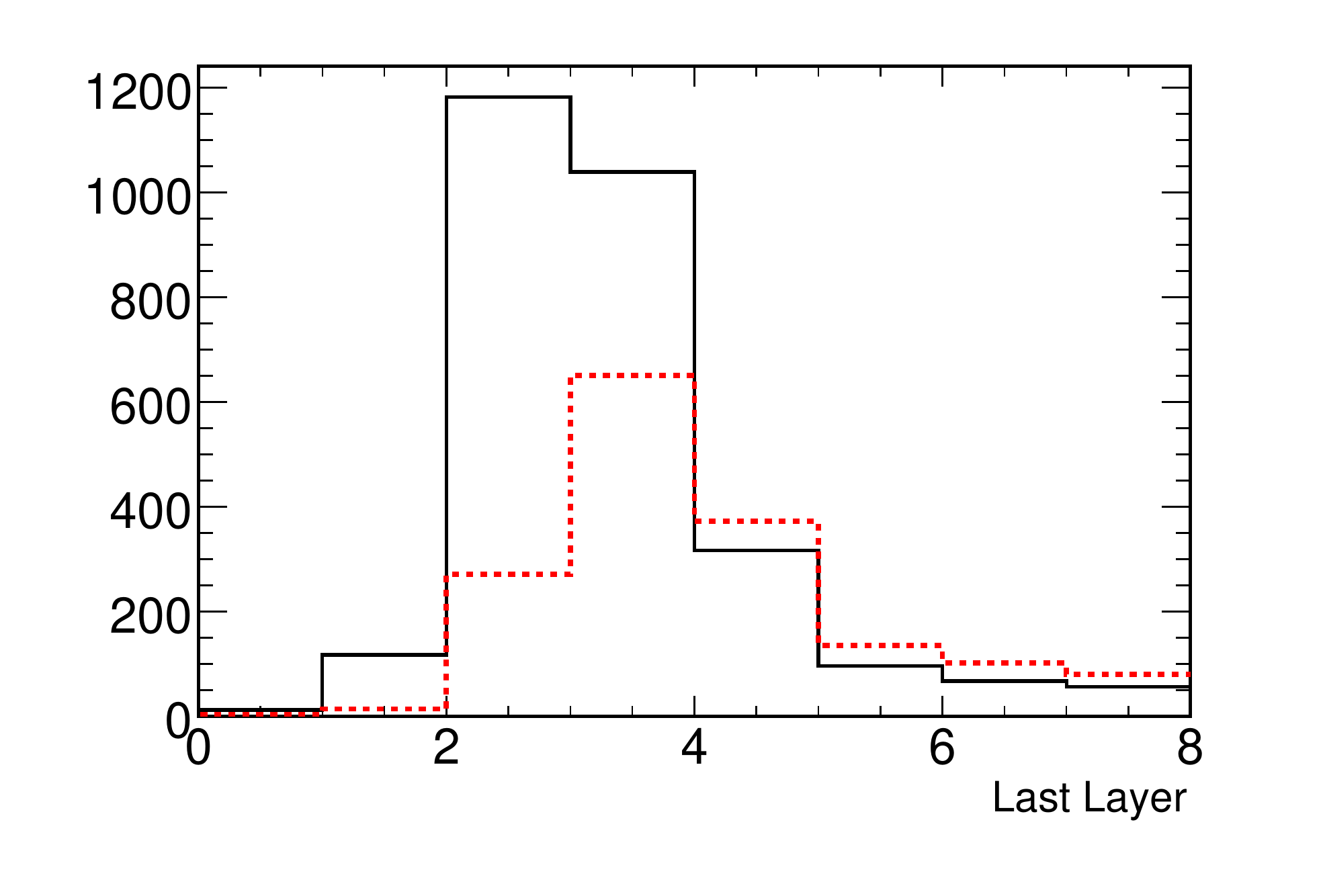}\\
\end{tabular}
\caption{Distribution of the last-layer reached for a sample of $K_L$ in the momentum
range 0--3\gevc for $K_L$ that interact in the EMC (solid) or after the EMC (dashed).} 
\label{fig:klong-lastl}
\end{figure}

The distribution of the last layer hit is on average higher when the
$K_L$ does not interact with the EMC, as can be seen in
Figure~\ref{fig:klong-laslmom}.  There is a correlation between
last-layer and the $K_L$ momentum, but the distribution is quite wide
due to the large fluctuations in the hadron interactions, and cannot
be used to infer the $K_L$ momentum.

\begin{figure}[!htb]
\begin{tabular}{c}
\includegraphics[width=\linewidth]{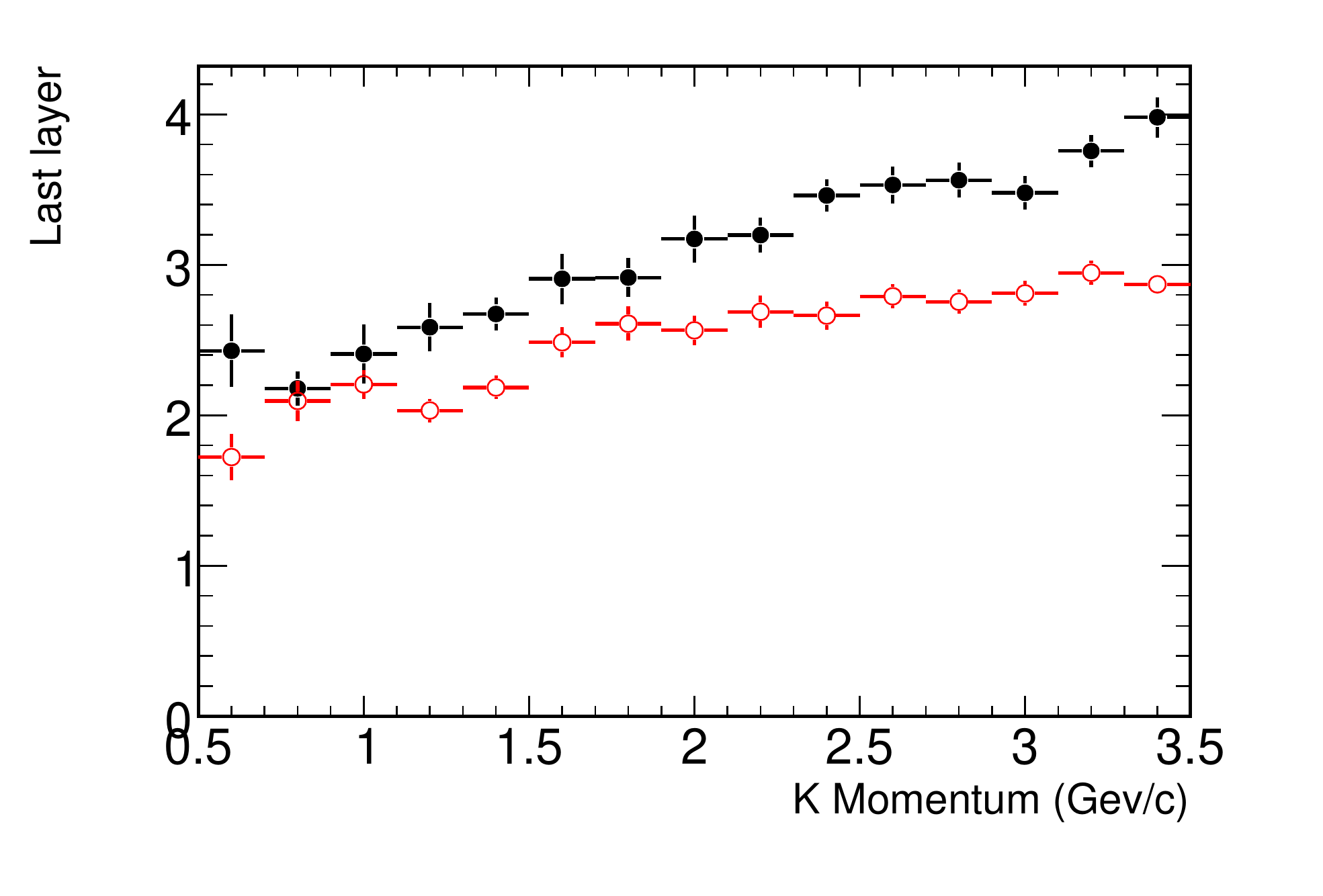}\\
\end{tabular}
\caption{The last-layer variable distribution as a function of the
  $K_L$ momentum.  On average, the last-layer is larger when the $K_L$
  does not interact in the EMC ($\circ$) than when it interacts in the
  EMC ($\bullet$).}
\label{fig:klong-laslmom}
\end{figure}

\begin{figure}[tbp]
\begin{tabular}{c}
\includegraphics[width=\linewidth]{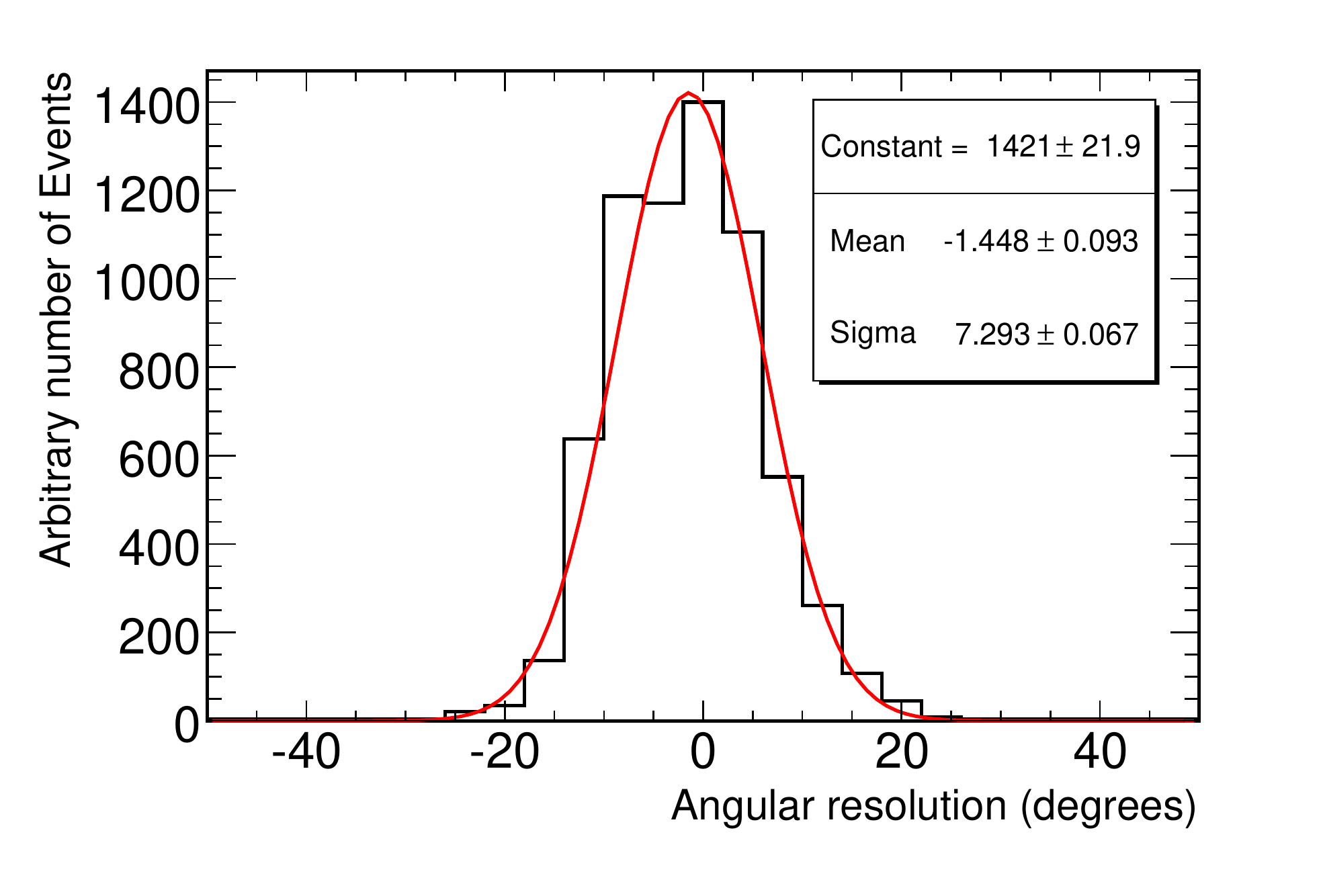}\\
\end{tabular}
\caption{$K_L$ angular resolution.} 
\label{fig:klong-res}
\end{figure}

The information that can be extracted by combining measurements from
EMC and IFR is limited to the $K_L$ direction.  This is very useful
when one needs to reconstruct exclusive decays with a $K_L$ in the
final state, like the $B\to J/\psi K_L$. The knowledge of the $K_L$
direction allows to constrain the decay kinematics. Moreover, the
reconstruction of the $K_L$ in the event is very important for
identifying rare decays whose signature is a large missing momentum
(e.g., $B\to \tau\nu_\tau$ or $B\to \pi\tau\nu_\tau$).  The angular
resolution is about $7.3^\circ$ for $K_L$ generated with a flat
momentum distribution between 0 and 4
\gevc~(Figure~\ref{fig:klong-res}).

The impact of the machine background, that affects mainly the first
layers in both the forward and the barrel regions, can be important
for the $K_L$ identification efficiency and background reduction.
These studies require full pattern recognition that has not been
performed yet.


\tdrsec{Detector R\&D}

	\tdrsubsec{Module Tests and Results}
	 
 The IFR detector technology is the result of an extensive R\&D
 program, carried out in the last years and devoted to the choice of
 the most effective component for each detector component. In more
 detail, the initial studies were mainly dedicated to:
\begin{itemize} 
\item   scintillators;
\item  fibers;
\item photodetectors. 
\end{itemize} 
For practical reasons, due to the many possible combinations, our R\&D
studies proceeded by comparing two different configurations in which
only one element was changed. At the end, the decision was made by
balancing the performance and reliability of the detection system with
the cost and complexity of the mechanical implementation.

\subsubsection*{Scintillators}
Given the large amount of scintillator needed ($\simeq$ 20 tons), our
attention was, since the beginning, focused on rather inexpensive
scintillators produced in large quantities by the FNAL-NICADD facility
at Fermilab~\cite{ifr:nicadd}.  They are fabricated by extrusion and a
thin layer of $TiO_2$ is co-extruded around the active core.  This
production method suits particularly well our needs to provide long
and thin scintillator strips in the desired shape.

We tested several samples of scintillator strips of external
dimensions $1.0 \times 4.5\cma$ (already available at the FNAL -NICADD
facility) and $2.0 \times 4.0\cma$, considering two options on the
positioning of the fibers: those placed in an embedded hole or in
surface grooves. The difference in light yield has been measured to be
about 10\pct higher for the embedded hole option. This relatively
small improvement in light yield is offset by the greater difficulty
of filling the embedded holes with optical glue. We have therefore
chosen the surface grooves as the baseline option.

For the barrel readout, larger strips of dimensions $1 \times 10\cma$
and $1\times5\cma$, produced by the Vladimir factory near
Moscow~\cite{ifr:vladimir}, have been studied. In terms of light yield
and attenuation length the Russian scintillators are very similar to
the above-mentioned FNAL-NICADD ones.

\subsubsection*{Optical fibers} 
Since the attenuation length of the scintillator is rather short
($\simeq$ 35\cm), the light produced by the particle interaction has
to be collected and transported to the photodetectors efficiently by
Wave Length Shifting (WLS) fibers, In our application the fibers
should have a good light yield to ensure a high detection efficiency
for fiber lengths in the range between 0.6 and 3\m. The time response
was also studied, since originally a TDC readout option was considered
(extracting one coordinate from the hit arrival time). While the TDC
option could match the required performance, it was deemed to be not
reliable enough for implementation on the detector.

We tested WLS fibers from Saint-Gobain (BCF92)~\cite{ifr:bicron} and
from Kuraray (Y11-300)~\cite{ifr:kuraray} factories.  Both companies
produce multiclad fibers with good attenuation length ($\lambda \simeq
\, 3.5\m$) and trapping efficiency ($\varepsilon \simeq 5\pct$). The
fibers from Kuraray have a higher light yield (about 40\pct more),
while Saint-Gobain fibers have a faster response ($\simeq\;2.7\ns$
versus $\simeq\;9\ns$ of the Kuraray), which ensures a better time
resolution.


\subsubsection*{Photodetectors}
For the IFR detector the photodetectors converting the light signal
from the WLS fibers need to work efficiently in tight spaces and in a
high magnetic field environment. Geiger mode avalanche photodiodes
(GMAPDs) have been selected as they meet these needs. GMAPDs have high
gain ($\simeq\,10^{5}$), low bias voltage ($<100$ V), good detection
efficiency ($\simeq\, 30\pct $), fast response (risetime below ns),
and are very small (few mm$^2$) and insensitive to magnetic fields.
On the downside they have a relatively high dark count rate (few 100
kHz/mm$^{2}$ at 1.5 p.e.) and non-negligible sensitivity to radiation.
Among several GMAPDs which are now on the market we concentrated our
efforts on the following devices: Silicon Photomultipliers (SiPMs)
from IRST-FBK~\cite{ifr:SIPM}, and Multi Pixel Photon Counters (MPPCs)
from Hamamatsu~\cite{ifr:MPPC}.

At first, small ($1\times 1\mma$) photodetectors had been considered.
However, given the small amount of light extracted out of the
scintillator with just one fiber, it was soon realized that larger
devices are needed to couple more readout fibers, while keeping the
active surface (and so the noise) as low as possible. The studies,
that led to the prototype design and construction were then performed
with $2 \times 2\mma$ FBK devices.  The detection efficiency of such
devices is related to the operating parameters and the tolerable dark
noise.  FBK devices are in general noisier and have a better time
resolution than MPPCs. MPPCs have a better light yield and are more
sensitive to bias and temperature variations.  Additional important
parameters to be taken into account for photodetectors are the rate of
aging and radiation hardness.

Custom devices from the FBK company with a rectangular active area
optimized for our needs ($1.2\, \times \,3.2\mma$, and $1.4\, \times
\,3.8\mma$) were produced and tested for the IFR prototype together
with the TDC readout. After we dropped the TDC readout, the Hamamatsu
devices seemed to guarantee the best performance and are currently our
baseline. Since this technology is rapidly evolving the final choice
will depend on the state of the art when this choice is made.

\subsubsection{Detection module R\&D}

\paragraph{Number of fibers and geometry.} The light collected with
cosmics data using one, two or three fibers placed on a scintillator
bar has been studied in order to determine the most effective number
of fibers.  As shown in Figure \ref{ifr:fig:fibers}, going from one to
two fibers provides the gain in light yield of the order of $100\pct$.
The application of three instead of two fibers increases the light
yield by another $\sim 50\pct$, a total gain of three times with
respect to one fiber. Using more than three fibers per scintillator
bar results in only in small increases of light that will not
compensate the increase of the dark count and of the cost. The studies
of this issue are still ongoing. From Table \ref{ifr:tab:fibers},
which summarizes all the results, we can also deduce that the
scintillator geometry has a sizable impact on the light output. \\

\begin{table}[htb]
\begin{center}
\caption{Light yield (the number of photoelectrons) for different readout configurations.}
\begin{tabular}{lcc} 
\hline
	& $1\times5\cma$ strip & $1\times10\cma$ strip\\
\hline
1 fiber &  37$\pm$3    &19$\pm$3  \\
2 fibers & 71$\pm$4   & 34$\pm$3  \\
3 fibers & 105$\pm$5 & 60$\pm$4 \\
\hline
\end{tabular}
\label{ifr:tab:fibers}
\end{center}
\end{table}

\begin{figure}[tbp]
  \begin{center}
     \includegraphics[width=0.5\textwidth]{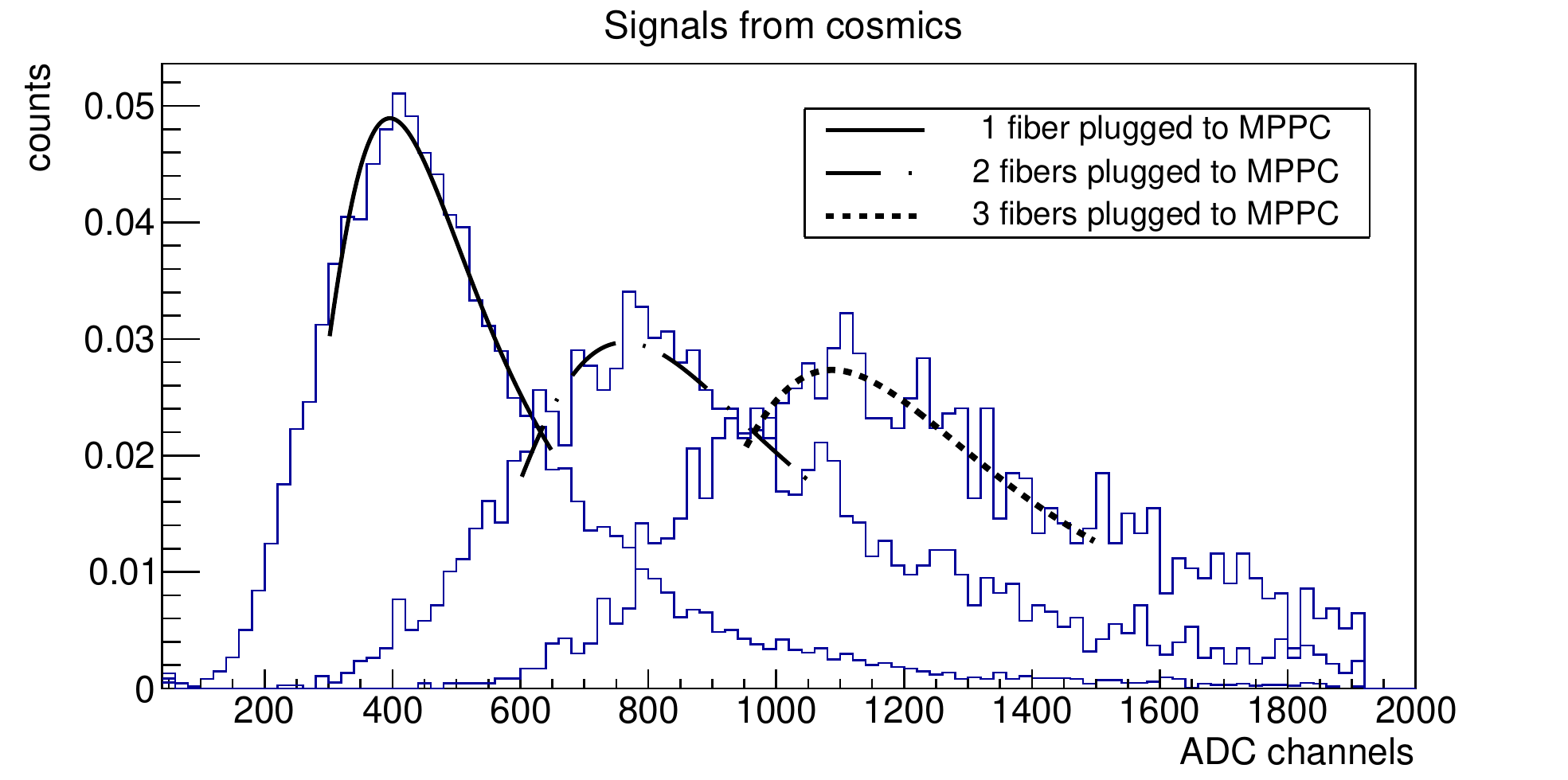}\\
     \includegraphics[width=0.5\textwidth]{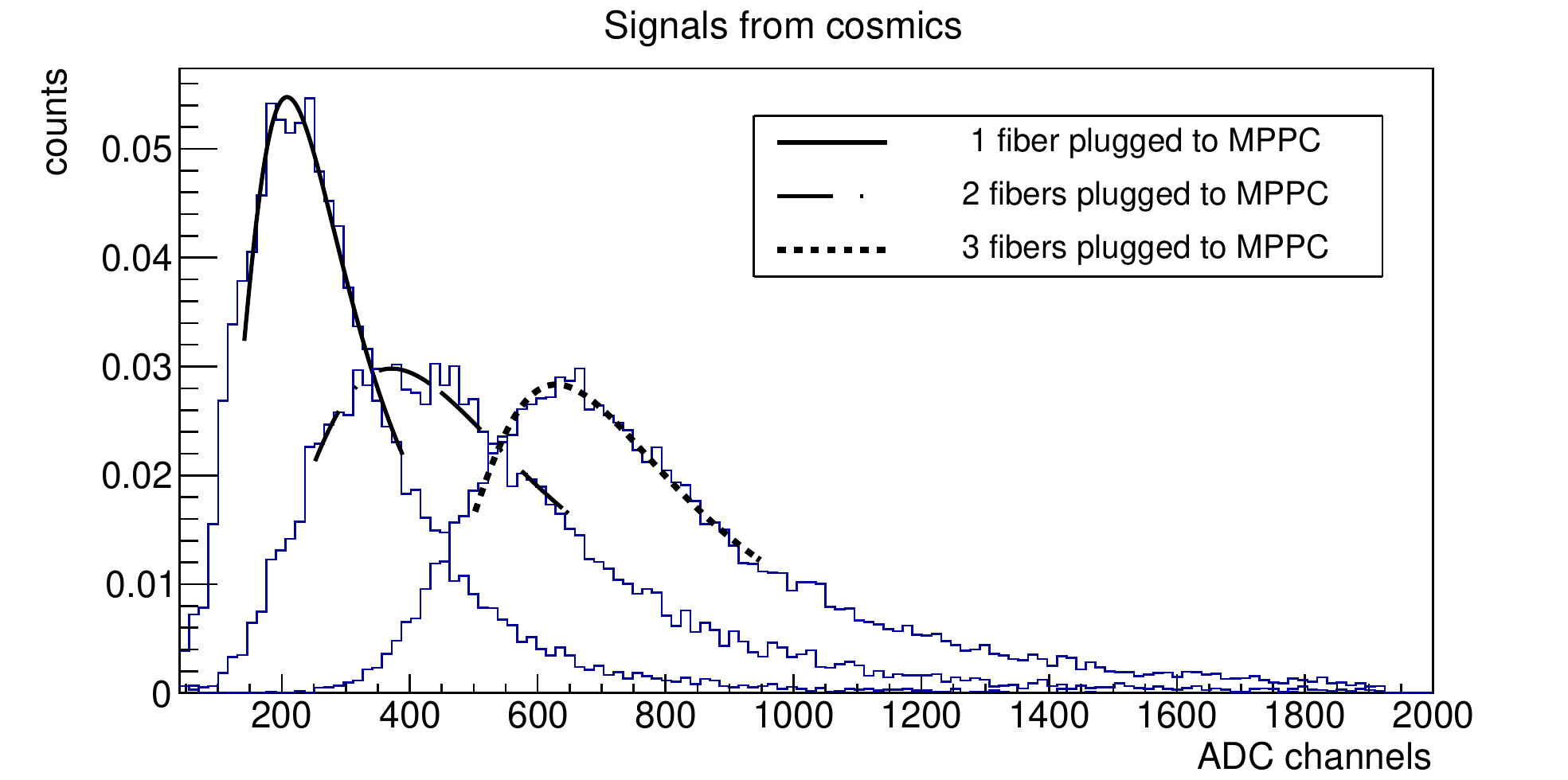}
     \caption{Comparison of the light yield with one, two and three fibers readout for 5 cm strip (top) and for 10 cm strip (bottom). 
     All the above plots correspond to data collected with  cosmic rays.}
     \label{ifr:fig:fibers}
  \end{center}
\end{figure}

\paragraph{Module length.} The studies described above have been
performed with the cosmic trigger located at about 40\cm from the
photodetector.  On the other side the amount of light collected by the
SipM is dependent on the path traversed by photons in the fibers
(Figure \ref{ifr:fig:lightdist}). The dependence of the light yield
on the distance of the muon impact point from the photodetector is the
result of the light attenuation inside the WLS fiber and the photon
propagation and collection along the scintillator strip. The
attenuation inside the WLS fibers has been extensively studied
in~\cite{ifr:fiberlen}. It can be modeled by the superimposition of
two exponentials, one with a short attenuation length of about 0.8\m
and another of about 10\m. The long attenuation length is typical for
polystyrene fibers and it is rather independent of the wavelength. The
component with short attenuation length is due a mechanism of self
absorption.  In this case the absorption spectrum of the fiber
overlaps with the emission spectrum.  In particular for the Kuraray
fiber this corresponds to relatively low wavelengths from the range
between 450\nm and 470\nm. The attenuation due to the self absorption
mechanism is dependent on the wavelength and it is greater in the
blue-violet region. This mechanism has to be taken into account when
coupling the WLS fiber to the photodetector. Generally, devices with
more sensitivity in the blue part of the spectrum are more affected by
the fiber attenuation length.

\begin{figure}[tbp]
  \begin{center}
     \includegraphics[width=0.5\textwidth]{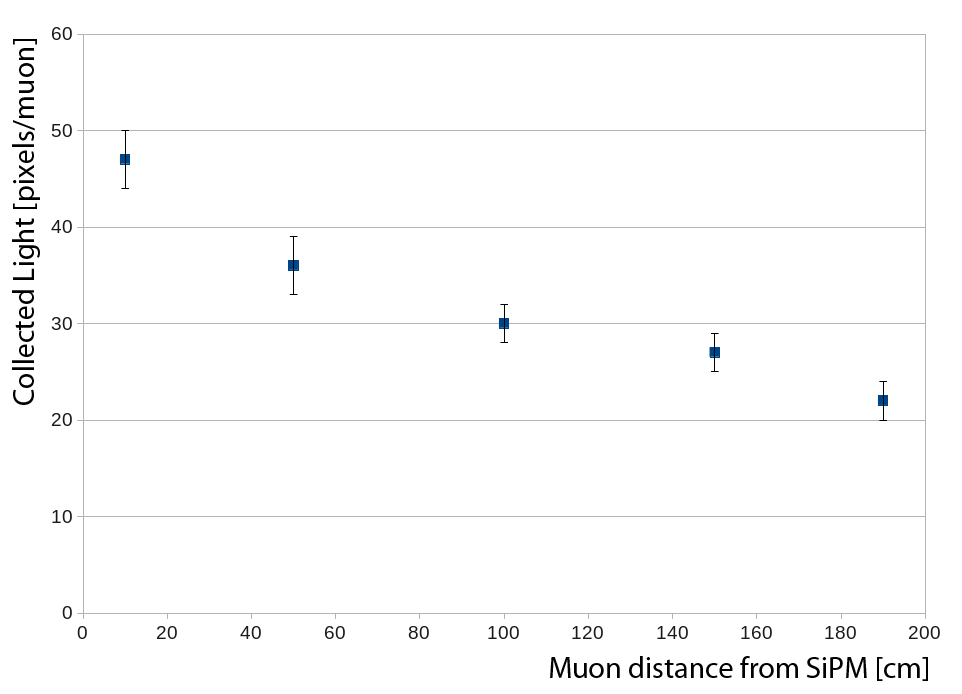}
     \caption{Light output as a function of the distance of the cosmic trigger from the photodetector.}
     \label{ifr:fig:lightdist}
  \end{center}
\end{figure}

\paragraph{Other related studies.} Performance required several other
studies which will be briefly discussed below.  Among them is {\bf
  gluing} of the fiber to the scintillator groove.  It would certainly
improve the light collection, but at the price of additional
mechanical complications in the production of the modules.  The
studies revealed that the usage of a Bicron optical epoxy resin
provides the gain in light output at the level of almost 50\pct higher
with respect to the non glued module.

The {\bf polishing} of the fiber surface was investigated in order to
understand which material and type of blade would be optimal in order
to obtain the best quality of the surface and optimal light
transmission.  We compared the quality of fiber polishing for natural
and synthetic diamond blades; the tests showed a $10\pct$ higher light
transmission for the natural diamond option.

Studies have been also performed on the coverage of the free end of
the fiber with aluminum (through a sputtering process) in order to
reflect back the light which would be lost otherwise. It was
estimated that the above technique yields a reflection coefficient
of the order of $50\pct$.  Thus we are considering the possibility to
aluminize the free end of the fibers to recover part of the light
and to reduce the light yield loss as a function of the length. \\

\subsubsection{Radiation damage of photodetectors} 

Similar to most other solid state devices, SiPMs are rather sensitive
to radiation. In the \superb\ environment we expect neutrons to be the
main source of background that can cause the aging of IFR
photodetectors.  An extensive program, started in 2009 with a first
test at the ENEA FNG (Frascati Neutron Generator)~\cite{ifr:FNG} is
currently ongoing to understand the effects of intense neutrons fluxes
on SiPMs. A test with low energy neutrons ($\le$ keV) has just been
carried out at the GELINA (GEel LINear Accelerator)~\cite{ifr:gelina}
facility in Belgium. The next test with higher energy neutrons is
foreseen in the following months.  The first irradiation
test~\cite{ifr:ifrsipm2} was carried out at the Frascati Neutron
Generator facility with neutrons of 2.5\mev energy~\cite{ifr:FNG}.  In
this test we observed an increase in dark current by a factor of 30
after a dose of $7.3\times 10^{10}\neqpcma$(1\mev equivalent). At the
same time, the dark count rate increased by a factor of 10 and the
average output signal (charge) was reduced to one third.  While the
increase in the dark current is not a major issue by itself, the
increase in dark count and the reduction of the output signal might
cause a significant deterioration of the detection efficiency.  This
reduction was studied by comparing the SiPM signal from cosmics before
and after the irradiation and it was estimated to be about $15\pct$ in
the worst case (see Figure~\ref{ifr:fig:FNGadc}). These results are
not conclusive yet because the SiPM techology is rapidly improving
with time and new tests of recently produced devices (e.g.  with an
initial dark count rate $\simeq$ a factor 10 lower) are needed.

\begin{figure*}[tbp]
\begin{center}
\includegraphics[width=\textwidth]{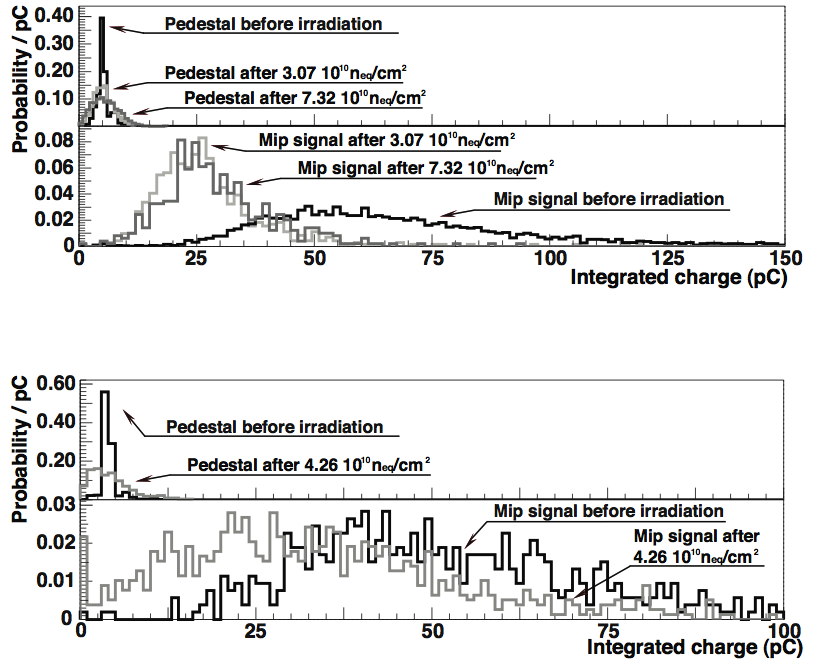}
\caption{Comparison of the cosmic signal distribution  before and after the irradiation, for a FBK 1 mm$^2$ (top) and
a MPPC 1 mm$^2$ (bottom) device.}
\label{ifr:fig:FNGadc}
\end{center}
\end{figure*}

Since an important fraction of neutrons expected at \superb\ would be
in the range of thermal energies, a test at the GELINA facility, where
a neutron beam is available with energy ranging from tens of meV to
about one MeV~\cite{ifr:gelina}, has been recently carried out on a
set of about 25 SiPM devices from AdvanSiD, Hamamatsu and SenSL.  For
each device, the dark rate and dark current were measured every minute
and I-V ({\it i.e.} current versus voltage) curve and dark rate
threshold scan were performed every hour. For some of the devices the
dark noise charge spectra have also been acquired. The data analysis
is ongoing and preliminary results indicate that the dark current and
dark count increase almost linearly starting from about 10\,$^8$
n/cm$^2$. Figure \ref{ifr:fig:gelina1} reports the different behavior
of the three technologies we tested: the Hamamatsu device seems to be
the most sensitive to the radiation.  In addition to standard devices,
two Hamamatsu radiation hard devices have been tested. The results are
still very preliminary, but while the current behavior is not so
different with respect to the standard devices, the charge spectra
seem to be less affected by the radiation.

\begin{figure*}
\begin{center}
\includegraphics[width=0.9\textwidth]{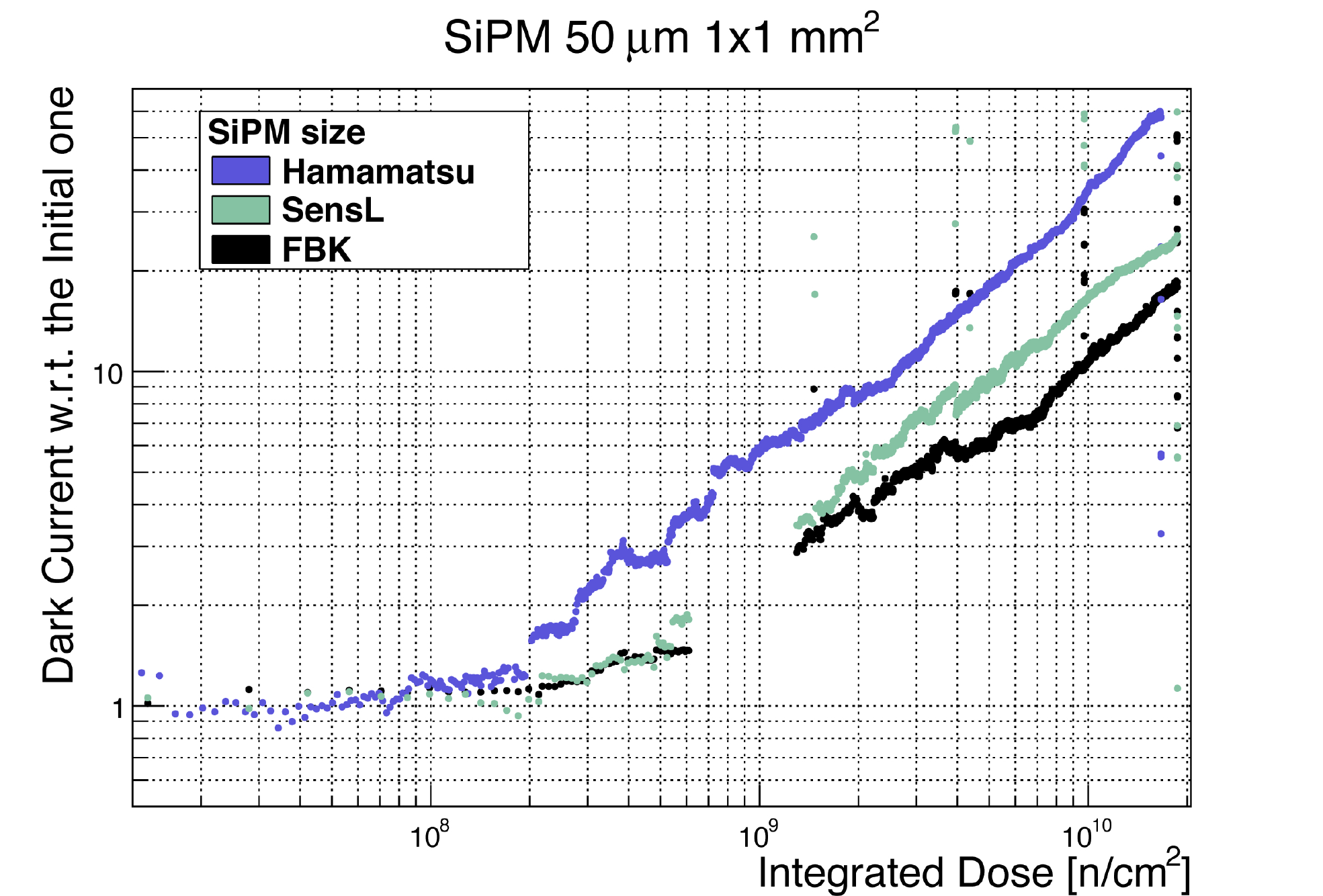}
\caption{Dark current as function of the integrated neutron dose during the GELINA test.}
\label{ifr:fig:gelina1}
\end{center}
\end{figure*}

	\tdrsubsec{Prototype Test and Results}
	
Once a baseline design of the IFR detector was established, a full
depth prototype was built and beam-tested at the Fermilab Test Beam
Facility~\cite{ifr:fnaltest} using muon and pion beams in the range of
interest for \superb.  The main goal of the beam test was to measure
the detector performance: mainly the detection efficiency, the time
resolution and the particle identification capability as a function of
the beam momentum.  The test was also the source of useful experience
concerning the future strategy of the construction and assembly of the
full detector.  In addition, the prototype was used to test two
different readout modes for the detection modules. The first one is
the ``Binary ReadOut'' (BiRO), where the SiPM signals, once
discriminated, are sampled and digitized, so that the hit coordinate
is given only by the scintillator bar position. The BiRO detection
modules are composed of two orthogonal layers of strips. The second
readout is the ``Time Readout'', where the discriminated SiPM signals
are read by a TDC system: this way one coordinate is given by the
strip position and the other by the hit arrival time. Therefore, only
one layer of strips is needed for the Time Readout.

\subsubsection{Design and construction of the IFR prototype} 

Figure~\ref{fig:ifr-prototype} shows a picture of the prototype.  It
represents a section of $60\times60\cma$ of the IFR detector, with the
full depth, and an iron structure designed to have the possibility of
moving the active layers in different positions, in order to change
the scintillator-iron segmentation and determine the configuration
being the most effective for particle identification.

\begin{figure}[tbp]
\begin{center}
\includegraphics[width=\linewidth]{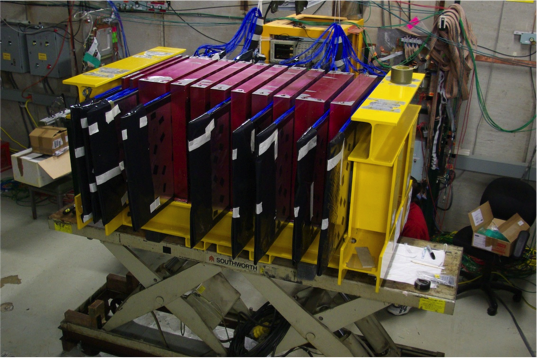}
\caption{A picture of the IFR prototype during the beam test a Fermilab.}
\label{fig:ifr-prototype}
\end{center}
\end{figure}
 
Twelve active modules were assembled in the prototype (as summarized
in Table \ref{ifr:table:protomodule}): four of them were ``standard''
in terms of the readout (BiRO and TDC) and designed according to our
baseline approach.  The remaining four were ``special'' {\it i.e.}
designed to estimate the impact of individual components (larger
fibers, different photodetector, etc.).


\begin{table}[tbp]
\begin{center}
\caption{Parameters of the modules built for the IFR prototype.}
\begin{tabular}{cclcl} 
\hline

{\small Id.} & {\small Scint.} & {\small Fibers }& {\small SiPM }&{\small Readout} \\
\#      & {\small (mm$^2$)}   & {\small type - }              & {\small $\calA$(mm$^2$)} &         \\
        &                     & {\small $\phi$ (mm)}           &                          &          \\
\hline
1& $20\times40$& Bc - 1.0 & $1.2\times3.2$ & TDC \\
2& $20\times 40$ &Bc - 1.0  &$1.2\times3.2$ &TDC \\
3&  $20\times 40$& Bc - 1.0 &$1.2\times3.2$ & TDC\\
4&  $20\times 40$& Bc - 1.0 &$1.2\times3.2$ & TDC\\
5&  $20\times 40$&Bc - 1.2 &$1.4\times3.8$ & TDC\\
6&  $20\times 40$& Bc - 1.2& round& TDC\\
7&  $20\times 40$& Bc - 1.0& MPPC & TDC\\
8&  $10\times 45$&Ku - 1.2& $1.4\times3.8$& BiRO\\
9& $10\times 45$& Ku - 1.2&$1.4\times3.8$&BiRO \\
10& $10\times 45$ & Ku - 1.2& $1.4\times3.8$& BiRO\\
11& $10\times 45$& Ku - 1.2& $1.4\times3.8$& BiRO\\
12 &$10\times 45$ &Ku - 1.2 & round& BiRO\\
\hline
\end{tabular}
\label{ifr:table:protomodule}
\end{center}
\end{table}

Figure \ref{pizzabox-biro} shows the general internal structure of an
active module of the prototype. It is composed of two layers of
orthogonal scintillating bars, 5\cm wide and 50\cm long with three
fibers in each bar. One of the fibers is housed in surface grooves and
the remaining two in embedded holes.  For the TDC modules, the
scintillator bars were 2\cm thick, so a single layer was present.  The
fibers are collected on cylindrical supports and then coupled to the
SiPM by means of custom made plastic (plexiglas) couplers.  Their
lengths range from 45 to 370\cm for BiRO readout; the fibers for the
TDC readout have a fixed length of 4\m. 

\begin{figure}[tbp]
\begin{center}
\includegraphics[width=\linewidth]{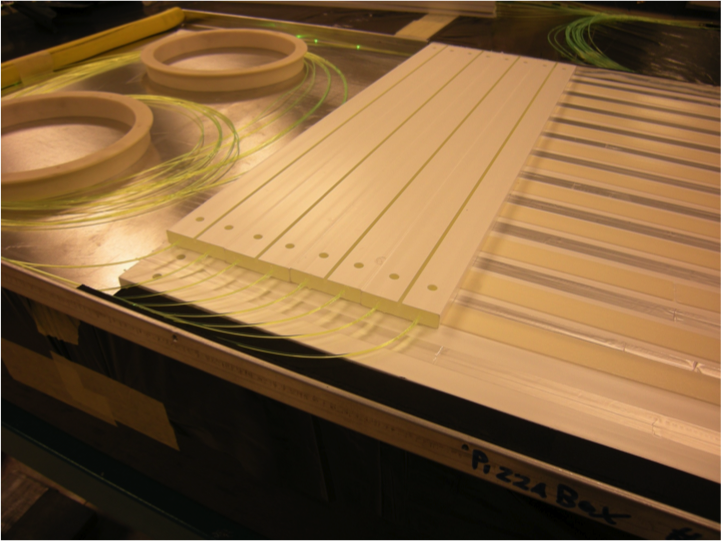} \\
\includegraphics[width=\linewidth]{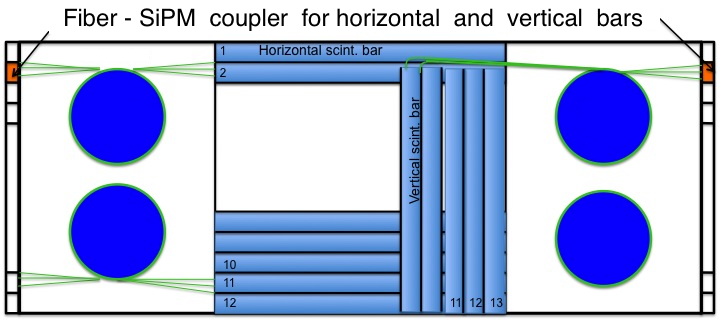} 
\caption{A picture and a scheme of a BiRO active module of the prototype.}
\label{pizzabox-biro}
\end{center}
\end{figure}

\noindent

The front-end electronics providing the bias to the SiPMs and
amplification and discrimination of signals was based on commercial
amplifiers MMIC BGA2748 and BGA 2716. The data acquisition and online
control systems were custom-designed specifically for the experimental
setup of the test beam.

\subsubsection{Beam Tests}

\paragraph{Beam Test Setup and data taking.} 

To assess the detection and identification performance the prototype
has been extensively tested with a muon and pion beam delivered by the
Fermilab Test Beam Facility at FNAL~\cite{ifr:fnaltest}. Muons and
pions were produced by means of a proton beam impingingon an Aluminum
target. The range of particle momenta was selected by two dipoles with
a spread of about 5-10\pct at 10\gevc and getting worse at lower
momenta.  In the momentum range we explored (1--10\gevc) the resulting
beam is composed mainly of electrons and pions, with a small admixture
of muons (less than 5\pct). Electrons were removed based on the veto
trigger from the Cherenkov counter (Figure \ref{ifr:fig:beamtest}).
Muons and pions with momenta above 2 \gevc were selected using another
Cherenkov signal. For momenta below 2\gevc no external particle
identification information was available, so these data have been used
only for cross checks.  The tagging capabilities of the Cherenkov
counters were deteriorating at smaller momenta. This was caused mainly
by the increase in the beam diameter from a few cm at 8\gevc to
20--30\cm at 1\gevc as well as by the spread in the beam momentum.
As a result it was hard to select clean muon and pion samples below 5\gevc.

Four beam tests have been performed between December 2010 and March
2012. While the basic concept, shown in Figure~\ref{ifr:fig:beamtest},
remained stable, the apparatus was modified for each test, taking into
account the experience gathered in the previous runs.

The setup encompassed:

\begin{itemize}
\item Cherenkov counters for particle identification, placed about
  20\m upstream the prototype;
\item a slab of 16\cm of iron upstream of the apparatus to
  ``simulate'' the material in front of the IFR detector in the
  \superb\ experiment and to get rid of most of the electrons;
\item two or three scintillator detectors with photomultipliers (PMT)
  readout placed between the iron and the IFR prototype as trigger;
\item the IFR prototype;
\item another set of scintillator detectors placed behind the
  prototype to select tracks going through the prototype.
\end{itemize}

All the signals from PMTs and Cherenkov counters were acquired with
the prototype TDC system in order to refine trigger requests offline.

Table~\ref{ifr:table:beamtestdata} shows the data samples collected at
different energies during the four beam tests along with the trigger
used to select the events.

\begin{figure*}[tbp]
\centering
\includegraphics[width=0.95\linewidth,height=0.30\textwidth]{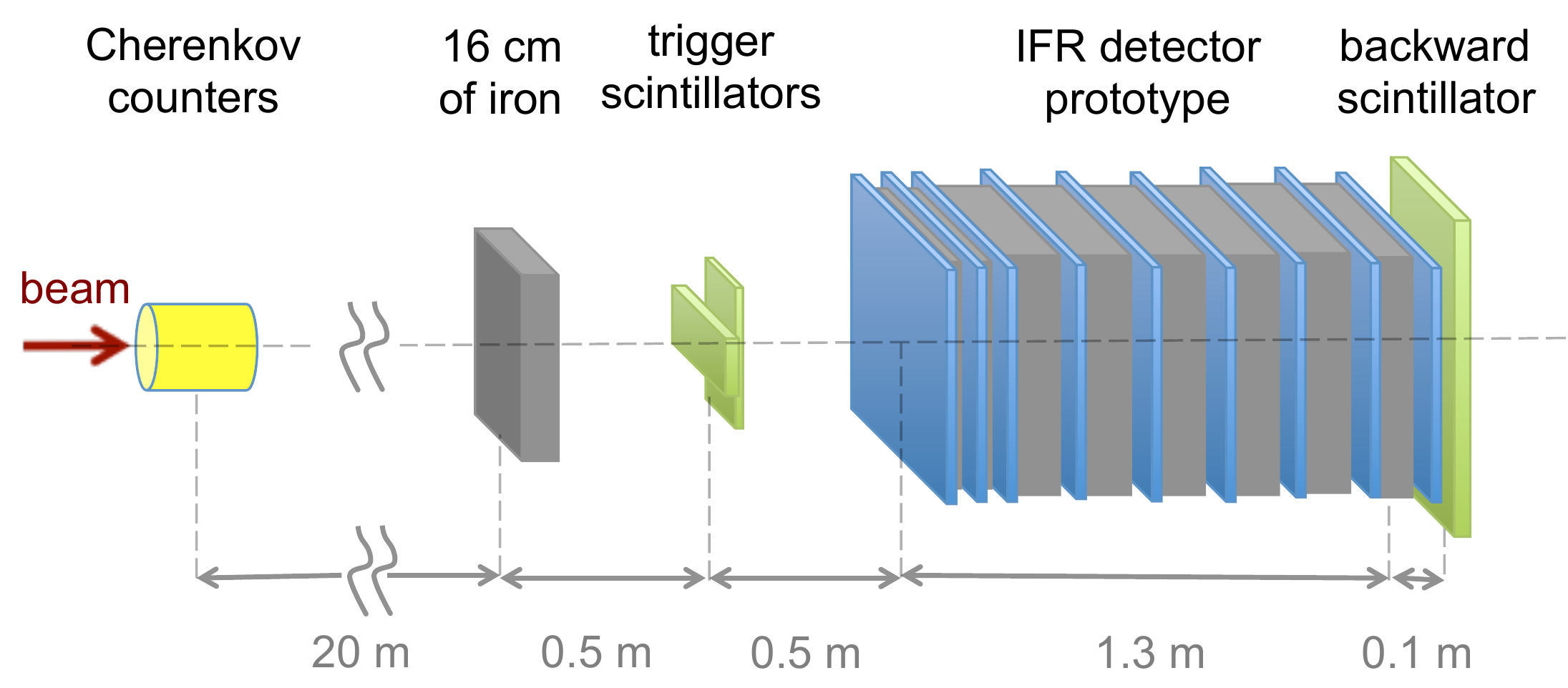}
\caption{The layout of the Fermilab beam test setup.}
\label{ifr:fig:beamtest}
\end{figure*}

\begin{table}[htb]
\begin{center}
\caption{Beam test data sample.}
\begin{tabular}{ccccc} 
\hline
beam            & min. &   muon   &   pion      & $\mu/\pi$  \\
mom.  & bias            &  trigger  &   trigger  & trigger  \\
(\mevc)        & (k evt)  &  (k evt)  & (k evt)  & (k evt)  \\
\hline
1  &  50 & -  & - & 386  \\
2  &  60 & -  & - & 236  \\
3  & 126 & 134 & 89 & 155  \\
4  & 101 & 106 & 76 & 129  \\
5  & 262 & 87   & 158 & 25  \\
6  & 230 &218  & 230  & 100  \\
8  &  236& 375 & 413 & 125  \\
\hline
\end{tabular}
\label{ifr:table:beamtestdata}
\end{center}
\end{table}

\subsubsection{Tests Results}
\label{sec:protoanalysis}

Data from different runs have been analyzed in order to study the
detection performance and the muon identification capability of the
IFR prototype. The resuts are described in detail in the following
paragraphs.

During all four tests the full setup, including the prototype detector
was working reliably with almost 100\pct data taking efficiency. After
almost 18 months of tests (both with the accelerator beam and cosmics)
only three channels out of 237 failed; their damage was caused by
shipping and handling of the detector during installation and
dismounting.

\paragraph{Detection Performance.}
One of the major goals of beam tests was to confirm the R\&D results
on a larger scale to assure that the system is capable not only to
detect particle hits but also to reconstruct three dimensional tracks
passing through it.  Therefore, not only the high detection
efficiency, but also the spatial resolution is of a paramount
importance; correspondingly for the TDC readout modules the time resolution 
has also been also evaluated. 

The detection efficiency has been evaluated using muon events passing
through the entire prototype and selected using the backward
scintillators. Results are shown in Figure~\ref{fig:ifr:deteff} both
for the binary and time readout modules.  The detection efficiency is
clearly dependent on the length of the light path inside the fibers,
which varies from module to module, but its value exceeds 95\pct in
almost any case. The only exception is for the module in layer number
two for which the light path amounted to almost 4 meters, however,
such long light paths are not foreseen in the final detector design.
The detection efficiency depends also on the electronic threshold.
Figure~\ref{fig:ifr:effthr} shows that, except for the module with the
long fibers (layer number two), there is room to raise the threshold
from the current nominal value, that is 3.5 photoelectrons, to 4.5 in
order to reduce the SiPM dark count with very small efficiency loss.
All these conclusions are consistent with the R\&D results.

\begin{figure}[h!]
\centering
\includegraphics[width=\linewidth]{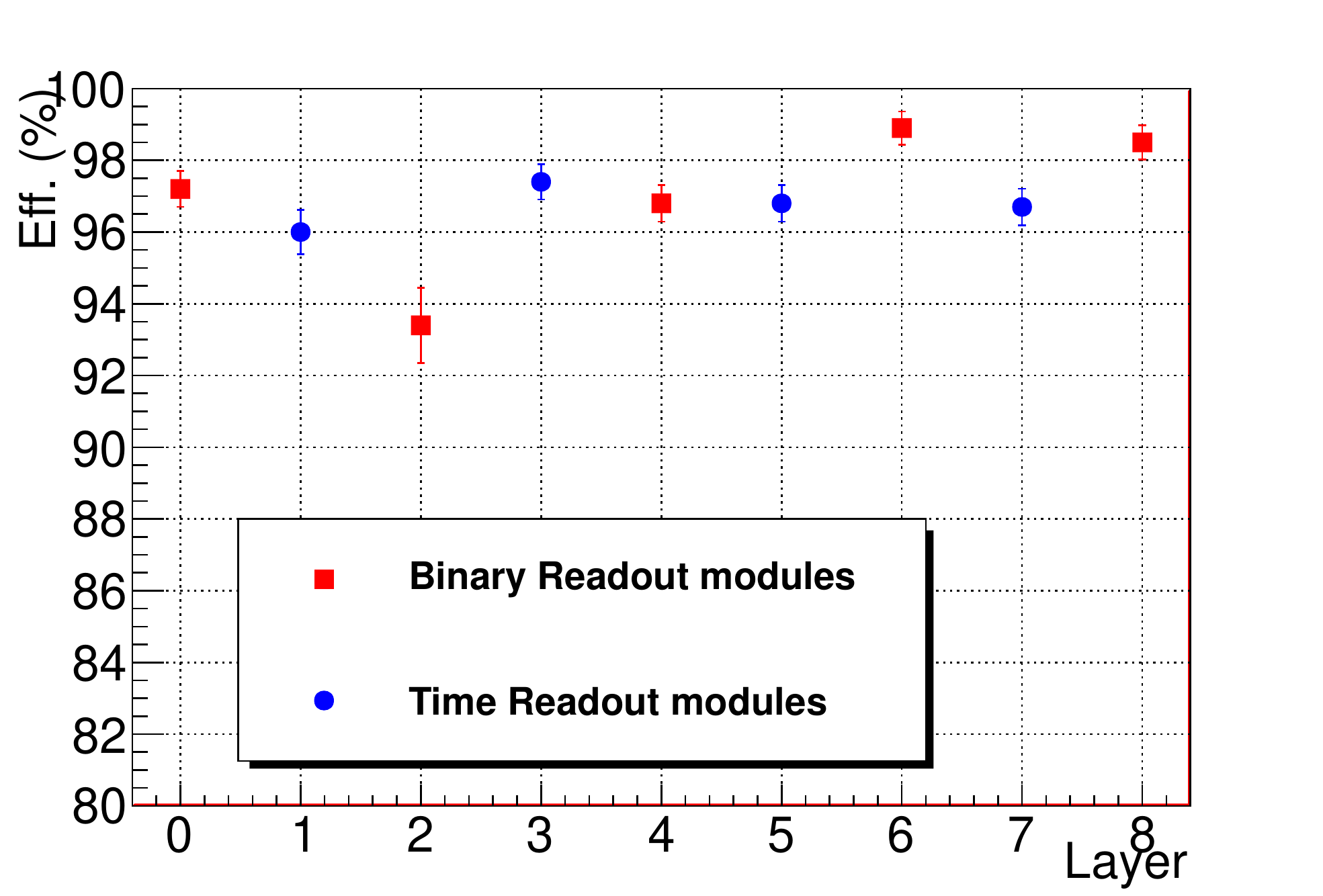}
\caption{The detection efficiency of each individual layer of the prototype as measured during the first beam test at Fermilab
in December 2010.}
\label{fig:ifr:deteff}
\end{figure}

\begin{figure}[h!]
\centering
\includegraphics[width=\linewidth]{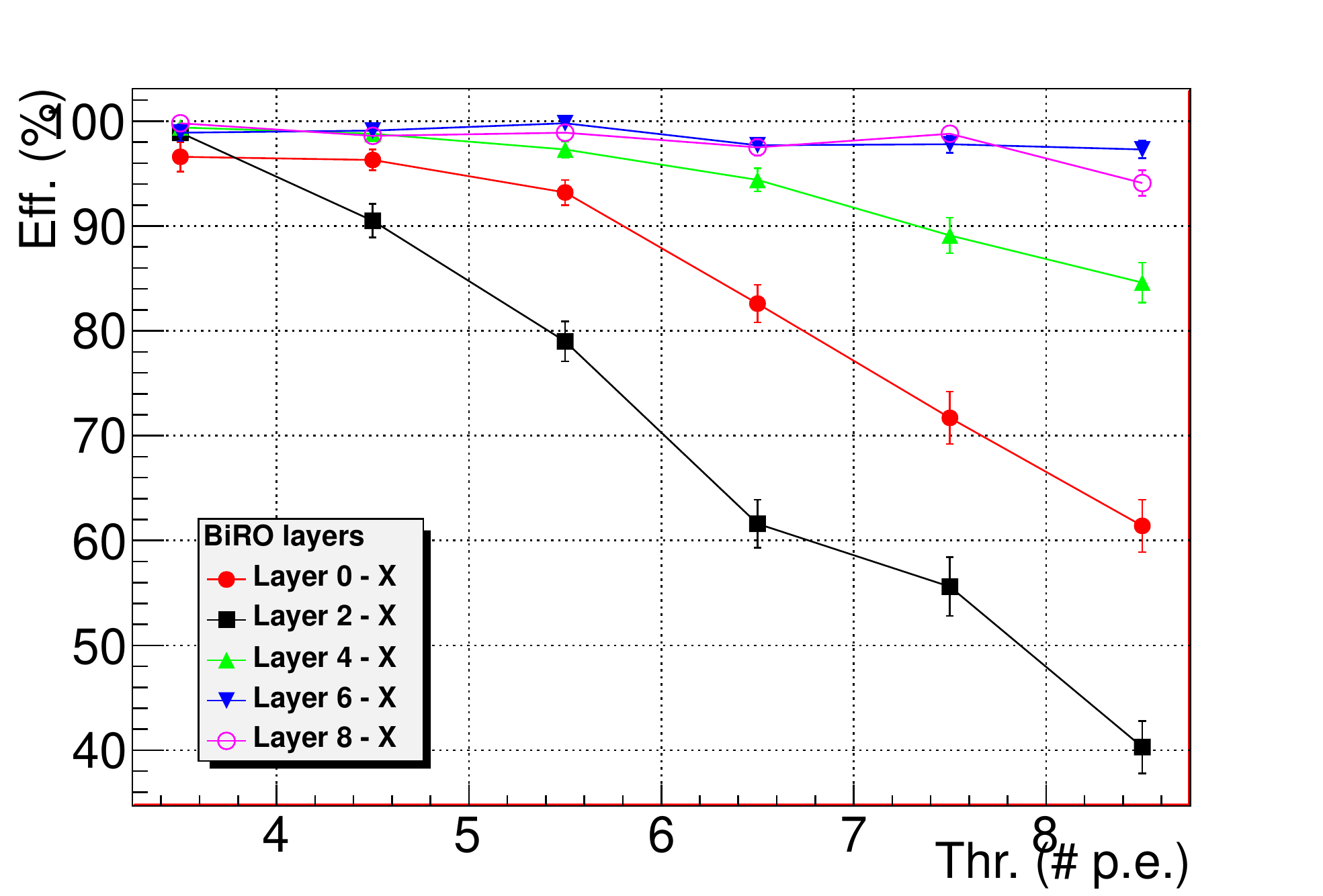}
\caption{The detection efficiency of each individual layer of the
  prototype as a function of the electronic threshold (measured in the
  number of photoelectrons) for the BiRO modules.}
\label{fig:ifr:effthr}
\end{figure}

The beam tests have also shown that the time resolution linearly
depends on the fiber length, in agreement with our R\&D results. The
average time resolution was measured to be about 1.2\ns, which would
be sufficient to guarantee the spatial resolution needed to match the
physics performance requirements.  On the other hand, there is very
little space for improvements in the time resolution and moreover the
risk of worsening the performance with the aging of the photodetectors
is rather high.


The last important factor estimated during beam tests was the average
strip occupancy caused by the SiPM dark noise. It was evaluated to be
about 1.7\pct both with a random trigger and using out-of-time hits.

\paragraph{Muon Identification Results}

The IFR muon identification performance described in
section~\ref{sec:muonid} is based on studies with Monte Carlo (MC)
events.  The prototype data have been compared with the respective MC
samples and the simulation parameters have been tuned in order to
ensure that the MC reproduces the prototype behavior (hit and particle
multiplicity, noise, detection efficiency and time response).  The
extensive data-MC comparisons and tuning have been done in order to
reproduce the prototype particle identification performance with the
MC.  For the reasons explained above, only the subsample of data
collected at 8\gevc have been used in the tuning procedure (to
improve the purity of the sample offline cuts on the hits arrival time
in the Cherenkov and in the trigger scintillators have been
performed).

Therefore a complete simulation of the prototype has been implemented
within the \superb\ Full Simulation (see Sec.
\ref{sec:Full_Simulation}) framework. The simulation included the
Cherenkov counters and the scintillators used in the trigger, the
prototype iron structure and the position of the momentum-selecting
magnet (about 70\m upstream with respect to the prototype) --- important
for taking into account the particles decaying in flight.

\begin{figure}[t]
\centering
\includegraphics[width=\linewidth]{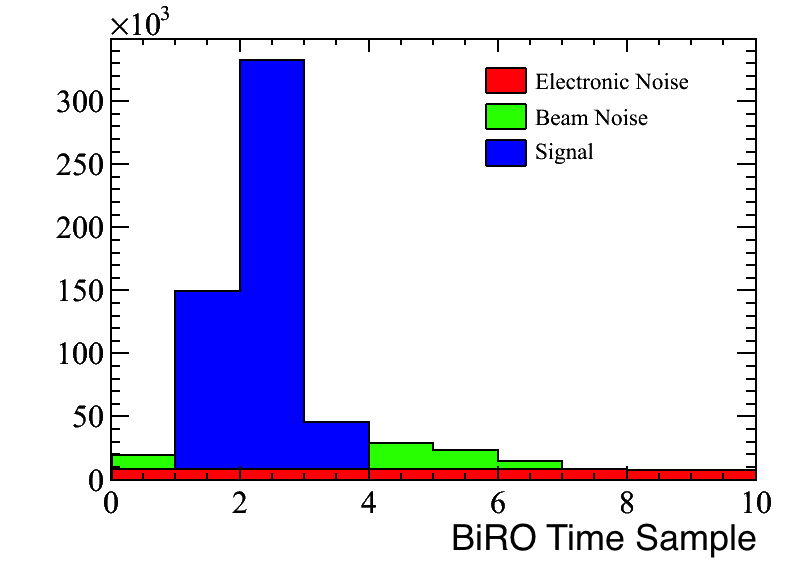}
\caption{Hit occupancy for muon events as a function of the binary readout sample; each sample corresponds to 12.5 ns.
Muon hits peak around sample number 2, the noise due to the SiPM dark count is flat over entire region while 
the remaining component is ascribed to beam background.}
\label{fig:ifr:birobkg}
\end{figure}

Beam background and SiPM dark count are parameters of paramount
importance to reproduce the beam test conditions. To estimate their
contribution we used the sampling feature of the BiRO readout with
muon events to disentangle the signal, beam and electronic components
of the hit occupancy as shown in Figure~\ref{fig:ifr:birobkg}: each
sample corresponds to a 12.5\ns interval.  The electronic component,
due to SiPM dark rate, is flat and is estimated using the last three
samples.  Muon hits arrive within few nanoseconds and the hit time
depends also on the light path into the fibers: the signal component
peaks between sample number 1, 2 and 3; the beam-related background
(mainly tails from electron showers) is distributed around the signal
peak, for its estimation we considered samples 0 and from 4 to 6; Beam
and electronic background have been evaluated strip by strip on all
the prototype channels and their occupancy has been added to the
simulation.

\begin{figure}[b!]
  \centering
        \begin{subfloat}
        \centering
                \includegraphics[width=\linewidth]{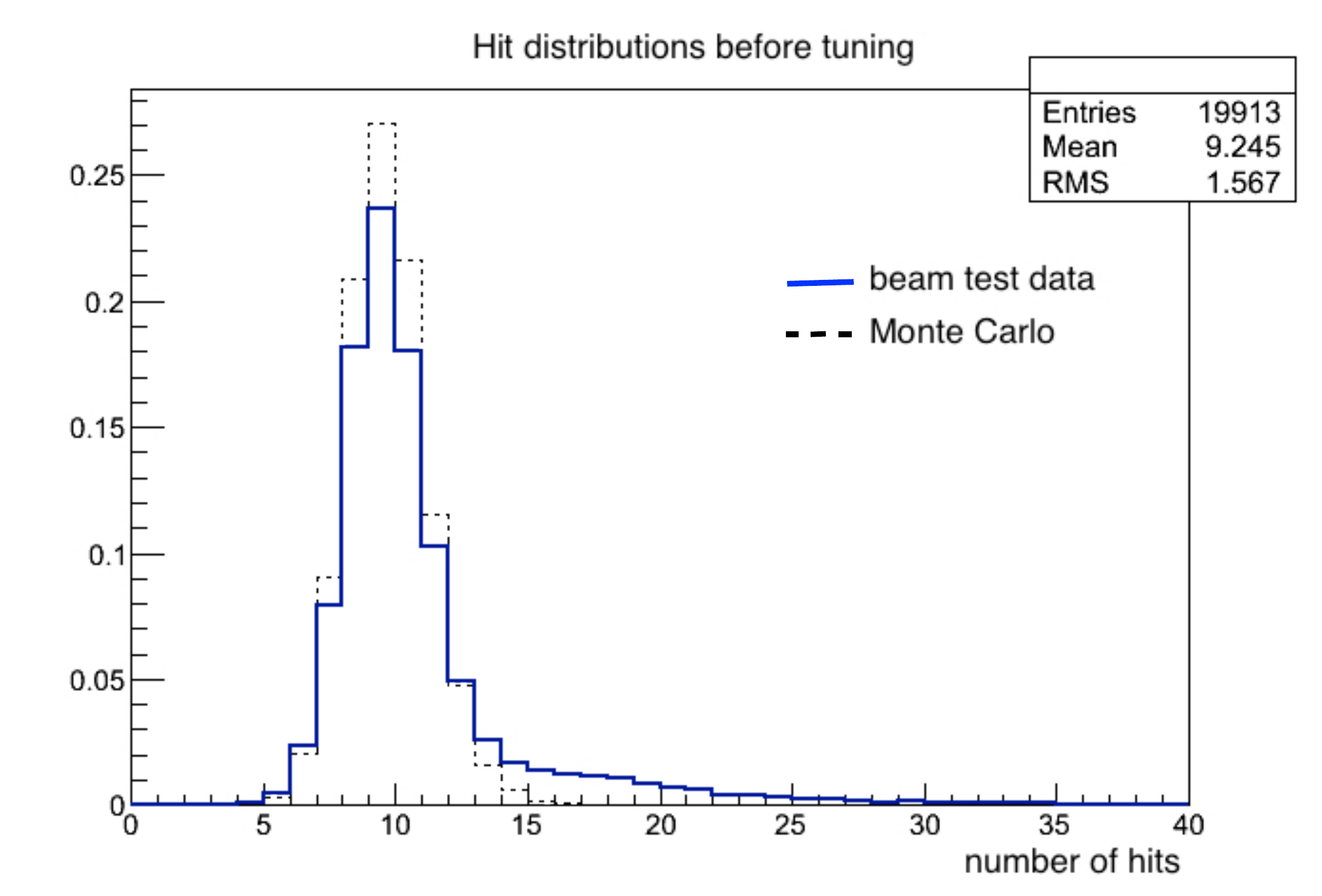}
        \end{subfloat}
        \begin{subfloat}
        \centering
                \includegraphics[width=\linewidth]{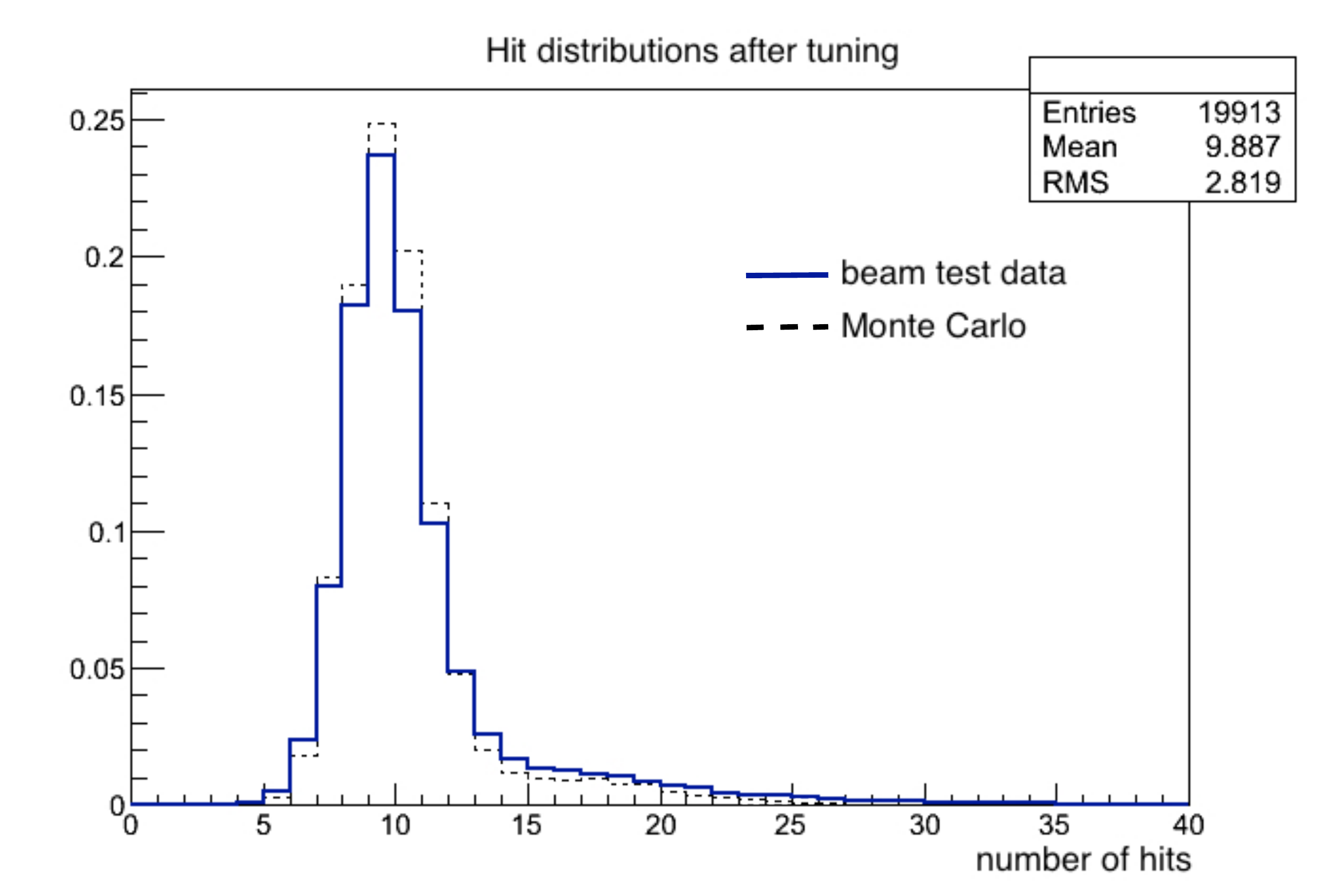}
        \end{subfloat}
  \caption{Data - Monte Carlo comparisons for the hit multiplicity before (top) and after (bottom) the addition of a secondary 
  particle to the Monte Carlo sample. }
  \label{fig:ifr:tuning}
\end{figure}

Similarly, the detection efficiency has been studied layer by layer
using 8\gev muon data; the layer inefficiency has been added
to the simulation by tuning a cut on the MC hit energy loss.

As last part of the tuning procedure we simulate a multi-particle
component: the beam composition, given by the beam test
facility~\cite{ifr:fnaltest}, is made of about 10\pct of
multi-particle events, in our case mostly pions; we randomly added
pion events to the 10\pct of each MC sample. The effect of this
addition is shown in Figure~\ref{fig:ifr:tuning}.

After the tuning, the data-MC agreement of all the variables we used
as BDT input for muon and pion samples was good, see
Section~\ref{sec:muonid} for a complete description of the BDT
selector.  The BDT was then trained separately on data and MC
sub-samples and applied to the remaining datasets; the signal
efficiency - background rejection curves (ROC) are reported in
Figure~\ref{ifr:fig:roc-prototype}: the consistency between data and
MC is within 5\pct in the worst case. Therefore, we can conclude that
the tuned MC at 8\gev is a good description of the beam test data and
we can use the tuning information to extend particle identification
studies with MC to increase the statistics and using the \superb\
geometry.

\begin{figure}[hb]
\centering
\includegraphics[width=\linewidth]{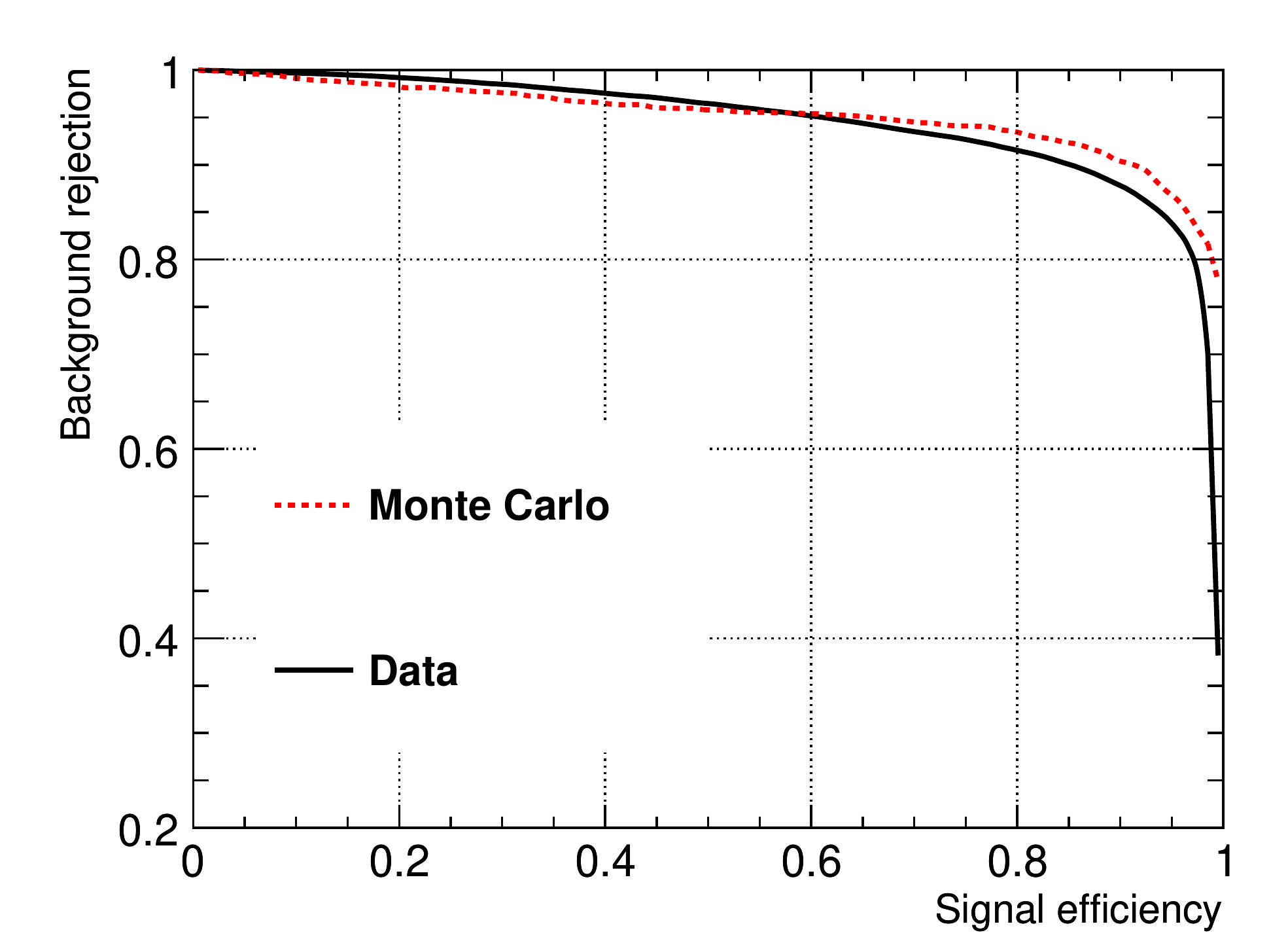}
\caption{ROC curve for BDT selector with prototype data (solid line) and Monte Carlo (dashed line).}
\label{ifr:fig:roc-prototype}
\end{figure}


\tdrsec{Baseline Design of the IFR Detector}

	\tdrsubsec{Detector Layout	}
	\label{ifr:sec:systlay}

The layout of the IFR detector takes into account several requirements
and limitations. They will be underlined below before the detailed
discussion of the detector design.

The IFR system must have high efficiency for selecting penetrating particles such as muons, 
while at the same time rejecting charged hadrons (mostly pions and kaons) in a wide momentum
range, from approximately 0.5\gevc to 5\gevc.

The possibility of reusing the \babar\ iron offers a cheap solution
for the flux return structure, but it introduces some mechanical
constraints which have a significant influence on the detector design.
In particular, the available space for the detection modules is
limited by the dimension of the existing gaps (25\mm). Also, the
thickness of the absorber material is restricted to about 90\cm of
iron and the clearances and paths for cables and other utilities must
follow the existing structure of the flux return.

The design of the IFR detector should also take into account the
presence of the accelerator-induced backgrounds that can potentially
reduce the quality of particle identification as well as speed up
aging of the system components. These two issues are discussed in the
next paragraphs.

The maximum rate of the accelerator-induced background is expected in
the forward endcap of the IFR.  A detailed Monte Carlo simulation
yielded that the above background should not exceed $1\kHzpcma$ and
that it decreases rapidly with the distance from the beam axis. It was
checked that with the above background estimation multiplied by a
safety factor of five, the performance of muon identification is still
acceptable.

For what concerns the aging of the photodetectors the main hazard
comes from the high flux of neutrons originating from radiative Bhabha
events. Recent neutron irradiation tests yielded \cite{ifr:ifrsipm2,
  ifr:musienko} that SiPM devices continue to work at least up to a
fluence of $ \sim 10^{11} - 10^{12}$\,n$_{eq}/$cm$^2$.  However, they
suffer a considerable increase of dark rate and dark current and an
associated loss in the light yield lower than 20\pct, depending on the
cell size.  The neutron flux in the part of the IFR where the SiPMs
will be located was estimated from Monte Carlo simulations to be in
the range from 50\Hzpcma to 500\Hzpcma. We use such information to
identify the safest place for the SiPM. 

R\&D results have revealed that using $1\times5\cma$ ($1\times10\cma$)
strips, and with three fibers for each strip, it is feasible to build
a detection module of 2.5\m (1\m) length.  For such a module the
detection efficiency would exceed 95\pct and the increase of the dark
count rate due to the neutron irradiation damage could be kept under
control by applying a sufficiently high threshold.  Moreover, beam
tests on a large scale prototype have demonstrated that the muon
identification capability and the reliability of the system are fully
acceptable.

The design of the IFR detector should also take into account its
abilities to provide a reasonable reconstruction of $K^0_L$s mesons.
Monte Carlo simulations and previous experience with the \babar\
experiment led to the conclusion that maintaining the existing fine
segmentation of the three internal layers is sufficient for this goal.

\tdrparagraph{Module concept.}  The basic element of the IFR detector
is composed of an extruded plastic scintillator with three WLS fibers
located in machined grooves and a SiPM installed as shown in Figure
\ref{fig:ifr:sipm-fib-coupl}. The scintillator will be the one
produced by the FNAL-NICADD facility and the fibers will be produced
by the Kuraray company (Y11-300, with a diameter of 1.2\mm).  The
$3\times3\mma$ MPPCs produced by Hamamatsu with a cell size of 50\mum
have been chosen as photodetectors.

\begin{figure*}[htb]
  \begin{center}
    \subfloat[Layout of the module, showing details of the fiber
    geometry and of the SiPM-PCB
    connection.]{\label{fig:ifr:sipm-fib-coupl}\includegraphics[width=0.45\textwidth,
      height=0.45\textwidth]{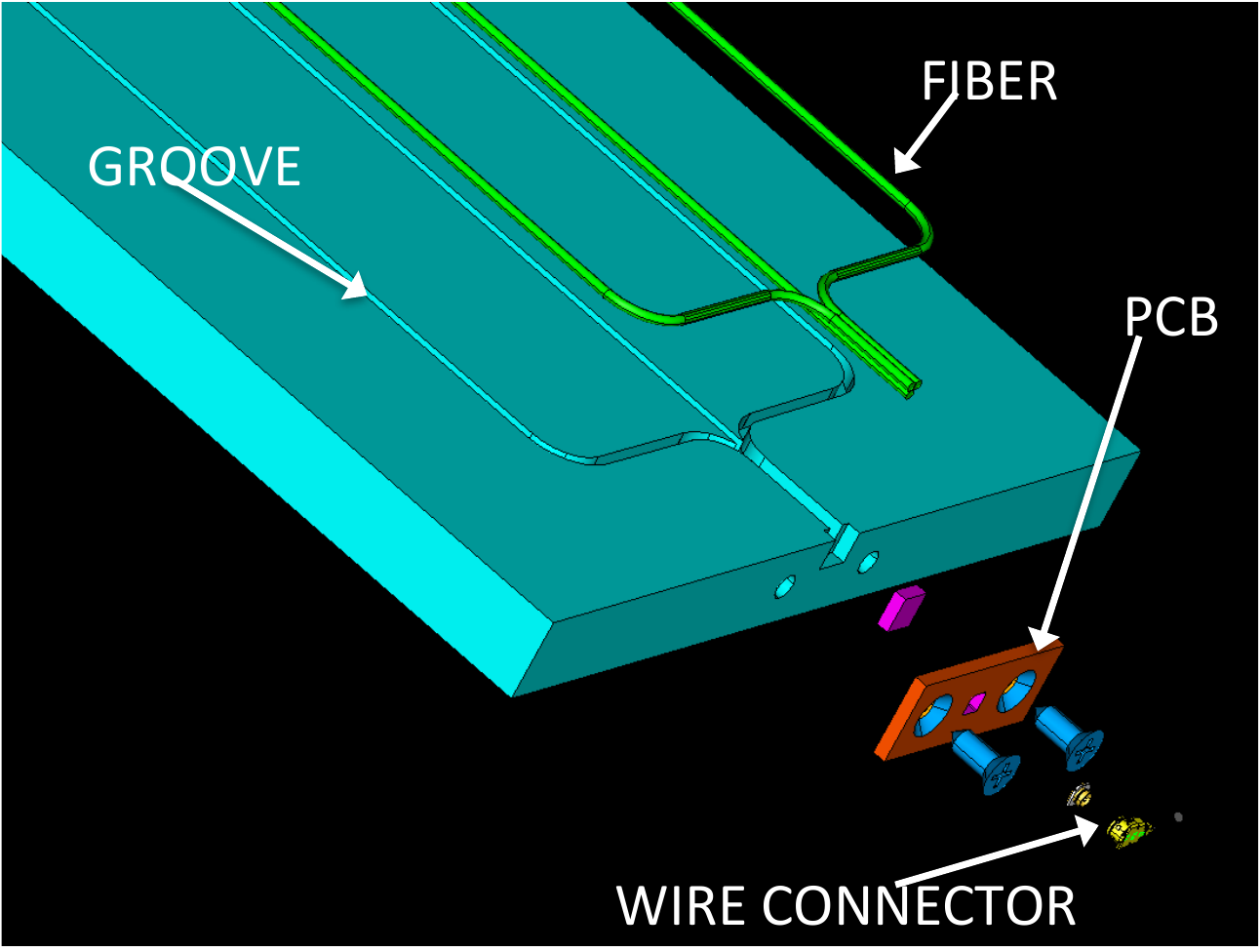} }
    \subfloat[Location of the scintillator bars and SiPMs inside the
    IFR barrel. The innermost layer is shown with the modules'
    enclosure
    removed.]{\label{ifr:fig:sipmloc}\includegraphics[width=0.45\textwidth,
      height=0.45\textwidth]{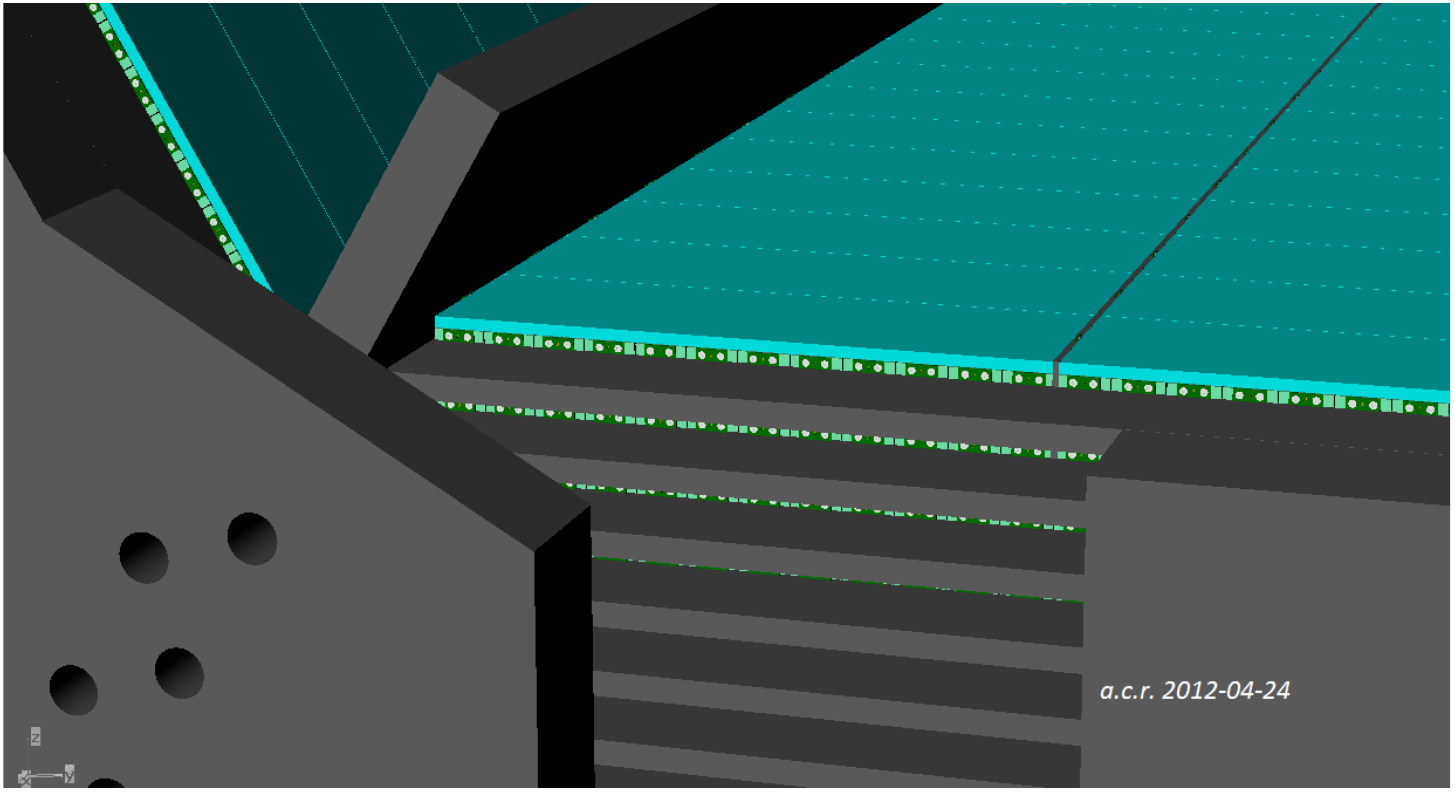}}
  \end{center}
  \caption{Concept of the IFR module design.}
  \label{ifr:fig:design}
\end{figure*}

From one end of each scintillator bar the WLS fibers are brought in
front of a SiPM mounted on a PCB.  A coaxial cable then connects the
PCB with the front-end stage of the IFR readout.  The scintillator is
wrapped with an aluminum foil reflecting the light toward the fibers.
The detector is composed of two perpendicular layers of scintillators.
The scintillator bars, together with the SiPM carrier PCBs and the
signal cables, will be enclosed in aluminum dark boxes.  Their main
purpose will be to shield the assembly from the ambient light and to
provide the necessary mechanical robustness. The above structure will
be, in the following, referred to as ''module''.



The modules will be inserted, in nine gaps of the iron structure, both
in the barrel and endcap sections of the IFR. Each of these gaps
instrumented with modules will be referred as an "active layer".
Figure \ref{ifr:fig:sipmloc} shows a few modules installed in the
barrel section of the IFR, with the aluminum ''envelopes'' removed in
order to show the scintillator's orientation.  The tight coupling of
the SiPM to the WLS fibers has been decided as it maximizes the number
of converted photoelectrons. This choice determines in turn that the
coupling between the SiPMs and the first stage of the IFR signal
processing chain must be done through coaxial cables and connectors.

\tdrparagraph{The barrel modules} will be of rectangular shape and
each module will be composed of a layer of $\phi$-strips superimposed
on a layer of z-strips. The $\phi$-strips will be 10\mm thick, 50\mm
wide, and 1800\mm long. The number of these strips will vary from
layer to layer. The z-strips will be 10\mm thick, 106 mm wide and
their length varies from 650\mm to 800\mm.  The number of strips for
each module is 17 except for the two outermost layers that will be
shorter and will contain 15 strips.  The number of modules of each
active layer is 6 or 8, depending on the gap width. Since each module
covers half of the gap length along the beam direction, we will have
half of the modules facing forward and half on the backward side. As a
result the cables will be routed on both sides of the barrel.
Figure~\ref{ifr:fig:barmodule} shows the innermost and the outermost
layers instrumented with forward modules. Modules of the outermost
layer will have a special design due to the details of the mechanical
connection between the last iron plate and the external structure.

\begin{figure}[h!]
\centering
\includegraphics[width=0.45\textwidth, height=0.45\textwidth]{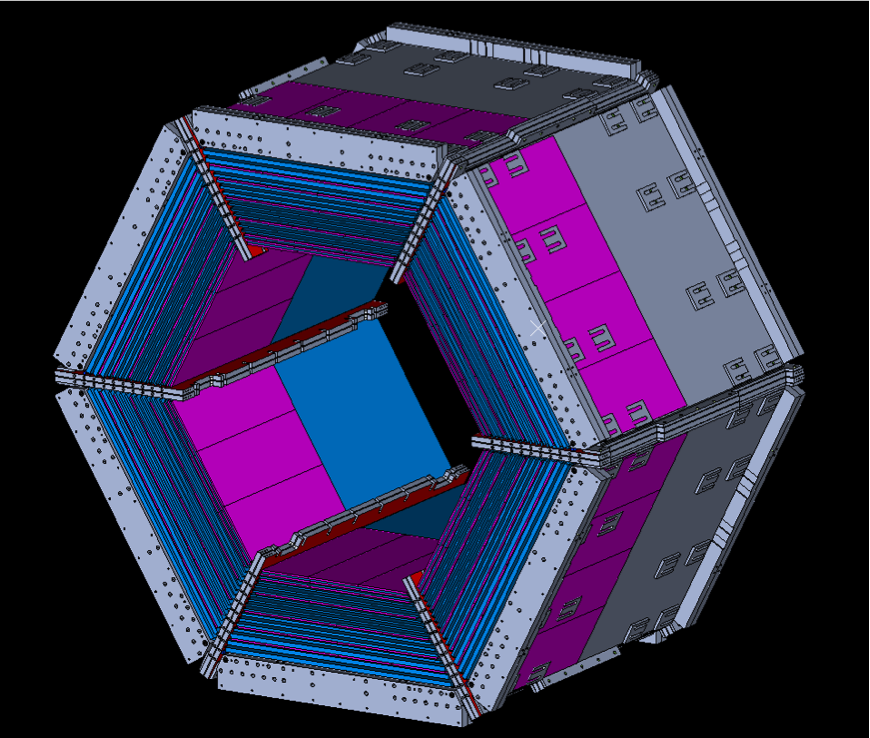}
\caption{Layout of the barrel modules of the IFR detector.}
\label{ifr:fig:barmodule}
\end{figure} 

Gap dimensions as well as the number of channels and modules for each
barrel layer are collected in Table~\ref{ifr:tab:barrel}.

\begin{table}[htb]
\begin{center}
\caption{The numbers of modules and strips per modules in layers of the barrel part of the IFR detector.}
\begin{tabular}{cccc} 
\hline
Layer  &   n. of            & $\phi$-strips      & z-strips\\
            &    modules    & per modules    & per modules \\
\hline
1 & 6 & 13 & 17 (65\cm) \\
2 & 6 & 13 & 17  (65\cm) \\ 
3 & 6 & 13 &  17 (65\cm) \\
4 & 6 & 15 &  17 (75\cm) \\
5 & 8 & 12 &  17 (60\cm)\\
6 & 8 & 13 &  17 (65\cm)\\
7 & 8 & 14 &  17 (70\cm)\\
8 & 8 & 15 &  15 (75\cm)\\
9 & 8 & 16 &  15 (80\cm) \\
\hline
\end{tabular}
\label{ifr:tab:barrel}
\end{center}
\end{table}

\tdrparagraph{The endcap modules.}  There will be six wide endcap
modules with each active layer as dictated by the reuse of the \babar\
flux return (Figure~\ref{ifr:fig:fwdmodule}). While top and bottom
modules will have the same size, the dimensions of the middle ones
will depend on the internal radius of the hole, that is conical for
the forward and cylindrical for the backward endcap.  Each module will
be instrumented with $x$ and $y$ strips of the same thickness (10\mm)
and width (50\mm). The strip lengths will range from about 500\mm up
to about 2500\mm. Table~\ref{ifr:tab:endcap} provides the information
about the number of strips in the endcap modules.

\begin{figure}[t]
\centering
\includegraphics[width=0.45\textwidth, height=0.45\textwidth]{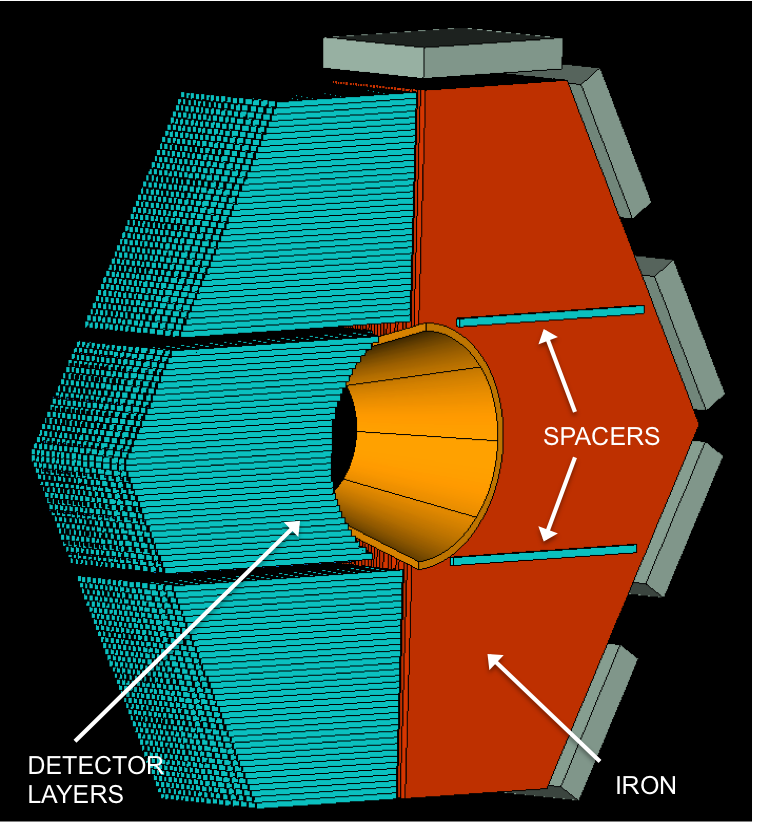}
\caption{Layout of the endcap modules of the IFR detector.}
\label{ifr:fig:fwdmodule}
\end{figure} 

\begin{table}[htb]
\begin{center}
\caption{The numbers of strips in endcap modules.}
\begin{tabular}{cccc} 
\hline
            &  Top           & Middle      & Bottom   \\
            &  modules   & modules  & modules \\
\hline
$x$ strips   & 51  & 64  & 51  \\
$y$ strips   & 37  & 36  & 37  \\
\hline
Total strips & 88 & 100 & 88 \\
\hline
\end{tabular}
\label{ifr:tab:endcap}
\end{center}
\end{table}

	\tdrsubsec{Chamber Construction, Assembly and Quality Control}
	
\subsubsection*{Module assembly procedure}

The assembly of the modules requires a careful production plan, 
defining, in particular, the requirements for the place, timing,
tooling and quality control. The procedure can be split into different
operations, some of which are sequential, while others will occur in
parallel.  The main steps of the assembly will be described below.

\tdrparagraph{Manufacturing of the scintillators} encompasses their
cutting the scintillator to the proper length, labeling and stocking.
It can be divided into the following operations:
\begin{enumerate}
\item positioning of the scintillator on the machine;
\item machining the linear grooves;
\item machining the curved grooves;
\item cleaning of the grooves and the scintillator; 
\item checking the grooves quality; 
\item stocking the scintillator.
\end{enumerate}
The operations 2 and 3 are to be done in parallel on the same machine
equipped with two heads; the first one with a cutting disk making the
linear grooves, and the second one with a cylindrical milling cutter
making the curved grooves.

\tdrparagraph{Preparation of the WLS fibers:} the three fibers
belonging to the same strip are to be inserted and clamped into a
plexiglass connector, machined and polished.  All the fibers and
connectors comprising a single module are to be polished at the same
time. The manufacturing of the scintillators and the preparation of
the fibers have to be done in parallel.

\tdrparagraph{Scintillators and fibers assembly.} This operation
encompasses :
\begin{enumerate}
\item the positioning of the scintillator on the machine, inserting
  the plexiglass connector and fibers into the grooves;
\item glueing the fibers into the grooves;
\item checking the quality of the glueing (diffusion of the resin and
  air bubbles);
\item stocking in the curing station.
\end{enumerate}

A two component optical epoxy resin is used for the glueing processes.
The resin is characterized by low viscosity and by a liquid
consistency.  These features facilitate the diffusion and avoids the
air bubbles due to the gluing not under vacuum. The curing time lasts
three days and when the production is started the steps which will be
described in the following are to be shifted by this time period.

\tdrparagraph{Covering scintillators and fibers with thin aluminum
  foils.}  The scintillators with the fibers are to be wrapped with a
thin reflecting aluminum foil in order to maximize the light collected
by the fibers themselves.

\tdrparagraph{The module assembly.}  The wrapped scintillators are
accommodated into the box defining the module. The scintillators are,
one by one, secured by double sided ribbons into the box comprising
both the longitudinal and the transversal layer. The PCB and the SiPM
are assembled in front of the polished surfaces of the fibers by
screws. Finally, the read-out cables are connected to the PCB
connector. Thus, the module assembly will proceed as follows:
\begin{enumerate}
\item fixing the ribbon into the inner surfaces of the box;
\item positioning the scintillator;
\item mounting the PCB and SiPM;
\item checking the reference position;
\item connecting the read-out cable;
\item fixing the ribbon between the two layers;
\item closing the box;
\item checking the module.
\end{enumerate} 

\tdrparagraph{Tooling for module assembly.}  Proper tooling will be
developed to automate some of the steps of the module preparation, in
particular for:
\begin{itemize}
\item manufacturing of the scintillators;
\item polishing of the fibers;
\item dispensing the resin on the fibers.
\end{itemize}

\subsubsection*{Quality Control procedures}

Quality Control (QC) encompasses a set of tests to be performed in
scheduled phases during production and assembly of the modules. These
are:

\begin{enumerate} 
\item {\bf QC on finished scintillators} to validate the quality of
  fiber grooves by visual inspection in search for any macroscopic
  defects.

\item {\bf QC test on finished fibers} to check the quality of
  polished surfaces. This plays a crucial role as far as the
  efficiency of the detected light is concerned. The quality of
  samples of polished fibers will be tested versus the aging time of
  the polishing tool. The sample to be tested will be inserted in a
  black box and will be illuminated with a source of light like a
  pulsed LED or a laser. Then, in order to determine the status of the
  polished surface and of the tool, the efficiency of light collection
  will be measured using the photodetectors coupled to fibers.

\item {\bf QC test on finished bars} to verify a global status of
  various couplings like the ones between the following pairs of
  components: scintillator-glued fibers, photodetector-fibers and
  photodetectors-PCB-cables.  In order to isolate the bars from
  external light, a cap will be placed on the test table and the QC
  will be performed on a group of bars. During the test the
  photodetector will be powered on and the single rate and dark
  current will be monitored in order to determine the status of
  channels. In addition a scan test will be done by passing a
  radioactive source orthogonally over a set of scintillator bars
  close to the PCB side.

\item {\bf QC test on finished modules} performed in groups of
  predefined number of them in a cosmic ray station. Taking into
  account the planned production rate, five modules per week would be
  a reasonable number. Then a QC test lasting a week would allow to
  obtain precise information about the status of the modules. A cosmic
  ray station is equipped with two extra layers for triggering cosmic
  events and allows for tests of of the status of each channel
  together with the verification of their efficiency and uniformity,
  based on reconstructed tracks.

\end{enumerate}


\tdrsec{Photodetectors and PCB}
     
The baseline design for the IFR detector foresees that the photodetector would be applied at the end of each bar. A suitable PCB will be designed to support the photodetector as well as the miniature connector to a thin coaxial cable.\\


\subsection{Photodetector PCB and optical coupling to fibers} 
The details of the assembly to be installed at the end of each scintillator bar are shown in Figure \ref{sipm-fib-coupl-1}.
The SiPM is viewed 
(from the solder pad side) suspended in front of the machined notch where it is lodged once the PCB assembly is fastened. The WLS fibers are guided by grooves machined at the scintillator surface. The WLS fibers are glued in the position in which their diamond-cut ends are as close as possible to the SiPM active surface. As the SiPM should guarantee an efficient collection of light from all three fibers, its size will be chosen to meet this requirement. Suitable SiPMs with an area of up to 9\mma have been tested during the detector R\&D in order to evaluate their efficiency and radiation tolerance properties.

\begin{figure}[tbp]
\centering
\includegraphics[width=0.45\textwidth]{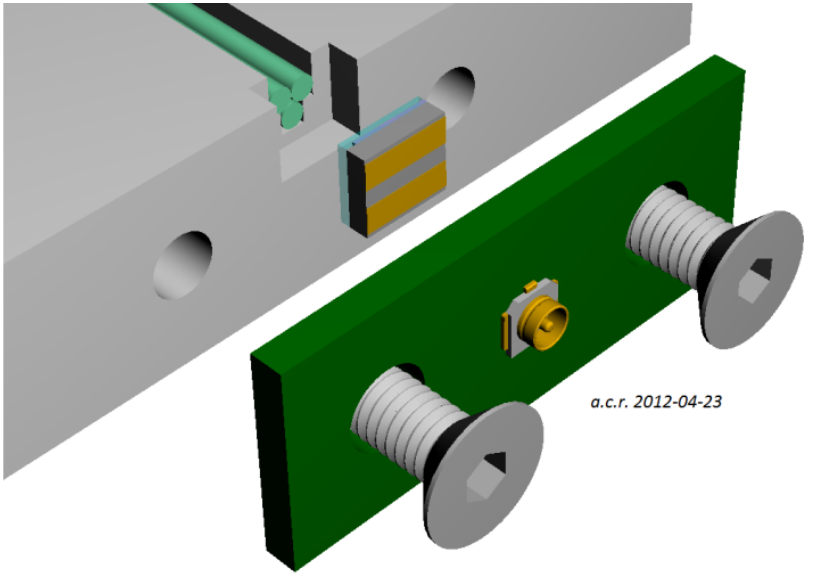}
\includegraphics[width=0.45\textwidth]{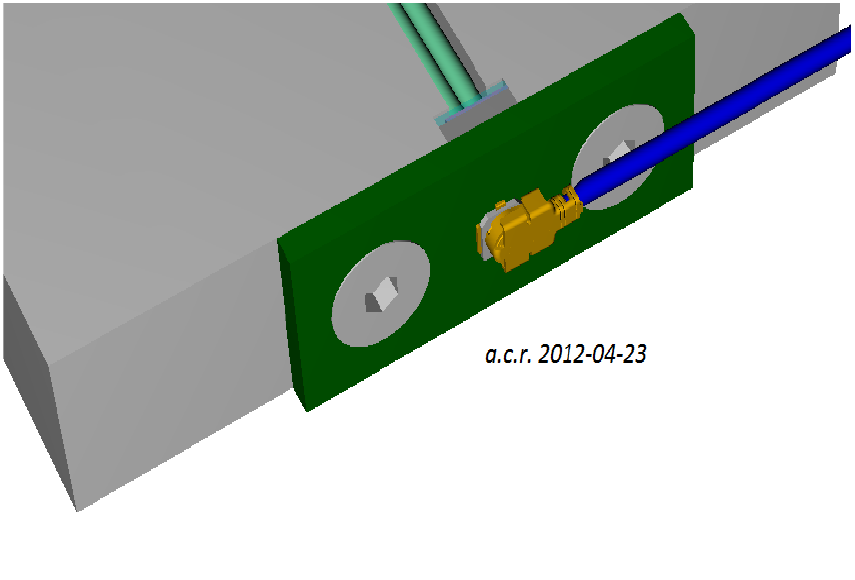}
\caption{Detailed view of the three WLS fibers and the PCB.}
\label{sipm-fib-coupl-1}
\end{figure} 

Figure \ref{sipm-fib-coupl-1} shows also the assembly in operation: a micro coaxial jack such as the Amphenol A-1 JB, soldered on the PCB, allows for connection of a small-diameter coaxial cable. All coaxial cables are routed out of the detector assembly  enclosure and reach the front end cards to which they are connected via a mass-terminated high density connector as shown in Figure \ref{conn-pcb}.

\begin{figure}[t]
\centering
\includegraphics[width=0.2\textwidth]{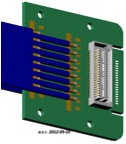}
\caption{Detailed view of the multi coaxial connector assembly}
\label{conn-pcb}
\end{figure} 

As shown in Figure \ref{sipm-fib-coupl-1},  the coupling of the WLS fibers to the SiPM is defined by means of machined grooves for the fibers, of a machined notch for the SiPM and of threaded holes for the PCB mounting screws (and eventually with a hole for precision positioning pins). Based on the experience acquired during the R\&D and prototype construction it was estimated that the relative positions of the fiber ends and the surface of the photodetector can be controlled  with a precision in the order of 100\mum. The optical grease or silicon pads can be used to improve the optical matching.\\

\subsection{Photodetector location}
As described before, the scintillator bars, fitted with the photodetectors, will be enclosed in light-tight sheet-aluminum boxes, which will also provide mechanical rigidity to the assemblies, or modules. 


It is to be said that once the modules are installed inside the IFR it will be impossible to perform any maintenance on the majority of them without a major overhaul of the detector; this is especially true for the barrel part of the detector.  In Figure \ref{ifr:fig:sipmloc} we have shown a view of the barrel section of the IFR with a few modules of the innermost layer (the modules' envelopes are not drawn, to show the locations of the SiPMs for the barrel). A similar arrangement is foreseen for the active layers of the endcaps. 
The SiPMs are distributed throughout the entire surface of an active layer, they are expected to operate reliably. 
This assessment of the radiation tolerance of the SiPMs has already been carried out for the devices available at the time the IFR prototype was assembled. More R\&D and irradiation tests have to be planned before the final SiPM choice is made, to characterize and possibly select devices built with the latest and most promising technologies. For the modules that will be subjected to the highest intensities of background radiation  a special layout is also being investigated in which the SiPMs are relocated at more suitable positions by extending the length of WLS fibers.\\

\subsection{Photodetector choice}
Different solid-state single photon detectors have been and are being evaluated for the IFR application. The technology of these devices is rapidly evolving and it would be reasonable to commit to a specific device as late as possible. 

\begin{figure}[tbp]
\centering
\includegraphics[width=0.3\textwidth]{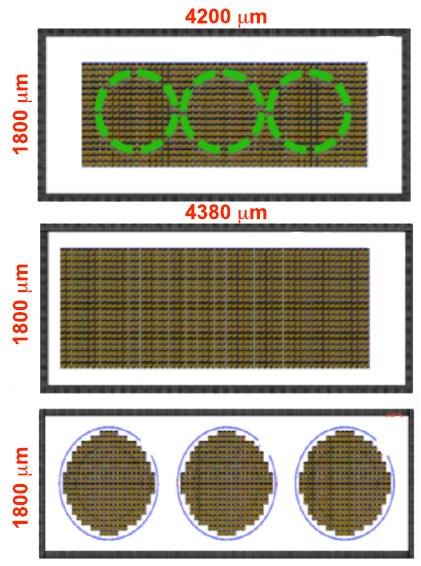}
\caption{The custom SiPM manufactured by FBK company for the IFR prototype.}
\label{sipm-FBK}
\end{figure} 

Devices tested so far by the \superb collaboration include SiPM manufactured by the Fondazione Bruno Kessler (FBK), Hamamatsu MPPC and SiPM by SensL  \cite{ifr:ifrsipm4, ifr:ifrsipm5, ifr:ifrsipm6, ifr:ifrsipm7, ifr:ifrsipm8}.
The key parameters describing the performance of a single photon detector are:

\begin{itemize}
  \item the photon detector efficiency at the WLS fiber characteristic wavelength;
  \item the fill factor;
  \item the optical cross-talk  ;
  \item the dark count rate;
  \item the sensitivity of breakdown voltage to temperature variations;
  \item the gain and its sensitivity to temperature variations.
\end{itemize} 

The SiPM suited for the IFR application should guarantee a detection efficiency better than 95\pct when installed in a detector assembly and should yield a dark count rate at 0.5 p.e. threshold in the range of a few 100\kHzpmma at room temperature and nominal bias voltage. SiPM devices satisfying these requirements and showing, after irradiation, the least perturbation of their original performance will be finally selected for the instrumentation of the IFR detector. 

\begin{figure}[t]
\centering
\includegraphics[width=0.45\textwidth]{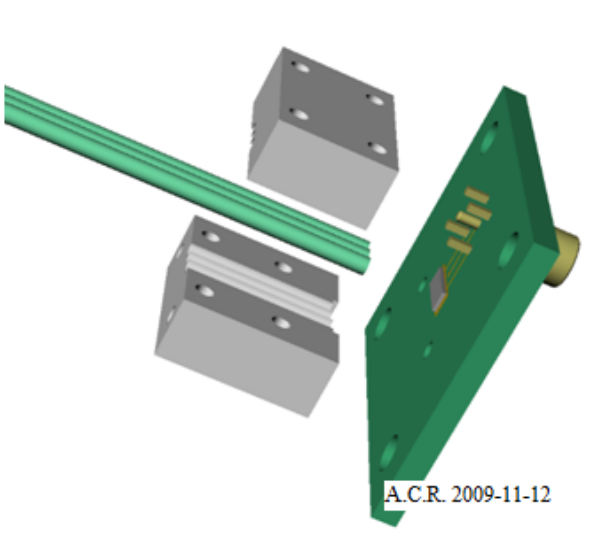}
\caption{A detail of the SiPM carrier PCB with the MMCX connector.}
\label{sipm-FBK-coupl}
\end{figure} 

Another parameter which will be considered for the SiPM device selection would be the availability of detector geometries which  minimize the unused surface area to optimize the ratio of dark count rate over signal rate. For the instrumentation of the IFR prototype a special device run was commissioned to the FBK and devices tailored to the IFR prototype needs were manufactured (see Figure \ref{sipm-FBK} and \ref{sipm-FBK-coupl}). The IFR prototype had the majority of the channels equipped with the FBK SiPMs, but also had channels equipped with MPPCs and both types of silicon photomultipliers performed very well. \\

\subsection{Aging and background issues}
The SiPMs manufactured by FBK and the MPPC by Hamamatsu have been irradiated with neutrons at the Laboratori Nazionali di Legnaro \cite{ifr:ifrsipm1}  and at the Frascati Neutron Generator to evaluate the effect of radiation on the devices' key parameters \cite{ifr:ifrsipm2}. These tests have shown that for neutron fluences above a few $10^8$\,n$_{eq}/$cm$^2$ the dark current and the dark count rate begin to increase. It was determined that after exceeding a few $10^{10}$ n$_{eq}/$cm$^2$ the number of photons detected by an irradiated silicon photomultiplier instrumenting a cosmic muon detector may drop to about 80\pct of the initial value \cite{ifr:ifrsipm2}. As a reference, the fluence of 1\,$\rm MeV$-equivalent neutrons expected~\cite{ifr:ifrsipm3} at the hottest  location of the IFR barrel is about $3 \times 10^9$\,n$_{eq}/$cm$^2$ per year.
The outcome of the irradiation tests results would seem to advise against installing the SiPM inside the IFR steel, where a possible replacement would require a major overhaul of the detector and a long downtime. On the other hand the positioning of the silicon photomultipliers in accessible locations would require to carry scintillation light to them via long clear fibers spliced to the WLS fibers. The latter would yield increasing costs, increasing dead space and loss of the photons available at the SiPM, with similar effects on the overall efficiency.
Other topic to consider is that the signal processing foreseen for the binary mode readout of the detector is more tolerant to device performance degradation  than the alternative ''timing mode'' readout \cite{ifr:ifrsipm8} and that new silicon photomultiplier devices are being introduced, which feature an improved radiation tolerance. The dose received by the sensors could also be reduced by inserting multiple and shielding layers within the IFR, which would thermalize the neutrons from the beam halo, capture the thermal neutrons and finally absorb the energetic photons resulting from these processes. These considerations lead the IFR collaboration to maintain the baseline design described above, while scheduling  more irradiation tests to evaluate the radiation tolerance of the newest SiPMs and the effectiveness of the shielding techniques. \\


\subsection{Temperature requirements}

\begin{figure}[h!]
\centering
\includegraphics[width=0.5\textwidth,height=0.35\textwidth]{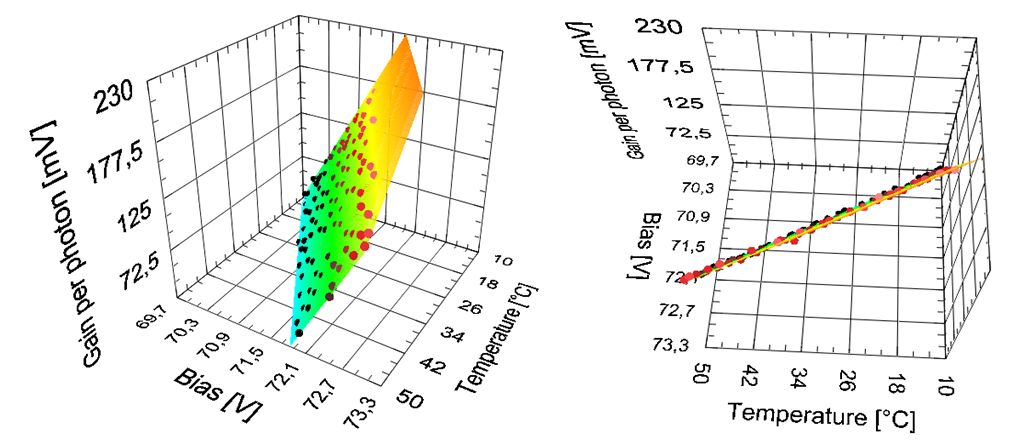}
\caption{The measured points and the fitting plane representing the SiPM gain vs. bias voltage and temperature relationship}
\label{temp-comp-krak-a}
\end{figure} 

Both the operating point and the parameters of SiPMs are influenced by temperature. In the case of the IFR detector it is not foreseeable to refrigerate the devices by means of thermoelectric coolers. As result we assume that the devices will operate at the temperature of the flux return steel, which was controlled, in \babar, by means of a chiller system. Rather than controlling the temperature of the silicon photomultiplier devices, the IFR environmental control system will periodically correct the bias point of each device to compensate for local temperature variations. This technique was used in the latest beam tests of the IFR prototypes \cite{ifr:ifrsipm9}.

The results of extensive studies on the compensation of temperature induced SiPM gain variations \cite{ifr:ifrsipm10}. 
Figure \ref{temp-comp-krak-a} illustrates the distribution of measurements points in the (gain, bias, temperature) space, while Figure \ref{temp-comp-krak-b} shows the effectiveness of the compensation techniques against temperature variation for two types of silicon photomultipliers.\\

\begin{figure}[tbp]
\centering
\includegraphics[width=0.45\textwidth]{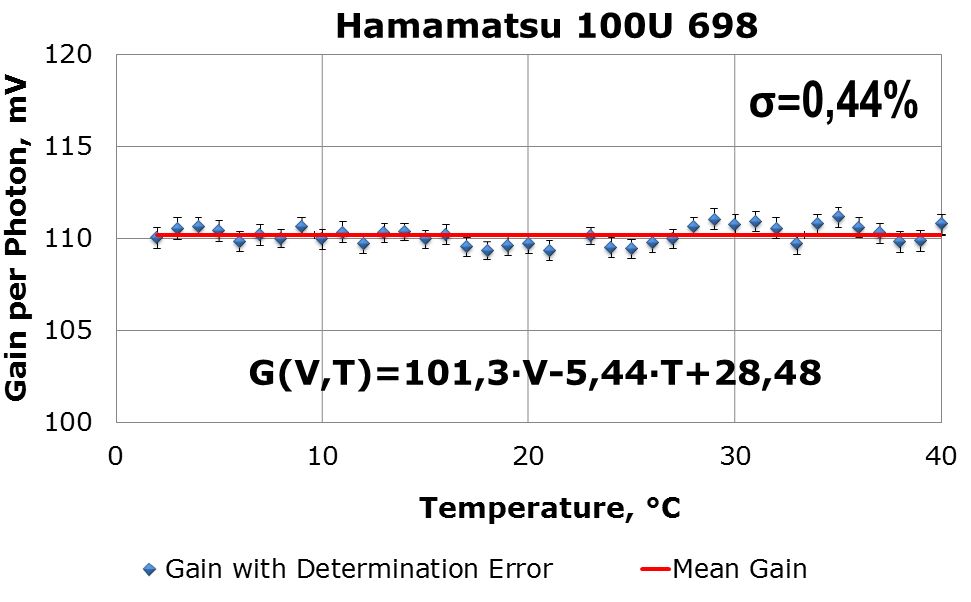}
\includegraphics[width=0.45\textwidth]{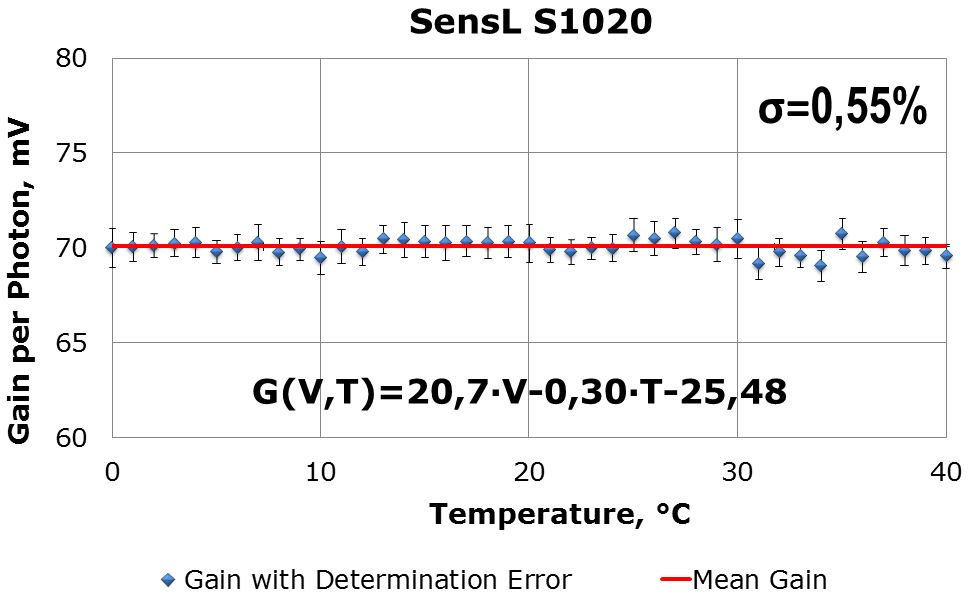}
\caption{Gain stabilization results obtained at the AGH Krakow for Hamamatsu (top) and SensL (bottom) devices.}
\label{temp-comp-krak-b}
\end{figure}


\tdrsec{Front-End Electronics}
	
This section describes the features and the design of the IFR readout
system.  At first, an estimation of the number of electronic channels,
of the event size and the data bandwidth at the nominal \superb\
trigger rate are presented. The design constraints determined by the
expected background radiation are then reviewed, along with the
results of irradiation tests on existing ASICs and off-the-shelf
devices suited for the implementation of some functional blocks of the
readout chain. This chapter describes, finally, the baseline IFR
readout system, going from the basic requirements of a dedicated IFR
front end ASIC to the proposed installation locations of the IFR
electronics building blocks and to the services required in the
detector hall.

\tdrsubsec{IFR channel count estimation}

The active layers of the IFR detectors are equipped with modules in
which the detector bars (of different widths for the $\phi$ and the
$z$ views in the barrel) are assembled in two orthogonal layers.
Table~\ref{ifr:tab:chnum} summarizes the current number of electronic
channels based on the figures reported in Table~\ref{ifr:tab:barrel}
end Table~\ref{ifr:tab:endcap}; the result is a total of 11640
channels for the IFR barrel and 9936 for the endcaps, considering 9
equipped gaps.

\begin{table}[htb]
\begin{center}
\caption{Estimation of the electronics channels count for the IFR detector.}
\begin{tabular}{cc|cc} 
\hline
Barrel            &             & Endcaps      &   \\
\hline
$\phi$ ch per sxt  &  884  & $x$ ch per door  & 1494  \\
$z$ ch per sxt       & 1056  & $y$ ch per door  & 990  \\
Ch. per sxt   &  1940  &  Ch. per door   &  2484  \\
\hline
Tot. ch.      &  11640   &  Tot. ch.      &   9936 \\
\hline
\end{tabular}
\label{ifr:tab:chnum}
\end{center}
\end{table}

\tdrsubsec{Estimations of the IFR event size and data bandwidth}

\begin{figure} [tbp]
  \centering
\includegraphics[width=0.48\textwidth, height=0.4\textwidth]{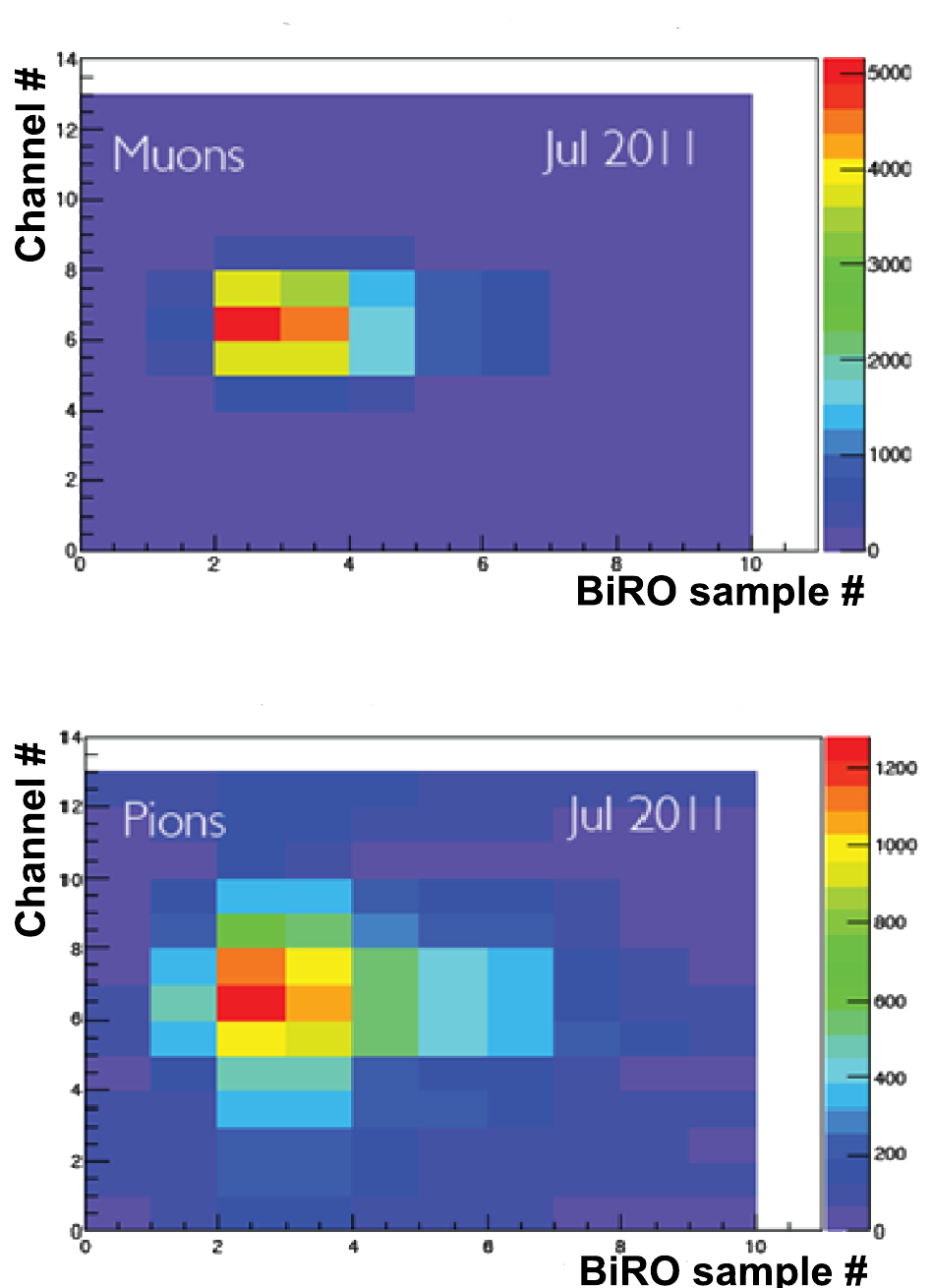}
\caption{Time evolution of muons and pions evaluated across the 10
  samples taken for each BiRO event (plots from
  \cite{ifr:ElexIFRref1} corresponding to a sampling period of 12.5\ns).}
\label{ElexIFRfig5}
\end{figure} 

The IFR detector will be read out in binary mode: the output of each
SiPM device will be amplified, shaped and compared to a threshold; the
binary status of each comparator output being the variable to be
recorded to reconstruct the particle tracks within the IFR detector.
The downstream stages of the IFR electronics are the digitizers which
sample the comparators outputs at \superb clock rate and store the
samples into local ``on-detector'' circular memories. These buffers
are designed to keep the data for a time interval equal at least to
the \superb trigger command latency. The last stage of the
``on-detector'' readout system is based on Finite State Machines (FSM)
which extract, from the local latency buffers, the data selected by
the trigger command coming from the Fast Control and Timing System
(FCTS). The IFR channels will be processed by functional blocks with a
modularity of 32 (``MOD32'').

Such a straightforward scheme has been successfully applied to the IFR
prototype, which has been tested with cosmic muons and with the beam
provided by the Fermilab Test Beam Facility~\cite{ifr:ElexIFRref1,
  ifr:ElexIFRref2}. The beam tests have shown that the time window for
the extraction of data matched to a trigger should be about 120\ns
wide (see Figure~\ref{ElexIFRfig5}), in order to recover the entire
signal from the shower initiated in the detector by an impinging
hadron.

Tables~\ref{ifr:tab:Tab_t1} and \ref{ifr:tab:Tab_t2} contain all
information relevant for the computation of the expected IFR event
size and data bandwidth at the nominal \superb trigger rate. In these
tables the number of samples has been set to 16 to account also for a
trigger jitter of the order of 100\ns. The event size and required
bandwidth could be reduced by applying the data reduction scheme
proposed by the ETD group. In this approach, if a new trigger would
select data that has already been sent in response to a previous one,
the repeating data is not transmitted and a pointer to the start of
the repeating data within the previous packet would be sent instead.

\begin{table}[htb]
\begin{center}
\caption{Estimation of the event size and data bandwidth for the Barrel.}
\begin{tabular}{lc} 
\hline
Max channel count per module    & 32 \\
\# of MOD32 units per module  & 1   \\
& \\
Tot. \# of barrel modules      & 384 \\
Tot. \# of MOD32 units          & 384 \\
& \\
Sampling period ($\frac{1}{FCTS_{clk}}$\ns)  & 17.86 \\
\# of samples in the trigger window    & 16 \\
Barrel event size (\kbyte) 				& 24.580 \\
& \\
Trigger rate (\kHz) 		& 150 \\
& \\
Total barrel bandwidth (\gbitps)  & 36.860 \\
\# of data links for the barrel & 24 \\
Bandwidth per link (\gbitps) 		& 1.536 \\
\hline
\end{tabular}
\label{ifr:tab:Tab_t1}
\end{center}
\end{table}

\begin{table}[htb]
\begin{center}
\caption{Estimation of the event size and data bandwidth for the Endcap.}
\begin{tabular}{lc} 
\hline
Max channel count per module    & 92 \\
\# of MOD32 units per module  & 3   \\
& \\
Tot. \# of endcap modules       & 108 \\
Tot. \# of MOD32 units          & 324 \\
& \\
Sampling period ($\frac{1}{FCTS_{clk}}$\ns)  & 17.86 \\
\# of samples in the trigger window    & 16 \\
Endcap event size (\kbyte) 				& 20.736 \\
& \\
Trigger rate (\kHz) 		& 150 \\
& \\
Total endcap bandwidth (\gbitps)  & 31.104 \\
\# of data links from the endcap & 16 \\
Bandwidth per link (\gbitps) 		& 1.944 \\
\hline
\end{tabular}
\label{ifr:tab:Tab_t2}
\end{center}
\end{table}

\tdrsubsec{Background radiation and electronics design constraints}

The current knowledge of the radiation environment at \superb\ comes
from detector simulations---constantly improving as far as the details
of the structure being simulated and thus the accuracy of the results,
are concerned.  The simulations~\cite{ifr:ElexIFRref3,
  ifr:ElexIFRref3_bis} allowed, in particular, for the analysis of the
known sources of radiation background in \superb\ and for the
evaluation of the doses in silicon at the locations of the SiPMs and
at the positions of the FEE.  The largest fraction of the total dose
absorbed by the IFR electronics is deposited by neutrons originating
back to the radiative Bhabha and the Touschek processes. The neutron
energy spectrum shown in Figure~\ref{fig:nspectrum} in
Sec.~\ref{neutronsection} is quite wide-spread and so different types
of interaction processes have to be considered to assess the effects
of the background neutrons on the performance of the sensors and of
the readout components. As a reference location for the neutron rate
estimation one could take innermost layer of the barrel where the
neutron rate is shown in Figure~\ref{fig:nrateL0}. The rates presented
in this plot are reported separately for the different background
sources, are normalized to 1\mev-equivalent neutrons in silicon and
include a {\it safety factor} that takes into account the fact that
the simulation may not reproduce perfectly the reality (as explained
in Sec.~\ref{ifr:sec:bkg}). A reference flux of 500\neqpcmaps would
result in a fluence of about $1.6\times10^{10}\neqpcma$ per solar
year.

The background simulations provided also information on the total dose
in silicon at locations around the IFR steel where ``on detector''
components of the electronic readout system could be located.
Figure~\ref{ElexIFRfig8} gives information on the dose absorbed at the
different locations in evidence. The highest dose is found at the C2
locations of the forward endcap and it is about 140 krad/year.

\begin{figure}[h!]
  \centering
        \begin{subfloat}
        \centering
                \includegraphics[width=0.5\textwidth,height=0.3\textwidth]{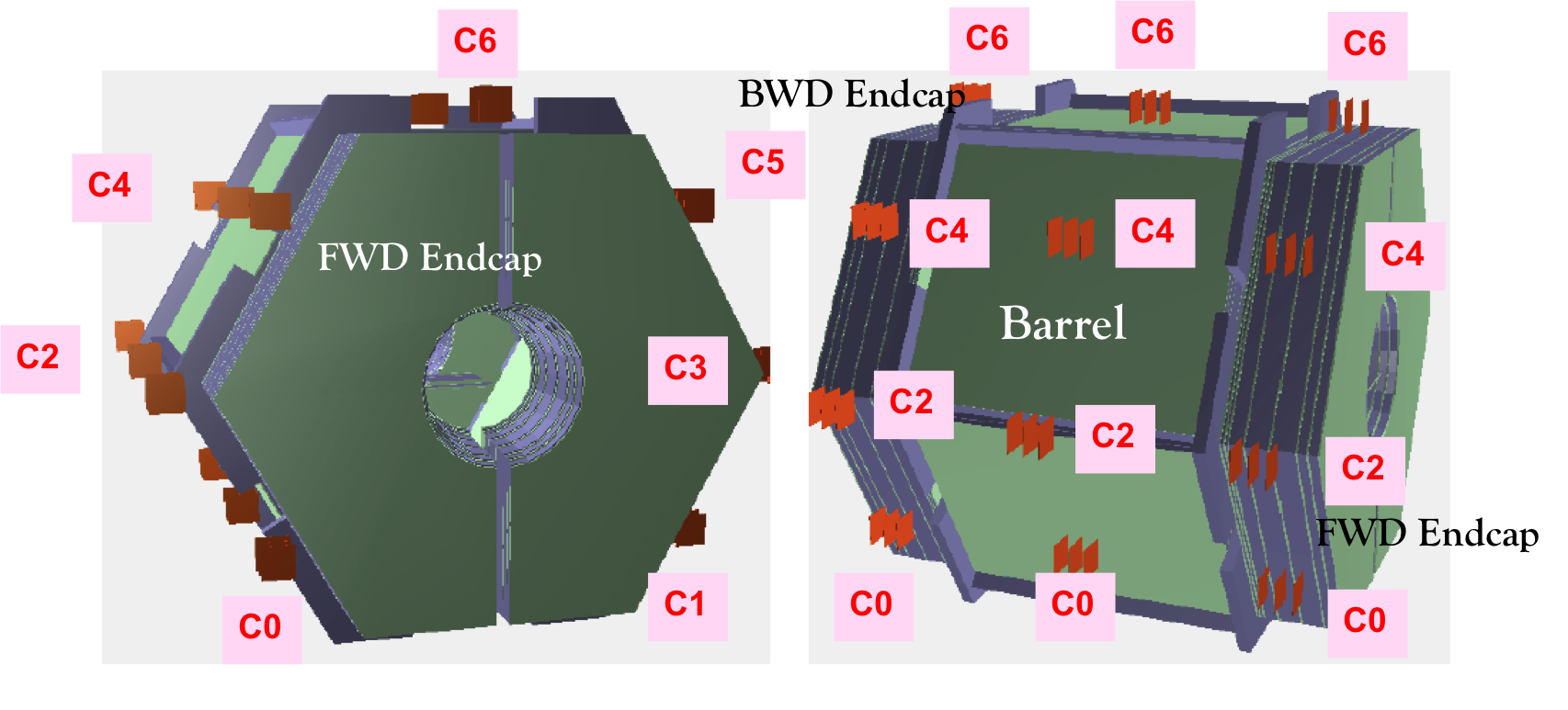}
        \end{subfloat}
        \begin{subfloat}
        \centering
                \includegraphics[clip=true,trim=0 0 30 0,width=0.5\textwidth]{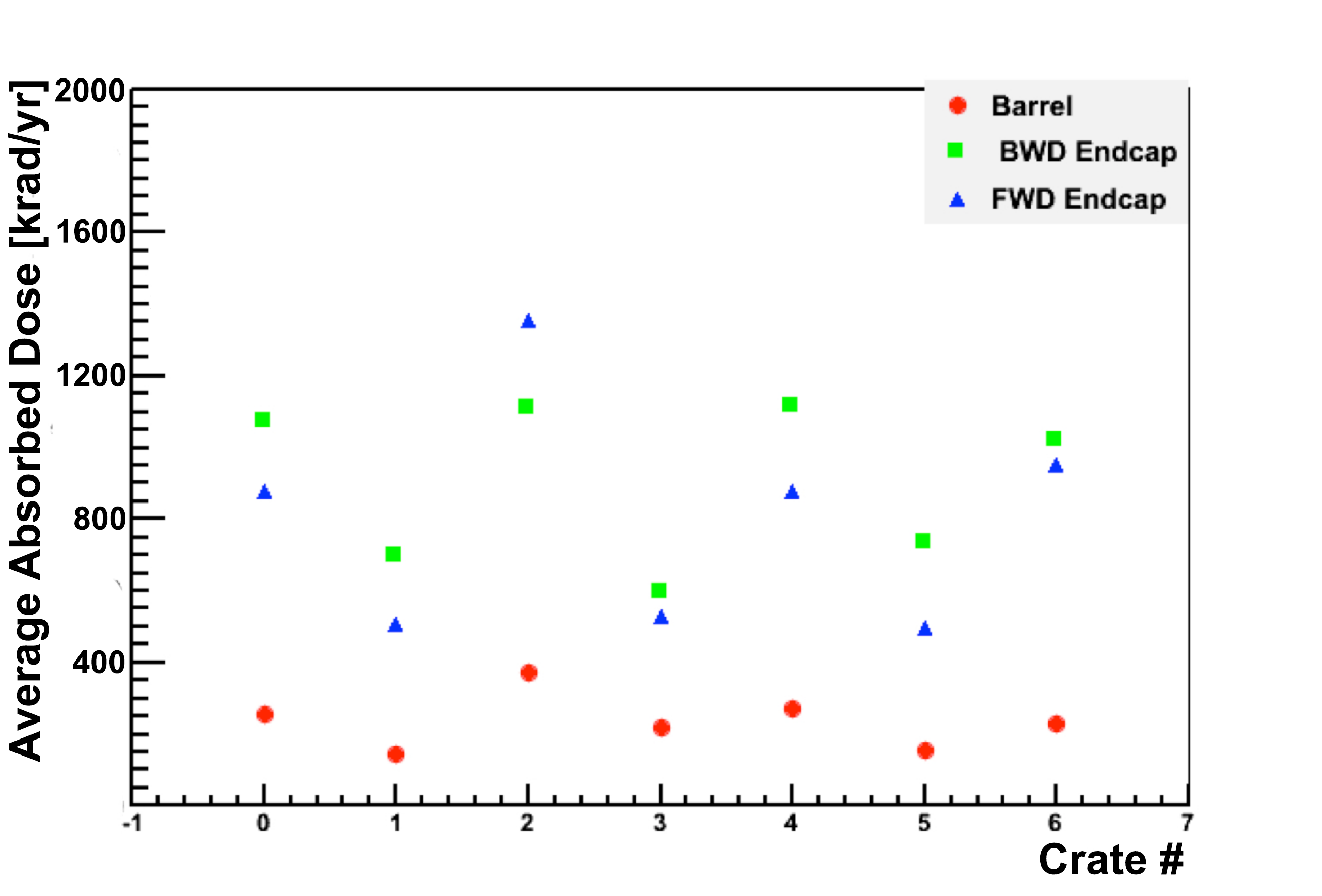}
        \end{subfloat}
  \caption{Location of the IFR FEE crates as simulated (top). Doses (krad/year) measured at some relevant locations around the detector (bottom).}
  \label{ElexIFRfig8}
\end{figure}

Electronic devices and systems, which must reliably and durably
operate under these radiation conditions, must be designed or selected
according to guidelines which have already been drawn, among others,
by the LHC community. As an example one could refer to the ``ATLAS
Radiation Tolerance Criteria'' for electronics~\cite{ifr:ElexIFRref6}
to see that the neutron fluence expected for the \superb IFR is of the
same order of magnitude simulated for the ATLAS MDT muon spectrometer.

We could then, if not exploit exactly the same technical solutions
adopted there, at least follow the design guidelines established by
the cited document and by similar papers. A more general approach to
the subject of radiation effects in silicon devices is described
in~\cite{ifr:ElexIFRref4, ifr:ElexIFRref5, ifr:ElexIFRref7,
  ifr:ElexIFRref23}. The R\&D activity for the preparation of the
\superb TDR has included the evaluation of the radiation effects on
SiPM~\cite{ifr:ElexIFRref8, ifr:ElexIFRref9, ifr:ElexIFRref10,
  ifr:ElexIFRref11, ifr:ElexIFRref12} and on electronics devices used
to implement the basic functions of the readout system. Irradiation
tests were performed at the CN facility of the INFN Laboratori
Nazionali di Legnaro (LNL) on samples of:

\begin{itemize}
\item low power, current feedback, operational amplifiers (op amp), to
  be used, if necessary, for ``on detector'' buffers, active summing
  junctions and so on;
\item ACTEL proASIC3E FPGA (backup solution for the implementation of
  digital functions of the IFR readout chain);
\item the ``EASIROC'' ASIC developed by the Omega group of the LAL,
  Orsay, France~\cite{ifr:ElexIFRref13, ifr:ElexIFRref14};
\item the ``RAPSODI'' ASIC developed by W. Kucewicz and his group at
  the AGH University, Krakow, Poland~\cite{ifr:ElexIFRref19};
\item the ``CLARO'' ASIC developed by G. Pessina and his group at INFN
  Milano-Bicocca, Milano, Italy~\cite{ifr:ElexIFRref26}.
\end{itemize} 

The CN facility of the LNL has a beam line delivering 4\mev
\textsuperscript{2}H$^+$ ions interacting on a beryllium target. The
energy spectra of the neutrons produced in the
\textsuperscript{9}Be(d,n)\textsuperscript{10}B reaction are shown in
Figure \ref{ElexIFRfig9}. Figure \ref{ElexIFRfig10} shows the stack of
boards carrying, in the first of the irradiation tests performed, the
EASIROC, the FPGA and the op amps installed right in front of the
beampipe, and a thermally isolated SiPM box located behind it, at
about 4\cm from the beam pipe. Some results of this irradiation tests
are presented in Tab.~\ref{ifr:tab:easi-lnl}. The OMEGA EASIROC ASIC,
manufactured in the 0.35\mum SiGe technology of Austria Micro Systems,
was exposed to an estimated total of $1.7 \times
10\textsuperscript{11}$ neutrons, equivalent to a few year's
exposition to the \superb IFR operating conditions.

\begin{figure*}[!]
\centering
\includegraphics[width=0.8\textwidth]{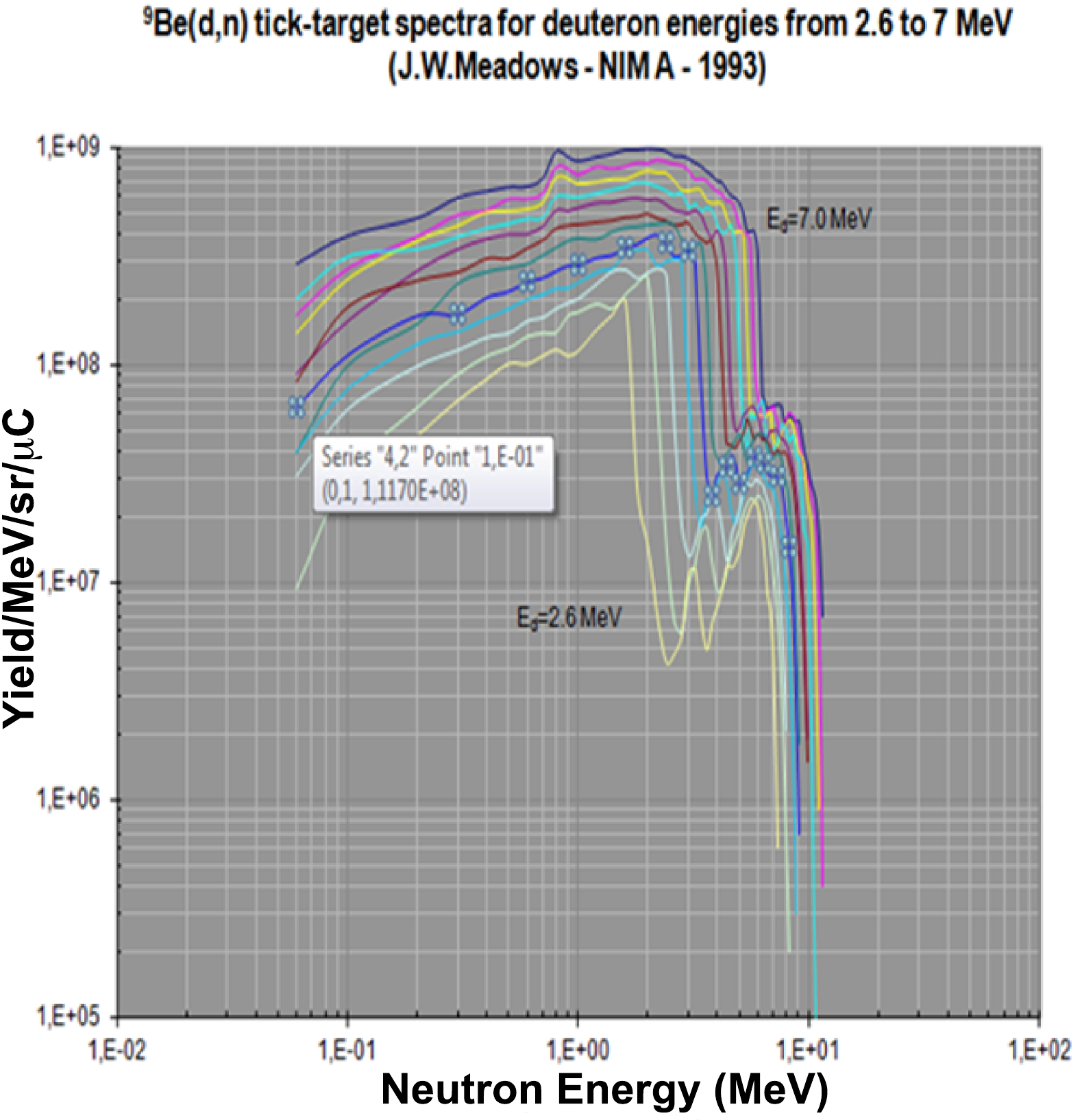}
\caption{Energy spectra of the neutron beam delivered by CN at LNL).} 
\label{ElexIFRfig9}
\end{figure*}

\begin{table*}[htb]
\begin{center}
\caption{Results of the EASIROC irradiation tests at CN-LNL.}
\begin{tabular}{lccc|c} 
\hline
total integration time (sec) & \multicolumn{4}{c}{60683}\\
total charge ($\mu$C)        & \multicolumn{4}{c}{4104}\\
\hline
 & \multicolumn{3}{c|}{{\bf ACTEL A3PE250-PQ208}} & {\bf EASIROC} \\
distance from source (mm)    & \multicolumn{3}{c|}{ 7}   & 20   \\
fluence at target (n/cm$^2$)& \multicolumn{3}{c|}{$6.12\times10^{12}$} & $1.03\times10^{12}$\\
fluence at DUT (n)& \multicolumn{3}{c|}{$4.95\times10^{12}$} &  $1.69\times10^{11}$  \\
\hline
& Configuration   &   SRAM   &  Flip  & Flip \\
& memory bits         & bits       &  flop  & flop \\
total memory elements monitored& N.A. & 32768 & 5824 & 456 \\
number of SEU detected & 0 & 752& 26& 0\\
number of SEL detected & 0 & 0 & 0 & 0\\
\hline
\end{tabular}
\label{ifr:tab:easi-lnl}
\end{center}
\end{table*}

The outcome of the test was that, as expected, while the FPGA
configuration memory contents was not corrupted, the on-chip SRAM
blocks and the registers in the FPGA fabric were subjected to upsets.
Therefore, any critical part of the FPGA design should include
suitable protection against Single Event Effects (SEE), which include
Single Event Upset (SEU) and Single Event Latchup (SEL). It is worth
to notice that no SEU afflicted the EASIROC configuration registers,
no latch-up occurred and finally that no evidence of Total Ionizing
Dose (TID) damage was found while comparing the current consumption
and the analog performance before and after the irradiation
test~\cite{ifr:ElexIFRref12}. The op amps to be tested had been
fastened to the top of the FPGA and have thus been irradiated with a
fluence of the order of $10^{12}\neqpcma$ with no noticeable effect on
their current consumption or DC offset.

\begin{figure}[!]
\includegraphics[width=0.47\textwidth]{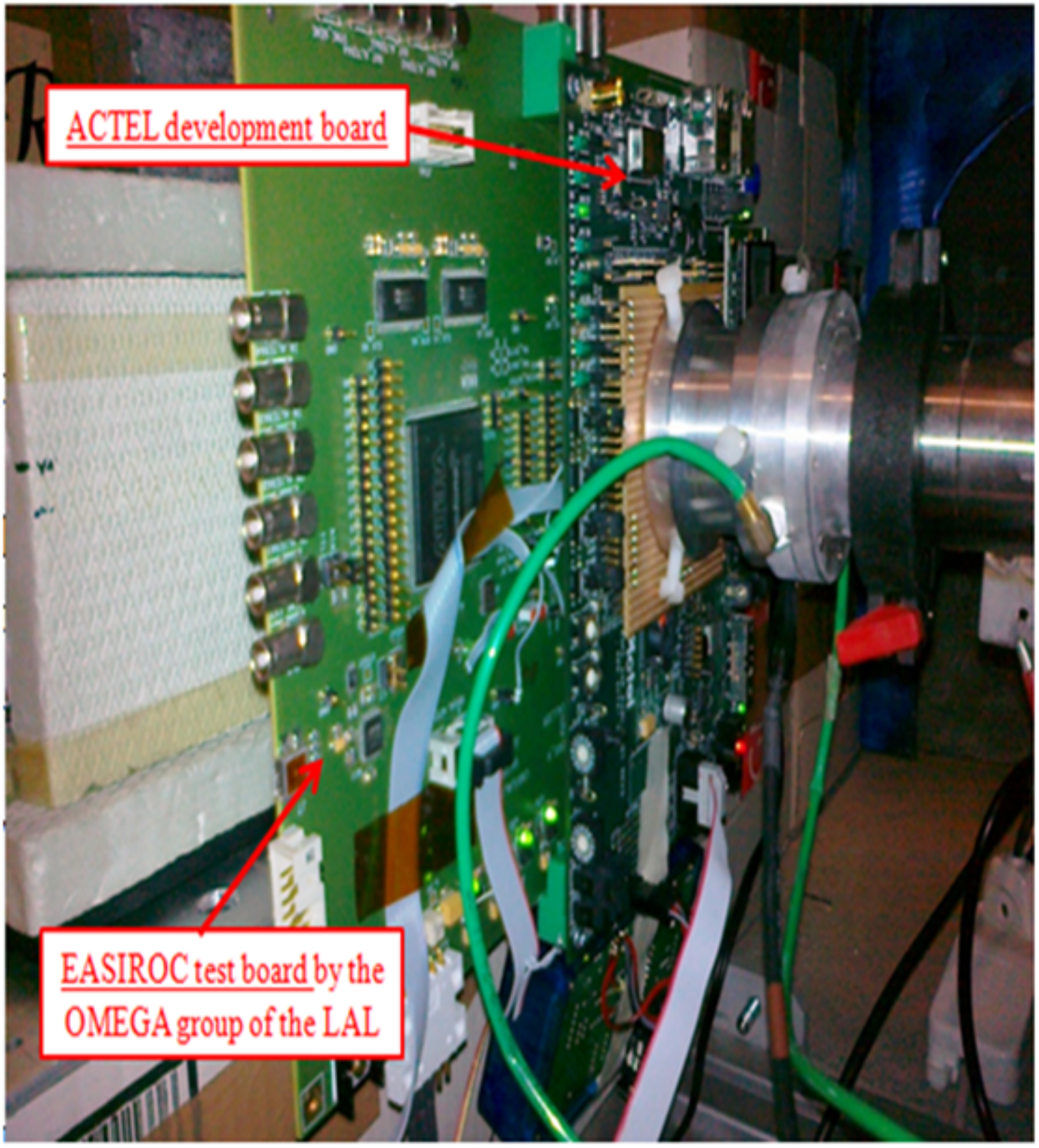}
\caption{Irradiation test setup at the CN-LNL facility.}
\label{ElexIFRfig10}
\end{figure}

On a later occasion, in June 2012, the RAPSODI ASIC-2 and the CLARO
ASICs were also irradiated at the CN facility, using similar setups as
the one illustrated above. Figure \ref{ElexIFR_CL_RAP_LNL} sketches
the relative positions of the neutron source and the devices under
test (DUTs). No SEL were detected both for the CLARO and the RAPSODI
ASIC-2, which were powered at nominal operating voltage during the
irradiation. Also a little effect was measured on the ASIC performance
such as DAC transfer functions, gain and comparator
offset~\cite{ifr:ElexIFRref27, ifr:ElexIFRref28}. As an example, the
comparison between transfer function measurements performed before and
after irradiation are shown in Figure \ref{CLARO_after_DAClin} for one
of the CLARO DACs. Also, the comparison between linearity measurements
performed before and after irradiation is presented in Figure
\ref{RAPSODI_LIN_AFTER} for one of the irradiated RAPSODI chips.

\begin{figure}[tbp]
  \centering
  \includegraphics[width=0.47\textwidth,  height=0.2\textwidth]{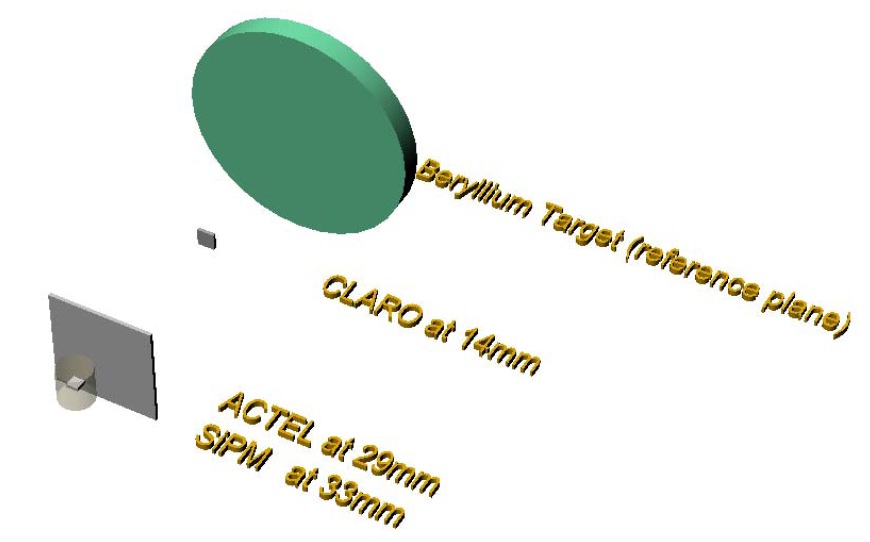}
  \includegraphics[width=0.47\textwidth,  height=0.2\textwidth]{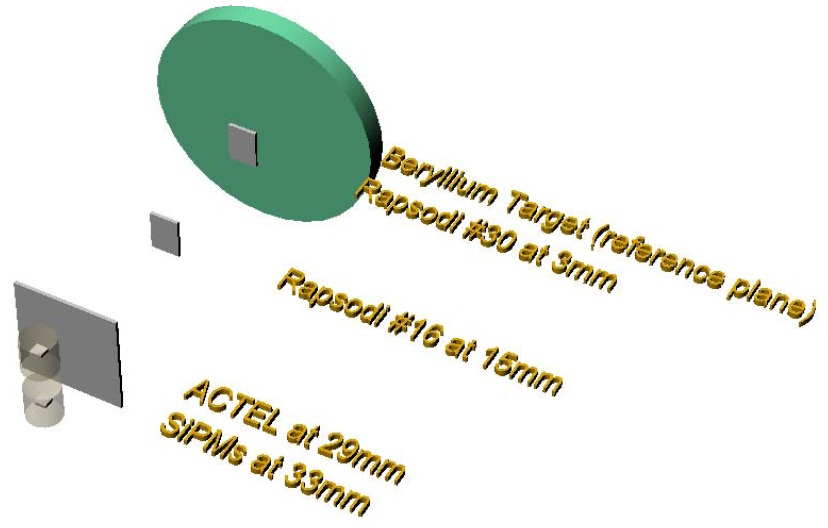}
\caption{Schematic representation of the CLARO and RAPSODI irradiation test setup at INFN-LNL.}
\label{ElexIFR_CL_RAP_LNL}
\end{figure}

\begin{figure}
\centering
\includegraphics[width=0.5\textwidth]{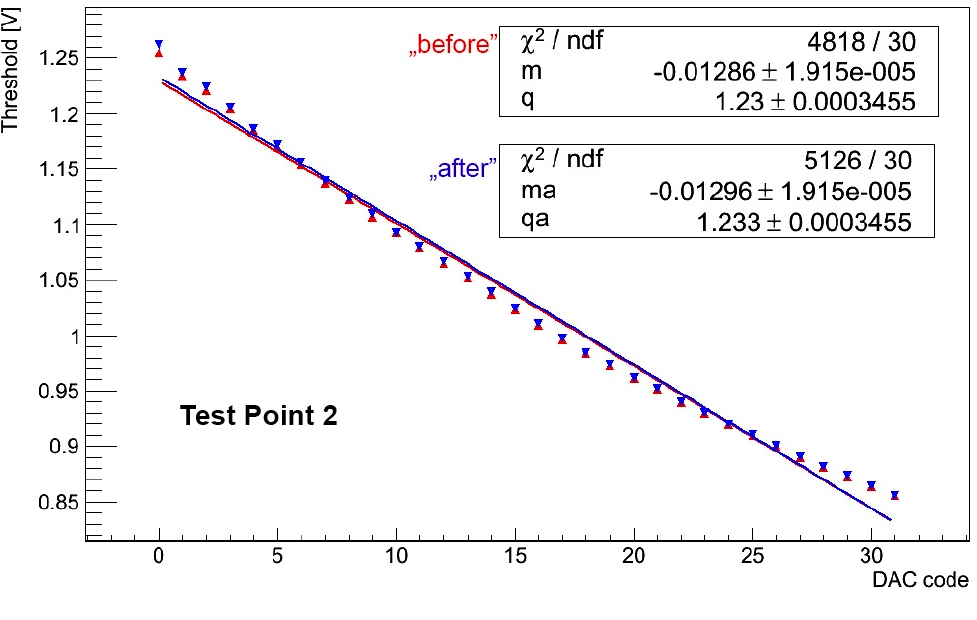}
\caption{CLARO DAC test performed before and after the irradiation.}
\label{CLARO_after_DAClin}
\end{figure}

The neutron fluences delivered to the devices under test, {\it i.e.} two samples of the CLARO ASIC and two samples of the RAPSODY ASIC-2, were:
\begin{itemize}
\item CLARO sample n. 1: $1.28 \times 10^{12}\neqpcma$;
\item CLARO sample n. 2: $1.32 \times 10^{12}\neqpcma$;
\item RAPSODY sample n. 1: $3.3 \times 10^{13}\neqpcma$;
\item RAPSODY sample n. 2: $1.32 \times 10^{12}\neqpcma$.
\end{itemize}

The results of the irradiation tests supported the decision on how to build and where to install the front end stages of the IFR readout system. As more neutron irradiation tests on SiPM are planned to assess the radiation tolerance of the latest generation devices, there will be more occasions to test, at the same time, more samples of ASICs built in the AMS 0.35\mum technology at different neutron energies and with different beam orientations with respect to the device under test. The 0.35\mum CMOS technology from AMS is being tested because it represents the preferred choice for the development of a dedicated ASIC, as described in the next chapter.

\begin{figure*}[!]
\centering
\includegraphics[width=0.8\linewidth]{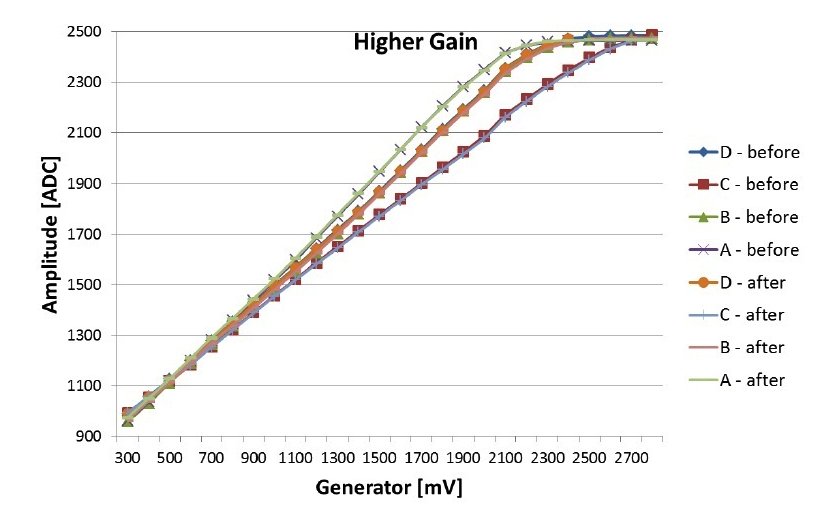}
\caption{RAPSODY linearity test performed before and after the irradiation.}
\label{RAPSODI_LIN_AFTER}
\end{figure*}

\tdrsubsec{The IFR readout system}

\tdrparagraph{IFR readout basics: outline of the IFR prototype readout.}
The basic principles of the IFR readout system design have been tested with the IFR prototype, which was read out in the binary mode described above. For the prototype readout, a dedicated front end board, the ABCD, was built from off-the-shelf components (COTs). Figure \ref{ElexIFRfig12} shows the ABCD block diagram.

The main functions which were implemented on the 32 channel ABCD board are:
\begin{itemize}
\item individual ``high side'' regulation (with 12 bit resolution) of the SiPM bias voltage; bias voltage and SiPM signal are carried by the inner conductor of the coaxial cable, whose outer conductor is at true ground potential;
\item wideband (-3 dB frequency of 1.5 GHz) amplification of the SiPM signal;
\item discrimination of each SiPM signal to an individually programmable (with 12 bit resolution) threshold voltages; the discriminators (two per channel) used AC coupled feedback to stretch the output pulse to a width of about 20\ns;
\item sampling of the discriminators
outputs and storing of samples onto local memories  to accommodate for the latency of the trigger signal;
an ALTERA Cyclone III FPGA was used to sample the discriminated SiPM signals and store them in an internal latency pipeline running at 80\MHz;
\item trigger processing: with each trigger a set of 10 samples was extracted from the latency pipeline and transferred to an output buffer (along with suitable framing words) from where they reached the ``BiRO Trigger Logic Unit (TLU) Interface board''~\cite{ifr:ElexIFRref16} acting as a crate-wide data-collector and interface to the DAQ PC and to the experiment TLU.
\end{itemize}

The DAQ system developed for the IFR prototype readout performed also functions which would, in \superb, be carried on by the Experiment Control System (ECS). Among the latter there are: room temperature acquisition, calculation and download to the ABCD DACs of new set points for the bias voltages (needed to stabilize the SiPM gain against variations of the operating temperature)~\cite{ifr:ElexIFRref1}. The Fermilab beam tests of the IFR prototype have demonstrated, among other things, that connecting the SiPMs to the front end cards by means of long (4 m) coaxial cables, although not optimal, was possible and resulted in a reliably operating system.
This result supported the current baseline choice of locating the SiPMs near the scintillator bars and far from the front end cards, so that the latter can be installed at more convenient (better accessible) locations. \\

\begin{figure} [t]
\centering
\includegraphics[width=0.47\textwidth, height=0.5\textwidth]{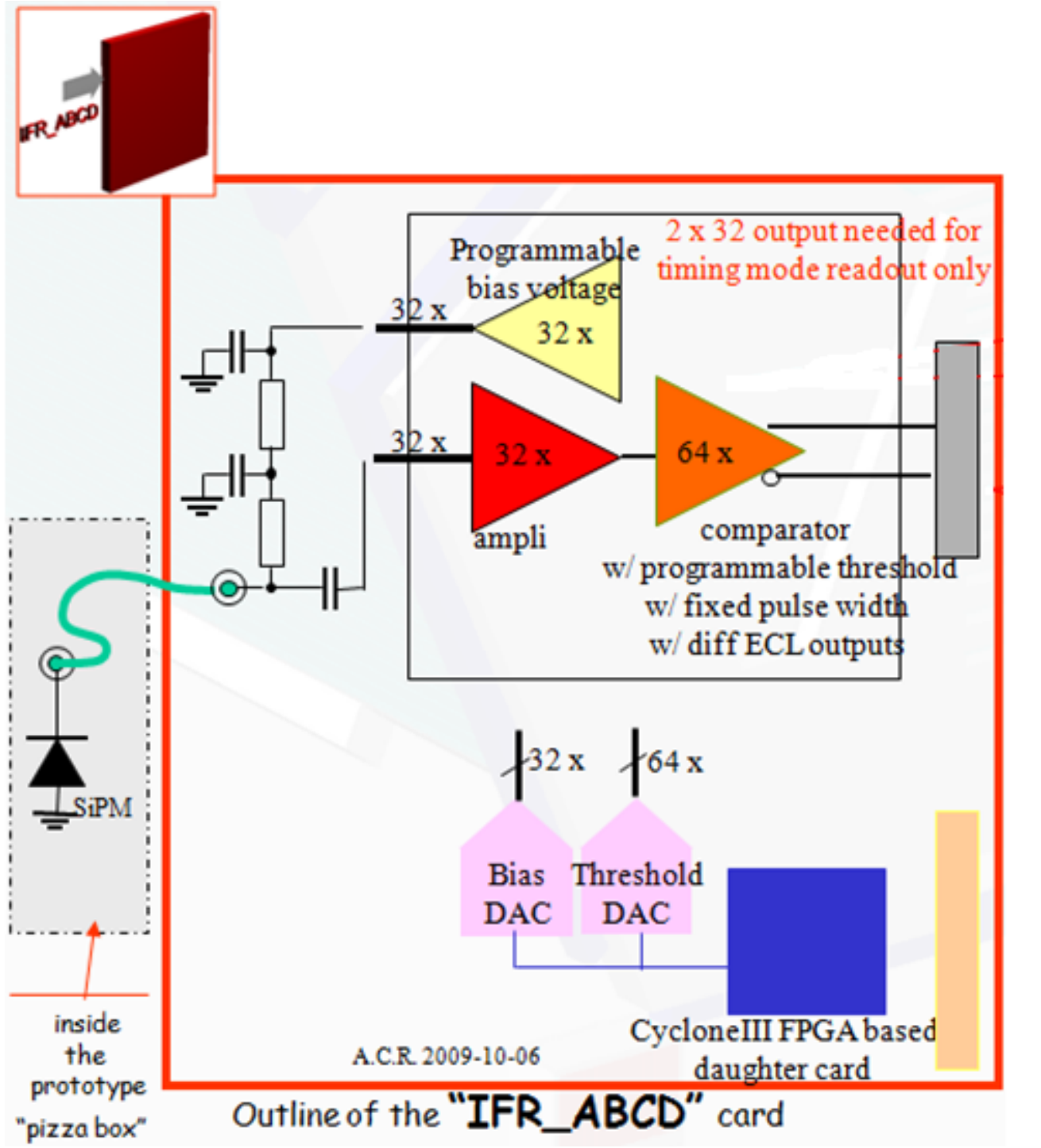}
\caption{Basic features of the ABCD card designed for the IFR
  prototype readout.}
\label{ElexIFRfig12}
\end{figure}

\tdrparagraph{ASIC based front end readout.} 
The COTS-based design of the ABCD card met goals such as short development time and flexibility (especially needed for tuning the parameters of the channels to be read out in the ``timing'' mode~\cite{ifr:ElexIFRref17}) but it wouldn't be convenient for large volume production. A search for existing ASICs suited for binary mode processing of SiPM signals was started and a good candidate was found in the ``\emph{Extended Analogue Silicon PM Integrated Read Out Chip}'' or EASIROC which had many features already suited for the IFR readout application~\cite{ifr:ElexIFRref13, ifr:ElexIFRref14}.
The EASIROC has been used, thanks to the test board and utilities provided by the LAL/OMEGA group, to test different SiPM technologies and coupling schemes~\cite{ifr:ElexIFRref18}. Figures \ref{ifr:ElexIFRfig14_1} and 
\ref{ifr:ElexIFRfig14_2} show, for instance, the pulse height histograms obtained from a 1\mma SensL SiPM connected to the EASIROC via:
\begin{itemize}
\item different lengths of coaxial cables, up to 12 m (Figure \ref{ifr:ElexIFRfig14_1});
\item one differential pair of a high density, double shielded, multi twisted pair cable 8 m long 
(Figure \ref{ifr:ElexIFRfig14_2}).
\end{itemize}

The features of the EASIROC which would suite the IFR readout application are:
\begin{itemize}
\item the integration of 8\bit DACs, one per channel, for ``low side'' adjustments of the SiPM bias voltage;
\item one fast (15\ns peaking time) shaper stage for each channel driving the trigger comparators;
\item the integration of a 10\bit DAC, common to all channels, for setting the threshold of the fast trigger comparators;
\item the availability of all 32 trigger outputs at the I/O pins (with single ended LV-TTL level) of the EASIROC;
\item low power consumption;
\item no sign of performance degradation or SEE detected during or after the irradiation tests described above.
\end{itemize}

These suitable features are partially offset by other characteristics which make the EASIROC not quite usable as it is. The most important is the 15 ns peaking time of the fast shaper: a lower value is necessary to cope with the increasing dark count rate for SiPMs operating in the \superb radiation environment. As the IFR collaboration has been growing in the last year to include groups experienced in VLSI design~\cite{ifr:ElexIFRref23}\cite{ifr:ElexIFRref19}\cite{ifr:ElexIFRref26}
\cite{ifr:ElexIFRref20, ifr:ElexIFRref21, ifr:ElexIFRref22, ifr:ElexIFRref25}, 
the development of an ASIC implementing the front end stages of the IFR readout system has become within reach. The EASIROC and an auxiliary flash based FPGA could still be considered as building blocks for a backup solution. \\

\begin{figure*}
  \centering
\includegraphics[width=\textwidth]{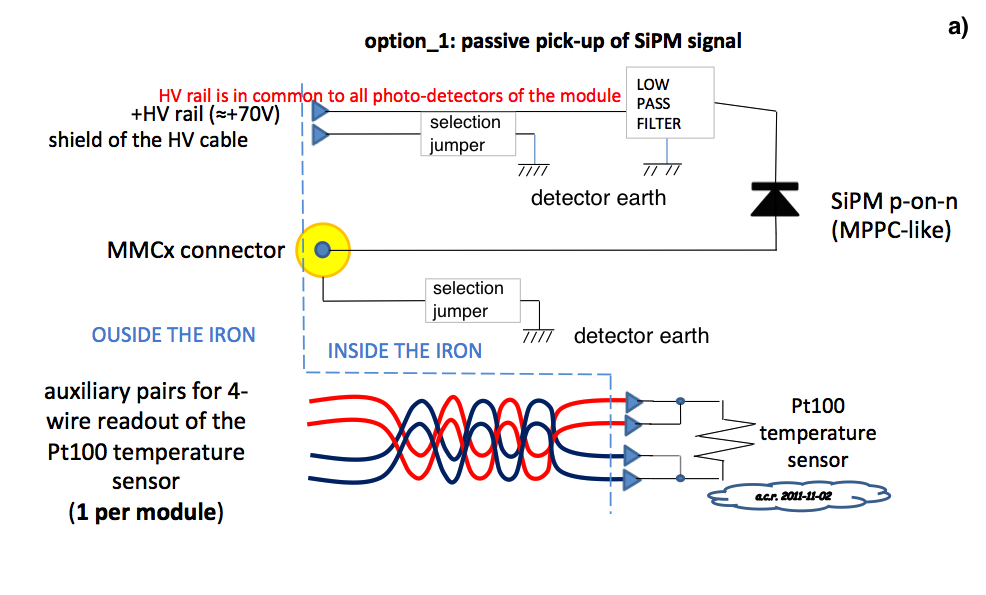}
  \includegraphics[width=0.50\textwidth]{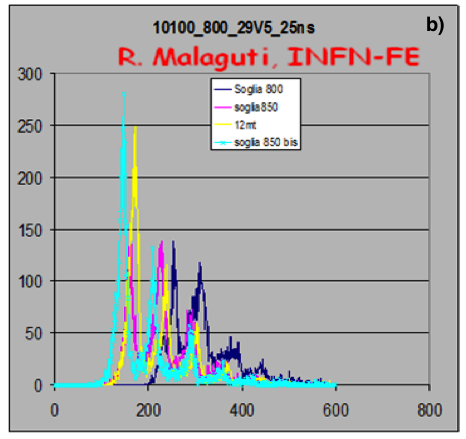}
\caption{a) Schematic view of the ``passive pick-up'' coupling. b) Histogram (counts vs. ADC bins) of the signal amplitude at the monitor output of the EASIROC. The SiPM was connected with coaxial cables up to 12\m.}
\label{ifr:ElexIFRfig14_1}
\end{figure*}
\begin{figure*}
  \centering
  \includegraphics[width=\textwidth]{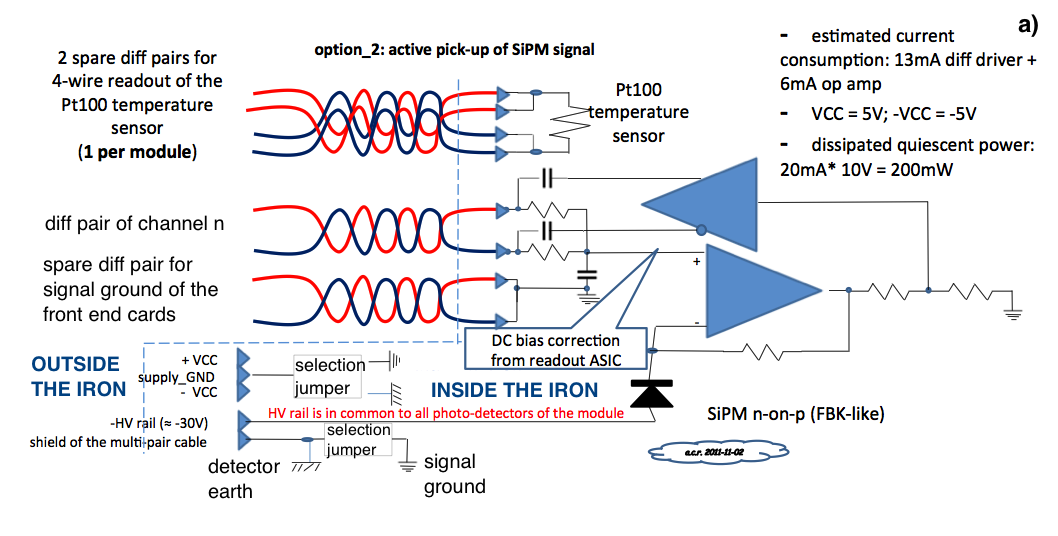}
  \includegraphics[width=\textwidth]{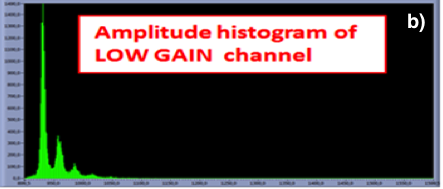}
\caption{a) Schematic view of the ``active pick-up'' coupling. b) Histogram (counts vs. ADC bins) of the signal amplitude at the monitor output of the EASIROC. The SiPM was connected with one differential pair of a high density cable 8 \m long.}
\label{ifr:ElexIFRfig14_2}
\end{figure*}

The new IFR front end ASIC should implement a set of basic features as follows:
\begin{itemize}
\item a preamplifier and shaper chain suited for positive and negative input signals and characterized by a dynamic range of about 100 times the single photon response signal;
\item a shaper peaking time not larger than 10 ns to minimize the pulse pile-up effects at high input rates;
\item individual DACs to set, with a few mV resolution, the DC level of each input and thus the ''low side'' bias voltage for the DC-coupled SiPM; the current compliance of the DAC should be specified considering that the dark current of a SiPM could increase two orders of magnitude during its operational life in \superb;
\item a fast comparator design, possibly differential; 
\item one threshold setting DAC with a resolution of at least one fourth of the single photon response signal and a linear range one fourth of the shaper linear dynamic range;
\item a configurable test pulse injection circuits;
\item a slow control interface logic: a simple serial protocol should be implemented in order to perform write and non destructive readback to/from all register controlling the ASIC's programmable features;
\item a clock interface unit: to avoid the need of on-chip PLLs the ASIC will receive an LVDS clock with a multiple of the \superb clock frequency; the timing unit will derive from it all on-board timing signals; a suitable fast reset signal input should also be foreseen;
\item a trigger primitives generator: a Lookup Table (LUT) based block combining signals from a programmable set of the on-chip comparators and driving one or two trigger outputs;
\item SEU protection through TMR or Hamming coding for all key registers and state machines in the ASIC.
\end{itemize}
The IFR front end ASIC should also possess a set of application specific features such as:
\begin{itemize}
\item a configurable latency buffer: a dual ported memory whose width would be equal to the number of channels and whose depth would be determined by the \superb trigger latency time interval. The constant (but configurable) offset between the write pointer and the read pointer would equal the trigger latency time expressed in terms of \superb clock periods;
\item a trigger interface: a trigger matching logic would detect a trigger pulse from the experiment, then wait for the relevant data to be extracted from the latency buffer and forward the trigger-matched data packet to the output serializer;
\item a set of low power serializers clocked by the input clock (which has a pace multiple of the sampling clock) needed to transfer the trigger matched data to the downstream data collector units; the serial output data could be 8b/10b encoded to allow the usage of AC coupled transmission links requiring DC-balancing.
\end{itemize}

\begin{figure*}[pt]
\centering
\includegraphics[width=\textwidth]{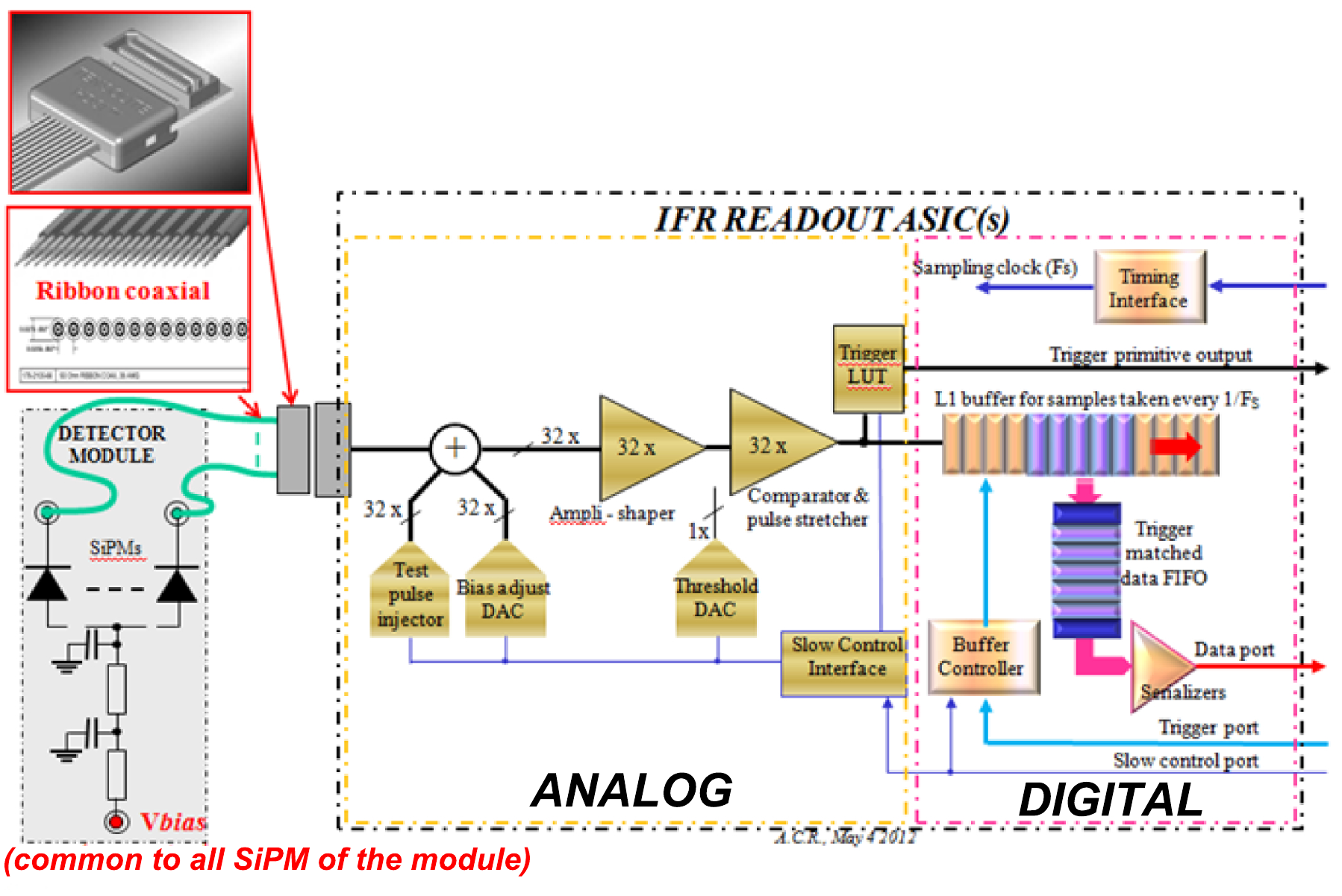}		
\caption{Block diagram of the IFR readout ASIC.}
\label{fig:ElexIFRfig15}
\end{figure*}

The features listed above are differentiated into a first set, mainly analog, and a mainly digital second set to point out that it might be convenient to implement the two lists of functions in two different ASICs, to increase the flexibility of the overall system. Figure \ref{fig:ElexIFRfig15} shows a block diagram of the new IFR readout ASIC and its connection to the SiPMs in the detector module. An IFR ASIC with a modularity of 32 channels would be able to readout all SiPMs in one module of the barrel. As shown in the inlay of Figure \ref{fig:ElexIFRfig15} a ribbon-coaxial cable, mass terminated to a high density connector, could be used to carry the signals from the sensors to the IFR readout ASICs and the ancillary components on the IFR front end cards. At the right side of the ASIC diagram, Figure  \ref{fig:ElexIFRfig15} shows the input and output ports for the connections between the IFR ASIC and the downstream data merger cards.\\

\tdrparagraph{Location of the front end stages of the IFR readout system.} 
The SiPMs would be installed, according to the baseline design, directly on the scintillator bars and thus distributed at different locations inside the modules. The ASIC based front end stages would be installed outside the modules, at the closest accessible locations. The reason behind this choice is that even if we were to install the front end ASICs inside the modules, we would not still be able to directly connect the SiPMs to the ASICs (in order to maximize the signal/noise ratio). This, in turn, would make it more difficult to remove the heat dissipated by the front end and it would make it impossible to access the front end cards, to replace faulty ones for instance, without disassembling large parts of the IFR system. \\

\begin{figure} [tbp]
\centering
\includegraphics[width=0.47\textwidth, height=0.5\textwidth]{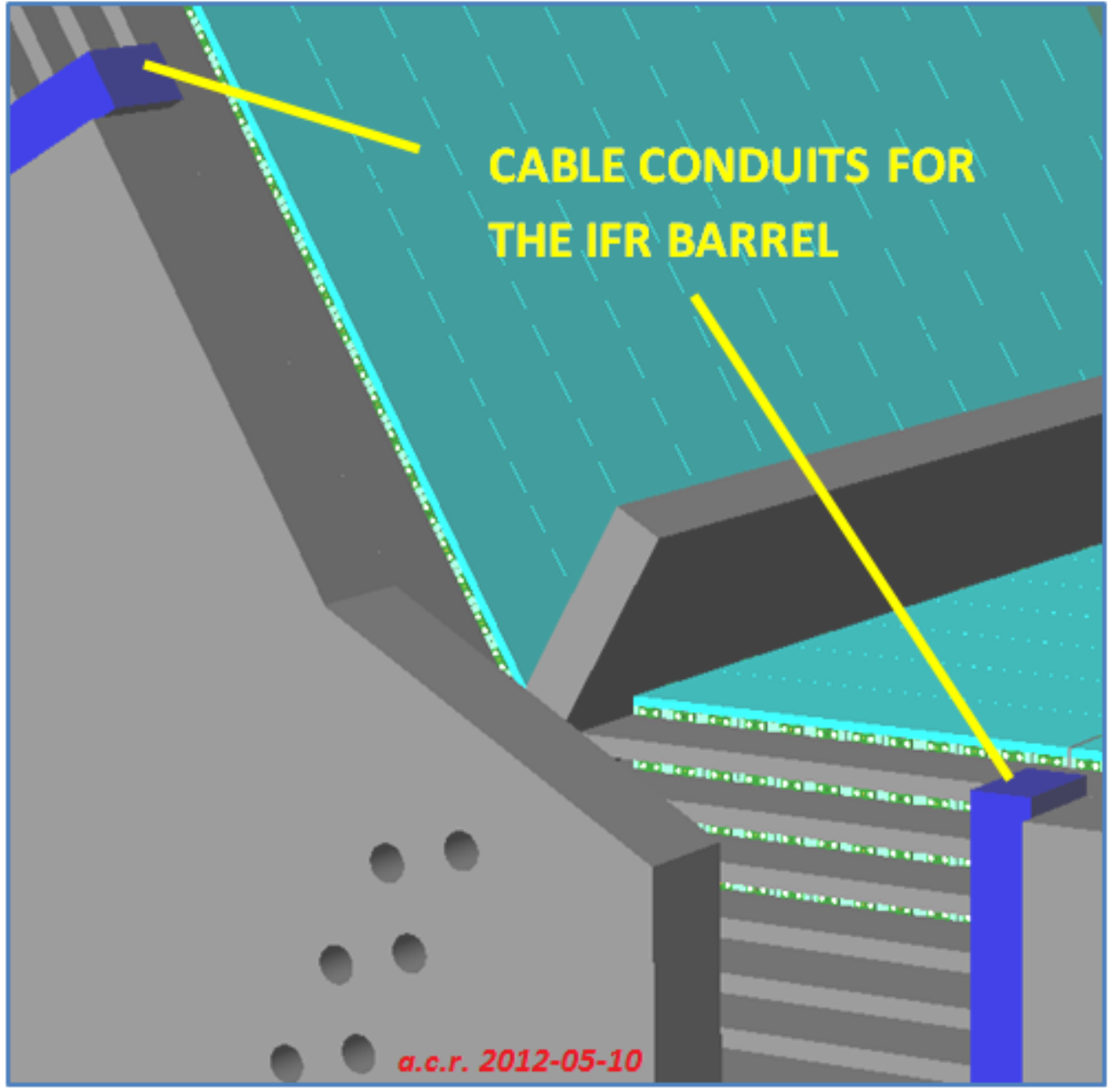}
\caption{The IFR cable conduits for two sextants.}
\label{fig:ElexIFRfig16}
\end{figure}

The position of two of the $5"\times3"$ conduits through which the IFR detector signal and power cables are routed are
shown in Figure \ref{fig:ElexIFRfig16}.
The printed circuit boards for the front end stages of the IFR signal processing chain could be located in these cable conduits. They would be accessible, if needed, without removing structural elements of the IFR structure. Figure \ref{fig:ElexIFRfig17} shows a section of one IFR cable conduit with a stack of 4 boards in evidence. Each IFR front end board could host 2 ASICs with 32 input channels each and since the IFR modules in the barrel are equipped with 32 SiPM at most, a stack of three to four IFR front end cards can handle all SiPMs installed in an active layer; 9 active layers are foreseen in the IFR detector. All digital signals to and from the two front end ASICs on a board are coming from the data merger cards downstream, physically located as close as possible to the end of the cable conduit emerging from the IFR, as illustrated in the next paragraph. The interconnection cables are shown on the right side of the Figure \ref{fig:ElexIFRfig17}: they are double shielded, 17 pairs, Amphenol SpectraStrip part number 425-3006-034 fitted with connectors by KEL (part no. KEL 8825E-034-175D). The mating connectors on the front end and the data merger PCBs are KEL 8831E-034-170LD. The power cables for the SiPM bias (one common bias voltage per detector module) and for the front end card supply voltage are not shown in Figure \ref{fig:ElexIFRfig17}.

\begin{figure}[tbp]
  \centering
\includegraphics[width=0.47\textwidth, height=0.5\textwidth]{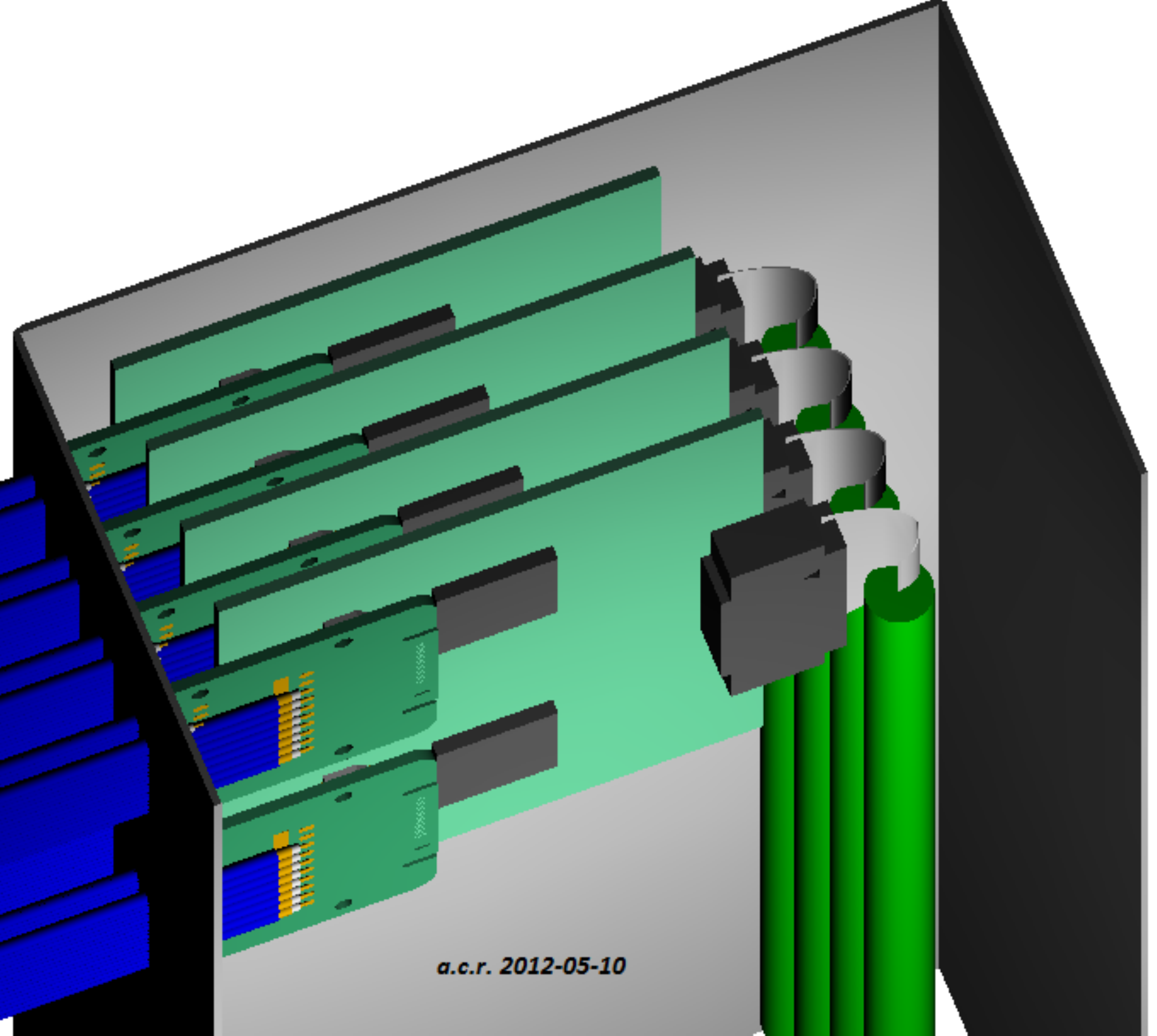}
\caption{Detailed view of the front end cards installed in the IFR cable conduit.}
\label{fig:ElexIFRfig17}
\end{figure}

The cable conduits have a big enough cross-section to accommodate all boards and cables needed for the 9 active layers of each barrel sextant. Not shown in Figure \ref{fig:ElexIFRfig17} are also the copper pillars which are foreseen to build a thermal conduction path from the IFR front end card to the IFR steel, to which the cable conduit is fastened. If individual coaxial cables were used to carry the SiPM signals instead of ribbonized coaxial cables, then the signal cables from the module could be soldered onto a suitable interface PCB, as it was shown in Figure \ref{conn-pcb}. The multi coaxial assembly, shown in this 3D drawing, features a SAMTEC QSE-20-01-F-D high density connector which mates to a QTE-020-01-F-D installed on the IFR front end PCB.

The above description concerned the instrumentation of the barrel section of the IFR detector. Figure \ref{fig:ElexIFRfig19} shows a perspective view of a pair of endcaps. Here the yellow callouts indicate the openings in the side lining steel through which the signal and power cables for the detector modules can be routed. The signal cables emerging from these openings are routed to the crates marked by the red callouts. For the endcap section of the IFR, the front end cards carrying the IFR read out ASIC could be installed directly in the crate and connected to the data merger boards through the crate's backplane interconnections. \\

\begin{figure}[b]
\centering
\includegraphics[width=0.47\textwidth, height=0.6\textwidth]{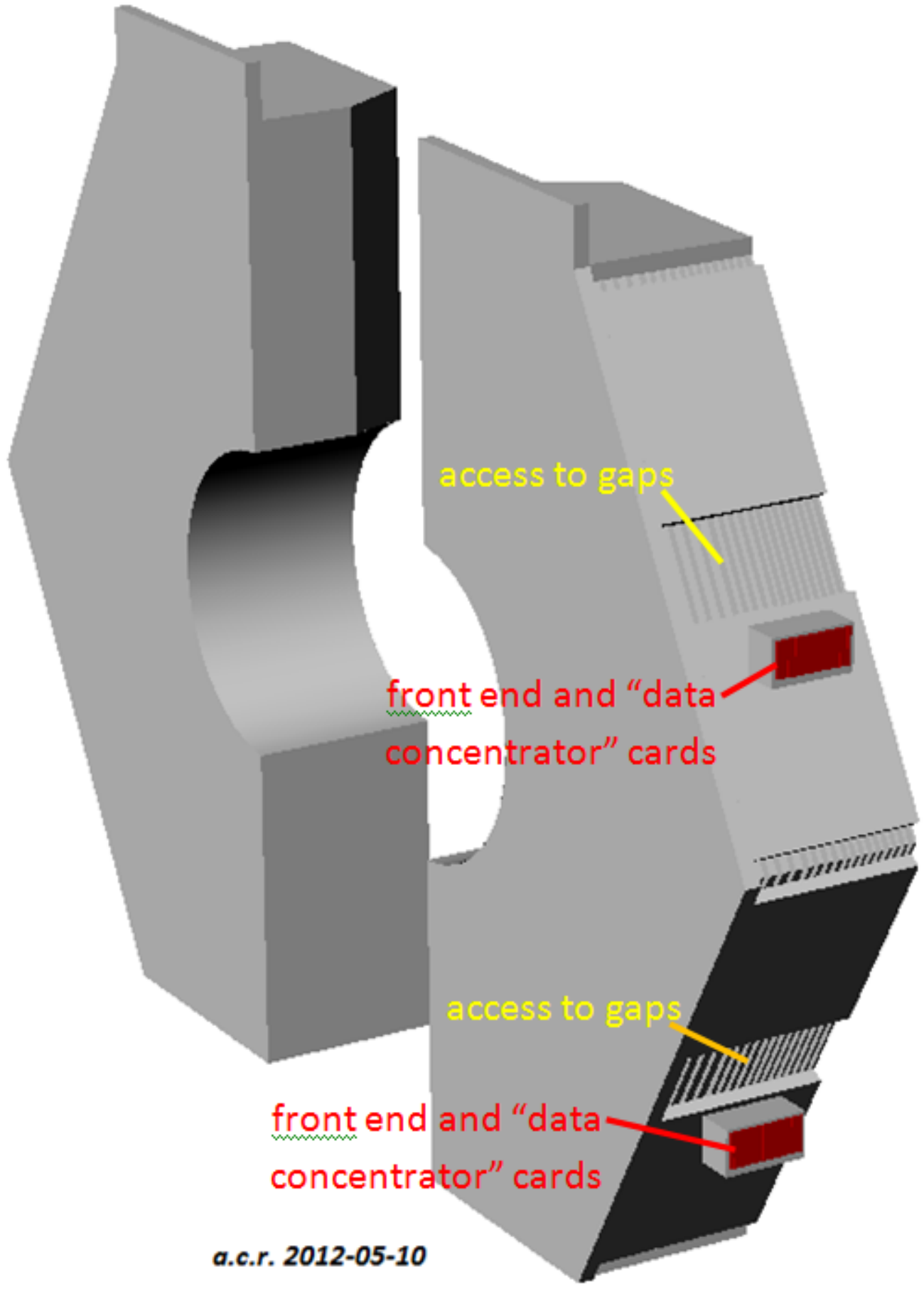}
\caption{Perspective view of IFR endcaps.}
\label{fig:ElexIFRfig19}
\end{figure}

\tdrparagraph{The IFR data merger crates.} The IFR data merger crates are interfaced to the \superb FCTS and ECS from which they receive and execute the commands controlling the data acquisition and the detector configuration, respectively. The data merger units also collect the trigger matched data from the front end cards and merge the input streams into the output packets sent to the \superb\ ROM (ReadOut Modules) via the high speed optical data links. Figure \ref{fig:ElexIFRfig20} shows the main functions performed by the units installed in a data-merger crate:

\begin{itemize}
\item FCTS interface: the key element of this functional unit is the FCTS receiver module which is linked via optical fiber to the \superb Fast Control and Timing System; the FCTS interface unit fans out to the IFR front end cards the timing (clock and reset) and the trigger commands decoded by the ``receiver of the FCTS protocol'' block;
\item muon trigger module: this unit receives the trigger primitives generated by the IFR front end cards and combines them to generate a  muon trigger for local debugging purposes;
\item data link: each data-merger crate is supposed, according to the baseline design, to receive and combine the serial streams of trigger matched data coming from the front end cards connected to it and to drive a suitable number of high speed optical data links (four for the barrel section, two for the endcap ones) to the ROMs. This functional unit will exploit the common TX link driver module to reliably send data to the ROM, even in the radiation environment of the IFR;
\item ECS interface: the key element of this functional unit is the ECS receiver module, which is linked via optical fiber to the \superb Experiment Control System. The ECS unit translates commands and status information back and forth between the ECS system and the serial configuration controllers implemented in the front end ASICs.
\end{itemize}

For the endcap section of the IFR the physical links between the front end and the data merger boards are composed of the differential lines provided by the backplane in which the front end and the data-merger units are installed. An ATCA or Micro-TCA crate~\cite{ifr:ElexIFRref23} provides a large number of such interconnection resources and the IFR data merger crates could conveniently conform to one of these TCA specifications.

\begin{figure}[htb]
\centering
\includegraphics[width=\linewidth]{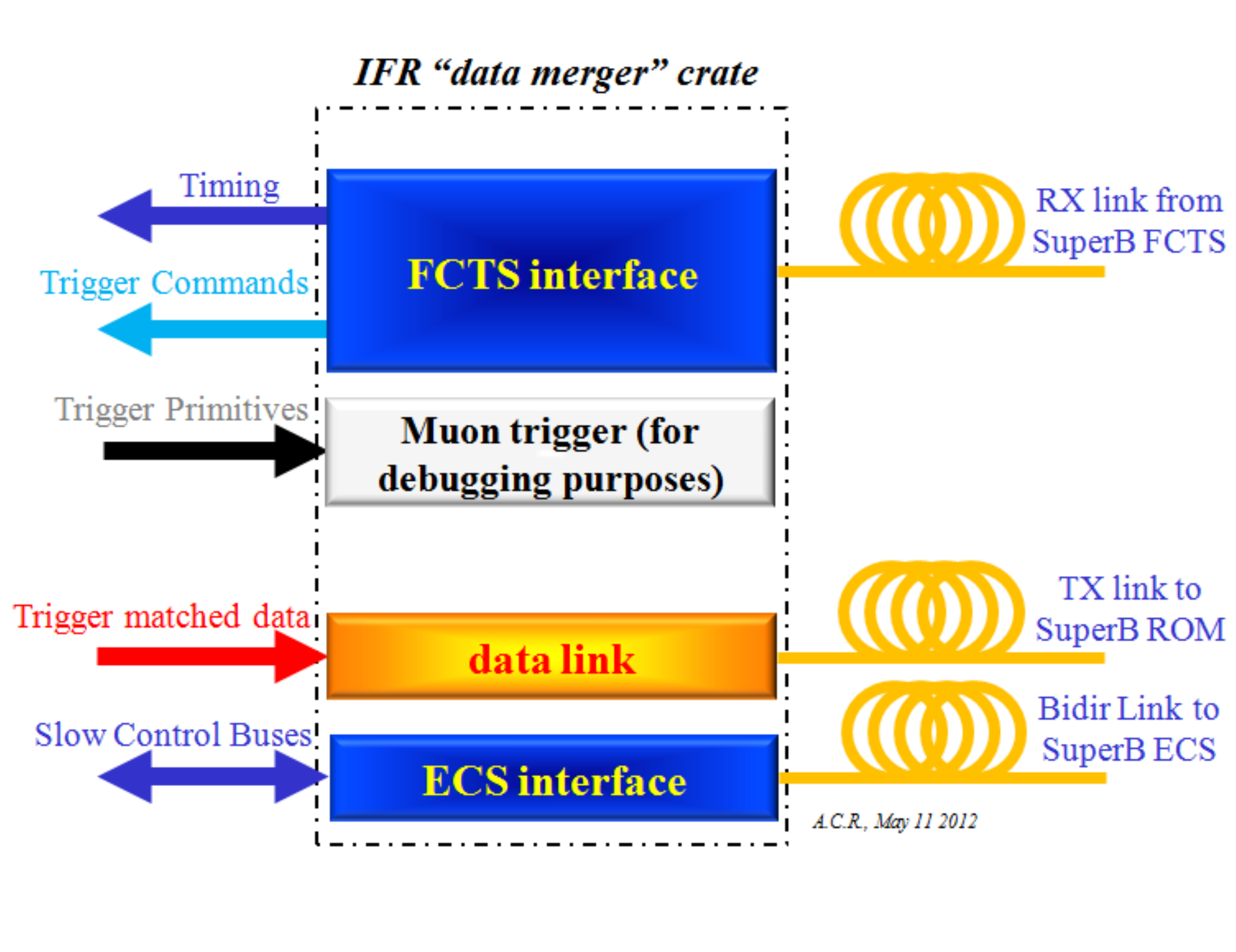}
\caption{Illustration of the main functions performed by the electronic units in the data-merger crate.}
\label{fig:ElexIFRfig20}
\end{figure}

Because of the \superb radiation environment it is likely that in the IFR data merger crates the standard xTCA IPM (``Intelligent Platform Management'') controller will be replaced by some ad-hoc interface card connected to the \superb ECS. The power supply units of the IFR data merger crate would have to be radiation tolerant or simply to be installed at a larger distance from the crates. The power entry module integrated into a xTCA crate are fit to both options.  
The xTCA
fans will have to be tested for operation in a radiation environment or replaced with qualified units. On the other hand, the convection cooling specification could be somewhat relaxed, since the data merger crates will most likely dissipate much less than the allowed 150--200\Watt per slot. Figure \ref{fig:ElexIFRfig21} shows the proposed location for one of the barrel data merger crates.

\begin{figure}[tbp]
\centering
\includegraphics[width=0.47\textwidth, height=0.5\textwidth]{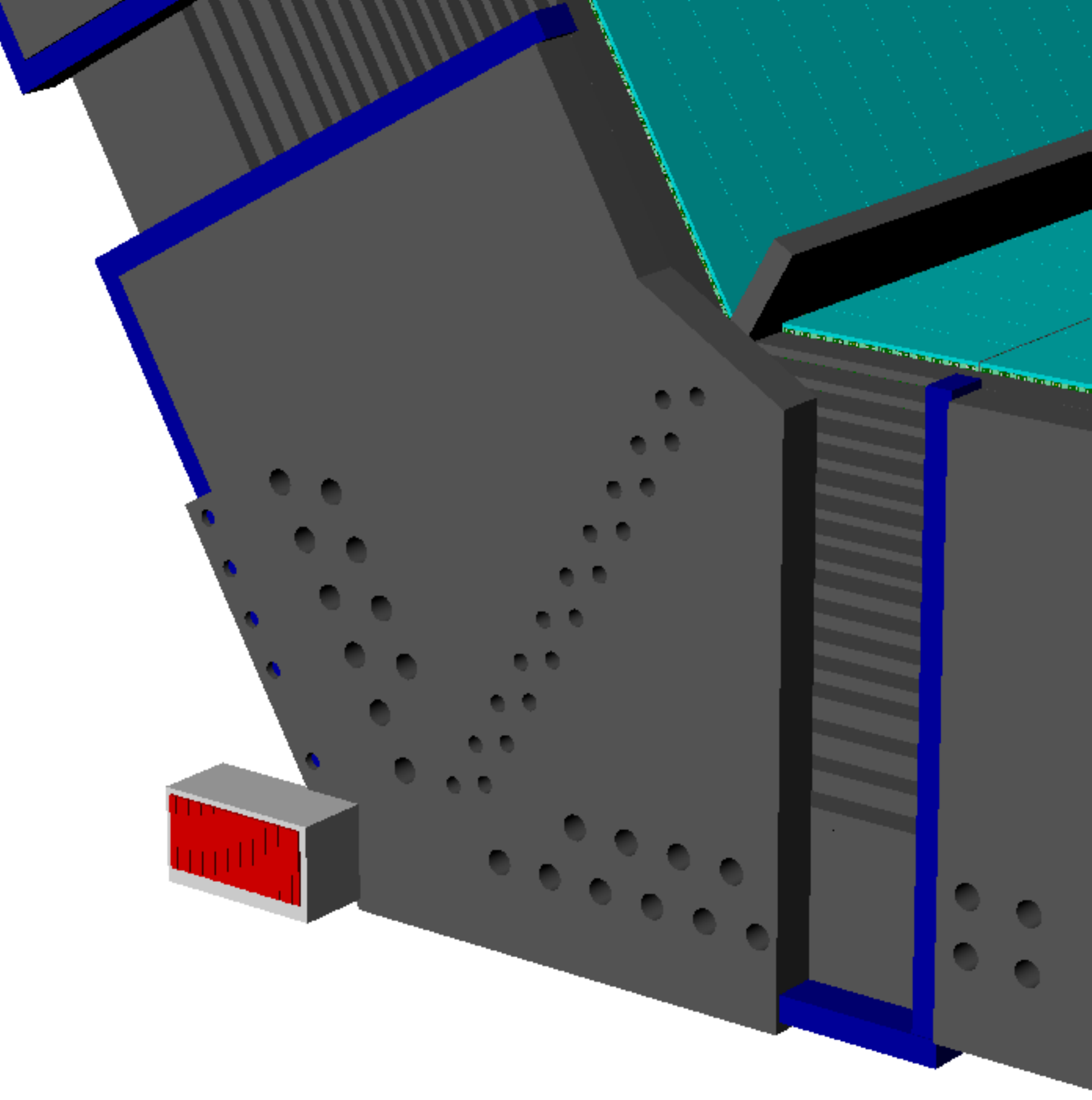}
\caption{The location of one of the six data merger crates for the Barrel part of the detector}
\label{fig:ElexIFRfig21}
\end{figure}

Two other such crates would be placed at 120\degrees on the forward end of the barrel while three crates will be located in a similar arrangement at the backward end of the barrel. Each data merger crate serves two sextants.\\

\tdrparagraph{Services needed for the IFR readout in the experimental hall.} 
The elements of the IFR readout system need to be properly powered, monitored and refrigerated. This paragraph summarizes the main requirements for a readout system built according to the baseline design:

\begin {itemize}
\item HV SiPM supply: since the SiPMs are handled in groups of 32 by the front end ASICs which set the individual correction to the SiPM bias, it is preferable to bias each group of 32 SiPMs from a single ``high side'' supply channel; the IFR needs then 384 programmable HV supply channels for the barrel and 324 for the endcaps, for a total of 708 HV independent supply voltages. An HV supply channel should feature: a programmable (with at least 10\bit resolution) output voltage ranging from 0 to 100\V (actual range and  polarity depend on the final SiPM technology chosen) and a current compliance of 25\mA. The power dissipated by the SiPMs in the whole IFR under nominal operating conditions is of the order of a few tens of Watts;
\item LV front end cards supply: in the baseline design a front end card hosts two ASICs, powered at 3.3\V and drawing 50\mA each and, eventually, one FPGA powered at 3.3\V and drawing about 150\mA. If optional ancillary op amp stages were installed, these would draw, for each group of 32 channels, a current of 350\mA from both the positive and negative 3.3 V supply rails. In summary the current consumption for a front end card would be: 600\mA at +3.3\V, 350\mA at -3.3\V, resulting in a total power consumption of about 2.7 W for a 32 channel group. The IFR would then need 354 + 354 LV supply channels with 1\amp compliance and +3.3\V and -3.3\V output voltage, respectively. Lower core voltages for the ASIC could be derived on-board by means of low-drop out regulators; the radiation tolerance of suitable LDO regulators is being characterized;
\item cooling: the temperature of the IFR steel should be controlled to within $\pm2^{\circ}$, as it was for \babar; the design of the barrel cooling system should then take into account the estimated 1140\Watt dissipated in total by the front end stages of the IFR barrel and endcap electronics. \\
\end{itemize}

\begin{figure}[t]
\centering
\includegraphics[width=0.47\textwidth]{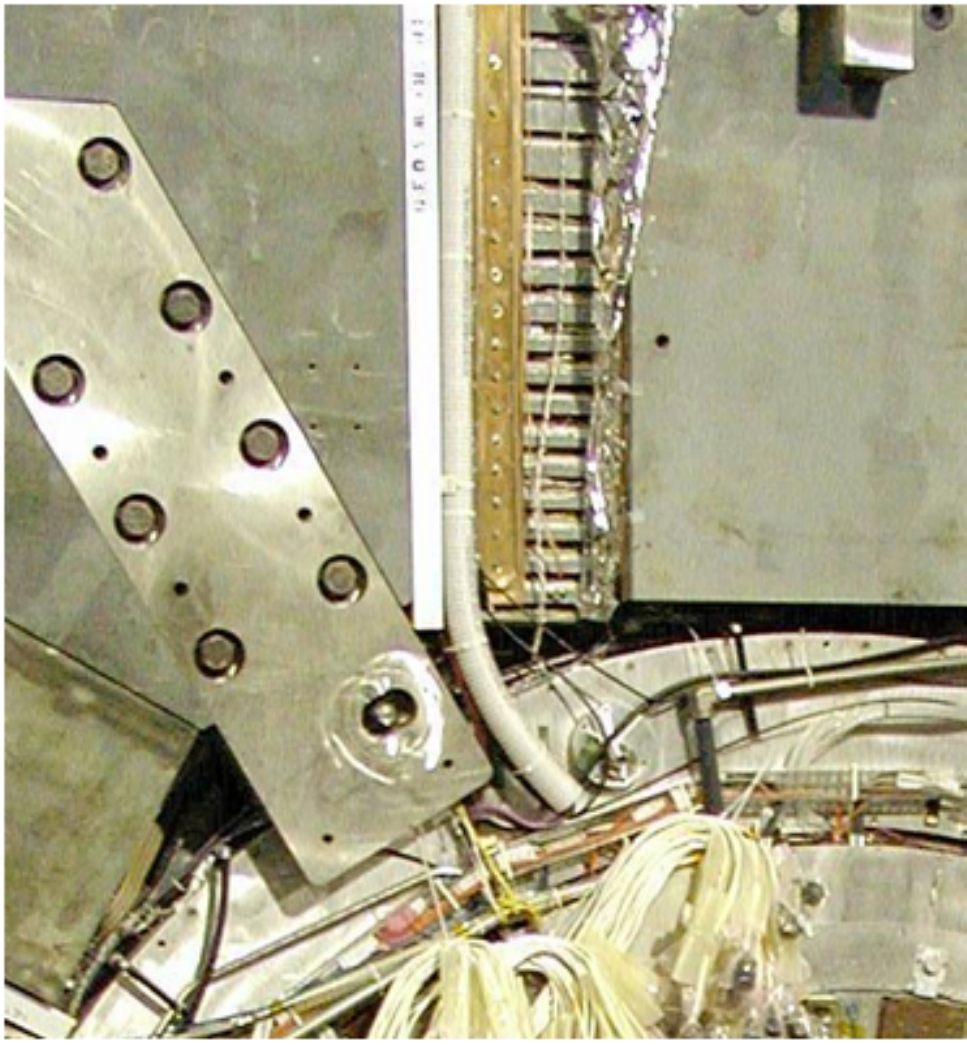}
\caption{A picture from \babar showing one of the elements of the
  cooling system for the IFR steel}
\label{fig:ElexIFRfig22}
\end{figure}


\tdrsec{Final assembly and installation}
	The assembly of the detector modules in the experimental area has to take into account the space needed for the modules, the installation tooling and the people involved. Safety rules have to be followed. The average weight is 50\kg for the barrel modules and 100\kg for the endcap modules.
For the insertion of the absorber plates into the gaps the structure used in \babar could be adapted.
Following the IFR structure, the installation is divided between the barrel and the end caps.\\

\tdrparagraph{Barrel installation.} 
The barrel is composed of six wedges called sextants. The top and bottom sextants are horizontal, while the lateral sextants are inclined at an angle of 60$^{\circ}$. For the assembly a lifting structure will be used which will allow the positioning of the module in front of the gap and the insertion of the module itself.  The module covers half the length of the barrel and the insertion will take place on both sides.
The plane of the lifting structure supporting the modules can be rotatated in order to be parallel to the gaps of the angled sextants. 
The barrel is supported by arches connected to the external plate of the sextants. This reduces the plates deformation and leaves the dimension of the gaps within the mounting tolerance. The external layers of the detector have to be sized in order to cover the maximum detector area. The external layers of the three bottom sextants need a supporting frame connected to the barrel structure. Figure~\ref{fig:ifr:inst-scheme} and~\ref{fig:ifr:inst-picture} 
show a schematic and a picture
of the \babar muon detector disassembly: the same kind of tooling can be used to install the IFR modules into the \superb detector.\\

\tdrparagraph{Endcap installation.}
The lifting structure used for the installation of the modules in the barrel will be used for the installation in the endcaps.
The endcap modules are heavier than the barrel modules and bigger in size: the lifting connections have to be chosen in a way to avoid deformation and damages during the handling and the insertion. \\

Installation tests both for the barrel and the endcaps have to be done in order to optimize the procedure
and the schedule.

\begin{figure}[t]
\centering
\includegraphics[width=0.45\textwidth]{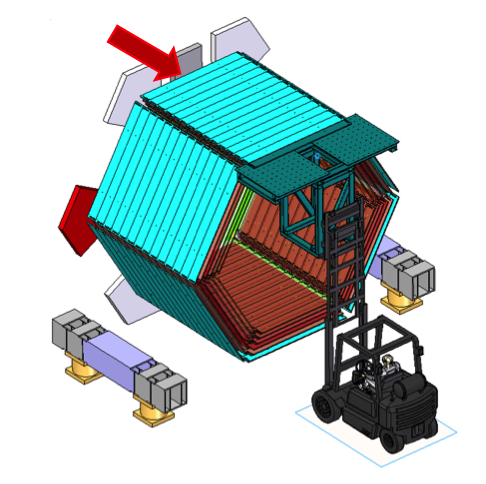}
\caption{Scheme of the \babar brass removal tooling that can be reused for the IFR module installation.}
\label{fig:ifr:inst-scheme}
\end{figure} 

\begin{figure}[tbp]
\centering
\includegraphics[width=0.45\textwidth]{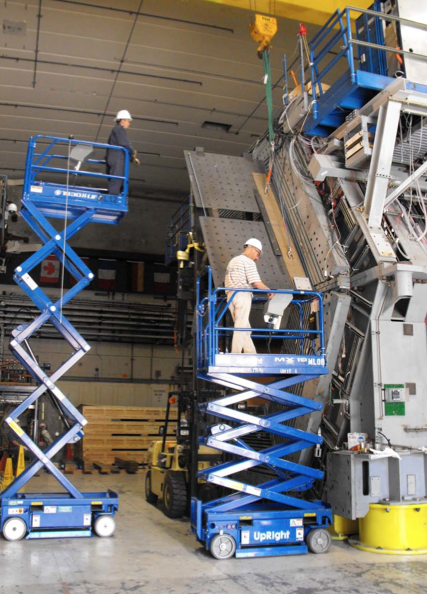}
\caption{Picture taken during the removal of the brass from a diagonal sextant of the \babar barrel.}
\label{fig:ifr:inst-picture}
\end{figure}








\bibliographystyle{\tdrbibstyle}
\balance
\bibliography{IFR/IFR-bib}



\renewcommand{\ChapLabel}{chap:Magnet} 
\tdrchap{Magnet and Flux Return}

The magnet for the \SuperB\ experiment is a thin superconducting
solenoid mounted inside a hexagonal flux return.  It was initially
designed and built in the 1990s for the \babar\
experiment~\cite{Fabbricatore:1996mc, Bell1999559} where it was
successfully operated with a high degree of reliability for
approximately ten years.  Because of this it has been chosen as the
preferred option for the \SuperB\ detector. In this chapter the main
characteristics of the magnet, the modifications required for its
integration into the \SuperB\ detector, and a plan for its transport
from SLAC to the Tor Vergata site, reassembly and installation are
described.

\tdrsec{Magnet main characteristics}

The \SuperB\ magnet, for which a schematic cross section is shown in
Figure~\ref{fig:mag1}, includes:

\begin{enumerate}
\item The superconducting solenoid;
\item The laminated barrel flux return, composed of 18 steel plates of
  different thickness (from 20\mm to 100\mm);
\item The two doors of the flux return with 19 steel plates of
  different thickness;
\item An iron end plug shield in the forward end door;
\item An iron end plug shield in the backward end door.
\end{enumerate}

\begin{figure*}[tbh]
\centering
    \includegraphics[width=0.8\textwidth]{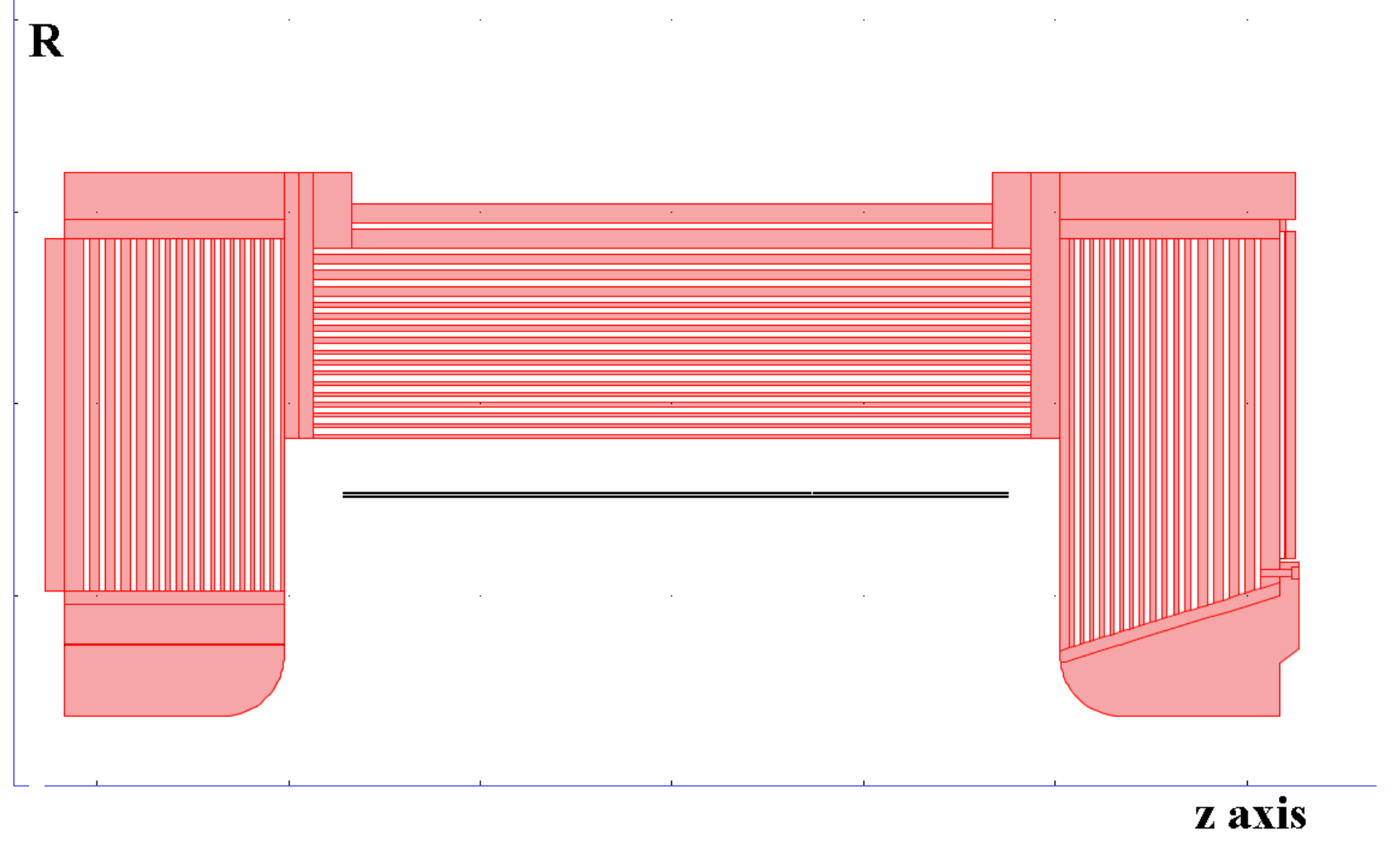}	
    \caption{Simplified cross-sectional view of the magnet in the
      plane z-r (cylindrical symmetry with respect the axis shown at
      the bottom of the figure). This model is used for 2D magnetic
      computations. The thin black line represents the superconducting
      solenoid (just the coil, the cryostat is not shown). It is
      included in the flux return composed of the barrel and two end
      caps (the doors) including two end plugs. The asymmetry is due
      to the reuse of the \babar\ equipment. }
    \label{fig:mag1}
\end{figure*}

\begin{figure}[tbh]
 \centering
    \includegraphics[width=0.45\textwidth]{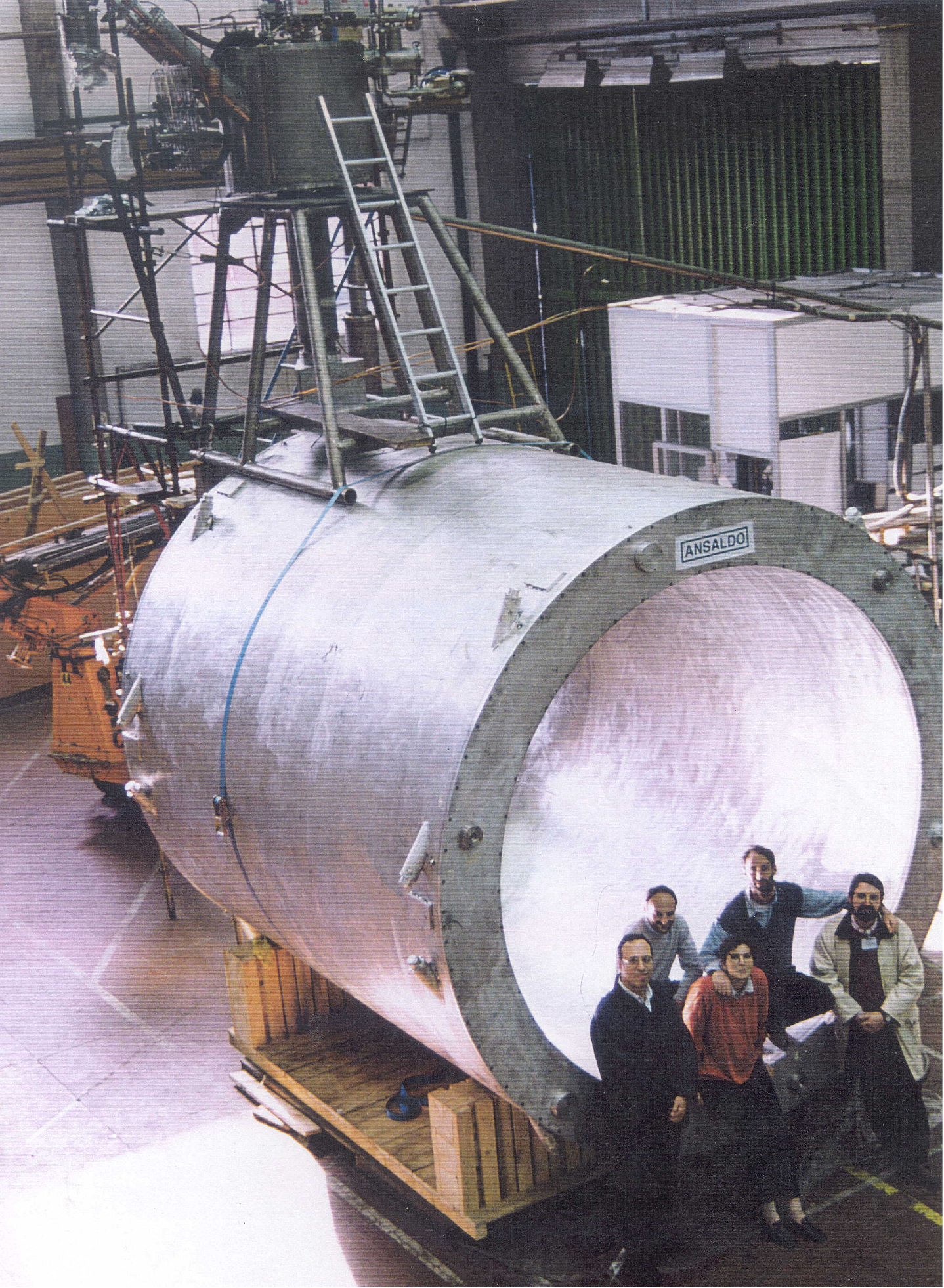}	
    \caption{The superconducting solenoid after the cryogenic test at
      Ansaldo. One can see the annular cryostat enclosing the
      superconducting solenoid and the turret with proximity
      cryogenics and current leads. }
    \label{fig:mag2}
\end{figure}

\begin{figure}[tbh]
\centering
    \includegraphics[width=\linewidth]{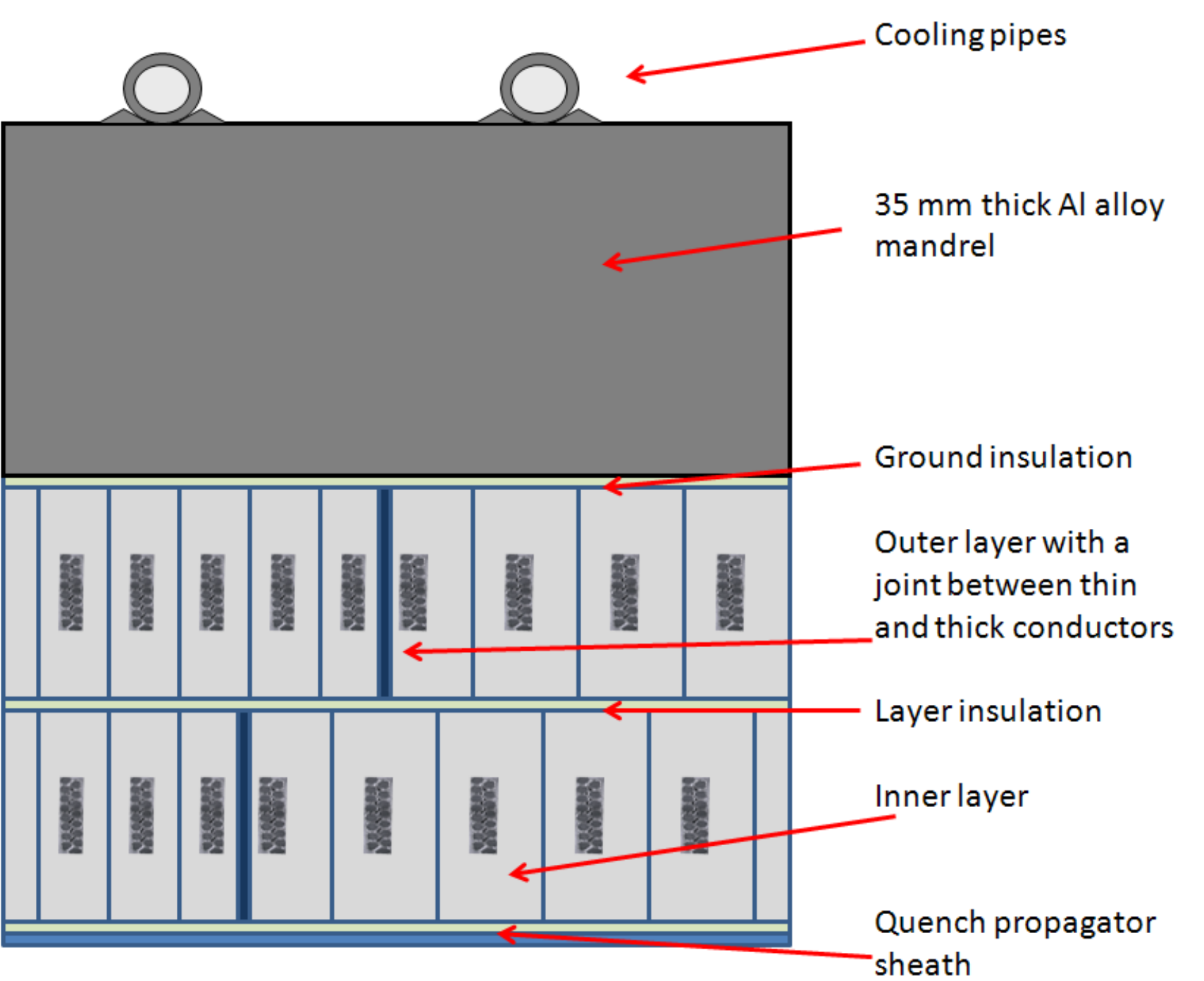}	
    \caption{Cross section of the cold mass. One can see the two
      layers made of a two different conductors for grading the axial
      current density. The two layers were directly wound inside an
      external mandrel in aluminum alloy 5083. The liquid helium is
      circulated in pipes directly attached to the external surface
      of the mandrel. }
    \label{fig:mag3}
\end{figure}

Compared to \babar, the design of the forward end plug shield has been
modified to make it as symmetric as possible with respect to the
backward end plug. The reason for the asymmetry of the \babar\ magnet
was that the magnetic field had to be shielded from the Q2 final focus
quadrupole. This quadrupole is no longer present in in the \superb\
accelerator lattice, allowing for a more symmetric detector magnet
configuration. In the present design the forward door configuration
remains the same as in \babar.  However, a future option for making the
forward door completely symmetric with respect to the backward door is
under consideration.  Another difference from the \babar\ magnet is
the addition of a 50\mm plate at the external ends of the backward
door and of a 100\mm plate at the external side of the forward door.
More detailed information about the flux return can be found in
Chapter~\ref{chap:IFR}.

The core of the magnet is a thin superconducting solenoid with
as-built characteristics shown in Table~\ref{tab:mag1}.
Figure~\ref{fig:mag2} shows the solenoid just after the preliminary
test at Ansaldo in Genova (Italy)~\cite{ansaldo}, where
it was built. The design of the solenoid was based on criteria
developed and tested with many superconducting detector magnets using
aluminum-stabilized thin solenoids~\cite{Yamamoto:2002ja:mag:ref3},
including the magnets for the ALEPH~\cite{Decamp1990121} and
DELPHI~\cite{tagkey1991233} experiments at CERN.

\begin{table}
\caption{Main characteristics of the \SuperB solenoid, as built.
(R.T.\ = room temperature). }
\label{tab:mag1}
\begin{center}
\begin{tabular}{p{4cm}p{3cm}}
\hline
Central Induction	& 1.5\Tesla \\
Conductor peak field	& 2.3\Tesla \\
Winding structure  &	2 layers graded current density\\
Uniformity in the tracking region &	$\pm$ 3\% \\
Winding axial length	& 3512\mm at R.T. \\ 
Winding mean radius	& 1530\mm at R.T. \\
Operating current	& 4596\amp\\
Inductance	  & 2.57\Henry \\
Stored Energy &	27\MJ \\
Total turns	  & 1067 \\ 
Total length of conductor &	10300\m \\
\hline
\end{tabular}
\end{center}
\end{table}

The coil is composed of two layers of aluminum-stabilized conductors
internally wound in a 35\mm-thick aluminum (alloy 5083) support
mandrel.  Cooling pipes welded to the outside of the support mandrel
form part of a cryogenic thermo-siphon system. Electrical insulation
consists of dry wrap fiberglass cloth and epoxy vacuum impregnation.
Figure~\ref{fig:mag3} shows a schematic cross section of the
solenoid's cold mass.  The conductors are 20\mm wide and are composed
of a superconducting 16-strand Rutherford-type cable embedded in a
pure aluminum matrix through a co-extrusion process, which ensures
good bonding between the aluminum and the superconductor.  In order to
achieve a field homogeneity of $\pm 3\,\%$ in the large volume
required by \babar\ (and confirmed for \superb), the current density
in the winding is graded: It is lower in the central region and higher
at the ends.  This gradation is obtained by using conductors of two
different thicknesses: 8.4\mm in the central region and 5\mm towards
the ends. Table~\ref{tab:mag2} shows the main characteristics of the
conductor.  In total six conductor lengths were employed in the
construction.  The electrical joints between the thin and the tick
conductors are integrated into the winding, as it can be seen in
Figure~\ref{fig:mag3}.

\begin{table*}
\caption{Summary of strands, Rutherford, and conductor characteristics.}
\label{tab:mag2}
\begin{center}
\begin{tabular}{lcc}

\bf{Component} & \bf{Characteristic} & \bf{Value} \\
\hline
Strand & Material & NbTi \\
&Composition & Nb $46.5 \pm 1.5\%$ wt \\
&Filament size& $< 40 \mum$ \\
&Twist pitch & 25 \mm \\
&Cu/NbTi ratio& > 1.1 \\
&RRR & Final >100 \\
&Wire diameter& $0.800 \pm 0.005 \mm $\\
\hline
Rutherford&Transposition pitch&$< 90 \mm$\\
&Number of strands&16\\
&Final size& $1.4 \times 6.4 \mm^2$\\
\hline
Conductor&Al-RRR&>1000 \\
&Thin conductor dimensions & $(4.93 \times 20)\pm 0.02 \mm$ \\
&Thick conductor dimensions& $(8.49 \times 20)\pm 0.02 \mm$ \\
&Rutherford-Al bonding&> 20\MPa \\
&Thin conductor Al/Cu/NbTi ratio &23.5:1.1:1 \\
&Thick conductor Al/Cu/NbTi ratio& 42.4:1.1:1\\
&Edge curvature radius& $ > 0.2 \mm$\\
&Critical current $(T=4.2\degk; B=2.5\Tesla)$&12680\amp\\
\hline
\end{tabular}
\end{center}
\end{table*}

\tdrsec{Magnetic forces on the solenoid}

The coil is located inside an asymmetric flux return yoke, a heritage
of the \babar\ configuration. The forward plug design has been
modified to make it more symmetric with respect to the backward plug,
but axial offset forces are still present. For the \babar\ magnet,
in order to have an offset force in one direction only (no inversion
during the ramp up), the coil was positioned with a 30\mm axial
displacement in the forward direction and held in place by three
Inconel 718 tie rods on the backward end. These tie rods were designed
to hold forces as high as 250\kN with a safety factor of 4. During
magnet operation at SLAC the axial force was continuously monitored
and measured be no more than 80\kN at 3800\amp magnet current. The
three similar tie rods on the forward end were never strained.

The modification of the iron yoke introduced for \superb\ requires an
adjustment of the axial coil position inside the cryostat. A
preliminary magnetic computation, for which the magnetic field map is
shown in Figure~\ref{fig:mag4}, indicates that the same conditions for
the axial forces can be obtained in \SuperB\ with a displacement of
20\mm in the forward direction. It has also been determined that the
gradient of the axial force is about 15\kNpmm.

\begin{figure*}[tbh]
  \begin{center}
    \includegraphics[width=0.8\textwidth]{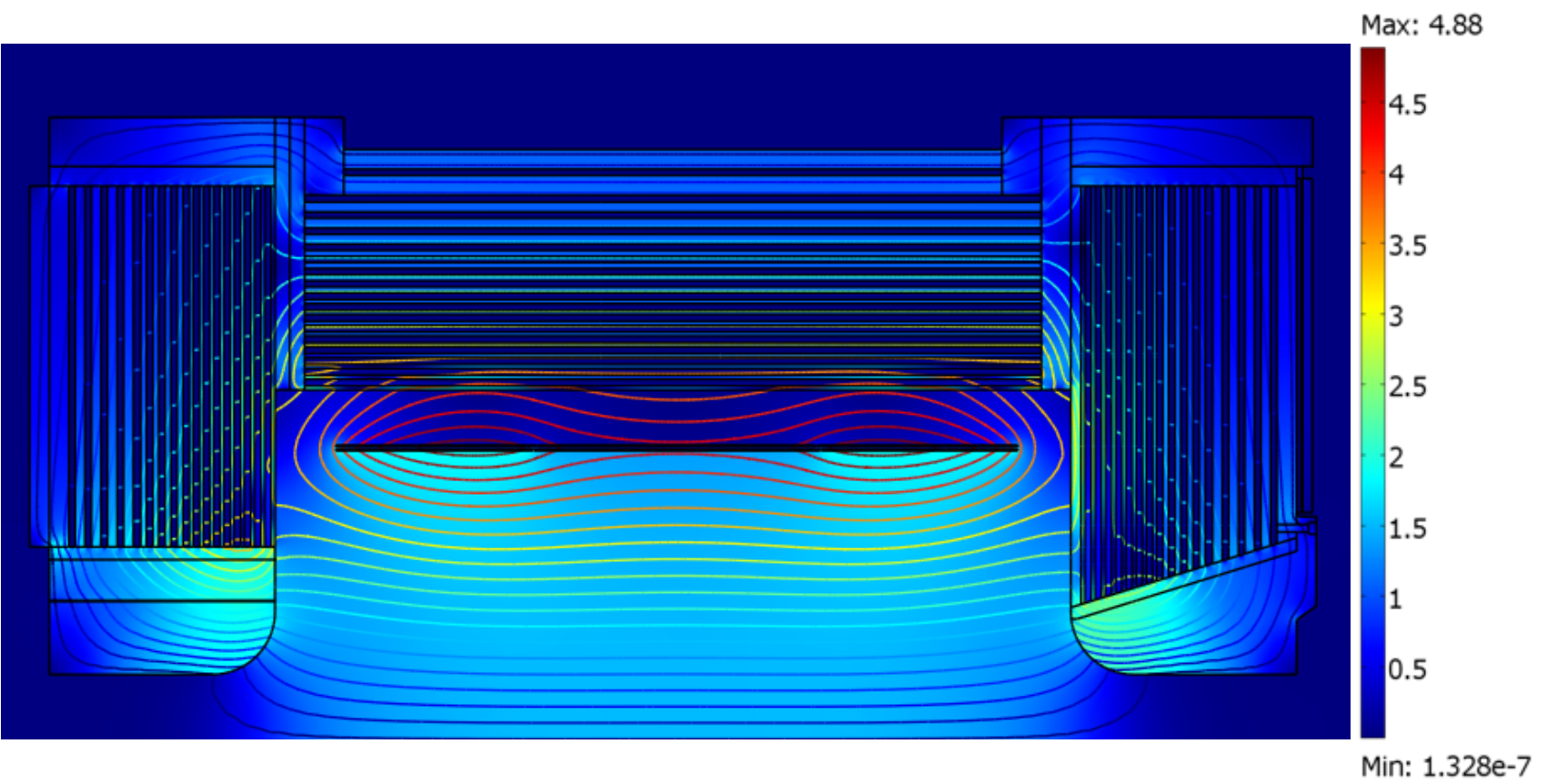}	
    \caption{ Magnetic field map (in T) in the \superb\ detector
      region showing that the field in proximity of the magnetic axis
      will be partially shielded by the anti-solenoids of the final
      focus.  In this simulation the superconducting coil is axially
      displaced by 20\mm in the forward direction (to the right in the
      figure). }
    \label{fig:mag4}
  \end{center}
\end{figure*}

\tdrsec{Cryogenics} 

The coil is indirectly cooled at an operating temperature of 4.5\degk
using the thermo-siphon technique with the liquid helium being
circulated in pipes welded to the support cylinder
(Figure~\ref{fig:mag3}). The piping was designed for a steady-state
cooling flow of 30\gmps. In the \babar\ magnet, cool down and cryogenic
supply to the coil and 40\degk radiation shields was accomplished by a
modified Linde TCF-200 liquefier/refrigerator~\cite{Burgess:1998bm}.
Liquid helium and cold gas from the liquefier/refrigerator and a
4\cubicm storage vessel was supplied to the coil and shields through a
60\m-long, coaxial, return gas-screened, flexible transfer line. A
similar, albeit updated system, will be implemented for \superb; the
proximity cryogenics integrated with the magnet and hosted in the
turret will however remain unchanged.  The schematic of the cooling
system is shown in Figure~\ref{fig:mag5}. It is possible to cool down
the coil by a mixture of warm and cold helium gas or by supplying gas
at increasingly lower temperature through the refrigerator. The
shields are cooled by part of the gas coming back from the coil. The
cool-down at SLAC took about a week.

\tdrparagraph{Refrigerator.} The refrigerator consists of a 4\cubicm
supply dewar, a helium liquefier, distribution valve box and a
compressor facility with three screw compressors. Descriptions of the
system design and function can be found in
References~\cite{Boutigny:1995ib} and~\cite{Thompson:2005dj}.

\tdrparagraph{Heat load.} In \babar\ a heat load measurement at
4.5\degk was performed by closing the input valve to the 4\cubicm
Dewar and then measuring the liquid helium consumption in that vessel. With a
1.58\Tesla magnetic field the mass flow rate per lead (at 40\mV across
each lead) was 90 NLP/m corresponding to a heat load of 5\Watt per lead.
Using these data, and considering that a 3\Watt loss can be attributed to
the transfer line, we can calculate that the magnet heat load was
between 19\Watt and 24\Watt at a flow rate of 14 l/h.  Cooling the shields
with cold helium gas from the liquid helium reservoir in the Valve Box
contributed to keeping the loss rate very low: Shield temperatures
ranged from from 37\degk to 49\degk with a mass flow rate of
0.35\gmps. Since the enthalpy variation of helium gas at atmospheric
pressure between 4.5\degk and 45\degk (average shield temperature) is
about 250\Jpgm, the total load at the shield was 87\Watt. For the magnet
in the \superb\ configuration we can assume the same heat load values.

\begin{figure*}
  \begin{center}
    \includegraphics[width=0.9\textwidth]{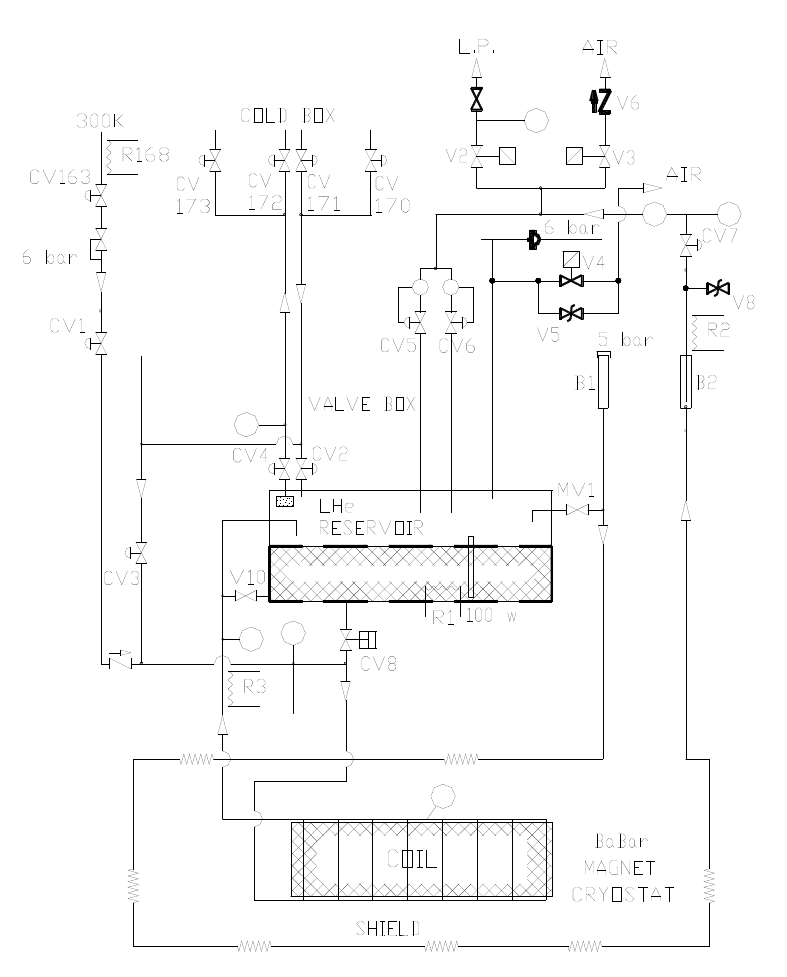}	
    \caption{ Schematic of cryogenic circuit with proximity cryogenics, which is completely integrated with the magnet. The reservoir, the valves and the current leads are hosted in the turret visible in Figure~\ref{fig:mag2}  }
    \label{fig:mag5}
  \end{center}
\end{figure*}

\tdrsec{Current supply and protection system}

The \superb\ magnet will be electrically operated with the same
circuit used in the \babar\ magnet (Figure~\ref{fig:mag6}).
A 5\kA\ - 20\V two-quadrant power supply supplies current to the
solenoid, which is protected by a resistor electrically connected in
parallel. If a quench is detected (50\mV unbalance signal between the
two voltages in two layers), the breakers open, allowing the coil to
discharge its stored energy into the dump resistor. During this
process the peak voltage at the coil ends can reach up to 340\V.
The fast discharge of the nominal current causes a quench of the whole
coil due to the heating of the supporting cylinder (\emph{Quench
  Back}) and its temperature increases to 37\degk uniformly. About
five hours are needed after a quench to cool down the coil and fill
the reservoir before the magnet is ready again for ramping up the
current.

\begin{figure*}
  \begin{center}
    \includegraphics[clip=true,trim = 10 10 10 20,width=0.8\textwidth]{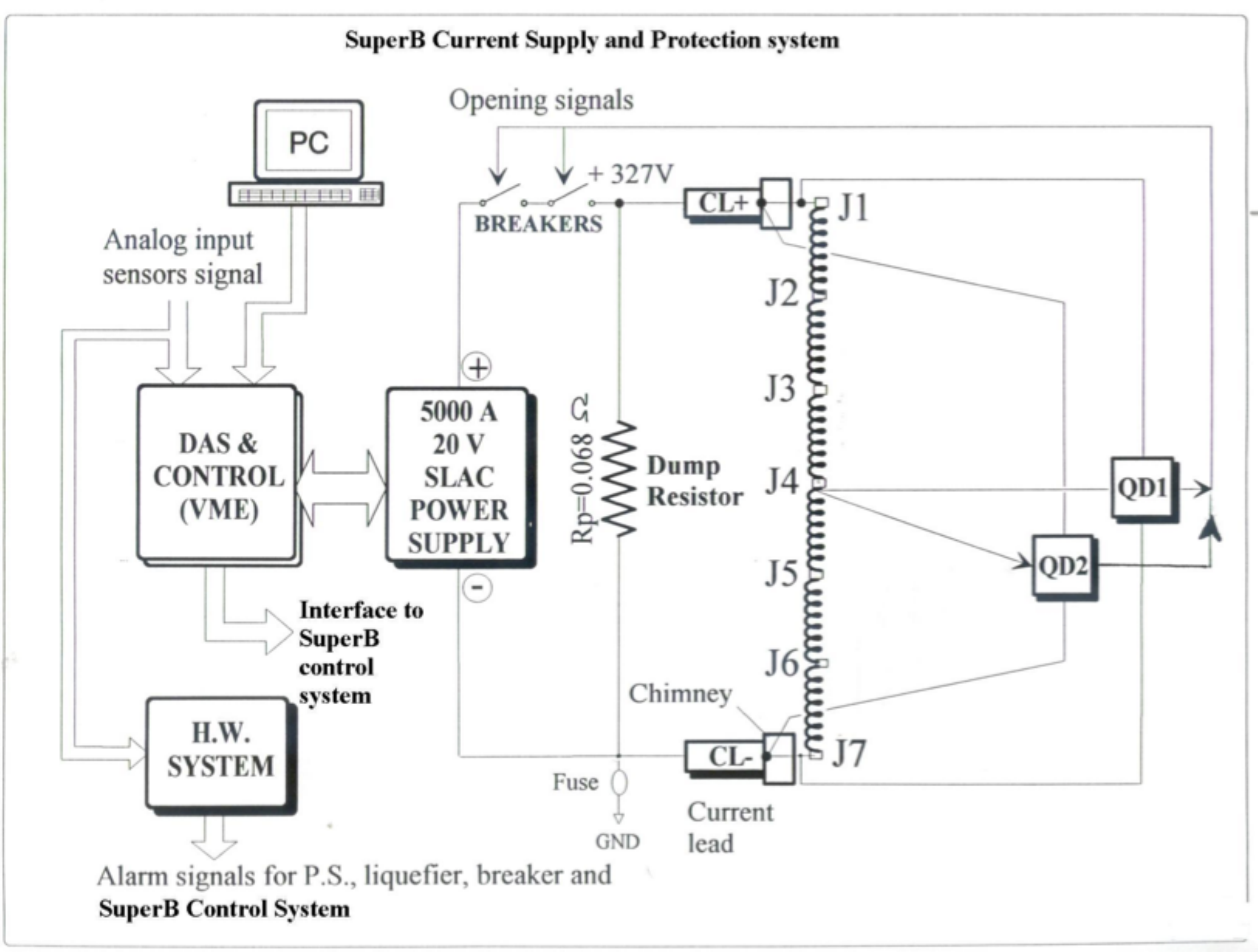}	
    \caption{Electrical scheme for current supply and protection system  }
    \label{fig:mag6}
  \end{center}
\end{figure*}

\tdrsec{Plan for reusing the \babar\ magnet equipment.}

With appropriate refurbishing, the superconducting solenoid and its
ancillaries from \babar\ can largely be reused for \SuperB.  The
following items are involved:
\begin{enumerate}
\item The superconducting coil composed of the cold mass (mass: 7.8\ton), the
cryostat (3.0 t), and the valve box (0.8\ton); 
\item  the dump resistor (1.2\ton); 
\item  the power supply (5\kA, 20\V); 
\item  two racks with controls; 
\item  the vacuum pumps and related systems: 
\item  the transfer line and quench line; 
\item  the breaker for 5\kA; 
\item  cables and miscellaneous hardware; 
\item  the cold box, the compressors, and the refrigerator controls.  
\end{enumerate}

Although some options are still under investigation, the most credible
scenario is described below. 

\tdrparagraph{Superconducting solenoid.}  The coil has been
disassembled (the turret has been removed) and is currently mounted
inside the transportation tool. For transport, the cold mass still
needs to be blocked with respect to the cryostat as it was during
delivery to SLAC. The supporting system should be able to hold the
expected transportation loads (3g vertical, 2g axial and 1.5 g
lateral). Prior to transportation the flanges will be removed and the
transportation blocks reapplied. The same consideration applies to the
valve box enclosed in the turret. A preferred option for the
transportation is a dedicated flight as it was done for the delivery
to SLAC, because there is much less risk of damage. Once in Italy the
solenoid will be reassembled in an area close to the \superb\ site to
perform a cryogenic test before the final installation in the detector.
Since the sensors on the tie rods (strain gauges and thermometers) are
no longer working, they will need to be replaced before the
realignment of the cold mass. Reassembly will also include the
reconnection of the current leads and restoration of the electrical
insulation.

\tdrparagraph{Dump resistor, quench detectors and breakers.} These
components are reusable after a check of their conditions. The
breakers might require maintenance.

\tdrparagraph{Power supply and control system.} The reuse of these
components is problematic, partly due to their old age, and partly due
to different electrical equipment standards and codes. The plan is to
procure them as new.

\tdrparagraph{Cryogenic devices.} The refrigerator could in principle
be reused, but its reinstallation and commissioning in Italy would
require the replacement of the screw compressors and the involvement
of personnel from both the manufacturer (Linde) and SLAC, resulting in
high costs with large uncertainties.  Furthermore, there are
significant risks in reusing a 15-year-old machine and expecting it to
operate for another 15 years. Similar considerations also apply to all
the other cryogenic components with the notable exception of the
flexible transfer lines.  The option of a new refrigerator, better
sized for the solenoid and final focus superconducting magnets, has
been studied.

\bibliographystyle{\tdrbibstyle}
\balance
\bibliography{Magnet/Magnet-bib}


\renewcommand{\ChapLabel}{chap:ETD} 
\tdrchap{Electronics, Trigger, Data Acquisition and Online}


\newcommand{\ktilde}{$~_{\widetilde{}}\,$}



The \superb~\cite{bib:etd-SuperBCDR} Electronics, Trigger, Data
acquisition and Online system (ETD) comprises the Level-1 trigger, the
event data chain, and their support systems. Event data corresponding
to accepted Level-1 triggers move from the Front-End Electronics (FEE)
through the Read-Out Modules (ROMs) and the network event builder to the
High Level Trigger (HLT). Events accepted by the HLT are streamed to a
data logging disk buffer from where they are handed over to offline
for further processing and archival. ETD also encompasses the hardware
and software components that control and monitor the detector and data
acquisition systems, and perform near-real-time data quality
monitoring and online calibration.

\tdrsec{Trigger Strategy and System Requirements}


The \babar\ and Belle~\cite{bib:etd-BELLENIM} experiments both chose
to use ``open triggers'' that preserved nearly 100\% of \BB\ events of
all topologies, and a very large fraction of \tautau\ and \ccbar\
events.  This choice enabled very broad physics programs at both
experiments, albeit at the cost of a large number of events that
needed to be logged and reconstructed, since it was so difficult to
reliably separate the desired signals from the \qqbar\ ($q=u,d,s$)
continuum and from higher-mass two-photon physics at trigger level.
The physics program envisioned for \superb\ requires very high 
efficiencies for a wide variety of \BB\ , \tautau, and \ccbar events,
and depends on continuing the same strategy, since few classes of the
relevant decays provide the kinds of clear signatures that allow the
construction of specific triggers.


All levels of the trigger system are designed to permit the
acquisition of prescaled samples of events that can be used to measure
the trigger performance.

The trigger system consists of the following components \footnote{
While at this time we do not foresee a ``Level~2'' trigger that acts
on partial event information in the data path, the data acquisition
system architecture would allow the addition of such a trigger stage
at a later time, hence the nomenclature.}:

\tdrparagraph{Level~1 (L1) Trigger:}

A synchronous, fully pipelined L1 trigger receives continuous data
streams from the detector independently of the event readout and
delivers readout decisions to the FCTS with a fixed latency. Like in
\babar, the \superb\ L1 trigger operates on reduced-data streams from
the drift chamber and the calorimeter.

\tdrparagraph{High Level Triggers (HLT) --- Level~3 (L3) and Level~4 (L4):}

The L3 trigger is a software filter that runs on a commodity computer
farm and bases its decisions on specialized fast reconstruction of
complete events. An additional ``Level~4'' filter may be implemented
to reduce the volume of permanently recorded data if needed. Decisions
by L4 would depend on a more complete event reconstruction and
analysis. If the worst-case per-event performance of the L4
reconstruction algorithms does not meet the near-real-time
requirements of L3, it might become necessary to decouple L4 from L3
-- hence, its designation as a separate trigger stage.

\tdrsubsec{Trigger Rate and Event Size Estimation}


The \superb\ L1-accept rate design standard of 150\kHz takes into
consideration the experience with \babar.  The \babar\ Level-1 physics
configuration produced a trigger of approximately 3\kHz at a
luminosity of $10^{34}\ {\rm cm}^{-2} {\rm sec}^{-1}$, however changes
in background conditions produced large variations in this rate. The
\babar\ DAQ system performed well, with little dead time, up to
approximately 4.5\kHz. This 50\% headroom was very valuable for
maintaining stable and efficient operation and will be retained in
\superb. In spite of the different energy asymmetry, the \superb\
geometrical acceptances are very similar to those in \babar, due to
changes in detector geometry.

In \babar\
the
output of the offline physics filter corresponded to a cross-section
of approximately 20\nb and included a highly efficient Bhabha veto. We
take this as an irreducible baseline for an open hardware trigger
design; in fact this is somewhat optimistic since the offline filter
used results from full event reconstruction. The accepted
cross-section for Bhabhas in \superb\ is also similar to \babar, and
approximately 50\nb. At the \superb\ design luminosity of $10^{36}\
{\rm cm}^{-2} {\rm sec}^{-1}$ these two components add up to 70\kHz
L1-accept rate without backgrounds. The maximum rate of
background-related L1 accepts is estimated to be 30\kHz, resulting in
a total L1-accept rate of 100\kHz. Without a Bhabha veto at L1 (which
is not part of the \superb\ baseline trigger design) this is the
minimum rate the readout system has to handle. By adding a \babar-like
reserve of 50\% to both accommodate the possibility of higher
backgrounds than design (e.g. during machine commissioning), and the
possibility that the machine exceeds its initial design luminosity, we
set the the \superb\ design rate at 150\kHz.

The event size estimate still has some uncertainties. The size of raw
events that have to be transferred from the FEEs through the ROMs to
the HLT is understood well enough to determine the number of data
links required, and to size the event building network. Based on
initial estimates from the SVT and the EMC we predict a raw event size
of 500\kbyte. As part of the event processing in the HLT farm, the
event size will be reduced by applying zero suppression or feature
extracting algorithms. The details of these algorithms and their
specific performance for data size reduction are not known yet, so we
make a conservative event size estimate of 200\kbyte for permanent
logging.


%

\tdrsubsec{Dead Time and Buffer Queue Depth Considerations}


The event data chain has been specified to handle the maximum average
rate of 150\kHz, and to absorb the expected instantaneous rates, both
without incurring dead time of more than 1\% under normal operating
conditions at design luminosity. Note that this 1\% dead time
specification does not include event loss due to individual
sub-detector's intrinsic dead times or the L1 trigger's limitations in
separating events that are very close in time. Dead time is generated
and managed centrally in the FCTS based on feedback (``fast
throttle'') from the FEE: the FCTS drops valid L1 trigger requests
when at least one FEE indicates that its rate capabability has been
exceeded. The ROMs and the event builder also provide feedback (``slow
throttle'') to slow down the trigger if they cannot keep up with the
L1 accept rate.

The required ability to handle the maximum average rate determines the
overall readout system bandwidth; the instantaneous trigger rate
requirement affects the FCTS, the data extraction capabilities of the
front-end-electronics, and the depth of the de-randomization buffers.

The minimum time interval between bunch crossings at design luminosity
is about 2.1\ns---so short in comparison to detector data collection
times that we assume ``continuous beams'' for the purposes of trigger
and FCTS design. In this model the inter-event time distribution is
exponential and the instantaneous rate is only limited by the
capability of the L1 trigger to separate events in time. Therefore,
the burst handling capability of the system (i.e. the derandomization
buffer size in the FEEs) is determined by the average L1 trigger rate,
minimum time between L1 triggers, and the average data link occupancy
between the FEEs and the ROMs.  Figure~\ref{fig:etd-minbuf} shows the
result of a simple simulation of the FCTS together with the FEE of a
typical subdetector: The minimum buffer depth required to keep the
event loss due to a full derandomization buffer below 1\% is shown as
a function of the average data link utilization.

In \superb, we are targeting a link utilization of 90\%. According to
Figure~\ref{fig:etd-minbuf}, this requires the derandomizer buffers to
be able to hold a minimum of 10 events. It is very important to
understand that there is no contingency in the target link utilization
--- all system-wide contingency is contained in the trigger rate
headroom.

Additional derandomizer capacity to what is shown in
Figure~\ref{fig:etd-minbuf} is required to absorb triggers generated
while the fast throttle signal is propagated from the CFEE to the
FCTS. Therefore, we estimate the total derandomizer depth to be at
least 16 events.

\begin{figure}[tbp]
\includegraphics[width=0.5\textwidth,clip=true,trim=70 0 66 0]{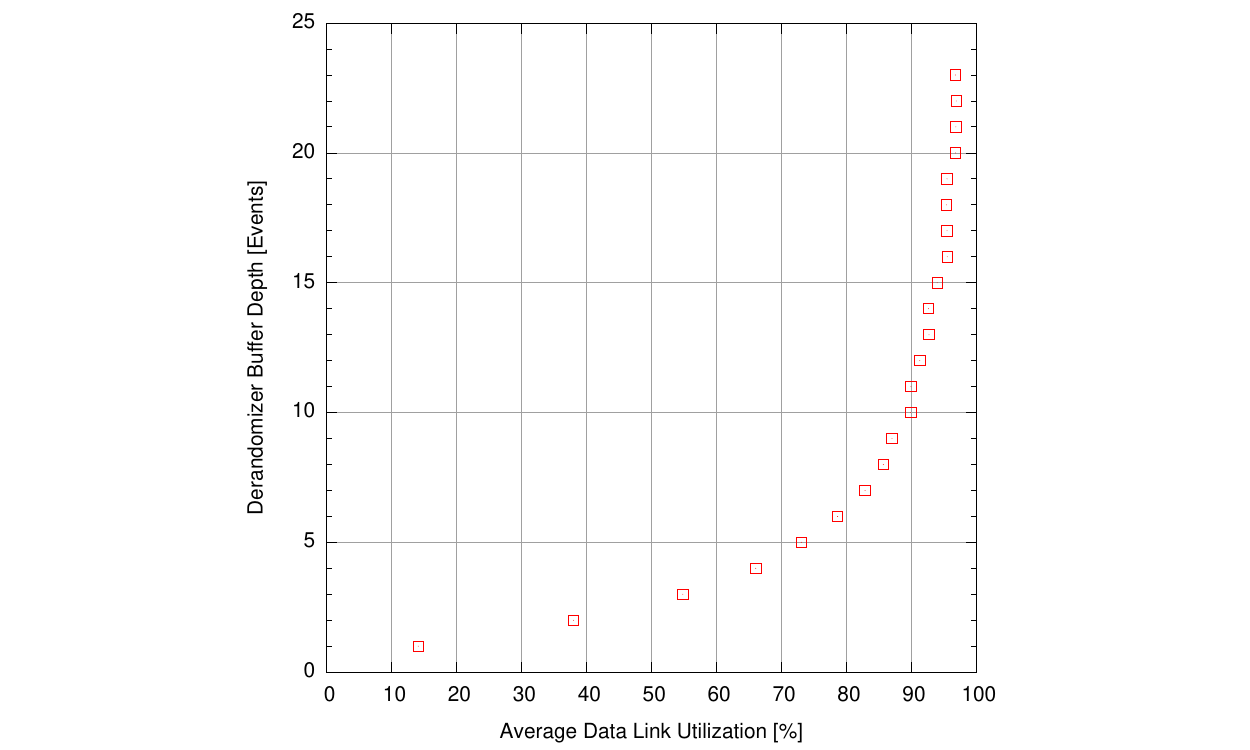}
\caption{Minimum derandomizer buffer depth required to keep the event loss due to ``derandomizer full'' below 1\% as a function of the average data link utilization.}
\label{fig:etd-minbuf}
\end{figure}

\tdrsubsec{Choice of Global Clock Frequency}

The \superb\ trigger and data acquisition system requires a global
clock that can be distributed to all components that operate
synchronously, i.e. FCTS, the L1 trigger and the sub-detector FEEs.

To allow features like selective blanking of triggers for a
well-defined part of the revolution of the stored beams, it is
critically important that the global clock is in well defined relation
with the machine RF of 476\MHz and the revolution phase of the stored beams.

In \babar the global clock was generated by dividing the RF by eight,
yielding a 59.5\MHz clock that was distributed to all components of
the experiment. In addition, a revolution (``fiducial'') signal was
fed to the FCTS, giving the system knowledge of the revolution phase
of the stored beams. 

The baseline for \superb\ is to use the same parameters and distribute
a 59.5\MHz clock throughout the ETD system.

We are investigating to change the RF clock divisor from 8 to 12. This
would result in a globally distributed experiment clock of
$39\frac{2}{3}\MHz$, allowing us to reduce \superb-specific R\&D and
cost by reusing components and system designs developed and qualified
for the 40\MHz\ clock frequency of the LHC experiments.

\tdrsec{Architecture Overview}



\begin{figure*}[t]
\begin{center}
\includegraphics[width=\textwidth]{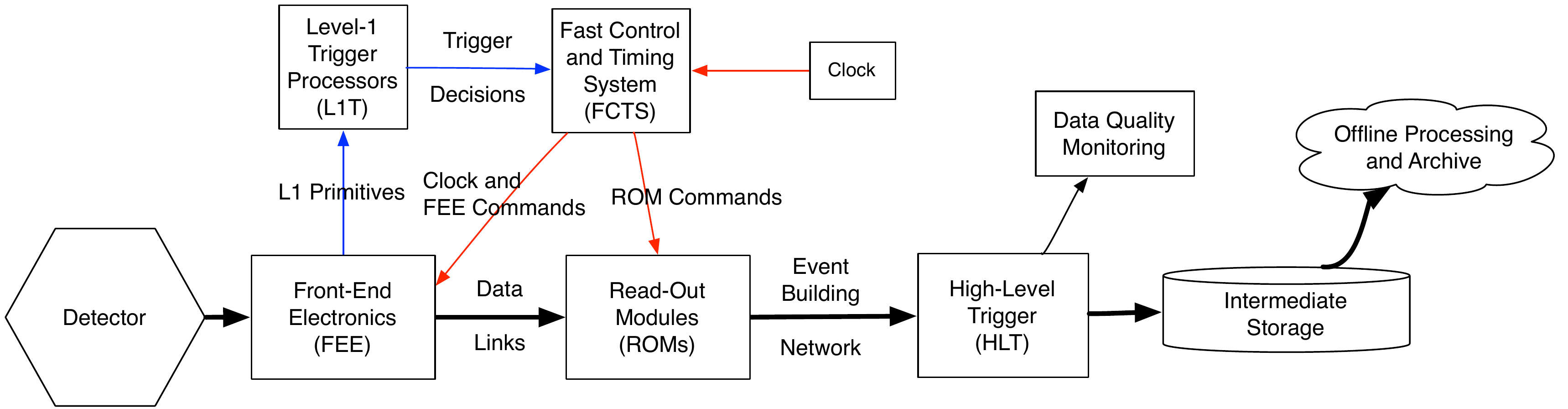}
\caption{Control and data flow in the event data chain}
\label{fig:etd-datachain}
\end{center}
\end{figure*}

\begin{figure*}[t]
\begin{center}
\includegraphics[width=\textwidth]{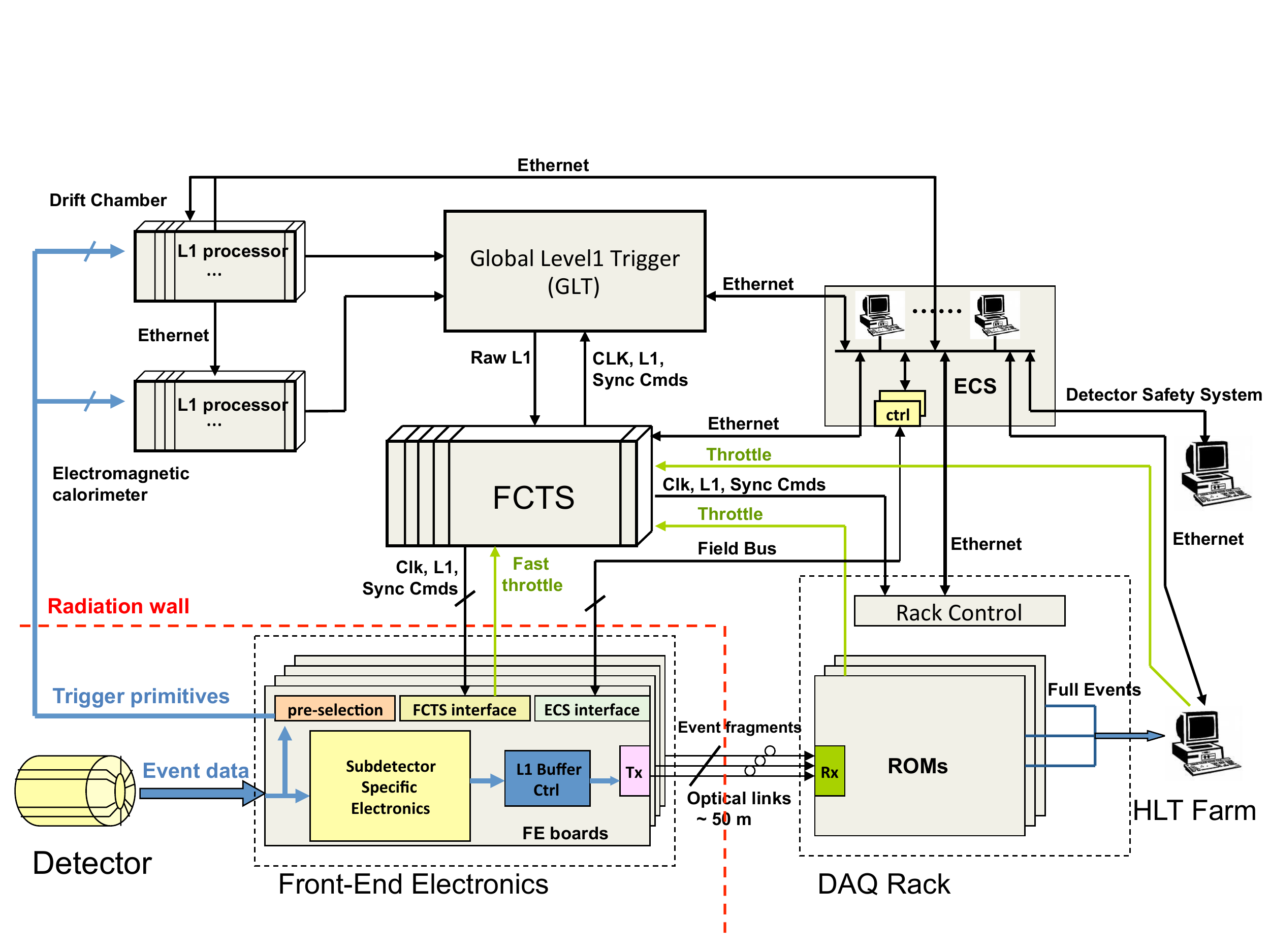}
\caption{Overview of the ETD system architecture}
\label{fig:etd-physical}
\end{center}
\end{figure*}

The system design takes into account the experience from running
\babar~\cite{bib:etd-BaBarNIM} and building the LHC
experiments~\cite{bib:etd-AtlasTDR},~\cite{bib:etd-CMSTDR},
~\cite{bib:etd-LHCBTDR}. To minimize the complexity of the FEEs and
the number of data links, the detector side of the system is
synchronous, and the readout of all sub-detectors is triggered by a
fixed-latency first-level trigger. Custom hardware components (e.g.
specialized data links) are only used where the requirements cannot be
met by commercially available off-the-shelf (COTS) components such as e.g.
Ethernet.

Radiation levels are significantly higher than in \babar, making it
mandatory to design radiation-tolerant on-detector electronics and
links (see also Chapter~\ref{chap:mdi}).  Figure~\ref{fig:etd-datachain}
shows the control and data flow in the trigger and event data chain,
Figure~\ref{fig:etd-physical} gives an overview of the ETD system
architecture.

A first-level hardware trigger processes dedicated data streams of
reduced primitives from the sub-detectors to provide trigger decisions
to the Fast Control and Timing System (FCTS).

The FCTS is the central ``bandmaster'' of the system; it distributes
clock and commands to all elements of the architecture. It initiates
the readout of the events by broadcasting ``L1-accept'' commands to
the detector FEEs at a fixed time period (``trigger latency'') after
the activity in the detector that caused the trigger to be asserted.
In response to a L1-accept, the FEEs read out a window of samples
(``readout window'') from the latency buffer, format the data and
transfer these ``event fragments'' via derandomizer buffers and
optical links to the ROMs. The ROMs then perform a first stage
internal event build and send the partially constructed events to the
HLT farm where they are combined into complete events, and processed
by the HLT which reduces the amount of data to be logged permanently
by rejecting uninteresting events. Note that the readout window has to
be large enough to contain the sub-detector information pertaining to
an event, and to accommodate the time uncertainty of the trigger
(``trigger jitter'') caused by the limited time resolution of the
detectors feeding the trigger.

The trigger, data acquisition and support components of the ETD system
are described in this chapter, sub-detector electronics, power
supplies, grounding and shielding, and the cable plant are described in
the next chapter.

\tdrsec{Trigger and Event Data Chain}

The systems in the trigger and event data chain manage event
selection, event filtering and end-to-end data flow from the detector
to the intermediate buffer.

\tdrsubsec{Level-1 Trigger}


The baseline for the L1 trigger is to re-implement the \babar\ L1
trigger~\cite{bib:etd-BaBarTRG} with state-of-the-art technology.  It
is a synchronous pipeline running at the 59.5\MHz system-wide clock
(or multiples of 59.5\MHz) that processes primitives produced by
dedicated electronics located on the front-end boards or other
dedicated boards of the respective sub-detector.  The raw L1 decisions
are sent to the FCTS which applies a throttle if necessary and then
broadcasts them to the FEEs and the ROMs. The standard chosen for the
trigger crates will most likely be either ATCA for Physics (Advanced
Telecommunications Computing Architecture) for the crates and the
backplanes, or a fully custom architecture.

\begin{figure*}[tbp]
\begin{center}
\includegraphics[width=\textwidth]{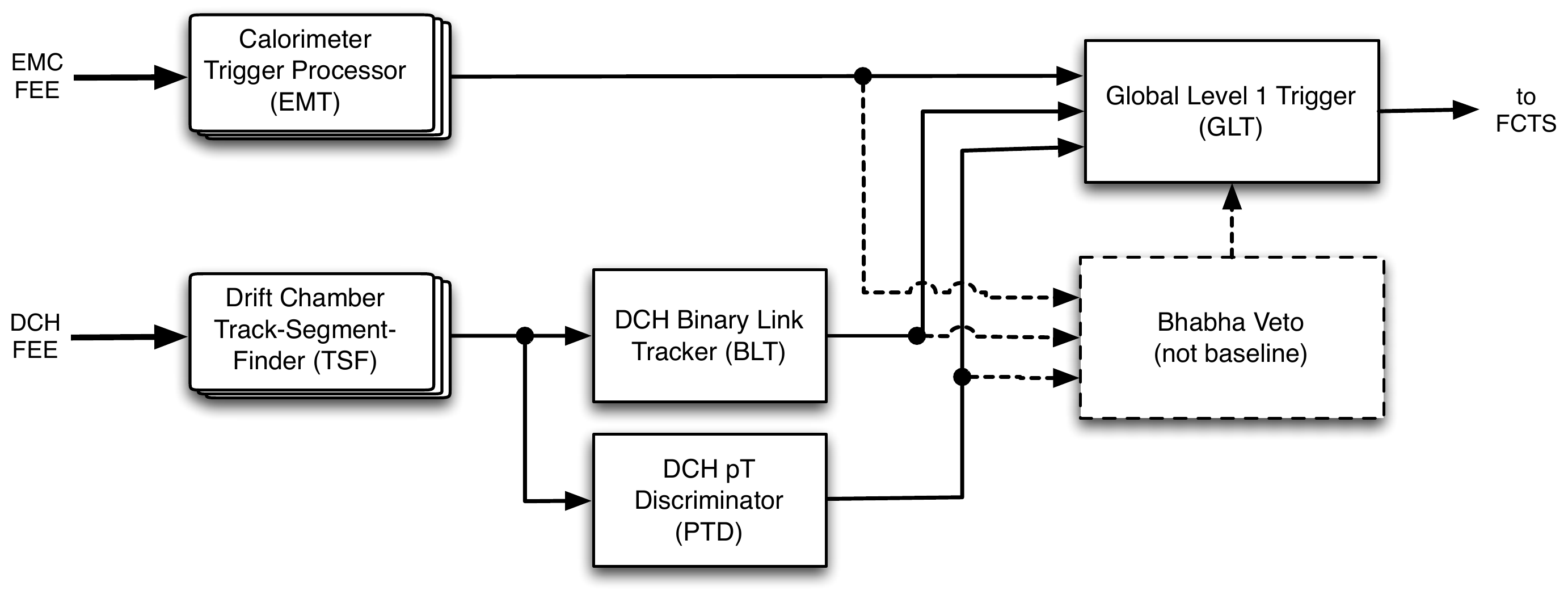}
\end{center}
\caption{Level~1 trigger overview}
\label{fig:etd-trigger-l1}
\end{figure*}

The main elements of the L1 trigger are shown in
Figure~\ref{fig:etd-trigger-l1} and are described below.

\tdrsubsubsec{Drift chamber trigger (DCT)}
\label{subsubsec:DCT}
The DCT consists of a track segment finder (TSF), a binary link
tracker (BLT) and a $p_{t}$ discriminator (PTD). The DCT also
extrapolates tracks to the interaction point (IP); this allows to
remove backgrounds that do not originate from the IP.

The layout of the DCT is shown in Figure~\ref{fig:etd-dct}. Track
finding is implemented in two stages: in the first stage, the
algorithm locally searches for track segments exploiting the data
available at the front-end level. The second stage links the track
segments and searches for complete tracks.

\begin{figure*}
\begin{center}
\includegraphics[width=\textwidth]{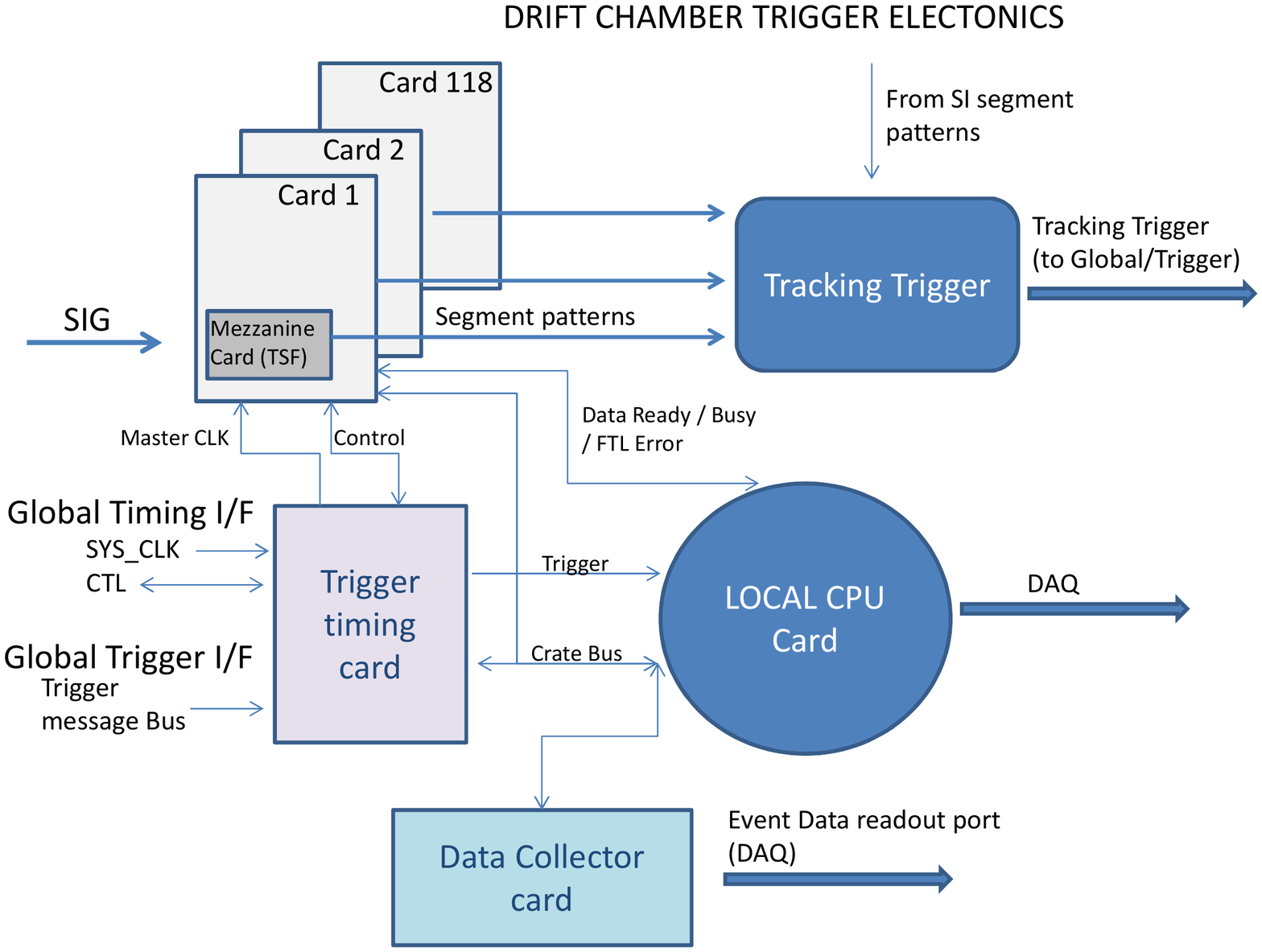}
\caption{DCT Overview}
\label{fig:etd-dct}
\end{center}
\end{figure*}

\tdrparagraph{Local Segment Finding:} 

Two options are under discussion for the DCH FEE. In the first option
time and charge are measured using standard FADCs and TDCs.  A second
option is instead based on a cluster counting technique that requires
very high-speed digitization of the signals. Granularity is 64
channels per front-end board in the first case and 16 channels per
front-end board in the second.  In the first case data delivered by
the front-end boards will be gathered by a single board on the
front-end crate and an optical link will connect the crate with the
trigger crate.  In the other option the DCH electronics has a
modularity of 64 channels distributed on 4 contiguous radial planes.
This geometry defines the super layer. Table ~\ref{table:multi} shows
the super-layer composition for the DCH and the front-end boards
necessary for its data acquisition. The track segment finder is
partially integrated in the DCH front-end.

\begin{table*}[t]
\caption{ DCH superlayer (SL) and wire readout }
\label{table:multi}
\begin{center}
\begin{tabular}{c|c|c|c|c|c|c|c|c|c|c|c}
\hline
   & {\bf Sl1} & {\bf SL2} & {\bf SL3} & {\bf SL4}& {\bf SL5} & {\bf
   SL6} & {\bf SL7} & {\bf SL8} & {\bf SL9}& {\bf SL10} & {\bf TOT}\\
\hline
 {\bf Planes} & 4 & 4 & 4 & 4 & 4 &4 & 4 & 4 & 4 & 4 &\\ 
 {\bf Type} & A & A & U & V & U & V & U & V & A & A & \\ 
 {\bf n. wires} & 736 & 864 & 496 & 560 & 624 & 688 & 752 & 816 & 896 & 960 & 7392 \\ 
 {\bf n. TSF} & 12 & 14 & 8 & 9 & 10 & 11 & 12 & 13 & 14 & 15 & 118\\ 
 {\bf n. BCD} & 1 & 1 & 1 & 1 & 1 & 1 & 1 & 1 & 1 & 1 & 10 \\
\hline
\end{tabular}
\end{center}
\end{table*}

The TSF will be implemented either at the crate level or at the
mezzanine level; the number of optical links and TSFs will be 118 in
the case of ADC/TDC-based DCH electronics or equal to the number of
DCH front-end crates in the case of cluster counting. With cluster
counting, data delivered by the FEE will be gathered by a single board
in the front-end crate and sent to the trigger processor via an optical link.
Each card reads a set of super cells. In order to define a
coincidence and define a track segment we need to stretch the signal
delivered by every wire for at least one drift time.

The TSF takes decisions at a programmable rate which can be as high as
the system clock frequency.  A possible track segment is shown in
Figure~\ref{fig:etd-track}.

\begin{figure*}
\begin{center}
\includegraphics[width=0.95\textwidth]{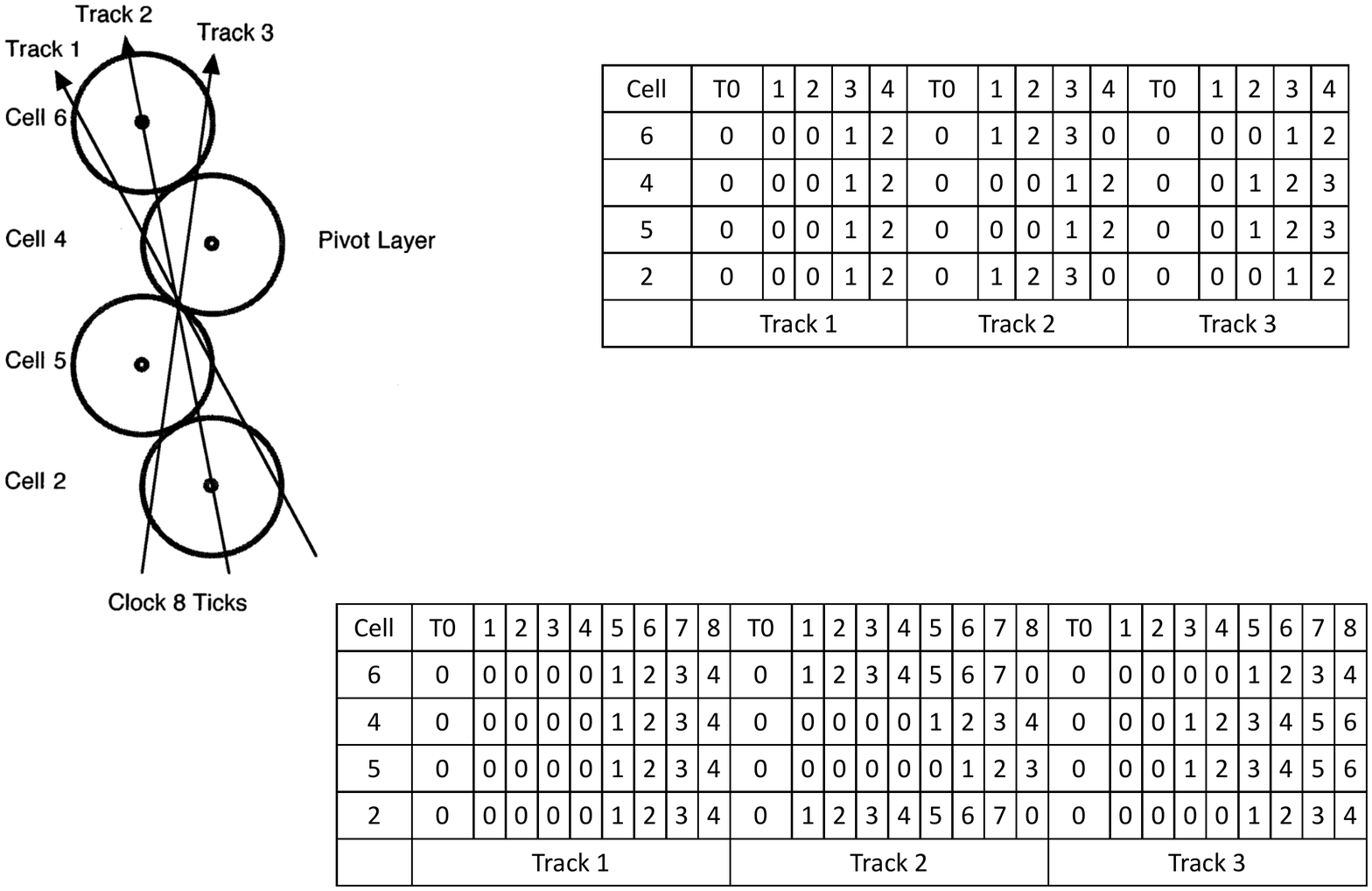}
\caption{Track segment: \babar\ vs \superb\ output bitstream. In this
  example we assumed to double the \babar\ sampling frequency.}
\label{fig:etd-track}
\end{center}
\end{figure*}

A pattern needs 3 hits to be taken in consideration. The TSF delivers
a bit stream to the following DCH stage, whose dimension depends on
sampling frequency. This bit stream represents the $\phi$ of the
segment in the super cell.  These data are delivered to the trigger
crate. To avoid efficiency losses a small number of contiguous
channels is collected by all the neighboring cards. It is relevant in
this scheme to optimize the sampling frequency as a function of track
efficiency and exploited bandwidth.

\tdrparagraph{Transmission links:} 

We will use high speed serial links to deliver trigger data to the
trigger crate. Therefore signal aggregation delivered by the TSF is made
possible, and a single card in the trigger crate has all the data from
the pertaining superlayer.

\tdrparagraph{Global Tracking:} A simple and efficient tracking
algorithm is the Binary Link Track Finder (BLT) developed within the
CLEO collaboration. This method starts from superlayer (SL) 2 and
moves radially outward combining the TSFs if in the following SL one
of the three contiguous TSF is active. The track length varies from
the first to the last super layer.  Track charge does not affect the
algorithm and it can be implemented on programmable logic.  DCT
primitives are received by the DCH L1 trigger box. In this crate the
optical links exploited by the DCH trigger lines are collected in such
a way that each board stores the information of a single SL. A bus
with a switch topology interconnects the single boards where the SL
information is present with the master where all the DC TSFs are
present.

\tdrparagraph{Particle Counter:} The BLT outputs are used to count
different tracks using a multiplicity logic exploiting also isolation
criteria.  To define a track we can use associative memory based
techniques so that in one clock cycle a pattern can be identified. The
very same associative memory can in principle be used to compute the
transverse momentum and the perigee parameters of the track. Track
counting takes into consideration 4 SL long tracks.

\tdrparagraph{Transverse Momentum:} Using TSF positions and their
$\phi$ angle it is possible to measure the multiplicity of tracks
above a predefined momentum threshold and their perigee parameters.
The data are delivered to the GLT which on the basis of the trigger
tables asserts the trigger.

\tdrsubsubsec{Electromagnetic Calorimeter Trigger (EMT)}
 
The EMT (Figure~\ref{fig:emctrigger}) processes the trigger output from
the calorimeter to find energy clusters. From the trigger point of
view the electromagnetic calorimeter in the barrel is composed of
``towers'' of $8(\theta)\times3(\phi)$ CsI(Tl) crystals, as shown in
Figure~\ref{fig:emctower}.

\begin{figure*}
\begin{center}
\includegraphics[width=\textwidth]{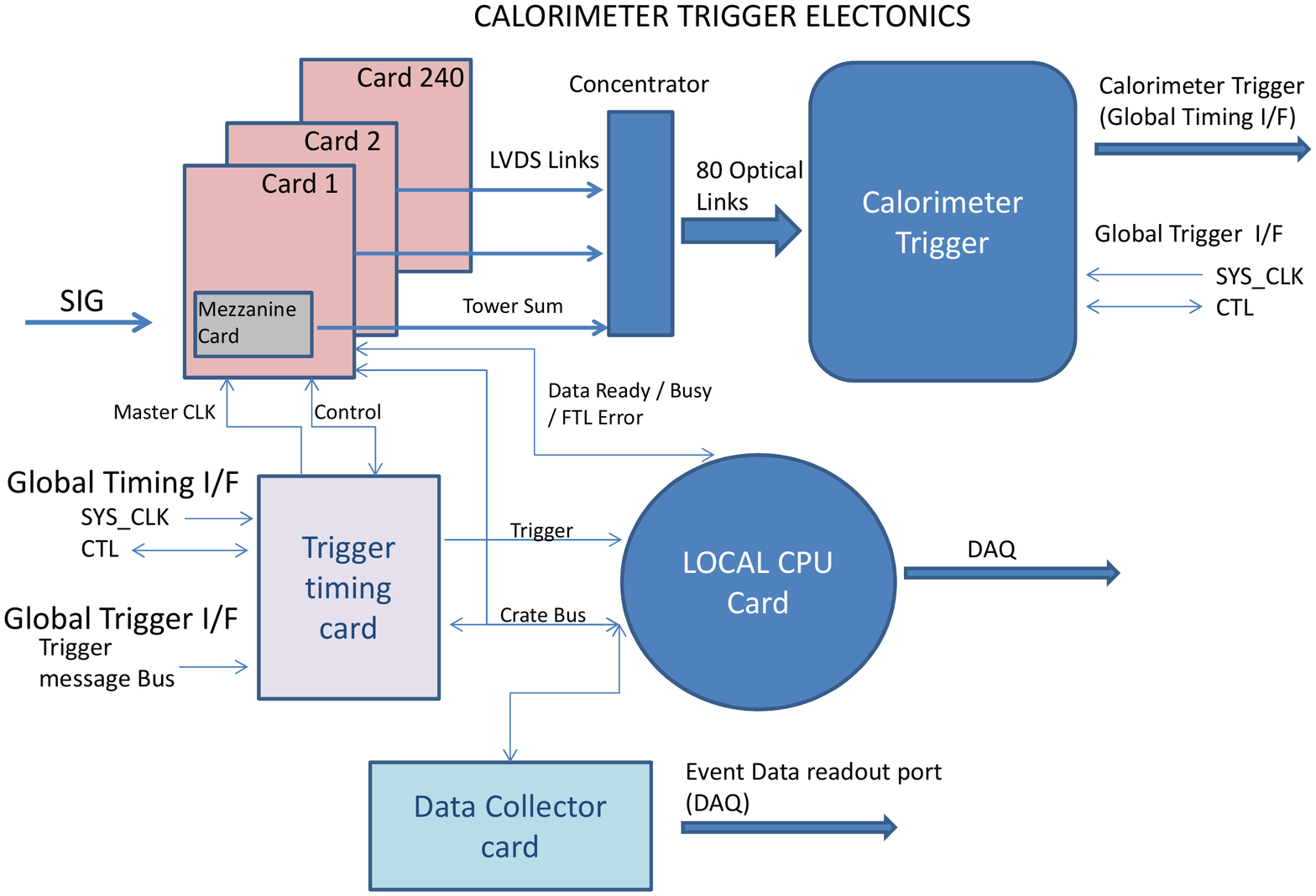}
\caption{EMT overview}
\label{fig:emctrigger}
\end{center}
\end{figure*}

\begin{figure}[tbp]
\includegraphics[clip=true,trim=30 100 60 100,width=0.5\textwidth]{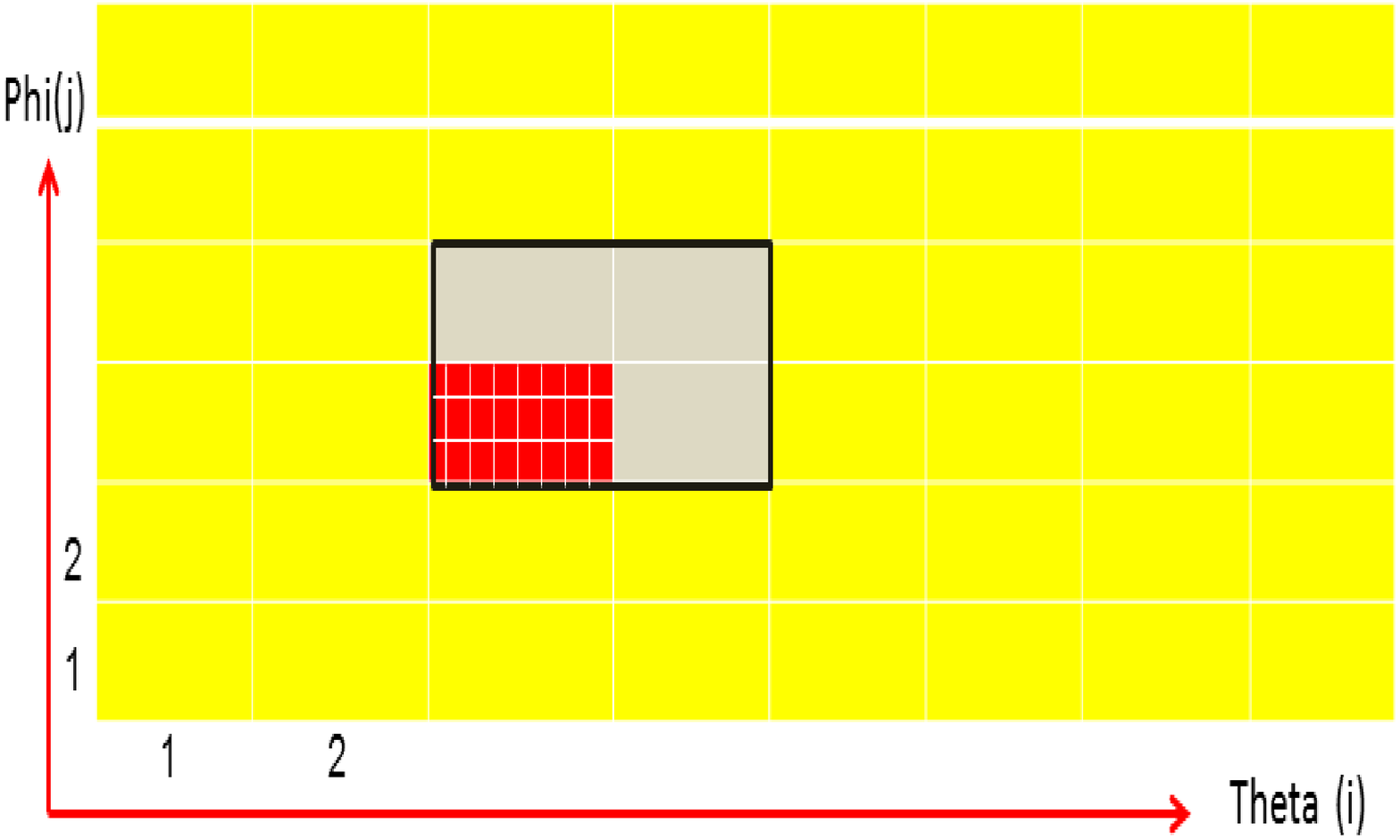}
\caption{EMT trigger tower layout. A tower (shown in red) in the
  barrel calorimeter is comprised of 24 crystals. The black rectangle
  shows an example of 4 neighboring towers that need to be considered
  when calculating cluster energies. }
\label{fig:emctower}
\end{figure}

The number of crystals in the barrel is 5760; this implies that the
trigger box will handle 240 towers in the barrel.  The trigger-level
configuration of the end-cap has not been finalized yet, but we expect
a maximum of 60 towers in the end-cap.  Two different strategies are
under study, both generating analog sums at the tower level.  The
first strategy uses the signals from the readout pre-amps, the other
strategy foresees to equip each crystal with independent photon
detectors (such as SiPM).

The tower-level sums are calibrated using look-up tables, encoded,
and sent via LVDS links to fiber-optic transceivers shielded from
radiation.

\tdrparagraph{Particle Finders and global cluster formation:} 

The energy deposit of a shower can be spread over adjacent towers,
therefore, neighboring towers will need to be considered in the
calculation of the cluster energy. Neighboring towers are scanned,
and if their energy deposit is above a threshold of interest, it is
added to the cluster energy.


As in \babar, there will be 3 separate energy thresholds for clusters
in the trigger: M cluster (above a low energy consistent with a
minimum ionizing), G cluster (above 500-600 MeV) and cluster (electron
Bhabha). The thresholds are all programmable. Bhabhas can be vetoed or
pre-scaled.  This algorithm will be implemented by the L1 EMT
processor.

\tdrparagraph{Particle Counters:} 
Since clusters can be dynamically defined this allows to apply
isolation cuts on particle definition.

\tdrsubsubsec{L1 DCT and EMT trigger processors}

A crate based on VME or ATCA technology will host the L1 processors.
They will be based on FPGA technology, will have the same optical and
electrical interfaces for both DCT and EMT, and will only differ in
firmware. Track finding and dynamic super-cluster definition will be
implemented on the L1 processor backplane.

\tdrsubsubsec{Global Trigger (GLT)}

The GLT processor combines the information from DCT and EMT (and
possibly other inputs such as a Bhabha veto) and forms a final trigger
decision that is sent to the FCTS. The GLT receives the information
from the sub-detectors participating in the trigger decision through an
optical link.

The trigger can be asserted by considering the information delivered
by the DCH and the EMC either separately or combined. This module
combines the information of the detectors and compares them to
pre-defined criteria. Through the use of an FPGA the trigger can be
fully programmable and upgradable.

\tdrsubsubsec{L1 Trigger Latency}

The \babar\ L1 trigger had 12\mus latency. However, since size and
cost of the L1 latency buffers in the sub-detectors scale directly with
trigger latency, it should be substantially reduced, if possible.  L1
trigger latencies of the much larger, more complex, ATLAS, CMS and
LHCb experiments range between 2 and 4\mus, however these experiments
only use fast detectors for triggering. Taking into consideration that
the DCH adds an irreducible latency of about 1\mus due to drift-time
and read-out, and adding some latency reserve for future upgrades, we
currently estimate a total trigger latency of no more than 6\mus.
More detailed engineering studies are required to validate this
estimate.

\tdrsubsubsec{Monitoring the Trigger}

To debug and monitor the trigger, and to provide cluster and track
seed information to the higher trigger levels, data information
supporting the trigger decisions is read out on a per-event basis
through the regular readout system.  In this respect, the low-level
trigger acts like just another sub-detector.

\tdrsubsubsec{L1 Trigger R\&D}
We will study the applicability of this baseline design at \superb\
luminosities and backgrounds, and will investigate improvements, such
as adding a Bhabha veto to the L1 trigger.  We will also study faster
sampling of the DCH and the impact of the forward calorimeter design
choice. For the barrel EMC we will need to study how the L1 trigger
time resolution can be improved and the trigger jitter can be reduced
compared to \babar. In general, improving the trigger event time
precision should allow a reduction in readout window and raw event
size. Other improvements might include the improvement of tracking and
clustering algorithms, or by exploiting better readout granularity in
the EMC.

\tdrsubsec{Fast Control and Timing System}


\begin{figure*}[t]
\begin{center}
\includegraphics[width=0.95\textwidth]{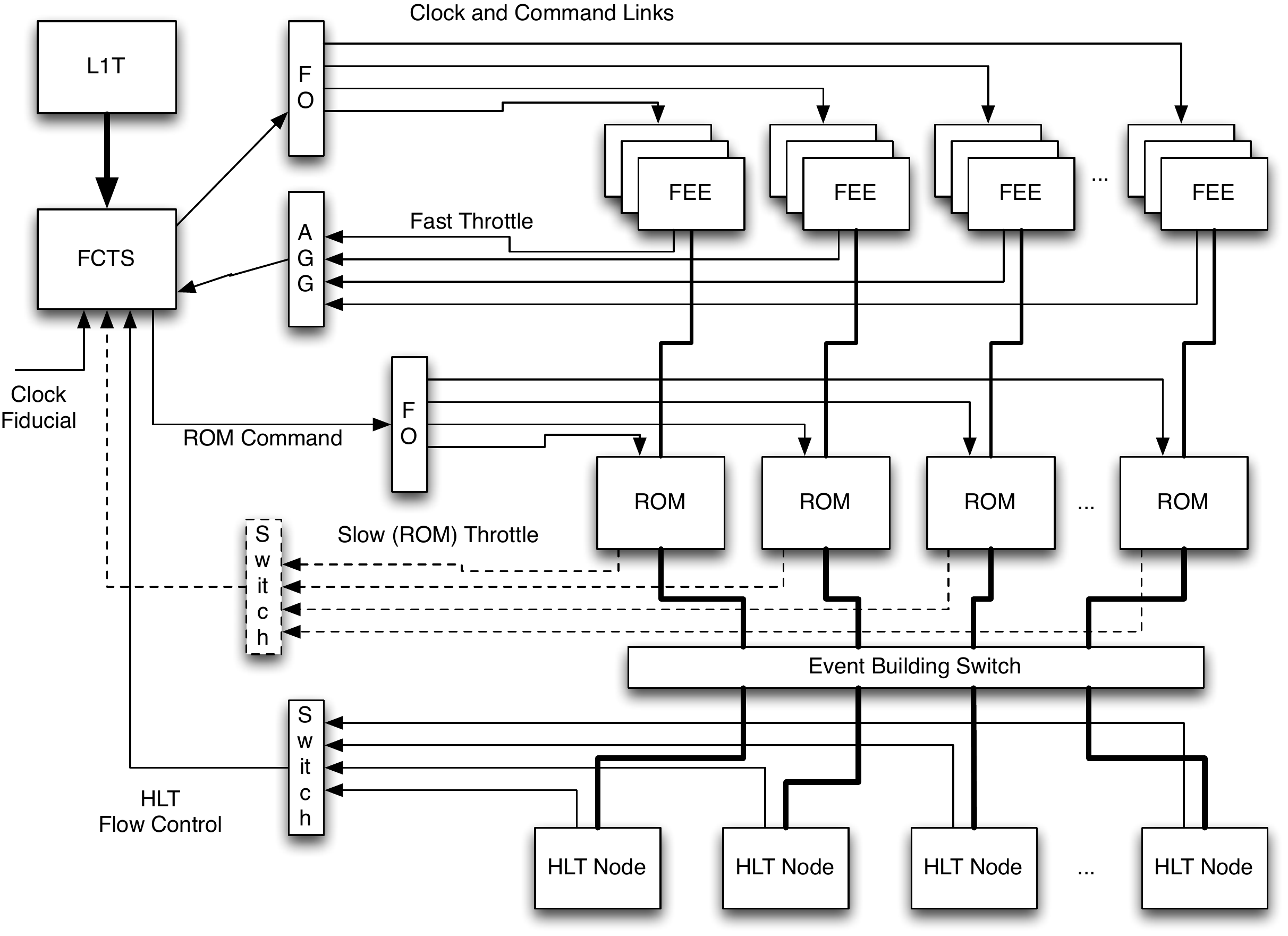}
\caption{Overview of L1T/FCTS/ROM/HLT integration}
\label{fig:etd-fcts-overview}
\end{center}
\end{figure*}

The Fast Control and Timing System (FCTS) manages all elements linked
to clock, trigger, and event readout, and is responsible for
partitioning the detector into independent sub-systems for testing and
commissioning. Figure~\ref{fig:etd-fcts-overview} shows how the FCTS is
connected to the L1 trigger, FEE, ROMs and HLT.

\begin{figure*}[t]
\begin{center}
\includegraphics[width=0.95\textwidth]{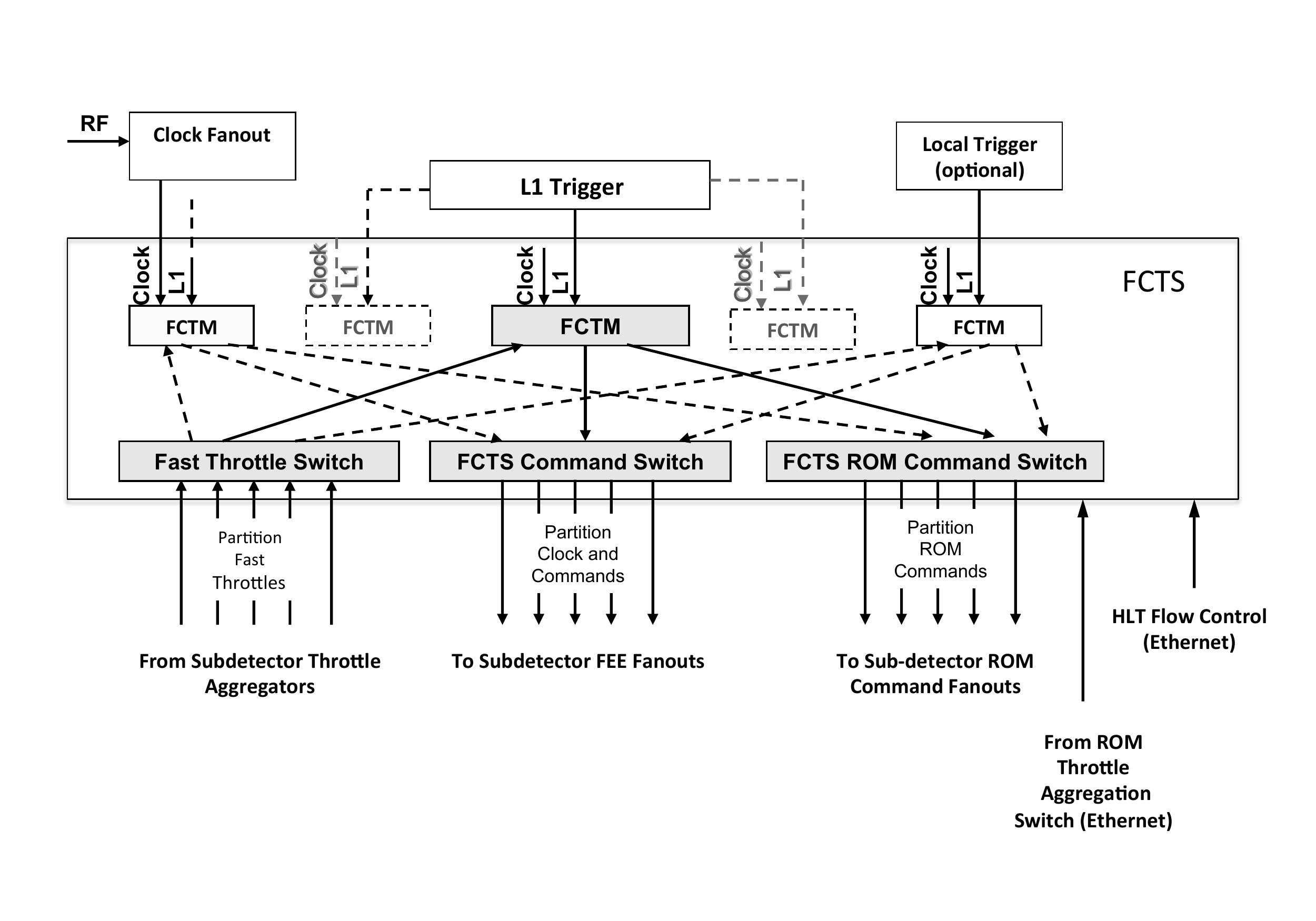}
\caption{The main functional blocks in the FCTS crate}
\label{fig:etd-fcts-crate}
\end{center}
\end{figure*}

The FCTS will be implemented in a crate where the backplane can be
used to distribute all the necessary signals in point-to-point mode.
This permits the delivery of very clean synchronous signals to all
boards --- avoiding the use of external cables.
Figure~\ref{fig:etd-fcts-crate} shows the main functional blocks of the
FCTS crate. The Fast Control and Timing Module (FCTM, shown in
Figure~\ref{fig:etd-fctm}) provides the main functions of the FCTS.

\tdrparagraph{Clock and Synchronization:} The FCTS synchronizes the
experiment with the machine and its bunch timing, buffers the clock
and distributes it throughout the experiment, and generates
synchronous reset commands.

The baseline configuration is to distribute the clock through the FCTS
command links. The deserializers will extract the clock from the
serialized bitstream and report any loss of the clock signal or
lock. Possible mechanisms to recover from loss-of-clock conditions are
still under investigation and might include the capability of
individual FEE to regain synchronization without major interruptions
of the data taking. Should reliability concerns arise during the final
design stage, we will consider a separate, dedicated clock
distribution network as a backup solution.

To globally synchronize the system, all FEE must support the ability
to reset all clock dividers, state machines and pipelines at system
initialization or system reset to ensure that all counters and divided
versions of the global clock are properly synchronized. This can be
achieved through a global reset command broadcast by the FCTS.
Alternatively, individual FEEs may be instructed by the ECS to
resynchronize.

\tdrparagraph{Trigger Handling and Throttling}

The FCTS receives the raw L1 trigger decisions, the fast throttle
signals from the sub-detector FEE and slow throttle signals from the
ROMs, generates readout commands and broadcasts them to the FEEs and
the ROMs. The fast throttle is needed because during a spike in
instantaneous rate or event size, data generated by the FEE can exceed
the bandwidth of the data links and fill up the derandomizer buffers. An
FEE generates a throttle signal when the free space in the
derandomizer buffer falls below a certain threshold determined by the
throttle latency. The throttle signal is removed when enough space in
the derandomizer buffer has been freed up.

Assertion of a throttle signal temporarily inhibits the generation of
L1-accept commands in the FCTS and thus stops the readout of new
events. Any L1-accepts that are already in-flight when a throttle is
asserted will still need to be absorbed by the derandomizer buffers.
In addition to acting as a rate limiting mechanism, the fast throttle
can also be used by the FEEs to request an ``emergency stop'' of the
event readout.

As a baseline, the fast throttle will be implemented as a 1-bit level
without any framing or encoding. This minimizes the complexity of
generating, aggregating, distributing and monitoring the signals. It
also allows to design a simple throttle aggregator device that can
easily be standardized across all subsystems.

Every throttle line is monitored by measuring the fraction of time a
throttle signal is asserted, and by counting the number of throttle
on/off transitions. Detailed monitoring of the throttle-aggregation
devices and detailed sub-detector specific diagnosis of the reasons
for a throttle/stop can be performed through the ECS.

Since there is no further local flow control mechanism on the data
links, the ROMs cannot assert backpressure through the FEEs but need a
separate path to slow down or stop the trigger when they are unable to
offload the data through the event builder. Since there is plenty of
buffer space in the ROMs, this throttle (``slow'' throttle) does not
have stringent latency requirements and can be implemented using
Ethernet.

\tdrparagraph{L1-accept Command Format and Distribution}

The commands broadcast to the FEEs have a strict fixed-latency
requirement and are as simple and short as possible to minimize the
achievable temporal inter-command spacing during transmission and the
amount of decoding required in the FEEs. We foresee a command word of
at least 16 bits, that in addition to the front end command code and
parameters contains an event tag of at least 8 bits to allow the ROM
to match the event fragment with its corresponding ROM command
word. The exact number of bits in the event tag (and thus the FEE
command word) will depend on the details of the clock and command link
protocol for framing and error detection.  The FEEs are required to
include the FEE command word with each event fragment they send to the
ROMs.

The commands distributed to the ROMs are not subject to a
fixed-latency requirement. In addition to a copy of the command word
sent to the FEEs they contain at least an event time\-stamp (minimum
of 56 bits), the full trigger word (32 bits) and the HLT destination
node (10-12 bits). Additional information such as a per-run event
counter or the time since the last trickle injection pulse in the HER
or LER might be included in the ROM command word as well. Since
latency is variable, the ROM commands can be derandomized by queuing
them before transmission; this means that the ROM command links only
need to sustain the average rate (150kHz) of ROM commands.

FEE and ROM commands are sent in the same order (parallel pipelines).
Suitable ways of handling lost or corrupted FEE commands will have to
be developed, however this will depend on the detailed failure modes
of clock distribution and command links.

\tdrparagraph{Event Management:}

The FCTS generates unique event identifiers, manages the assignment of
events to nodes in the HLT farm and uses a per-HLT-node sliding window
protocol to load-balance the HLT farm and to stop sending events to
unresponsive HLT farm nodes. For this, the FCTS manages a per-node
counter of events it is still allowed to send to this node. A node's
counter gets decremented with each event sent to that node. The FCTS
determines the next destination node by searching for the next
non-zero counter. Every HLT node asynchronously sends ``generic event
requests'' with the number of events it is willing to take --- when
the FCTS receives such a request from a node it updates the
corresponding counter with the value contained in the request. The
ROM/HLT/FCTS protocol is described in more detail in the network event
builder section below. Handling the event distribution with the help
of the FCTS minimizes the complexity of the ROMs and the event
building protocol.

\superb\ events are large enough that in an Ethernet-based event
builder we will not need to batch events into multi-event-packets
(MEPs); the system, however, does not preclude the addition of a MEP
scheme.  Such a scheme might be required to send events that are very
close in time to the same HLT node or to accommodate a non-Ethernet
event building network that necessitates transmission of data in
significantly larger units.

To generate a MEP, the FCTS would simply send the same HLT
destination address for subsequent events which would then be packed
into the same MEP (the packing can in fact be determined by the
individual ROM). Sending a ROM command with a new HLT destination will
then close the current MEP and open a new one.

The FCTS also keeps track of all its activity, including an accounting
of triggers lost due to FEE throttling or other sources of trigger
inhibits. 

\tdrparagraph{Calibration and Commissioning:} The FCTS can trigger the
generation of calibration pulses and flexibly programmable local
triggers for calibration and commissioning. To support the EMC source
calibration where a meaningful global trigger can not be constructed,
the system can run in a special mode in which the readout of
individual EMC channels is self-triggered, and the resulting waveforms
are aggregated and collected under the control of the FCTS.

\begin{figure*}[tbp]
\begin{center}
\includegraphics[width=\textwidth]{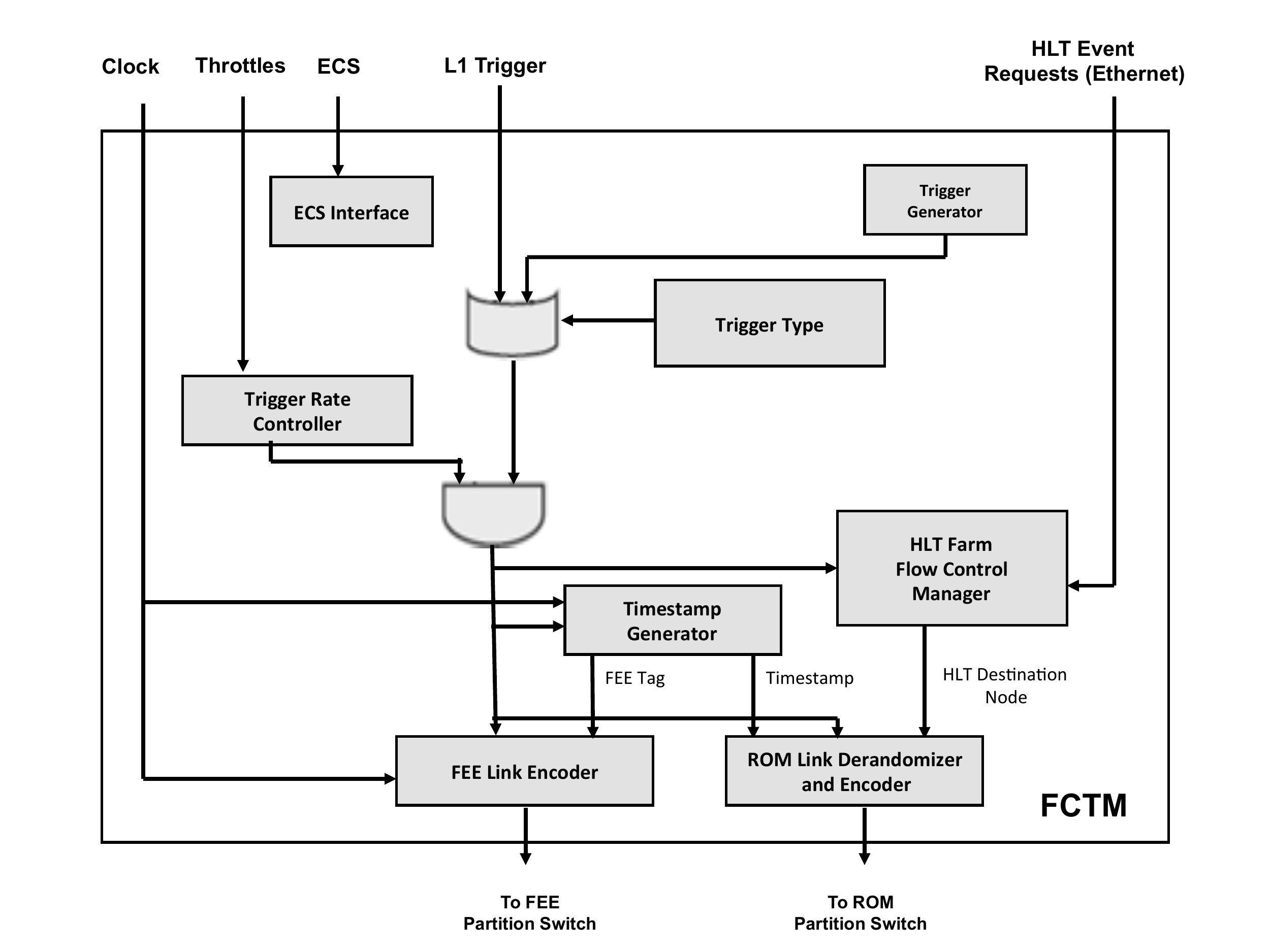}
\end{center}
\caption{Fast Control and Timing Module}
\label{fig:etd-fctm}
\end{figure*}

\tdrparagraph{Partitioning:}
The FCTS crate includes as many FCTM boards as required to cover all
partitions. One FCTM will be dedicated to the unused sub-systems in
order to provide them with the clock and the minimum necessary
commands.

As shown in Figure~\ref{fig:etd-fctm}, three dedicated switches are
required for partitioning the system into independent sub-systems or
groups of sub-systems: One switch each for FEE clock and commands, ROM
commands, and fast throttle feedback from the FEEs.

\tdrsubsec{Control and Data Links}

\tdrparagraph{Requirements}

High-speed optical links will be used for data transfer, triggering
and fast control distribution. The clock recovered from the serial
streams will be used to synchronize the whole experiment.
  
Considerations based on event size, average expected trigger rate,
topology, and cost of the link components, suggest to choose a
per-link throughput in the 1-3 \gbitps range.

\superb\ policy explicitly requires the use of COTS components where
available, avoiding the design of application specific integrated
circuits (ASIC) where possible. 

Since the whole ETD is designed to work synchronously with a
frequency-divided machine clock, the latency and the recovered clock
phase of serial links deployed for trigger and fast control
distribution should be fixed and predictable, at a level below 1 unit
interval. This requirement does not apply to read-out links, where the
latency can be variable.

Moreover, the line coding should be DC-balanced in order to allow
AC-coupled transmissions where needed and in such a way not to stress,
and consequently not reduce the lifetime of, optical transmitters.

\tdrparagraph{Radiation Tolerance Issues}

Because the high-speed optical links are a key element of the ETD,
their electrical and optical components must withstand the expected
radiation levels.

According to preliminary simulations of the luminosity-scaling
background component (see Table~\ref{tab:fullsim_samples}), and under
the assumption that no links will be installed in the high-radiation
regions closer to the beam than the innermost layer of the DCH, the
worst-case Total Ionizing Dose (TID) is 7.6 Gy(Si), and the neutron
and hadron (E>20\mev) fluences are $3\cdot10^{11}$cm$^{-2}$ and
$7.6\cdot10^{10}$cm$^{-2}$, respectively.

R\&D has been performed to identify the most reliable devices in terms
of radiation tolerance among those compatible with the \superb\
requirements. These included the irradiation testing (with 62\mev
protons) of Serializer-Deserializers (SERDESes), both standalone
(Texas Instruments TLK2711A \cite{TLK2711_ds}, DS92LV18
\cite{DS92LV18_ds} and DS92LV2421 -serializer only-
\cite{DS92LV2421_22}), and embedded in SRAM-based FPGAs (Xilinx
Virtex-5 \cite{GTP_ug} and Virtex-6 \cite{GTX_ug}). The DS92LV18 has
also been tested with gamma rays from a $^{60}Co$ source.  For the
optical part of the link, a VCSEL-based (Avago AFBR-57R5APZ
\cite{AFBR_ds}) transceiver has been tested.

\begin{figure*}[t]
\begin{center}
\includegraphics[width=0.95\textwidth]{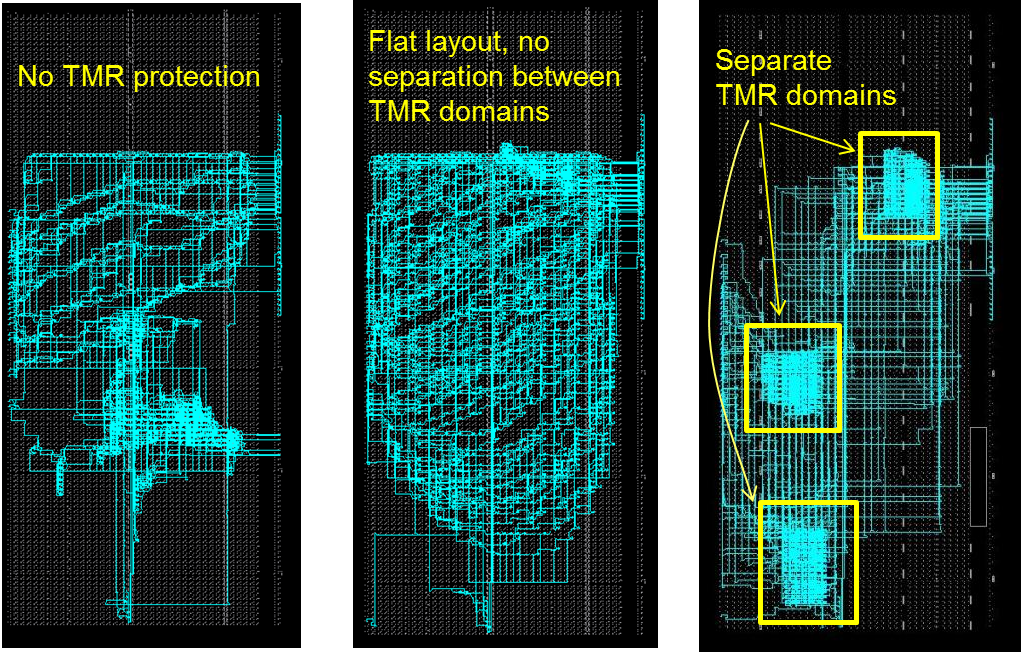}

\caption{\label{fig:link_Layouts}Layouts of the serial link implementation
in a Xilinx FPGA V5LX50T. Left: link implementation without TMR
protection.  Center: TMR-protected version with flat layout. Right:
TMR-protected version with separation of the TMR domains in distinct
areas.}
\end{center}
\end{figure*}

The links implemented with SRAM-based FPGAs have also been protected
with triple modular redundancy (TMR) and configuration scrubbing
mitigation techniques. The layout of the TMR-protected version has
been optimized with placement-hardening algorithms in order to produce
different circuit layouts (Figure \ref{fig:link_Layouts}), which have
been tested.

Before discussing the irradiation results, it is worth remarking that
the optical links are a chain of different devices (SERDES, optical
transceiver, fiber) and the failure of any component impacts the whole
chain. In general the irradiation with 62\mev protons generates both
total ionizing dose and single event effects.

For the different devices the following dose effects were observed:

\begin{itemize}
\item {\bf TLK2711A}: failed persistently (not recoverable by cycling
power) after absorbing a dose of $\sim$ 400 Gy (Si).

\item {\bf DS92LV18}: did not fail either during testing with protons, or
with gamma rays. For a total dose of 2.5 kGy(Si) the only effect
observed was a moderate ($\sim$ 0.5\%) variation in power dissipation.

\item {\bf DS92LV2421/22}: did not fail persistently during the testing and
did not show any measurable current variation due to TID.

\item {\bf Xilinx V5 and V6 FPGAs}: did not exhibit TID-related variations of
the power dissipation, however, their power consumption increased
gradually during the accumulation of Single Event Upsets
(SEUs). Reconfiguration of the devices restored the initial power
consumption.

\item {\bf Avago AFBR-57R5APZ}: failed persistently at 400 Gy (Si).
\end{itemize}

\begin{figure}[tbp]
\begin{center}
\includegraphics[clip=true,trim=0 0 10 0,width=0.5\textwidth]{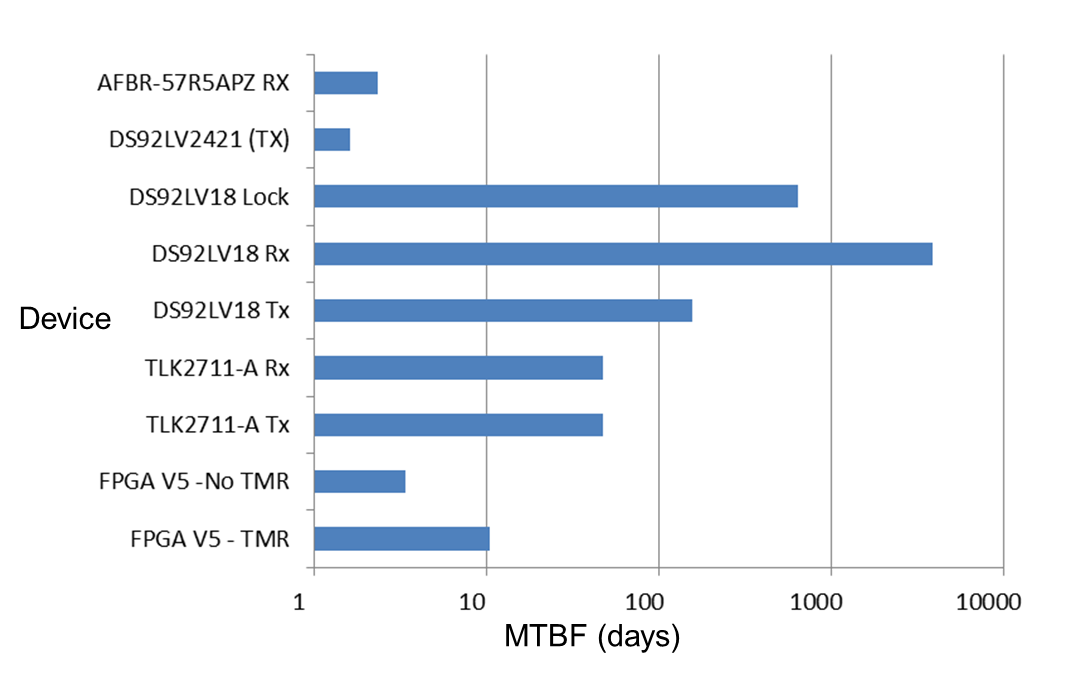}
\caption{\label{fig:Expected-failure-rates}Expected failure rates at
  \superb\ for a fluence of $3\cdot10^{11}$cm$^{-2}$ 62\mev protons in
  one effective year ($10^{7}$s ). }
\end{center}
\end{figure}

Concerning transmission errors related to single event effects, Figure
\ref{fig:Expected-failure-rates} shows the expected mean time between
failures (MTBF) for each tested device, considering a reference fluence
of $3\cdot10^{11}$cm$^{-2}$ 62\mev protons per effective year ($10^{7}$
s). In the following, we use the word {}``failure'' to refer to
the following: 
\begin{enumerate}
\item for a standalone SERDESes, to a single bit error either in the transmitter
or in the receiver (for the DS92LV18 also to a loss of lock);
\item for a FPGA-embedded SERDES, a persistent failure of the link requiring
the reconfiguration of the device.
\end{enumerate}

The separate MTBFs for the transmitter and receiver parts of the
TLK2711A and of the DS92LV18 are reported. For the DS92LV18, the rate
of losses-of-lock is also shown. The Virtex-6 FPGA MTBF is not shown
since it is lower than $10^{-1}$days.

These results show that optical receivers are expected to be the main
source of errors (we did not find any error in the transmitters during
testing). For the reference fluence of $3\cdot10^{11}$cm$^{-2}$62\mev
protons, a MTBF of 2.3 days is expected for errors (either single or
burst) in the optical receiver. Among the SERDESes, the most reliable
are the DS92LV18 and the GTP embedded in V5 FPGAs, when protected with
TMR (we do not consider the TLK2711, since it has a low tolerance to
TID). The DS92LV18 is limited to 1.32 \gbitps and it adopts a
proprietary line encoding, therefore only the payload, rather than the
full frame, can be protected by means of external EDAC components. The
line protocol of the device includes a 20-bit symbol consisting of a
start bit, an 18-bit payload and a stop bit.  When one of these bits
is incorrect, the receiver loses the lock to the stream and the
recovered clock stops toggling.  Since the errors in the optical
receiver are uniformly distributed on the parallel symbol, there is a
2/20=0.1 chance of hitting a start or a stop bit and losing the lock
at the deserializer level.

Due to the line protocol of the DS92LV18, on average 10\% of the
incoming bit errors from the optical receiver will result in a
loss-of-lock from the stream, also with the loss of the recovered
clock. This means that the effective mean time between two losses of
lock for the DS92LV18 will be lowered to nearly 23 days (from the
value of 640 days related to the SERDES alone). On the other hand, the
FPGA-based links do not lose the lock due to single or (short) bursts
errors in the optical receiver and for the TMR-protected versions the
failure rates are in the order of $10$ days. The results in terms of
MTBF suggest that link failures are unavoidable and therefore the
whole ETD should be implemented as a fault-tolerant machine. This is
further discussed in each ETD component subsection.

However, in the light of the present irradiation results the
electrical part of the links should be based either on Virtex-5 FPGAs
or on the DS92LV18. For both these devices the fixed-latency operation
required by the trigger and fast control links is guaranteed. Although
tested at a line rate of 2 \gbitps, the links in V5 FPGAs can run up
to 3.125 \gbitps (limited by the embedded SERDES throughput) and the
coding, including also error detection and correction (EDAC) schemes,
can be completely customized by implementing the required logic in the
fabric.

\tdrparagraph{Error Detection and Correction}

Two common solutions are error \emph{correcting} codes (ECC) or the
simpler error \emph{detecting} codes (EDC). The ECC solution, for
example using a Reed Solomon corrector, appears to be too intensive
with regard to the computational effort of both encoding and decoding,
resulting in increased hardware complexity, longer link latency and
reduced reliability.  The expected low error rate will permit a
simpler and faster solution in the form of a proper EDC code. This
will simply flag the occurrence of the error in the transmitted block,
so that the block can just be discarded. As shown below, the expected
frequency of this kind of error is extremely low, thus, bit errors
will not affect data quality and event reconstruction. 

As far as the choice of the error detection code is concerned, both
cyclic redundancy check and check-sums have been evaluated. Since CRC
coding has a complexity that is comparable to a full ECC approach, it
is not a viable option.

Checksums have a lower error detection ability than CRCs but have the
advantage of a lower hardware complexity that meets the latency
requirements and allow for a widespread use in the experiment,
increasing the overall reliability. A parameter to be evaluated is the
overhead in terms of ratio of the number of check-sum bits added to
the useful message bits.

Let us suppose to deploy the DS92LV18 SERDES for the optical links. In
this case, the 18 bits block managed as a whole by the SERDES will be
buffered in groups of 3 blocks to be protected as a single 54 bit
super-block. In order to achieve the best possible DC balance, the
super-block will then be scrambled, checksum bits will be added and
the data will be transmitted by the SERDES and the optical link. On
the receiving end, the inverse process will be applied and possible
errors will be flagged.

Tools have been developed in order to evaluate error detecting
efficiencies of different checksums, such as \emph{Parity},
\emph{Modular Sum}, \emph{One Complement Sum} and \emph{Fletcher Sum}.
The choice will be the adoption of Fletcher check-sum which provides
detection of all 1 bit errors and two bits burst errors. Less frequent
2 or 3 bits errors are also detected even if not with the same 100\%
efficiency. Hardware complexity is limited to simple binary adders
using the so called one's complement addition.  Using the 54 bits
block including a 6 bits check-sum, overhead is as low as 12.5\%.
Error detection probability, evaluated by direct simulation mixed with
proper analytical approach, provides the following figures for the
detection efficiency: 1 bit errors 100\%, 2 bits errors 98.7\%, 3 bit
errors 98.0\%, 4 bit errors 98.1\%. At the same time complete coverage
is provided for burst errors encompassing 2-3-4 bits, an invaluable
feature in our case foreseeing these kind of errors. The above
figures, for a reference Bit Error Ratio BER=$10^{-10}$ will deliver a
probability of undetected error as low as $1.5\cdot10^{-19}$.  This
undetected error rate, in turn, for a reference 2.5 \gbitps
transmission speed, will deliver 8 years average time between
undetected errors in a single link; for 1000 links this will deliver 3
days average time between undetected errors in the whole apparatus.

\tdrparagraph{Physical Implementation }
Serial links could be implemented on a plug-in mezzanine board (a board which plugs onto a carrier mainboard in order to extend its functionality). One of the main advantages of this approach is an easier maintenance and possibility of upgrading the link performance (data rate, power, jitter) by simply replacing mezzanine cards. This also guarantees protection against obsolescence of components. A single design group would be in charge of characterizing performance and testing of the links, while other ETD boards might be designed by other groups independently in the meanwhile.
The link-related issues will be decoupled from the logic-related ones, leaving critical issues (latency, jitter) to be solved at the mezzanine level. An higher design efficiency will be achieved since the mezzanine design group will focus on a 'critical but single\textquoteright{} board. A single printed circuit board (PCB) will be customized with different components to fit different needs (e.g. jitter cleaners or rad-hard SERDES for on detector nodes, copper and/or optical lanes where needed).

The implementation of serial links, as separate systems, by means of
mezzanine cards is a widespread solution. Examples of this trend are
the timing trigger and control (TTC) \cite{TTC} system of the LHC, the
S-Link~\cite{SLink} deployed in several CERN experiments, and the SODA
\cite{SODA} timing distribution system adopted in the PANDA
experiment. In the TTC system, the implementation on a mezzanine
allowed to upgrade the receiver with a VCXO-based PLL for improved
output jitter ($\sim$ 20 ps rms), compatible with the requirements of
the reference clock of \gbitps SERDESes and other jitter sensitive
devices.

Due to the limited space in the FCTS crate, the mezzanines for the FEE
clock, command and throttle links will be housed in dedicated fan-out
crates that will also contain the logic to aggregate and monitor the
throttle signals. 

The ROM PCIe boards have similar space constraints, so a mezzanine
solution for the data links will require the mezzanines and circuitry
to convert the signals to e.g. LVDS to be housed in external crates.

\tdrparagraph{Backup Solutions}

In order to optimize the design effort, control and data links should
be based on the same components and mezzanine printed circuit boards.
Should the tested devices be found not to be compatible with the radiation
tolerance requirements of the experiment, the gigabit optical link
(GOL) \cite{GOL} transmitter could be considered as a backup solution
for data links. The radiation-hard GBT \cite{GBT} transceivers could
also be considered, if available within a time scale compatible with
the links realization phase.

\tdrsubsec{Common Front-End Electronics}


\begin{figure*}
\begin{center}
\includegraphics[width=\textwidth]{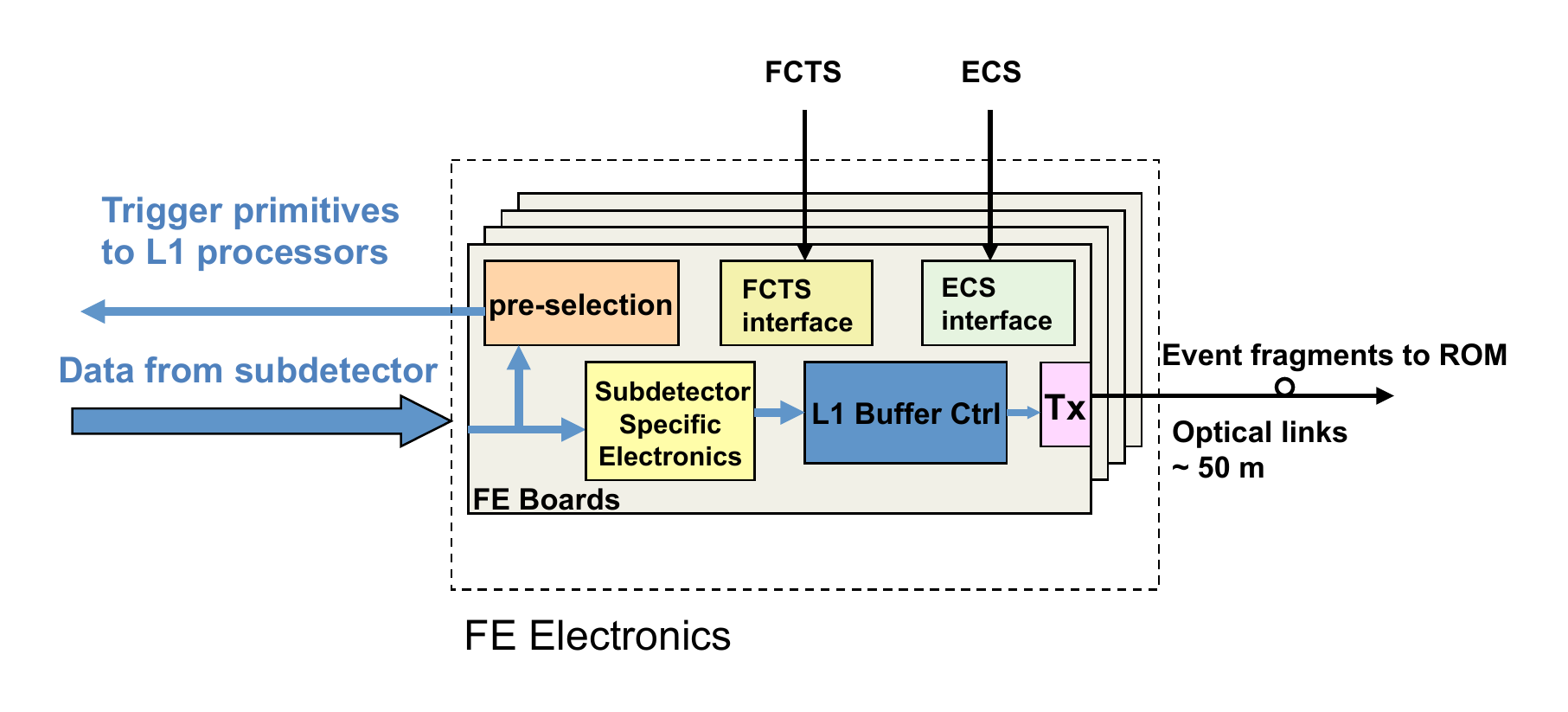}
\end{center}
\caption{Common Front-End Electronics}
\label{fig:etd-fex}
\end{figure*}

It will be beneficial to implement the different
functions required to drive the front-end electronics as independent
modular elements. These elements can for example be implemented as
mezzanines or as circuits directly mounted on the front-end modules.
For instance, as shown in Figure~\ref{fig:etd-fex}, one mezzanine can
be used for FCTS signals and commands decoding, and one for the ECS
management.

Depending on the implementation of the FEE, it is also possible to
decode the FCTS and ECS signals on one mezzanine board, and to
distribute them to neighboring boards, permitting a great reduction in
the number of links. Driving the L1 buffers may also be implemented in
a dedicated common control circuitry inside a radiation-tolerant FPGA.
This circuitry would handle the L1 accept commands and provide the
signals necessary to control the data transfer from the latency
pipeline into the derandomizer, and finally, the transmission of event
fragments to the serializer, as shown in
Figure~\ref{fig:etd-cfee-detail}.

 The latency buffers can be implemented either in the same FPGA
or directly on the carrier boards, and one such single circuit could
be able to drive numerous data links in parallel, thus reducing the
amount of electronics on the front-end boards.

The control circuit will also handle events that have overlapping
readout windows because they were triggered very close in time. The
strategy for this is that data shared between events with overlapping
readout windows is only read out and transmitted once.  Complete
events will then be reconstituted in the ROM by duplicating the data
from the overlapping region as necessary.

It is also the responsibility of the CFEE to manage the derandomizer
buffer and the fast multiplexer that feeds the data link serializer.


Figure~\ref{fig:etd-fex} shows a possible implementation of the L1
buffers, their control electronics and the outputs towards the optical
readout links. The control electronics may be located within a
dedicated FPGA.

Another important requirement is that all (rad-tolerant) FPGAs in the
FEEs have to be reprogrammable without dismounting a board. While this
could be done through dedicated front panel connectors which might be
linked to numerous FPGAs, the preferred solution is to program FPGAs
through the ECS system without any manual intervention on the detector
side.

\begin{figure*}
\begin{center}
\includegraphics[clip=true,trim = 0 95 0 20,width=\textwidth]{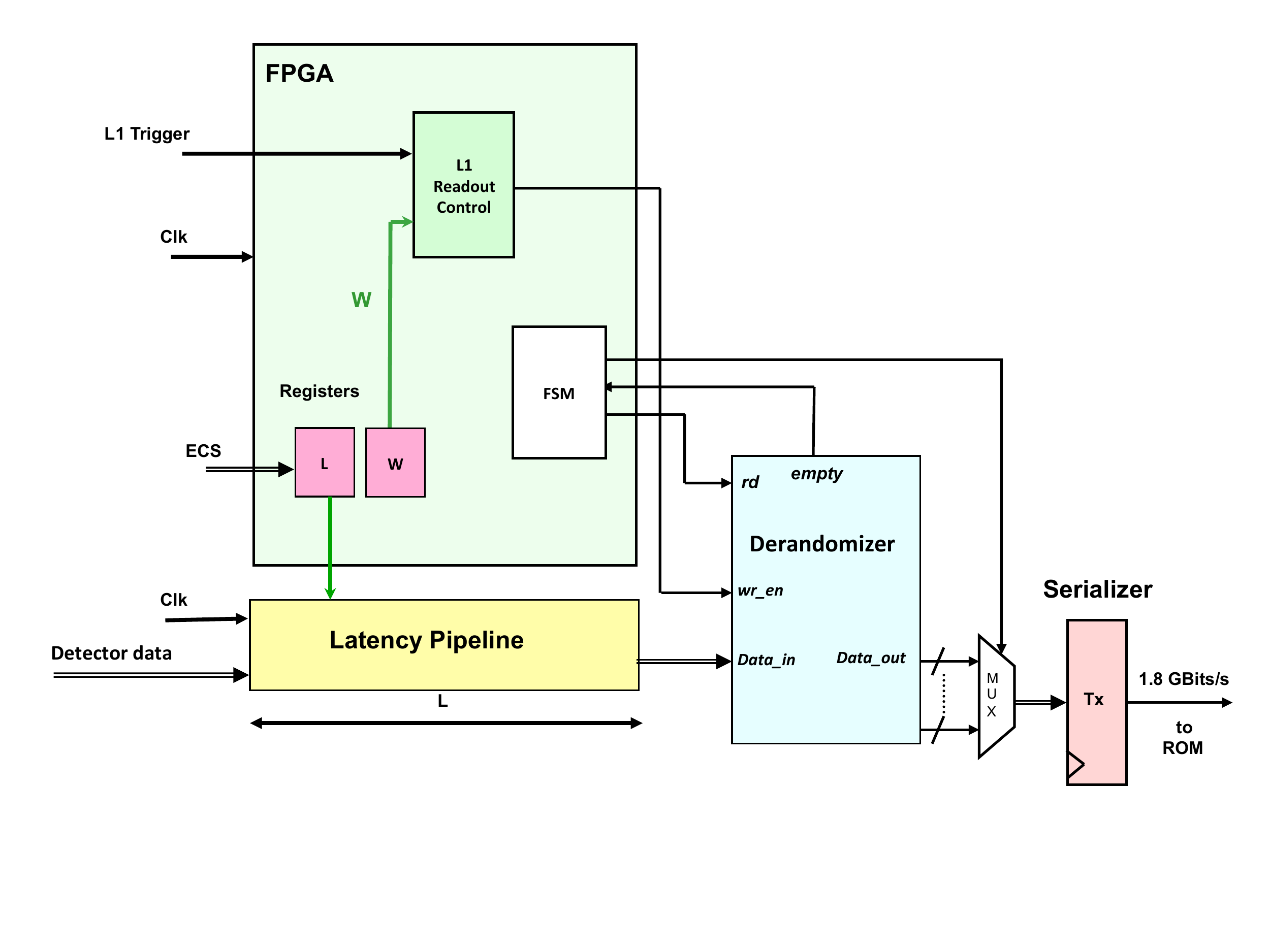}
\end{center}
\caption{Control of latency pipeline and derandomizer in the CFEE}
\label{fig:etd-cfee-detail}
\end{figure*}

\tdrsubsec{Read-Out Modules}


\begin{figure}[tbp]
\center{\includegraphics[width=0.45\textwidth]{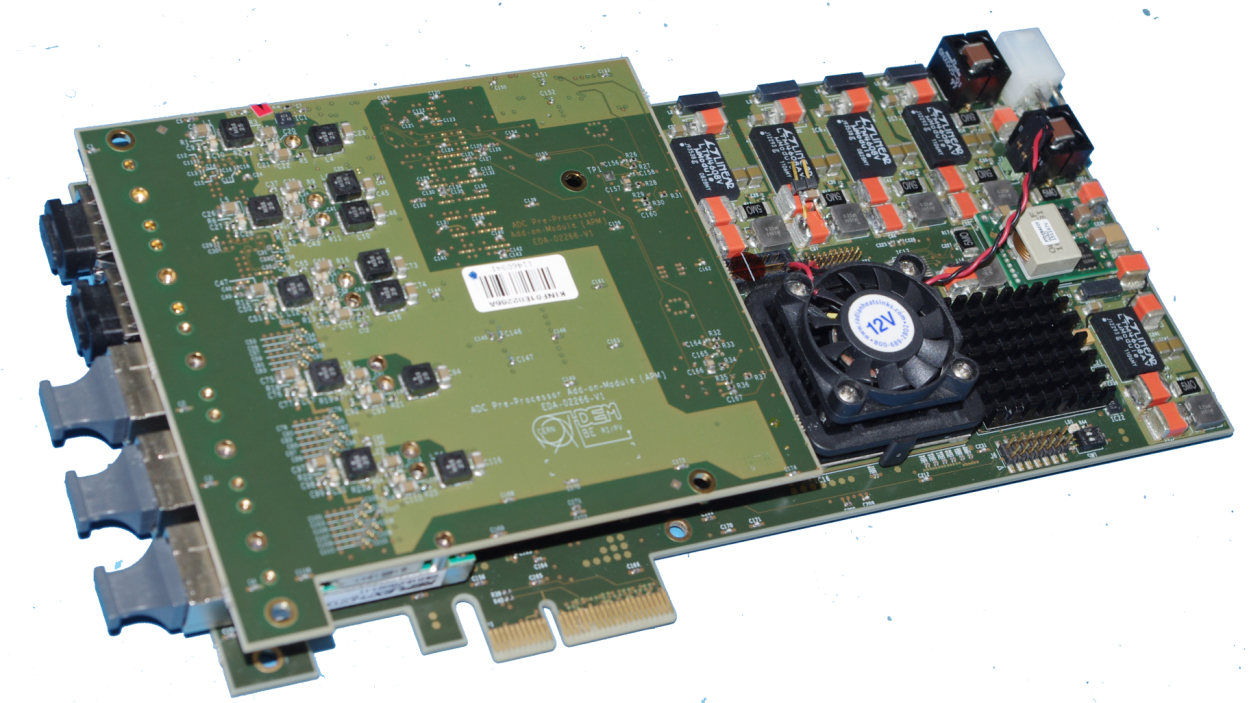}}
\caption{Photo of ROM prototype developed and tested for \superb.}
\label{fig:etd-rom-photo}
\end{figure}

The Readout Modules (ROMs) are the hardware interfaces between the FEE
and the event builder. The processing flow in the ROM starts with
receiving event fragments from the sub-detectors' front-end
electronics and the corresponding ROM command words from the FCTS.
Processing then continues by cross-checking front-end identifiers
and absolute time-stamps, buffering the event fragments in
derandomizing memories, performing processing (still to be defined) on
the fragment data, and eventually injecting the formatted fragment
buffers into the event builder and HLT farm. Processing includes
duplicating data that are shared between events because of overlapping
readout windows.

Connected to the front-end electronics via optical fibers, the ROMs
will be located in an easily accessible, no-radiation area. ROM
hardware is identical for all sub-detectors, however, if the need
arises, the peculiarities of each sub-detector can be addressed by
customization of the ROM firmware.

A ROM will have 10 or more optical input channels operating at a rate
of 1\gbitps and at least one output channel at 10\gbitsps. To exploit
flexibility and processing power of computer technology, the ROMs will
be manufactured as PCI Express (PCIe) add-on cards for servers in the
event-building farm. This approach helps to keep production costs low
(when compared to a field-bus based solution), allows the processing
of event fragment data with a combination of hardware (FPGA on the
ROM) and software (host CPU), and provides the flexibility to address
unforeseen changes and upgrades with a low impact on the overall
architecture.  It is important to note that the space on a PCI express
card is rather limited, both for connectors and mezzanines, so the
strategy of using mezzanines and the mechanical design has to be
reviewed carefully.

To assess the suitability of this approach, a prototype ROM has been
manufactured and qualified. It is a PCIe 2.0 add-on card
(Figure~\ref{fig:etd-rom-photo}) built around a Xilinx Virtex6-250T FPGA
with 48 high speed serial transceivers. The plug-in module
accommodates 3 SNAP12 optical receivers and 2 QSP optical transceivers
for a total of 44 receivers and 8 transmitters each operating at max
6.2\gbitps. An 8\mbyte true dual-port RAM is used to derandomize
incoming event fragments before they are streamed to host server
memory.

\begin{figure}[tbp]
\center{\includegraphics[width=0.5\textwidth]{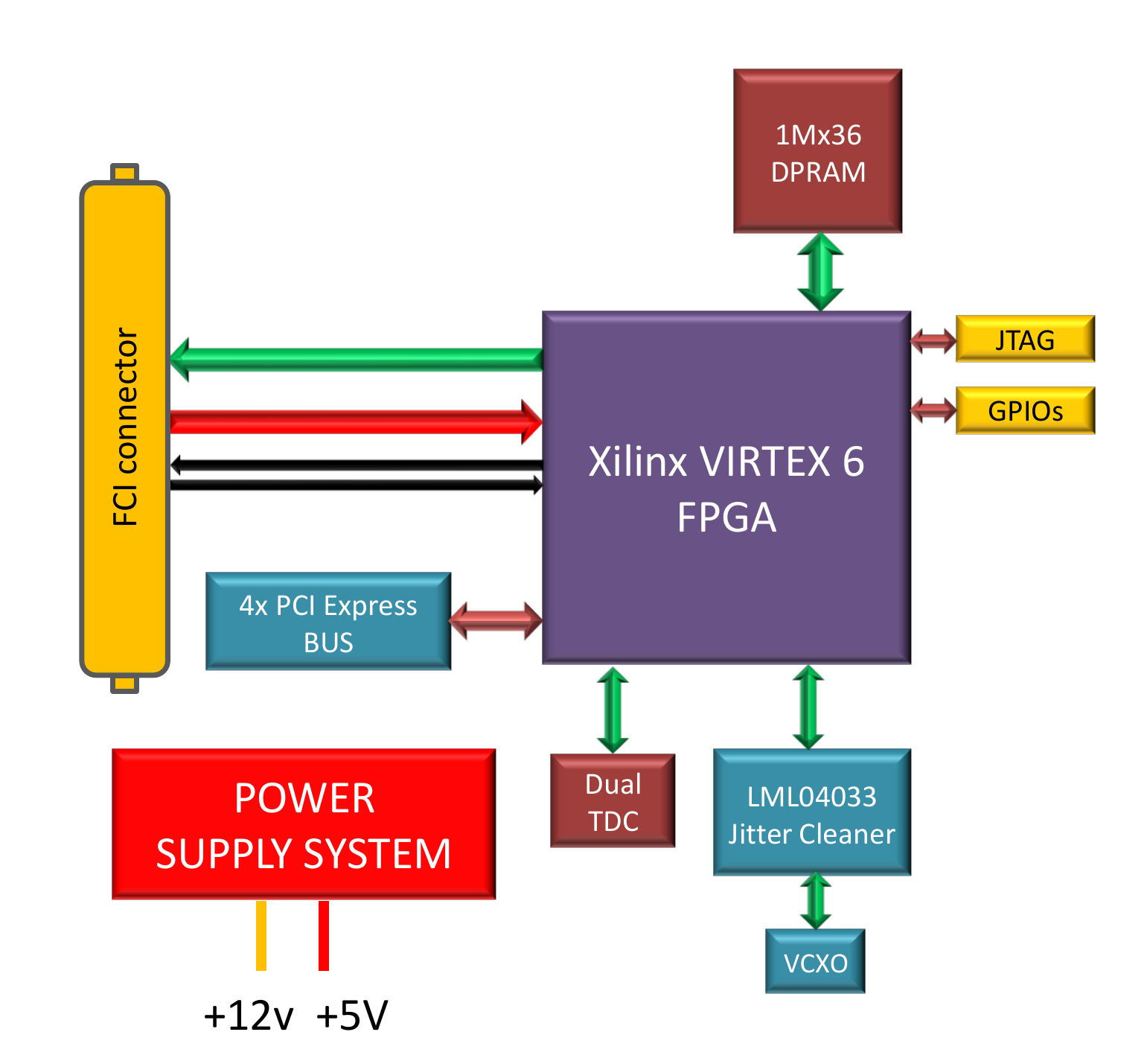}}
\caption{Block diagram of a ROM. The fiber-optic receivers are not shown.}

\label{fig:etd-rom-diagram}
\end{figure}

The prototype ROM features a TDC and a ultra-low jitter PLL for the
benefit of the FPGA (Fig~\ref{fig:etd-rom-diagram}). A 4x PCIe 2.0
interface with an aggregate bandwidth of 20\gbitsps per direction
allows fragment data to be moved to host memory efficiently.  To
measure the worst-case power consumption, the FPGA was configured with
38 input channels at 2.0\gbitsps, eight FIR filters per input channel
for data processing, a 10\gbitsps PCI-express interface, a
scatter-gather DMA engine, and a timing unit, resulting in a total of
92\% occupancy of logic resources. With this configuration the board
uses 60 W, and does not require any special cooling measures when
housed in a low-cost 2U Dell server.

Data transfers from the ROM's dual-port RAM to the host memory were
exercised through a custom-developed Linux driver. This driver uses a
custom designed scatter-gather DMA engine in the FPGA to efficiently
move the fragment data from the ROM directly into user-space buffers.
By handling the mapping between virtual pages and physical memory
regions, it allows the direct transfer of data from the dual-port RAM
into data vectors defined in user space, obviating the need for an
additional kernel space to user space copy (``zero-copy'').

Bandwidth tests show that with 4x PCIe 1.1, fragment buffers larger
than a few hundred bytes can be moved at a rate of approximately
950\mbytesps. With 4x PCIe 2.0 (using a 256 byte PCIe payload), this
rate almost doubles. Since these test results by far exceed the
requirements for the \superb\ ROMs, we are confident that the
PCIe-based ROM approach can be safely adopted.

\tdrsubsec{Network Event Builder}


The ROMs receive event fragments in parallel from the sub-detector
front-end-electronics, perform a first stage event-build, store the
event fragments in memory buffers, and add the corresponding ROM
command received from the FTCS on the dedicated ROM-command links.
The resulting partially-built events are then encapsulated in UDP
datagrams and sent to the HLT node determined by the destination node
field in the ROM command word. Thus, all fragments of an event are
sent to the same HLT node which completes the event building process
by combining them. Full events are passed to the HLT through
a queue. To utilize the multiple cores and CPUs on a HLT node, the
HLT can be implemented as multithreaded application or, alternatively,
as multiple independent processes working in parallel.

As described earlier, the FCTS determines the HLT destination node for
every event using a round-robin algorithm that takes into
consideration a simple flow control scheme based on a per-node sliding
window maintained by the FCTS.  Each HLT node maintains a queue of
fully built events that is filled by the event building process and
drained by the HLT process(es). As long as the event building process
is running and the number of events in the queue stays below a
high-water mark, the HLT node periodically sends requests for more
events to the FCTM that is responsible for the partition. These
requests are sent as UDP packets over Ethernet; their aggregate rate
can be limited to a fraction of the L1-accept rate, since each request
is for multiple events.

If the HLT processes on a node cannot keep up with the incoming
events, the node stops sending requests and after the FCTM has
exhausted the node's window of outstanding events, no more events are
directed to the node. No events are lost in this case, since the node
still queues all event fragments that are still in-flight. Only after
the FCTM receives a new event request from this HLT node, it considers
it as a valid event destination again. This provides a flow-control
and load balancing mechanism for the HLT farm. In the case of a
complete HLT node failure (or the failure of the event building
process) the event loss is less or equal to the number of events last
requested.

The overall process of event building is inherently parallel and its
rate can be scaled up as needed, up to the bisection bandwidth of the
event building network. The baseline technology for the event builder
network is standard 10\gbitps Ethernet. We will investigate the
suitability of end-to-end flow control mechanisms (such as IEEE
802.1Qbb) at the Ethernet layer for avoiding packet loss in the event
building network. We will also investigate alternative network
technologies and protocols (such as RDMA or Infiniband).

With a L1 trigger rate of 150\kHz and a pre-HLT event size of
500\kbytes, the bandwidth in the network event builder is about
75\gbyteps, corresponding to about 750\gbitps with network overhead
included. To avoid packet loss and to maintain stability, the event
building network cannot be operated at 100\% link utiliziation, and we
need to retain an additional safety factor of \ktilde1.5. Therefore,
the minimum bisection bandwidth of the event building network is
1200\gbitsps. When implemented with 10\gbitsps Ethernet, this means
120 ``source'' network interfaces on the ROMs and at least 120
``destination'' interfaces on the HLT nodes. Network switches that
provide the necessary bandwidth and can host at least $120+120=240$
10\gbitsps Ethernet ports are commercially available at the time of
the writing of this document.

\tdrsubsec{High-Level Trigger Farm}


The HLT farm needs to provide sufficient aggregate network bandwidth
and CPU resources to handle the full Level~1 trigger rate on its input
side. The Level~3 trigger algorithms should operate and log data
entirely free of event time ordering constraints and be able to take
full advantage of modern multi-core CPUs (the simplest implementation
would be to run multiple identical HLT threads or processes which get
their input from a single queue of built events).  Extrapolating from
\babar, we expect 10\ms core time per event to be more than adequate
to implement a software L3 filter, using specialized fast
reconstruction algorithms. With such a filter, an output cross-section
of 25\nb should be achievable.  Using contemporary (as of Summer 2012)
hardware and taking into account the CPU overhead for event building,
logging and data transfer, a suitable farm could be implemented with
approximately 150 nodes.

\tdrparagraph{Level-4 Option} To further reduce the amount of
permanently stored data, an additional filter stage (L4) could be
added that acts only on events accepted by the L3 filter. This L4
stage could be an equivalent (or extension) of the \babar\ offline
physics filter---rejecting events based either on partial or full
event reconstruction. If the worst-case behavior of the L4
reconstruction code can be well controlled, it could be run in near
real-time as part of, or directly after, the L3 stage.  Otherwise, it
may be necessary to use deep buffering to decouple the L4 filter from
the near real-time performance requirements imposed at the L3 stage.
The discussion in the \superb\ CDR ~\cite{bib:etd-SuperBCDR} about
risks and benefits of a L4 filter still applies.

\tdrsubsec{Data Logging}

\label{sec:etd-logging}

The output of the HLT is written to disk storage. With a cross-section
of about 25\nb accepted by the HLT, the event rate is 25\kHz at a
luminosity of $10^{36}\ {\rm cm}^{-2} {\rm sec}^{-1}$. With an event
size of 200\kbyte, this corresponds to an aggregate logging rate of
5\gbytesps, or \ktilde430\tbyte per day. In a farm of 150 HLT nodes,
the per-node rate is approximately 33\mbyteps.

The most likely mode of operation is that every HLT farm node writes a
single file at a time and that data from multiple farm nodes is not
aggregated into larger files. Instead, the individual files from the
farm nodes are maintained in the downstream system. This means that
the bookkeeping system and data handling procedures have to manage
these files, and also deal with missing or damaged run contribution
files.

There will be least a few \tbytes of usable space per HLT farm node,
implemented as directly attached low-cost disks in a redundant (RAID)
configuration, a storage system connected through a network or SAN, or
as a combination of the two.

A network separate from the event builder one is used to transfer
data asynchronously to archival storage and/or near-online farms for
further processing.  It has not yet been decided where such facilities
will be located, or if they will even be on site, since a distributed,
virtual Tier-0 computing facility is the \superb\ computing
infrastructure baseline. In any case, network connectivity with with
adequate bandwidth and reliability is required. At the experiment,
enough local storage must be available to allow to continue taking
data during periods of link downtime. Many considerations will
determine the exact amount of local storage, including cost, expected
data collection efficiency, link downtime, and the availability of
backup links and facilities.

\tdrparagraph{Data Format}While the format for the raw data has yet to
be determined, many of the basic requirements are clear, such as
efficient sequential writing, compact representation of the data,
portability, long-term accessibility, and the freedom to tune file
sizes in order to optimize storage system performance.

\tdrsubsec{Data Quality Monitoring System}

Event data quality monitoring is based on quantities calculated by the
HLT, as well as quantities calculated by a more detailed analysis of a
subset of the data. A distributed histogramming system collects the
monitoring output histograms from all sources and makes them available
to automatic monitoring processes and operator GUIs.

\tdrsec{System Integration and Error Handling} 

Due to the radiation environment, SEUs and the corresponding rate of
link and FEE failures cannot be neglected, as there will likely be at
least a few failures per day. For this reason it is very important
that detection of and recovery from these failures is reliable and
fast.  Because we cannot predict all possible failure modes, the
system design uses a generic approach to recovery. The FCTS and all
FEEs implement mechanisms that allow individual front-ends to detect
and report the loss of synchronization, and to resynchronize without
the need to restart the whole system. This includes the periodic
broadcasting of synchronization frames on the clock and command links,
the reporting of FEE problems through the fast throttle, and the
ability to diagnose, reset and synchronize individual front-ends under
the control of the ECS.  Other components in the system that
participate in the detection of FEE problems are the ROMs, the HLT,
and the data quality monitoring system.

Recovery from failures will be orchestrated by the global control
system software and the ECS, this provides maximum flexibility in
handling unforeseen failure modes.



\tdrsec{Control Systems} 

A major lesson learned during \babar\ operations was that achieving
high operational efficiency and true ``factory-mode'' data taking
required a high degree of automation and control system integration.
The traditional approach of separate control systems for different
aspects of the experiment (detector control, run control, farm and
logging control) designed and implemented with completely separate
tools and very limited capability to communicate with each other,
greatly limited the amount of automation and automatic error detection
and recovery.

In \superb, all routine operations will need to be orchestrated across
subsystem boundaries.  For example, performing a simple calibration
might require high voltages to be ramped to a calibration set-point,
the FEE and the FCTS to be configured for calibration, farm nodes to
be allocated and configured, the calibration run to be performed,
calibration data to be analyzed and after completion the system to be
returned to its normal data taking configuration.

\begin{figure*}
\begin{center}
\includegraphics[width=0.7\textwidth]{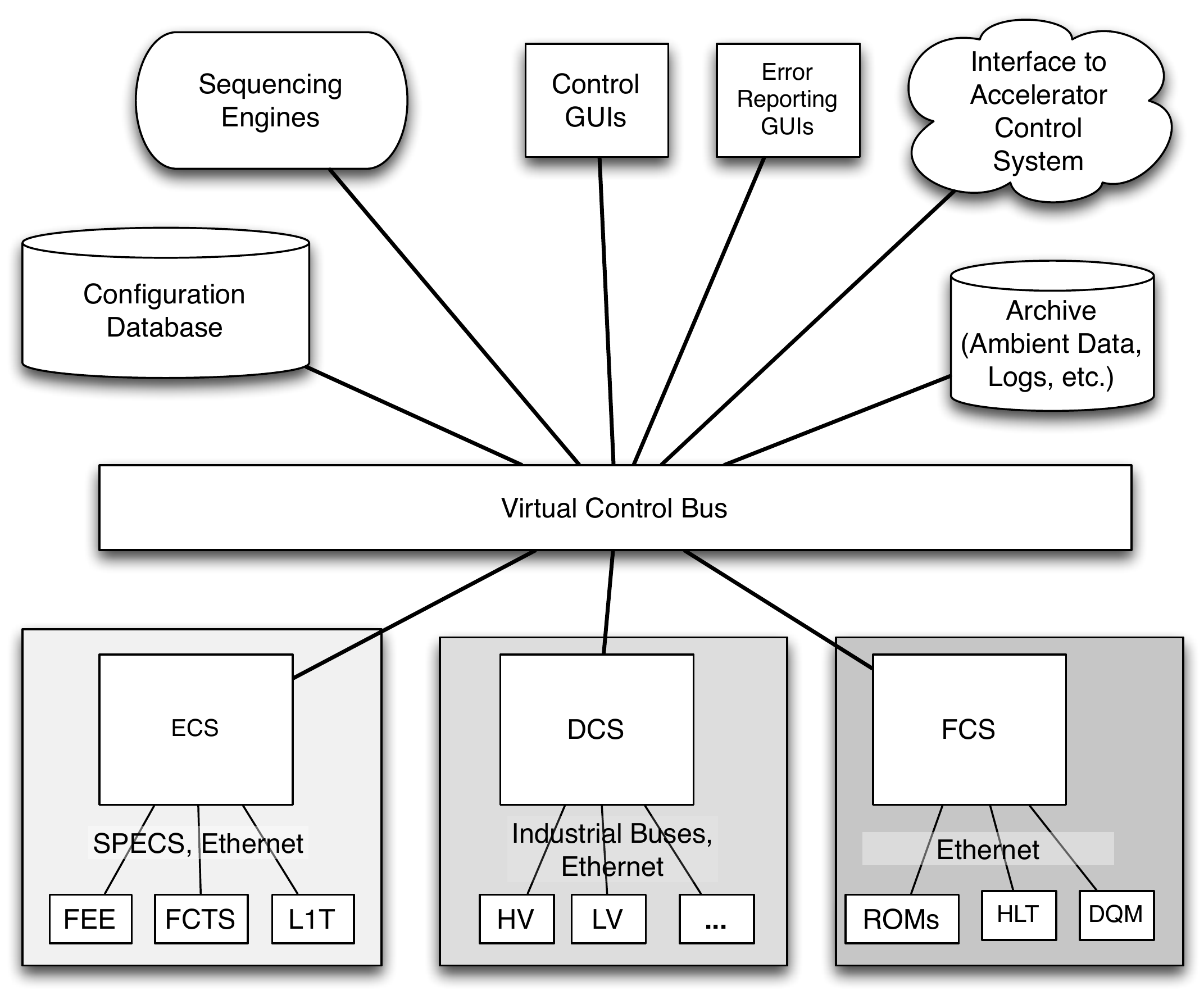}
\end{center}
\caption{The \superb\ unified control system architecture}
\label{fig:etd-control-systems}
\end{figure*}

For \superb\ we therefore foresee a unified control system that can
automatically perform all routine operations. The system comprises\
the Electronics Control System (ECS), the Detector Control System
(DCS), the Farm Control System (FCS), the configuration database,
sequencing engines to implement distributed state machines, an
archiving and logging system, operator GUIs, and the interface to the
accelerator control system. All components are connected by a central
virtual control bus (``Global Control System'').

All operations are driven by the configuration database and executed
under the control of one or more sequencing engines.

A large part of the system, including the central virtual control bus,
will be implemented using the !CHAOS~\cite{bib:etd-CHAOS} control
system toolkit which is currently under development for the \superb\
accelerator. !CHAOS is a state-of-the-art scalable distributed control
system framework that combines high-performance data acquisition and
archiving capabilities with a plug-in architecture to provide
low-level controller interfaces and graphical user interfaces (GUIs).

An overview of the unified control system architecture is shown in
Fig~\ref{fig:etd-control-systems}; its major components are described
below.


\tdrsubsec{Electronics Control System}
\label{sec:etd-ecs}
The Electronics Control System (ECS) controls and monitors the FEE,
the FCTS and the Level~1 trigger. Its main responsibilities are:

\tdrparagraph{Configuring the Front-ends:} Many front-end parameters
must be initialized before the system can work correctly. The number
of parameters per channel range from a only a few to large per-channel
lookup tables. The ECS may also need to read back parameters from
registers in the front-end hardware to check the status or verify that
the contents have not changed. For a fast detector configuration and
recovery turnaround in factory mode, it is critical to not have
bottlenecks either in the ECS itself, or in the ECS' access to the
front-end hardware.

\tdrparagraph{Error recovery:} The ECS plays a central role in
recovering from temporary system failures due to radiation (such as
loss-of-clock). It provides the interface to the dead time monitoring
system, allows detailed diagnostics of FEE failures and can reset and
resynchronize indidividual FEEs without the need for issuing a global
reset or resynchronization command.

\tdrparagraph{Calibration:} Calibration runs require extended
functionality of the ECS. In a typical calibration run, after loading
calibration parameters, event data collected with these parameters are
sent through the DAQ system and analyzed. Then the ECS loads the
parameters for the next calibration cycle into the front-ends and
repeats the operation.  

\tdrparagraph{Testing the FEE:} The ECS is also used to remotely test
all FEE electronics modules using dedicated software. This obviates
the need for independent self-test capability for all modules.

\tdrparagraph{Monitoring the Experiment:} The ECS continuously
monitors FEE boards, FCTS and the L1 Trigger to ensure that they
function properly. This might include independent spying on event data
to verify data quality, and monitoring operational parameters on the
boards (such as voltages, currents, temperatures and error flags). By
monitoring these parameters, the ECS also participates in protecting
the experiment from a variety of hazards. An independent
hardware-based detector safety system, which is part of the DCS,
protects the experiment against equipment damage in case the
software-based ECS is not operating correctly.

ECS support must be built into all electronics modules that are to be
controlled by the ECS -- this includes the FEE.

\tdrparagraph{Programming the FPGA firmware:} Where applicable, the
ECS can also be used to program or update FPGA configurations. It is
highly desirable to be able to perform such operations without
physical access to the boards. Depending on the FPGA this may require
appropriate programming voltages to be made available on the boards.

\tdrparagraph{}The specific requirements that each of the sub-systems
makes on ECS bandwidth and functionality must be determined (or at
least estimated) as early as possible so that the ECS can be designed
to incorporate them.  Development of calibration, test, and monitoring
routines must be considered an integral part of sub-system
development, as it requires detailed knowledge about sub-system
internals.

\tdrparagraph{ECS Implementation:}

The field bus used for the ECS has to be radiation tolerant on the
detector side and provide very high reliability. Such a bus has been
designed for the LHCb experiment: it is called SPECS (Serial Protocol
for Experiment Control System) \cite{bib:etd-SPECS}. It is a
bidirectional 10\mbitsps bus that runs over standard Ethernet Cat5+
cable and provides all possible facilities for ECS (like JTAG (Joint
Test Action Group) and I2C (Inter IC)) on a small mezzanine. It can be
easily adapted to the \superb\ requirements.  Though SPECS was
initially based on PCI boards, it is currently being translated to an
Ethernet-based system, as part of an LHCb upgrade, also integrating
all the functionalities for the out-of-detector elements.  For the
electronics located far from the detector, Ethernet will be used for
ECS communication. The \superb\ ECS will be implemented using SPECS;
an interface to !CHAOS will be developed.





\tdrsubsec{Detector Control System}


\label{sec:detector_control_system}

The Detector Control System (DCS) is responsible for ensuring detector
safety, controlling the detector and its support system, and
monitoring and recording detector and environmental conditions. The
DCS also provides the primary interface between the accelerator and
the detector.

Efficient detector operations in factory mode require high levels of
automation and automatic recovery from problems. Here, the DCS plays a
key role and a tight integration with the Accelerator Control System
(ACS) is highly desirable. The DCS in conjunction with the ACS
manages the accelerator-detector interlocks and beam and detector
states and has access to all information from the accelerator and the
detector that is needed to fully automate data taking operations.

The DCS-ACS connection is also used to provide the accelerator with
beam measurements performed by the detector (such as beam spot
positions or bunch-by-bunch luminosities). Operational experience from
\babar\ has shown that mutual access between machine and detector to
their respective archived control system data (such as records of
background levels, detector currents, trigger rates on the detector
side and vacuum pressures, temperatures and stored currents on the
machine side) are invaluable for improving the accelerator and
detector performance.

The DCS will be implemented with !CHAOS. Due to its distributed nature
and modular storage design, !CHAOS will allow us to federate the
independent instances of DCS and ACS and provide a unified query
interface for data archived by the respective systems. 

Low-level components and interlocks responsible for detector safety
(Detector Safety System, DSS) will be implemented as simple circuits
or with programmable logic controllers (PLCs).

\tdrsubsec{Farm Control System}


Processes on the ROMs and on the HLT farm will be started, controlled
and monitored by the Farm Control System (FCS) and will be implemented
using the !CHAOS framework or a traditional network inter-process
communication system such as DIM~\cite{Online:DIM}.


\tdrsec{Support Systems}


A number of systems and infrastructure components are needed to
support detector and data taking operations, and ETD/Online software
development. These include:

\tdrparagraph{Software Infrastructure:}The data acquisition and online
system is a distributed system built with commodity hardware
components and a substantial software design and implementation
effort.  Taking a homogeneous approach in both the design and
implementation phases will help to keep the effort under control and
also facilitate maintenance and improvement of the system during the
operations phase. An Online software infrastructure framework will be
of great help.  It should provide basic memory management,
communication services, and the environment to execute the Online
applications.  Specific Online applications will make use of these
general services to simplify the performance of their functions.
Middleware designed specifically for data acquisition exists, and may
provide a simple, consistent, and integrated distributed programming
environment.

\tdrparagraph{Electronic Logbook:} A web-based logbook, integrated
with all major Online components, allows operators to keep an ongoing
log of experiment status, activities and changes.

\tdrparagraph{Databases:} Online databases such as configuration,
conditions, and ambient databases are needed to track, respectively,
the intended detector configuration, calibrations, and actual state
and time-series information from the DCS.

\tdrparagraph{Configuration Management:} The configuration management
system defines all hardware and software configuration parameters, and
records them in a configuration database.

\tdrparagraph{Software Release Management:} Strict software release
management is required, as is a tracking system that records the
software version (including any patches) that was running at a given
time in any part of the ETD/Online system. Release management must
cover FPGAs and other firmware as well as software.

\tdrparagraph{Performance Monitoring:} The performance monitoring
system monitors all components of ETD/Online.

\tdrparagraph{Computing Infrastructure:} The Online
computing infrastructure (including the specialized and
general-purpose networks, file, database and application servers,
operator consoles, and other workstations) must be designed to provide
high availability, while being self-contained (sufficiently isolated
and provided with firewalls) to minimize external dependencies and
downtime.

\tdrsec{R\&D for Electronics, Trigger and Data Acquisition and Online}



The baseline design presented in this chapter can be implemented with
technology and components available at the time of writing of this
document. However, we expect that by the times when we have to freeze
various aspects of the design to start construction or purchasing,
components that are significantly more performant and/or cost
effective might be available. In order to take advantage of these
developments, we will need to develop a detailed plan on when we have
to finalize the parts of our design.


\tdrparagraph{Data Links:}
The data links for \superb\ require further R\&D in the following
areas: (1) studying jitter related issues and filtering by means of
jitter cleaners; (2) coding patterns for effective error detection and
correction; (3) radiation qualification of link components; and (4)
performance studies of the serializers/de-serializers embedded in the
new generation of FPGAs (Virtex6, Xilinx, etc.). We will also closely
follow developments that are in progress for other experiments
(e.g. LHC experiment upgrades) to understand their applicability in
the \superb\ system.

\tdrparagraph{Readout Modules:}
Readout Module R\&D requires further investigation of 10\gbitsps
Ethernet and alternative networking technologies, and detailed studies
of the I/O sub-system performance in the ROMs.

\tdrparagraph{Trigger:}
For the L1 trigger, the time resolution, algorithms and their physics
performance, pile-up handling, and the achievable mimimum latency will
need to be studied in detail. We will also need to investigate the
feasibility of a L1 Bhabha veto. For the HLT, studies of achievable
physics performance and rejection rates need to be conducted,
including understanding the risks and benefits of a possible L4
option.



\tdrparagraph{Event Builder and  HLT Farm:}

The main R\&D topics for the Event Builder and HLT Farm are (1) the
applicability of existing tools and frameworks for constructing the
event builder; (2) the HLT farm framework; and (3), event building
protocols and how they map onto network hardware.

\tdrparagraph{Software Infrastructure:}

To provide the most efficient use of resources, it is important to
investigate how much of the software infrastructure, frameworks and
code implementation can be shared with Offline computing. This
requires us to determine the level of reliability-engineering required
in such a shared approach. We also must develop frameworks to take
advantage of multi-core CPUs.



\tdrsec{Conclusions} 

The architecture of the ETD system for \superb\ is optimized for
simplicity and reliability at the lowest possible cost. It builds on
substantial in-depth experience with the \babar\ experiment, as well
as more recent developments derived from building and commissioning
the LHC experiments.  The proposed system is simple and safe. Trigger
and data readout are fully synchronous---allowing them to be easily
understood and commissioned.  Safety margins are specifically included
in all designs to deal with uncertainties in backgrounds and radiation
levels. Event readout and event building are centrally supervised by a
FCTS system which continuously collects all the information necessary
to optimize the trigger rate.  The hardware trigger design philosophy
is similar to that of \babar\ but with better efficiency and smaller
latency.  The event size remains modest.

The Online design philosophy is similar---leveraging existing
experience, technology, and toolkits developed by \babar, the LHC
experiments, and commercial off-the-shelf computing and networking
components---leading to a simple and operationally efficient system to
serve the needs of \superb\ factory-mode data taking.



\bibliographystyle{\tdrbibstyle}
\balance
\bibliography{ETD/ETD-bib}

\onecolumn
\begin{subappendices}
\section{Appendix: ETD design rules for the FEE}
\label{app:etd-rules}

This appendix outlines the rules for a correct implementation of the
FEE.

\begin{enumerate}
\item Average and instantaneous rate handling capabilities
  \begin{enumerate}
  \item Every FEE must be able to handle an average rate of at least
    150\kHz
    \begin{enumerate}
    \item At 150\kHz with equidistant L1-accepts it must not generate
      any dead time
    \item At 150\kHz with the standard exponential inter-event time
      distribution, it must not generate more than 1\% dead time
    \end{enumerate}
  \end{enumerate}

\item Properties of L1-Accepts
  \begin{enumerate}
  \item L1-accepts are synchronous with the system clock and have fixed latency
    \begin{enumerate}
    \item L1-accepts arrive at a time $L\pm(J/2)$ after the activity
      in the detector that caused the trigger to be asserted.
      \begin{enumerate}
      \item $L$ is the trigger latency (approximately 6\mus in the
        current design)
      \item $L=N/f$ where $N$ is the trigger pipeline depth in clock
        cycles, and $f$ the system clock frequency (59.5\MHz in the
        current design). In other words, $L$ is an integer multiple of
        the system clock period.
      \item $J$ is the peak-to-peak trigger jitter, caused by the
        limited time resolution of the detectors feeding the trigger,
        and by the resynchronization of detector signals to the
        (global) clock running the trigger system.
      \item $J/2$ is expected to be in the order of few tens of \ns.
      \end{enumerate}
    \end{enumerate}

    \begin{enumerate}
    \item L1-accepts have no minimum inter-command spacing, so the
      rate of sending L1-accepts is only limited by the control link
      wire speed and protocol.
      \begin{enumerate}
      \item Every FEE must be able to process back-to-back L1-accepts,
        even if data has to be shared between the successive events
        (pile-up events).  See the discussion about dead time and
        pile-up below for more details about this requirement.
      \item We expect the time length of an L1-accept command word to
        be no more than ~60\ns.
      \end{enumerate}
    \end{enumerate}
  \end{enumerate}
\item Latency buffer and read-out windows
  \begin{enumerate}
  \item The latency buffers in the FEE hold the samples during the
    trigger latency.
    \begin{enumerate}
    \item They constitute pipelines that run parallel to the trigger
      pipeline.
    \item They are filled synchronously and read-out at the same
      frequency, which can be the global clock or multiples thereof
    \item When an L1-accept arrives, a time window $W$ of samples
      around the end of the latency buffers is read out, formatted and
      transferred to the derandomizer buffers
    \item $W$ contains all the information pertaining to the
      particular event
      \begin{enumerate}
      \item $W$ has to be large enough to also accommodate the trigger
        jitter $J$
      \end{enumerate}
    \item Alternative implementations are permissible, as long as they
      do not violate the principle of parallel pipelines and
      back-to-back L1-accept handling (see item \ref{item:deadtime}
      below for a discussion of dead time management)
    \end{enumerate}
  \end{enumerate}
\item Management of overlapping read-out windows (pile-up)
  \begin{enumerate}
  \item In case of close consecutive triggers, data can be shared
    between consecutive events if their readout windows overlap
    \begin{enumerate}
    \item In this case, the first event is sent in full, for
      subsequent overlapping events, only the new, non-overlapping
      data are sent.
    \item Full events are restored in the ROMs by replicating the
      shared data.
    \end{enumerate}
  \end{enumerate}

\item Dead time management \label{item:deadtime}
  \begin{enumerate}
  \item Dead time is centrally managed by the FCTS.
    \begin{enumerate}
    \item Should the instantaneous rate exceed an FEE's capacity, the
      FEE must assert a ``fast throttle'' (FT) signal to the FCTS to
      temporarily stop L1-accepts
    \item The FEE must be able to handle all L1-accepts that have
      already been generated, and are in flight at the time it asserts
      FT, and absorb any L1-accepts generated during the latency
      period from asserting the FT signal to stopping the L1-accepts
    \item The preferred implementation is to only use an ``almost
      full'' signal from the derandomizer buffers to generate the
      throttle.  This makes it easy to calculate the derandomizer
      depth based on the maximum link utilization
      \begin{enumerate}
      \item Alternative implementations are permissible, however,
        great care must be taken that none rules outlined in this
        appendix are violated.
      \end{enumerate}
    \end{enumerate}
  \end{enumerate}

\item Derandomizer buffers and data link utilization
  \begin{enumerate}
  \item In the standard design, the derandomizer buffers absorb spikes
    in the instantaneous rate
  \item To be able to catch up after a rate spike, the data link
    utilization corresponding to the average L1-accept rate must not
    exceed 90\% for every individual link.
  \item Alternative implementations for derandomizing are permissible,
    however, great care must be taken that no rules outlined in this
    appendix are violated.
  \end{enumerate}
 
\item Radiation
  \begin{enumerate}
  \item Each element of the FEE has to be validated to work safely and
    reliably in its expected radiation environment
  \end{enumerate}
\end{enumerate}

\end{subappendices}
\twocolumn


\renewcommand{\ChapLabel}{chap:ELEX} 
\tdrchap{Subdetector Electronics and Infrastructure}

The general design approach is to standardize components across the
system as much as possible, implementing common functions on mezzanine
boards, and using commercially available common off-the-shelf (COTS)
components where viable. In this chapter we will discuss and summarize
the aspects of the front-end electronics (FEE) that relate to the
central fast control and timing system (FCTS) and data acquisition
system. We will also discuss the general infrastructure for power
distribution to the FEE.

\tdrsec{Subsystem-specific Electronics}

\tdrsubsec{SVT Electronics}
\label{sec:ELEX:svt}

The full details of the SVT electronics are given in
Sect.~\ref{svt:electronicsreadout}. Here we recall the main features
relevant for data collection, trigger distribution, system programming
and monitoring.

In the baseline SVT option, all layers will be equipped with double
sided silicon strips (or alternatively, striplets). Custom front-end
chips will be used to read the analog information of a particle
traversing the detector and digitize it as soon as possible, yielding
position (layer, strip), energy deposit (signal time over threshold)
and time. The different requirements of all SVT layers will be handled
by a single chip design with programmable features and operating
parameters that are adjustable over a wide range. A possible upgrade
of the SVT internal layers might require a pixelated
detector and a different custom front-end chip design.  The digital
architecture of the pixel chips has already been partially developed;
it will be based on the same general readout architecture, and
will---from the point of view of the trigger and DAQ system---share
the same interface.

Common features of all chips will include the capability to work both
in data-push and data-pull mode, the presence of internal buffers to
allow for a trigger latency of up to 10\mus, the use of a periodic
signal to time-tag the recorded hits, a serialized hit output, and the
chip programmability via two digital lines.

The full SVT data chain will therefore be able to provide and
distribute all the signals, clocks, triggers, and time-tagging signals to
all the front-end chips in a system-wide synchronous way.

\begin{figure}[b]
\includegraphics[width=0.45\textwidth]{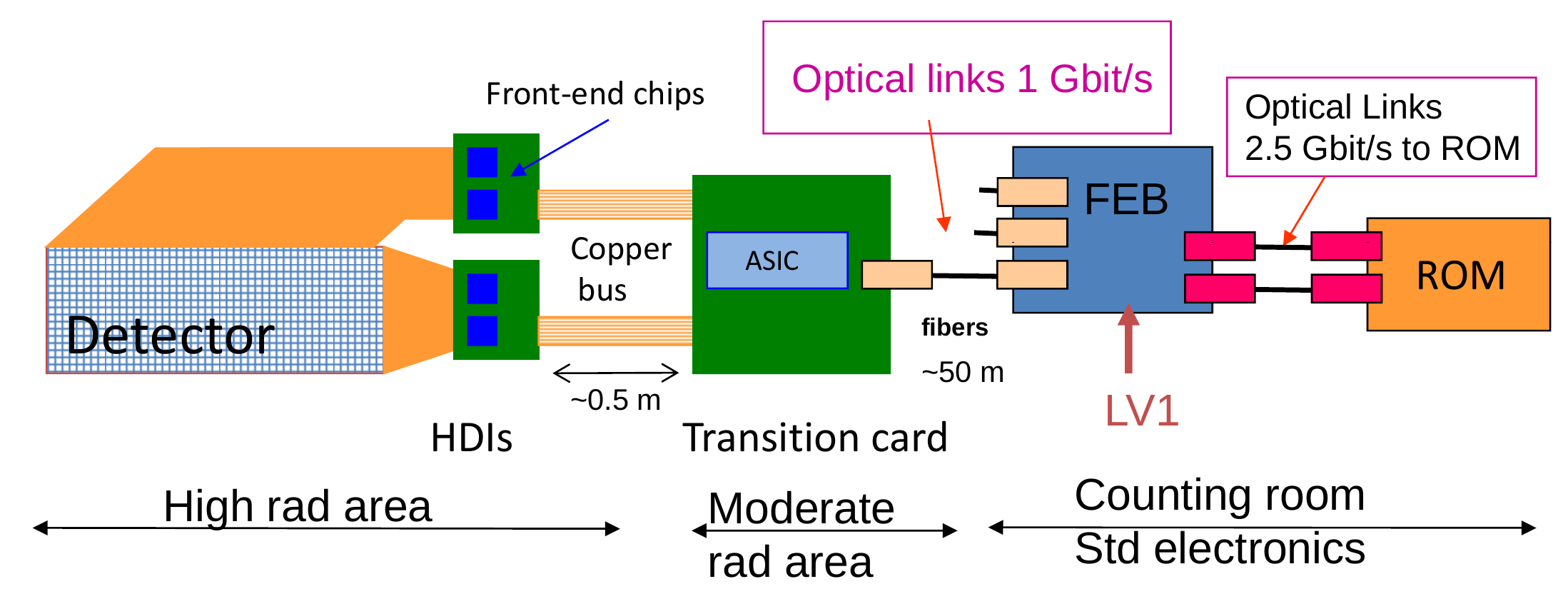}
\caption{Sketch of the SVT Electronics and data chain.}
\label{fig:etd-svt}
\end{figure}

A sketch of the data chain is given in Figure~\ref{fig:etd-svt}.
Starting from the detector and going to the ROM boards, the chain
contains: a) the front-end chips mounted on an HDI placed immediately
at the end of the sensor modules; b) wire connections to a transition
card (signals in both directions and power lines); c) a transition
card, placed about 50\cm from the end of the sensors, hosting the
wire-to-optical conversion; d) a bidirectional optical line running at
or above 1\gbitsps; e) a programmable Front-End-Board (FEB).

Since the silicon sensors are double sided, the front-end (FE) chips
are mounted on both sides of the HDIs.  Each HDI side will host from 5
to 10 FE chips servicing a side of the silicon strip module, a
read-sut section (ROS).  All the chips will share the same
input lines (\textit{reset}, \textit{clock},
\textit{fastclock}, \textit{trigger}, \textit{time-stamp},
\textit{registerIn}) and at least a \textit{registerOut} line.  The
chips can either be programmed individually (by addressing a single
chip) or in bulk by sending broadcast commands to all chips in an HDI.
Hits will be serialized on a programmable number of lines (1 or 2)
using the \textit{fastclock} signal. Each HDI will have a maximum
number of 14 data output lines running at the \textit{fastclock} rate.

The role of the transition card is threefold: a) it will distribute
the power to the HDI; b) it will receive all control signals for the
front-end chips via the optical line connected to the FEB; and c) it
will ship the data to FEB for data acquisition. For the inner layers
(0--3) there will be a transition card for each HDI. For the outer
layers (4--5) it will be possible to group the data of the two HDIs
servicing the two sides of the same silicon sensor into the same
transition card, reducing the total number of optical links needed
(Table~\ref{tab:SVT:DataTransmission}).

The FEBs will handle all the communications with the front-end chips,
the FTCS, the ECS and the downstream DAQ system. Each FEB will host up
to 12 optical links: up to 4 of them could be run at 2.5\gbitsps and
used for communication with the ROMs. The other links will be
connected to the transition cards and run at 1\gbitsps. Important
functions of the FEB are clock distribution, trigger handling, and
data collection. There are multiple clock signals that need to be
distributed in a sub-detector-wide synchronous manner: The experiment
clock (59.5\MHz), a fast clock (120--180\MHz), and a time-stamping
clock (up to 30\MHz).  In order to synchronize the system at the
sensor (or front-end) level, the latencies of all serializers and
deserializers in the signal and DAQ chain will be measured at each
power-up; the phases of the signals will then be adjusted according to
the measurements.

The time-stamping clock is used to time-tag the hits; its frequency
can be higher for the inner layers, where the high track rates require
short signal shaping times and short DAQ time windows. In the outer
layers, a lower frequency could be used to adapt to the longer shaping
times (800--1000\ns) and correspondingly larger readout time
windows.

Data volumes have been estimated assuming a time-stamping clock period
of 30\ns in the inner and outer layers. The acquisition window will be
defined as a time window centered around the L1 trigger time, and
lasting at least 10 time-stamp clock cycles (300\ns) for the inner
layers, and 33 time-stamp clock cycles (990\ns) or more for the outer
layers. The FEB forwards trigger requests via the transition boards to
the FE chips, and also collects their data, both for the main data
acquisition and for monitoring. Data are deserialized in the board,
redundant information is stripped, and optionally, further data
compression algorithms can be applied. Finally, the data will be sent
to a ROM via an optical link.

A possible change in the global clock frequency from 59.5 to 39.66\MHz
has already been considered in the design; only minor changes in the
DAQ chain would be required: In this scenario, a \textit{fastclock}
running at 120 or 160 MHz would be used to collect the data, and a
time stamping clock at 39.66\MHz would be used to time-tag the hits.
No changes would be necessary in other parts of the SVT system, in
particular, the number or types of optical links, and the number of
transition cars and FEB boards would remain the same. 

\begin{table*}[t]
 \caption{
Electronic load on each layer (assuming a $\times 5$ safety factor on background levels), 
readout sections and optical links.}
\begin{center}
\begin{tabular}{|l|c|c|c|c|c|c|c|c||} 
\hline\hline
      & Layer& chips/ & available & Backgnd              &  Gbits/trig & FE      & Event \\ 
Layer & side & ROS    & channels & (MHz/cm$^2$) &  (per GROS)  & Boards  & Size (kHits)\\ 
\hline\hline
0  & u &  6  &  768  &  100  &  0.66 &   2 &   3.5 \\
0  & v &  6  &  768  &  100  &  0.66 &   2 &   3.5 \\
1  & z    &  7  &  896  &  14.0 &  0.56 &   2 &   2.2 \\
1  & phi  &  7  &  896  &  16.0 &  0.64 &   2 &   2.6 \\
2  & z    &  7  &  896  &   9.6 &  0.54 &   2 &   2.2 \\
2  & phi  &  7  &  896  &  10.3 &  0.58 &   2 &   2.3 \\
3  & z    & 10  & 1280  &   4.2 &  0.50 &   2 &   2.0 \\
3  & phi  &  6  &  768  &   3.0 &  0.36 &   2 &   1.5 \\
4a & z    &  5  &  640  &   0.28&  0.26 &   1 &   1.4 \\
4a & phi  &  4  &  512  &   0.43&  0.38 &   1 &   2.0 \\
4b & z    &  5  &  640  &   0.28&  0.26 &   1 &   1.4 \\
4b & phi  &  4  &  512  &   0.43&  0.40 &   1 &   2.1 \\
5a & z    &  5  &  640  &   0.15&  0.17 &   1 &   1.0 \\
5a & phi  &  4  &  512  &   0.22&  0.24 &   1 &   1.5 \\
5b & z    &  5  &  640  &   0.15&  0.17 &   1 &   1.0 \\
5b & phi  &  4  &  512  &   0.22&  0.24 &   1 &   1.5 \\
\hline
\hline
\end{tabular}
\end{center}
\label{table:ELEXSVT:Rates}
\end{table*}

\tdrparagraph{Data volumes.}

As discussed in the SVT chapter (see sect.~\ref{svt_background}), the
data rates and volumes are dominated by the background. For the SVT
front-end chip and DAQ design, the latest background simulations at
nominal luminosity were used, and a safety factor of $5$ was applied
on the results. Due to the strong non-uniformity of the particle rate
in the sensors, the front-end chip characteristics have been chosen
according to the peak rates, while the data volumes have been
calculated from the mean rates for each layer.

To evaluate the data rate, a trigger rate of 150\kHz with 10\mus
maximum latency and 100\ns of time jitter has been used. A 16-bit hit
size is used as the FE chip output, which becomes 20 bits when
serialized with an 8b/10b protocol.  The bandwidth needed by the
layer-0 ROS in a data-push configuration is of the order of 20\gbitsps
per ROS. This rate is high enough and difficult enough to handle, so
we considered a fully triggered SVT.  In Table
\ref{table:ELEXSVT:Rates} the mean expected data load for each layer
type, evaluated assuming a safety factor of $\times 5$, is shown.

For events accepted by the L1 trigger, the bandwidth requirement is
only 1\gbitsps, and data from each ROS can be transferred on optical
links to the FEBs and from there to the ROMs through the 2.5\gbitsps
optical readout links at an average 90\% link utilization.

\tdrsubsubsec{SVT Summary}

In total, the SVT electronics requires 24 FEBs, 56 optical links at
2.5\gbitsps for connection to the ROMs and 172 bidirectional links at
1\gbitsps (radiation hard) for command, trigger and data shipping.
The average SVT event size is 79\kbyte, 22\% coming only from the
layer0.  Table~\ref{table:SVTsummary} summarizes the SVT specification
concerning the DAQ interface requirements.

\begin{table}[h]
\caption{SVT summary}
\label{table:SVTsummary}
\begin{tabular}{|l|r|}
\hline\hline
Number of Control Links & 172 \\
Number of Data Links &  56\\
Total Event Size & 79\kbyte \\
\hline\hline
\end{tabular}
\end{table}

\tdrsubsec{DCH Electronics}
\label{sec:ETD_DCH}

\tdrsubsubsec{Design Goals} 

\begin{figure*}[t]
\begin{center}
  \includegraphics[angle=-0, scale = 0.60]
  {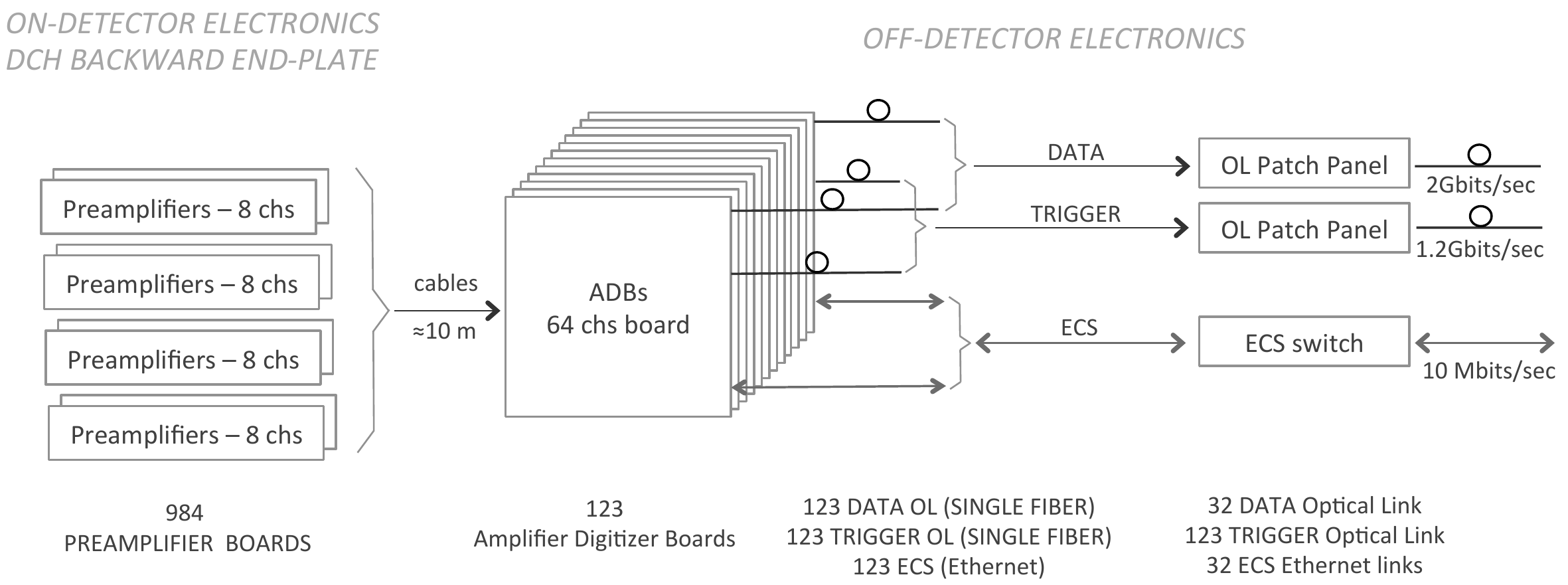}
\caption{DCH front end system (Standard Readout)}
 \label{fig:FeBlockDiaTM}
 \end{center}
\end{figure*}

\begin{figure*}[t]
\begin{center}
  \includegraphics[angle=0, scale =
  0.45]{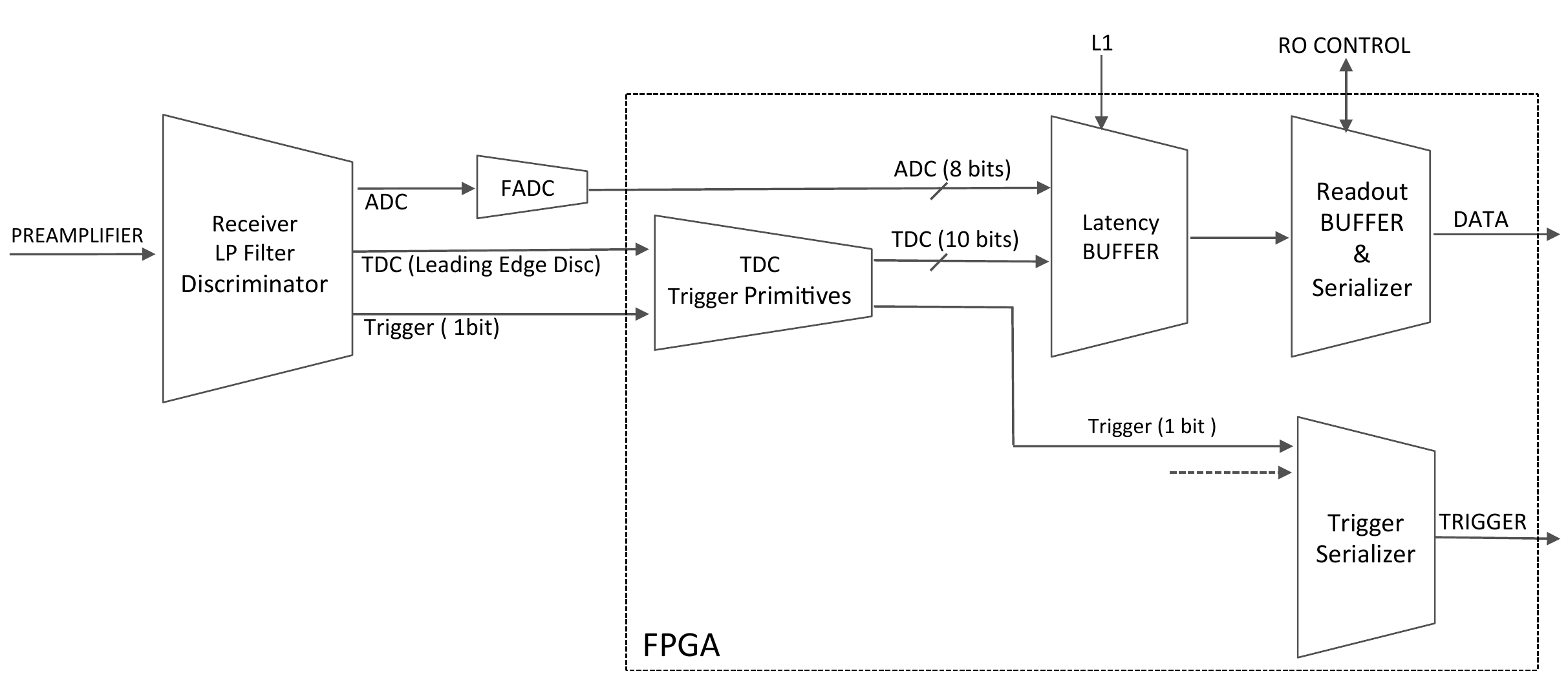}
\caption{Off-Detector Electronics: ADB board (Standard Readout)}
\label{fig:ReadoutBoardBlockDiagram}
\end{center}
\end{figure*}

The \superb\ Drift Chamber (DCH) Front End Electronics (FEE) extracts
and processes approximately 8000 sense wire signals for tracking and energy
loss measurement purposes. Moreover it provides information (trigger primitives) to
the L1 trigger processor.

For the $dE/dx$ measurements, two possible scenarios will be investigated. 
The first one is based on the measurement of the integrated charge on each sense wire 
(Standard Readout), while the second one is based on the detection of
electron clusters in a given cell (Cluster Counting)
A description of the different requirements for the two scenarios can be found in 
Section~\ref{sec:DCHmain}.

\tdrsubsubsec{DCH FEE (overall design)}

The DCH FEE block diagram is very similar for the two options. It is shown in 
Figure \ref{fig:FeBlockDiaTM} for the Standard Readout case. It consists of two
subsystems:

\begin{itemize}
\item{On-Detector Electronics: the HV distribution boards 
located on the front endplate and the preamplifier 
cards located on the back endplate.} 
\item{Off-Detector Electronics: Amplifier Digitizer Boards (ADBs) and
    Concentrator boards, located approximately 10~m away.}
\end{itemize}

\tdrsubsubsec{Off Detector Electronics - Standard Readout}

\tdrparagraph{ADBs - Overall Design} Each ADB will host up to 64
channels and will consist of three stages as shown in
Figure~\ref{fig:ReadoutBoardBlockDiagram}. The first stage receives
signals from the preamplifiers and generates analog and discriminated
outputs.  The second stage provides digitization for charge and time
measurements and includes the logic for trigger primitives generation
as well. Finally, the third stage contains the latency and readout
buffers and the dedicated control logic. The ADBs will also include an
interface (not shown in the diagram) that will be used for parameters setting/readout 
and to test the entire FEE chain. 

Input connections will use twisted or mini-coaxial cables, while output connections 
will use different types of cables depending on the chain. DAQ and Trigger 
chain will use optical links while the ECS chain will use copper links 
(see section \ref{sec:etd-ecs}).

\tdrparagraph{ADBs - Receiver Section}

This stage is designed to amplify and to split the preamplifier output
signal to feed a low-pass filter for charge measurement, and a leading
edge discriminator for time measurement.

\tdrparagraph{ADBs - Digitization Section}

The low-pass filter output signal is routed to an 8-bits, 30
megasamples/s FADC whose outputs are sent to a section of the latency buffer
implemented in a FPGA. Upon a L1-accept, the system extracts 32 FADC
output samples (corresponding to a time window of about 1\mus,
enough to span the full signal development) from the
latency buffer and transfers them to a readout buffer. 

The comparator output is routed to a FPGA where it is split along two paths. 
One signal is sent to a TDC, implemented in the FPGA itself (using the
oversampling method) for time measurement. The other signal, 
synchronized with the system clock and stretched appropriately
to remove redundant information, is sent to the DCH Trigger
Segment Finder (TSF) modules (Figure~\ref{fig:etd-dct}). The TDC
outputs are also sent to the latency buffer, to be readout in the presence of a L1-accept 
trigger signal. 

Normally, the event data structure will have different lengths
since L1 triggers, spaced less than a single event readout 
time, will extend the time window to include the new event. However, during Feature EXtraction 
(FEX) implementation (extraction of information from the digitized 
data), the transferred data stream will have a fixed length. An example of a possible readout
data structure is shown in Table \ref{tab:DataStreamFex}.

In both instances the first, non zero, FADC output signal can be used to
correct the  discriminator output jitter, thereby minimizing the input
signal slewing effects on the timing mesurement.

\begin{table}[h]
\centering
\caption{\small{FEX based data stream (fixed length structure)}}
\begin{tabular}{c}
\hline\hline
\bf{Data stream example} \\
\hline
Digitizer Module Address (2 bytes)  \\
\hline
Flag (1 byte)\\
\hline
Trigger Tag (1 byte) \\
\hline
Counter (1 byte)\\
\hline
Charge (2 bytes)\\
\hline
Time (2 Bytes) \\
\hline
1st ADC sample (different from baseline) for  \\
time walk correction (1 byte)\\
\hline\hline
\end{tabular}
\label{tab:DataStreamFex}
\end{table}

\tdrsubsubsec{Off Detector Electronics - Cluster Counting}

The Cluster Counting technique for $dE/dx$ measurements is based on
the detection of electron clusters (primary ionization measurements).
This requires high bandwidth devices and digitizers with high sampling
frequencies of, at least, 1 gigasamples/s (GSPS). Very fast processing
is required to sustain the 150\kHz average L1-accept rate foreseen in
\superb. With current technology, this results in a very large power
dissipation, and as a consequence, the number of channels per ADB
board has to be limited in the present design.

If we can detect the electron clusters with an efficiency of 100\%, the
system can also be used for tracking purposes; all that is needed is
to measure and store the cluster arrival times, in addition to
counting them.

The DCH FEE system block diagram for Cluster Counting is
similar to the one shown in Figure~\ref{fig:FeBlockDiaTM}.  However,
because of the smaller number of channels per board (16 instead of
64), the number of boards and crates increases significantly 
(Table~\ref{tab:LinksBoardsCrates}).

\tdrparagraph{ADBs} The ADBs will have high-sampling-rate ( $\ge$ 1
GSPS) digitizers. To meet the power requirements, only 8 channels
working at 1 GSPS can be packaged in a single VME 6U board at this
time. However, in the future, we expect to be able to increase the
number of channels per board to 16, or even 24, with only a small
increase in power requirements.

The circuit structure is very similar to the one shown in Figure
~\ref{fig:ReadoutBoardBlockDiagram}. The digitizing section is
different, since no TDC is required for the time measurements, and the
trigger primitives are generated from the FADC output instead.

Because of the large amount of data per L1-accept and channel (about
1\kbyte), we cannot transfer raw data to the DAQ. This means that the
FEX must be implemented on the ADB, that is, for every L1-accept, all event
samples must be examined to identify possible clusters.  While the
time required for cluster identification using current technology is
compatible with the L1-accept rate expected at the nominal \superb\
luminosity, this could become a limiting factor at higher L1-accept
rates. Another potential problem may be the radiation sensitivity of
the high performance RAM-based FPGA used in the present design.

\tdrsubsubsec{FEE Crates}

Each FEE Crate will host up to 16 ADBs, a Power Supply board for the
preamplifiers, Data Concentrators and, possibly, a Trigger Patch Panel
or Trigger Concentrator. Custom backplanes will be designed to:

\begin{itemize}
\item {distribute power and common signals to the ADBs}
\item{ make use of Interconnection Boards (IB) to gather the
    preamplifier cables (Fig.~\ref{fig:InterconnectionsBoards})}
\item {send trigger signals to adjacent boards (Section \ref{subsubsec:DCT} )}
\end{itemize}

\begin{figure}[t]
\begin{center}
  \includegraphics[angle=0, scale =
  0.6]{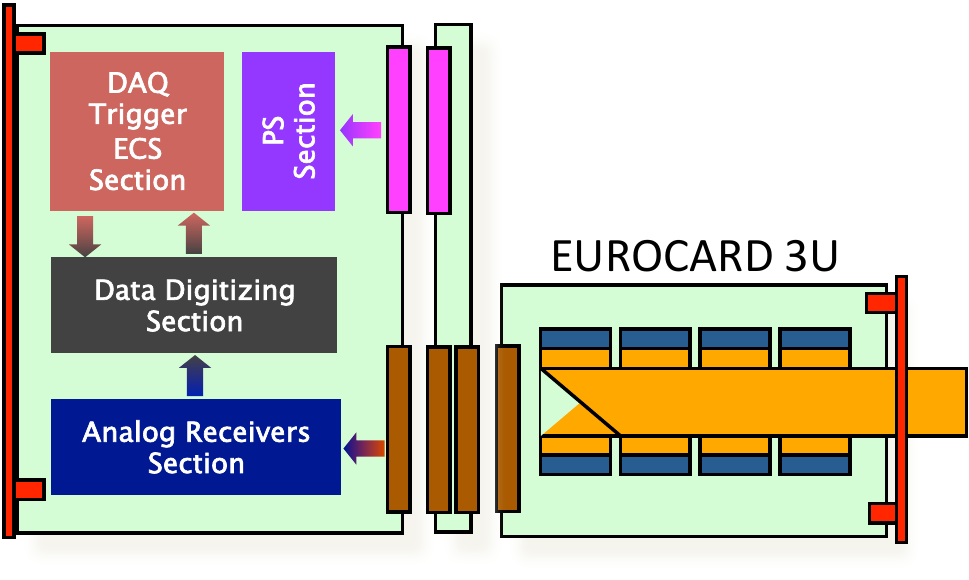}
  \caption{On-Off Detector FEE connections (ADB - Custom Backplane -
    IB)}
  \label{fig:InterconnectionsBoards}
\end{center}
\end{figure} 

\tdrsubsubsec{Number of Crates and Links}

Table~\ref{tab:LinksBoardsCrates} shows the estimate of the number of
links, boards and crates required for both the DAQ and Trigger FEE
chains.  As shown in the table, the number of trigger signals is the
same for both options even though the number of ADBs for the Standard
Readout scenario is not the same as for the CC scenario.
The reason for this is that we intend to also use, in the CC option, a
Trigger Concentrator board for the trigger chain to compensate for the limited
number of channels per ADB. This board will generate a single trigger
output from the trigger inputs coming from several ADBs and will send
this single output to the TSF modules.

The estimate has been done assuming about 1\mus sampling window and:

\begin{itemize}
\item{150 $kHz$ L1 trigger rate, 7872 sense wires (subdivided in 10
super-layers)}
\item{ $10\%$ chamber occupancy in 1\mus time window}
\item{SR - NO FEX option: 50 bytes per channel data transfer (charge: 32 bytes; time: 10 bytes; information: 8 bytes)}
\item{SR - FEX option: 11 bytes per channel data transfer (Table \ref{tab:DataStreamFex})}
\item{CC - FEX option (18 clusters - 2 bytes/cluster): 44 bytes per channel data transfer (data: 36 bytes; event information: 8 bytes)}
\item{single link bandwidth of 2
Gbits/sec for the DAQ data path and 1.2 Gbits/sec for the Trigger data path}
\end{itemize}

\begin{table}[t]
\centering
\caption{Number of links (Data, ECS, Trig), ADBs and crates (Cr) for 64 channels Standard Readout (SR) and 16 channels Cluster Counting (CC) boards }
\begin{tabular}{|lcccccc|}
  \hline\hline
  & chs & Data & ECS & Trig & ADBs & Cr\\
 \hline
  SR & 64 & 32 & 32 & 123  & 123 & 8  \\
  CC & 16 & 32 & 32 & 123 & 492 & 31\\
  \hline\hline
\end{tabular}
\label{tab:LinksBoardsCrates}
\end{table}

\tdrsubsubsec{ECS} Each ADB will have an interface section to manage ECS
communications. The interface will control the ADB and
will provide the capability of data readout for debugging purposes.
ECS details can be found in section \ref{sec:etd-ecs}.

\tdrsubsubsec{Cabling} Because of the large number of channels
involved, the DCH cable layout must be carefully designed. The main
requirement is the possibility to replace failed preamplifier boards
without having to disconnect too many cables. Thus, signal and HV
cables should be routed to the chamber's outermost layer to minimize 
the overlap of cables. Table~\ref{tab:Cables} gives the expected number 
of cables and a rough estimate of their sizes.

\tdrsubsubsec{Power Requirements}

A very preliminary power requirement estimate is shown in Table
\ref{tab:PowerRequirement}. The estimate for On Detector Electronics
is based on preamplifier simulation and prototype tests, while the
estimate for Off Detector Electronics comes from the latest available
state of the art digitizing boards. Local (preamplifier) voltage
regulation is supposed to be implemented by means of linear low-drop
rad-hard regulators.

\begin{table}[h]
\centering
\
\caption{Estimate of the number and dimension of DCH signal cables (7872 sense wires - Preamplifiers of 8 channels each)}
\begin{tabular}{|lcccc|}
\hline\hline
& LVPS &HV &Signal &Signal\\
&                &            &  (coax)  & (twist)\\
\hline
Quantity & 123 & 16 &  7872 & 984 \\
\hline
Number & 16 & 28 &  1  & 16  \\
of cores &&&& \\
\hline
Core area & 0.5  & 0.07 & & \\
(\mma) & & & &  \\
\hline
Cable diam. & 12 & 14 & 1.8 & 6.5 \\
(\mm) & & & & \\
\hline\hline
\end{tabular}
\label{tab:Cables}
\end{table}

\begin{table}[h]
\centering
\caption{Power requirement estimate for SR and CC  (DC = Data Concentrators; TC = Trigger Concentrators)}
\begin{tabular}{|l|c|c|}
\hline\hline
& \bf{Boards} & \bf{Power Req.} \\
\hline
SR Preamplifier & 984 & 0.25\kW \\
SR ADB & 123 & 5\kW \\
SR DC & 8 & 0.24\kW \\
\hline
CC Preamplifier & 984 &  1.2\kW \\
CC ADB & 492 & 20\kW \\
CC DC & 31 & 1\kW \\
CC TC & 31 &  1\kW \\
\hline\hline
\end{tabular}
\label{tab:PowerRequirement}
\end{table}

\tdrsubsubsec{DCH Summary}

Table~\ref{tab:DchSummary} summarizes the DCH FEE main
specifications concerning the DAQ interface and power requirements.  As
already stated, the number of links is the same for both SR and CC
options, while the power requirement is significantly different
for the two options.

\begin{table}[h]
\centering
\caption{DCH FEE summary}
\begin{tabular}{|l|r|}
\hline\hline
Number of TriggerLinks & 123 \\
Number of Control Links & 32 \\
Number of Data Links & 32 \\
\hline
SR Total Event Size (NO FEX) &  40\kbyte\\
SR Total Event Size (FEX) &  9\kbyte\\
SR On-Detector Power Req. & 0.25\kW \\
SR Off-Detector Power Req. &  5\kW \\
\hline
CC Total Event Size &   35\kbyte\\
CC On-Detector Power Req. & 1.2\kW \\
CC Off-Detector Power Req. &  20\kW \\
\hline\hline
\end{tabular}
\label{tab:DchSummary}
\end{table}

\tdrsubsec{PID Electronics}
\label{subsec:PIDElectronics}

The PID electronics will read out the 18,432 channels of the 12
sectors of the FDIRC. The electronics chain is based on a high
speed TDC, a time-associated charge measurement
with 12-bit precision, and an event data packing block, sending event
data frames to the data acquisition system (DAQ). The target
performance of the overall electronics chain is a time precision of
200\ps rms and a per-channel count rate capability of up to 500\kHz,
yet dealing with the global system constraints (average L1-accept rate
of up to 150\kHz and a minimum spacing between triggers of about
50\ns).

The estimated radiation level is expected to be about 1 Gray per
year, so it is mandatory to use radiation-tolerant or off-the-shelf
radiation-qualified components. Latch-up effects are very unlikely at
the expected particle energies, so the design has only to take into
account Single Event Upsets. The ACTEL family FPGA components have
been chosen here for their non-volatile configuration memories based
on flash technology, which are well adapted to this kink of radiation 
environment.

The electronics consists of a front-end analog part and a TDC both integrated 
in a 16 channel ASIC called CATS. The analog part contains an amplifier, a peak
detector, an analog pipeline for the charge measurement and a fast comparator
to drive the TDC inputs. The TDC part provides a time 
measurement with both a high resolution (~100\ps RMS) and a large dynamic 
range (53 bits). Thanks to an external 12-bit ADC, the charge measurement can 
be used for electronics calibration, monitoring and survey purposes. 
The front-end (FE) board FPGA synchronizes the process, associates the time 
and charge information coming from CATS and finally packs them into a data 
frame which is sent via the backplane to the FBLOCK control board (FBC). 
The FBC is in charge of distributing signals from the FCTS and ECS, 
packing the data received from the FE boards to an n-event frame including 
control bits, and transferring it to the DAQ.

\tdrsubsubsec{The Front-end Crate}

\begin{figure}[ht]
\centering
\includegraphics[width=0.52\textwidth]{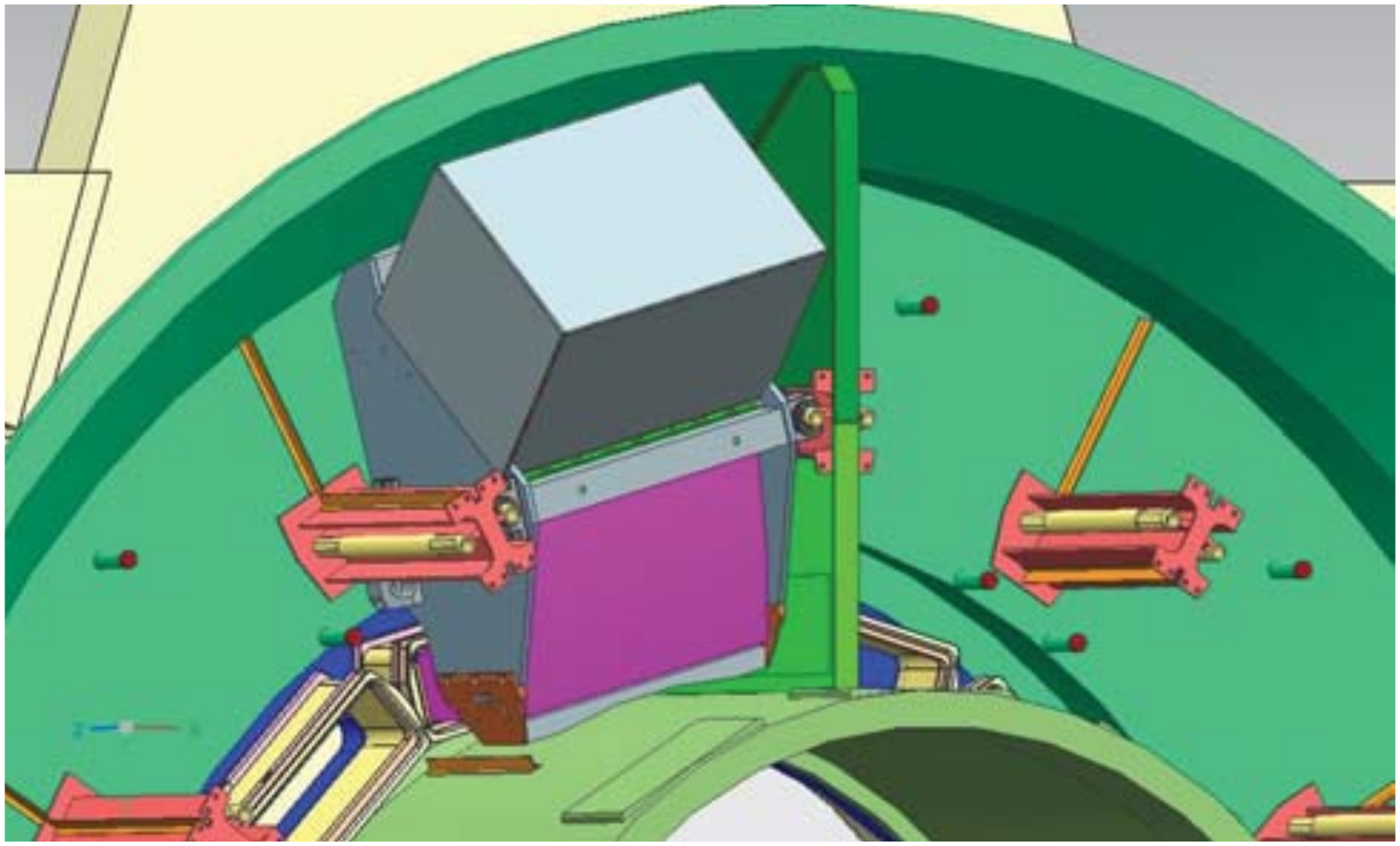}
\caption{The FBLOCK equipped with the boards and fan tray}
\label{fig:ELEX_PID_FBLOCK}
\end{figure}

\begin{figure*}
\centering
\includegraphics[width=0.85\textwidth]{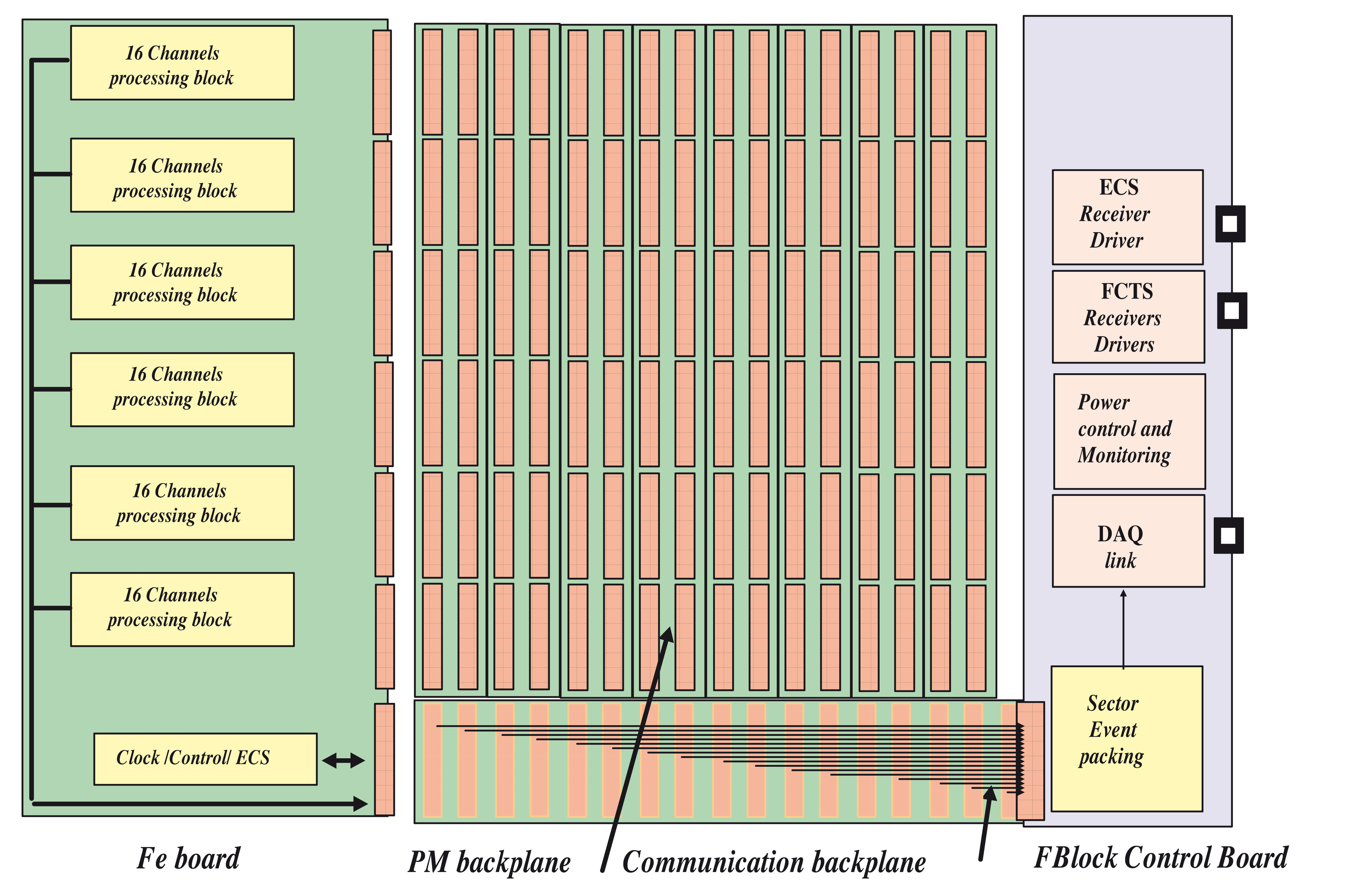}
\caption{The Front end Crate}
\label{fig:ELEX_PID_Frontend}
\end{figure*}

The board input will match the topological distribution of the PMTs on
the FBLOCK. The PMTs are arranged in a 6 (vertical) by 8 (horizontal)
matrix.  Each column of 6 PMTs will fit to one FE board. One vertical
backplane (PMT Backplane) will interface between the 4 connectors of
each PMT base to one connector of the FE board. The PMT Backplane also
distributes the High Voltage, thus avoiding HV cables to pass over the
electronics. In order to minimize the need for custom-designed
elements, such as board guides, the FB crates will use as many
commercially available elements as possible.

\tdrsubsubsec{The Communication Backplane}

The communication backplane distributes the ECS and FCTS signals from
the FBC to the 16 FE boards thanks through point to point LVDS
links. It connects each FE board to the FBC for system control and
data transfer.

\tdrsubsubsec{The PMT Backplane}

The PMT backplane is an assembly of 8 motherboards, each corresponding
to a column of 6 PMTs.

\tdrsubsubsec{Cooling and Power Supply}

The cooling system must be designed in order to maintain the
electronics located inside the FBLOCK at a constant temperature close
to the optimum of 30\degc. 

The air inside the volume will be extracted by dedicated fans while
dry, clean, and temperature controlled air will replace it. Like
commercial crates, each FB crate will have its own fan tray.
4000\,${\rm m^3/h}$ can be considered as the baseline value for the
whole detector. The crate and boards will be cooled by a water based
cooling system.

\tdrsubsubsec{The Front-end Board}

\begin{figure*}
\centering
\includegraphics[width=0.85\textwidth]{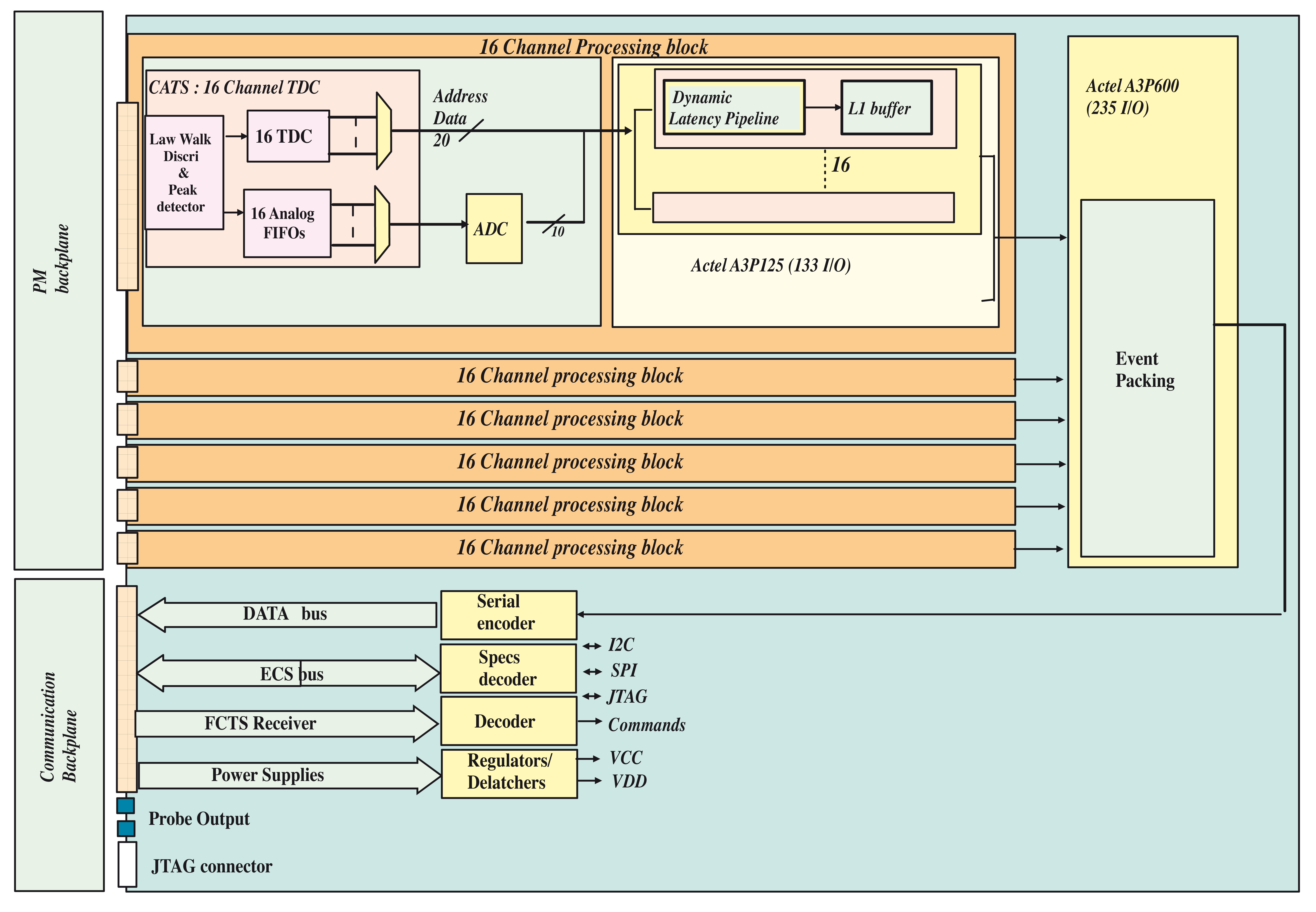}
\caption{The PID Front-end Board }
\label{fig:ELEX_PID_Frontend_Board}
\end{figure*}

One FE board (Figure \ref{fig:ELEX_PID_Frontend_Board}) handles a total
of 96 channels and actually comprises 6 channel-processing blocks.

A channel-processing block consists of one CATS chip, one ADC, one
ACTEL FPGA, and the associated ancillary logic. The FPGA receives event
data from the CATS and the associated charge data from the ADC.  The
16-bit data bus coming out of the CATS is de-serialized and split into
16 separate data buses, each corresponding to one channel.  A
dedicated mechanism inside the FPGA inserts the events into the
latency buffer at their proper position with respect to the hit
time. From there on, the design perfectly corresponds to the common
ETD proposal for the latency buffer and derandomizer (CFEE). Event
data is stored in the event buffer where it is kept until either been
thrown away if it is not selected, or sent to the derandomizer located
in the master FPGA upon the reception of a L1-accept from the FCTS.

\tdrsubsubsec{The Crate Controller Board (FBC)} 

The master FPGA receives event data from the 6-channel processing
blocks, packs it in a frame with parity bits for data checking and
transmits it in a serial differential LVDS format to the FBC via the
communication backplane.

\tdrsubsubsec{PID Summary}

There is thus one link per sector, leading to a total of 12 readout
links for the whole FDIRC. Each link has a mean occupancy of 15\pct,
far below the limit fixed to 90\pct by ETD official rules. The 
parameters of the PID system concerning the DAQ interface and power consumption
are summarized in Table~\ref{table:PIDsummary}.

\begin{table}[h]
\caption{PID summary}
\label{table:PIDsummary}
\begin{tabular}{|l|r|}
\hline\hline
Number of Control Links & 12 \\
Number of Data Links &  12 \\
Total Event Size & 5\kbyte\\
On-Detector Power Consumption & 6.1\kW  \\
\hline\hline
\end{tabular}
\end{table}

\tdrsubsec{EMC Electronics}

\begin{figure*}
\includegraphics[width=\textwidth]{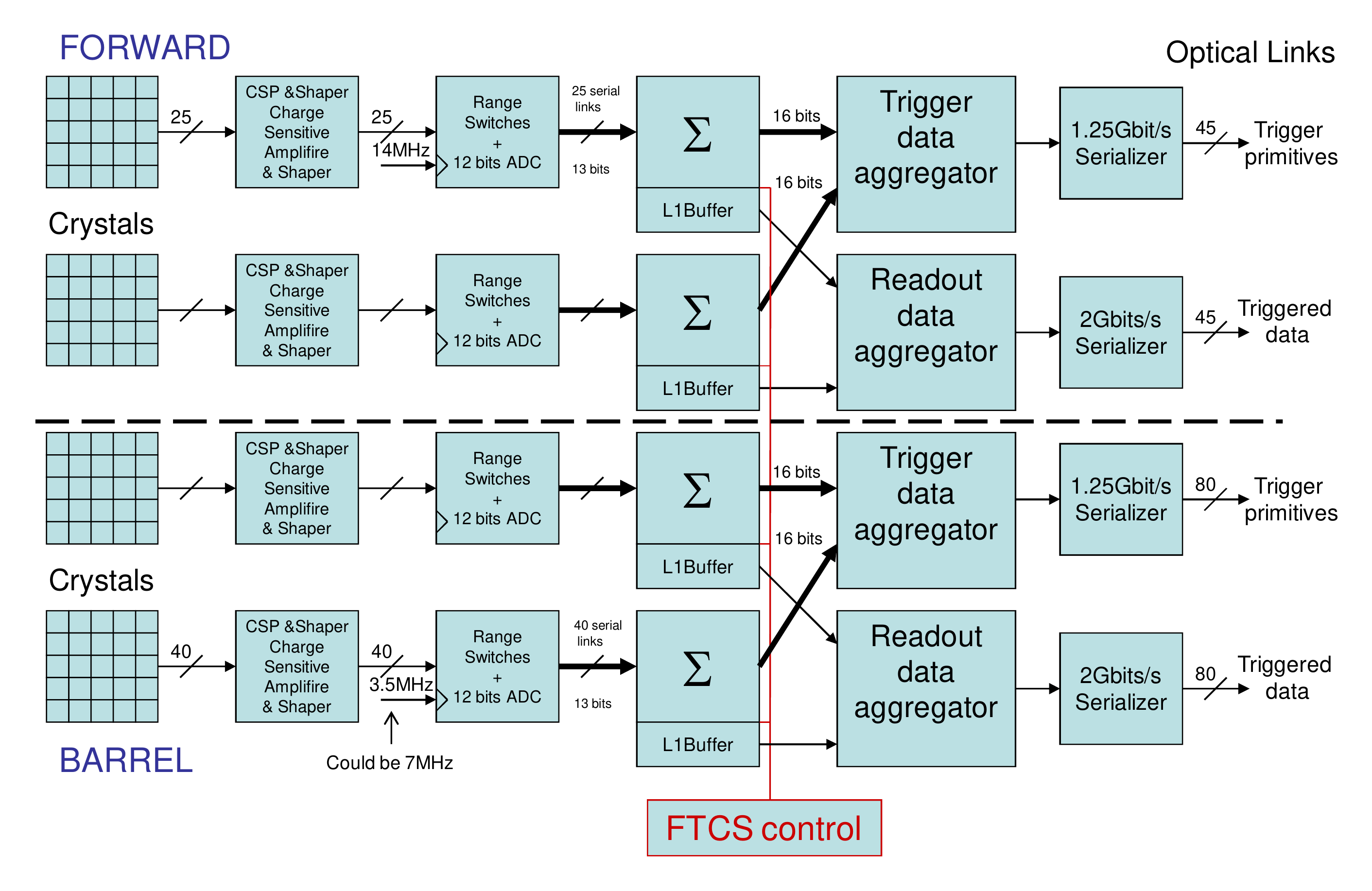}
\caption{EMC Electronics}
\label{fig:etd-emc}
\end{figure*}

The EMC is fully triggered in order to minimize the number
of optical links between FEEs and ROMs, and trigger primitives are
sent to the L1 trigger processor on a separate, independent data
path. Compared to the untriggered EMC readout system in \babar, the
new system keeps the same number of optical links for the
barrel EMC, and requires only a very moderate increase in the link
count for the new hybrid forward calorimeter design, even with a
large increase in trigger rate.  The separate trigger data path is
mandated by the need for a fast calorimeter trigger. To support the
neutron-activated liquid-source calibration (see
Sec.~\ref{sec:EMC:source_calibration}), where no central trigger can
be provided, both the barrel and the end-cap readout systems need to
support a free running ``self-triggered'' mode where only samples
associated with an actual signal are sent to the ROM. This may require
digital signal processing to suppress noisy channels.

\tdrparagraph{Barrel Calorimeter}

While the 5760 crystals and PIN diodes from \babar\ are reused in the
barrel EMC, the charge-sensitive preamplifiers and shapers that
convert the charge into voltage will be replaced with faster devices
that have a 500\ns shaping time and provide a pulse with a FWHM of
1180\ns.  Each shaped signal is then amplified with two different
gains ($\times 1$ and $\times 32$) and digitized. An auto-range
circuit decides on-the-fly which of the two gains will be digitized by
a 12-bit pipeline ADC running at 9.915\MHz ($1/6$ of the \superb\
global clock frequency). Note that a change of the global system
frequency to 39.66\MHz can easily be handled by using a divisor of 4
instead of 6. The 12 bits from the ADC plus an additional bit for the
selected gain covers the full scale from 10\mev to 10\gev with
a resolution better than 1\%. Storing these 13-bit samples in 16-bit
(2 byte) words leaves an additional 3 spare bits per sample that could
be used for bit error detection or correction.

In order to optimize the energy scale for the 6\mev photons produced
by the calibration source, separate programmable gain amplifiers are
used during source calibrations.

Each channel is sampled continuously, and the samples are stored in
latency buffers in the Data Unit Concentrators (DUTs). Upon a
L1-accept, a window of 32 samples (corresponding to approximately
3.3\mus) around the trigger time is transferred from the latency
buffers to the derandomizer buffers that feed the optical links to the
ROMs. The mean time interval between two L1-accepts is approximately
6.6\mus. The triggered system requires only approximately half of the
bandwidth of a comparable untriggered system, albeit at the price of
moving the latency and derandomizer buffers into the FEE. Additional
gains will result from transferring data shared between overlapping
events only once.

The \superb\ EMC barrel FEE reuses the \babar\ topology, mechanics and
mechanical specifications: A single digitizer board hosts 12 channels.
Each DUT hosts latency buffers, readout logic and derandomizer buffers
for 6 digitizer boards, and feeds 2 optical links to the ROMs. This
configuration allows complete reuse of the \babar\ mechanics and form
factors, but requires new boards implementing the new electronics
specification. There are a total of 80 DUTs (one for each of the 40
$\phi$ sectors on each end of the barrel), resulting in a total of 160
optical links for the barrel EMC. 

Each L1-accept generates 64 bytes of waveform data per channel (32
samples $\times$ 2 bytes/sample). 36 channels feed one optical link,
generating a total of 2304 bytes per L1-accept. At a mean trigger rate
of 150\kHz, this corresponds to an average bandwidth of 1.38\gbitsps.

\tdrparagraph{Forward Calorimeter}

For the forward calorimeter, the proposed baseline configuration
consists of a hybrid detector with 360 CsI(Tl) crystals and 2160 LYSO
crystal; it was chosen as a trade-off between cost and performance of
the forward EMC. The readout scheme for the CsI(Tl) crystals follows
the EMC barrel scheme, and uses 5 DUTs and 10 optical links.  For the
LYSO crystals used in the high-background area, the faster response
time requires the shaping time to be shortened to 100\ns, and the
sampling frequency to be increased correspondingly. We expect to use
the same digitizer boards as in the EMC barrel, increase the sampling
frequency to 29.75\MHz ($1/2$ the baseline \superb\ global system
frequency), and to use a readout window of 32 samples,
corresponding to 1.08\mus. To read the 2160 LYSO crystals, 30 DUTs
with 60 optical links are required. To accommodate a change of the
global system clock to 39.66\MHz, one would instead sample at the rate
of the system clock, and the readout time window would become 807\ns.

\tdrparagraph{Backward Calorimeter} The backward EMC has 1152 readout
channels.
The inputs
are the MPPC signals from the scintillator strips amplified by a
charge-sensitive preamplifier. 
We have been using SPIROC boards in tests and are investigating whether
a fast enough version of this technology can be defined. However, our baseline
electronics for the detector is the same as for the PID system, with
12 bit ADC and 200~ps TDCs. Further details are in section~\ref{subsec:PIDElectronics}.
To match the PID electronics, the preamplifier will produce a negative pulse. Seventy-two of the
16-channel PID ASICs are required for the backward EMC. The circuit boards
are placed behind the backward EMC.
For calibration and monitoring we use two calibration boards, each
housing 12 LEDs and 12 PIN diodes. Each LED distributes light via four
clear fibers that have 12 notches each and span the full calorimeter
length. A fifth fiber is coupled to a PIN diode. Thus, with four
fibers two full layers (24 tiles each) can be illuminated, while the
fifth fiber is used to monitor the LED light.

\tdrsubsubsec{EMC Summary}

The parameters of the EMC concerning the DAQ and power consumption are summarized in Table~\ref{table:EMCsummary}.

\begin{table}[h]
\caption{EMC summary}
\label{table:EMCsummary}
\begin{tabular}{|l|r|}
\hline\hline
Number of Control Links &  266 \\
Number of Data Links &   496\\
Total Event Size & 277\kbyte\\
On-Detector Power Consumption & 16.6\kW\\
\hline\hline
\end{tabular}
\end{table}

\tdrsubsec{IFR Electronics} 

The full description of the IFR readout electronics, from the design
constraints determined by the features of the IFR detector, to the
details of the adopted baseline design, is given in the subdetector
chapter. In this section, only the features of the IFR electronics
related to the common ETD and ECS infrastructure, are described.

The active layers of the IFR are equipped with modules in which the
detector elements (of different widths for the $\phi$ and the Z views
in the barrel) are assembled in two orthogonal layers. Assuming that
the IFR is instrumented with 9 active layers, the total channel count
amounts to 11604 channels for the barrel section, and 9540 channels
for the endcaps.

\begin{table}[h]
\caption{Estimation of the event size and data bandwidth for the Barrel}
\label{Tab_t1}

 \begin{tabular}{|l|c|} 
\hline
\hline
Maximum channel count per  module: & 32  \\  
\hline 
Number of MOD32 processing & \\ units per module & 1 \\
 \hline 
Total number of modules & \\ in the barrel & 384 \\
 \hline 
Total number of MOD32 & \\ processing units & 384 \\  
 \hline 
 Sampling period (ns)  & 17.86 \\ 
 \hline 
 Number of  samples in the trigger & \\ matching window & 16 \\
 \hline 
Barrel Event Size (kB)  &  24.58 \\
 \hline 
 Trigger Rate (kHz) & 150 \\
 \hline 
 Total Barrel Bandwidth (Gbps), & \\ including  8b\/10b encoding  overhead & 36.86 \\
 \hline 
Number of data links from the barrel & \\ detectors & 24 \\ 
 \hline 
 Bandwidth per link (Gbps) & 1.536  \\ 
 \hline 
\end{tabular}
\end{table}

The IFR detector will be read out in ``binary mode'': the output of
each SiPM device will be amplified, shaped and compared to a
threshold. The binary outputs of the comparators are then sampled by
``digitizers'' at the \superb\ clock rate, and the samples are stored
into local ``on-detector'' circular memories. These \textit{latency
  buffers} are designed to keep the data for a time interval that is
at least the L1 trigger latency. The last stage of the
``on-detector'' readout system is based on Finite State Machines
(FSMs); upon arrival of a L1-accept command, it transfers the data
corresponding to the L1 trigger from the latency buffers to the
derandomizer FIFOs.




\begin{table}[h]
\caption{Estimation of the event size and data bandwidth for the Endcaps}
\label{Tab_t2}

 \begin{tabular}{|l|c|} 
\hline
\hline
Maximum channel count per  module: & 92 \\  
\hline 
Number of MOD32 processing & \\ units per module & 3 \\
 \hline 
Total number of modules & \\ in the endcaps & 108 \\
 \hline 
Total number of MOD32 & \\ processing units & 324 \\  
 \hline 
 Sampling period (ns)  & 17.86 \\ 
 \hline 
 Number of  samples in the trigger &  \\ matching window   & 16 \\
 \hline 
Endcap Event Size (kB)  &  20.74 \\
 \hline 
 Trigger Rate (kHz) & 150 \\
 \hline 
 Total Barrel Bandwidth (Gbps), & \\ including  8b\/10b encoding  overhead & 31.10\\
 \hline 
Number of data links from the barrel & \\ detectors & 16 \\ 
 \hline 
 Bandwidth per link (Gbps) & 1.944  \\ 
 \hline 

 \end{tabular}
\end{table}

In this baseline version of the IFR electronics, the digitizer blocks
have 32 input channels. In response to a L1-accept, the digitizers
send the data corresponding to that trigger via copper serial links to
the ``data merger'' units. The data merger units assemble the data
from a number of digitizers into event data packets of suitable size
and forward these via the derandomizer buffers and optical links to
the ROMs. Parts of this binary readout scheme have already been tested
on an IFR prototype detector, allowing the determination and optimization
of the readout system parameters, such as the sampling frequency and the
number of samples. 

Tables \ref{Tab_t1} and \ref{Tab_t2} show the estimated
event size and data bandwidth required for the IFR readout at the
nominal trigger rate of 150\kHz. These tables assume a
readout window that can accommodate a trigger jitter of 100\ns, but do
not take into account the data reduction that can be achieved by not
sending the shared data of events with overlapping time windows twice.

\begin{figure*}[ht]
  \begin{center}
\includegraphics [width=0.95\textwidth] {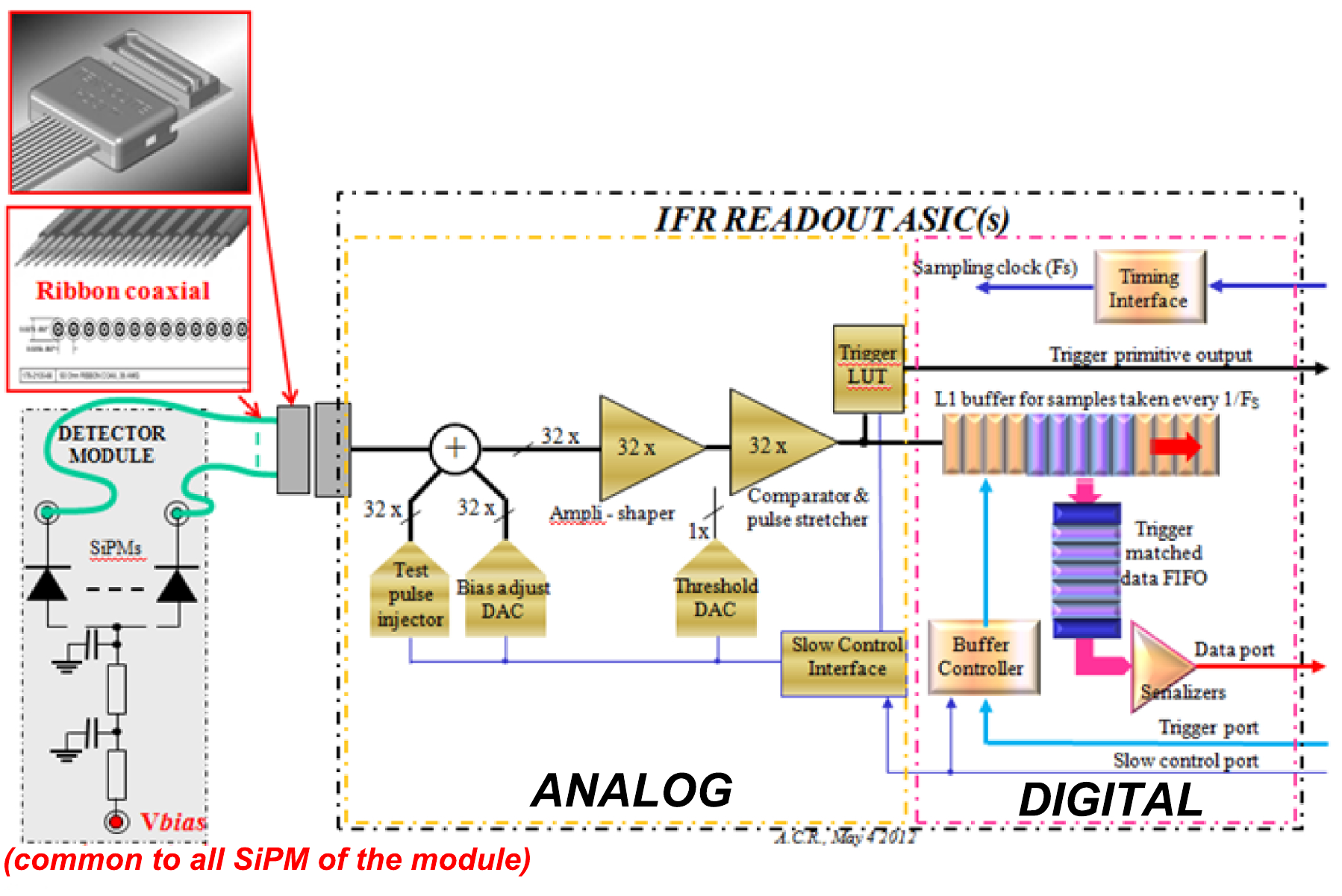}
\caption{Block diagram of the IFR readout}
\label{Fig_f3}
\end{center}
\end{figure*}

While a readout system based on COTS was designed and successfully
used in the IFR prototype, the front end stages of the readout chain
for the final IFR detector will be implemented in an ASIC
whose block diagram is shown in Figure~\ref{Fig_f3}.

The IFR scintillation detectors are each equipped with a silicon
photomultiplier (SiPM) capable of counting single photons. They are
grouped in modules which are inserted in selected gaps of the flux
return steel or around it. The signals from all the SiPMs of a module
are carried by thin coaxial cables. These cables exit the module's
aluminum enclosure and are mass-terminated to a high density connector
on a carrier PCB (printed circuit board). The average length of the
coaxial cable bundles is in the order of a few meters to allow the
carrier boards to be located in an area conveniently accessible for
maintenance.

\begin{figure}[ht]
\begin{center}
\includegraphics [width=0.48\textwidth, height=0.5\textwidth] {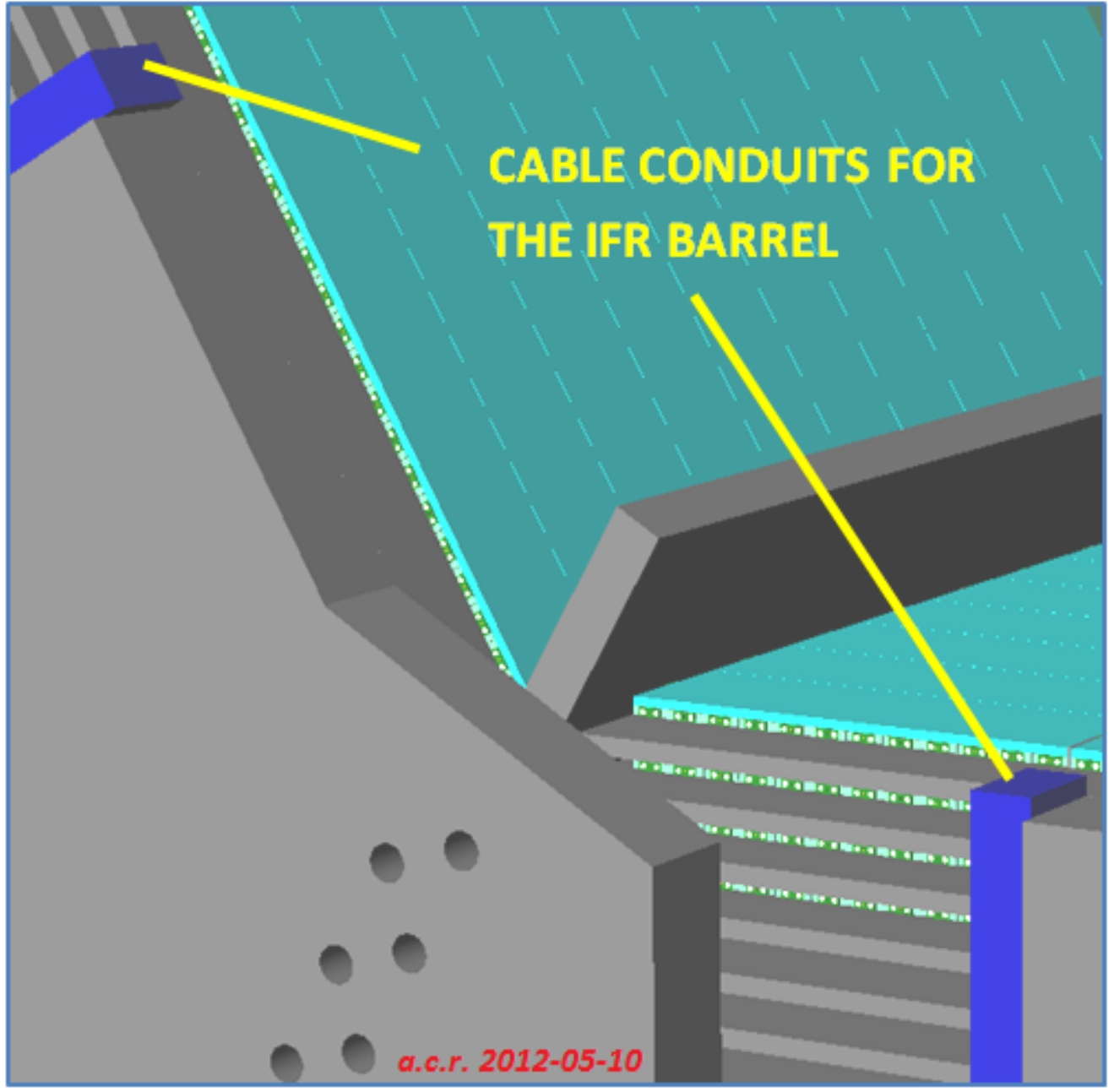}
\caption{The IFR cable conduits for 2 sextants}
\label{Fig_f4}
\end{center}
\end{figure}

Figure \ref{Fig_f4} shows the situation for the IFR barrel: in this
representation the metal enclosures of the modules are not shown; the
closest convenient locations for the ASICs carrying boards are
indicated by the yellow callouts. The front end cards are installed
inside the 3'' x 5'' cable conduits as indicated in Figure
\ref{Fig_f5}. While the figure only shows the double shielded, multi
differential-pair cables (green jacket) that are needed to connect the
front end cards' output serial lines to the data merger units, the
cable conduits are also meant to host cables for the distribution of
the SiPM bias voltages (one bias voltage common to all SiPMs of a
module), of the fast (FCS) and slow (ECS) commands and, finally, of
the supply voltages for the front end cards.

\begin{figure}[ht]
\begin{center}
\includegraphics [width=0.48\textwidth, height=0.5\textwidth] {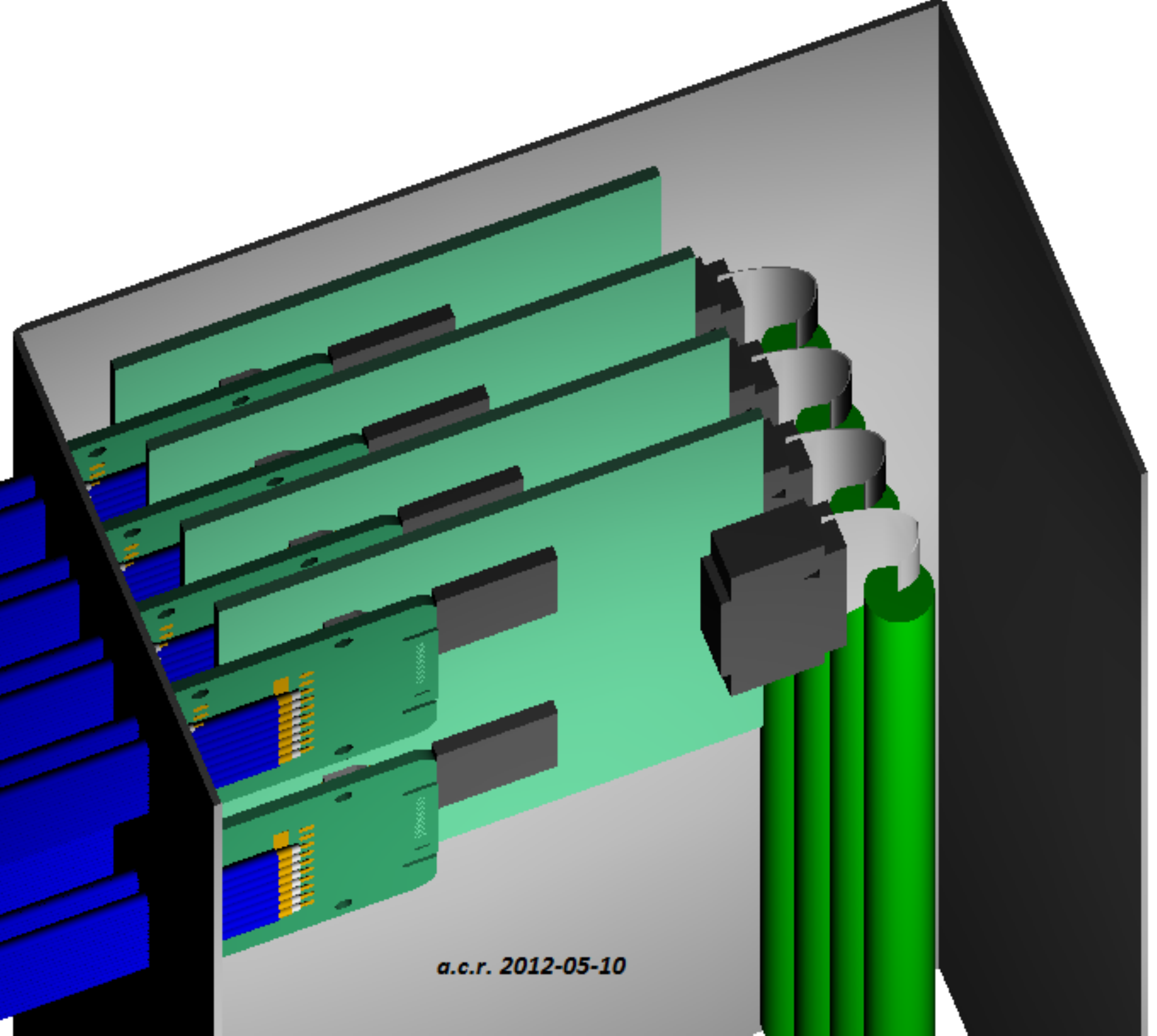}
\caption{Detail of front end cards installed in the IFR cable conduit}
\label{Fig_f5}
\end{center}
\end{figure}

Figure \ref{Fig_f6} shows the perspective view of the two ``doors'' of
an endcap: the yellow callouts indicate the openings in the
side-lining steel through which the data, control and power cables for
the detector modules will be routed. The signals emerging from these
openings are routed to the crates indicated by the red callouts. For
the endcap section of the IFR, the front end cards carrying the IFR
read out ASICs could be installed directly in the crate and connected
to the data merger cards through the crate's backplane
interconnections.

The data merger crates receive the global clock and fast commands
from the FCTS and send their merged data to the ROMs, both via optical
links.  They also have a connection to the ECS for data acquisition
system configuration and diagnostics.

\begin{figure}[ht]
\begin{center}
\includegraphics [width=0.48\textwidth, height=0.5\textwidth] {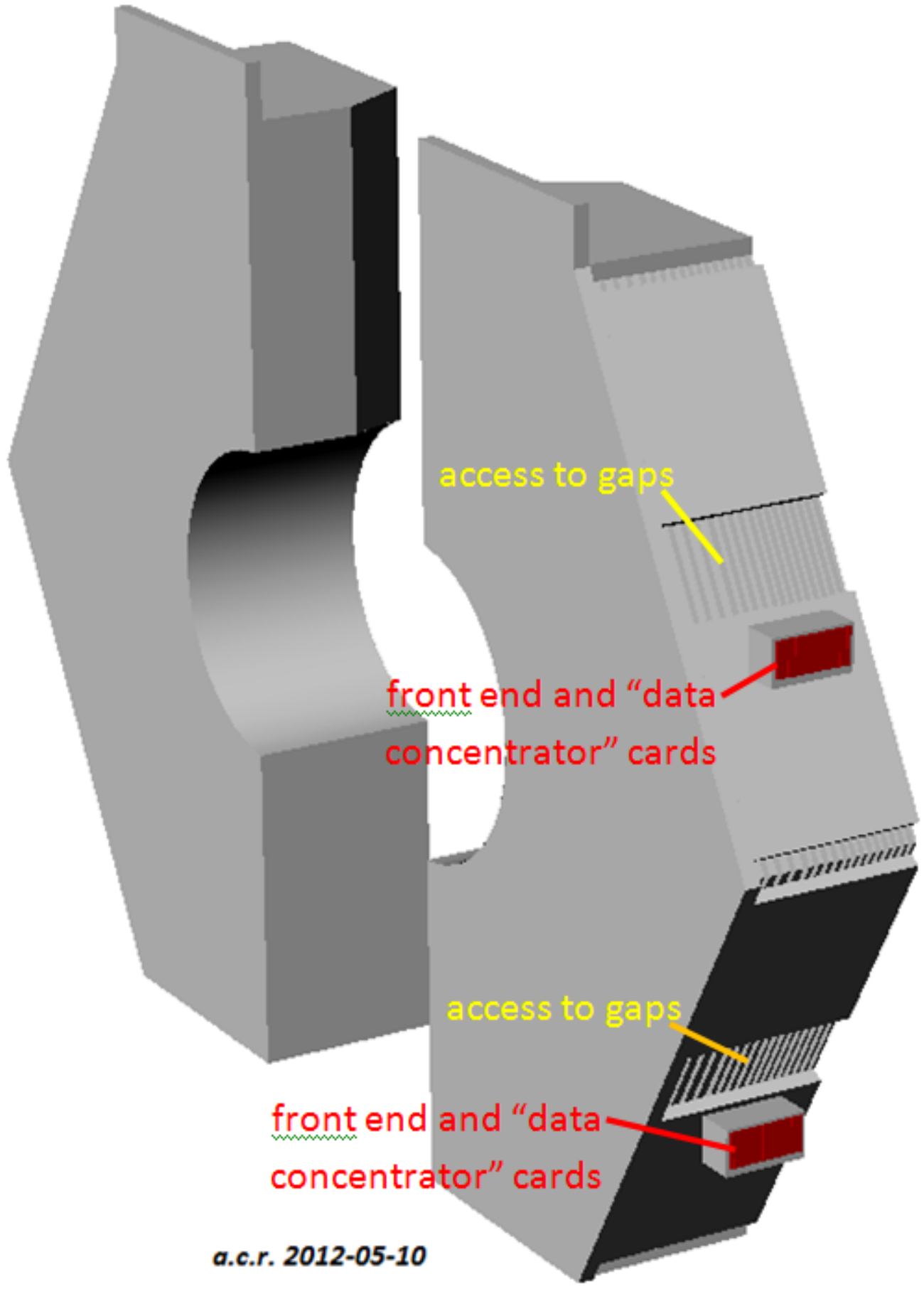}
\caption{Perspective view of IFR endcaps}
\label{Fig_f6}
\end{center}
\end{figure}

The IFR barrel will be equipped with a total of 6 data merger
crates while 8 crates are foreseen for the endcaps. Further details on
partitioning, location and construction of the data merger crates
are given in the IFR subdetector chapter, which also provides a
description of the services needed by the IFR readout in the
experimental hall:
\begin{itemize}

\item SiPM supply voltages: 384 (barrel) + 324
(endcaps) = 708 ``HV'' channels with: maximum output voltage 100\V, 10\bit
 resolution, maximum output current 25\mA. The power dissipated by
the SiPMs in the whole IFR under nominal operating conditions is of
the order of a few tens of Watts. 

\item Front end cards supply voltages:
354 + 354 ``LV'' supply channels with: +3.3\V or -3.3\V output voltage
respectively, maximum output current 1\amp, IR drop compensation. The
power dissipated by the front end stages for each 32 SiPM group is
around 1.6\Watt and thus the power dissipated by the front end stages is
about 620\Watt for the barrel and 520\Watt for the endcap.

\item Cooling: the
temperature of the IFR steel should be controlled to within a few
degrees, as it was for \babar; the design of the barrel cooling system
should then take into account the estimated 1140\Watt dissipated in total
by the front end stages of the IFR barrel and endcap electronics.

\end{itemize}

\begin{figure}[ht]
\begin{center}
\includegraphics [width=0.48\textwidth, height=0.5\textwidth] {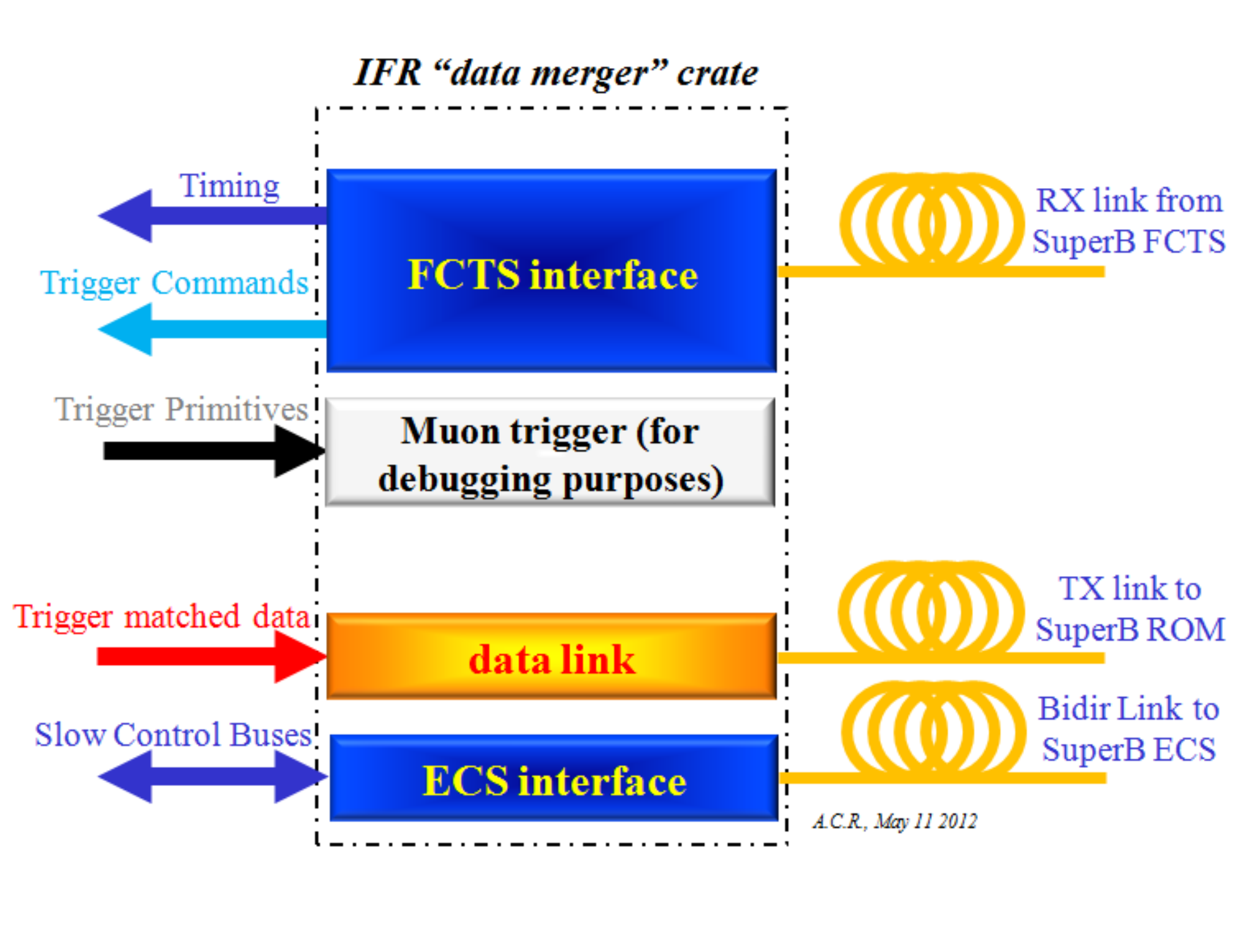}
\caption{Main functions performed by the electronic units in the data merger crate}
\label{Fig_f7}
\end{center}
\end{figure}

\tdrsubsubsec{IFR Summary}

Table~\ref{table:IFRsummary} summarizes the IFR specification concerning the DAQ interface and power requirements. 
\begin{table}[h]
\caption{IFR summary}
\label{table:IFRsummary}
\begin{tabular}{|l|r|}
\hline\hline
Number of Control Links & 20  \\
Number of Data Links & 40  \\
Total Event Size & 45\kbyte\\
On-Detector Power Consumption & 1.2\kW\\
\hline\hline
\end{tabular}
\end{table}

\tdrsec{Electronics Infrastructure}

\tdrsubsec{Power supplies, grounding and cabling}

\def \ad {AC/DC\xspace}
\def \dd {DC/DC\xspace}

The infrastructure for grounding, shielding and powering the detector
is an integral part of the electronics design. As a general rule, the
ground has to be as equipotential as possible across the experiment.
When constructing the experimental hall, it should be foreseen to
connect the iron reinforcement bars sunken in the concrete ground to
the experiment's electrical ground in order to create the most perfect
possible mesh.  Shielding is not limited to cables carrying signals
that can cause or are sensitive to electrical interference, but must
also extend to electronics sensitive to any kind of interference.
Interference can be electric, magnetic or radiative. If possible,
power supplies should remain outside of the radiation area, because
radiation-tolerant power supplies are more expensive and less reliable
than standard ones. This implies long cables. Therefore, power
consumption on the detector has to remain reasonable in order to limit
the power loss along the cables.

\tdrsubsubsec{Grounding and shielding}

Noise in the front-end analog electronics, typically located in or on
the detector, is the most dangerous one for the event data quality. For this
reason, we need the best possible ground on the detector. One has to
avoid high currents flowing back to the power supplies through
unforeseen paths, especially through other subdetectors. Ground
connections have to be strong everywhere, but the currents they carry
have to be very limited. Subdetectors have to be fully equipotential
internally; between subdetectors, this requirement can be somewhat
relaxed, because there are no analog data connections between them.
The choice of having a ``perfect'' equipotential ground has to be made
very early in the experimental hall and detector design, because it is
very difficult to retrofit later.

\tdrsubsubsec{Rules for cables} Cable interconnection can use DC or AC
coupling. DC coupling requires a perfect grounding, and is sensitive
to any kind of currents, whereas AC coupling is only sensitive to AC
currents. In all cases, the goal is to try to avoid any kind of parasitic
currents. Therefore, the best solution is again to ensure a strong
common ground. Moreover, it is strongly advised to also have a common
analog and digital ground on electronics boards.

Cables have to be carefully defined and have to be perfectly matched
to the required currents and the noise levels tolerated. Such cables
have already been designed for other HEP experiments. Shielding mainly
concerns remote interconnections. As a rule, in order to be effective,
shields must be connected to ground on both ends. Within the
electronics house, high-rate serialized data can be transmitted on
shielded Cat6 cables with shielded RJ45 connectors on both ends,
covering distances of up to about 15 meters. 

Generally, we will try to avoid any power consumption local to the
receiver side, and favor differential signalling because it offers
constant power consumption. In cases where unipolar links are
absolutely necessary, serial adaptaion at the source will be used to
make the signal return path the same as the forward path, thus
avoiding current flow in the cables.



\tdrsubsubsec{Power Supply to the Front-end}

Ground is not supposed to be the return path for supply currents,
therefore power cables have to offer a very low impedance path for
current returns. Moreover, the best solution to avoid having return 
currents flowing into ground is to let the power supplies float
and to fix the ground potential on the sole detector side of the power
cables. To avoid voltage drop fluctuations, it is normally better to
regulate the power near to its use; this also permits to use cheaper
DC/DC main supplies. In the \superb\ environment, however, this would
require radiation-tolerant supplies for many components of the
on-detector electronics. Since radiation-tolerant supplies are
expensive and less reliable, we will try to avoid them as much as
possible by using alternative schemes to power the front-ends. 

The voltage supply system is normally composed by a cascade of \ad,
\dd and linear regulators. An example of a scheme for powering
front-end electronics is shown in Figure~\ref{fig:etd-pws-0}. The main
power supplies are located outside the detector, and linear regulators
are used to regulate the power locally. Such as system has of course
to be optimized for the power dissipation and noise requirements. In
\superb, the presence of a strong magnetic field and large particle
fluences poses additional constraints, as they can have an impact on
the aging and behavior of electronic equipment. Radiation is addressed
by adopting only radiation-hardened technology, and by using suitable
layouts recipes for monolithic circuits. Magnetic fields generally
have less impact, except for \ad and \dd converters that use
inductances and/or transformers with ferromagnetic cores.

\tdrparagraph{Power Supply outside the detector area}

In the following we will describe our solution that is able to meet
the requirements outlined above. The basic strategy is to minimize the
amount of power electronics in the detector area with a system design that
allows the distance between the main power source and the front-ends to be a
few tens of meters. In such as system, the energy stored in the
inductive component of a cable can be large, and attention must be
paid to protect the electronic equipment from damage in case of an
accidental short circuit to ground or a sudden change of the current
draw (e.g. in the case of loss and recovery of clock in digital
electronics).

\begin{figure}[ht]
\centering
\includegraphics[width=0.47\textwidth]{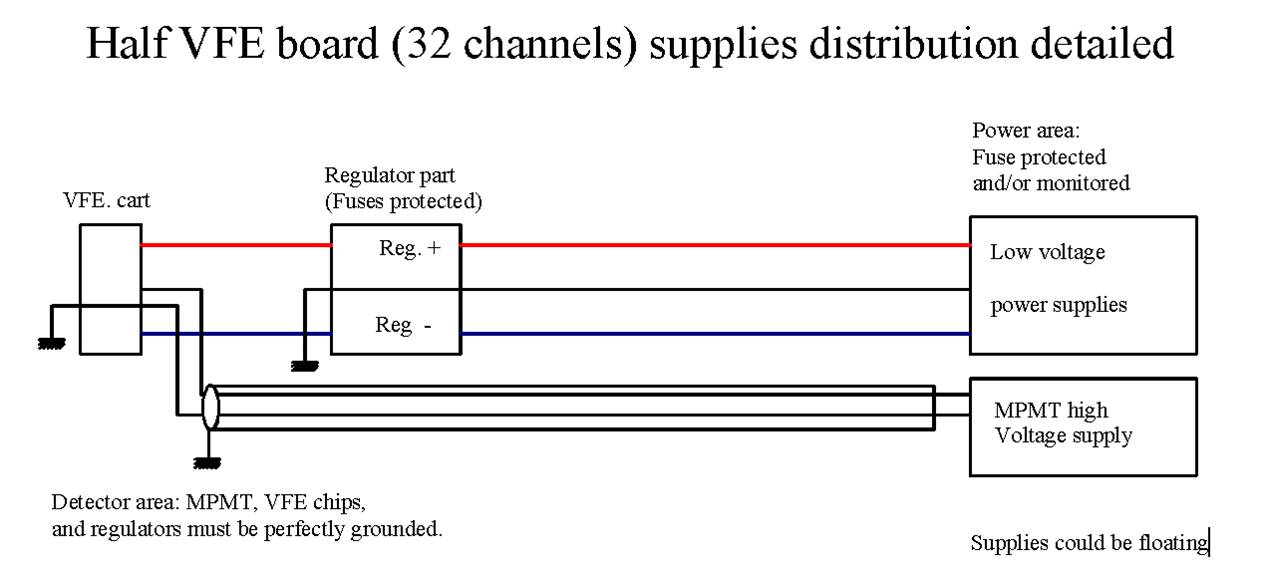}
\caption{Example of implementation for front-end electronics powering. }
\label{fig:etd-pws-0}
\end{figure}

We have found that most cables with 3 or 4 wires and cross-sections
between 1.5\mma\ and 4\mma\ have an inductance per unit length, $L_M$,
of the order of 0.7\muH/m from DC to few hundreds of \kHz. To limit
the voltage at a safe level we add a capacitance in parallel to the
load that is able to store the released energy.  For instance, we
calculated that with 50\m of cable length, $I_{short}=$20\amp and
$C_{lim}=$160\muF, the maximum overvoltage excursion would be less
than 9.5\V.

\begin{figure}[ht]
\centering
\includegraphics[width=0.47\textwidth]{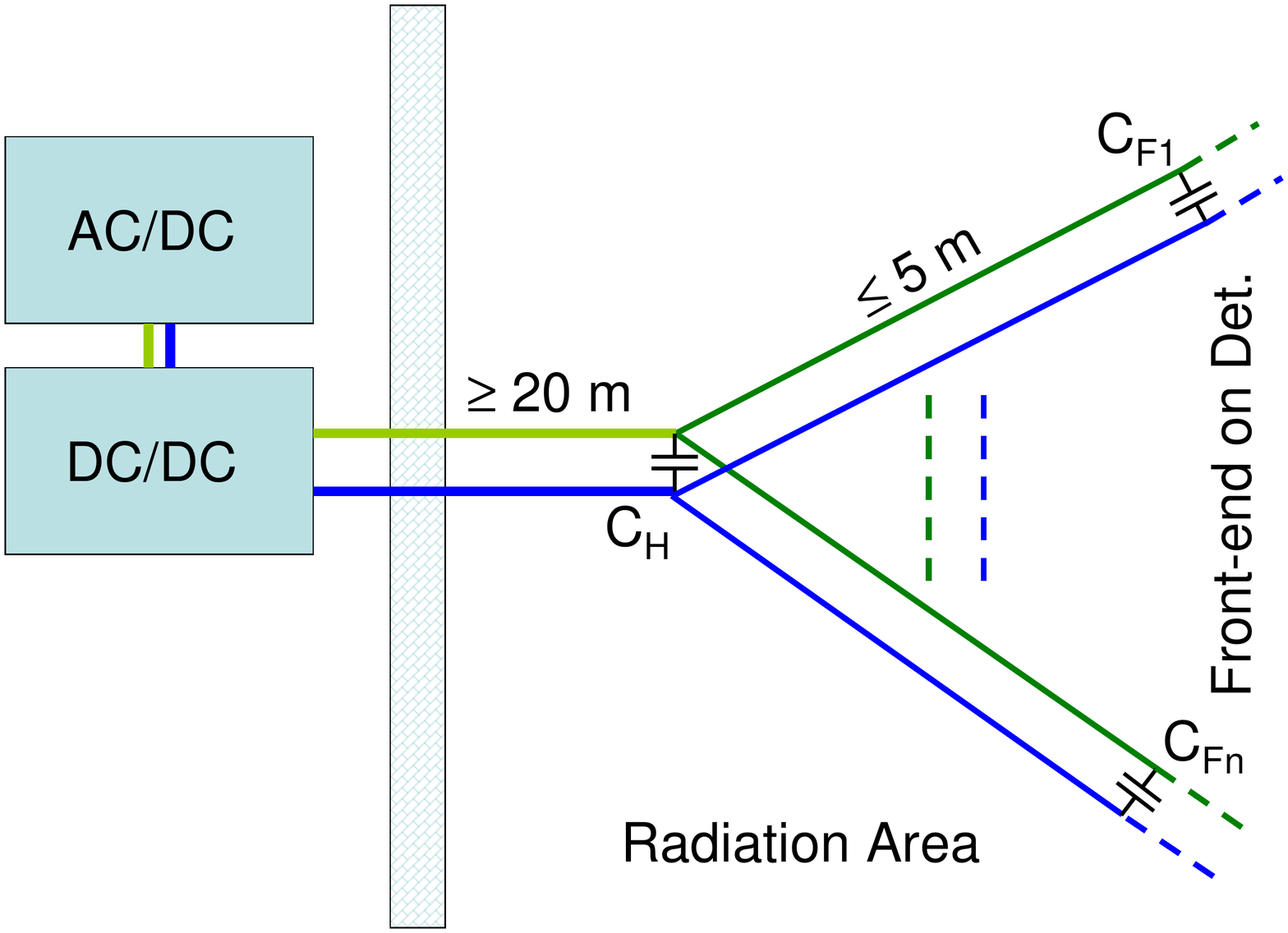}
\caption{Possible voltage supply layout for a sub-detector}
\label{fig:etd-pws-3}
\end{figure}

Adopting this technique, a hub for distribution to several shorter
cables can also be implemented as shown in Figure~\ref{fig:etd-pws-3}.
A large value capacitance, $C_H$, is at the end of the cable that
connects the \dd regulators from the outside to the inside of the
detector area. In our example we continue to assume 50\m cable length
and $C_H = 160\muF$. From the distribution hub several shorter cables,
or stubs, connect the various parts of the detector or sub-detector
front-end. To save space, the cross sections of the stubs can be
smaller, since they each only have to carry a smaller current.

At the end of each of these stubs, 5\m in this example, a smaller
value capacitance, $C_{Fx}$, (33\muF) is connected. In case of short
at the end of a stub all the current flows into it. But as soon as the
short is opened, capacitance $C_H$ absorbs the energy of the longer
cable, while the energy of the shorted stub is managed by the
corresponding capacitance $C_{Fx}$ (Figure~\ref{fig:etd-pws-3}).

The suppression capacitance must have a very small series resistance
and inductance.  Capacitors with plastic dielectric such as Metalized
Polypropylene Film satisfy this condition. As an example the 160\muF
capacitor we have adopted for the test has only 2.2\mOhm of series
resistance, but, being big in volume, it shows a series inductance of
a few tens of \nH. To compensate for this latter effect, a smaller
value (and volume) capacitance (1\muF), able to account for the fast
part of the rising signal, is put in parallel. Metalized Polypropylene
Film capacitances have a range of values limited to a few hundreds of
\muF.  As a consequence, a limited current per cable, 10\amp to 20\amp, results in
a good compromise. Many commercial regulators, also in the form of the
so called bricks and half-bricks layouts, are available off the shelf
at low cost. This strategy is particularly useful for minimizing the
dropout along the cable and it is of particular concern when a low
voltage is needed.


An overvoltage can also happen due to a possible
malfunctioning of the regulator. To reject rapidly and with good
precision this effect a stack of fast diodes is a good choice. For
instance with a voltage supply of 10\V, about 20 diodes would allow to
maintain a safe operating condition for a while, provided that they
are in contact with a heat sink. The diode stacks can be located close
to the regulator where space constraints are not an issue.


\tdrparagraph{Noise cabling and shielding} The combination of the
inductance component of the wires and the suppression capacitance has
an additional benefit, as it behaves also as a low pass filter.
Figure~\ref{fig:etd-power-supply-7} shows the noise at the end of the
50\m cable with two 5\m stubs when 160\muF plus 2$\times$33\muF
capacitances load the combination. The applied supply voltage was 10\V
and the load 3.3\Ohm. In this case a standard commercial regulator was
used. Very low noise \dd regulators have been designed \cite{OurDCDC}
and Figure~\ref{fig:etd-power-supply-8} shows the noise performance
under the same conditions. Even better performances can be obtained by
cascading the \dd to a linear regulator of very good quality
\cite{MySupply}.

Low noise results can be obtained by careful choice of the type of
cable.  In all our measurements described so far we used only shielded
cables. This protects the supply voltage from outside
interferences, and also avoids creating disturbances for the
outside world. We intend to adopt this layout for the final \superb\
setup. In addition, where needed, we intend to add another shield by
inserting the cables into a tubular copper mesh.

The connection scheme of Figure~\ref{fig:etd-pws-3} is, in a natural
way, also suitable to route ground. Assume that the ground of
every detector or sub-detector to which the cables are routed is
isolated. We can then route a tinned copper wire (or a copper bar)
very close to the power supply cables to reduce the area sensitive to
EMI interferences. Such a routing scheme allows a ``star'' connection
with only one ground contact node (recall that \ad and \dd
regulators are floating) which is the standard requirement.

Shielding is considered for those regions where the electric or
magnetic field can affect performance. The shields can be considered
for the whole sub-detector or individually on a channel by channel
basis. This is particularly true for the effect of magnetic field on
those detectors that extend on a large volume, such as photomultiplier
tubes (PMTs).  Past experience shows that in these cases a local
shield implemented with mu-metal around every PMT is essential.

\begin{figure}[ht]
	\centering
	\includegraphics[width=0.47\textwidth]{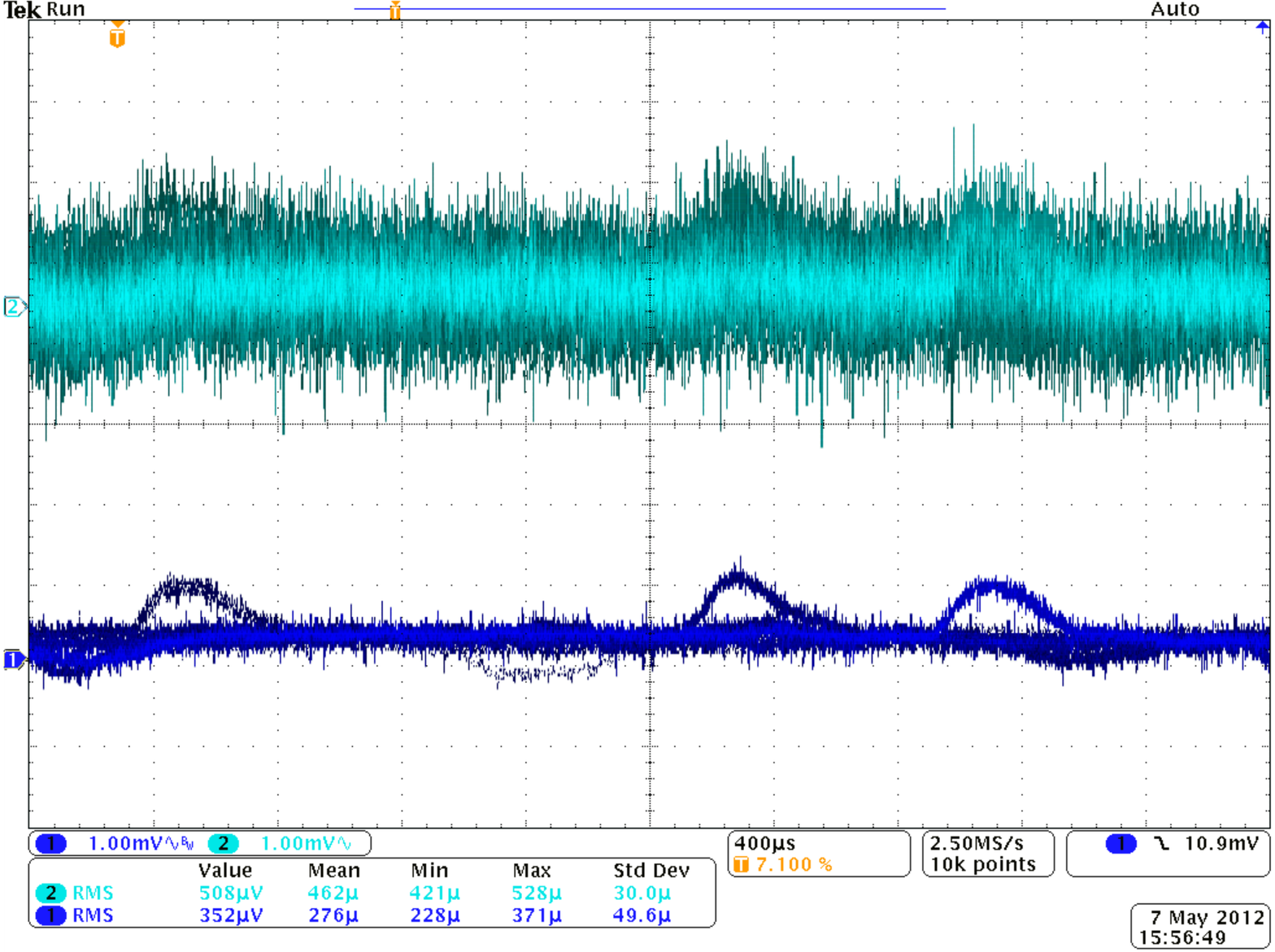}
	\caption{Noise after 50\m cable loaded with 160\muF plus
	2$\times$33\muF capacitance at 10\V and with 3.3\Ohm load. The
	upper noise trace is measured with the full 350\MHz bandwidth
	of the oscilloscope, the lower noise trace has has the scope
	bandwidth limited to 20\MHz.}  \label{fig:etd-power-supply-7}
\end{figure}

\begin{figure}[ht]
	\centering
	\includegraphics[width=0.47\textwidth]{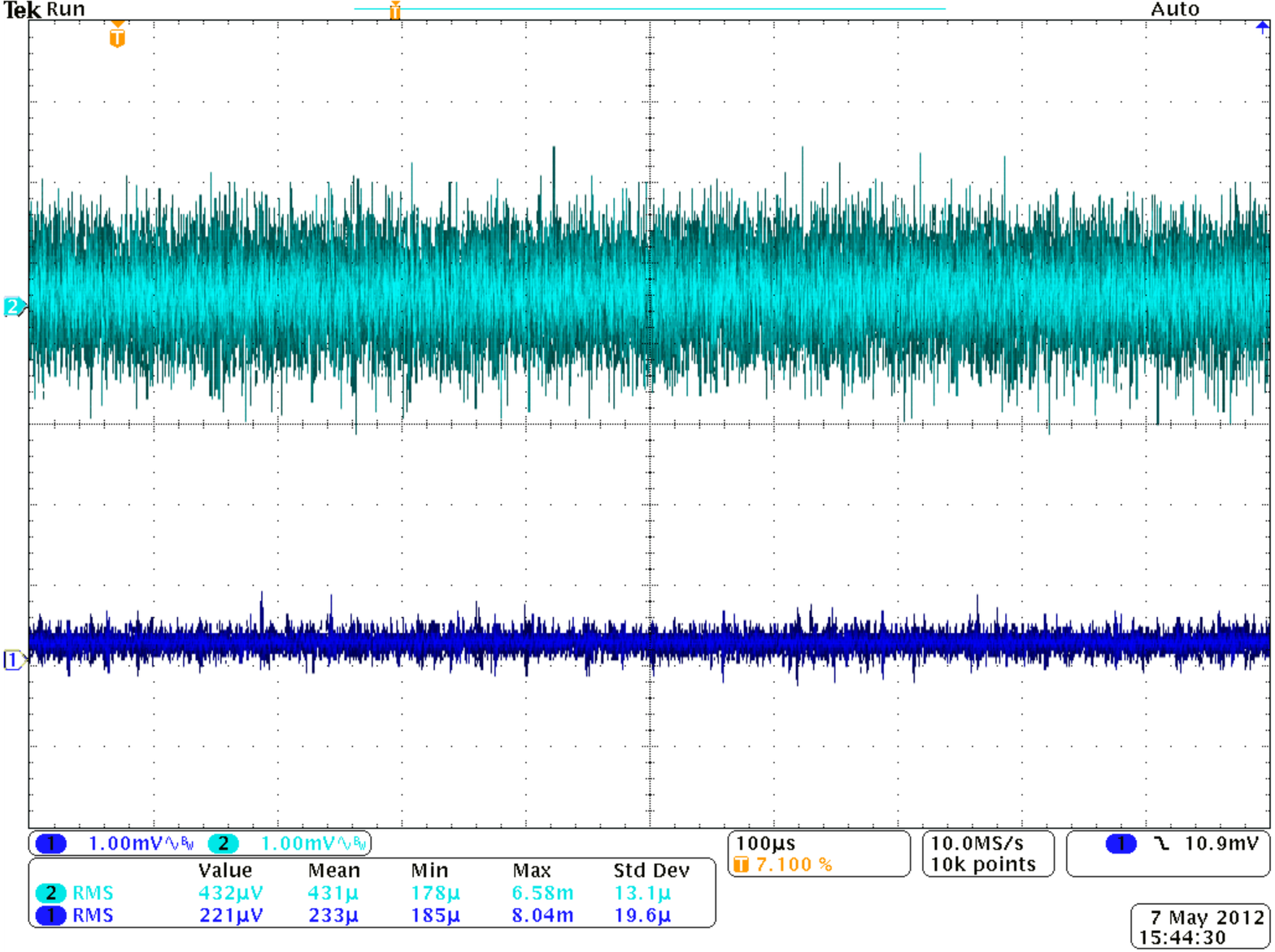}
	\caption{Very low noise \dd regulator
	\cite{OurDCDC}. Measurements condition and setup as for
	Figure~\ref{fig:etd-power-supply-7}}
	\label{fig:etd-power-supply-8}
\end{figure}

\tdrparagraph{Power Supply in the detector area} We are considering to
use both \dd and linear regulators inside the detector area where
inductances and transformers cannot use ferromagnetic coils.  As a
consequence these are limited in their range of values and the
switching speed of the \dd must be very large. This is the case for
the monolithic \dd regulator we are considering \cite{Faccio_1},
developed in 0.35\mum CMOS technology based on a rad-hard layout and
components. Its modulator has a switching frequency of a few \MHz,
which allows the use of a coil-free inductance. A linear regulator
based on the same technology is also available. \cite{Faccio_2}.

\tdrsubsubsec{High Voltage Power Supply for the Detectors} 

High voltage power supplies suffer of similar problems as \ad and \dd
regulators. As a consequence these regulators must be located outside
the detector, sub-detector area (we do not know about any commercial
rad-hard high voltage regulator). The energy released to the load in
case of accidental short circuit would be not an issue thanks to the
fact that such regulators normally have their driving current limited
to a few hundreds of \muA. For instance, if we load the line with a
1\nF high voltage capacitor and assume 1\amp as the short circuit
current, we expect an overvoltage of about 0.2\V. Finally, commercial
overvoltage protectors based on gas discharge tubes are very efficient
and fast.



\bibliographystyle{\tdrbibstyle}
\balance
\bibliography{ELEX/ELEX-bib,IFR/IFR-bib}


\renewcommand{\ChapLabel}{chap:COMP} 
\tdrchap{Software and Computing}
\label{sec:computing}  

During 5 years of data taking, the \superb\ detector will produce more
than 500\pbyte of raw data. To this, event reconstruction and Monte Carlo
simulation will add another 300\pbyte.

To cope with these large data volumes, \superb\ will rely on
predictable progress in computing technology to provide cost reduction
and performance increase. The effective exploitation of computing
resources on the Grid has become well established in the LHC era, and
will enable \superb\ to access a huge pool of world wide distributed
computing resources. 

Crucial issues for \superb\ are the ability to efficiently use modern
CPU architectures, and to efficiently and reliably access very large
amounts of data spread over multiple geographically distributed sites.

So far, the \superb\ computing effort has been devoted to the
development and support of simulation software tools and computing
infrastructure needed to design and validate the detector, to the
initial definition of a computing model, and to computing R\&D.

In the first part of this chapter, the existing software tools are
presented. The second part describes the current baseline computing
model, gives an estimate of the \superb\ computing requirements, and
presents a possible implementation of the computing infrastructure.
The third part describes the dedicated computing R\&D program that has
been initiated to investigate possible technical solutions to the
\superb\ computing problems.

\tdrsec{Tools to support detector studies}

The \superb\ collaboration has developed a set of tools to perform
fairly detailed and sophisticated detector and physics studies. This
toolset includes a detailed Monte Carlo simulation (Bruno) based on
\geantFour~\cite{ref:geant4}, a fast parametric simulation (\fastsim)
which directly leverages the existing \babar\ analysis code base, and
a production system that exploits the computing resources available on
the European and US Grids to perform very large--scale simulation
productions.

In addition, the computing group is providing and supporting the
infrastructure and tools to support and coordinate the day-to-day
activities of the \superb\ collaboration: source code
management, a portal for document management, and tools to enable and
support collaborative work and manage collaboration membership.

A description of the tools made available to the collaboration and
their capabilities is presented in the following.

\tdrsubsec{Full Simulation}    
\label{sec:Full_Simulation}

The availability of reliable full simulation tools is crucial for
designing both the accelerator and the detector. First of all, the
background rates in the sub-detectors need to be carefully assessed
for each proposed accelerator design.  Secondly, for a given
background scenario, the design of the sub-detectors must be optimized
to obtain the best possible performance.  Moreover, full simulation
software can be used to improve the results of the fast simulation in
some particular cases, as discussed below.

\tdrparagraph{Bruno: the \superb\ full simulation software}

To better profit from both the \babar\ legacy, and the experience
gained in the development of the full simulation softwares for the LHC
experiments, the choice was made to rewrite from scratch the core
simulation software. Therefore, \geantFour\ and the C++ programming
language were the natural choice for the underlying technology. 

After some years of development, the \superb\ full simulation software
is, in its present state, usable (and, indeed, used) to assist in the
design of the accelerator and detector. The basic functionality is in
place and well tested; more features are added based on user requests.
We will give in the following a short overview of the main
characteristics, emphasizing areas where future developments are
planned.

\tdrparagraph{Geometry description}

The need to reuse as much as possible the existing geometrical
description of the \babar full simulation, called for some
application-independent interchange format to describe the geometry
and materials of the sub-detectors.  In the present phase of the
\superb\ experiment, a human-readable format is preferred, so that it
is easy to change the geometry details in order to model different
design
alternatives.\\

From the formats currently used in HEP applications, the Geometry
Description Markup Language (GDML) was chosen for the following
reasons:

\begin{itemize}
\item its XML-based structure allows for relatively easy human inspection and
editing, while still providing (by construction) a certain level of robustness.
\item it provides some level of modularization, by allowing to split the global
geometry into smaller files.
\item its specification is an independent project, not linked to any specific
application
\item \geantFour\ provides native interfaces to read and write GDML files. This
allowed, for example, to export the \babar\ original geometry, modify it as
necessary, and read it back into Bruno.
\item ROOT can process GDML geometry as well, allowing for example to easily
visualize some simulation result together with the underlying detector geometry
using ROOT embedded geometry viewer
\end{itemize}

The present implementation relies on GDML as the only source of
geometry information. The file structure reflects the natural division
in sub-detectors, with one global file defining only sub-detector
envelopes, each filled in turn (using a different file) with all
needed details. Space allocation is thus centrally managed at the
level of the top file, while the sub-detector communities have the
freedom of experimenting with different layouts without interfering
with each other.

The consistency of the geometry (volume overlaps, volume
overshoots) is periodically checked with standard \geantFour\ tools.
The choice of GDML also brings in some limitations. One example is the
limited support for loops and volume parameterization. In the longer
term, after the detector layout has finalized, and the ease of
inspection is no longer as crucial as it is now, the GDML-based
approach might be replaced with a custom solution that allows more
flexibility at the cost of reduced human readability.

\tdrparagraph{Simulation input: Event generators}

Bruno can be interfaced to an event generator in two ways: either by
directly embedding the generator code, or by using an intermediate
exchange format. In the latter case, the event generator is run as a
different process and its results are saved in a file, which is then
used to feed the full simulation job.  Bruno presently supports two
interchange formats: a plain ASCII file and a purely ROOT-based format
that uses persisted instances of the TParticle class.

\tdrparagraph{Simulation output: Hits and Monte Carlo Truth}

Hits in the different sub-detectors are created by specific
SensitiveDetectors. The transient representation is managed through
instances of custom implementations of the G4VHit interface which are
translated into instances of ROOT-based classes for persistency.
Presently, all sub-detectors are producing hits, which are then saved
in the output (ROOT) file for further processing and analysis. In
addition to the simulated event as seen from the detector, the Monte
Carlo Truth (MCTruth) describes the event as seen by the simulation
engine itself, i.e. with full detail. The same ROOT-native class
(TParticle) is used for the transient and persistent representations
of MCTruth.  Examples of MCTruth information presently handled by
Bruno include:

\begin{itemize}
\item Full snapshots of the simulation status at different scoring
  volumes. All particles crossing a given scoring volume are persisted
  and saved into the output file, allowing easy estimation of the particle
  flux. Predefined scoring volumes include all sub-detector
  boundaries, and the list is extensible/modifiable at run time without
  need to recompile.

\item Detailed lists of secondaries produced within a given volume.
  The secondaries can be filtered based on the particle type or energy

\item Full particle trajectories, i.e. the list of steps a particle
  makes while propagating through the detector. Due to its very high
  level of detail, this information is of little use without advanced
  filtering. The present implementation allows, for example, to save
  the trajectories only for particles of selected type/energy, being
  produced in a given volume, under the condition that the particle
  leaves the volume it was created in. This is specifically tailored
  for studying the effect of shielding components.
\end{itemize}

\tdrparagraph{Simulation optimization}

In order to properly optimize the simulation results, several
parameters need to be adjusted. The most important parameter is
probably the physics list to be used. A conservative choice was made
by using QGSP\_BERT\_HP, which from other experiments is known to
provide a good description of both electromagnetic and hadronic
physics (including low-energy neutron interactions), but at the cost
of more computing time. Tests with different physics lists are
foreseen for the near future.

On top of the default physics lists, the user can choose to apply some
\superb-specific tunings, such as simulation of optical photons or
different models of Multiple Coulomb Scattering.

Another crucial aspect is the choice of the production threshold for
different particles in different parts of the detector, since this has
impact on both computing and physics performance. The present policy
is to use the default \geantFour\ production cut (0.7mm) unless specific
studies from sub-detectors require different settings. Such studies
are already ongoing, and results will be reflected in the default
configuration.

\tdrparagraph{Staged simulation}

Especially in the design phase, a very common use case is where
the layout of a detector is modified and full simulation is used to
evaluate the effect of the change, requiring the production of an
appropriate number of events. When all sub-detectors are working in
parallel on their own implementations, this may result in sub-optimal
use of computing resources. One way to ease this situation is the use
of staged simulation. 

The basic idea is that if the layout of a detector changes, there is
no need to re-simulate all the interactions taking place
\textit{before} particles hit that detector: the simulation must be
redone only for that detector (and the ones downstream). For example,
if the IFR wants to test a new layout, all existing simulated events
can be re-used up to the electromagnetic calorimeter, and only the
very last simulation step needs to be redone.

The infrastructure for this kind of application is already implemented
and validated, and it uses features already described above: 

\begin{itemize}
\item All simulated events save snapshots of the simulation status at
  sub-detector boundaries in the output ROOT file.
\item The resulting TParticles can be read back by Bruno as if they
  were the output from an event generator, thus seeding the new
  simulation process.
\end{itemize}

In the concrete IFR example, one can use the existing simulated events
as input to the new simulation, instructing Bruno to consider the
snapshot at the exit of the electromagnetic calorimeter as a seed.

\tdrparagraph{Interplay with fast simulation}

As already mentioned in the introduction, full simulation can also be
used in conjunction with the fast simulation programs in certain contexts.

The design of the interaction region in particular has a large impact
on the background rates seen by the detector. Simulating such a
complex geometry with the required level of detail is beyond the
capability of the fast simulation. On the other hand, full simulation
is not fast enough to generate the high statistics needed for signal
events.

\superb\ uses a hybrid approach to this problem. Bruno is used to
simulate background events up to (and including) the interaction
region and a snapshot of the simulation status is saved. In order to
save time, the full simulation of the event can optionally be aborted
once the relevant information has been saved. The result of this
procedure is a set of \textit{background frames}, which can be read
back in the fast simulation program which propagates these particles
through the simplified detector geometry and overlays the resulting
hits with the hits coming from signal events. This approach allows to
combine the two simulations and to use each one only for the tasks it
performs best.

The interplay between fast and full simulation is also very useful for
the evaluation of the neutron background. The idea here is to make
Bruno handle all particle interactions within the interaction region,
as explained above, \textit{plus} all neutron interactions afterwards.
Neutrons are tracked in full simulation until they decay, and the
decay products are saved in the output file, as part of the background
frame. Fast simulation can then include these interactions in the
overlaying procedure.  All these functionalities are presently
implemented, and have been used in the recent productions.

\tdrparagraph{Long term evolution of the full simulation software}

While the full simulation software already provides adequate
functionality for the design phase of the experiment, it will need to
be constantly updated and improved to be ready to support the needs of
the data taking period.

Studies of the readout electronics will require a digitization
framework and digit persistency. Since digitization is generally not
very demanding in terms of CPU, it will be implemented as a separate
step from the full simulation and possibly be run in a different job,
starting from the persisted hits. 

Particular attention will be devoted to performance issues:

\textbf{Physics performance} must be improved, both through a finer
tuning of Bruno's internal parameters and by following closely the
evolution of the \geantFour\ code itself, in order to better profit from
improved modeling of the underlying physics processes. 

Optimizing the \textbf{computing performance} is also of paramount
importance, in particular in the context of highly distributed and/or
parallelized computing. The present implementation is already capable
of running on distributed computing resources, while it still lacks
intra-event parallelism. This is mainly due to limitations of the
\geantFour\ toolkit itself, which will hopefully be overcome in a
timescale compatible with the lifespan of the \superb\ experiment.  In
order to take advantage of new parallelization features, Bruno will
most likely need to be heavily restructured. Similar considerations
apply to the optimal exploitation of the new (yet already widespread)
computing architectures, such as multi/many core CPUs and GPUs.

\tdrsubsec{Fast Simulation}\label{sec:fastsim}

\fastsim\ relies on simplified models of the detector geometry,
materials, response, and reconstruction to achieve an event generation
rate a few orders of magnitude faster than is possible with a
\geantFour-based detailed simulation, but with sufficient
precision and detail to allow realistic physics analyses.

In order to produce more exact results, \fastsim\ incorporates to some
degree the effects of expected machine and detector backgrounds.  It
is easily configurable and allows different detector options to be
selected at run time. It is also compatible with the \babar\ analysis
framework, allowing sophisticated analyses to be performed with
minimal software development.

A diagram summarizing the structure of \fastsim\ is shown in
Figure~\ref{fig:fastsim_diag}.  The simulation proceeds through four
main steps: particles generation, detector configuration and response,
particle reconstruction and analysis of the event.

\tdrparagraph{Event generation} Since \fastsim\ is compatible with the
\babar\ analysis framework, we can exploit the same event generation
tools used by \babar .  On-peak events ($e^+e^-\to\Upsilon(4S)\to
B\bar{B}$), with the subsequent decays of the $B$ and $\bar B$ mesons,
are generated with the EvtGen package~\cite{ref:evtgen}.  EvtGen
also has an interface to JETSET for the generation of continuum
$e^+e^-\to q\bar q$ events ($q=u,d,s,c$) and the generic hadronic
decays that are not explicitly defined in EvtGen. The \superb\ machine
design includes the ability to operate with a $70-80\%$ longitudinally
polarized electron beam, which is especially relevant for $\tau$
physics studies. Events $e^+e^-\to\tau^+\tau^-$ with polarized
electron beam are generated using the KK generator and
Tauola\cite{ref:kkgen}.

Machine backgrounds are superimposed on top of the physics event at the event
generation stage.

\begin{figure}[tbp]
\centering
\includegraphics[width=\linewidth]{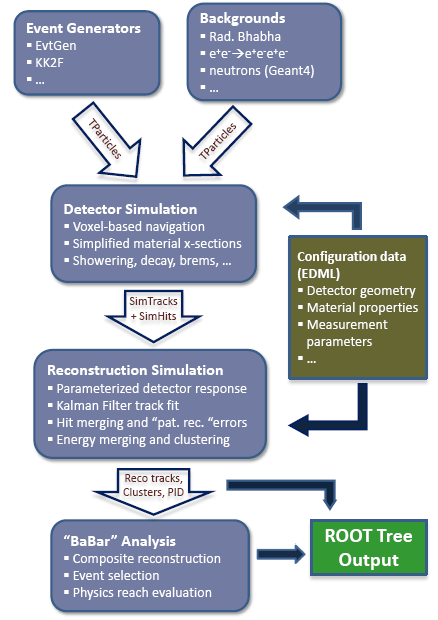}
\caption{Simulation diagram of \fastsim. }
\label{fig:fastsim_diag}
\end{figure}

\tdrparagraph{Detector description} \fastsim\ models \superb\ as a
collection of detector elements that represent medium-scale
pieces of the detector.  The overall detector geometry is assumed to
be cylindrical about the solenoid $\vec{B}$ axis ($z$ axis), which
simplifies the particle navigation.  The detector's volumes are
divided into voxels, defined as cells in the $z$, $\rho$, $\phi$ space
with sizes specified in the XML configuration files. Particles are
tracked through these voxels. When a particle crosses a voxel, it
interacts with all elements inside the voxel without any assumption
about the order of these elements.

Individual detector elements are described as sections of
two-dimensional surfaces such as cylinders, cones, disks and planes;
the effect of physical thickness is modeled parametrically. For
example, a barrel layer of Si sensors is modeled as a single
cylindrical element. Thick detector components, such as the EMC
crystals, are modeled by layering several elements and summing their
effects.  Gaps and overlaps between the real detector pieces within an
element are modeled statistically.  Density, radiation length,
interaction length, and other physical properties of common materials
are described in a simple database.  Composite materials are modeled
as mixtures of simpler materials. A detector element may be assigned
to be composed of any material, or none.  

Sensitive components are modeled by optionally adding a measurement type 
to an element.  Measurement types describing Si
strip and pixel sensors, drift wire planes, absorption and sampling
calorimeters, Cherenkov light radiators, scintillators, and TOF are
available.  Specific instances of measurement types with different
properties (resolutions) can co-exist.  Any measurement type instance
can be assigned to any detector element or set of elements.
Measurement types also define the time-sensitive window, which is used
in the background modeling.  The geometry and properties of the
detector elements and their associated measurement types are defined
through a set of XML files using the EDML (Experimental Data Markup
Language) schema invented for \superb.

\tdrparagraph{Interaction of particles with matter} \fastsim\ models
particle interactions using parametric functions.  Coulomb scattering
and ionization energy loss are modeled using the standard
parameterization in terms of radiation length and particle momentum
and velocity.  Moli\`ere and Landau tails are included.  Bremsstrahlung
and pair production are modeled using simplified cross-sections.
Discrete hadronic interactions are modeled using simplified
cross-sections extracted from a study of \geantFour\ output.
Electromagnetic showers are modeled using an exponentially-damped
power law longitudinal profile (a simplified version of the gamma
distribution in \cite{ref:longo_sestili}\cite{ref:PDGch27}) and a
double Gaussian transverse profile\cite{ref:grindhammer}, which
includes the logarithmic energy dependence and electron-photon
differences of shower-max.  Hadronic showers are modeled with a simple
exponentially-damped longitudinal profile~\cite{ref:had_shower1}\cite{ref:had_shower2} tuned
using \geantFour\ output.

Unstable particles are allowed to decay during their traversal of the
detector.  Decay rates and modes are simulated using the \babar\
EvtGen code and parameters.

\tdrparagraph{Detector response} All measurement types for the
detector technologies relevant to \superb\ are implemented.  Tracking
measurements are described in terms of the single-hit and two-hit
resolution, and the efficiency. Si strip detectors and (optional)
pixel detectors are modeled as two independent orthogonal projections,
with the efficiency being uncorrelated (correlated) for strips
(pixels) respectively.  Wire chamber planes are defined as a single
projection with the measurement direction oriented at an angle,
allowing stereo and axial layers.  Ionization measurements ($dE/dx$)
used in particle identification (PID) are modeled using a Bethe-Bloch
parameterization.

Cherenkov rings are simulated by using a lookup table to define the
number of photons generated based on the properties of the charged
particle when it hits the radiator.
Timing detectors are modeled based on their intrinsic resolution.

In the EMC the energy deposits are distributed across a grid
representing the crystal or pad segmentation, taking into account
profile fluctuations and the energy resolution of the crystals.  In
the IFR the hadronic shower profile produced by charged pions or the
ionization energy loss by muons are used to determine the
2-dimensional distribution of hits over the scintillator planes,
taking into account the intrinsic spatial resolution. The detector
response for charged pions and $K_L$ is also simulated.

\tdrparagraph{Reconstruction} On one hand, full reconstruction based
on pattern-recognition is beyond the scope of \fastsim, on the other
hand, a simple smearing of particle properties is insensitive to
important effects like backgrounds.  As a compromise, \fastsim\
reconstructs high-level detector objects (tracks and clusters) from
simulated low-level detector objects (hits and energy deposits), using
the simulation truth to associate detector objects.  Pattern
recognition errors are introduced by perturbing the truth-based
association, using models based on observed \babar\ pattern
recognition algorithm performance.

In tracking, hits from different particles within the two-hit
resolution of a device are merged, the resolution degraded, and the
resulting merged hit is assigned randomly to one particle.  Hits
overlapping within a region of `potential pattern recognition
confusion' are statistically mis-assigned based on their proximity.
The final set of hits associated with a given charged particle are then
passed to the \babar\ Kalman filter track fitting algorithm to obtain
reconstructed track parameters at the origin and the outer detector.
Outlier hits are pruned during the fitting, based on their
contribution to the fit $\chi^2$.

Ionization measurements from the charged particle hits associated with
a track are combined using a truncated-mean algorithm; this is done
separately for the SVT and DCH hits.  The truncated mean and its
estimated error are used in PID algorithms.  Figure~\ref{fig:dedx_dch}
shows the reconstructed $dE/dx$ in the drift chamber as a function of
the particle momentum for different particle types. The measured
Cherenkov angle from the DIRC is smeared according to its intrinsic
resolution and the Kalman filter track fit covariance at the radiator.

\begin{figure}[tbp]
\centering
{\includegraphics[trim=40 0 90 0,clip=true,width=\linewidth]{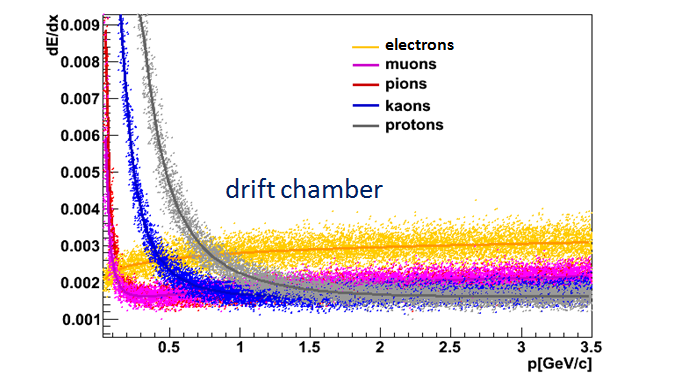}}
\caption{
Reconstructed $dE/dx$ in the drift chamber as a function of the 
track momentum for different charged particle types. 
}
\label{fig:dedx_dch}
\end{figure}

In the EMC, overlapping signals from different particles are summed
across the grid.  A simple cluster-finding algorithm based on a local
maxima search is run on the grid of calorimeter response.  The
energies deposited in the cluster cells are used to define the
reconstructed cluster parameters (cluster energy and position).
Simple track-cluster matching based on proximity of the cluster
position to a reconstructed track is used to distinguish charged and
neutral clusters.

Hits in IFR are combined into clusters, which are then fitted with a
straight line ($\vec{B}$ is zero outside the coil). A number of
quantities useful to distinguish pions from muons are computed, such
as the number of interaction lengths crossed by the particle. The
distribution of this quantity for charged pions crossing the \babar\
IFR is shown in Figure~\ref{fig:ifr_intlength}. The fast simulation
agrees reasonably well with the \babar\ detailed simulation.

\begin{figure}[!t]
\centering
\includegraphics[trim=40 0 90 0,clip=true,width=\linewidth]{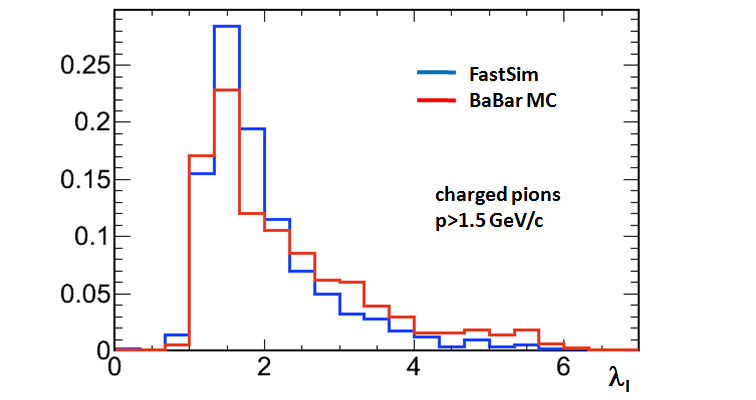}
\caption{
Number of interaction lengths for $\pi^\pm$ crossing the \babar\ IFR.
}
\label{fig:ifr_intlength}
\end{figure}

\fastsim\ has its own event display. Figure~\ref{fig:fastsim_display}
shows a reconstructed $e^+e^-\rightarrow\Upsilon(4S)\rightarrow
B\bar{B}$ event with the $B$ mesons decaying to hadrons. The
subsystems have cylindrical symmetry. Both charged particles and
photons (straight segments) are visible.  The green spots are
reconstructed clusters in the EMC.

\begin{figure}[!t]
\centering
\includegraphics[width=3.0in]{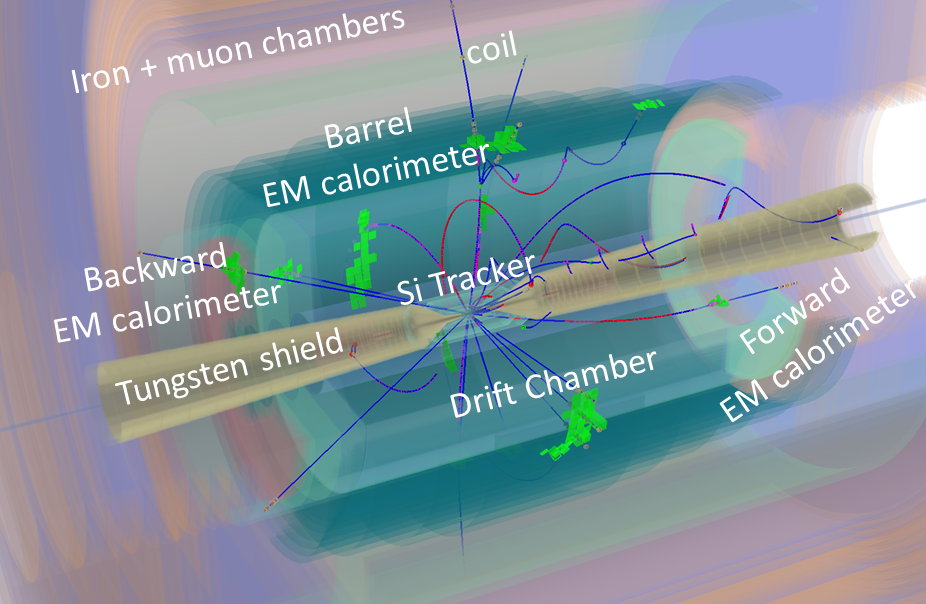}
\caption{Reconstructed $e^+e^-\rightarrow\Upsilon(4S)\rightarrow B\bar{B}$ 
event with $B\rightarrow$~hadrons represented using the \fastsim\ event display.}
\label{fig:fastsim_display}
\end{figure}

\tdrparagraph{Machine backgrounds}\label{sec:machine_bkg} The main
contributions to machine background at \superb\ are radiative Bhabhas,
Pair production ($e^+e^-\to e^+e^-e^+e^-$), Touschek background and
beam-gas interactions. Background events are generated in dedicated
full simulation (based on \geantFour) runs, and then superimposed on the
\fastsim\ physics events. \geantFour\ is needed to model the effect of
background showers in the elements of the interaction region, since
the detailed description of these elements and the processes involved
are beyond the scope of \fastsim.
Background events are stored as lists of the generated particles ({\em
  TParticles}~\cite{ref:root}), these lists are then filtered to save
only those particles which enter the sensitive detector volume.

Background events from all sources are overlaid on top of each
generated physics event during \fastsim\ simulation.  The time origin
of each background event is assigned randomly across a global window
of 0.5~$\mu$s (the physics event time origin is defined to be zero).
Background events are sampled according to a Poisson distribution
whose mean is the rate of the background process times the global time
window.  Particles from background events are simulated exactly as
those from the physics event, except that the response they generate
in a sensitive element is modulated by their different time origin.
In general, background particle interactions outside the
time-sensitive window of a measurement type do not generate any
signal, while those inside the time-sensitive window generate nominal
signals.  The EMC response to background particles is modeled based
on waveform analysis, resulting in exponentially-decaying signals
before the time-sensitive window, and nominal signals inside.  The
hit-merging, pattern recognition confusion and cluster merging
described earlier are also applied to background particle signals, so
that fake rates and resolution degradation can be estimated from the
\fastsim\ output.  The mapping between reconstructed objects and particles
is kept, allowing analysts to distinguish background effects from
other effects.

Background effects on electronics (hit pileup) and sensors (saturation
or radiation damage) are also crucial for \superb, but are best
studied using the full simulation and other tools. Their impact on the
measurement performance can be implemented parametrically at the
detector response or reconstruction level.

\tdrparagraph{Analysis tools}\fastsim\ has been designed to be
compatible with the \babar\ analysis framework, so that existing
\babar\ analysis tools such as vertexing and combinatorics engines can
be used in \fastsim\ with small modifications.
PID selectors are also available. They provide lists of identified
particles (pions, kaons, electrons, etc.) selected by combining the
PID information measured in the relevant subsystems and are key
ingredients in most of the \superb\ physics analyses.

User packages have been developed to support the reconstruction and
analysis of specific decay modes. For instance, packages are available
for the selection of tau decays in $\tau^+\tau^-$ events or the
reconstruction of semi-inclusive hadronic and semileptonic $B$ decays
for the recoil analysis of rare $B$ decays.

The standard tool used in \babar\ to store analysis information into a
ROOT tuple has been adapted to work in \fastsim, allowing even complex
analyses to be run in \fastsim\ approximately as in \babar.

Typical per-event processing times on a dual quad core Intel(R)
Xeon(R) E5520 CPU (2.27\GHz) are 1\ms for particle generation, 10\ms
for propagation of particles through the detector, 100\ms for
reconstruction, and 100-1000\ms for composites selection, depending on
the event complexity.

\tdrparagraph{Simulation validation and detector studies} Many aspects
of \fastsim\ have been validated either by simulating the \babar\
detector and comparing the output with the full simulation or real
data of the \babar\ experiment, or by comparing with the \superb\ full
simulation. In general the outcome of these comparisons is quite
satisfactory. The tracking simulation works particularly well, despite
the lack of a real pattern recognition.

\fastsim\ is being used extensively to optimize the \superb\ detector
design by studying the performance of different layouts. Since the
complete simulation and reconstruction chain is available, including
the tools to perform a complete physics analysis of the simulated
events, it is possible to compare different detector configurations
directly in terms of the physics reach of benchmark channels.

An example is illustrated in Figure~\ref{fig:btoknunu}. The top plot
shows the reconstructed mass of the tag $B$ mesons selected in
hadronic decays, with the \fastsim\ result superimposed to the \babar\
full simulation. The bottom plot shows how $S/\sqrt{S+B}$ ($S$ and $B$
are the signal and background yields, respectively) scales as a
function of the integrated luminosity and detector layouts.

\begin{figure}[!]
\centering
\includegraphics[trim=0 0 40 0,clip=true,width=\linewidth]{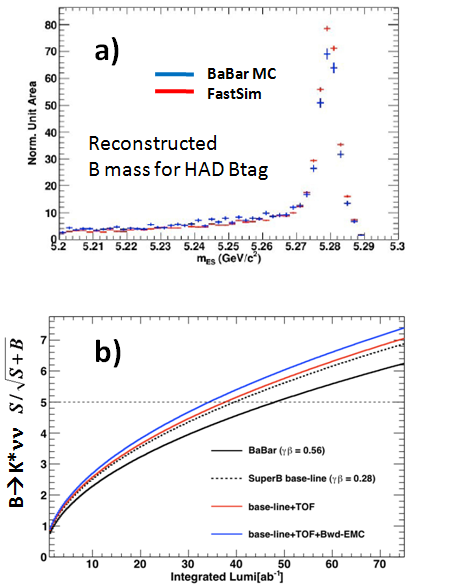}
\caption{Top: Mass of the reconstructed $B$ decays selected into hadronic final states. 
Bottom: precision of the $B\rightarrow K^*\nu\bar\nu$ measurement as a function of the integrated 
luminosity in four different detector configurations.}
\label{fig:btoknunu}
\end{figure}

\tdrsubsec{Distributed computing tools}


Even at this early stage in its lifetime, the \superb\ experiment
needs very large samples of Monte Carlo simulated events to evaluate
the machine background, to optimize the detector design, and to
estimate the physics analysis performances. Since producing these
samples is beyond the capacity of a single computing farm, the
\superb\ collaboration developed a suite of tools to fully exploit the
existing HEP world wide Grid computing infrastructure.
\cite{ref:egi}\cite{ref:osg}\cite{ref:lcg-tdr}. The \superb\
distributed computing tools are built upon the LHC Computing Grid
(LCG)~\cite{ref:lcg-tdr} which is widely adopted in the HEP community
and has long term support by the Grid initiatives.  The \superb\
Virtual Organization (VO) is currently enabled at several sites located in
countries participating in the project; both Grid flavors,
EGI~\cite{ref:egi} and OSG~\cite{ref:osg} are in use within the
\superb\ VO, so all services need to be able to support multi-flavor
Grid middlewares.

The \superb\ distributed tools use the following
LCG services and applications:
\begin{itemize}
   \item The \emph{Workload Manager System} (WMS)~\cite{ref:wms} for
     job brokering, resource matching and job submission;
   \item the \emph{Virtual Organization Membership System}
     (VOMS)~\cite{ref:voms} to manage user authentication and
     authorization;
   \item the \emph{LCG File Catalog} (LFC)~\cite{ref:lfc} to provide
     logical to physical file mappings;
   \item the \emph{SRMV2 resource manager layer}~\cite{ref:srm} to
     share, access and transfer data among heterogeneous and
     geographically distributed data centers; and
   \item the \emph{LCG-Utils}~\cite{ref:lcg} to perform data handling
     tasks compliant with LCG and SRMV2.
\end{itemize}

\begin{figure}[!h]
\centering
\includegraphics[width=\linewidth]{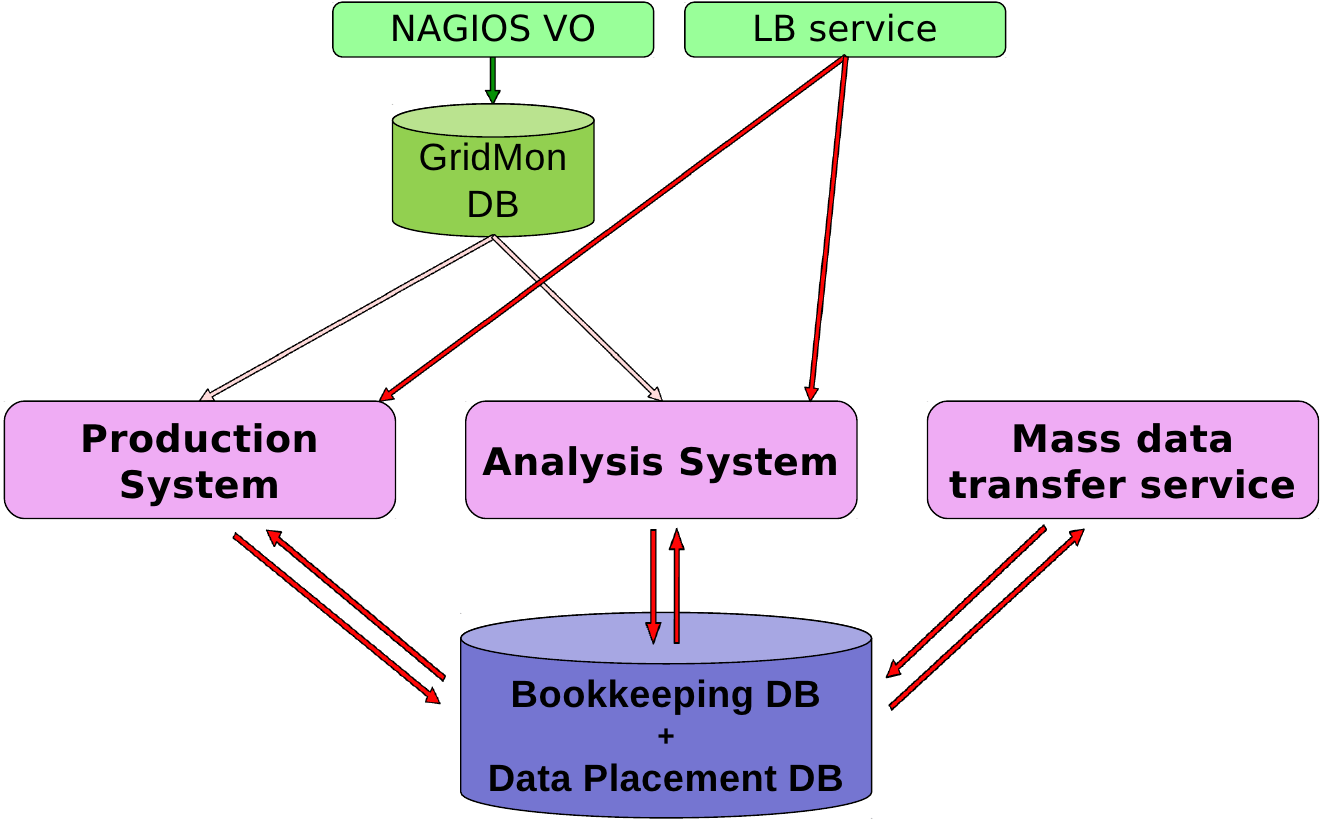}
\caption{Distributed systems bird's-eye view.}
\label{fig:dist_systems}
\end{figure}

Figure~\ref{fig:dist_systems} shows the important elements of the
\superb\ distributed computing system: actual processing and data
movement are performed in the \textit{Simulation Production System},
the distributed \textit{Analysis System} and the \textit{Mass data
transfer service}.  The \textit{Bookkeeping and Data Placement
databases} hold metadata related to
\fastsim and FullSim, and information about dataset structures and data
placement on Grid resources. The monitoring system uses the
\textit{Nagios}~\cite{ref:nagios} tools suite to implement a
\textit{Service Availability Monitoring} (SAM)~\cite{ref:sam} to
maintain a per-VO status of the Grid elements in the GridMon database.
Finally, the \textit{Logging and Bookkeeping} (LB) service provides
detailed job status information from the Grid-internal processes.

\tdrparagraph {The Simulation Production System} has been developed to
manage the production of very large Monte Carlo samples in a
distributed environment.  It supports both \fastsim and FullSim, is
tightly integrated with the bookkeeping database, permits a fine
tuning of operational tasks via web portal, and hides the intrinsic
complexity of the Grid infrastructure from the user by providing
automated submission procedures and rich monitoring features. Job
submission is performed by the Ganga~\cite{ref:ganga} system.  A Ganga-specific script
has been developed to permit run time customizations and the
generation of site-specific job files.

The web interface provides rich monitoring features by means of
querying the bookkeeping database and interacting with the Logging and
Bookkeeping gLite service.  The user can retrieve the list of jobs as
a function of their unique identifier (or range of them), their
specific parameters, the execution site, status, and so forth. The
monitor section provides a per job list of output files and a direct
access to the corresponding log files. The output file size, execution
time, computing resource load status, and the list of the last
finished jobs (successfully or with failures) are also provided.

A simple authentication and authorization layer, based on an
LDAP~\cite{IETF:LDAP} directory service permits to apply a user-role
based policy to access the system.  A VOMS proxy user authentication
has been added to be compliant with Grid Security Infrastructure (GSI)
standards.

For a detailed and technical description of the production system 
framework please refer to \cite{ref:simprod_chep12}.

\tdrparagraph{The data analysis system prototype}

Physicists involved in Monte Carlo simulation productions and in data
analysis need to perform job management and basic data transfer
operations in a user-friendly way and with minimal training.  Ganga
has been adopted as the interface layer of this distributed
analysis infrastructure and a \superb-specific Ganga-plugin has been
developed. Two use cases are currently implemented: a private Monte
Carlo production and data reduction and analysis of centrally (or
privately) produced Monte Carlo datasets. The main components of the
\superb\ data analysis suite are the job wrapper, the bookkeeping and
data placement database and the Ganga plug-in software layer.  The job
wrapper allows the management of stage-in, user application execution,
monitoring and stage-out phases. It interacts with the Ganga plug-in
for monitor purpose only. Bookkeeping and data placement databases
are the back end providing the Ganga plug-in with all the information
needed regarding distributed resources, dataset structure and
placement and output metadata. For a detailed and technical description 
of the distributed analysis prototype please refer to \cite{ref:ganga_chep12}.

\tdrparagraph{Bookkeeping and data placement database}

Both the distributed production and data analysis systems require a
way to identify selected data files and the Storage Elements (SEs)
that hold these files in the distributed system. The immediate access
to the execution status of jobs and their parameters is crucial for
users in order to plan their activities and summarize the results.
Therefore, a data bookkeeping system has been developed, that stores
the semantic information associated with data files and tracks the
relation between jobs, their parameters and their output data.

The bookkeeping database is extensively used by the production job
management system itself in order to schedule subsequent submissions,
and to keep the information about request completion and site
availability up to date. It was modeled after the general requirements
of a typical simulation production application; its design uses the
relational model, and the current implementation uses the PostgreSQL
RDBMS~\cite{ref:postgresql} in a centralized way.

The database connections from/to processes running in a Wide Area
Network environment are managed via web services, in particular, a
RESTful~\cite{ref:rest} service in front of the database has been
developed.  Database accesses to the central information systems from
jobs running at remote sites are authenticated by VOMS proxy
certificates. Local jobs may alternatively use direct connections to
PostgreSQL.

The feasibility of integrating and using a schema-free and/or
document-oriented information system is under study. Such a system
might be required for distributed metadata management and data
replication with bi-directional conflict detection and management.

\tdrsubsec{Collaborative tools}

The computing group supports a suite of computing tools to facilitate
the day-to-day activities of the \superb\ collaboration, to present
the project to the general public, to exchange information among
collaborators, to manage and store documents, and to coordinate
software development. A short description is presented in the
following. 

\tdrparagraph{Authorization} The need to manage authorization and
authentication for the number of people in a \superb-size
collaboration imposes the use of some kind of database. Since several
services (code repository, internal web pages, wiki, etc.) require
authentication/authorization, a directory service based on LDAP~\cite{IETF:LDAP} 
has been chosen to implement single-sign-on (SSO) access across
the \superb\ service suite, and to manage authorization information
such as access permissions and group membership.

\tdrparagraph{Web Portal} The web plays a central role for the
\superb\ collaboration. Web services and tools are both used for
outreach to the general public, and to provide collaboration space
where \superb\ members can interact, and share content and knowledge.
Initially, \superb\ started out with two separate web content
management systems (CMS), one for the general public, the other for the
collaboration website. Since then, these services have been
consolidated in a portal system. The \superb\ collaboration portal is
based on the \textit{Liferay} portal system \cite{ref:liferay} and
provides both an outreach website, and a collaboration website. For
the collaboration website it uses \textit{Jasig CAS}
\cite{ref:jasig_cas} to provide SSO to collaboration members; after
authenticating through the portal, they can use collaboration services
without the need to sign in to each service individually\footnote{Not
  all collaboration services have been converted to SSO yet.}.
Collaboration services include the \superb\ event calendar, a members
directory, and an administrative database to manage collaboration
membership, speakers, and service access privileges.

\tdrparagraph{Document Management} The \superb\ document management
system is implemented using the \textit{Alfresco} platform
\cite{ref:alfresco}, an enterprise class suite of applications for
document and content management. The repository is fully integrated
with the collaboration portal and is accessible both via web
interfaces or WebDAV. It provides document workflows, automated
actions on documents, user/group based document-level security,
\superb-specific data dictionaries, fully indexed documents (metadata
and content), and auditing and versioning of all documents. The
repository can be browsed or searched by category or keywords.

\tdrparagraph{Documentation} The document repository notwithstanding,
the \superb\ collaboration has realized from the beginning that it
also needs a tool to facilitate the creation and maintenance of web
pages to be used as internal documentation. For this, Mediawiki~\cite{ref:mediawiki}
has been chosen, because it combines easy creation and
modification of content in a web browser (WYSIWYG or simple markup
language) with basic content management features, integrates well with
the directory service, and can also handle content like \TeX\
formulas, plots, and pictures.


\tdrparagraph{Code repository} The \superb\ collaboration currently
uses \textit{Subversion} (SVN) \cite{ref:subversion} in conjunction
with \textit{Trac} \cite{ref:trac}\ to manage the source code for
\fastsim, FullSim, several component projects, and documentation. SVN
provides the management of source code and revisions, while Trac
provides add-on services such as code navigation via web browser, a
ticketing system, a wiki for the code documentation and a blog.

\tdrparagraph{Code packaging and distribution} The \superb\ software
release system is based on the \textit{RPM} \cite{ref:rpm} packaging
system. It has been chosen because it provides an easy way to install,
update and uninstall software packages, and also provides a mechanism
to declare explicit dependencies on other RPM packages. External
packages (e.g. ROOT, \geantFour\ and others) are also packaged and
distributed in RPM format. \superb\ code distribution is performed
using \textit{yum} \cite{ref:yum}, an updater and package
installer/remover which facilitates the use of RPM by automatically
computing and resolving dependencies and orchestrating the download
and installation/uninstallation of RPM packages.


\tdrsec{The \superb\ baseline computing model}

The computing models of the \babar\ and Belle experiments have proven
to be very successful in handling the data volume generated by a
flavor factory in the ${\cal L}=10^{34}\,\rm{cm}^{-2}\rm{s}^{-1}$
luminosity regime. The data volumes produced by \superb\ will be
two orders of magnitude larger, nevertheless, we expect that a
\babar-like computing model can still apply.  In this section we
introduce the \babar-inspired data processing strategy and computing
model, give an estimate of the extrapolated \superb\ computing
requirements, and finally present a possible computing infrastructure.

\tdrsubsec{Data processing strategy}

\begin{figure*}[t]
\begin{center}
\includegraphics[width=0.95\linewidth]{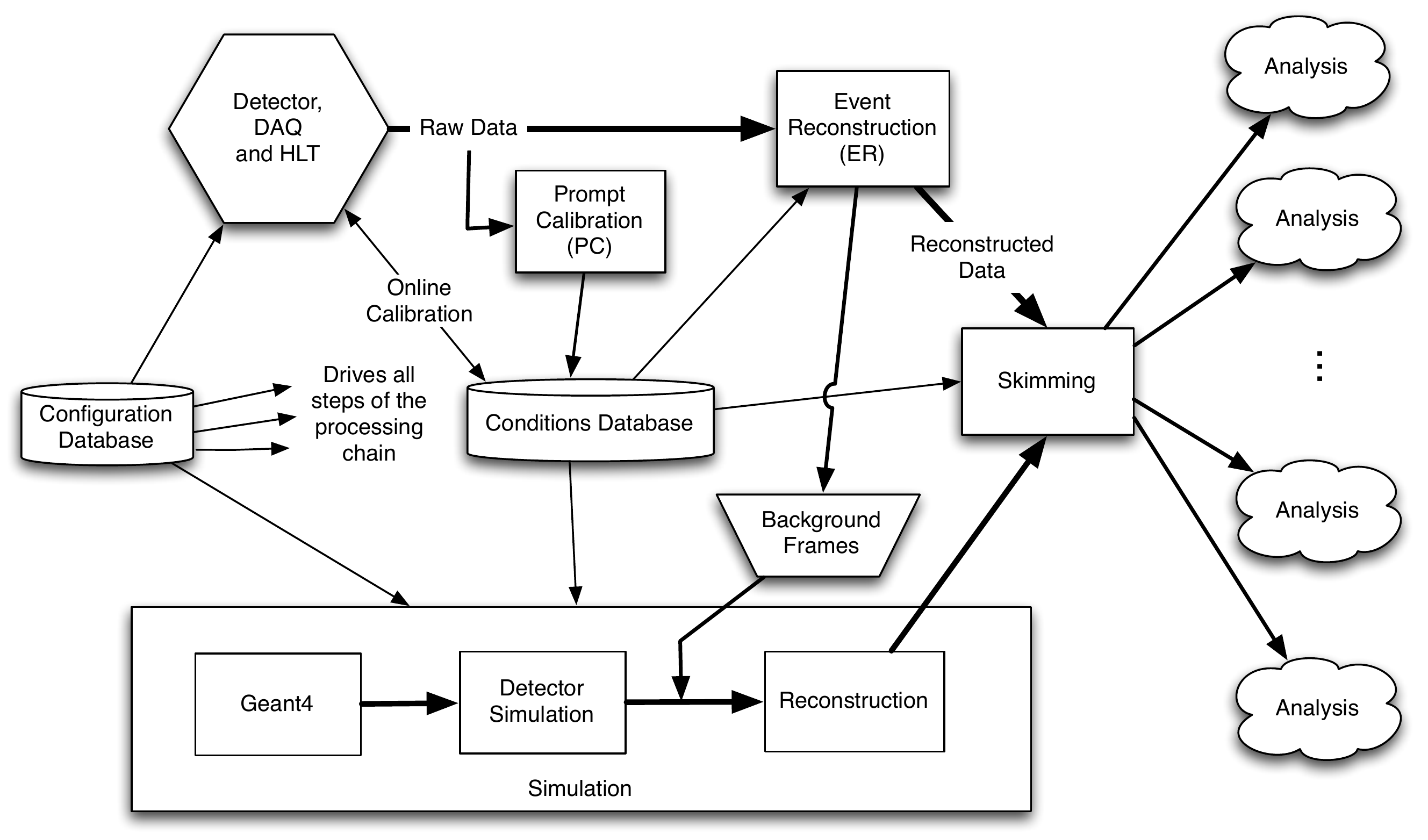}
\caption{Data and control flow in the \superb\ baseline computing model.}
\label{fig:cm-overview}
\end{center}
\end{figure*}

Data processing in \superb\ is envisaged to be similar to the strategy
used in \babar. An overview of the proposed computing model and data
flow is shown in Figure~\ref{fig:cm-overview}.

``Raw data'' coming from the detector are permanently stored and then
reconstructed in a two-step process:

\begin{enumerate}
\item a \textit{Prompt Calibration} (PC)  pass is performed on a subset of the
  events to determine various calibration constants.
\item a \textit{ full Event Reconstruction} (ER) pass that uses the
  calibration constants derived in the first step is then performed on
  all events.
\end{enumerate}
Reconstructed data are permanently stored and data quality is
monitored at each step of the process. 

A comparable amount of Monte Carlo simulated events are also produced
in parallel and processed in the same way. The Monte Carlo simulation
incorporates the trigger configuration, the calibration constants, and
\textit{background frames} derived from ``random'' triggers recorded
by the data acquisition system.

Reconstructed data (both from the detector and from the simulation) are
stored in two different formats: The \textit{Mini} contains reconstructed
tracks and energy clusters in the EMC, as well as information from the
other subdetectors. Efficient packing of data and noise suppression
makes this format relatively compact. The \textit{Micro} contains only
information that is essential for physics analyses.

Detector and simulated data are made available for physics analysis in
a convenient form through the process of ``skimming''. This involves
the production of selected subsets of the data, the ``skims'',
designed for specific areas of analysis.  Skims are very convenient
for physics analysis, but they increase the storage requirement
because the same events can be present in more than one skim.

From time to time, as improvements in constants, reconstruction code,
or simulation are implemented, the data may be ``reprocessed'', or new
simulated data may be generated. When a set of new skims become
available, an additional skim cycle can be run on all the
reconstructed events.

At the core of the \superb\ computing model there are two databases, the
\textit{configuration database} that is the authoritative source for the
detector, data acquisition system and trigger configuration, and the
\textit{conditions database} that records calibration information and
provides it to all components of the computing model.

\tdrsubsec{Resource estimate}

Computing resource estimates for \superb\ are based on the extensive
experience with \babar\ computing. As described above, the baseline
\superb\ computing model is very similar to the \babar\ one. This
allows us to estimate the \superb\ computing resource needs by
extrapolating from the well-understood \babar\ computing parameters.
Table \ref{tab:comp-bbrtosb} shows the basic parameters of the \babar\
and \superb\ computing models. 


\begin{table*}[t]
\begin{center}
\begin{tabular}{lrrlr}
Parameter & \babar &  \superb & & Factor \\
\hline
Raw event size & 35 & 200 & \kbyte & $\sim 6$\\
Raw data size  & 875 & 5250 & \tbpab & $\sim 6$\\
Mini/Micro event size & 42 & 84 & \tbpab & 2 \\
\hline
CPU for Reconstruction  &88.4 & 354 & \khspab & 4 \\
CPU for Simulation Production& 280 & 1120 & \khspab & 4 \\
CPU for Skimming (Data)& 40.8& 122 & \khspab & 3 \\
CPU for Skimming (MC) & 40 & 120 & \khspab & 3 \\
CPU for Physics Analysis (Data) & 4.8 & 14.4 & \khspab & 3 \\
CPU for Physics Analysis (MC) & 5.2 & 15.6 & \khspab  & 3 \\
\hline
Number of raw data copies & & 2 & &\\
Number of Mini copies & & 1 & & \\
Number of Micro copies & & 4 & &\\
Skim expansion factor & & 5 & & \\
Fraction of Mini on disk & &  $100\%\rightarrow10\%$ (Year 5+) & & \\
\hline
Reprocessing cycles / year & & 1 & & \\
\hline
\end{tabular}
\end{center}
\caption{Basic \superb computing resource estimate parameters.}
\label{tab:comp-bbrtosb}
\end{table*} 

Compared to \babar, the \superb\ raw data size increases by about a
factor of 6, mostly due to detectors with higher granularity and
sampling rates, and the projected need to permanently store raw EMC
waveforms to recover detector performance during reconstruction. After
reconstruction, the event sizes in Mini and Micro are expected to be
about twice the corresponding sizes in \babar.

We expect the CPU needs to increase by a factor of 4 for
reconstruction and simulation production, and by a factor of 3 for all
other activities. The factor of 3 is an estimate, and is based on the
assumption that skimming and analysis will require more CPU compared
to \babar\ to handle increased combinatorics and for advanced
algorithms to handle the larger backgrounds. The factor of 4 reflects
the additional penalty imposed by the increased event size. 

For storage, we anticipate two copies of the raw data stored at
different geographical locations.

\begin{table*}
\begin{center}
\begin{tabular}{llrrrrrl}
& {\bf Year} & {\bf 1} & {\bf 2} & {\bf 3} & {\bf 4} & {\bf 5} \\
\hline\hline
\multirow{3}{*}{Luminosity}
&Peak & {\bf0.25} & {\bf0.7} & {\bf1.0} & {\bf1.0} & {\bf1.0} & \designlumi\\
&per Year & 3.75 & 10.51 & 15.01 & 15.01 & 15.01 & \invab \\
&integrated & 3.75 & 14.26 & 29.26 & 44.27 & 59.28 & \invab \\
\hline\hline
\multirow{4}{*}{Data Sets}
&Raw Data & 39.4 & 149.7 & 307.3 & 464.9 & 622.4 & \pbyte \\
&Mini & 1.2 & 4.5 & 9.2 & 14.0 & 18.7 & \pbyte \\
&Micro & 2.8 & 10.6 & 21.8 & 33.0 & 44.1 & \pbyte \\
&User & 0.4 & 1.3 & 2.8 & 4.2 & 5.6 & \pbyte \\
\hline\hline
\multirow{2}{*}{Storage}
& Disk & 9.4 & 31.4 & 51.3 & 68.5 & 77.1 & \pbyte\\
& Tape & 43.7 & 168.0 & 348.2 & 530.7 & 713.1 & \pbyte \\
\hline\hline
\multirow{5}{*}{CPU}
& Reconstruction & 0.09 & 0.34 & 0.69 & 1.04 & 1.40 & \mhs \\
& Simulation & 0.28 & 1.06 & 2.18 & 3.30 & 4.42 & \mhs \\
& Skimming & 0.06 & 0.26 & 0.59 & 0.95 & 1.32 & \mhs \\
& Analysis & 0.11 & 0.43 & 0.88 & 1.33 & 1.78 & \mhs \\
& {\bf Total} & {\bf 0.54} & {\bf 2.09} & {\bf 4.34} & {\bf 6.63} & {\bf 8.91} & \mhs \\ 
\hline\hline
\end{tabular}
\end{center}

\caption{Computing resource estimate for the first 5 years of \superb\ data 
  taking.}
\label{tab:comp-res-est}
\end{table*}

Table \ref{tab:comp-res-est} shows the computing resource estimate for
the first 5 years of \superb\ data taking.  The data set and storage
sizes include replication factors, contingency, on-disk fractions and
support storage. Raw data is processed at the same rate at which it is
produced, with no more than 48 hours of latency; the corresponding
simulated data sets are produced within a month. As a consequence, the
CPU required for these activities is proportional to the peak
luminosity. The CPU required to reprocess the data collected in the
previous years scales with the integrated luminosity.

\tdrsubsec{Computing Infrastructure}

\superb\ will exploit distributed computing resources using Grid or
eventually Cloud technologies. At this early stage in the experiment
life, the \superb\ VO is already enabled on more than 26 sites and
actively accessing the available hardware.  At data taking time we
foresee to access pledged resources in Tier-0/1/2 computing centers
located in Italy and the other participating countries. In addition to
the INFN/CNAF computing center in Bologna, four new centers are under
construction in Bari, Catania, Cosenza, and Napoli, funded by the PON
ReCaS~\cite{ref:ponrecas}. As a consequence, only minimal
computing resources will be needed at the experiment site to perform
the online and near-online computing tasks that are essential for data
acquisition, calibration (PC), data quality monitoring, and providing
feedback about the detector status and performance.

Raw data will be sent to CNAF and to another center for permanent
storage on tape, and to the ReCaS centers for ER. As discussed in
\ref{sec:etd-logging}, sufficient disk space at the experimental site
(Cabibbo Lab) will be provided to allow data collection to temporarily
continue during wide area network failures. The ReCaS sites will
provide the disk storage to host the raw data for the initial
processing, and, after the first year of data taking, for the
reprocessing of data collected in the previous years. In this scheme,
Cabibbo Lab, CNAF and the ReCaS centers will together act as a
distributed Tier-0 center.

In the other participating countries there are already Tier-1/2-sized
sites primarily used by the LHC experiments. Pledged resources for
\superb\ are expected to become available and to provide about 50\% of
the total \superb\ computing needs.

Monte Carlo production can be performed at all tiers, while skimming
will be done only at the Tier-0 and Tier-1 facilities.

Multiple copies of real and Monte Carlo events in Mini and Micro
format together with the collections of skimmed events will be stored
on disk at Tier-1 and Tier-2 sites for physics analysis. The
possibility of also performing physics analysis at the Tier-0 is under
study. At this stage it is difficult to estimate the amount of data
replication needed to mitigate network failures, and to accommodate the
analysis load; in our estimate we have used a conservative replication
factor of 4.

The network bandwidth required for each member of the distributed
Tier-0, and the Tier-1 sites is estimated to be 100\gbitsps. The
GARR-X network~\cite{ref:garr} and similar networks in other countries
will provide that bandwidth.

\tdrsec{The \superb\ Computing R{\&}D program}

Computing technology is evolving in several areas. The number of cores
per CPU and the amount of storage per disk are rapidly increasing, as
are the resilience and bandwidth of Wide Area Network (WAN) links. In
addition, gains in computing performance are increasingly being
achieved by offloading some computing tasks to specialized
co-processors such as graphical processing units (GPUs), or by
adopting radically different, highly parallel computing paradigms.

In order to be cost-effective, computing solutions adopted or
developed by the \superb\ collaboration will need to take advantage of
this evolution as much as possible.

For example, the \superb\ event reconstruction and analysis framework
will very likely need to implement an approach based on parallelism as
the programming paradigm. Since legacy and current HEP software is not
optimized for this, current frameworks and many familiar algorithms
will need to be redesigned.

Another requirement for the \superb\ computing system is to be able to
reliably and efficiently access large amounts of data that are spread
over multiple, geographically disjoint sites. This requires mechanisms
to mass-transfer and directly access distributed data in a
user-friendly, robust and efficient manner. 

In the following, we describe the R\&D efforts that are underway to
investigate possible technical solutions to these problems.

\tdrsubsec{Exploiting parallelism} 

Parallelism is one of the most promising programming paradigms to
improve the efficiency of data processing and analysis tools for
future HEP experiments. The main ways to achieve software
parallelization in HEP are to parallelize the \textit{algorithms}, and
to parallelize the \textit{data flow}. The computing R\&D program aims
at understanding the different strategies and techniques to obtain
parallelism, and how they can be combined and applied to the \superb\
computing problems.

Algorithm parallelization can be achieved by using architectures that
provide a large number of computing units that are tightly connected
through low-latency links and execute the same algorithm on a
sub-sample of data in each computing unit. Graphical processing units
(GPUs) providing several hundred computing units with a very
restricted set of operations are currently the most common
representatives of this architecture. New architectures, like Intel's
MIC~\cite{ref:MIC}, that could in principle be more user friendly than
GPUs, are also emerging. Typically, all these architectures offer very
good internal communication between their computing units, but have
relatively high latency and limited bandwidth for the communication
with the host CPU and memory.

Obviously, GPU-like approaches are only useful where algorithms are
executed in parallel on subsets of the same data. 

Although our studies of \textit{MIC} and \textit{GPUs} have shown
promising results for some algorithms, they also have shown that
neither the programming tools nor the hardware itself are mature.
Intel MIC hardware is in a beta test stage and not generally available
yet, and can only programmed with a proprietary compiler that still
needs optimization. nVidia hardware is more mature, but the
combinations of hardware and software we have used so far, still show
large latencies in memory handling and data transfer from/to the GPU
board.

Another approach is to parallelize the processing of different parts
of the data with different and independent software ``modules''. Here,
an HEP program can be thought as being made of modules that operate on
sets of data which are generally encapsulated in a container called
\textit{event}. Independent parts of the same physical event can then
be processed concurrently by independent modules. 

Yet another approach that has already being used widely in HEP for a
long time, is the concurrent processing of complete events. This
approach has the disadvantage of inefficient use of data and
instruction caches, especially on modern multi-core CPUs, and in many
cases a very large memory footprint.

Almost certainly, the \superb\ software will use a combination of
these methods.

\tdrparagraph{GPU R\&D}

GPUs are many-core architectures that provide \textit{Single
  Instruction, Multiple Thread} (SIMT) parallelism. They allow for
interesting scenarios but at the same time require extensive studies
and major changes to the existing HEP programming paradigms. Several
groups in the HEP community are working on exploiting GPUs for a
multitude of use cases. In \superb, the GPU-related activities started
in 2011 and are ongoing. They have been mainly related to offline
analysis software with the goals to assess the effectiveness of GPUs
as co-processors for specific analysis tasks, to evaluate the
cost-benefit ratio of porting sequential code to GPUs, and to create
a prototype that can be integrated in the current \superb\ code.
 
A first phase of inspection and profiling of the existing physics
reconstruction algorithms that are part of the \superb\ fast
simulation has started. The focus is on the \superb\ physics software
that provides the basic tools for composition of particle hypotheses,
fitting of four-momenta, and track vertexing. These high-level
reconstruction procedures are performed during central productions
when the complete data are filtered and processed into overlapping
subsets optimized for particular analyses (skimming).  This process is
combinatorial in nature and computationally expensive, but has the
potential of great gains from parallelization.

A second line of activity aims at the parallelization of maximum
likelihood fits of complex physics models, evaluating the probability
density functions on the GPU, while the MINUIT algorithm is run on the
CPU as usual. The first tests of fitting on simple, simulated datasets
already show encouraging results with speed-up factors up to several
hundreds with respect to running on a single CPU.

\tdrparagraph{MIC Architecture R{\&}D}

The Many Integrated Core (MIC) architecture is a multiprocessor
architecture developed by Intel, that provides a co-processor composed
of many simple x86-compatible cores.Two prototypes have appeared so
far, code-named Knights Ferry and Knights Corner, the latter providing
up to about 60 CPU cores and a few GBs of memory. Each core supports 4
threads of in-order execution, has 512-bits wide vector units and has
its own L2 cache. The L2 caches are nonetheless coherent among all the
cores, which are connected through a wide high-speed ring bus.

The above characteristics and the fact that the system runs the Linux
operating system make it an appealing target for parallel
applications, since at least in principle, very few changes are needed
to adapt software to MIC. The \superb\ computing group is
collaborating with the COKA~\cite{ref:COKA} project to understand whether
and how the MIC architecture might be usable in the HEP computing
environment.

The first results of the attempt to port an event generator (written
in C++) to the MIC architecture show that the changes to the code are
not very intrusive, and that the difference of the MIC platform can
easily be handled in the build process. Once the commercial product
becomes available, the next steps will be to conduct performance
measurements, to identify more code that is suitable for running on
MIC, and to investigate how offloading of computing to a MIC
co-processor could be transparently integrated into the \superb\
Framework.

\tdrparagraph{Framework Architecture and R{\&}D}

The efforts on understanding GPUs and MIC are tightly coupled with an
investigation of how module-parallelism and these new computing
architectures could be exploited in the \superb\ reconstruction and
analysis software framework. In the \superb\ Framework (inherited from
\babar) modules operate on data generally encapsulated in an
event container. Modules communicate through
messages and put the result of their computations inside the
event container. In general, to produce something, a module needs some
input (possibly taken from the container) which is processed by the
module's algorithms; the output is stored back inside the container.
A common feature of all modules is that there is \textit{required
  input} and \textit{provided output}. This allows to build an
execution schedule based on dependencies which can be represented by a
directed graph.

To understand the impact of module-parallelism, performance
measurements on a typical configuration of \fastsim\ were performed to
establish a baseline, and data and execution dependencies between
\fastsim\ modules were studied in detail. Theoretical speedup was
determined by developing a dependency-aware scheduling algorithm and
using it to simulate code execution and measure the performance gains.

As a next step, the collaboration has decided to build a
proof-of-concept parallel Framework prototype using the \textit{Intel
  Thread Building Block} (TBB)~\cite{ref:tbb} library. A major benefit
of TBB is that it has native support for dependency graphs, allowing a
near 1:1 mapping of module dependencies to the TBB execution schedule.
In parallel to the development of the Framework, there is ongoing work
to migrate some analysis modules from the legacy serial environment to
the parallel prototype, mainly focusing on improving parallelism at
the module level. Efforts are also spent on aspects of the Framework
not related to parallelism, for example analysis configuration, so
that the finished prototype can be used as a complete test bed.

At the time of writing of this document, work is still in progress,
but it is expected to have preliminary measurements of the prototype
performance in multi-core environments in the near future.

\tdrsubsec{Distributed computing R\&D: DIRAC framework evaluation}

As part of the distributed computing R\&D, the \superb\ collaboration
is evaluating the DIRAC~\cite{ref:dirac} framework. DIRAC
(Distributed Infrastructure with Remote Agent Control) was initially
developed as a simulation production tool using pilot jobs for the
LHCb experiment. Since then data management functionality has been
added, making it a complete Grid-based data and workload management
solution for HEP experiments. DIRAC aims to provide a complete
distributed computing environment, capable of supplying all features
needed in a small VO.

Two realistic use cases have already been successfully tested within
\superb: distributed analysis and Monte Carlo production. Further
integration of \superb-specific functions into DIRAC will take place
in a new DIRAC module called SuperBDIRAC. Integration between DIRAC
and the \superb\ bookkeeping database is currently under development,
and the DIRAC web portal will be extended to include the functions
provided by the \superb\ distributed computing system prototype. 

Due to the modular and distributed architecture of DIRAC, it should be
straightforward to build up an infrastructure for \superb. The minimal
set of functionalities included in a DIRAC installation are the DIRAC
framework, the configuration service, the data and workload management
services, accounting functionalities, the bookkeeping service, log and
sandbox storage and the web interface.  While DIRAC components can be
deployed on several servers, a single well equipped machine can be
configured as head node to run all previously cited functionalities.

Future work on DIRAC will be focused on completing the evaluation,
testing advanced features, and implementing more \superb-specific use
cases. For a detailed and technical description of the Dirac evaluation 
process please refer to \cite{ref:dirac_chep12}.

\tdrsubsec{Data management and distributed storage R{\&}D}

One of the key drivers for \superb\ computing R{\&}D activities is the
study of data management in a distributed resource environment. Since
\superb\ is at least a few years away, it is important that we
investigate the next generation of tools in this field. The main
interests are Wide Area Network (WAN) data access, mass data transfer
and file catalogs.

\tdrparagraph{WAN data access} 

In this section we will discuss the state of the art in the data
management field. For WAN data access our current focus is on the use
of HTTP or xrootd as the underlying data transport protocols, and the
optimization of remote data access through mechanisms such as
pre-fetch and caching.

During the next 2-5 years we expect to be able to exploit remote data
access solutions in order to simplify the management of a typical
\superb\ computing center, and to increase the flexibility of data
movement and placement. 

This approach is supported by the trends in the computing models of
other HEP experiments. The LHC experiments are very interested in
dynamic and remote data access to augment the data placement models
already in place. The Alice~\cite{ref:alice} experiment has been fully
working with this paradigm for the last two years; CMS and ATLAS have
recently implemented such solutions for specific use cases
\cite{ref:pd2p}\cite{ref:cmsdp}.

Neither the current grid-based data management solutions in EGI and
OSG, nor the SRM protocol provide a solid infrastructure for direct
remote WAN data access; their model still relies on a data driven
computing paradigm.

The ALICE data access strategy uses the xrootd/ROOT suite, which
natively provides name spaces and data brokering. 

The following use cases will greatly benefit from the adoption of a
reliable direct WAN data access design:

\begin{itemize}
\item Interactive usage of \superb\ data, such as event display and
  single event browsing
\item Analysis code development and debugging
\item Opportunistic analysis tasks executed on resources not dedicated
  to \superb, for example non-\superb\ grid computing centers or
  dynamically allocated cloud resources.
\item Job execution on small sites where the experiment does not
  have an allocation of storage resources (Tier-3-like)
\item Improved reliability and availability, especially when
  recovering from temporary or partial storage failure at \superb\
  sites
\end{itemize}

Suitable network protocols for remote data access will have to provide
at least the following functionality:

\begin{itemize}
\item Support of a minimal Posix-like API (open, read, seek, close)
\item Ability to work through routers and firewalls
\item Caching and pre-fetching mechanisms to improve performance over
  high-latency networks (``latency-hiding'')
\item Native support by ROOT~\cite{ref:root} framework
\end{itemize}

At the present time, the two viable protocol candidates that could
fulfill these requirements are xrootd~\cite{ref:xrootd} and HTTP.

Xrootd was developed and is being used primarily within the HEP
environment, and provides a high level of functionality useful in HEP.
In contrast, HTTP is the dominating protocol on the Internet, is very
mature, and is collecting huge interest also outside the HEP
environment for use as large-scale data distribution and access
protocol.

\tdrparagraph{Data access library}

Like all the other HEP experiments, \superb is developing its own
software framework. As part of this framework, it will be important to
implement a library and APIs to standardize and facilitate end user
access to data, hiding the complexity of the underlying hardware and
software storage implementations, and network contexts.

The development of this general file/data access library is in
progress, and it will implement the following features:

\begin{itemize}
\item Intelligent pre-fetching and buffering, using the time
  spent in processing events to read the data from remote
  storage
\item Mapping of logical file names to different and multiple
  storage URIs. 
\item Enabling the support of storage protocols not already supported
  by ROOT
\item Read-ahead and caching mechanisms in order to match the
  performance requirements of different network, application, and
  storage solution
\end{itemize}

\tdrparagraph{File Transfer Service evolution}

In addition to direct remote data access, \superb will also need
mechanisms for pre-determined data placement. Many other HEP
experiments use experiment-developed frameworks to move data between
sites, that rely on FTS (File Transfer Service, developed within the
gLite middleware) as an underlying transport. Currently, FTS can only
be used if sites provide an SRM interface for managing the files and a
gridftp server to transfer the data. As soon as new protocols like
xrootd and HTTP are widely used, it will be important to have an FTS
implementation that support these protocols. Within the EMI (European
Middleware Initiative) project, there are developments to improve the
FTS service, and to add these new protocols.

\tdrparagraph{Dynamic file catalog technology}

One of the most important components of a distributed data and computing
architecture is the File Catalog. Today, each HEP experiment has its
own implementation of this. \superb\ should consider reusing one of
the already available file catalog solutions, choosing the one that
best fits the needs of the experiment. Since HTTP does not provide
multi-site brokering functionality, its use will require the file
catalog to manage and provide lists of replicas in HTTP. The
next-generation LFC (LCG File Catalog~\cite{ref:lfc}) will provide this
functionality. As soon as it is released, we will start tests to
check whether it can fulfill the \superb\ requirements.

\tdrparagraph{Storage system evaluation}

Moving and accessing the large volumes of \superb\ data puts
significant requirements on the performance of the underlying storage
systems and necessitates dedicated storage system R{\&}D. This is
particularly important in order to provide adequate guidance to the
site admin community that has to manage the experiment's data.

{\bf Data Storage} During the last few years the evolution of
storage and file system technologies provided significant new features
in the areas of clustering and distributed data management. Today we
have a multitude of available file system solutions, ranging from
open-source file systems to commercial products. The large body of
experience in the HEP community will facilitate the selection of
working solutions that accommodate the most relevant requirements of
\superb\ computing.  The sites that are already running production
systems for \superb\ and other experiments constitute large test beds
for specific file systems like GPFS and Lustre. Their experiences
provide the starting point for our evaluation campaign.

The expected use of many-core CPUs has a major impact on storage
system capabilities; technology is moving towards a computational
model built upon clusters of nodes with large amounts of computing
power provided by many-core CPUs as well as GPUs. Such configurations
place much higher demands on the I/O subsystem and networks and can
easily introduce bottlenecks and scalability problems between CPU
nodes and storage systems. The availability of cost-effective
multi-10\gbitps networks in the near future will also affect how we
distribute the access to our data. 

These new developments could invalidate the current cluster and site
configurations and open up new ways of data distribution and analysis
techniques. The \superb\ storage R{\&}D program is expected to yield
feedback on the best use of these new technologies to support the
requirements of the experiment.

The Distributed Storage R{\&}D group is in the process of testing
several file systems. This activity has two main goals:

\begin{itemize}
\item Provide input to the \superb\ computing centers on the storage
  technologies that could be used to serve the storage requirements of
  the experiment.
\item Identify storage technologies that are suitable for building one
  or more ``virtual computing centers'' from loosely coupled
  federations of geographically distributed data centers and that can
  provide complete fail-over solutions for \superb\ data processing.
\end{itemize}

The main storage technologies currently under test are:
GlusterFS~\cite{ref:gluster}, HADOOP-FS, NFSv4.1~\cite{ref:nfs4} and
EOS/xrootd~\cite{ref:eos}. The tests primarily focus on:

\begin{itemize}
\item Fail-over characteristics;
\item Performance in data analysis and simulation production use
  cases, and;
\item Scalability.
\end{itemize}

The evaluation process will identify the best storage system solutions
that will fit within the different computing infrastructures and data
management requirements for the various use cases, such as buffering the
raw data at the detector site, archiving raw data on tape, and storage of
intermediate data products for simulation and analysis. 

For a detailed and technical description of the R{\&}D lines in distributed storage field please refer to \cite{ref:httpda_chep12}~\cite{ref:distr_tier1_chep12}.

\tdrsec{Outlook}

So far the effort of the \superb\ computing group has been focused on developing and 
mantaining the suites of tools to support the detector and physics studies and on investigating 
possible technical solution to strategic issues with an active R{\&}D program.
We expect to complete these R{\&}D activities and to have a first computing and data model 
design within a year. 

The next step will be the coding of a first version of the full set
of the software tools of the experiment and this should require two more years.
The plan is to have a group of computing experts designing and coding the core software under 
the leadership of a software architect. Detector and physics analysis experts will provide 
subsytem and analysis specific code (simulation and reconstruction modules for their subsystems,
physics analysis modules, etc.).

One of the challenges to be faced will be that the Reconstruction and Analysis Framework
is foreseen to be based on parallel computing and expertise on it is so far scarse 
in the HEP community and has to be build up.
A possible strategy is to have a group of core developers professionally trained.
They will then mentor other developers to increase the number of skilled programmers.
Some level of familiarity with parallel computing will be required to the core group of expert, 
but care has to be taken to isolate people contributing detector and analysis specific code
from the complexity of parallel computing.

The remaining time before the beginning of the data taking will be devoted to debugging and 
testing the code. Large scale data challenges are foreseen to stress test the system.

\bibliographystyle{\tdrbibstyle}
\balance
\bibliography{COMP/COMP-bib}



\renewcommand{\ChapLabel}{chap:Integration} 
\tdrchap{Facilities, Mechanical Integration and Assembly}


\tdrsec{Introduction}

The \babar\ detector was built to a design optimized for operations at
a high luminosity asymmetric $B$ meson factory.  The detector
performed well during the almost nine years of colliding beams
operation, at luminosities three times the \pepii accelerator
design. There are substantial cost and schedule benefits that result
from reuse of detector components from the \babar\ detector in the
\superb\ detector in instances where the component performance has not
been significantly compromised during the last decade of use. These
benefits arise from two sources: one from having a completed detector
component which, though it may require limited performance
enhancements, will function well at the \superb\ Factory; and the
other from requiring reduced interface engineering, installation
planning and tooling manufacture (most of the assembly/disassembly
tooling can be reused, with procedures that are mirrors of the
disassembly procedures). The latter reduces risk to the overall
success of the mechanical integration of the project. Though reuse is
very attractive, risks are also introduced. Can the detector be
disassembled, transported, and reassembled without compromise? Will
components arrive in time to meet the project needs?

The reuse of elements of the PID, EMC and IFR system, and associated
support structures have been described in previous chapters. In this
section, issues related to the magnet coil and cryostat, and the IFR
steel and support structure are discussed as well as the integration
and assembly of the \superb\ detector, which begins with the
disassembly of \babar, and includes shipping components to Italy for
reassembly there.

\tdrsubsec{Magnet and Instrumented Flux Return}

The \babar\ superconducting coil, its cryostat and cryo-interface box, and the helium compressor and liquefier plant will be reused in whole or in part. The magnet coil, cryostat and cryo-interface box will be used in all scenarios. Use of on-detector pumps and similar components may not be cost effective due to electrical incompatibility. The existing \babar\ helium liquefier plant, which is halfway through its forty year service life, has sufficient capacity to cool both the detector magnet and the final focus superconducting magnets. The final decision about reuse of  other external service components, such as compressors, etc., will take into account electrical compatibility, schedule and cost. 

The initial \babar\ design contained too little steel for quality $\mu$ identification at high momentum. Additional brass absorber was added during the lifetime of the experiment to compensate. This will be retained and augmented.  The flux return steel is organized into five structures: the barrel portion, and two sets of two end doors. Each of these is, in turn, composed of multiple structures. The substructures were sized to match the 50 ton load limit of the crane in the \babar\ hall.

Each of the end doors is composed of eighteen steel plates organized
into two modules joined together on a thick steel platform. This
platform rests on four columns with jacks and Hilman rollers. A
counterweight is also located on the platform. There are nine steel
layers of 20\mm thickness, four of 30\mm thickness, four of 50\mm
thickness, and one of 100\mm thickness. During 2002, five layers of
brass absorber were installed in the forward end door slots in order
to increase the number of interaction lengths seen by $\mu$ candidate
tracks. In the baseline, these doors will be retained, including the
five 25\mm layers of brass installed in 2002, as well as the outer
steel modules which can house two additional layers of
detectors. Additional layers of brass or steel will be added,
following the specification of the baseline design in the instrumented
flux return section. The aperture of the forward plug must be opened
to accommodate the accelerator beam pipe. Compensating modifications
to the backward plug are likely to be necessary. These modifications
to the steel may affect the central field uniformity and the centering
forces on the solenoid coil, and so must be carefully re-engineered.

The barrel structure consists of six cradles. These are each composed
of eighteen layers of steel. The inner sixteen layers have the same
thicknesses as the corresponding end door plates. The two outermost
layers are each 100\mm thick. The eighteen layers are organized into
two parts; the inner sixteen layers are welded into a single unit
along with the two side plates, and the outer two layers are welded
together and then bolted into the cradle. The six cradles are in turn
suspended from the double I-beam belt that supports the
detector. During the 2004/2006 barrel LST upgrade, layers of 22\mm
brass were installed, replacing six layers of detectors in the
cradles. In the \superb\ baseline, these brass layers will be
retained, as well as all the additional flux return steel attached to
the barrel in the gap between the end doors and barrel. As in the end
doors, additional layers of absorber will be placed in gaps occupied
in \babar\ by LSTs. In order to provide more uniform coverage at the
largest radius for the $\mu$ identification system, modifications to
the sextant steel support mechanism are likely to be needed. Finite
element analyses will be required to confirm that deflections of the
steel structure due to this alternative support mechanism do not
reduce inter-plate gaps needed for the tracking detectors.

\tdrsec{Component Extraction}

Extraction of components for reuse requires the disassembly of the
\babar\ detector. This process began after the completion 
of \babar\ operations in April 2008. The first stage of the project was to
establish a minimal maintenance state, including stand alone
environmental monitoring, that preserves the assets that have reuse
value. The transition was complete in August 2008. In order to disassemble
the detector into its component systems, it was necessary to move it off the
accelerator beam line where it was pinned between the massive supports
of accelerator beamline elements into the more open space in the IR2
Hall. This required removal of the concrete radiation shield wall,
severing of the cable and services connections to the electronics
house which contained the off detector electronics, and roll-back of
the electronics house 14\m. Electronics, cables and infrastructure
that were located at the periphery of the detector were
removed. Beamline elements close to the detector were removed to allow
access to the central core of the detector by July 2009, allowing
removal of the support tube, which contained the \pepii accelerator
final beamline elements as well as the silicon vertex detector, the
following month.

The core disassembly sequence was optimized after the Conceptual
Design Report. Completion of disassembly of the detector required
fewer steps, less time, and posed fewer risks, with the end doors
being disassembled while the detector was on beamline. Tooling that
was used in the initial installation of the detector at IR2 was
refurbished and additional tooling has been fabricated for the
disassembly effort. All of this tooling is available for use on
\superb. In 2010, the end doors were broken down into component parts,
and the EMC forward endcap and the drift chamber were removed. The
detector was rolled out for core disassembly in Fall 2010. The DIRC
was removed. In Winter 2011 the Barrel EMC and Superconducting
Solenoid were removed. Brass was removed from the Barrel Flux Return,
and the Barrel Flux Return disassembly completed in August 2011.

This work has been accomplished as a low priority project with a small
crew of engineers and technicians. Though the disassembly project is
behind schedule, due to laboratory priorities, the earned value
compared to actual costs indicates that the level of effort estimated
to perform the disassembly is very accurate. The same methodology used
to determine the level of effort needed for the \babar\ disassembly
has been applied in estimating the needs for \superb\ assembly.

The components from \babar\ that are expected to be 
reused in \superb\ were available for transport to Italy in mid-2011. 
In order to
minimize the possibility of damage to the DIRC bar boxes and their
fused silica bar content, disassembly proceeded with removal of the bar
boxes one by one from the bar box support structure inside the
DIRC. The twelve DIRC bar that were removed from the DIRC structure
during the disassembly of the core of the detector have been stored in
an environmentally controlled container.

\tdrsec{Component Transport}

The magnet steel components will be crated for transport to limit
damage to mating surfaces and edges. Most, if not all, of these
components can be shipped by sea.

The \babar\ solenoid was shipped via special air transport from Italy
to SLAC. It is hoped that this component can be returned to Italy in
the same fashion. The original transport frame has been refurbished.

The DIRC and barrel calorimeter present transportation challenges. In
both cases transport without full disassembly is preferred. In the
case of the DIRC, the central support tube will be separated from the
strong support tube and transported as an assembly. The bar boxes
storage container provides a dry environment, but is unsuitable
for bar box transport. Design and fabrication are required.

In the case of the EMC, there are two environmental constraints on
shipment of the device or its components. The glue joint that attaches
the photodiode readout package to the back face of the crystal has
been tested, in mock-up, to be stable against temperature swings of
$\pm 5^{\circ} \rm{C}$. During the assembly of the endcap calorimeter,
due to a failure of an air conditioning unit, the joints on one module
were exposed to double this temperature swing. Several glue joints
parted. The introduction of a clean air gap causes a light yield drop
of about 25\%. In order to avoid this reduction in performance,
temperature swings during transport must be kept small. Since the
crystals are mildly hygroscopic, it is best that they be transported
in a dry environment to avoid changes in the surface reflectivity, and
consequent modification in the longitudinal response of the
crystal. Individual \babar\ endcap modules constructed in the UK were
successfully shipped to the USA in specially constructed containers
that kept the temperature swings and humidity acceptably small.

Disassembly of the barrel calorimeter for shipment presents a
substantial challenge. Both the disassembly and assembly sites need to
be temperature and humidity controlled. The disassembly process
requires removal of the outer and inner cylindrical covers, removal of
cables that connect the crystals with the electronics crates at the
ends of the cylinder, splitting of the cylinder into its two component
parts and removal of the 280 modules for shipment. Though much of the
tooling is still in hand, the environmentally conditioned buildings
used in calorimeter construction at SLAC no longer exist so that
alternative facilities will need to be outfitted. The cooling and
drying units used in the module storage/calorimeter assembly building
continue to be available.

The clear preference is to ship the barrel calorimeter as a single
unit by air. With tooling support stand and environmental conditioning
equipment, the load is likely to exceed 30 tons. It is anticipated
that such a load could be transported in the same way as the
superconducting coil and its cryostat, but verification is
needed. Detailed engineering studies, which model accelerations and
vibrations involved with flight that might cause the crystal
containing carbon fiber modules to strike against one another, are
needed to determine if the calorimeter can be safely transported. It
may be that a module restraint mechanism will need to be engineered
and fabricated. A transport frame must be designed and its performance
modeled. Engineering studies have begun, however, it is not unlikely
that it will be necessary to transport the barrel calorimeter via air,
but disassembled into modules.

\tdrsec{Detector Assembly}

Assembly of the \superb\ detector is the inverse of the disassembly of
the \babar\ detector. Ease of assembly will be influenced by the
facilities which are available. In the case of \babar, the use of the
IR2 hall, which was ``too small'', led to engineering compromises in
the design of the detector. Assembly was made more complicated by the
weight restrictions posed by the 50 ton crane. Upgrades were made more
difficult because of limitations in movement imposed by the size of
the hall.

The preferred dimensions for the area around the \superb\ detector
when it is located in the accelerator housing are at least 16.2\m
transverse to the beamline, 20.0\m along the beamline, 11.0\m from the
floor to the bottom of the crane hook, 15.0\m from floor to ceiling,
and 3.7\m from the floor to the beamline. The increased floor to
beamline height relative to \babar-\pepii will require redesign
effort for the underpinnings of the detector cradle. However, this
will permit improved access for installation of cable and piping
services for the detector, as well as make possible additional IFR
absorption material for improved $\mu$ detector performance below the
beamline.  In order to facilitate detector assembly, the preferred
capacity for the two hook bridge crane is 100/25 tons.


\renewcommand{\ChapLabel}{chap:Management} 
\tdrchap{The \superb\ Collaboration and Project Management}

\label{sec:Management}  

During the pre-approval phase of the project,  the  \superb\ detector was organized as a ``Proto-collaboration'', with a structure modeled on that of the \babar\ detector. Leadership was centered in the ``Proto-technical'' Board which included  technical leadership for all subsystems (SVT, DCH, PID,  EMC, IFR); the general detectors systems (electronics, DAQ,  computing, and simulation), and the machine detector interface (MDI). The ``Proto-technical'' Board was led by  Detector Co-leaders and the Proto-Technical Coordinator who provided executive direction and decision making for the collaboration with the guidance of the Board. Subsystem status and progress were reviewed by the Board at regularly scheduled meetings. Major overall decisions about the detector were also reviewed by the Board,  aided in the more challenging cases by recommendations provided by ad-hoc review committees and a geometry working group appointed by the leadership.  

Following formal  project approval in Dec. 2010, the \superb\ detector ``Proto-Collaboration'' began the transition to the  \superb\ detector Collaboration. The inaugural Collaboration forming meeting was held in Elba in May-June 2011. Initial steps were defined to begin to organize the experimental collaboration, while allowing time for new groups to become involved from the worldwide community. The basic model was suggested by the long history of successful experimental collaborations in particle physics with particular reference to the successful experience of the  \babar\ detector.  Given the success of the \superb\ Proto-collaboration to date and its ability to deal with most executive and technical functions, this structure was retained during the interim period until the new executive and technical structures could be defined by the Collaboration and new leadership appointed. This bootstrapping process began by establishing a Proto-council composed of the PIs of the collaborating institutions to provide input to the process, approve default membership and interim leadership,  and provide monitoring of the working committees.  This Proto-council then approved the appointment of a Governance committee to  develop the bylaws of the collaboration, and a continuation of the Proto-speakers bureau to select speakers for conferences to the bridging period.

The Governance committee, in consultation with collaboration membership, Proto-collaboration leadership, and the Laboratory project leadership (now the Cabbibo Laboratory), drafted a Governance document for the collaboration. This document~\cite{Governance_2012} was unanimously approved at the March 2012 meeting of \superb\ and the Proto-council became the Council of the Collaboration. Following the procedures described in the Governance document, The Council chair was selected and a Council committee is now being appointed to select the leadership of the collaboration. The Council also appointed an Executive Board to deal with issues that go beyond those managed by the Proto-technical board. Given the fact that the collaboration will evolve substantially in the coming months, the initial terms for this Executive Board were set to one year.

In this chapter, we summarize the collaboration organization as detailed in the Governance document. 

\tdrsec{Collaboration Membership}

The \superb\ Detector Collaboration is the organization of scientists belonging to academic or research Institutions worldwide whose goal is the exploitation of the physics potential of the \superb\ electron-positron asymmetric collider.  Institutions who participate in the Collaboration are expected to make significant contributions to the construction and operation of the \superb\ detector, including its computing systems, both hardware and/or software. Participating Institutions are entitled to be represented in the Collaboration Council, and, to preserve their membership, they must actively participate in the \superb\ program and comply with the obligations that are defined by the Collaboration in terms of financial and service contributions. Ph.D. physicists, engineers, and Ph.D. thesis students who belong to a participating Institution can become members of the Collaboration if they intend to devote a substantial fraction of their research time to the \superb\ program over a period of several years. Participation of an individual in \superb\ is at the discretion of the Institution to which he or she is accredited. Members of the Collaboration are eligible to be included in the author list.

\tdrsec{The Collaboration Council}

The Collaboration Council, representing the full membership of the Collaboration, is the principal governing body of the Collaboration. Institutions with two or more Collaboration members who are Ph.D. physicists will be directly represented on the Collaboration Council. Members of the Collaboration from Institutions with fewer than two Ph.D. physicists may affiliate with another Institution for the purpose of representation, while larger institutions will be given additional representation.  The Council has an elected Chair and Vice-Chair. Council decisions are to be agreed by consensus wherever possible.  Should a situation arise where consensus cannot be reached, a time scale shall be set for the decision to be reached at a later date and the matter will then be decided at a subsequent meeting of the Council, by a vote if necessary.

The Council selects and may remove the executive leadership of the collaboration, particularly the Spokesperson and the Executive Board via specific defined procedures of the Governance document. The Spokesperson is the scientific leader of the Collaboration, responsible for financial, scientific, technical, and organizational decisions pertaining to the Collaboration. The Spokesperson is assisted by a Management Team.  The Executive Board is responsible for advising the Spokesperson on all scientific, financial, and organizational matters pertaining to the Collaboration. The details of the management structure, including the roles and responsibilities of individual managers, the roles and responsibilities of additional groups such as the Technical Board, the Management Team, the Publication Board, and Physics Analysis Groups will be established in a management plan, proposed by the Spokesperson, endorsed by the Executive Board and ratified by the Council.

The Collaboration Council will concern itself with issues related to the overall framework of the Collaboration. It will ratify the Collaboration by-laws and the organizational structure, insofar as it is specified in the Collaboration Constitution, and vote on proposed amendments, which may be introduced by any Council representative or by the Spokesperson.

The Chairperson, in consultation with the Collaboration, possibly through the appointment of a search committee, will select candidates for ratification by Council for the several boards and committees that are to be established by the Council. 

The Collaboration Council will appoint a Membership Committee, which will make recommendations to the Council regarding membership issues, including the admission of new Institutions. Ratification of such recommendations requires a two-thirds majority vote of the Council. 

The Collaboration Council will be responsible for developing a publications policy and for appointing a Publications Board to approve papers for publication. The Board will be responsible for maintaining a list of eligible authors for both instrumental and physics papers. The Collaboration Council will also be responsible for appointing a Speakers' Bureau to organize Collaboration presentations at conferences and workshops as appropriate and to determine who will represent the Collaboration at such meetings.

\tdrsec{The Spokesperson}

The Collaboration will be led by a single Spokesperson, who is the scientific leader of the Collaboration, responsible for financial, scientific, technical, and organizational decisions pertaining to the Collaboration.  The Spokesperson also serves, ex-officio, as Chair of the Executive Board and as an ex-officio member of the international body of agencies organized by the Cabibbo-Lab to support the  \superb\ detector experiment Collaboration through the construction and operation phases (indicated in the following as IFC, International Finance Committee). 

The Spokesperson, with the endorsement of the Executive Board, proposes a Management Plan to the Council for ratification, and a Management Team that will assist him/her in the execution of the \superb\ program. The Spokesperson is expected to devote all his/her research time to the position, and must be willing and able to take up residence at Cabibbo-Lab when detector commissioning begins.

During the construction phase, the initial term for the \superb\ Spokesperson is 3 years long and renewable, which will allow long-range planning and will provide continuity.  Renewal for additional terms will be discussed and endorsed by the Council at least 9 months before the end date. 
The construction phase, and the Spokesperson term, will end on the September 1 that follows the first year in which the experiment has accumulated at least three months of beam. During the data taking phase,  the Spokesperson will serve a single 2 year term, non-renewable, beginning September 1 of the relevant year, during which time he/she must be resident at the laboratory.  At the end of his/her term, the retiring spokesperson serves for a year in the management team by mutual agreement with the incoming spokesperson. 

\tdrsec{The Executive Board}

The Executive Board advises the Spokesperson on all scientific, financial, and organizational matters of the Collaboration. It consists of members distinguished by their scientific judgment, technical expertise, commitment to the experiment, and ability to speak knowledgeably and effectively for the regions they represent.The Spokesperson will chair the Executive Board. The Executive Board will meet at least eight times per year. The Board will also meet at least once each year with ex-officio members not present to review the performance of the management team.
By a two-thirds majority vote of the entire Executive Board, the Board may recommend to  the Collaboration Council that the Spokesperson be removed.

The number of members reflects the national composition of the collaboration. Initially, the Board will consist of one representative each from Canada, France, Russia, Poland, U.S., and the U.K. and six representatives from Italy.  In addition, the Spokesperson, the Spokesperson-Elect, and the Chairperson of the Collaboration Council will be nonvoting ex-officio members of the Executive Board. The Cabibbo-Lab Directorate will appoint a senior scientist who is also a member of the \superb\  Detector Collaboration to sit on the Executive Board as an ex-officio member. From time to time, the Board will review its regional composition, in consultation with the Cabibbo-Lab management, and will propose changes to the Collaboration Council that it may approve with a qualified majority.

Membership on the Collaboration Council does not preclude membership on the Executive Board. However, members of the Management Team may not serve simultaneously as members of the Executive Board. The normal term of office for the Executive Board will be three years, renewable, with one-third of the members being replaced or renewed each year.

\tdrsec{The Management Team and Management Plan}

The overall execution of  the \superb\ program is the responsibility of a Management Team led by the Spokesperson. The Management Team consists of: the Spokesperson, the Physics Analysis Coordinator, the Technical Coordinator, the Computing Coordinator, and the Spokesperson-elect when that office is filled. The Spokesperson may decide to designate at most two additional members to the  management team, such as a Deputy Spokesperson, Financial Coordinator, and/or a Senior Advisor. 

The Technical Coordinator is in charge of detector operations and hardware development. He/she is responsible for the common project construction and the technical integration of all \superb\ components. He/she provides oversight for the implementation of \superb\ engineering standards and procedures and the sub-system construction, commissioning and operations. He/she is assisted by managers responsible for the sub-systems.  The Physics Analysis Coordinator is in charge of coordinating the analysis infrastructure and planning for physics analyses leading to publication. He/she is assisted by physics-topic group conveners .
The Computing Coordinator is in charge of planning for and coordinating the software and computing activities and resources necessary for data acquisition and analysis.

The precise details of the management structure may evolve with time, including the roles and responsibilities of individual members of the Management Team and of additional groups such as the Technical Board, and Physics Analysis Groups will be established in a Management Plan, proposed by Spokesperson, approved by the Executive Board by a majority vote and ratified by the Council with a majority vote.  In particular, the roles of the Technical Board and its members will be at the heart of collaboration activities during the fabrication and commissioning phases of the experiment, but will become less central as time passes. The Technical Coordinator will be nominated with the agreement of the Cabibbo-Lab Management. 

\tdrsec{The International Finance Review Committee}

An International Finance Review Committee instituted by the Cabibbo Laboratory will monitor the financial aspects of the experiment as detailed in the management plan and agreed between the Cabibbo Laboratory, the collaboration, and the International Funding Agencies. This will be described in Memoranda of Understanding between the participating institutional partners and Cabibbo Laboratory.  

\tdrsec{Interaction with the Cabibbo-Lab}

The Spokesperson will plan regular meetings with Cabibbo-Lab management.  He/she will report to the International Cabibbo-Lab Scientific Committee on different aspects of the experiment in order to converge on the detector design, review the construction, installation and commissioning.  He/she will also propose for endorsement to the Cabibbo-Lab and International Finance Committee the annual construction, maintenance and operation budgets of the detector.

\tdrsec{Construction Responsibilities}

The design of the \superb\ detector described in this report began with the excellent \babar detector. It then went through an extended process of optimization of the design of individual systems and the detector as a whole against the required physics performance, the interests and technical capabilities of the collaborating institutions, and financial constraints. In cases where there were conflicting requirements or competing technologies, task forces were employed to evaluate alternatives and recommend the appropriate choice. These include task forces for the forward PID, the backward EMC, and the forward EMC. All design decisions have been informed by a long term simulation task force (the Geometry Working Group), and background simulation effort. 

Specific responsibilities for design and construction of the various detector subsystems have yet to be fully specified, and will be discussed more completely in a later document. We expect that they will be assigned through the traditional process of matching interests, capabilities, and resources.  Final  responsibilities will be detailed in Memoranda of Understanding. The collaboration expects to finance certain items through a common fund. Details remain under discussion. 

\bibliographystyle{\tdrbibstyle}
\balance
\bibliography{Management/Management-bib}


\renewcommand{\ChapLabel}{chap:Budget} 
\tdrchap{Cost and Schedule}


The \superb\ detector cost and schedule estimates, presented in this
chapter, rely heavily on experience with the \babar\ detector at
PEP-II. The reuse and refurbishing of existing components has been
assumed whenever technically possible and financially advantageous.

The costing model used here is similar to that already used for the
\superb\ CDR (2007) and White Paper (2010). In all cases where significant changes have been made to the design, the estimates has been updated to include design changes and new vendor quotes obtained since 2010, but if no such changes have occurred, the estimate from 2010 has sometimes been taken without escalation. Some  small cost escalation from 2010 to 2012 is therefore not
consistently included, but it is considered to be
a small perturbation, well within the contingency levels.  The
components are estimated in two different general categories; (1)
``LABOR'' and, (2) M\&S (Materials and Services).  M\&S cost estimates
are given in 2010 Euros and include 20\% of Value Added Tax (VAT). The
``LABOR'' estimates comprise two sub categories which are kept and
costed separately as they have differing cost profiles;(1) EDIA
(Engineering, Design, Inspection, and Administration) and (2) Labor
(general labor and technicians). Estimates in both categories are
presented in manpower work units (Man-Months) and not monetized, as a
monetary conversion can only be attempted after institutional
responsibilities have been identified and the project timescale is
known.  The total project cost estimate can be calculated, once the
responsibilities are identified, by summing the monetary value of
these three categories.

Given the long term nature of this multi-national project, there are
several challenging general issues in arriving at appropriate costs
including (1) fluctuating currency exchange rates, and (2) escalation.
M\&S costs and factory quotes that have been directly obtained in
Euros can be directly quoted. M\&S estimates in US Dollars are
translated from Dollars to Euros using the exchange rate as of Jun 1,
2010 (0.8198 Euros/US\$).  For costs in Euros that were obtained in
earlier years, the yearly escalation is rather small. For simplicity,
we use a cost escalation rate of 2\% per year which is consistent with
the long term HICP (Harmonized Index of Consumer Prices) from the
European Central Bank. Costs given in 2007 Euros are escalated by the
net escalation rate (1.061) for three years to arrive at the 2010 cost
estimates given here~\cite{bib:INFLATEEURO}.

For all items whose cost basis is \babar, we accept the procedure
outlined in the \superb\ CDR which arrived at the costs given there in
2007 Euros. This procedure escalated the corresponding cost (including
manpower) from the \pepii and \babar\ projects from 1995 to 2007 using
the NASA technical inflation index~\cite{bib:NASA} and then converted
from US Dollars to Euros using the average conversion rate over the
1999---2006 period~\cite{bib:USDEURO}. The overall escalation factor
in the CDR from 1995 Dollars to 2007 Euros is thus $1.21 = 1.295
\times 0.9354$.

Similarly, the replacement values (``Rep.Val.'') of the reused
components, \ie, how much money would be required to build them from
scratch, as presented in separate columns of the cost tables, have
been obtained by escalating the corresponding \babar\ project cost
(including manpower) from 1995 to 2007.  Though it is tempting to sum
the two numbers to obtain an estimate of the cost of the project if it
were to be built from scratch, this procedure yields somewhat
misleading results because of the different treatment of the manpower
(rolled up in the replacement value; separated for the new cost
estimate) and because of the double counting when the refurbishing
costs are added to the initial values.

Contingency is not included in the tables. Given the level of detail
of the engineering and the cost estimates, a contingency of about 35\%
would be appropriate.

\tdrsec{Detector Costs}

The costs, detailed in Table\,\ref{tab:budget}, are presented for the
detector subsystem at WBS level 3/4.  Although the \superb\ baseline
detector is essentially defined, some further options remain open:
some components, such as the forward PID, have overall integration and
performance implications that need to be carefully studied before
deciding to install them; for some other components, such as the SVT
Layer0, promising new technologies require additional R\&D before they
can be definitively used in a full scale detector. The cost estimates
list some of the different technologies separately, but the rolled-up
value only includes the baseline detector choice. Technologies that
are not included in the rolled-up value are shown in italics.


\onecolumn {
  \setlength{\tabcolsep}{4pt} 
  \begin{center} {\footnotesize
\begin{longtable}{lp{6cm}rrrr}
  \caption{\superb\ detector budget. }
\label{tab:budget} \\

\hline\hline
\textbf{ } & \textbf{ }  & \textbf{EDIA} & \textbf{Labor} & \textbf{M\&S} & \textbf{Rep.Val.} \\
\textbf{WBS} & \textbf{Item}  & \textbf{mm} & \textbf{mm} & \textbf{kEuro} & \textbf{kEuro} \\
\hline
\endfirsthead

\multicolumn{6}{c}%
{\tablename\ \thetable{} -- continued from previous page} \\[3mm]
\hline\hline
\textbf{ } & \textbf{ }  & \textbf{EDIA} & \textbf{Labor} & \textbf{M\&S} & \textbf{Rep.Val.} \\
\textbf{WBS} & \textbf{Item}  & \textbf{mm} & \textbf{mm} & \textbf{kEuro} & \textbf{kEuro} \\
\hline
\endhead

 \\ \hline
  \multicolumn{6}{c}{Continued on next page} \\
\endfoot

\endlastfoot

   {\bf 1} & {\bf SuperB detector} & {\bf 3749} & {\bf 2464} & {\bf 50840} & {\bf 48922} \\
\hline
 {\bf 1.0} & {\bf Interaction region} &   {\bf 21} &   {\bf 12} &  {\bf 860} &    {\bf 0} \\
     1.0.1 & Be Beampipe &         10 &          4 &        260 &          0 \\
     1.0.2 & Tungsten Shield &          9 &          6 &        540 &          0 \\
     1.0.3 & Radiation monitors &          2 &          2 &         60 &          0 \\
\hline
 {\bf 1.1} & {\bf Tracker (SVT + Strip + MAPS)} &  {\bf 258} &  {\bf 364} & {\bf 4870} &    {\bf 0} \\
{\bf 1.1.1} &  {\bf SVT} &  {\bf 222} &  {\bf 309} & {\bf 4328} &    {\bf 0} \\
   1.1.1.1 & Mechanical &         48 &        129 &        400 &          0 \\
   1.1.1.2 &    Cooling &          8 &         10 &        155 &          0 \\
   1.1.1.3 & Silicon Wafers and Fanout &         24 &        120 &       2642 &          0 \\
   1.1.1.4 & On-detector electronics &         72 &         42 &       1014 &          0 \\
   1.1.1.5 & Detector monitoring &          4 &          4 &         94 &          0 \\
   1.1.1.6 & Detector assembly &          6 &          4 &          0 &          0 \\
   1.1.1.7 & System Engineering &         60 &          0 &         24 &          0 \\
{\bf 1.1.2} & {\bf L0 Striplet option} &   {\bf 36} &   {\bf 55} &  {\bf 541} &    {\bf 0} \\
   1.1.2.1 & Mechanical &         12 &         30 &         60 &          0 \\
   1.1.2.2 &    Cooling &          3 &          3 &         48 &          0 \\
   1.1.2.3 & Silicon Wafers and Fanout &         16 &         18 &        326 &          0 \\
   1.1.2.4 & On-detector electronics &          5 &          4 &        107 &          0 \\
\textbf{\textit{1.1.3}} & \textit{\textbf{L0 MAPS option}} & \textit{\textbf{150}} & \textit{\textbf{78}} & \textit{\textbf{1576}} & \textit{\textbf{0}} \\
{\it 1.1.3.1} & {\it Mechanical} &   {\it 18} &   {\it 48} &   {\it 90} &    {\it 0} \\
{\it 1.1.3.2} & {\it Cooling} &    {\it 6} &    {\it 6} &   {\it 96} &    {\it 0} \\
{\it 1.1.3.3} & {\it MAPS Modules Components} &  {\it 126} &   {\it 24} & {\it 1390} &    {\it 0} \\
\textit{\textbf{1.1.4}} & \textit{\textbf{L0 Hybrid Pixel option}} & \textit{\textbf{156}} & \textit{\textbf{84}} & \textit{\textbf{1684}} & \textit{\textbf{0}} \\
{\it 1.1.4.1} & {\it Mechanical} &   {\it 18} &   {\it 48} &   {\it 90} &    {\it 0} \\
{\it 1.1.4.2} & {\it Cooling} &    {\it 6} &    {\it 6} &   {\it 96} &    {\it 0} \\
{\it 1.1.4.3} & {\it Hybrid Pixel Modules Components} &  {\it 132} &   {\it 30} & {\it 1498} &    {\it 0} \\
\hline
 {\bf 1.2} &  {\bf DCH} &  {\bf 169} &  {\bf 151} & {\bf 3481} &    {\bf 0} \\
     1.2.1 & System engineering &         24 &          0 &         60 &          0 \\
     1.2.2 &  Endplates &         16 &          6 &        660 &          0 \\
     1.2.3 & Inner cylinder &          8 &          2 &        200 &          0 \\
     1.2.4 & Outer cylinder &          6 &          2 &        120 &          0 \\
     1.2.5 &       Wire &          4 &          6 &        308 &          0 \\
     1.2.6 & Feedthroughs &          9 &         10 &        439 &          0 \\
     1.2.7 & Endplate systems &         12 &         12 &        445 &          0 \\
     1.2.8 & Assembly \& Stringing &         74 &         96 &        960 &          0 \\
     1.2.9 & Gas System &         10 &          8 &        240 &          0 \\
     1.2.A &       Test &          6 &          9 &         48 &          0 \\
\hline
 {\bf 1.3} & {\bf FDIRC Barrel (Focusing DIRC)} &   {\bf 97} &  {\bf 101} & {\bf 4129} & {\bf 7138} \\
     1.3.1 & Radiator Support Structure (new support disc) &          2 &          1 &         18 &       2516 \\
     1.3.2 & New magnetic door and inner cylinder &          4 &          1 &         96 &          0 \\
     1.3.3 & FDIRC photon camera background shielding &          4 &          1 &        120 &          0 \\
     1.3.4 & Background shield displacement system and rails &          4 &          1 &         30 &          0 \\
     1.3.5 & Bar box transport to Italy &          6 &          1 &        120 &          0 \\
     1.3.6 & Radiator box/Photon camera assembly &         23 &         12 &       1159 &       4515 \\
     1.3.7 & Photon Camera mechanical boxes &         24 &         44 &        282 &          0 \\
     1.3.8 & Photodetector assembly &         20 &         20 &       2004 &          0 \\
     1.3.9 & Calibration System &          4 &          0 &         72 &          0 \\
    1.3.10 & Temperature, water, nitrogen and helium leak interlock system &          4 &          2 &         12 &          0 \\
    1.3.11 & Alignment services &          2 &          4 &         36 &          0 \\
    1.3.12 & Mechanical Utilities &          0 &          4 &         60 &        107 \\
    1.3.13 & System Integration &          0 &         10 &        120 &          0 \\
\hline
 {\bf 1.4} &  {\bf EMC} &  {\bf 332} &  {\bf 578} & {\bf 10047} & {\bf 31574} \\
{\bf 1.4.1} & {\bf Barrel EMC} &  {\bf 110} &  {\bf 169} & {\bf 1802} & {\bf 30923} \\
   1.4.1.1 & Crystal Procurement &          0 &          0 &          0 &      21742 \\
   1.4.1.2 & Light Sensors \& Readout &          0 &          0 &          0 &       2654 \\
   1.4.1.3 & Crystal Support Modules &          0 &          0 &          0 &       2875 \\
   1.4.1.4 & Barrel Structure &          0 &          0 &          0 &       3419 \\
   1.4.1.5 & Transfer to Italy &         22 &         15 &       1318 &          0 \\
   1.4.1.6 & Refurbish and re-assemble &         24 &         42 &        240 &          0 \\
   1.4.1.7 & Electronics &         52 &         88 &        221 &          0 \\
   1.4.1.8 & Project management &         12 &         24 &         24 &        233 \\
{\bf 1.4.2} & {\bf Forward EMC} &  {\bf 127} &  {\bf 298} & {\bf 7268} &    {\bf 0} \\
   1.4.2.1 &   Crystals &         27 &        110 &       5990 &          0 \\
   1.4.2.2 & Light sensors and electronics &         59 &        106 &        654 &          0 \\
   1.4.2.3 & Transfer to Italy &          6 &         14 &        312 &          0 \\
   1.4.2.4 & Refurbish and assembly &         24 &         50 &        288 &          0 \\
   1.4.2.5 & Project management &         12 &         18 &         24 &          0 \\
{\bf 1.4.3} & {\bf Crystal Calibration Systems} &   {\bf 46} &   {\bf 19} &  {\bf 332} &  {\bf 650} \\
   1.4.3.1 & Source calibration &         10 &          7 &        116 &        463 \\
   1.4.3.2 & Light pulser system &         12 &         12 &         96 &        187 \\
   1.4.3.3 & Project Management &         24 &          0 &        120 &          0 \\
{\bf 1.4.4} & {\bf Backward EMC} &   {\bf 30} &   {\bf 54} &  {\bf 600} &    {\bf 0} \\
   1.4.4.1 & Scintillator &          2 &         12 &        246 &          0 \\
   1.4.4.2 &   Radiator &          1 &          5 &         11 &          0 \\
   1.4.4.3 &     Fibers &          4 &          9 &         19 &          0 \\
   1.4.4.4 & Photodetectors &          2 &          5 &         47 &          0 \\
   1.4.4.5 & Mechanical support &         17 &         15 &        146 &          0 \\
   1.4.4.6 & Electronics &          2 &          6 &        107 &          0 \\
   1.4.4.7 & Project Management &          2 &          2 &         24 &          0 \\
{\bf 1.4.5} & {\bf Monitoring} &    {\bf 4} &    {\bf 8} &    {\bf 8} &    {\bf 0} \\
{\bf 1.4.6} & {\bf Project Management} &   {\bf 15} &   {\bf 30} &   {\bf 36} &    {\bf 0} \\
\hline
 {\bf 1.5} &  {\bf IFR} &   {\bf 37} &  {\bf 188} & {\bf 2340} &    {\bf 0} \\
     1.5.1 & Scintillators &          0 &          0 &        516 &          0 \\
     1.5.2 & WLS fibers &          0 &          0 &        600 &          0 \\
     1.5.3 & Photodetectors and PCBs &          1 &          6 &        990 &          0 \\
     1.5.4 & Mechanics (Production and QC) &         16 &         62 &        234 &          0 \\
     1.5.5 & Module Installation &         20 &        120 &          0 &          0 \\
\hline
 {\bf 1.6} & {\bf Magnet} &   {\bf 93} &   {\bf 59} & {\bf 3767} & {\bf 10210} \\
     1.6.0 & System Management &         36 &          0 &          0 &        612 \\
     1.6.1 & Superconducting solenoid &          0 &          0 &          0 &       2421 \\
     1.6.2 & Mag. Power/Protection &          0 &          0 &          0 &        181 \\
     1.6.3 & Cryogenics &         34 &         36 &       1753 &          0 \\
     1.6.4 & Cryo monitor/Control &         17 &         11 &        214 &          0 \\
     1.6.5 & Flux return &          6 &         12 &       1800 &       6481 \\
     1.6.6 & Installation/test equipment &          0 &          0 &          0 &        515 \\
\hline
 {\bf 1.7} & {\bf Electronics} &  {\bf 719} &  {\bf 361} & {\bf 11476} &    {\bf 0} \\
     1.7.1 &        SVT &         32 &          8 &        470 &          0 \\
     1.7.2 &        DCH &         55 &         81 &       4248 &          0 \\
     1.7.3 & PID Barrel &        136 &         18 &        612 &          0 \\
     1.7.4 &        EMC &        110 &        164 &       2726 &          0 \\
     1.7.5 &        IFR &         32 &         48 &        644 &          0 \\
     1.7.6 & Infrastructure &          4 &         36 &        314 &          0 \\
     1.7.7 & Systems Engineering &         12 &          0 &          0 &          0 \\
     1.7.8 & Hardware Trigger &        120 &          0 &        677 &          0 \\
     1.7.9 & ETD (without Trigger) &        218 &          6 &       1784 &          0 \\
\hline
 {\bf 1.8} & {\bf Online System} &  {\bf 951} &   {\bf 27} & {\bf 2058} &    {\bf 0} \\
     1.8.1 & Event Data Chain &        285 &          3 &       1594 &          0 \\
     1.8.2 & Control Systems &        192 &          0 &        108 &          0 \\
     1.8.3 & Databases, Web, Logbook &        114 &          0 &         12 &          0 \\
     1.8.4 & Infrastructure &         48 &         12 &        246 &          0 \\
     1.8.5 & Software Triggers &        216 &          0 &          0 &          0 \\
     1.8.6 & Coordination and Commissioning &         72 &         12 &          0 &          0 \\
\hline
 {\bf 1.9} & {\bf Installation and integration} &  {\bf 353} &  {\bf 624} & {\bf 7596} &    {\bf 0} \\
     1.9.1 & Disassembly &         95 &        161 &        612 &          0 \\
     1.9.2 &   Assembly &        222 &        463 &       3984 &          0 \\
     1.9.3 & Structural analysis &         36 &          0 &          0 &          0 \\
     1.9.4 & Transportation &          0 &          0 &       3000 &          0 \\
\hline
{\bf 1.10} & {\bf Project Management} &  {\bf 720} &    {\bf 0} &  {\bf 216} &    {\bf 0} \\
    1.10.1 & Project engineering &        300 &          0 &        120 &          0 \\
    1.10.2 & Budget, Schedule and Procurement &        300 &          0 &         48 &          0 \\
    1.10.3 &    ES \& H &        120 &          0 &         48 &          0 \\
\hline

    \hline
\end{longtable}
}
\end{center}
}

\twocolumn

\tdrsec{Basis of Estimate} 

\tdrparagraph{Vertex Detector and Tracker:} System cost is estimated
based on experience with the \babar\ detector and vendor quotes. A
detailed estimate is provided for the cost of the main detector
(layers 1 to 5). The costs associated with Layer0 are analyzed
separately. The costing model assumes that a striplet detector will be
installed initially -- followed by a second generation upgrade to a
pixel detector, which could either be either a MAPs or hybrid pixel
device.  Substantial R\&D on these new technologies is needed in
either case before such a detector can be built.  The total SVT cost
is obtained by using the baseline option with striplets for Layer0,
while the MAPS and Hybrid Pixels Layer0 costs are shown as options.

\tdrparagraph{Drift Chamber:} The DCH costing model is based on a
straightforward extrapolation of the actual costs of the existing
\babar\ chamber to 2010, since, as discussed in Sec.~\ref{sec:DCHmain}
the main design elements are comparable, and many related components,
such as the length of wire, number of feedthroughs, duration of wire
stringing, etc., can be reliably estimated. Although the cell layout
is still being finalized, the total cell count will likely be about
25\% larger. The endplates will be fabricated from carbon fiber
composites instead of aluminum. Though this will require a somewhat
longer period of R\&D and engineering design, it is unlikely to result
in significantly larger production costs for the final endplates. The
DCH electronics on Table\,\ref{tab:budget} assumes the  cluster
counting readout option discussed in Sec.\,\ref{sec:ETD_DCH}.

\tdrparagraph{Particle Identification:} Barrel PID costs and
replacement values are derived from \babar\ costs as extrapolated to
2010, with updated quotes from vendors. The main new component of the
barrel FDIRC is its new camera. For each module, the optical portion
consists of the focusing block (FBLOCK), an addition to the wedge (the
New Wedge) and possibly a Micro-Wedge. We have contacted about twelve
optics companies and received four preliminary bids.  We use the
average bid in the present budget. We hope to continue to refine the
values through further R\&D. The photon detector cost estimate is
based on the Hamamatsu bid for 600 H-8500 MaPMTs.  No budget estimate
is included at this time for a forward endcap PID.

\tdrparagraph{Electromagnetic Calorimeter:} The electromagnetic
calorimeter (EMC) consists of a barrel calorimeter, a forward
calorimeter, and a backward calorimeter. The barrel and forward
calorimeters make substantial reuse of \babar\ components. The
backward EMC is a new lead/scintillator sandwich device.  As
described in the calorimeter chapter, there are some technical
alternatives for the forward EMC.  The present cost estimate is for
our baseline ``hybrid'' design. The cost estimate also includes the
backward EMC, although a final decision has not been taken on this
device; it could become a later upgrade.

There is no cost assigned for the value of the reused \babar\ components. 
They are assumed to be available as a scientific transfer or loan of equipment
at no cost to the project. However, there are significant costs associated with
preparing these items for transport, and for the transport. It is assumed that
both the barrel and endcap will be dis-assembled to the module level, although
some engineering is included to further study the possibility of shipping the
barrel as a whole.

Manpower and costs for engineering and tooling required for the
preparation for the move of the barrel EMC from SLAC are engineering
estimates. These incorporate substantial experience with the \babar\
installation as well as current dis-assembly operations of the \babar
detector. Engineering and labor provided by SLAC are included in the
M\&S cost estimate. Engineering and labor from the project are
provided separately in man months.  The similar estimates for the
endcap are not as well developed, but are scaled from the barrel
estimates. The labor for the barrel and forward endcap includes 21 man
months for module testing that might be provided by physicists (e.g.,
graduate students).

The transportation cost for the endcap and associated tooling is based
on a shipping vendor quote, while the transportation for the barrel is
obtained based on the endcap quote.

The main cost driver for the forward endcap is the cost of LYSO
crystals.  The estimate is based on guideline quotes from vendors,
where we assume the six rings of LYSO in the baseline design.  The
next largest element is the APD photodetectors, with a cost based on a
quote from the vendor.

The PIN diodes on the CsI(Tl) crystals will be reused from \babar, but
the preamps will all be replaced. The preamp costs for the CsI(Tl) and
the LYSO are based on engineering estimates and vendor quotes. The
replacement of the barrel preamps is complicated by accessibility, and
the labor estimate for this is based on an in situ test.

For the backward endcap, the scintillator, lead, wavelength-shifting
fiber, and readout MPPC costs, as well as some other minor materials,
are all based on vendor quotes. Other items are based on experience
and estimator judgment.

The source calibration system costs are estimated based on experience
with the system at \babar, which we anticipate reusing. Due to the
half-life of tritium, replacement of the DT generator is included in
the cost.  Monitoring is planned to be achieved largely with the
existing \babar\ sensors and readout. An estimate of costs for
refurbishing and augmentation for the LYSO is included.


\tdrparagraph{Instrumented Flux Return:} The IFR cost is based on
quotations received for the prototype construction appropriately
scaled to the real detector dimensions. While the active part of the
detector is quite inexpensive the total cost is driven by the
electronics and the photodetectors. The current baseline design allows
the reuse of the \babar\ iron structure with some modification that
needs to be taken into account.  Manpower and cost for engineering and
module installation is based on the \babar\ experience.

\tdrparagraph{Electronics, Trigger, DAQ and Online:} The cost for the
Electronics and Trigger subsystems is estimated with a combination of
scaling from the \babar\ experience and from direct estimates.  For
items expected to be similar to those used in \babar\ (such as
infrastructure, high and low voltage or the L1 trigger) costs are
scaled from \babar. The same methodology is used to estimate EDIA and
Labor costs for the Online system.  However, some modifications based
on ``lessons learned'' are applied.  In particular, we are including
costs for development work that, in our opinion, should have been
centralized across sub-detectors in \babar\ (but wasn't) and work that
should have been done upfront but was only done or completed as part
of \babar\ Online system upgrades.

The readout systems for which the higher data rates require redesigned
electronics are estimated from the number of different components and
printed circuit boards, and their associated chip and board counts.
This methodology is also used for the possible new detectors (forward
EMC, backward EMC, forward PID) and for the elements of the overall
system architecture that are very different from \babar.

The hardware cost estimates for the Online computing system (including
the HLT farm) are, very conservatively, based on the current prices of
hardware necessary to build the system, with the assumption that
Moore's Law will result in future systems with the same unit costs but
higher performance. This is justified by our observation that for COTS
components, constraints from system design, topology and networking
are more likely to set minimum requirements for the \emph{number} of
devices than for the per-device performance.




\tdrparagraph{Magnet} The costs were evaluated on the basis of the
experience and some market prices (for vacuum system). The cost for a
new refrigerator and cryogenic equipment are included. The preferable
option for managing the reuse of the solenoid should be a contract to
an expert firm, which takes the full responsibility of the operation.
An estimated cost for this contract is also included.

\tdrparagraph{Transportation, installation, and commissioning: }
Installation and commissioning estimates, including disassembling and
reassembling \babar, are based on the \babar\ experience, and
engineering estimates use a detailed schedule of activities and
corresponding manpower requirements. The transportation costs have
been estimated from costs associated with disassembling and
transporting \babar\ components for dispersal, if they were not to be
reused.

\tdrsec{Schedule}

The detector construction schedule has not been finalized since it
depends heavily on the availability of civil and accelerator
infrastructures.  The fabrication of the detector subsystems can
proceed in parallel while the the \babar\ detector is disassembled,
transported to the new site, and reassembled. The detector subsystems
will be installed in sequence. An extended detector commissioning
period, including a cosmic ray run, will follow to ensure proper
operation and calibration of the detector. The total construction and
commissioning time is estimated to be a little over five years.


\bibliographystyle{\tdrbibstyle}
\balance
\bibliography{Budget/Budget-bib}

\begin{thebibliography}{10}
\bibitem{bib:FC77}
{Fluorinert is the trademark for polychloro-trifluoroethylene, manufactured by
  3M corporation, St. Paul, MN}.
\end{thebibliography}


\cleardoublepage
\thispagestyle{empty}
\phantom{test}
\clearpage
\thispagestyle{empty}


\ThisCenterWallPaper{1.015}{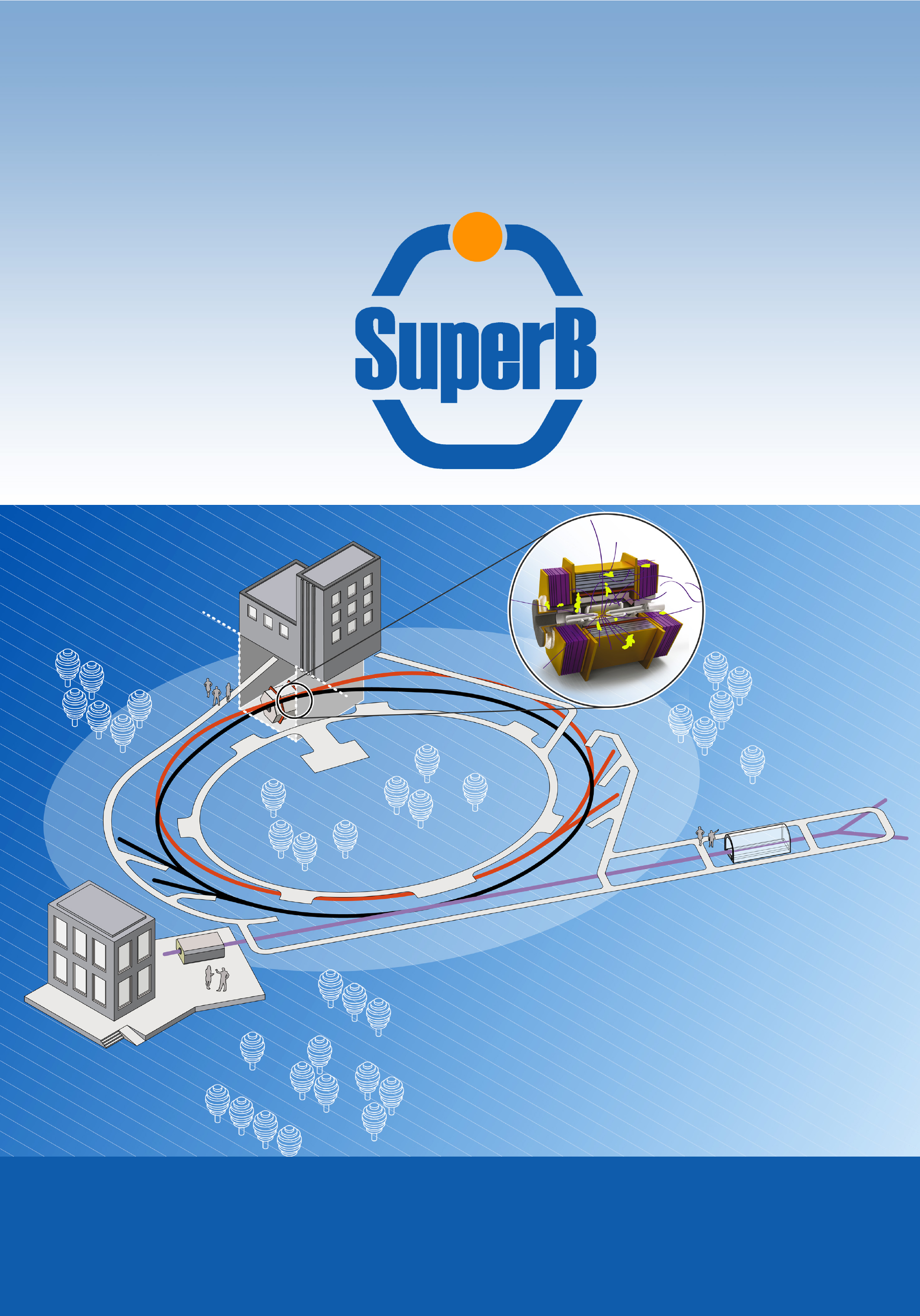}

\phantom{test}

\end{document}